\documentclass{raa_twocolumn} 
\usepackage{pdflscape}
\usepackage{natbib}
\usepackage{amssymb,amsmath}
\bibpunct{(}{)}{;}{a}{}{,}
\usepackage{graphicx}
\usepackage{grffile}
\usepackage{times} 
\usepackage{multirow}
\usepackage{makecell}
\usepackage[pagebackref=true]{hyperref}

\usepackage[final]{changes}

\hypersetup{pdftitle = The title of my PDF, pdfauthor = My name, pdfsubject= The subject, pdfkeywords = keyword1 keyword2 keyword3}
\hypersetup{colorlinks = true, linkcolor = blue, anchorcolor = blue, citecolor = blue, filecolor = blue,  urlcolor = blue}
\footnotesize
\newdimen\digitwidth    
\setbox0=\hbox{\rm0}
\digitwidth=\wd0
\catcode`!=\active
\def!{\kern\digitwidth}
\normalsize

\begin{document}

\title{FAST observations of an extremely active episode of FRB 20201124A: \\ 
I. Burst morphology}
\volnopage{ {\bf 2021} Vol.\ {\bf 21} No. {\bf 1}, A01}
\setcounter{page}{1}
\author{
    D. J. Zhou\inst{1,2},
    J. L. Han\inst{*1,2,3},
    B. Zhang\inst{*4},
    K. J. Lee\inst{5,1},
    W. W. Zhu\inst{1},
    D. Li\inst{1,2},
    W. C. Jing\inst{1,2},
    W. -Y. Wang\inst{5,6},
    Y. K. Zhang\inst{1,2},
    J. C. Jiang\inst{1,6},
    J. R. Niu\inst{1,2},
    R. Luo\inst{7},
    H. Xu\inst{1,5,6},
    C. F. Zhang\inst{5,1},
    B. J. Wang\inst{5,1},
    J. W. Xu\inst{5,1},
    P. Wang\inst{1},
    Z. L. Yang\inst{1,2},
    Y. Feng\inst{8}
   }
\institute{
National Astronomical Observatories, Chinese Academy of Sciences, 20A Datun Road, Chaoyang District, Beijing 100101, China;  {\it hjl@nao.cas.cn}\\
\and
School of Astronomy, University of Chinese Academy of Sciences, Beijing 100049, China; \\
\and
CAS Key Laboratory of FAST, NAOC, Chinese Academy of Sciences, Beijing 100101, China; \\
\and     
Department of Physics and Astronomy, University of Nevada, Las Vegas, NV 89154, USA;  {\it bing.zhang@unlv.edu}\\
\and
Kavli Institute for Astronomy and Astrophysics, Peking University, Beijing 100871, China; \\
\and
Department of Astronomy, Peking University, Beijing 100871, China; \\
\and
Research Center for Intelligent Computing Platforms, Zhejiang Laboratory, Hangzhou 311100, China 
\vs \no \\
{\small Received 2022 XXX; accepted 2022 XXX}
}

\abstract{
We report the properties of more than 600 bursts (including cluster-bursts) detected from the repeating fast radio burst (FRB) source FRB 20201124A with the Five-hundred-meter Aperture Spherical radio Telescope (FAST) during an extremely active episode on UTC September 25-28, 2021, in a series of four papers. The observations were carried out in the band of 1.0 - 1.5~GHz by using the center beam of the L-band 19-beam receiver. We monitored the source in sixteen 1-hour sessions and one 3-hour session spanning 23 days. All the bursts were detected during the first four days. In this first paper of the series, we perform a detailed morphological study of 624 bursts using the 2-dimensional frequency-time “waterfall'' plots, with a burst (or cluster-burst) defined as an emission episode during which the adjacent emission peaks have a separation shorter than 400 ms.
The duration of a burst is therefore always longer than 1 ms, with the longest up to more than 120 ms. The emission spectra of the sub-bursts are typically narrow within the observing band with a characteristic width of $\sim$277 MHz. The center frequency distribution has a dominant peak at about 1091.9~MHz and a secondary weak peak around 1327.9~MHz. Most bursts show a frequency-downward-drifting pattern. Based on the drifting patterns, we classify the bursts into five main categories: downward drifting (263) bursts, upward drifting (3) bursts, complex (203), no drifting (35) bursts, and no evidence for drifting (121) bursts. Subtypes are introduced based on the emission frequency range in the band (low, middle, high and wide) as well as the number of components in one burst (1, 2, or multiple). We measured a varying scintillation bandwidth from about 0.5~MHz at 1.0~GHz to 1.4~MHz at 1.5~GHz with a spectral index of 3.0.
}
\authorrunning{D. J. Zhou et al. }            
\titlerunning{FAST observations of FRB 20201124A: I. Burst Morphology}  
\maketitle
%
%
\section{Introduction}
Fast radio bursts (FRBs) are radio flashes with a short duration typically in milliseconds \citep{Lorimer2007Sci}. They have a high dispersion measure (DM), exceeding the maximum estimated from the electron column density model \citep{Cordes2002astro, YaoMW2017ApJ} for the Milky Way so that almost all of them are considered as extragalactic \citep{Thornton2013Sci, Petroff2019A&ARv, Cordes2019ARA&A, ZhangB2020Natur}. Since the discovery of the first event FRB 010724 \citep{Lorimer2007Sci}, hundreds of FRB sources have been discovered\footnote{\url{https://www.herta-experiment.org/frbstats/catalogue}}. A small number of sources are observed to emit repeated bursts. After the discovery of the
first repeater FRB 20121102A  \citep{Spitler2014ApJ},
a large number of repeated bursts have been detected from 20 FRB repeating sources ~\citep{Chime19b, Fonseca20, CHIME/FRB2021arXiv210604352T}.
Recently, a Galactic FRB (dubbed FRB 20200428) was detected from a Galactic magnetar, SGR J1935+2154 \citep{Bochenek2020Natur.587, CHIME2020Natur.587}. Even though its brightness is lower than most extragalactic FRBs, it is 
orders of magnitude brighter than single pulses of pulsars. The two peaks of the burst also exhibited narrow-band emission characteristics, which is typical for repeating FRBs.

The physical origin and emission mechanism of FRBs are unsolved mysteries \citep{Cordes2019ARA&A,Petroff2019A&ARv,ZhangB2020Natur}. Hints can be found in the different radiation patterns between the repeaters and non-repeaters. Repeater bursts typically have a narrower frequency band emission and a broader temporal width than non-repeater bursts (e.g., \citealt{Pleunis21}). While the lone bursts for non-repeaters prevent detailed studies of non-repeater sources, a large amount of data for repeaters have been collected and analyzed thoroughly. For example, the sub-pulse frequency drifting feature was discovered in many repeated bursts \citep[e.g.][]{Chime19a, Chime19b, Hessels19, Josephy19, Chime2020Natur, Fonseca20, Day20, LiDi2021Natur, Platts21}, which brings interesting constraints on theoretical models \citep[e.g.][]{WangWY2019ApJ,ZhangB2022ApJ}.
Sometimes precursors~\citep{Hardy2017MNRAS, Caleb2020MNRAS, Rajwade2020MNRAS} or postcursors~\citep{Scholz2017ApJ, Gourdji2019ApJ, Cruces2021MNRAS} have been detected from some repeaters, i.e. a faint emission component appears prior to or after the brighter primary burst within a very short time from sub-milliseconds to tens of milliseconds, which causes a deviation from the trend of frequency drifting of the bursts.

FRB 20201124A is an active repeater first discovered by \citet{Chime2021ATel14497}.  Similar to FRB 20121102A~\citep{Rajwade2020MNRAS} and FRB 20180916~\citep{Chime2020Natur}, a large number of bursts have been detected during the active episode. In the active phase from March to May 2021, extensive observations \citep{Lanman2021arXiv210909254L, Nimmo2021arXiv211101600N, XuHeng2022Natur} have been carried out. Its location has been well-determined within an arcsecond~\citep{Wharton2021ATel14538} by the European VLBI Network (EVN) \citep{Nimmo2021arXiv211101600N}, which is about 1.3 kpc from the optical center of its host galaxy, SDSS J050803.48+260338.0, a massive star-forming galaxy with a spectroscopic redshift of $z = 0.0979 \pm 0.0001$ \citep{Fong2021ApJ,Ravi2022MNRAS,Piro2021A&A,XuHeng2022Natur}.

Following-up observations of FRB 20201124A have revealed its various emission characteristics which are similar to other repeaters, such as downward frequency drifting, narrow band emission and scintillation. Based on the observations at 550 -- 750 MHz by the upgraded Giant Metrewave Radio Telescope (uGMRT) and in 1210 -- 1520 MHz by Effelsberg, \citet{Main2021MNRAS} worked out that the scintillation timescale is ${\tau}_{\rm GHz}$ = 0.31 $\pm$ 0.06 ${\mu s}$ with a best-fit power index of ${\gamma}$ = 3.5 $\pm$ 0.1, lower than that 4.0 or 4.4 of the Kolmogorov spectrum. \citet{Hilmarsson21} discovered circular polarization from one of the bursts 
and found frequency downward drifting structures of sub-bursts in FRB 20201124A at 1.36GHz. A persistent radio source, PRS 201124, was subsequently found at the location of this FRB \citep{Ravi2022MNRAS}, and the spectral energy distribution of the host galaxy is consistent with a star formation galaxy. However, this persistent radio source is extended rather than local to the source \citep{Piro2021A&A}. A bright burst of this repeater had a flux density of 0.7$\pm$0.01 mJy at 650 MHz as detected by uGMRT \citep{Wharton2021ATel14529} and 0.34$\pm$0.03 and 0.15$\pm$0.01 mJy at 3 and 9 GHz as detected by the VLA \citep{Ricci2021ATel14549}.

As the largest and most sensitive single antenna radio telescope in the world, the Five-hundred-meter Aperture Spherical radio telescope \citep[FAST,][]{NanRD2011} is an ideal facility for detecting weak radio signals from FRBs \citep{Luo2020Natur, LiDi2021Natur, NiuCH2021arXiv211007418N} and pulsars \citep{HanJL2021RAA}. It can track sources well with a pointing accuracy of 8 arcseconds \citep{JiangP2020RAA} and record polarization signals in the pulsar search mode \citep{HanJL2021RAA}. During the active period of FRB 20201124A from April 1 to June 11, 2021, \citet{XuHeng2022Natur} reported the detection of 1863 bursts. Based on the analysis of recorded polarized signals, they revealed rich features from this large sample of bursts, including detecting significant variations of the Faraday rotation measures and the oscillation features of the polarization properties with respect to wavelength in a small fraction of bursts.

\begin{figure}
\centering
\includegraphics[width=0.99\columnwidth]{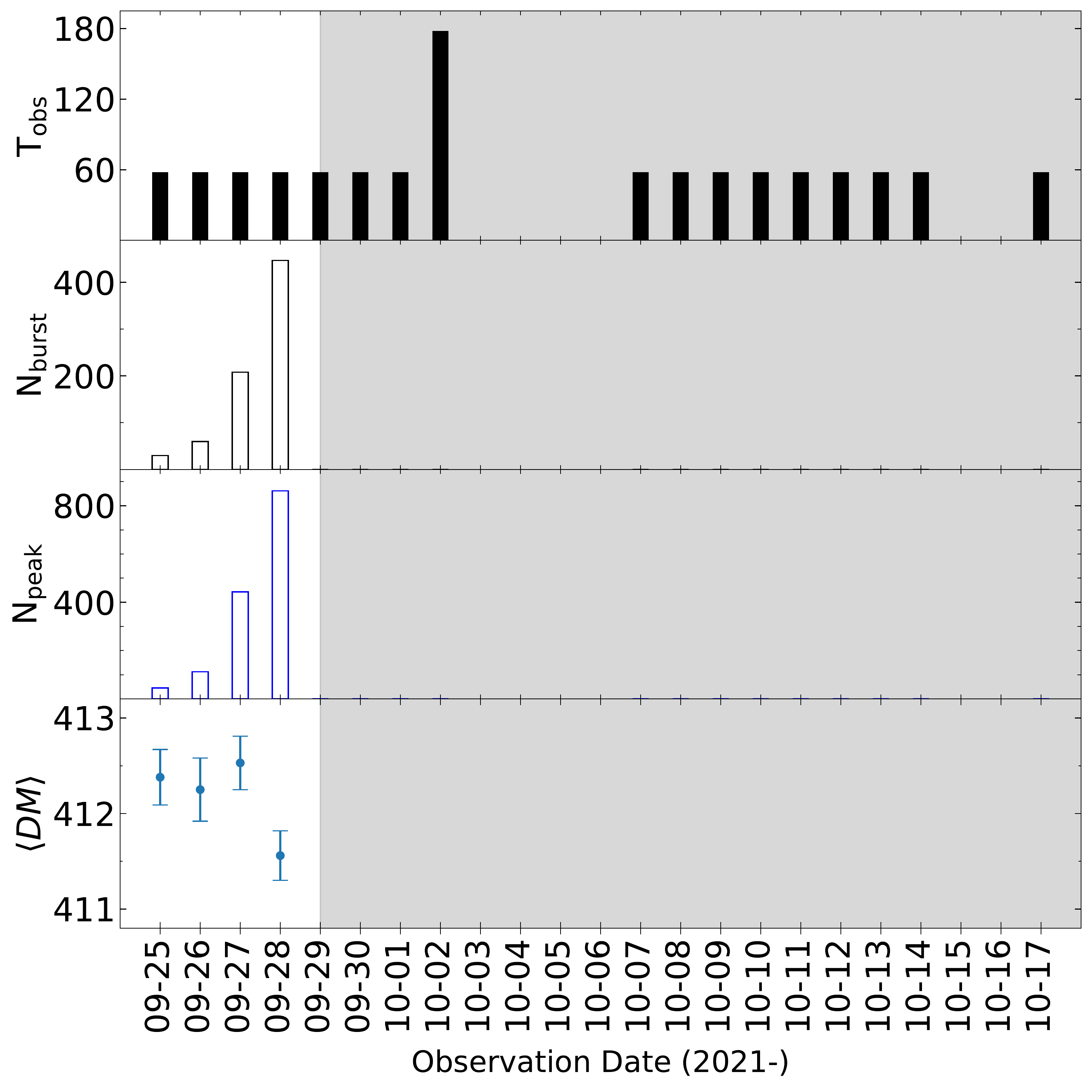}
\caption{
FRB 20201124A was monitored from September 25 to October 17, 2021 by FAST.
The panels from top to bottom show the observation duration (in minute) each day, numbers of detected bursts, numbers of fitted peaks and averaged DMs (in pc~cm$^{-3}$) respectively. The gray shaded area indicates no bursts detected from the FAST observations for many days.
See the relevant numbers in Table~\ref{tab:Obsparameters}.
}
\label{fig:Obsparameters}
\end{figure}

Triggered by the Canadian Hydrogen Intensity Mapping Experiment (CHIME) detection\footnote{\url{https://www.chime-frb.ca/repeaters/FRB 20201124A}} and the report by \citet{Main2021ATel14933}, we started to monitor FRB 20201124A with FAST on September 25th, 2021 and continued to monitor the source almost daily until October 17. A large number of bursts have been detected in the first 4 days (see Fig.~\ref{fig:Obsparameters}) with very diverse emission properties. In this paper, we focus on the burst morphology and taxonomy of more than 600 detected bursts or cluster-bursts. Their energy distribution is discussed in Paper II (Y.K. Zhang et al. 2022); the polarization properties are analyzed in Paper III (J.C. Jiang et al. 2022); and the bursts arrival time is analyzed in Paper IV (J.R. Niu et al. 2022). In Section~\ref{2Observation}, we briefly introduce the FAST observations and burst detection. The detailed analysis methods are presented in Section~\ref{3DataAnallysis}. DM measurements and parameter determination for emission frequency peak and emission bandwidth, sub-burst width and fluence are presented in Section~\ref{DMdetermination} and~\ref{param}. The statistics of burst parameters and morphology classifications are presented in Section~\ref{sect:BMclassify}. Finally, the summary is presented in Section~\ref{4Conclusions} and the comparison of observational results with other repeating sources are discussed there.

\section{Observations and Burst Detection}
\label{2Observation}

\begin{table}
\centering
\caption{Observation sessions and detected bursts}
\label{tab:Obsparameters}
\setlength{\tabcolsep}{4.0pt}
\begin{tabular}{lrrrrr}
\hline\noalign{\smallskip}
Date    & T$_{\rm obs}$ &Burst  & Burst Rate & Peak     & $\langle DM \rangle$ ($\sigma$)  \\
        & (min)     &     No.   &(hr$^{-1}$) &     No.      & (cm$^{-3}$pc) \\
\hline
20210925& 58        & 29        & 30.0       & 44           & 412.4(3) \\
20210926& 58        & 57        & 60.0       & 111          & 412.2(3) \\
20210927& 58        & 169       & 174.8      & 441          & 412.5(3) \\
20210928& 58        & 369       & 381.7      & 865          & 411.6(3) \\
20210929& 58        & 0         & 0          & 0            &          \\
20210930& 58        & 0         & 0          & 0            &          \\
20211001& 58        & 0         & 0          & 0            &          \\
20211002& 178       & 0         & 0          & 0            &          \\
20211007& 58        & 0         & 0          & 0            &          \\
20211008& 58        & 0         & 0          & 0            &          \\
20211009& 58        & 0         & 0          & 0            &          \\
20211010& 58        & 0         & 0          & 0            &          \\
20211011& 58        & 0         & 0          & 0            &          \\
20211012& 58        & 0         & 0          & 0            &          \\
20211013& 58        & 0         & 0          & 0            &          \\
20211014& 58        & 0         & 0          & 0            &          \\
20211017& 58        & 0         & 0          & 0            &          \\
\hline
\end{tabular} 
\end{table}

After FRB 20201124A was found to be active again by the CHIME and Effelsberg Telescope \citep{Main2021ATel14933},
we started the monitoring program by FAST at the coordinate of RA = $\rm{05^{h}08^{m}03.5077^{s}}$, Dec = $\rm{+26^{\circ}03^{\prime}38.504^{\prime \prime}}$ for FRB 20201124A obtained by the European Very Long Baseline Interferometry Network (EVN) \citep{Marcote2021ATel14603}, and scheduled 17 effective FAST observations of FRB 20201124A from September 25 to October 17, 2021.

The central beam of the L-band 19-beam receiver is used to cover the frequency range from 1.0~GHz to 1.5~GHz \citep{JiangP2020RAA,li18IMMag}. Our FAST observations were carried out for one hour each day, except for three hours on October 3rd, 2021. The calibration signals of periodic noise are injected for 1 minute at the beginning and the end of each observation, so that data on the source is effectively 58 minutes. The 4 polarization channels (XX, YY, XY*, X*Y) are sampled with a time resolution of 49.152~$\mu$s by the digital backend in pulsar searching mode, using 4096 frequency channels (0.122070 MHz each channel) to cover 1.0 to 1.5 GHz. The data are stored in PSRFITS format~\citep{Hotan2004PASA}, and each FITS file records data of about 12.88~s. The observation setup details are as same as the first FAST monitoring session in 2020 April to May ~\citep{XuHeng2022Natur}.

\subsection{Data Analysis and burst detection}
\label{3DataAnallysis}

An independent data analysis for searching single pulses is carried out by the first author of this paper. The discovered bursts are verified by members of other groups in the author list, as are also reported in other papers of this series. 

Before the single pulse search, all small FITS files, of which each consists of 12.88-s observation data, are merged to a single FITS file in the chronological order. To reduce the file size, 256 out of 4096 frequency channels at each edge of the L-band that have a very low gain are discarded, thus a 31.25 MHz band on each side of the band is cut off. The power of XX and YY of other frequency channels are then combined, and the channel number and sampling time are reduced by a factor of 8 and 4, respectively. The total power for a total of 448 (rather than 512, reduced by a factor of 8) frequency channels with a sampling time of 49.152*4~$\mu$s (i.e. downsampling by a factor of 4) are prepared for the following single pulse detection. If a pulse is detected, the original raw data for all channels are analyzed for pulse properties.

Specifically for FRB 20201124A, we search pulses in a few steps. First of all, we dedisperse data in the range of 3 to 1000 cm$^{-3}$pc in steps of 1.0~cm$^{-3}$~pc to form 2-dimensional images on the time versus DM. An artificial intelligence (AI) approach is carried out in the GNU ${Parallel}$ \citep{tange2020} to identify dedispersed pulses. Some frequency channels always with strong Radio frequency interference (RFI) are discarded during this process. To improve the pulse detection, we divide 448 channels into two parts, the upper half and lower half of the total band, and perform the single pulse search separately. We also search pulses in the upper and lower quarter of 448 channels. Any pulse candidates with signal-to-noise ratio $\rm SNR > 7$ are manually viewed and checked further in the two-dimensional frequency-time water-fall plots.

\begin{figure*}[!ht]
\centering
\includegraphics[width=0.64\columnwidth]{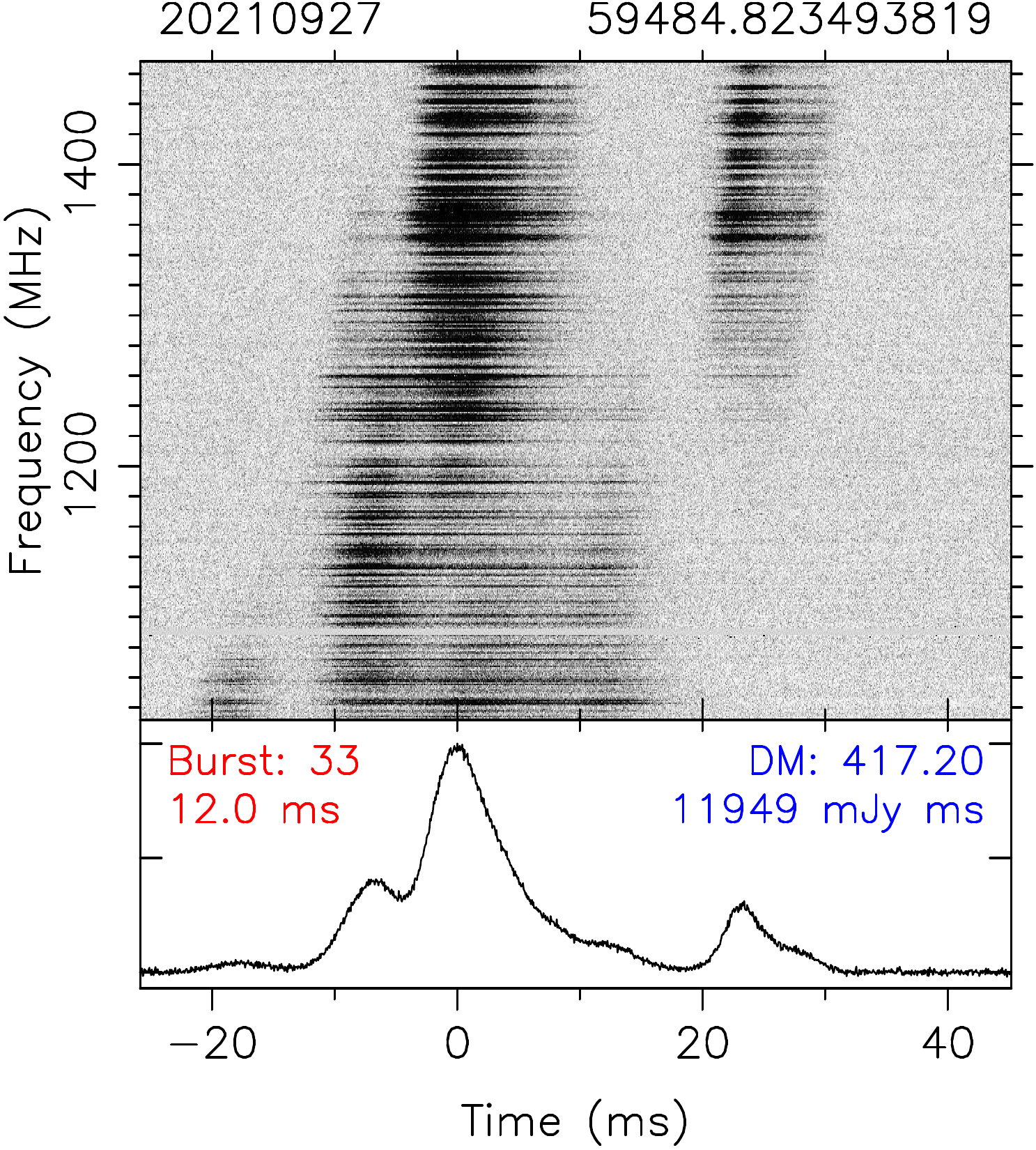}
\includegraphics[width=0.64\columnwidth]{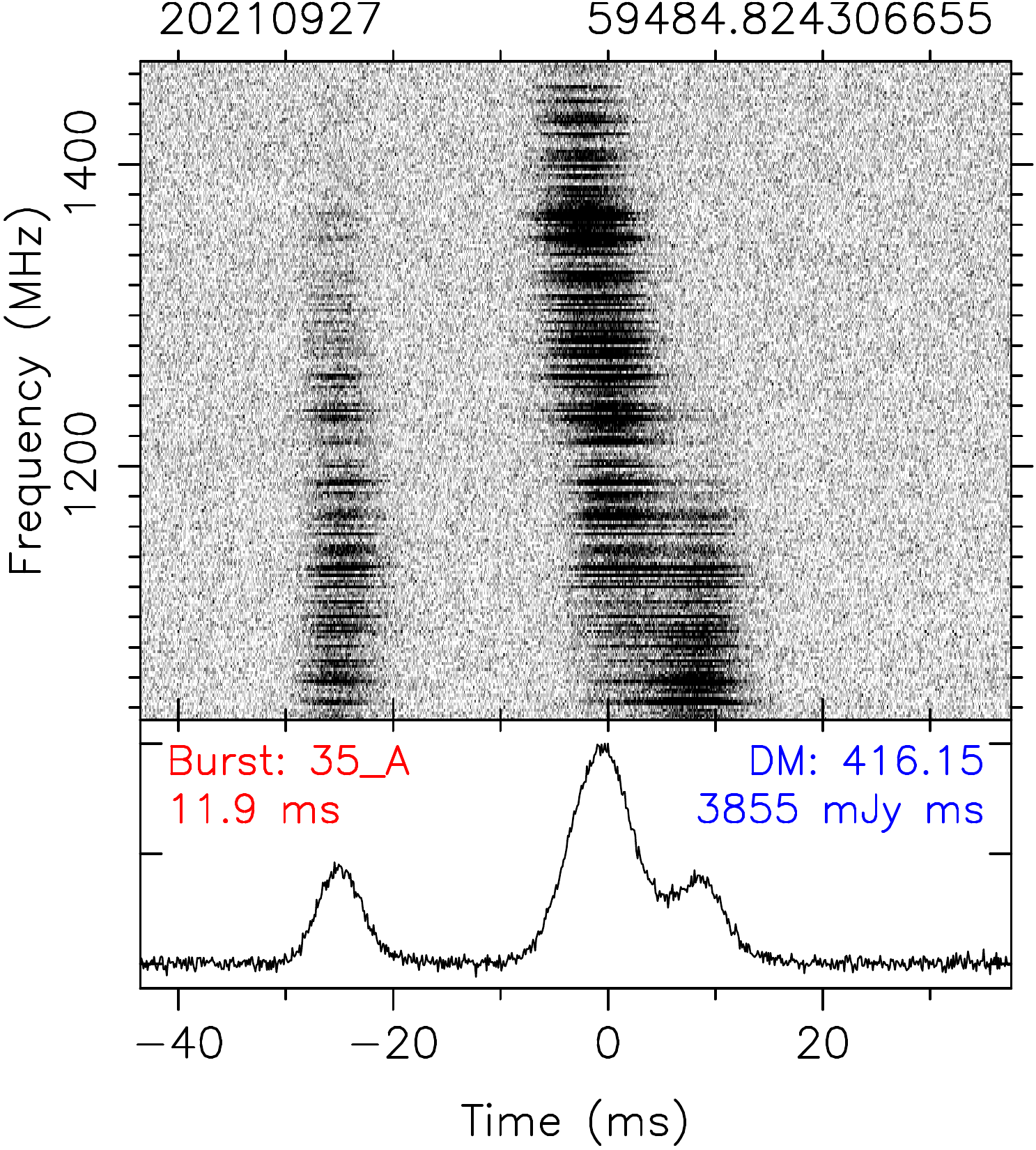}
\includegraphics[width=0.64\columnwidth]{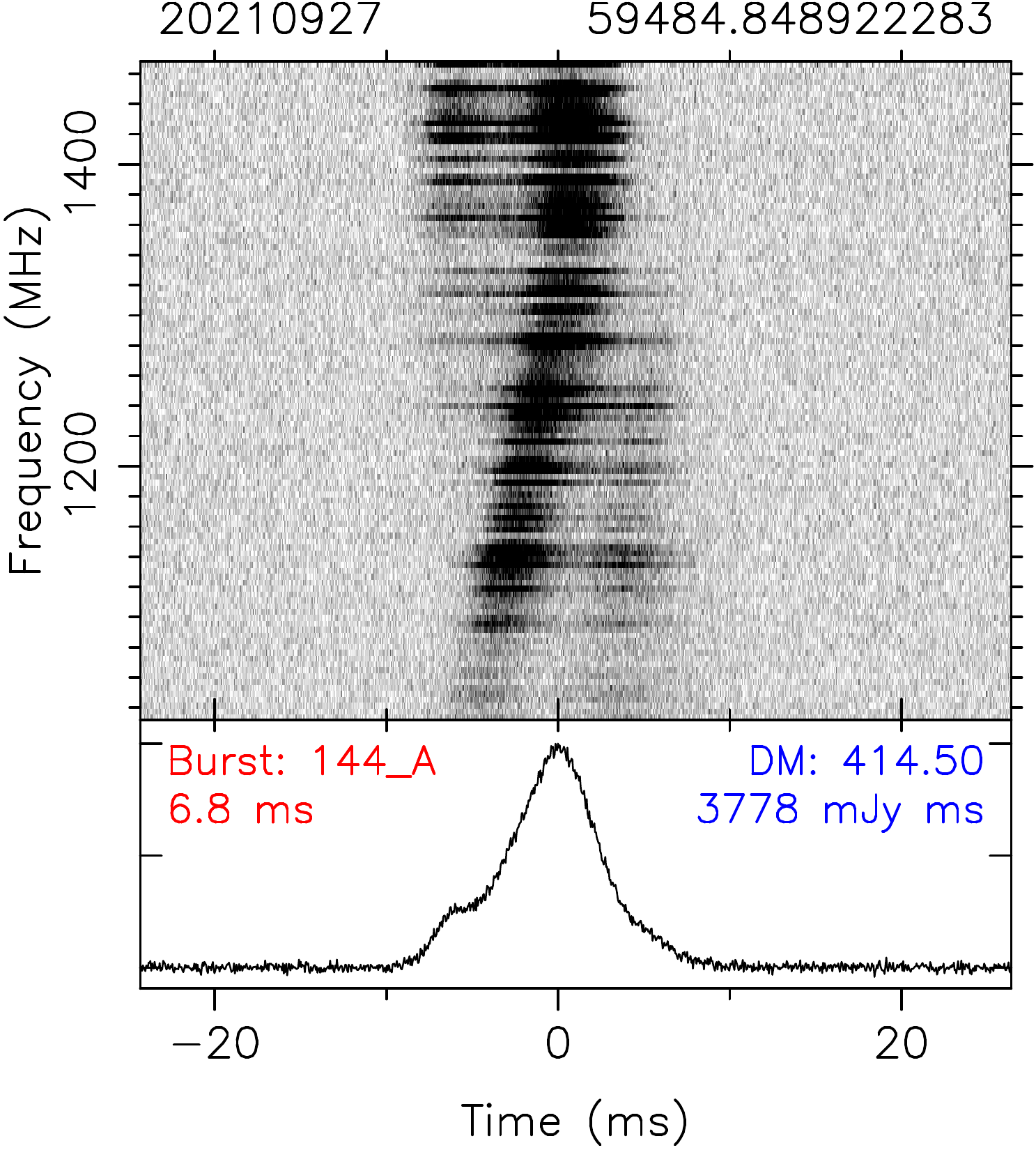}
\includegraphics[width=0.64\columnwidth]{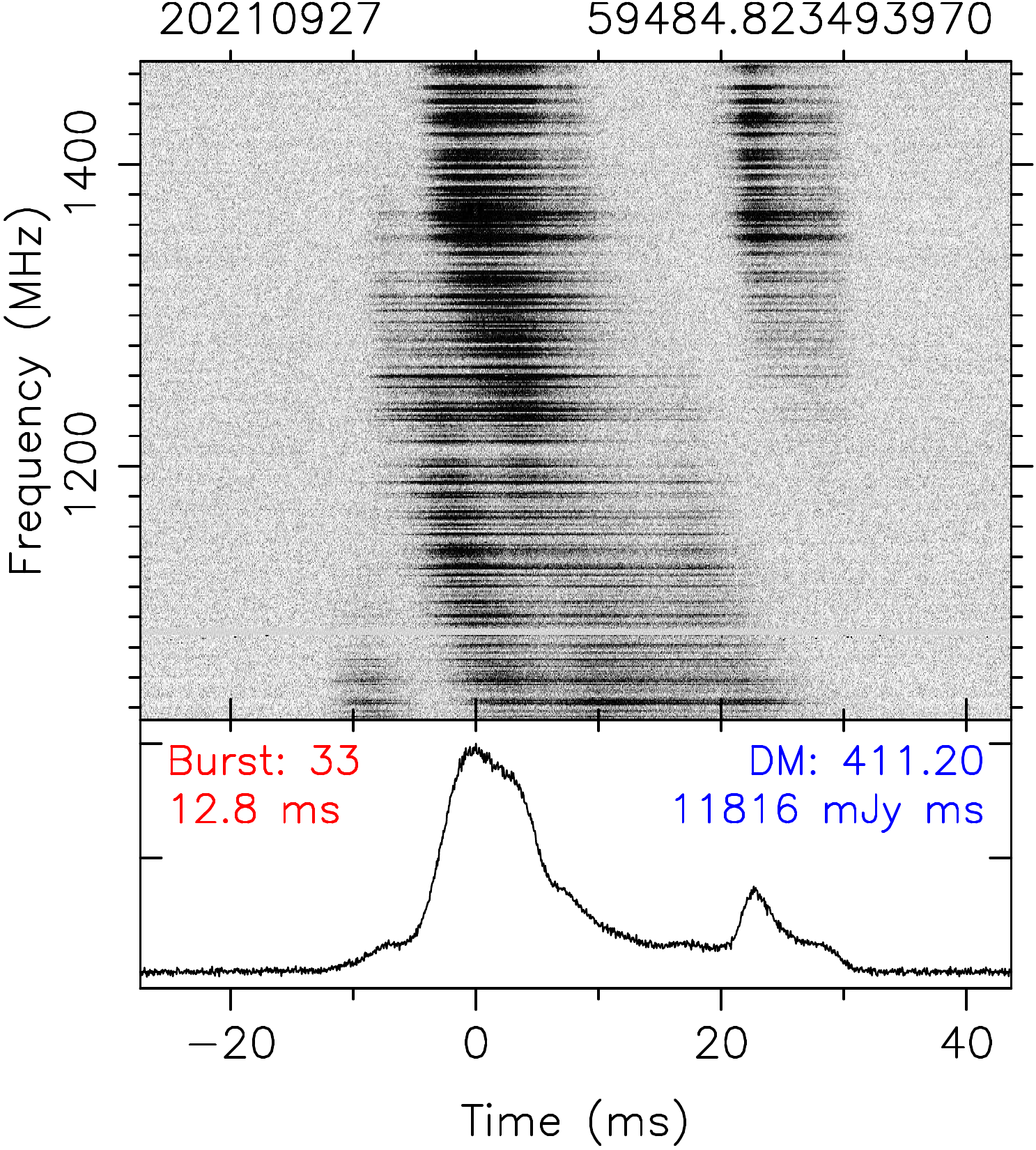}
\includegraphics[width=0.64\columnwidth]{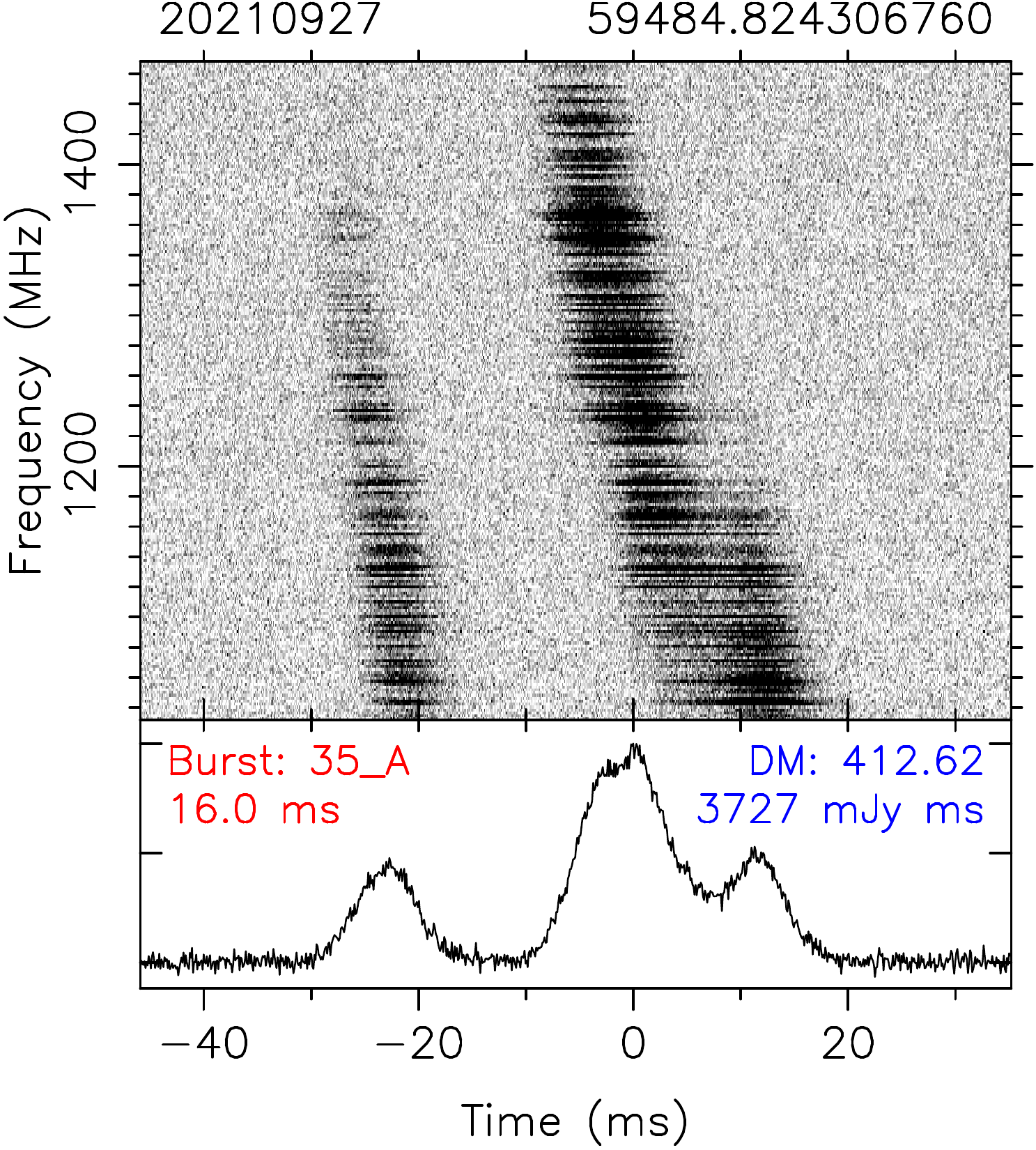} 
\includegraphics[width=0.64\columnwidth]{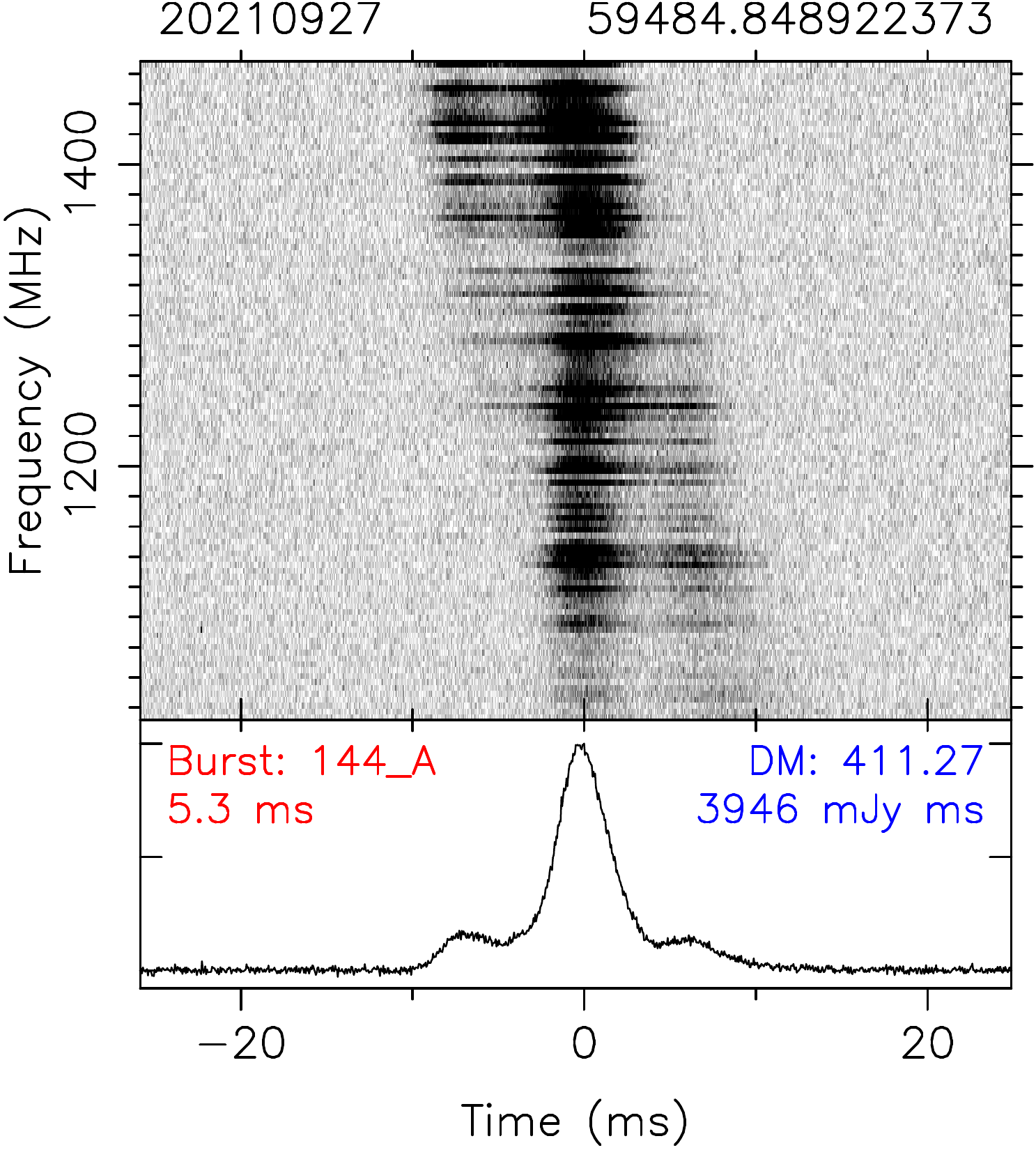}
\includegraphics[width=1.925\columnwidth]{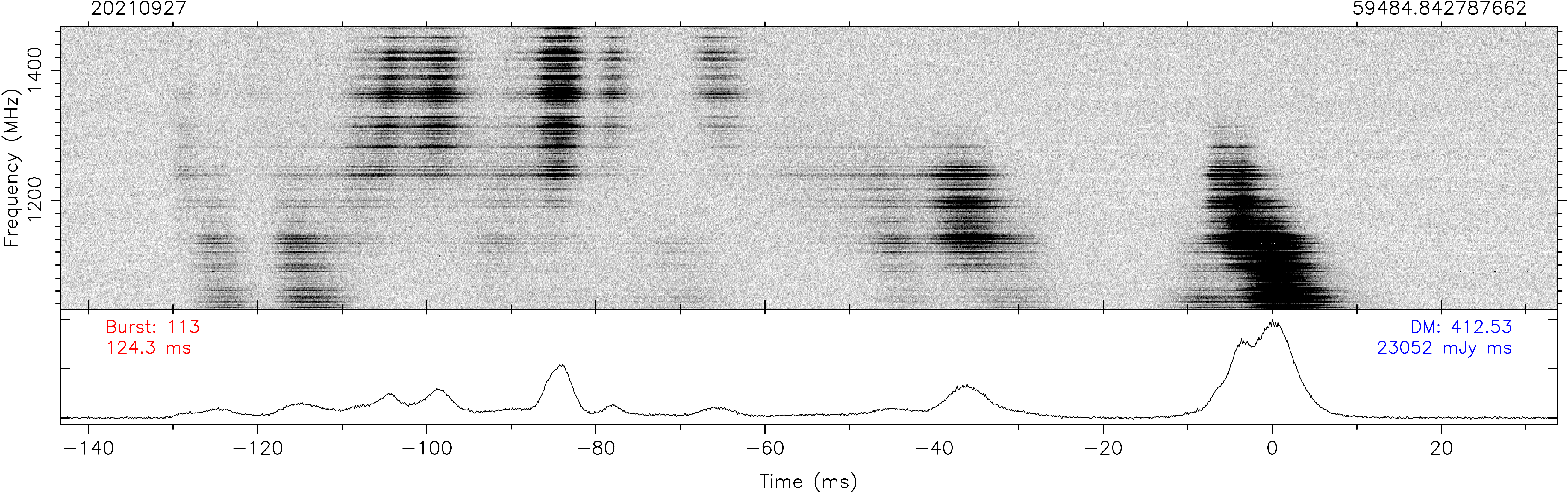}
  \caption{The dynamic spectra of four bursts illuminating the difficulty to determine DMs. In the upper sub-panel of each plot is the frequency-time water-fall plot, and the average burst profile over all frequency channels is shown in the bottom sub-panel. The observational date and the time of arrival (TOA) in MJD are marked on the top of each plot. The burst number on that day and burst width in ms, together with the DM value used and the total fluence of this burst are marked in the lower-sub-panel. In the top three panels, the DM values of bursts No.~33 and No.~35$\_$A and No.~144$\_$A on 20210927 are determined from burst emission at the upper half band of the FAST observations, and the middle panels are the same data but with DMs determined from the emission at the lower half band. The wide panel in the bottom shows a complex, No.~113 on 20210927, which have many sub-bursts so that it is difficult to obtain one DM to align with all the burst emission. We then have to compromise by using the averaged DM of the day to generate this plot.
}
\label{fig:difDM}
\end{figure*}

In an earlier FAST paper reporting the previous active episode of the source \citep{XuHeng2022Natur} and that in the Paper II (Y.K. Zhang et al.2022), the waiting time distribution of dedispersed emission peaks from FRB 20201124A exhibits two wide peaks, one from a few ms to a few tens of ms, and the other from a few seconds to a few tens of seconds. 
The first peak may indicate several emission components occuring in one burst, while the second peak probably stands for the interval between two adjacent independent bursts. The valley between the two peaks of the waiting time distribution can be used to distinguish independent bursts. Based on the waiting time distribution, we take the separation of 400~ms between any of two emission components as indication for the another independent burst.
This allows us to better characterize and compare the emission morphology of the emission peaks within the same emission episodes. More specifically, we adopt the following definitions for the terminology:  
\begin{itemize}
    \item A burst is defined as all emission components with the adjacent peak separation not longer than 400~ms, according to the waiting time distribution of the emission peaks (\citet{XuHeng2022Natur}, Niu et al 2022). 
    \item A sub-burst stands for one of several more or less connected components the frequency-time waterfall plots (see Figure~\ref{fig:difDM}) generally within a few tens of milliseconds, and it have a distinguished peak in the dedispersed burst profile.
    \item A cluster-burst is defined as a collection of several somehow independent bursts with a separation less than 400~ms, but there is no bridge emission in between since the burst intensities come back to the system noise level.
\end{itemize}
We note that the definitions of the bursts are somewhat different in other papers in the series. This is because different papers have different scientific purposes and the lead authors defined their ``bursts'' for the convenience of conducting their respective analyses. Therefore, the numbers of  ``bursts'' reported in different papers are somewhat different, even though the same data set has been analyzed. 

\begin{figure*} 
 \centering
 \includegraphics[height=58mm]{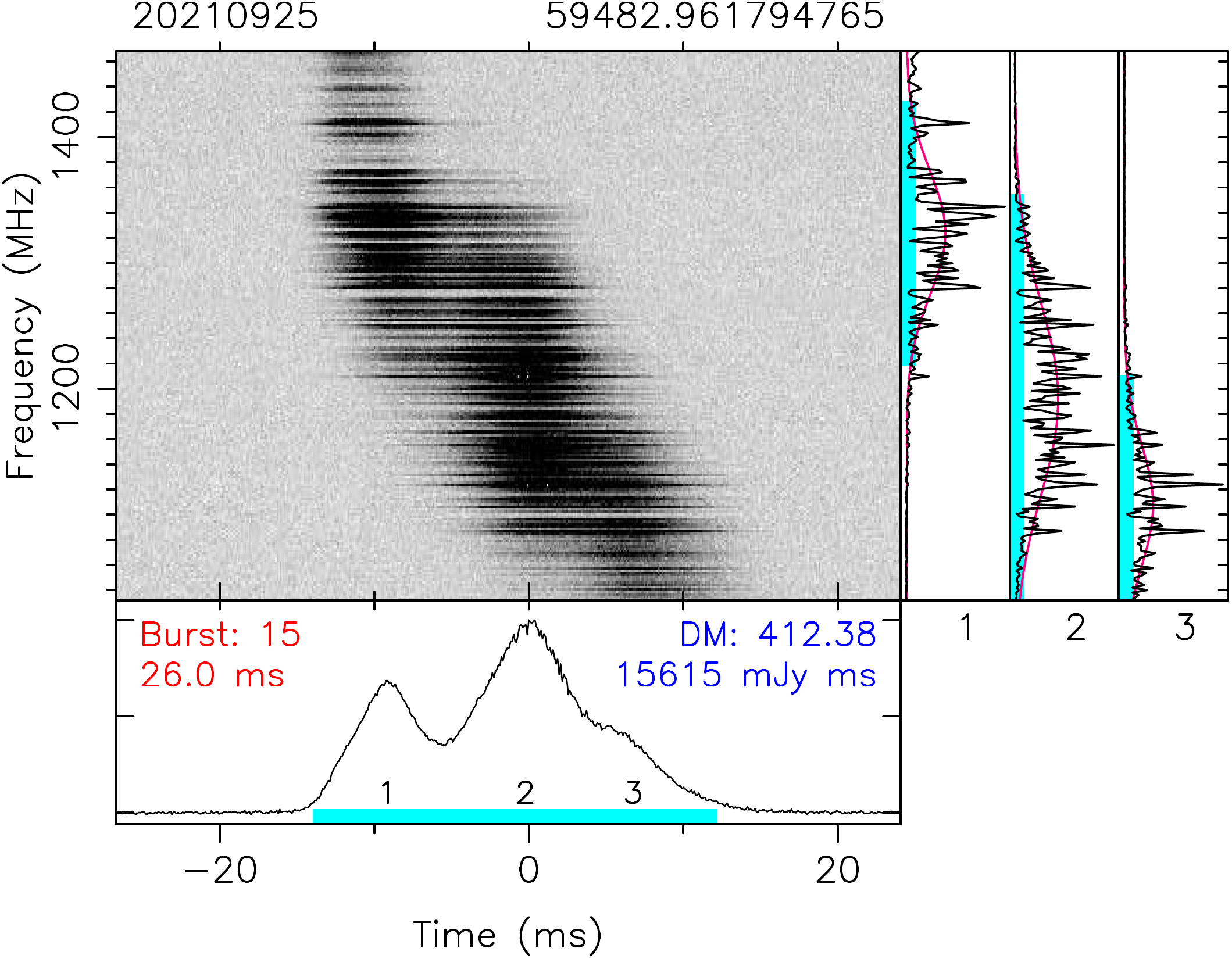}
 \includegraphics[height=58mm]{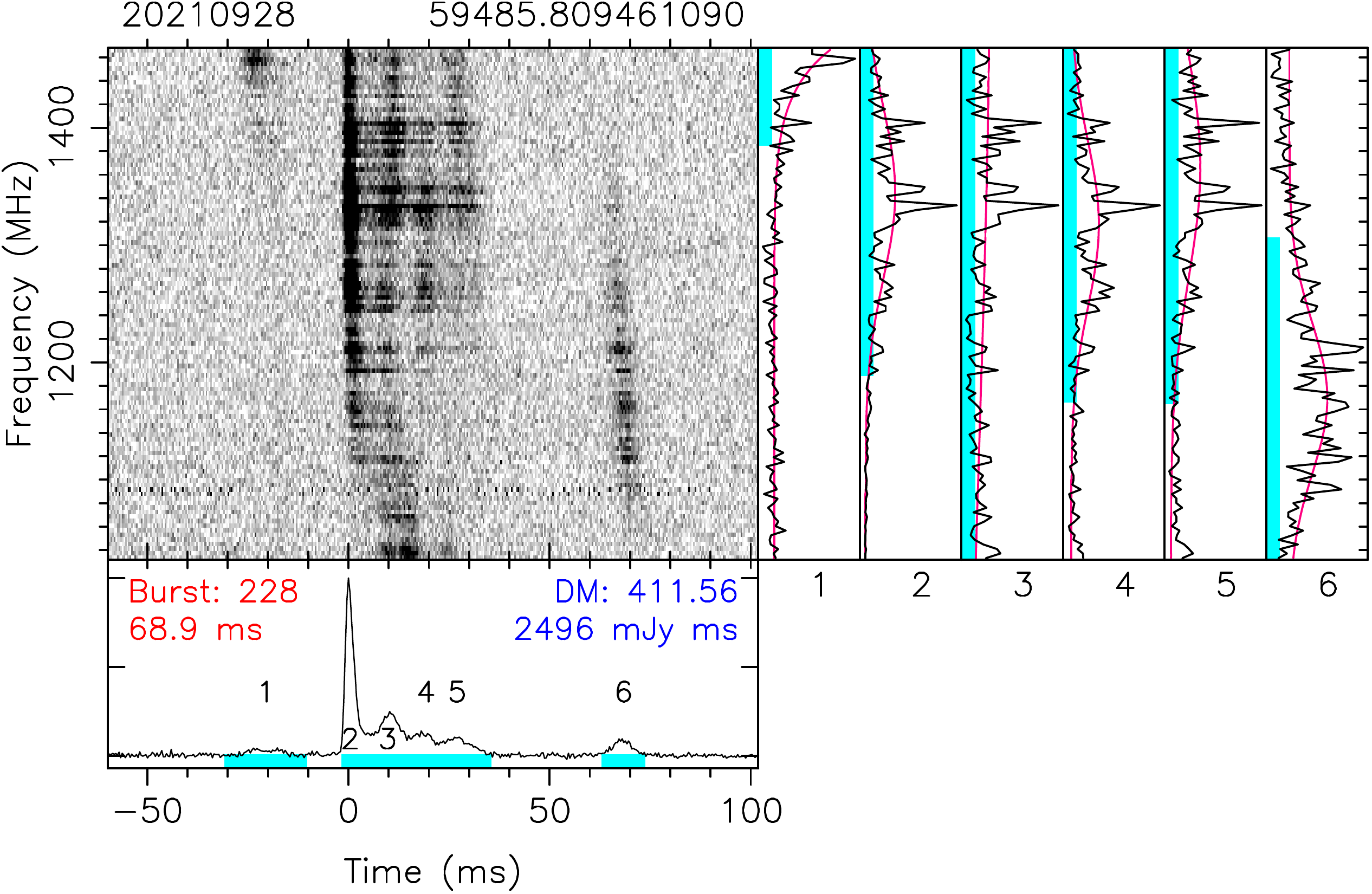}
\caption{The dynamic spectra of two bursts with DM values well determined by the {\sc DM\_phase} package. In the bottom sub-panel, the sub-burst numbers are marked. Their energy distributions over frequency are shown in the right sub-panels and fitted with a Gaussian function to obtain the emission peak frequency $\nu_{0}$ and emission bandwidth $\rm BW_{e}$.}
\label{fig:gooddm}
\end{figure*}

\begin{figure}
  \centering
  \includegraphics[width=0.99\columnwidth]{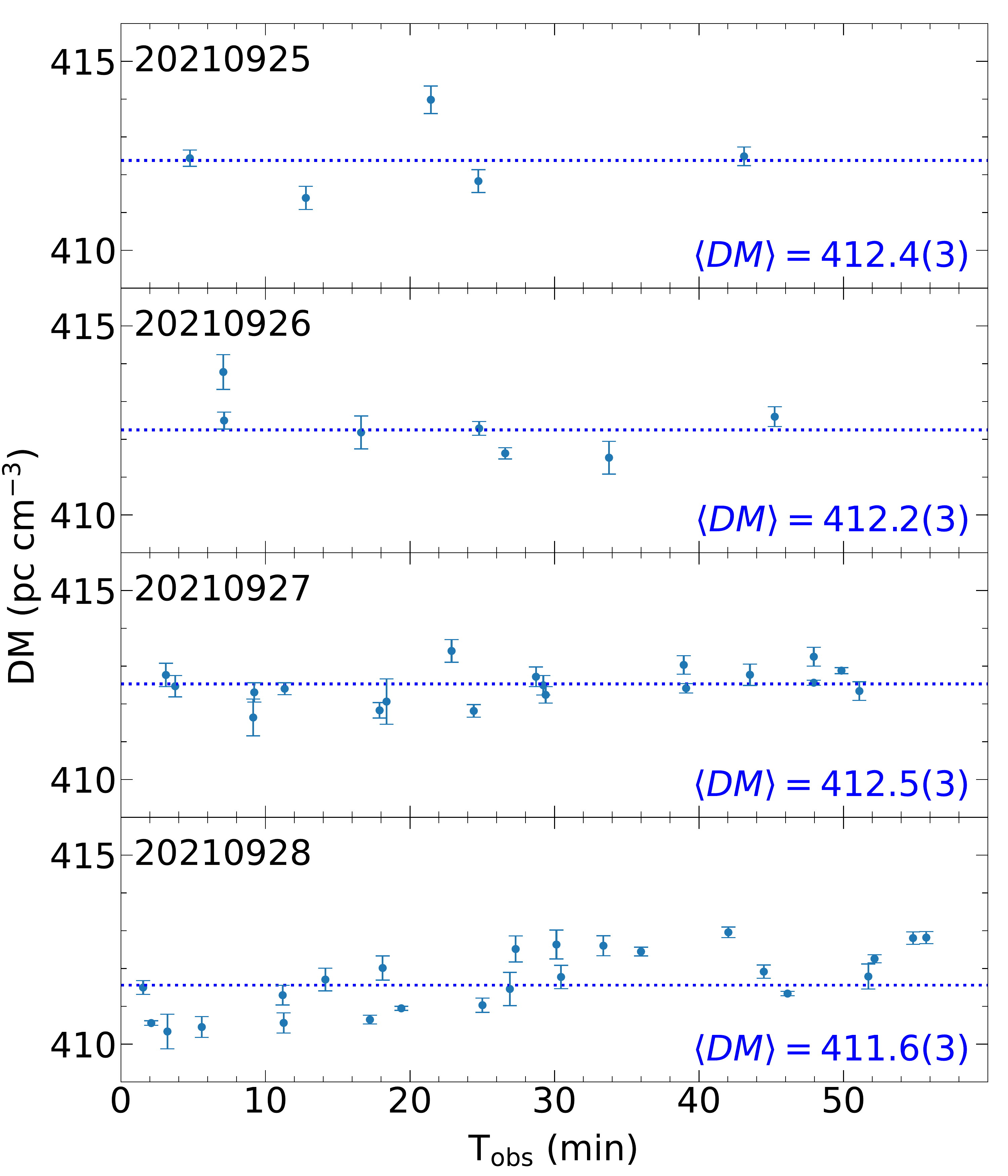}
  \caption{DM values are confidently determined for a number of bursts using the package {\sc DM\_phase}, with the mean DM values marked for each day.}
  \label{fig:timeDM}
\end{figure}

We obtained 29, 57, 169, and 369 bursts in the first 4 days, respectively, and in total 624 bursts from FRB 20201124A during this active episode. A full list of the bursts (see Table~\ref{tab:appendix}) together with plots for each bursts (see Figure~\ref{fig:appendix:D1L} to Figure~\ref{fig:appendix:NEH}) are presented in the Appendix. The total numbers of detected bursts in different observing sessions are listed in Table~\ref{tab:Obsparameters} and presented in Figure~\ref{fig:Obsparameters}. The burst rate rapidly increases on 20210927 and 20210928, reaching 381.7 bursts per hour (hr$^{-1}$) and about one magnitude higher than 45.8 bursts per hour as reported in the April-May observation sessions \citep{XuHeng2022Natur}. The highest burst rate for FRB 20201124A on September 28, 2021, is nearly four times of the previously known highest rate of FRBs, which is 122 hr$^{-1}$ for FRB 20121102A \citep{LiDi2021Natur}. It is worth mentioning here that our most inclusive definition of bursts has reduced the total number of bursts, which means that the burst rate is even higher if the same burst definition is adopted as \cite{LiDi2021Natur} or \citet{XuHeng2022Natur}. More intriguing is that the FRB source was suddenly quenched after 20210928 so that no plausible bursts were detected in many following days. It is the first time to see such a dramatic change in the burst rate, and it is puzzling how an FRB source increases the burst rate exponentially, rushes to a peak rate and then is suddenly quenched.

\subsection{The most probable DM value}
\label{DMdetermination}

The bursts of FRB 20201124A detected by the FAST show various dynamic spectra structures. They are detected first with a given dispersion measure (DM) value. However, when the detailed structures are studied, a proper DM is the key to determine the physical properties of the bursts (see Figure~\ref{fig:difDM}). 

Following the approach for other FRBs in the literature \citep{Chime19b, Marthi2021MNRAS,XuHeng2022Natur}, we first use the package {\sc DM\_phase}\footnote{\url{https://www.github.com/DanieleMichilli/DM_phase}} to find the best DM for every burst. This handy program can find the best frequency alignment of de-dispersed sub-burst features. 

For bursts of FRB 20201124A, however, the outcomes of {\sc DM\_phase} need to be taken with caution. It gives a very reasonable result for some bursts, but not for all, mainly because of the complicated burst properties of FRB 20201124A. After examining the determined DMs and the waterfall plots for all bursts, we found that it is hard to make a good and uniform standard to determine the DMs for various bursts. Figure~\ref{fig:difDM} shows some examples of the dilemma. The burst No.~33 on 20210927 shows a more reasonable DM determined from the upper half-band by using {\sc DM\_phase}, while it is reversed for the bursts No.~35\_A and 144\_A on 20210927. If one takes the leading component of the burst No.35\_A as an independent burst (see Figure~\ref{fig:difDM}), {\sc DM\_phase} would derive a DM that is so large that this burst appears to have its lower part distorted toward earlier time in the water-fall plot, rather than simply vertically aligned. We get many similar cases. This fact suggests that the bursts of FRB 20201124A almost always have an intrinsic downward frequency drifting.

For this session of FRB 20201124A, we have to determine the DMs only for a small number of selected bursts which we feel confident in the DM results from  {\sc DM\_phase},  having either a significant inter-structural gap as No.~15 on 20210925 in Figure~\ref{fig:gooddm} or a sharp leading edge as No.~228 on 20210928. With some well-determined DMs of a small fraction of bursts in Figure~\ref{fig:timeDM}, we see the insignificant DM variation in about one hour each day, and the averaged DM values are consistent with each other on the first three days. A slightly smaller averaged DM value and a weaker trend of DM increasing is seen on the last day 20210928.

In the following analyses, the averaged DM for each day is adopted for data analyses. There is no question that a different DM would cause a different drifting rate, but we do not have a better choice at present. 

\subsection{Burst parameters} 
\label{param}

We measured the observational parameters of 624 bursts, such as the TOA expressed in MJD for the peak of the each sub-burst, the emission peak frequency  ($\nu_{\rm 0}$, in MHz), the sub-burst emission bandwidth ($BW_{\rm e}$, in MHz), the sub-burst width ($W_{\rm sb}$, in ms), the detection signal-to-noise ratio (SNR), and the fluence of a bursts or a sub-burst ($F_{\nu}$, in mJy\,ms). All these parameters are listed in Table~\ref{tab:appendix} in the Appendix.

For each burst, we dedisperse the data by using the average DM value each day and obtain a burst profile first from an obvious frequency range manually selected, because most bursts have emission merely in some parts of the observation band of 500~MHz of the FAST L-band receiver, see examples in Fig.~\ref{fig:gooddm} and plots for all bursts in the Appendix. Guided by the primary burst profile, the observed burst energy over a frequency range is then fitted by a Gaussian function, so that the emission peak frequency $\nu_{\rm 0}$ and the emission bandwidth $BW_{\rm e}$ is determined. Here the $BW_{\rm e}$ is defined as the full frequency width
at the 10\% of emission peak (FWTM) obtained from the fitted Gaussian function. The burst profile is finally obtained from the integrated data of these frequency channels. Note that the fluctuations of emission strength over the frequency channels is not caused by random noises but by the scintillation which we will discuss later. For bursts brightening at an edge of the observed band, it is difficult to have $\nu_{\rm 0}$ and $BW_{\rm e}$ well determined, so we may get a $\nu_{\rm 0}$ outside the range of the FAST band, or we may take the frequency boundary of the L-band receiver as the replacement. For a burst with many sub-bursts, we take the emission band to cover all these sub-bursts to get the burst profiles. The such determined emission bandwidth leads to a more accurate estimation of burst energy, because many bursts of FRB 20201124A have emission in the limited band only, rather than in all 500~MHz band of the FAST L-band 19-beam receiver.

For each sub-burst, the TOA is defined as the arrival time of the emission peak at the infinity frequency. The peak is obtained from the Gaussian function fitted to the de-dispersed burst profile. There are 44, 111, 441 and 865 sub-burst peaks in the first 4 days, respectively, and in total 1459 peaks. The TOA is then converted to the Solar barycentric center using the DE438 ephemeris.

The sub-burst width $W_{\rm sb}$ is defined as the full width of 10\% of maximum (FWTM) measured from a fitted Gaussian function to the dedispersed profile of a sub-burst. The burst width $W_{\rm b}$ is then defined as the overlapping width of these fitted multi-Gaussian functions for all sub-bursts with the 10\% of maximum of two Gaussian functions for the two outermost components. The signal-to-noise ratio (SNR) is calculated from the summed energy of the fitted Gaussian profile relative to the standard deviation ($\sigma$) obtained from the nearby off-pulse ranges with a similar width of a burst. 

The fluence of a burst or sub-bursts, $F_{\nu}$ in units of mJy\,ms, is estimated from the above-defined sub-burst parameters and the system characteristics, via
\begin{equation}
F_\nu = \frac{\rm \sum{S_{i}} * T_{\rm sys} * t_{samp} * 10^{3}}{\rm \sigma * G_0 * \sqrt{\rm n_p * t_{samp} * BW_{e}}}
\end{equation}
here $\rm{\sum S_i}$ is the summed value of on-burst bins, ${\rm T_{sys}}=25$K is the system noise temperature, and $\rm G_0 = 16.1$ K/Jy is the effective gain of the telescope \citep{JiangP2020RAA}, $\rm n_p=2$ is the number of polarization summed, $\rm t_{samp}$ is sampling time (s), and $\rm BW_{e}$ is emission frequency bandwidth (MHz) obtained above.

\begin{figure*}
  \centering
  \includegraphics[width=0.9\textwidth]{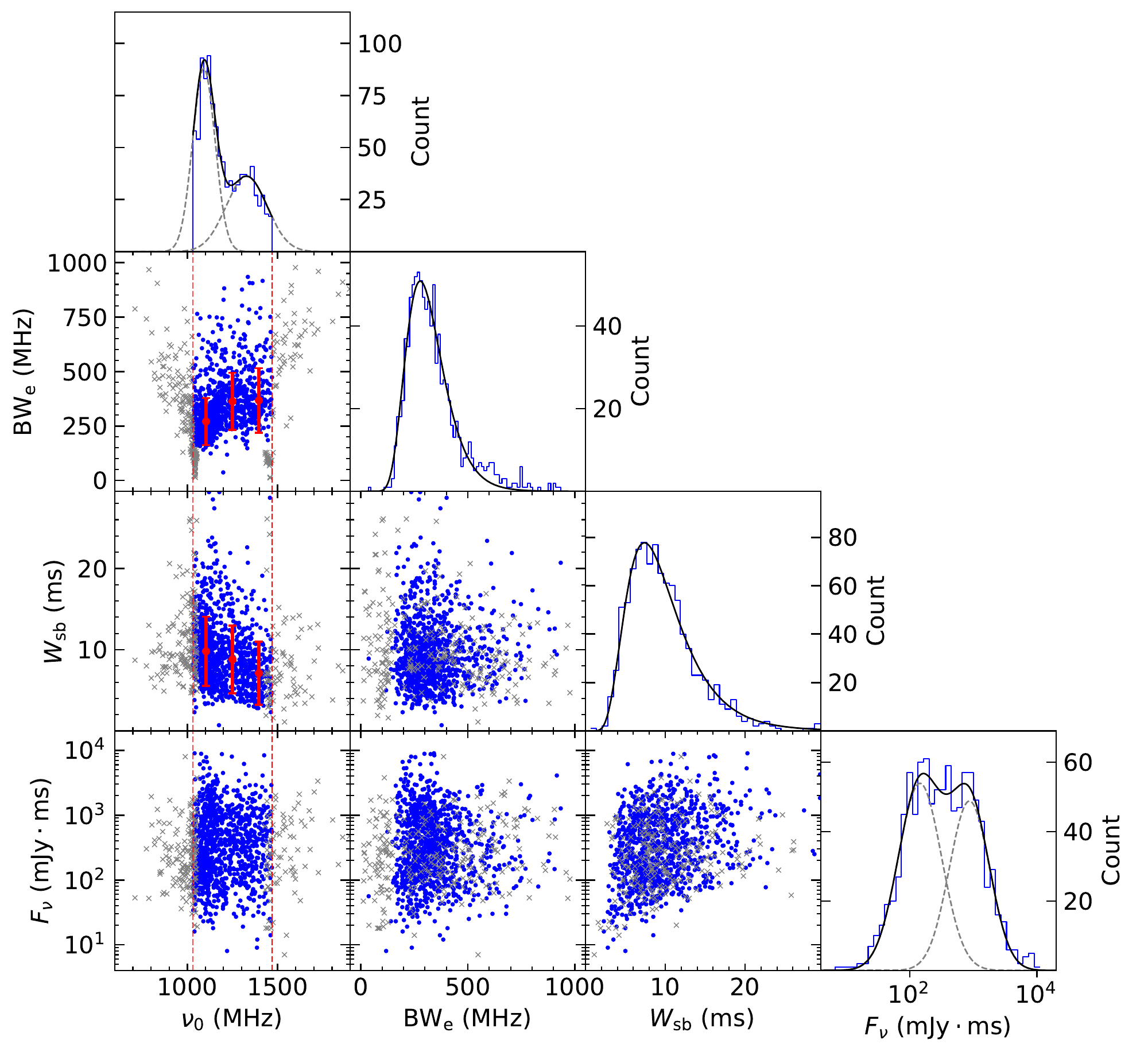}
  \caption{Data distributions of emission peak frequency ($\nu_{\rm 0}$), burst emission bandwidth ($\rm BW_{e}$), sub-burst width $W_{\rm sb}$ and specific fluence $F_{\nu}$ of all sub-bursts of FAST detected bursts. The values are estimated via Gaussian fittings, with a relative uncertainty less than 10\%. The blue dots stand for good values obtained from the fittings (see Sect.~\ref{param}). The gray crosses stand for data with $\nu_{\rm 0}$ outside the frequency range or that $\rm BW_{e}<150$~MHz but $\nu_{\rm 0}$ lower than 1080 MHz or higher than 1420 MHz.  The vertical dashed lines indicate the effective observational frequency range of the FAST from 1031.25 MHz to 1468.75 MHz. A few points with $W_{\rm sb}>30$~ms are not included. The histogram of $\nu_{\rm 0}$ distributions shows two peaks, which are fitted with two Gaussian functions peaking at 1901.9~MHz and at 1327.9~MHz. The histogram of $\rm BW_{e}$ and $W_{\rm sb}$ distributions can be fitted by a log-normal function, and they peak at 277~MHz and 7.4~ms, respectively. The histogram of the log~$F_{\nu}$ distribution has a dip near the peak but can be fitted with two Gaussian functions with peaks at log~${F_\nu= 2.1\pm0.3}$ and log~${F_\nu= 2.9\pm0.4}$, respectively. It is noticed that the sub-burst width $W_{\rm sb}$ tends to be larger for the bursts emerging at the lower part of the band, while the emission band widths  $\rm BW_{e}$ of these bursts tend to be smaller than the bursts emerging at high frequency part of the band, as illustrated by the median and the standard deviation for the three sub-bands in the first column of data distribution of $\nu_{\rm 0}$ vs $\rm BW_{e}$ and distribution of $\nu_{\rm 0}$ vs $W_{\rm sb}$. }
\label{fig:FRB_freq_w_flux}
\end{figure*}

\section{Burst Morphology and Classification}
\label{sect:BMclassify}

As shown in the plots for 624 bursts in Figures~\ref{fig:appendix:D1L} to \ref{fig:appendix:NEH}, the burst emission detected by the FAST always has a narrow band, sometimes detected only in the lower or upper edges of the FAST band of 500~MHz, occasionally have a wide-band covering all the band. Another interesting phenomenon, as burst No.15 in Figure~\ref{fig:gooddm} shows, is that a burst emission starts to appear in the upper band, and slightly later a new component emerges in the central band, and then another new component comes later in the lower band. This is the so-called downward frequency drifting of burst emission \citep{Hessels19}. With such a large sample of detected bursts and their characteristics, the bursts can be classified according to their morphology in the time-frequency waterfall plots. In this section, we first analyze the frequency distribution of the burst emission, then discuss the drifting patterns, and finally classify the bursts based on morphology.

\subsection{Burst parameters distributions}
\label{subsect:spectrum}

After obtaining the burst parameters, i.e.,  emission peak frequency $\nu_{\rm 0}$ and the emission bandwidth $BW_{e}$, sub-burst width $W_{\rm sb}$ and fluence $F_{\nu}$ of every sub-burst of all bursts, we perform a statistics quantitatively for these parameters in Figure~\ref{fig:FRB_freq_w_flux}. Due to the limitation of the FAST observation bandwidth, some fitted $\nu_{\rm 0}$ for the sub-bursts emerging near the band edges have the best fitted value outside the FAST frequency range or just  near 1080 or 1420 MHz but with a $\rm BW_{\rm e}< 150$ MHz. These account for 21\% of the total sub-bursts. 
The fitted parameters for these 21\% sub-bursts are unusual and have a large uncertainty and are therefore dropped. The remaining approximately 79$\%$ of data as indicated by the blue symbols in Fig.~\ref{fig:FRB_freq_w_flux} are used for the statistical analysis.

First of all, we notice that there are two peaks in the histogram of the $\nu_{\rm 0}$ distribution, which are at 1091.9 MHz and 1327.9 MHz as the Gaussian fitting gives, and the standard deviation $\sigma$ of the Gaussian distribution is 62.0 MHz and 113.0 MHz, respectively. The number of sub-bursts distributed in the lower-frequency component is larger, suggesting that this FRB source preferably emits at the lower part of the FAST band. 

One may question if the two peaks of the $\nu_{\rm 0}$ distribution in Fig.~\ref{fig:FRB_freq_w_flux} are caused by removing the strong Radio frequency interference (RFI) around 1250~MHz within the observation bandwidth of FAST. Because RFI appears mostly in a few tens of MHz in the central part of the observation band, the emission strength of the bursts from channels in the two sides of the RFI bands are not affected. Because the emission bandwidth is as wide as more than 200 MHz,  discarding some channels with RFI does not affect the fitting results for both $\nu_{\rm 0}$ and $\rm BW_{\rm e}$. This has also been verified by simulations (not presented in this paper). 

The emission bandwidth $\rm BW_{\rm e}$ in Fig.~\ref{fig:FRB_freq_w_flux} follows a log-normal distribution with a peak at about 277 MHz. The borderlines of the full width half maximum of the emission bandwidth are at about 193 MHz and 399 MHz, narrower than the FAST observed bandwidth. 

The sub-burst width $W_{p}$ in Fig.~\ref{fig:FRB_freq_w_flux} also follows a log-normal distribution concentrated in the range of 1 to 30 ms, with a peak at about 7.4$_{-3.0}^{+5.0}$~ms. We notice that sub-burst width is wider for the bursts emerging at the lower part of the band. We divide the band into three sub-bands centered at 1104.2~MHz, 1250.0~MHz and 1395.8~MHz,  and found that sub-burst widths have median values of $W_{\rm sb}$ of 9.8$\pm$4.3~ms, 8.8$\pm$4.2~ms and 7.1$\pm$3.9~ms, respectively, for the sub-bursts peaking in these three sub-bands. On the other hand, the emission bandwidth becomes wider for the sub-bursts emerging at the higher part of the band, i,e, $\rm BW_{\rm e}$ =~270.4$\pm$108.3~MHz, 362.8$\pm$131.2~MHz and~367.3$\pm$148.2~MHz for the three sub-bands. The correlation coefficient between $\nu_{\rm 0}$ and $W_{\rm sb}$ is -0.25 and that between $\nu_{\rm 0}$ and $\rm BW_{\rm e}$ is 0.36.

The distribution of sub-burst specific fluence $F_{\nu}$ has a dip near the peak but can be roughly fitted with two Gaussian functions in the logarithmic scale, peaking at log${F_{\nu}}$=2.2$\pm$0.3 and log${F_{\nu}}$=2.9$\pm$0.3. Details of the energy distribution of FRB 20201124A are discussed in Paper II (Y.K. Zhang et al. 2022).

\begin{table}
\centering
\caption{Drifting rate $R_d$ of single-component bursts and multi-component bursts.}
\label{tab_downdrifting}
\setlength{\tabcolsep}{4.0pt}
\begin{tabular}{lrccrr}
\hline\hline\noalign{\smallskip}
Date    & Burst &No.&$\nu_{\rm 0}$&$BW_{\rm e}$& $R_d$(err)\\
        & No.    & Comp.       & (ms)           &(MHz)     & (MHz~ms$^{-1}$)\\
\hline
\multicolumn{6}{c}{Single component bursts}\\
\hline
20210925& 8   & 1       & 1153.8   &   89.5  &     -58(7)  \\
20210926& 9   & 1       & 1188.8   &   118.8 &     -64(8)  \\
20210927& 54  & 1       & 1151.3   &   73.1  &     -95(12)  \\
20210927& 80  & 1       & 1124.9   &   81.9  &     -59(7)  \\
20210927& 87  & 1       & 1127.5   &   101.5 &     -54(7)  \\
20210927& 125 & 1       & 1250.7   &   118.4 &     -77(10)  \\
20210927& 138 & 1       & 1057.3   &   70.8  &     -55(7)  \\
20210928& 3   & 1       & 1146.2   &   80.4  &     -58(7)  \\
20210928& 39  & 1       & 1280.2   &   118.8 &     -68(9)  \\
20210928& 46  & 1       & 1367.0   &   69.1  &    -166(18)  \\
20210928& 61  & 1       & 1249.7   &   118.7 &     -36(4)  \\
20210928& 112 & 1       & 1383.2   &   105.5 &     -58(7)  \\
20210928& 125 & 1       & 1342.4   &   118.8 &     -78(10)  \\
20210928& 143 & 1       & 1151.2   &   79.3  &     -41(5)  \\
20210928& 154 & 1       & 1413.9   &   89.0  &     -62(7)  \\
20210928& 162 & 1       & 1204.3   &   118.8 &     -104(13)  \\
20210928& 179 & 1       & 1165.0   &   74.0  &     -52(6)  \\
20210928& 187 & 1       & 1129.1   &   73.8  &     -93(12)  \\
20210928& 189 & 1       & 1260.1   &   117.5 &     -62(8)  \\
20210928& 197 & 1       & 1174.1   &   81.9  &     -61(5)  \\
20210928& 210 & 1       & 1192.6   &   98.0  &     -66(8)  \\
20210928& 220 & 1       & 1269.4   &   91.2  &     -128(17)  \\
20210928& 236 & 1       & 1160.2   &   81.9  &     -28(3)  \\
20210928& 275 & 1       & 1217.6   &   107.4 &     -104(14)  \\
20210928& 276 & 1       & 1135.2   &   77.2  &     -54(7)  \\
20210928& 280 & 1       & 1333.2   &   73.5  &     -74(9)  \\
20210928& 286 & 1       & 1215.1   &   99.9  &     -50(6)  \\
20210928& 291 & 1       & 1256.5   &   118.8 &     -86(11)  \\
20210928& 298 & 1       & 1351.3   &   84.0  &     -74(9)  \\
20210928& 355 & 2       & 1398.0   &   74.7  &     -42(5)  \\
20210928& 369 & 1       & 1298.2   &   92.6  &     -90(12)  \\
\hline
\multicolumn{6}{c}{Multiple component bursts}\\
\hline
20210925& 15  &1-3      & 1323.6   &  89.1  & -14(2) \\
20210926& 5   &1-4      & 1440.3   & 104.9  & -18(2) \\
20210926& 22  &1-4      & 1254.7   &  51.2  &  -6(4) \\
20210927& 59  &1-4      & 1426.3   &  72.3  & -17(7) \\
20210927& 75  &1-4      & 1356.7   &  96.0  & -29.0(6)  \\
20210927& 111 &1-3      & 1473.0   & 127.7  & -28(6) \\
20210927& 165 &1-4      & 1407.4   &  85.5  & -18(2) \\
20210928& 3   &4-6      & 1153.1   &  34.5  &  -5(3) \\
20210928& 182 &1-3      & 1361.7   & 118.4  & -23(5) \\
20210928& 186 &1-4      & 1387.2   &  86.9  & -28(3) \\
20210928& 267 &1-3      & 1411.0   &  90.6  & -21(2) \\
20210928& 296 &1-4      & 1402.9   & 101.8  & -19(5) \\
\hline
\hline
\end{tabular} 
\end{table}

\subsection{Frequency drifting of bursts}
\label{subsect:drifting}

\begin{figure}
  \centering
  \includegraphics[width=0.85\columnwidth]{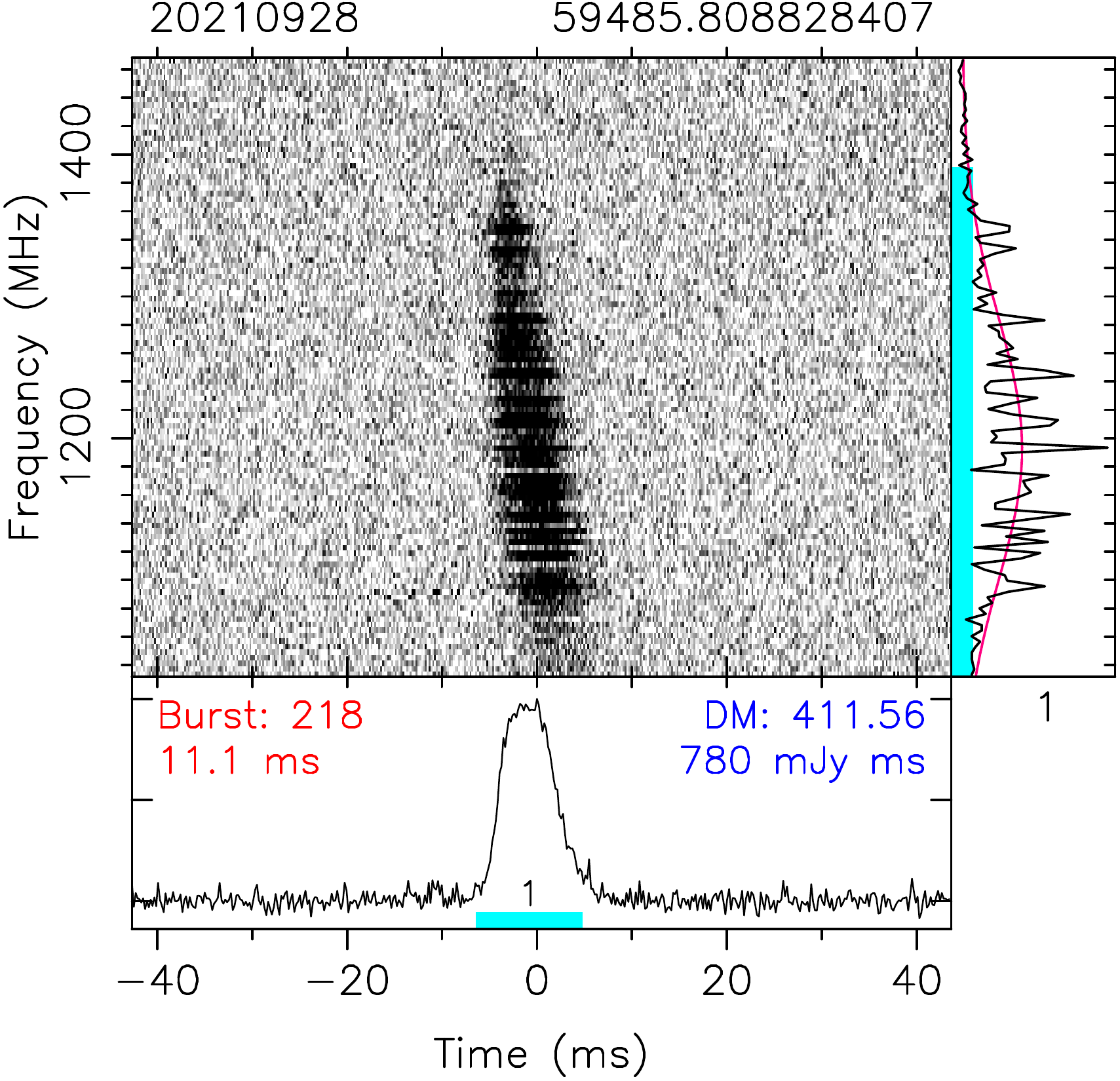}
  \includegraphics[width=0.60\columnwidth]{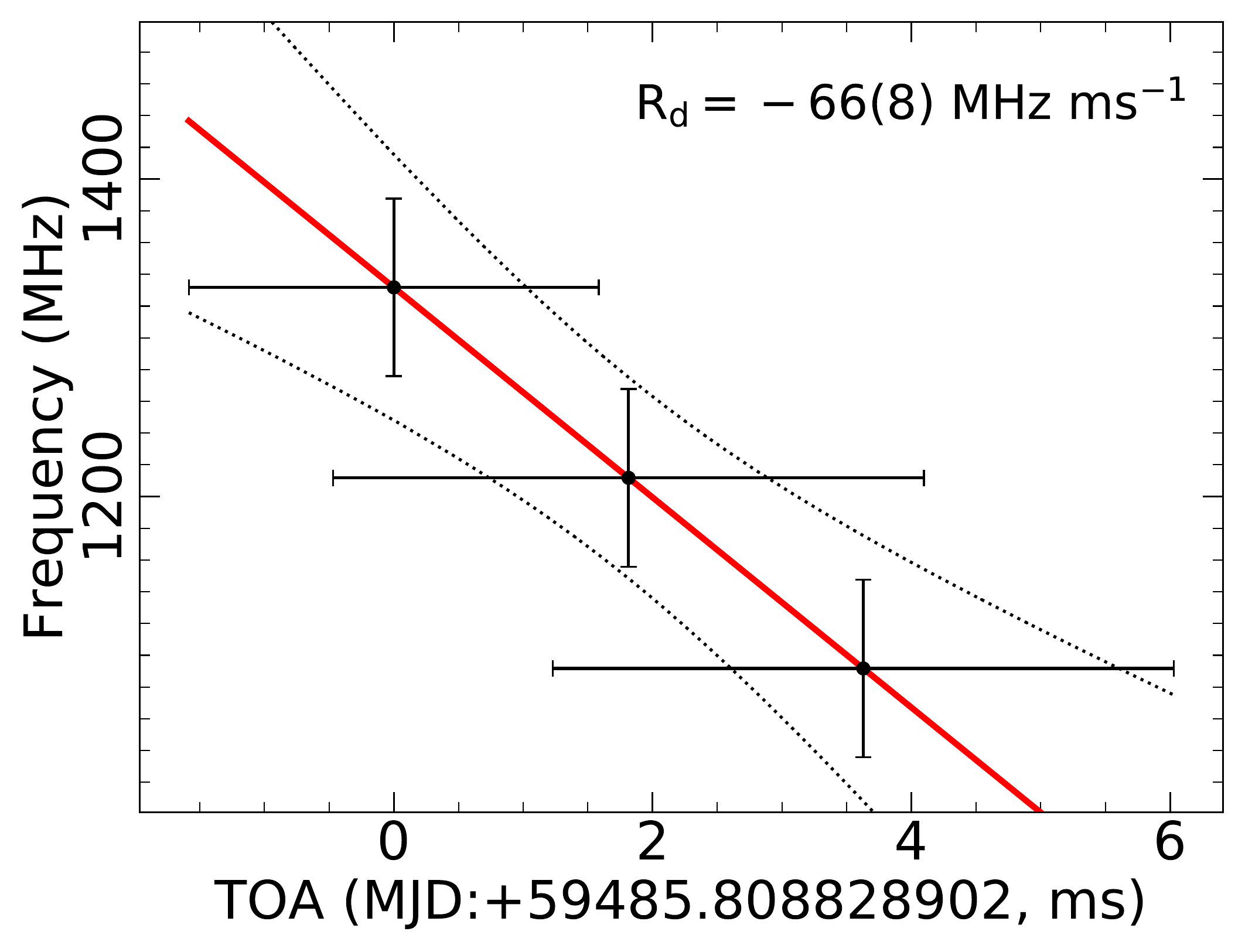} 
  \caption{
    An example of frequency drifting for a single-component burst, i.e. burst No.~218 detected on 20210928. The top panel in Figure~\ref{fig:gooddm} is the dynamic spectrum. The front edge of this burst is aligned well with the average DM of the day. The emission bandwidth of the burst $\rm BW_{\rm e}$, as marked in the top-right sub-panel, is cut to three sub-bands, and the relative TOAs of the burst are obtained respectively for the frequency-scrunched profiles of the three sub-bands, as shown in the bottom panel. The downward frequency drifting rate is then found to be $R_d = -66\pm8\rm~MHz~ms^{-1}$. The thick solid line stands for the best-fitting result, and the dotted lines in black indicate the range of uncertainty.
}
\label{fig:gooddrifting}
\end{figure}

An important feature of FRB 20201124A is the time-frequency drifting pattern clearly shown in most bursts, not only for bursts with multiple components, but also for single component bursts. In the literature  \citep{Gajjar18,Chime19a,Hessels19,Hilmarsson21}, frequency drifting was always discussed for the bursts with multiple components and the authors defined the shift of sub-bursts in the 2-D frequency and time domain. Our observations of FRB 20201124A in Figures~\ref{fig:appendix:D1W} to \ref{fig:appendix:D1L} show that even the bursts with a single component also have a clear drifting pattern when the front edge of burst is aligned with the average DM, as shown in the example presented in Figure~\ref{fig:gooddrifting}.

\begin{figure}
  \centering
  \includegraphics[width=0.75\columnwidth]{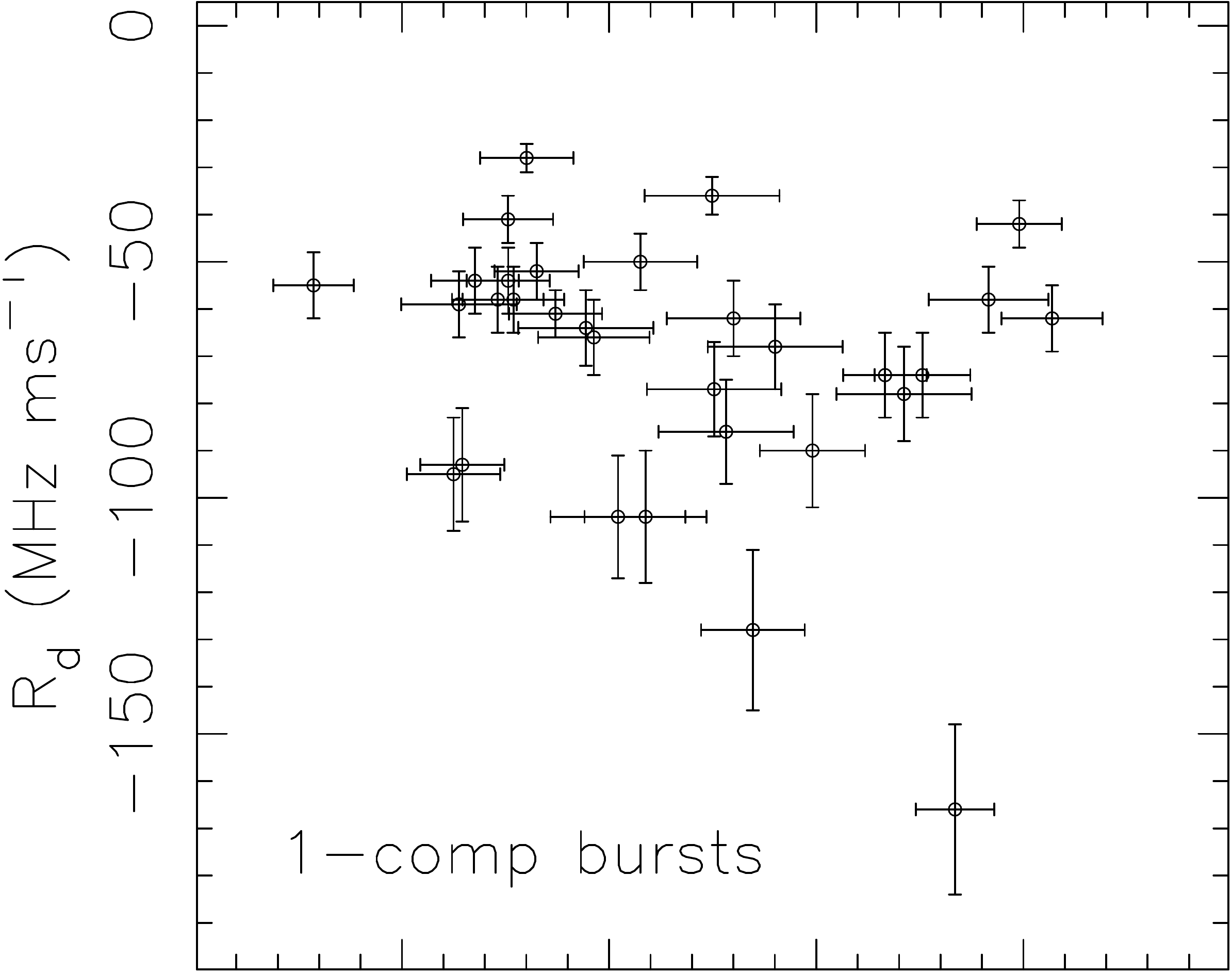}
  \includegraphics[width=0.75\columnwidth]{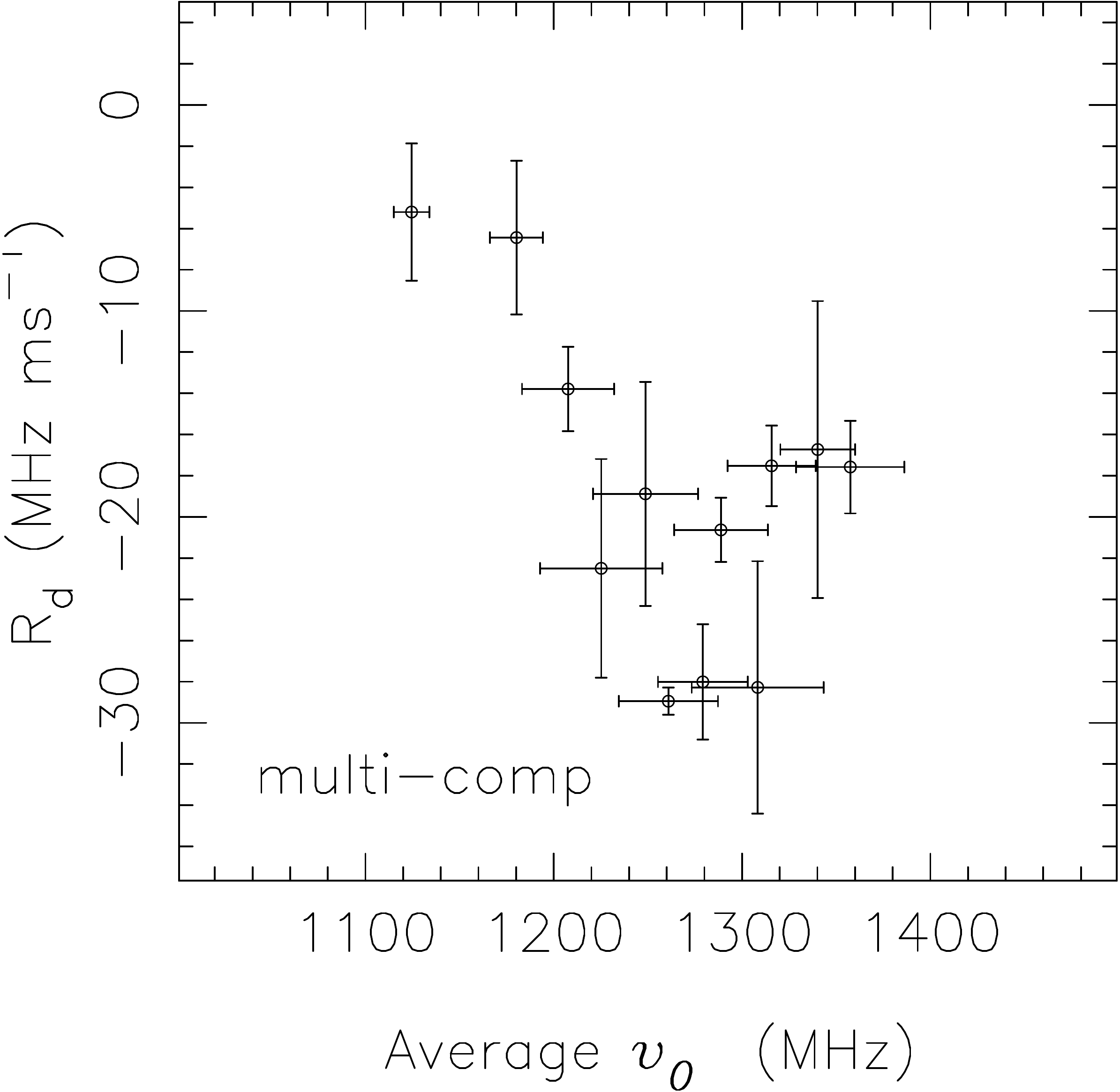}
  \caption{ 
The drifting rate distribution of single-component bursts (top panel) and multi-component bursts (bottom panel) listed in Table~\ref{tab_downdrifting}, plotted against the emission peak frequency. 
}
\label{fig:downdrifting}
\end{figure}

First we look at drifting of single component bursts. We cut the emission band of the burst $\rm BW_{\rm e}$ to three sub-bands, then get three averaged burst profiles from each sub-band, and obtain the TOAs of sub-burst peaks with Gaussian fitting. The drifting rate, $R_d$ in MHz/ms, is obtained as the slope of the least square fit to TOAs for the three sub-bands, as demonstrated in Figure~\ref{fig:gooddrifting}. The drifting rate $R_d$ depends on the implemented DM value for the burst, and the data in Table~\ref{tab_downdrifting} are obtained with the average DM of the day. The uncertainty of $R_d$ depends on the uncertainty of the central frequency and the TOAs in each sub-band. In principle, we can measure $R_d$ for each sub-burst. Nevertheless, to examine how the bursts drift, we choose only 31 single-component bursts with $\rm BW_{\rm e} > 250$ MHz and burst detection SNR $>$50. Their drift rates $R_d$ are listed in Table~\ref{tab_downdrifting} and plotted against the emission peak frequency in Figure~\ref{fig:downdrifting}.

For multi-component bursts, we selected 12 bursts with at least three sub-bursts. Each sub-burst is taken as a component to obtain the central emission frequency $\nu_{\rm 0}$ and the TOA of the frequency scrunched profiles, e.g. the multi-component burst No.~15 on 20210925 in the left panel of Figure~\ref{fig:gooddm}. The precursor and post-cursor near the band edges are excluded. We confidently obtain the frequency drifting rate for these 12 multi-component bursts listed in Table~\ref{tab_downdrifting} and their values are also plotted against the emission peak frequency in Figure~\ref{fig:downdrifting}. 

\begin{figure*}
\flushleft
\includegraphics[height=46mm]{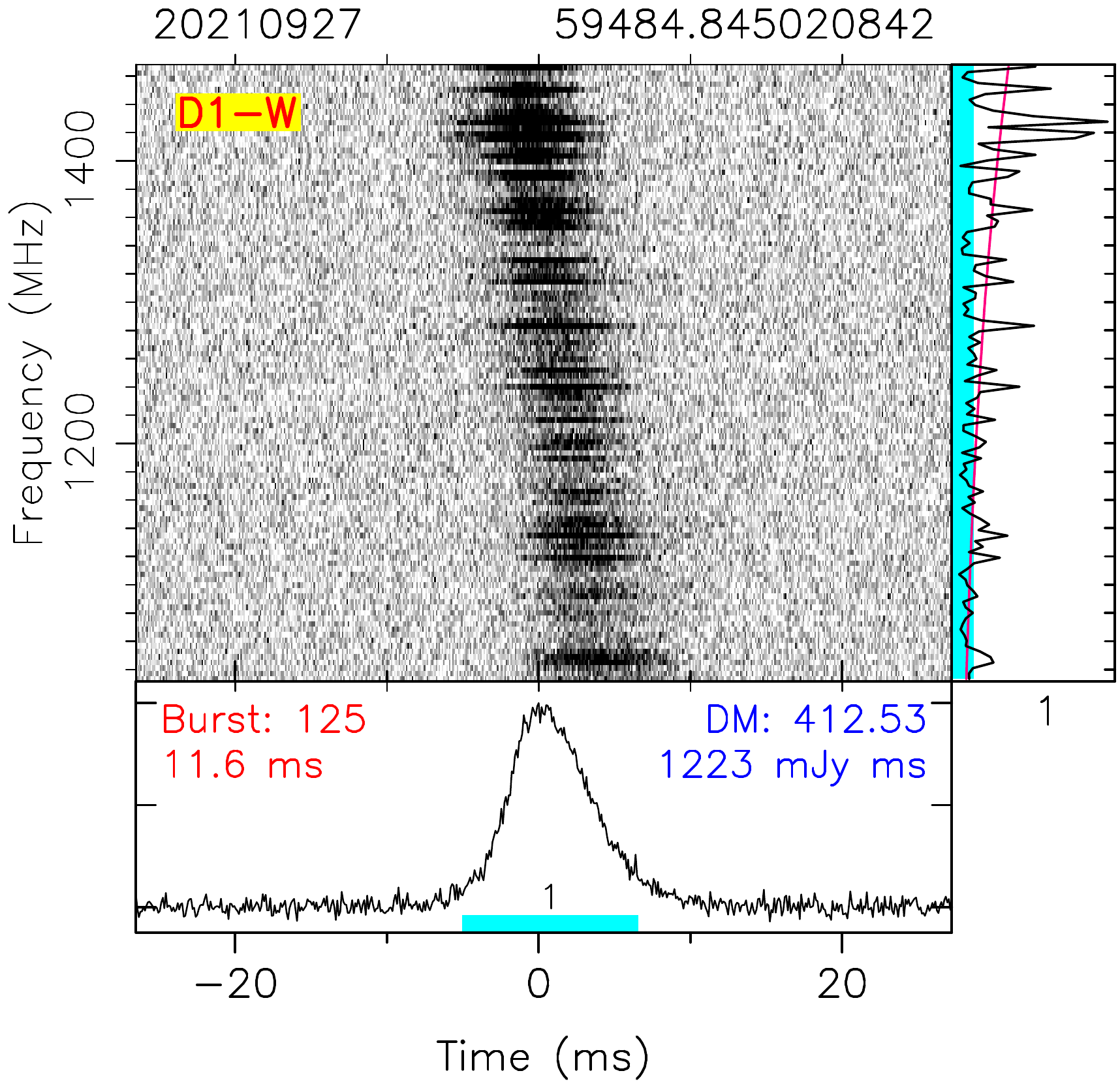} 
\includegraphics[height=46mm]{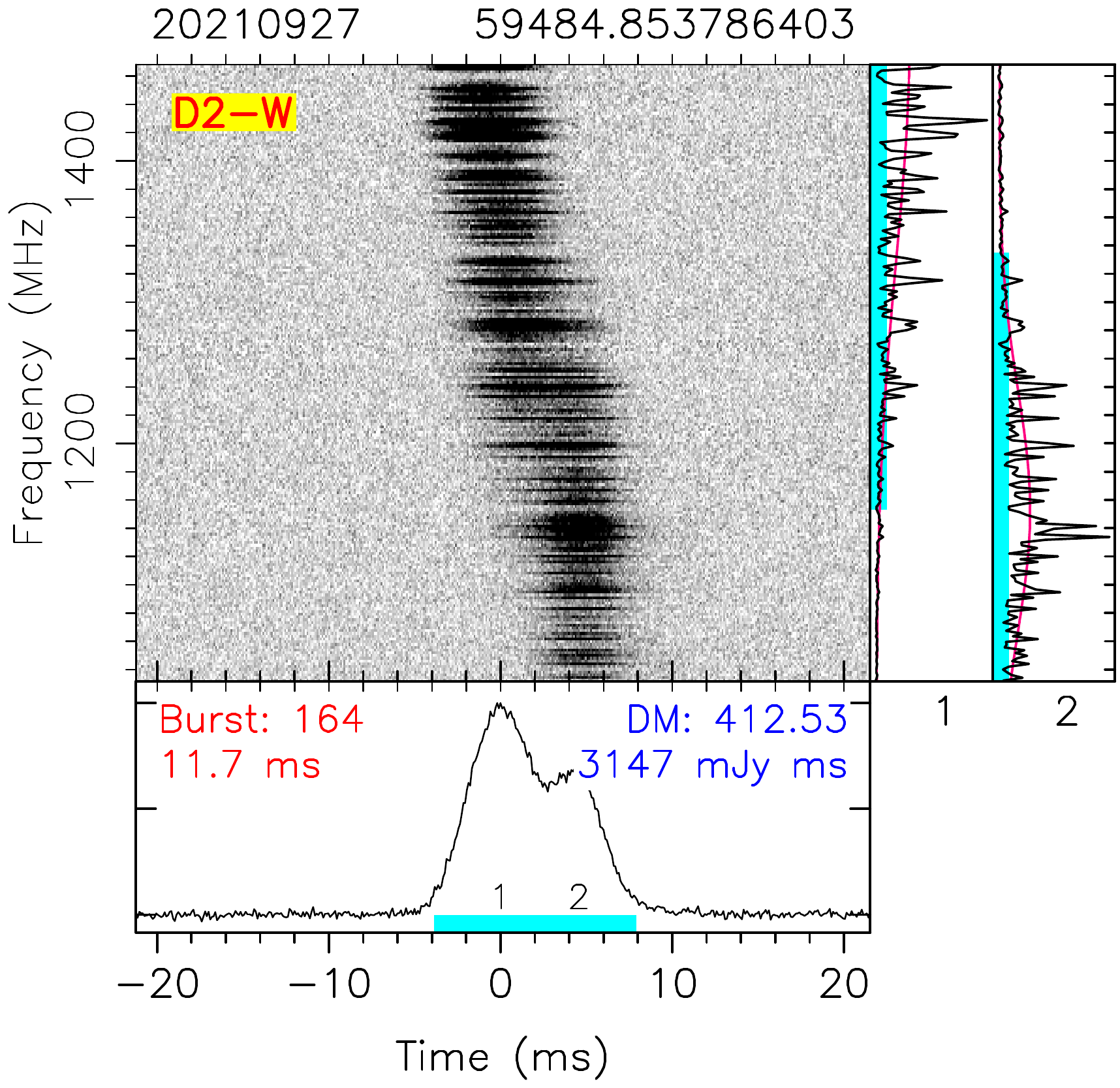}   
\includegraphics[height=46mm]{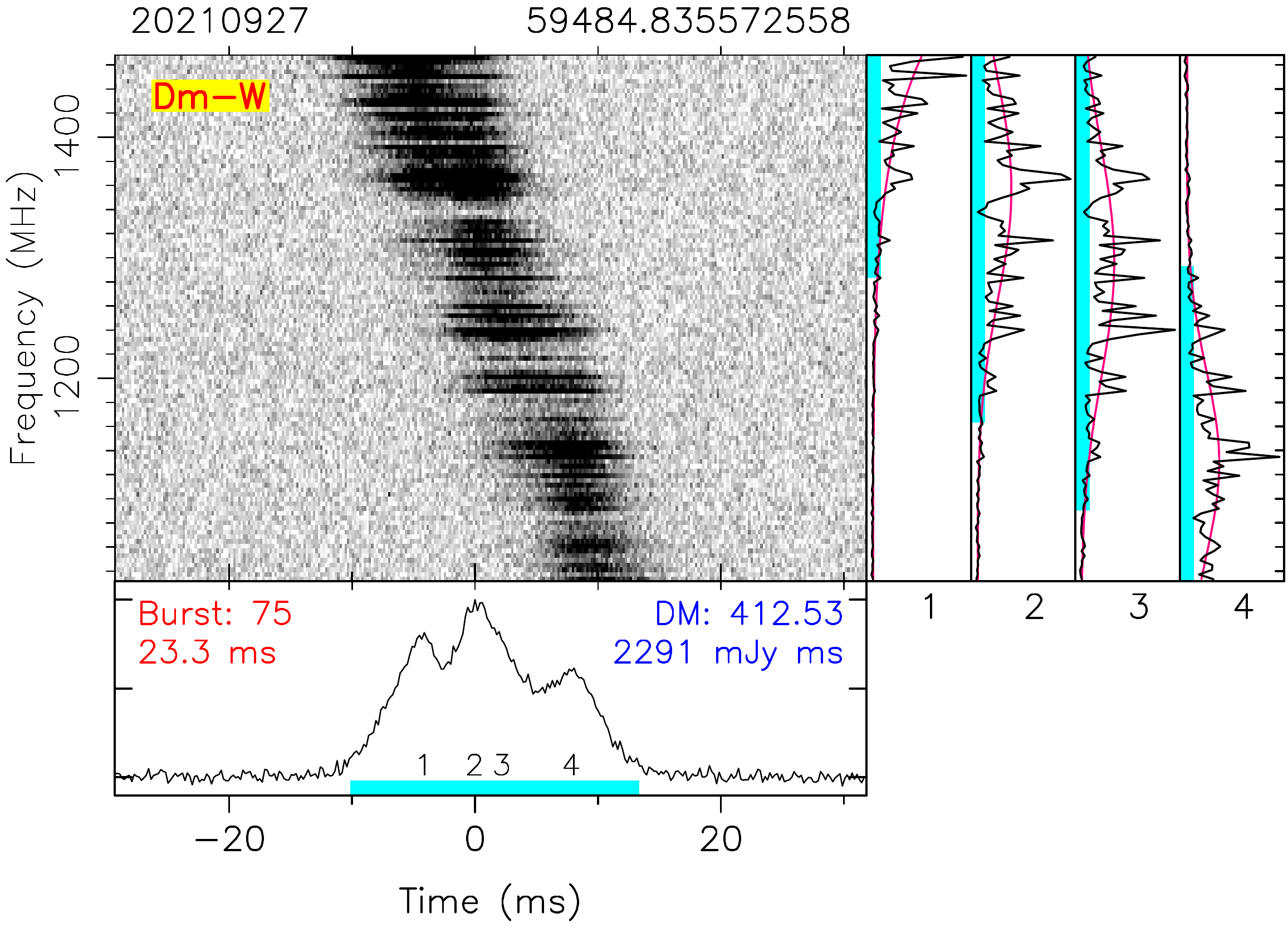}\\[2mm] 
\includegraphics[height=46mm]{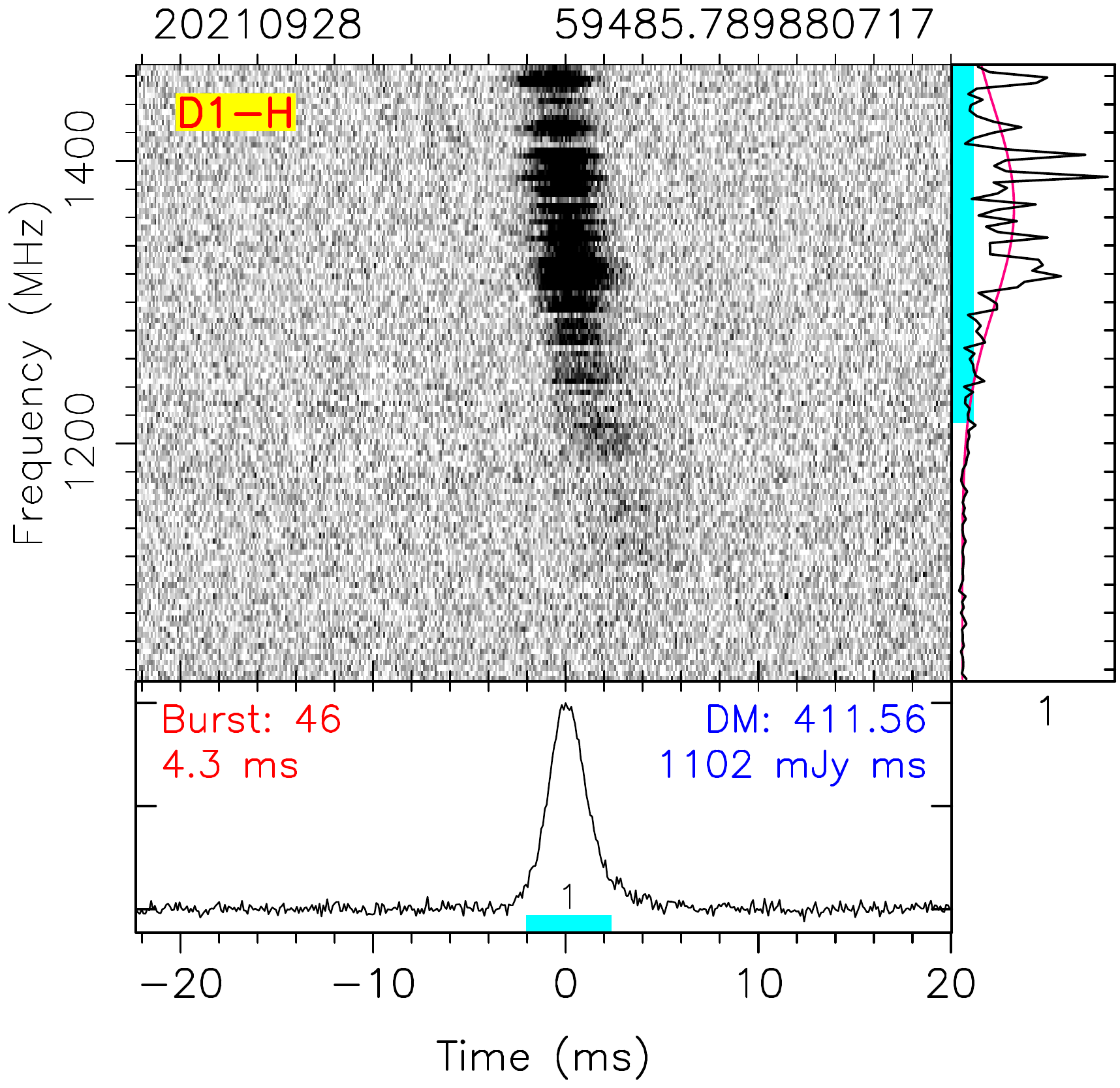}  
\includegraphics[height=46mm]{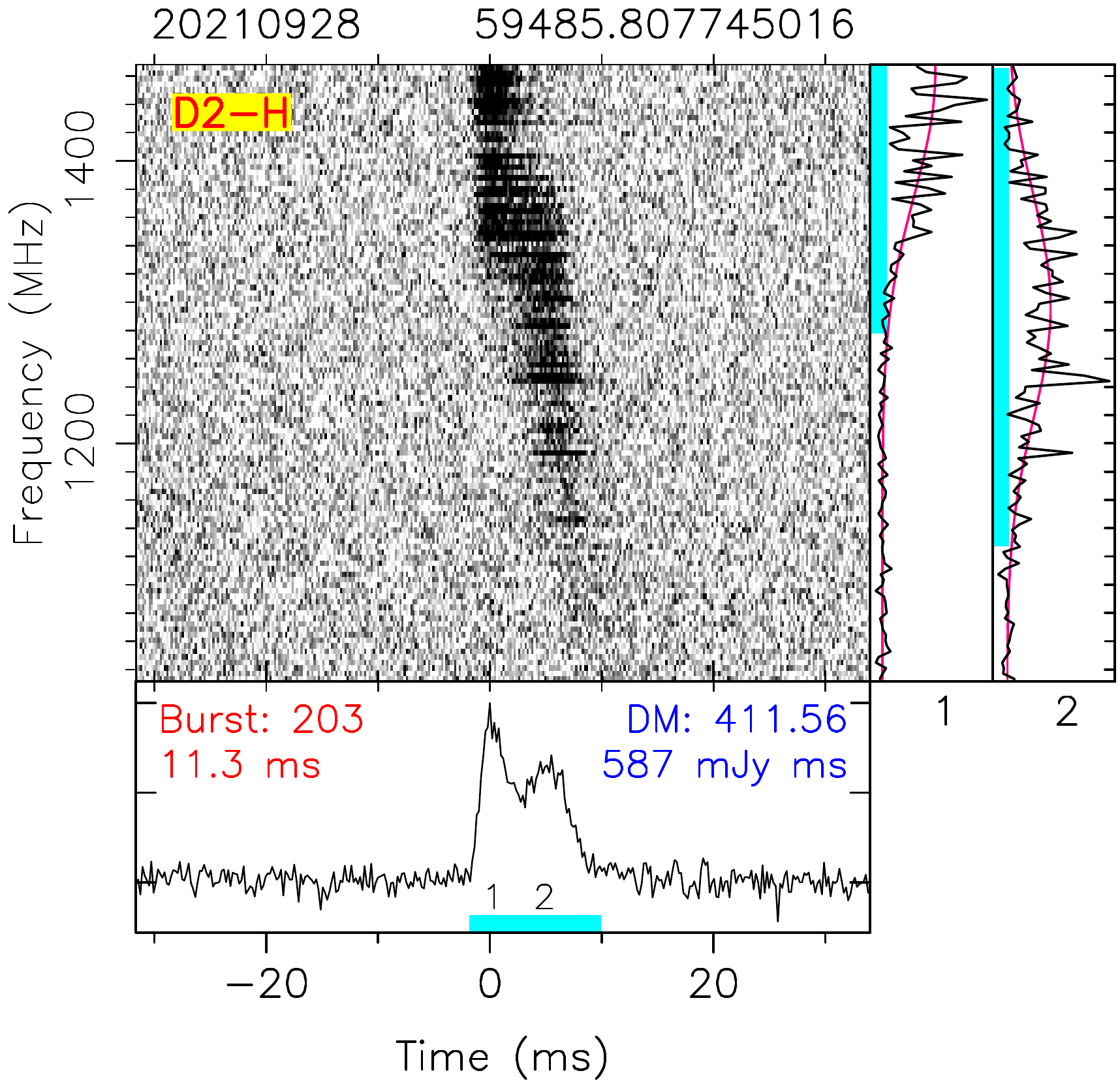}
\includegraphics[height=46mm]{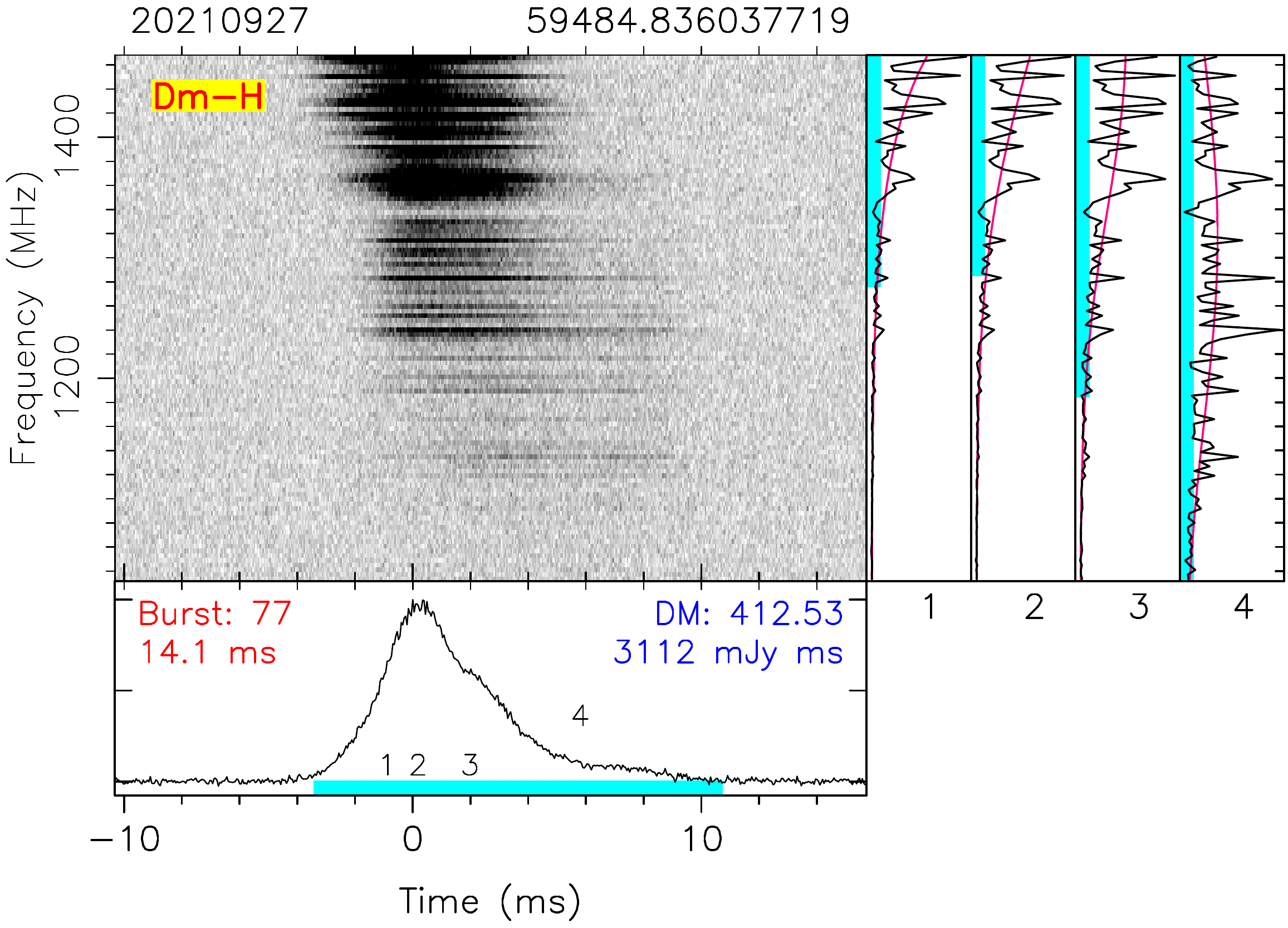}\\[2mm] 
\includegraphics[height=46mm]{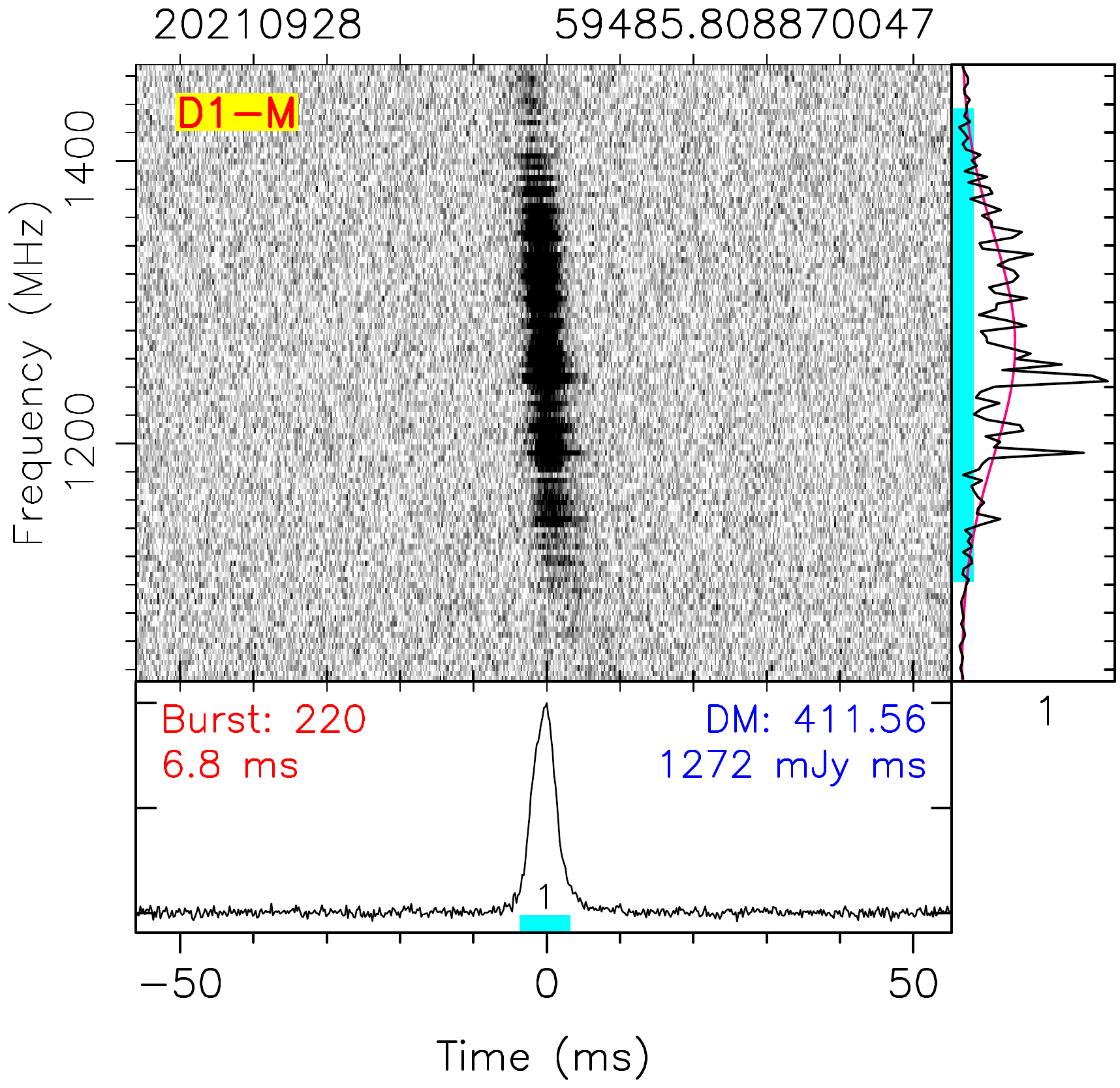} 
\includegraphics[height=46mm]{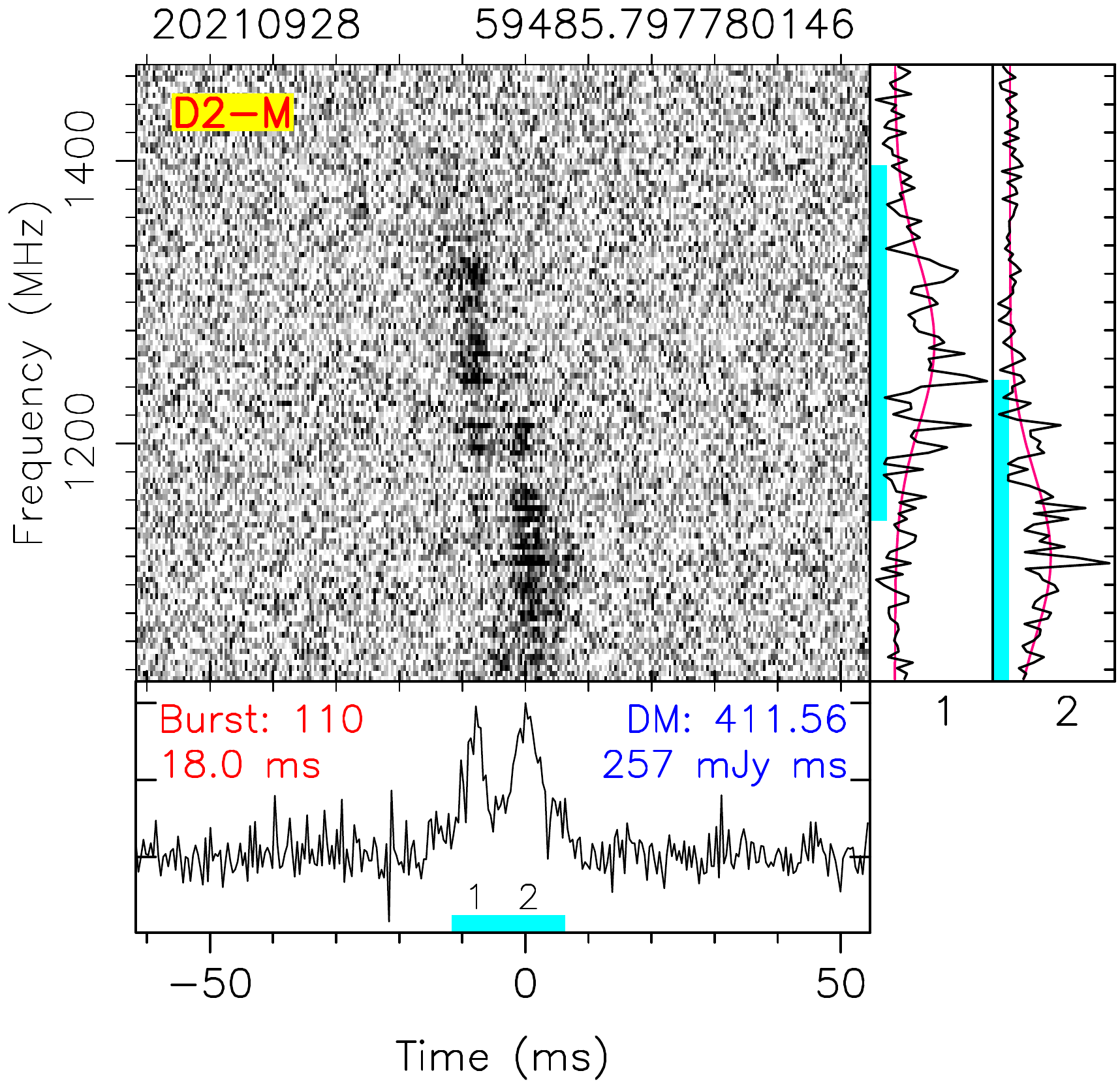}
\includegraphics[height=46mm]{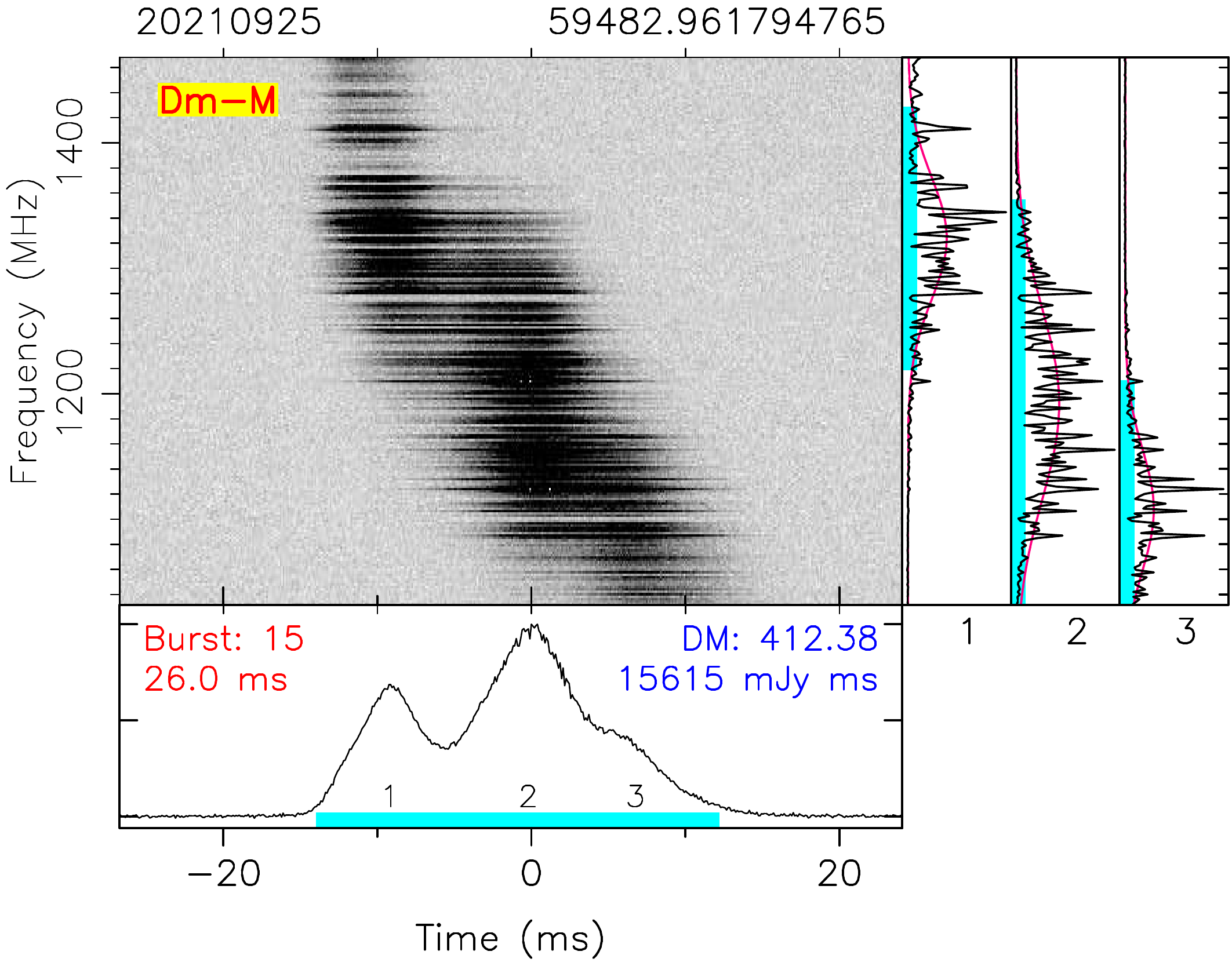}\\[2mm]
\includegraphics[height=46mm]{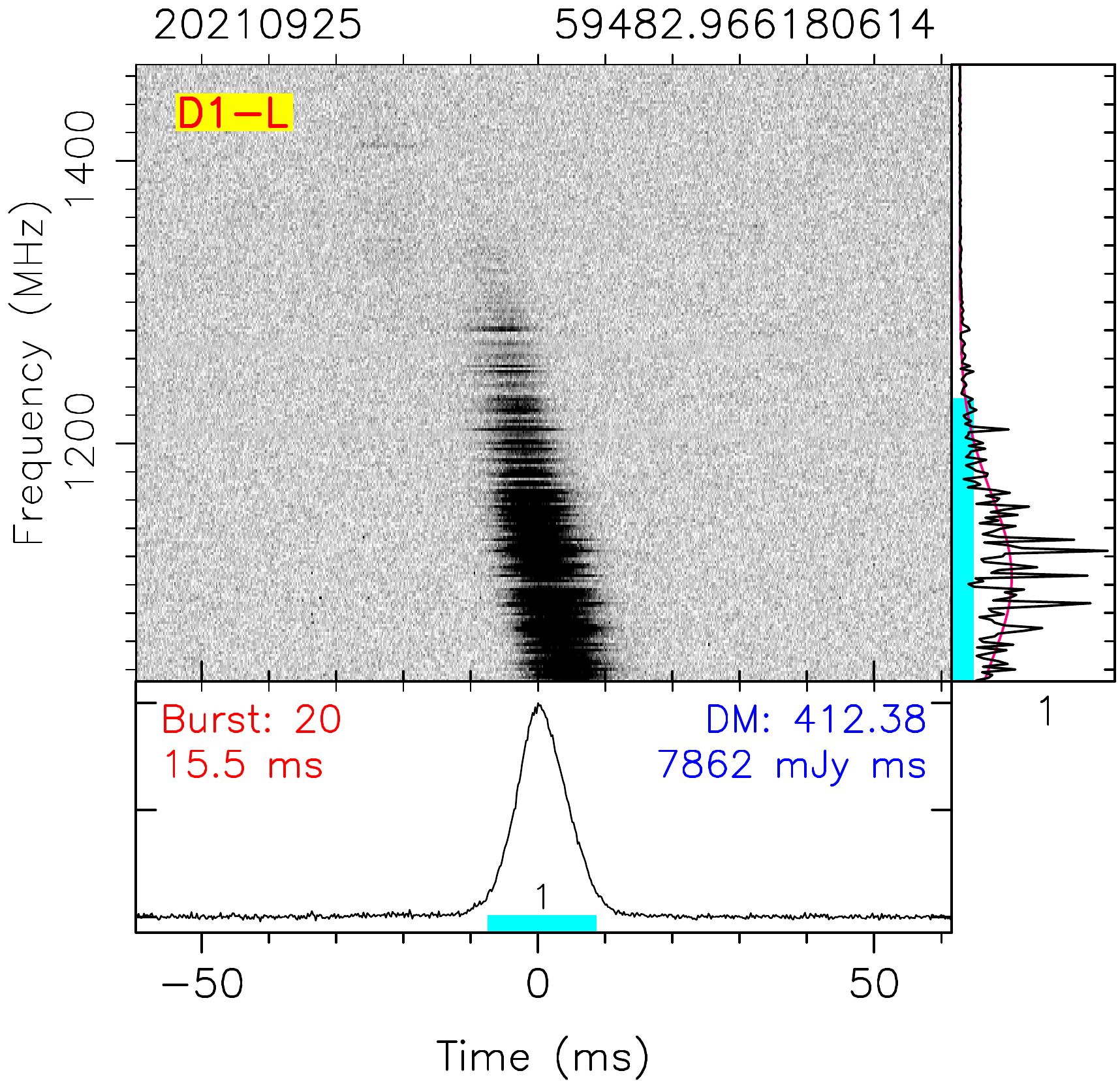} 
\includegraphics[height=46mm]{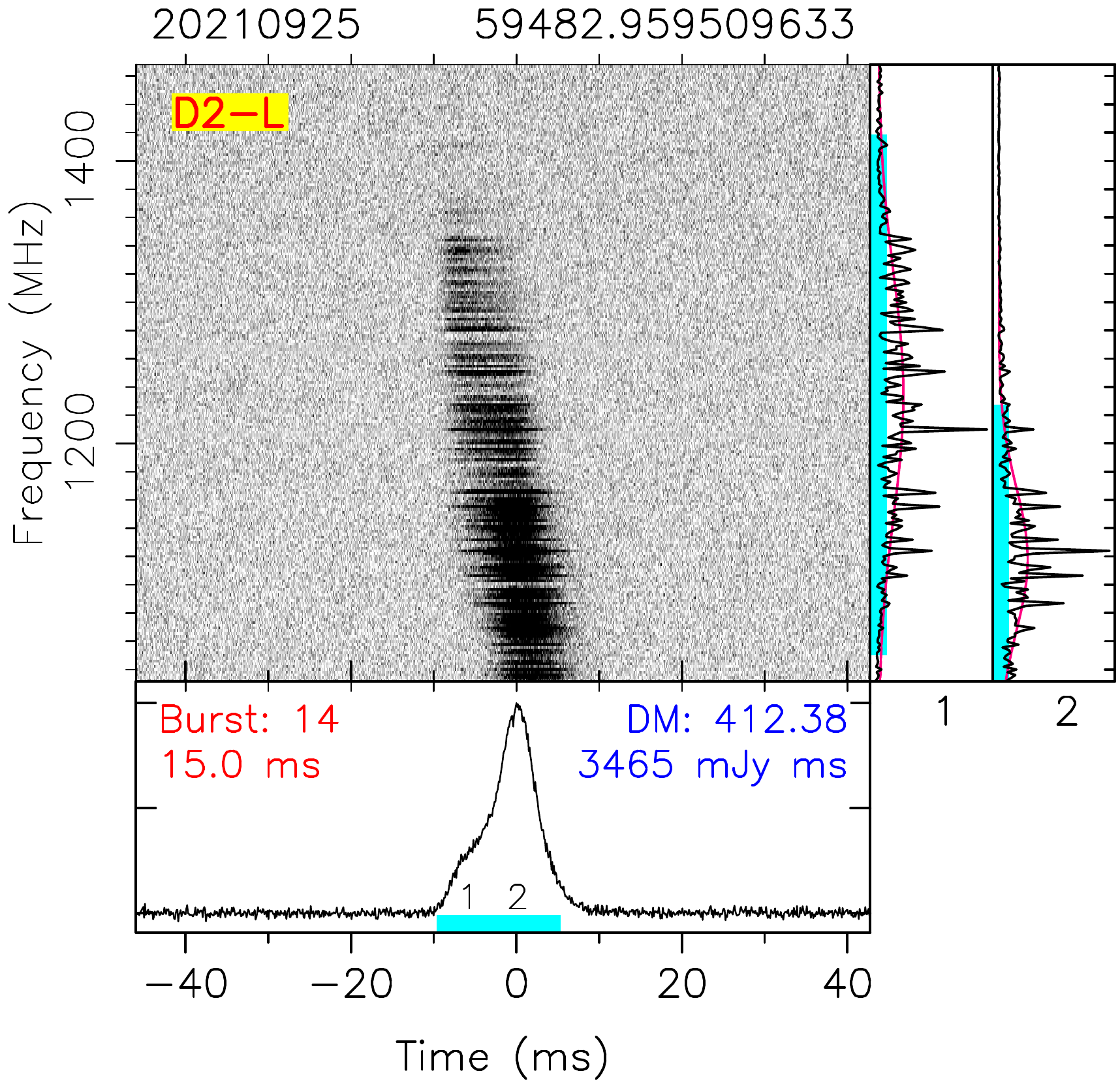} 
\includegraphics[height=46mm]{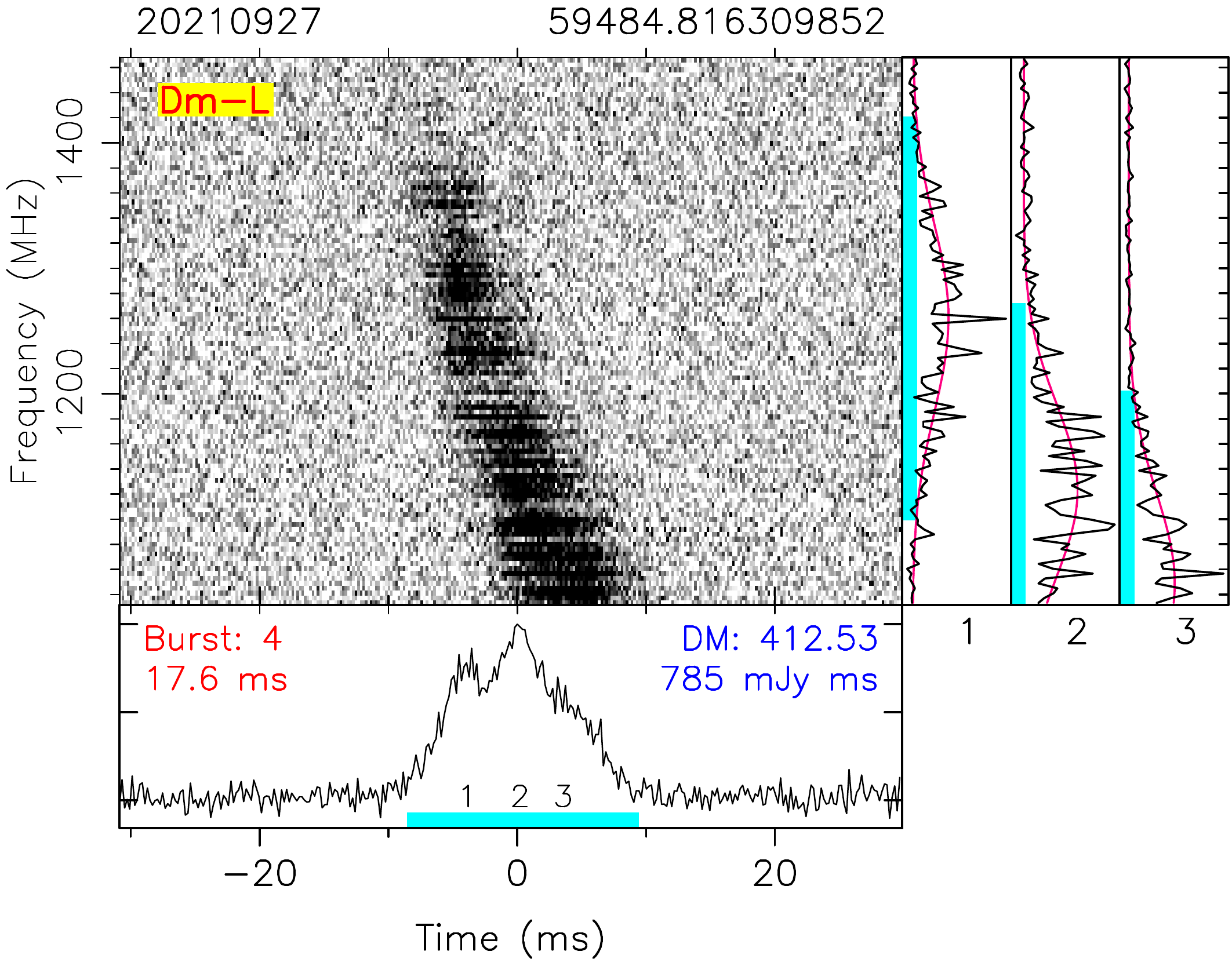}
\caption{
Dynamic spectra for frequency downward-drifting bursts with one (in the left column panels), two (in the middle column panels) and multiple components, and their emission is shown in the wide-band, the high part, the middle part, or the lower part of the FAST band (from top to bottom). The plot for each burst includes not only the water-fall plot but also the frequency-scrunched burst profile in the bottom sub-panel, with the number of the profile components and burst width marked. In the left sub-pannels are the burst energy distribution over frequency for each component, fitted with Gaussian functions, with the effective emission bandwidths are marked. The observation date and the TOA of the burst peak are marked on the top of the water-fall plot, and the classification of each burst is marked on the plot. The burst number and burst width, as well as the DM and fluence are marked in the lower sub-panel. }
\label{fig:class:D-drift}
\end{figure*}

\begin{table*}
\renewcommand{\arraystretch}{1.2}
\centering
\caption{Classification of the bursts and the detected burst numbers for each category for the burst detected in this 2021-09-(25-28) active episode.}
\label{tab:ClassProp}
\begin{tabular}{|c|c|l|l|l|l|}
\hline
\multirow{2}{*}{Drifting mode} & Component &\multicolumn{4}{c|}{Burst emission emerging in the frequency band}  \\
\cline{3-6}
    & No.  & Wide band   & High part   & Middle part   & Lower part    \\
\hline 
\multirow{3}{*}{\makecell[c]{Downward: 263}}  
    & one       & D1-W: 18 & D1-H: 16 & D1-M: 31 & D1-L: 68 \\
\cline{2-6}
    & two       & D2-W: 22 & D2-H: 14 & D2-M: 11 & D2-L: 48 \\
\cline{2-6}
    & multiple  & Dm-W: 14 & Dm-H: 6  & Dm-M: 8 & Dm-L: 7 \\
\hline
\multicolumn{2}{|c|}{Upward: 3}
          &    & U-H: 1  &    & U-L: 2\\
\hline 
\multicolumn{2}{|c|}{No evidence: 121} 
          &    & NE-H: 10  &    & NE-L: 111\\ 
\hline
\multicolumn{2}{|c|}{No Drifting: 35} & \multicolumn{4}{c|}{ND: 35} \\
\hline \hline
\multicolumn{2}{|c|}{Complex: 203}  & \multicolumn{4}{c|}{C: 203 (including 157 cluster-bursts)}   \\
\hline
\end{tabular} 
\end{table*}

As  listed  in Table~\ref{tab_downdrifting} and seen in Figure~\ref{fig:downdrifting}, all confidently and quantitatively derived frequency drifting rates have negative values, which indicates the downward frequency drifting of bursts. The drifting rate is plainly understandable in the frequency-time waterfall plot, though the exact values depend on the implemented DM value. As seen in Table~\ref{tab_downdrifting}, the drifting rates for very carefully selected single-component bursts are in the range between $-166$ and $-28$ MHz~ms$^{-1}$, and their average value is $-61\pm9$ MHz ms$^{-1}$. The drifting rates for multi-component bursts represent the average delay of several components in different parts of the frequency band, hence they have much lower values than those of single-component bursts.
The drifting rates for multi-component bursts are in the range from $-29$ to $-5$ MHz~ms$^{-1}$, with an average value of $-21\pm4$ MHz~ms$^{-1}$. This is only about one third of the drifting rate for single-component bursts. The bursts emerging in the higher half band tend to have a larger drifting rate than those detected only at the lower part of frequency band.

\subsection{Burst morphology classification}

With the implemented daily-average DM to all the bursts, one can systematically study the burst morphology. One immediate observation is that the bursts show a diverse morphology. The primary features of the bursts are the frequency drift and the limited emission band. The number of sub-bursts, i.e. the component of burst profiles, is another interesting feature. We therefore can classify the detected bursts from FRB 20201124A based on these key features. 

First of all, based on the frequency drifting, more than half of 624 bursts have shown a downward frequency drifting feature in the waterfall plots. Only a few cases show an upward drifting feature. The remaining bursts are either complex, show no drifting with the given DM, or have no evidence for drifting due to the limited emission bandwidth caught by FAST. Second, only a small fraction of bursts, which we categorize as ``wide-band'', are detected in the entire FAST observation band. Most of the bursts are detected only in a narrow part of the FAST L-band of 1.0-1.5 GHz, either in the high frequency part, the middle part, or the lower part. Therefore, the sub-classes can be distinguished according to the emission band where the bursts appear. Moreover, for the downward drifting bursts, the sub-classes can be further grouped according to the number of burst components. See Table~\ref{tab:ClassProp} for the number of bursts in each class or sub-class. Examples are shown in Figure~\ref{fig:class:D-drift} for the dynamic spectrum of frequency downward-drifting bursts with one (in the left column panels), two (in the middle column panels) and multiple components, with their emission seen in  wide-band,  high-frequency, middle-frequency band, or low-frequency part of the FAST band (from top to bottom). 

In the following, we briefly discuss each class and sub-class of bursts.

\subsubsection{One-component bursts with downward frequency drifting: D1-W, D1-H, D1-M and D1-L}

D1W: Only a few percent of bursts show wide-band emission, with an emission bandwidth larger than 500~MHz of the FAST observations. The integrated burst profile has a single component. If the waterfall plot is forced to be aligned to get a DM value, the emission at the lower part of the band would be distorted due to a slightly larger DM from the alignment (see the separated leading component of burst No.~35 in Figure~\ref{fig:difDM}). With the average DM of the day, we get 18 such one-component downward-drifting bursts, as shown in Figure~\ref{fig:appendix:D1W}. 

D1H: There are 16 one-component bursts in Figure~\ref{fig:appendix:D1H} which have emission only in the higher frequency part of the FAST band, showing the downward frequency drift pattern.

D1M: There are 31 one-component bursts in Figure~\ref{fig:appendix:D1M} which have emission only in the middle part of the FAST observation band.

D1L: There are 68 one-component bursts in Figure~\ref{fig:appendix:D1L} which have emission only in the lower part of the band, with the emission peak frequency near or below the FAST lower band limit. Some bursts have much extended emission towards the band lower limit, which also indicates the trend of frequency downward drifting, even though a quantitative analysis would be difficult. 

\subsubsection{Two-component bursts with downward  frequency  drifting: D2-W, D2-H, D2-M and D2-L}

The two-component bursts are practically the same as the one-component burst, except that two sub-pulses are observed within 50 ms. For a longer separation, they would be considered as independent bursts or a burst with precursor or postcursor. The burst profiles can be fitted with two Gaussian functions. There are 22, 14, 11 and 48 bursts in the sub-classes of D2-W, D2-H, D2-M and D2-L, respectively. See Figure~\ref{fig:appendix:D2W} to Figure~\ref{fig:appendix:D2L} for their plots. The longer tails of D2-L bursts cannot be well distinguished from the scattering tail of some bursts.  

\subsubsection{Multi-component bursts with downward frequency drifting: Dm-W, Dm-H, Dm-M and Dm-L}

Multi-component bursts have 3 or more  components in the burst profiles, and the downward drifting occurs between these components, as discussed above. These bursts can appear in the wide-band, or in the high part,  middle part, or the lower part of the FAST band, as shown in Figure~\ref{fig:appendix:DmW} to Figure~\ref{fig:appendix:DmL}. There are 14, 6, 8 and 7 bursts in each of these sub-classes, respectively.

\begin{figure}
    \flushleft
    \includegraphics[height=38mm]{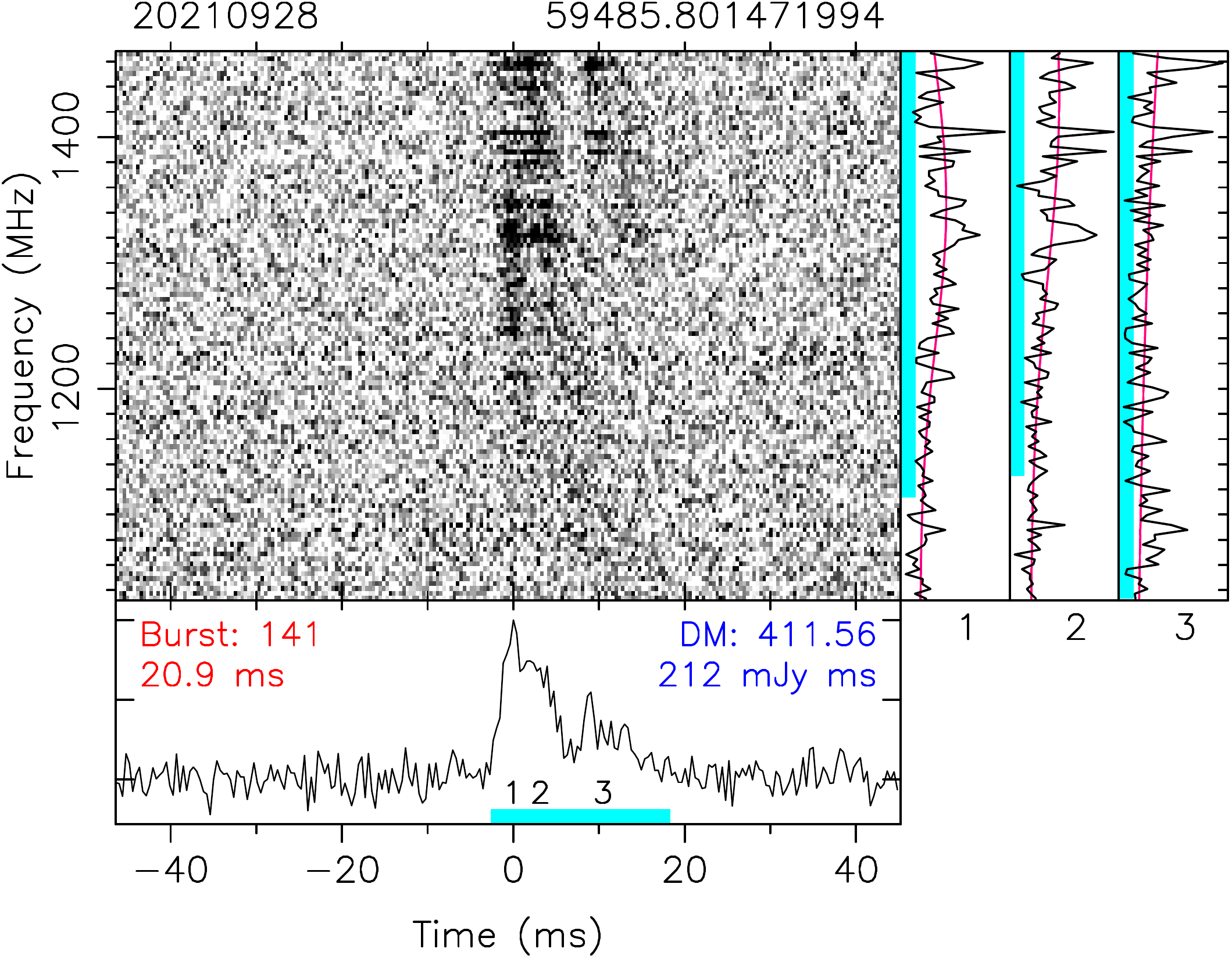} 
    \includegraphics[height=38mm]{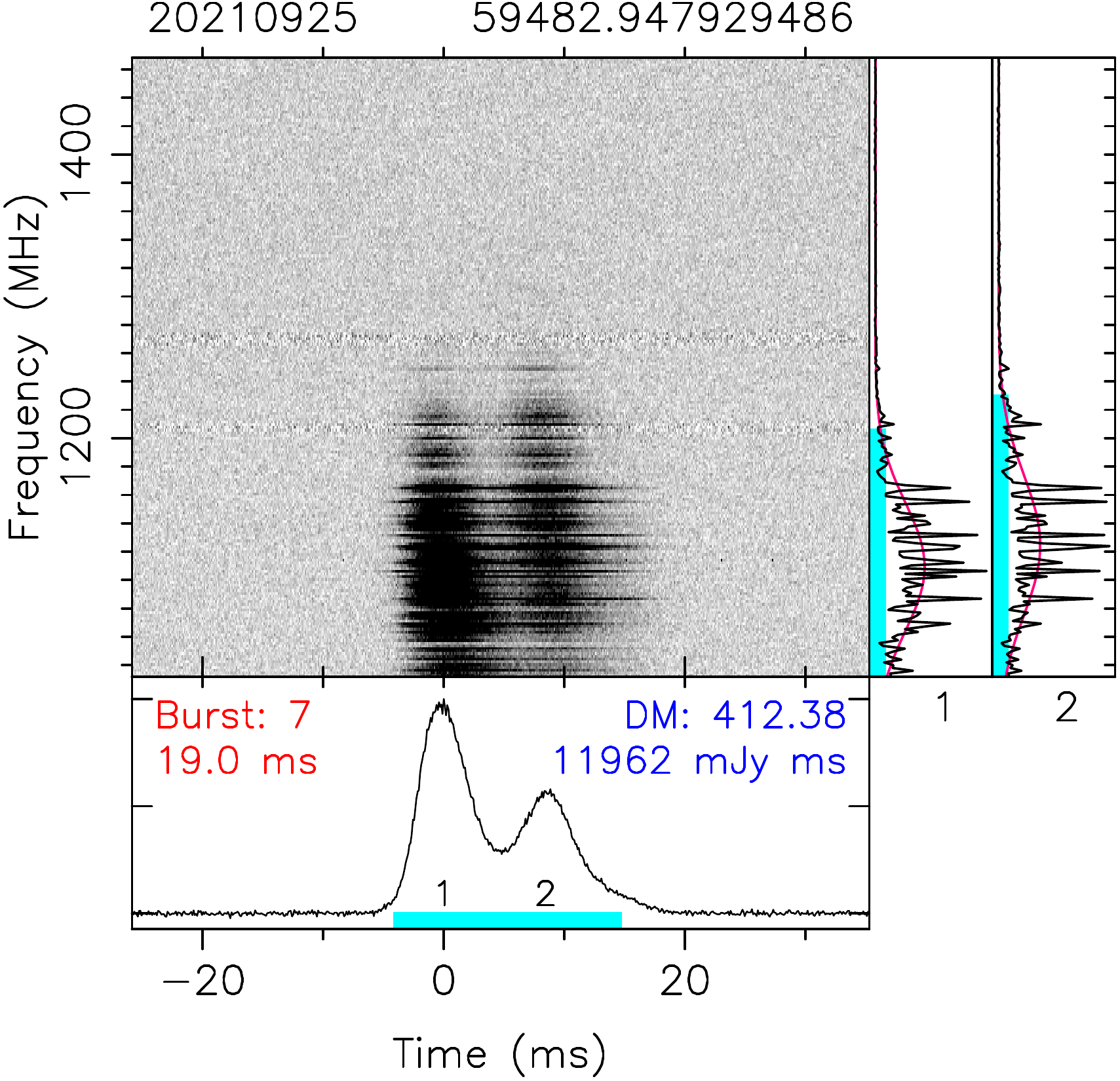}
    \includegraphics[height=38mm]{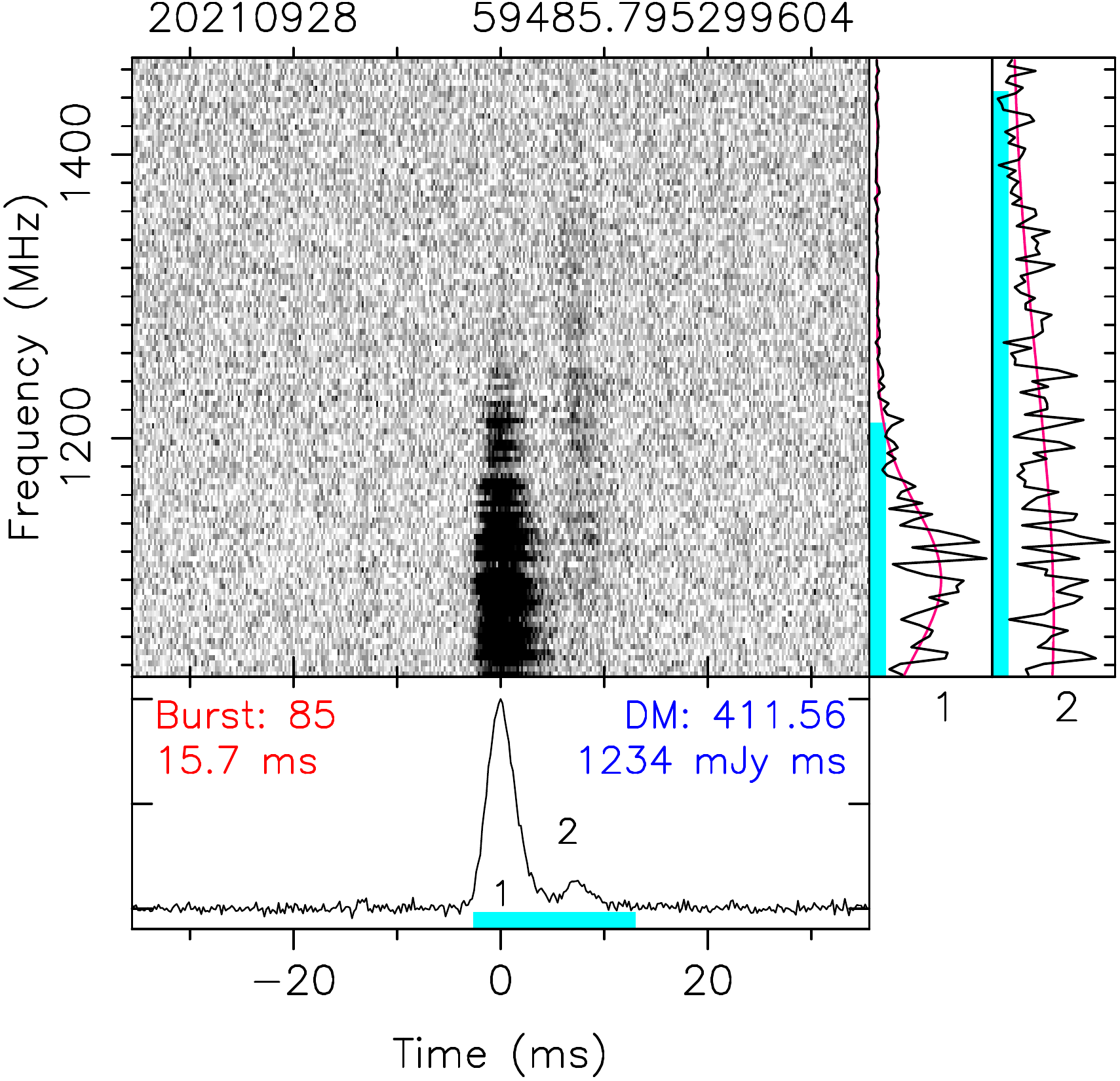}
\caption{Same as Figure~\ref{fig:class:D-drift} but for upward drifting of bursts with two or three components, i.e., the later component has a higher emission peak frequency than the first one. 
}
\label{fig:upward} 
\end{figure}

\subsubsection{Upward frequency drifting bursts}

Three bursts have the second or the third component appearing at a higher frequency band than the first one, so that they look like an upward frequency drifting as shown in Figure~\ref{fig:upward}. One of them have emission in the higher part of the band, denoted as subclass U-H, and two of them have emission in the lower part of the band, denoted as subclass U-L. 

\subsubsection{Bursts with no evidence for drifting}
Some bursts emerge near the edges of FAST observation band, either the higher or lower edge. The burst emission band is too narrow to determine whether there is a frequency drifting pattern. We classify these bursts as "No evidence" for drifting. They are further classified as NE-L (111 bursts) or NE-H (10 bursts) as their emission emerges in the lower or higher part of the FAST frequency band,  respectively. See Figure~\ref{fig:appendix:NEH} and Figure~\ref{fig:appendix:NEL} for the plots.

\subsubsection{No drifting bursts}
Dynamic spectra of 35 bursts (see Figure~\ref{fig:appendix:ND})  manifest a very vertically pattern, with their emission band wide enough to clearly judge that there is no frequency drifting (ND), regardless whether the emission appears at high, middle or low part of the frequency band.

\subsubsection{Complex}

There are 203 bursts, including 157 cluster-bursts (see Figure~\ref{fig:appendix:C}), which have many interesting burst components showing complex structures in the waterfall plots, either a mixture of upward and downward drifts or a mixture of individual downward drifts with precursors or postcursors in a short time or a cluster burst. It is really worth reading all plots in  Figure~\ref{fig:appendix:C}. The burst No.~113 on 20210927 in  Figure~\ref{fig:difDM} is an extreme case of complex bursts, which has the longest continuous emission duration (120~ms) with at least 10 burst components one after another. Some of them show upward drifting, but some others show downward  drifting patterns.

\section{Discussion and Conclusions}
\label{4Conclusions}

We report above the FAST detections of 624 bursts from FRB 20201124A in a star-forming galaxy at $z=0.0979$ during an extremely active episode in the end of September 2021. On September 28, 2021, the burst rate was 381.7 hr$^{-1}$, which is the highest among known FRB repeaters. The source was then suddenly quenched, with no bursts detected in the following three weeks. In our morphological study, we define a burst as an emission episode during which the adjacent emission peak  separation is shorter than 50~ms. The sub-bursts coming in such a duration are then counted as components of a burst. If all the components are counted as independent bursts, we would get 1461 bursts in total, which almost double the number of the bursts claimed in this paper.

The morphology of detected bursts of FRB 20201124A is diverse and intriguing. Most bursts are emitted in a relatively narrow frequency range inside the FAST observation band, and their energy distribution over frequency can be fitted with a Gaussian function. The typical emission bandwidth is $\rm BW_e = 277_{-84}^{+122}$ MHz, as described in Section~\ref{subsect:spectrum}. The emission peak frequency $\nu_{\rm 0}$ has a distribution in the FAST observation band, which can be fitted with two Gaussion functions, one with  1091.9$\pm$62.0~MHz and the other with 1327.9$\pm$113.0~MHz. Some bursts have emission detected only in the higher frequency part of the band, some in the middle, and some others in the lower frequency part of the observation band. A small fraction of bursts have wide band emission and are detected in the entire FAST band. The sub-burst widths ($W_{p}$) of the bursts have a wide distribution of 7.4$_{-3.0}^{+5.0}$~ms. Downward frequency drifting is observed from more than half of the detected bursts, including both one-component and multi-component bursts. Based on the burst features in the dynamic spectra of the frequency-time waterfall plots, the bursts of FRB 20201124A are classified into 18 groups. The complex features of some bursts are caused by the intrinsic emission properties of the FRB, rather than by its environment. This is because the variation of morphology occurs among sub-bursts of one single burst and the time is too short to introduce a large DM variation due to varying free electron column density along the line of sight.

In the following, we compare the frequency drifting properties of FRB 20201124A with other FRBs in detail, and then discuss the scintillation-induced emission intensity fluctuations. 
 
\subsection{Frequency drifting and radiation mechanisms}

\begin{figure}
  \centering
  \includegraphics[width=0.99\columnwidth]{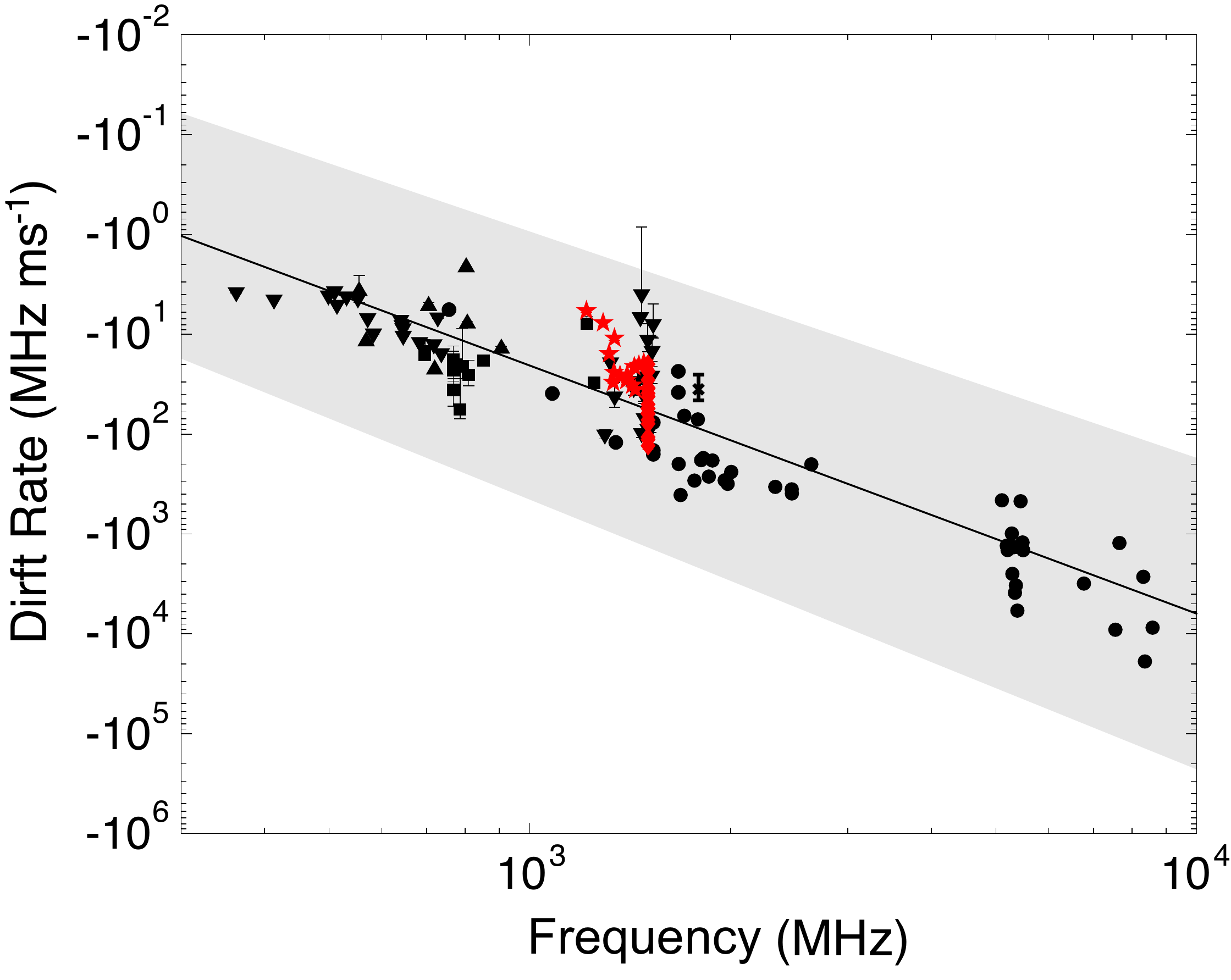}
  \caption{
Drifting rate vs frequency for FRB 20201124A (red) as compared with other bursts. Both drift rates and frequencies are transferred to the rest frame of the host galaxy. Different symbols stand for: FRB 121102 (circles), FRB 180814.J0422+73 (up triangles), FRB 180916.J0158+65 (down triangles), FRB 190711 (crosses), CHIME bursts (squares) and FRB 20201124A (diamonds for~\citet{Hilmarsson21} and pentagrams for this work).
The black solid lines are the best fitting line for the power law $\dot{\nu} = -10^{-6.12}\nu^{2.48}$.
The grey zone is the 1$\sigma$ region of the best fitting.
}
\label{fig:frbsRd}
\end{figure}

Frequency drifting structure have been observed in many FRB bursts, e.g. FRB 121102, FRB 180814.J0422+73, FRB 180916.J0158+65 and FRB 190711 \citep{Gajjar18,Michilli18, Chime19a,Chime19b, Hessels19,Josephy19,Caleb2020MNRAS,Day20,Fonseca20, Pastor-Marazuela20, Chamma21, Chawla21,Hilmarsson21,Platts21, Pastor2021Natur}. The drifting rate varies in a wide range for each FRB, and it may be related to other parameters. For example, \citet{Chamma21} found that the drifting rate is inversely correlated with the widths of the sub-bursts. Our results of sub-burst width of 7.4 ms and drifting rate of multi-component burst $R_d=-21$ MHz~ms$^{-1}$ of FRB 20201124A  are slightly higher but consistent with the result of \citet{Hilmarsson21}. \citet{WWY2021arXiv211111841W} investigated the relationship between the drifting rate and the burst emission frequency and found an anti-correlation. Our mean drifting rate $R_d=-21$ MHz~ms$^{-1}$ at 1250 MHz, as well as the drifting rate of bursts with multi-component, which is shown in Figure~\ref{fig:frbsRd} with parameters that have been converted to the
rest frame of the host galaxy, are consistent with the predication given by \cite{WWY2021arXiv211111841W}.

Frequency drifting and narrow-band emission features as shown from FRB 20201124A have been also observed from other radio sources. For example, PSR J0953+0755 has emission in a narrow band at a low frequency band of 18 - 30~MHz and shows a sub-pulse frequency drifting structure  \citep{Ulyanov2016MNRAS}. \citet{Bilous2021arXiv210908500B} observed the sub-pulse drifting structure of PSR J0953+0755 at several tens of MHz. The giant pulses of Crab \citep{Thulasiram2021MNRAS} often show narrow-band emission at various center frequencies with much pulse broadening at relatively lower frequencies which are not caused by scattering. These features are somewhat similar to the drifting behavior of the FRB 20201124A bursts.

\subsection{Scintillation}

\begin{figure}
  \centering
  \includegraphics[width=0.99\columnwidth]{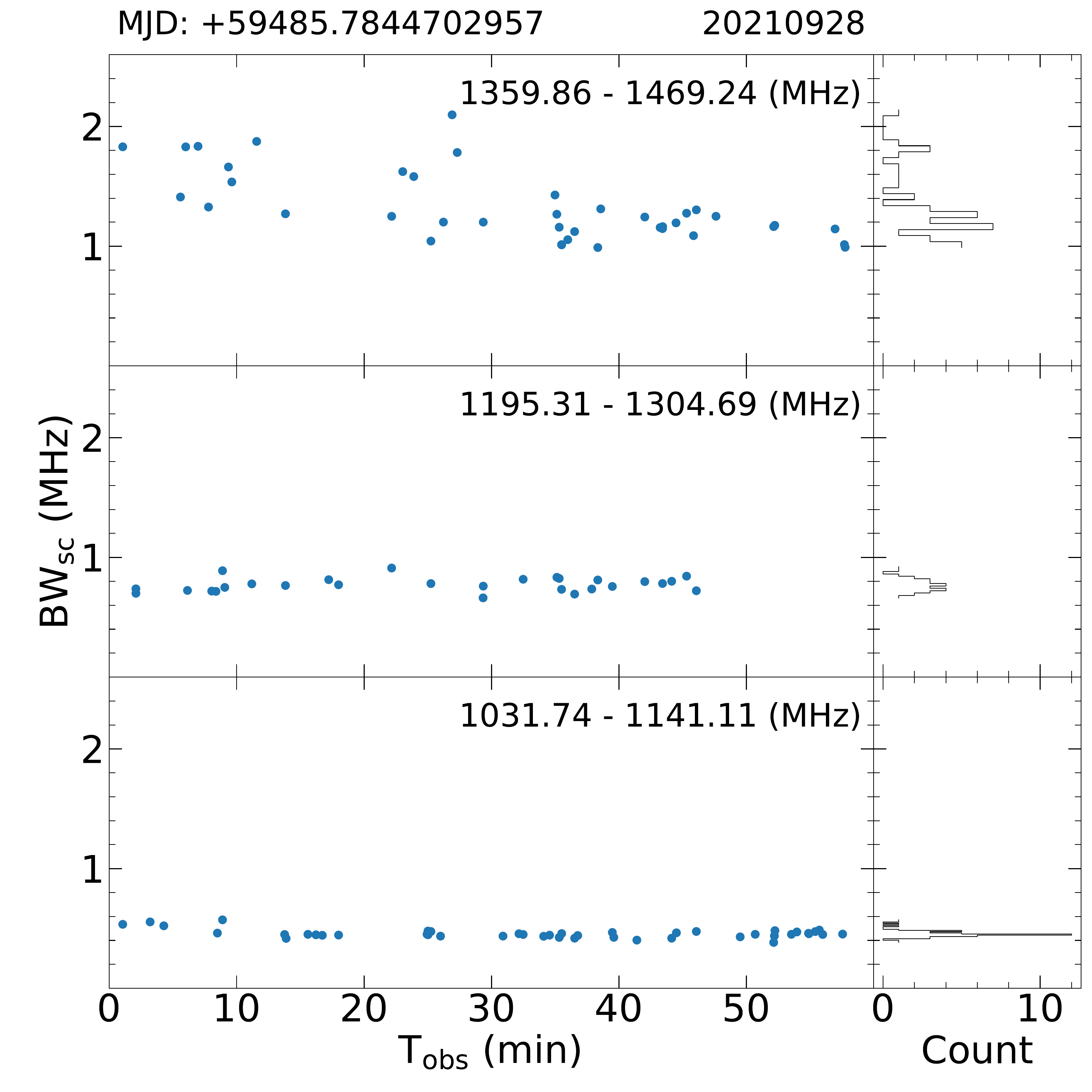}
  \caption{
Scintillation bandwidth BW$_{\rm sc}$ derived from some strong single bursts of FRB 20201124A  on 2021-09-28 in three sub-bands plotted against the observation time T$_{\rm obs}$. Their distributions are shown in the right sub-panels. 
}
\label{fig:timeScintBW}
\end{figure}

Scintillation is a distinct feature of FRB 20201124A. In order to test if the scintillation bandwidth changes within one hour observation, we calculate the scintillation bandwidths of strong bursts on 2021-09-28  using the autocorrelation function (ACF) method \citep{Cordes1986ApJ} independently in three sub-bands: the lower band of 1031.74 -- 1141.11 MHz, the medium band of 1195.31 -- 1304.69 MHz, and the higher band of 1359.86 -- 1469.24 MHz. The results are shown in Figure~\ref{fig:timeScintBW}. The mean values of scintillation bandwidths are  0.456$\pm$0.035 MHz, 0.772$\pm$0.057 MHz, and 1.325$\pm$0.285~MHz for the three sub-bands, respectively. Obviously, the scintillation bandwidths increase with the observation frequency. The variations of the calculated scintillation bandwidths are smaller than the width of one frequency channel (0.122070 MHz) in the lower and middle sub-bands. There is no systematic variation of scintillation bandwidth with time in one-hour sessions of FAST observations.

\begin{figure}
  \centering
  \includegraphics[width=0.99\columnwidth]{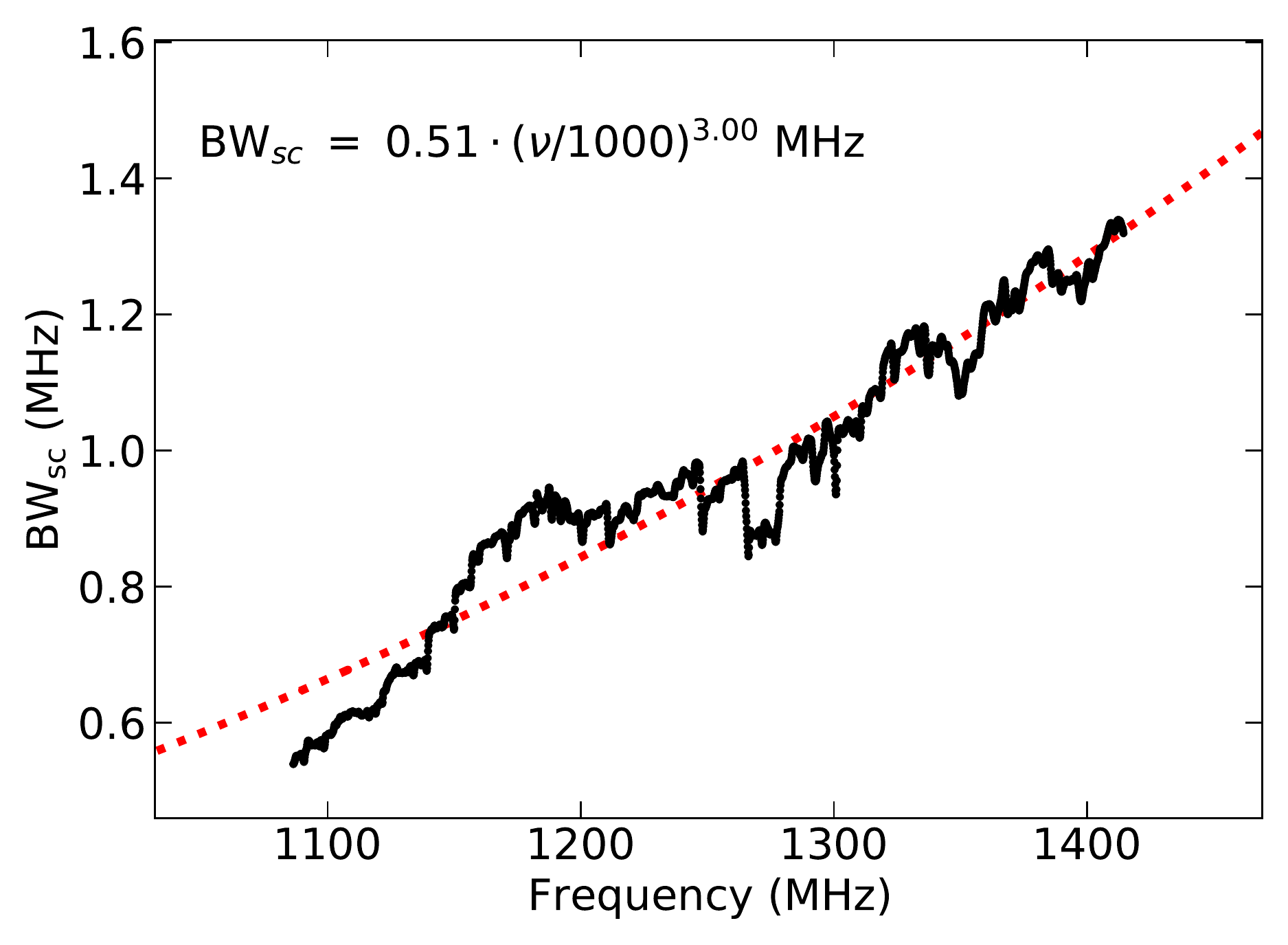}
  \caption{
Scintillation bandwidth of the integrated burst-energy spectrum of  FRB 20201124A  plotted against the central frequency of a car-box sub-band for scintillation calculation. The red dotted line denotes the power-law fitting.
}
\label{fig:ScintBW}
\end{figure}

To get a clear relation between scintillation bandwidths and the observing frequencies, we integrated the energy of all bursts for each frequency channels, and then analyzed scintillation in a car-box sub-band of 1/4 band. The results are shown in Figure~\ref{fig:ScintBW} for the center frequency ${\nu}$ and scintillation bandwidth $\rm{BW}_{\rm sc}$, which can be fitted by a power-law function $\rm {\nu}_{s} = {a\nu}^{\gamma}$, where $\gamma=3.0$ is the power-law index. 

The scintillation bandwidth increases from about 0.5~MHz at 1.05~GHz to 1.4~MHz at 1.45~GHz in our observation frequency, with a mean scintillation bandwidth of about 0.94 MHz. The power-law index $\gamma=3.0\pm0.2$ is smaller than the value of 3.5$\pm$0.01 obtained by~\citet{Main2021MNRAS}, or 4.9 reported by~\citet{XuHeng2022Natur}, or the theoretical value of 4.0 or 4.4 of Kolmogorov spectrum. The corresponding scattering timescale of about 0.31 $\rm \mu{s}$ is consistent with the results of \citet{Main2021MNRAS}, which is much smaller than the sampling time of 49.152 $\rm \mu{s}$ of our observation. Therefore  scattering has an almost negligible effect on the morphological study for the bursts of FRB 20201124A. For the large number of bursts we observed, the properties shown in this paper should be intrinsic to the FRB 20201124A source.

We emphasize at the end of this paper that the sensitive observations by a large radio telescope such as the FAST is very important to detect rich features of bursts and reveal their properties. The classification scheme and other results presented in this paper are hard to be achieved without such sensitive observations. We also realize that multi-epoch wide-band observations are fundamental to understand the environment and physical mechanisms of FRB emission.

\normalem
\begin{acknowledgements}
We thank the referee for helpful comments.
This work makes use of the data from the FRB key science project, as one of five key projects of FAST, a Chinese national mega-science facility, operated by National Astronomical Observatories, Chinese Academy of Sciences.
J.~L. Han is supported by the National Natural Science Foundation of
China (NSFC, Nos. 11988101 and 11833009) and the Key Research
Program of the Chinese Academy of Sciences (Grant No. QYZDJ-SSW-SLH021);
D.~J. Zhou is supported by the Cultivation Project for the FAST scientific Payoff and Research Achievement of CAMS-CAS.
Y. Feng is supported by the Key Research
Project of Zhejiang Lab no. 2021PE0AC0.
J.~C. Jiang, K.~J. Lee, H. Xu, C.~F. Zhang, B.~J. Wang, J.~W. Xu are supported by the National SKA Program of
China (2020SKA0120100), the National Key R\&D Program of China (2017YFA0402602), the 
National Nature Science Foundation grant No. 12041303, the
CAS-MPG LEGACY project, and funding from the Max-Planck Partner Group.
%

%
\end{acknowledgements}

\section*{Authors contributions}  

D.~J. Zhou developed the single pulse module and the related software for this work, and processed almost all data presented in this paper. 
J.~L. Han realized the diversity on radio emission morphology on this special FRB source observed by the FAST, and led the writing of this paper. 
B. Zhang and W.~W. Zhu proposed and chaired the FAST FRB key science project. 
B. Zhang, W.~W. Zhu, J.~L. Han, K.~J. Lee and D. Li coordinated the teamwork, the observational campaign, co-supervised data analyses and interpretations.
W.~C. Jing analyzed the burst and sub-bursts parameter distributions.
W. -Y. Wang performed the sub-pulse frequency analysis.
Y.~K. Zhang performed the analysis of energy distribution and the results are presented in the Paper II of this series.
J.~C, Jiang performed the analysis of the polarization properties and the results are presented in the Paper III.
J.~R. Niu performed the periodicity search and the results are presented in the Paper IV.
R. Luo improved the energy distribution analysis and make many suggestions to improve the paper.
H. Xu, C.~F. Zhang, B.~J. Zhang, J.~W. Xu, P. Wang, Z.~L. Yang and Y. Feng also performed the single-pulse search, DM search, and the analysis of the burst energy, polarization, and periodicity search. 
All authors contributed to discussions and the paper writing. 

\section*{Data availability}
All data for the plots in this paper, including these in appendix, can be obtained from the authors with a kind request.

\bibliographystyle{raa}
\bibliography{bibfile}{}

\appendix
\section{A complete list of detected bursts of FRB20201124A in 202109}

In the FAST monitor sessions of 20210925-20210928, we detected 30, 62, 208, and 447 bursts in these 4 days, respectively. In total, there are 747 bursts. For each burst and the associated sub-bursts, the TOA expressed in MJD, the frequency of emission peak ($\nu_{\rm 0}$, in MHz), the detected emit low and high frequency ($\nu_{\rm low}$ and $\nu_{\rm high}$ respectively, in MHz), the bandwidth ($BW_{\rm e}$, in MHz) of observed emission, the sub-burst width (W$_{\rm sb}$, in ms), the detection signal-to-noise ratio (SNR), and the fluence ($F_{\nu}$, in mJy~ms) together with the burst morphology classification, are all listed the Table~\ref{tab:appendix}.  

\begin{figure*}
    \flushleft
    \includegraphics[height=37mm]{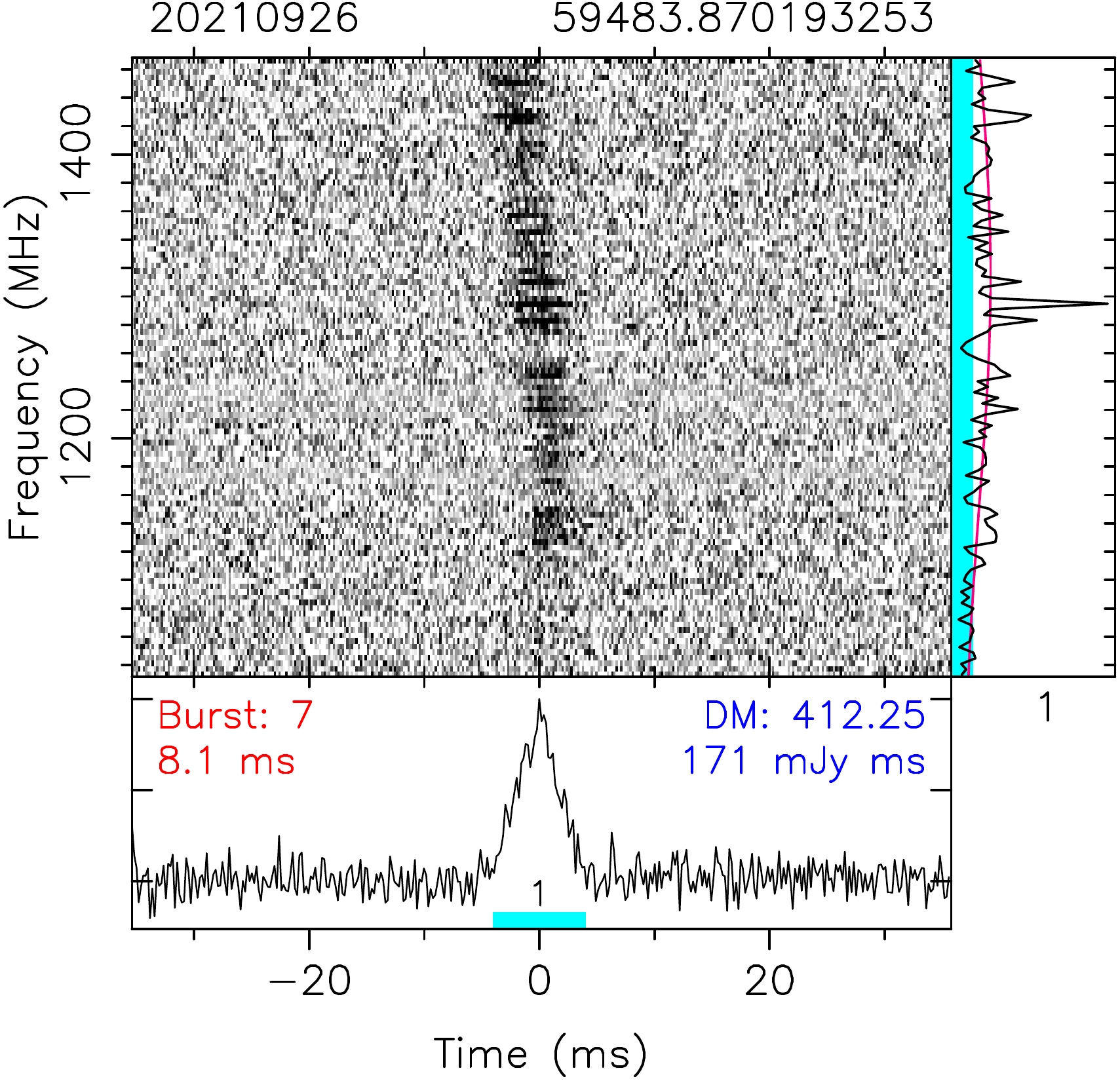}
    \includegraphics[height=37mm]{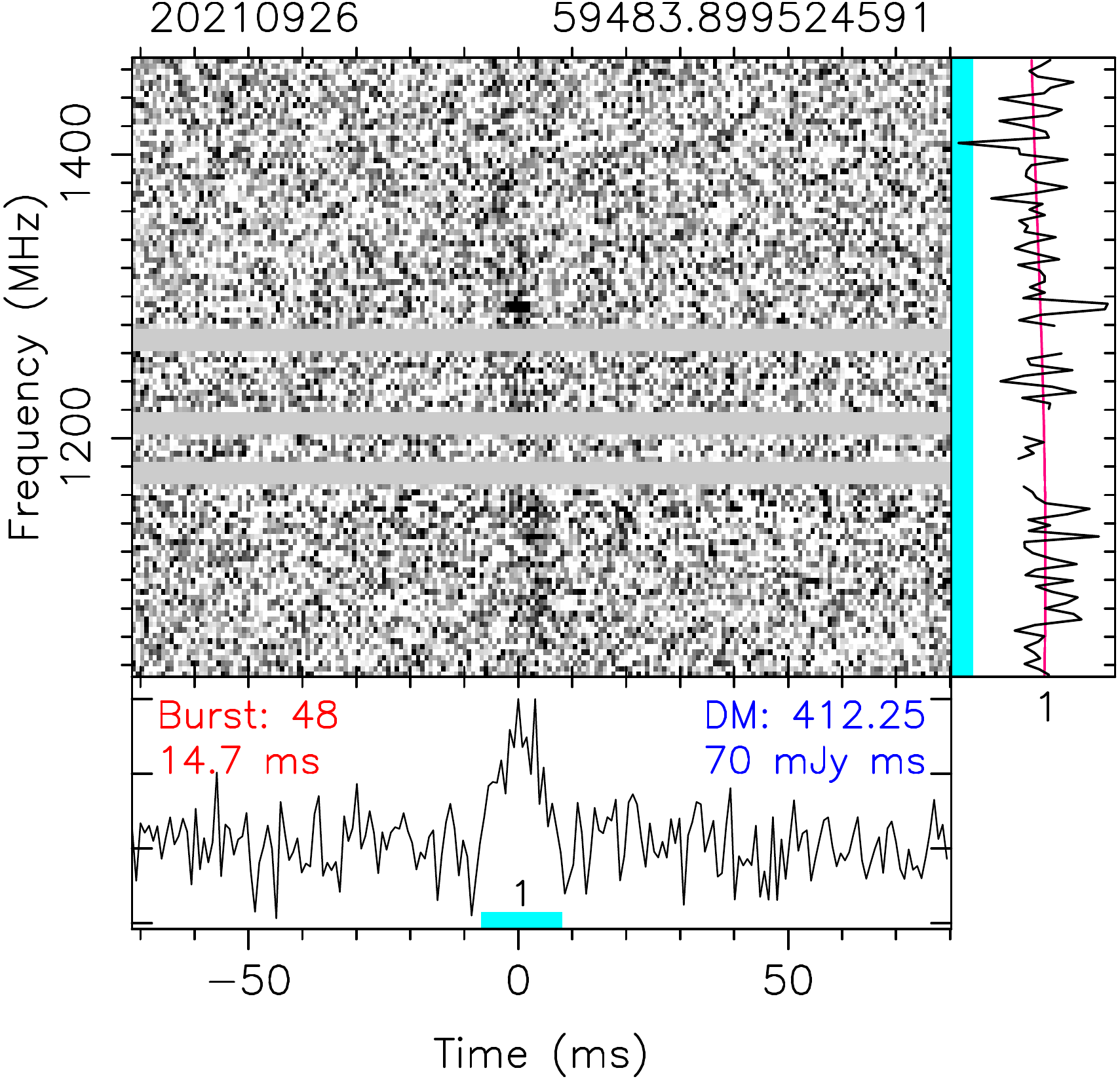}
    \includegraphics[height=37mm]{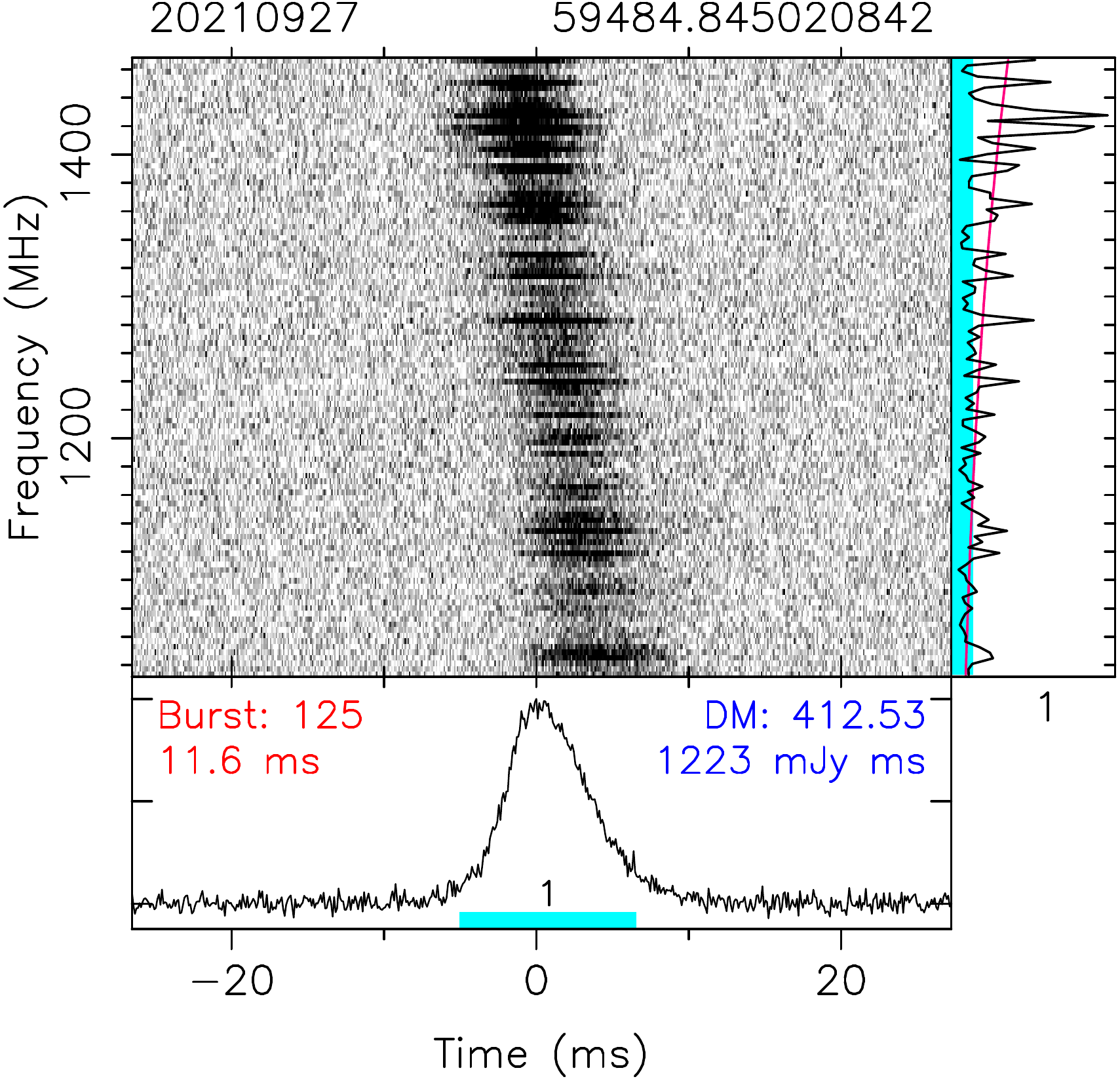}
    \includegraphics[height=37mm]{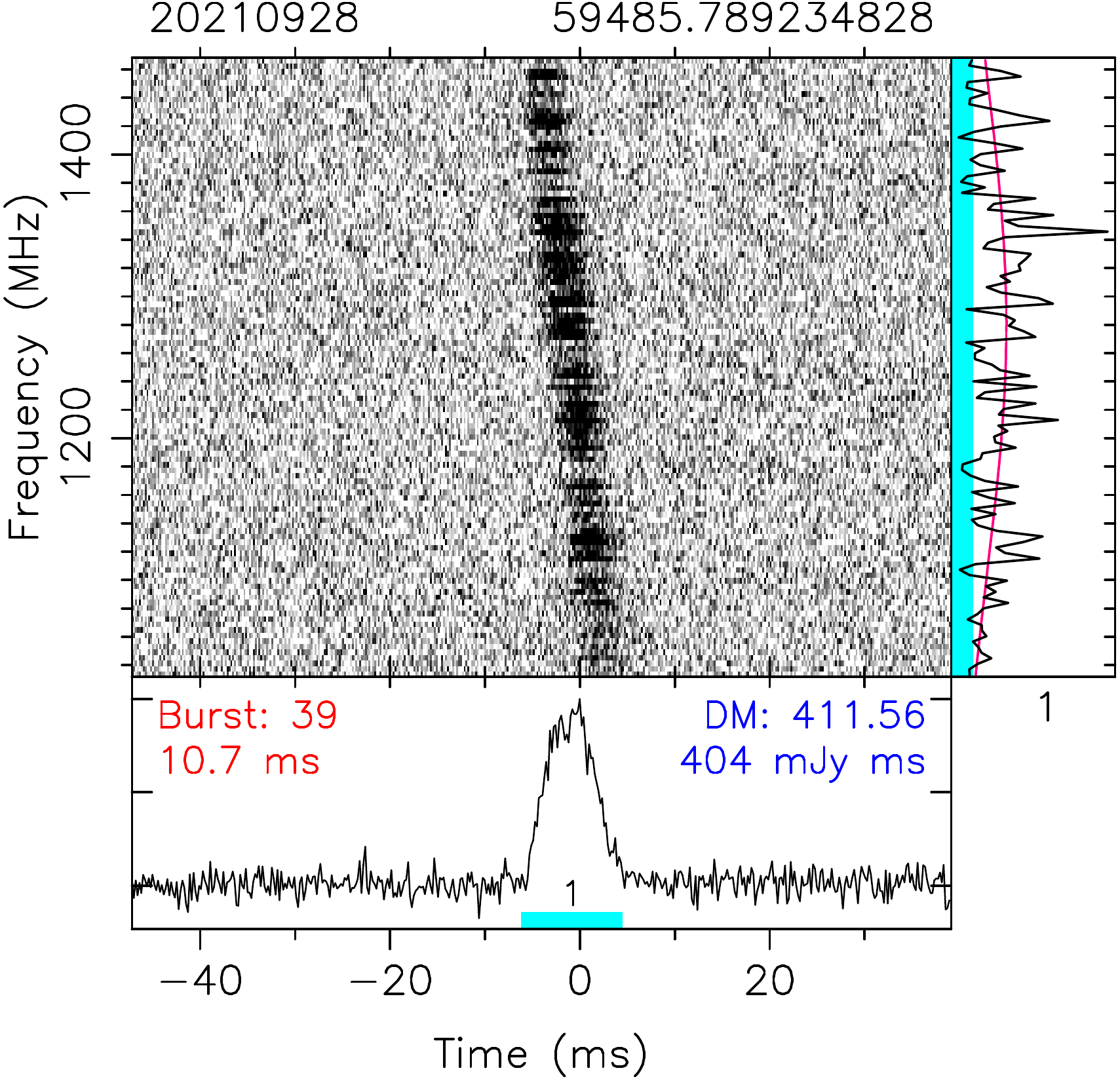}
    \includegraphics[height=37mm]{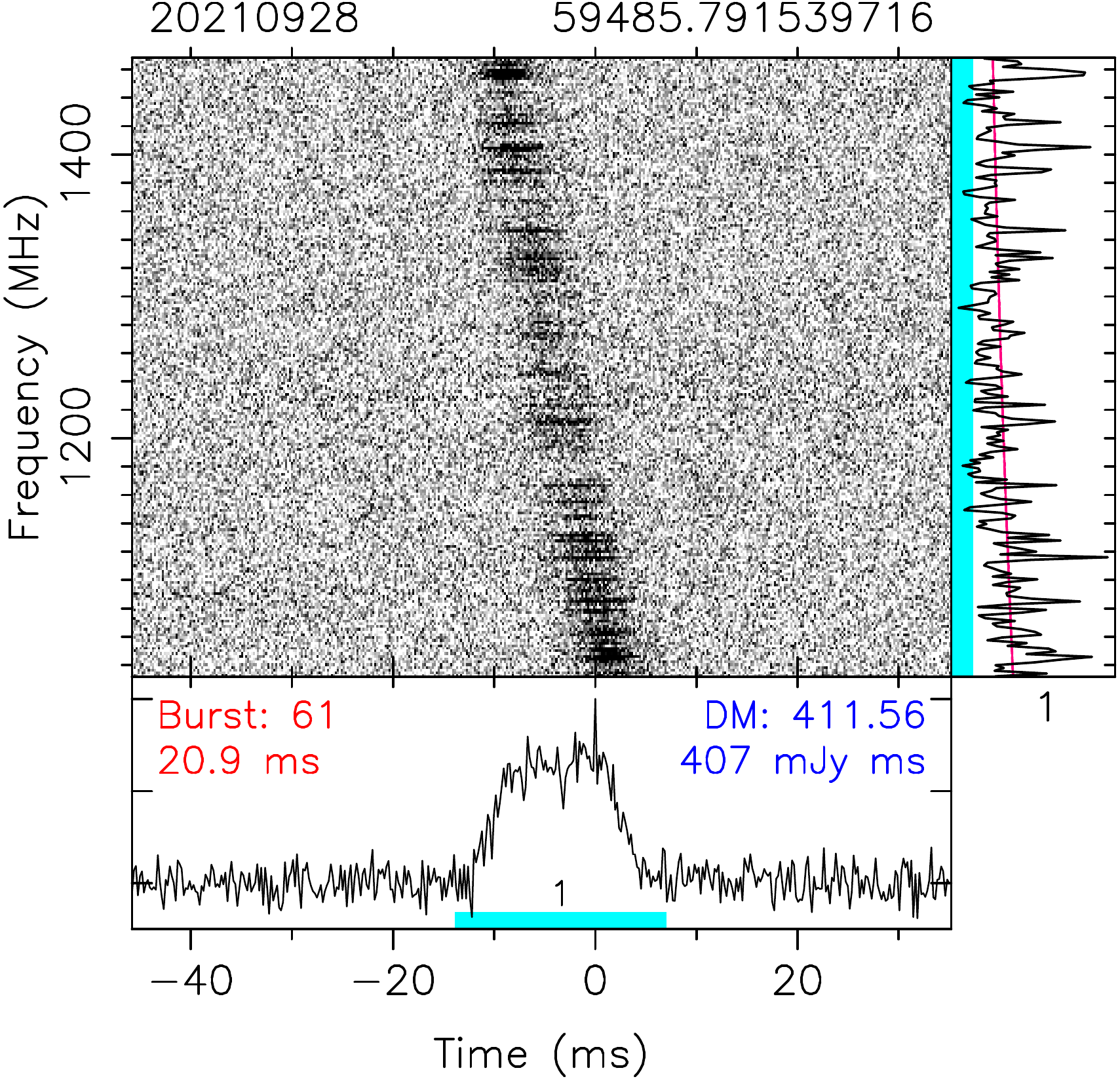}
    \includegraphics[height=37mm]{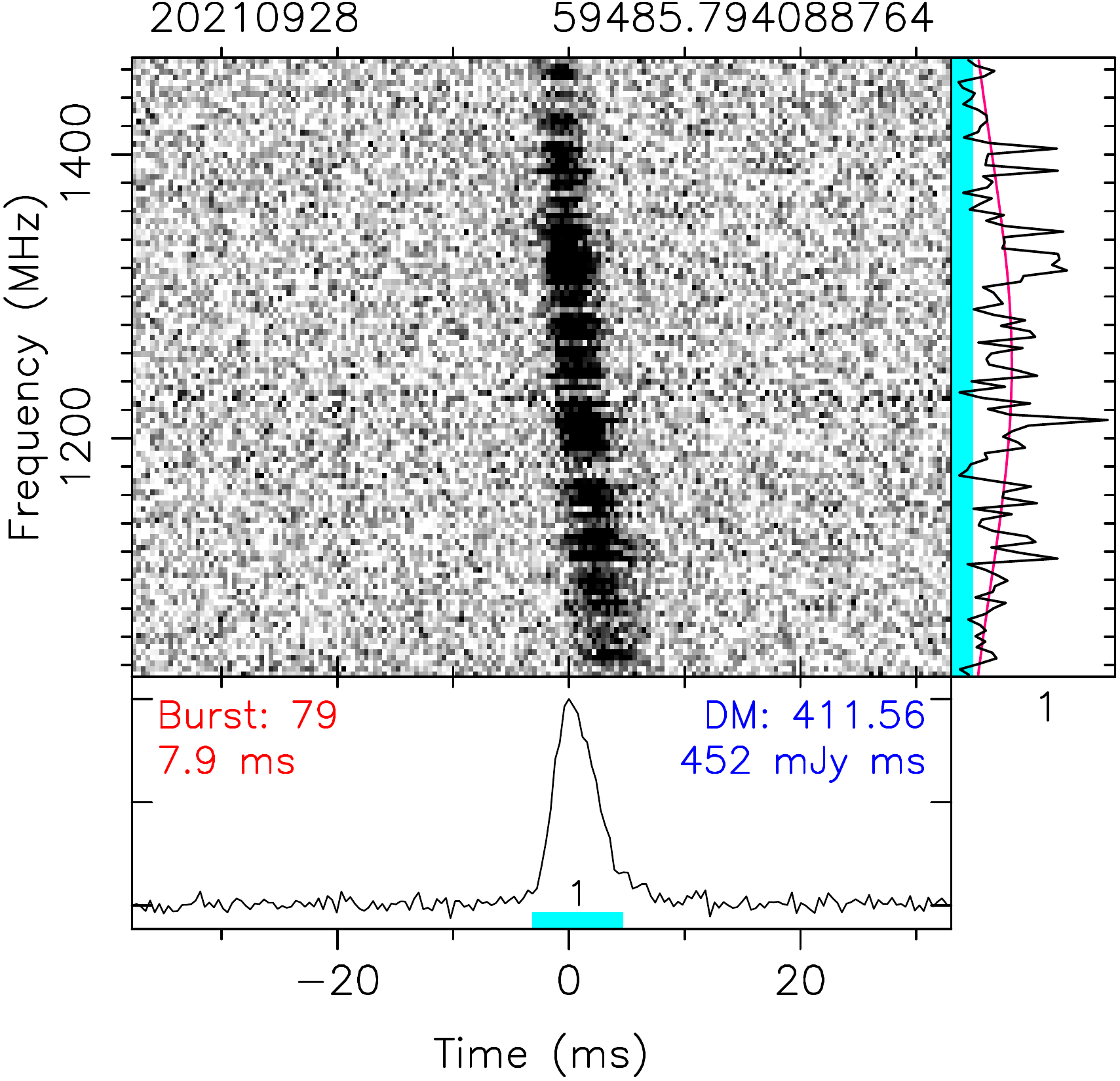}
    \includegraphics[height=37mm]{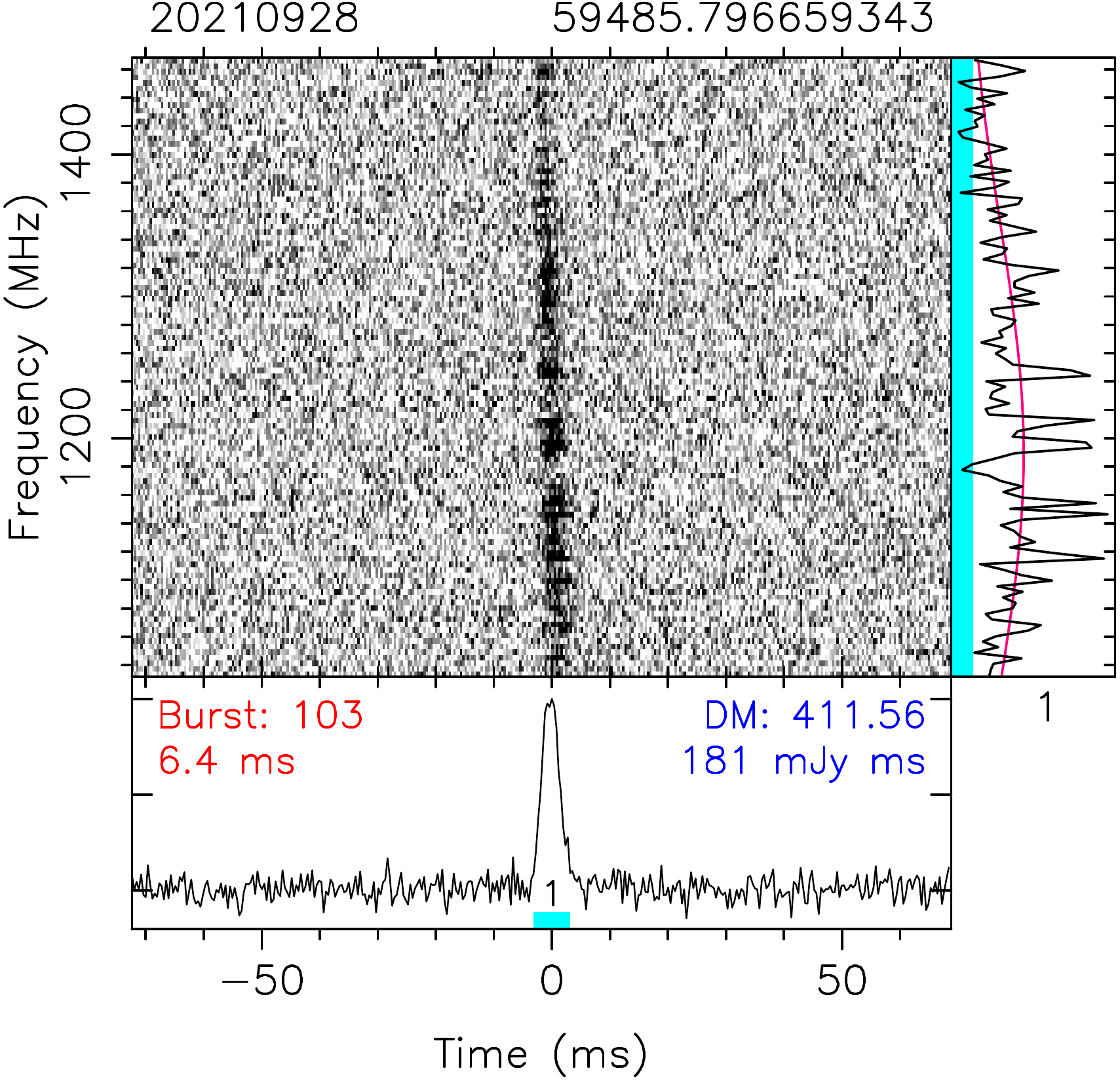}
    \includegraphics[height=37mm]{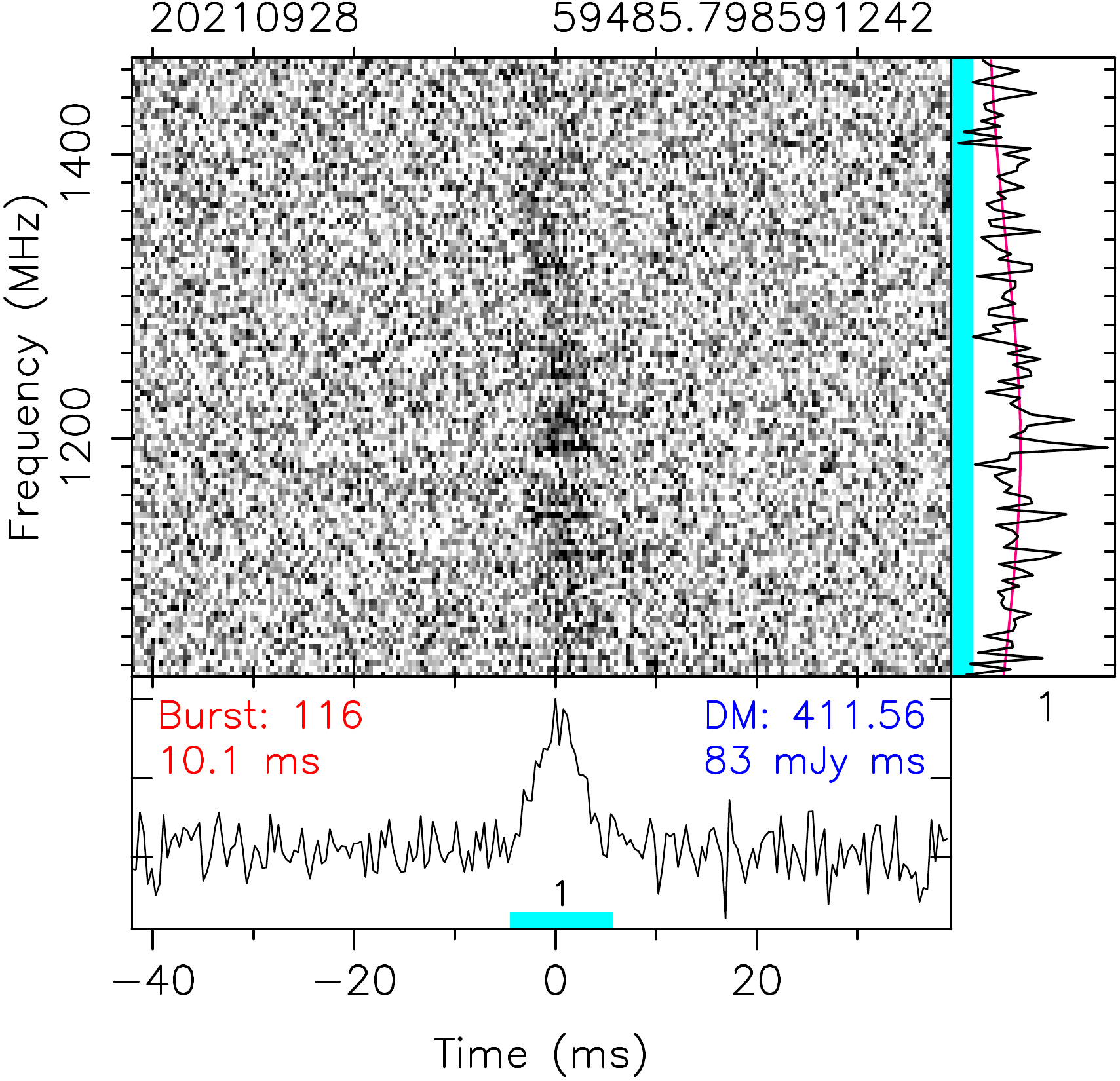}
    \includegraphics[height=37mm]{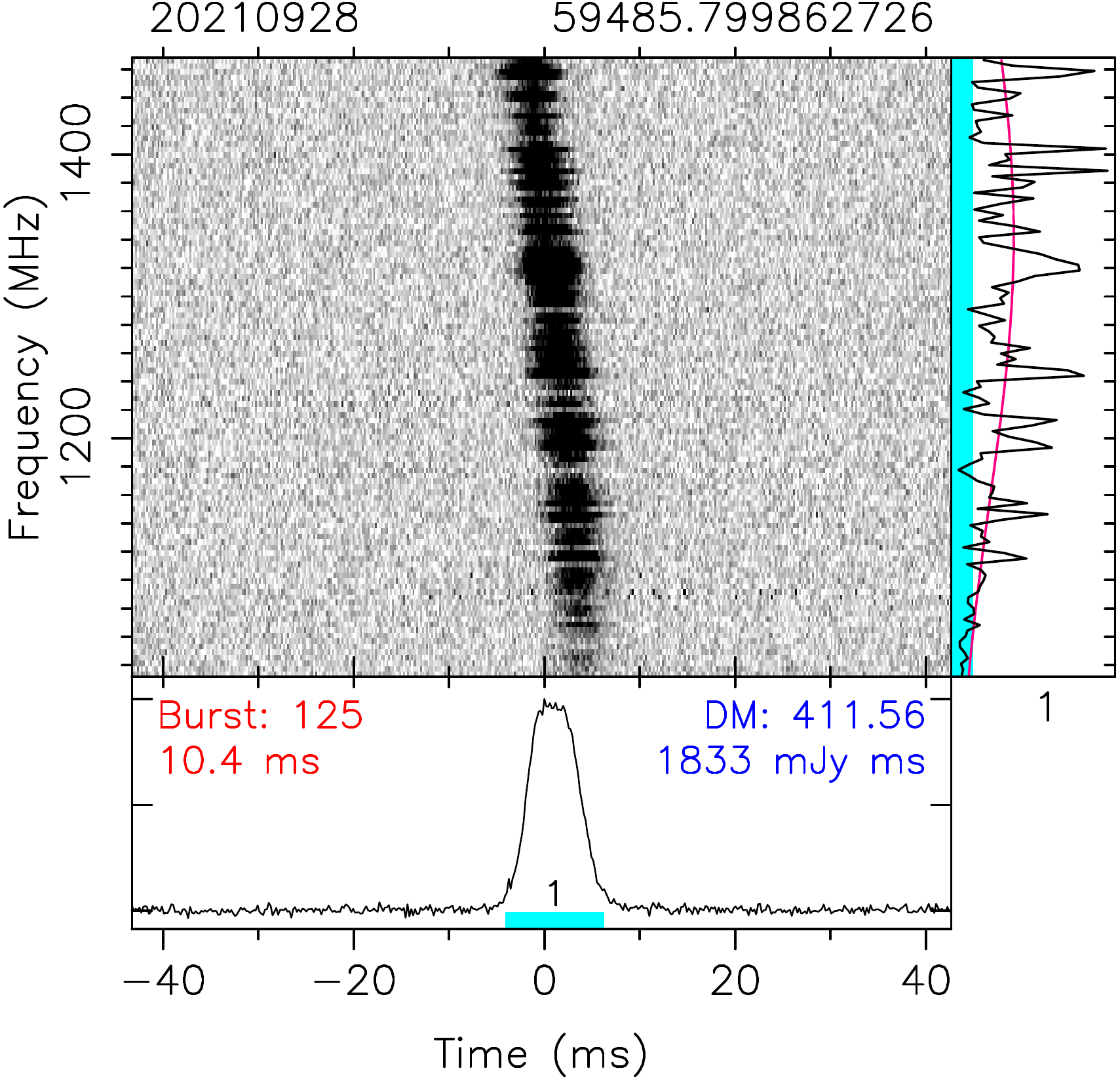}
    \includegraphics[height=37mm]{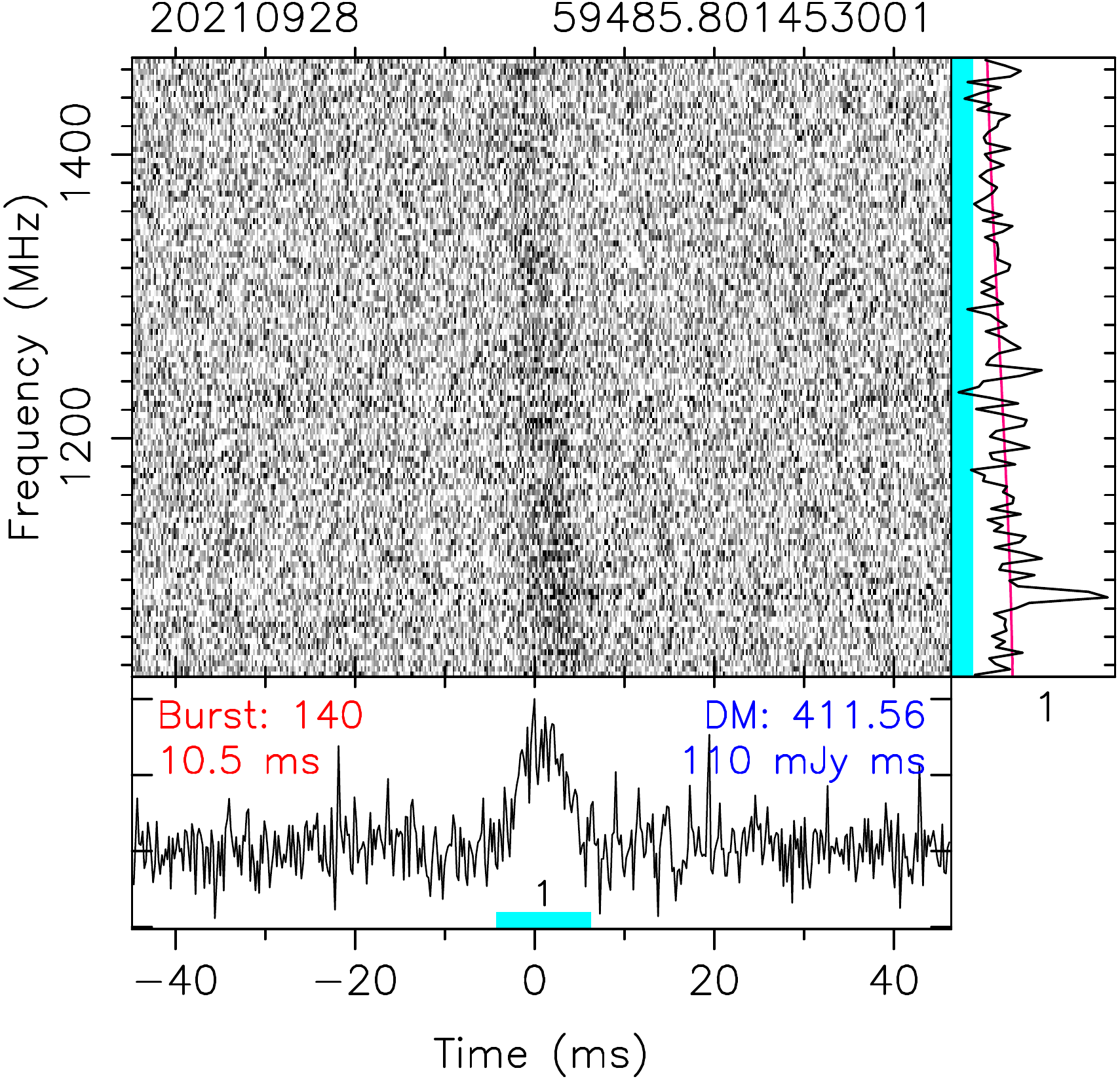}
    \includegraphics[height=37mm]{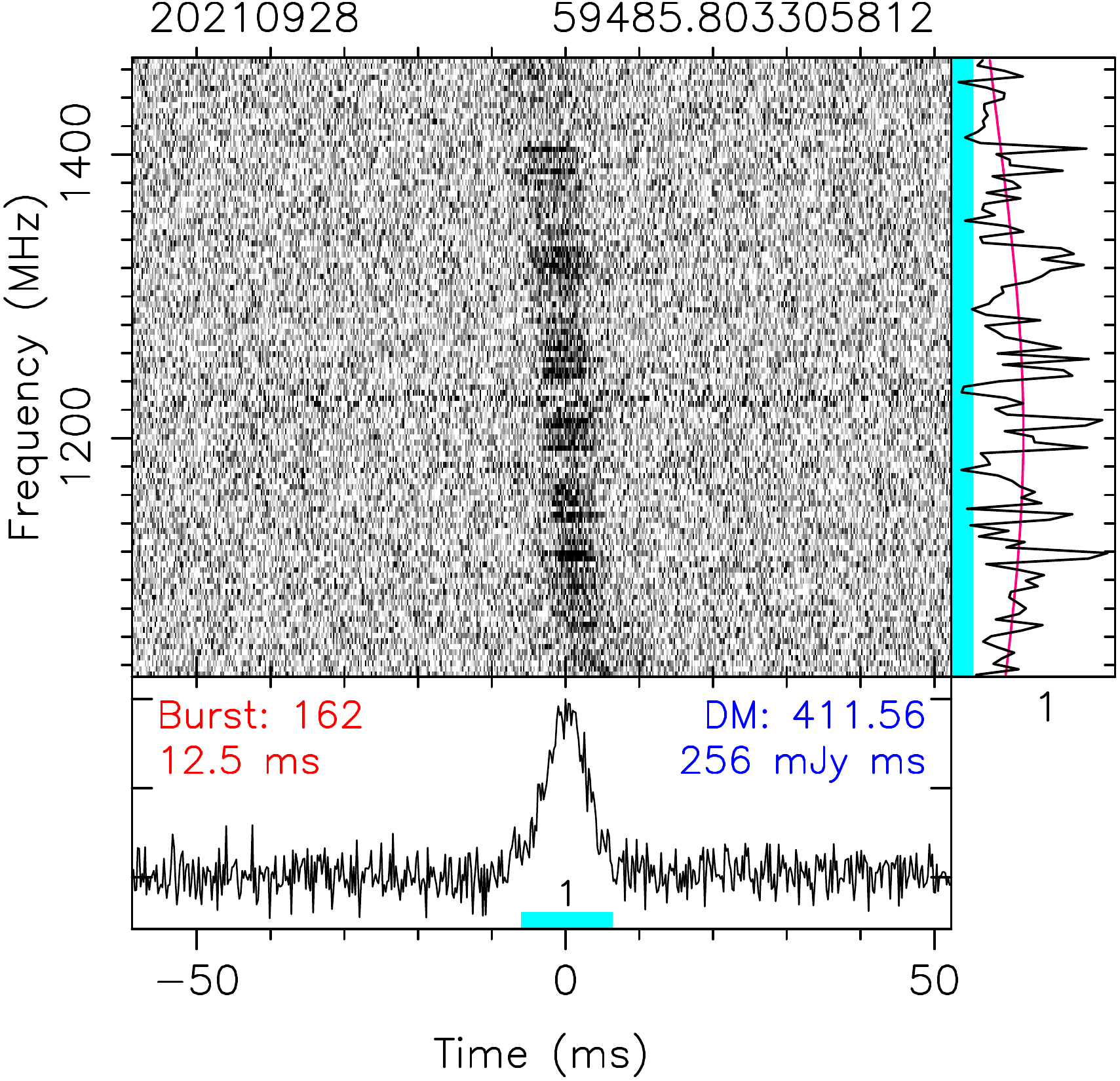}
    \includegraphics[height=37mm]{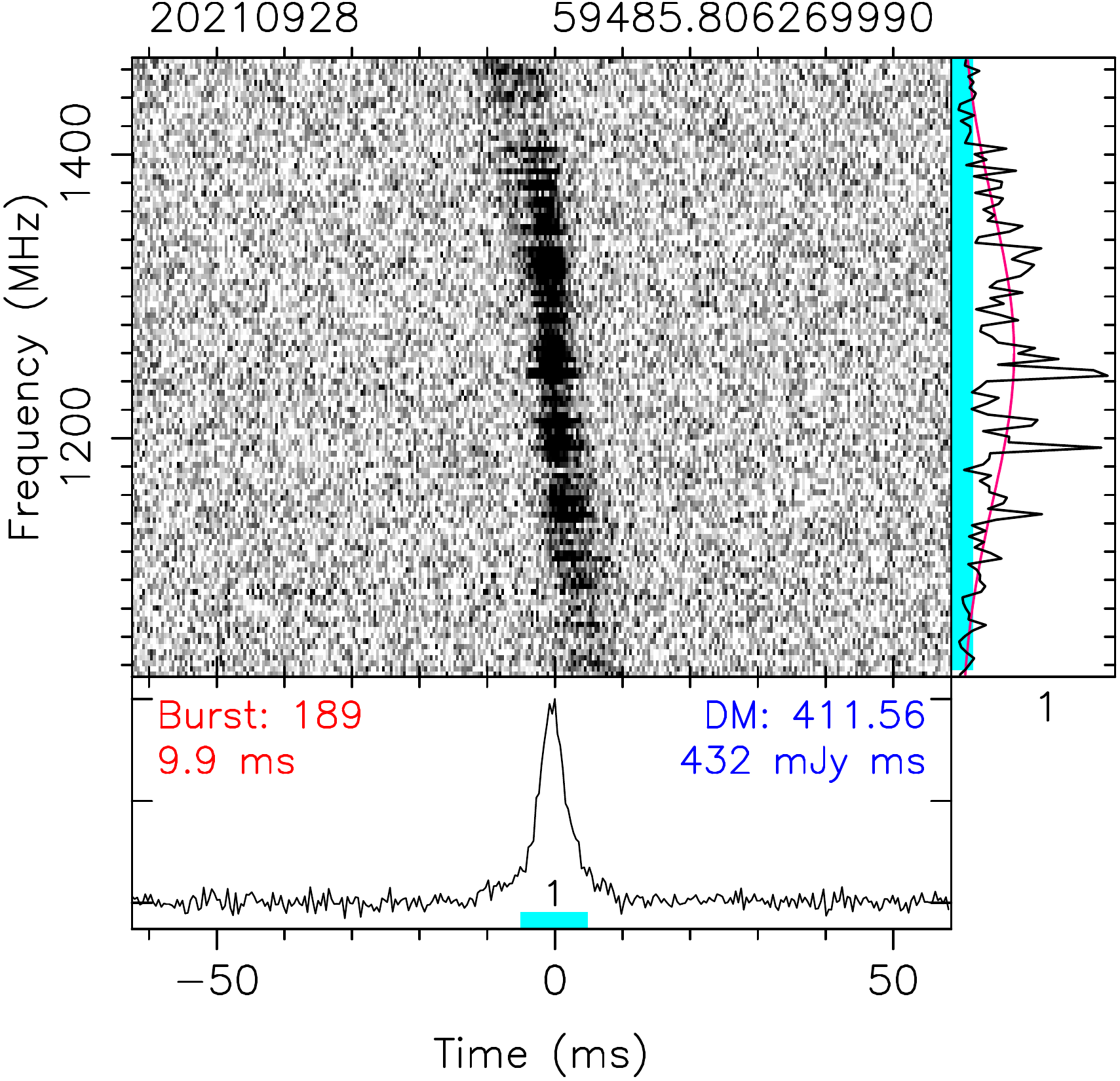}
    \includegraphics[height=37mm]{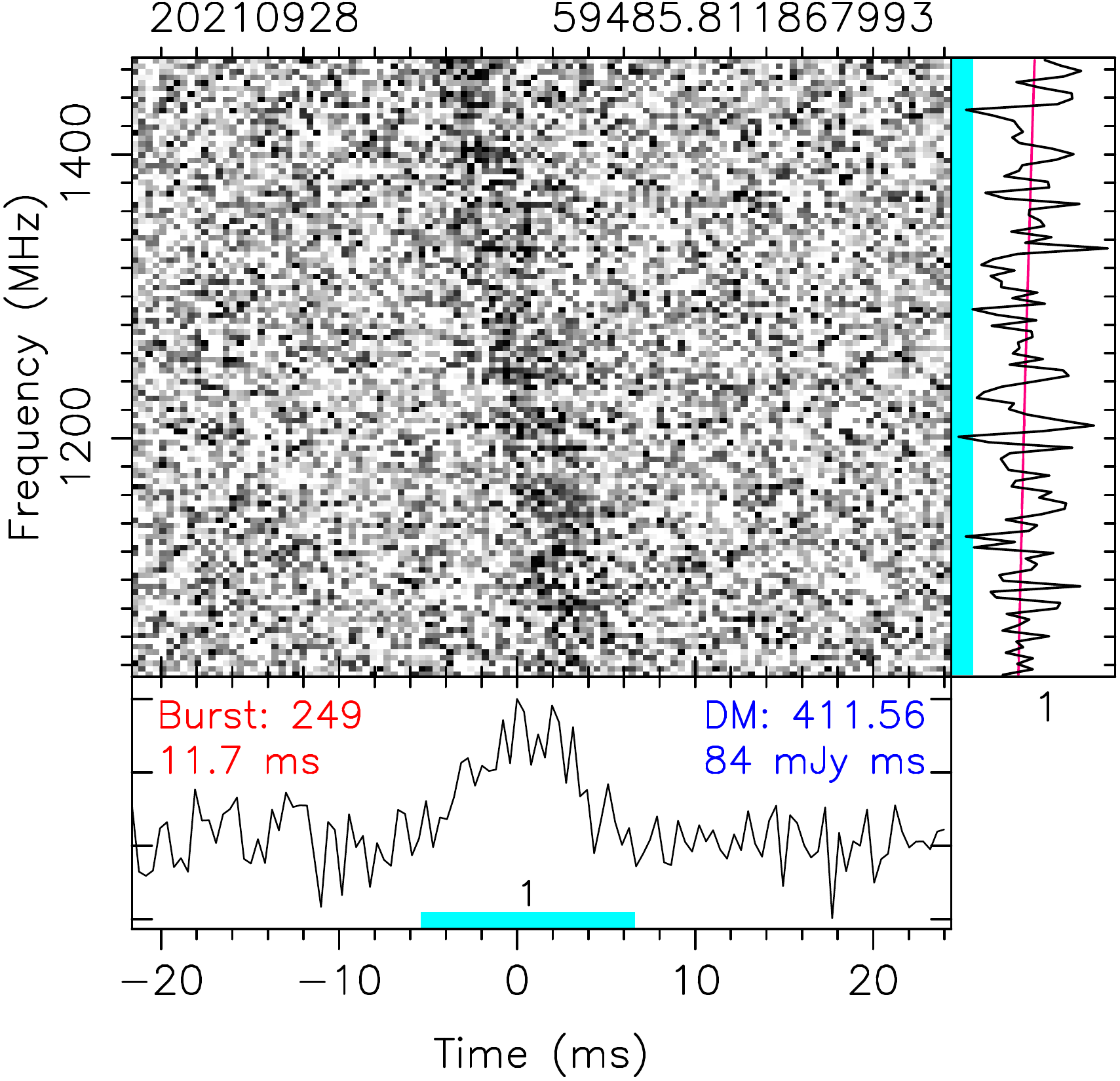}
    \includegraphics[height=37mm]{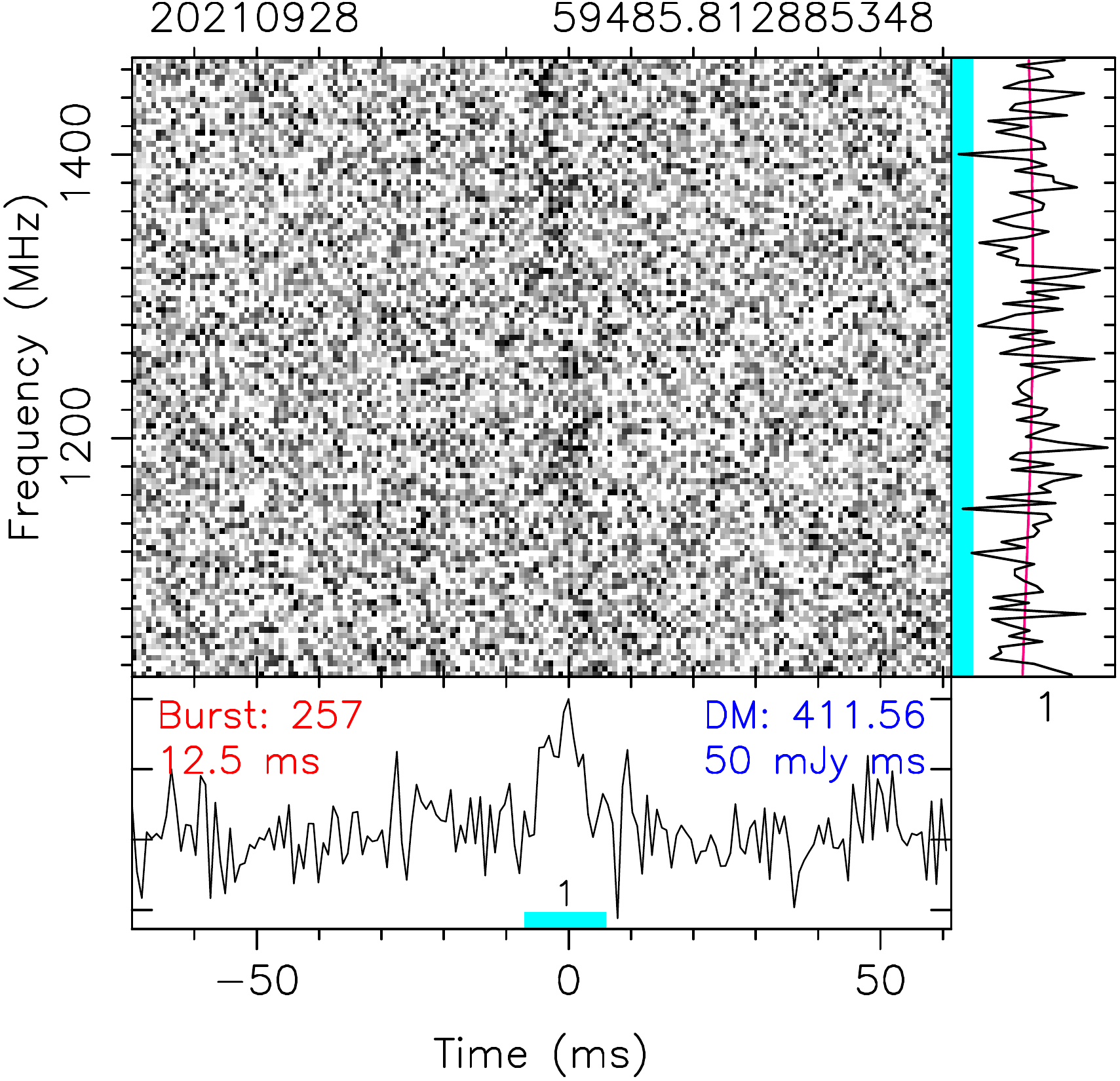}
    \includegraphics[height=37mm]{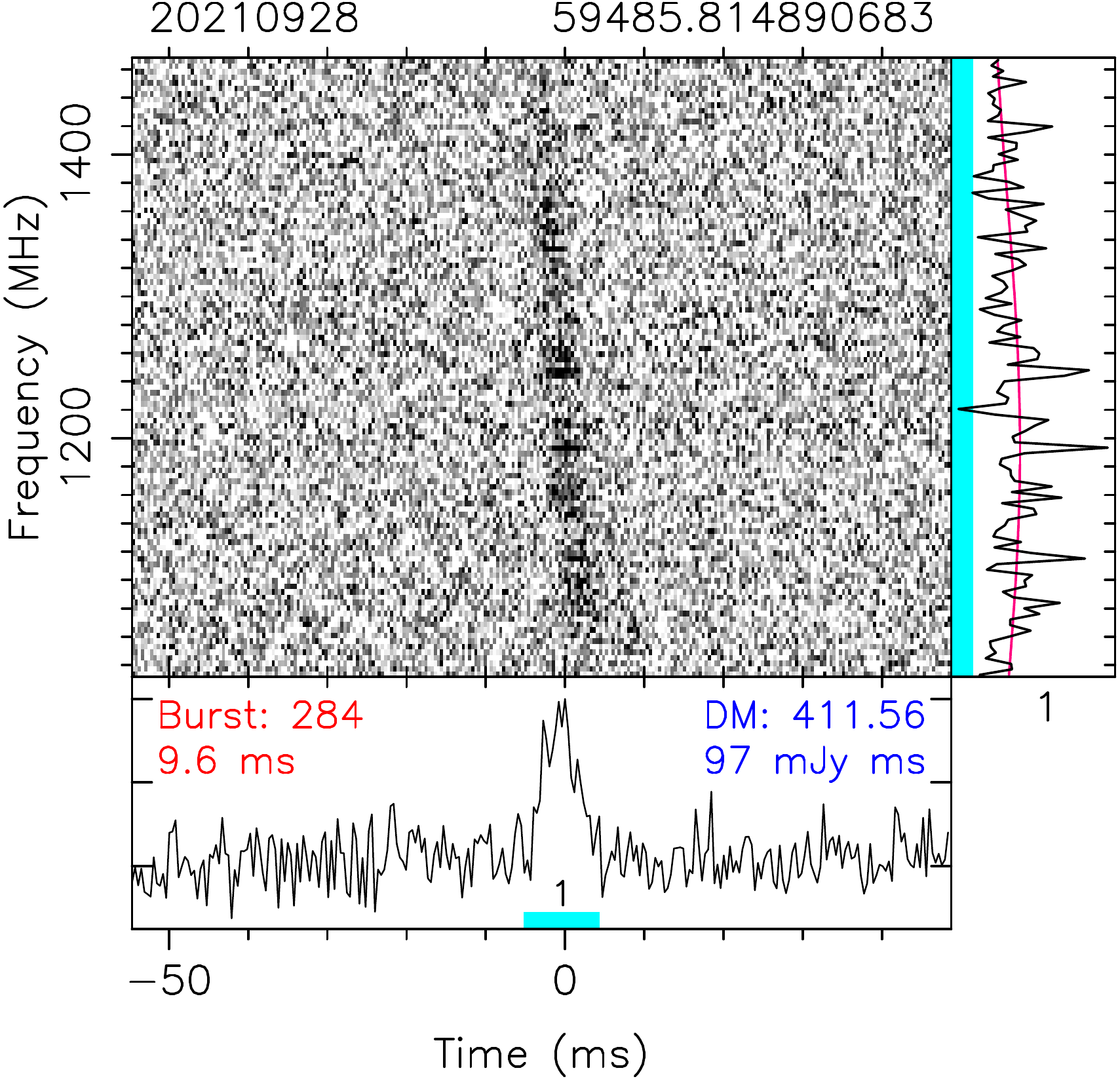}
    \includegraphics[height=37mm]{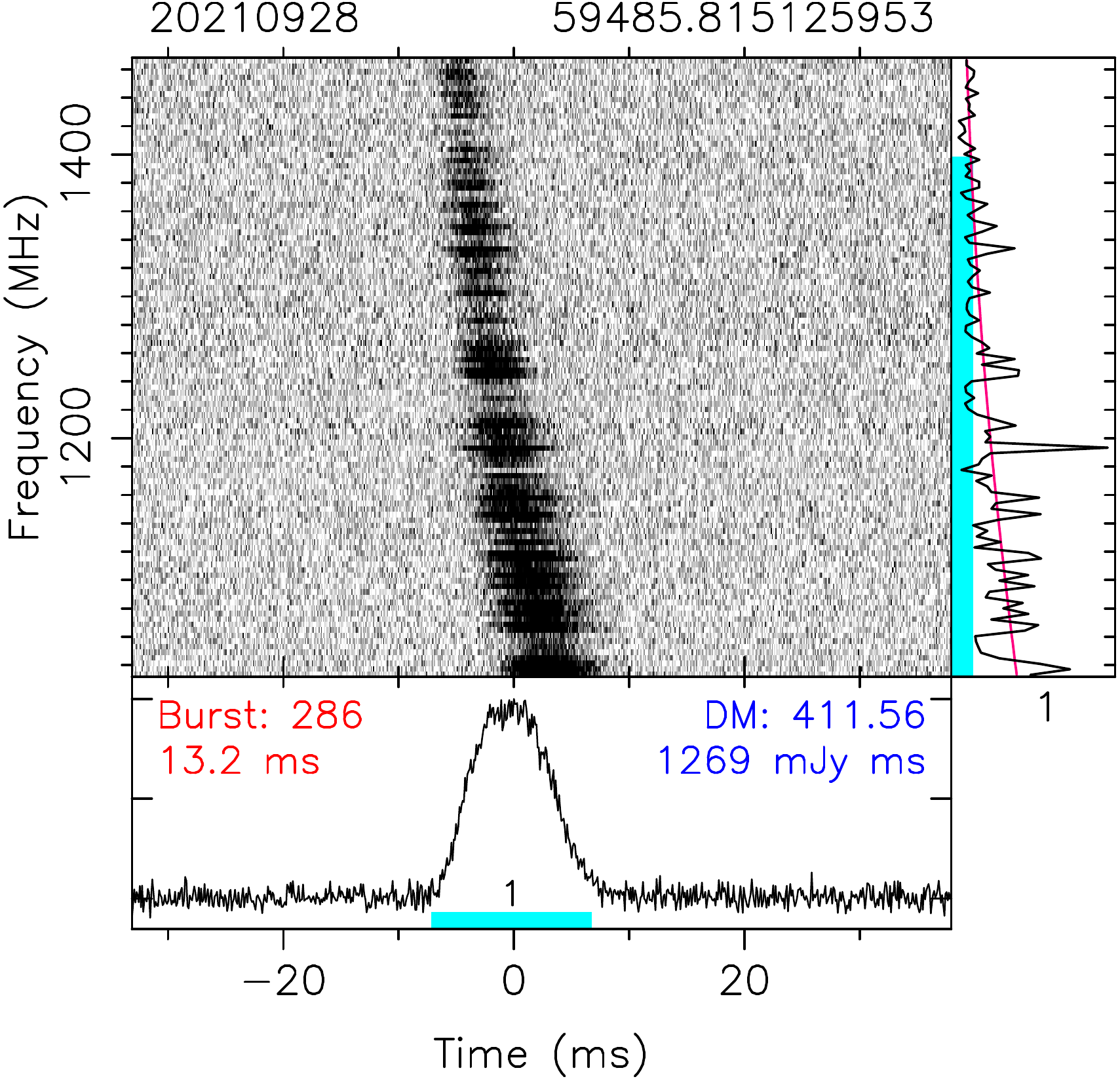}
    \includegraphics[height=37mm]{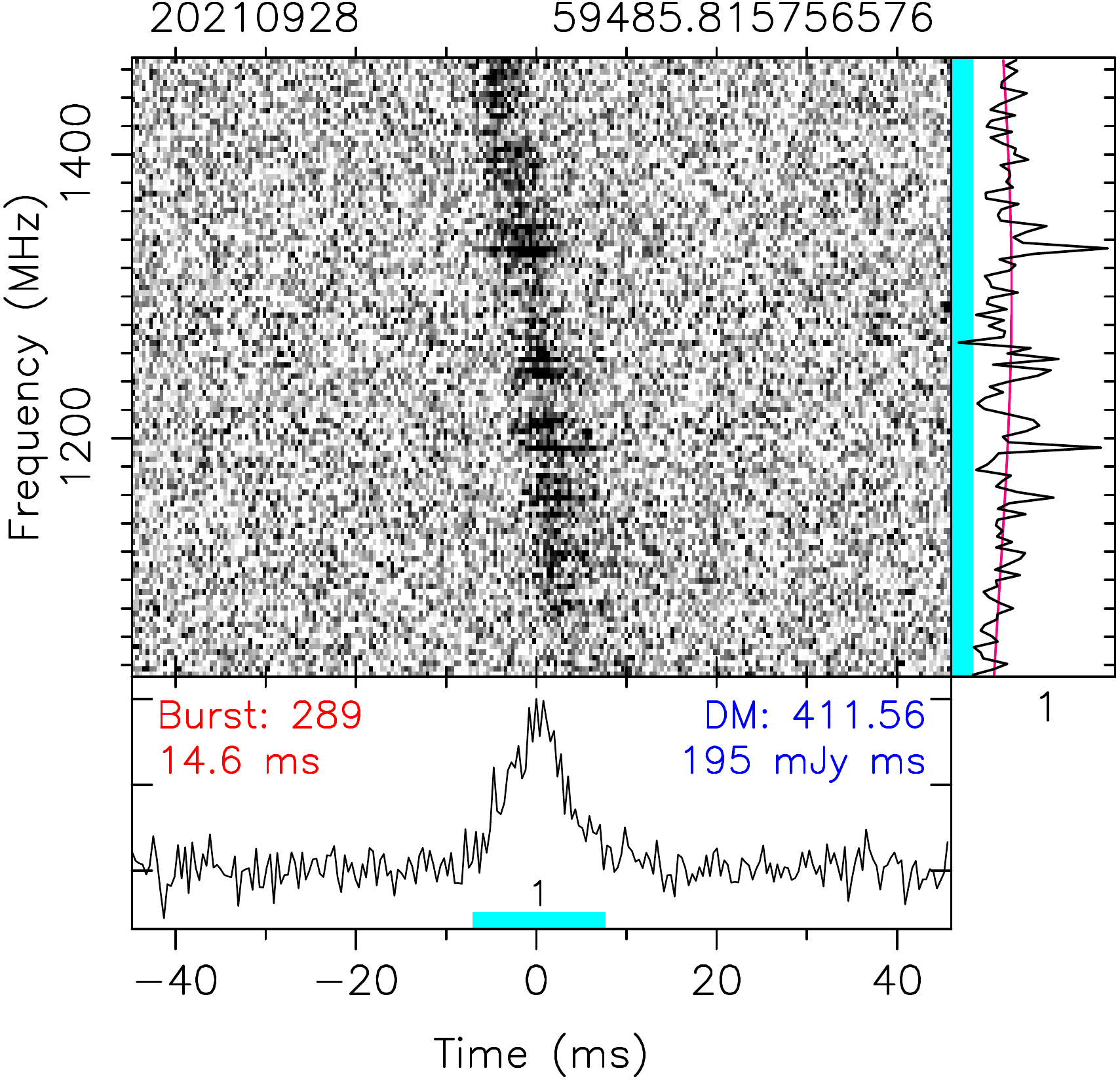}
    \includegraphics[height=37mm]{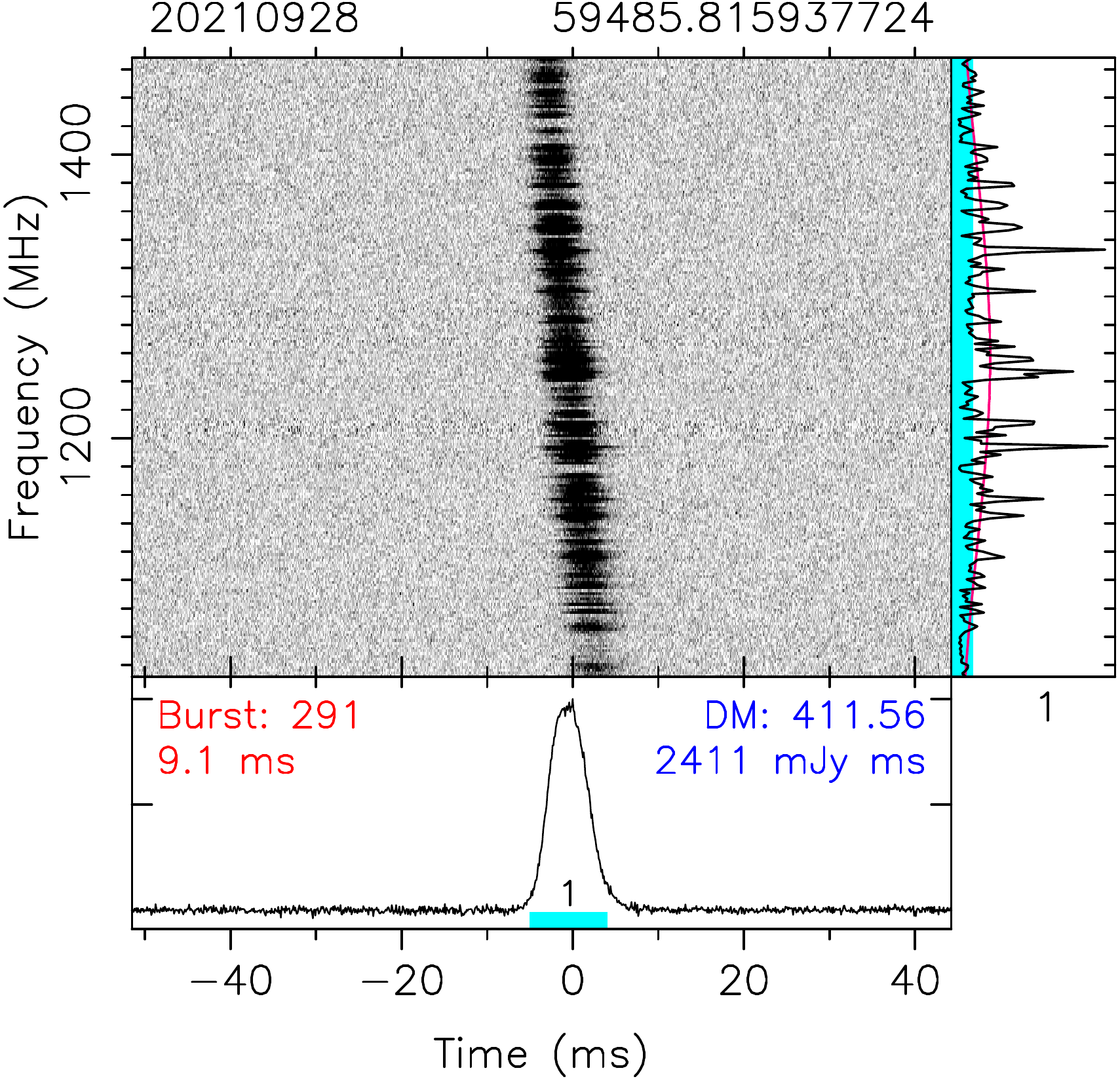}
    \includegraphics[height=37mm]{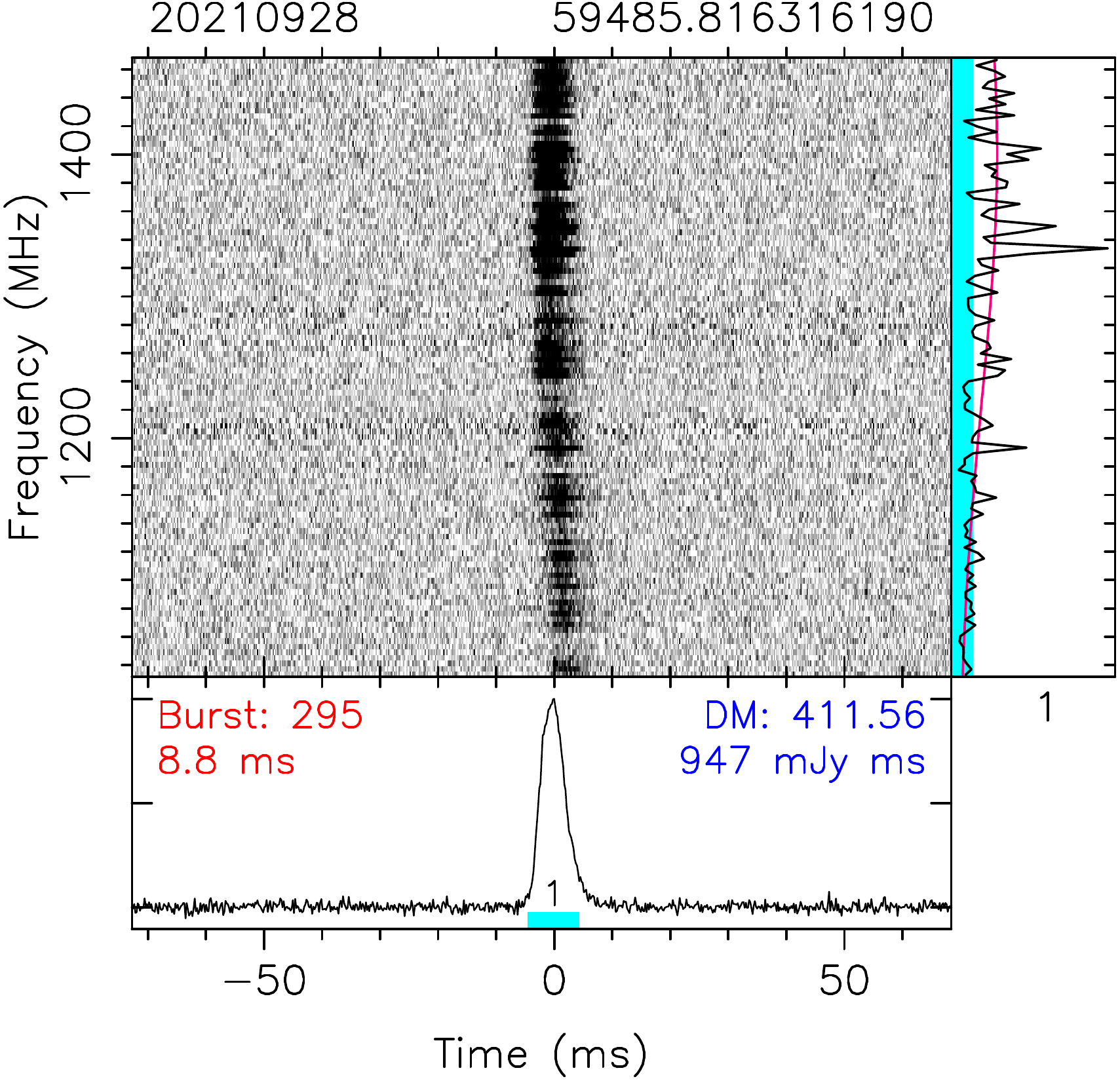}
\caption{The water-fall plot for D1-W bursts.
}\label{fig:appendix:D1W}
\end{figure*}

\begin{figure*}
    \flushleft
    \includegraphics[height=37mm]{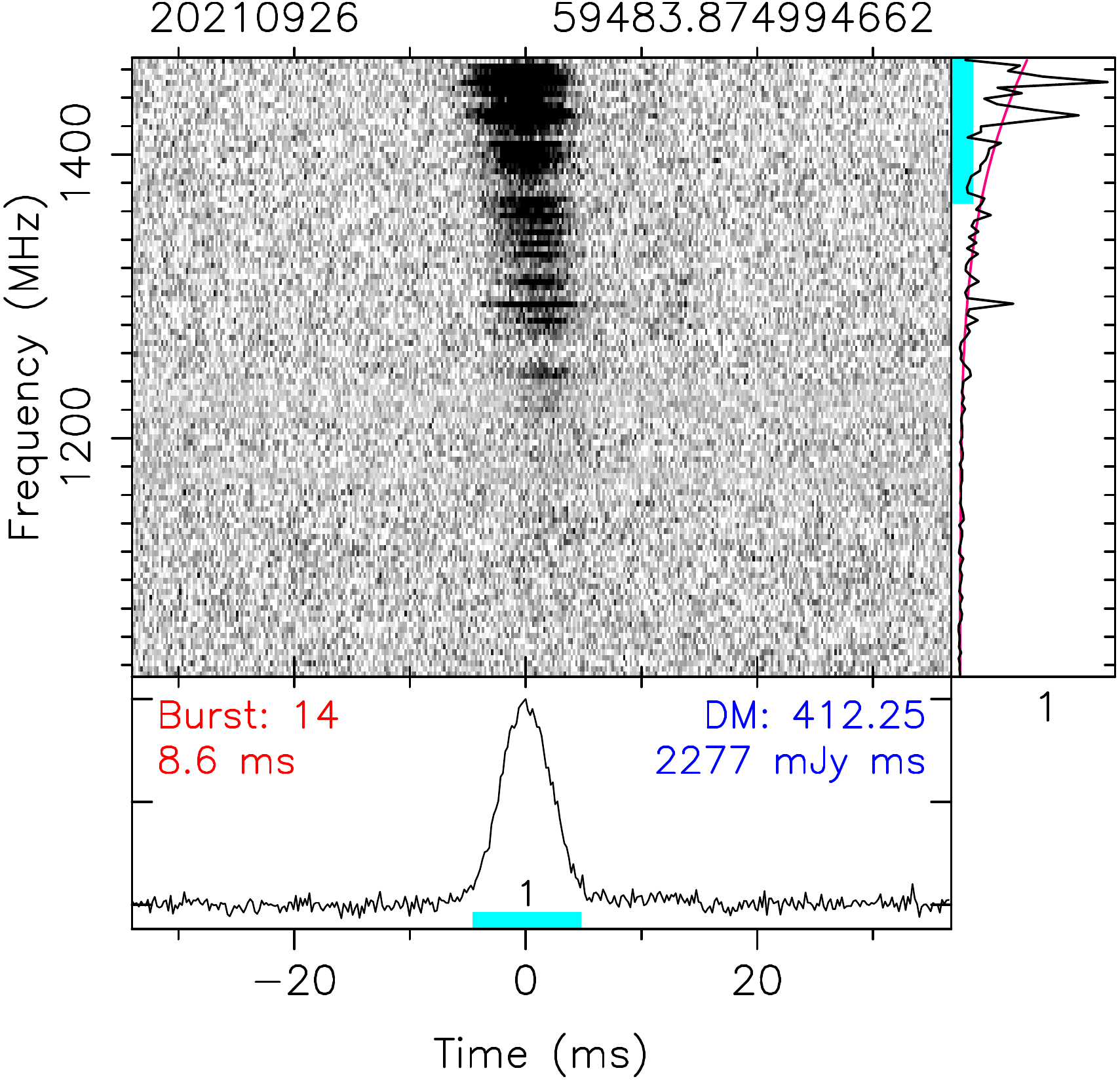}
    \includegraphics[height=37mm]{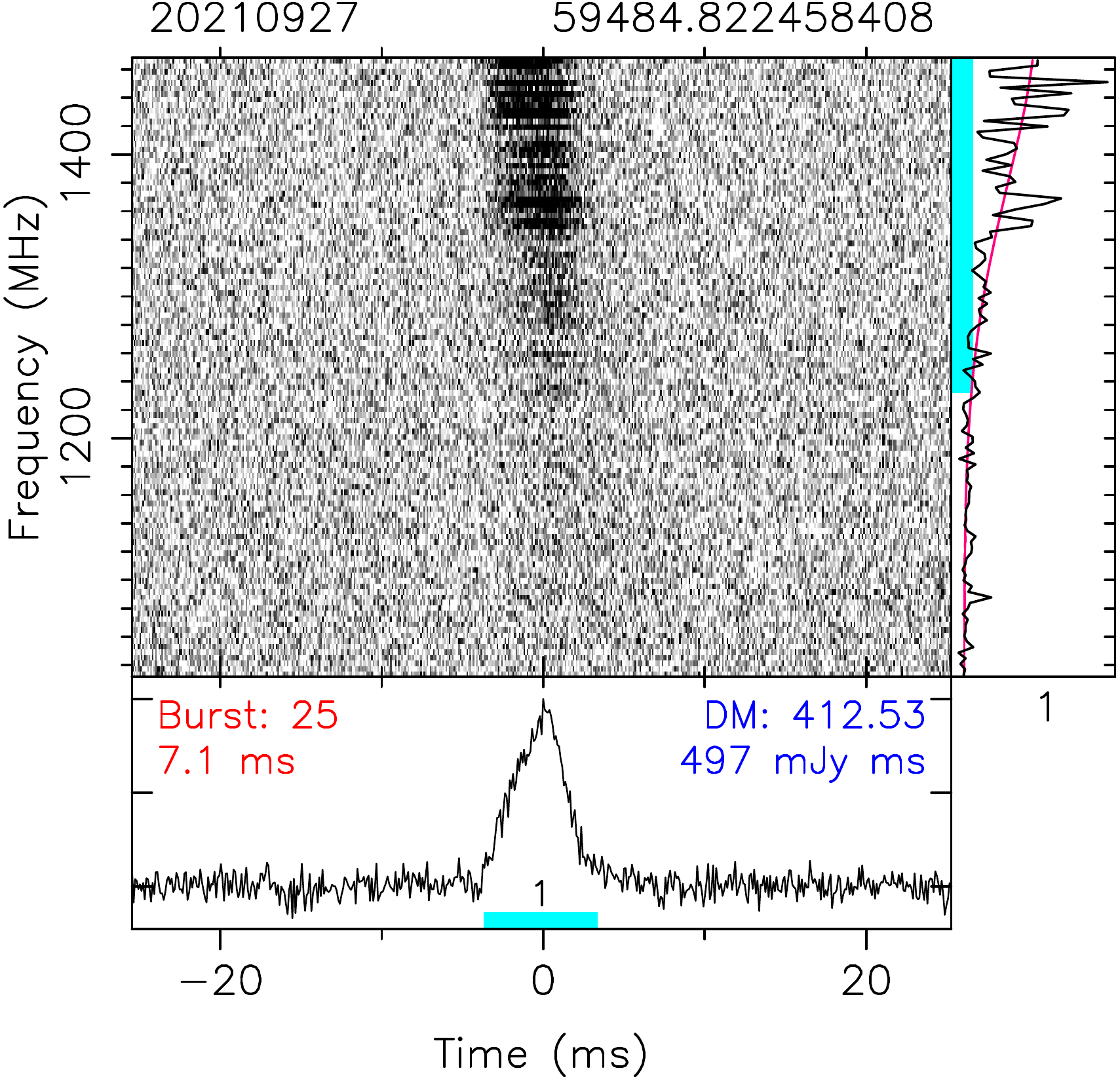}
    \includegraphics[height=37mm]{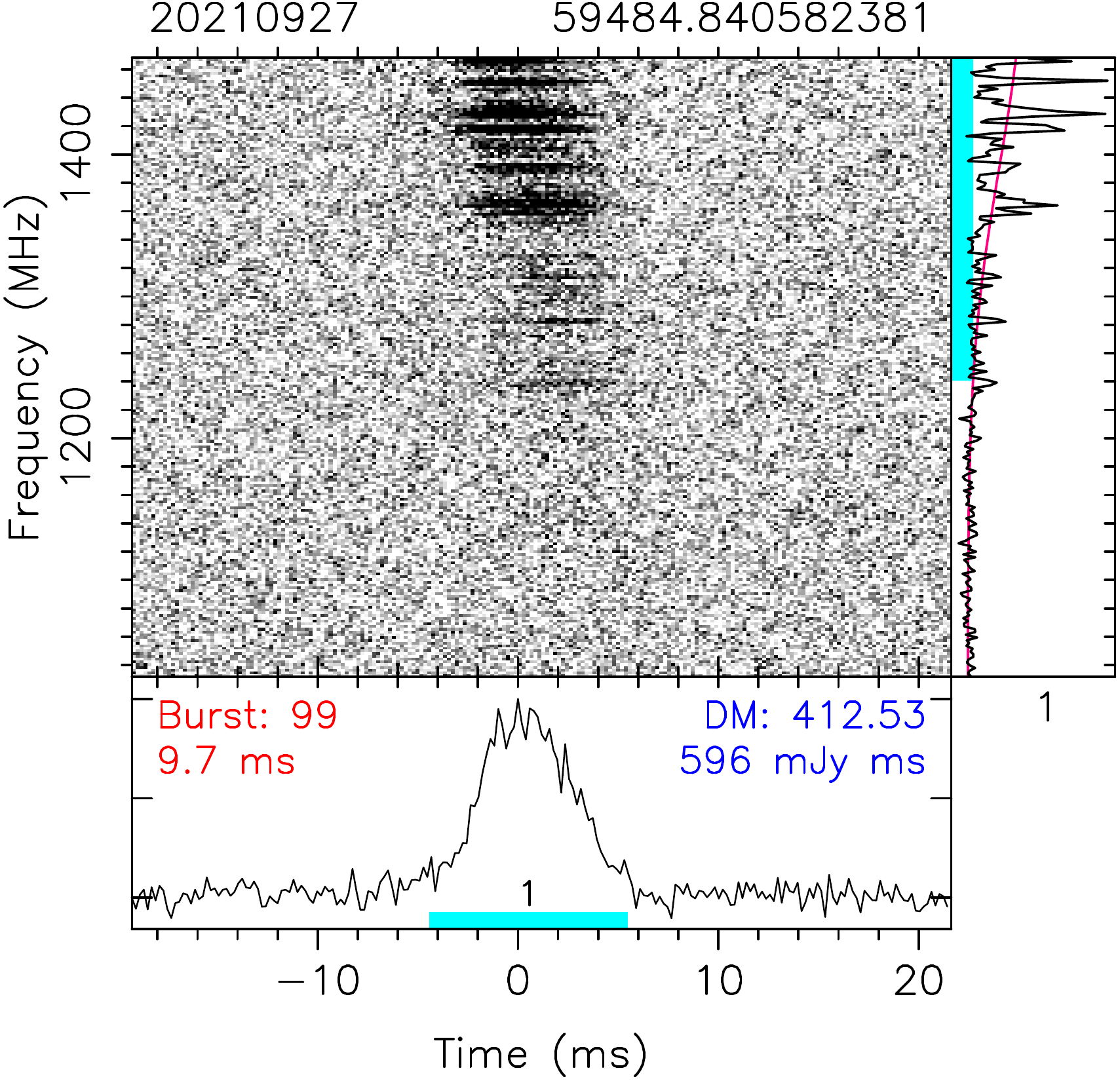}
    \includegraphics[height=37mm]{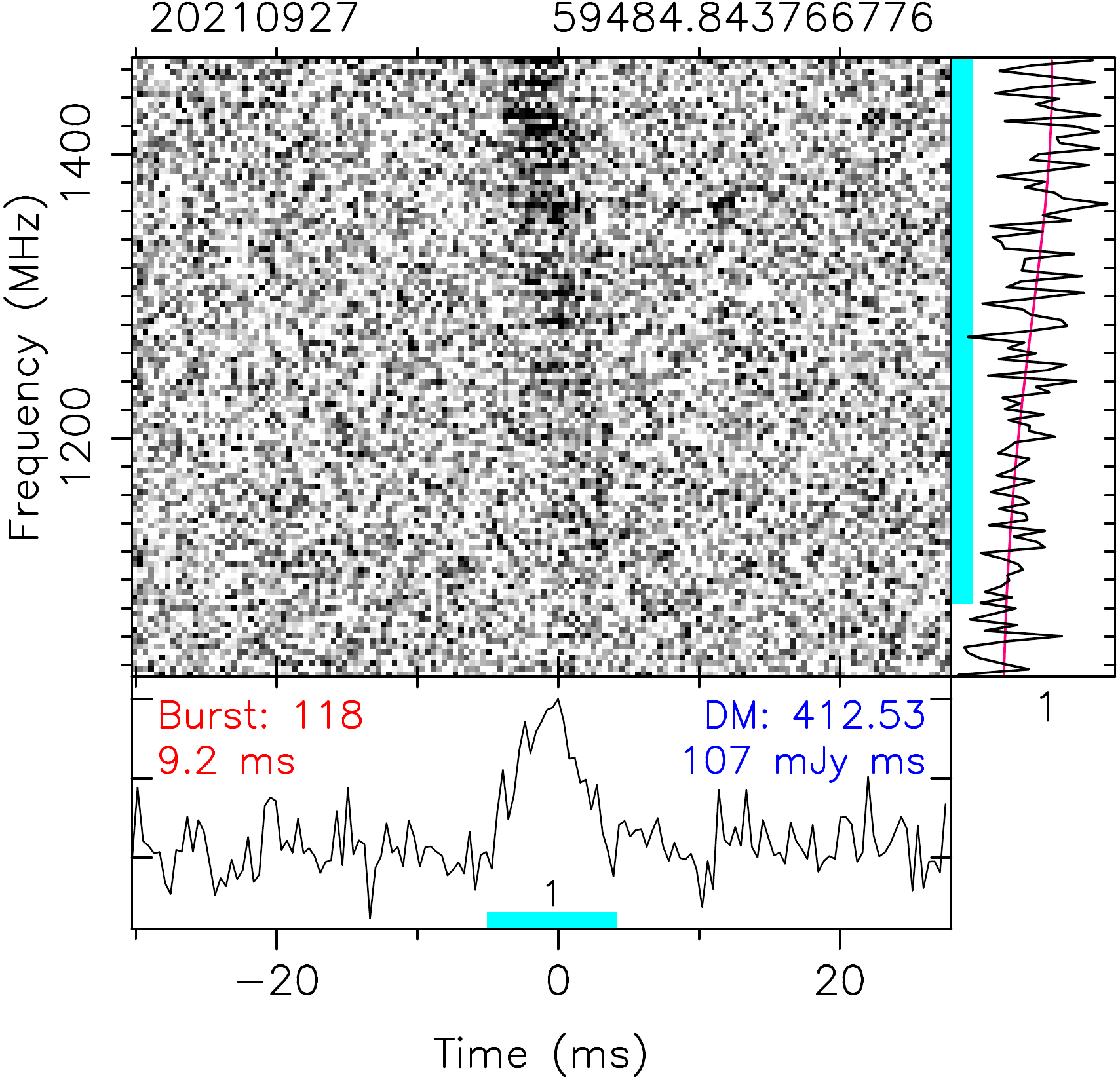}
    \includegraphics[height=37mm]{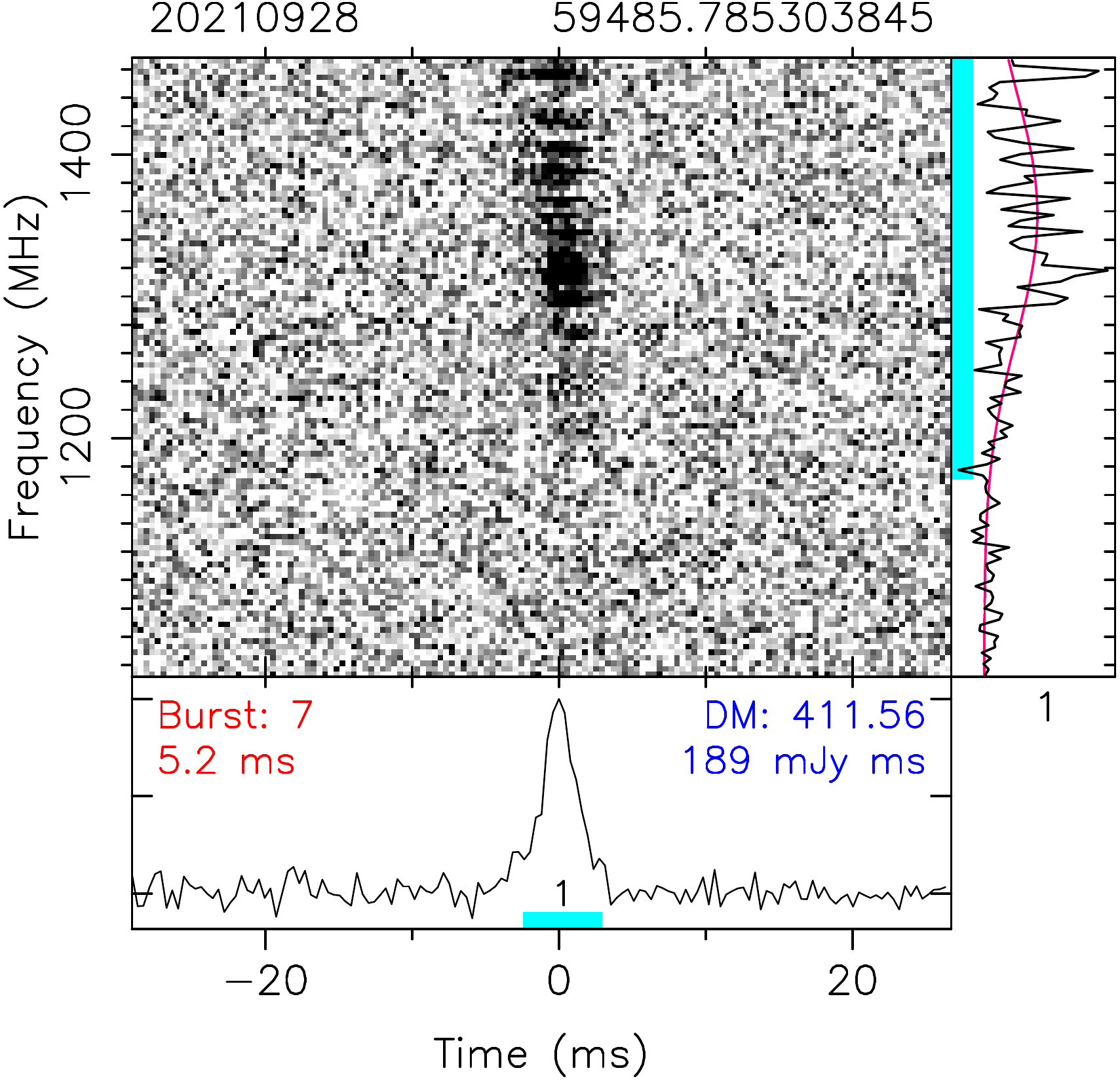}
    \includegraphics[height=37mm]{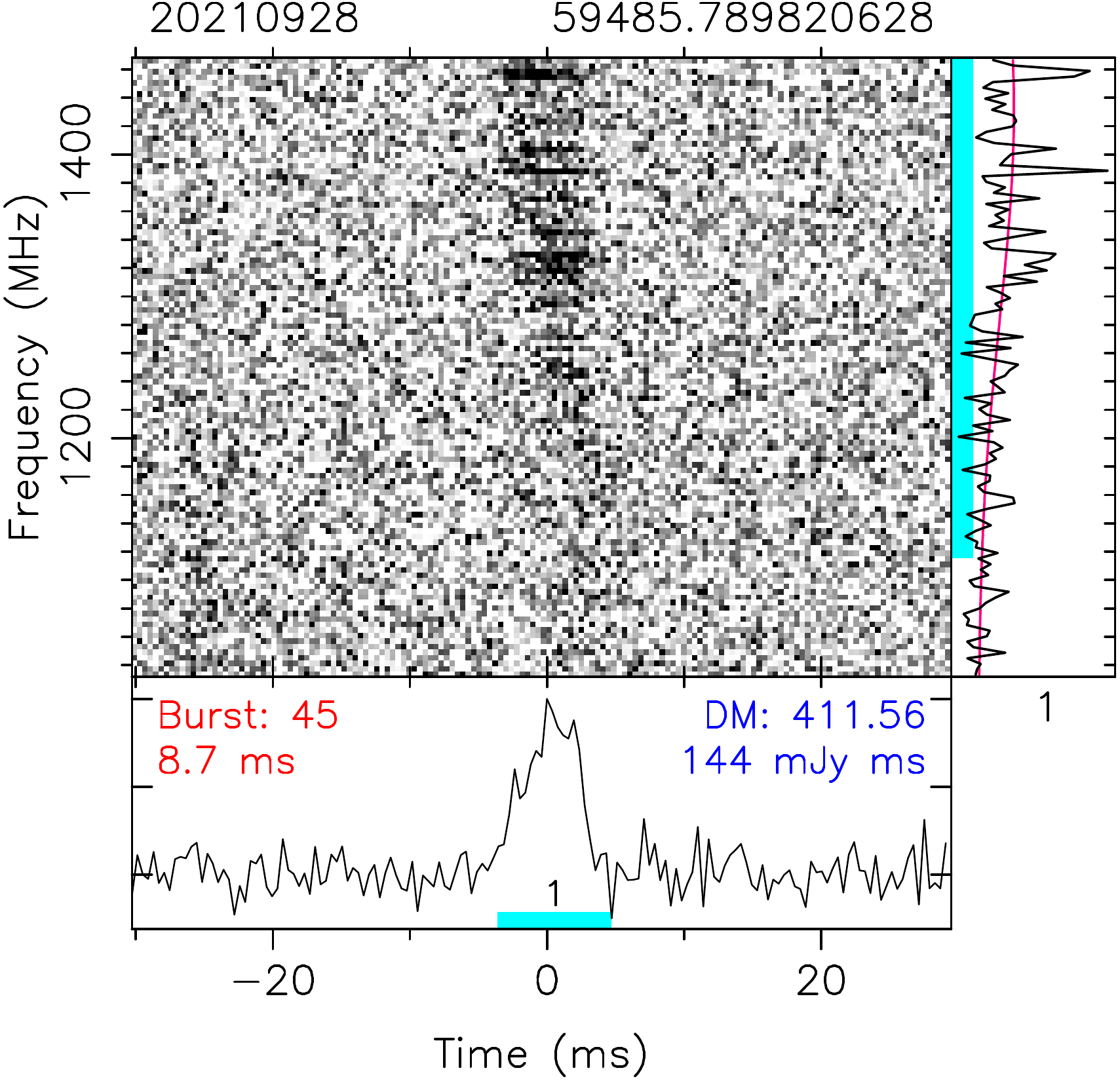}
    \includegraphics[height=37mm]{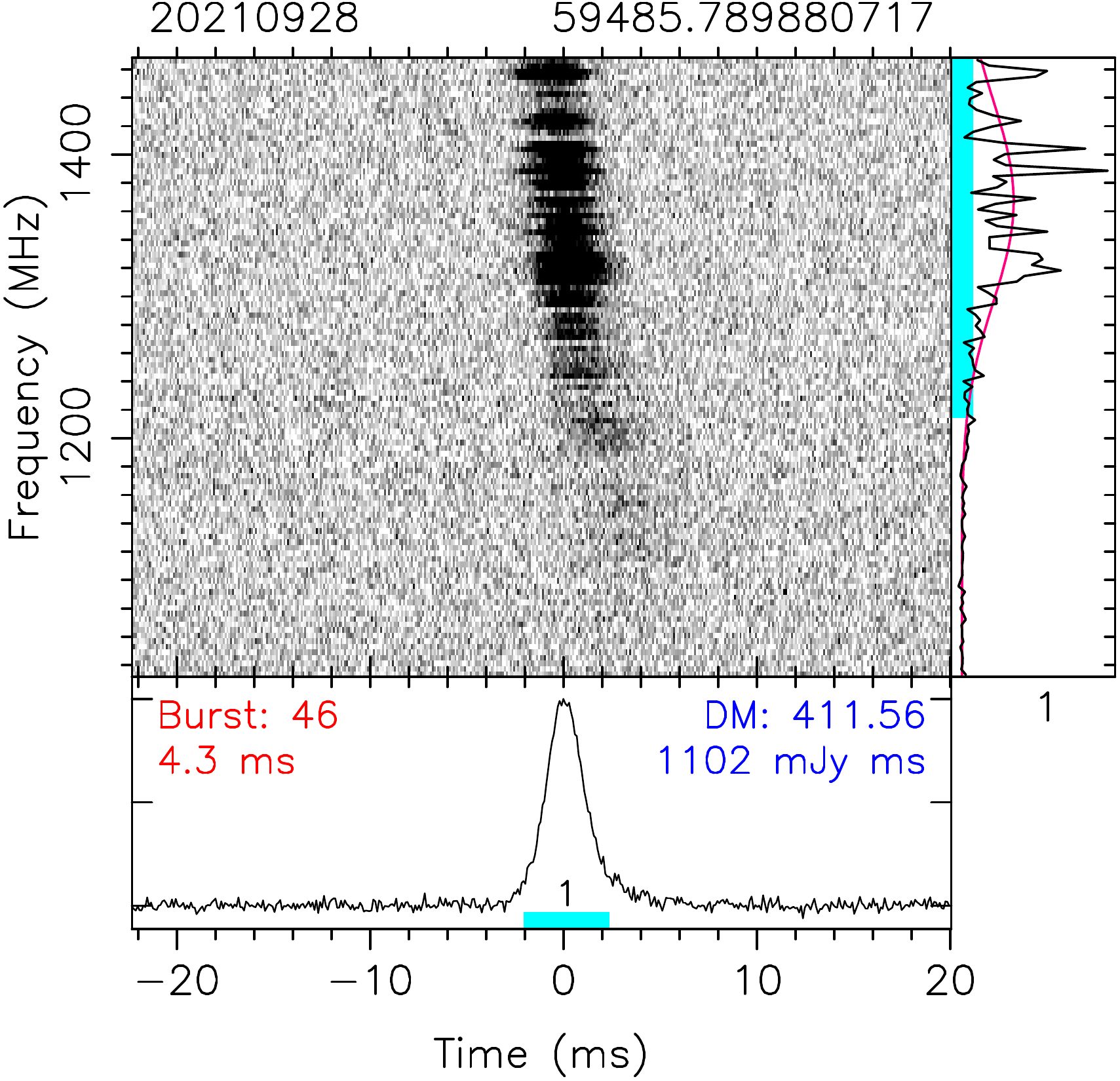}
    \includegraphics[height=37mm]{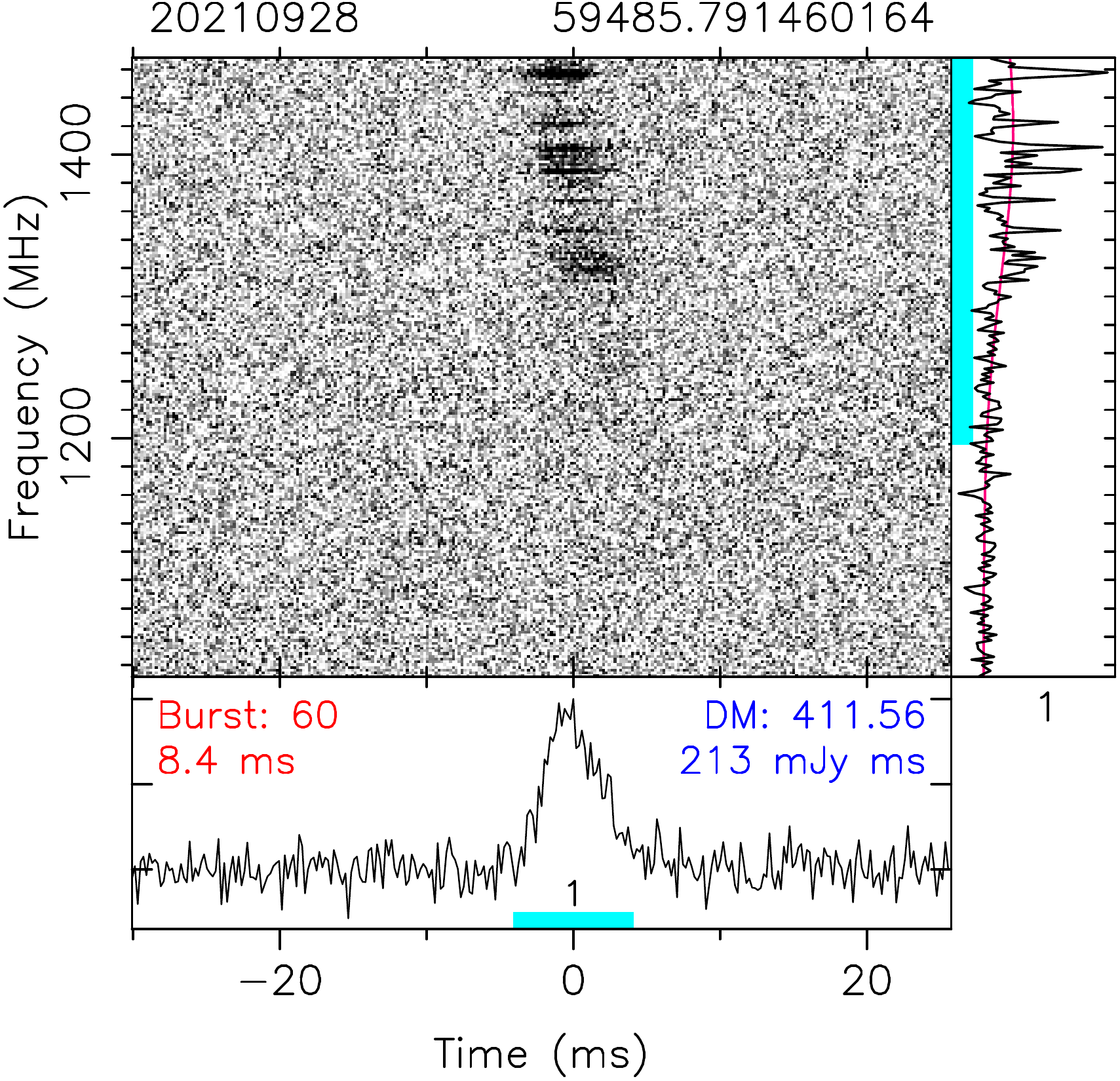}
    \includegraphics[height=37mm]{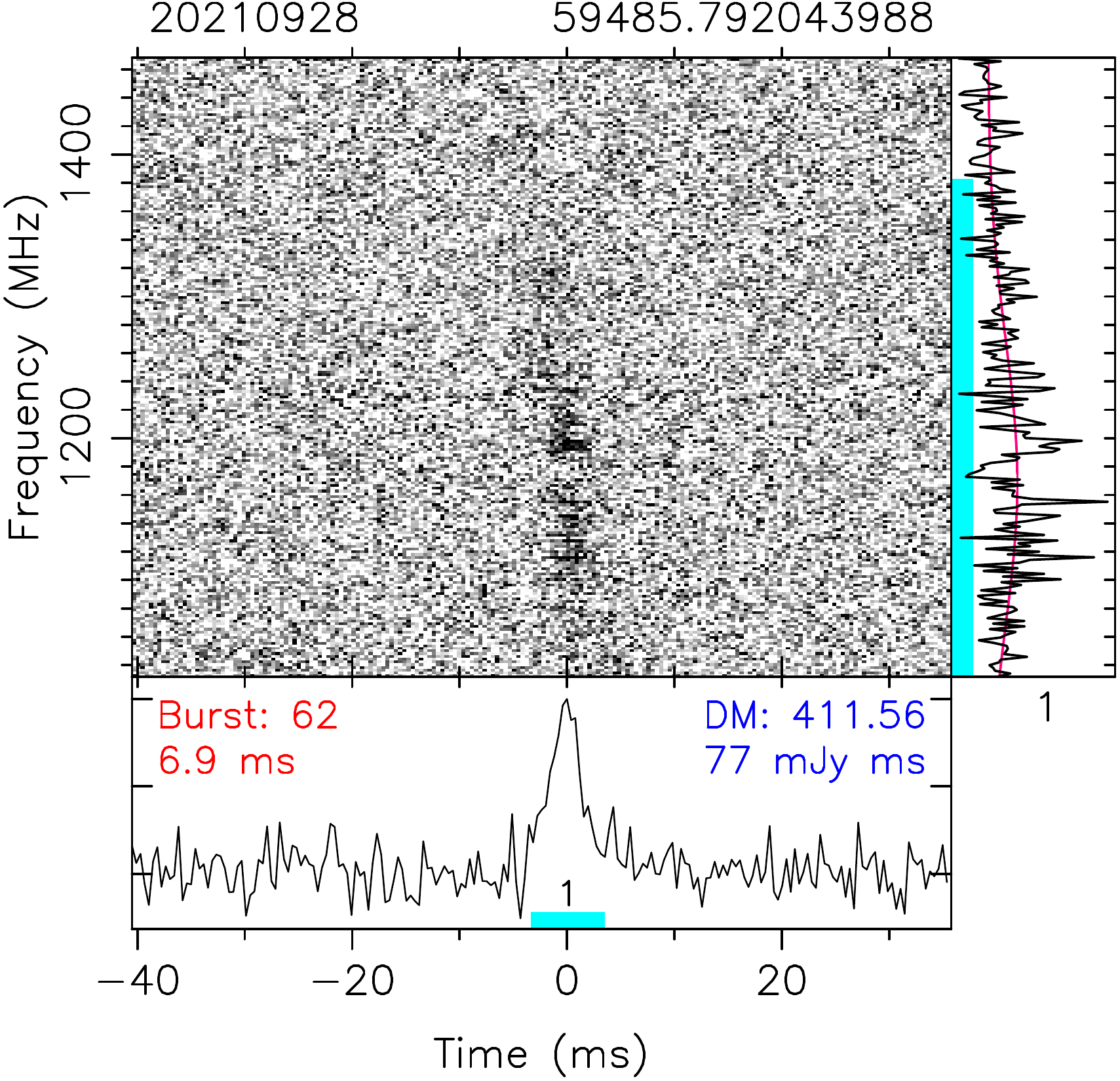}
    \includegraphics[height=37mm]{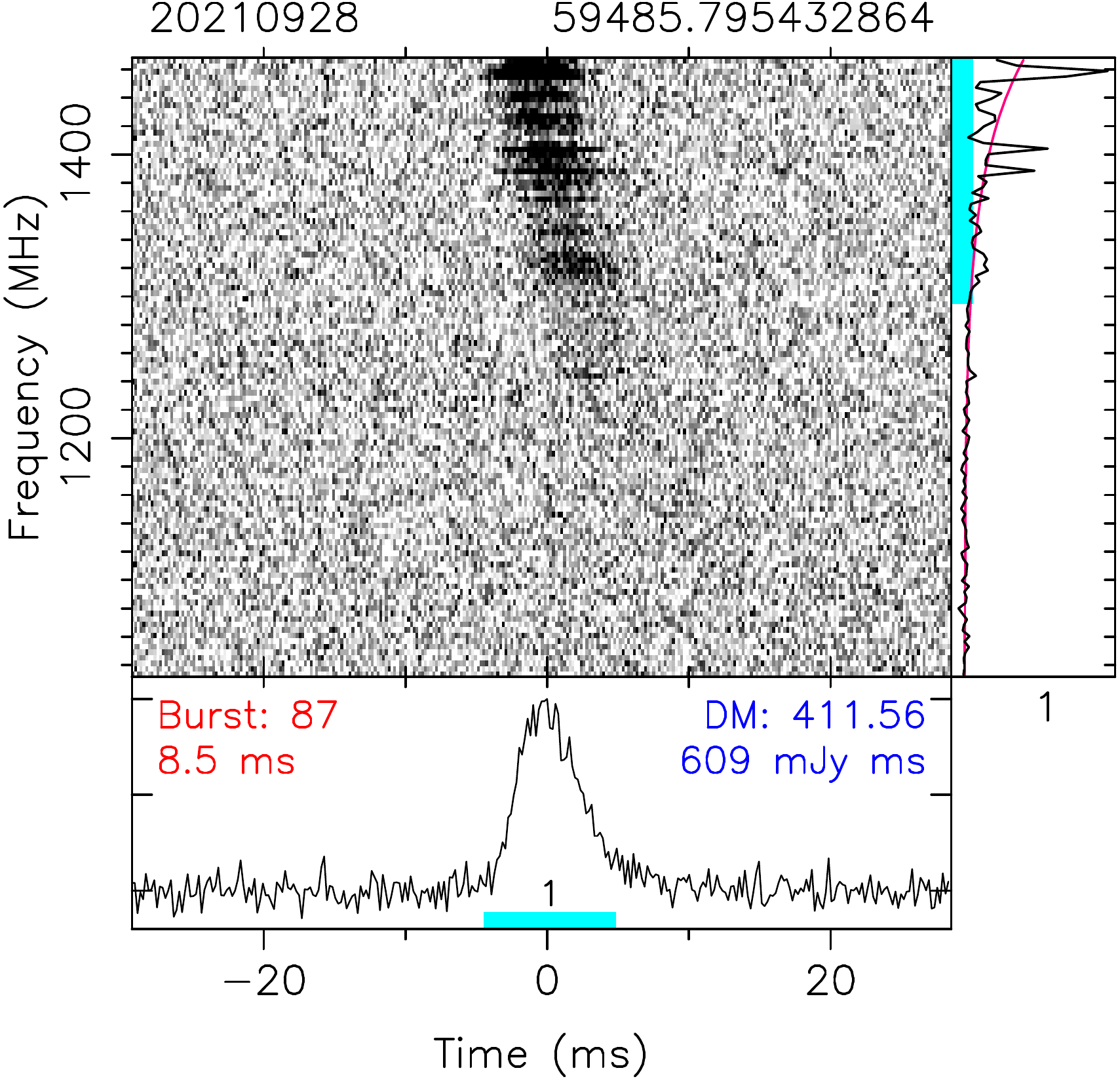}
    \includegraphics[height=37mm]{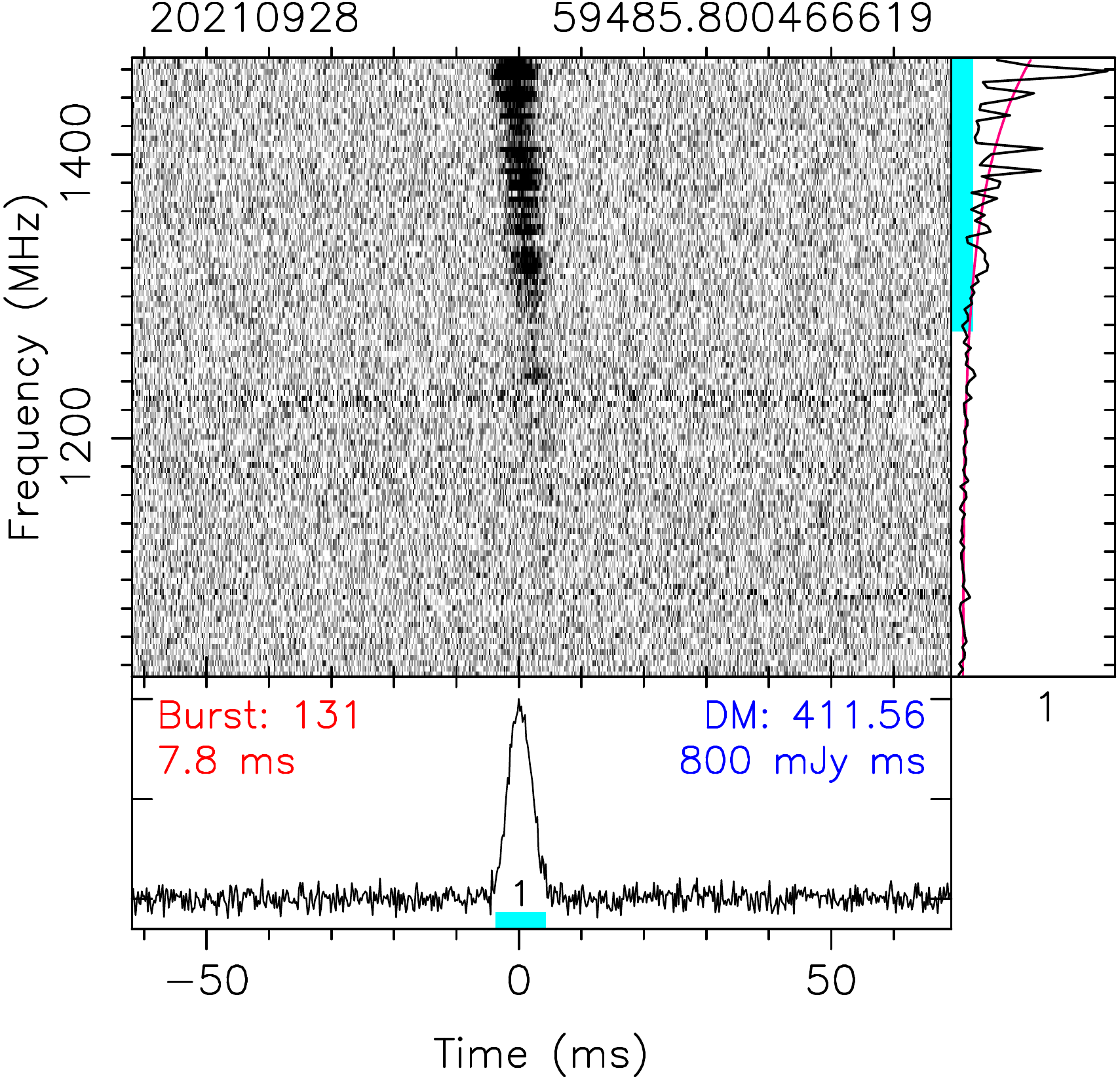}
    \includegraphics[height=37mm]{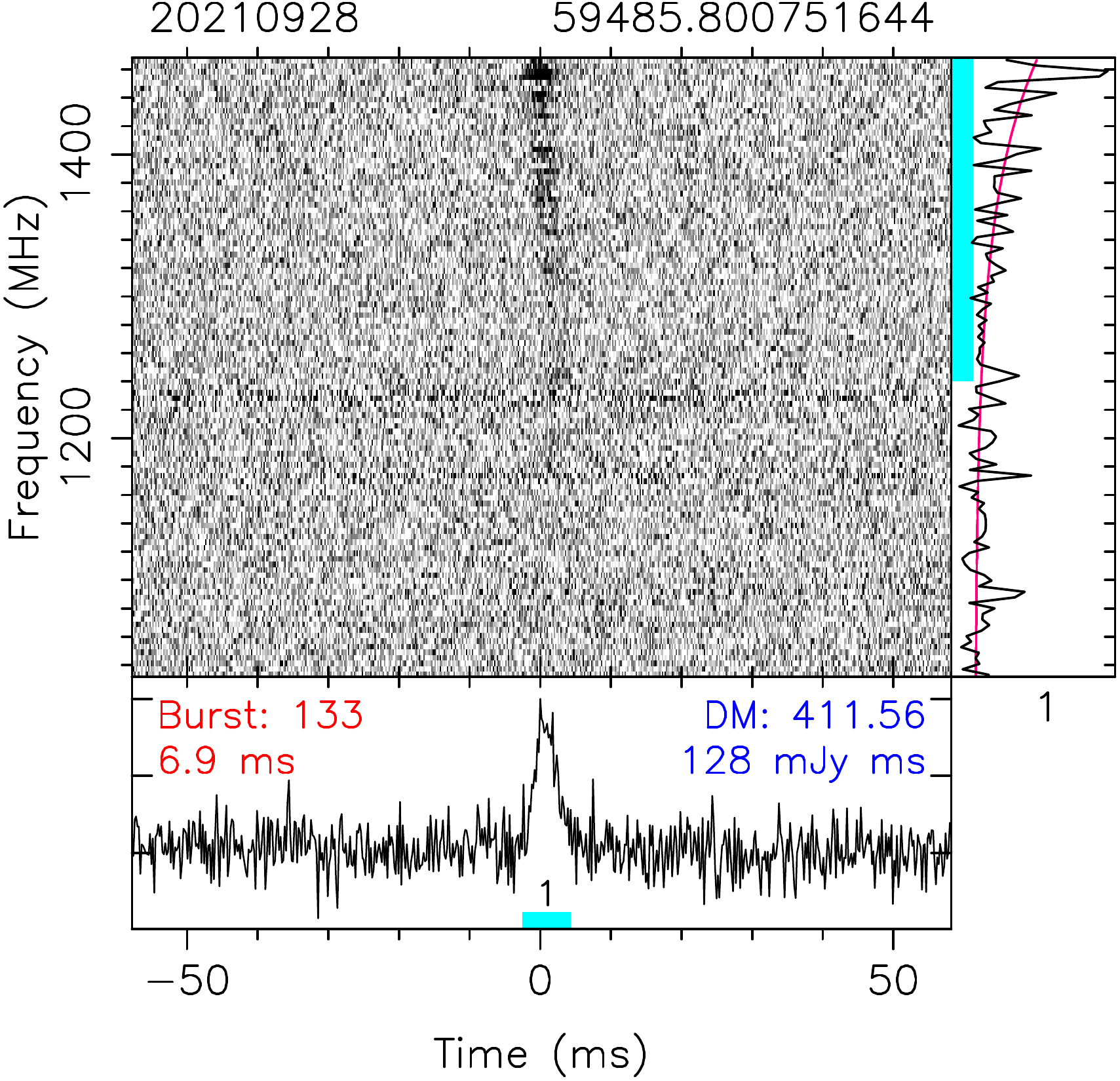}
    \includegraphics[height=37mm]{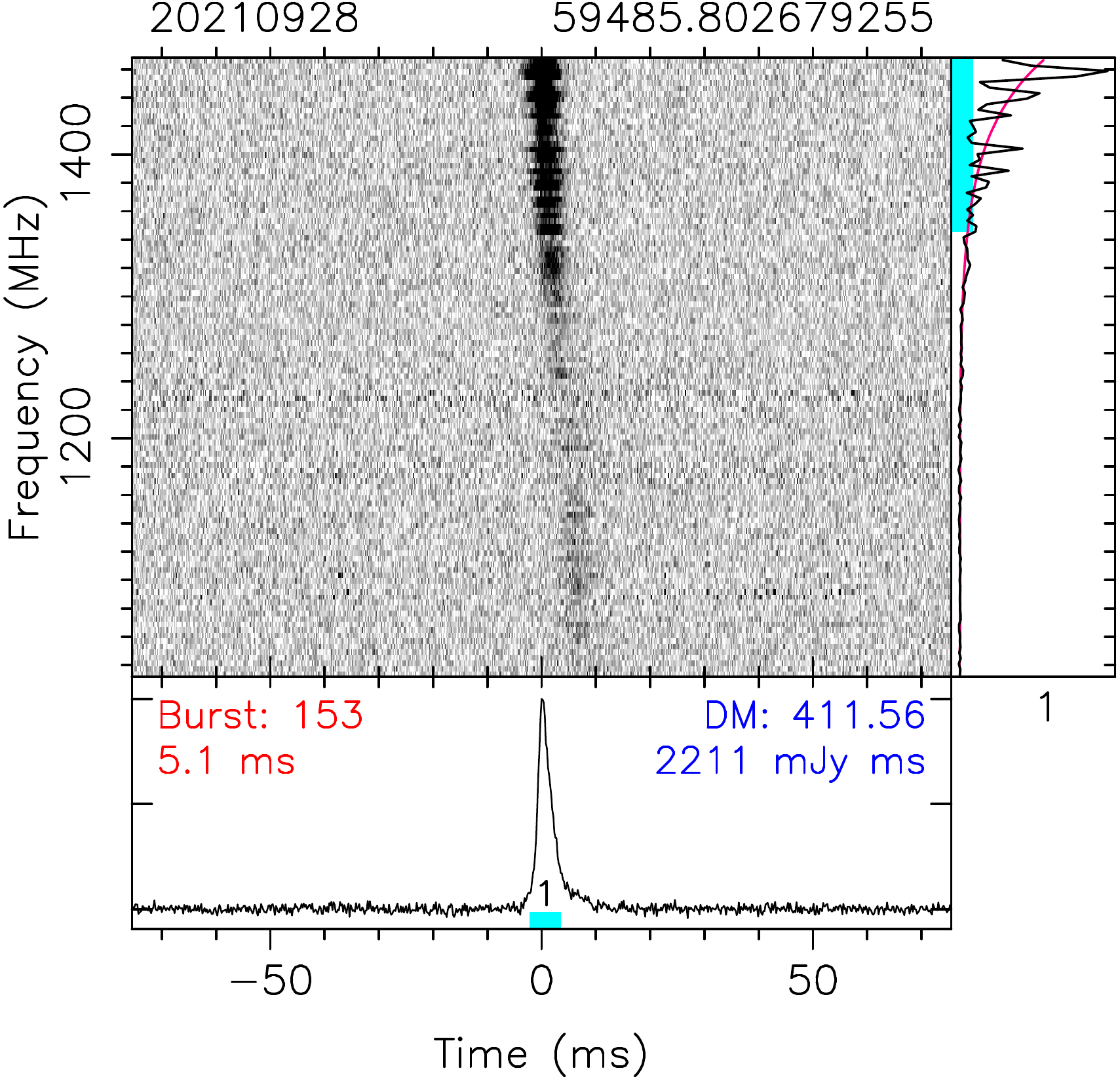}
    \includegraphics[height=37mm]{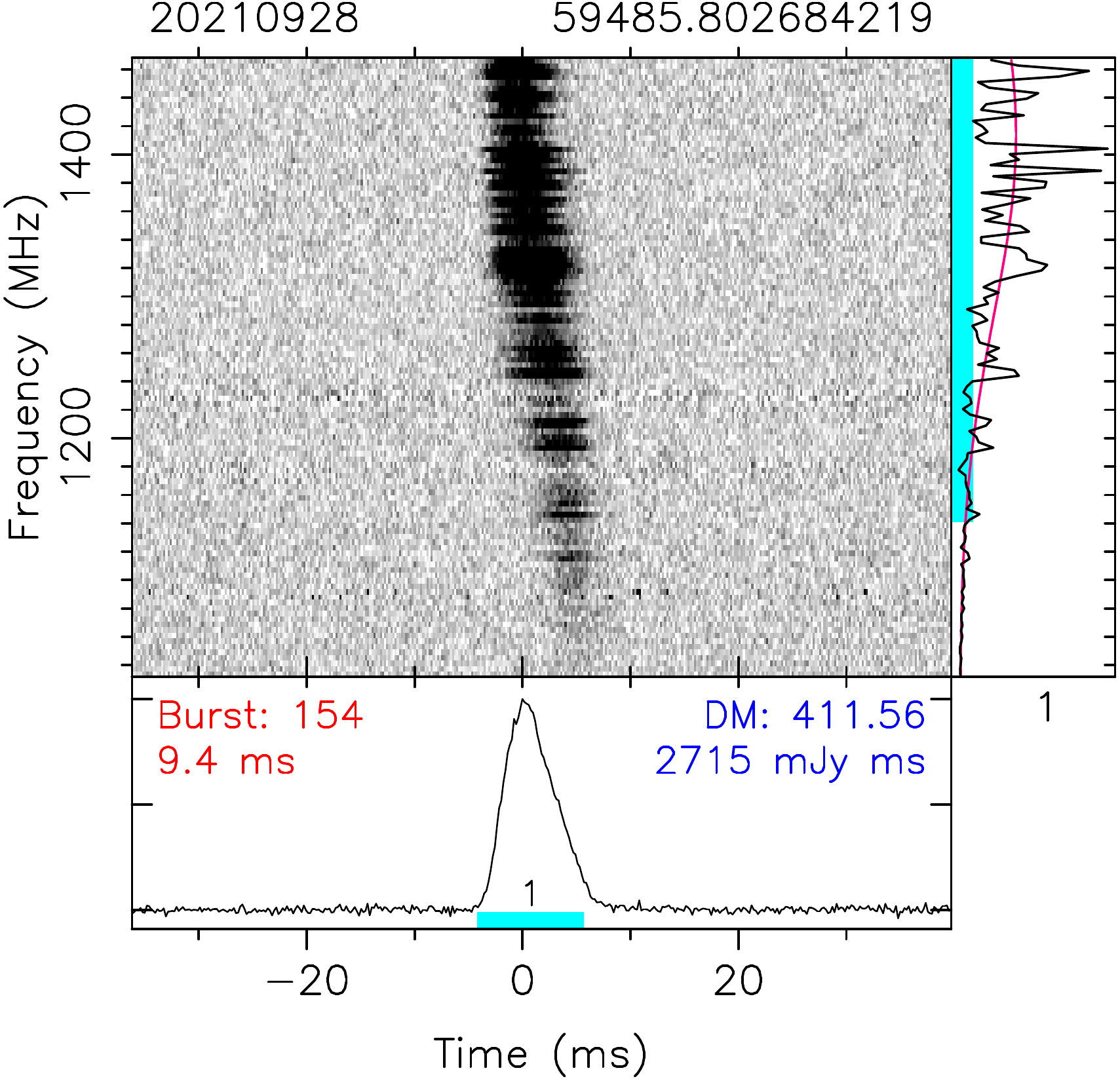}
    \includegraphics[height=37mm]{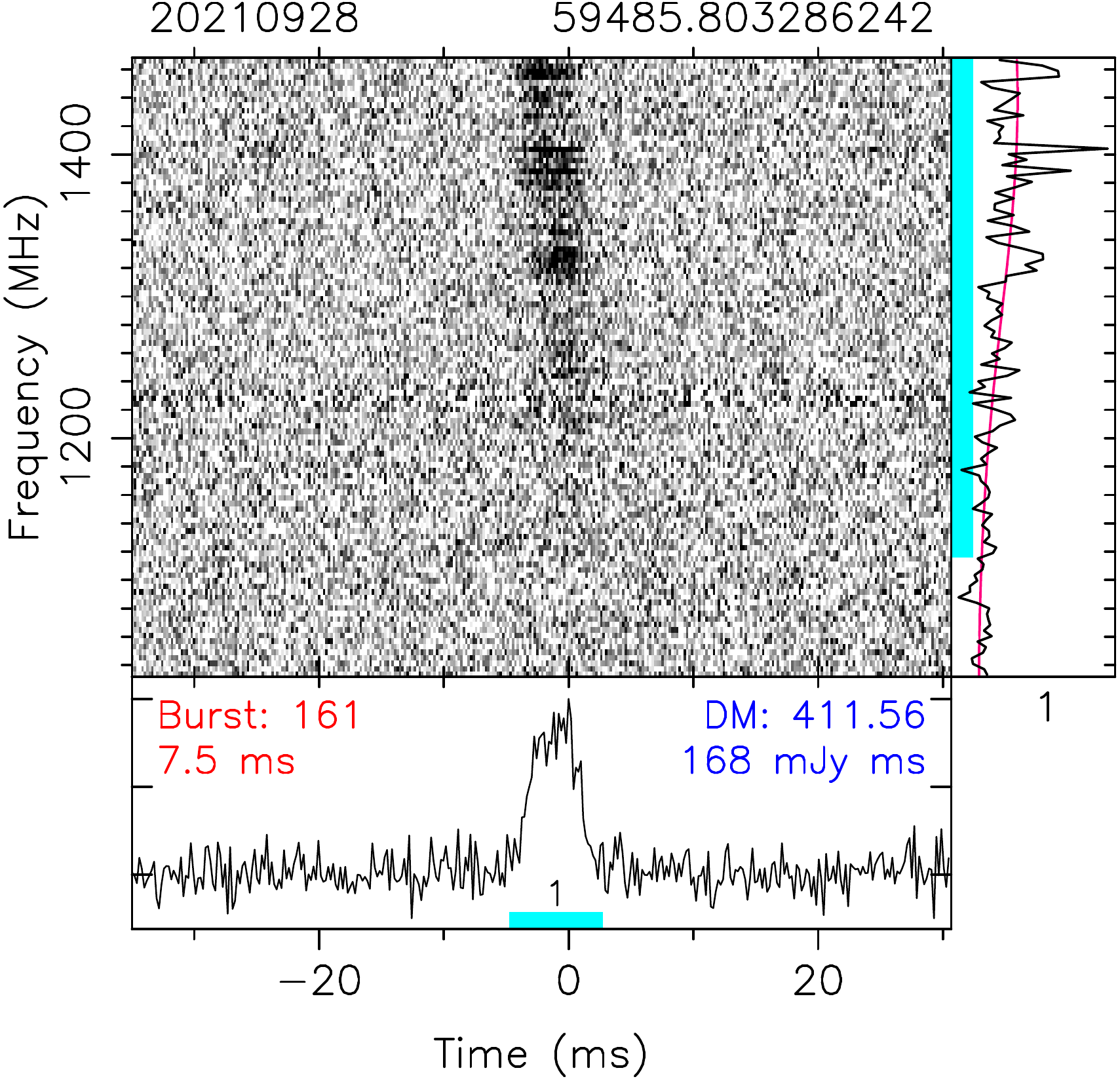}
    \includegraphics[height=37mm]{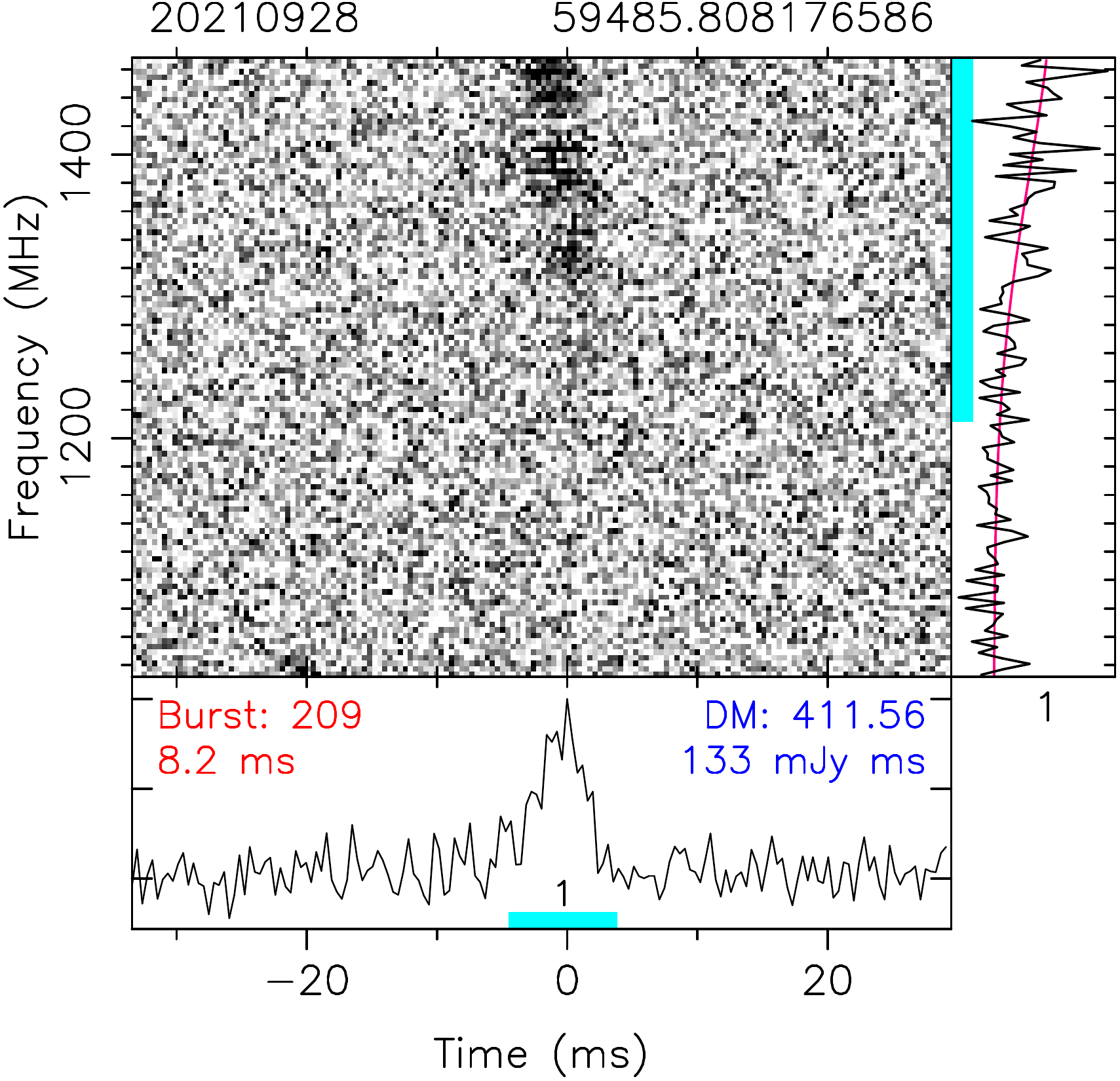}
    \includegraphics[height=37mm]{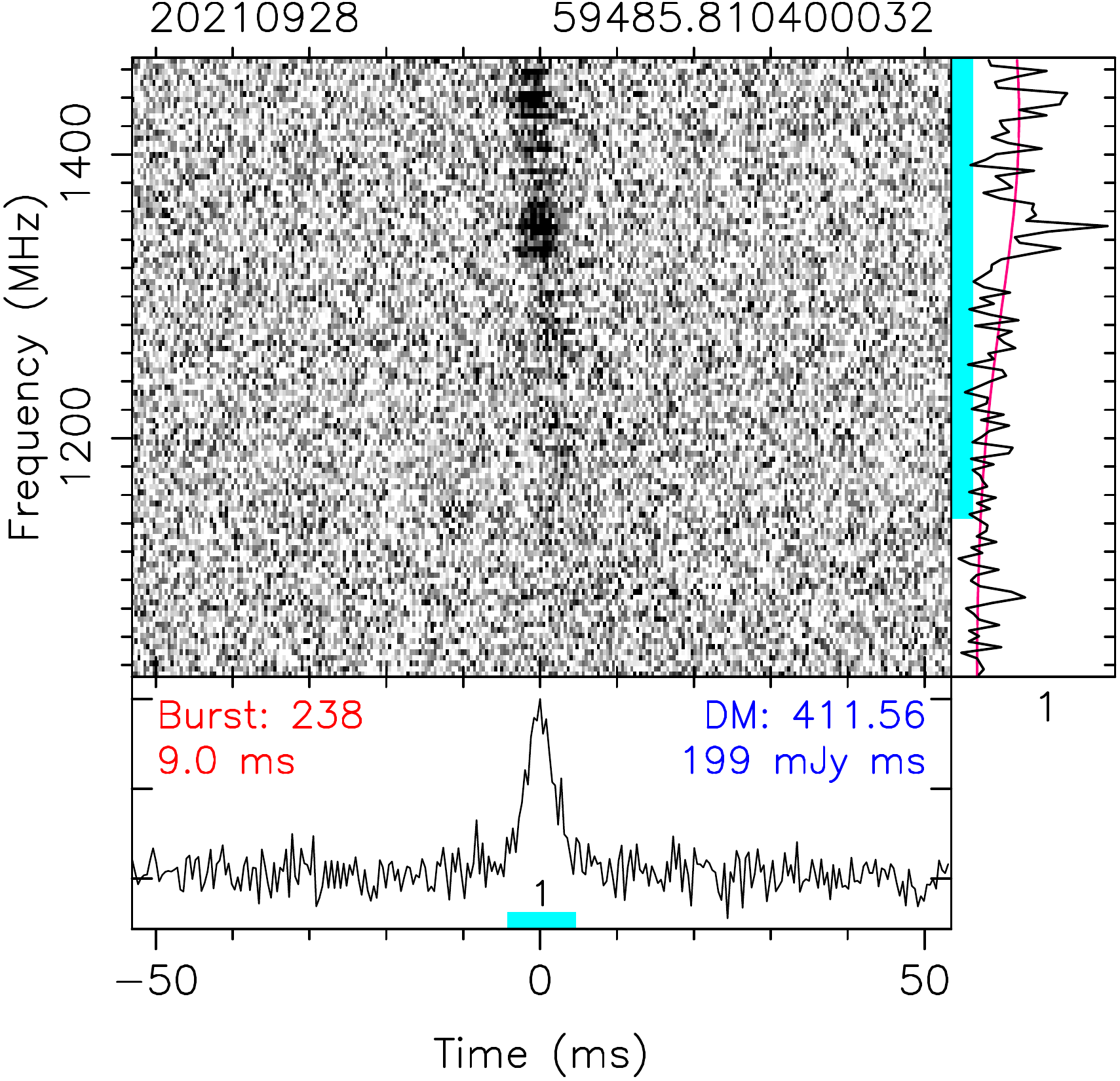}
    \includegraphics[height=37mm]{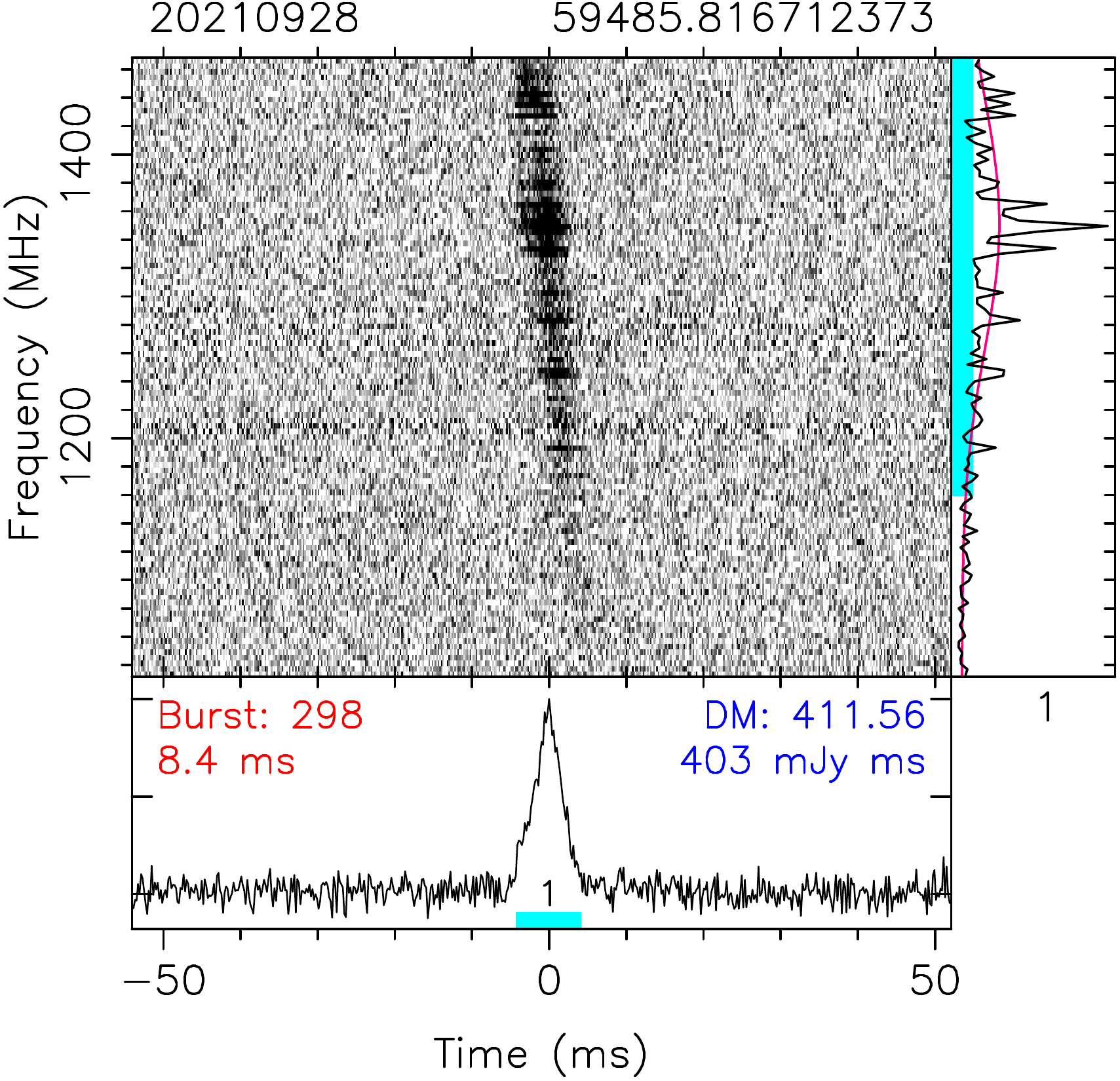}
    \includegraphics[height=37mm]{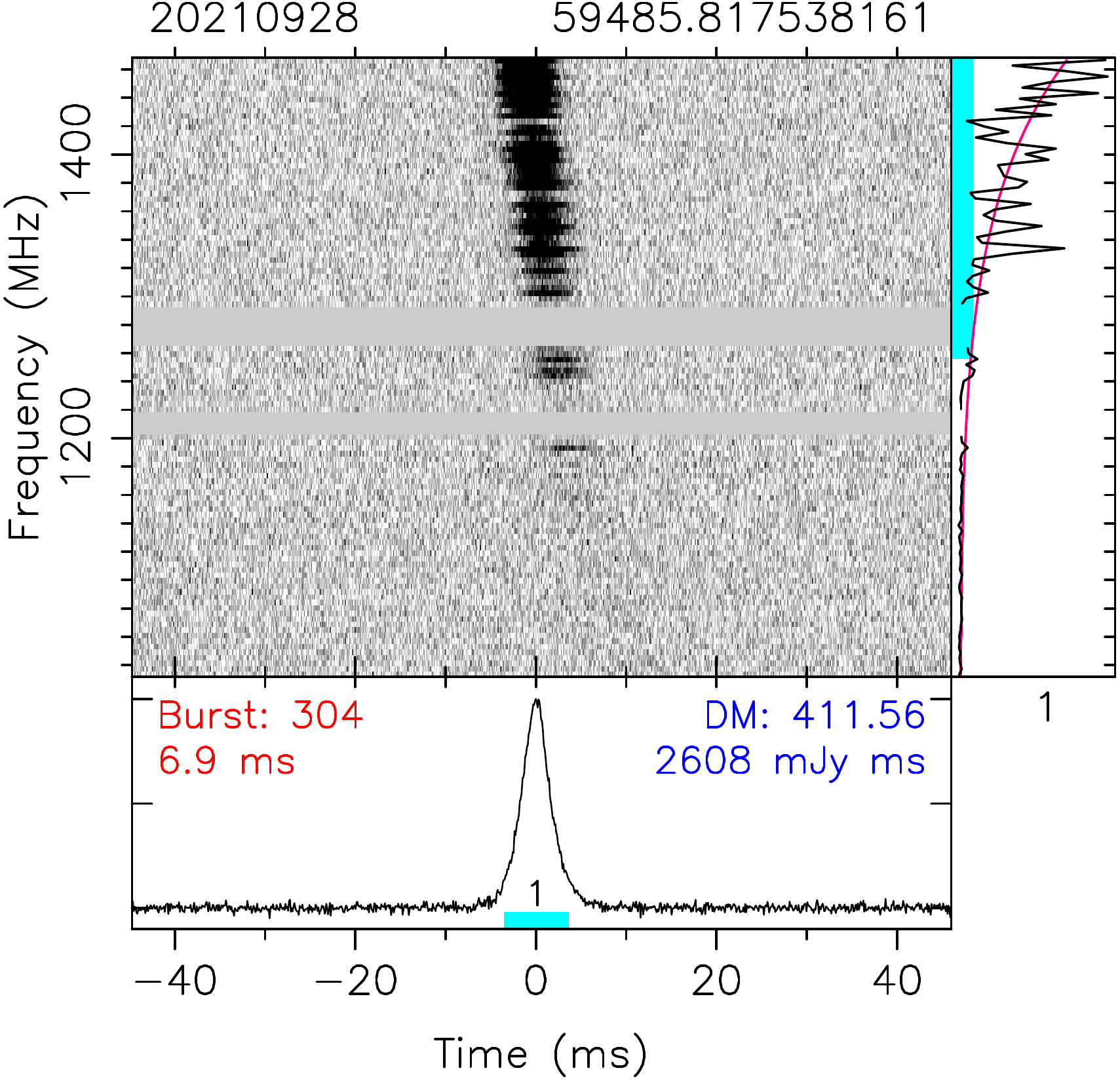}
    \includegraphics[height=37mm]{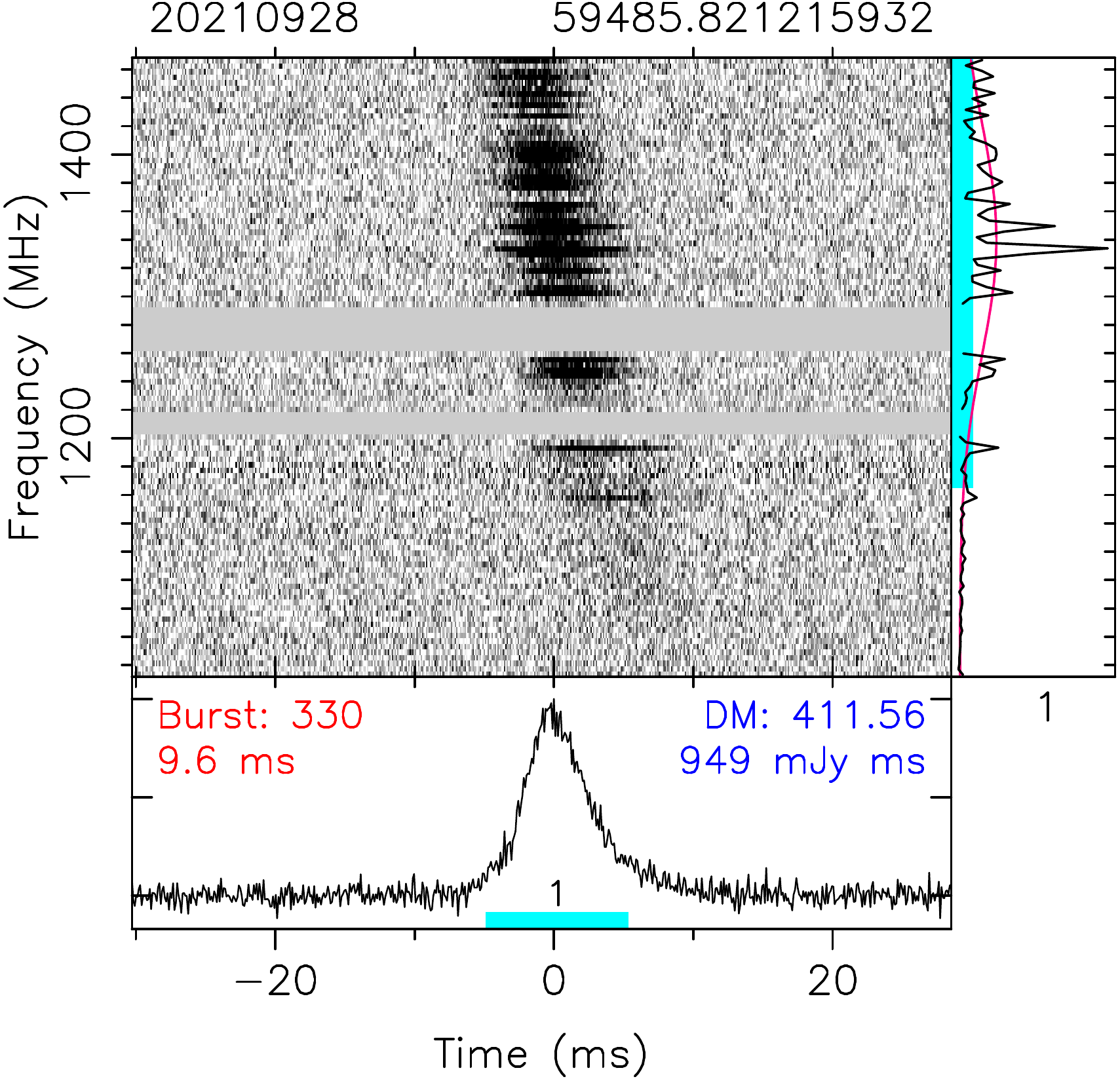}
    \includegraphics[height=37mm]{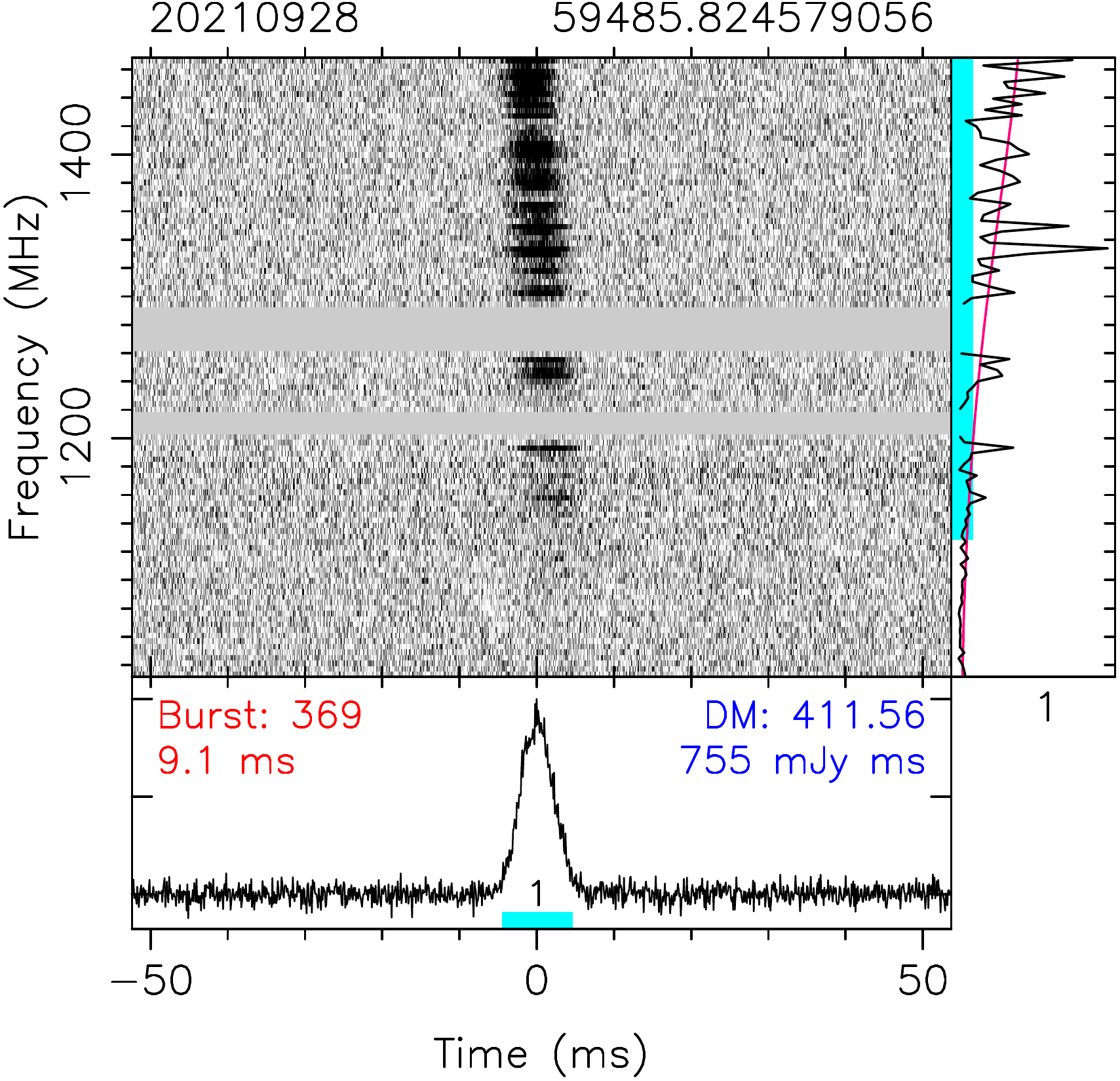}
\caption{The same as Figure~\ref{fig:appendix:D1W} but for bursts in D1-H.
}\label{fig:appendix:D1H}
\end{figure*}

We present the water-fall plot and burst profile for each burst, together with the energy distribution and the Gaussian fitting over the observational frequency. According to their morphology classification, the bursts can be found in  Figure~\ref{fig:appendix:D1W} to Figure~\ref{fig:appendix:C}.


\begin{table*}
\centering
\caption{Parameters of bursts and sub-bursts of FRB20201124A detected by the FAST in September 2021}
\label{tab:appendix}
\centering
\setlength{\tabcolsep}{6.0pt}
\footnotesize
\begin{tabular}{lcrrrrrrrl}
    \hline\noalign{\smallskip}
Burst  & TOA     &$\nu_{\rm low}$ &$\nu_{\rm 0}$&$\nu_{\rm high}$&$\rm BW_e$  & W$_{\rm sb}$ & SNR  &$F_{\nu}$          & Group  \\
No.    & (MJD)   &(MHz)        &(MHz)         &(MHz)         &(MHz)     & (ms)        &      &(mJy\,ms)      &    \\ 
(1) & (2) & (3) & (4) & (5) & (6) & (7) & (8) & (9) & (10) \\
\hline
\multicolumn{10}{c}{20210925: $ {\rm DM_{ mean}} $ = 412.38$\pm$0.93~cm$^{-3}$pc} \\
\hline
1       &59482.944753682& 1031.7& 1071.5& 1267.0&   390.8&    10.0&        20&        93& D1-L  \\
2       &59482.946971143& 1031.7&       & 1130.8&    99.1&     8.1&        52&       459& NE-L  \\
3       &59482.947246822& 1031.7&       & 1080.0&    48.3&    18.4&        15&       177& NE-L  \\
4       &59482.947360478& 1031.7&  984.7& 1098.6&   227.7&     8.6&        11&       115& NE-L  \\
\hline
\end{tabular} 
\end{table*}
\addtocounter{table}{-1}
\begin{table*}
\centering
\caption{${continued}$.}
\centering
\setlength{\tabcolsep}{6.0pt}
\footnotesize
\begin{tabular}{lcrrrrrrrl}
    \hline\noalign{\smallskip}
Burst  & TOA     &$\nu_{\rm low}$ &$\nu_{\rm 0}$&$\nu_{\rm high}$&$\rm BW_e$  & W$_{\rm sb}$ & SNR  &$F_{\nu}$          & Group  \\
No.    & (MJD)   &(MHz)        &(MHz)         &(MHz)         &(MHz)     & (ms)        &      &(mJy\,ms)      &    \\ 
(1) & (2) & (3) & (4) & (5) & (6) & (7) & (8) & (9) & (10) \\
\hline
\multicolumn{10}{c}{20210926: $ {\rm DM_{ mean}} $ = 412.3$\pm$1.6~cm$^{-3}$pc} \\
\hline
1       &59483.867443675&       &       &       &         &    36.7&        67&       633& C      \\
1-1     &59483.867443670& 1031.7& 1051.5& 1110.6&    118.2&     9.7&        30&       288&        \\
1-2     &59483.867443818& 1031.7& 1164.8& 1343.5&    357.4&     8.3&        10&        96&        \\
1-3     &59483.867443847& 1031.7& 1043.6& 1156.5&    225.8&    33.0&        26&       248&        \\
\hline
\end{tabular} 
\end{table*}
\addtocounter{table}{-1}
\begin{table*}
\centering
\caption{${continued}$.}
\centering
\setlength{\tabcolsep}{6.0pt}
\footnotesize
\begin{tabular}{lcrrrrrrrl}
    \hline\noalign{\smallskip}
Burst  & TOA     &$\nu_{\rm low}$ &$\nu_{\rm 0}$&$\nu_{\rm high}$&$\rm BW_e$  & W$_{\rm sb}$ & SNR  &$F_{\nu}$          & Group  \\
No.    & (MJD)   &(MHz)        &(MHz)         &(MHz)         &(MHz)     & (ms)        &      &(mJy\,ms)      &    \\ 
(1) & (2) & (3) & (4) & (5) & (6) & (7) & (8) & (9) & (10) \\
\hline
\multicolumn{10}{c}{20210926: $ {\rm DM_{ mean}} $ = 412.3$\pm$1.6~cm$^{-3}$pc} \\
\hline
24      &59483.881435345& 1031.7&       & 1114.2&     82.5&    15.9&       272&      4413& D1-L   \\
25      &59483.881832586&       &       &       &         &    24.7&        34&       153& Dm-W   \\
25-1    &59483.881832585& 1401.8& 1445.7& 1468.2&     87.8&     2.9&         4&        18&        \\
25-2    &59483.881832632& 1396.9& 1439.4& 1468.2&     85.0&    19.8&        25&       115&        \\
\hline
\end{tabular} 
\end{table*}
\addtocounter{table}{-1}
\begin{table*}
\centering
\caption{${continued}$.}
\centering
\setlength{\tabcolsep}{6.0pt}
\footnotesize
\begin{tabular}{lcrrrrrrrl}
    \hline\noalign{\smallskip}
Burst  & TOA     &$\nu_{\rm low}$ &$\nu_{\rm 0}$&$\nu_{\rm high}$&$\rm BW_e$  & W$_{\rm sb}$ & SNR  &$F_{\nu}$          & Group  \\
No.    & (MJD)   &(MHz)        &(MHz)         &(MHz)         &(MHz)     & (ms)        &      &(mJy\,ms)      &    \\ 
(1) & (2) & (3) & (4) & (5) & (6) & (7) & (8) & (9) & (10) \\
\hline
\multicolumn{10}{c}{20210926: $ {\rm DM_{ mean}} $ = 412.3$\pm$1.6~cm$^{-3}$pc} \\
\hline
50-1    &59483.900996833& 1290.0&       & 1468.2&    178.2&     5.2&        31&       146&        \\
50-2    &59483.900996856& 1142.3& 1573.5& 1468.2&    862.6&     8.3&        30&       141&        \\
51      &59483.901564891& 1031.7& 1087.7& 1410.4&    645.3&    11.0&        26&        64& ND     \\
52      &59483.902297238& 1031.7& 1035.3& 1100.9&    131.3&    13.6&        26&       250& NE-L   \\
\hline
\end{tabular} 
\end{table*}
\addtocounter{table}{-1}
\begin{table*}
\centering
\caption{${continued}$.}
\centering
\setlength{\tabcolsep}{6.0pt}
\footnotesize
\begin{tabular}{lcrrrrrrrl}
    \hline\noalign{\smallskip}
Burst  & TOA     &$\nu_{\rm low}$ &$\nu_{\rm 0}$&$\nu_{\rm high}$&$\rm BW_e$  & W$_{\rm sb}$ & SNR  &$F_{\nu}$          & Group  \\
No.    & (MJD)   &(MHz)        &(MHz)         &(MHz)         &(MHz)     & (ms)        &      &(mJy\,ms)      &    \\ 
(1) & (2) & (3) & (4) & (5) & (6) & (7) & (8) & (9) & (10) \\
\hline
\multicolumn{10}{c}{20210927: $ {\rm DM_{ mean}} $ = 412.5$\pm$1.9~cm$^{-3}$pc} \\
\hline
12-7    &59484.818234285& 1325.1& 1448.2& 1468.2&   246.2&     6.7&        79&      1007&        \\
12-8    &59484.818234346& 1294.7& 1434.5& 1468.2&   279.7&    14.9&       101&      1290&        \\
13      &59484.818453861& 1031.7&  975.3& 1155.4&   360.2&    16.5&        28&       121& NE-L   \\
14      &59484.818472324& 1031.7& 1052.3& 1159.1&   213.6&     8.4&        27&       199& NE-L   \\
\hline
\end{tabular} 
\end{table*}
\addtocounter{table}{-1}
\begin{table*}
\centering
\caption{${continued}$.}
\centering
\setlength{\tabcolsep}{6.0pt}
\footnotesize
\begin{tabular}{lcrrrrrrrl}
    \hline\noalign{\smallskip}
Burst  & TOA     &$\nu_{\rm low}$ &$\nu_{\rm 0}$&$\nu_{\rm high}$&$\rm BW_e$  & W$_{\rm sb}$ & SNR  &$F_{\nu}$          & Group  \\
No.    & (MJD)   &(MHz)        &(MHz)         &(MHz)         &(MHz)     & (ms)        &      &(mJy\,ms)      &    \\ 
(1) & (2) & (3) & (4) & (5) & (6) & (7) & (8) & (9) & (10) \\
\hline
\multicolumn{10}{c}{20210927: $ {\rm DM_{ mean}} $ = 412.5$\pm$1.9~cm$^{-3}$pc} \\
\hline
34-3    &59484.823628447& 1031.7& 1033.8& 1113.1&   158.5&     7.5&         9&       109&        \\
35      &59484.824306757&       &       &       &        &    60.4&       311&      3749& C      \\
35-1    &59484.824306496& 1031.7& 1121.0& 1319.7&   397.5&    12.3&        56&       679&        \\
35-2    &59484.824306710& 1218.0& 1351.7& 1468.2&   267.4&     6.8&        14&       166&        \\
\hline
\end{tabular} 
\end{table*}
\addtocounter{table}{-1}
\begin{table*}
\centering
\caption{${continued}$.}
\centering
\setlength{\tabcolsep}{6.0pt}
\footnotesize
\begin{tabular}{lcrrrrrrrl}
    \hline\noalign{\smallskip}
Burst  & TOA     &$\nu_{\rm low}$ &$\nu_{\rm 0}$&$\nu_{\rm high}$&$\rm BW_e$  & W$_{\rm sb}$ & SNR  &$F_{\nu}$          & Group  \\
No.    & (MJD)   &(MHz)        &(MHz)         &(MHz)         &(MHz)     & (ms)        &      &(mJy\,ms)      &    \\ 
(1) & (2) & (3) & (4) & (5) & (6) & (7) & (8) & (9) & (10) \\
\hline
\multicolumn{10}{c}{20210927: $ {\rm DM_{ mean}} $ = 412.5$\pm$1.9~cm$^{-3}$pc} \\
\hline
55-3    &59484.831110409& 1031.7& 1029.7& 1256.2&   452.9&    32.4&        27&       340&        \\
55-4    &59484.831110728& 1031.7& 1016.7& 1088.4&   143.4&     9.8&        12&       145&        \\
56      &59484.831273419&       &       &       &        &    37.8&       258&      2233& C      \\
56-1    &59484.831273084& 1031.7& 1110.8& 1260.3&   299.0&    11.2&        73&       629&        \\
\hline
\end{tabular} 
\end{table*}
\addtocounter{table}{-1}
\begin{table*}
\centering
\caption{${continued}$.}
\centering
\setlength{\tabcolsep}{6.0pt}
\footnotesize
\begin{tabular}{lcrrrrrrrl}
    \hline\noalign{\smallskip}
Burst  & TOA     &$\nu_{\rm low}$ &$\nu_{\rm 0}$&$\nu_{\rm high}$&$\rm BW_e$  & W$_{\rm sb}$ & SNR  &$F_{\nu}$          & Group  \\
No.    & (MJD)   &(MHz)        &(MHz)         &(MHz)         &(MHz)     & (ms)        &      &(mJy\,ms)      &    \\ 
(1) & (2) & (3) & (4) & (5) & (6) & (7) & (8) & (9) & (10) \\
\hline
\multicolumn{10}{c}{20210927: $ {\rm DM_{ mean}} $ = 412.5$\pm$1.9~cm$^{-3}$pc} \\
\hline
74-2    &59484.835442394& 1031.7&  883.6& 1042.3&   317.4&     7.9&        23&       233&        \\
75      &59484.835572558&       &       &       &        &    23.3&       276&      2291& Dm-W   \\
75-1    &59484.835572507& 1283.1&       & 1468.2&   185.1&    11.8&       104&       867&        \\
75-2    &59484.835572554& 1163.1& 1356.7& 1468.2&   387.2&     4.5&        27&       221&        \\
\hline
\end{tabular} 
\end{table*}
\addtocounter{table}{-1}
\begin{table*}
\centering
\caption{${continued}$.}
\centering
\setlength{\tabcolsep}{6.0pt}
\footnotesize
\begin{tabular}{lcrrrrrrrl}
    \hline\noalign{\smallskip}
Burst  & TOA     &$\nu_{\rm low}$ &$\nu_{\rm 0}$&$\nu_{\rm high}$&$\rm BW_e$  & W$_{\rm sb}$ & SNR  &$F_{\nu}$          & Group  \\
No.    & (MJD)   &(MHz)        &(MHz)         &(MHz)         &(MHz)     & (ms)        &      &(mJy\,ms)      &    \\ 
(1) & (2) & (3) & (4) & (5) & (6) & (7) & (8) & (9) & (10) \\
\hline
\multicolumn{10}{c}{20210927: $ {\rm DM_{ mean}} $ = 412.5$\pm$1.9~cm$^{-3}$pc} \\
\hline
94$\_$A2&59484.839509421& 1031.7& 1037.0& 1079.2&    84.4&    14.3&         8&       197&        \\
94$\_$B &59484.839510100&       &       &       &        &    16.3&       171&      1429&        \\
94$\_$B1&59484.839509714& 1062.7& 1216.0& 1369.4&   306.7&     4.8&         5&        45&        \\
94$\_$B2&59484.839510080& 1031.7&  872.6& 1152.1&   559.1&     7.7&        89&       745&        \\
\hline
\end{tabular} 
\end{table*}
\addtocounter{table}{-1}
\begin{table*}
\centering
\caption{${continued}$.}
\centering
\setlength{\tabcolsep}{6.0pt}
\footnotesize
\begin{tabular}{lcrrrrrrrl}
    \hline\noalign{\smallskip}
Burst  & TOA     &$\nu_{\rm low}$ &$\nu_{\rm 0}$&$\nu_{\rm high}$&$\rm BW_e$  & W$_{\rm sb}$ & SNR  &$F_{\nu}$          & Group  \\
No.    & (MJD)   &(MHz)        &(MHz)         &(MHz)         &(MHz)     & (ms)        &      &(mJy\,ms)      &    \\ 
(1) & (2) & (3) & (4) & (5) & (6) & (7) & (8) & (9) & (10) \\
\hline
\multicolumn{10}{c}{20210927: $ {\rm DM_{ mean}} $ = 412.5$\pm$1.9~cm$^{-3}$pc} \\
\hline
113$\_$A12&59484.842785907& 1250.4& 1394.5& 1468.2&   288.3&    15.1&        35&       652&      \\
113$\_$B  &59484.842787663&       &       &       &        &    52.5&       696&     16055&      \\
113$\_$B1 &59484.842787161& 1038.9& 1177.5& 1316.0&   277.1&    29.4&        80&      1840&      \\
113$\_$B2 &59484.842787243& 1053.3& 1177.6& 1301.8&   248.5&     8.3&        78&      1802&      \\
\hline
\end{tabular} 
\end{table*}
\addtocounter{table}{-1}
\begin{table*}
\centering
\caption{${continued}$.}
\centering
\setlength{\tabcolsep}{6.0pt}
\footnotesize
\begin{tabular}{lcrrrrrrrl}
    \hline\noalign{\smallskip}
Burst  & TOA     &$\nu_{\rm low}$ &$\nu_{\rm 0}$&$\nu_{\rm high}$&$\rm BW_e$  & W$_{\rm sb}$ & SNR  &$F_{\nu}$          & Group  \\
No.    & (MJD)   &(MHz)        &(MHz)         &(MHz)         &(MHz)     & (ms)        &      &(mJy\,ms)      &    \\ 
(1) & (2) & (3) & (4) & (5) & (6) & (7) & (8) & (9) & (10) \\
\hline
\multicolumn{10}{c}{20210927: $ {\rm DM_{ mean}} $ = 412.5$\pm$1.9~cm$^{-3}$pc} \\
\hline
129$\_$D &59484.845239907& 1032.1& 1044.3& 1056.6&    24.5&    15.0&         8&       205&        \\
129$\_$E &59484.845241875& 1031.7&       & 1464.8&   433.1&    10.4&        21&       117&        \\
129$\_$F &59484.845242388& 1031.7&       & 1173.8&   142.1&    10.2&        36&       311&        \\
130     &59484.845445281& 1200.5& 1400.3& 1468.2&   399.6&     5.5&        15&        55& NE-H   \\
\hline
\end{tabular} 
\end{table*}
\addtocounter{table}{-1}
\begin{table*}
\centering
\caption{${continued}$.}
\centering
\setlength{\tabcolsep}{6.0pt}
\footnotesize
\begin{tabular}{lcrrrrrrrl}
    \hline\noalign{\smallskip}
Burst  & TOA     &$\nu_{\rm low}$ &$\nu_{\rm 0}$&$\nu_{\rm high}$&$\rm BW_e$  & W$_{\rm sb}$ & SNR  &$F_{\nu}$          & Group  \\
No.    & (MJD)   &(MHz)        &(MHz)         &(MHz)         &(MHz)     & (ms)        &      &(mJy\,ms)      &    \\ 
(1) & (2) & (3) & (4) & (5) & (6) & (7) & (8) & (9) & (10) \\
\hline
\multicolumn{10}{c}{20210927: $ {\rm DM_{ mean}} $ = 412.5$\pm$1.9~cm$^{-3}$pc} \\
\hline
152-1   &59484.850811546& 1031.7& 1070.1& 1208.7&   277.3&     5.3&       601&      6124&        \\
152-2   &59484.850811582& 1031.7& 1023.5& 1203.8&   360.7&     7.4&        68&       691&        \\
152-3   &59484.850811624& 1031.7&       & 1137.6&   106.0&    12.5&        65&       662&        \\
153     &59484.851114302&       &       &       &        &    67.2&       729&     11794& C      \\
\hline
\end{tabular} 
\end{table*}
\addtocounter{table}{-1}
\begin{table*}
\centering
\caption{${continued}$.}
\centering
\setlength{\tabcolsep}{6.0pt}
\footnotesize
\begin{tabular}{lcrrrrrrrl}
    \hline\noalign{\smallskip}
Burst  & TOA     &$\nu_{\rm low}$ &$\nu_{\rm 0}$&$\nu_{\rm high}$&$\rm BW_e$  & W$_{\rm sb}$ & SNR  &$F_{\nu}$          & Group  \\
No.    & (MJD)   &(MHz)        &(MHz)         &(MHz)         &(MHz)     & (ms)        &      &(mJy\,ms)      &    \\ 
(1) & (2) & (3) & (4) & (5) & (6) & (7) & (8) & (9) & (10) \\
\hline
\multicolumn{10}{c}{20210928: $ {\rm DM_{ mean}} $ = 411.56$\pm$1.9~cm$^{-3}$pc} \\
\hline
1       &59485.784451366& 1031.7&       & 1103.5&    71.8&    12.1&        22&       181& NE-L   \\
2       &59485.784856798& 1031.7& 1074.9& 1220.9&   292.0&    13.8&        15&        75& D1-L   \\
3       &59485.785071425&       &       &       &        &    57.4&       221&      3316& C      \\
3-1     &59485.785070219& 1031.7& 1150.0& 1323.6&   347.2&    10.6&        37&       560&        \\
\hline
\end{tabular} 
\end{table*}
\addtocounter{table}{-1}
\begin{table*}
\centering
\caption{${continued}$.}
\centering
\setlength{\tabcolsep}{6.0pt}
\footnotesize
\begin{tabular}{lcrrrrrrrl}
    \hline\noalign{\smallskip}
Burst  & TOA     &$\nu_{\rm low}$ &$\nu_{\rm 0}$&$\nu_{\rm high}$&$\rm BW_e$  & W$_{\rm sb}$ & SNR  &$F_{\nu}$          & Group  \\
No.    & (MJD)   &(MHz)        &(MHz)         &(MHz)         &(MHz)     & (ms)        &      &(mJy\,ms)      &    \\ 
(1) & (2) & (3) & (4) & (5) & (6) & (7) & (8) & (9) & (10) \\
\hline
\multicolumn{10}{c}{20210928: $ {\rm DM_{ mean}} $ = 411.56$\pm$1.9~cm$^{-3}$pc} \\
\hline
23-1    &59485.787444292& 1031.7& 1116.5& 1220.7&   208.5&    16.7&         8&        90&        \\
23-2    &59485.787444709& 1031.7& 1099.2& 1201.7&   205.0&    14.3&         7&        81&        \\
23-3    &59485.787445463& 1124.7& 1213.3& 1301.8&   177.1&     4.5&         5&        50&        \\
23-4    &59485.787445645& 1031.7& 1053.4& 1160.9&   215.0&     7.2&         8&        92&        \\
\hline
\end{tabular} 
\end{table*}
\addtocounter{table}{-1}
\begin{table*}
\centering
\caption{${continued}$.}
\centering
\setlength{\tabcolsep}{6.0pt}
\footnotesize
\begin{tabular}{lcrrrrrrrl}
    \hline\noalign{\smallskip}
Burst  & TOA     &$\nu_{\rm low}$ &$\nu_{\rm 0}$&$\nu_{\rm high}$&$\rm BW_e$  & W$_{\rm sb}$ & SNR  &$F_{\nu}$          & Group  \\
No.    & (MJD)   &(MHz)        &(MHz)         &(MHz)         &(MHz)     & (ms)        &      &(mJy\,ms)      &    \\ 
(1) & (2) & (3) & (4) & (5) & (6) & (7) & (8) & (9) & (10) \\
\hline
\multicolumn{10}{c}{20210928: $ {\rm DM_{ mean}} $ = 411.56$\pm$1.9~cm$^{-3}$pc} \\
\hline
48      &59485.790191086& 1031.7& 1104.1& 1270.7&   333.2&    10.0&        63&       420& D1-L   \\
49      &59485.790287948&       &       &       &        &    43.2&       205&      1997& C      \\
49-1    &59485.790287623& 1216.1& 1358.0& 1468.2&   283.8&    16.2&        23&       228&        \\
49-2    &59485.790287864& 1196.8& 1322.8& 1448.8&   252.0&    11.0&        25&       241&        \\
\hline
\end{tabular} 
\end{table*}
\addtocounter{table}{-1}
\begin{table*}
\centering
\caption{${continued}$.}
\centering
\setlength{\tabcolsep}{6.0pt}
\footnotesize
\begin{tabular}{lcrrrrrrrl}
    \hline\noalign{\smallskip}
Burst  & TOA     &$\nu_{\rm low}$ &$\nu_{\rm 0}$&$\nu_{\rm high}$&$\rm BW_e$  & W$_{\rm sb}$ & SNR  &$F_{\nu}$          & Group  \\
No.    & (MJD)   &(MHz)        &(MHz)         &(MHz)         &(MHz)     & (ms)        &      &(mJy\,ms)      &    \\ 
(1) & (2) & (3) & (4) & (5) & (6) & (7) & (8) & (9) & (10) \\
\hline
\multicolumn{10}{c}{20210928: $ {\rm DM_{ mean}} $ = 411.56$\pm$1.9~cm$^{-3}$pc} \\
\hline
68-3    &59485.792505550& 1451.8& 1457.6& 1463.3&    11.6&     3.7&         5&        47&        \\
68-4    &59485.792505615& 1329.0&       & 1468.2&   139.2&     8.6&        57&       578&        \\
68-5    &59485.792505747& 1209.1& 1592.2& 1468.2&   766.3&     7.6&       124&      1258&        \\
68-6    &59485.792505806& 1031.7& 1155.6& 1468.2&   625.1&     7.0&        15&       152&        \\
\hline
\end{tabular} 
\end{table*}
\addtocounter{table}{-1}
\begin{table*}
\centering
\caption{${continued}$.}
\centering
\setlength{\tabcolsep}{6.0pt}
\footnotesize
\begin{tabular}{lcrrrrrrrl}
    \hline\noalign{\smallskip}
Burst  & TOA     &$\nu_{\rm low}$ &$\nu_{\rm 0}$&$\nu_{\rm high}$&$\rm BW_e$  & W$_{\rm sb}$ & SNR  &$F_{\nu}$          & Group  \\
No.    & (MJD)   &(MHz)        &(MHz)         &(MHz)         &(MHz)     & (ms)        &      &(mJy\,ms)      &    \\ 
(1) & (2) & (3) & (4) & (5) & (6) & (7) & (8) & (9) & (10) \\
\hline
\multicolumn{10}{c}{20210928: $ {\rm DM_{ mean}} $ = 411.56$\pm$1.9~cm$^{-3}$pc} \\
\hline
92-1    &59485.796146534& 1031.7& 1123.3& 1283.7&   320.8&     9.5&        67&       984&        \\
92-2    &59485.796146674& 1031.7& 1016.9& 1123.4&   213.0&    26.1&        26&       389&        \\
92-3    &59485.796147529& 1051.0& 1185.8& 1320.7&   269.8&    14.1&        64&       936&        \\
92-4    &59485.796147623& 1031.7& 1141.9& 1252.5&   221.2&    12.1&       224&      3289&        \\
\hline
\end{tabular} 
\end{table*}
\addtocounter{table}{-1}
\begin{table*}
\centering
\caption{${continued}$.}
\centering
\setlength{\tabcolsep}{6.0pt}
\footnotesize
\begin{tabular}{lcrrrrrrrl}
    \hline\noalign{\smallskip}
Burst  & TOA     &$\nu_{\rm low}$ &$\nu_{\rm 0}$&$\nu_{\rm high}$&$\rm BW_e$  & W$_{\rm sb}$ & SNR  &$F_{\nu}$          & Group  \\
No.    & (MJD)   &(MHz)        &(MHz)         &(MHz)         &(MHz)     & (ms)        &      &(mJy\,ms)      &    \\ 
(1) & (2) & (3) & (4) & (5) & (6) & (7) & (8) & (9) & (10) \\
\hline
\multicolumn{10}{c}{20210928: $ {\rm DM_{ mean}} $ = 411.56$\pm$1.9~cm$^{-3}$pc} \\
\hline
111-10  &59485.797944123& 1151.2& 1271.3& 1391.4&   240.2&     6.2&         7&        90&        \\
111-11  &59485.797944828& 1065.7& 1162.6& 1259.5&   193.8&     6.8&         3&        38&        \\
111-12  &59485.797945071& 1031.7& 1049.2& 1159.1&   219.9&    21.1&        32&       406&        \\
112     &59485.797999518&       &       &       &        &    21.7&        62&       395& C      \\
\hline
\end{tabular} 
\end{table*}
\addtocounter{table}{-1}
\begin{table*}
\centering
\caption{${continued}$.}
\centering
\setlength{\tabcolsep}{6.0pt}
\footnotesize
\begin{tabular}{lcrrrrrrrl}
    \hline\noalign{\smallskip}
Burst  & TOA     &$\nu_{\rm low}$ &$\nu_{\rm 0}$&$\nu_{\rm high}$&$\rm BW_e$  & W$_{\rm sb}$ & SNR  &$F_{\nu}$          & Group  \\
No.    & (MJD)   &(MHz)        &(MHz)         &(MHz)         &(MHz)     & (ms)        &      &(mJy\,ms)      &    \\ 
(1) & (2) & (3) & (4) & (5) & (6) & (7) & (8) & (9) & (10) \\
\hline
\multicolumn{10}{c}{20210928: $ {\rm DM_{ mean}} $ = 411.56$\pm$1.9~cm$^{-3}$pc} \\
\hline
139-4   &59485.801319175& 1111.0& 1263.3& 1415.6&   304.6&    10.2&        37&       359&        \\
139-5   &59485.801319314& 1031.7& 1077.3& 1189.0&   223.5&    13.6&        38&       366&        \\
139-6   &59485.801322622& 1031.7& 1054.0& 1159.7&   211.5&     9.9&        18&       179&        \\
140     &59485.801453000& 1031.7&  989.2& 1468.2&  1132.3&    10.5&        27&       110& D1-W   \\
\hline
\end{tabular} 
\end{table*}
\addtocounter{table}{-1}
\begin{table*}
\centering
\caption{${continued}$.}
\centering
\setlength{\tabcolsep}{6.0pt}
\footnotesize
\begin{tabular}{lcrrrrrrrl}
    \hline\noalign{\smallskip}
Burst  & TOA     &$\nu_{\rm low}$ &$\nu_{\rm 0}$&$\nu_{\rm high}$&$\rm BW_e$  & W$_{\rm sb}$ & SNR  &$F_{\nu}$          & Group  \\
No.    & (MJD)   &(MHz)        &(MHz)         &(MHz)         &(MHz)     & (ms)        &      &(mJy\,ms)      &    \\ 
(1) & (2) & (3) & (4) & (5) & (6) & (7) & (8) & (9) & (10) \\
\hline
\multicolumn{10}{c}{20210928: $ {\rm DM_{ mean}} $ = 411.56$\pm$1.9~cm$^{-3}$pc} \\
\hline
158     &59485.802911643&       &       &       &        &    19.4&        19&       212& C      \\
158-1   &59485.802911500& 1031.7&  921.9& 1133.1&   422.4&     9.3&         7&        81&        \\
158-2   &59485.802911640& 1031.7&  849.5& 1045.7&   392.3&    10.0&        12&       131&        \\
159     &59485.803134283&       &       &       &        &    37.3&       227&      2144& C      \\
\hline
\end{tabular} 
\end{table*}
\addtocounter{table}{-1}
\begin{table*}
\centering
\caption{${continued}$.}
\centering
\setlength{\tabcolsep}{6.0pt}
\footnotesize
\begin{tabular}{lcrrrrrrrl}
    \hline\noalign{\smallskip}
Burst  & TOA     &$\nu_{\rm low}$ &$\nu_{\rm 0}$&$\nu_{\rm high}$&$\rm BW_e$  & W$_{\rm sb}$ & SNR  &$F_{\nu}$          & Group  \\
No.    & (MJD)   &(MHz)        &(MHz)         &(MHz)         &(MHz)     & (ms)        &      &(mJy\,ms)      &    \\ 
(1) & (2) & (3) & (4) & (5) & (6) & (7) & (8) & (9) & (10) \\
\hline
\multicolumn{10}{c}{20210928: $ {\rm DM_{ mean}} $ = 411.56$\pm$1.9~cm$^{-3}$pc} \\
\hline
178-5   &59485.805052796& 1031.7& 1173.2& 1321.9&   297.4&    11.4&        79&       887&        \\
179     &59485.805159833& 1031.7& 1165.0& 1303.5&   277.0&    13.3&       166&      1457& D1-M   \\
180     &59485.805174373& 1031.7& 1092.6& 1203.3&   221.4&     7.6&        23&       163& D1-L   \\
181     &59485.805257276&       &       &       &        &    34.3&        83&       698& C      \\
\hline
\end{tabular} 
\end{table*}
\addtocounter{table}{-1}
\begin{table*}
\centering
\caption{${continued}$.}
\centering
\setlength{\tabcolsep}{6.0pt}
\footnotesize
\begin{tabular}{lcrrrrrrrl}
    \hline\noalign{\smallskip}
Burst  & TOA     &$\nu_{\rm low}$ &$\nu_{\rm 0}$&$\nu_{\rm high}$&$\rm BW_e$  & W$_{\rm sb}$ & SNR  &$F_{\nu}$          & Group  \\
No.    & (MJD)   &(MHz)        &(MHz)         &(MHz)         &(MHz)     & (ms)        &      &(mJy\,ms)      &    \\ 
(1) & (2) & (3) & (4) & (5) & (6) & (7) & (8) & (9) & (10) \\
\hline
\multicolumn{10}{c}{20210928: $ {\rm DM_{ mean}} $ = 411.56$\pm$1.9~cm$^{-3}$pc} \\
\hline
199     &59485.807660207& 1345.6&       & 1468.2&   122.6&     9.3&        18&       120& NE-H   \\
200     &59485.807673849&       &       &       &        &    13.7&        51&       297& Dm-H   \\
200-1   &59485.807673803& 1380.8&       & 1468.2&    87.4&     5.5&        12&        70&        \\
200-2   &59485.807673845& 1251.9&       & 1468.2&   216.3&     6.6&        34&       197&        \\
\hline
\end{tabular} 
\end{table*}
\addtocounter{table}{-1}
\begin{table*}
\centering
\caption{${continued}$.}
\centering
\setlength{\tabcolsep}{6.0pt}
\footnotesize
\begin{tabular}{lcrrrrrrrl}
    \hline\noalign{\smallskip}
Burst  & TOA     &$\nu_{\rm low}$ &$\nu_{\rm 0}$&$\nu_{\rm high}$&$\rm BW_e$  & W$_{\rm sb}$ & SNR  &$F_{\nu}$          & Group  \\
No.    & (MJD)   &(MHz)        &(MHz)         &(MHz)         &(MHz)     & (ms)        &      &(mJy\,ms)      &    \\ 
(1) & (2) & (3) & (4) & (5) & (6) & (7) & (8) & (9) & (10) \\
\hline
\multicolumn{10}{c}{20210928: $ {\rm DM_{ mean}} $ = 411.56$\pm$1.9~cm$^{-3}$pc} \\
\hline
224-3   &59485.809130189& 1221.3& 1421.7& 1468.2&   400.7&     4.2&         9&       103&        \\
224-4   &59485.809130237& 1227.8& 1415.0& 1468.2&   374.3&     7.4&        30&       355&        \\
224-5   &59485.809130290& 1140.3& 1326.4& 1468.2&   372.3&     9.3&        23&       270&        \\
224-6   &59485.809130391& 1031.7& 1132.2& 1306.8&   349.2&    20.1&        13&       154&        \\
\hline
\end{tabular} 
\end{table*}
\addtocounter{table}{-1}
\begin{table*}
\centering
\caption{${continued}$.}
\centering
\setlength{\tabcolsep}{6.0pt}
\footnotesize
\begin{tabular}{lcrrrrrrrl}
    \hline\noalign{\smallskip}
Burst  & TOA     &$\nu_{\rm low}$ &$\nu_{\rm 0}$&$\nu_{\rm high}$&$\rm BW_e$  & W$_{\rm sb}$ & SNR  &$F_{\nu}$          & Group  \\
No.    & (MJD)   &(MHz)        &(MHz)         &(MHz)         &(MHz)     & (ms)        &      &(mJy\,ms)      &    \\ 
(1) & (2) & (3) & (4) & (5) & (6) & (7) & (8) & (9) & (10) \\
\hline
\multicolumn{10}{c}{20210928: $ {\rm DM_{ mean}} $ = 411.56$\pm$1.9~cm$^{-3}$pc} \\
\hline
243-4   &59485.811097721& 1068.9& 1290.3& 1468.2&   442.9&    10.7&       186&      2035&        \\
243-5   &59485.811097807& 1031.7& 1137.2& 1278.7&   283.0&     6.4&        62&       680&        \\
243-6   &59485.811097812& 1031.7& 1149.6& 1335.5&   371.9&    27.4&       108&      1185&        \\
243-7   &59485.811098122& 1053.2& 1237.2& 1421.2&   368.0&    15.6&        11&       123&        \\
\hline
\end{tabular} 
\end{table*}
\addtocounter{table}{-1}
\begin{table*}
\centering
\caption{${continued}$.}
\centering
\setlength{\tabcolsep}{6.0pt}
\footnotesize
\begin{tabular}{lcrrrrrrrl}
    \hline\noalign{\smallskip}
Burst  & TOA     &$\nu_{\rm low}$ &$\nu_{\rm 0}$&$\nu_{\rm high}$&$\rm BW_e$  & W$_{\rm sb}$ & SNR  &$F_{\nu}$          & Group  \\
No.    & (MJD)   &(MHz)        &(MHz)         &(MHz)         &(MHz)     & (ms)        &      &(mJy\,ms)      &    \\ 
(1) & (2) & (3) & (4) & (5) & (6) & (7) & (8) & (9) & (10) \\
\hline
\multicolumn{10}{c}{20210928: $ {\rm DM_{ mean}} $ = 411.56$\pm$1.9~cm$^{-3}$pc} \\
\hline
273-1   &59485.814026355& 1031.7& 1192.3& 1373.7&   362.8&     8.4&        35&       276&        \\
273-2   &59485.814026437& 1031.7& 1154.1& 1291.9&   275.5&    12.1&       104&       817& NE-L   \\
274     &59485.814139859& 1031.7&       & 1138.6&   106.9&     7.1&        47&       387& NE-L   \\
275     &59485.814193864& 1031.7& 1217.6& 1426.1&   417.1&     7.3&        64&       343& D1-M   \\
\hline
\end{tabular} 
\end{table*}
\addtocounter{table}{-1}
\begin{table*}
\centering
\caption{${continued}$.}
\centering
\setlength{\tabcolsep}{6.0pt}
\footnotesize
\begin{tabular}{lcrrrrrrrl}
    \hline\noalign{\smallskip}
Burst  & TOA     &$\nu_{\rm low}$ &$\nu_{\rm 0}$&$\nu_{\rm high}$&$\rm BW_e$  & W$_{\rm sb}$ & SNR  &$F_{\nu}$          & Group  \\
No.    & (MJD)   &(MHz)        &(MHz)         &(MHz)         &(MHz)     & (ms)        &      &(mJy\,ms)      &    \\ 
(1) & (2) & (3) & (4) & (5) & (6) & (7) & (8) & (9) & (10) \\
\hline
\multicolumn{10}{c}{20210928: $ {\rm DM_{ mean}} $ = 411.56$\pm$1.9~cm$^{-3}$pc} \\
\hline
295     &59485.816316190& 1031.7& 1401.8& 1468.2&   746.9&     8.8&       177&       947& D1-W   \\
296     &59485.816469076&       &       &       &        &    26.1&       686&      5926& Dm-W   \\
296-1   &59485.816469075& 1252.1& 1400.9& 1468.2&   297.6&     6.1&       219&      1890&        \\
296-2   &59485.816469138& 1117.8& 1277.3& 1436.8&   319.0&    10.5&       281&      2430&        \\
\hline
\end{tabular} 
\end{table*}
\addtocounter{table}{-1}
\begin{table*}
\centering
\caption{${continued}$.}
\centering
\setlength{\tabcolsep}{6.0pt}
\footnotesize
\begin{tabular}{lcrrrrrrrl}
    \hline\noalign{\smallskip}
Burst  & TOA     &$\nu_{\rm low}$ &$\nu_{\rm 0}$&$\nu_{\rm high}$&$\rm BW_e$  & W$_{\rm sb}$ & SNR  &$F_{\nu}$          & Group  \\
No.    & (MJD)   &(MHz)        &(MHz)         &(MHz)         &(MHz)     & (ms)        &      &(mJy\,ms)      &    \\ 
(1) & (2) & (3) & (4) & (5) & (6) & (7) & (8) & (9) & (10) \\
\hline
\multicolumn{10}{c}{20210928: $ {\rm DM_{ mean}} $ = 411.56$\pm$1.9~cm$^{-3}$pc} \\
\hline
320-2   &59485.820385914& 1189.9& 1323.5& 1457.1&   267.2&     7.8&        96&      1002&        \\
320-3   &59485.820385959& 1124.7& 1285.5& 1446.4&   321.6&     7.1&        72&       744&        \\
320-4   &59485.820386016& 1051.5& 1223.8& 1396.1&   344.7&    17.6&       115&      1195&        \\
320-5   &59485.820386250& 1031.7& 1087.9& 1194.7&   213.7&     7.6&        52&       542&        \\
\hline
\end{tabular} 
\end{table*}
\addtocounter{table}{-1}
\begin{table*}
\centering
\caption{${continued}$.}
\centering
\setlength{\tabcolsep}{6.0pt}
\footnotesize
\begin{tabular}{lcrrrrrrrl}
    \hline\noalign{\smallskip}
Burst  & TOA     &$\nu_{\rm low}$ &$\nu_{\rm 0}$&$\nu_{\rm high}$&$\rm BW_e$  & W$_{\rm sb}$ & SNR  &$F_{\nu}$          & Group  \\
No.    & (MJD)   &(MHz)        &(MHz)         &(MHz)         &(MHz)     & (ms)        &      &(mJy\,ms)      &    \\ 
(1) & (2) & (3) & (4) & (5) & (6) & (7) & (8) & (9) & (10) \\
\hline
\multicolumn{10}{c}{20210928: $ {\rm DM_{ mean}} $ = 411.56$\pm$1.9~cm$^{-3}$pc} \\
\hline
340     &59485.821889646&       &       &       &        &    18.8&        32&       168& C      \\
340-1   &59485.821888484& 1373.0&       & 1468.2&    95.2&     3.0&         3&        14&        \\
340-2   &59485.821888509& 1302.7&       & 1468.2&   165.5&     9.7&        10&        53&        \\
340-3   &59485.821889647& 1031.7& 1044.6& 1237.8&   386.4&     9.1&        19&       102&        \\
\hline
\end{tabular} 
\end{table*}
\addtocounter{table}{-1}
\begin{table*}
\centering
\caption{${continued}$.}
\centering
\setlength{\tabcolsep}{6.0pt}
\footnotesize
\begin{tabular}{lcrrrrrrrl}
    \hline\noalign{\smallskip}
Burst  & TOA     &$\nu_{\rm low}$ &$\nu_{\rm 0}$&$\nu_{\rm high}$&$\rm BW_e$  & W$_{\rm sb}$ & SNR  &$F_{\nu}$          & Group  \\
No.    & (MJD)   &(MHz)        &(MHz)         &(MHz)         &(MHz)     & (ms)        &      &(mJy\,ms)      &    \\ 
(1) & (2) & (3) & (4) & (5) & (6) & (7) & (8) & (9) & (10) \\
\hline
\multicolumn{10}{c}{20210928: $ {\rm DM_{ mean}} $ = 411.56$\pm$1.9~cm$^{-3}$pc} \\
\hline
362-4   &59485.823667259& 1081.9& 1199.2& 1316.5&   234.6&    10.2&         8&        68&        \\
363     &59485.823856586& 1314.4&       & 1468.2&   153.8&     3.4&        58&       320& ND     \\
364     &59485.824029381&       &       &       &        &    14.0&       147&       914& Dm-H   \\
364-1   &59485.824029380& 1291.0&       & 1468.2&   177.2&     4.8&        70&       436&        \\
\hline
\end{tabular} 
\end{table*}
\clearpage

\begin{figure*}
    \flushleft
    \includegraphics[height=37mm]{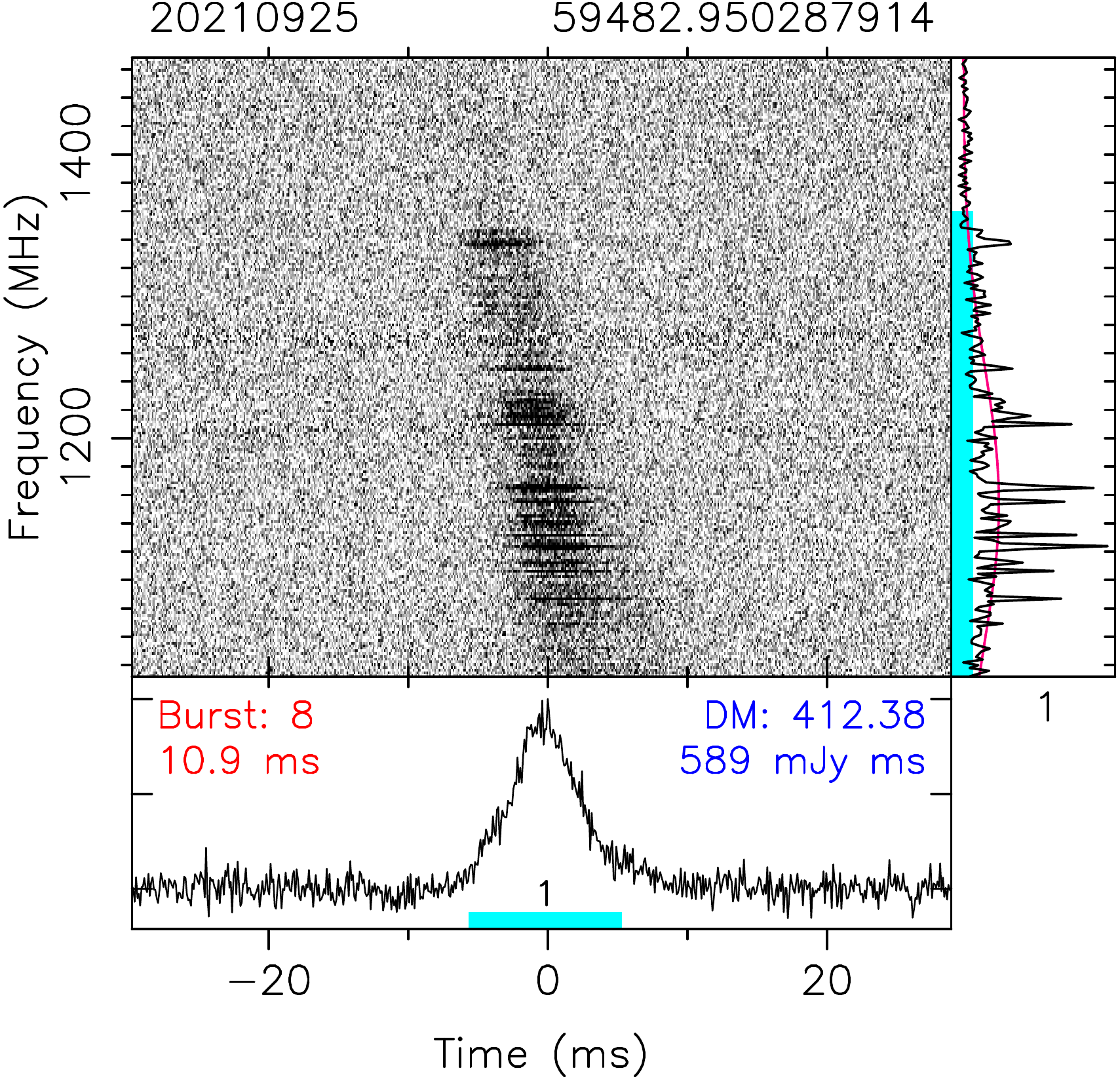}
    \includegraphics[height=37mm]{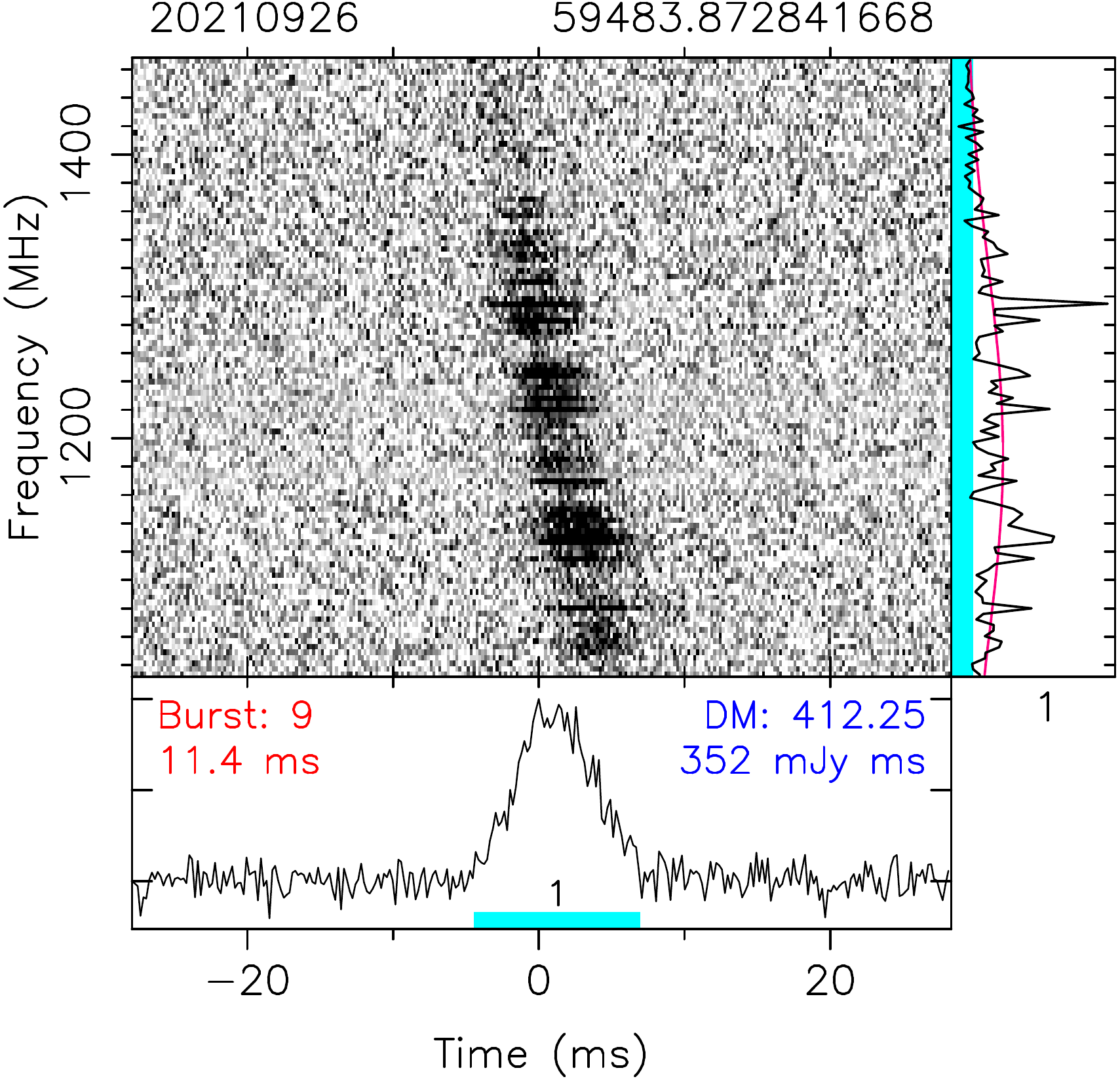}
    \includegraphics[height=37mm]{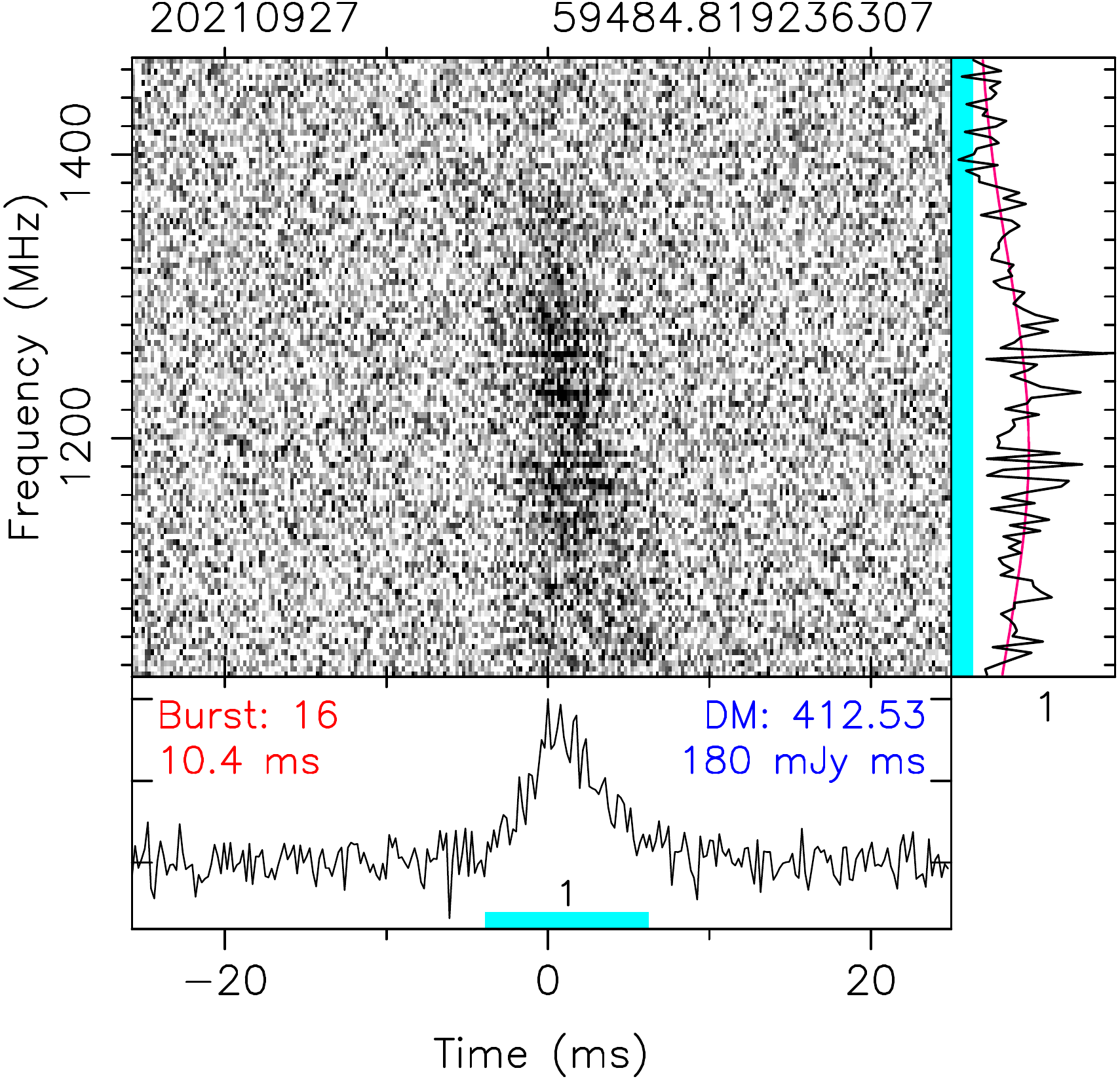}
    \includegraphics[height=37mm]{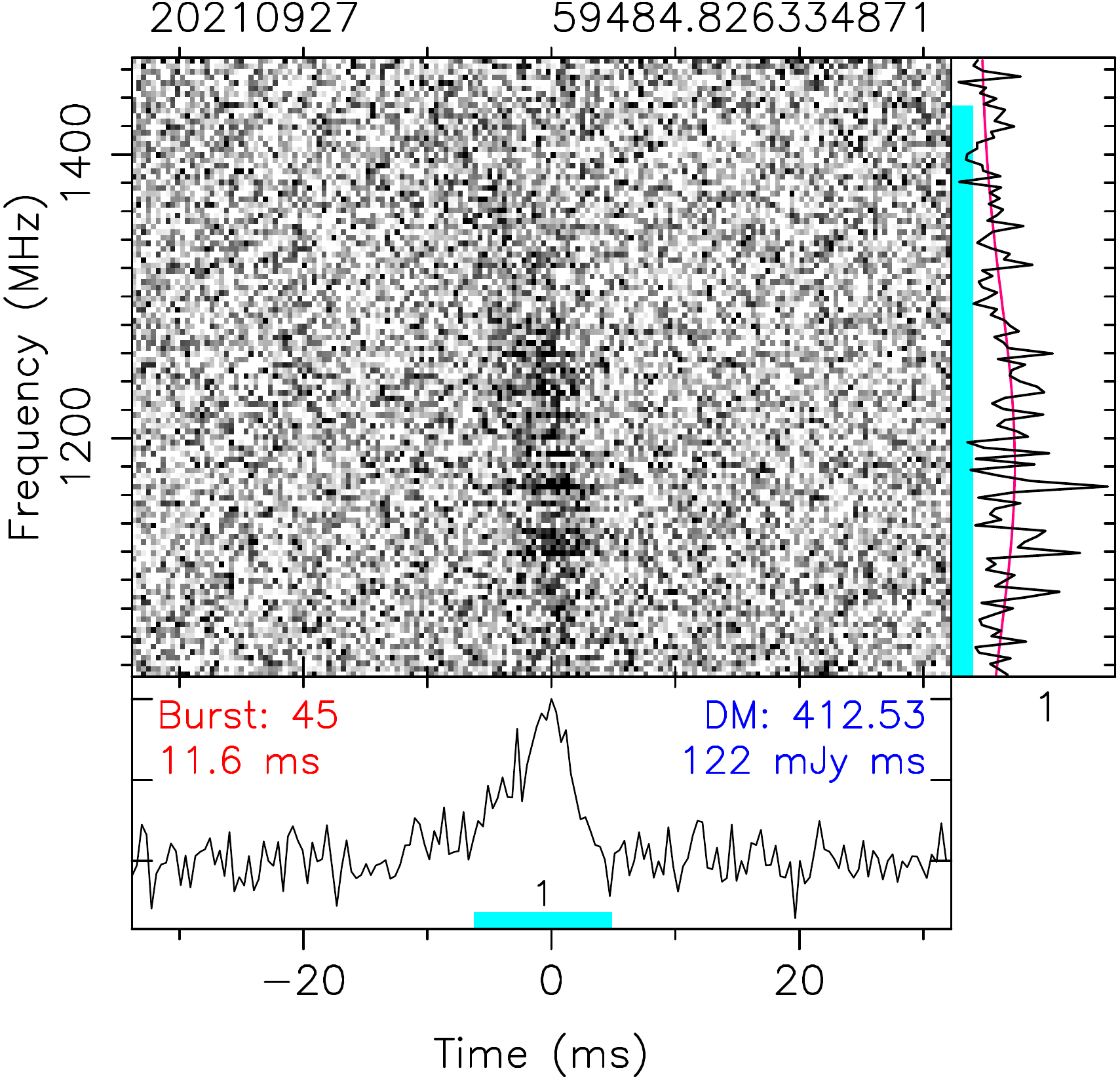}
    \includegraphics[height=37mm]{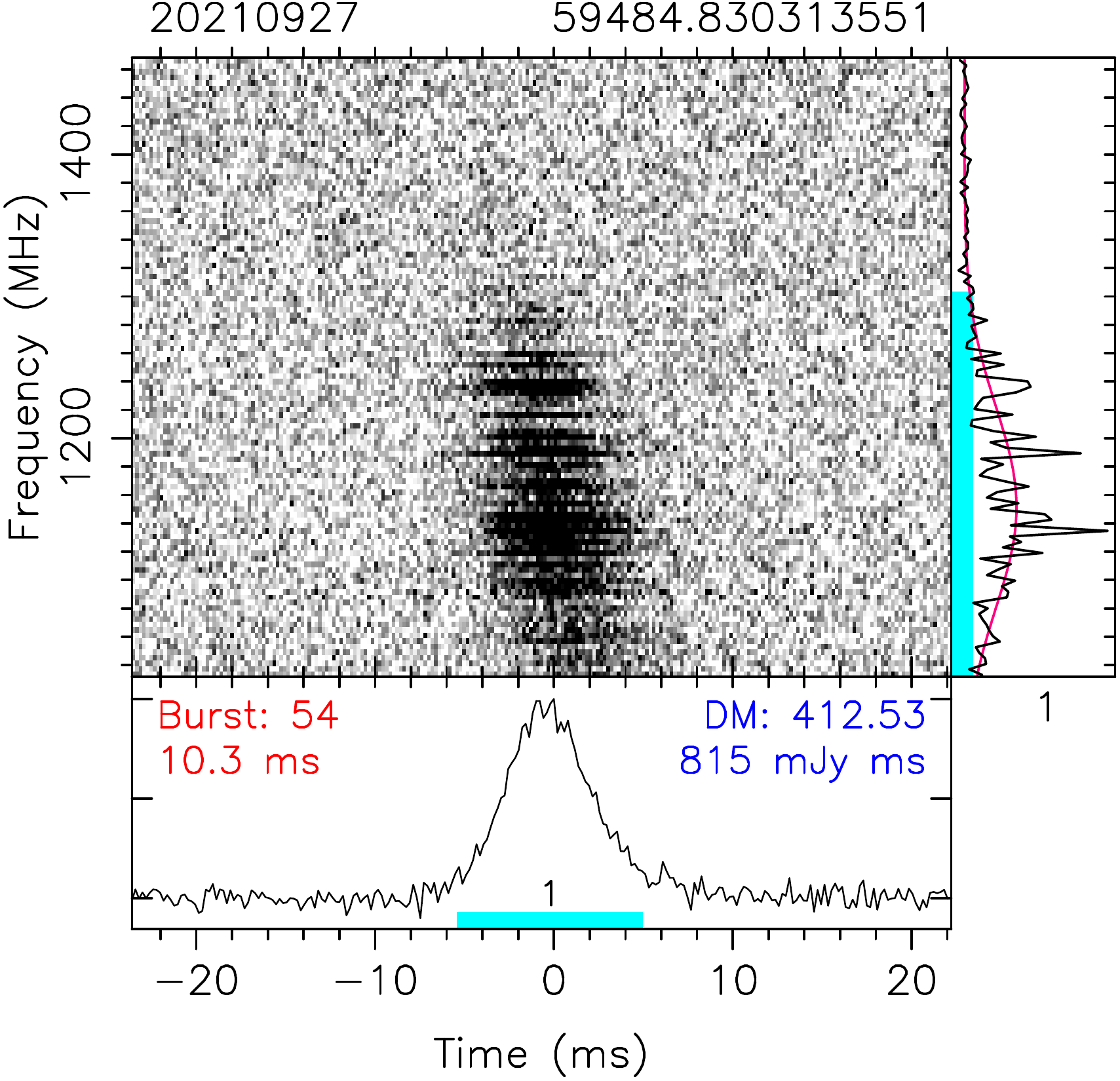}
    \includegraphics[height=37mm]{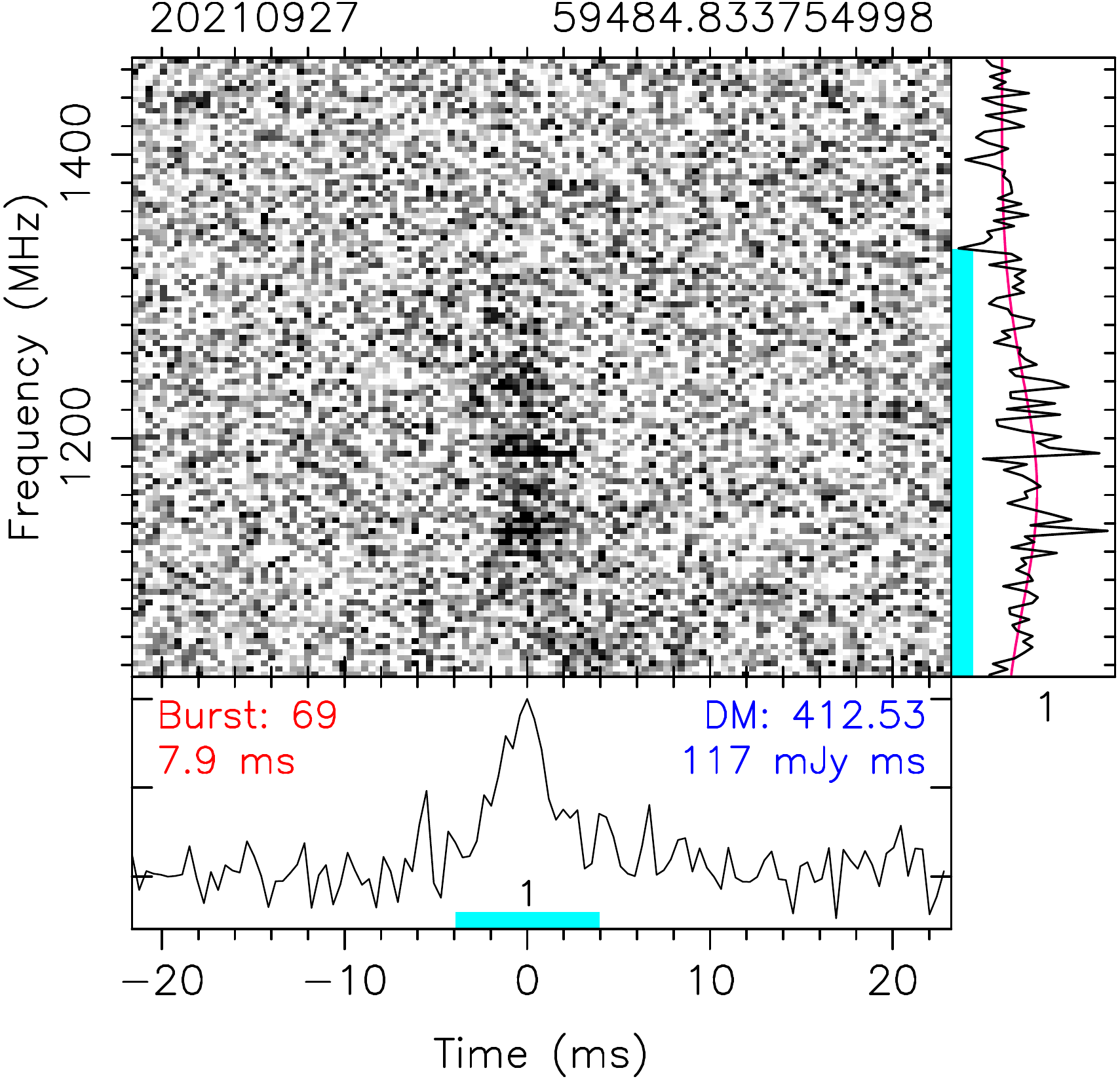}
    \includegraphics[height=37mm]{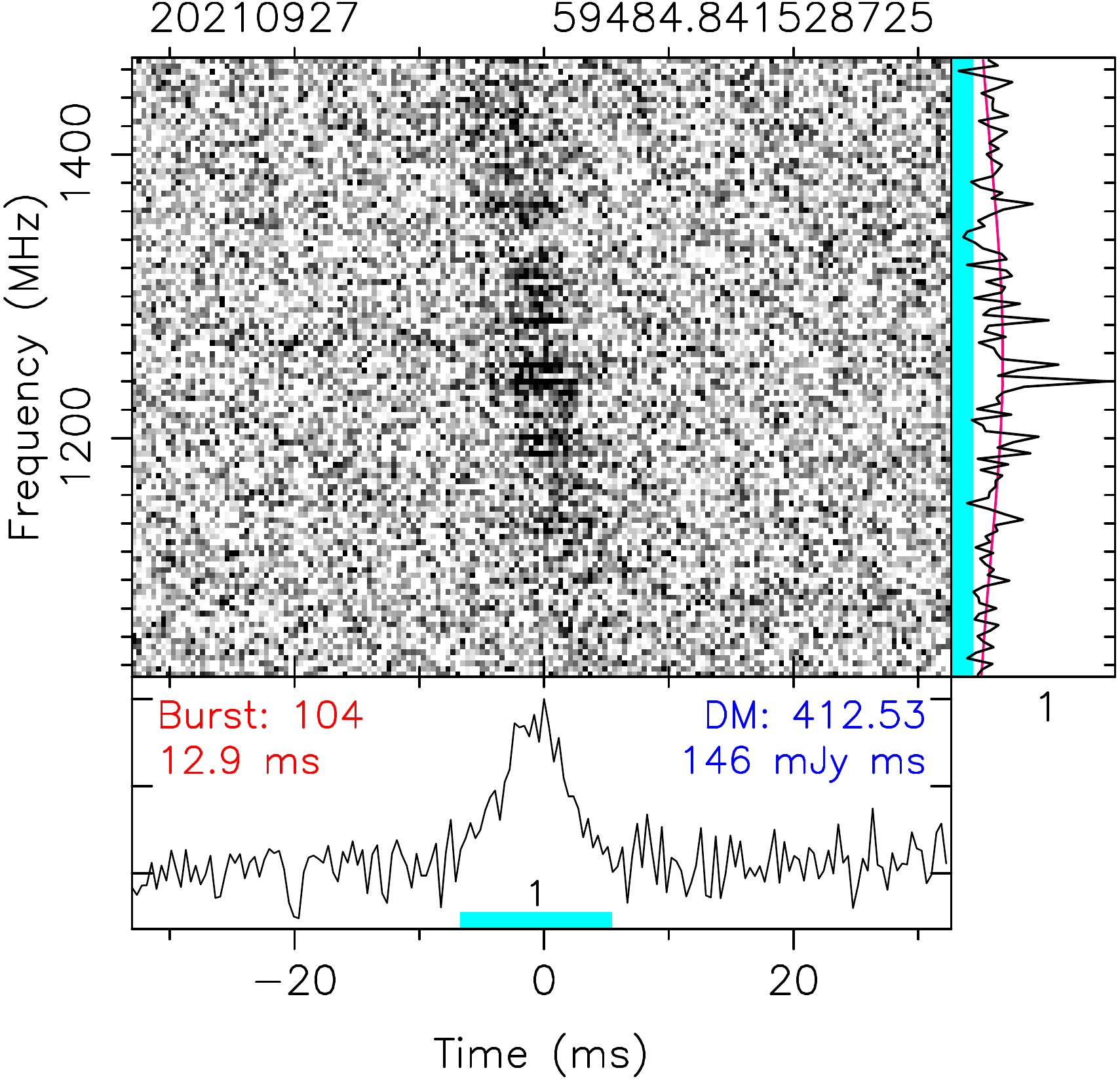}
    \includegraphics[height=37mm]{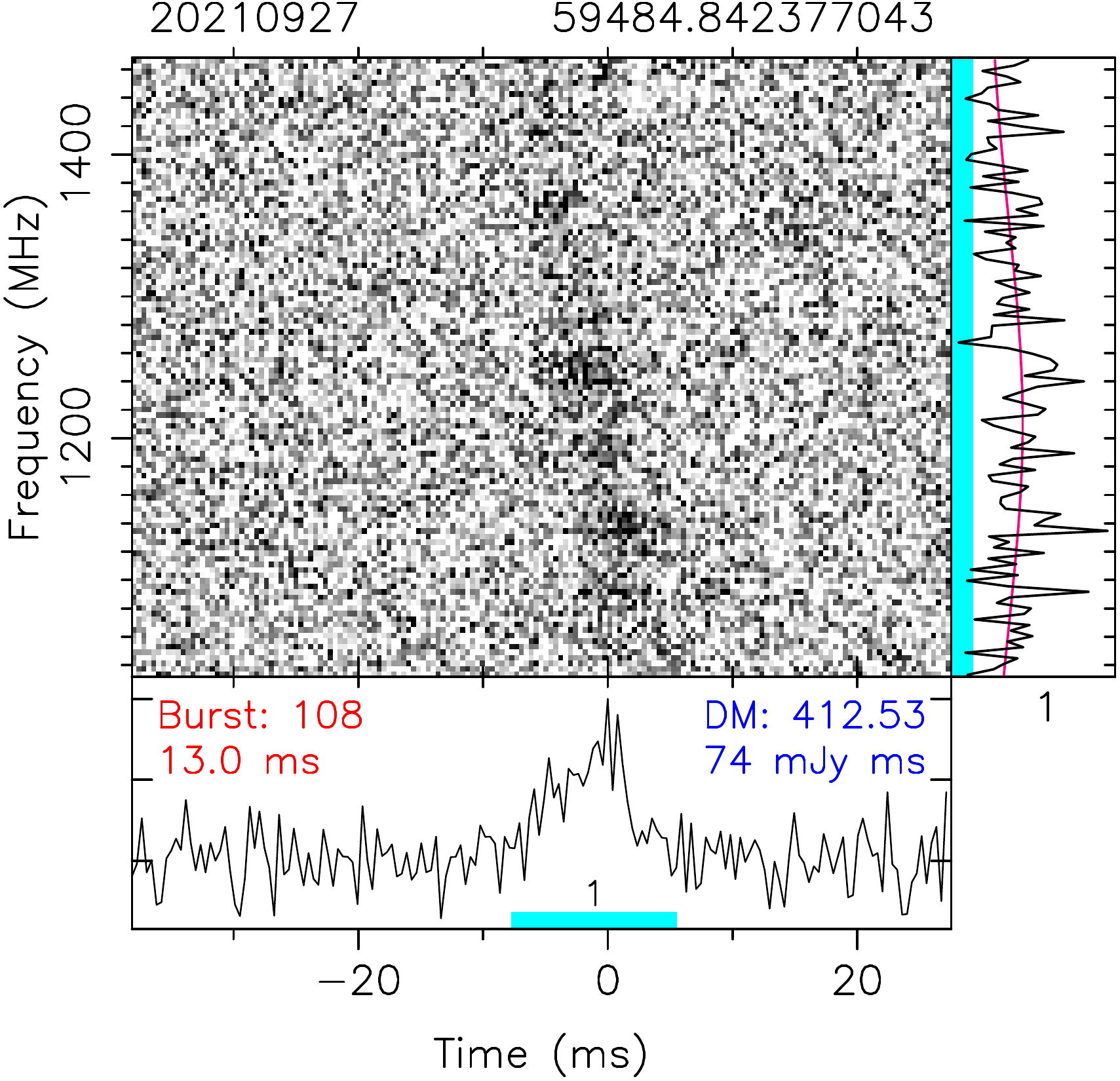}
    \includegraphics[height=37mm]{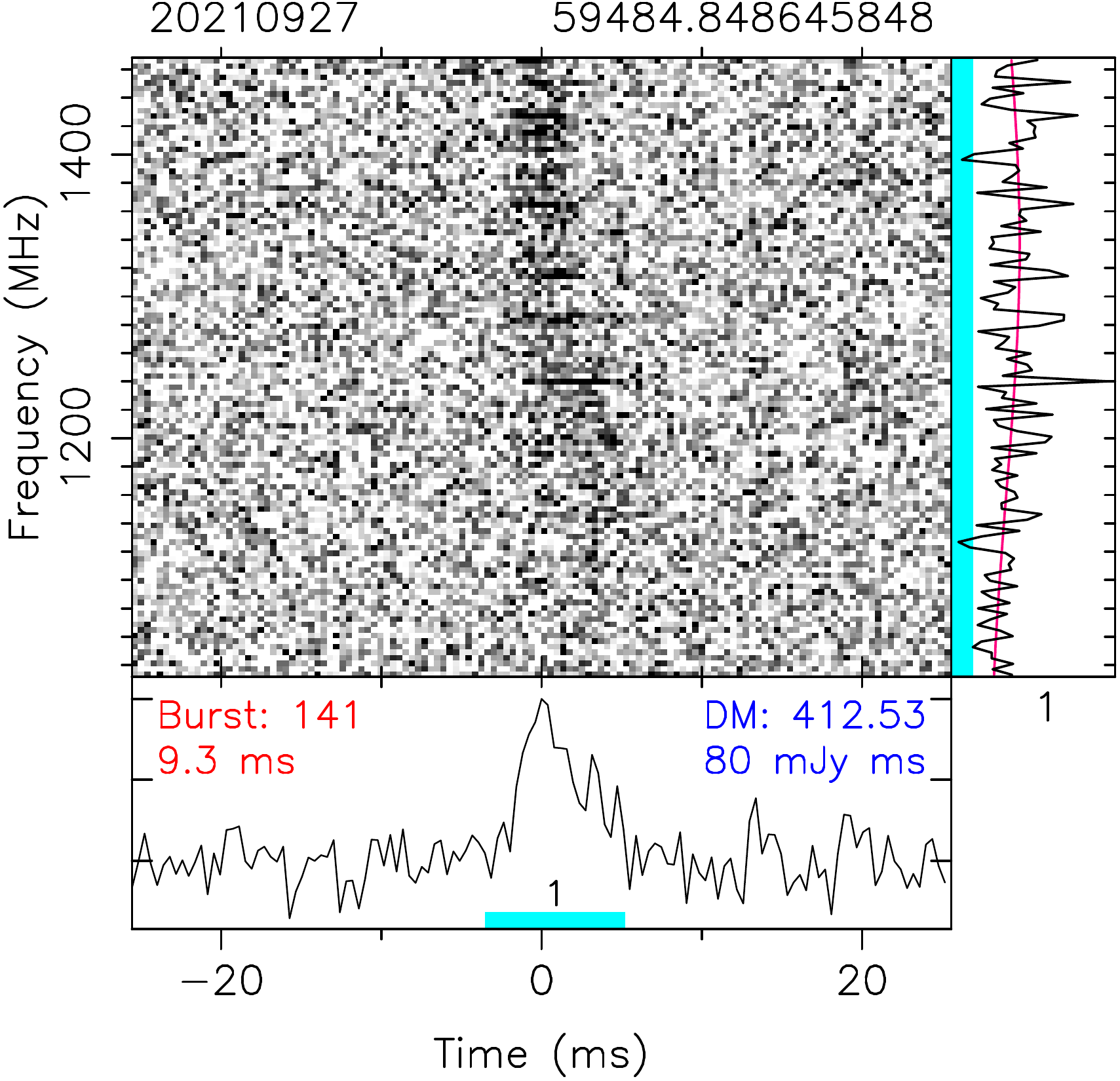}
    \includegraphics[height=37mm]{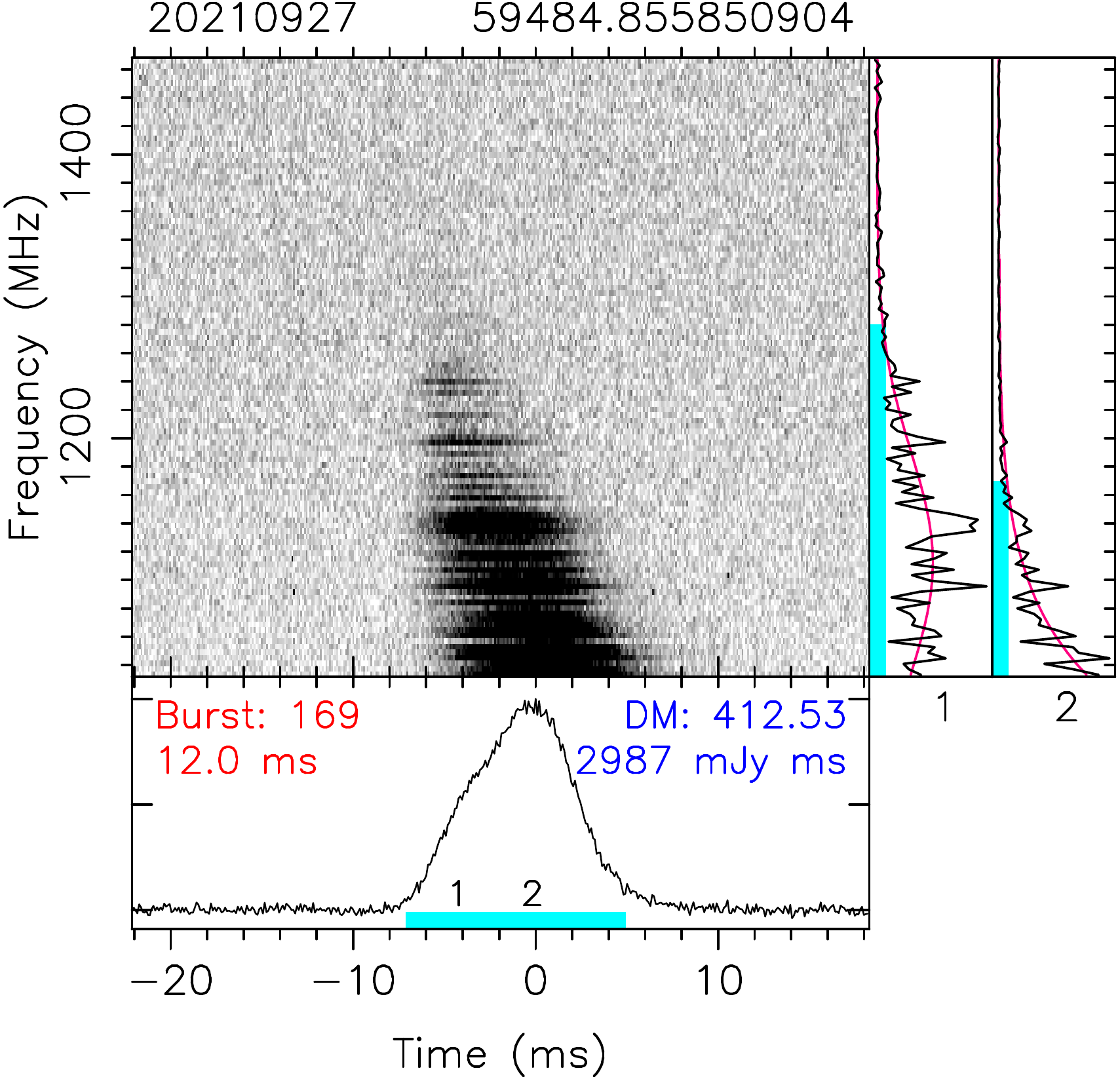}
    \includegraphics[height=37mm]{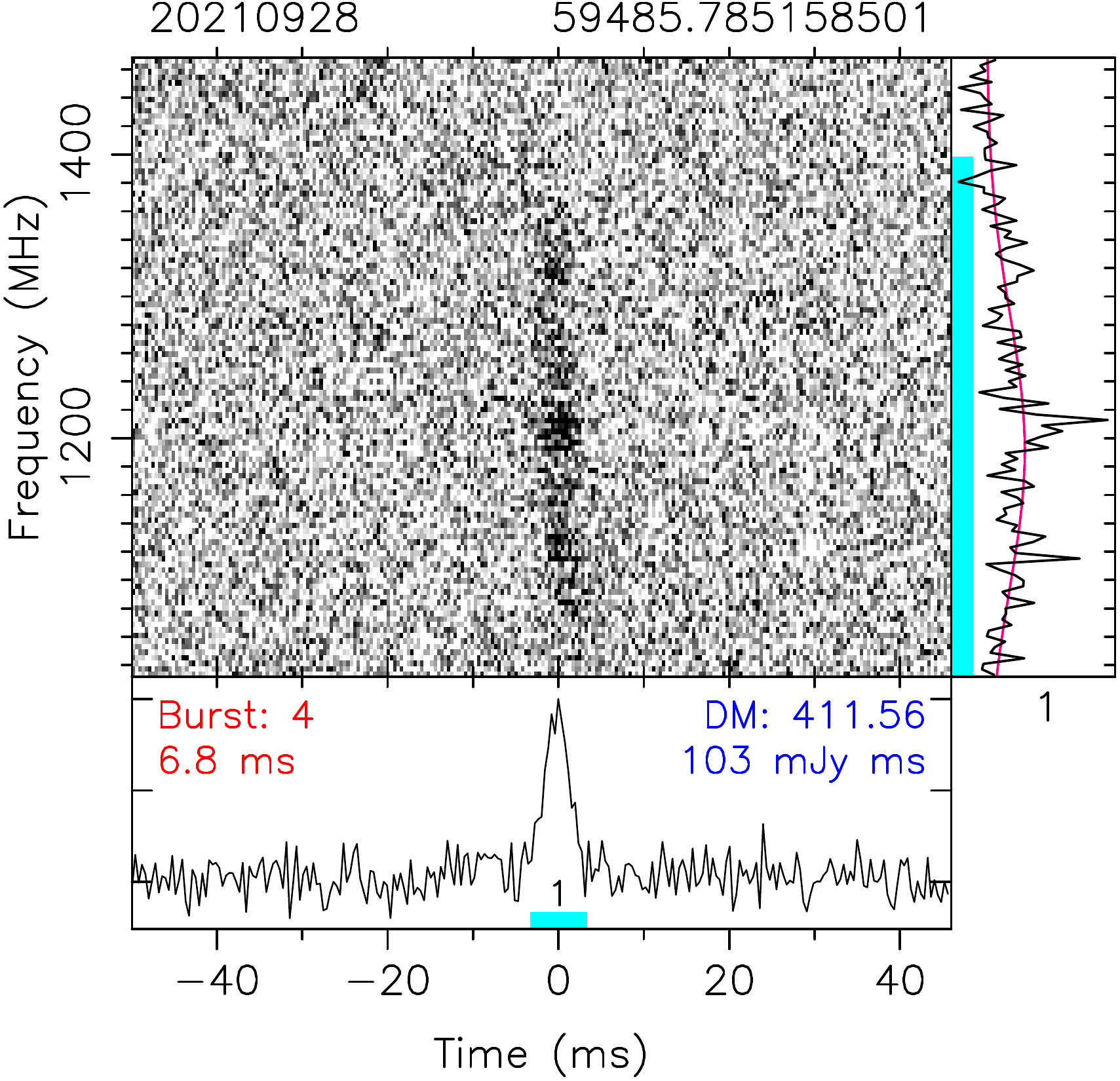}
    \includegraphics[height=37mm]{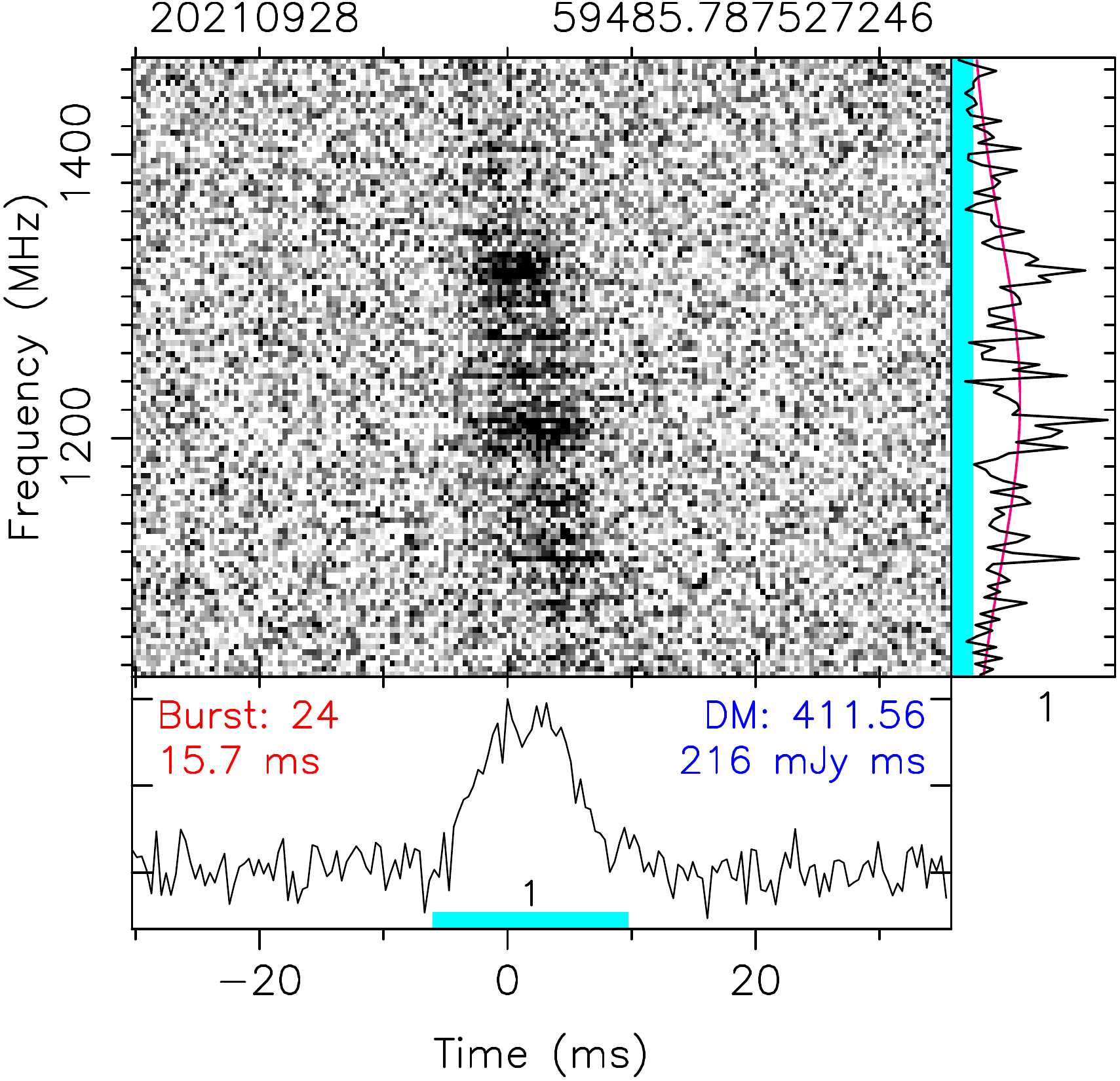}
    \includegraphics[height=37mm]{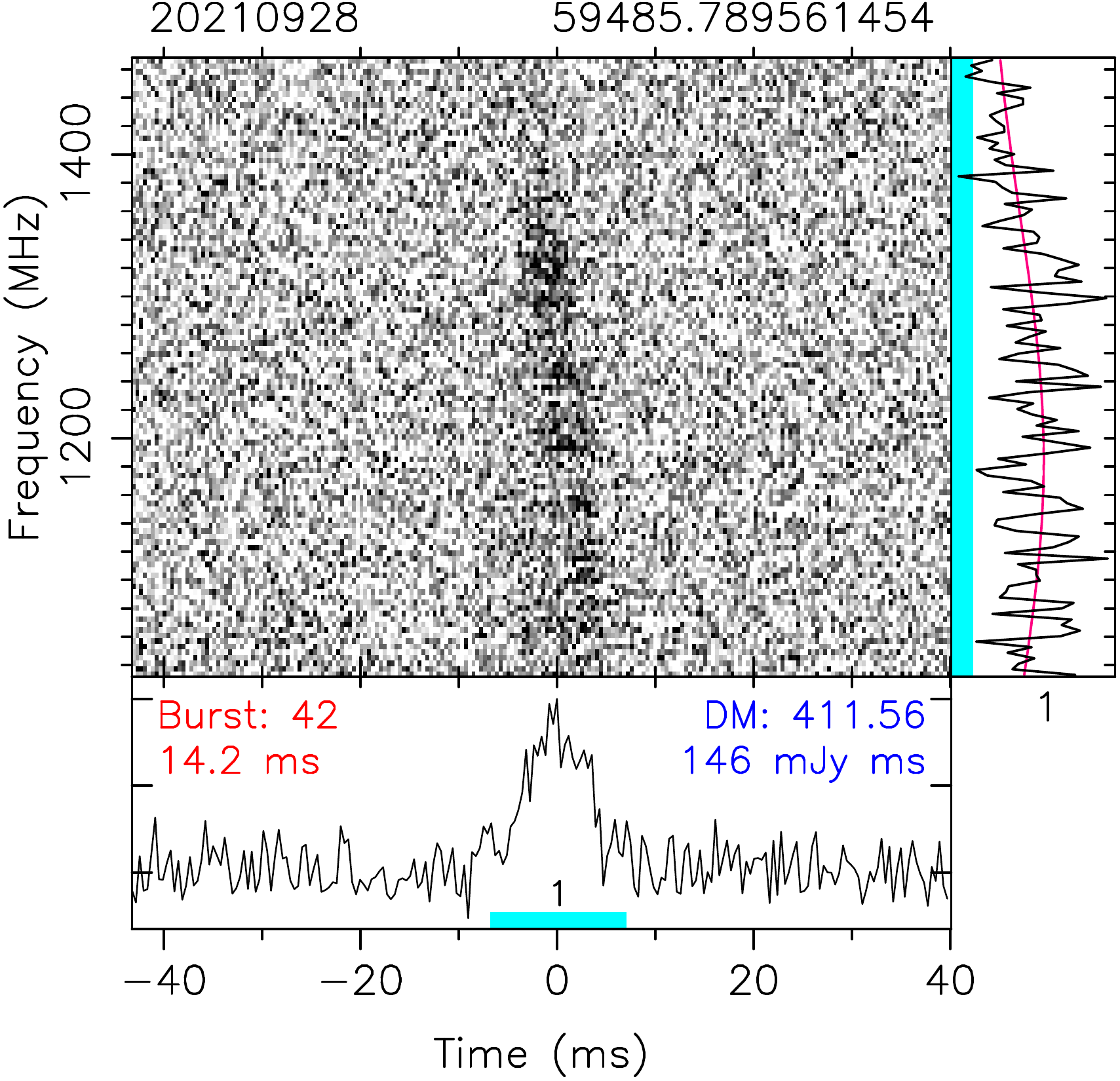}
    \includegraphics[height=37mm]{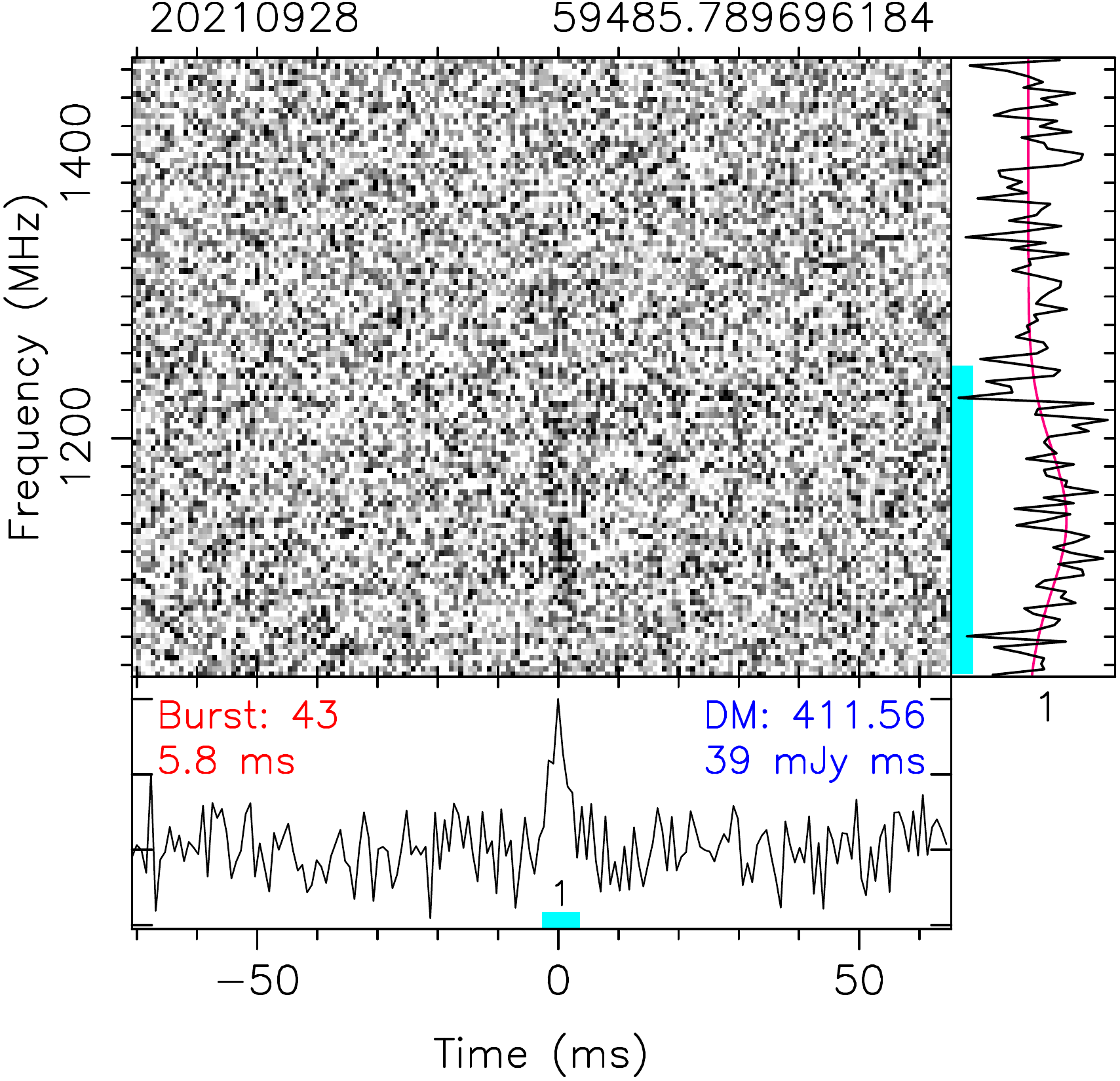}
    \includegraphics[height=37mm]{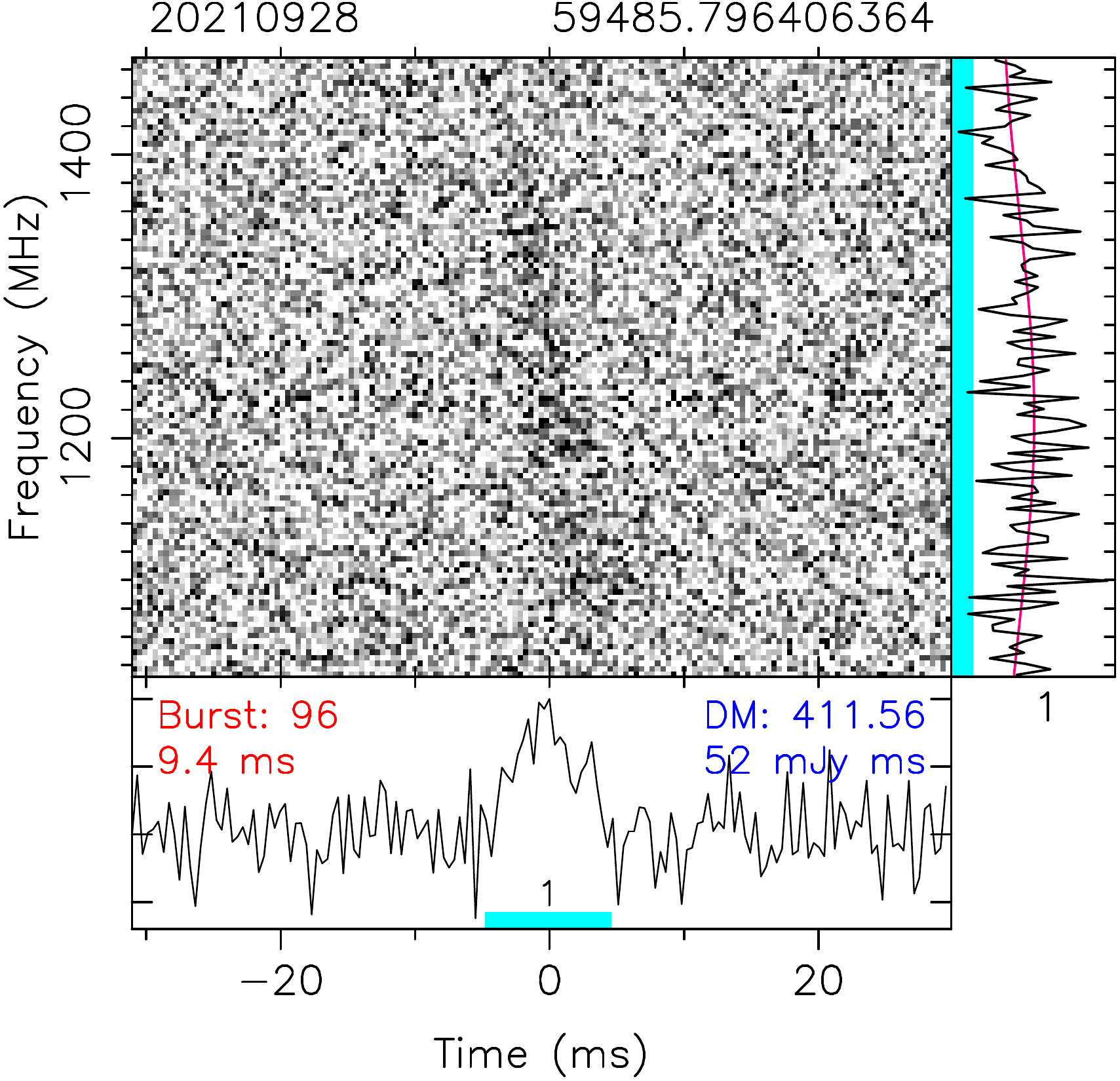}
    \includegraphics[height=37mm]{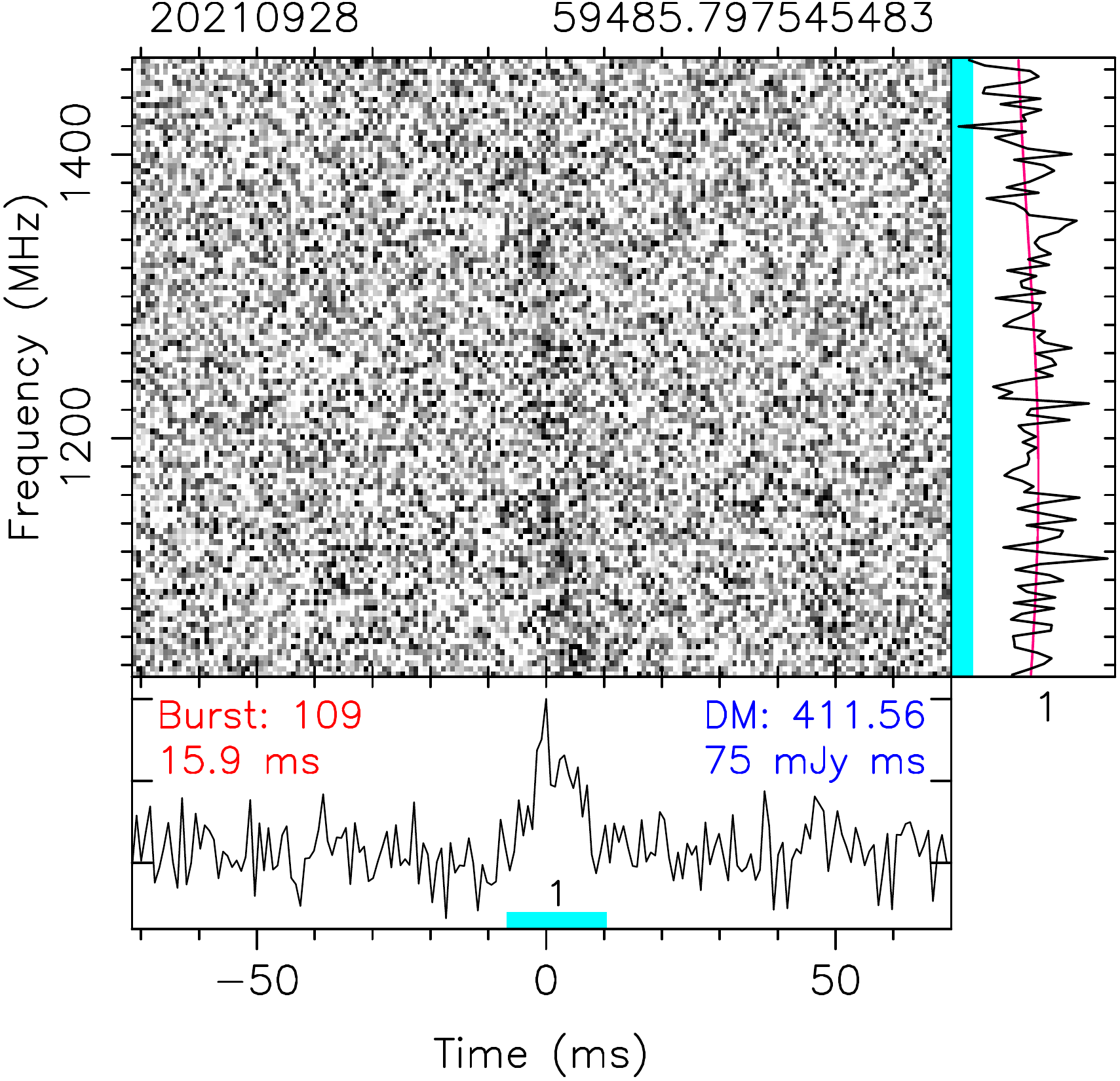}
    \includegraphics[height=37mm]{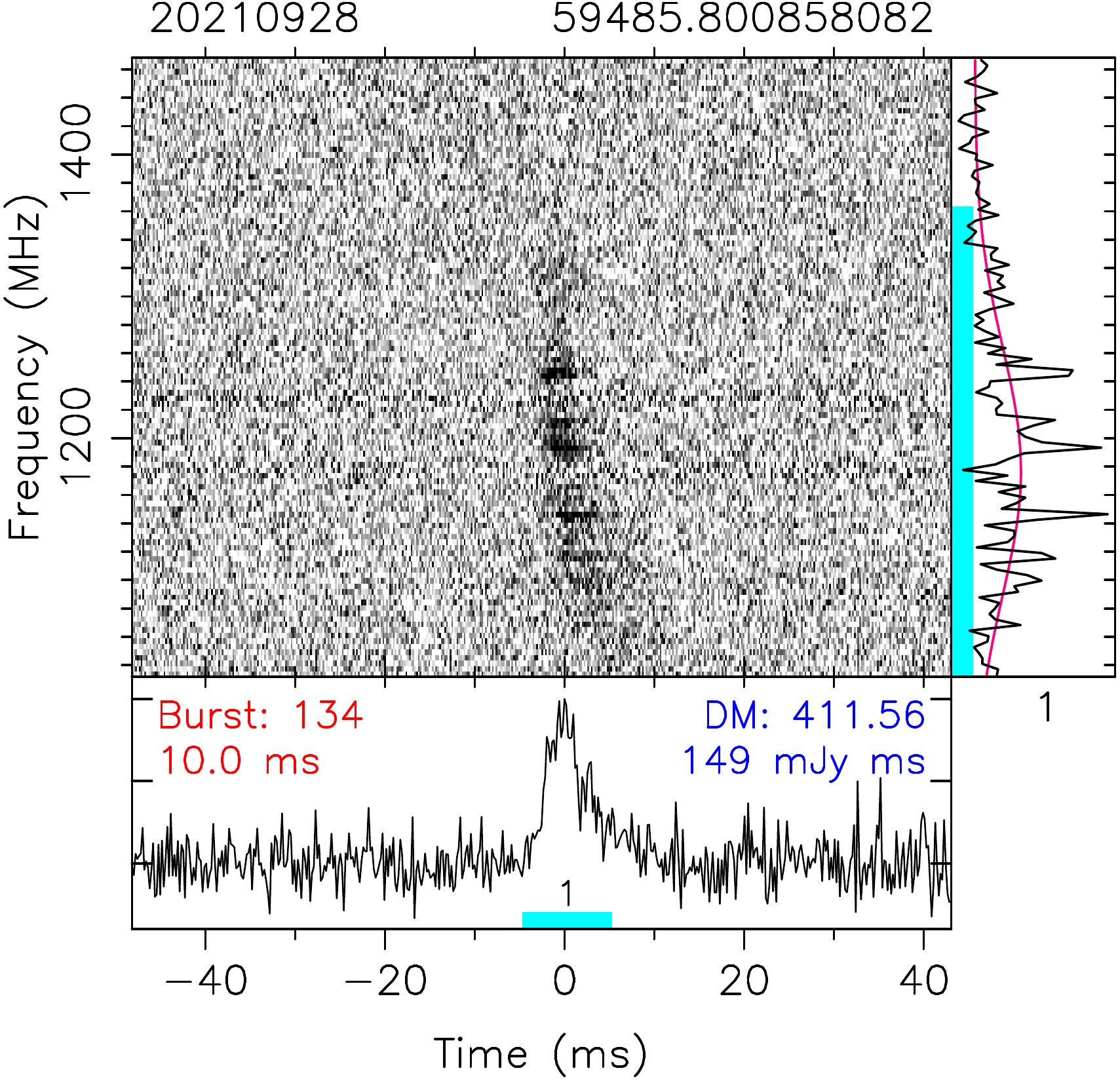}
    \includegraphics[height=37mm]{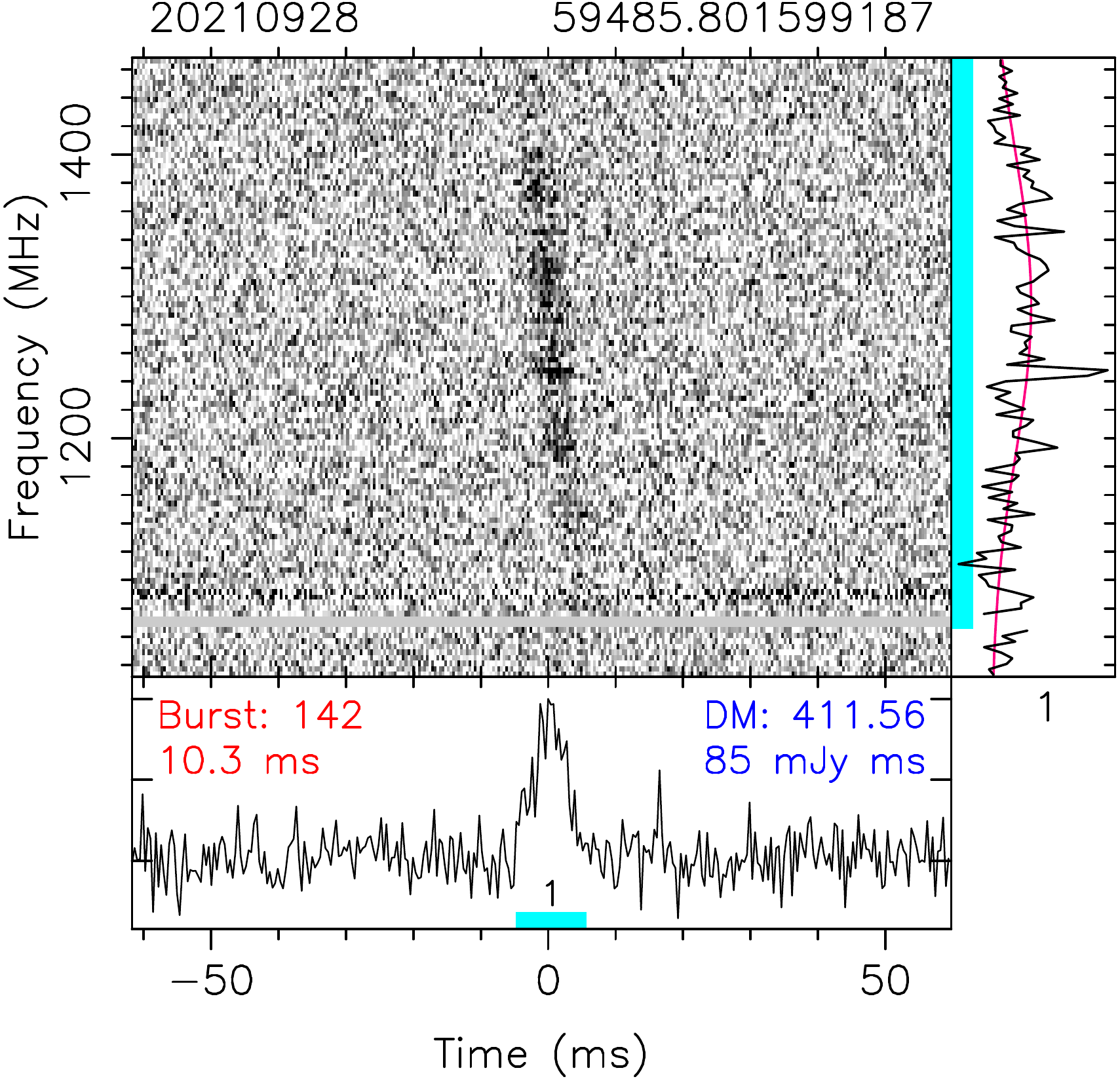}
    \includegraphics[height=37mm]{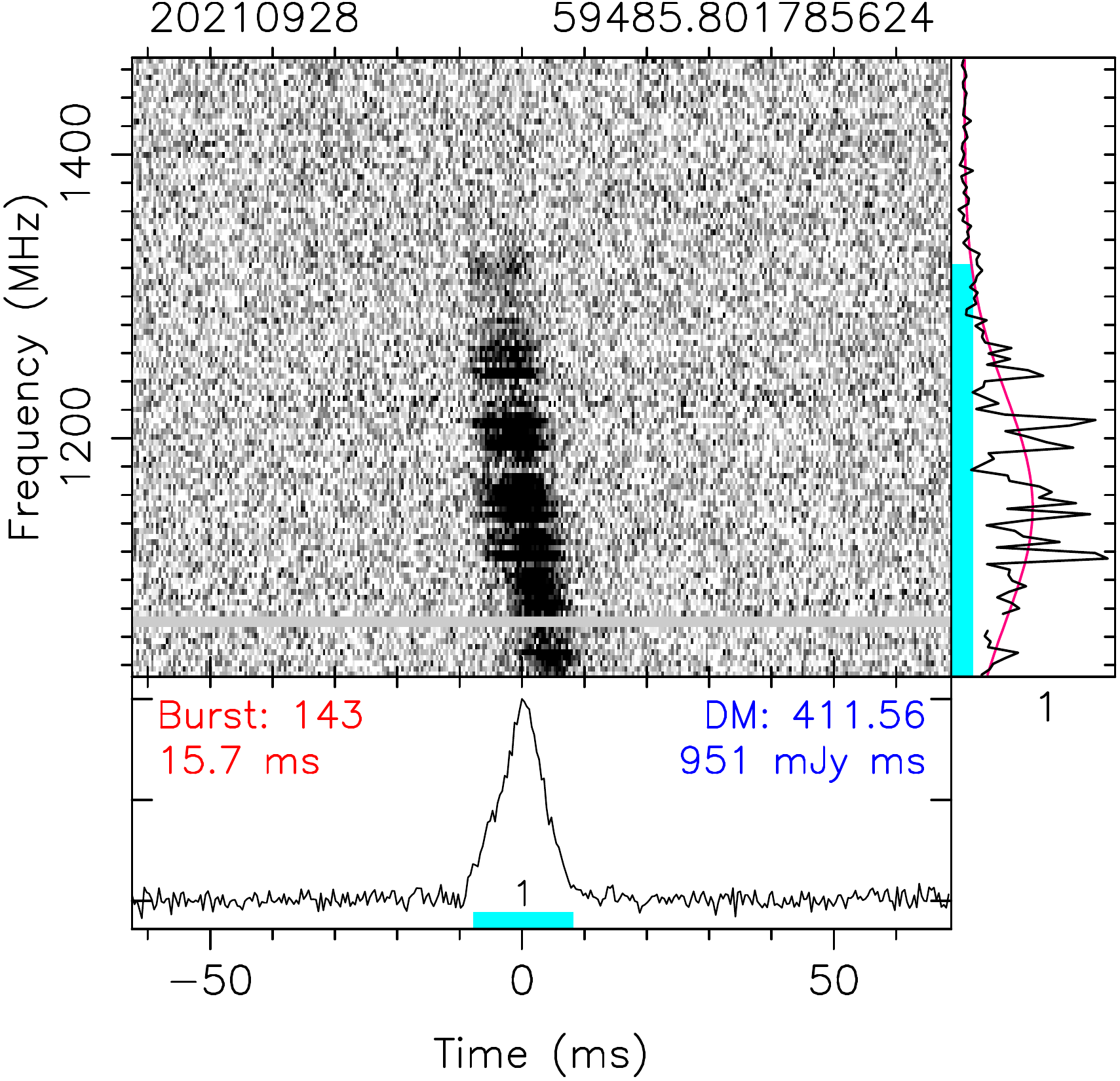}
    \includegraphics[height=37mm]{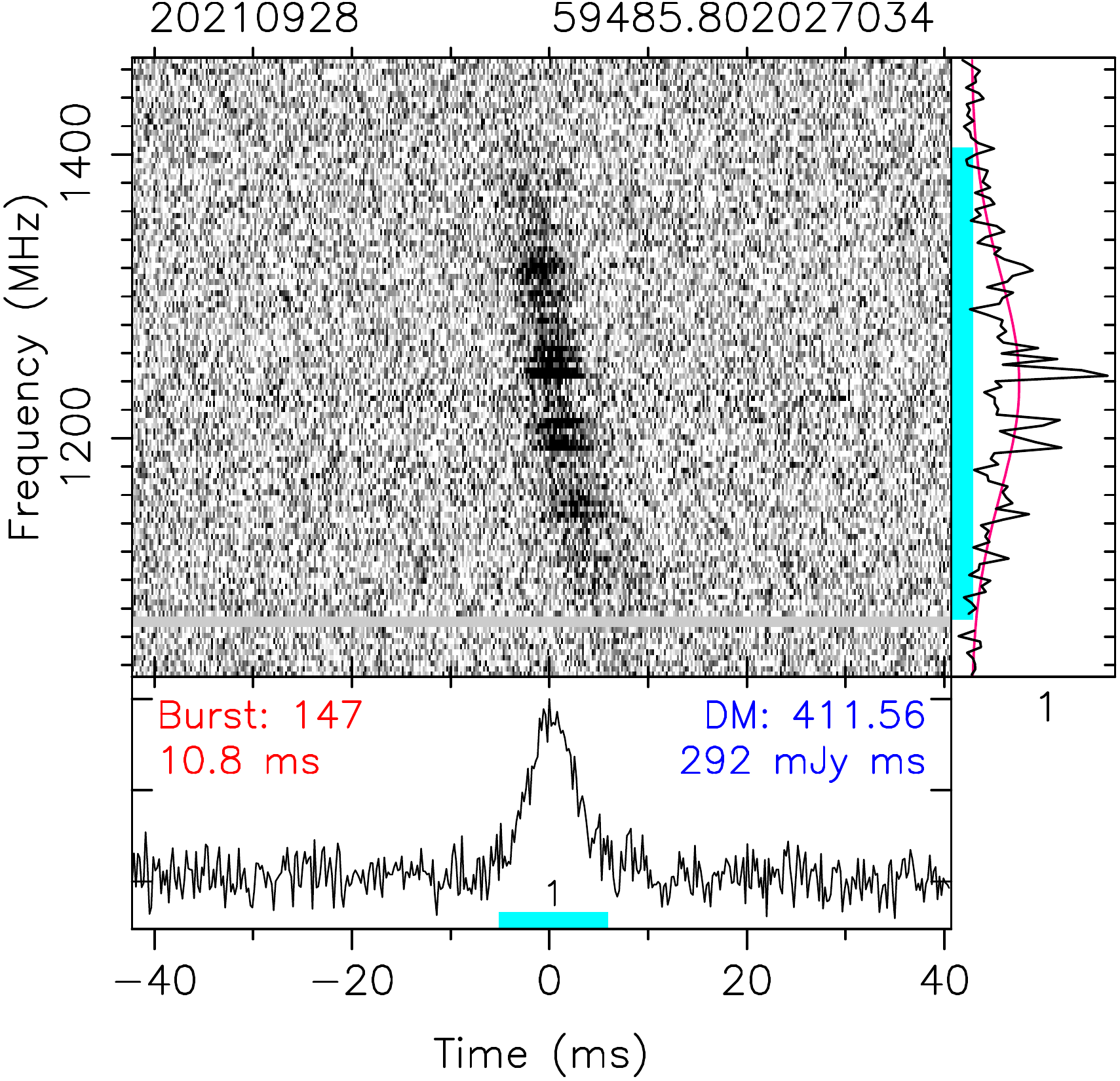}
    \includegraphics[height=37mm]{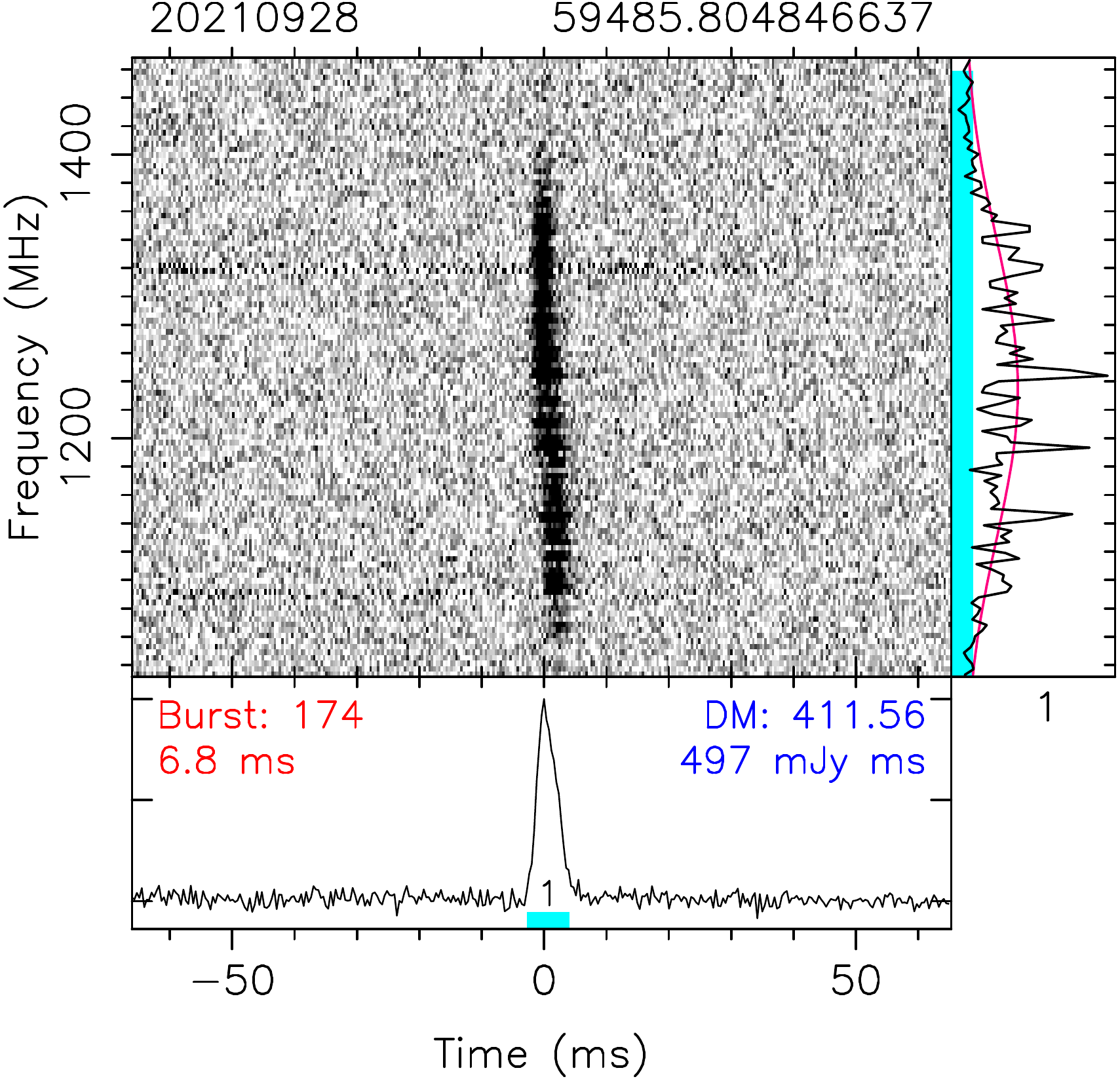}
    \includegraphics[height=37mm]{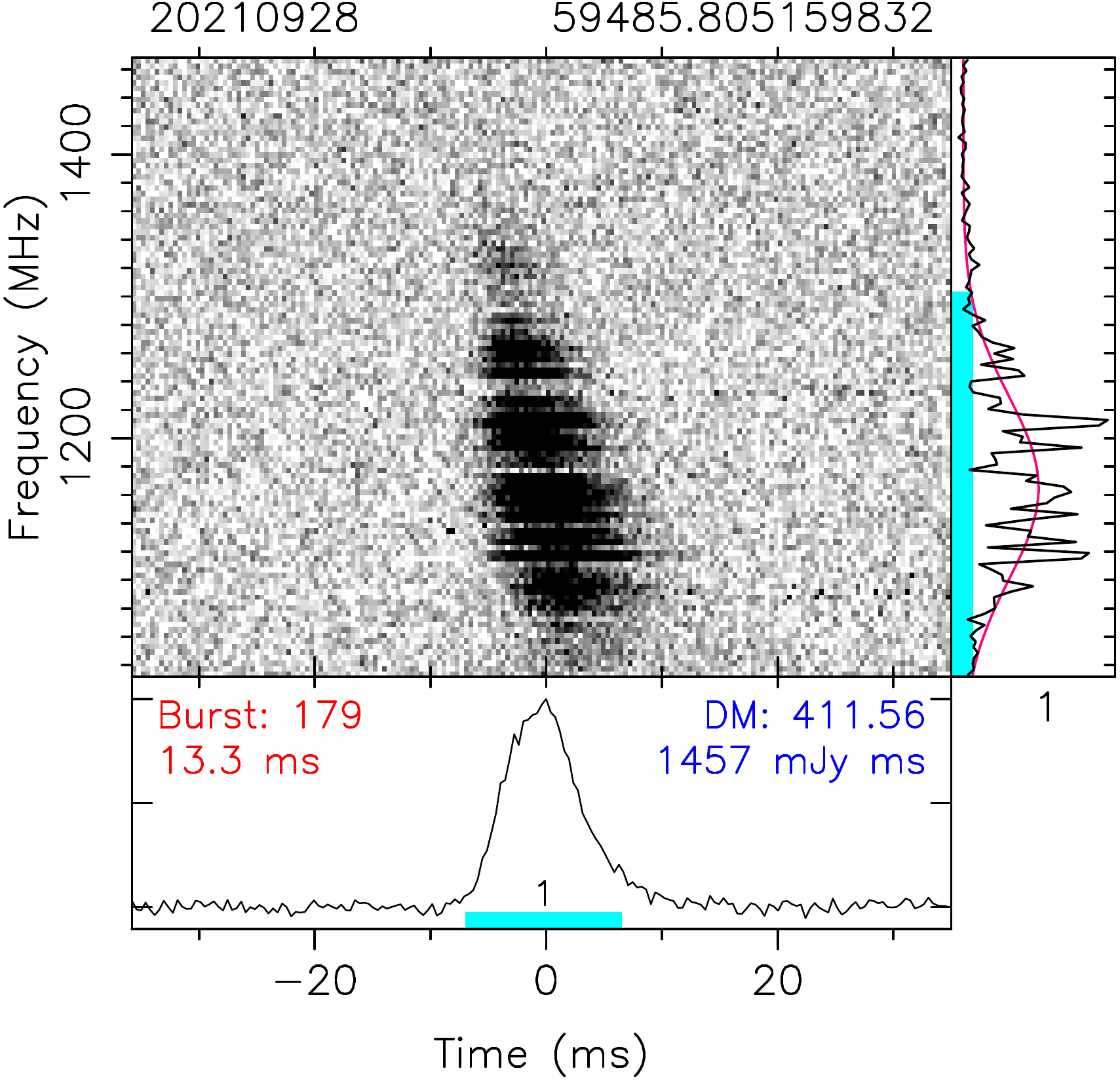}
    \includegraphics[height=37mm]{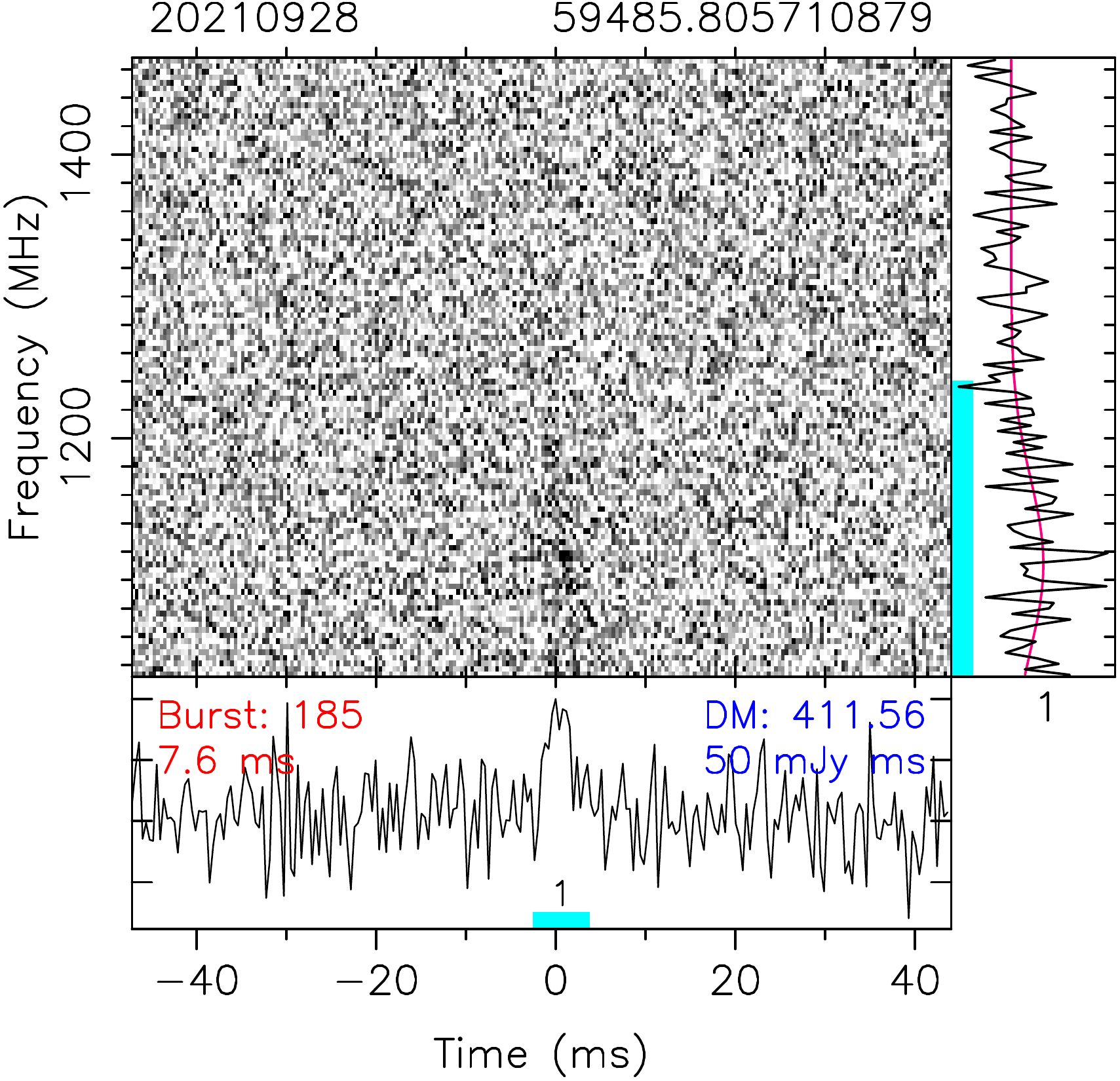}
    \includegraphics[height=37mm]{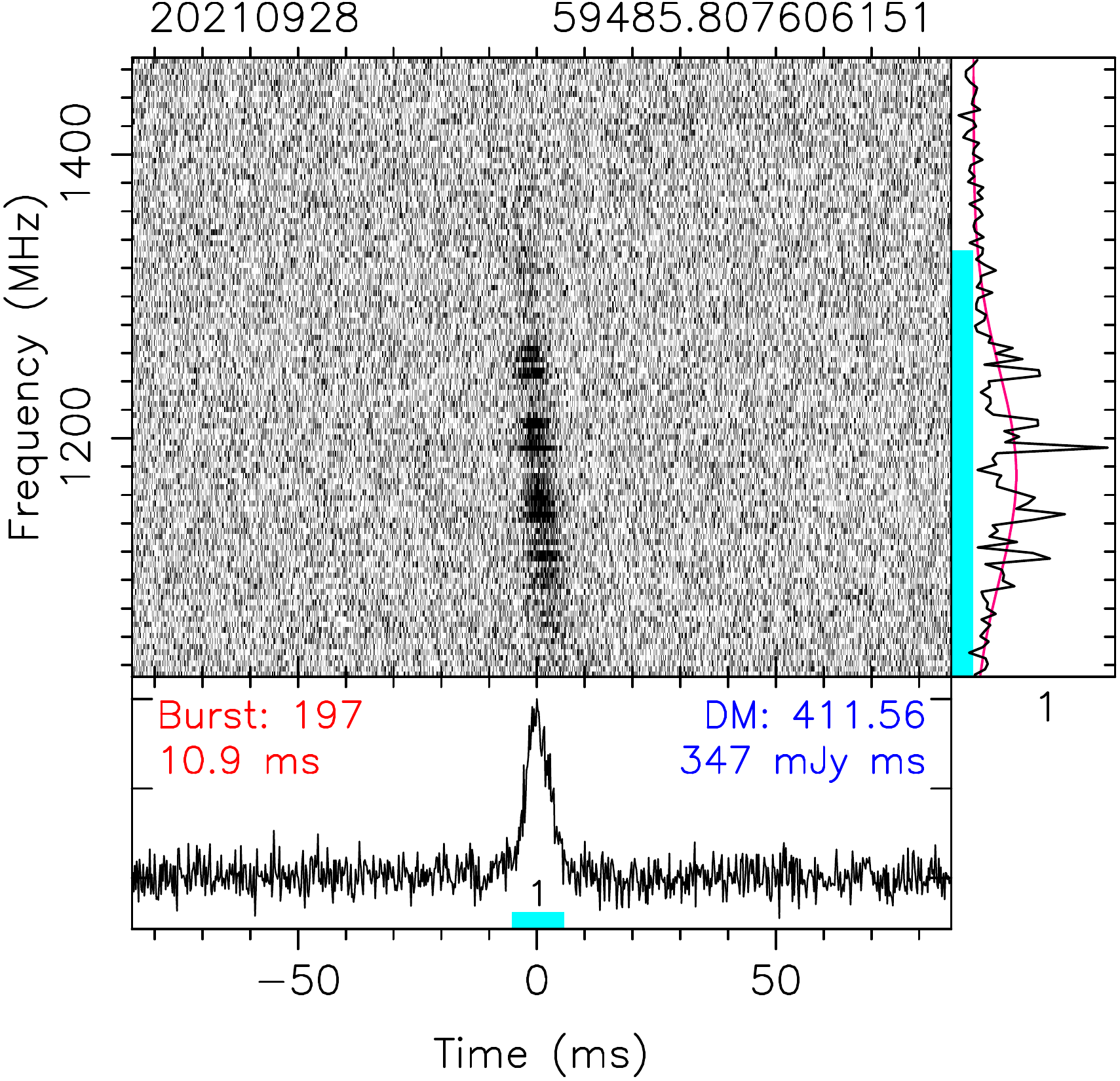}
\caption{The same as Figure~\ref{fig:appendix:D1W} but for bursts in D1-M.
}
\label{fig:appendix:D1M}
\end{figure*}
\addtocounter{figure}{-1}
\begin{figure*}
    \flushleft
    \includegraphics[height=37mm]{20210929/FRB20201124A_20210929_tracking-M01-P1-c512b1.fits-218-T-2165.136-2165.223-DM-411.6.pdf}
    \includegraphics[height=37mm]{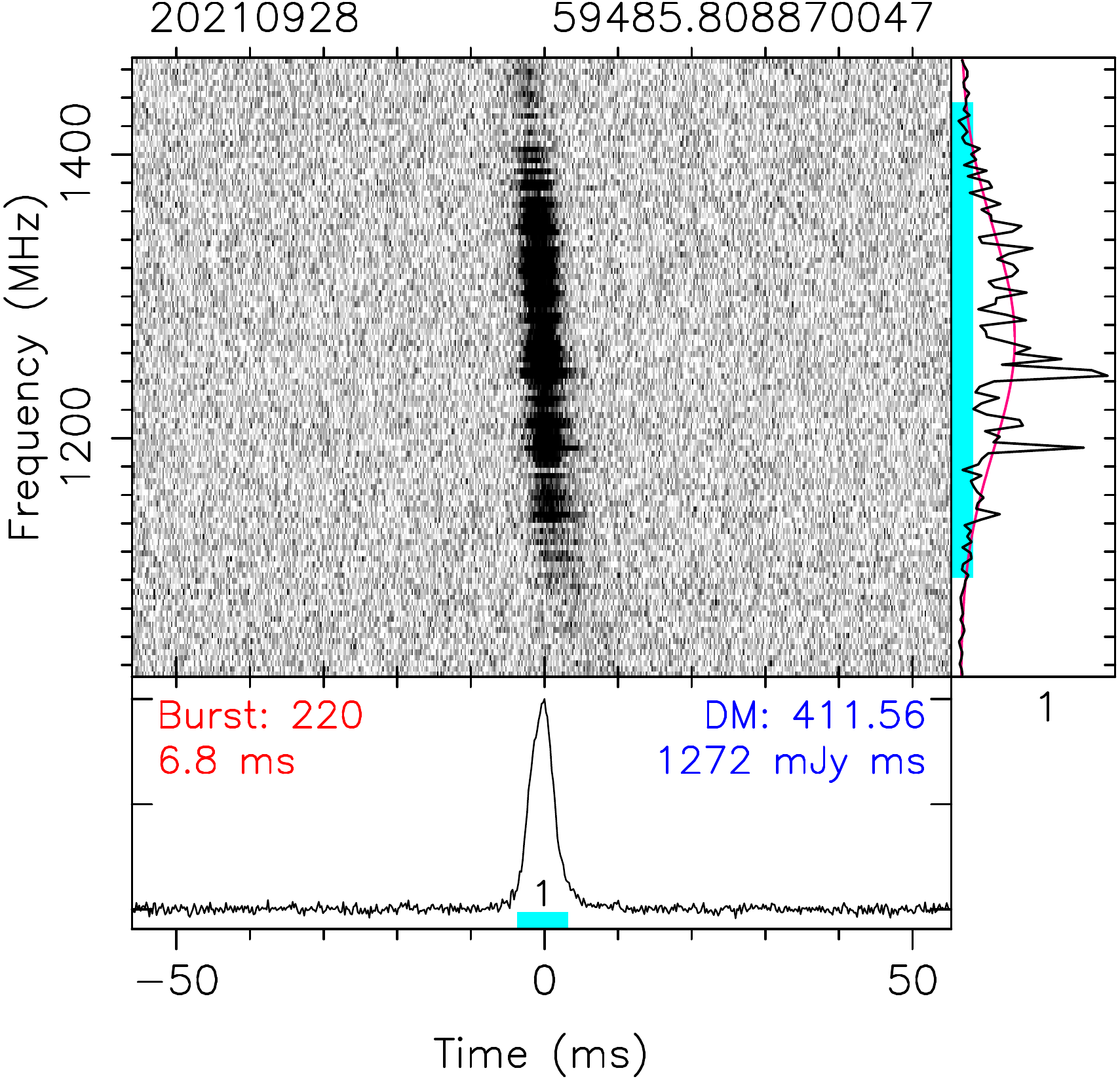}
    \includegraphics[height=37mm]{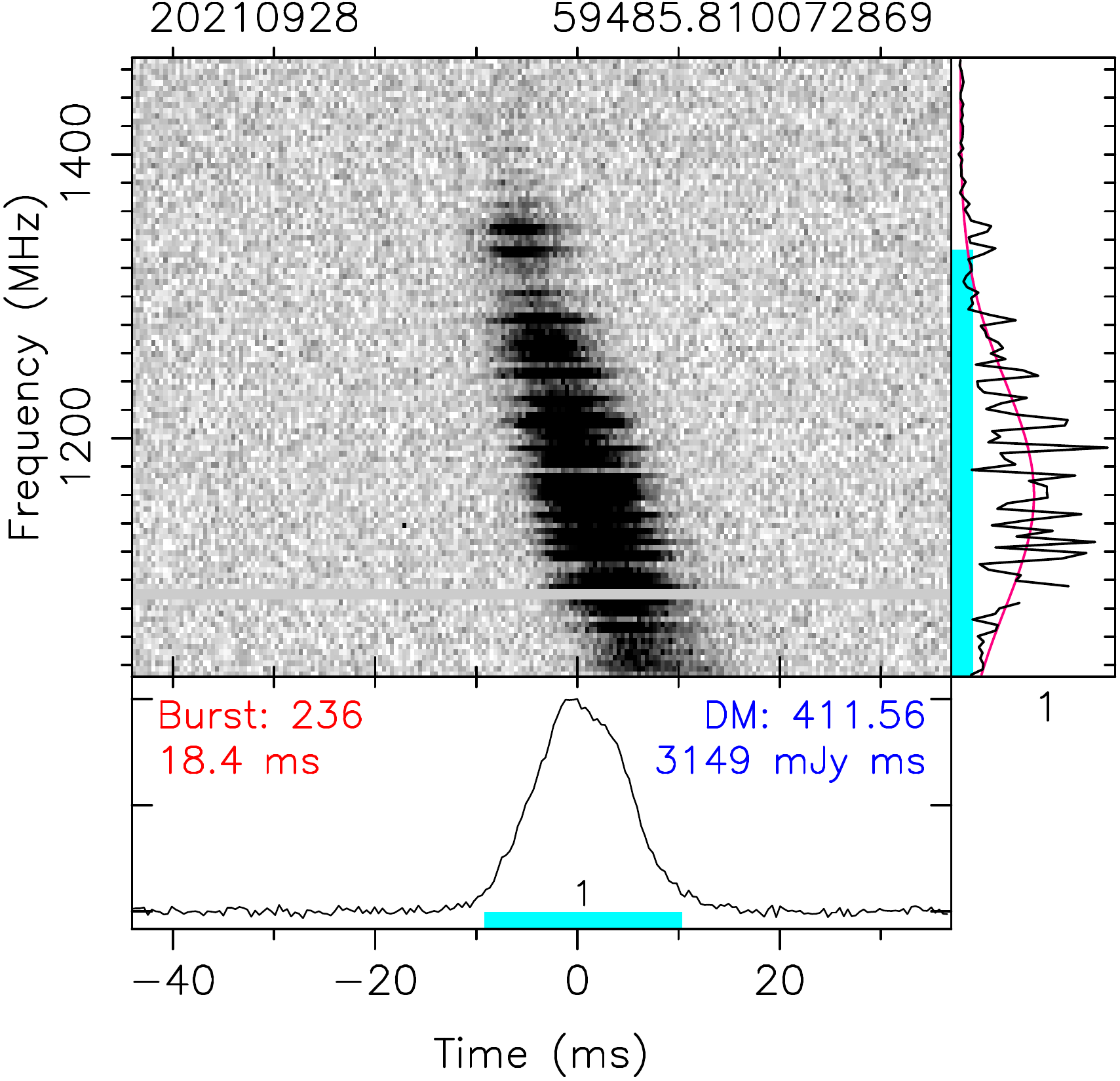}
    \includegraphics[height=37mm]{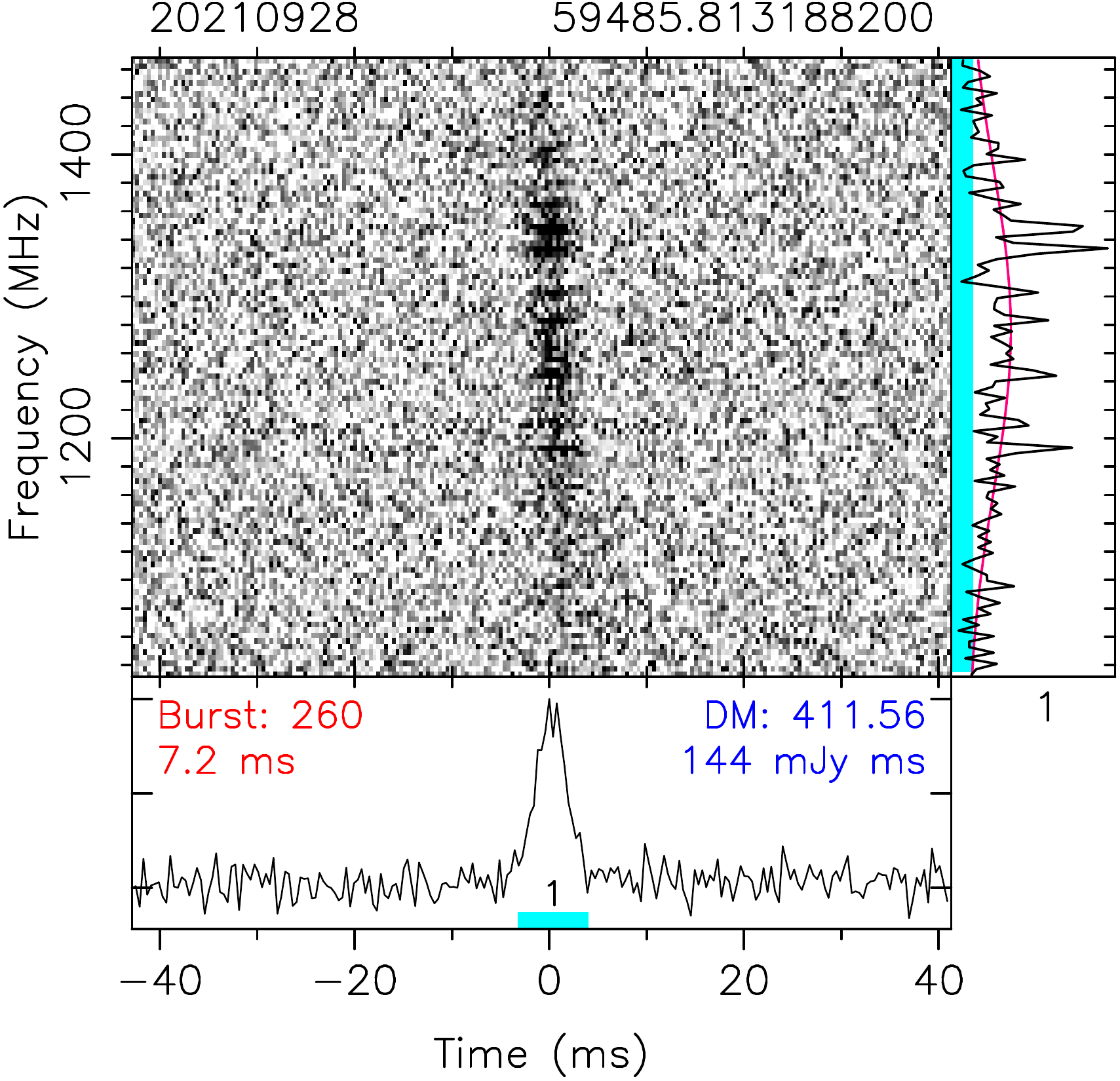}
    \includegraphics[height=37mm]{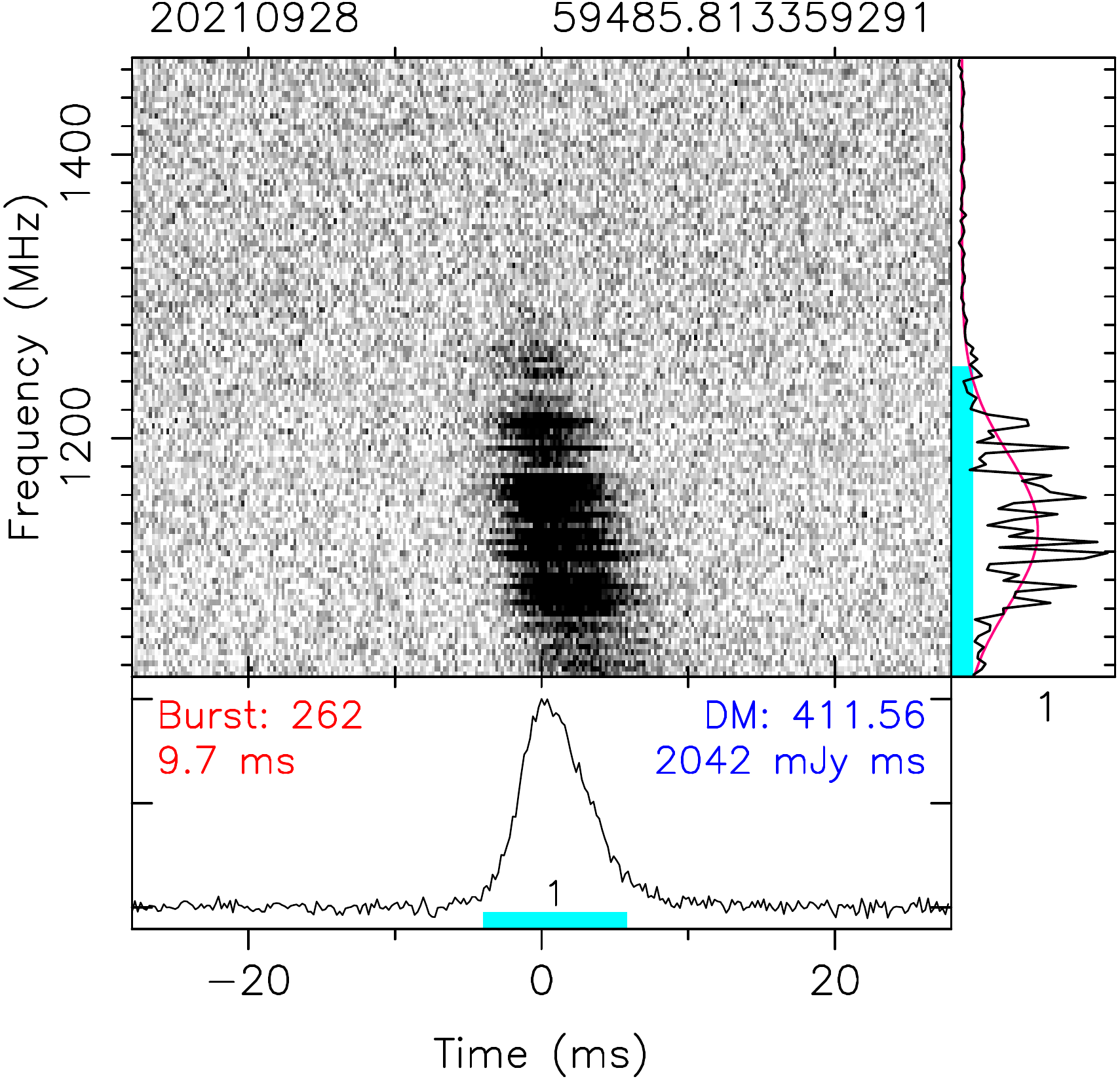}
    \includegraphics[height=37mm]{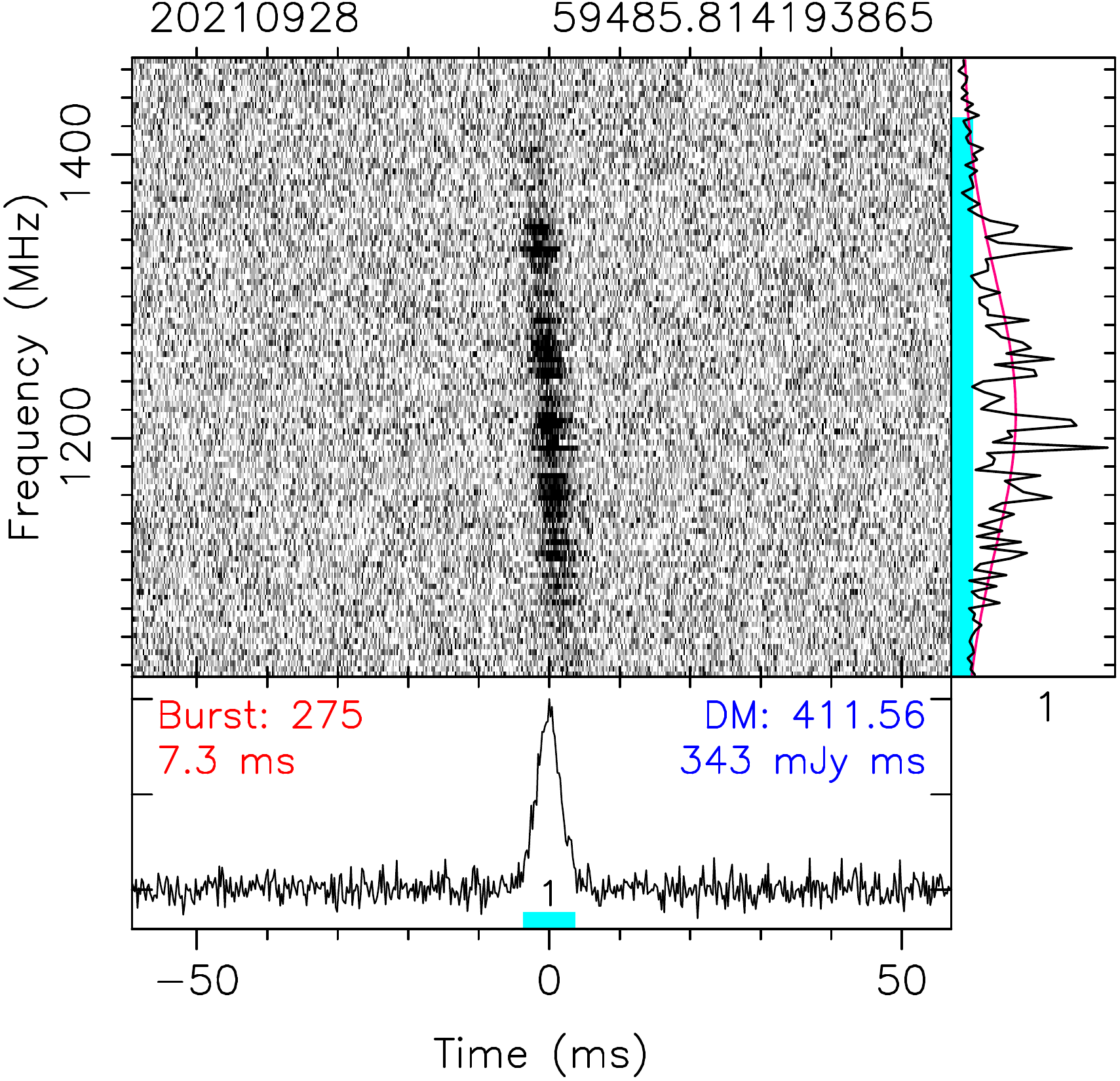}
    \includegraphics[height=37mm]{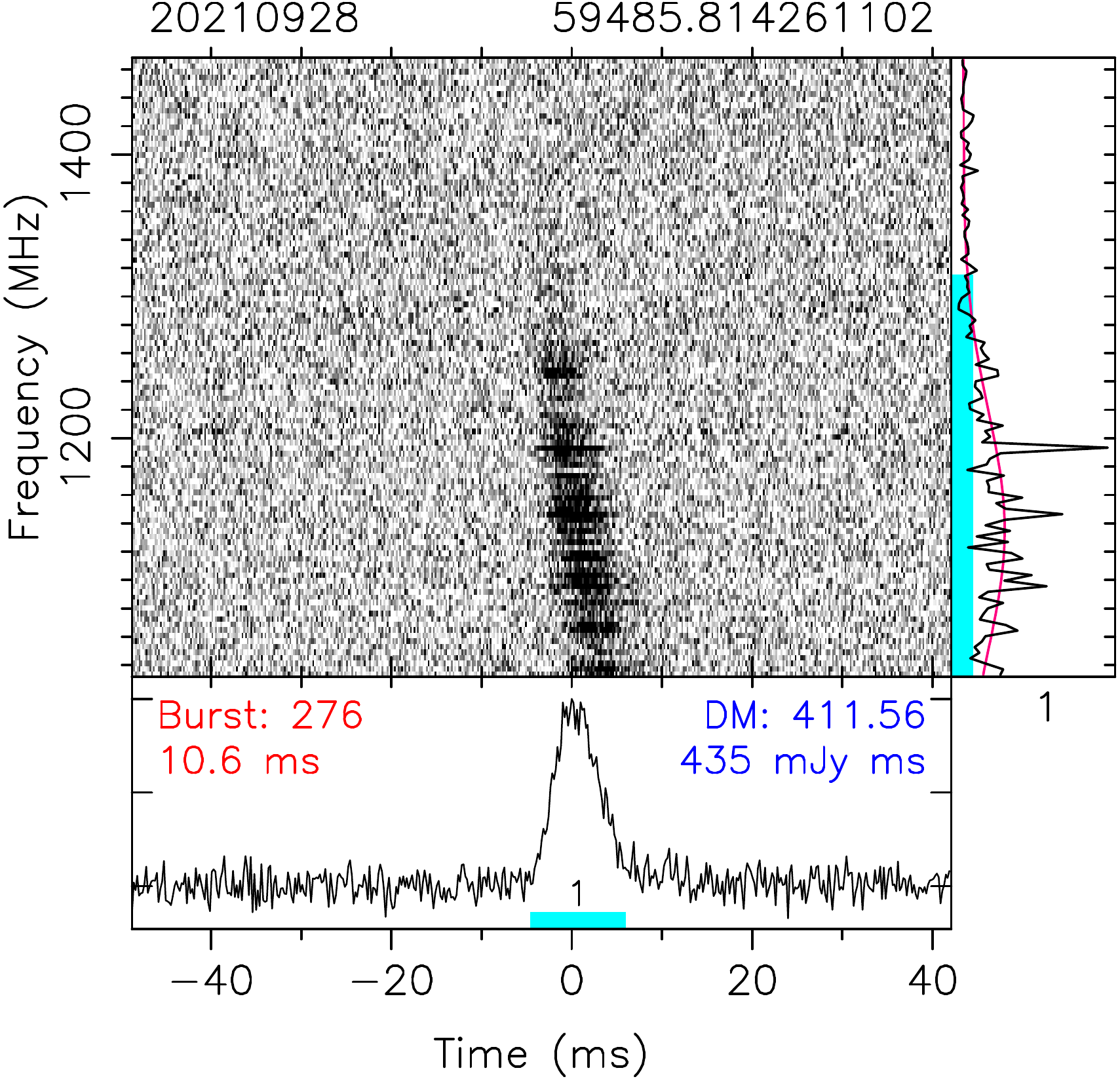}
    \includegraphics[height=37mm]{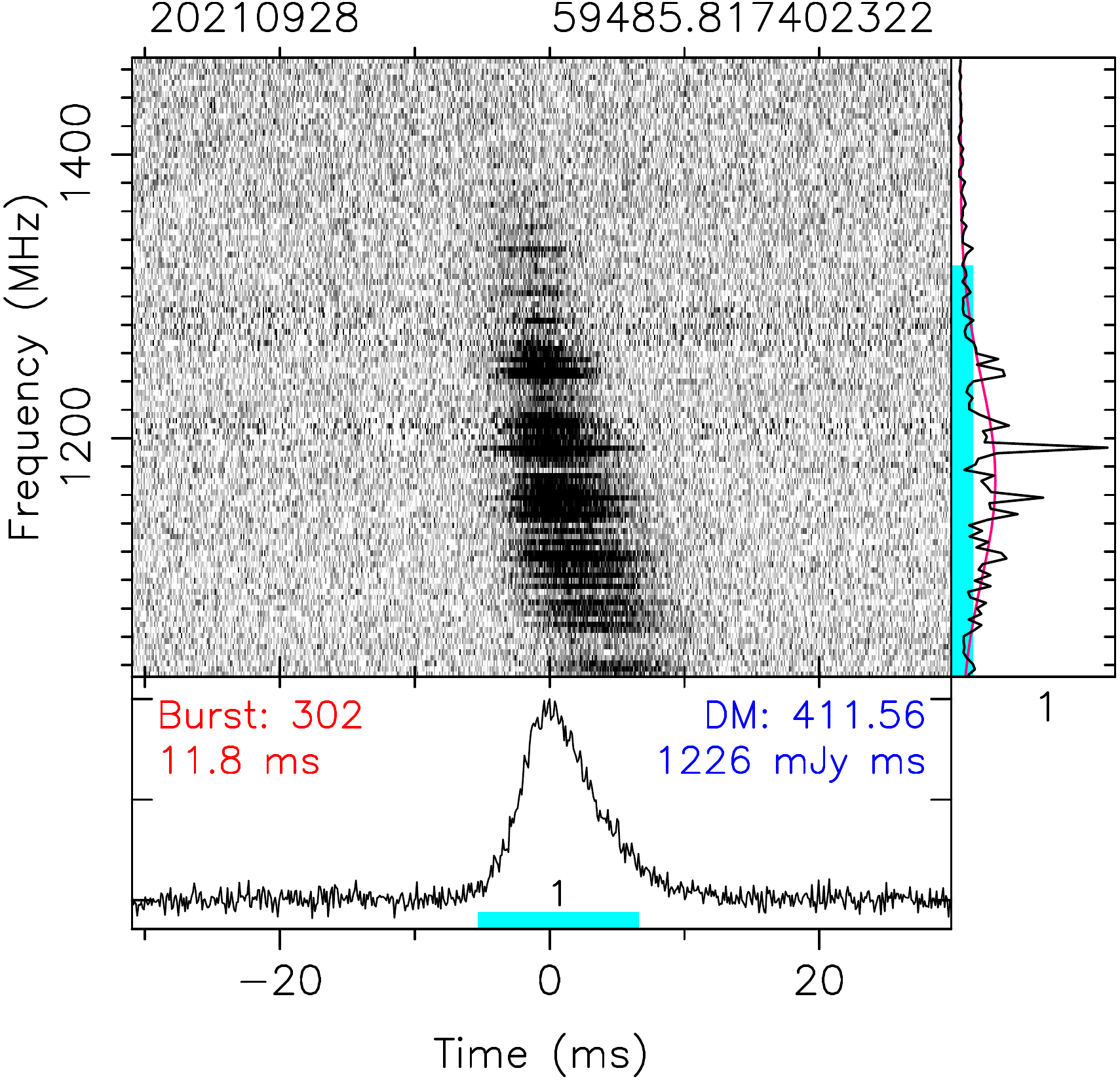}
    \includegraphics[height=37mm]{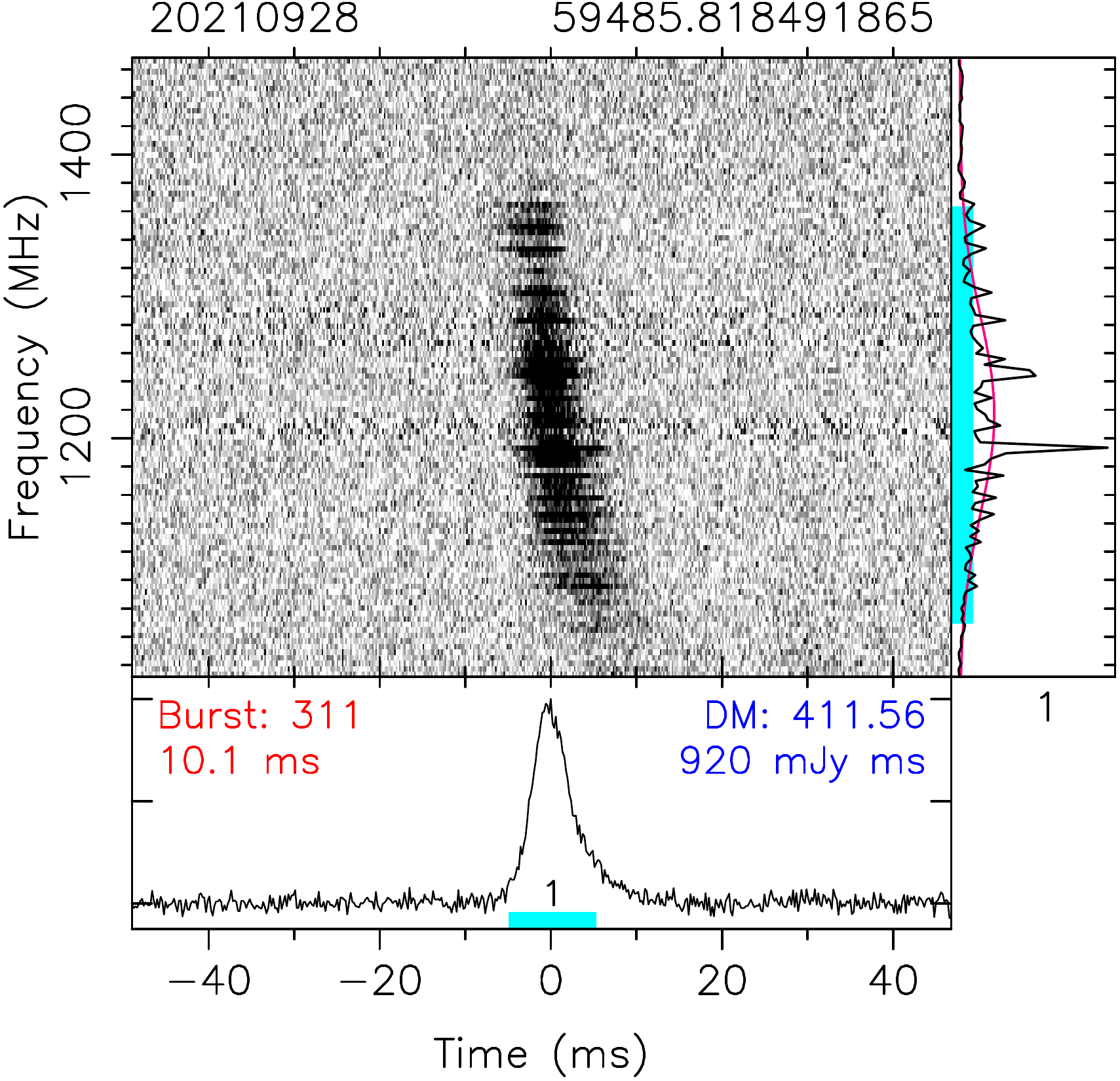}
    \includegraphics[height=37mm]{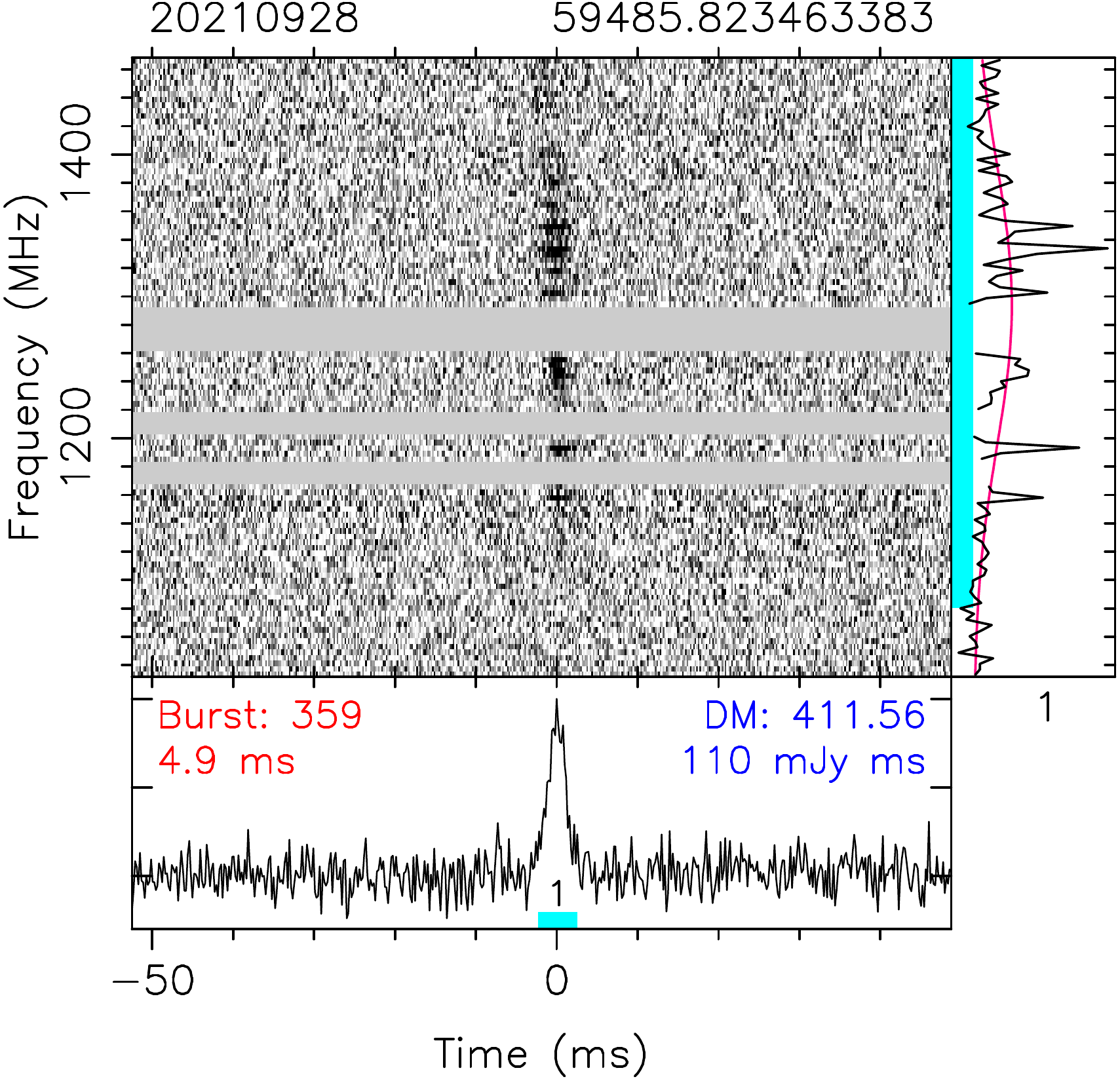}
\caption{\it{ -- continued and ended}.
}
\end{figure*}

\begin{figure*}
    \flushleft
    \includegraphics[height=37mm]{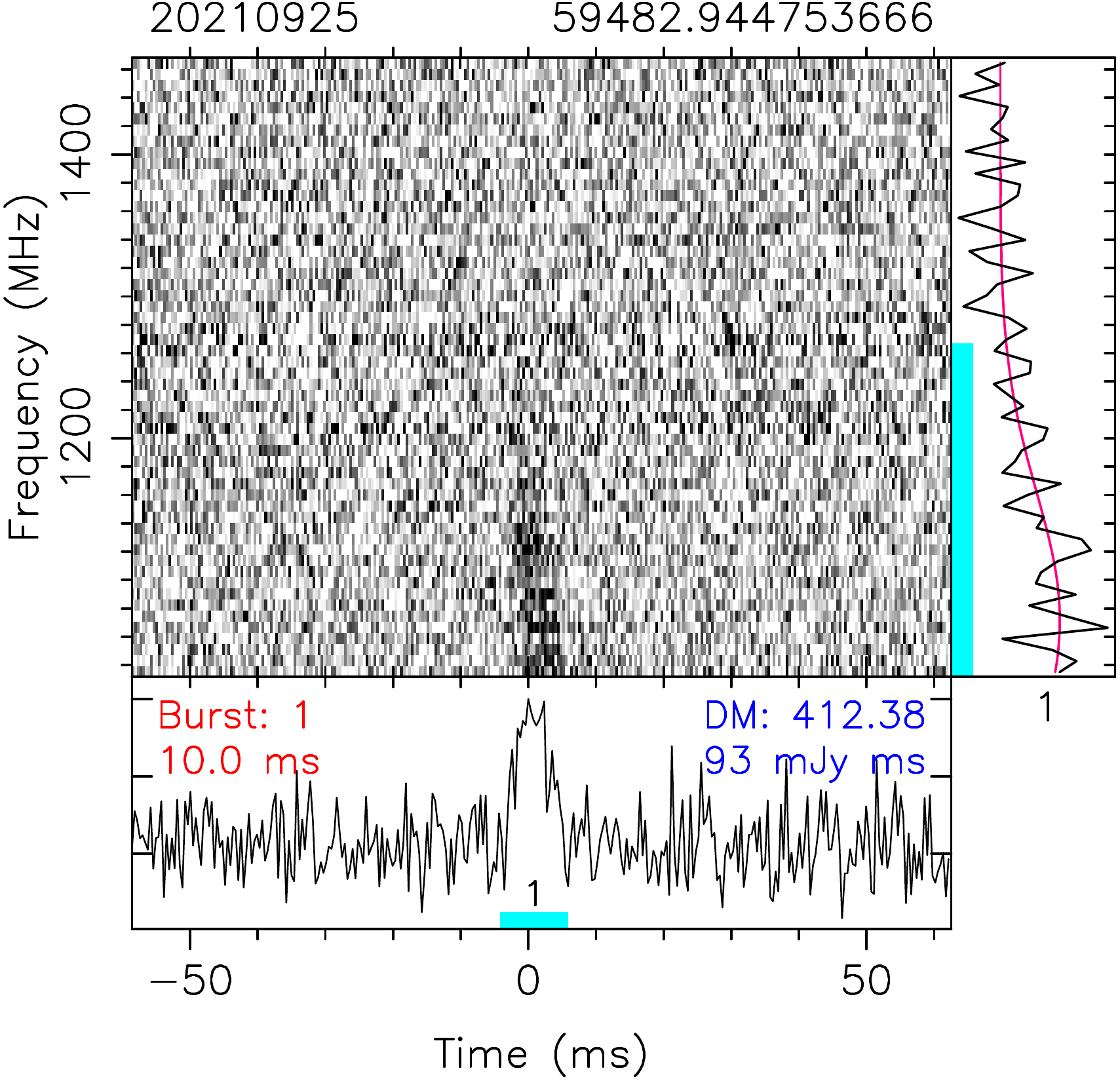}
    \includegraphics[height=37mm]{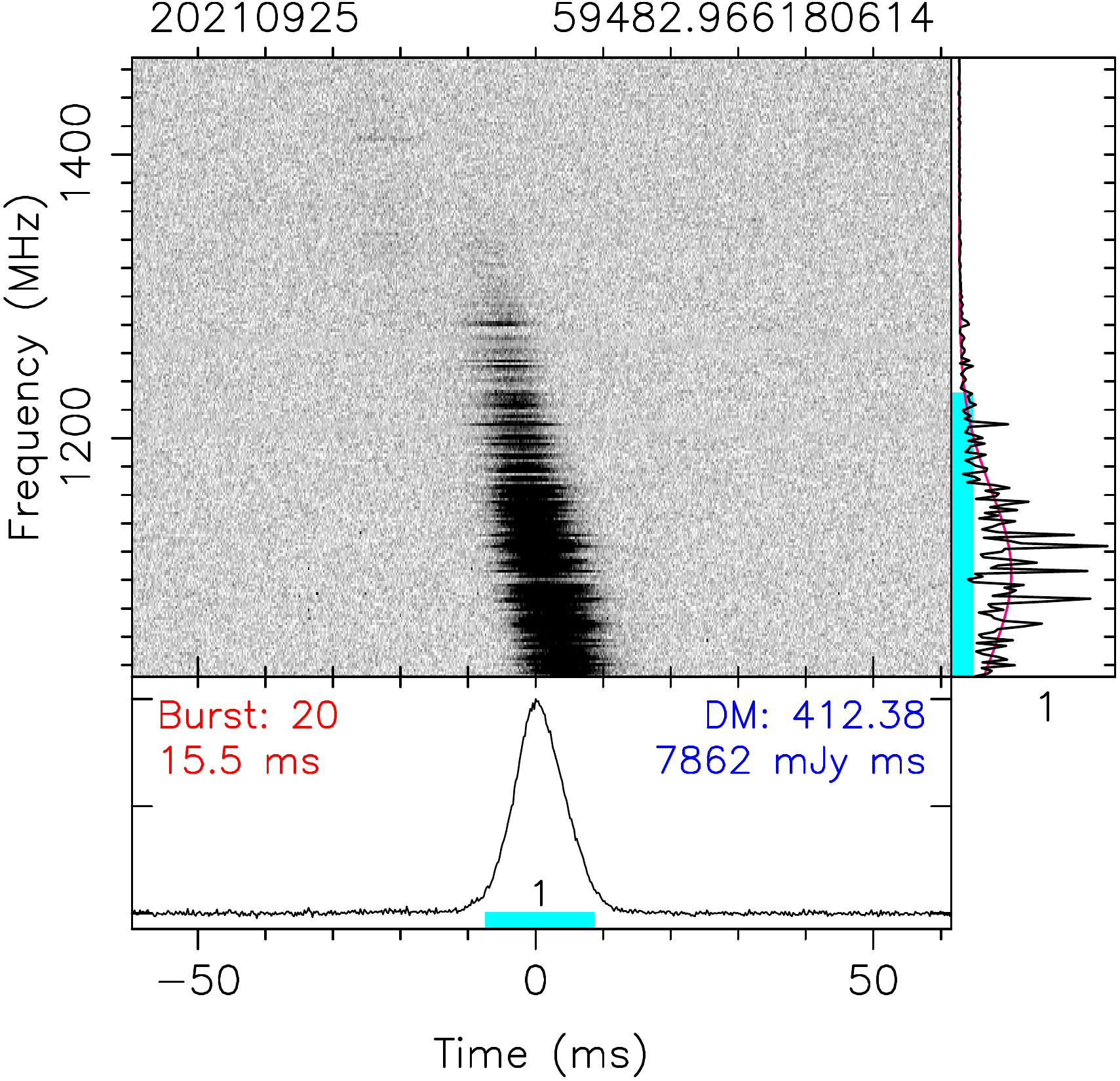}
    \includegraphics[height=37mm]{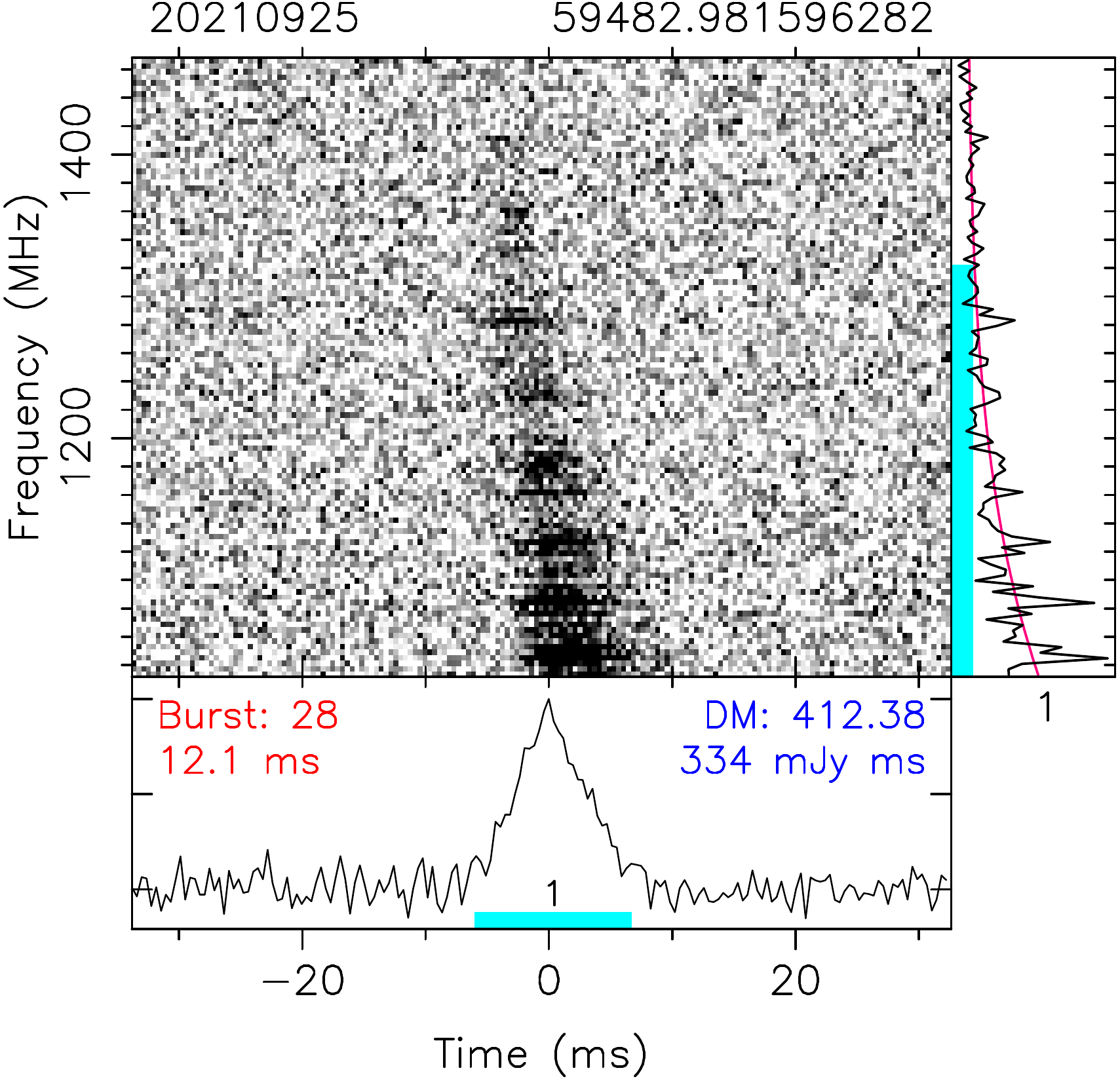}
    \includegraphics[height=37mm]{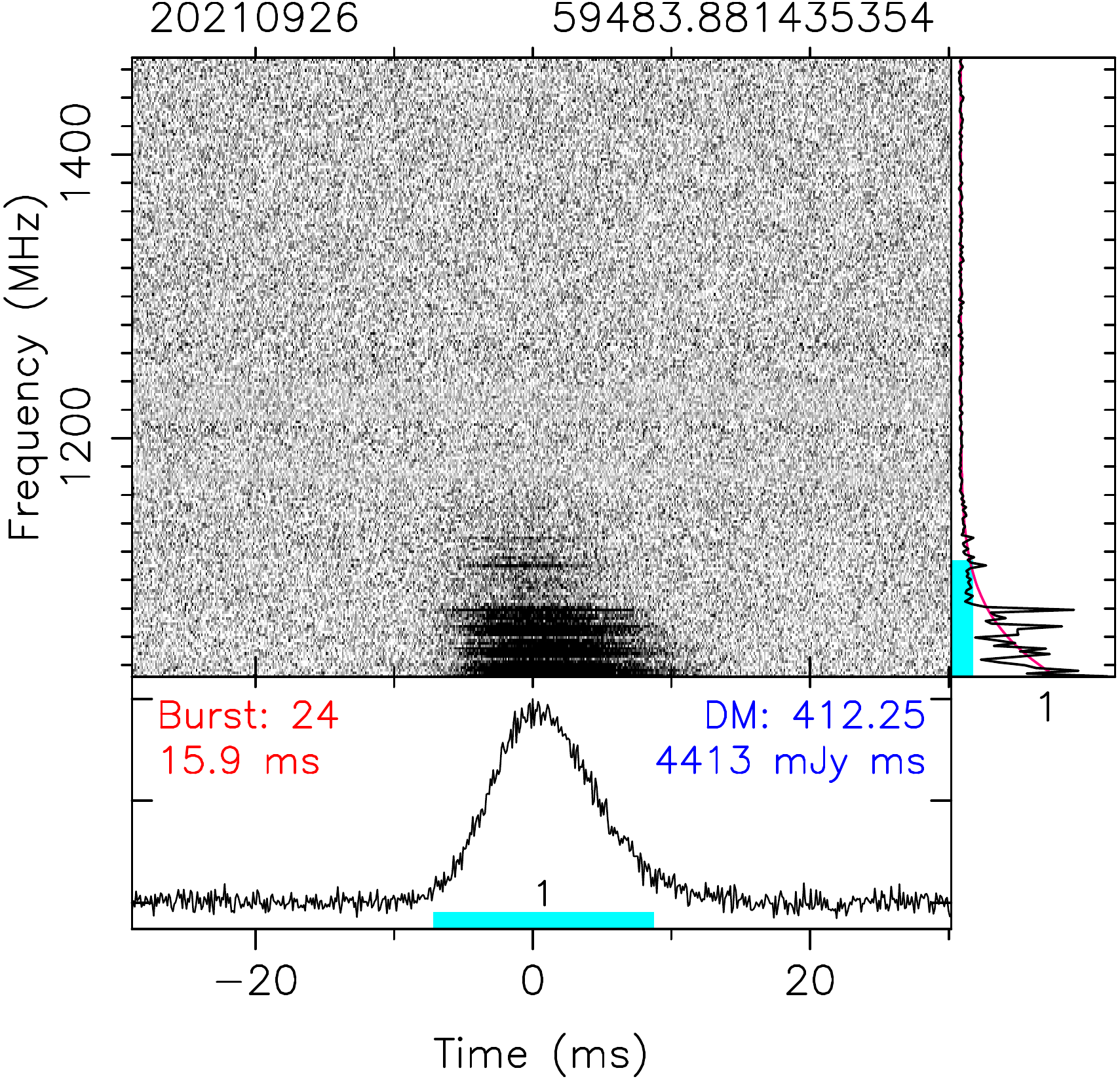}
    \includegraphics[height=37mm]{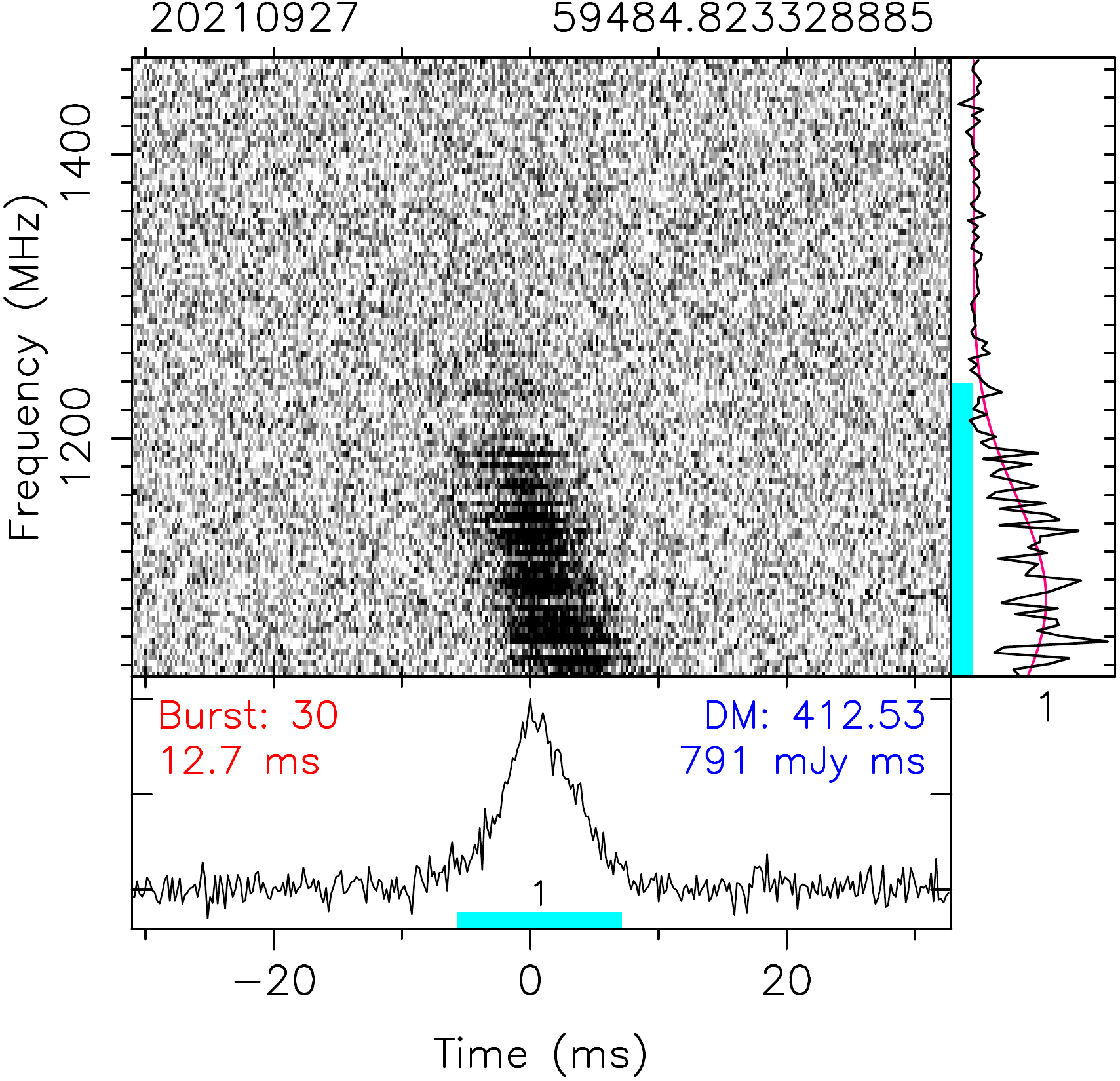}
    \includegraphics[height=37mm]{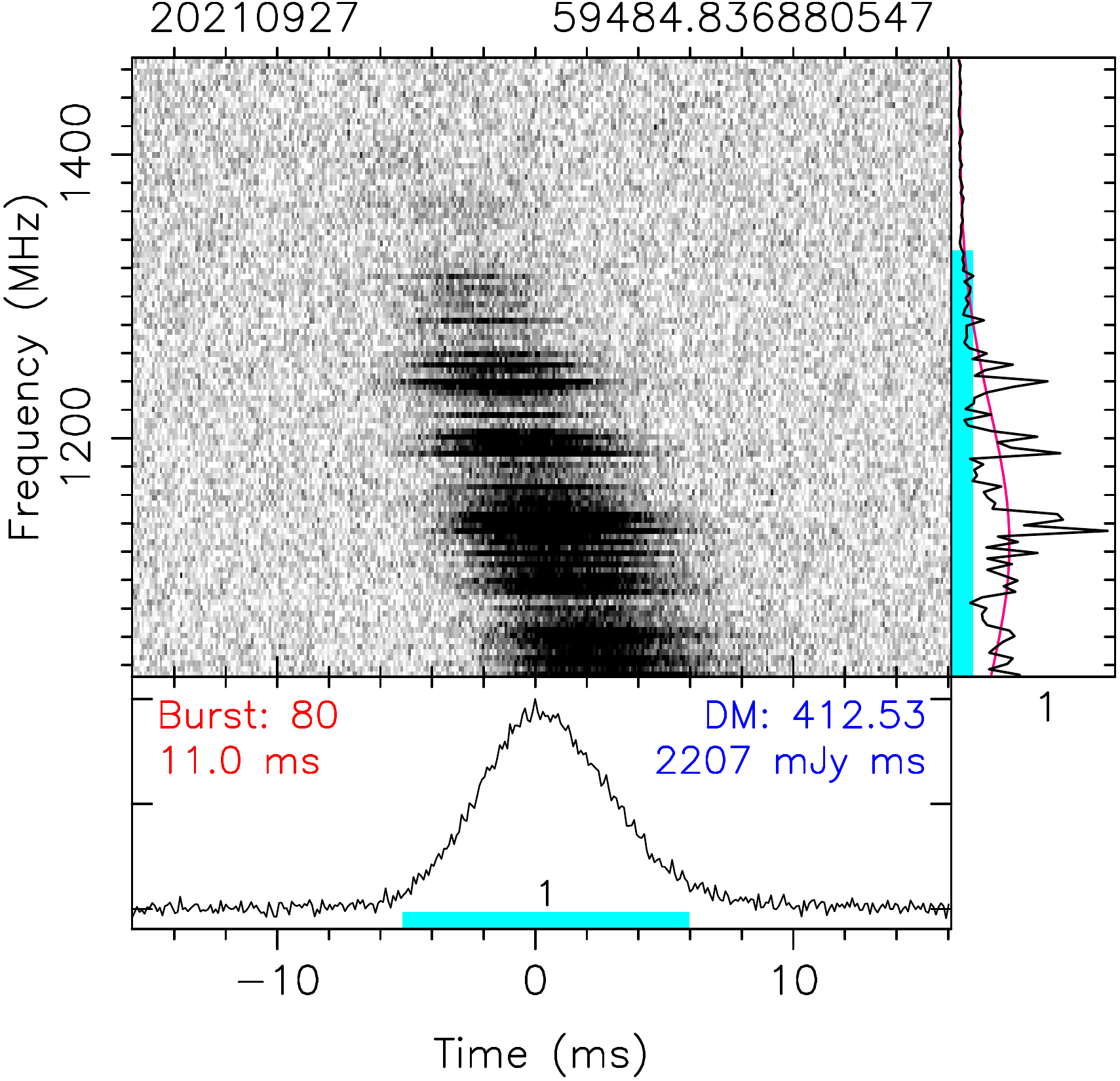}
    \includegraphics[height=37mm]{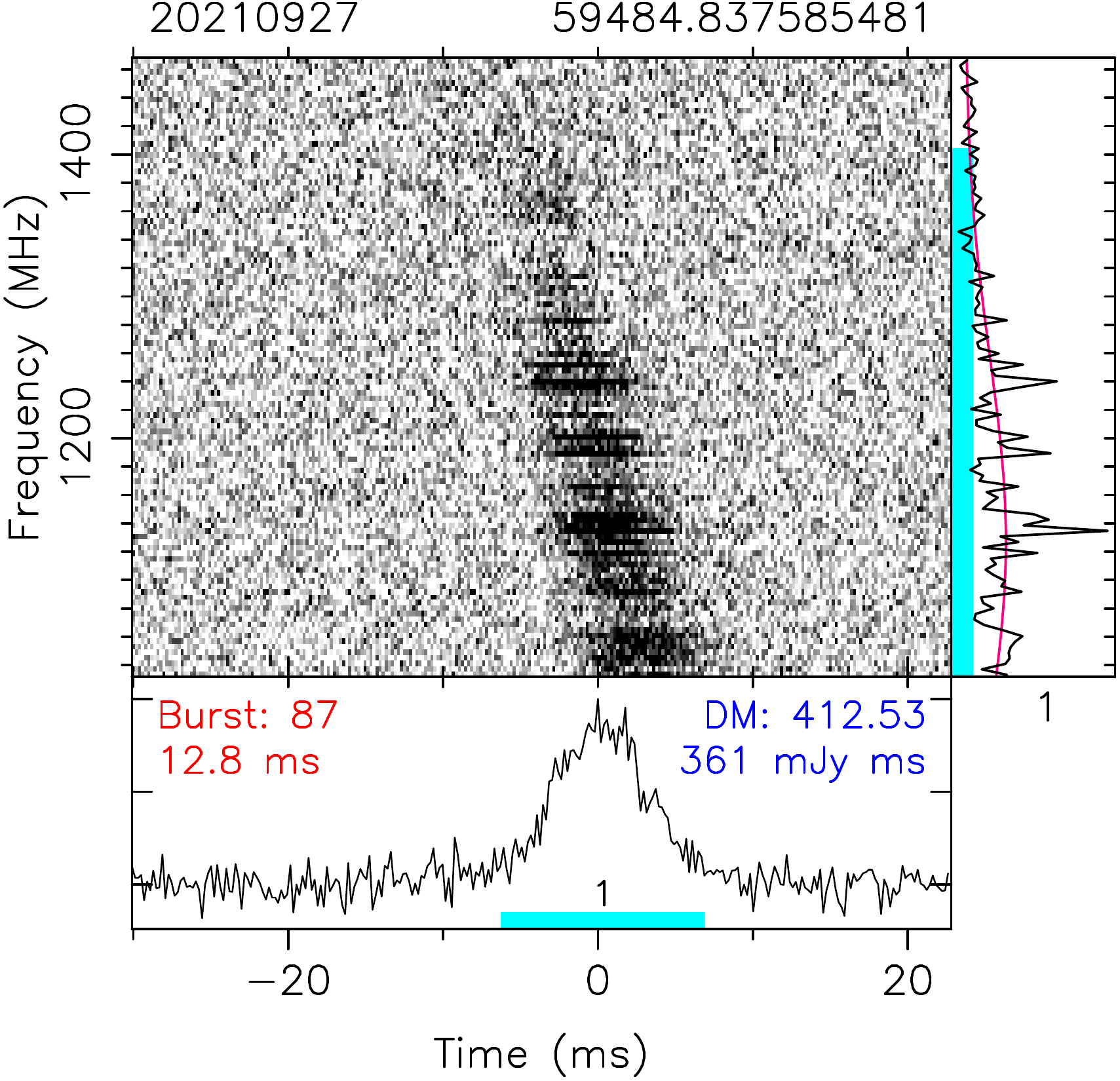}
    \includegraphics[height=37mm]{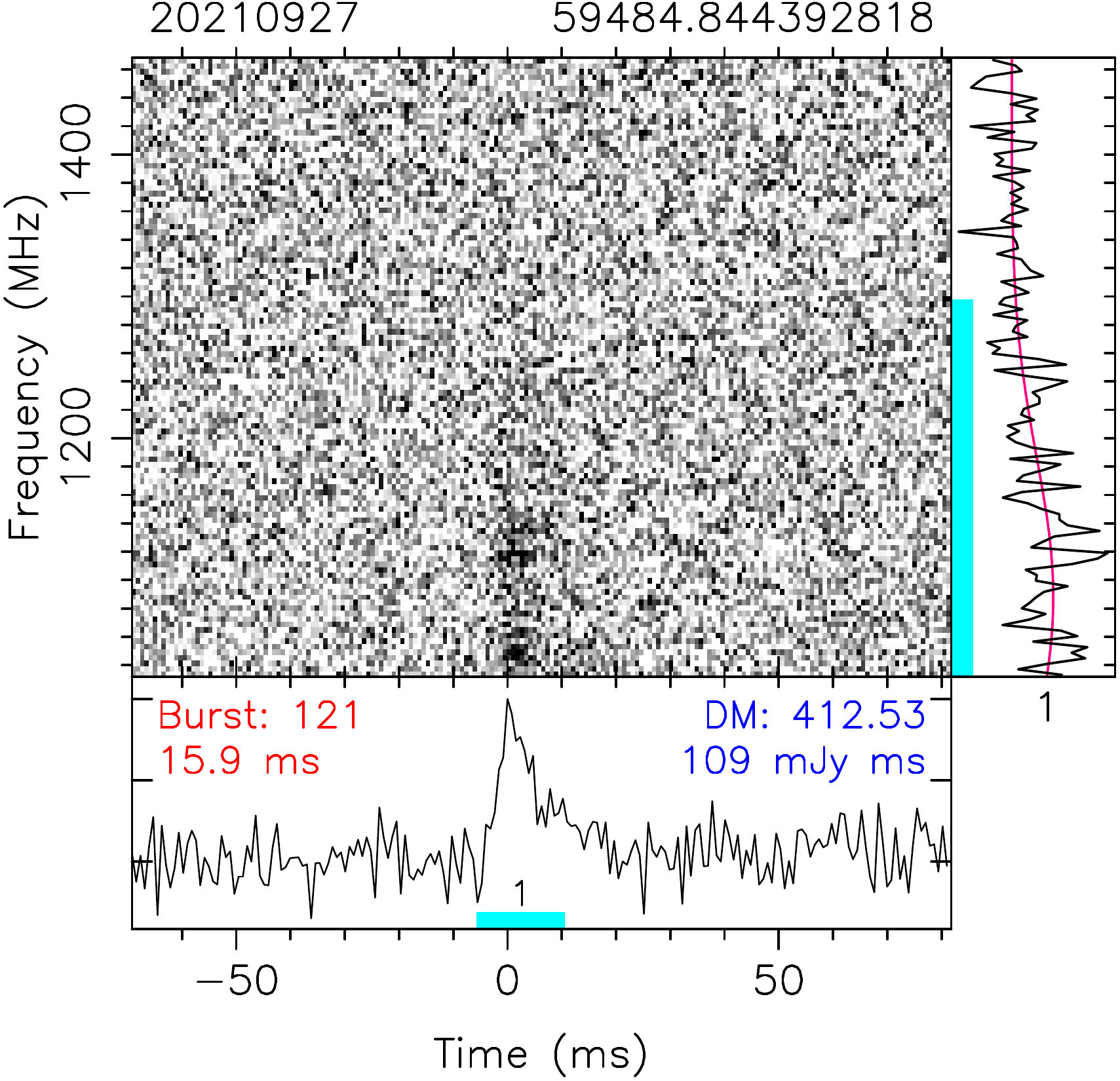}
    \includegraphics[height=37mm]{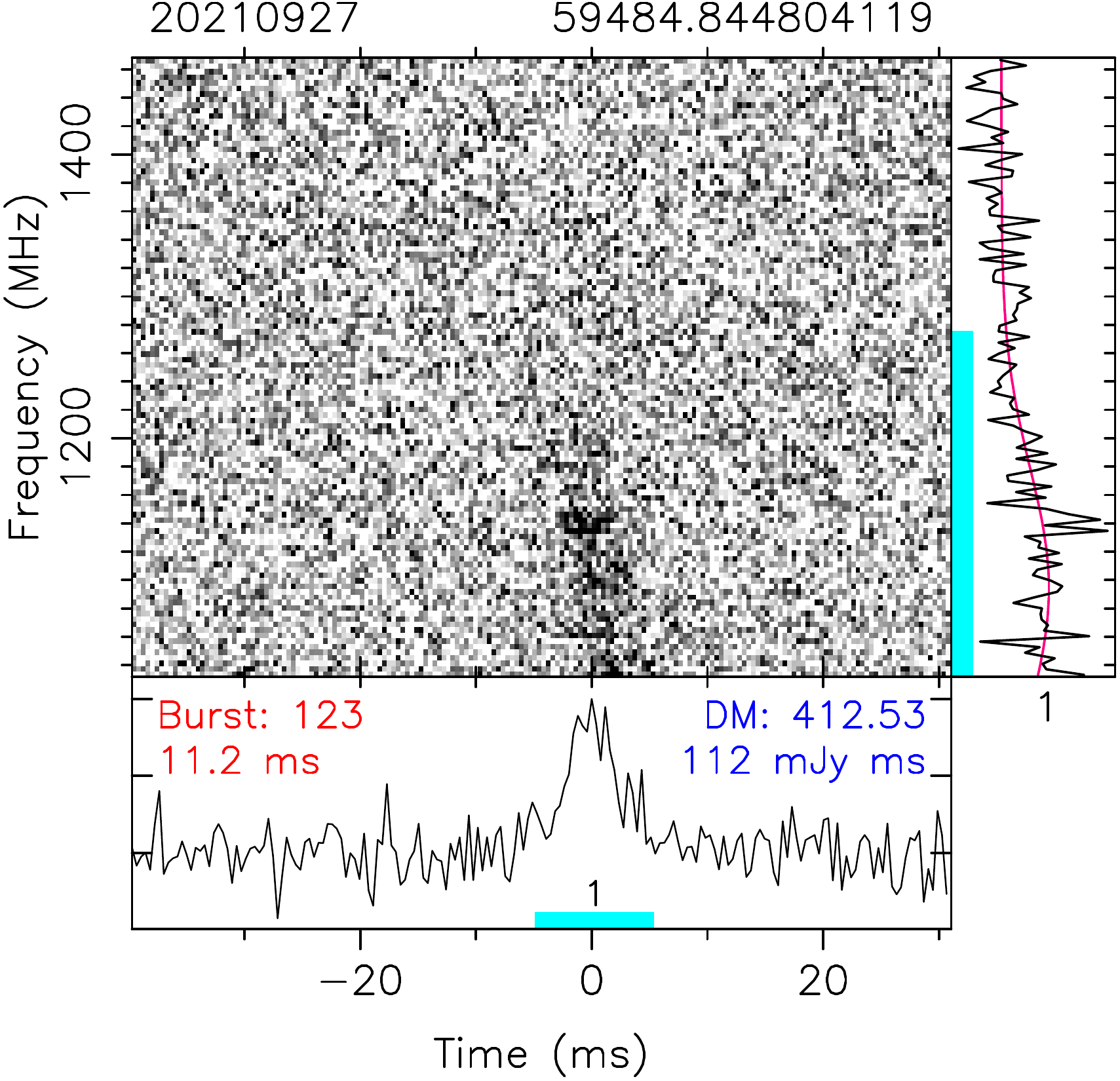}
    \includegraphics[height=37mm]{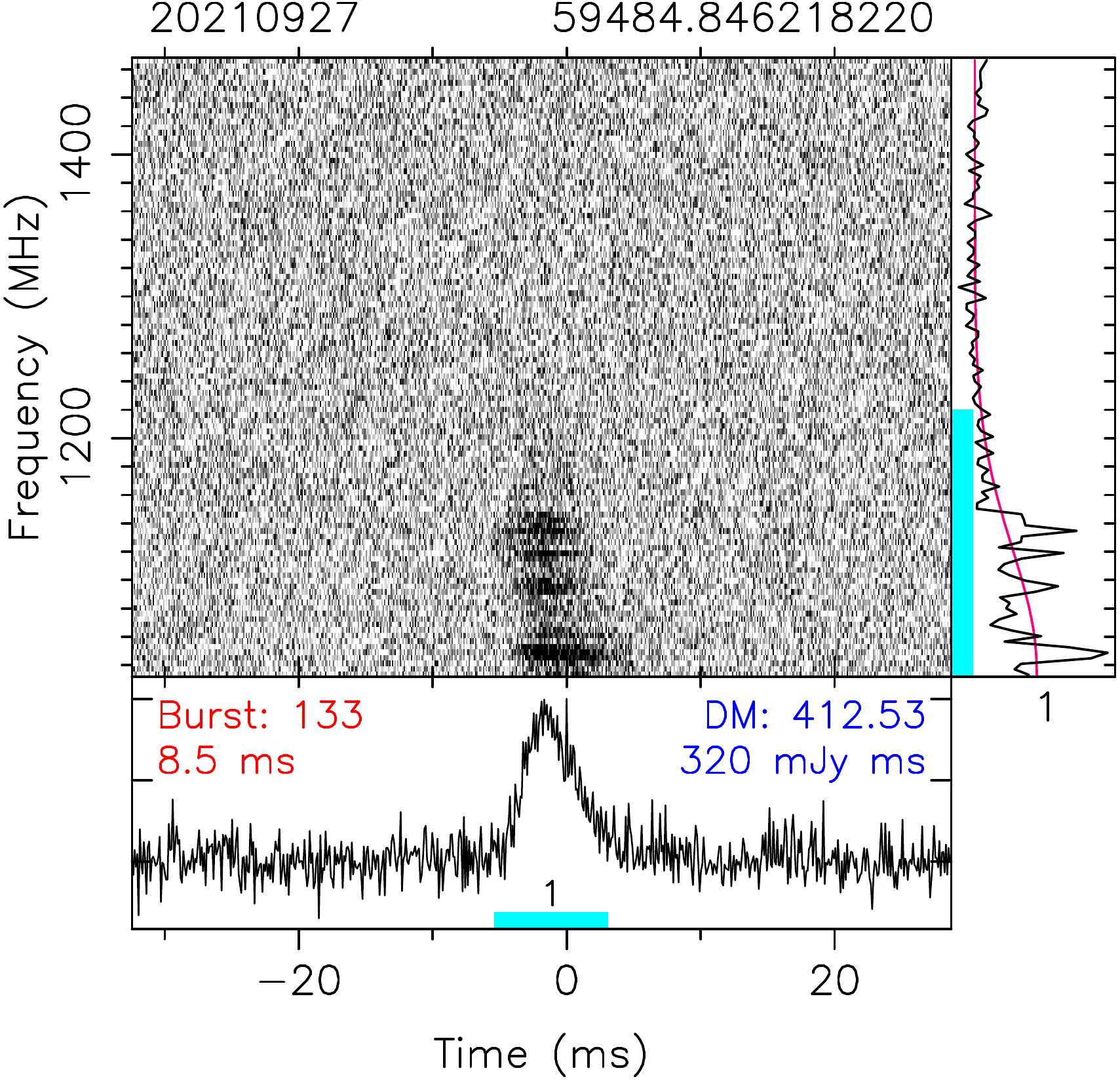}
    \includegraphics[height=37mm]{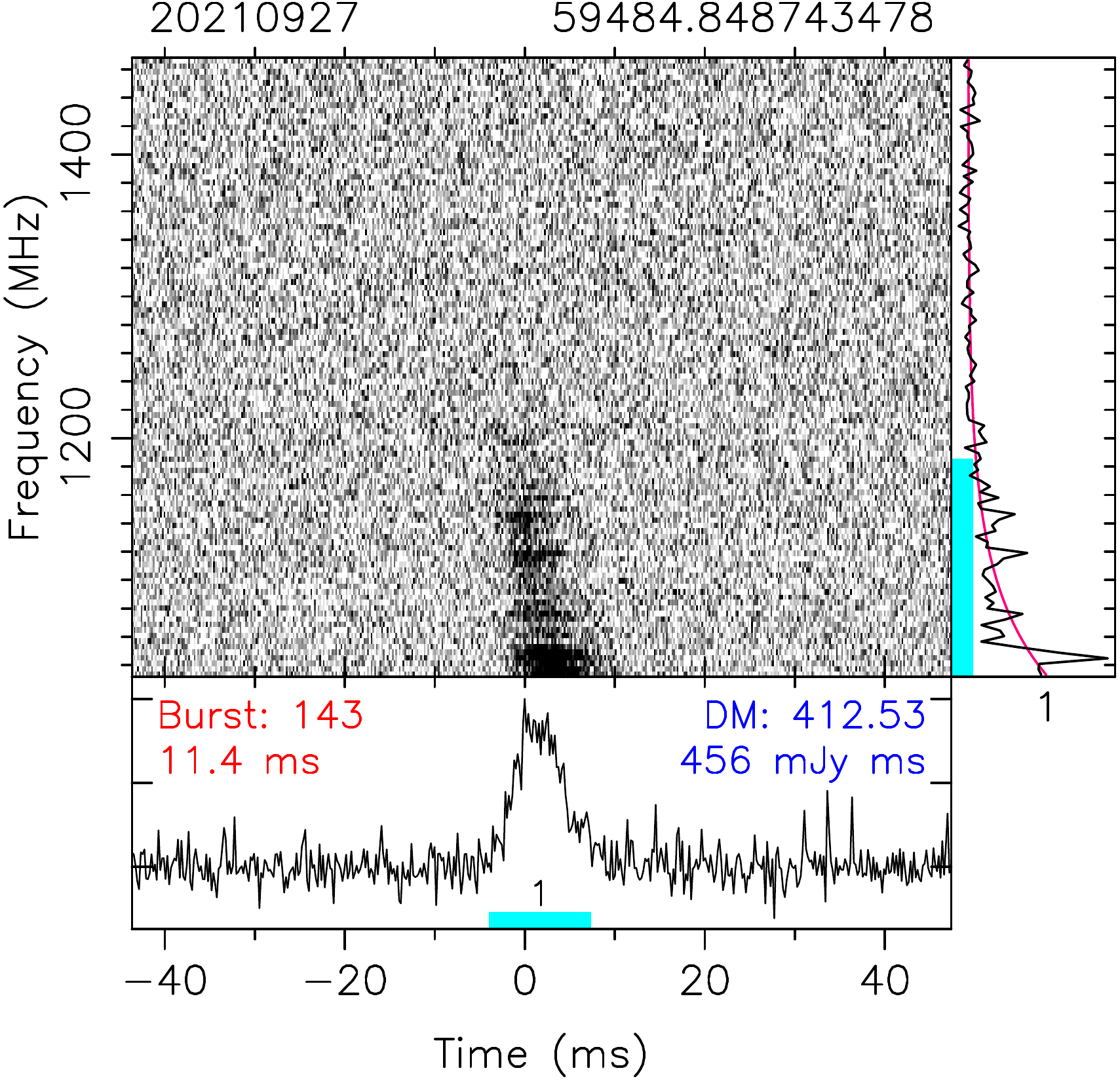}
    \includegraphics[height=37mm]{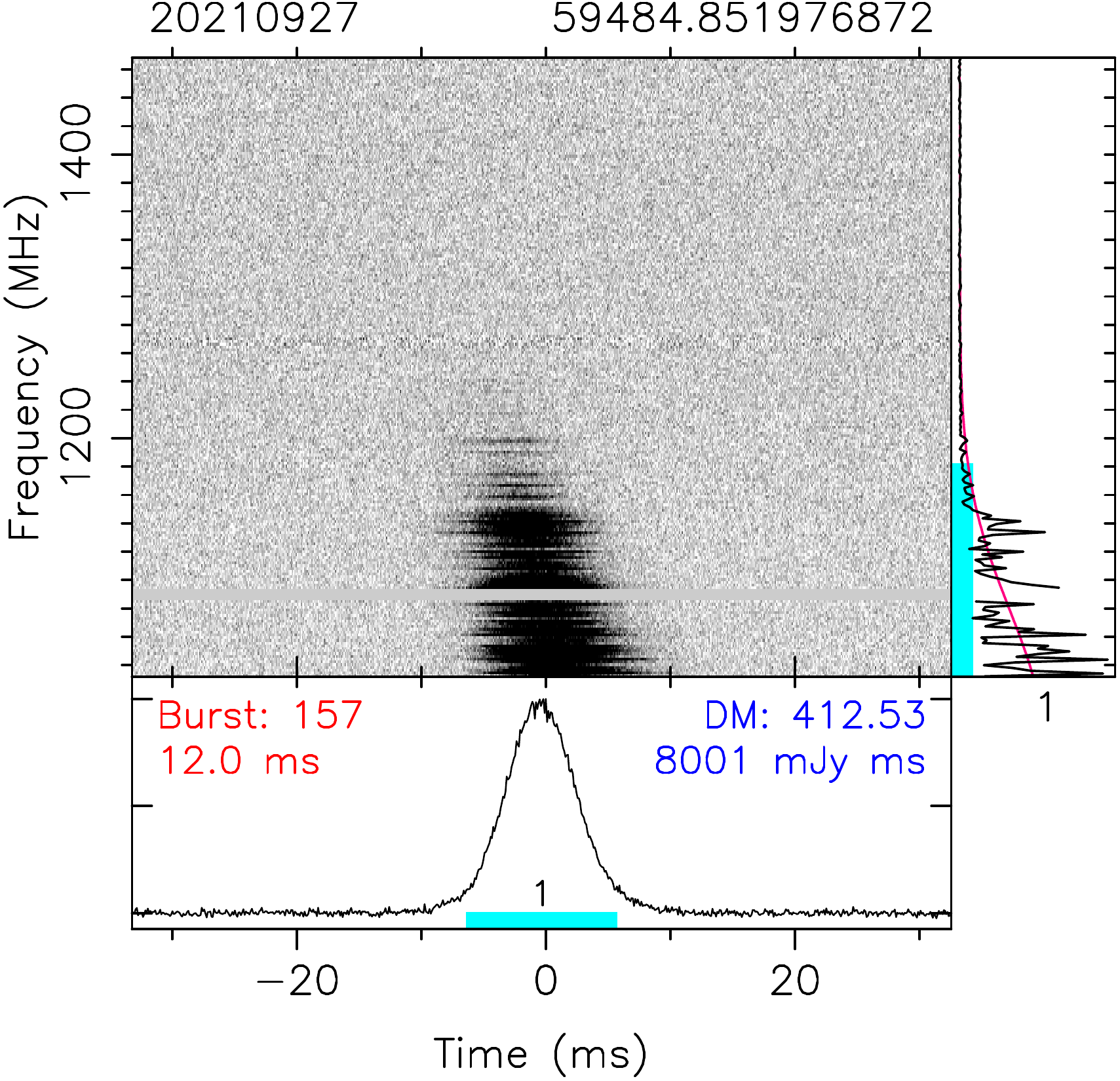}
    \includegraphics[height=37mm]{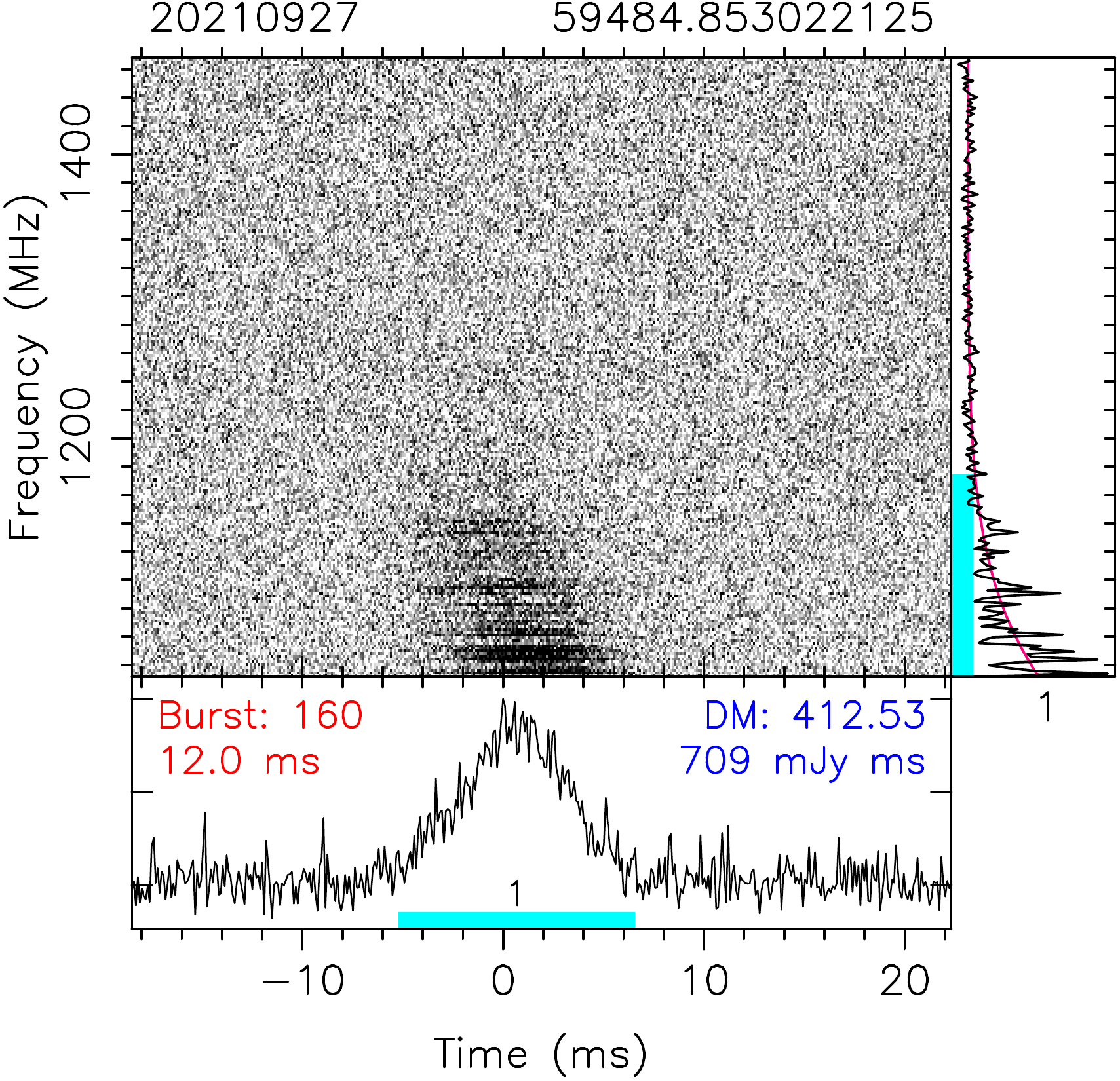}
    \includegraphics[height=37mm]{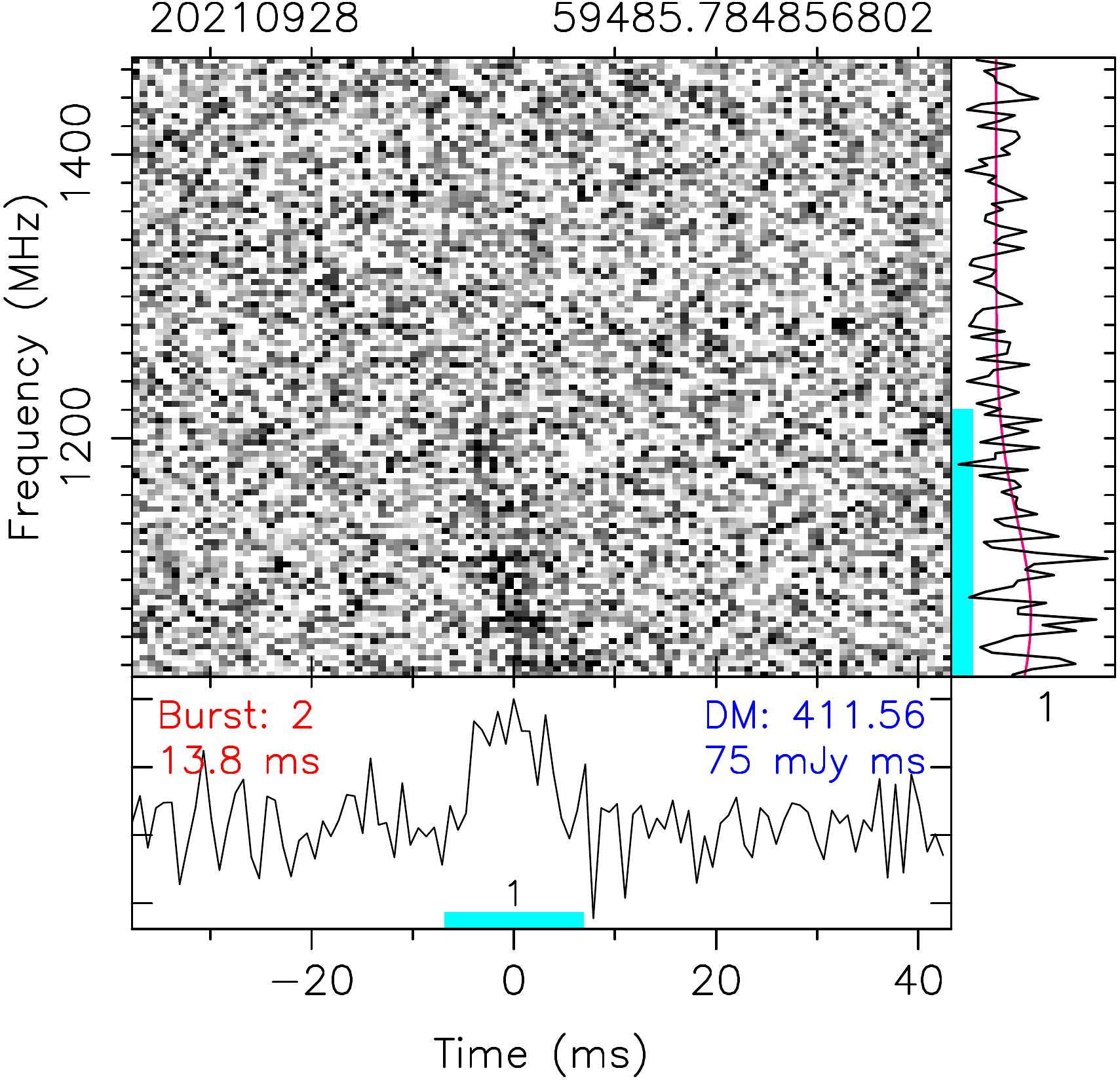}
    \includegraphics[height=37mm]{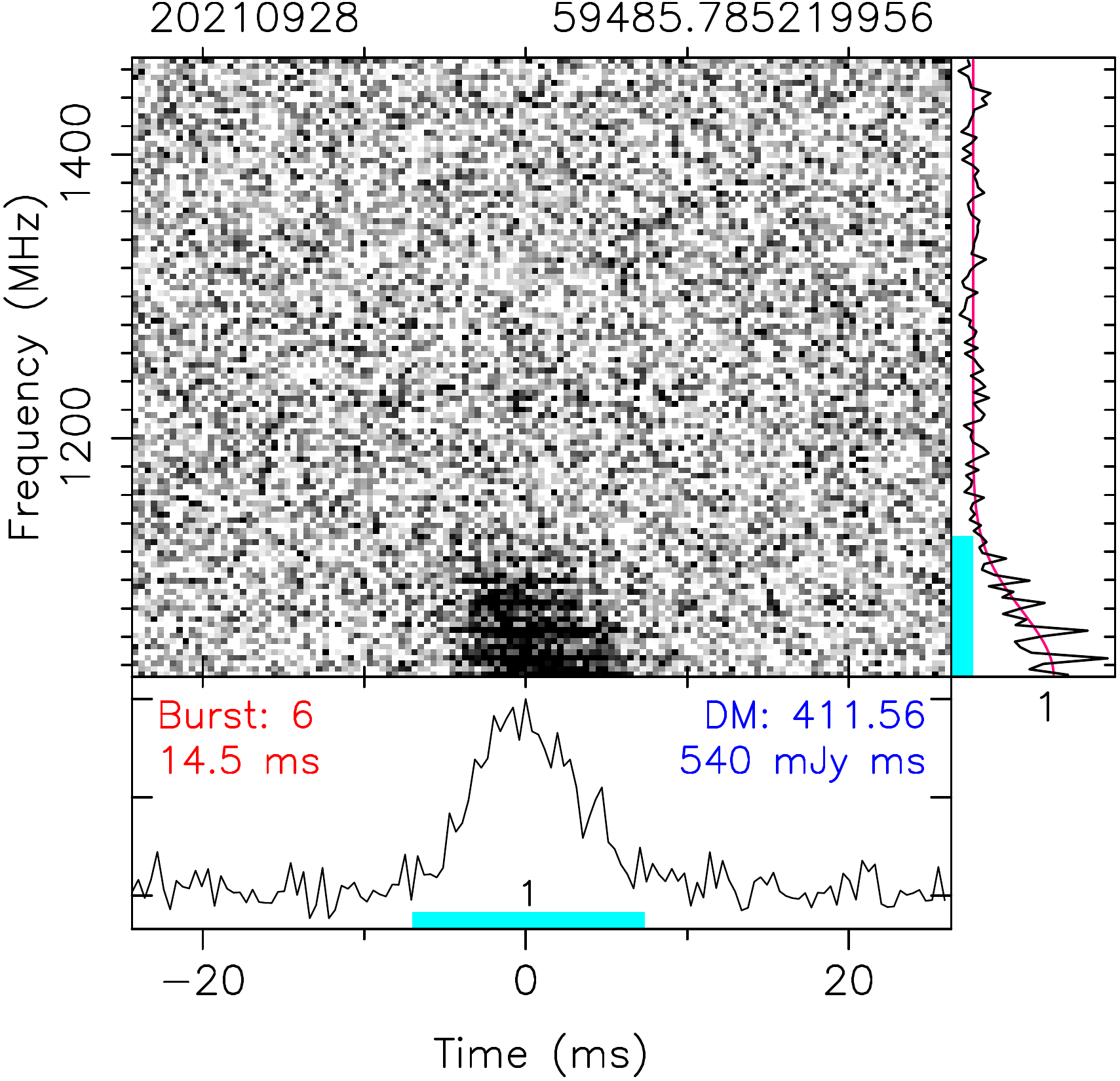}
    \includegraphics[height=37mm]{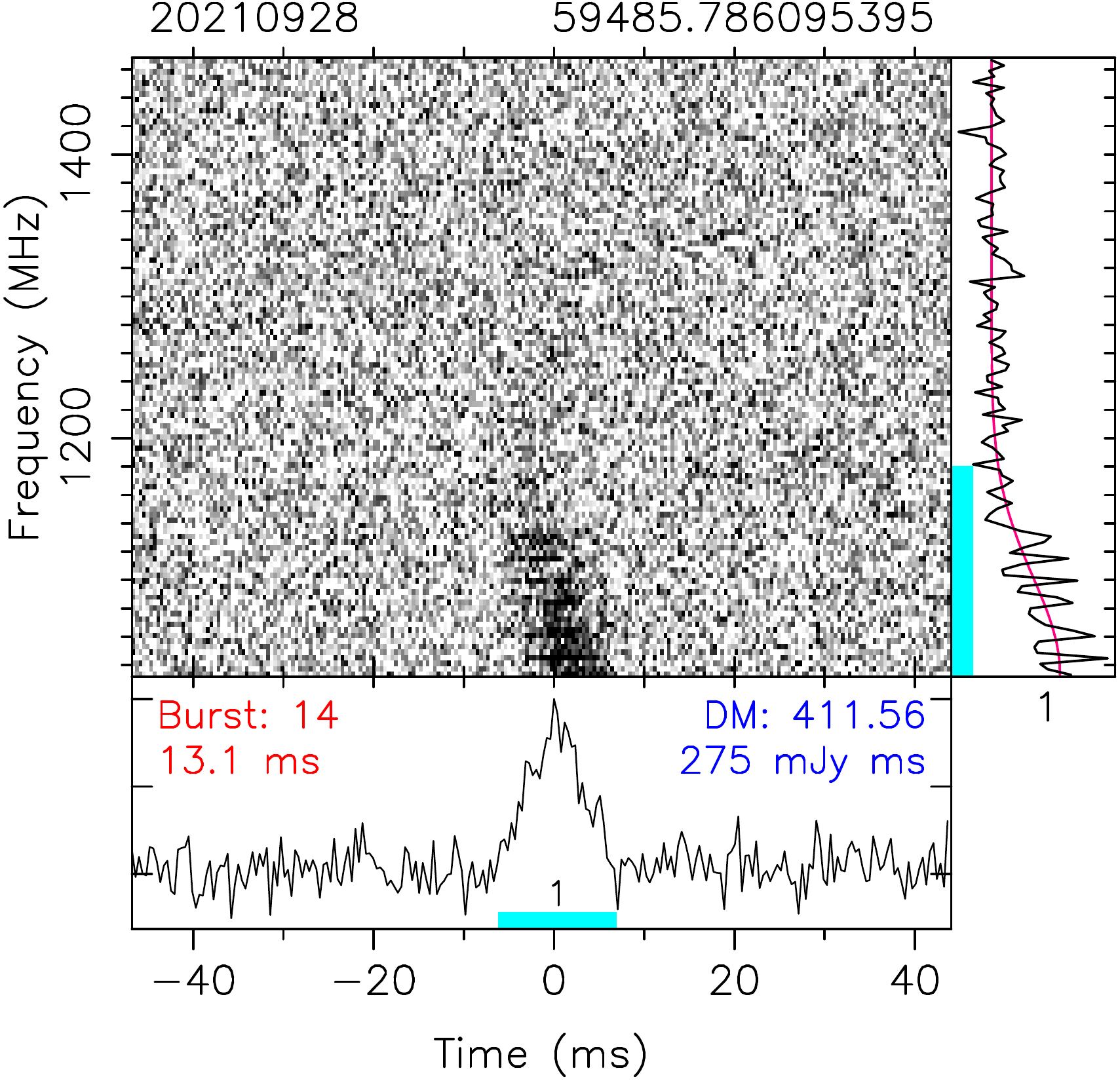}
    \includegraphics[height=37mm]{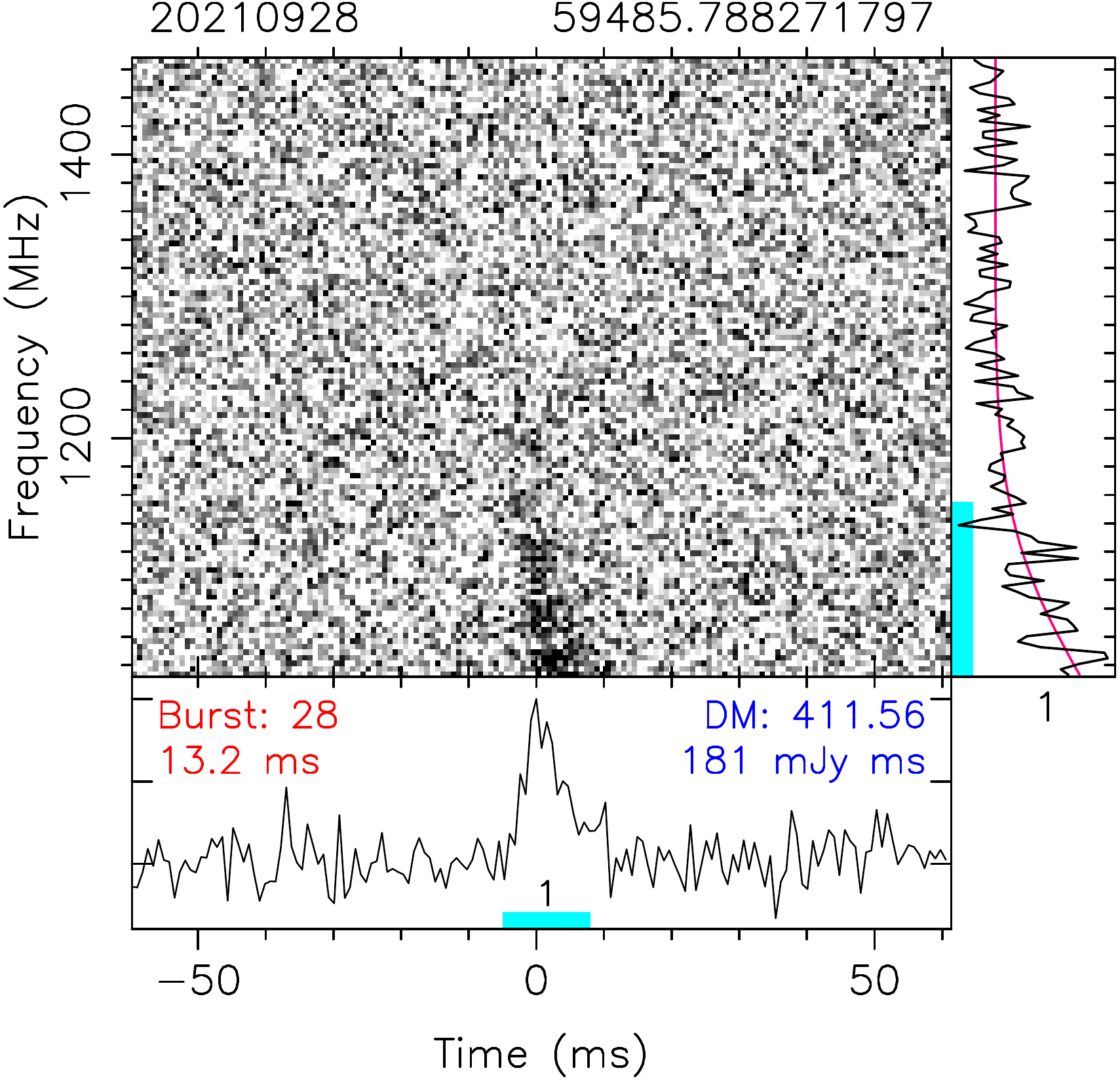}
    \includegraphics[height=37mm]{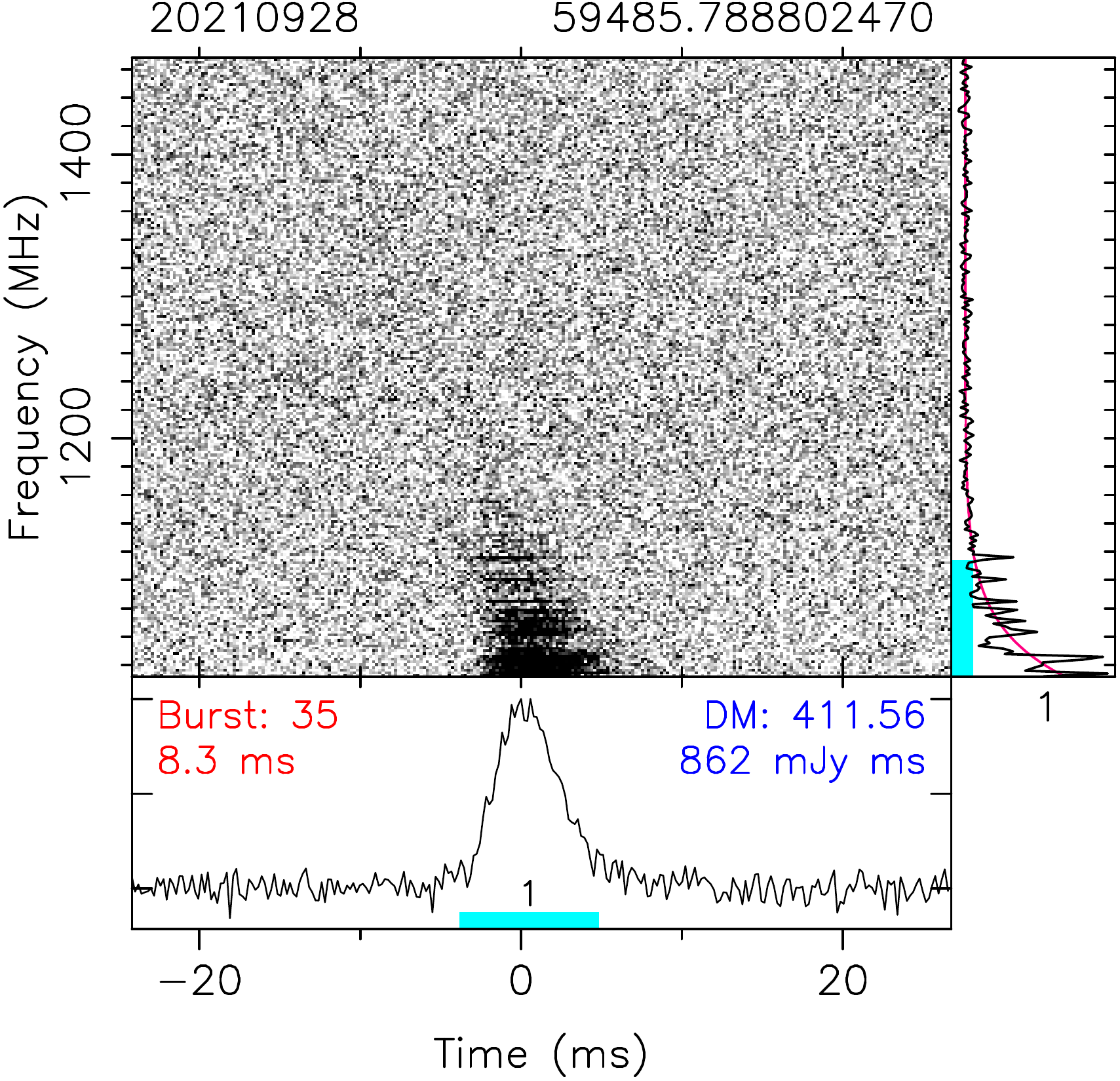}
    \includegraphics[height=37mm]{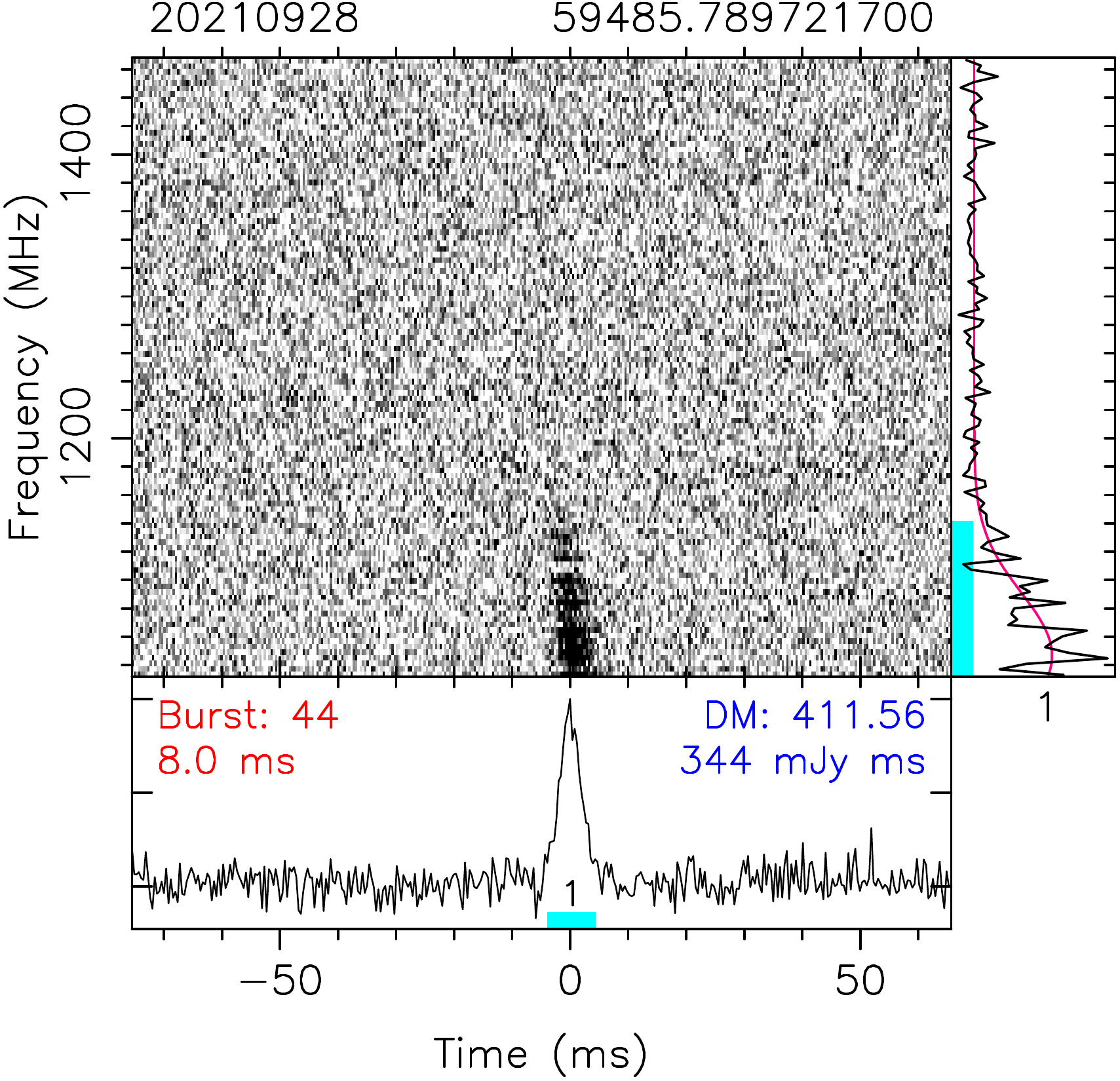}
    \includegraphics[height=37mm]{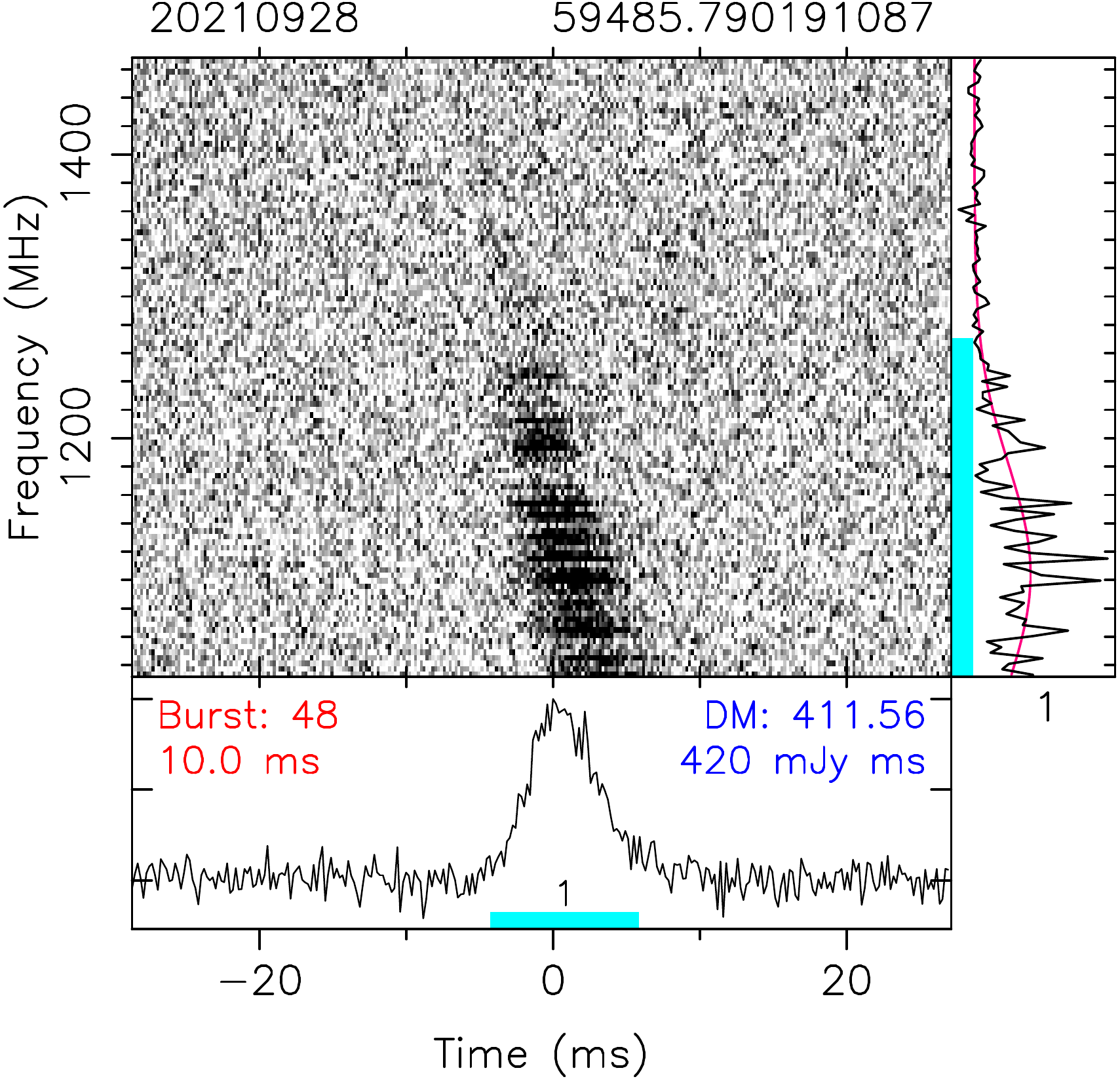}
    \includegraphics[height=37mm]{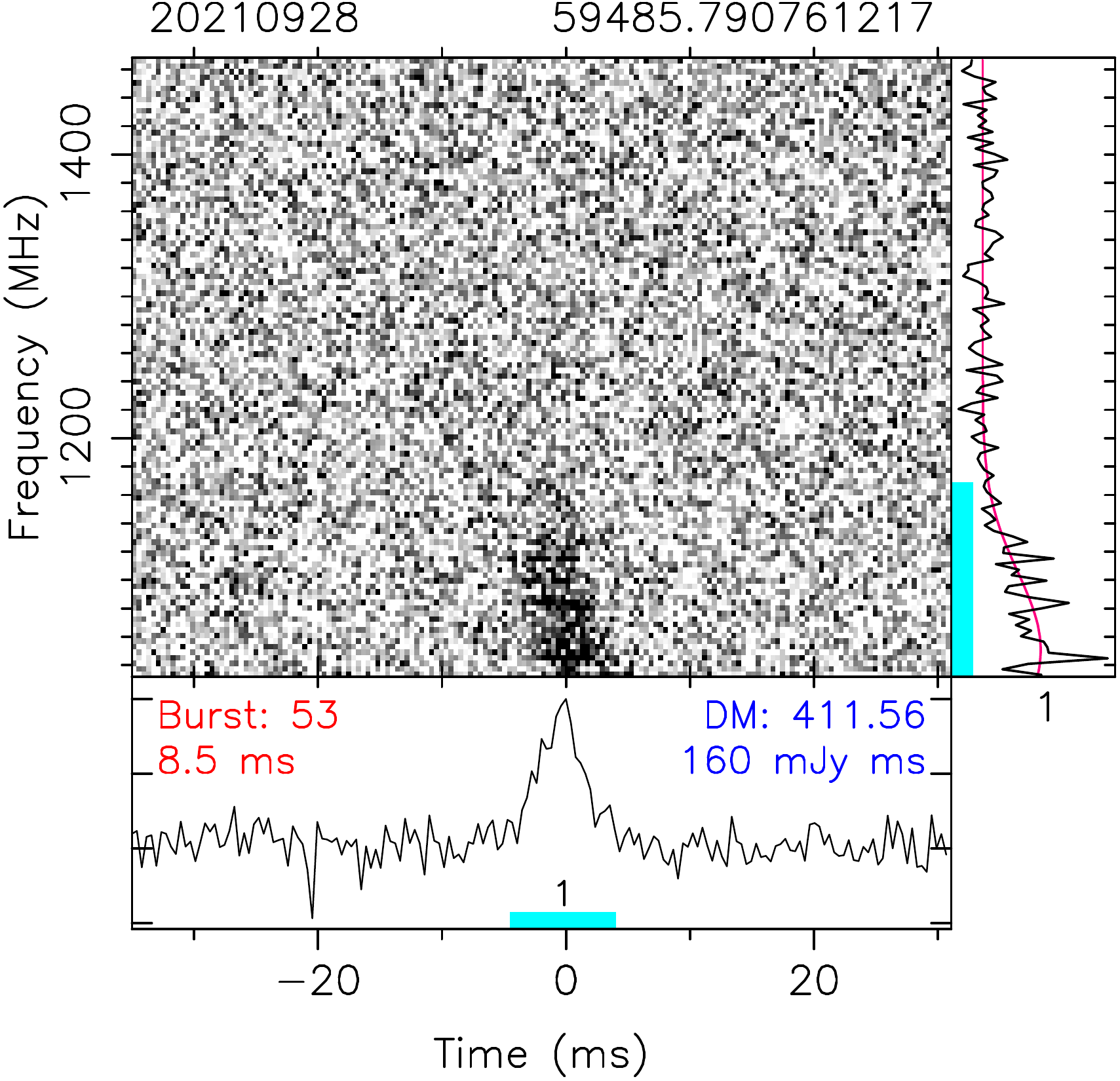}
    \includegraphics[height=37mm]{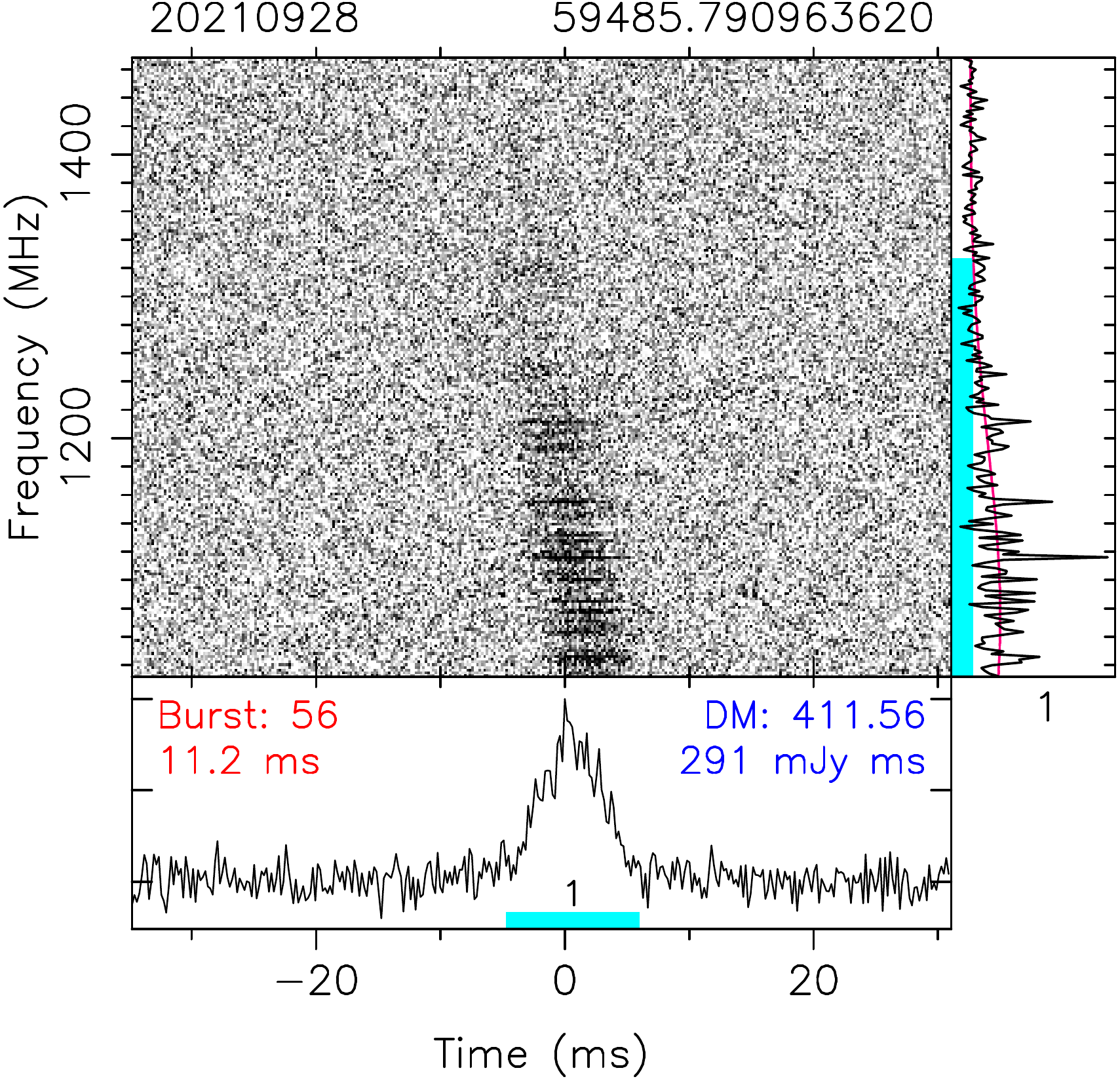}
    \includegraphics[height=37mm]{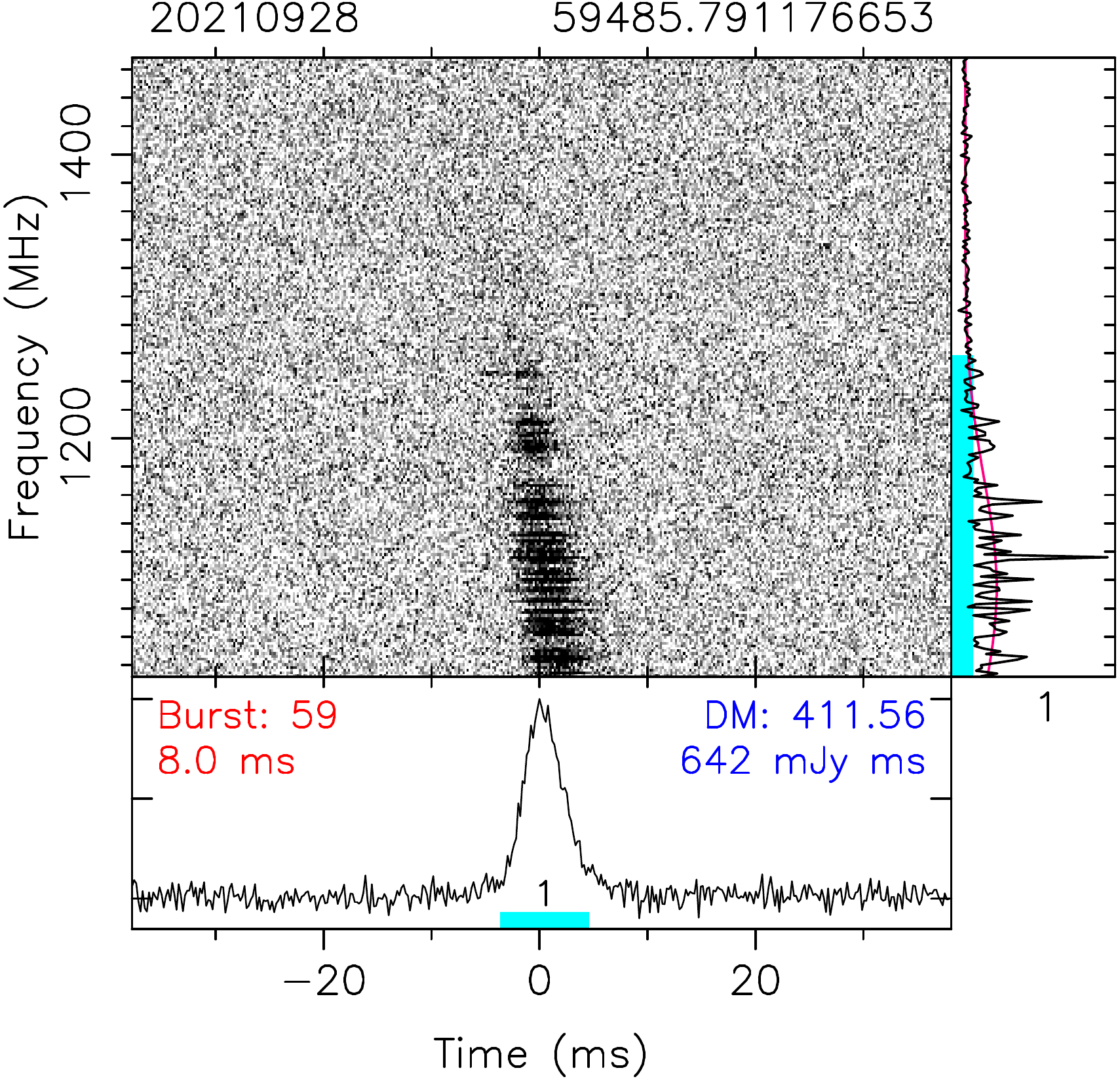}
    \includegraphics[height=37mm]{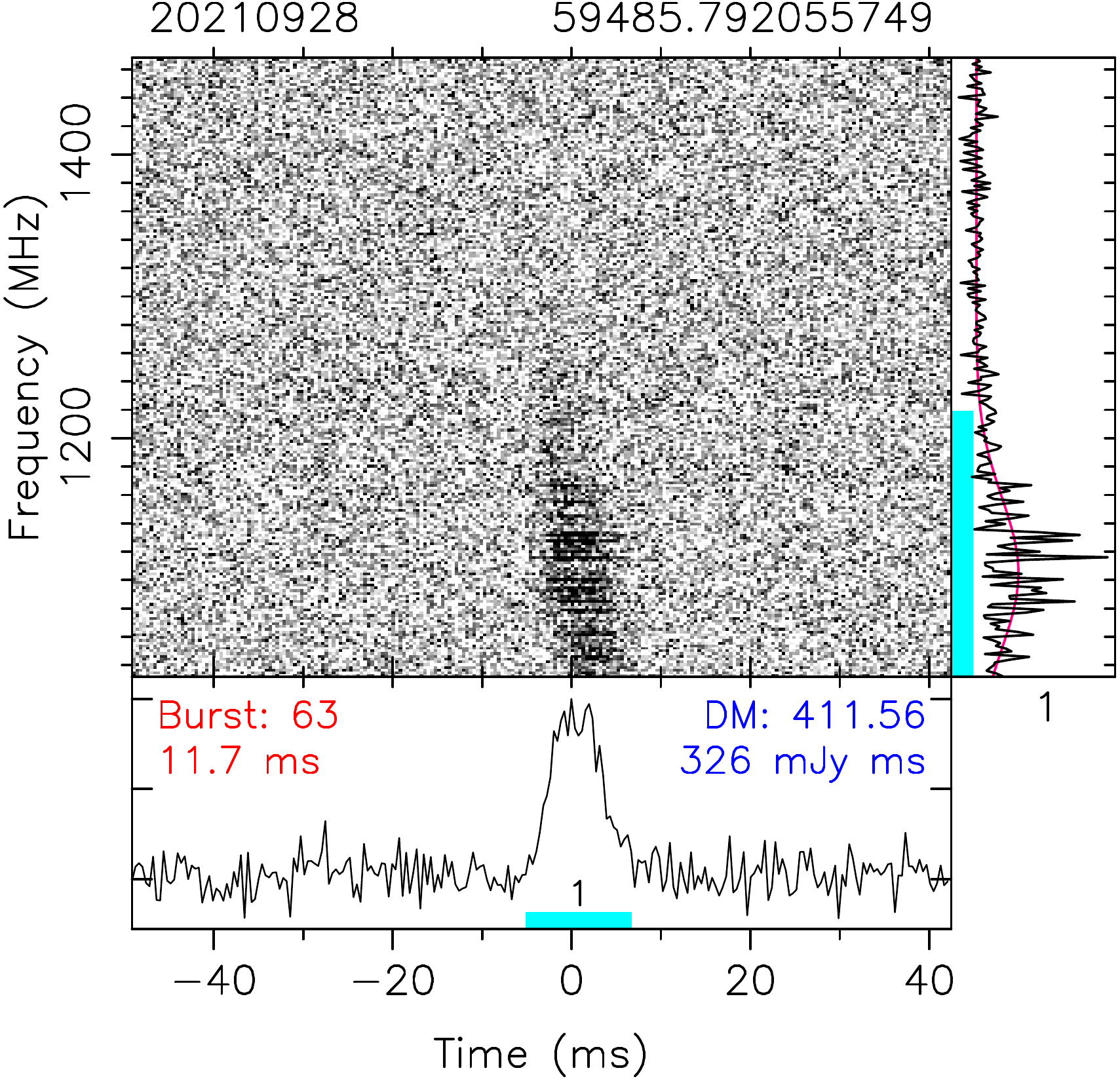}
  \caption{The same as Figure~\ref{fig:appendix:D1W} but for bursts in D1-L.
 }
\label{fig:appendix:D1L}
\end{figure*}
\addtocounter{figure}{-1}
\begin{figure*}
    \flushleft
    \includegraphics[height=37mm]{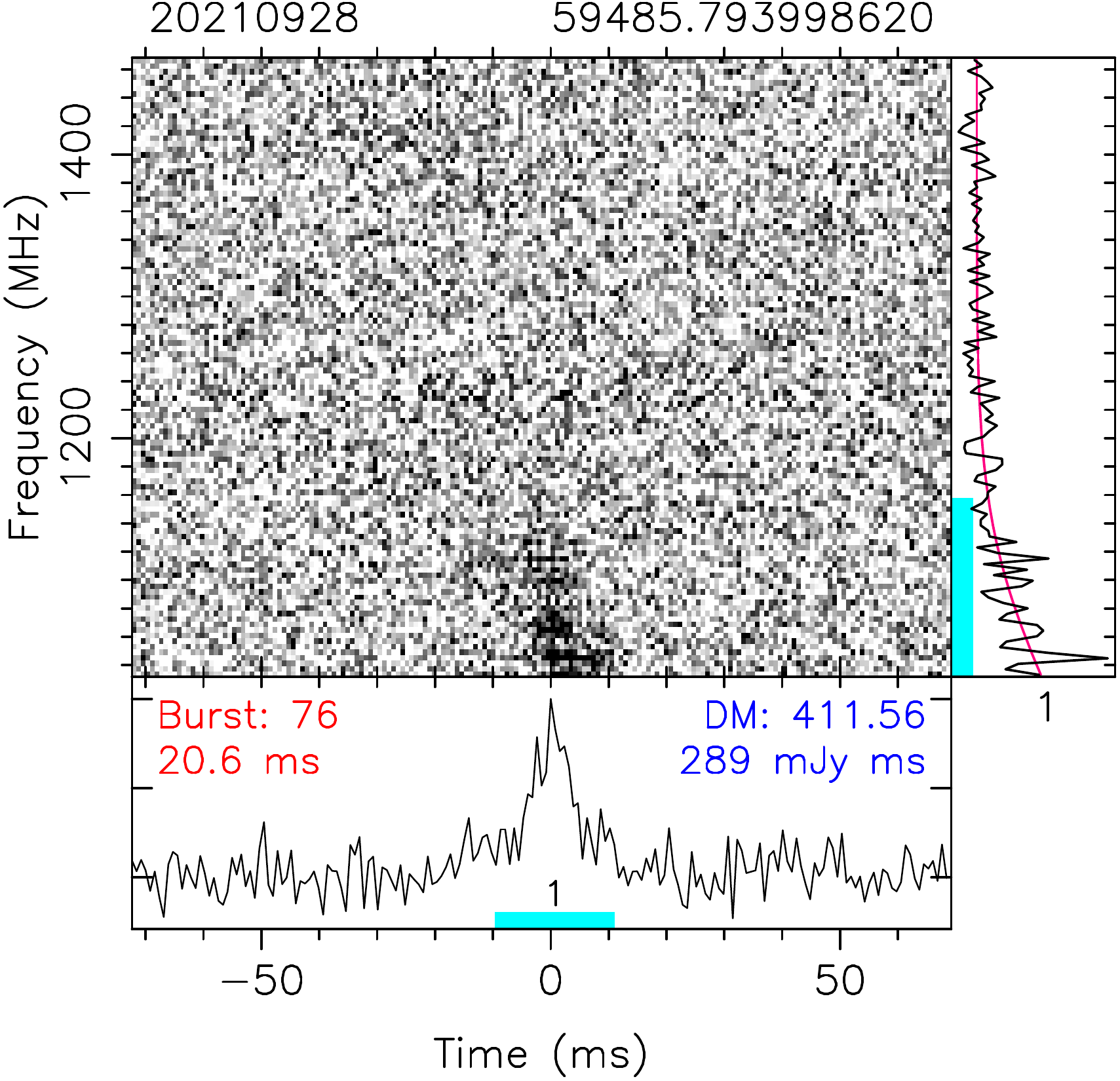}
    \includegraphics[height=37mm]{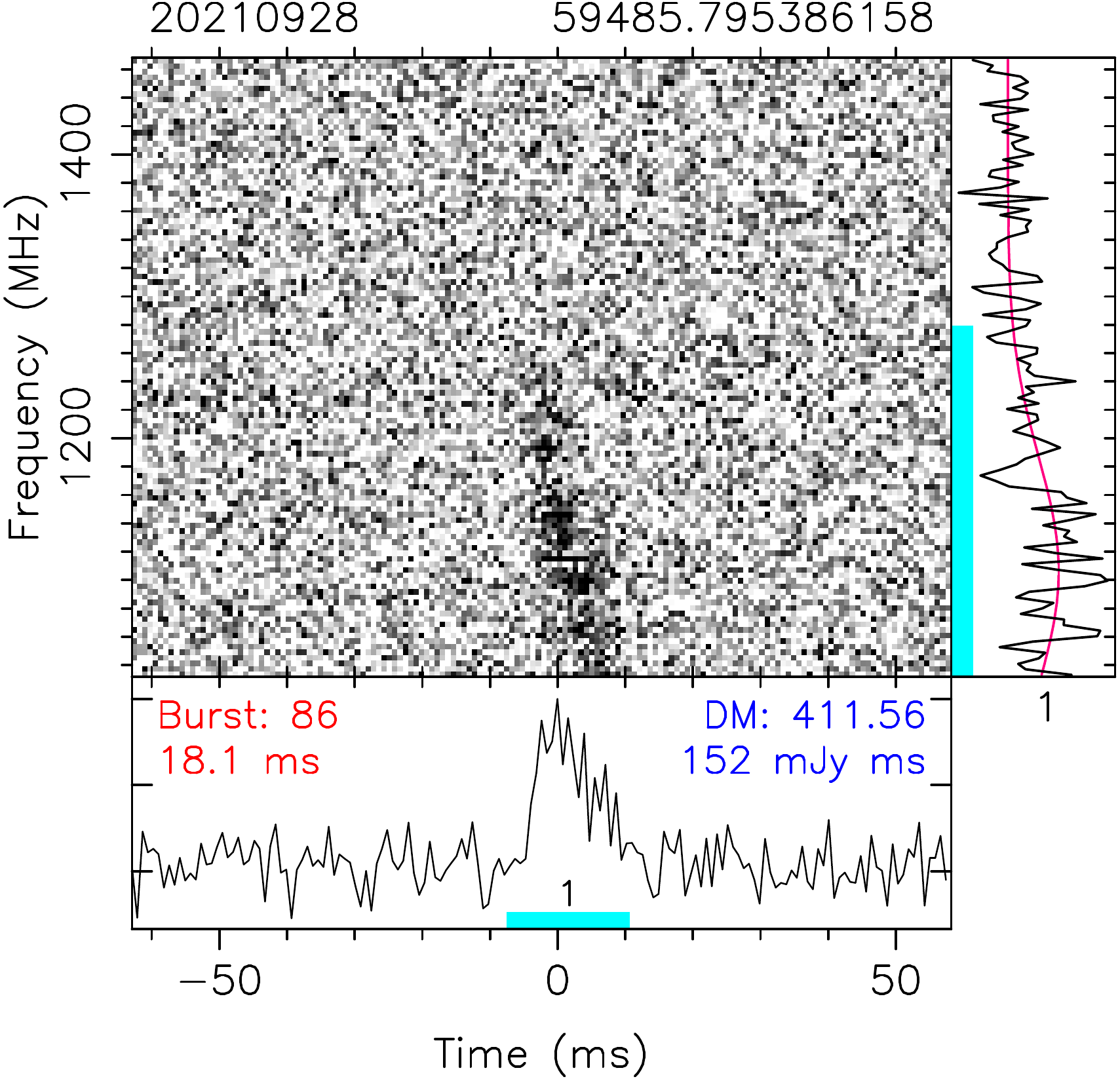}
    \includegraphics[height=37mm]{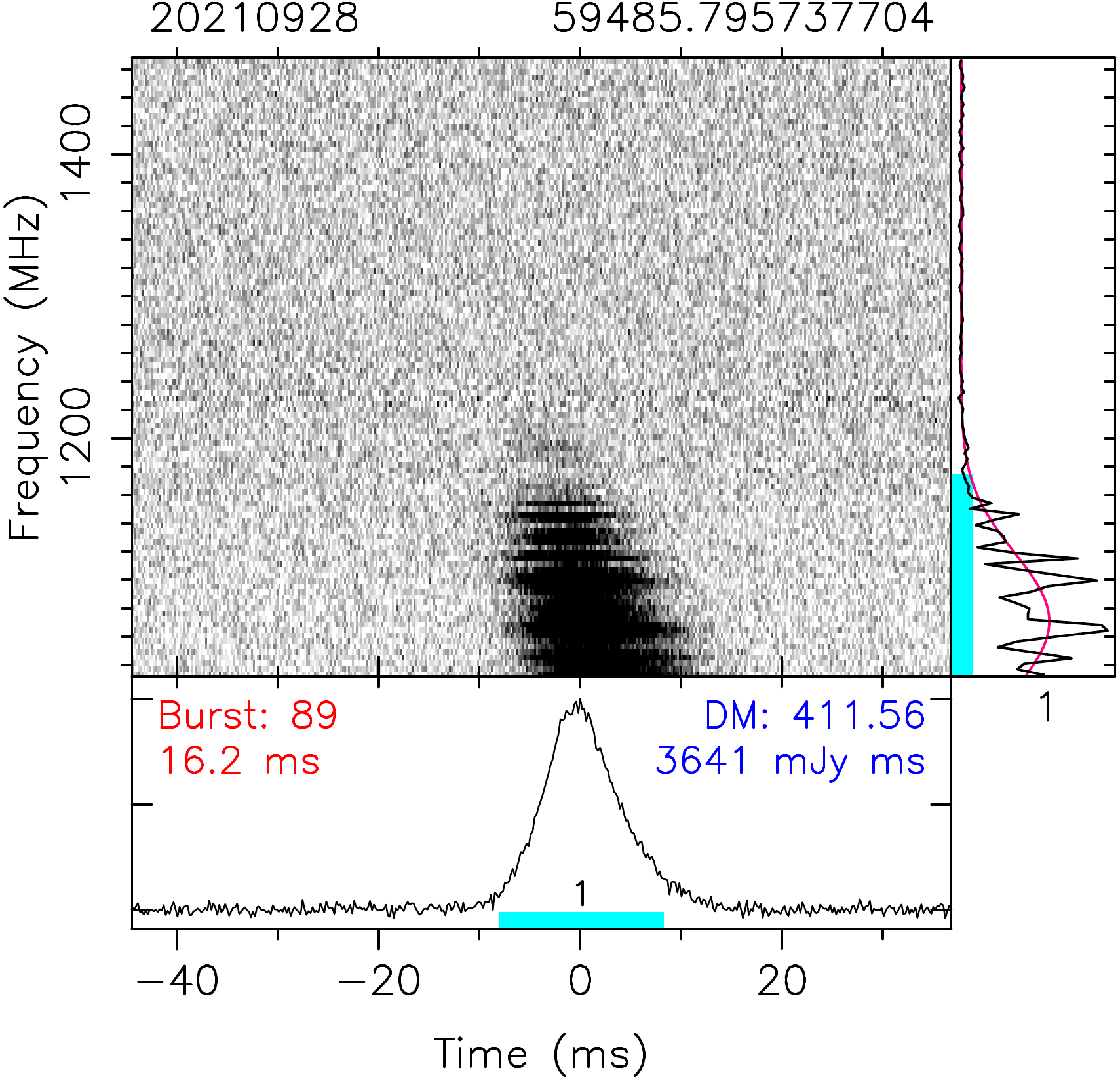}
    \includegraphics[height=37mm]{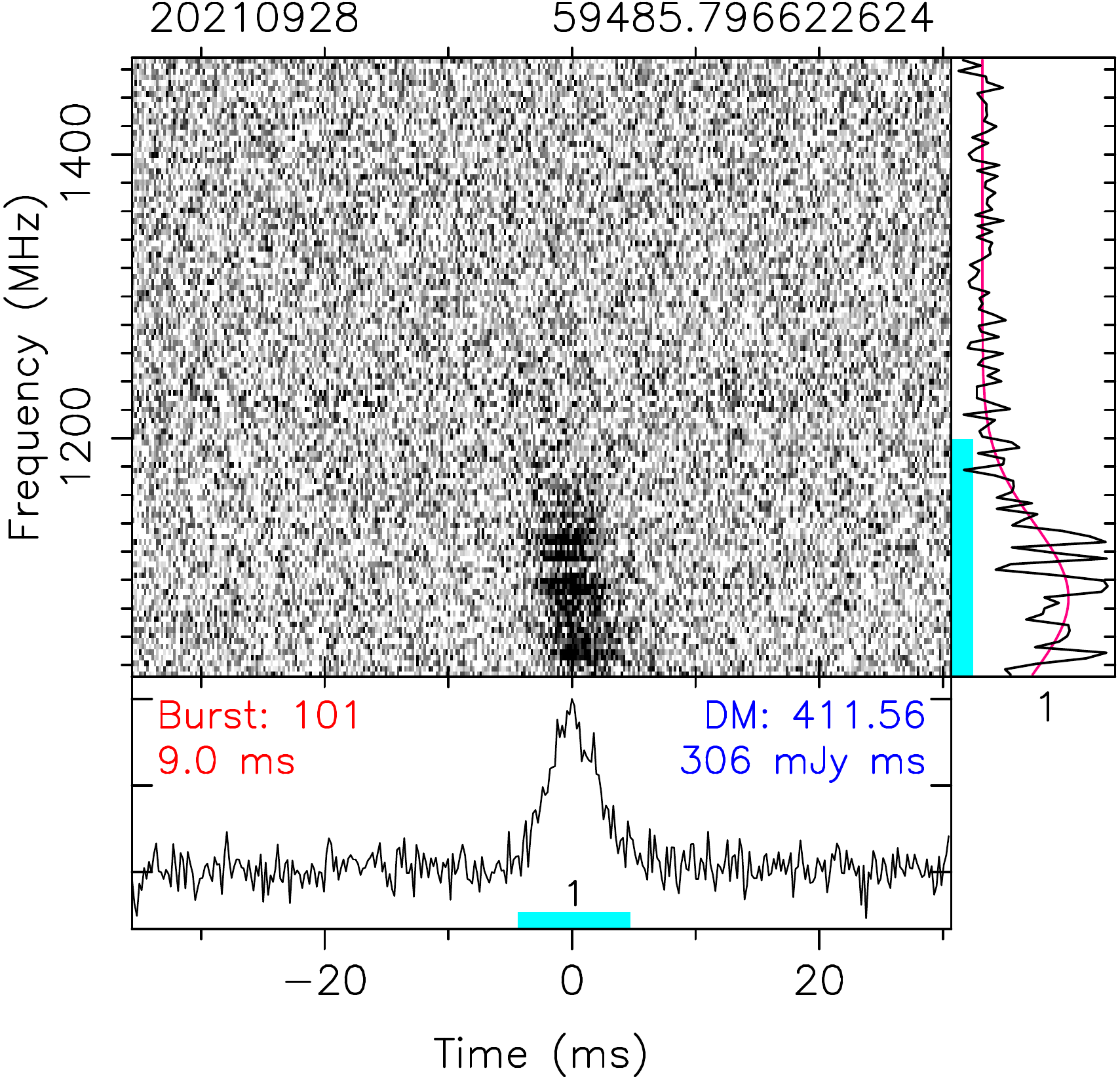}
    \includegraphics[height=37mm]{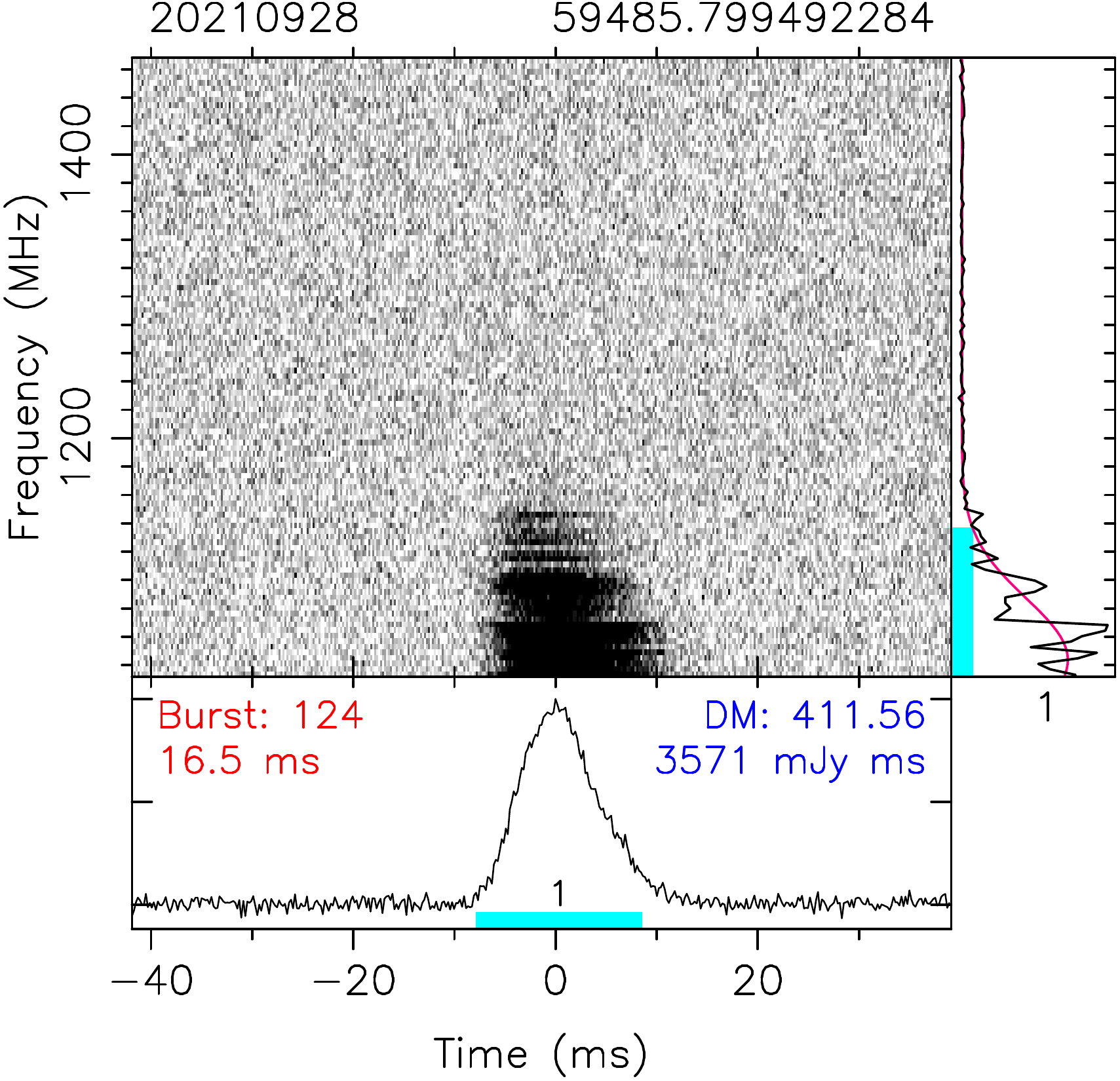}
    \includegraphics[height=37mm]{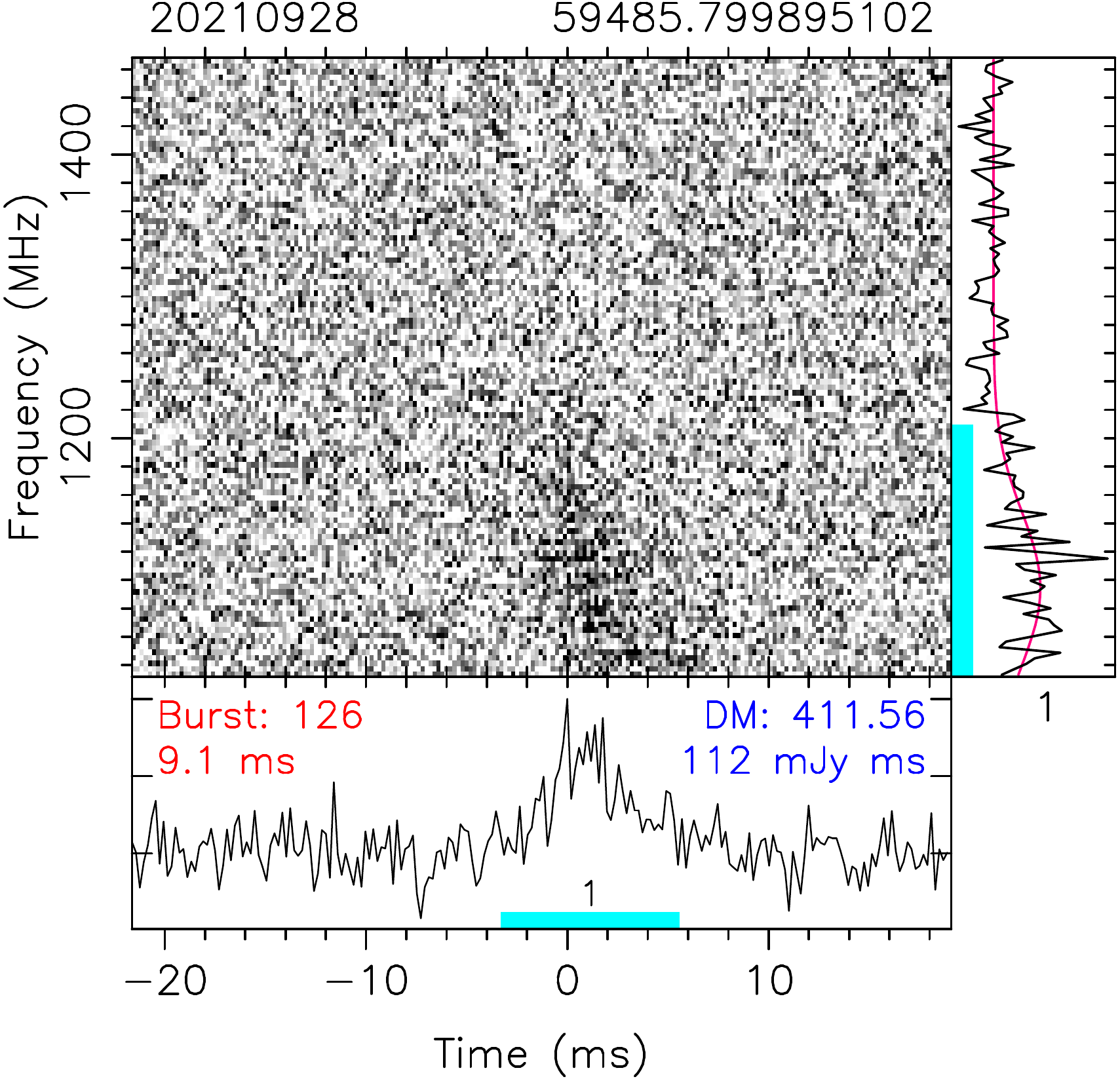}
    \includegraphics[height=37mm]{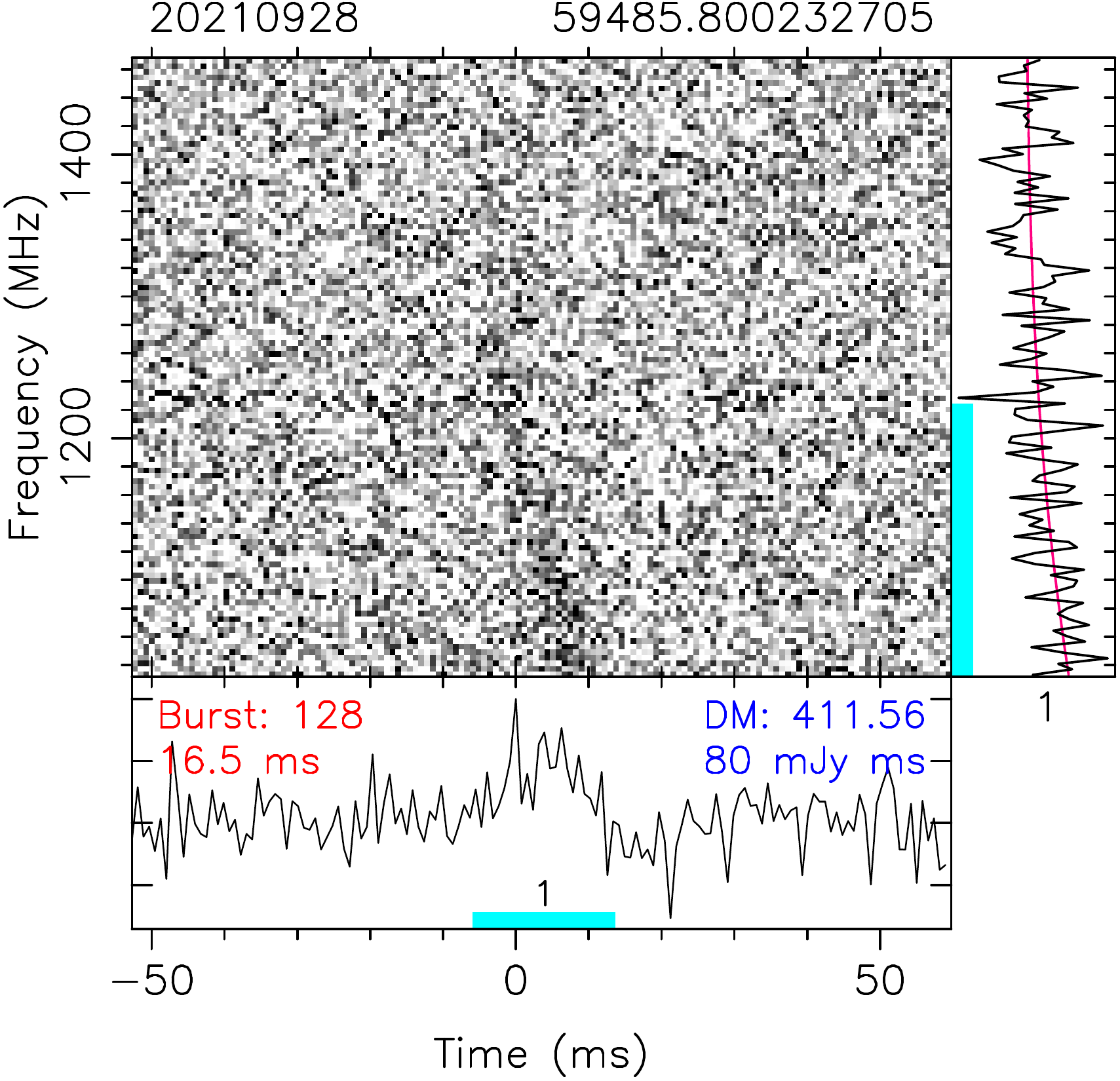}
    \includegraphics[height=37mm]{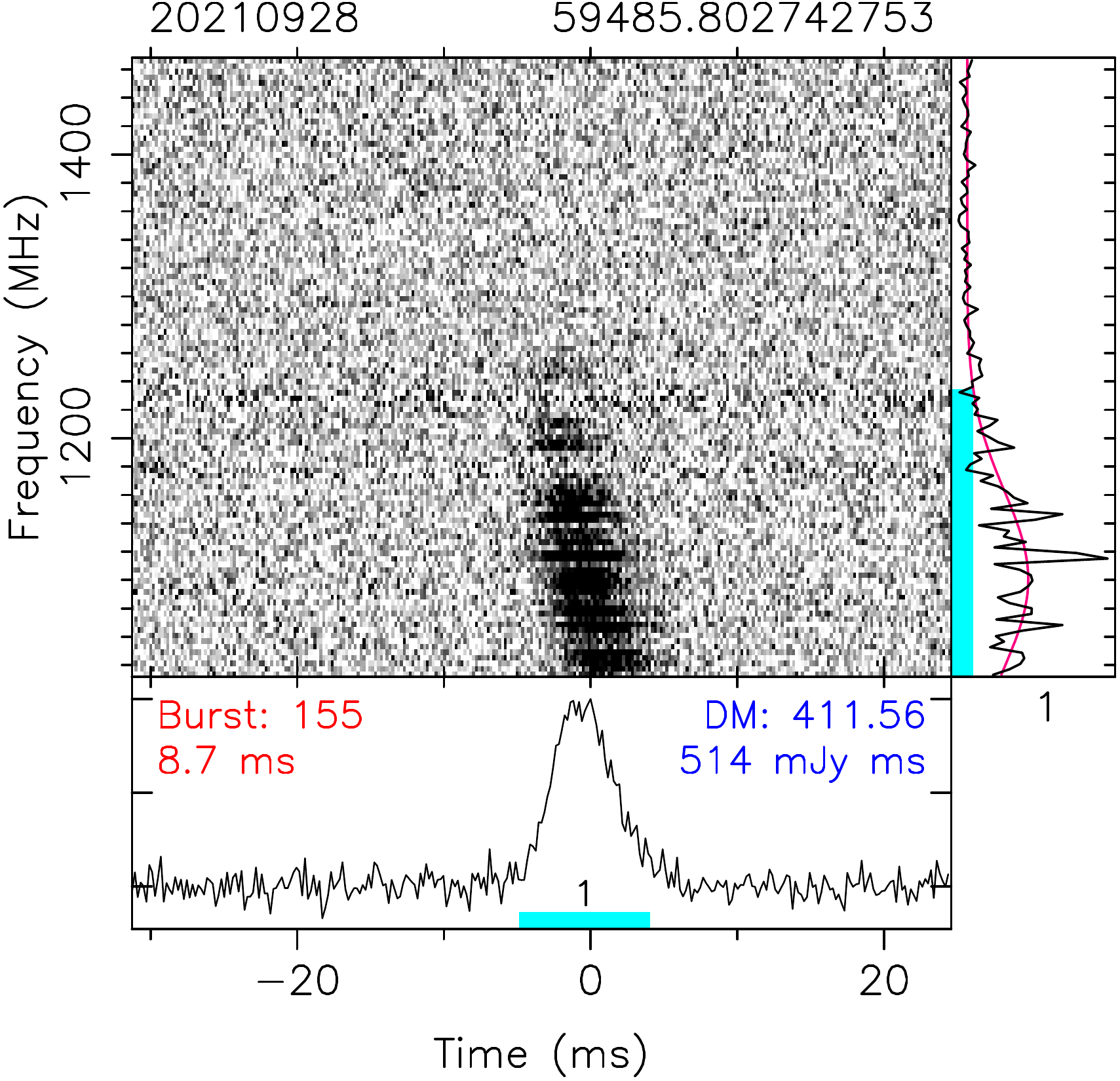}
    \includegraphics[height=37mm]{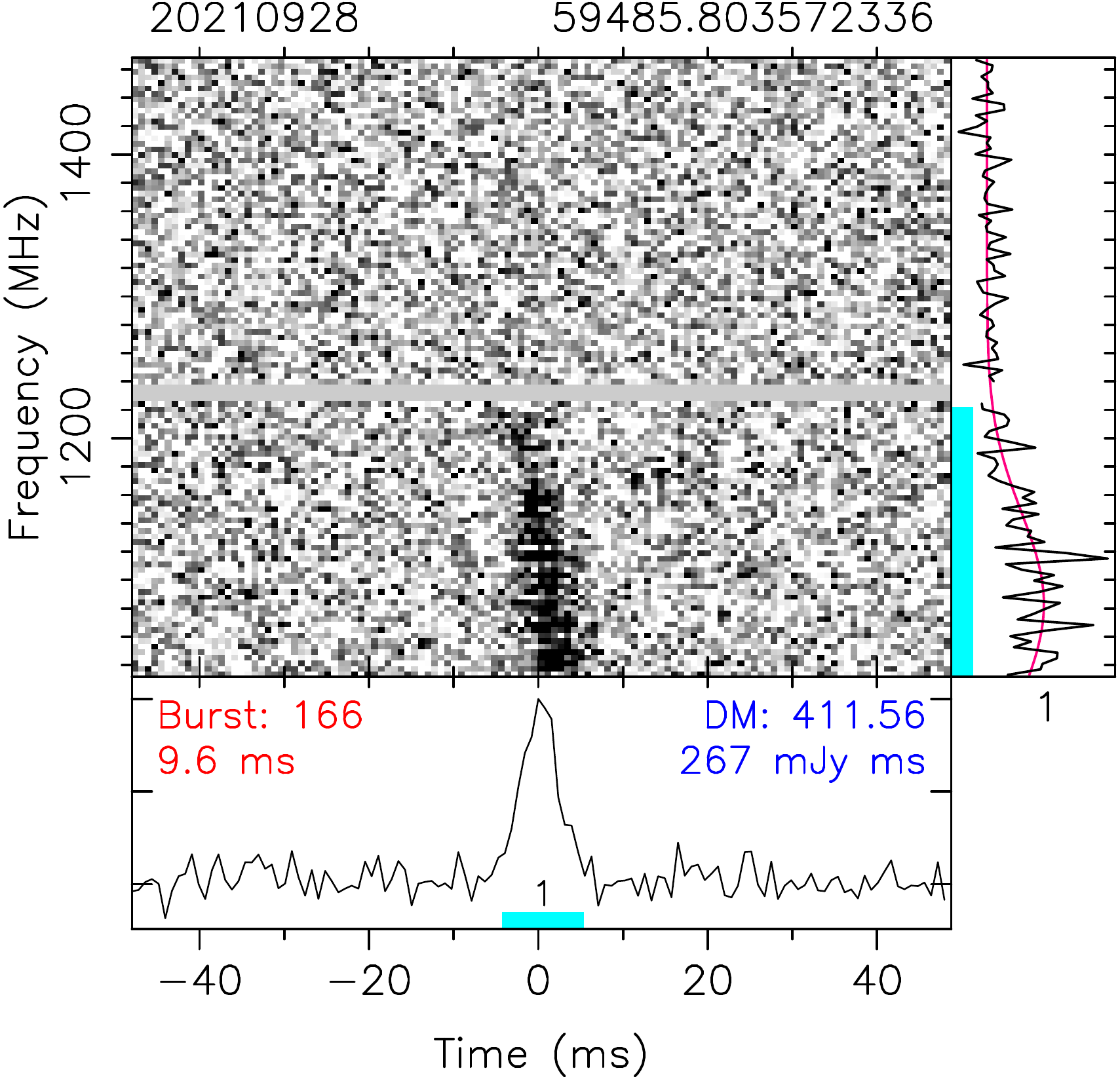}
    \includegraphics[height=37mm]{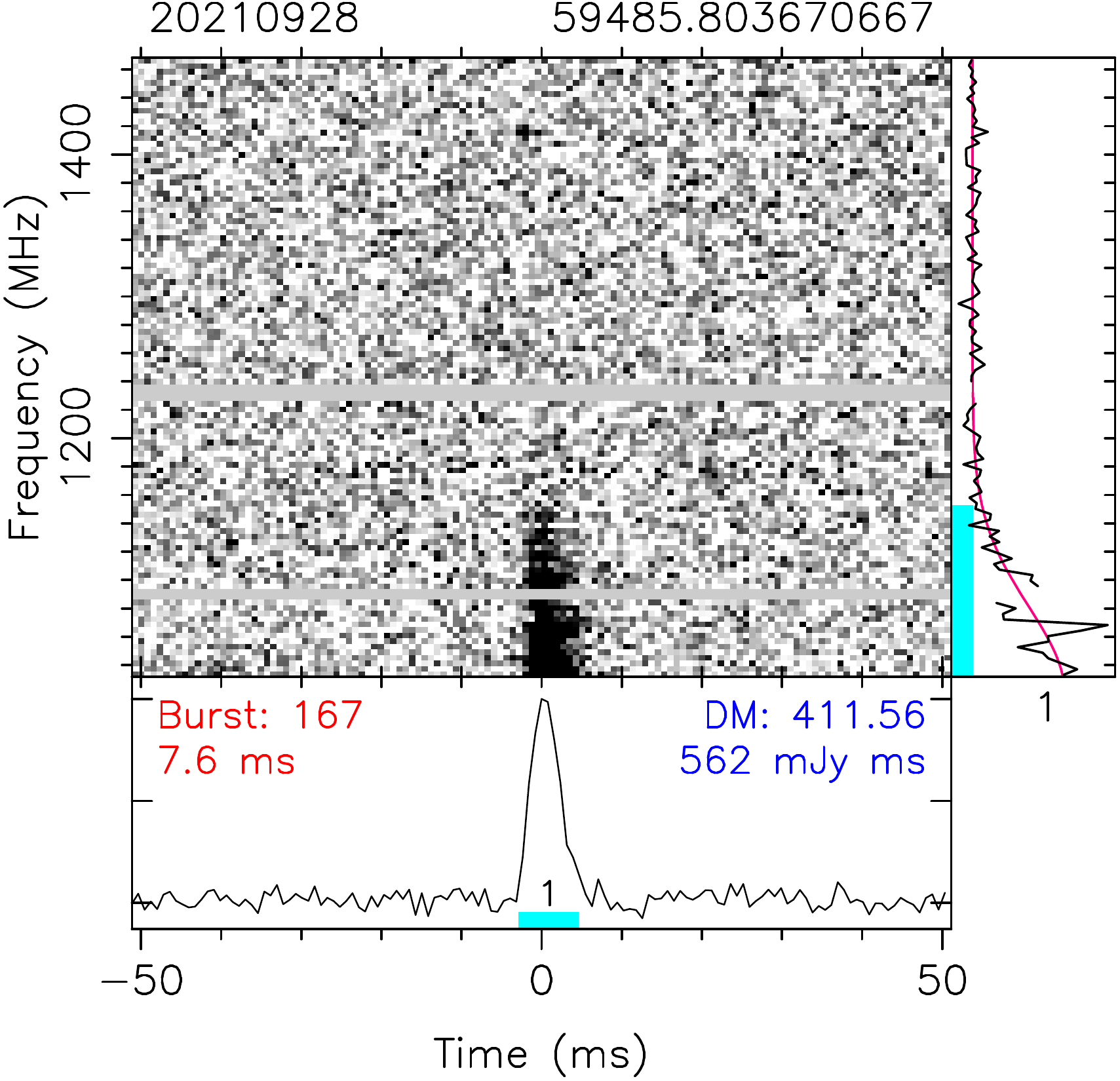}
    \includegraphics[height=37mm]{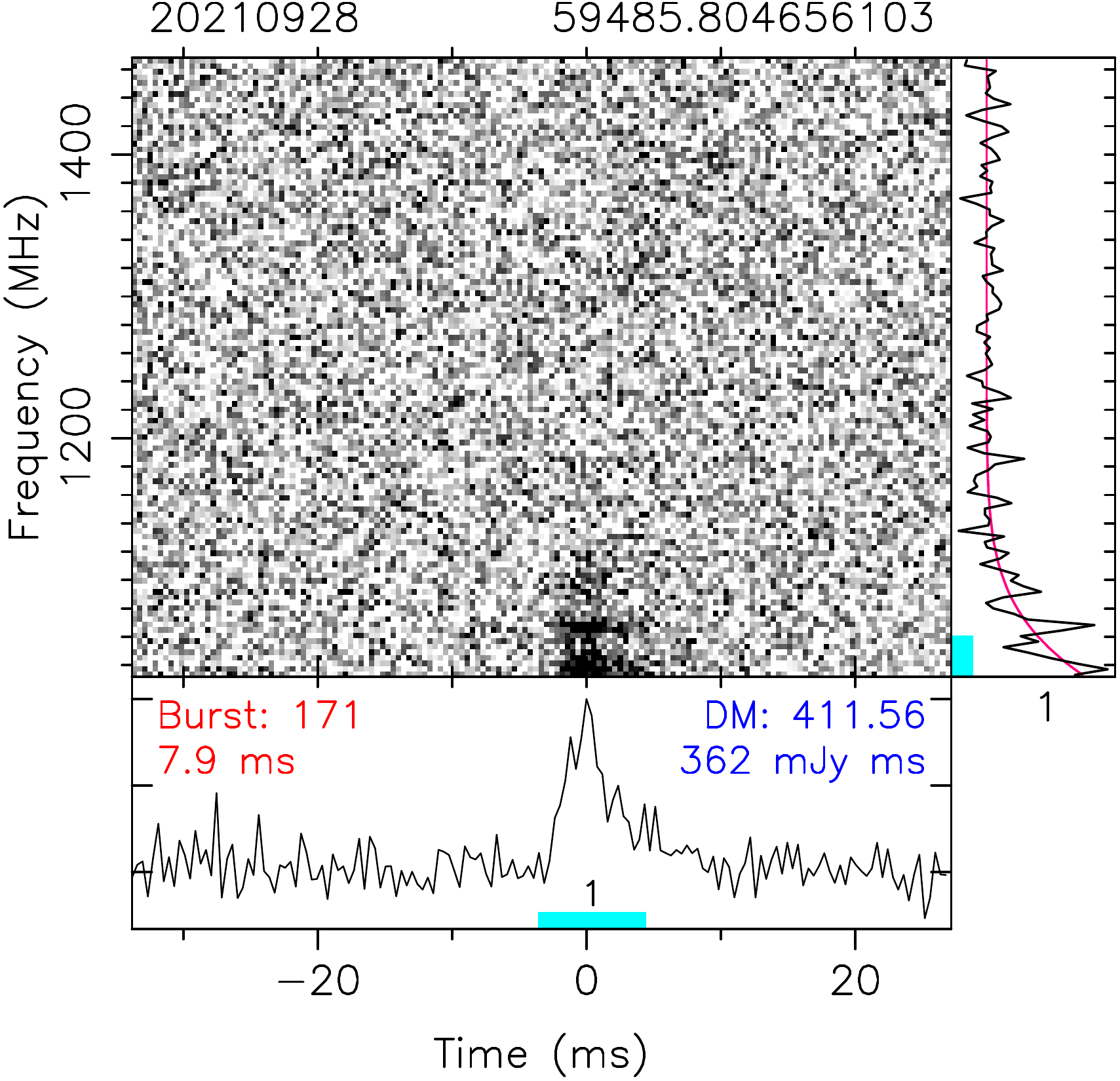}
    \includegraphics[height=37mm]{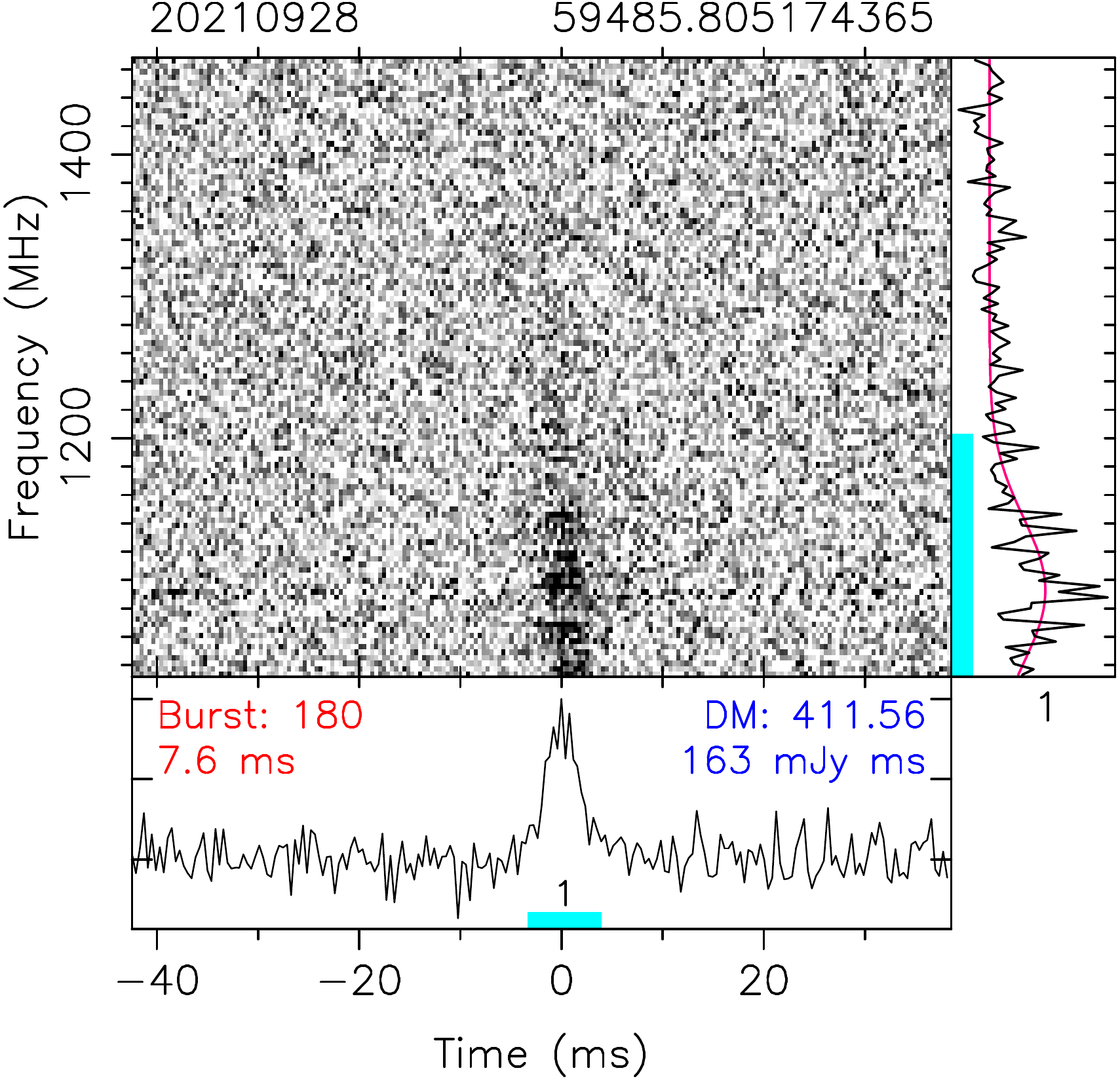}
    \includegraphics[height=37mm]{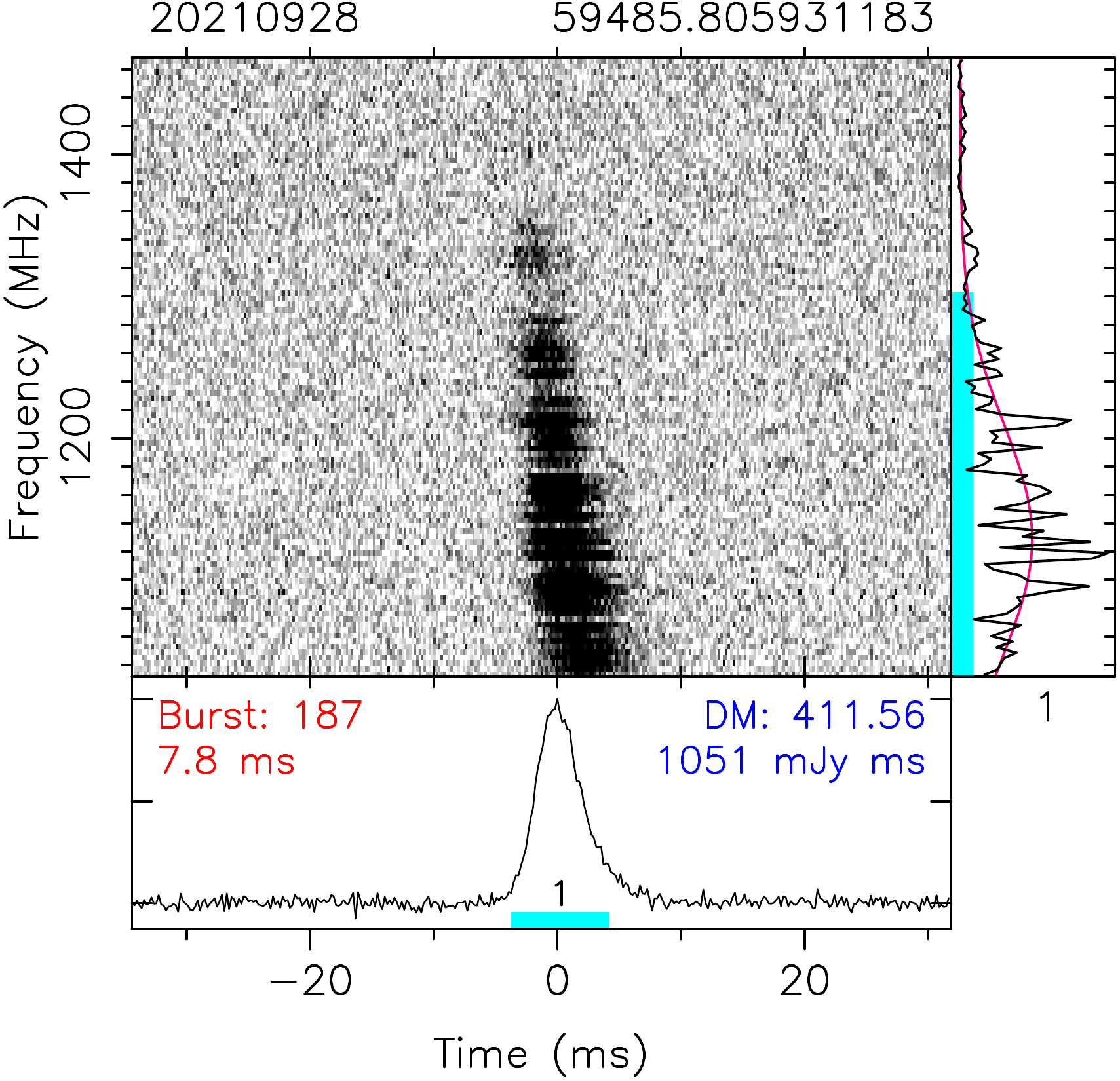}
    \includegraphics[height=37mm]{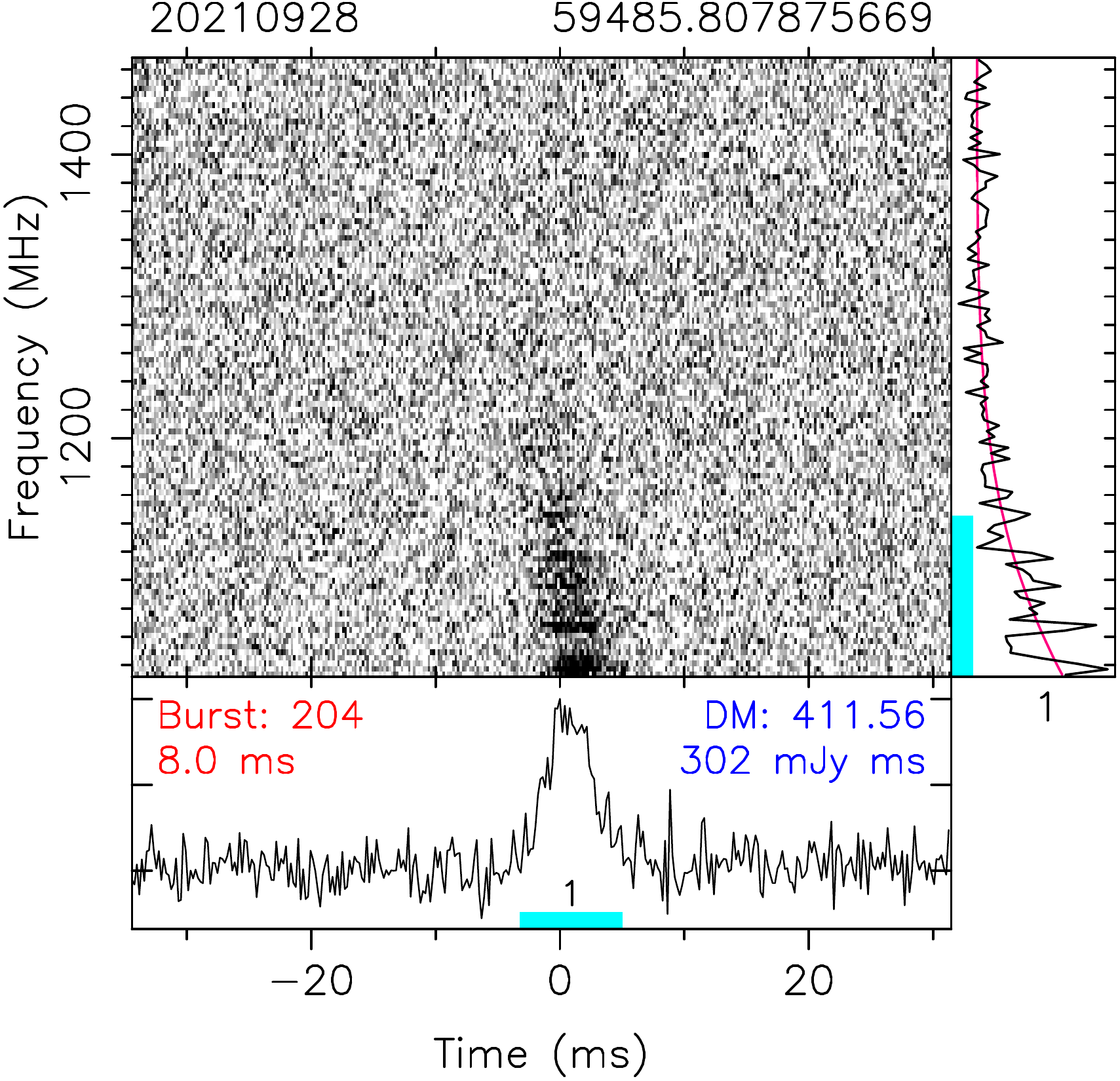}
    \includegraphics[height=37mm]{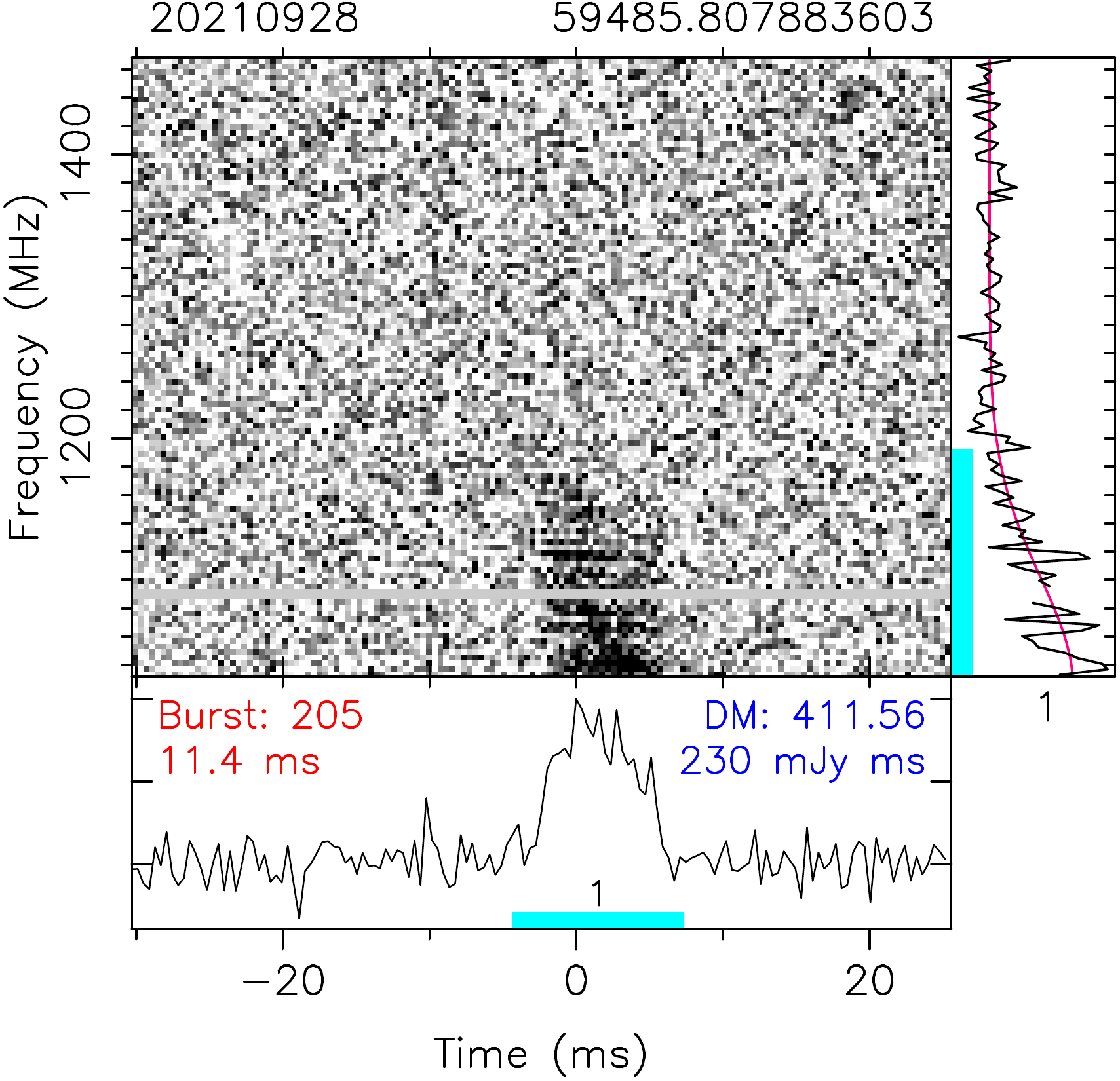}
    \includegraphics[height=37mm]{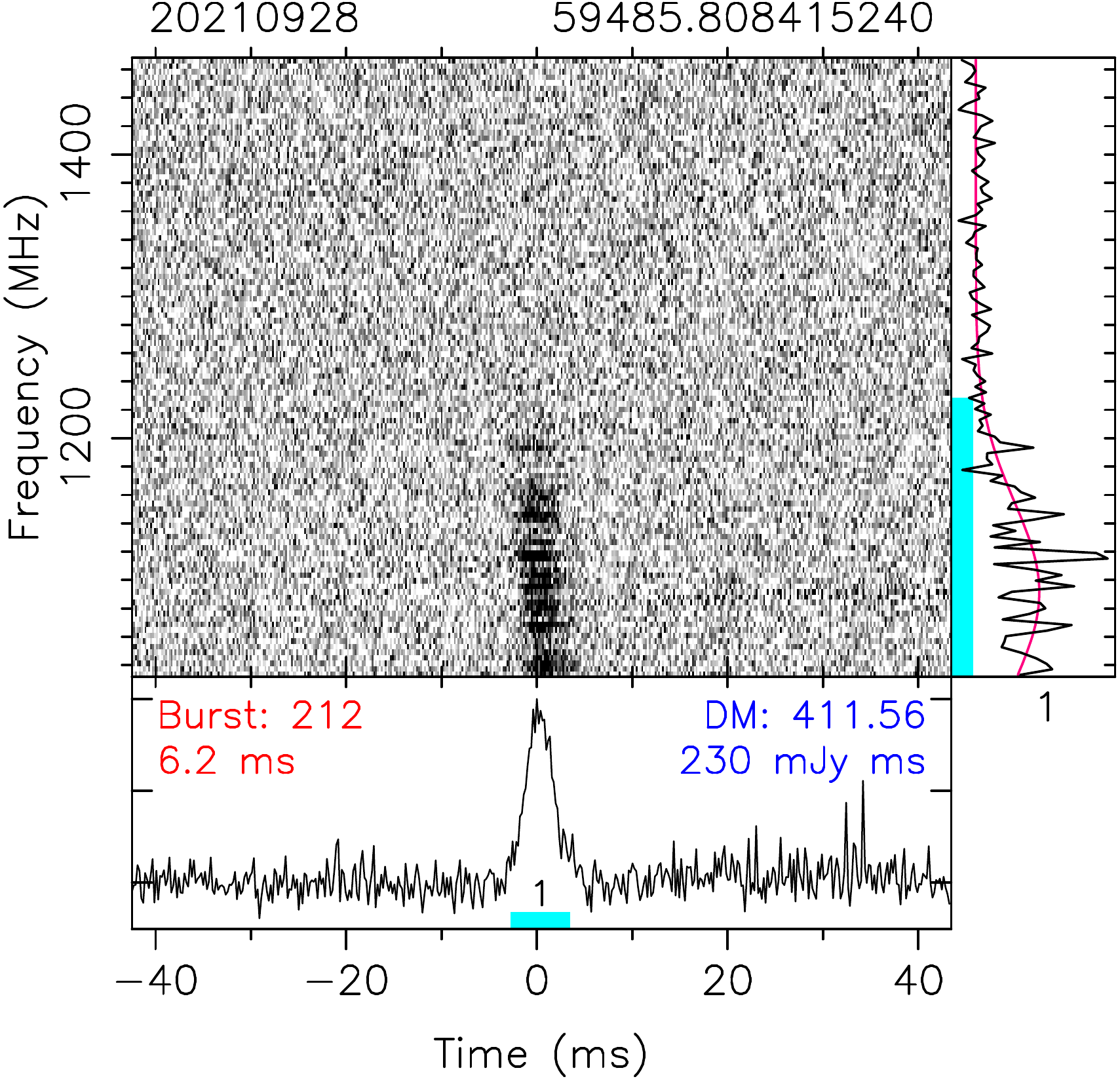}
    \includegraphics[height=37mm]{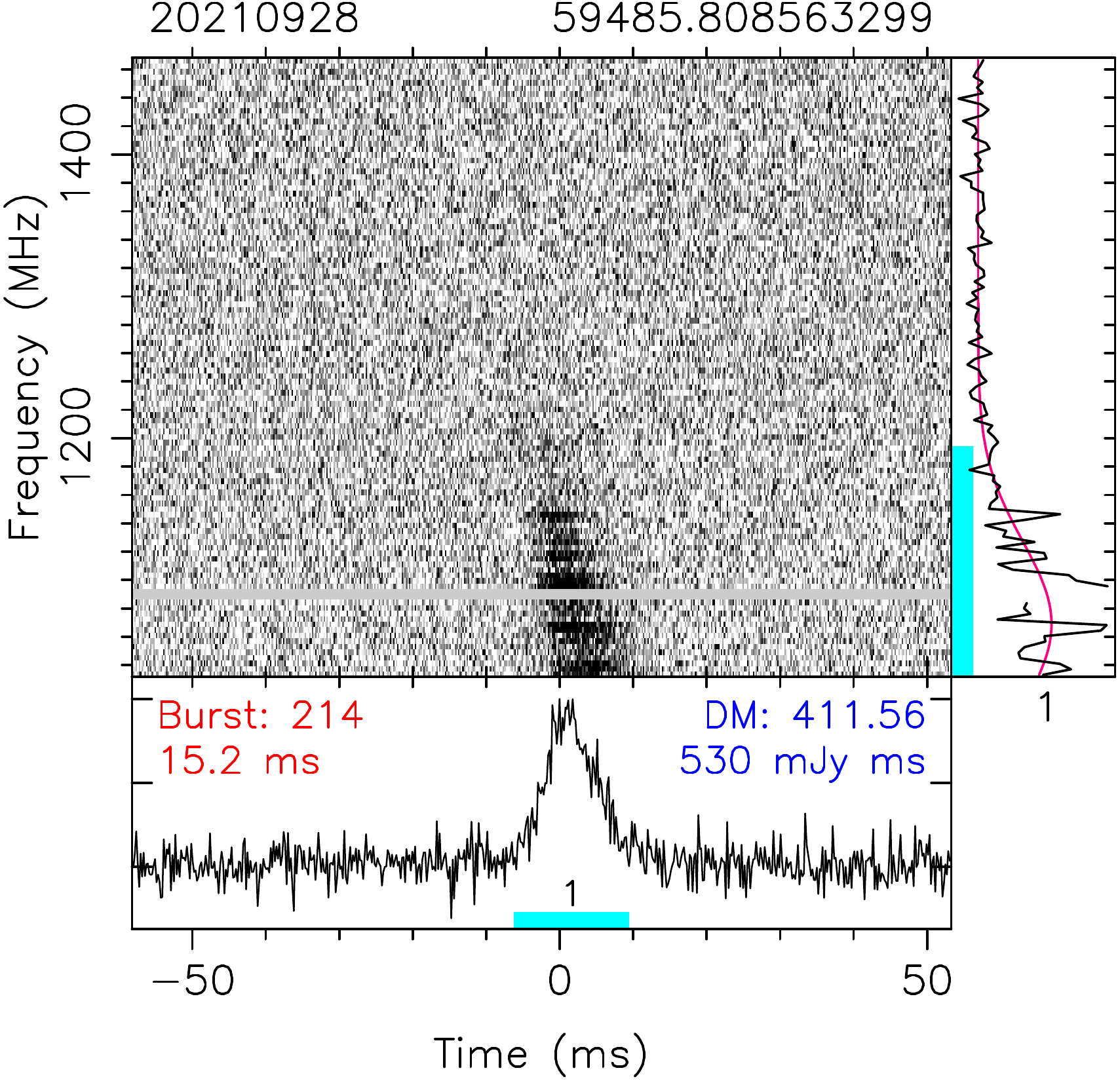}
    \includegraphics[height=37mm]{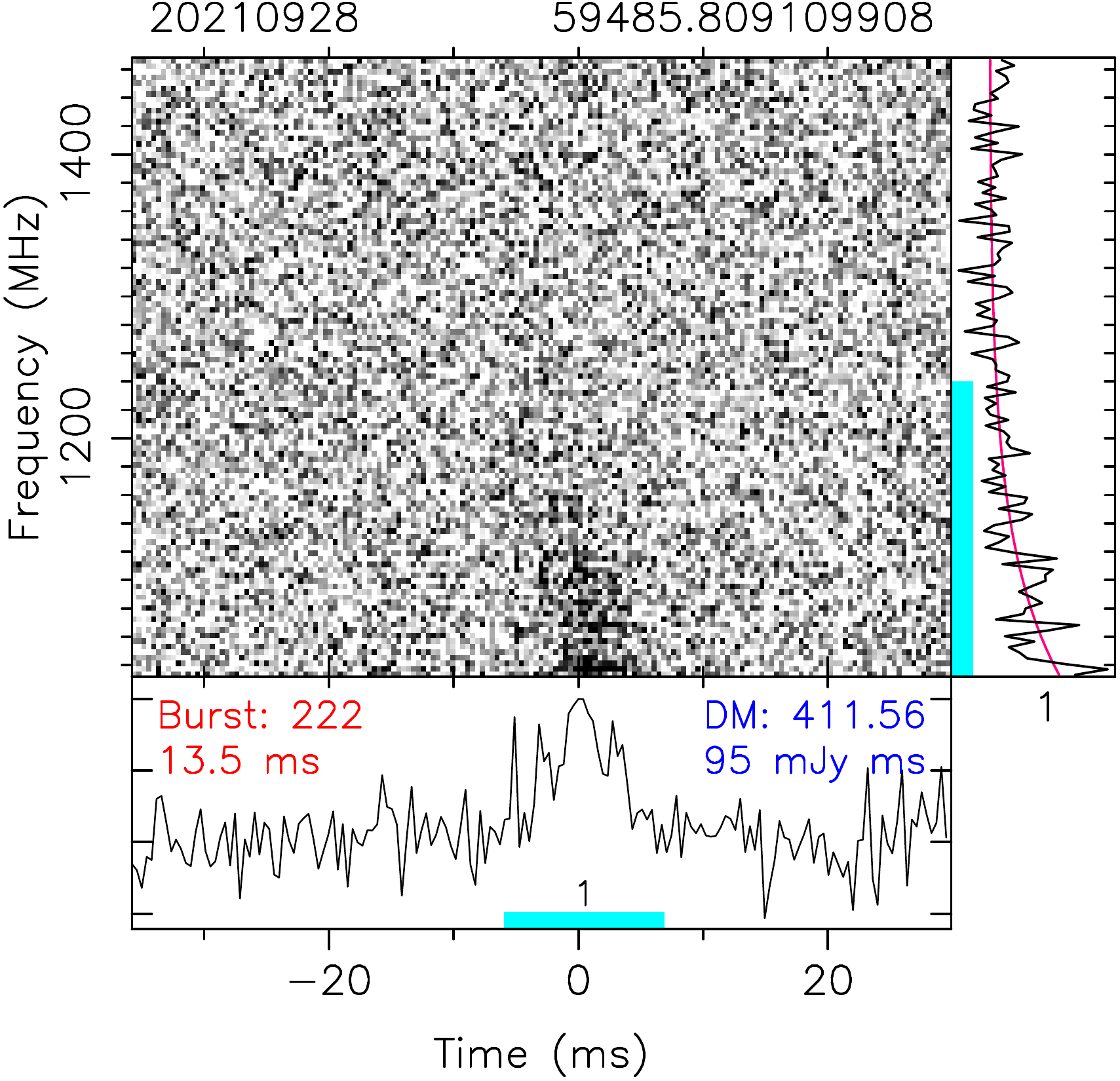}
    \includegraphics[height=37mm]{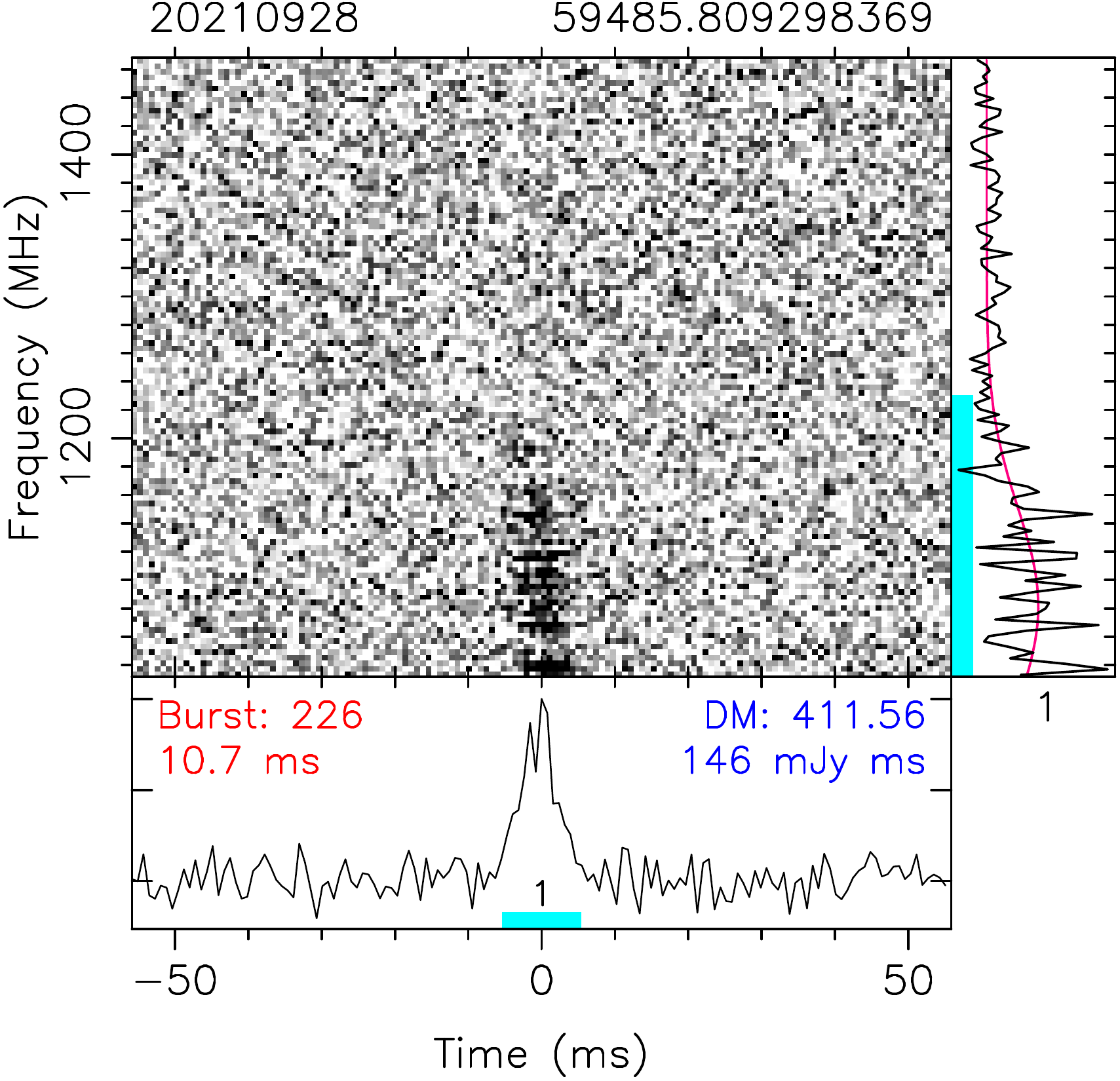}
    \includegraphics[height=37mm]{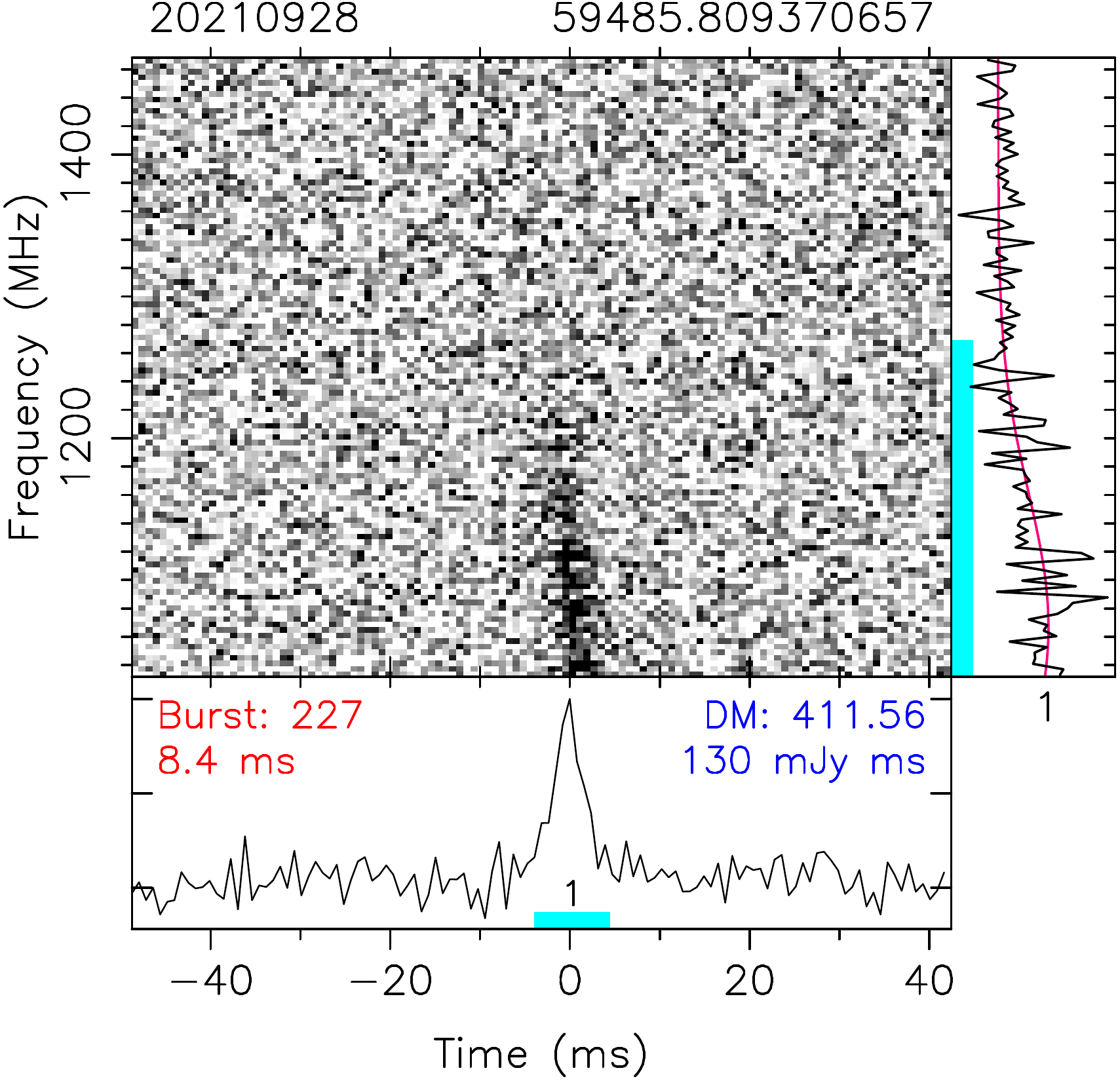}
    \includegraphics[height=37mm]{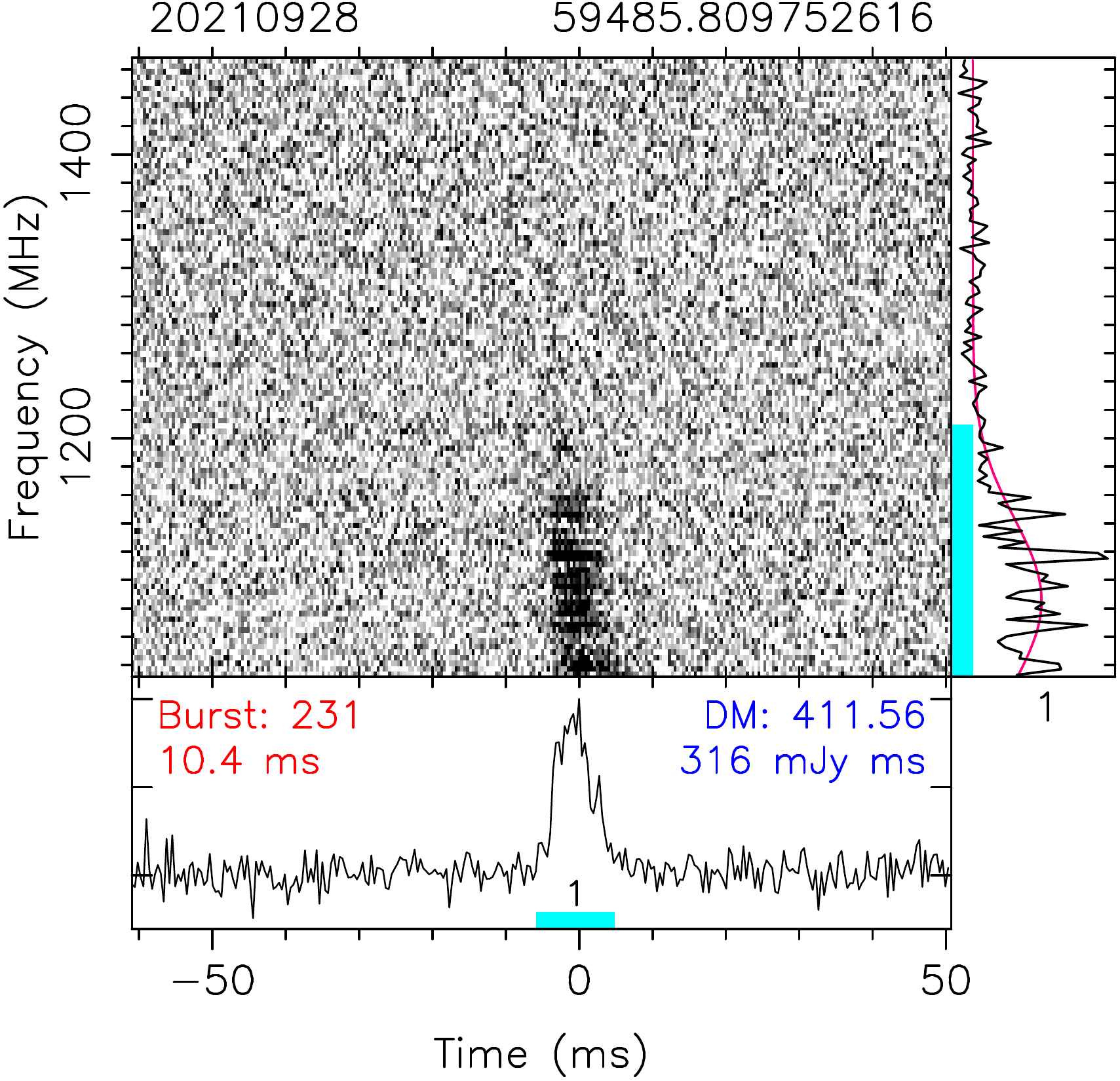}
    \includegraphics[height=37mm]{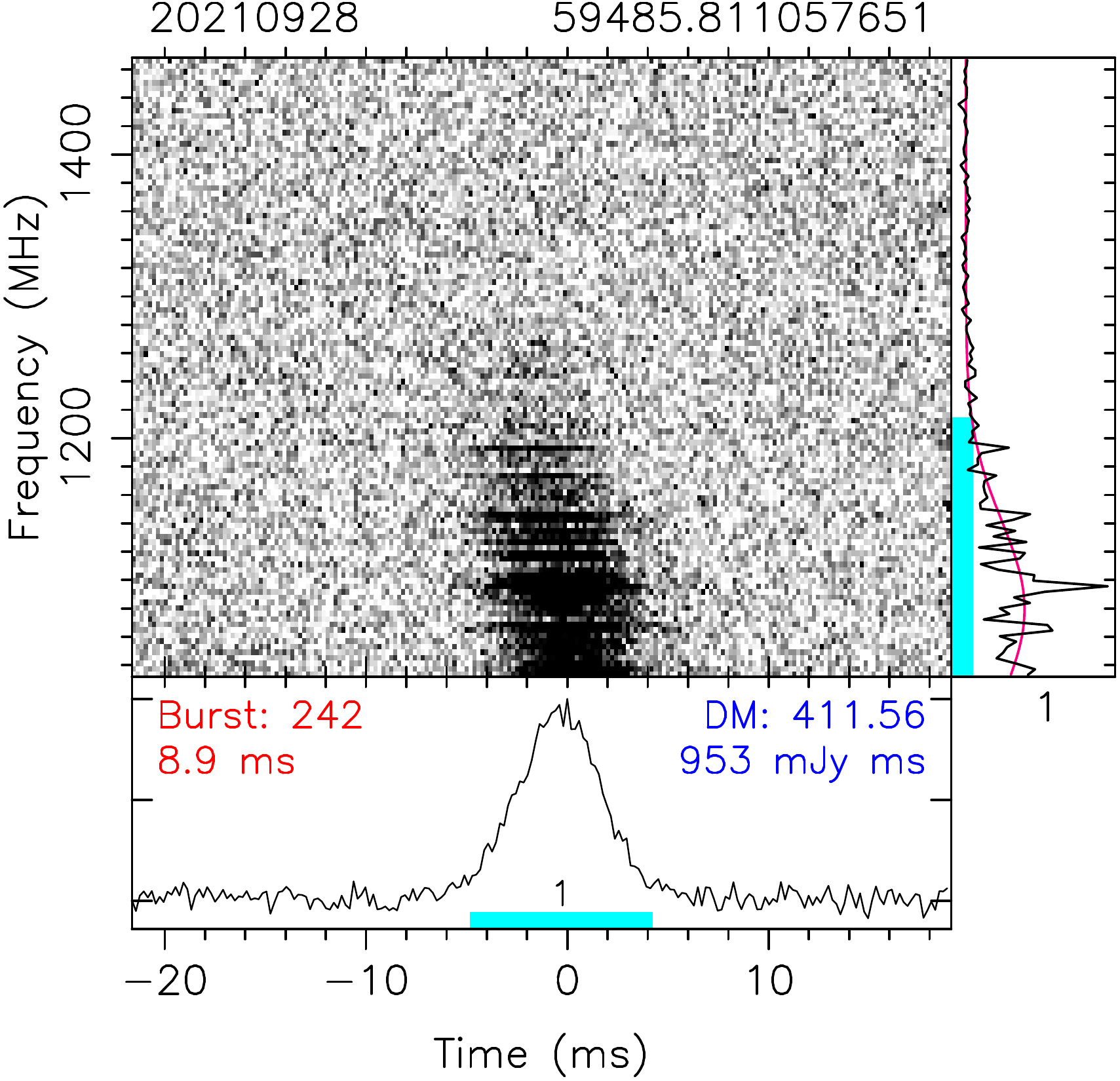}
    \includegraphics[height=37mm]{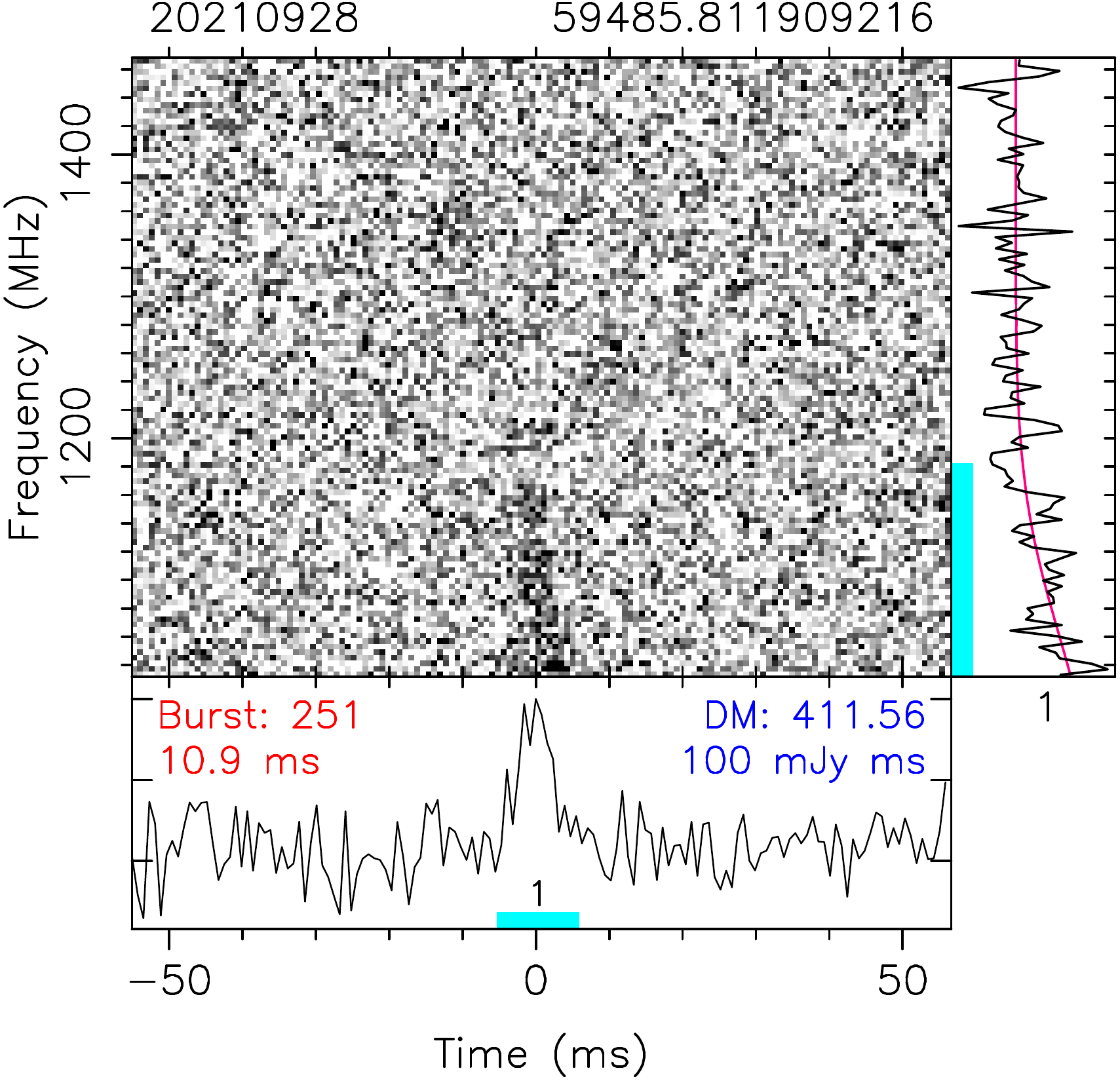}
    \includegraphics[height=37mm]{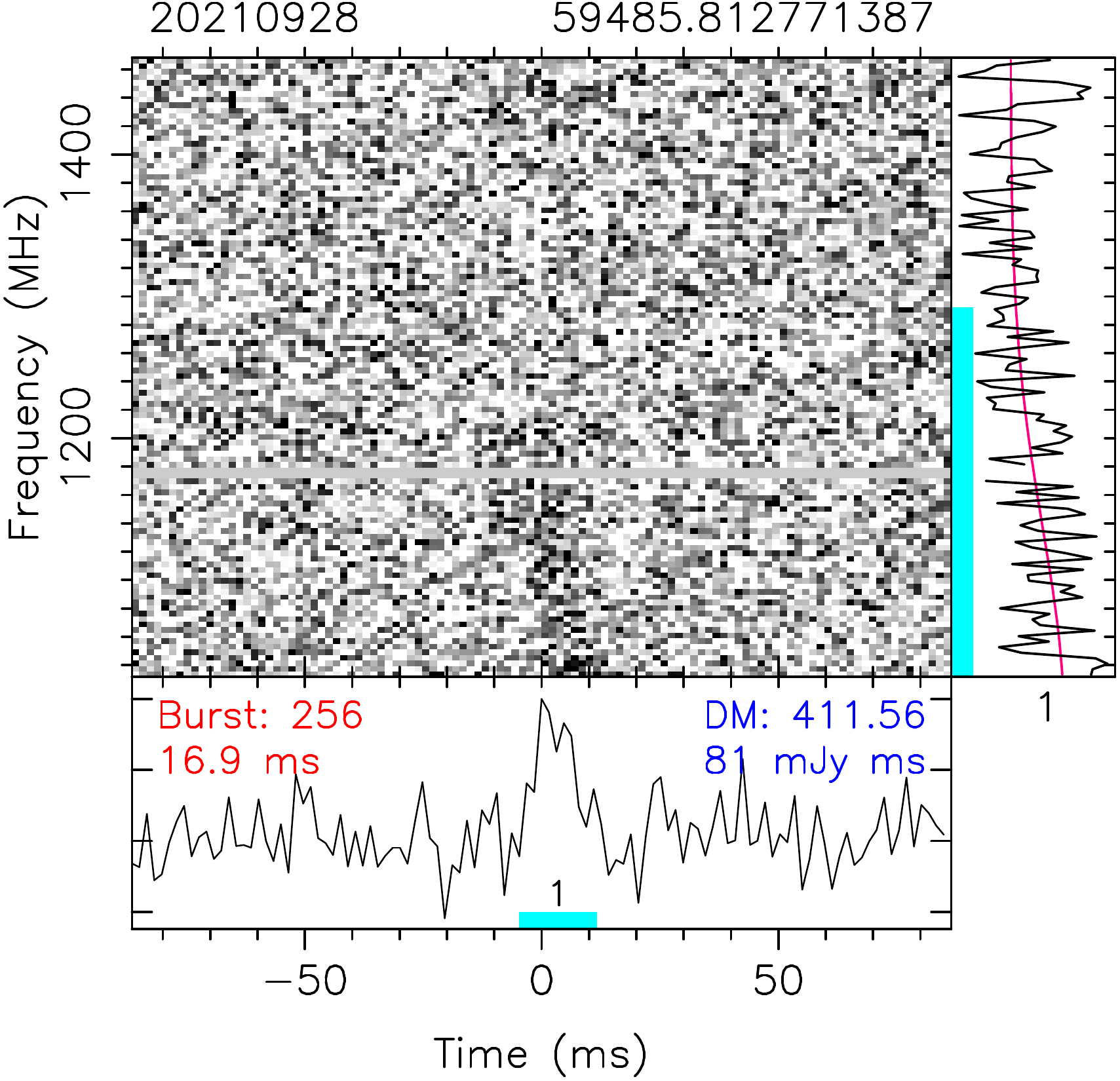}
\caption{\it{ -- continued}.
}
\end{figure*}
\addtocounter{figure}{-1}
\begin{figure*}
    \flushleft
    \includegraphics[height=37mm]{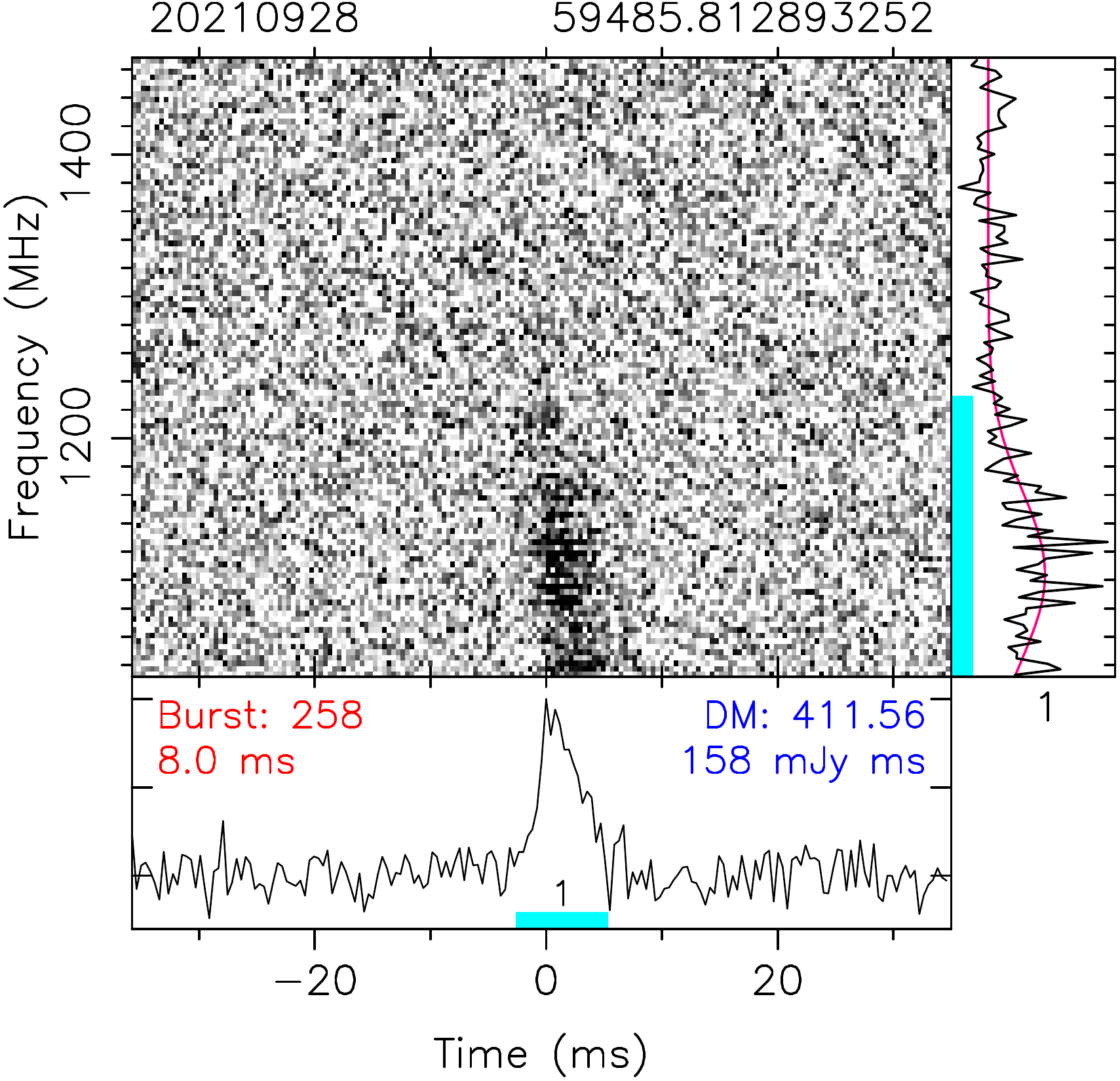}
    \includegraphics[height=37mm]{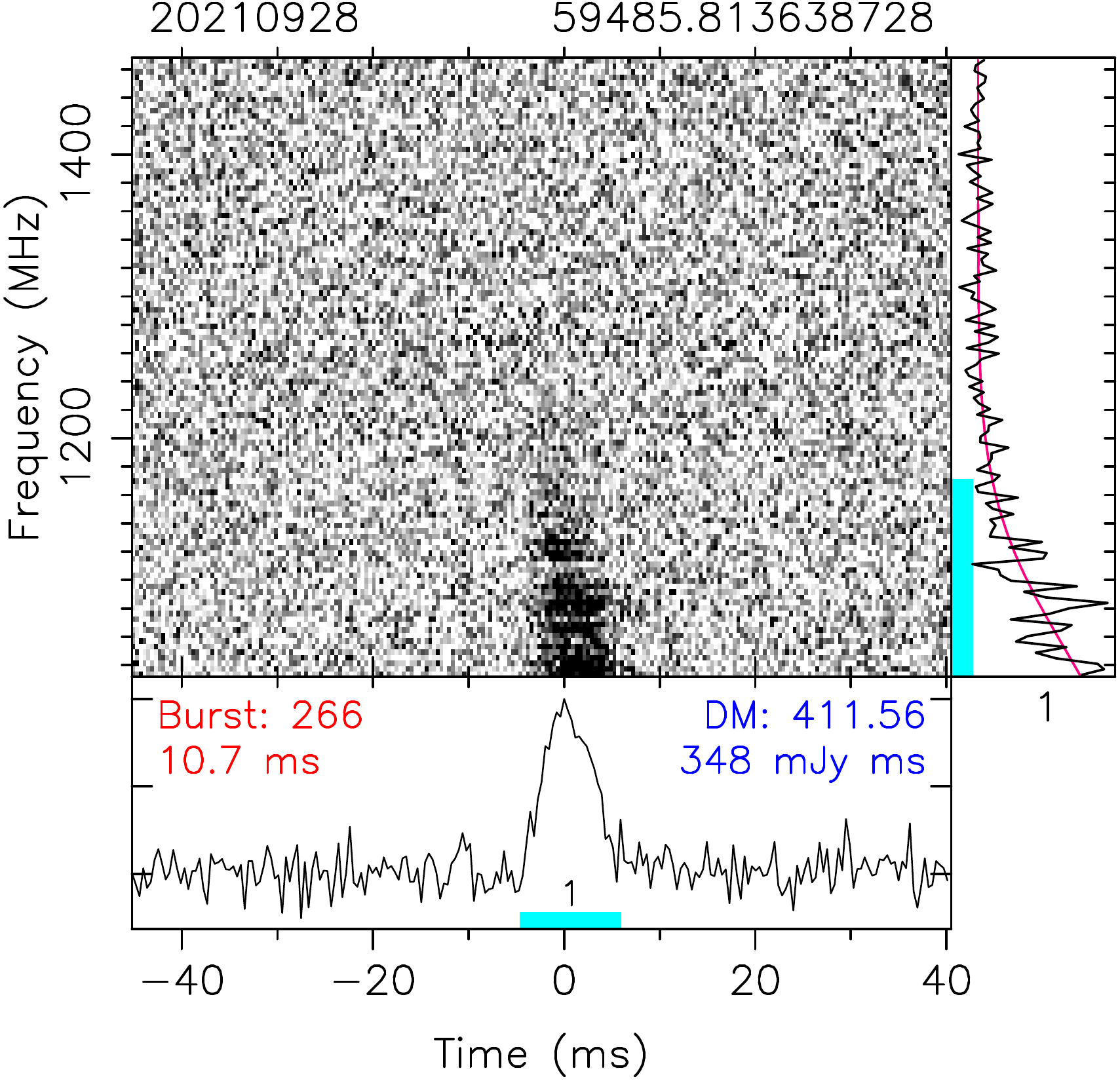}
    \includegraphics[height=37mm]{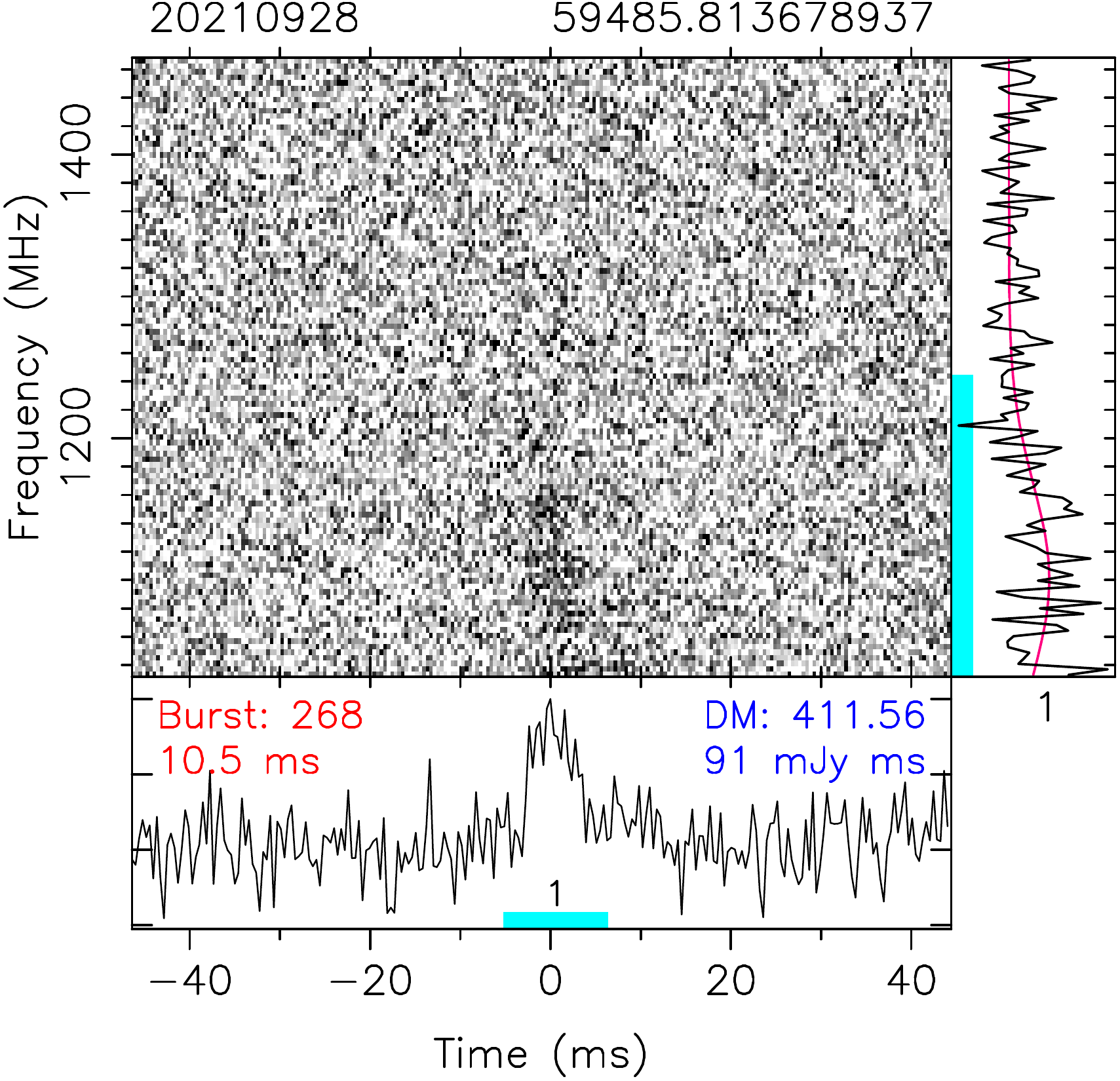}
    \includegraphics[height=37mm]{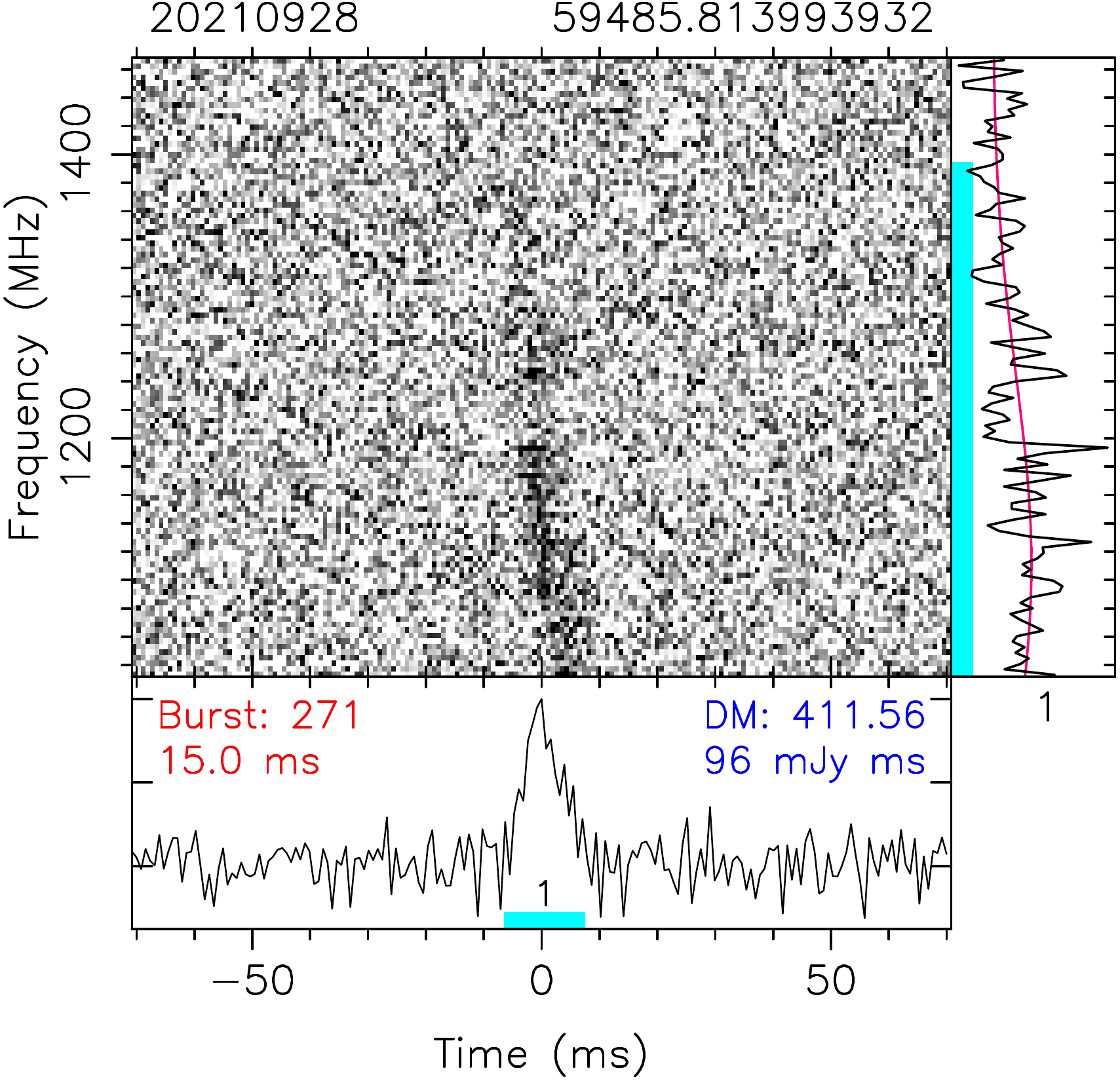}
    \includegraphics[height=37mm]{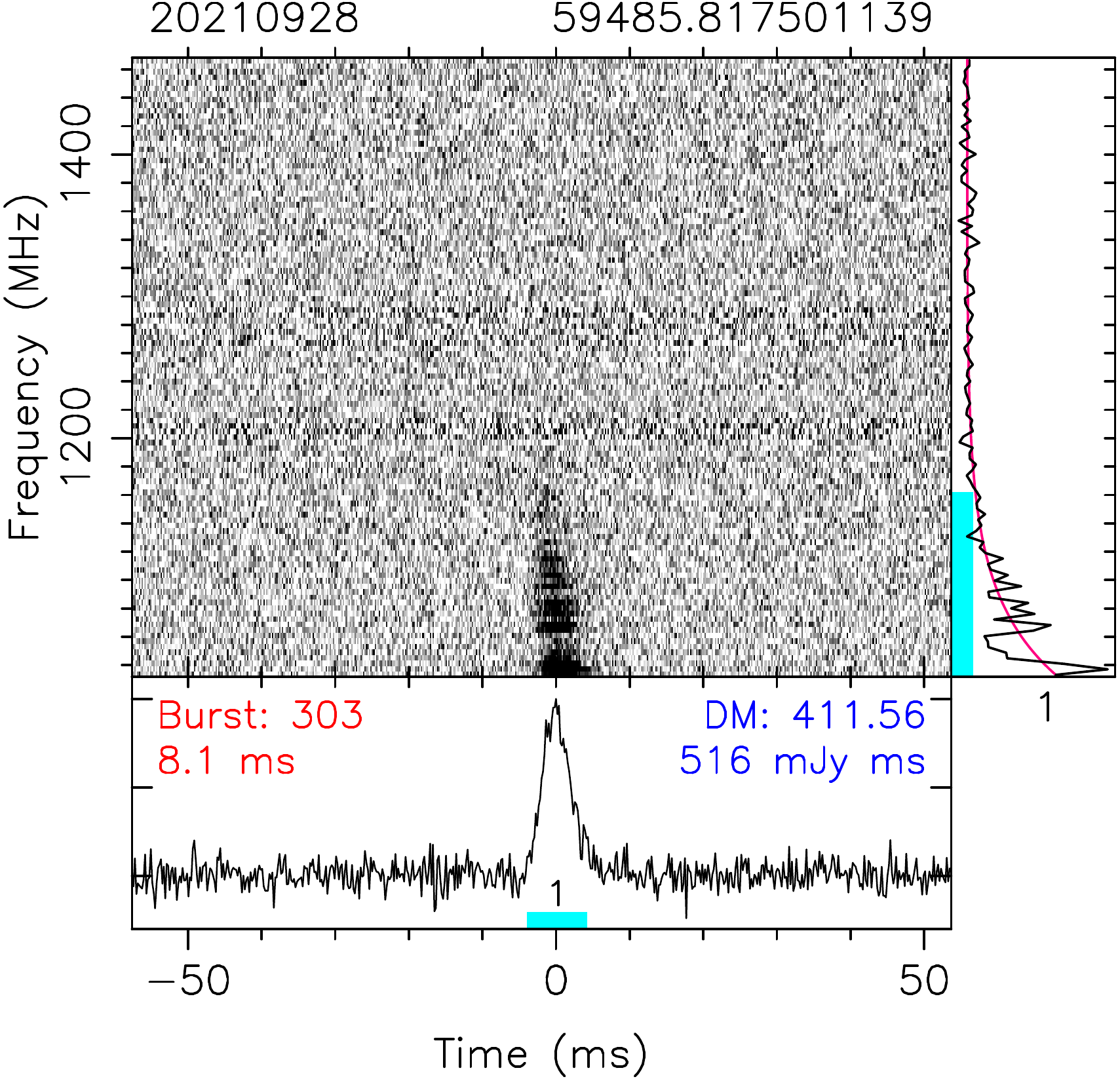}
    \includegraphics[height=37mm]{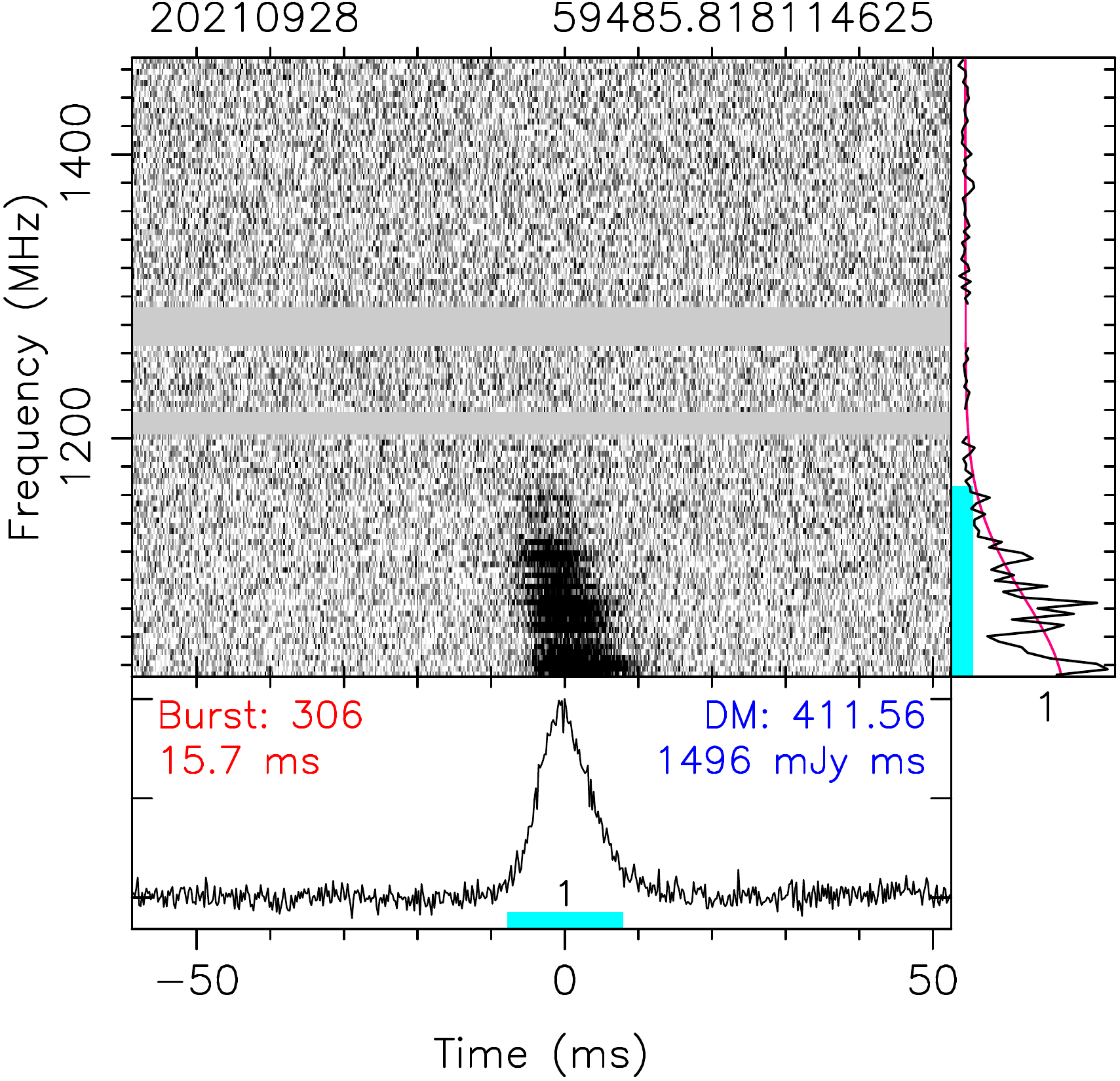}
    \includegraphics[height=37mm]{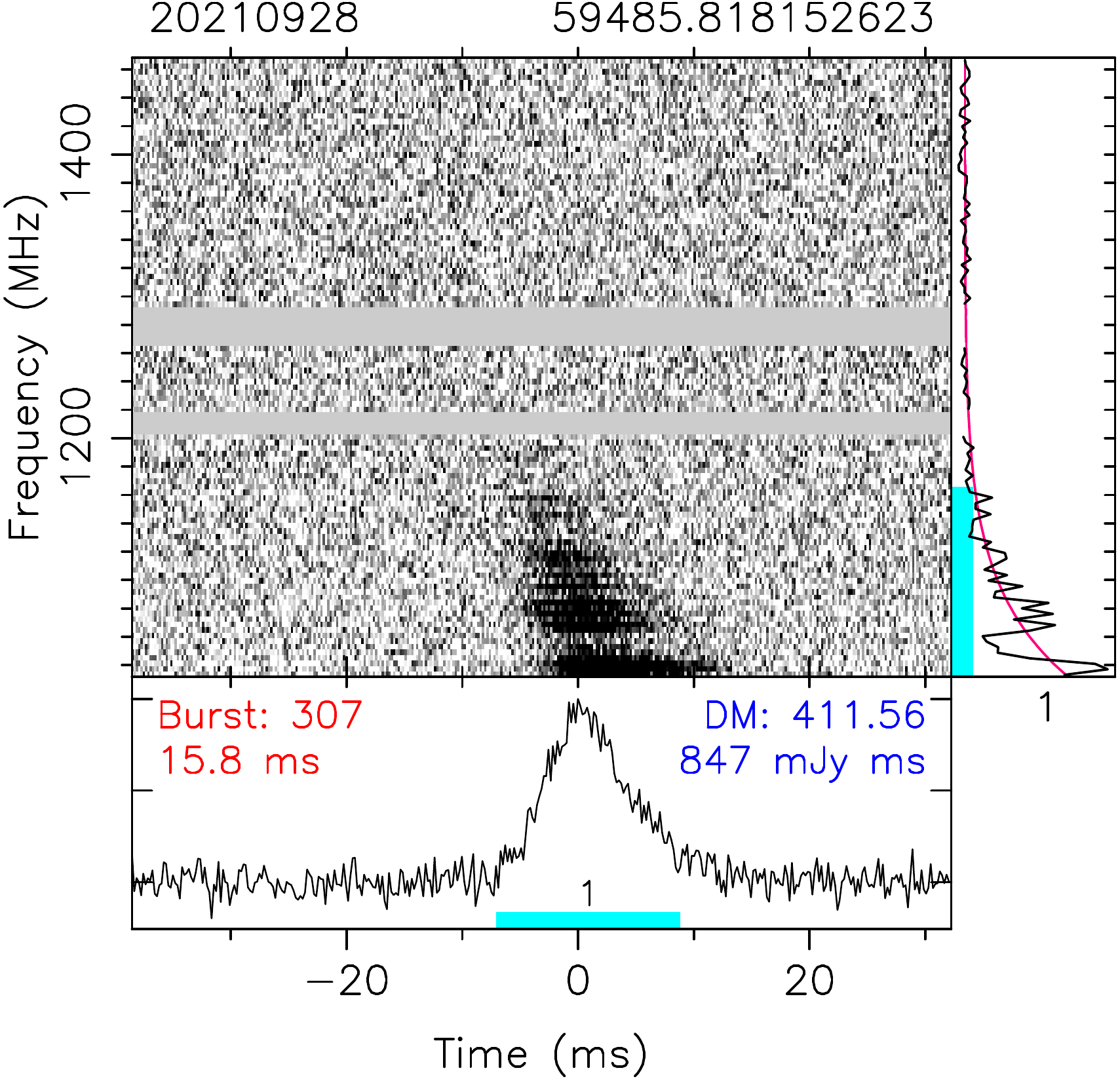}
    \includegraphics[height=37mm]{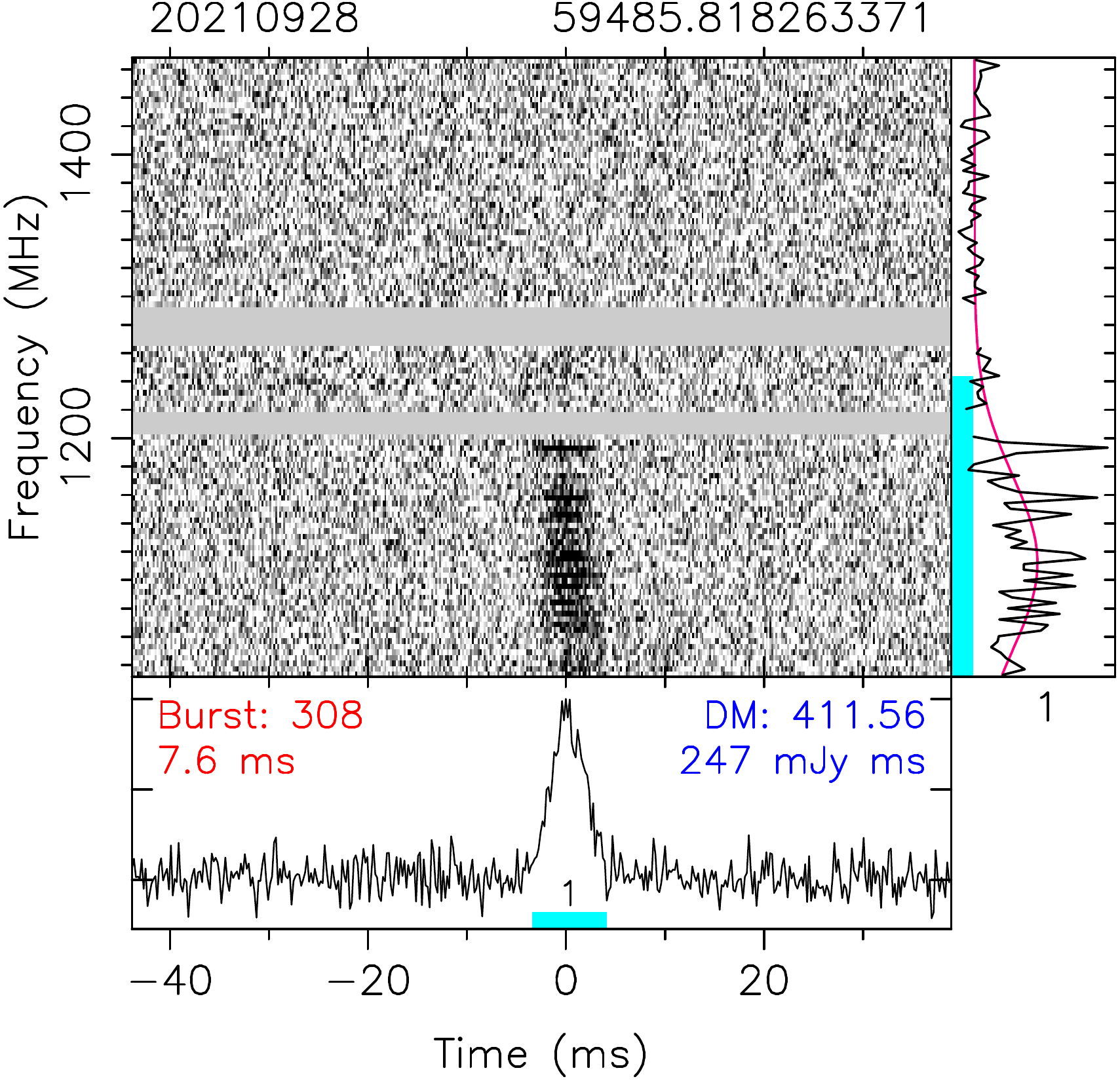}
    \includegraphics[height=37mm]{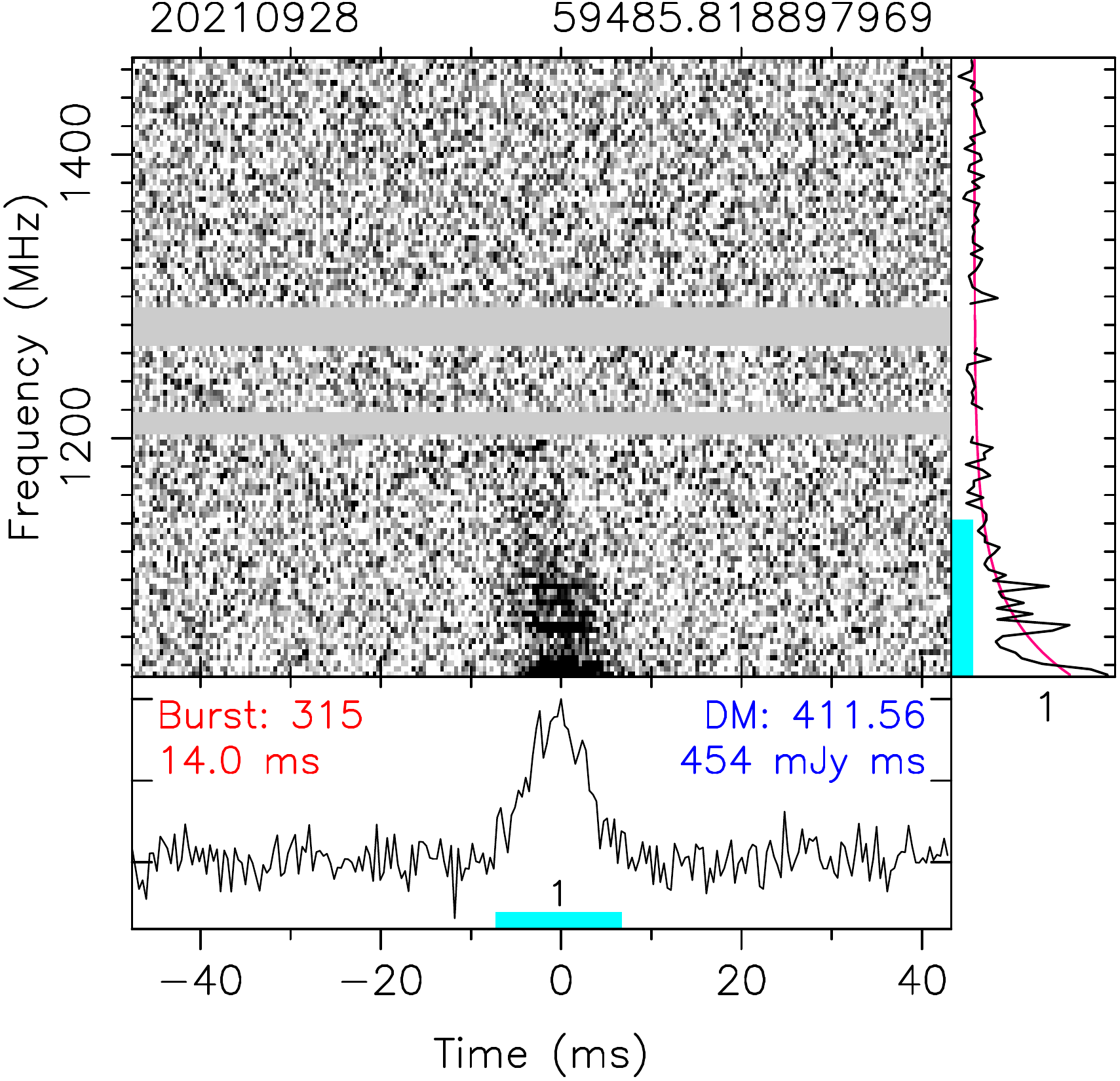}
    \includegraphics[height=37mm]{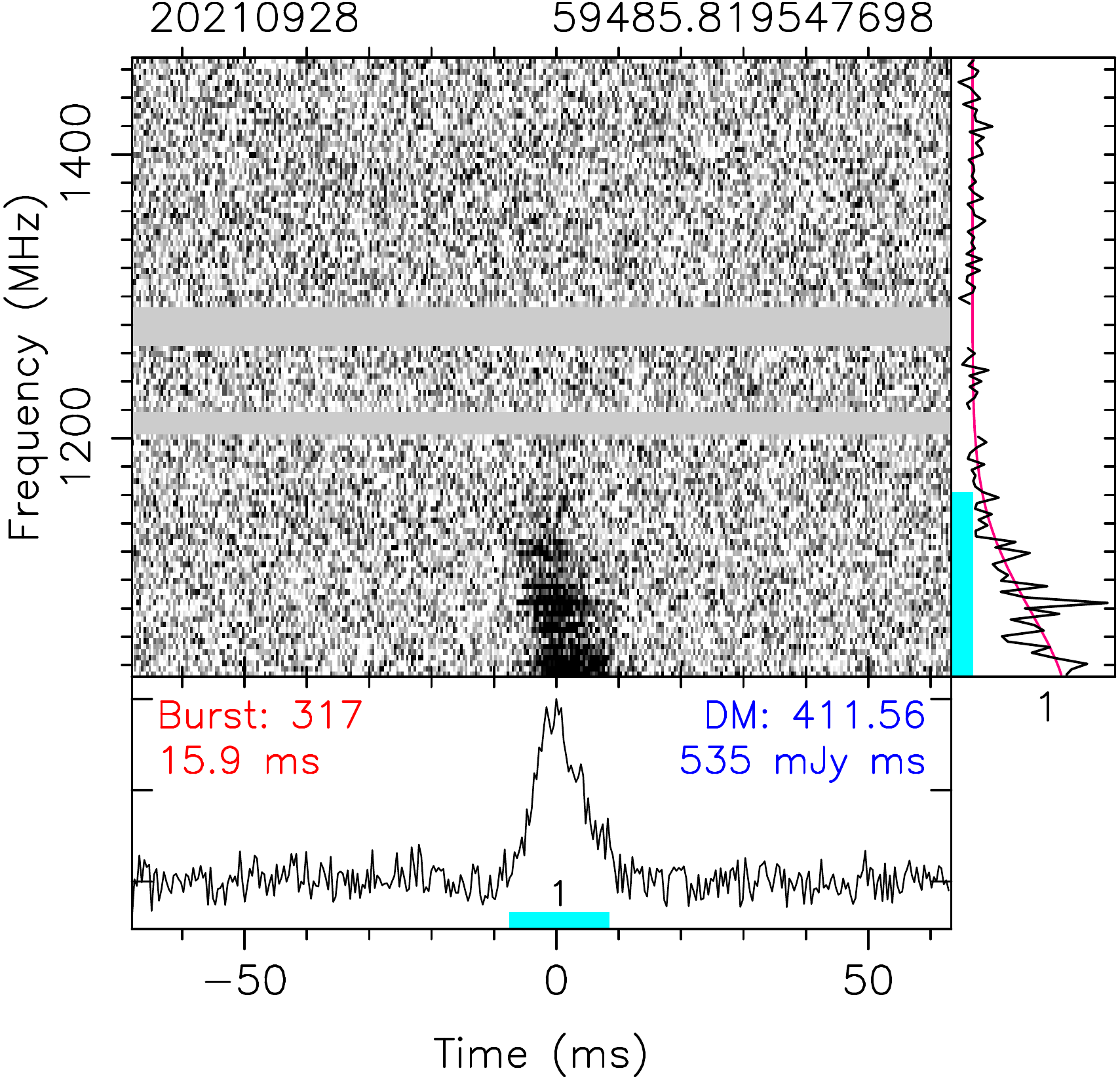}
    \includegraphics[height=37mm]{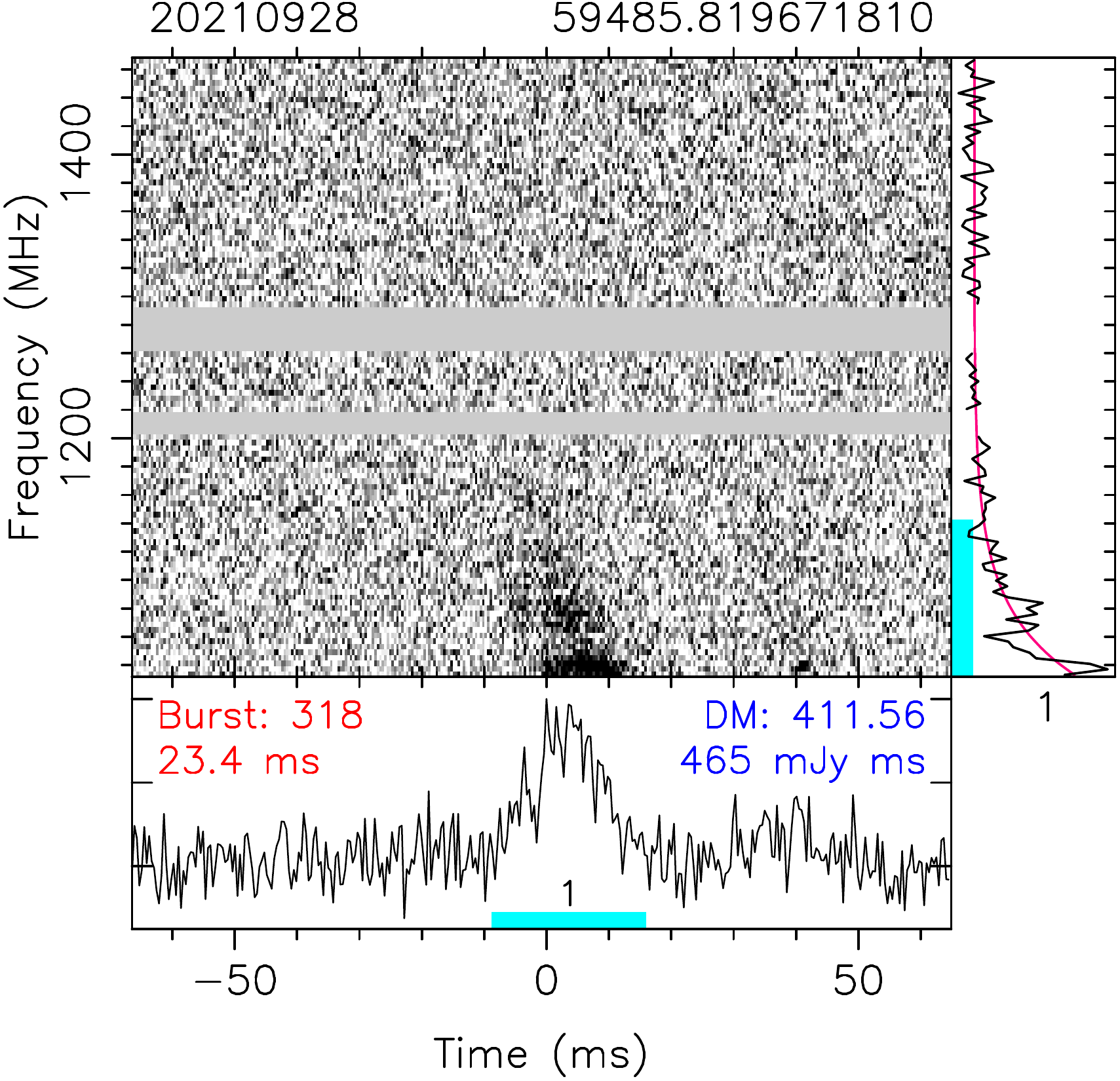}
    \includegraphics[height=37mm]{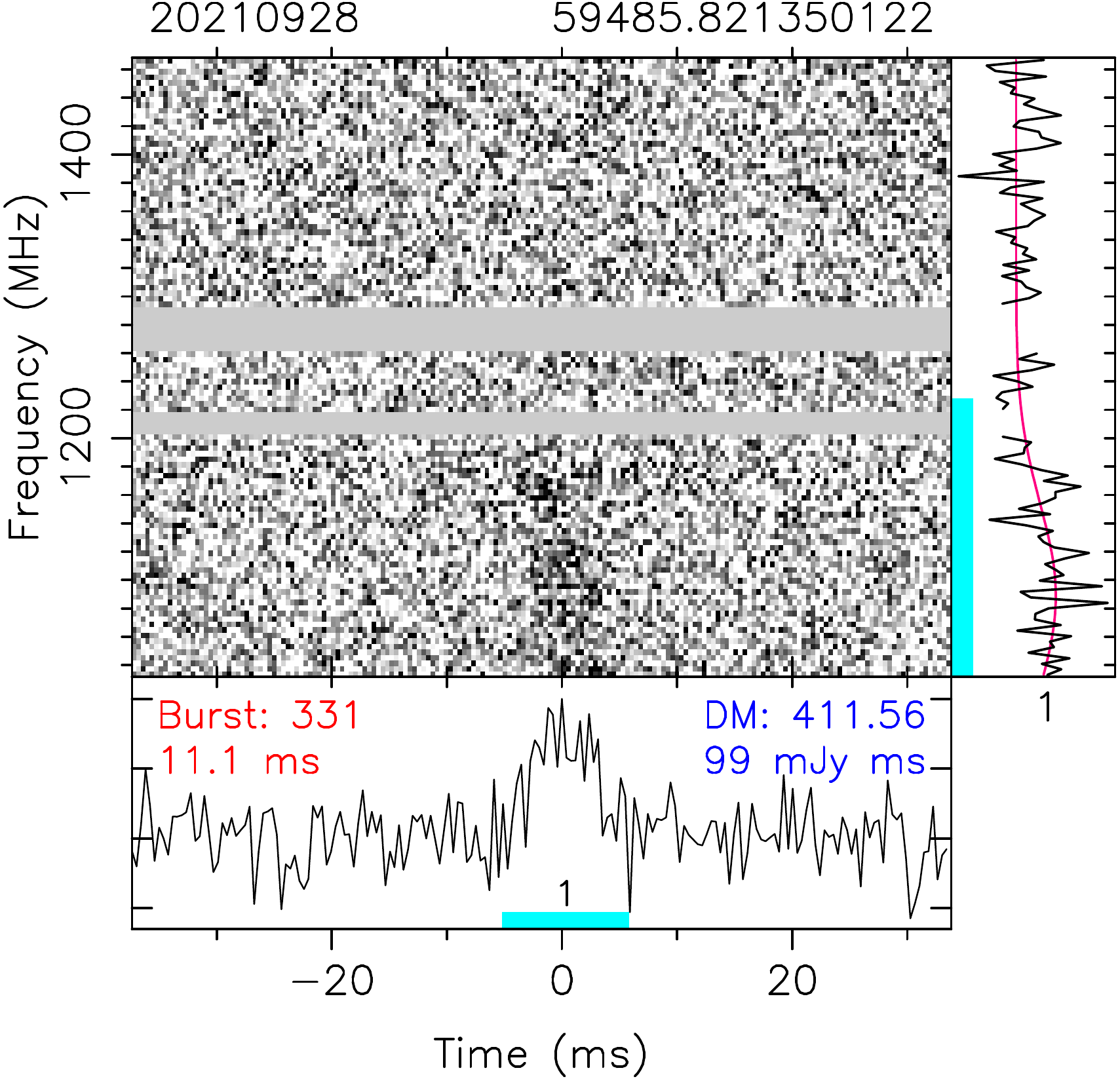}
    \includegraphics[height=37mm]{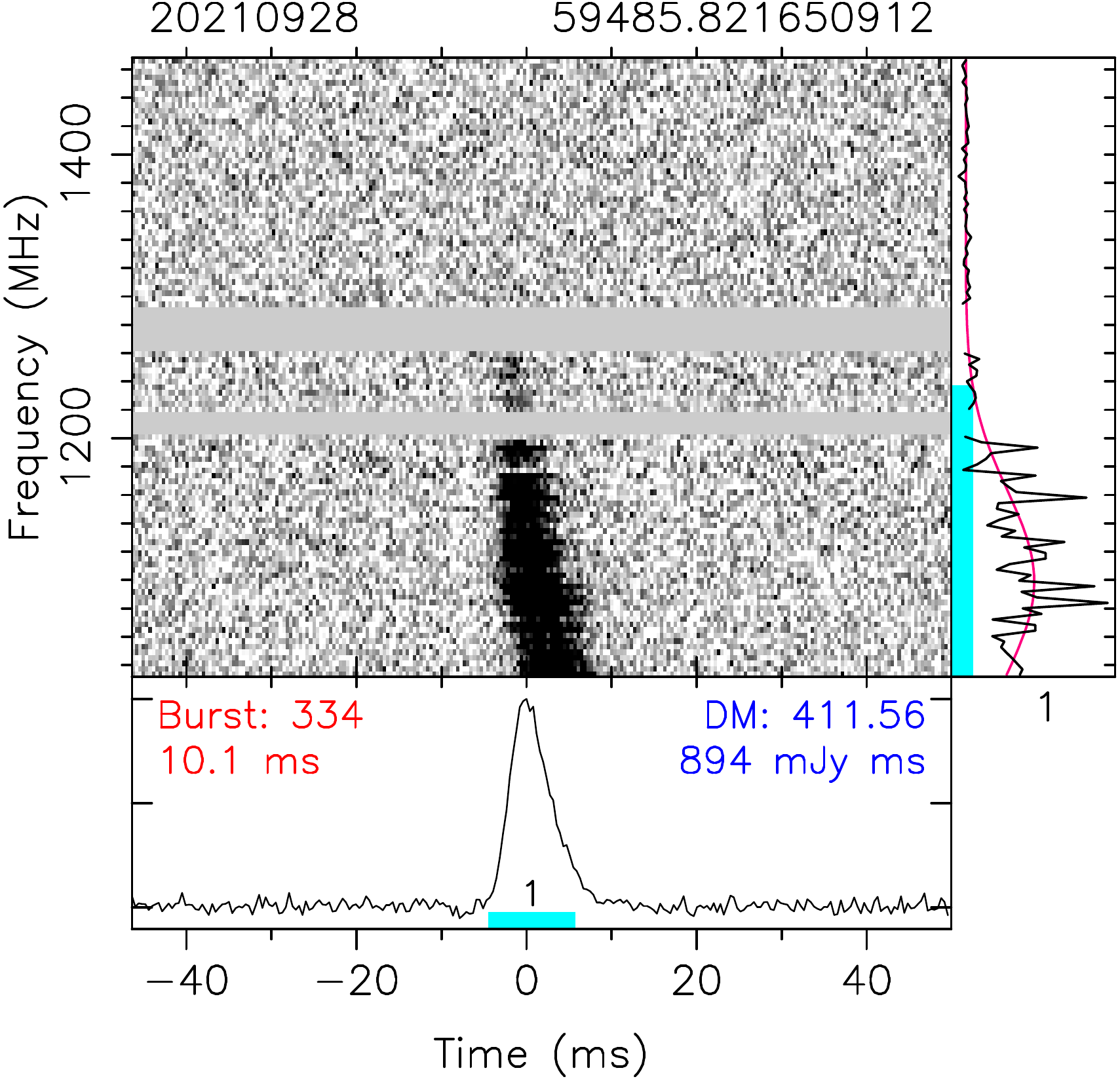}
    \includegraphics[height=37mm]{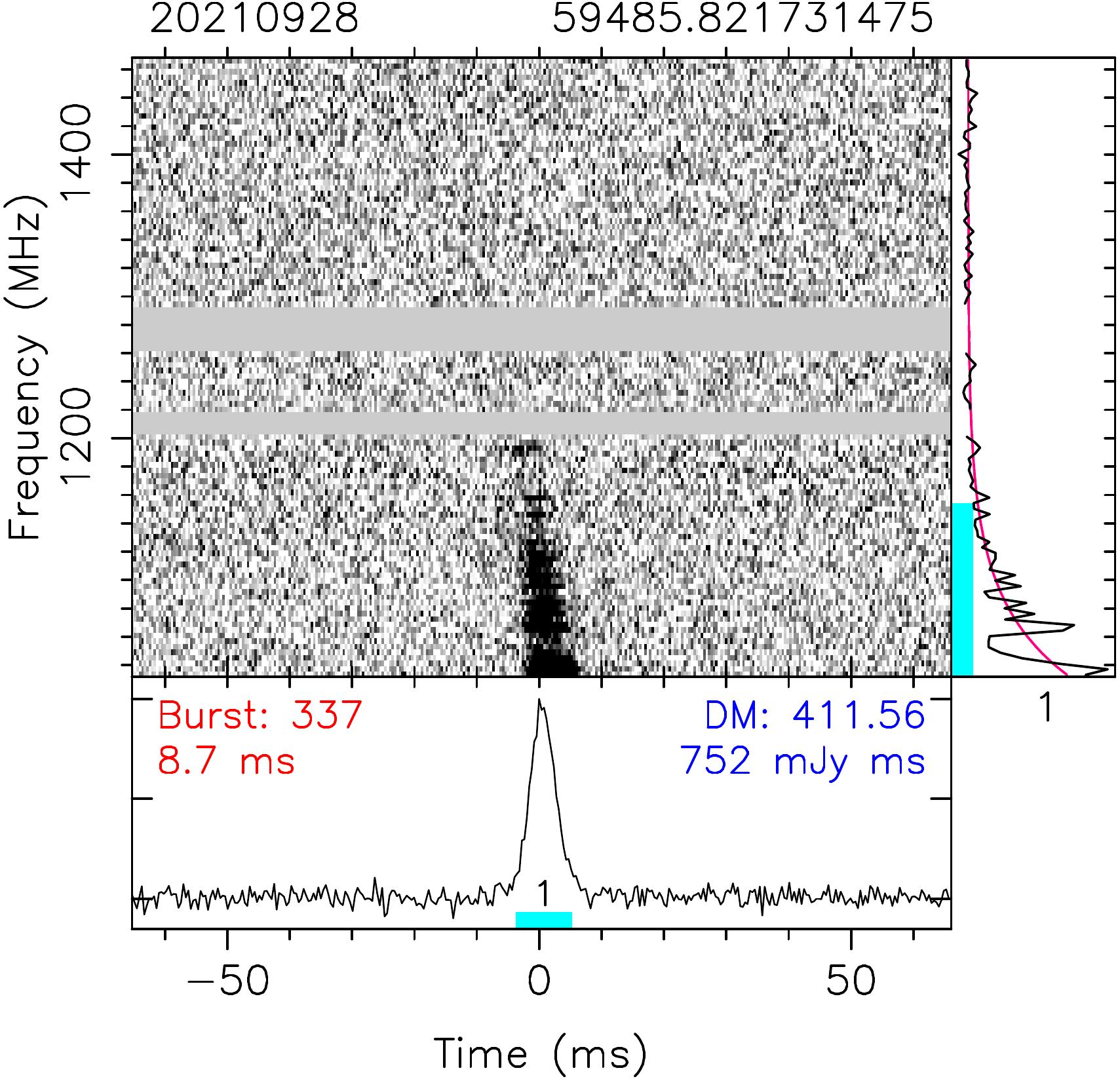}
    \includegraphics[height=37mm]{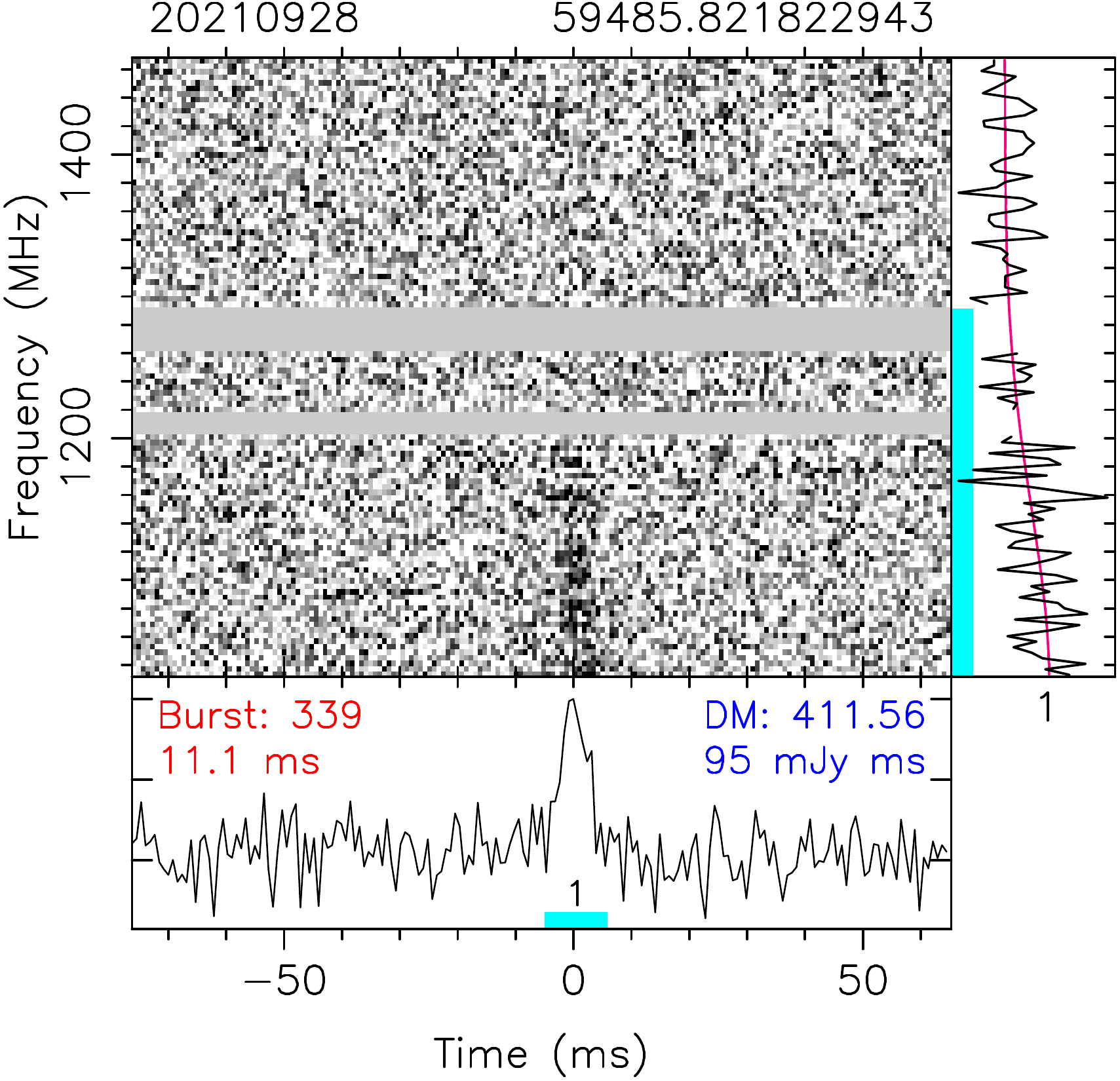}
    \includegraphics[height=37mm]{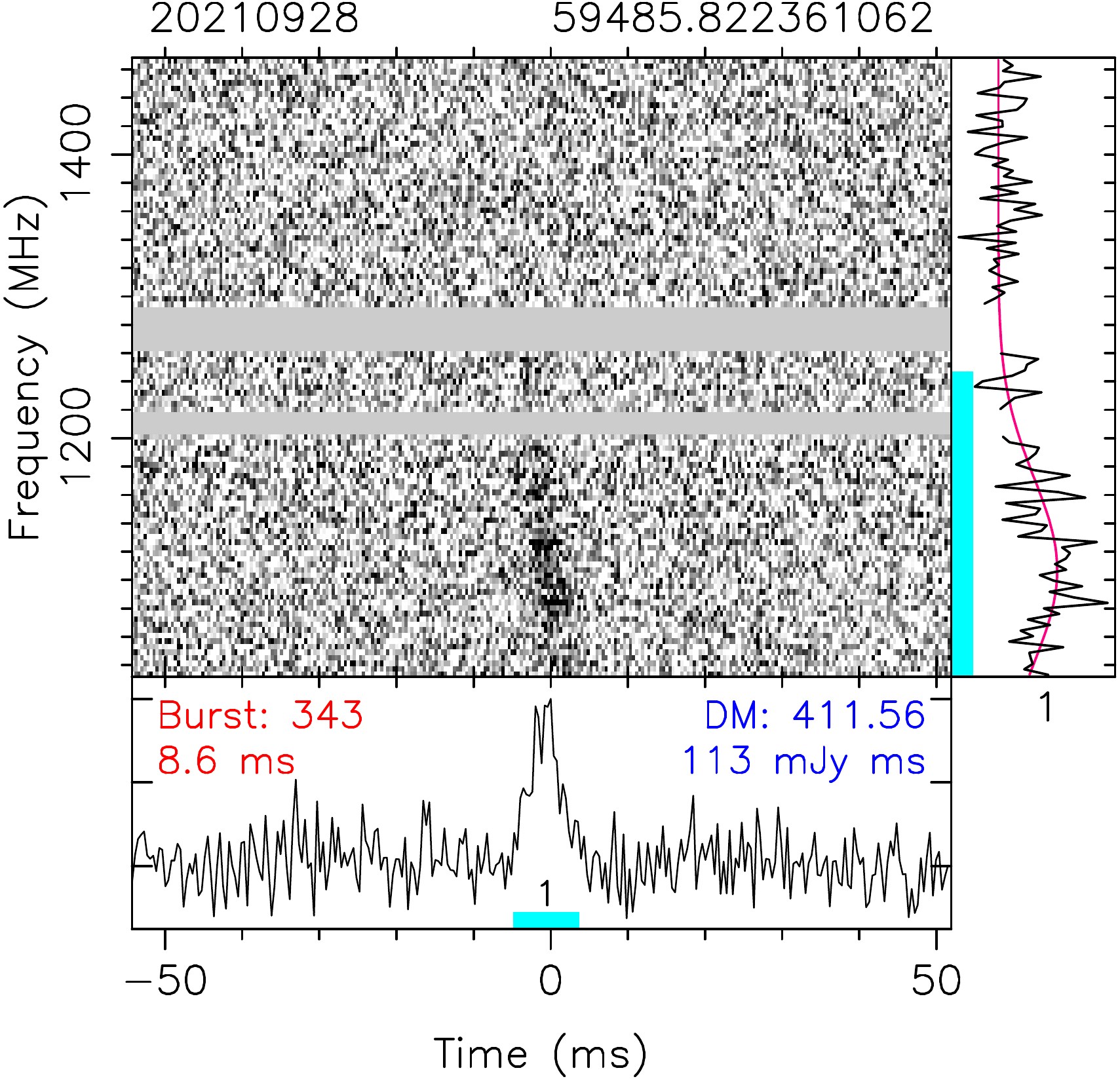}
    \includegraphics[height=37mm]{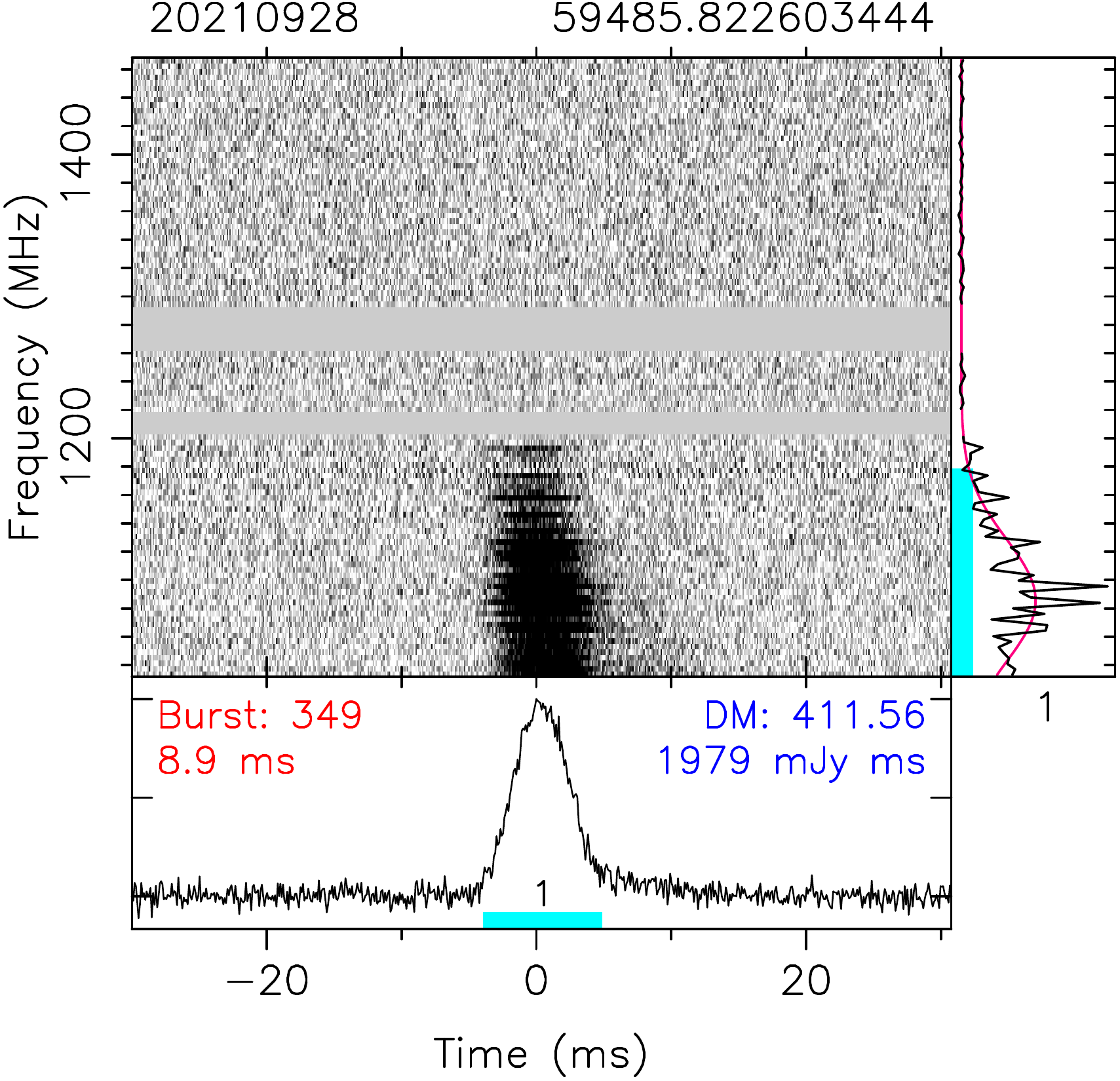}
    \includegraphics[height=37mm]{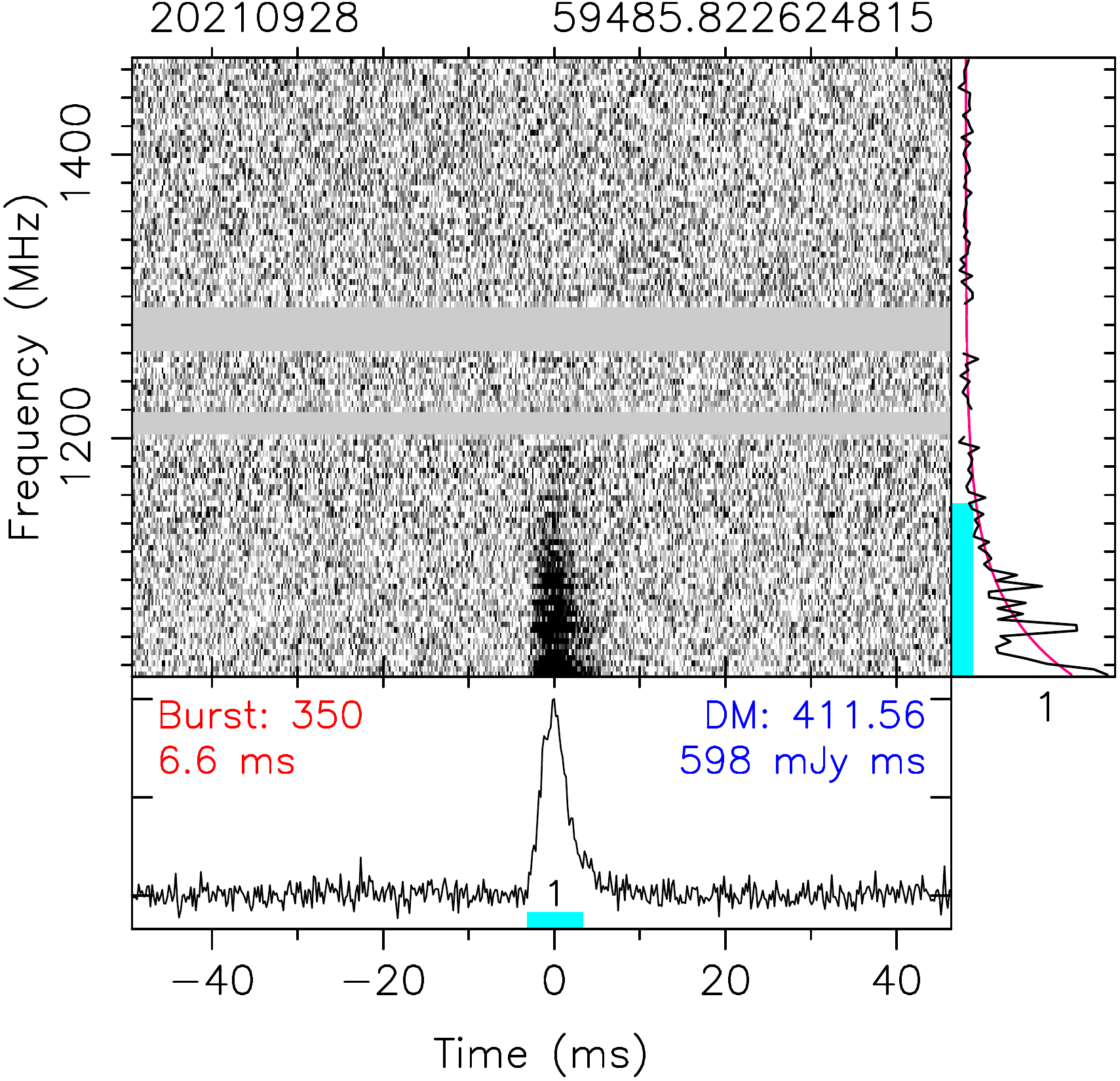}
    \includegraphics[height=37mm]{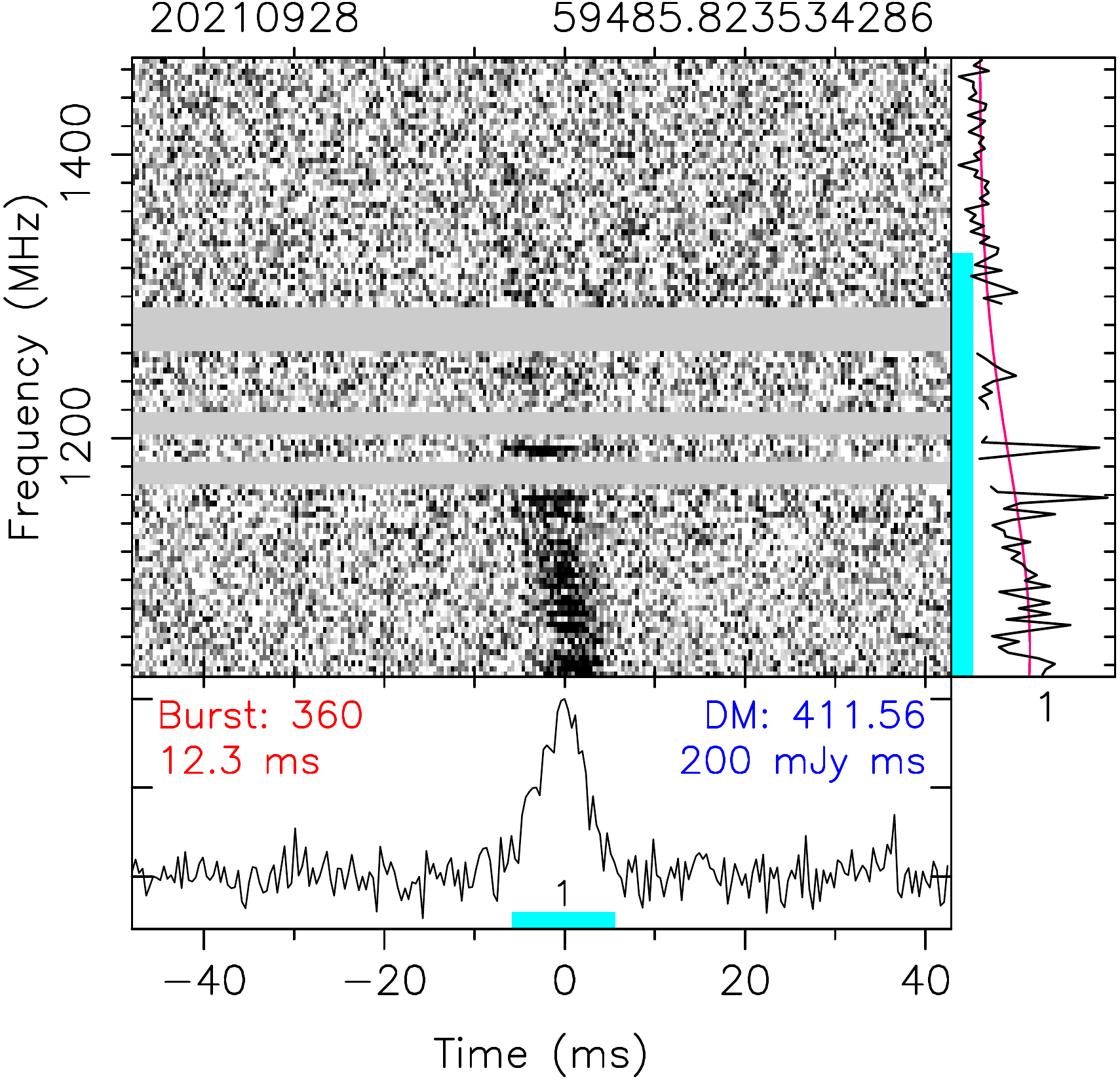}
    \includegraphics[height=37mm]{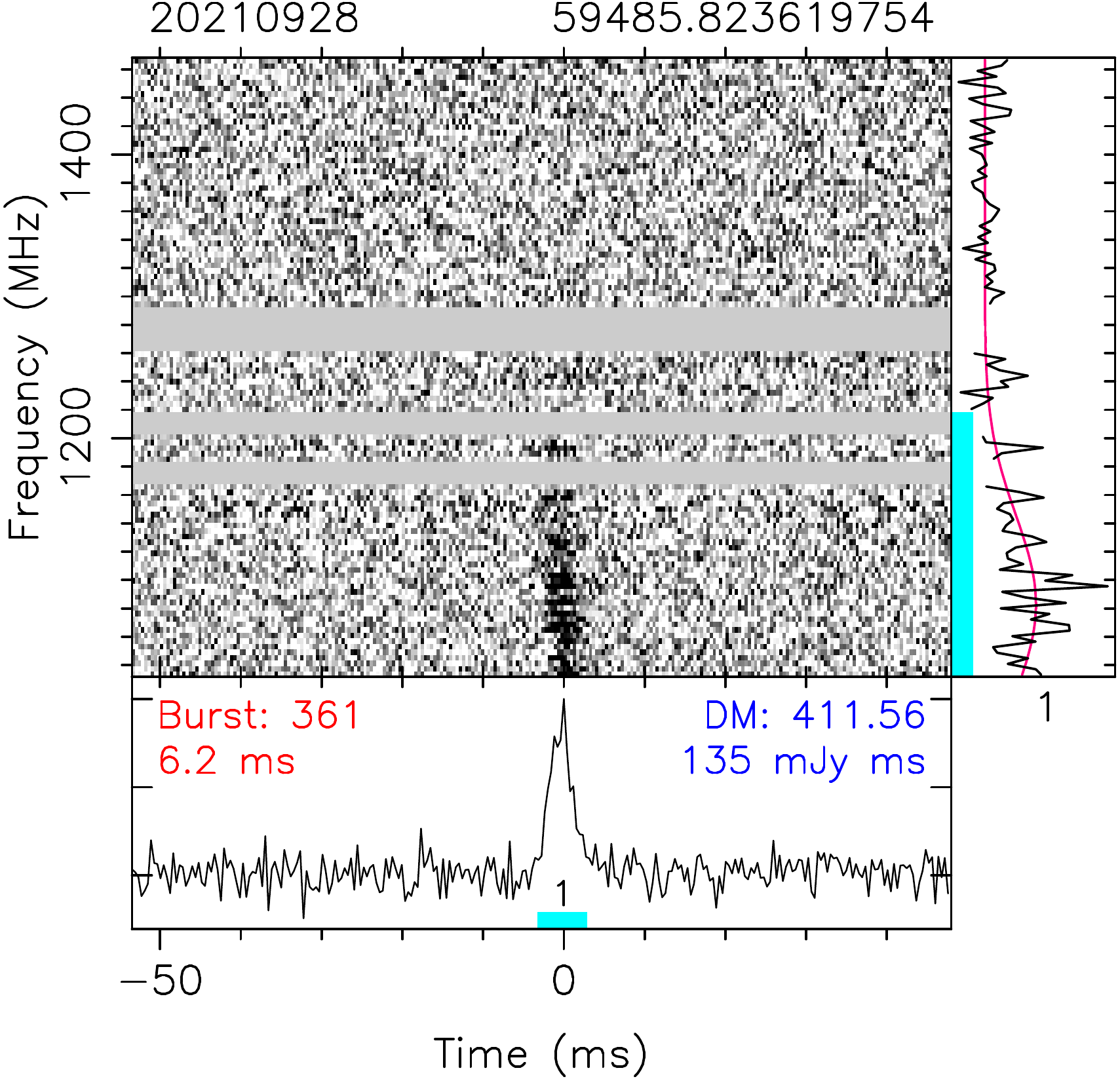}
    \caption{ \it{ -- continued and ended}.
}
\end{figure*}

\begin{figure*}
    \flushleft
    \includegraphics[height=37mm]{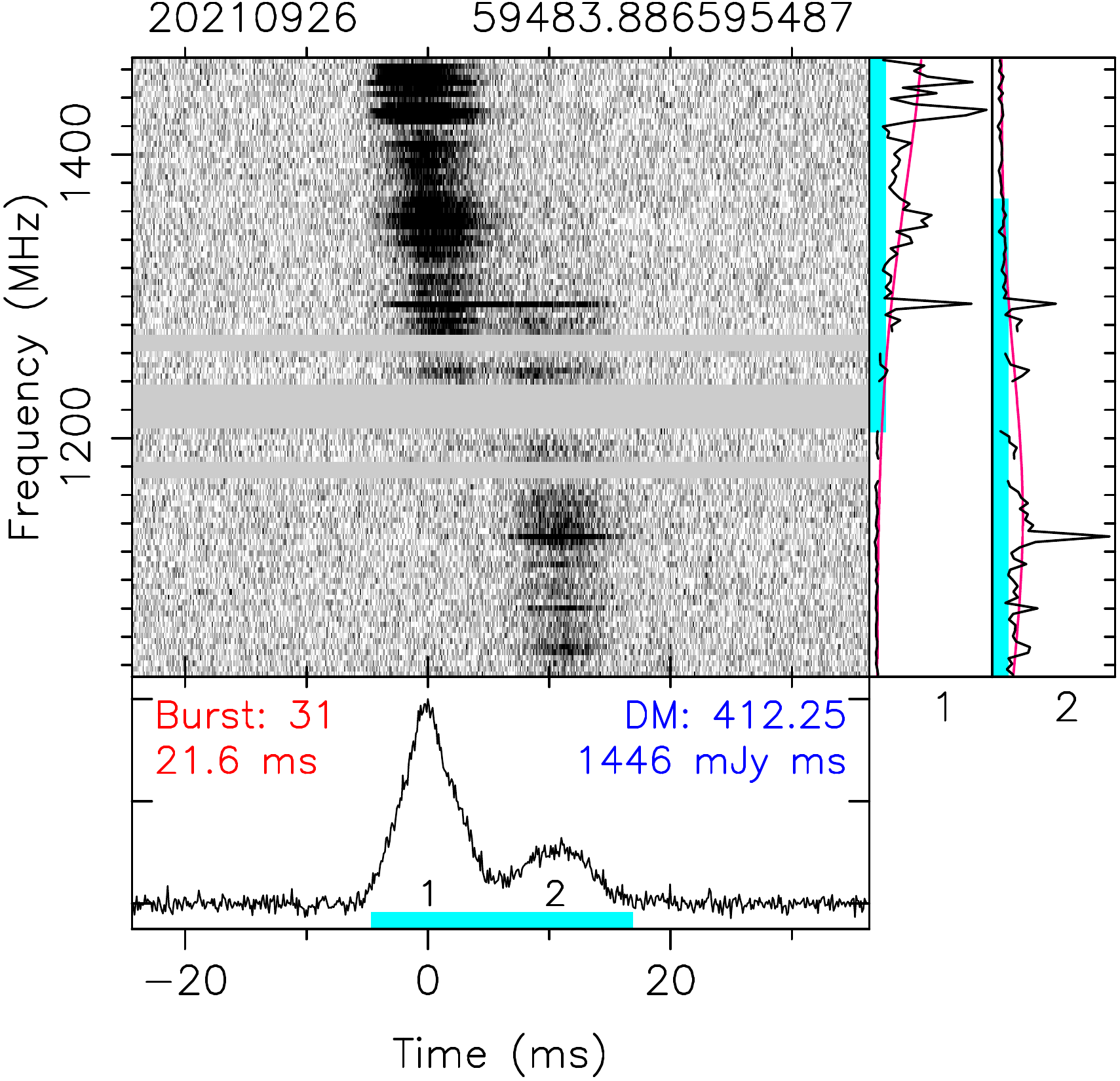}
    \includegraphics[height=37mm]{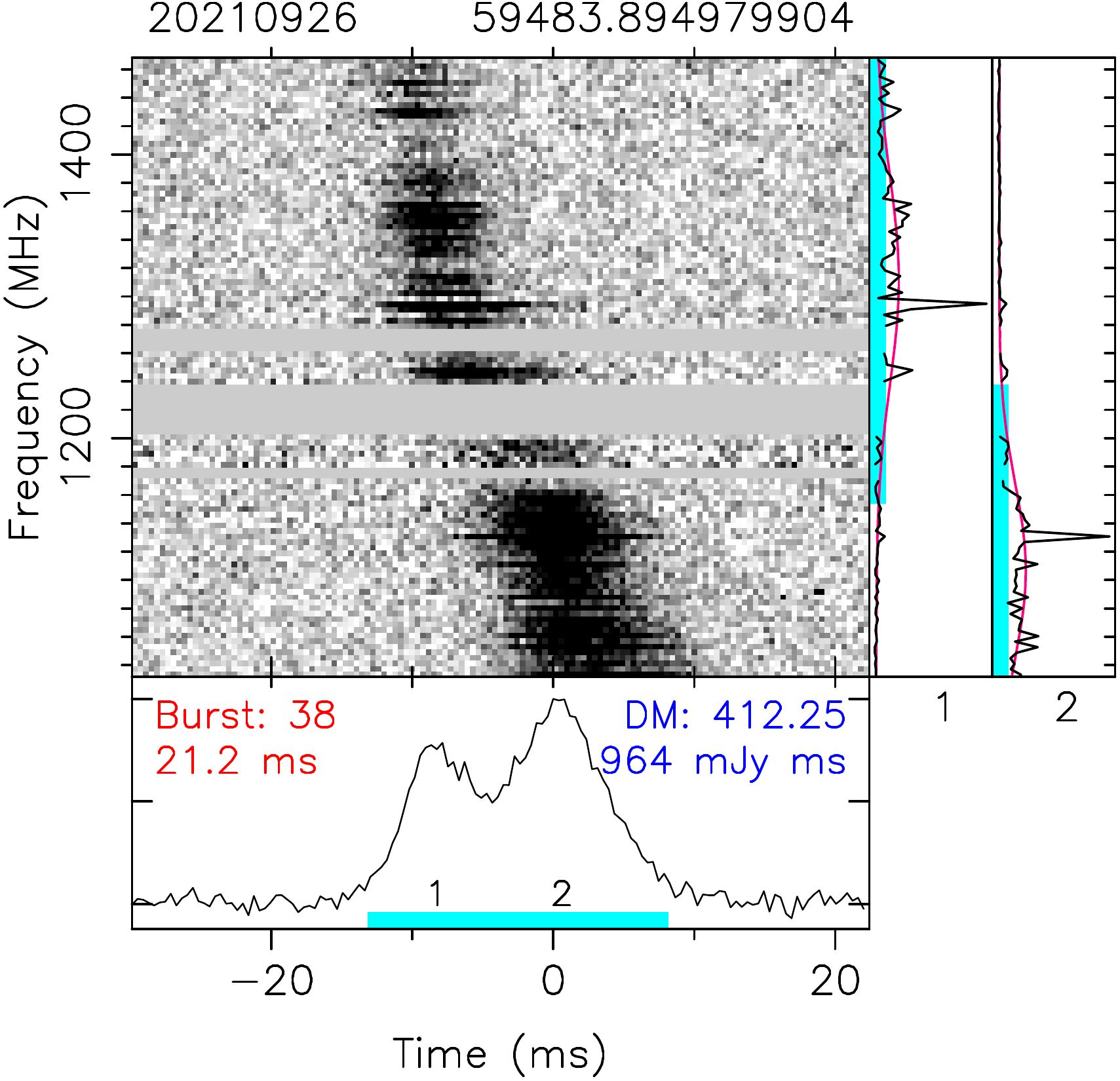}
    \includegraphics[height=37mm]{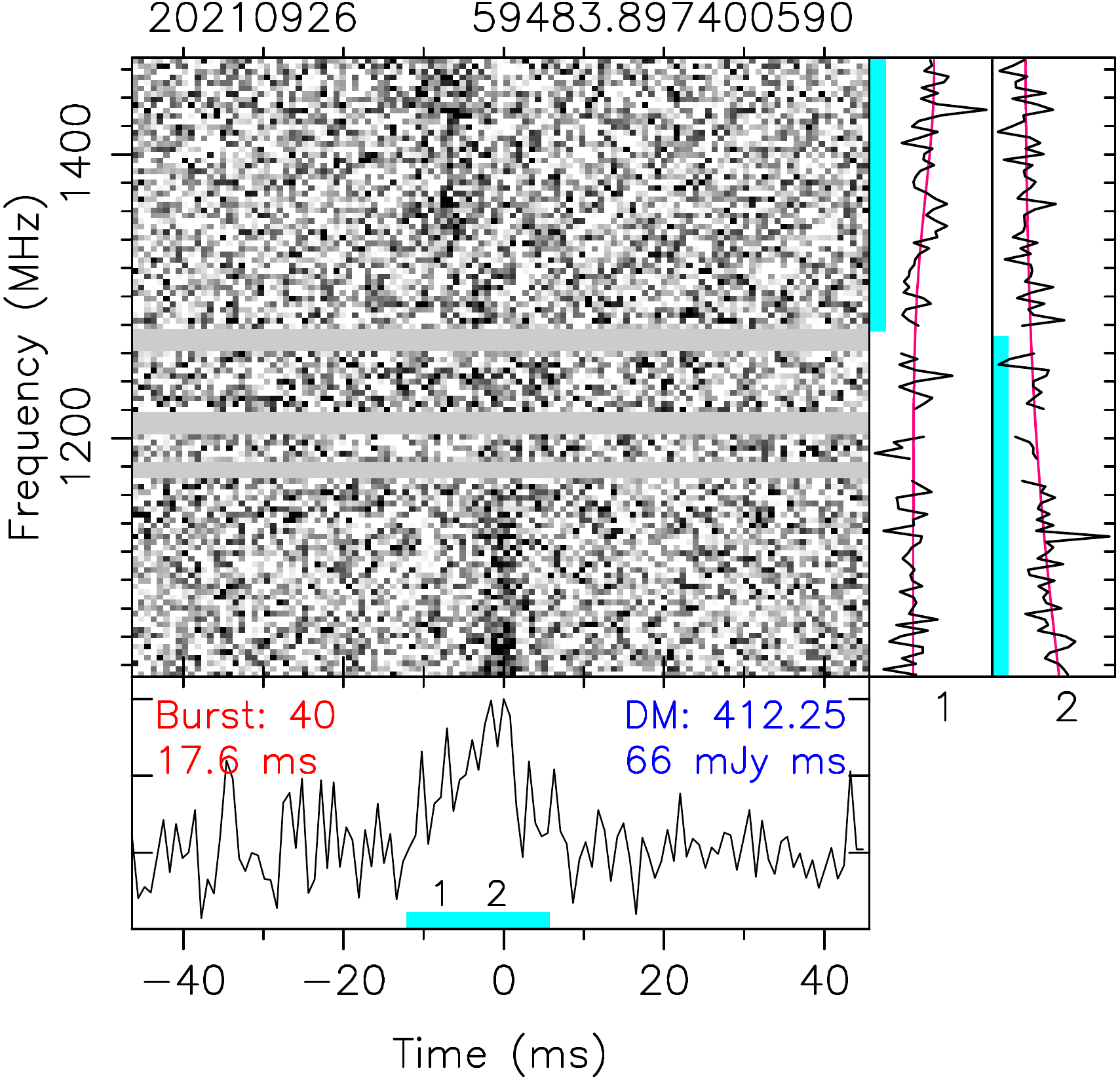}
    \includegraphics[height=37mm]{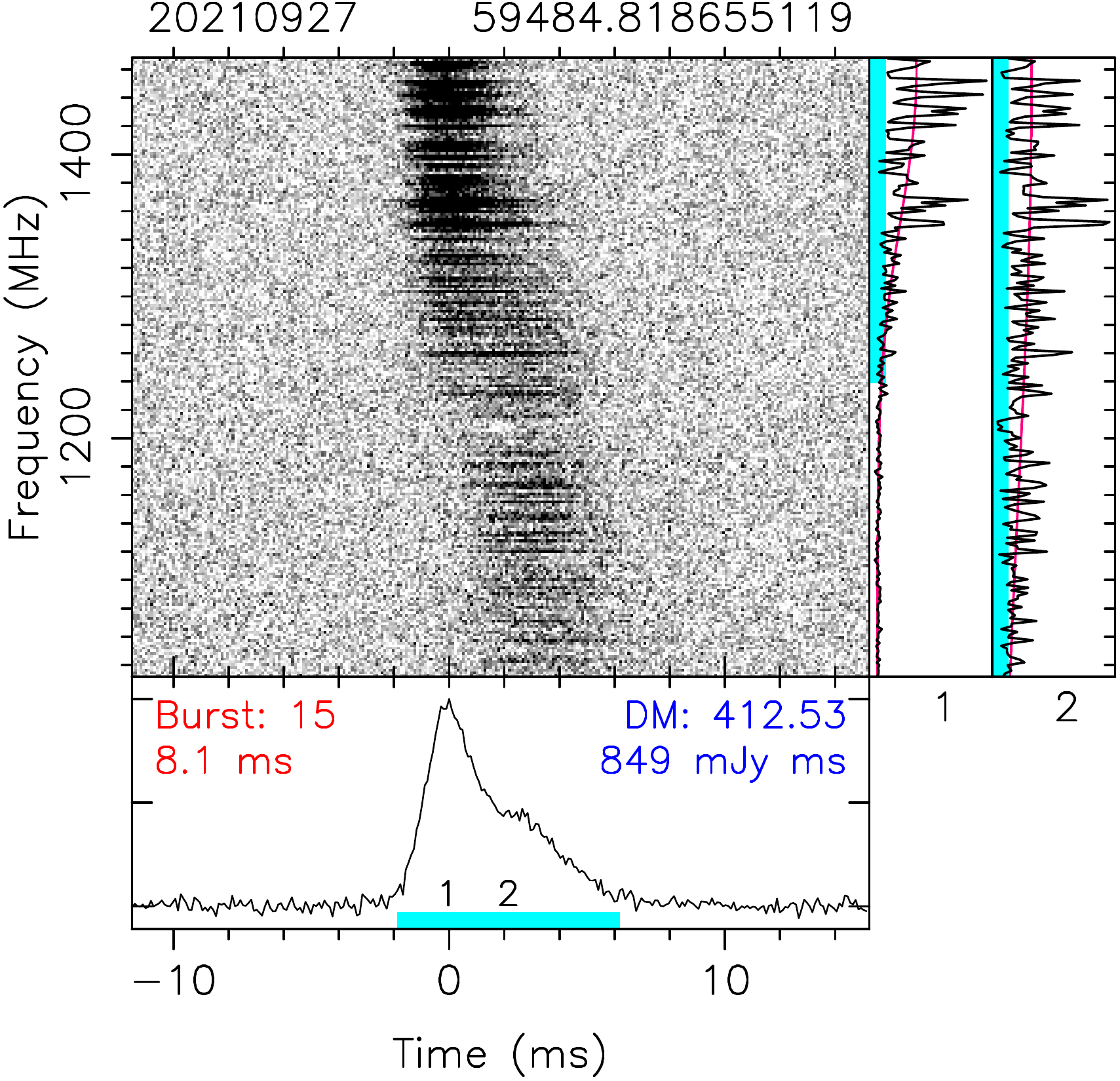}
    \includegraphics[height=37mm]{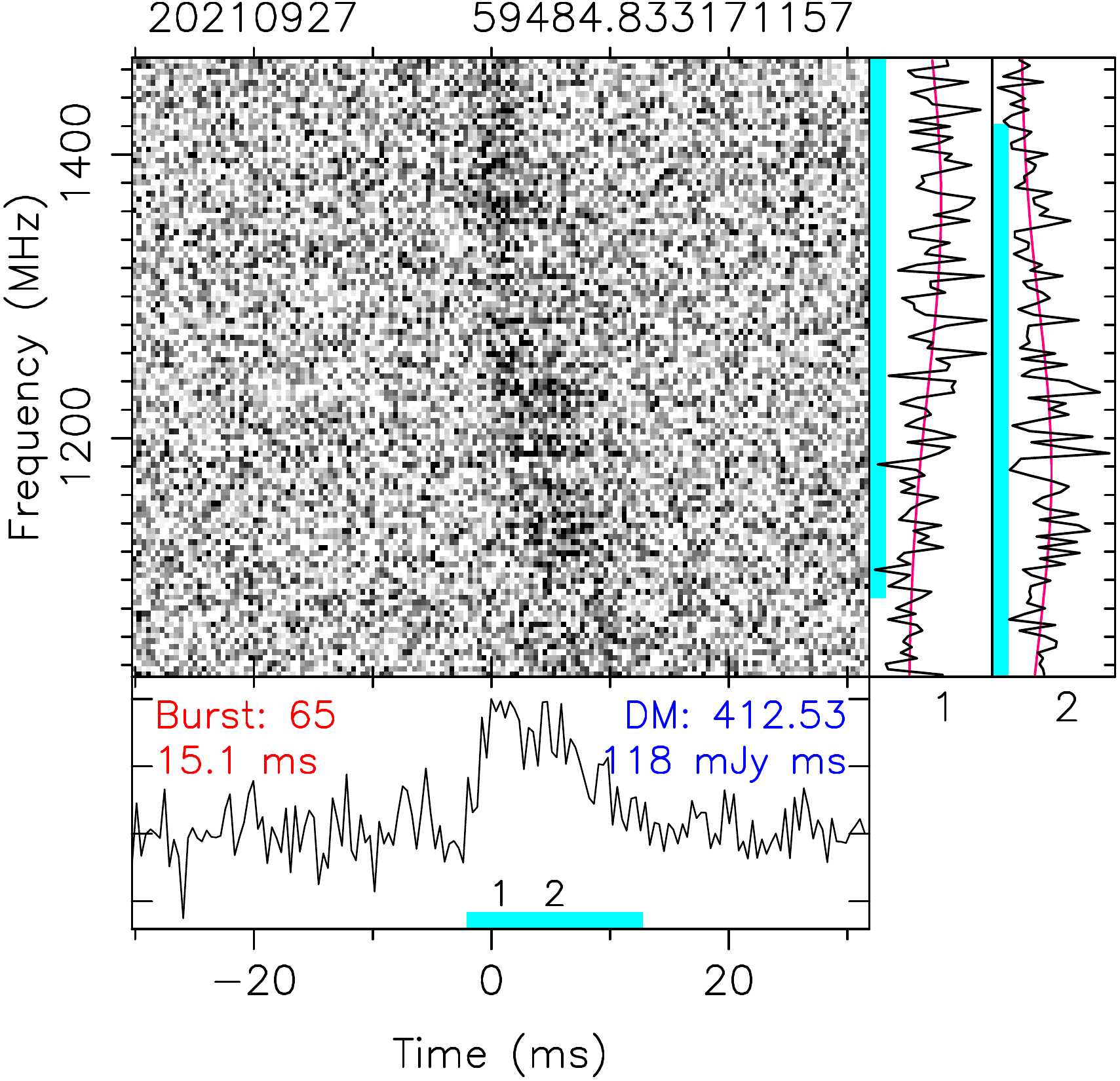}
    \includegraphics[height=37mm]{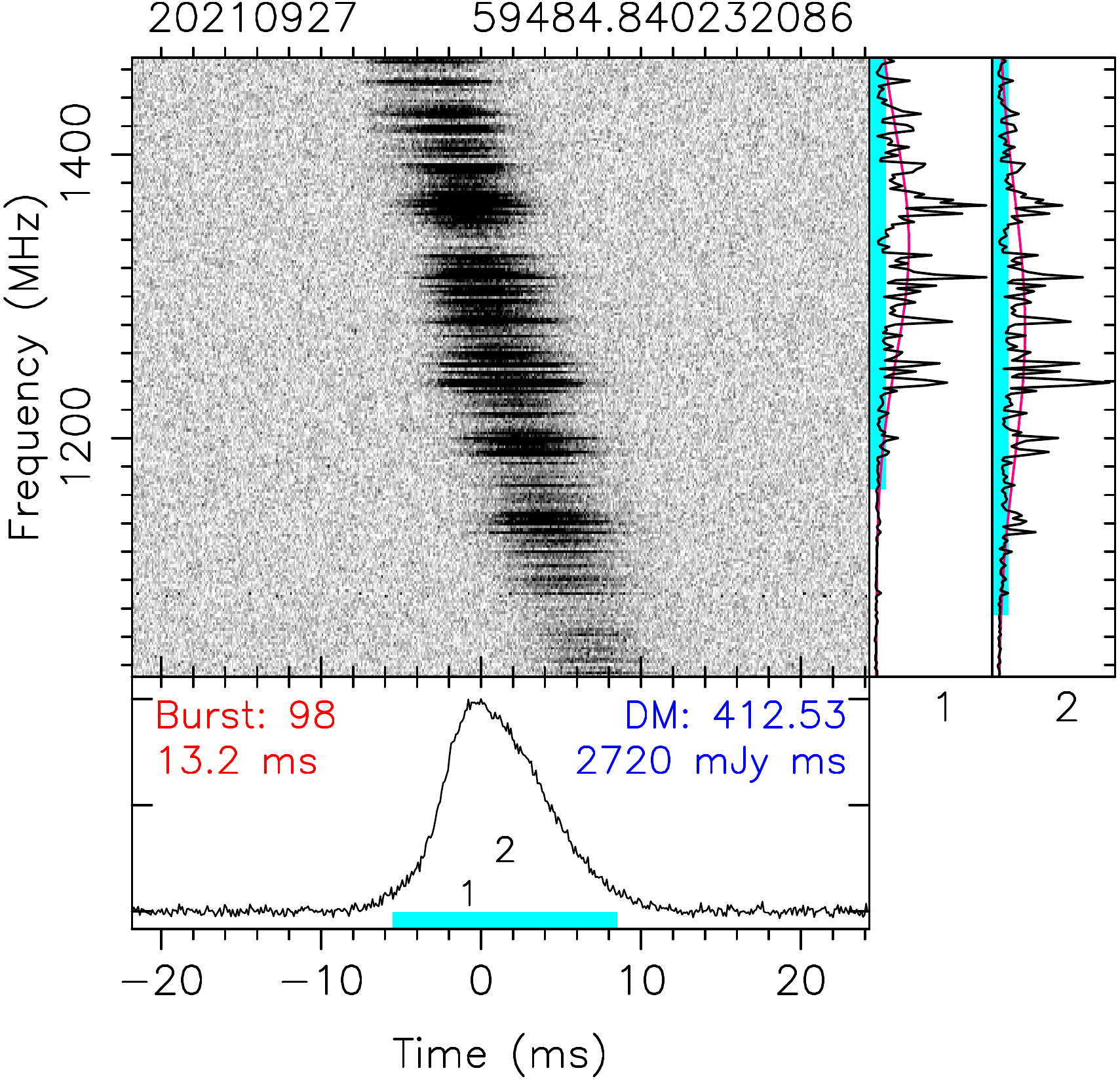}
    \includegraphics[height=37mm]{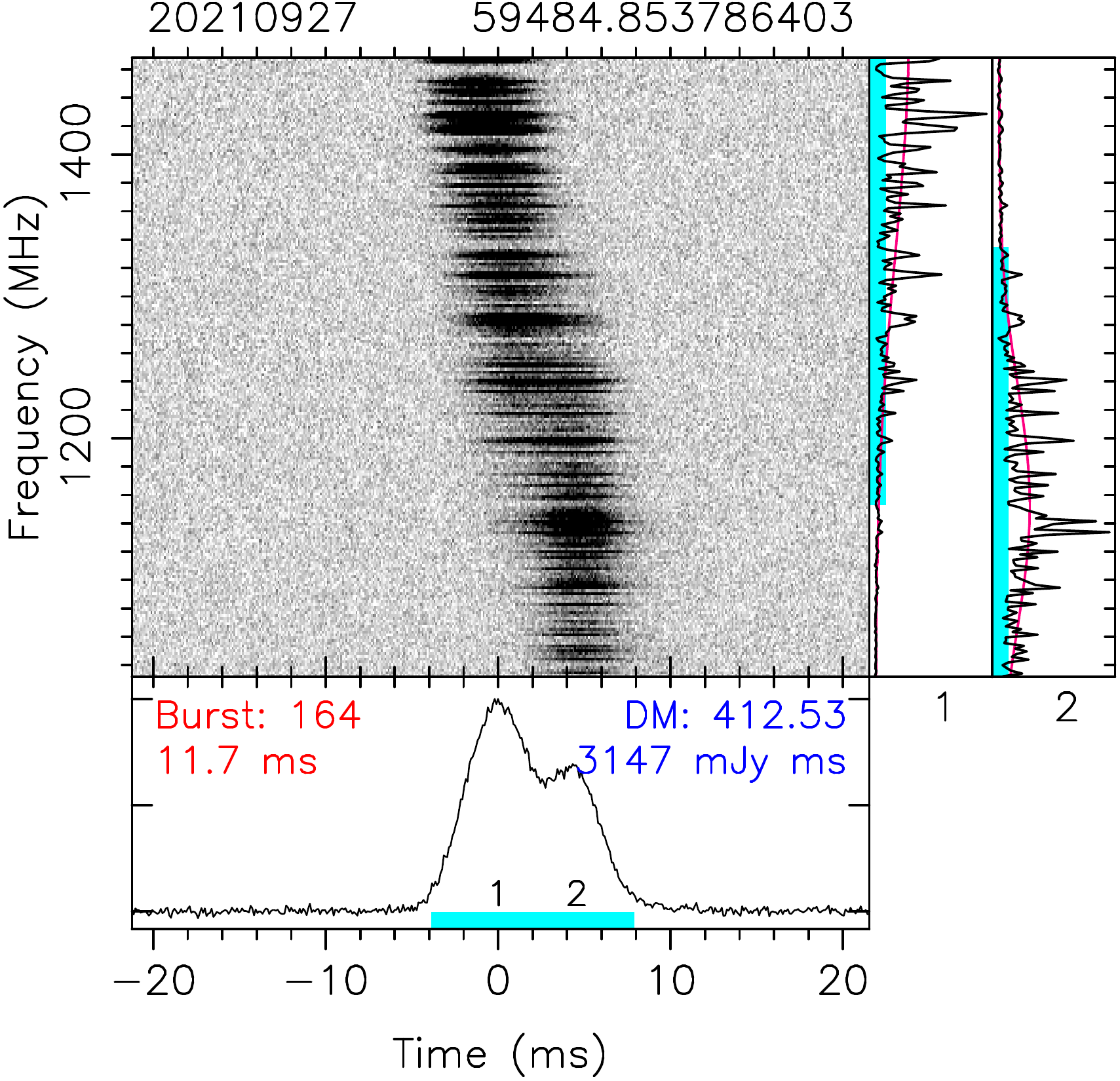}
    \includegraphics[height=37mm]{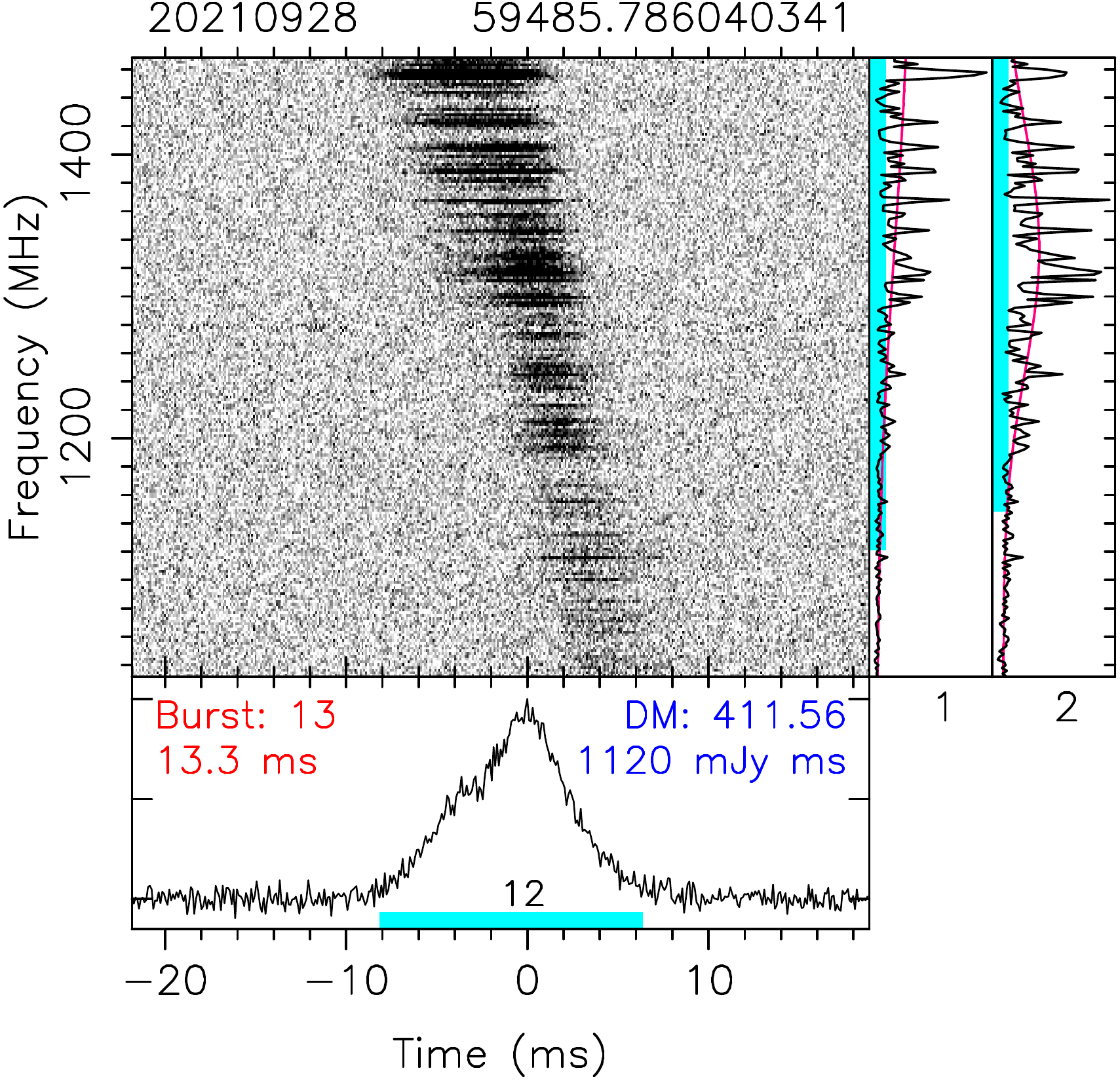}
    \includegraphics[height=37mm]{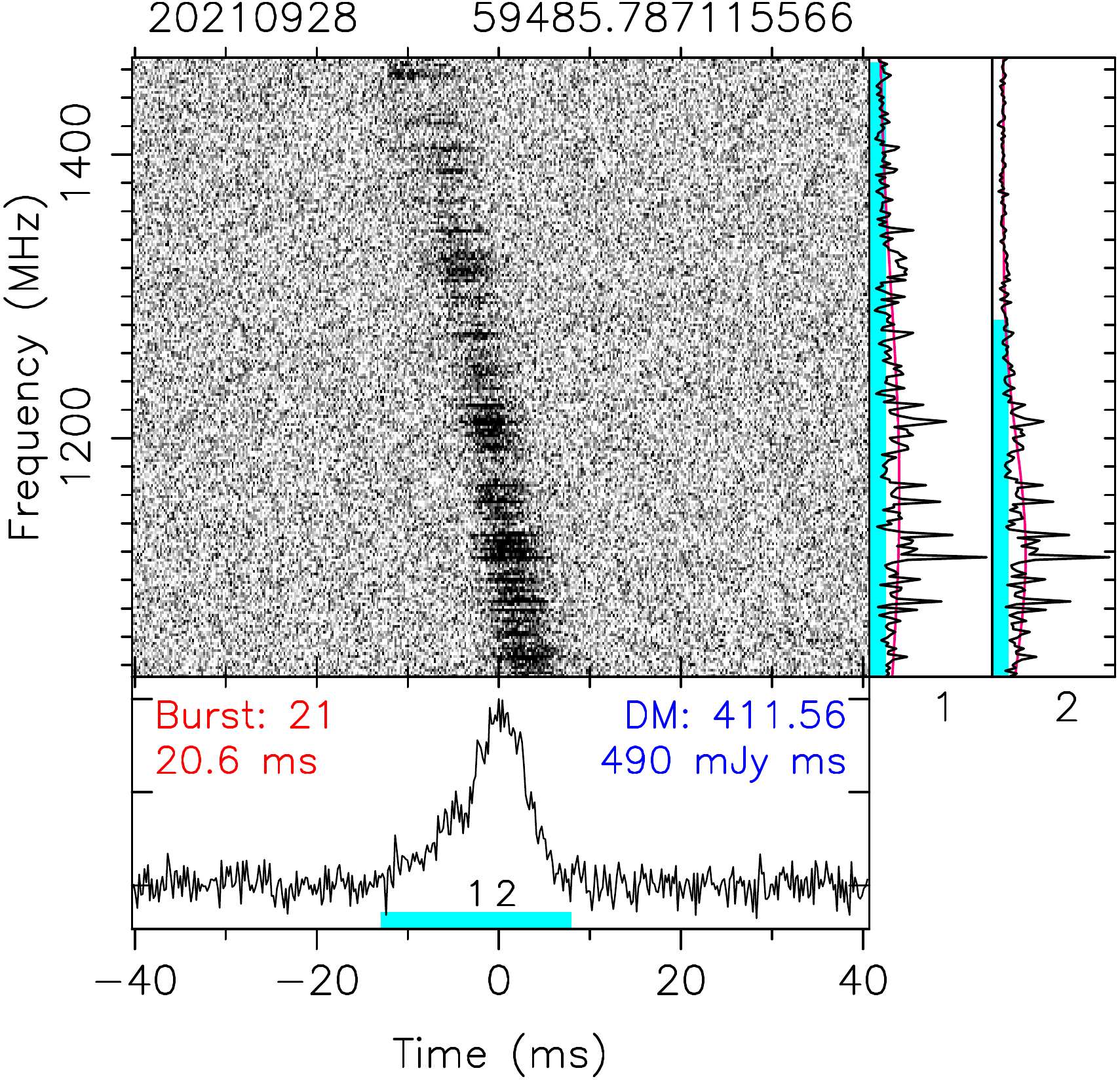}
    \includegraphics[height=37mm]{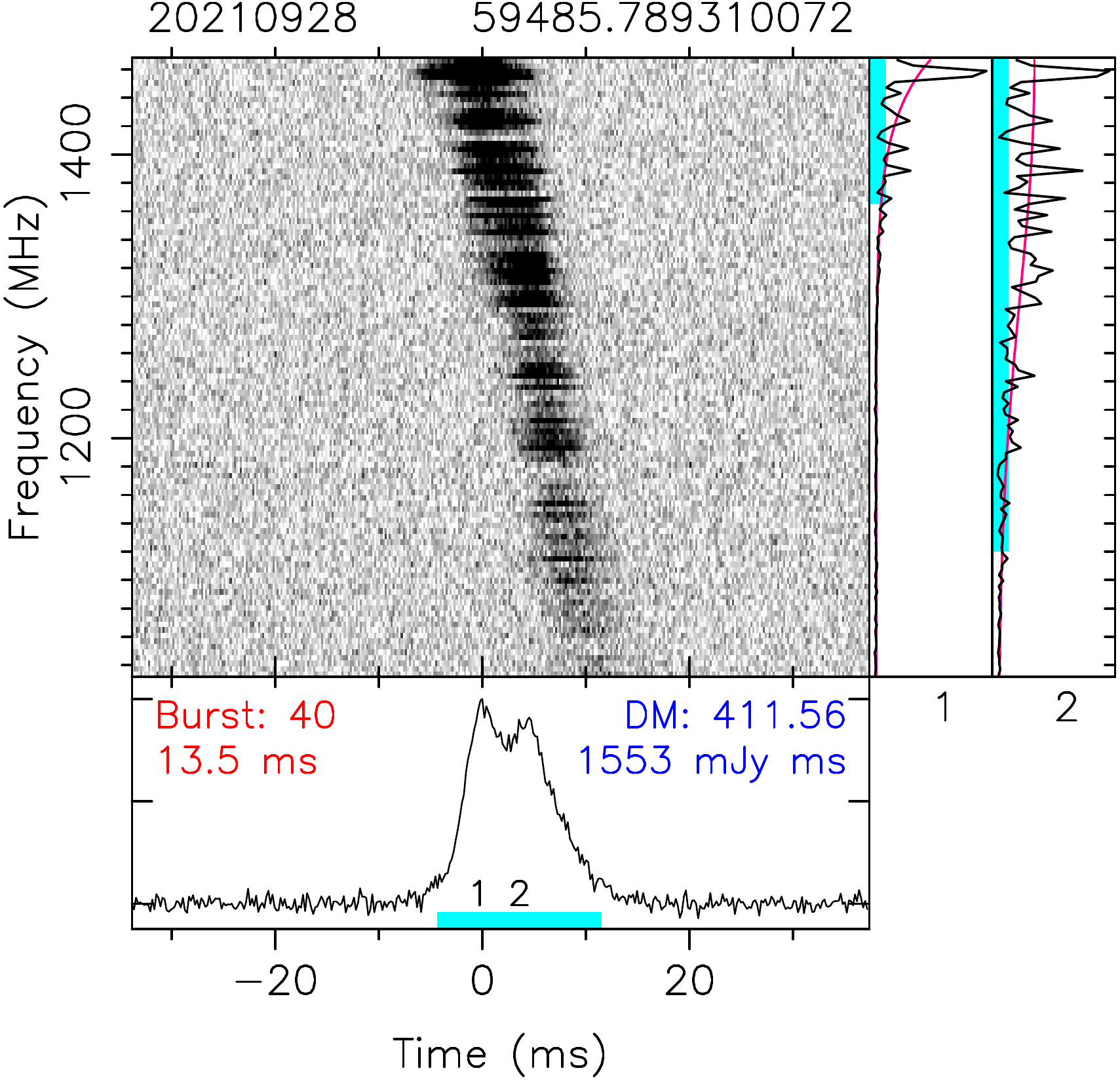}
    \includegraphics[height=37mm]{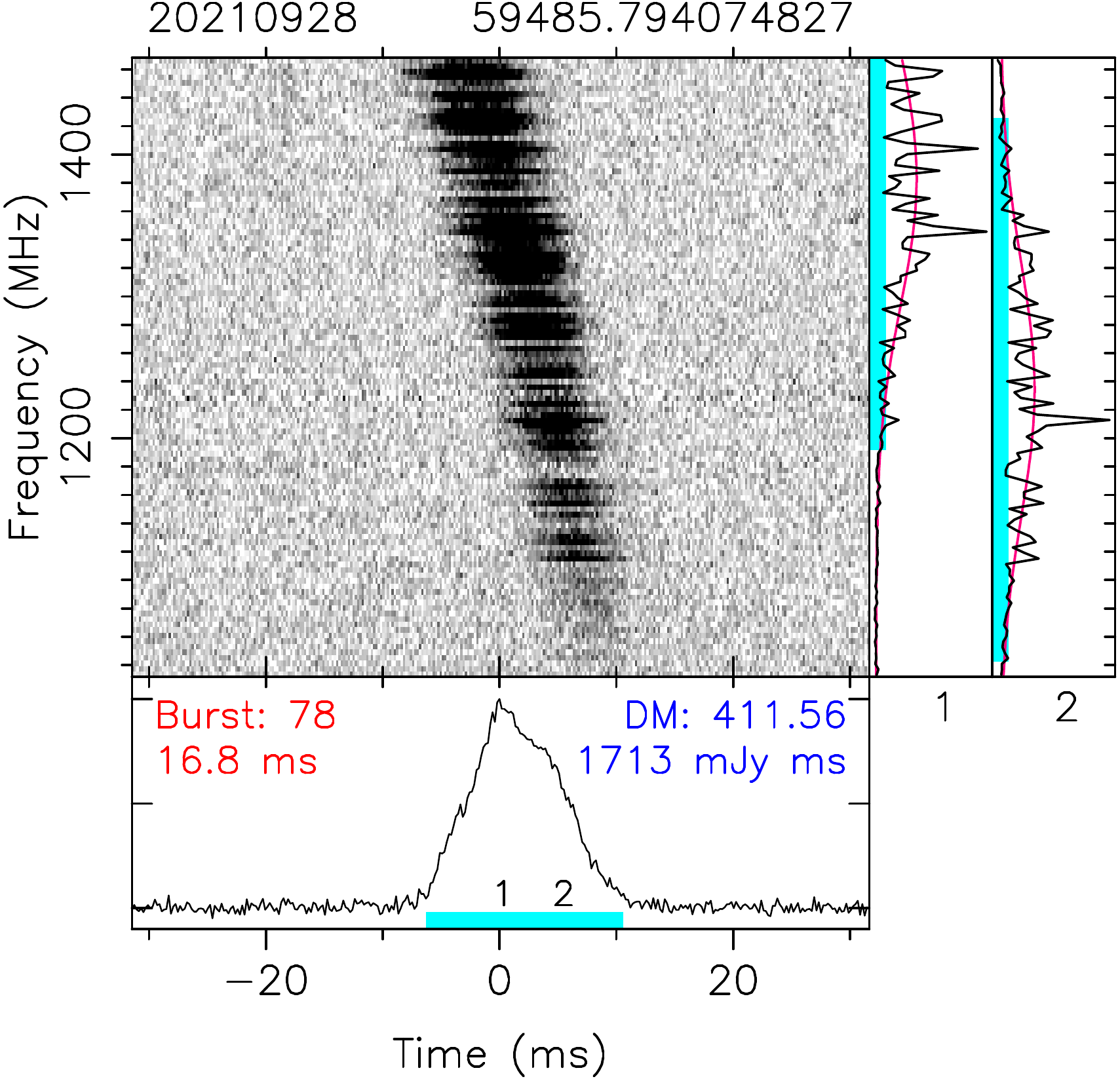}
    \includegraphics[height=37mm]{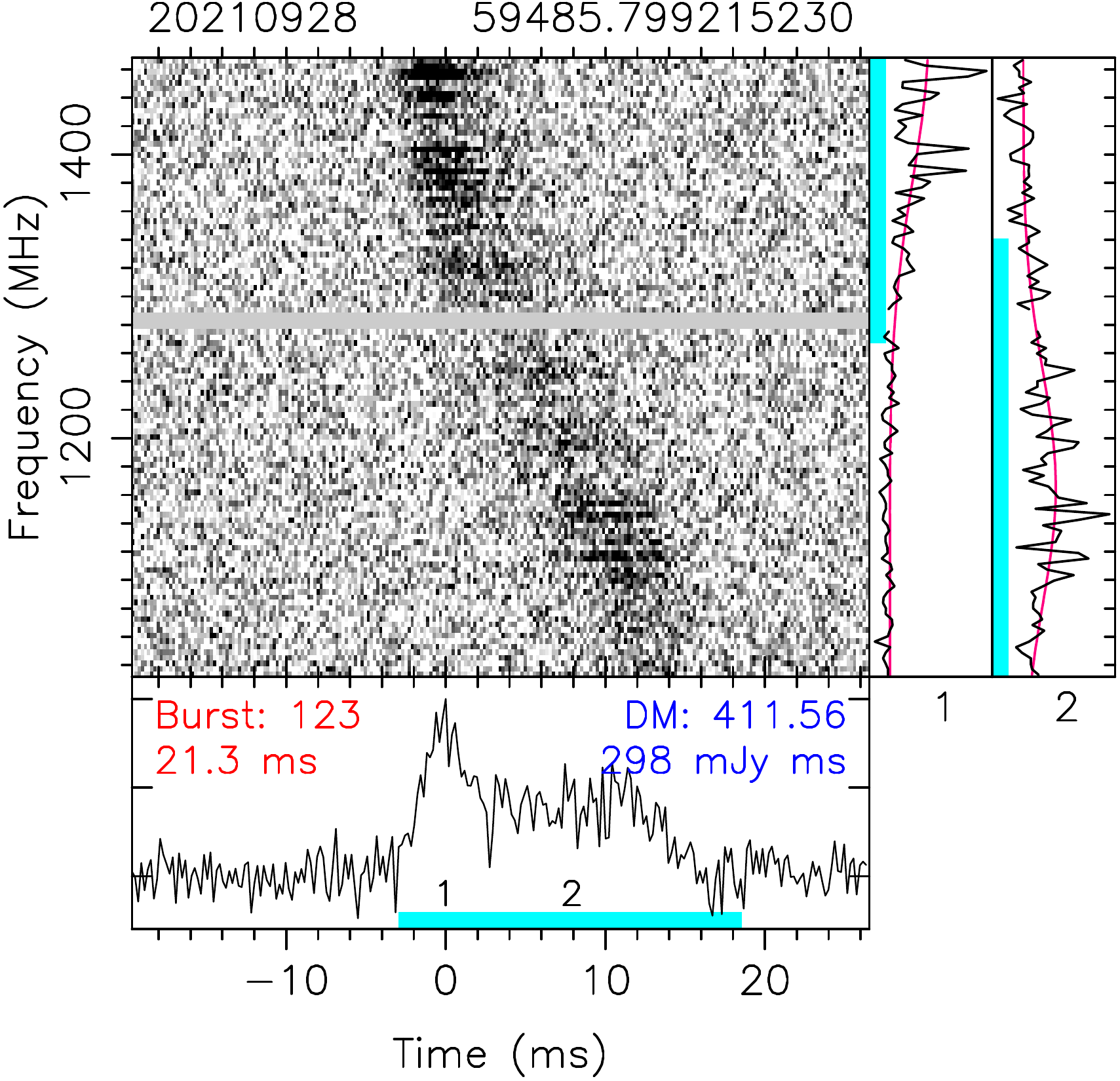}
    \includegraphics[height=37mm]{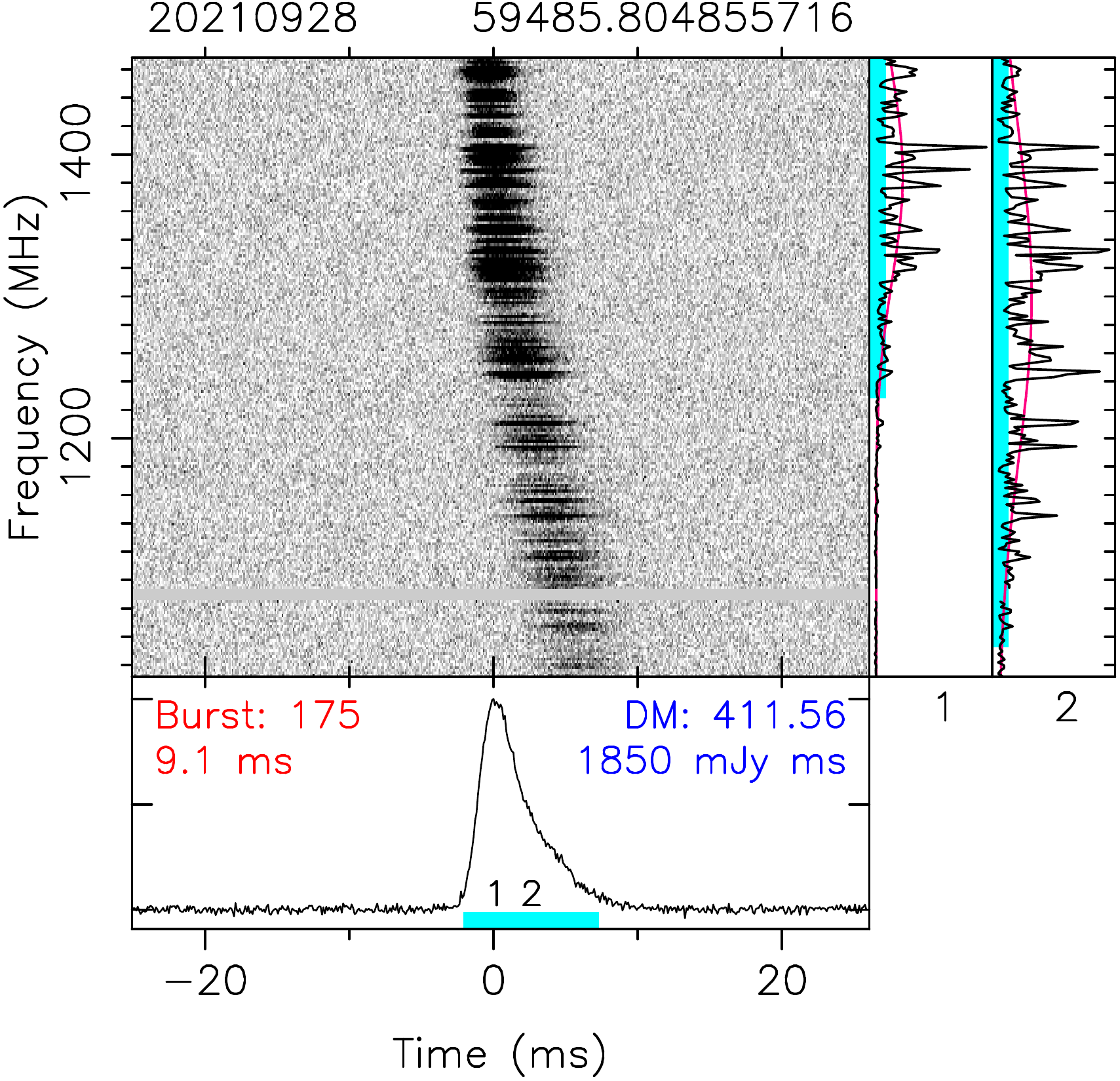}
    \includegraphics[height=37mm]{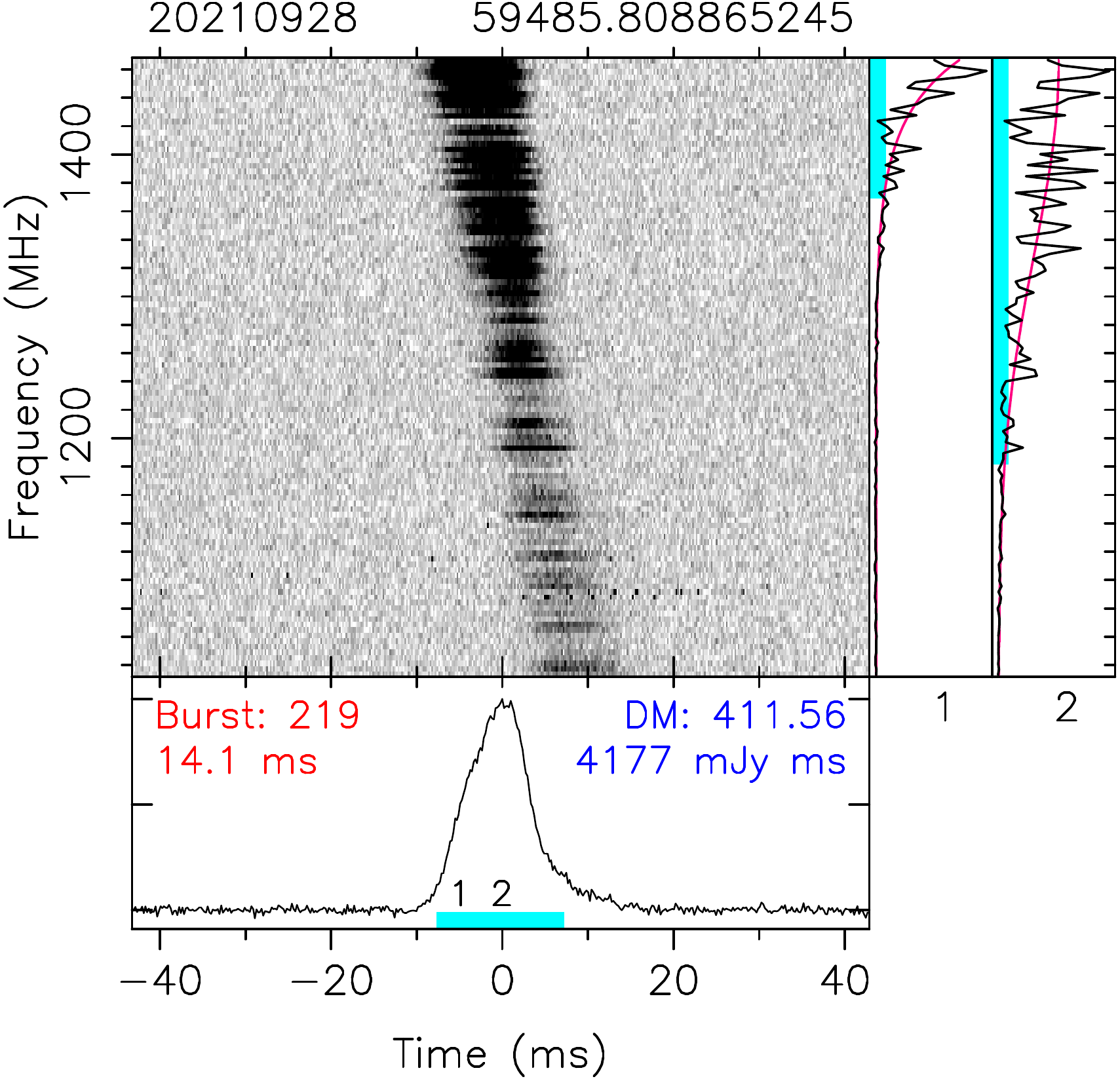}
    \includegraphics[height=37mm]{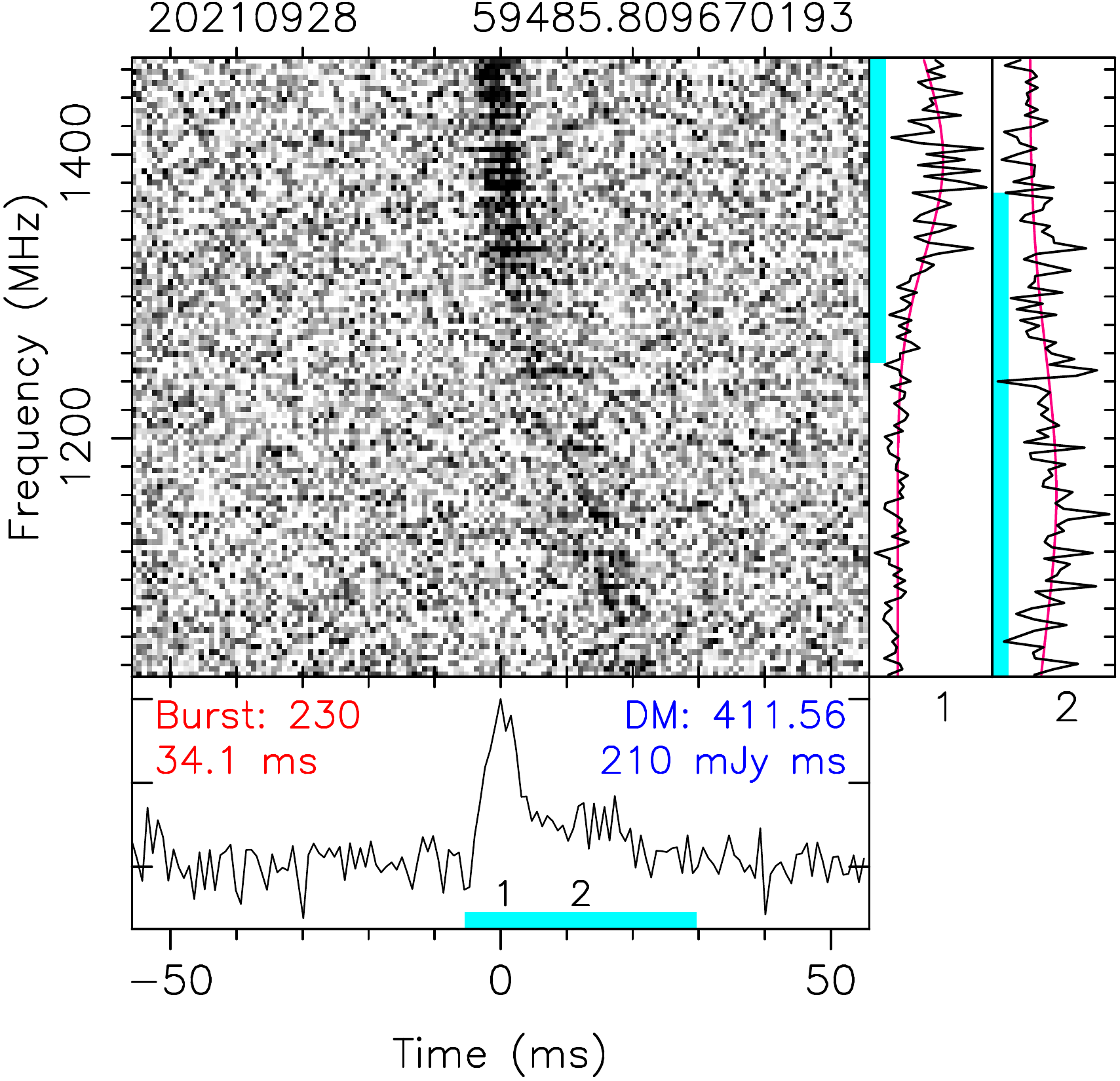}
    \includegraphics[height=37mm]{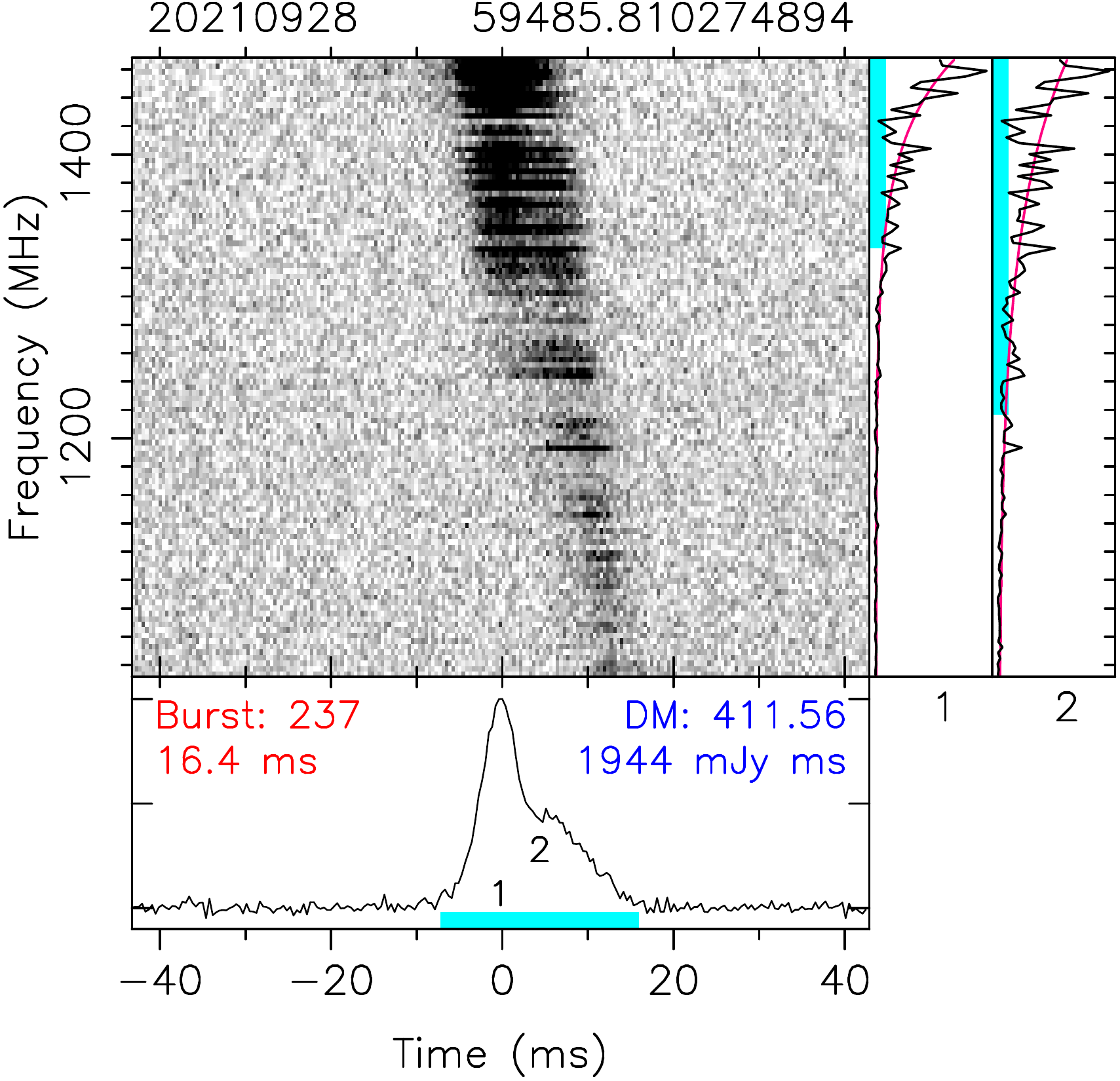}
    \includegraphics[height=37mm]{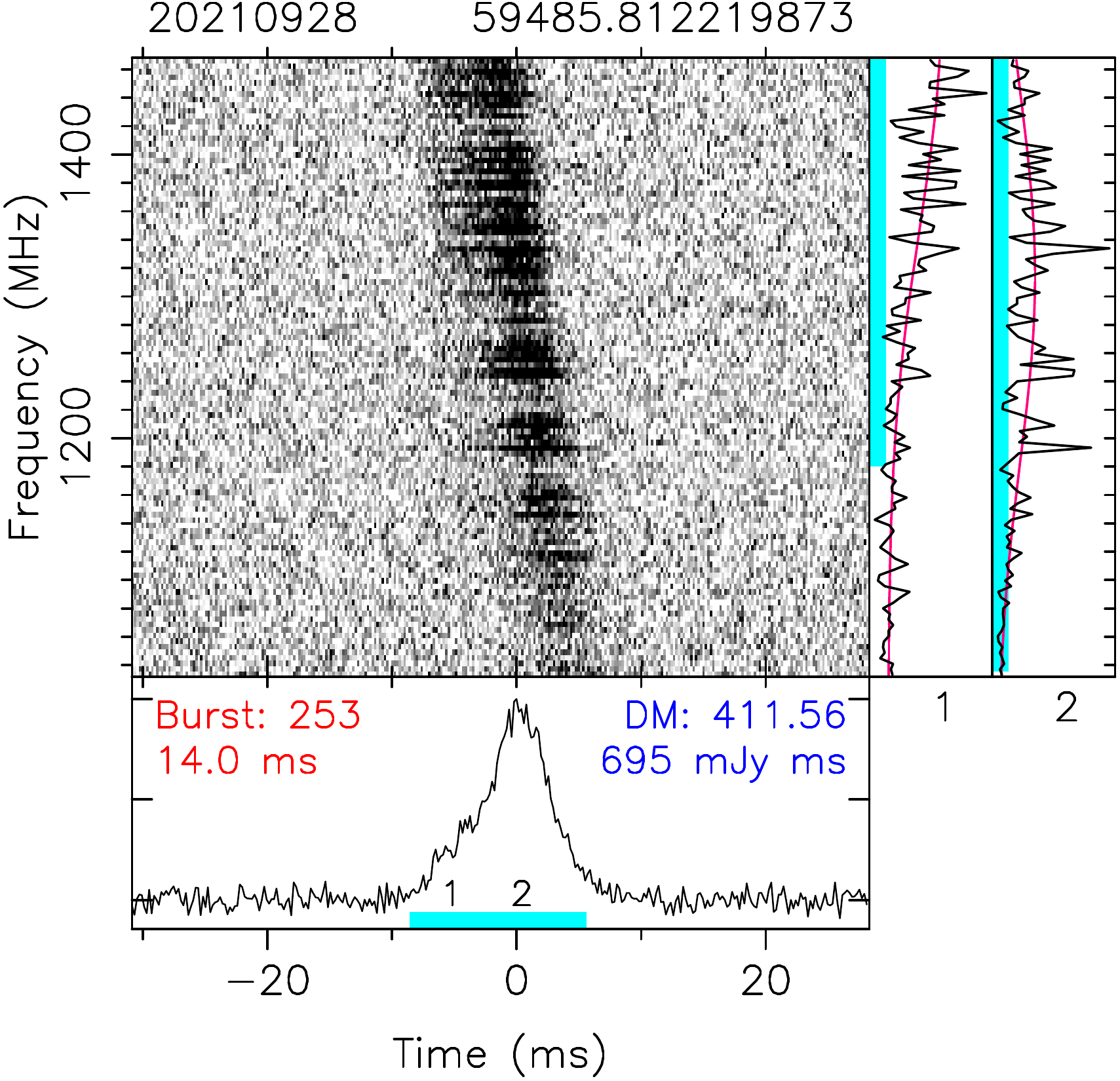}
    \includegraphics[height=37mm]{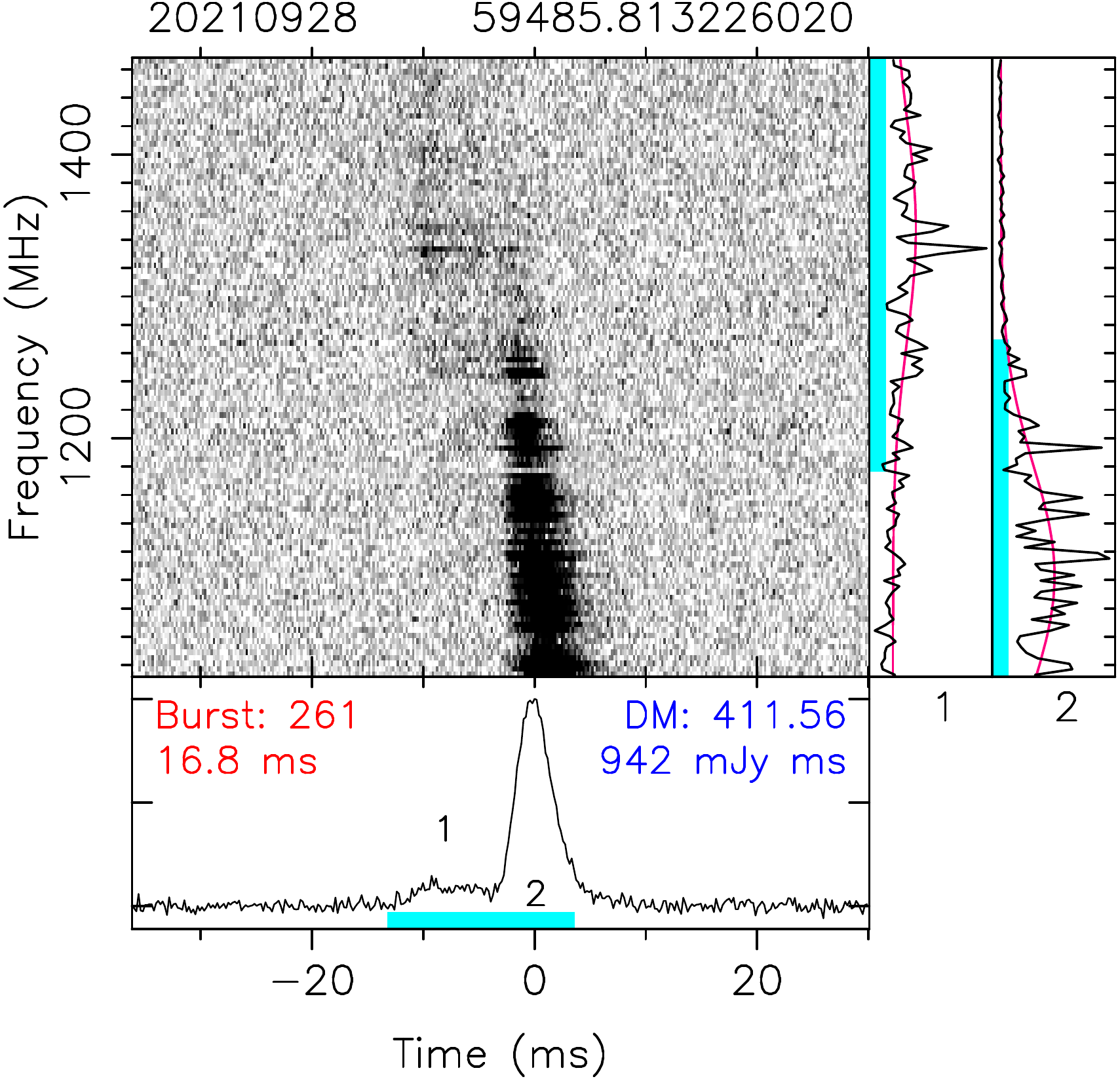}
    \includegraphics[height=37mm]{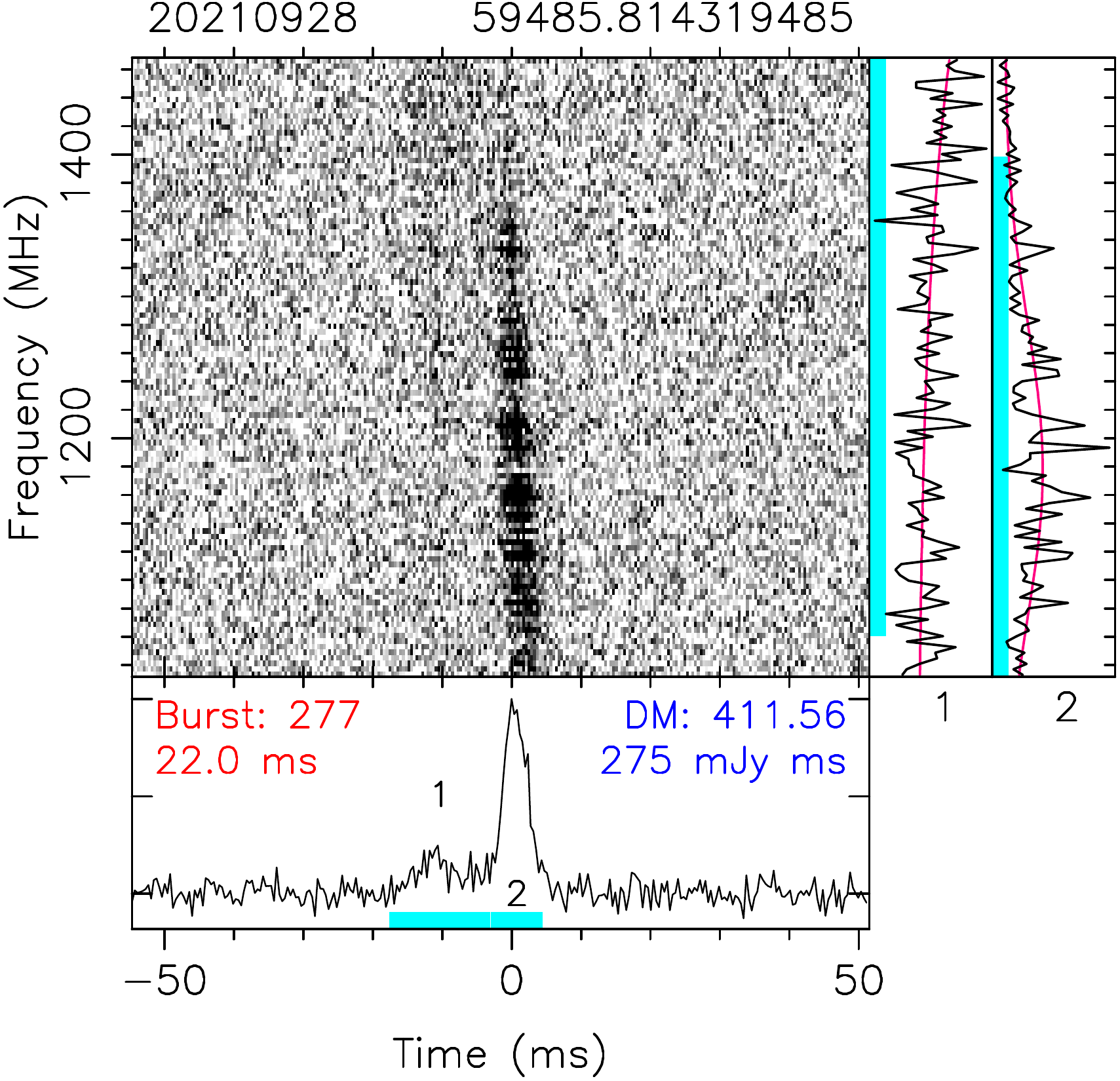}
    \includegraphics[height=37mm]{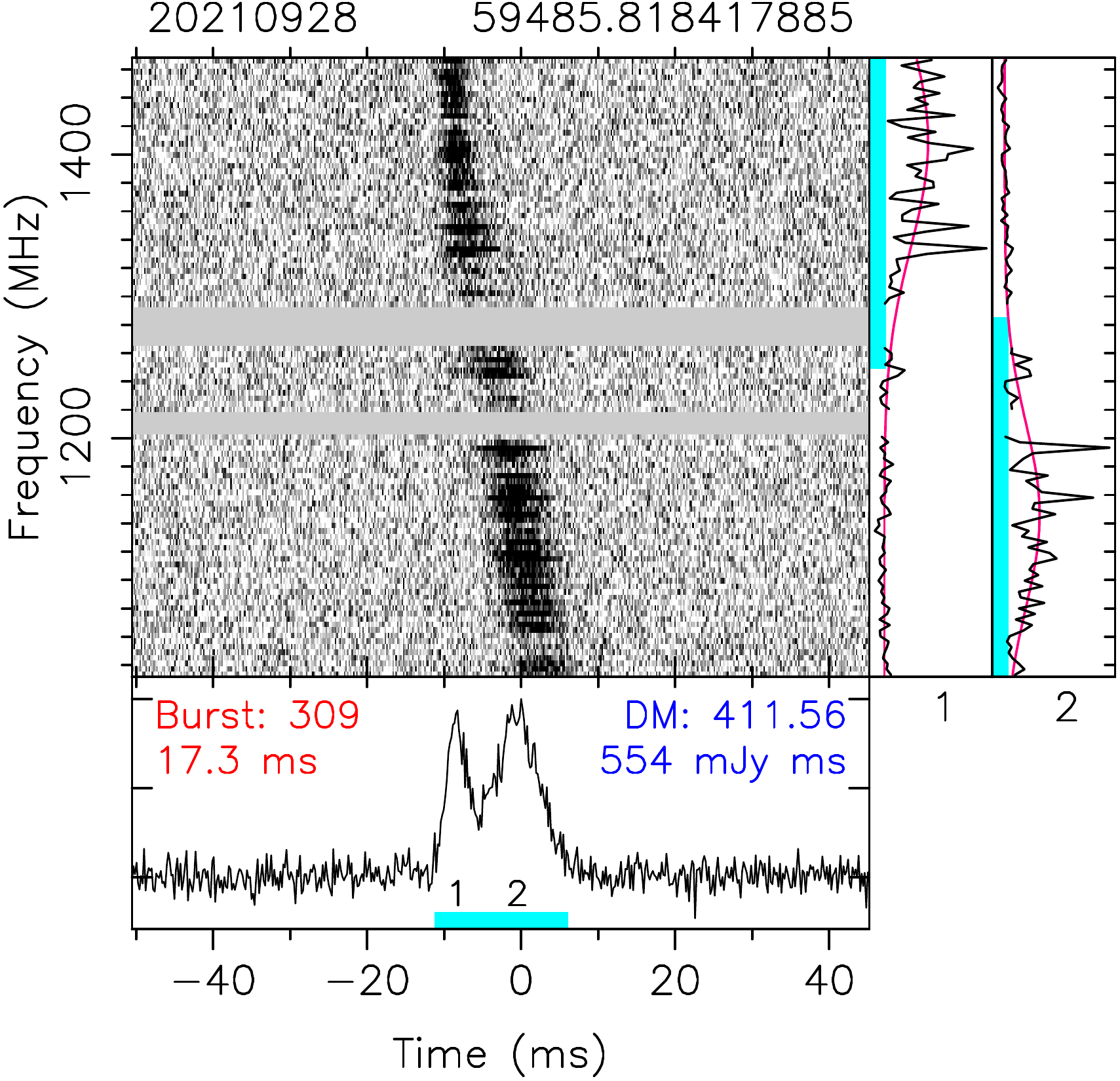}
    \includegraphics[height=37mm]{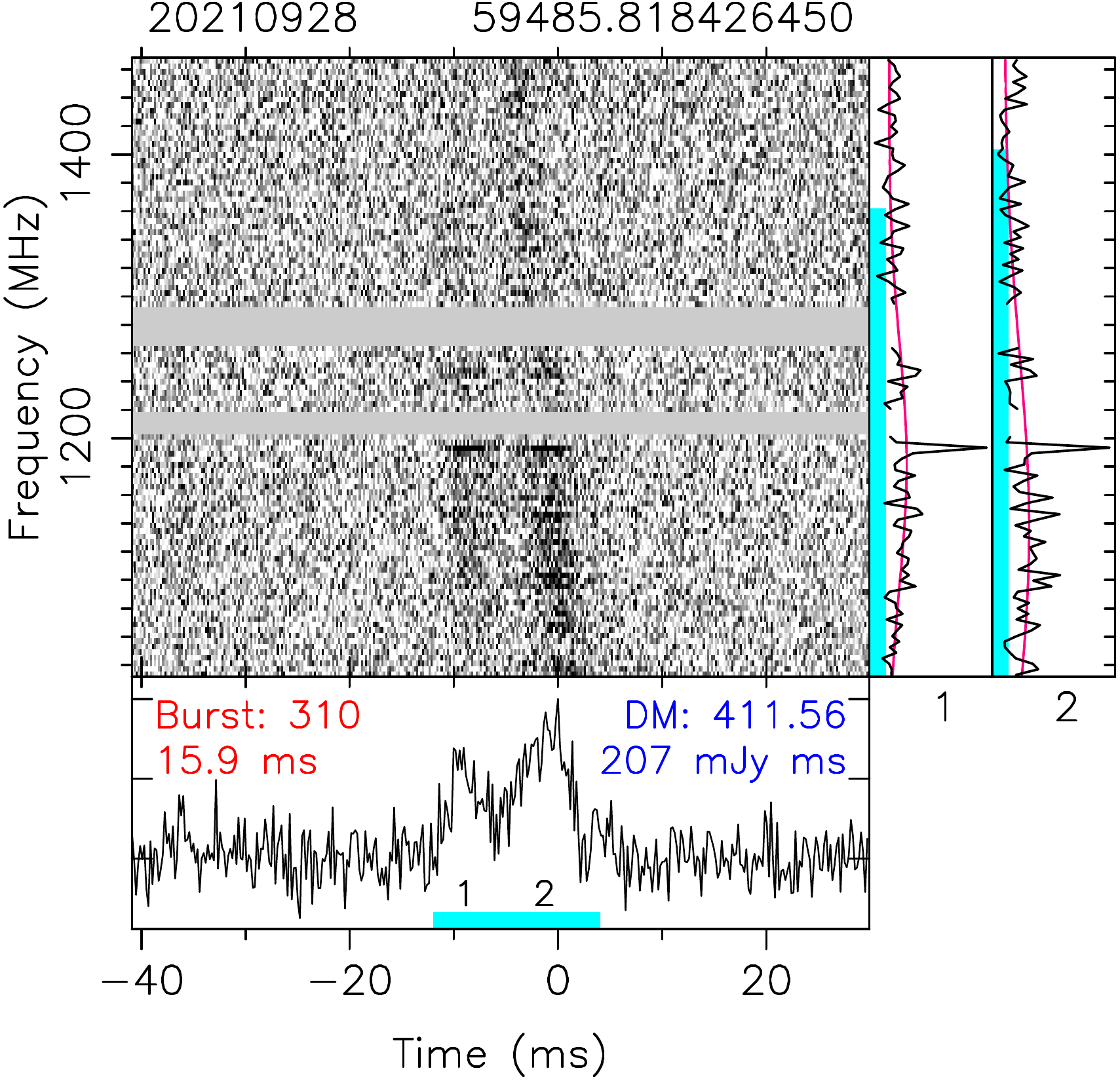}
    \includegraphics[height=37mm]{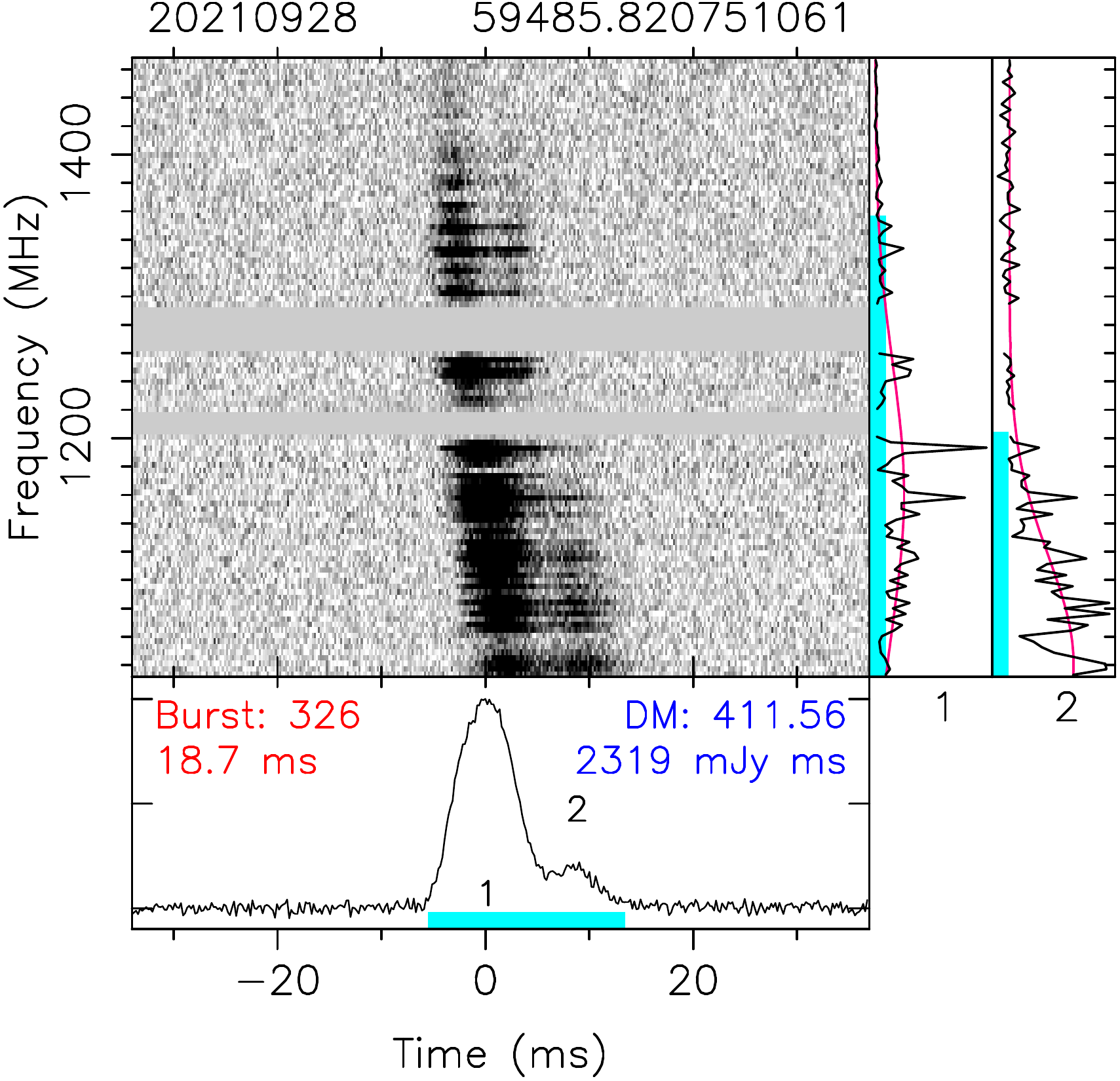}
    \includegraphics[height=37mm]{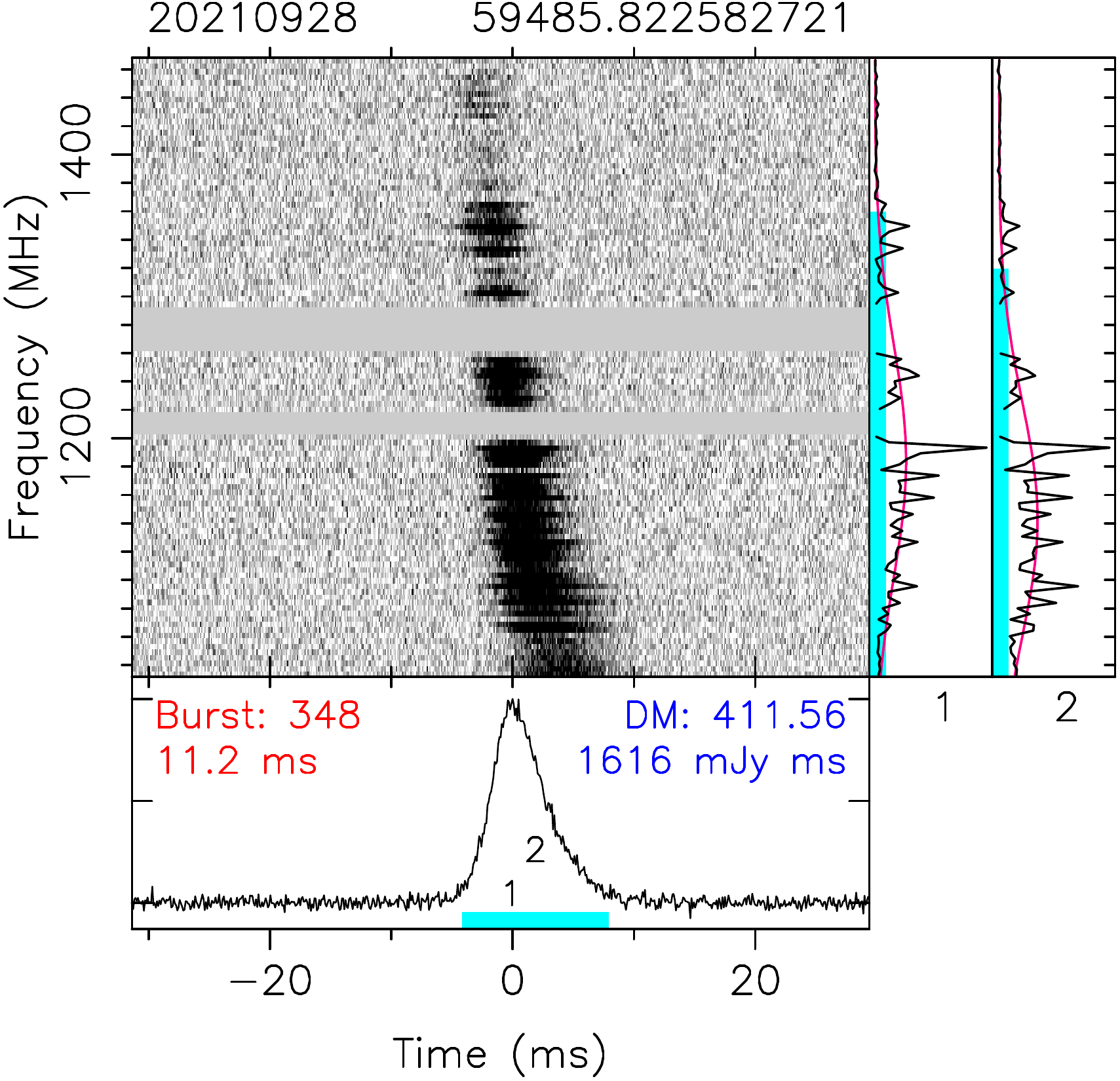}
\caption{The same as Figure~\ref{fig:appendix:D1W} but for bursts in D2-W.
}\label{fig:appendix:D2W} 
\end{figure*}

\begin{figure*}
    \flushleft
    \includegraphics[height=37mm]{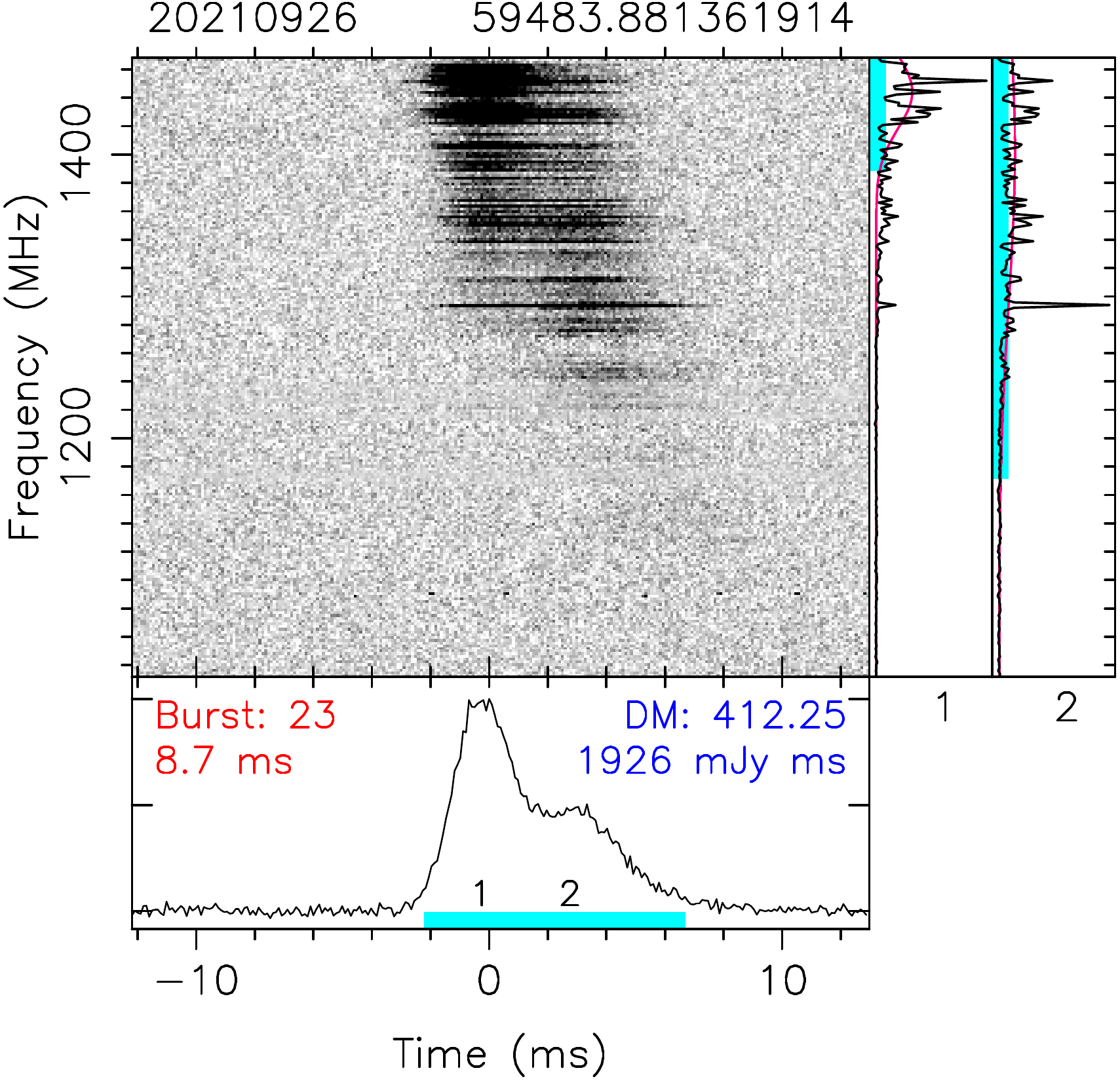}
    \includegraphics[height=37mm]{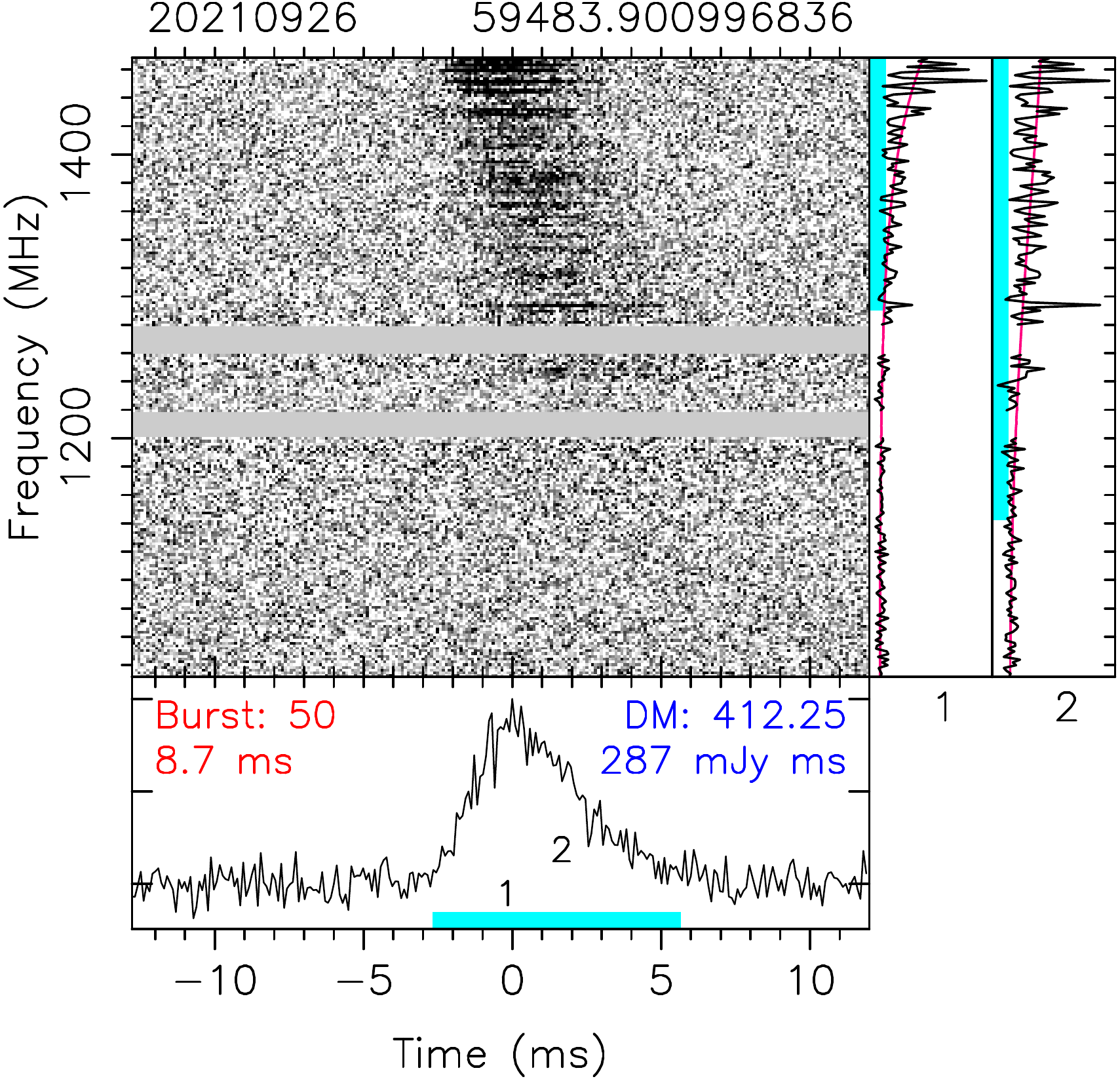}
    \includegraphics[height=37mm]{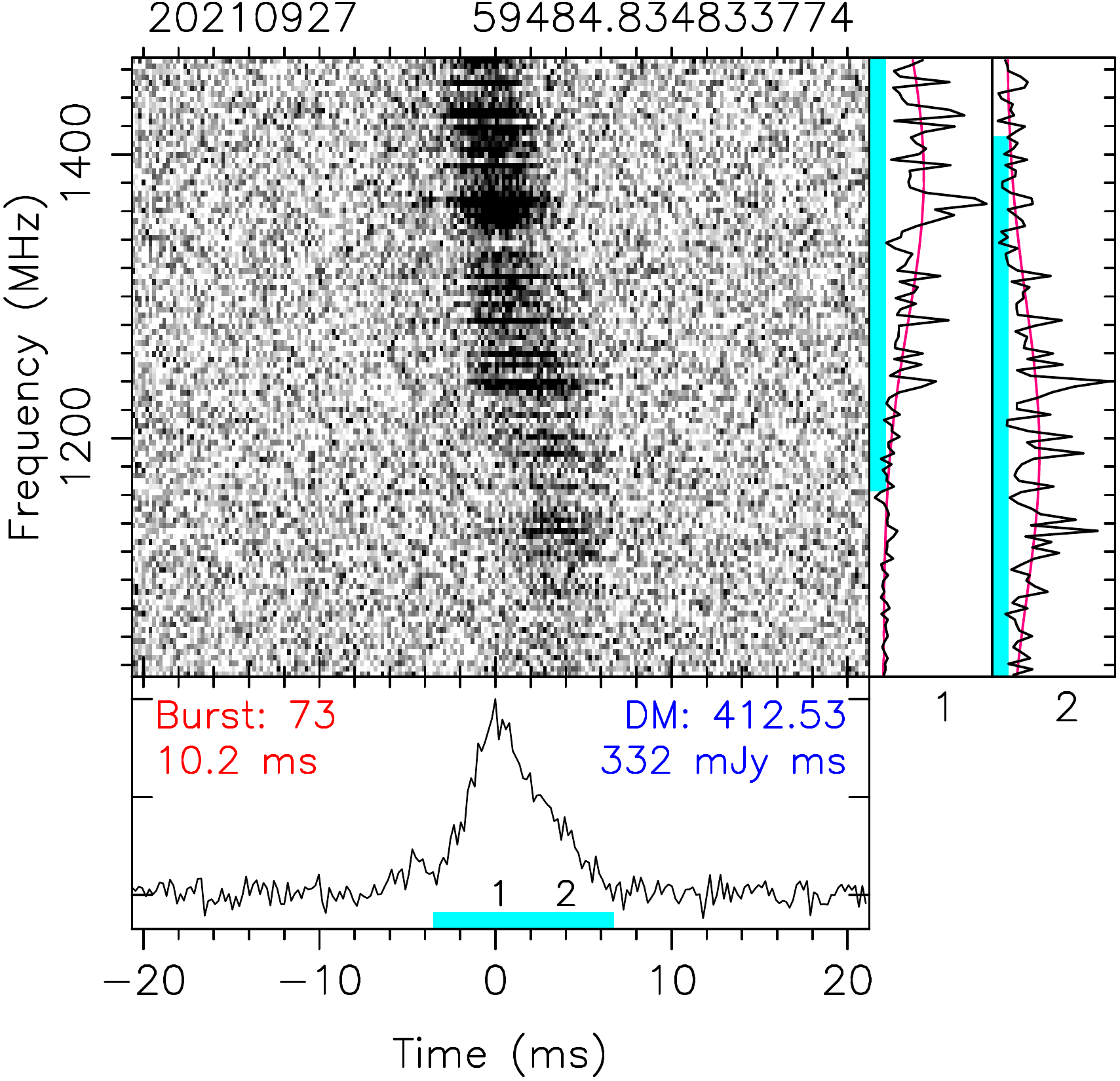}
    \includegraphics[height=37mm]{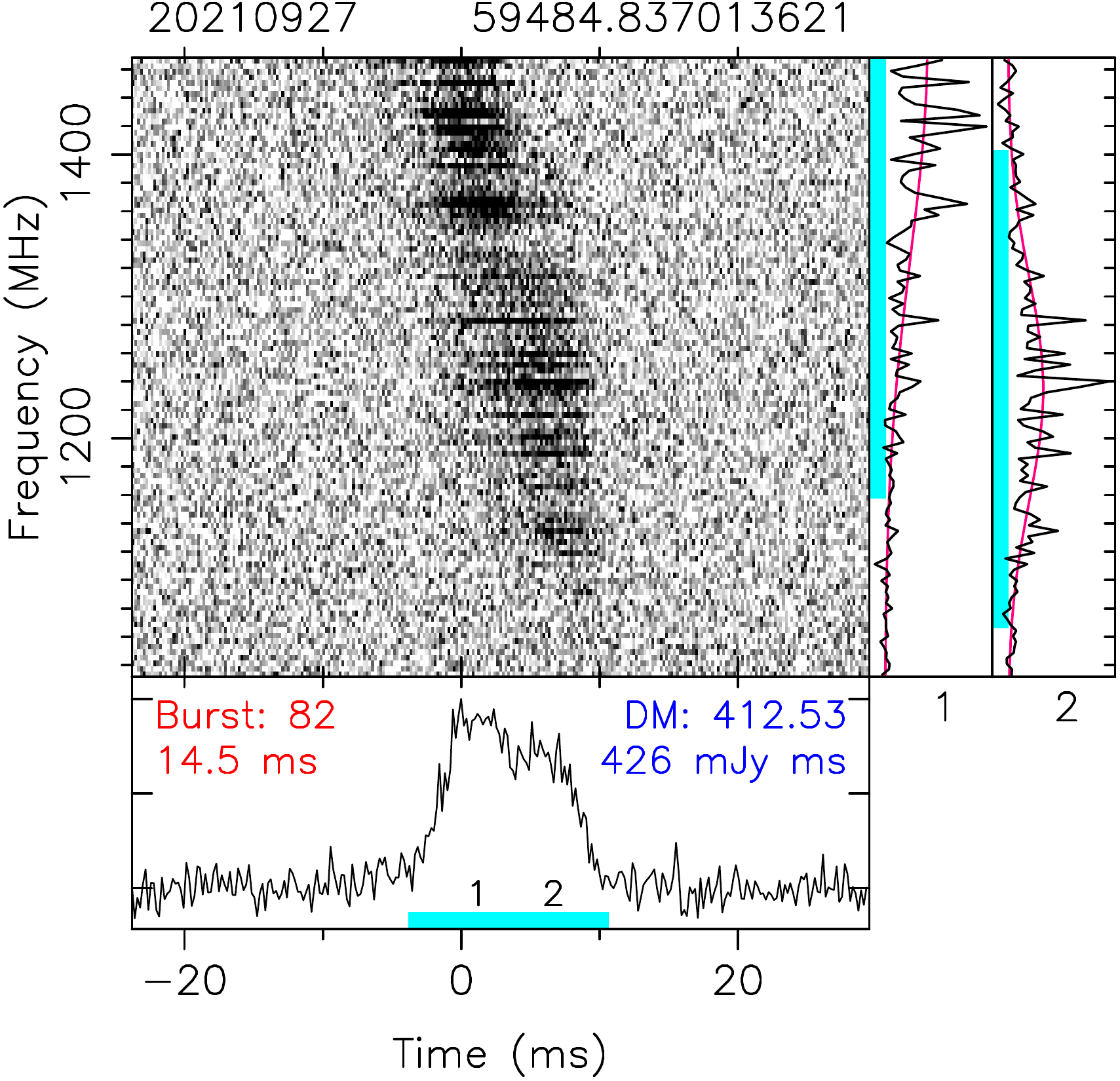}
    \includegraphics[height=37mm]{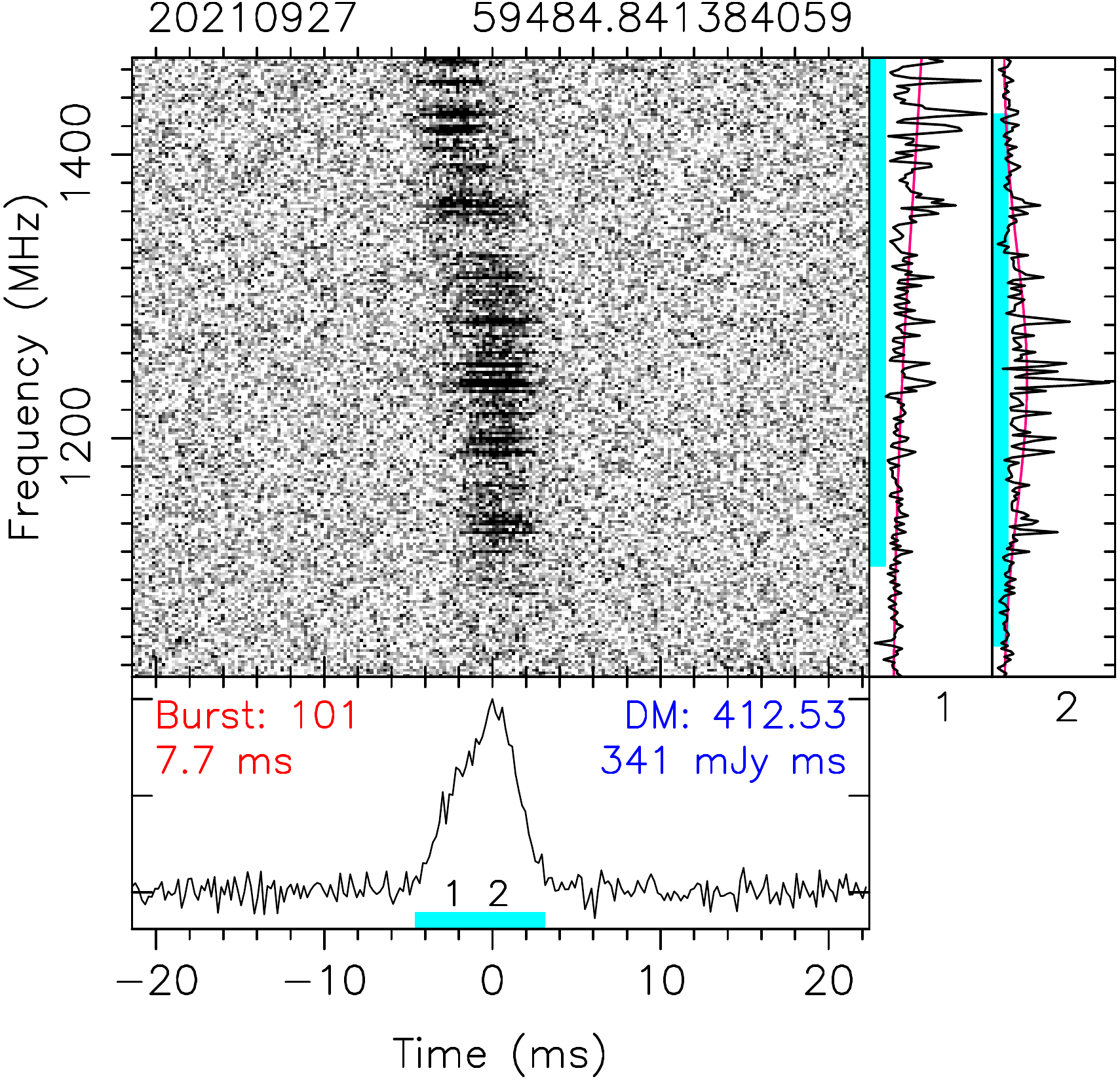}
    \includegraphics[height=37mm]{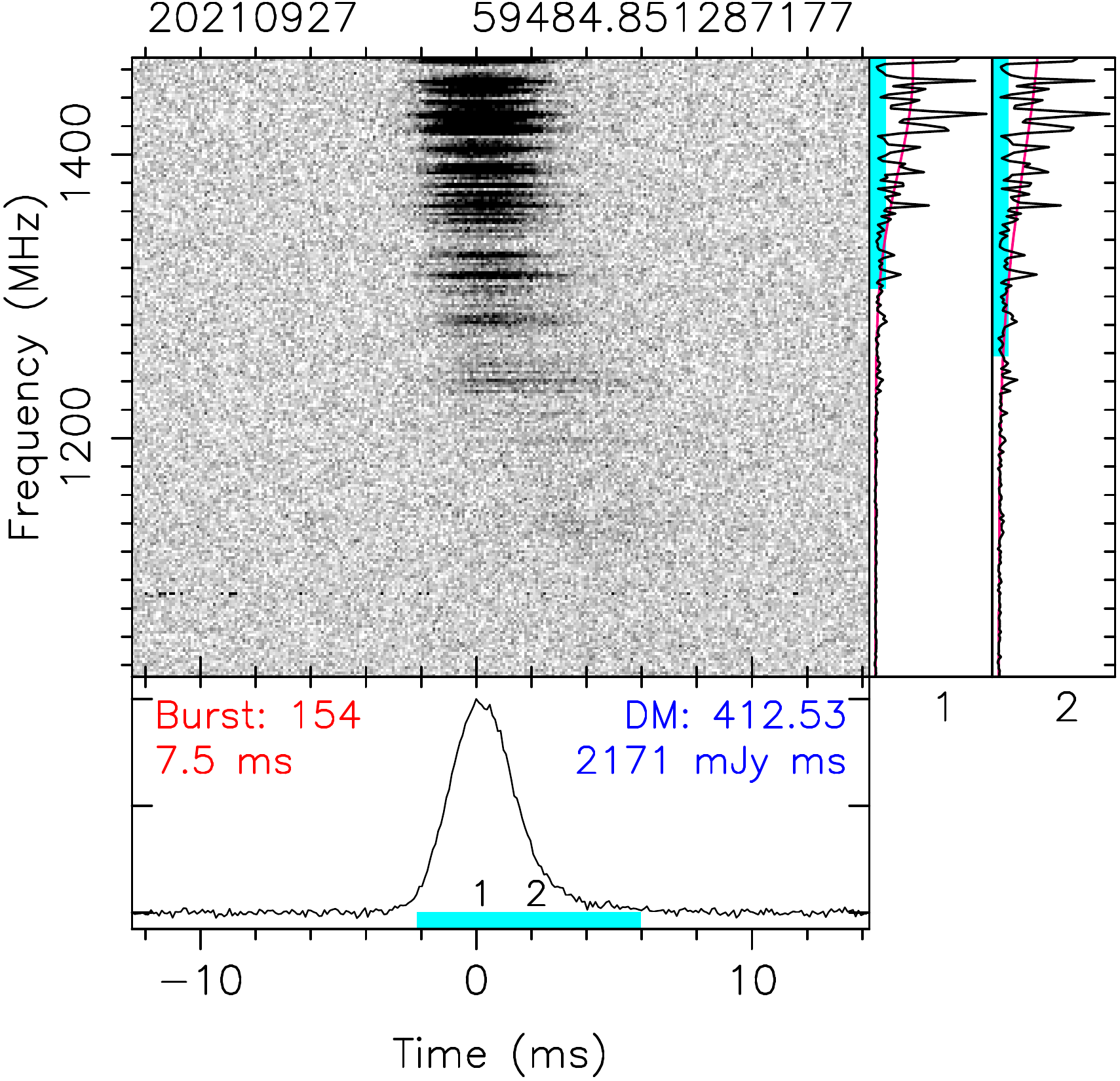}
    \includegraphics[height=37mm]{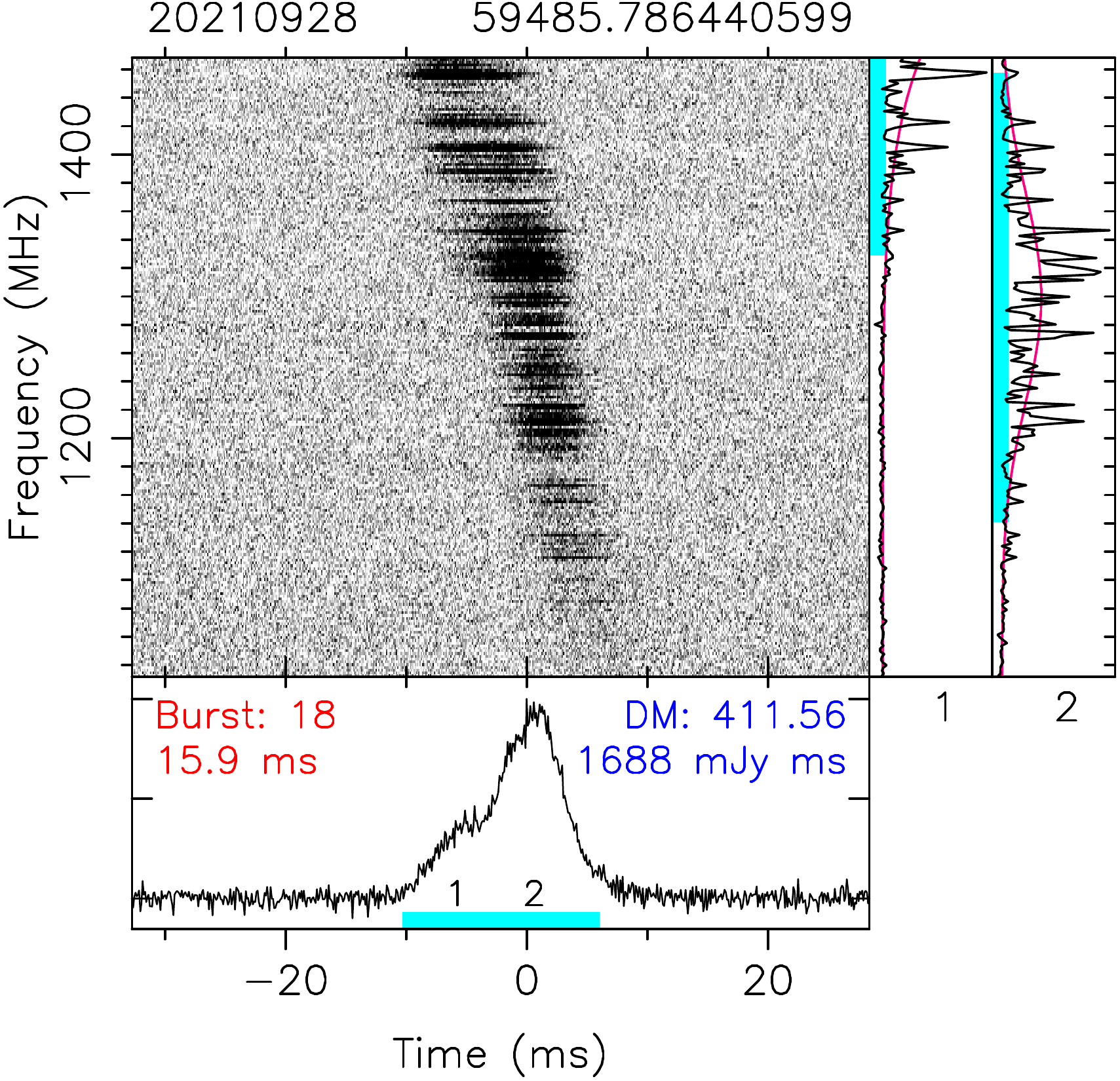}
    \includegraphics[height=37mm]{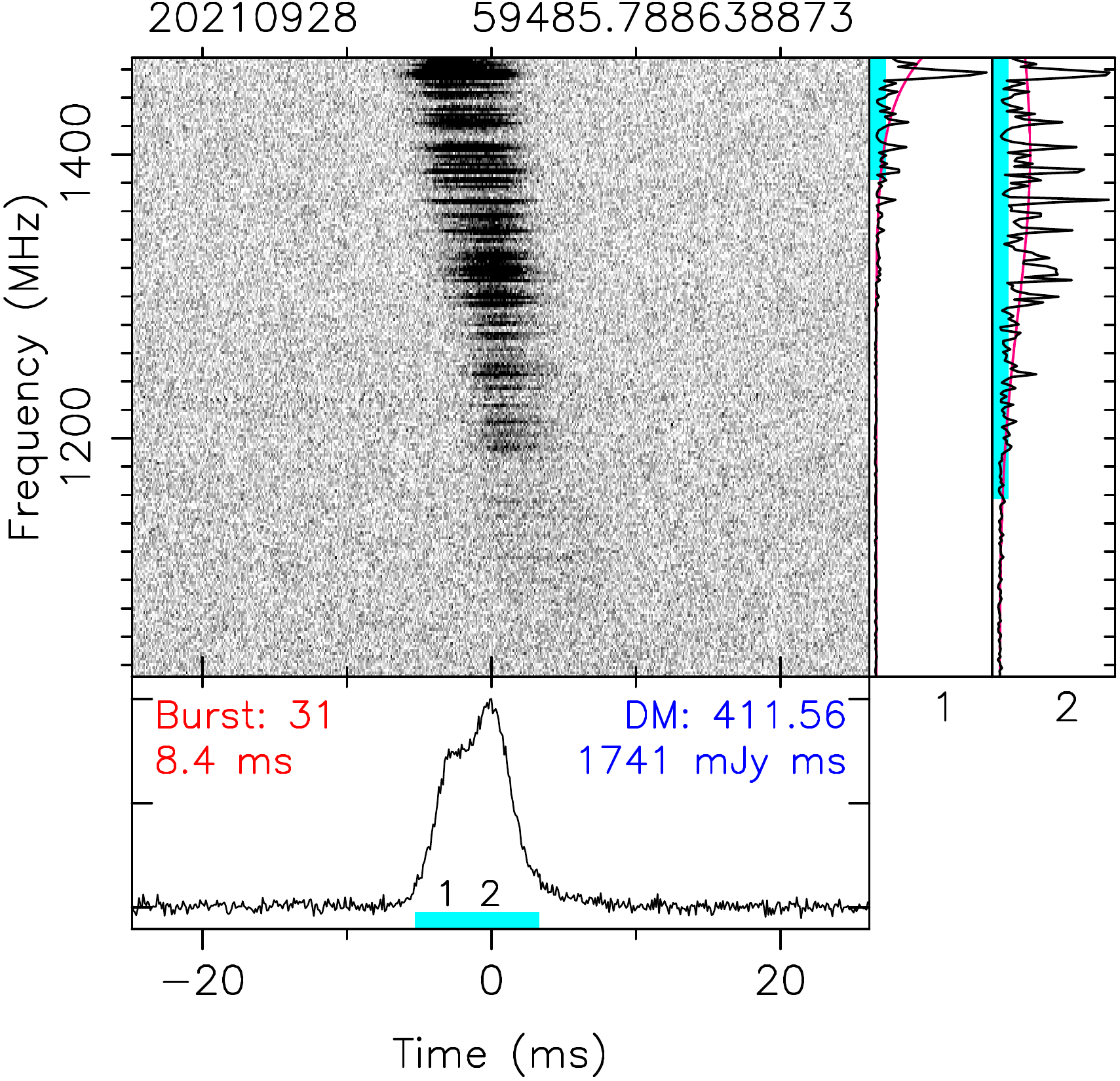}
    \includegraphics[height=37mm]{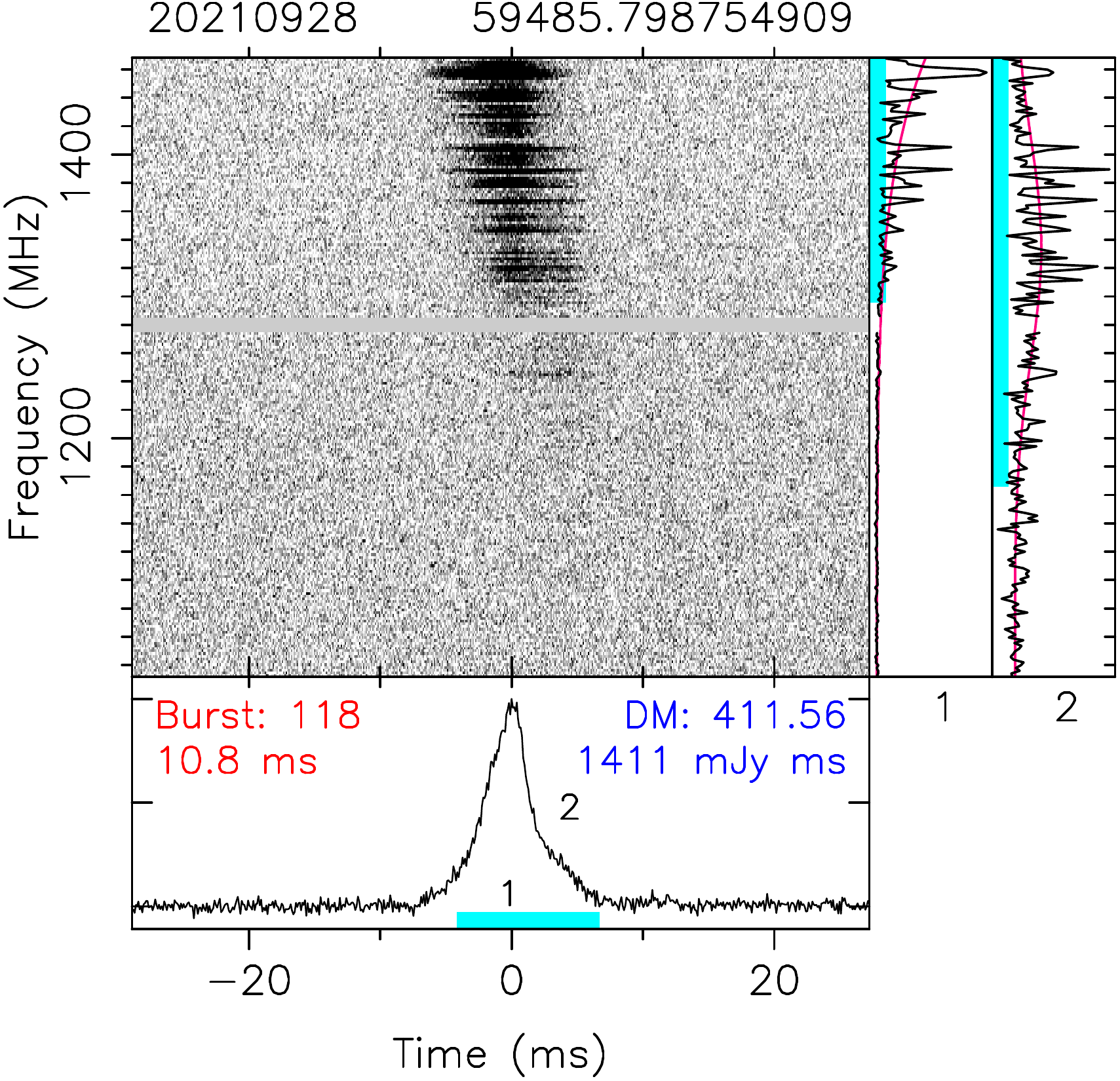}
    \includegraphics[height=37mm]{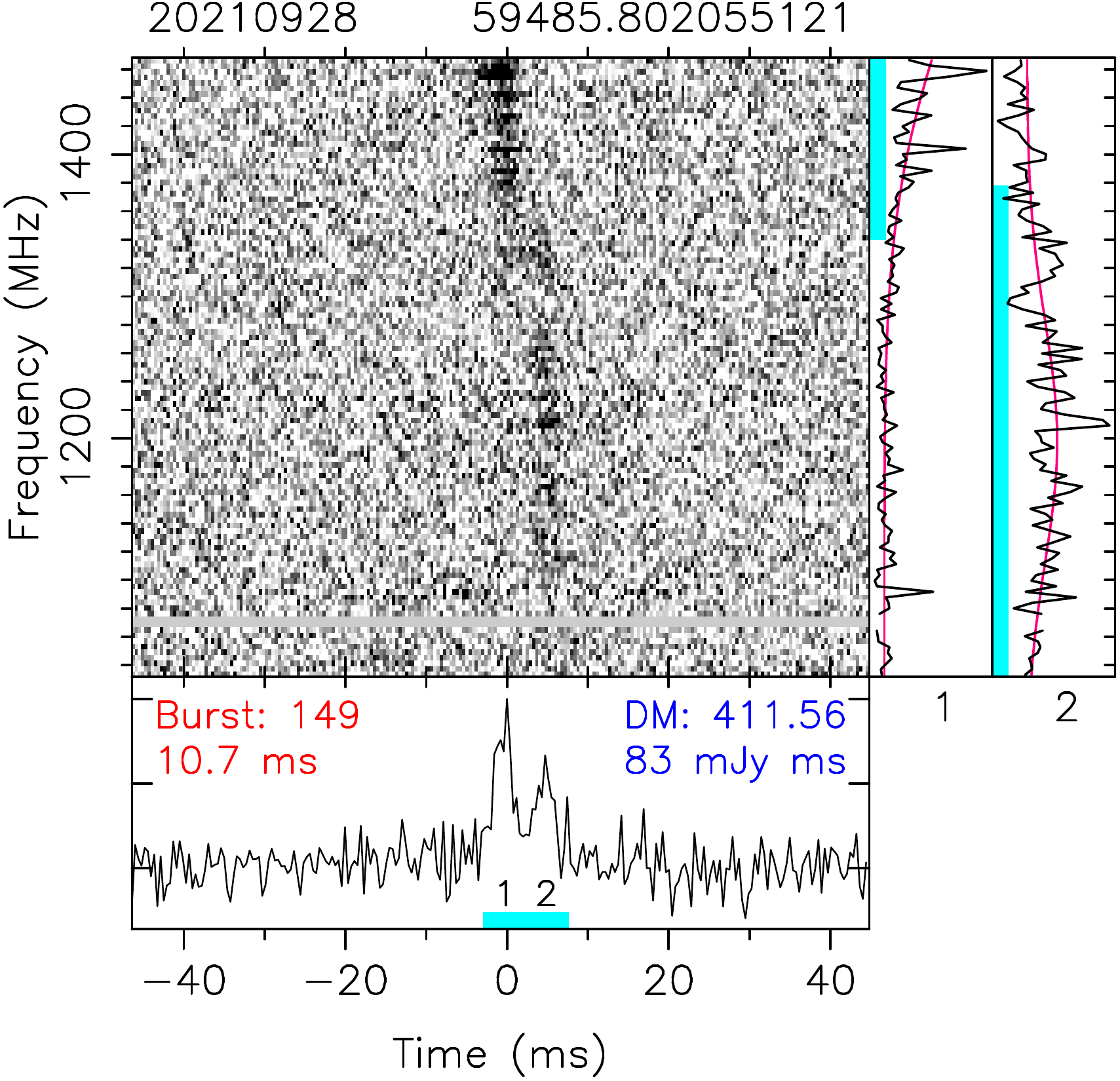}
    \includegraphics[height=37mm]{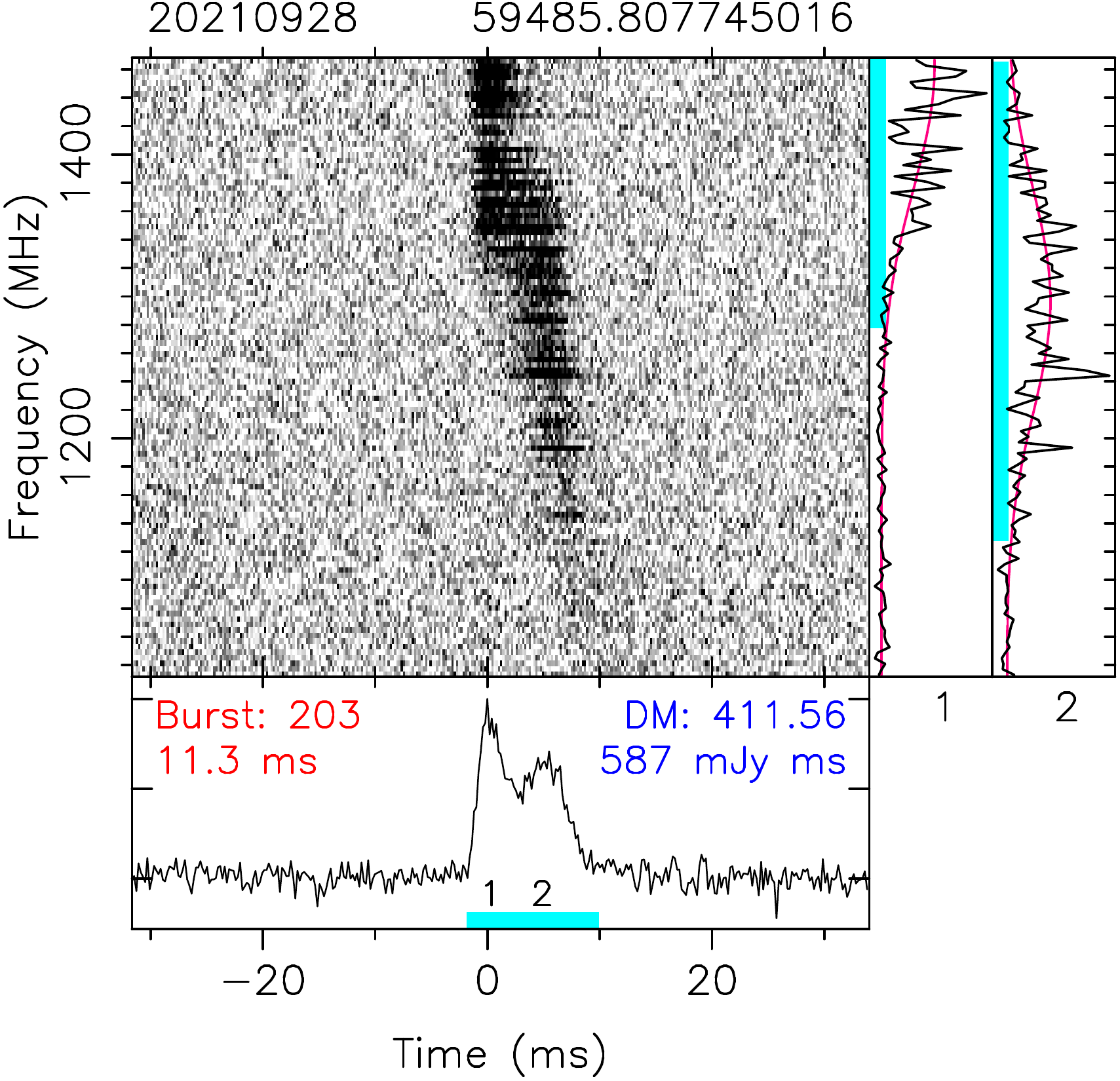}
    \includegraphics[height=37mm]{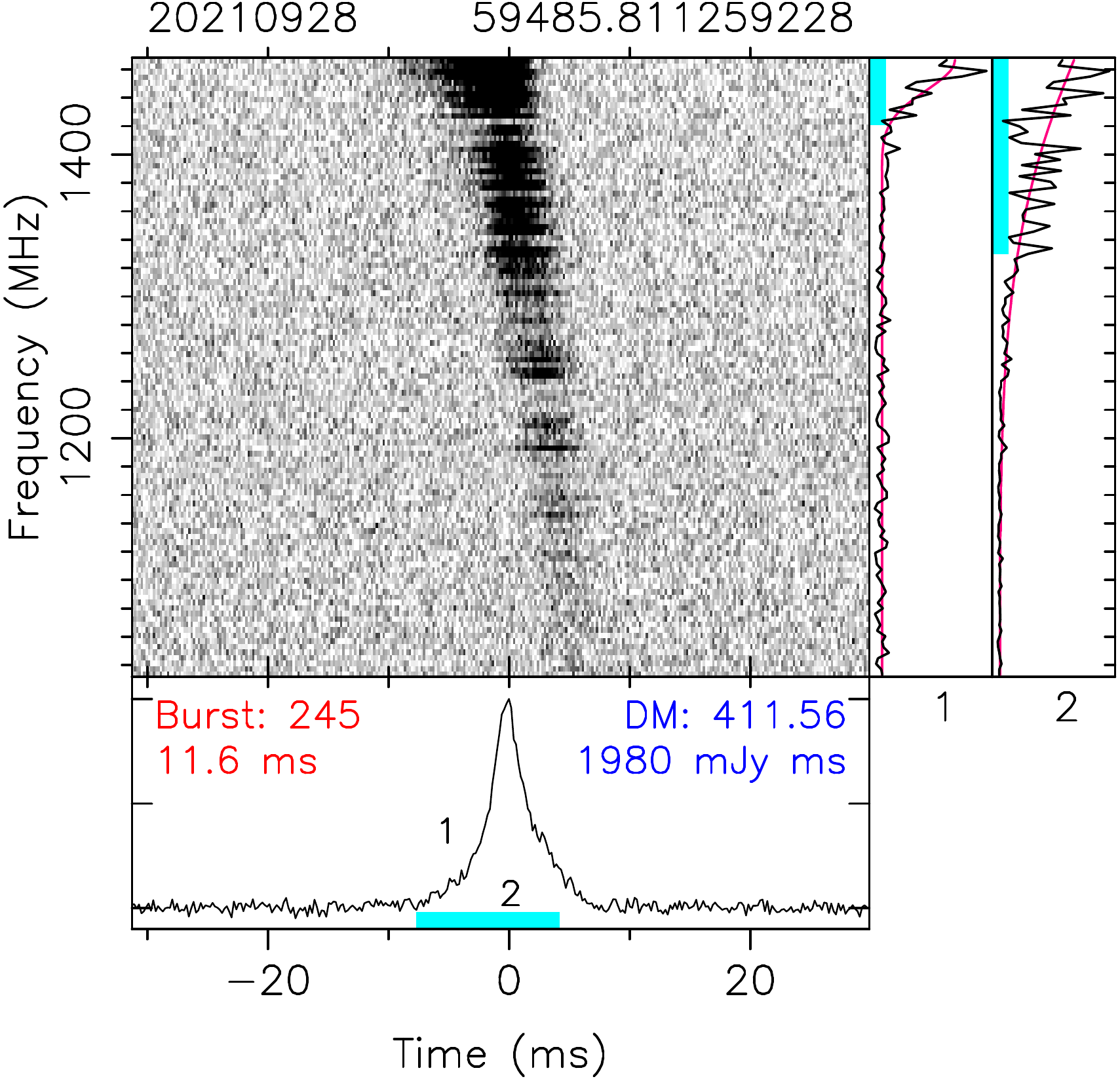}
    \includegraphics[height=37mm]{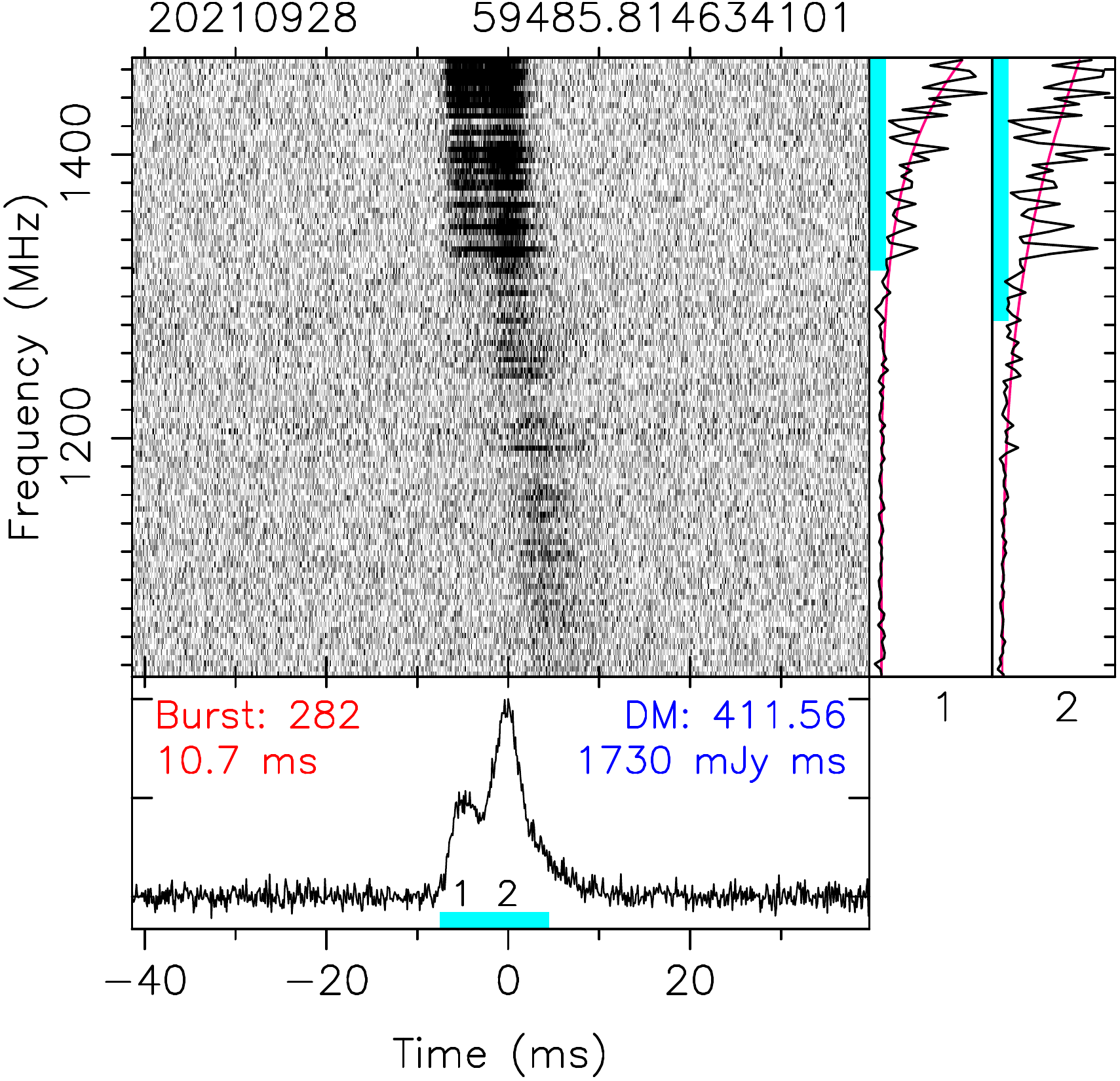}
    \includegraphics[height=37mm]{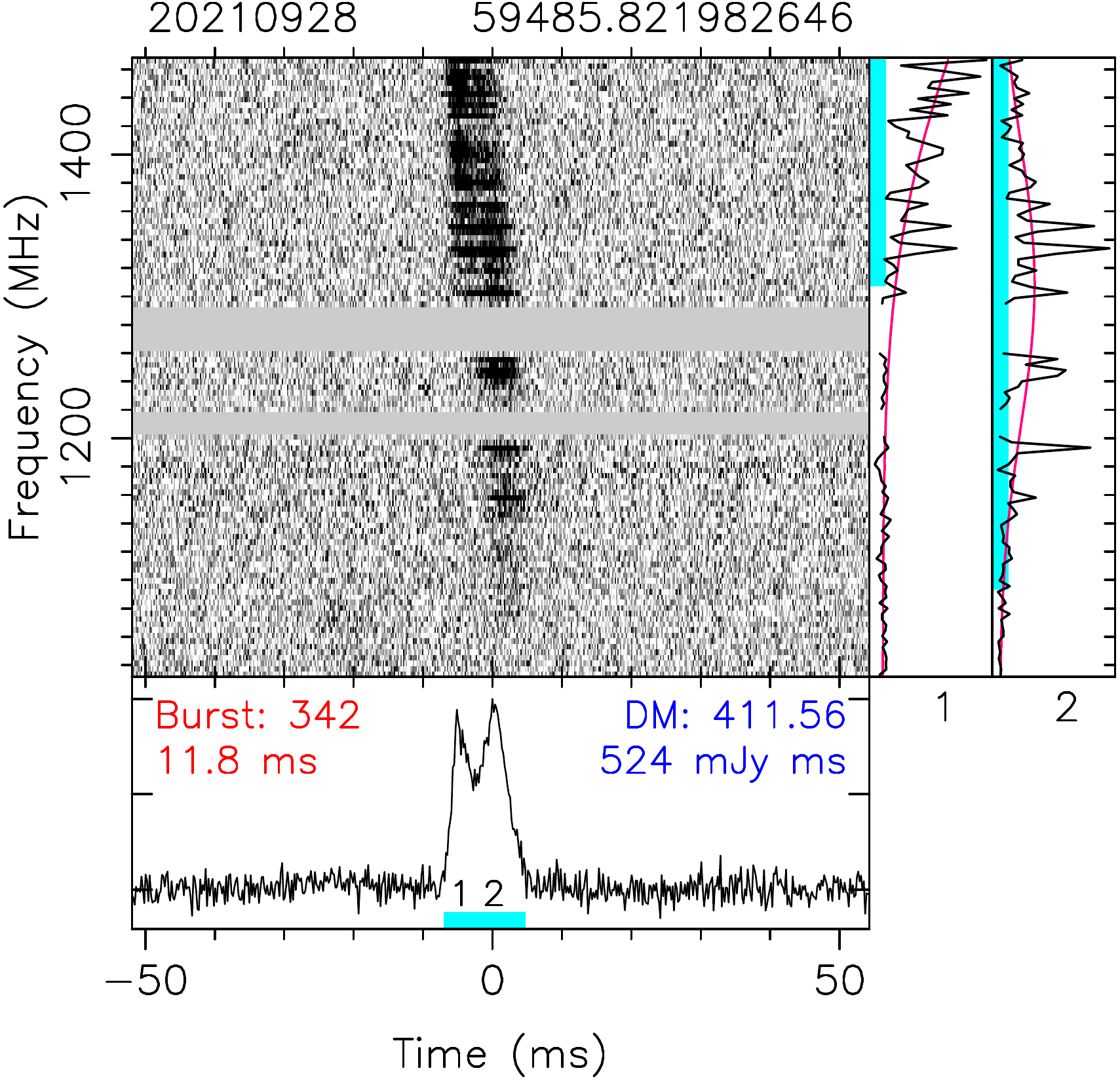}
\caption{The same as Figure~\ref{fig:appendix:D1W} but for bursts in D2-H.
}\label{fig:appendix:D2H}
\end{figure*}

\begin{figure*}
    \flushleft
    \includegraphics[height=37mm]{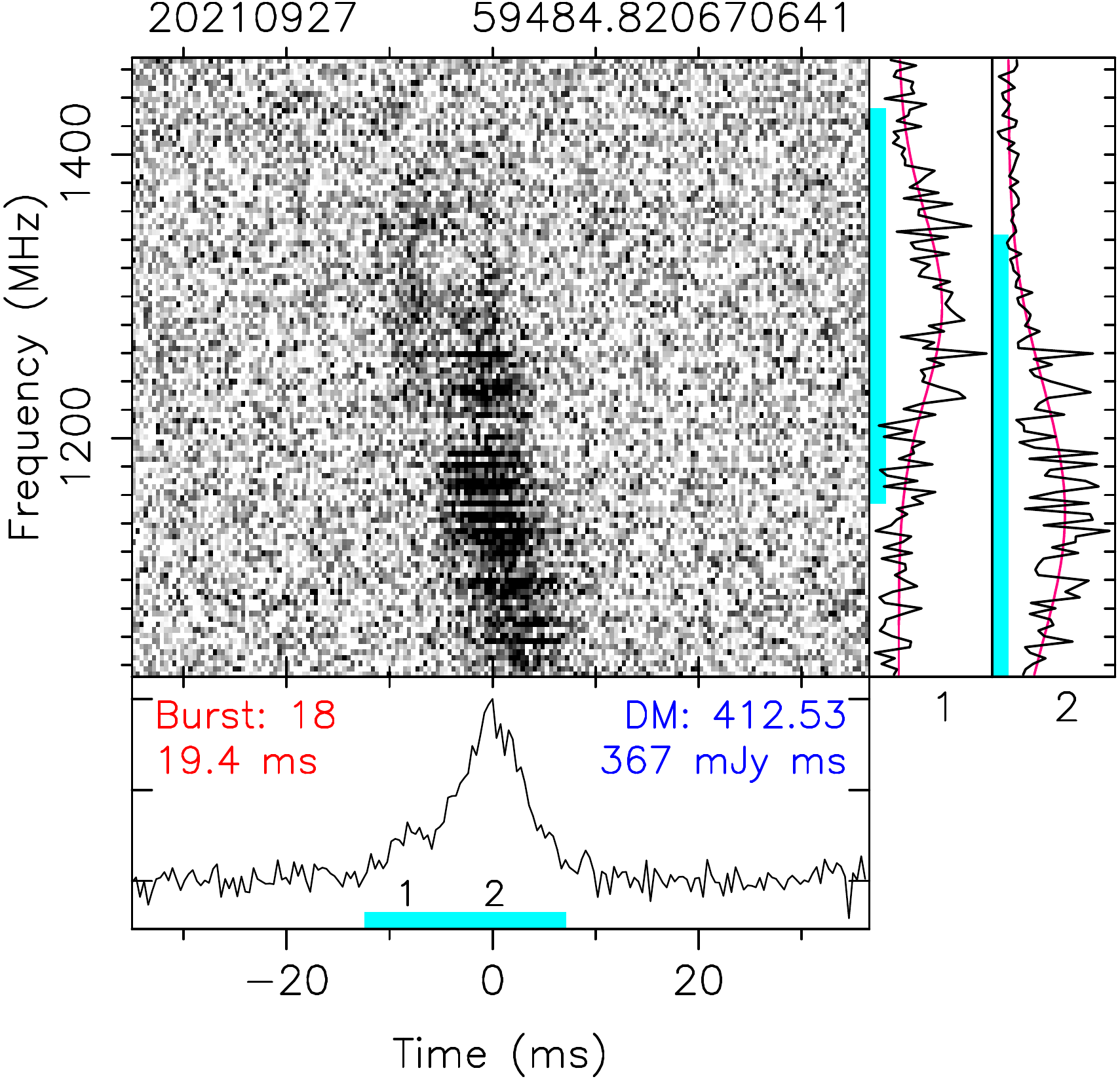}
    \includegraphics[height=37mm]{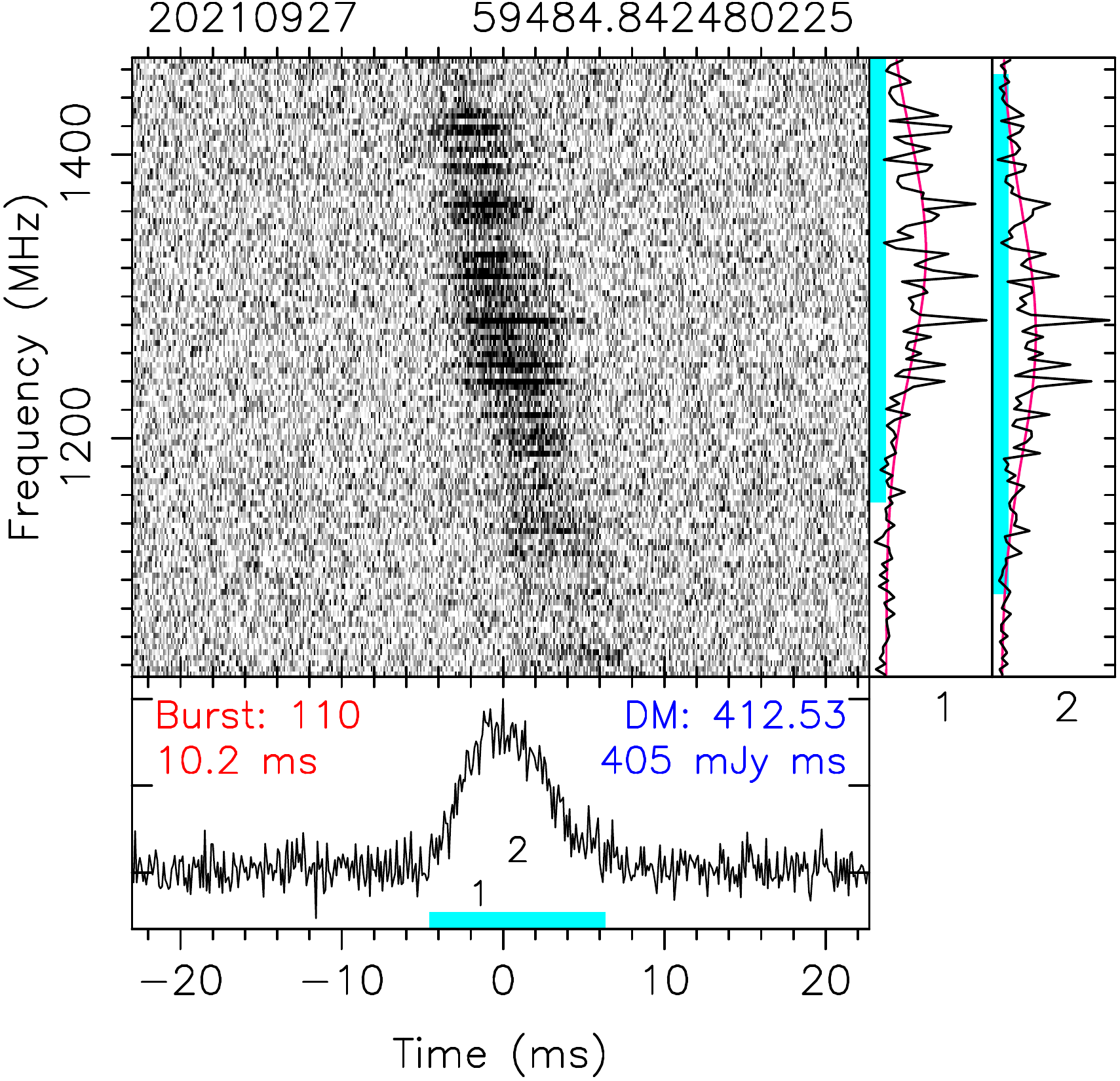}
    \includegraphics[height=37mm]{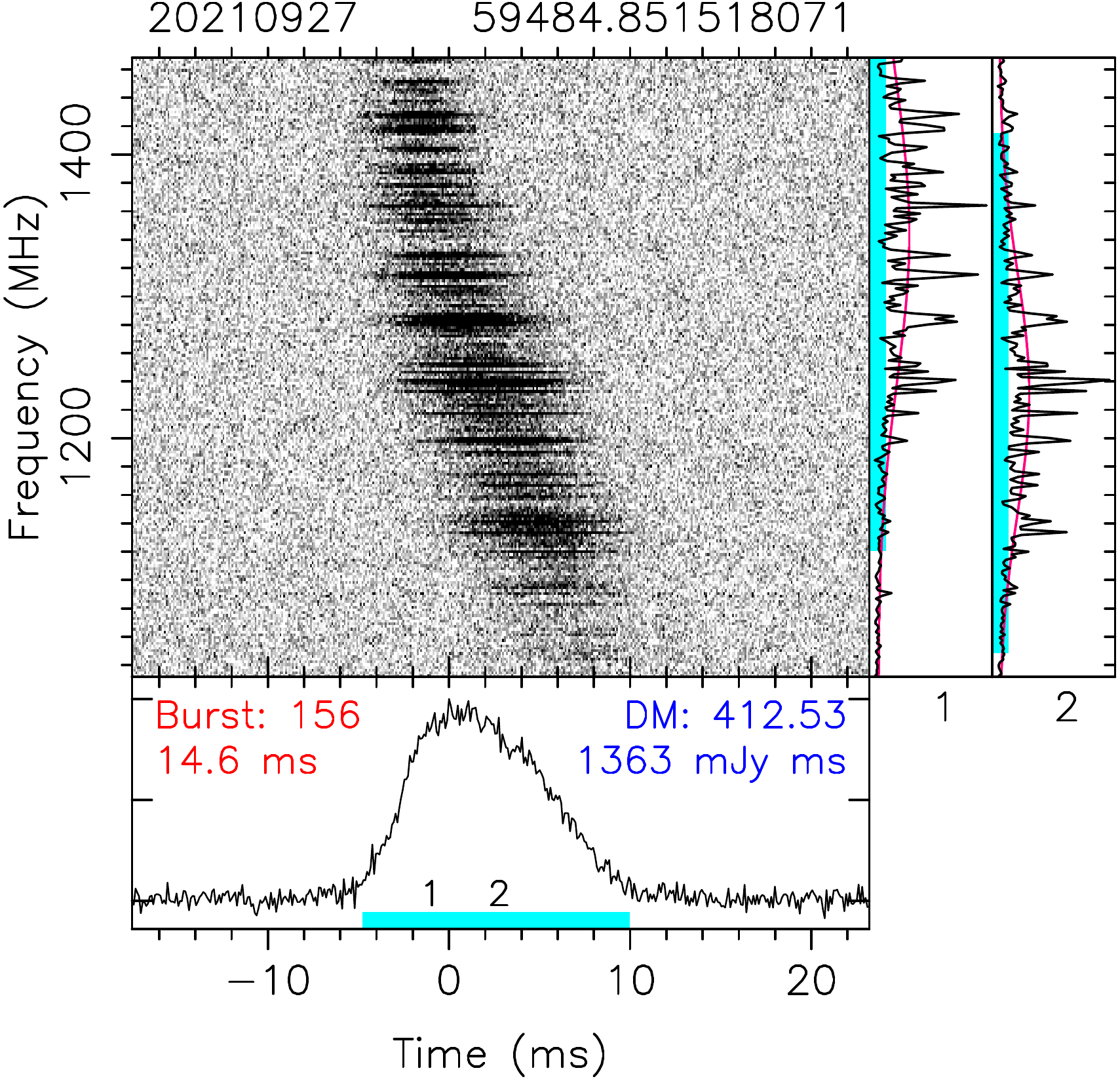}
    \includegraphics[height=37mm]{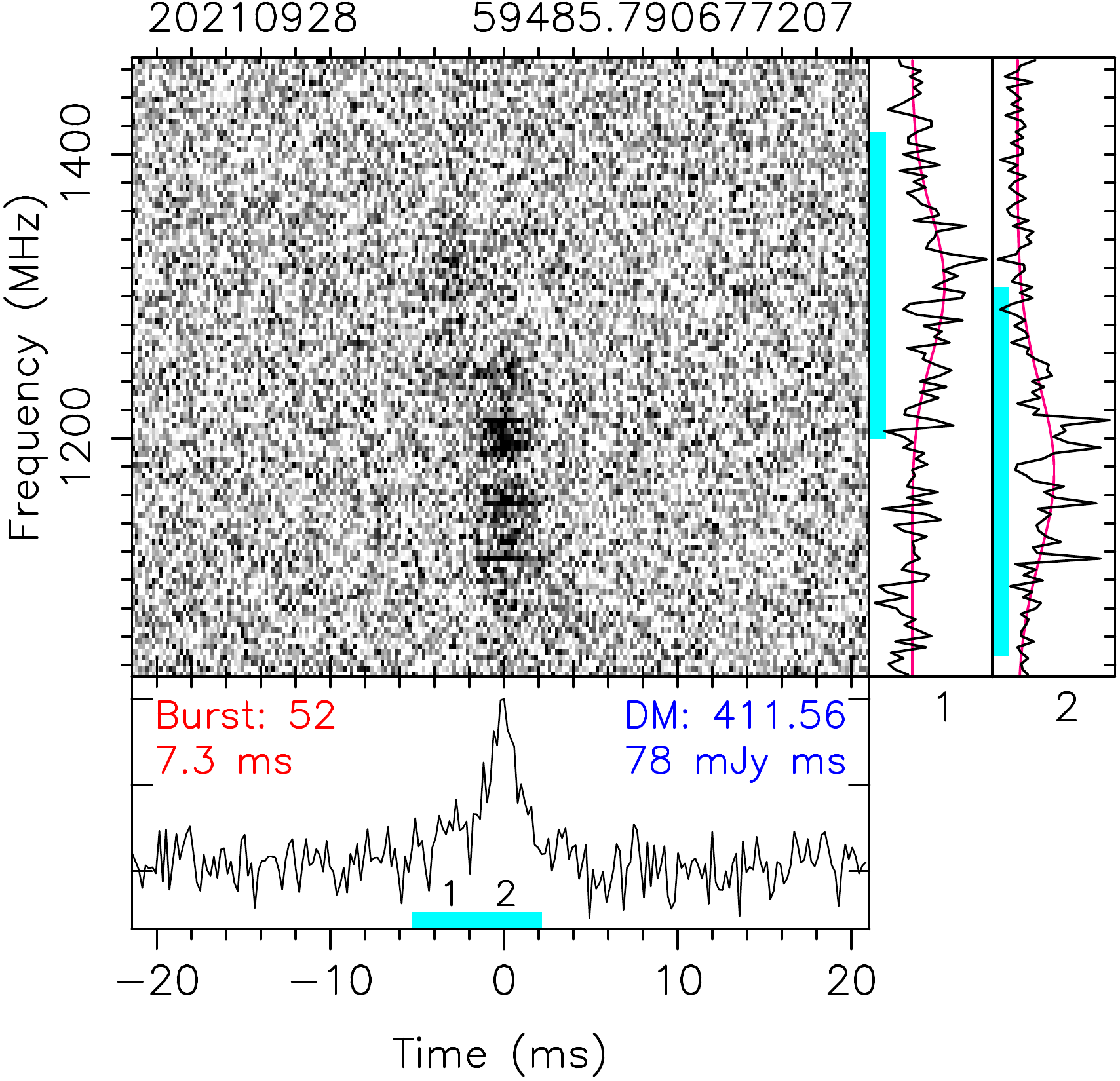}
    \includegraphics[height=37mm]{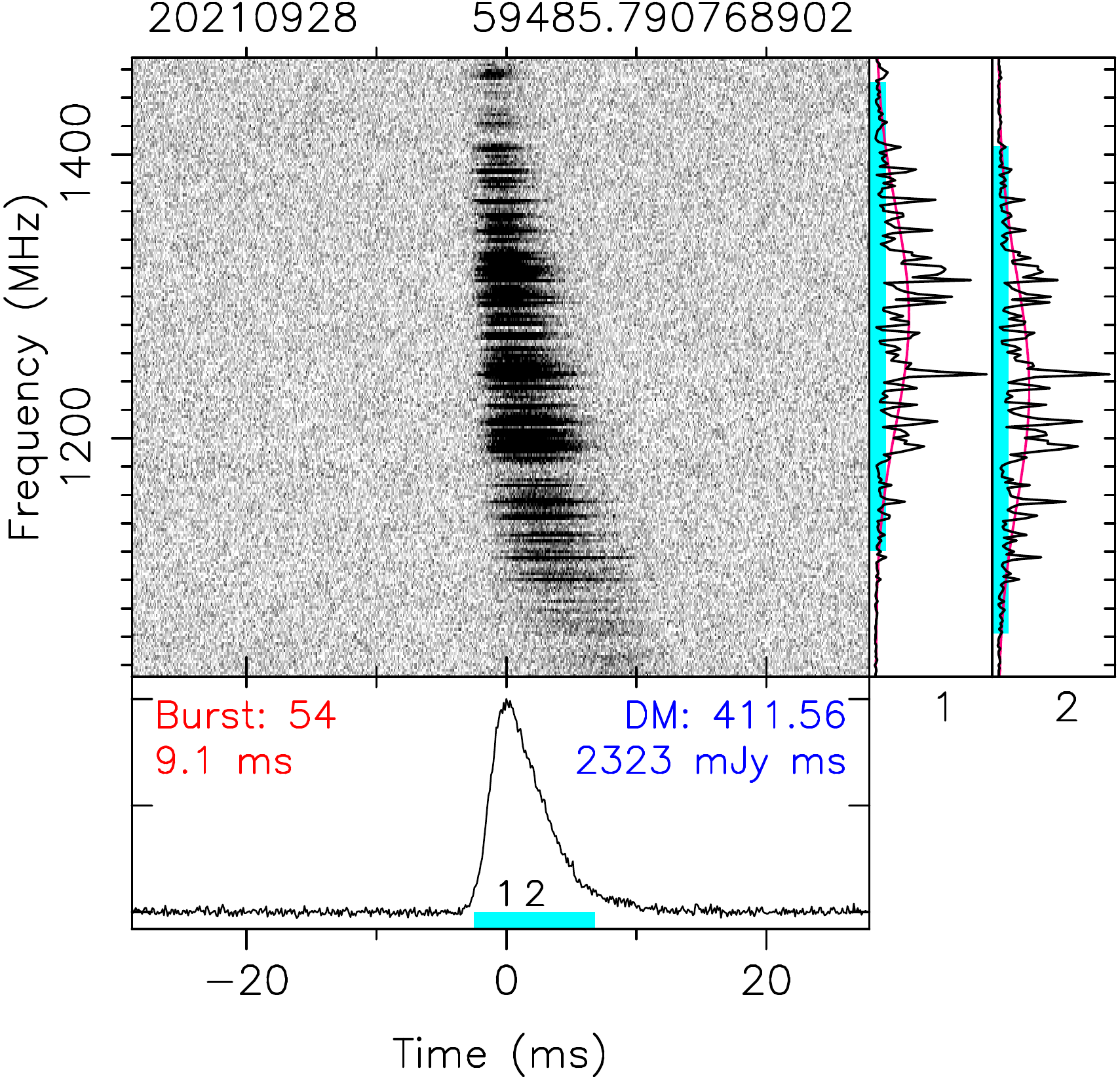}
    \includegraphics[height=37mm]{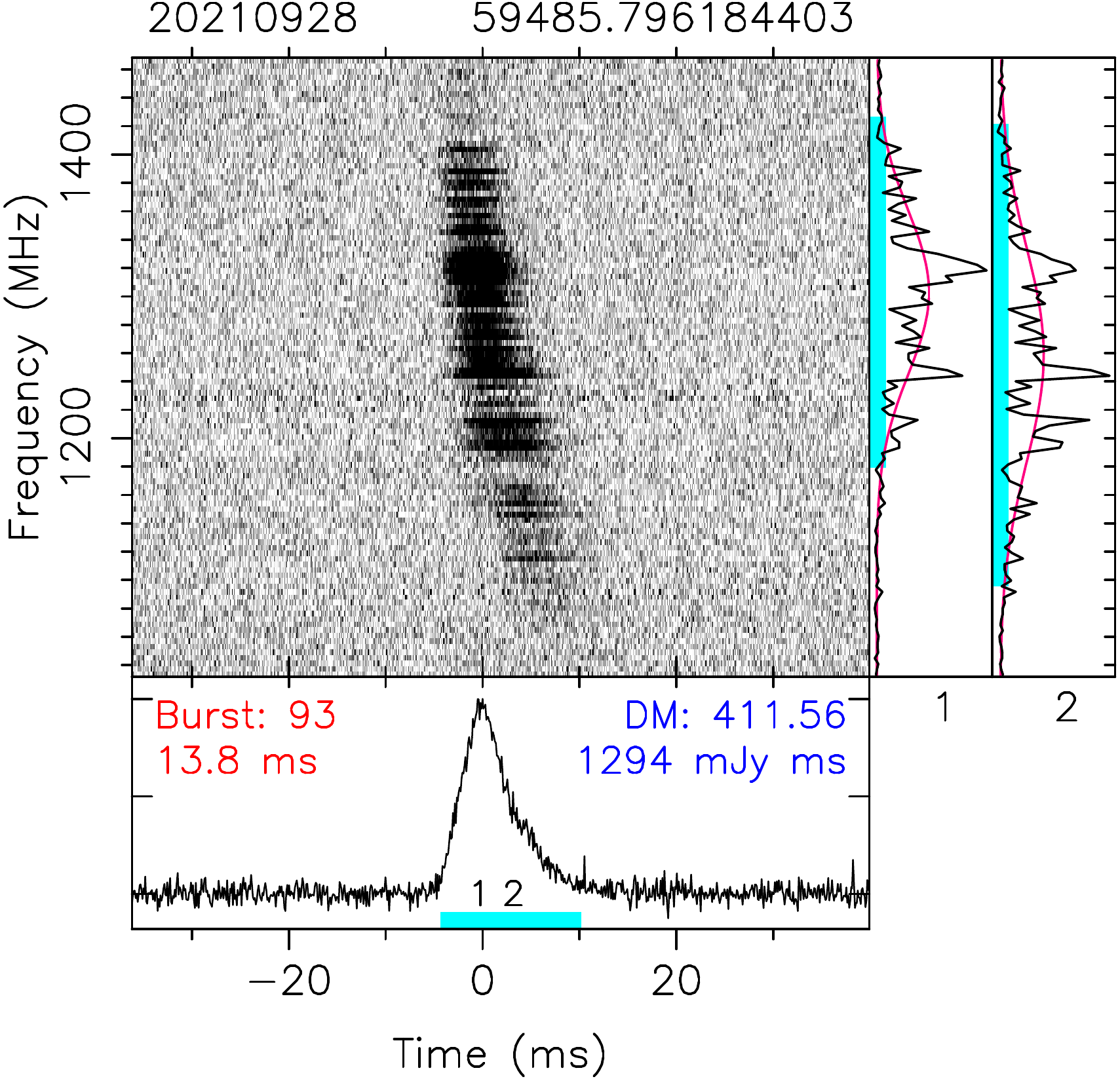}
    \includegraphics[height=37mm]{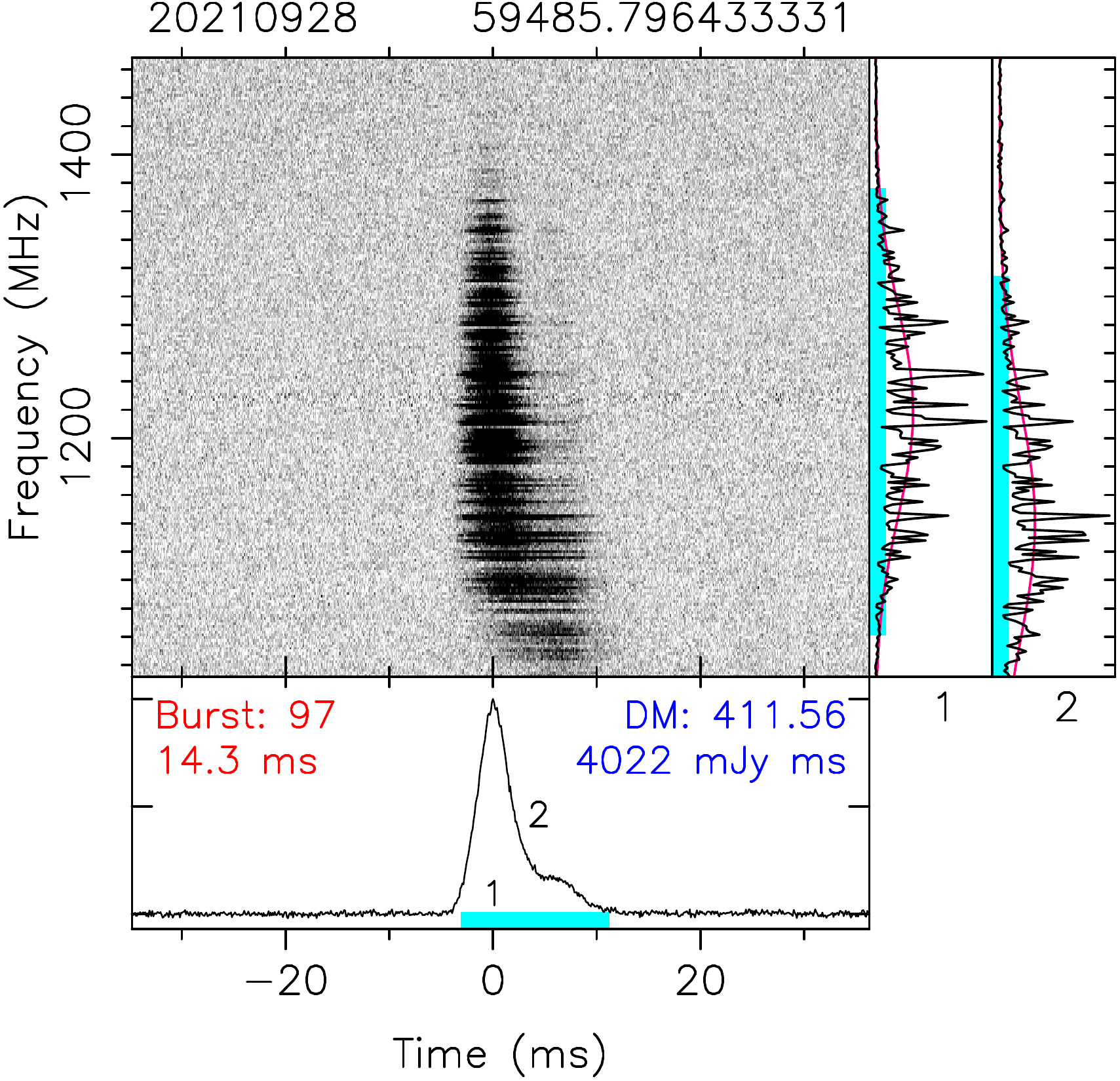}
    \includegraphics[height=37mm]{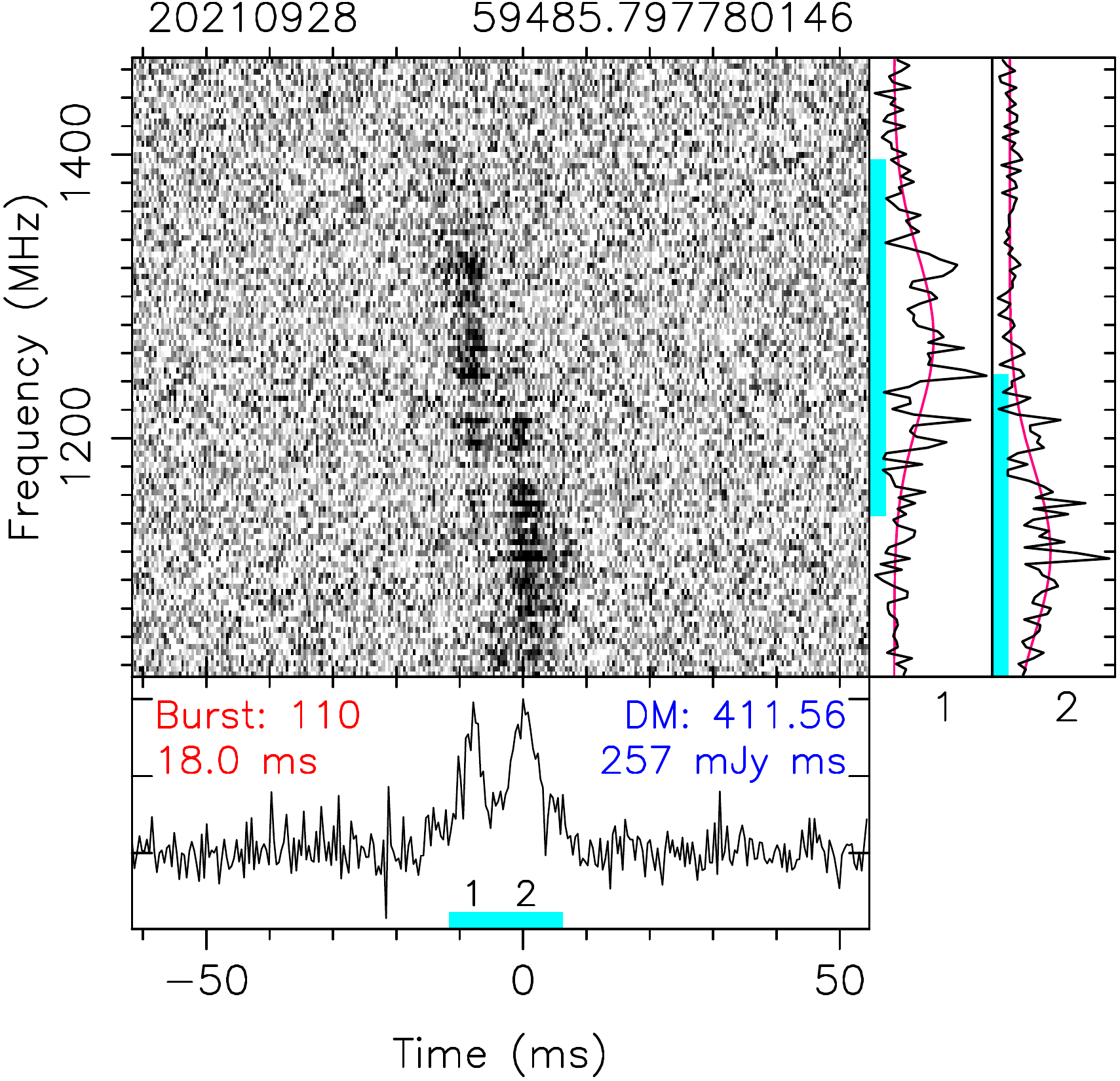}
    \includegraphics[height=37mm]{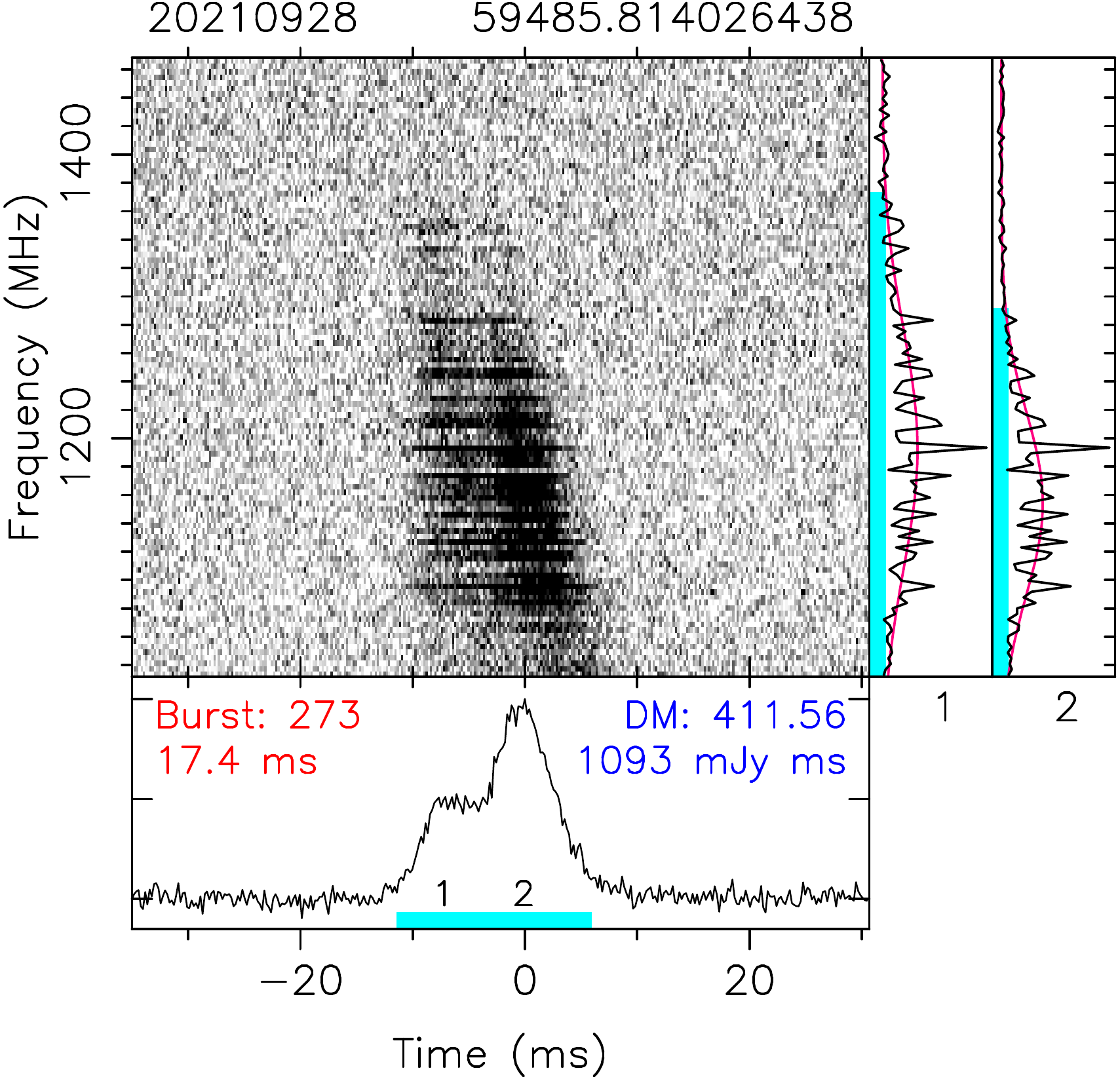}
    \includegraphics[height=37mm]{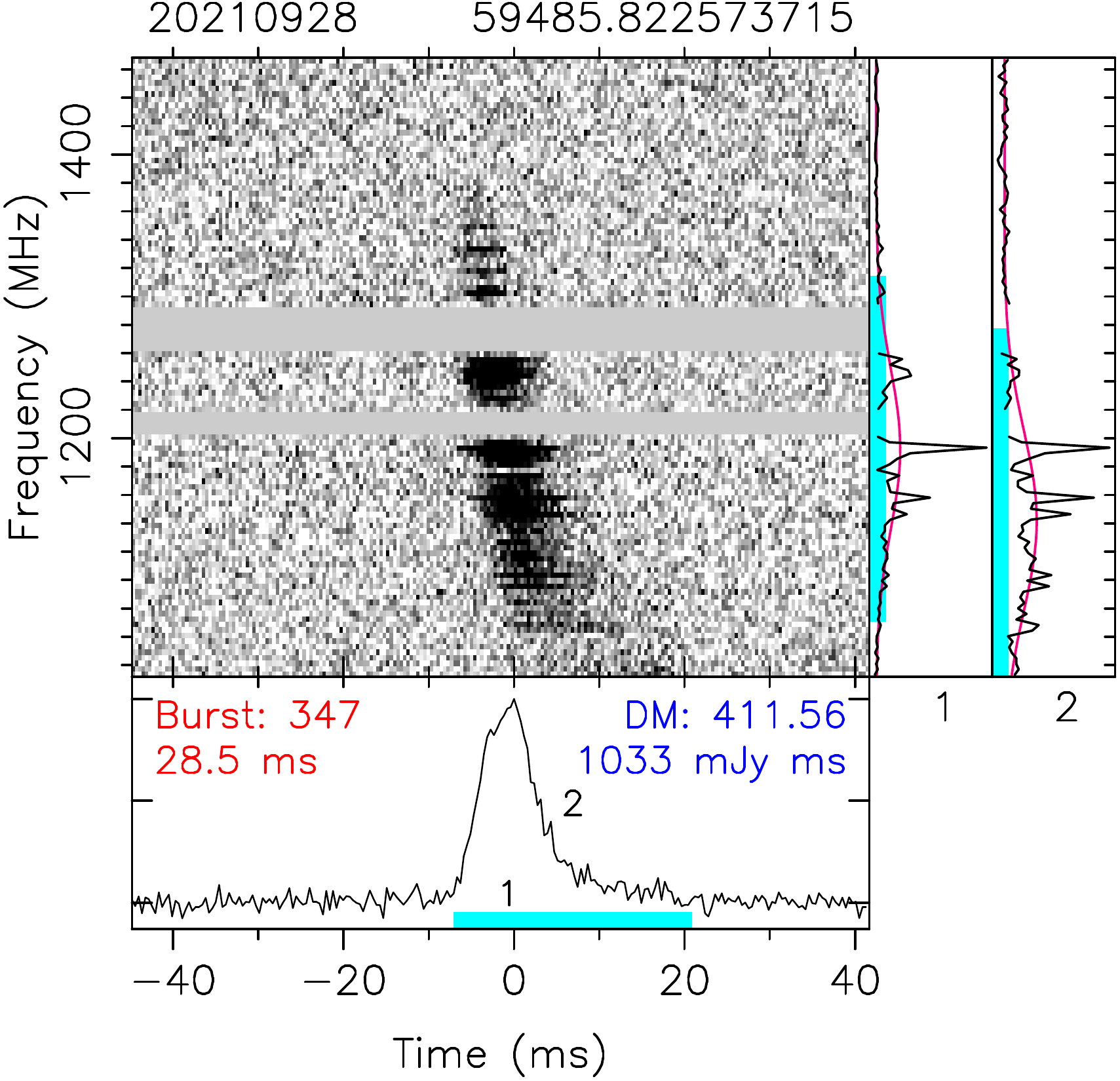}
    \includegraphics[height=37mm]{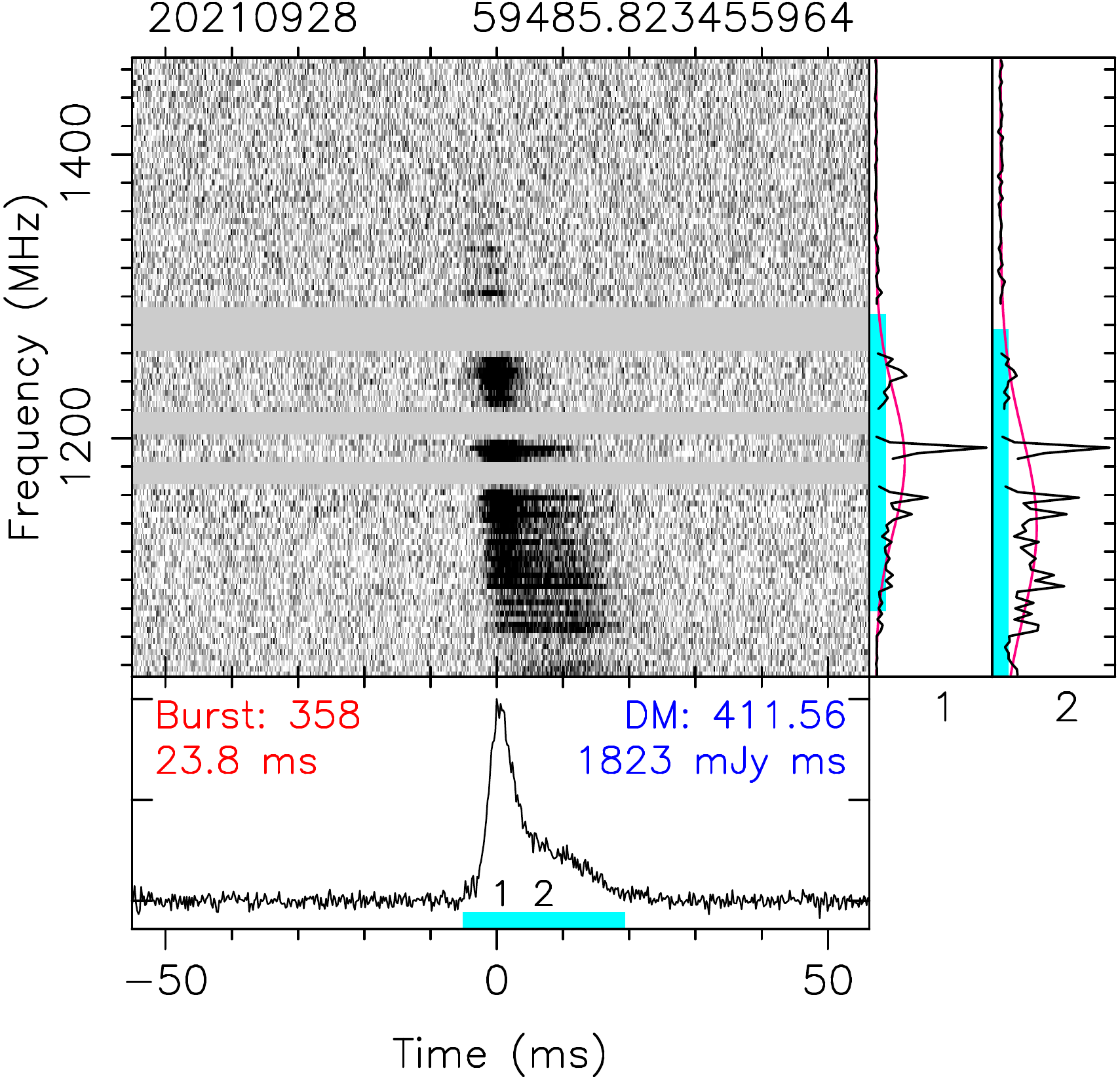}
\caption{The same as Figure~\ref{fig:appendix:D1W} but for bursts in D2-M.
}\label{fig:appendix:D2M}
\end{figure*}

\begin{figure*}
    \flushleft
    \includegraphics[height=37mm]{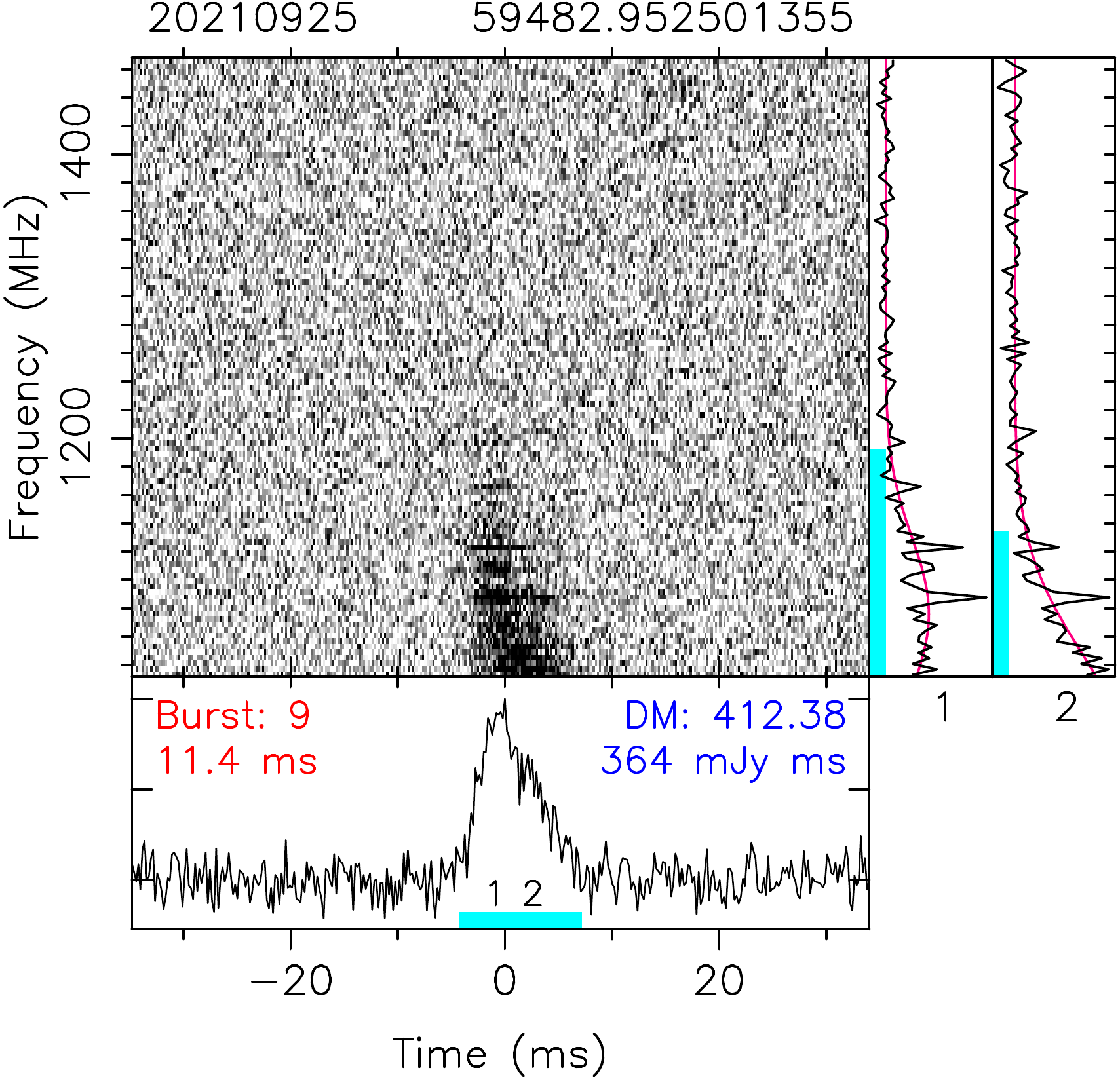}
    \includegraphics[height=37mm]{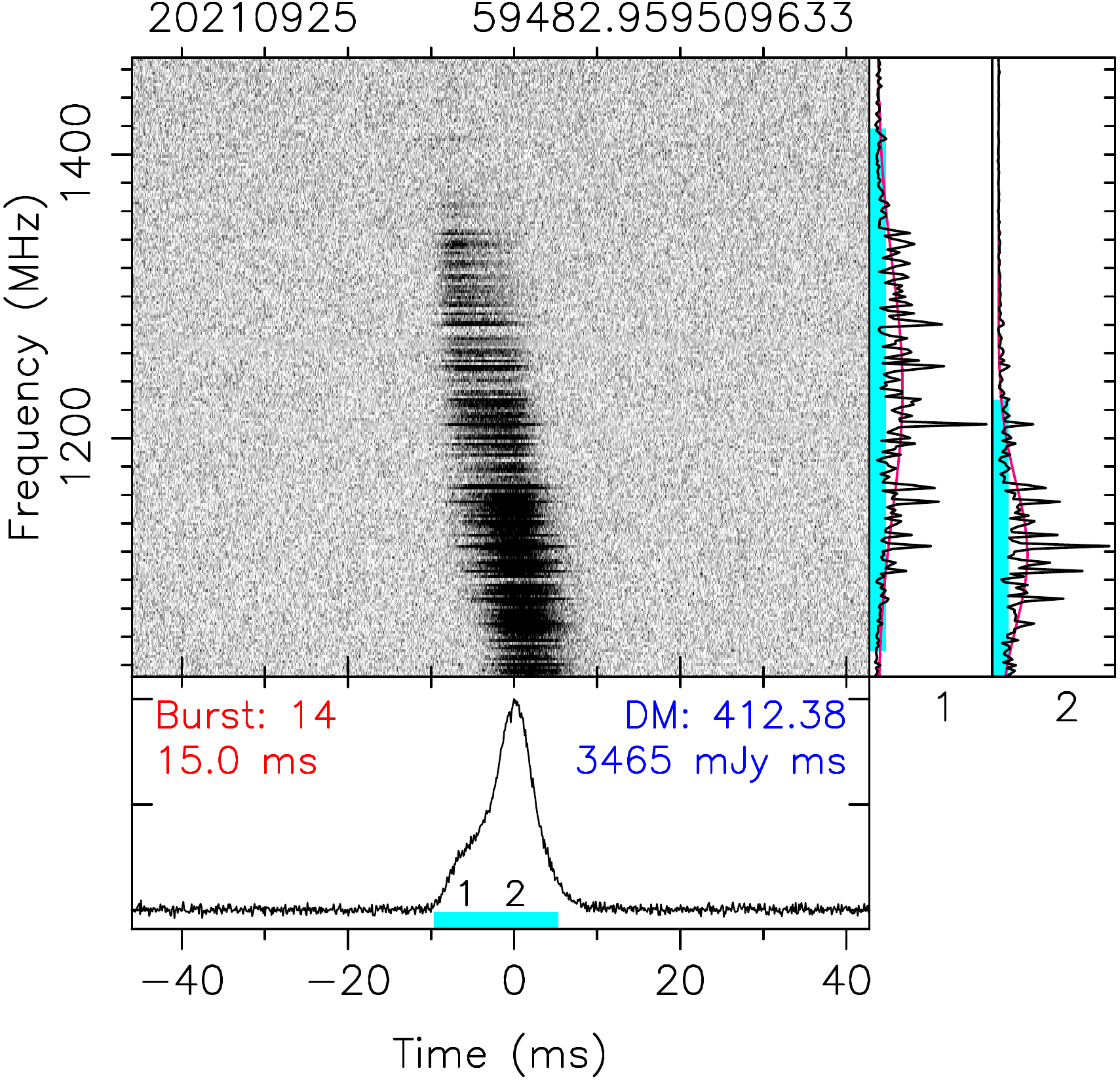}
    \includegraphics[height=37mm]{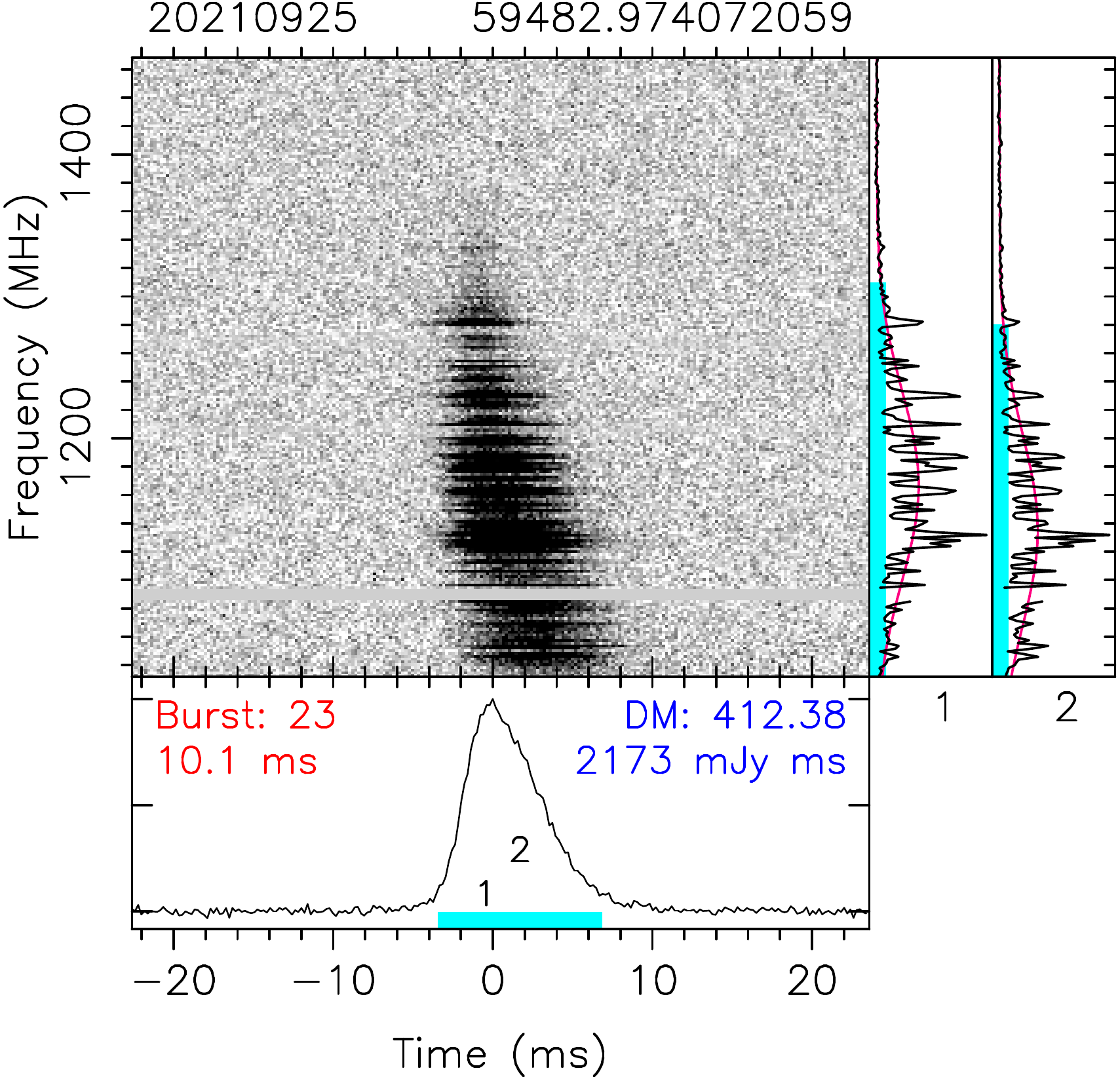}
    \includegraphics[height=37mm]{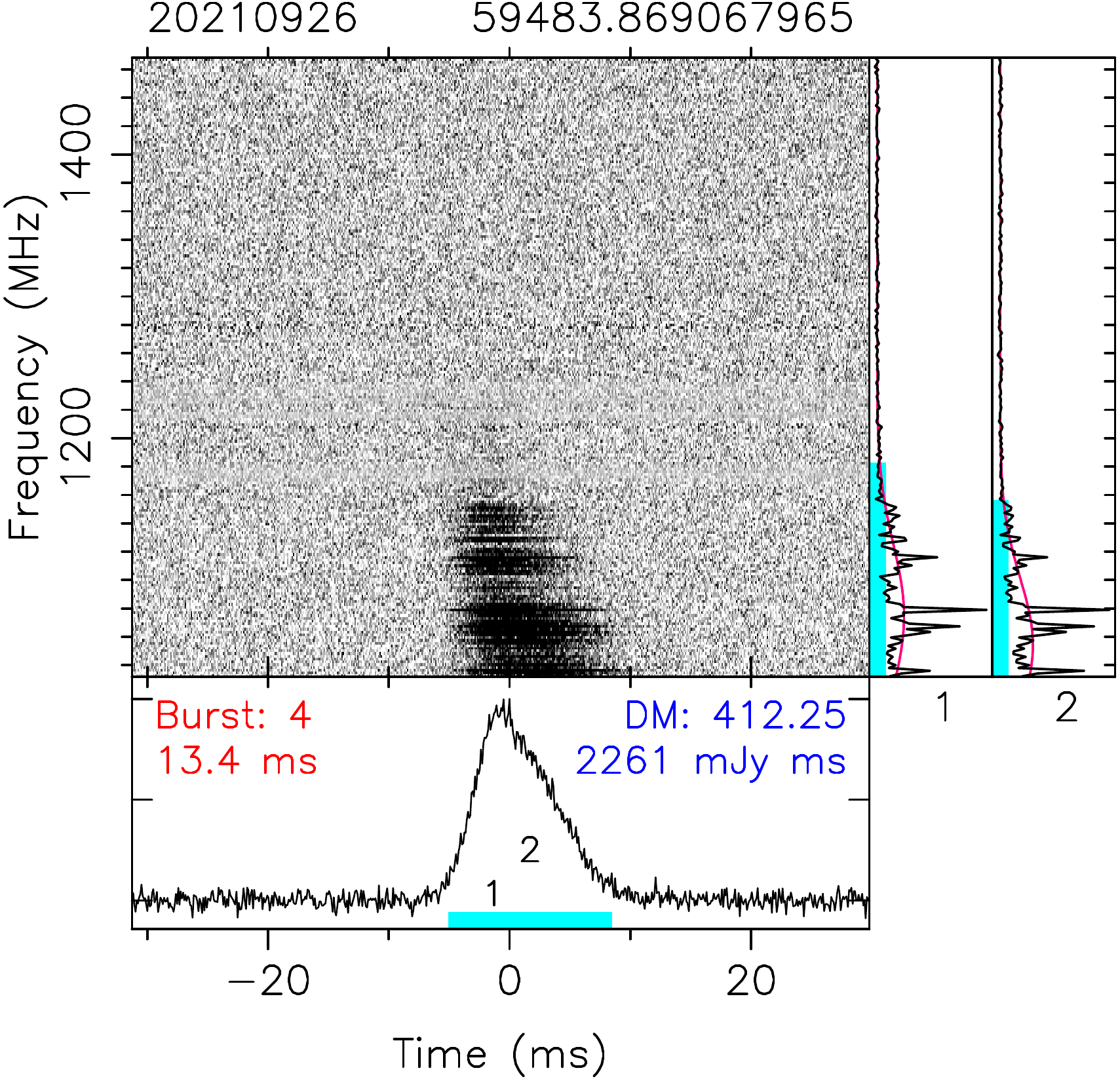}
    \includegraphics[height=37mm]{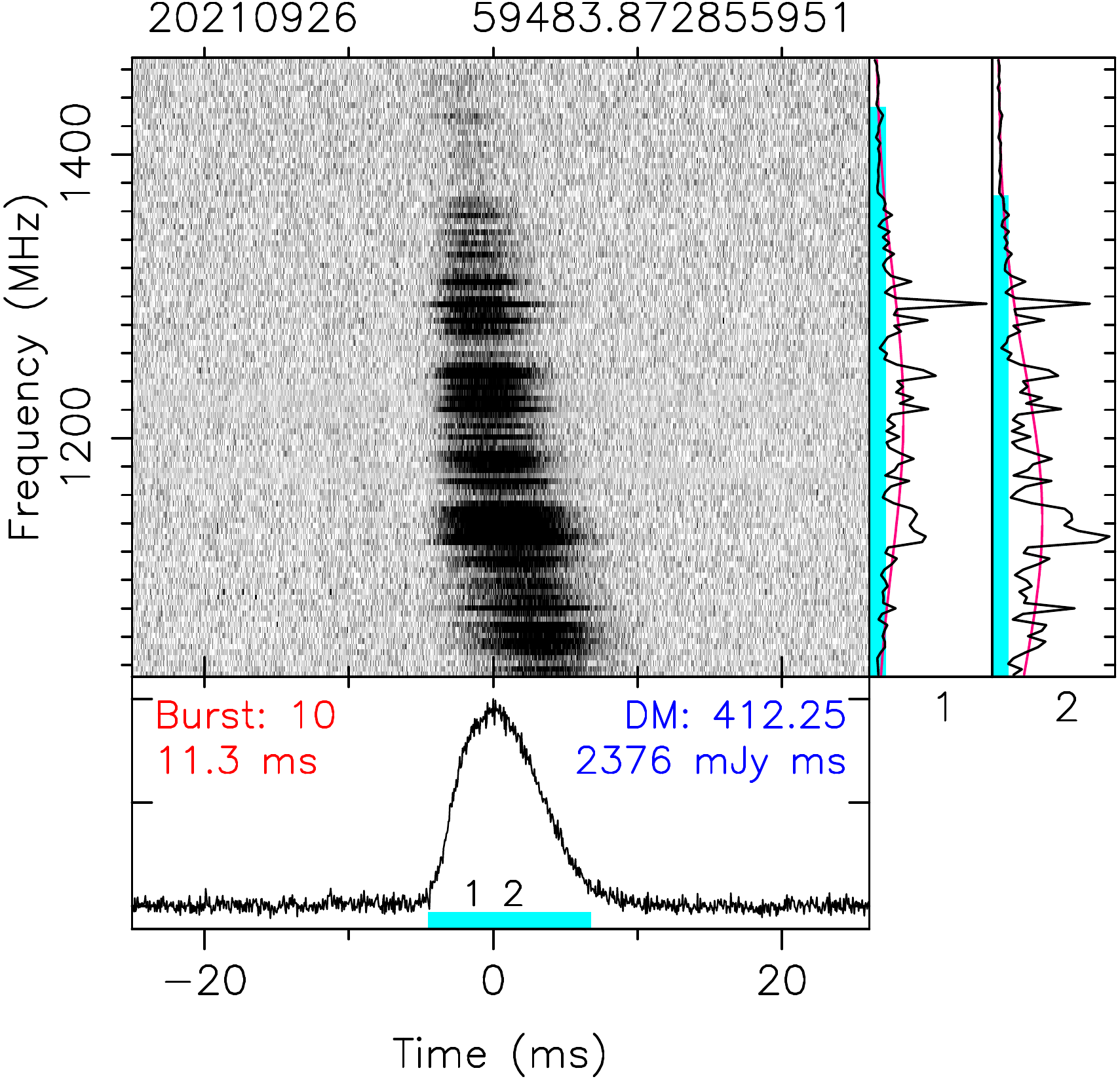}
    \includegraphics[height=37mm]{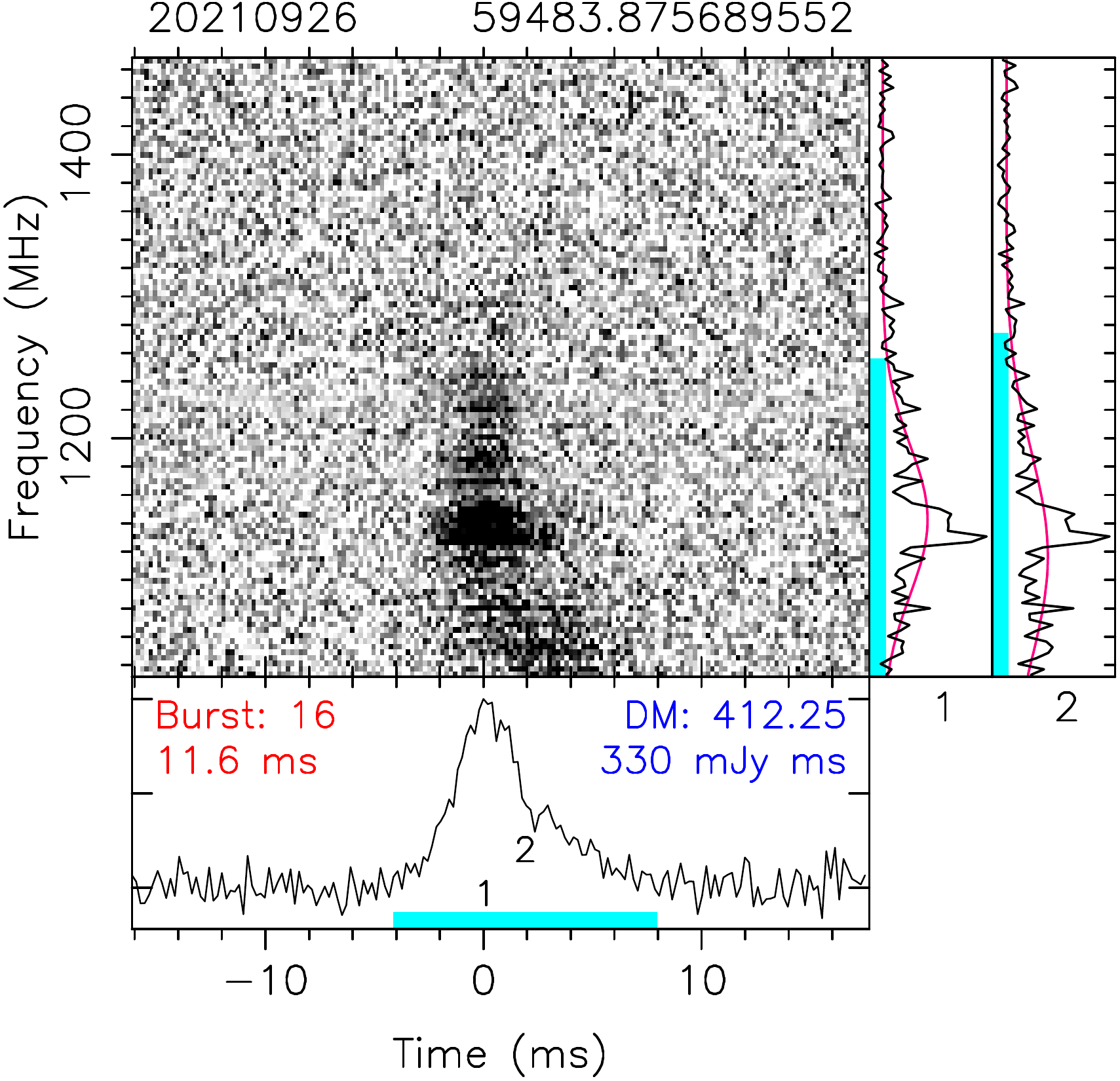}
    \includegraphics[height=37mm]{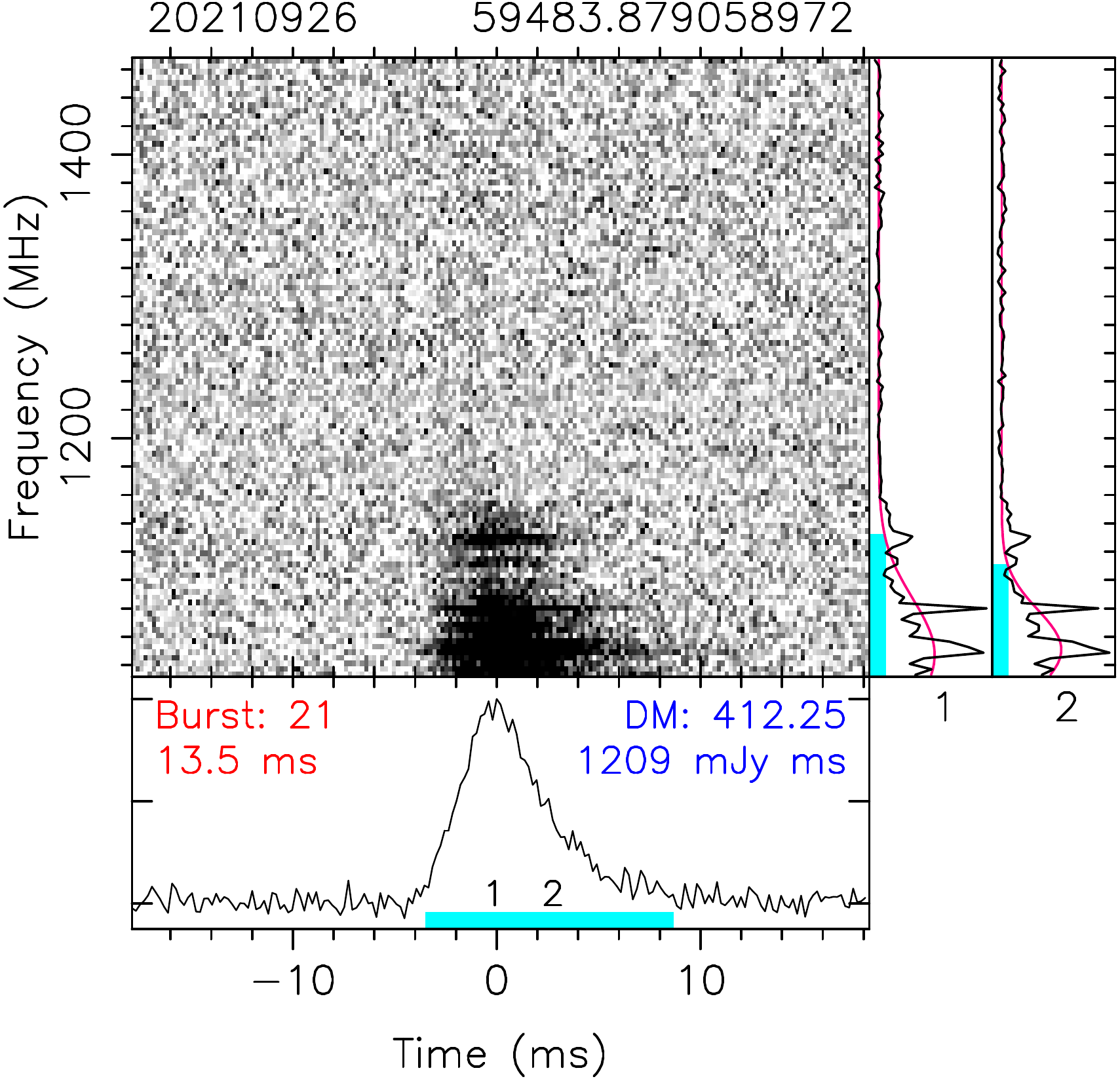}
    \includegraphics[height=37mm]{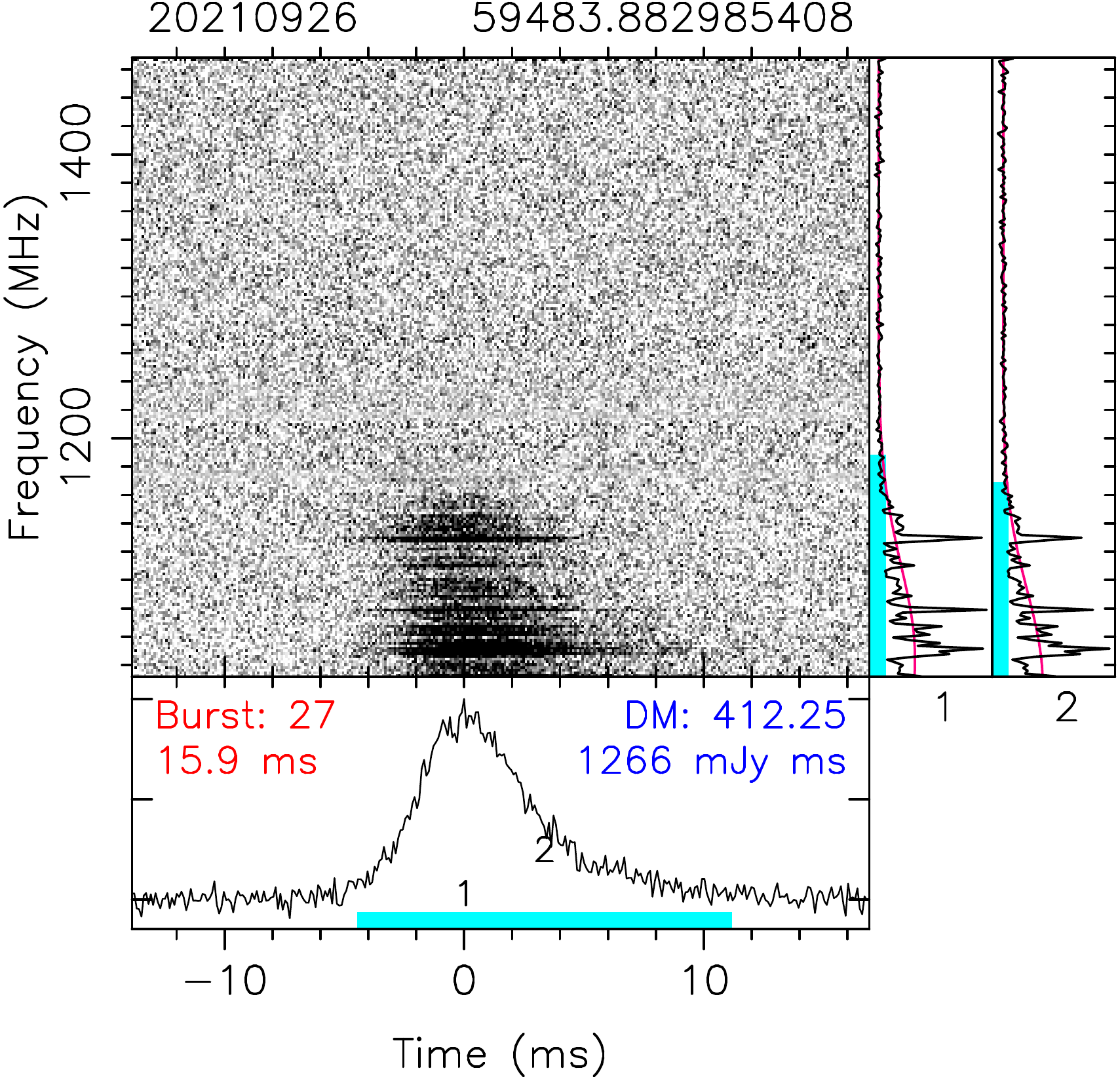}
    \includegraphics[height=37mm]{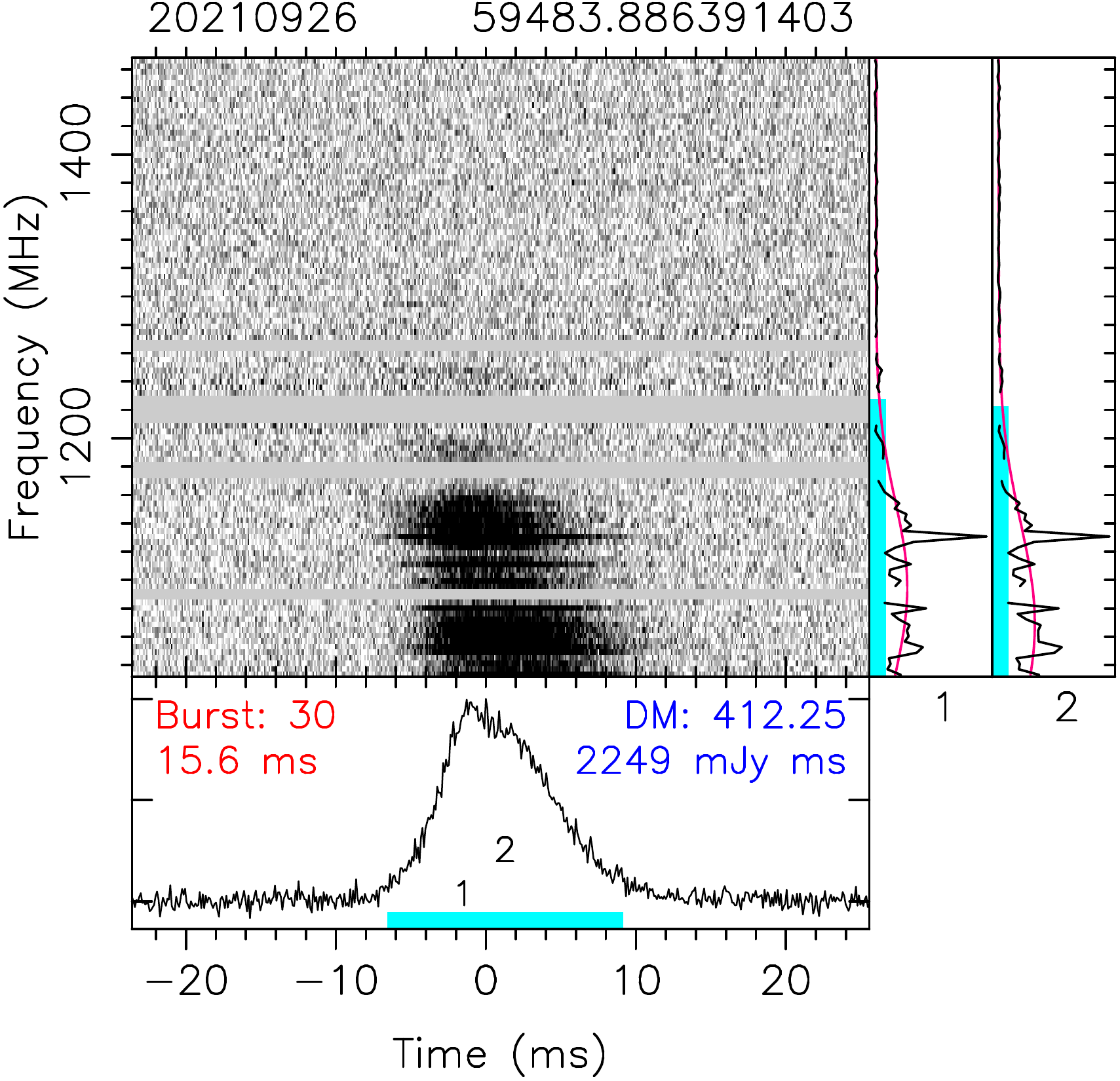}
    \includegraphics[height=37mm]{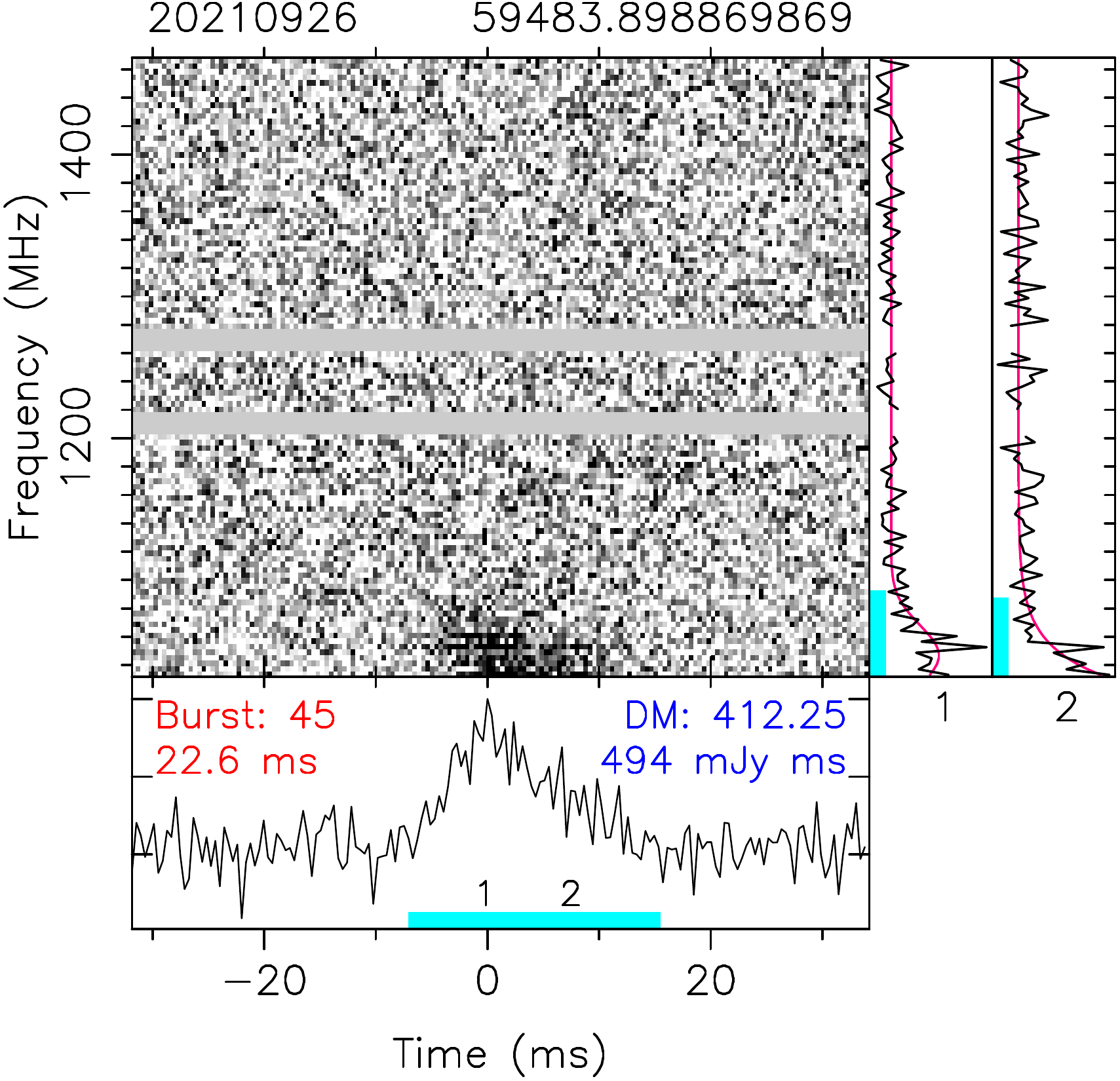}
    \includegraphics[height=37mm]{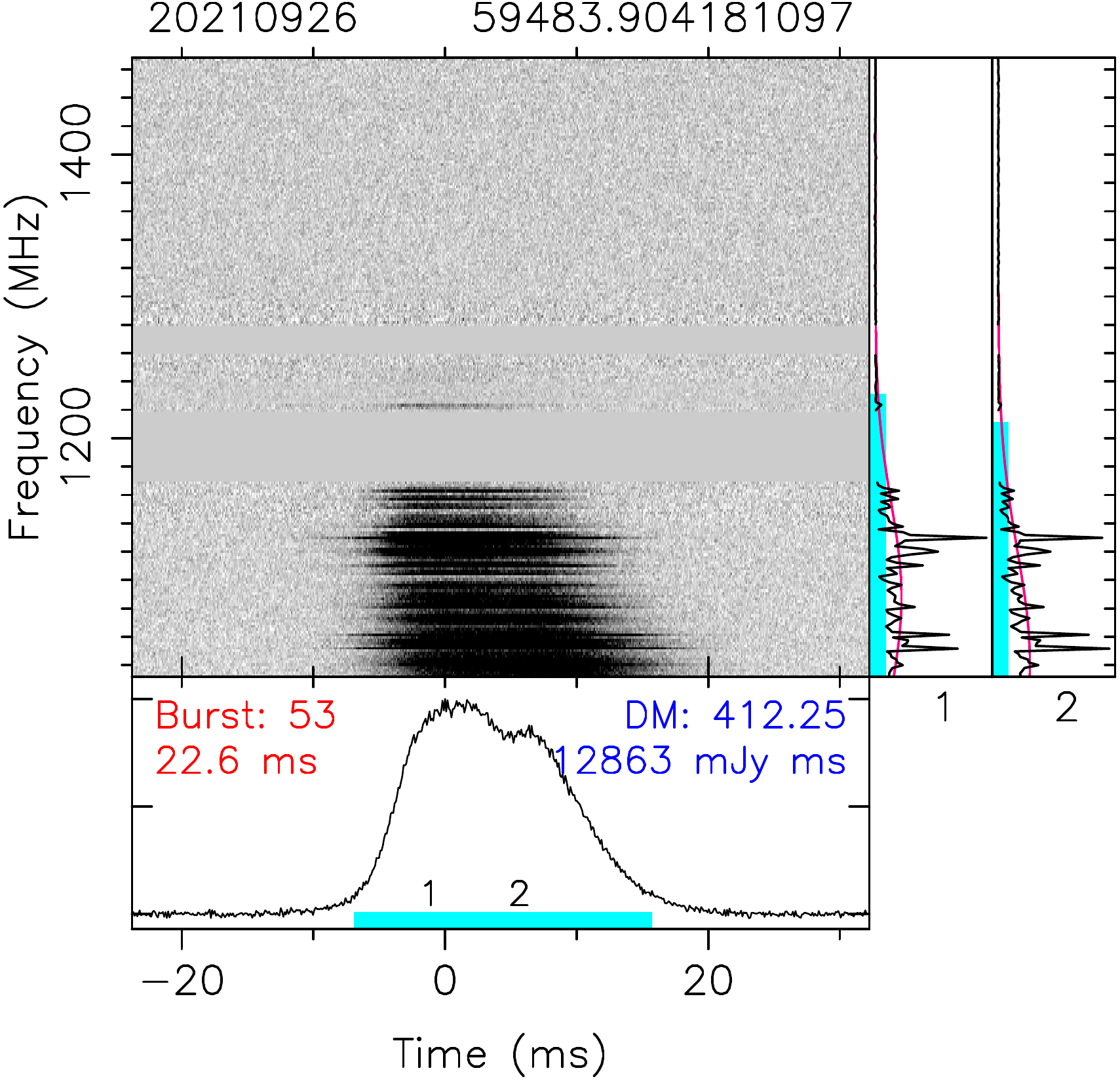}
    \includegraphics[height=37mm]{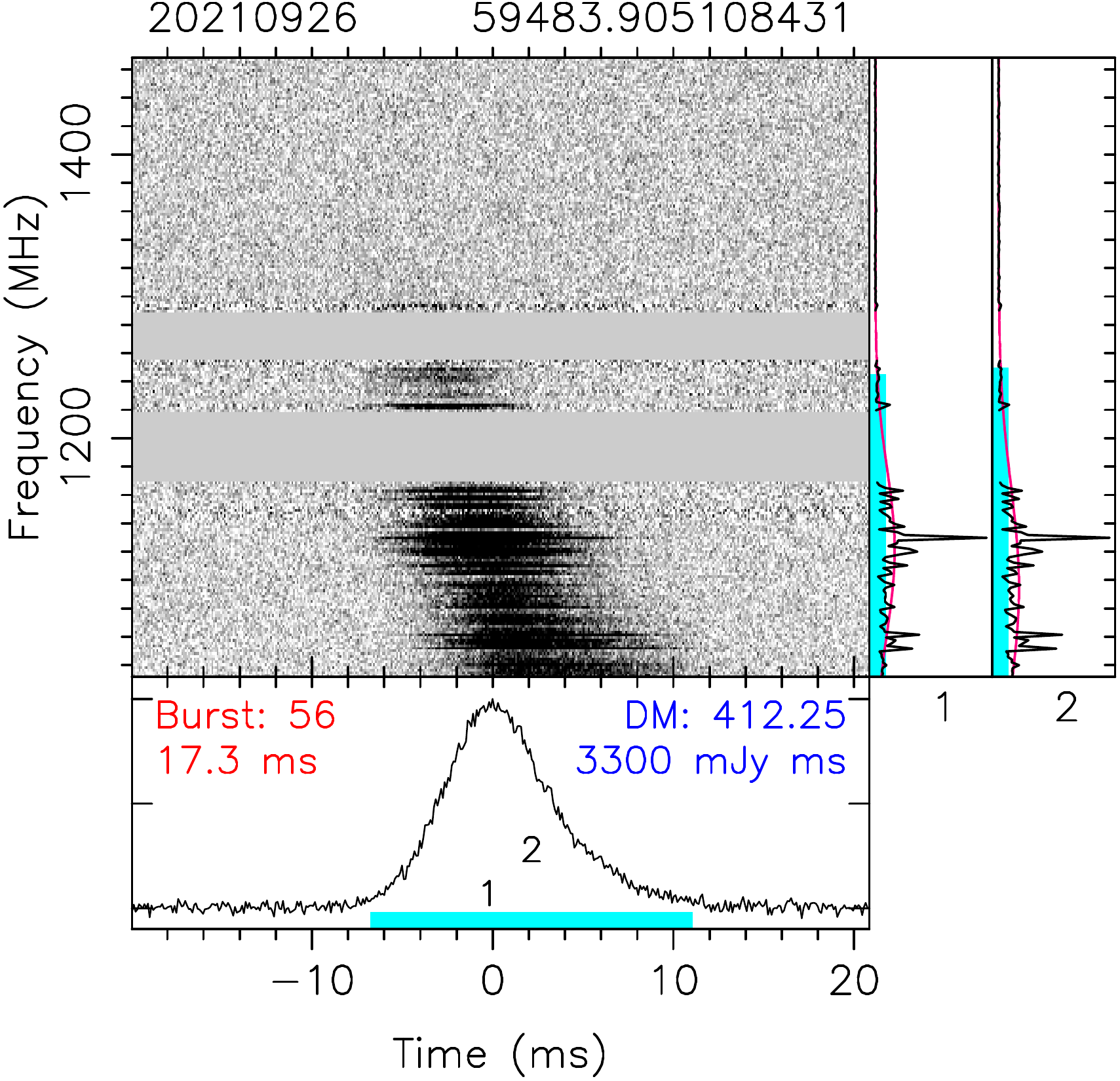}
    \includegraphics[height=37mm]{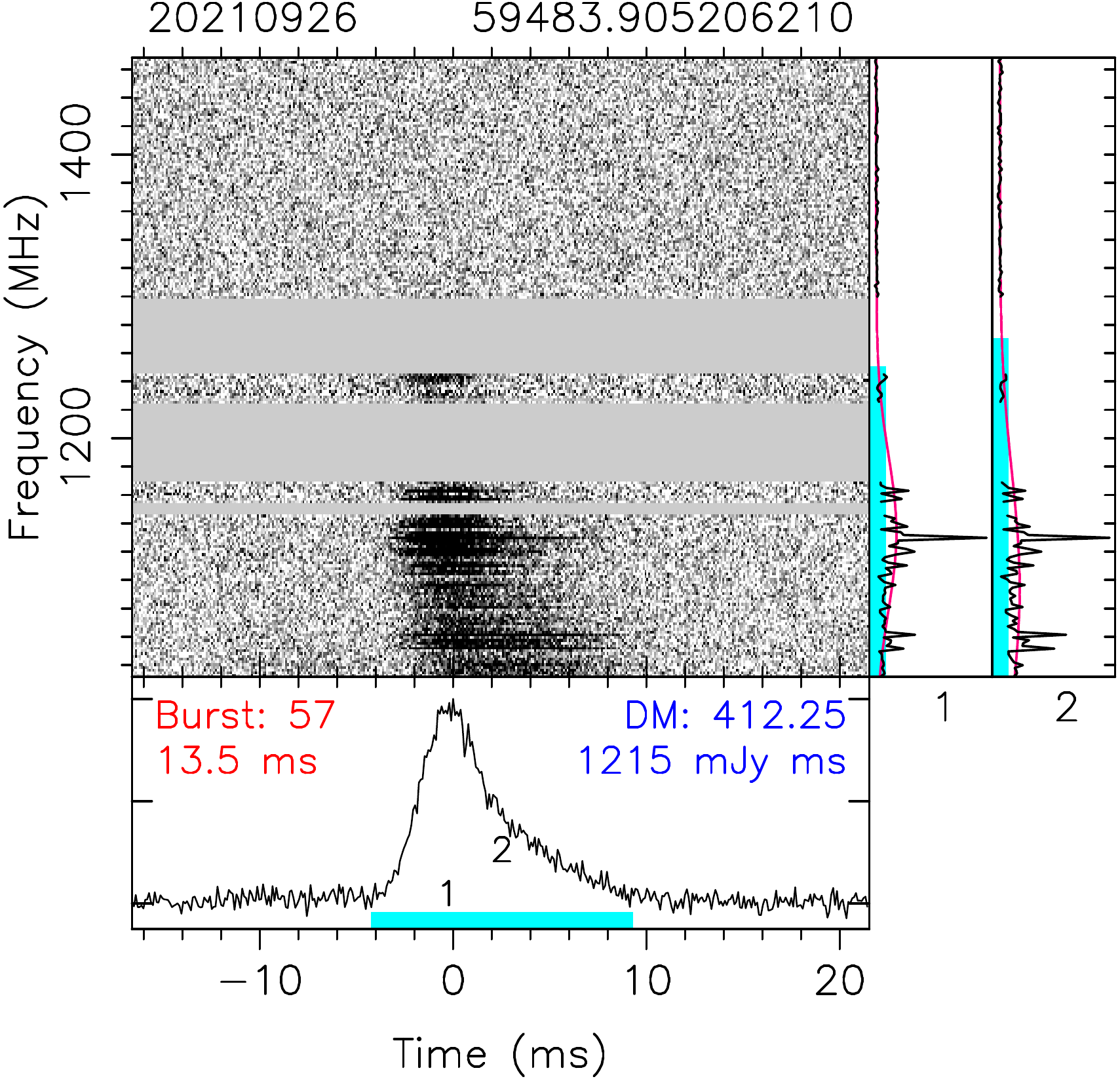}
    \includegraphics[height=37mm]{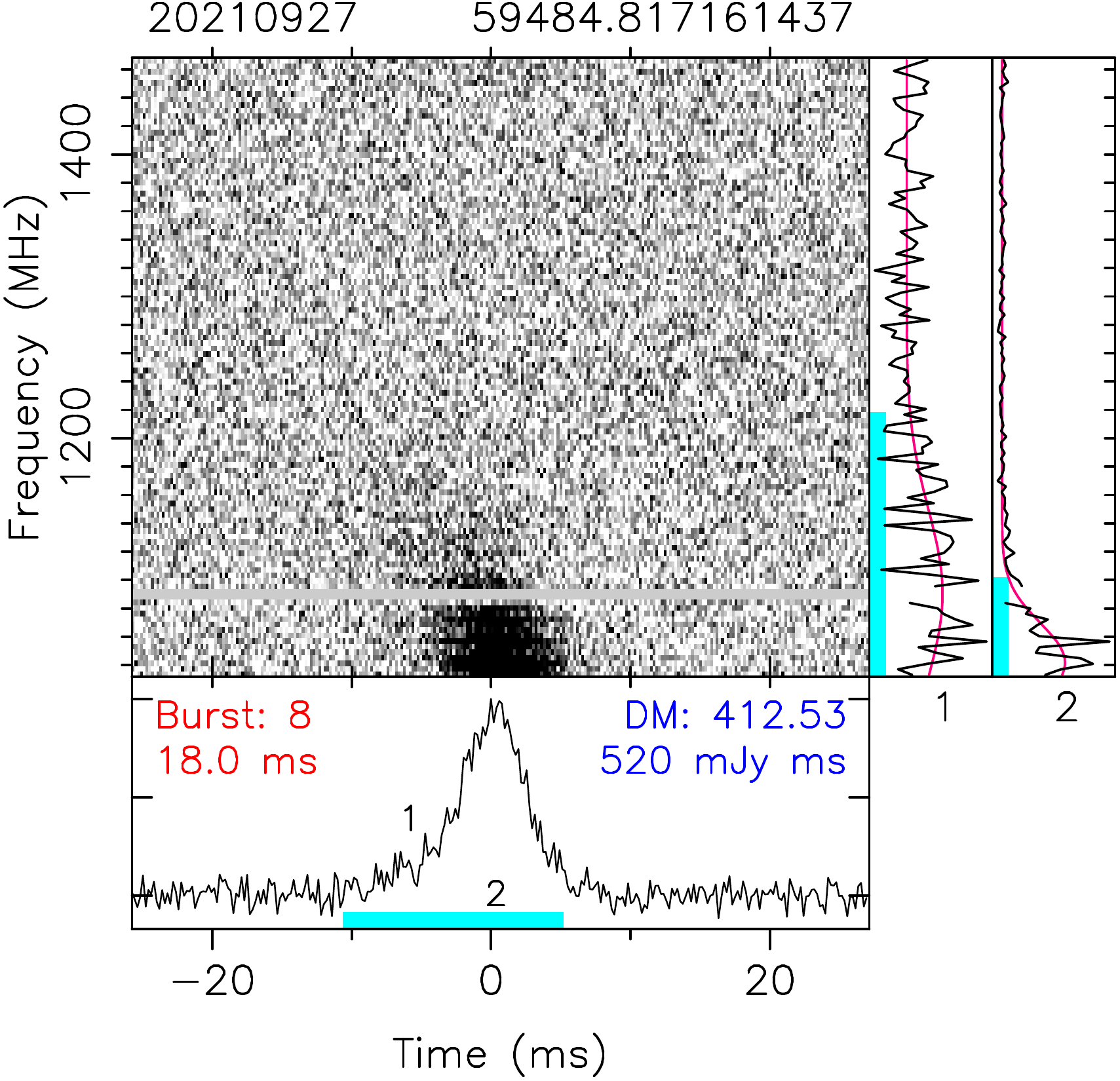}
    \includegraphics[height=37mm]{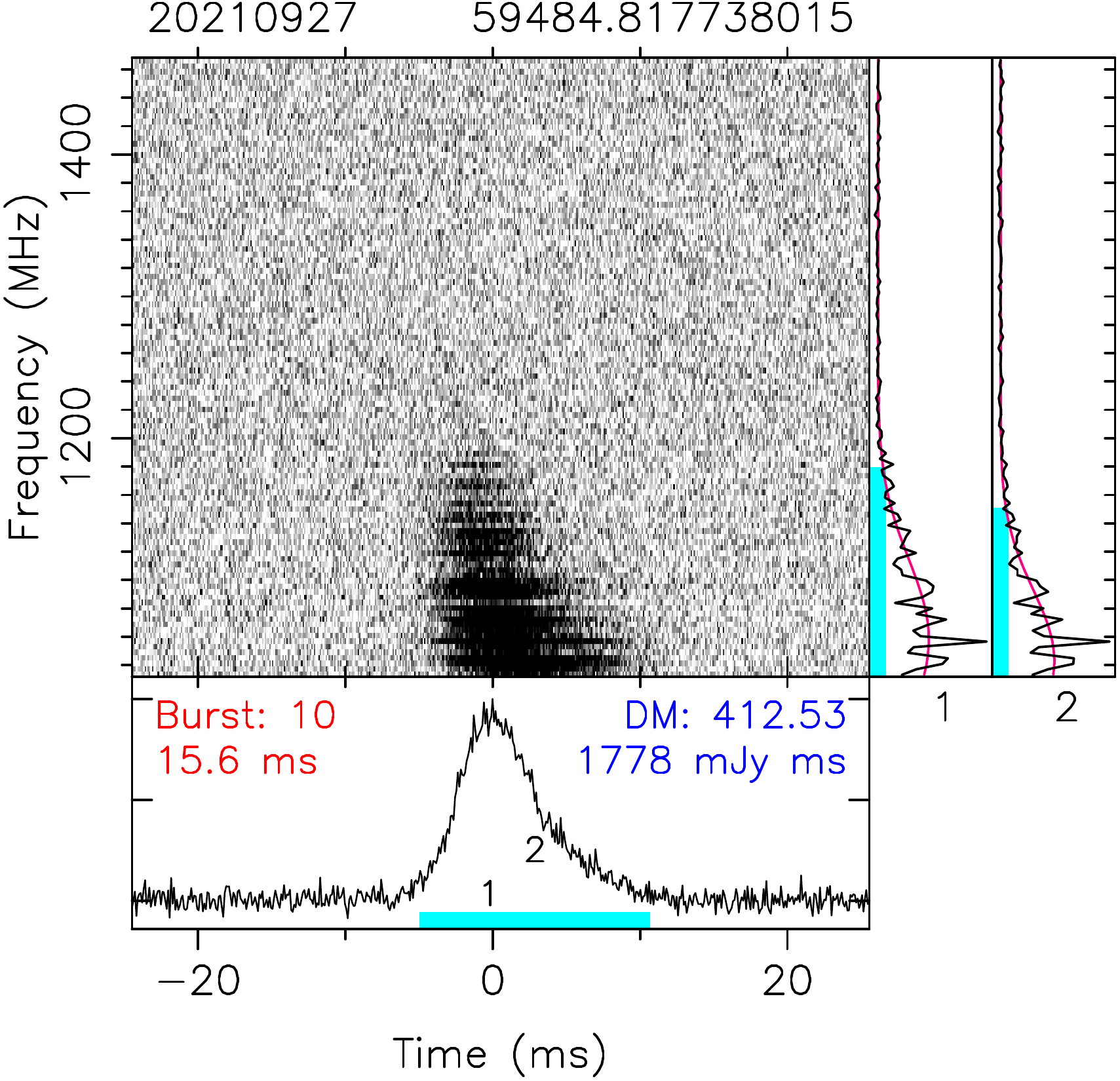}
    \includegraphics[height=37mm]{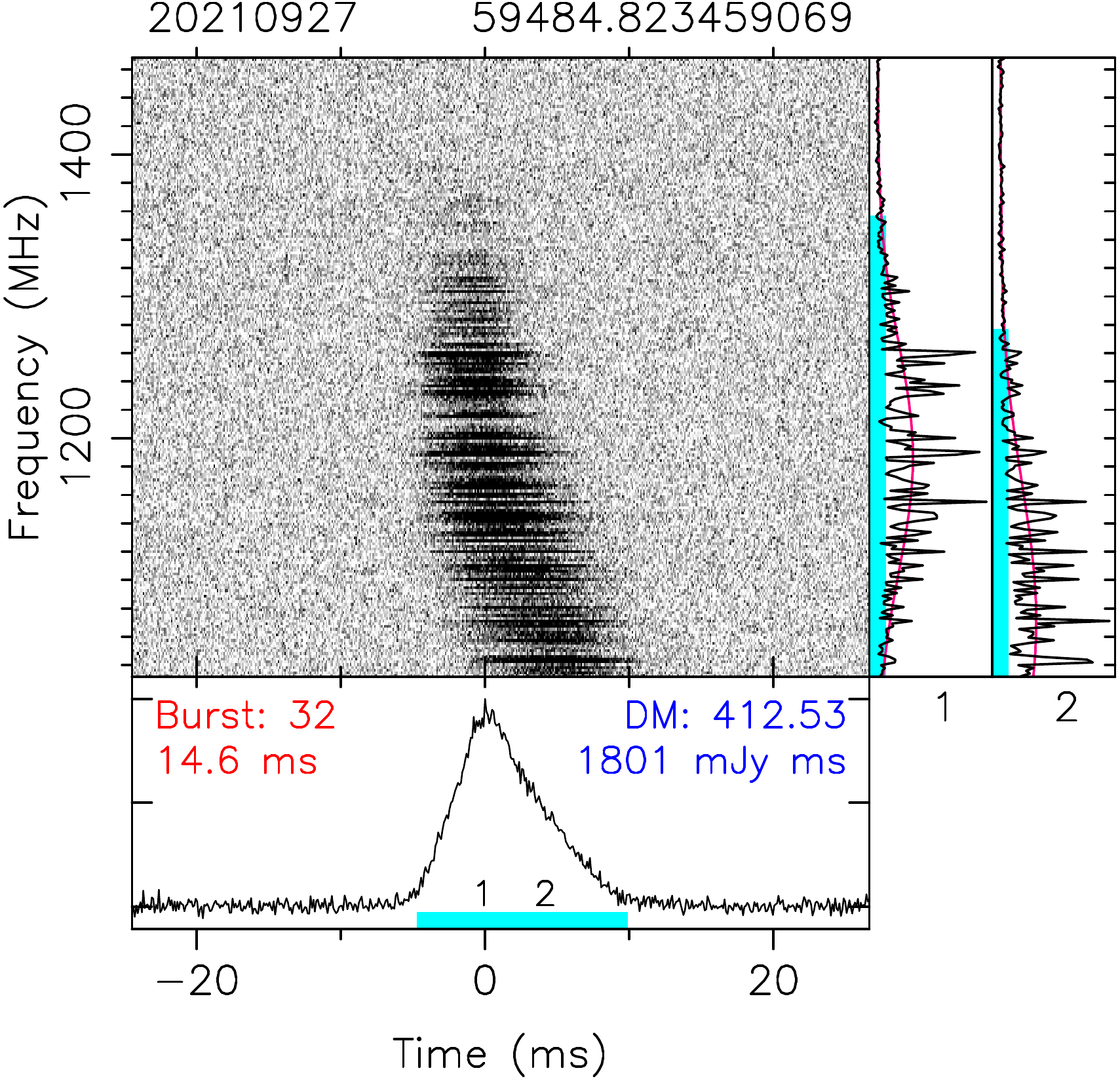}
    \includegraphics[height=37mm]{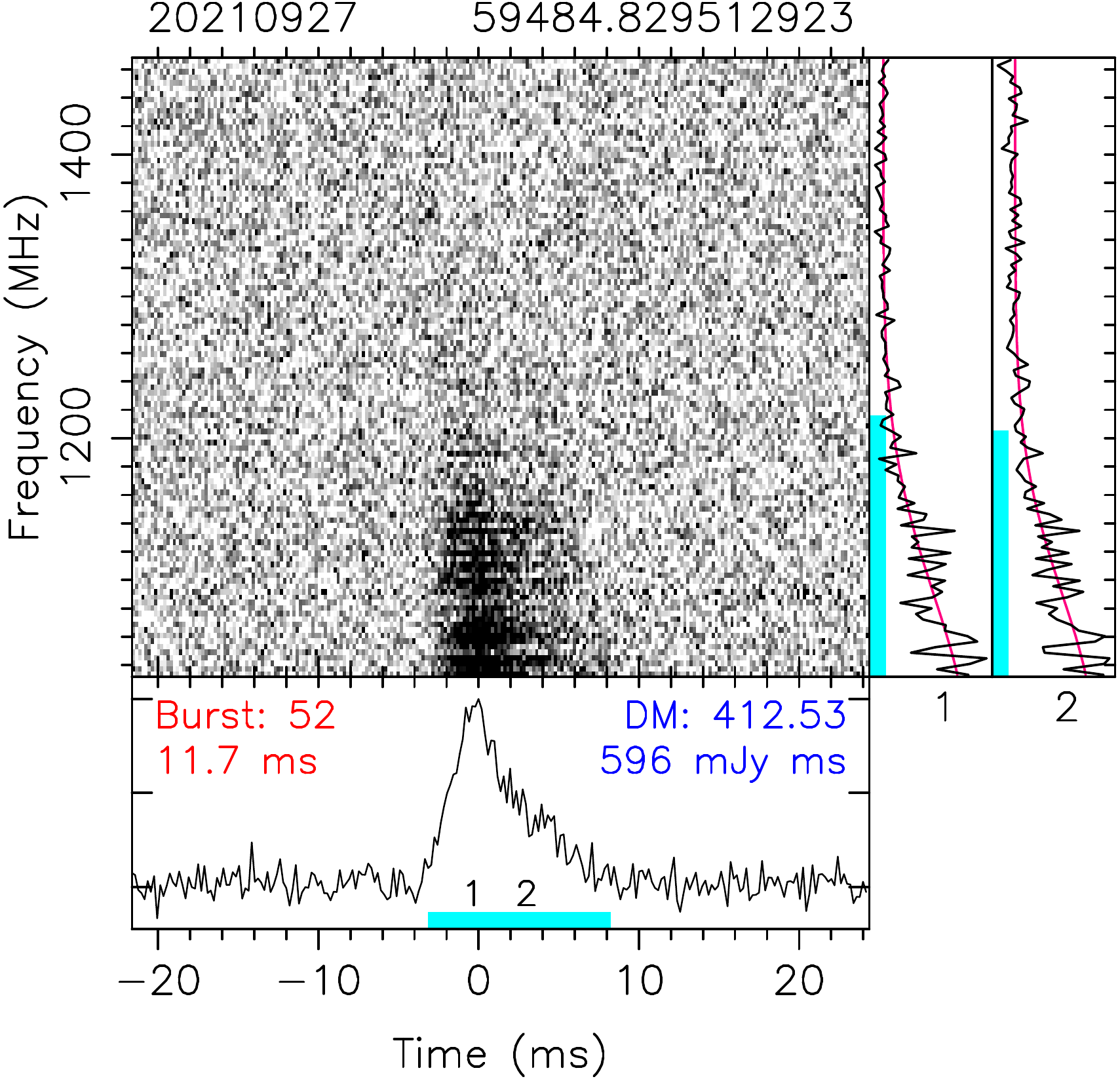}
    \includegraphics[height=37mm]{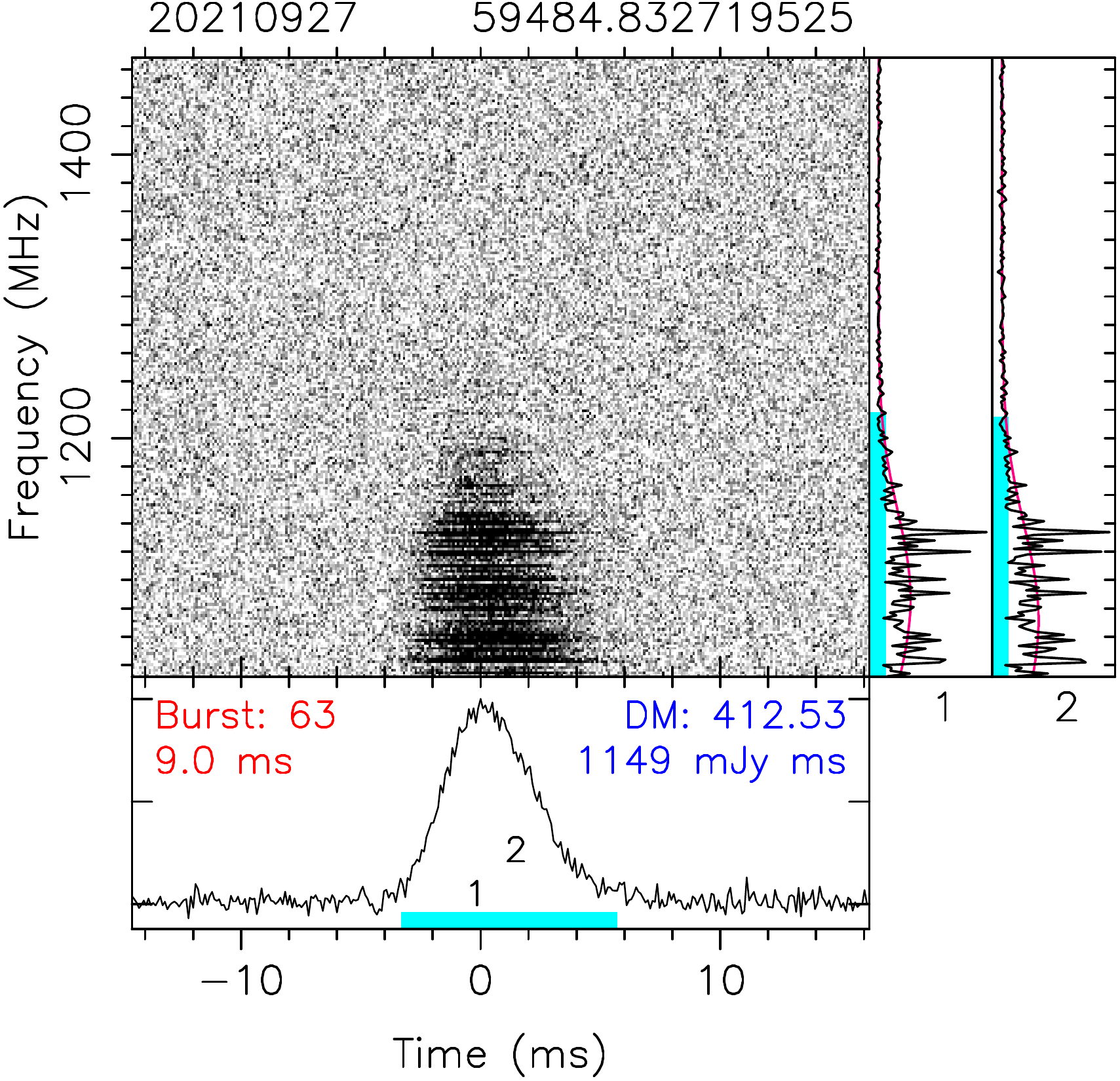}
    \includegraphics[height=37mm]{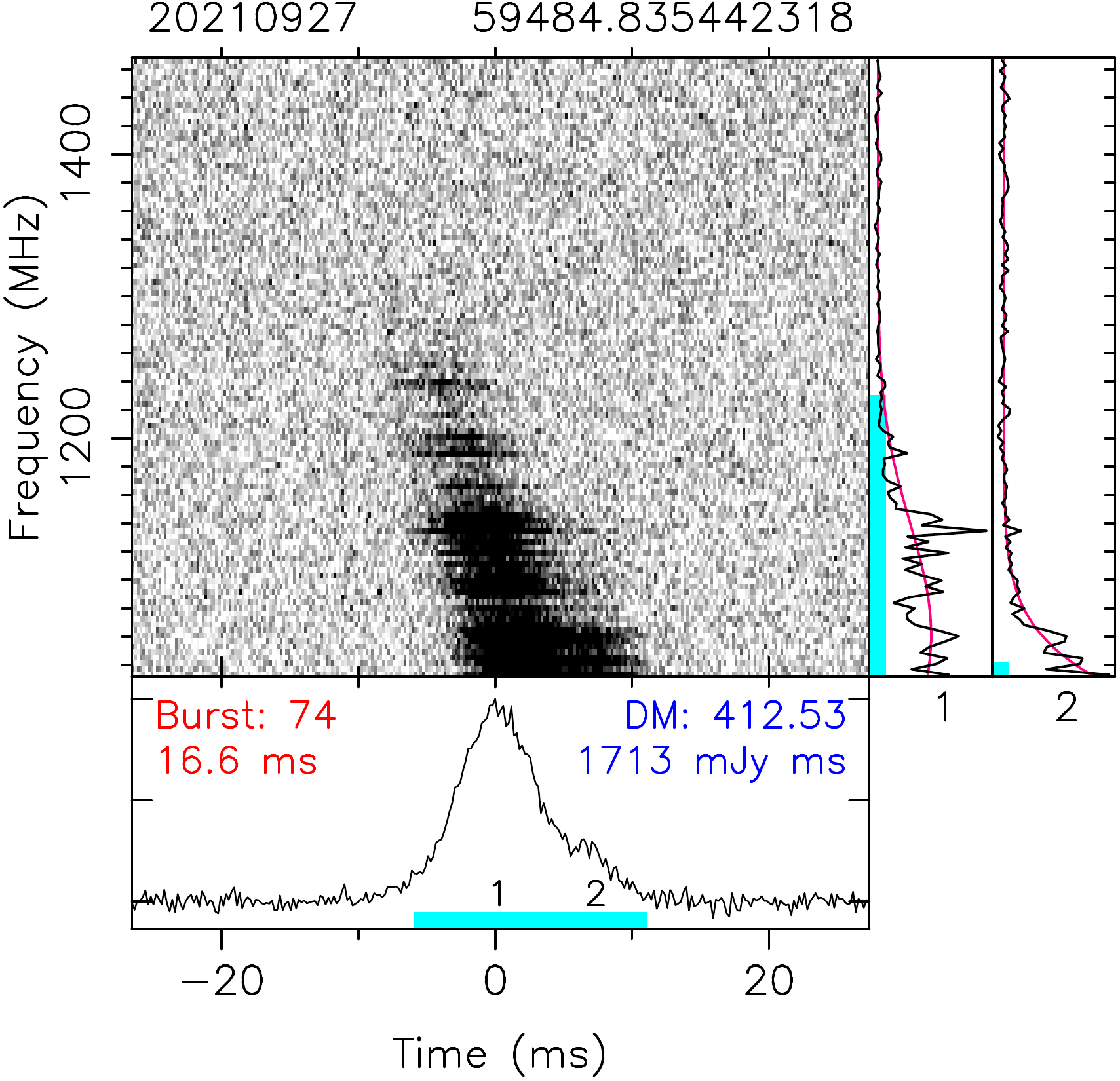}
    \includegraphics[height=37mm]{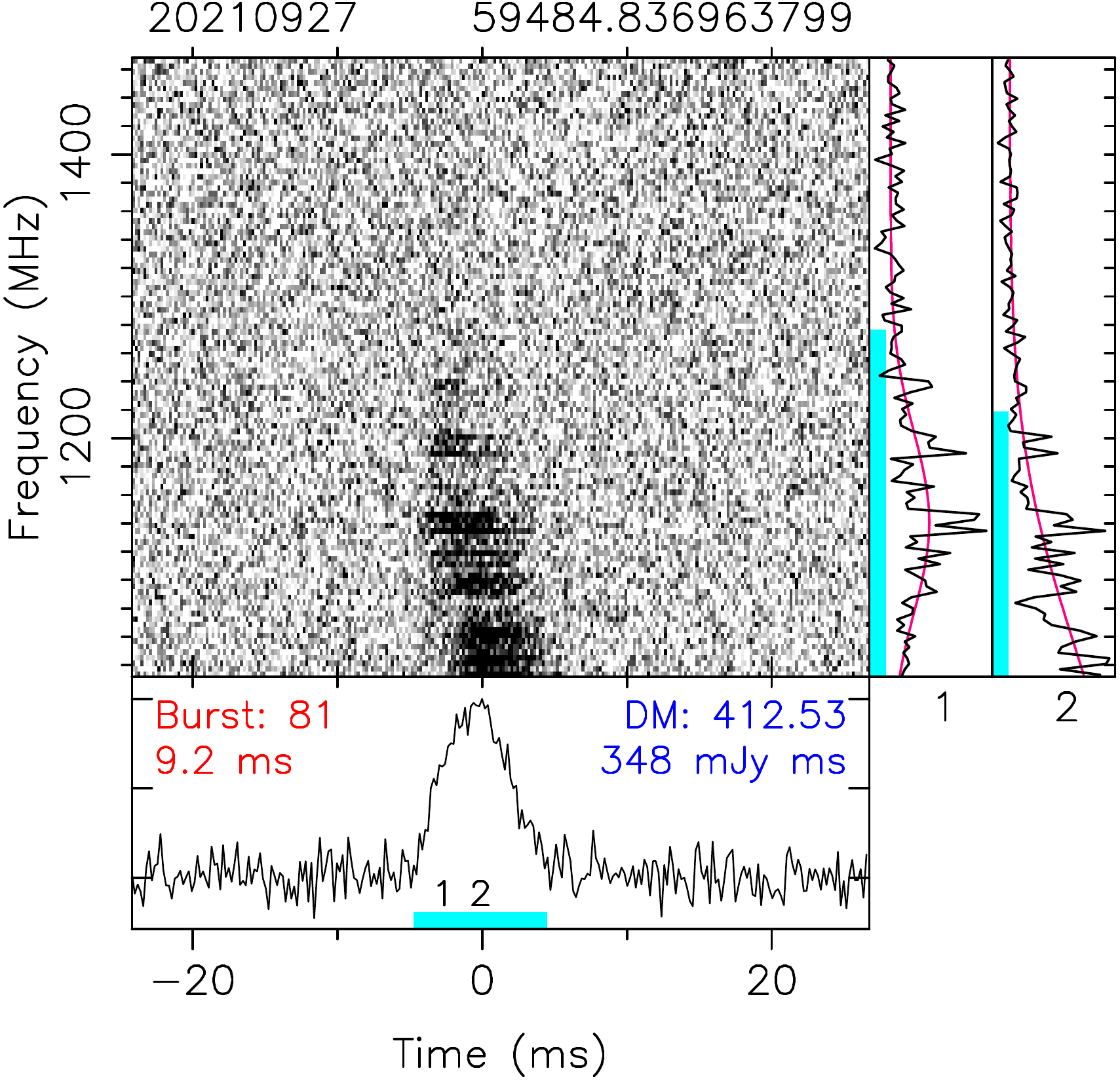}
    \includegraphics[height=37mm]{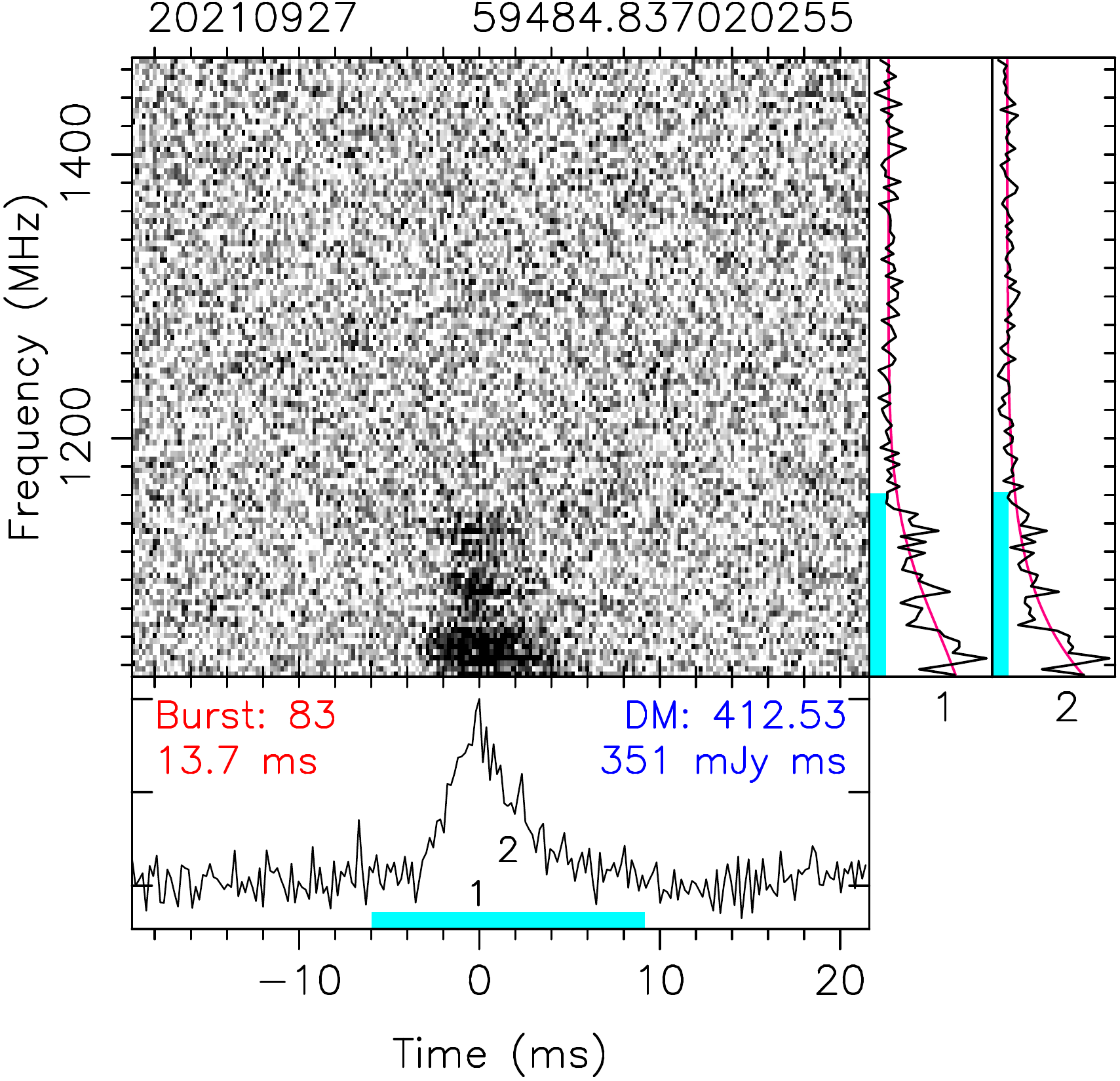}
    \includegraphics[height=37mm]{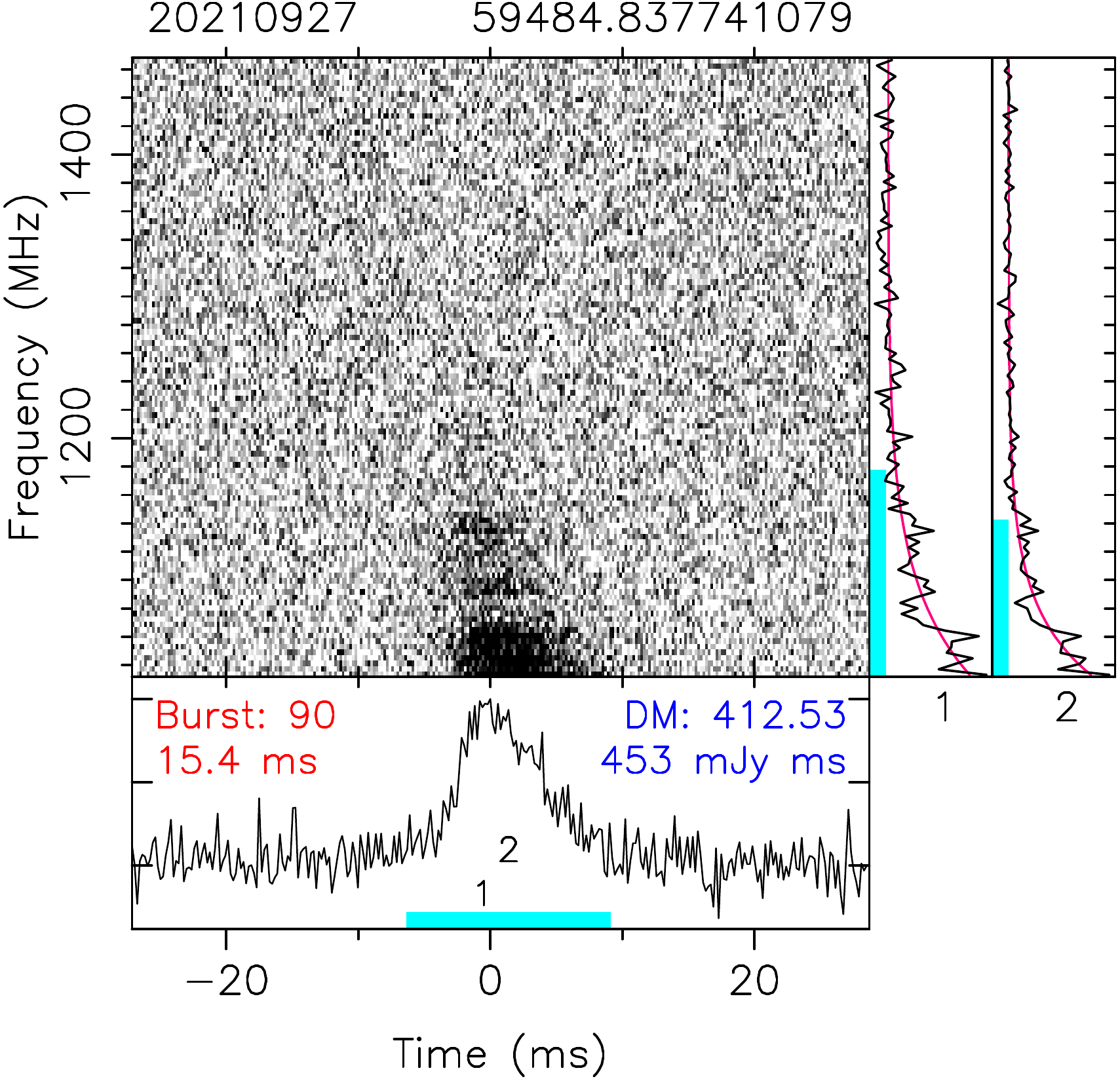}
    \includegraphics[height=37mm]{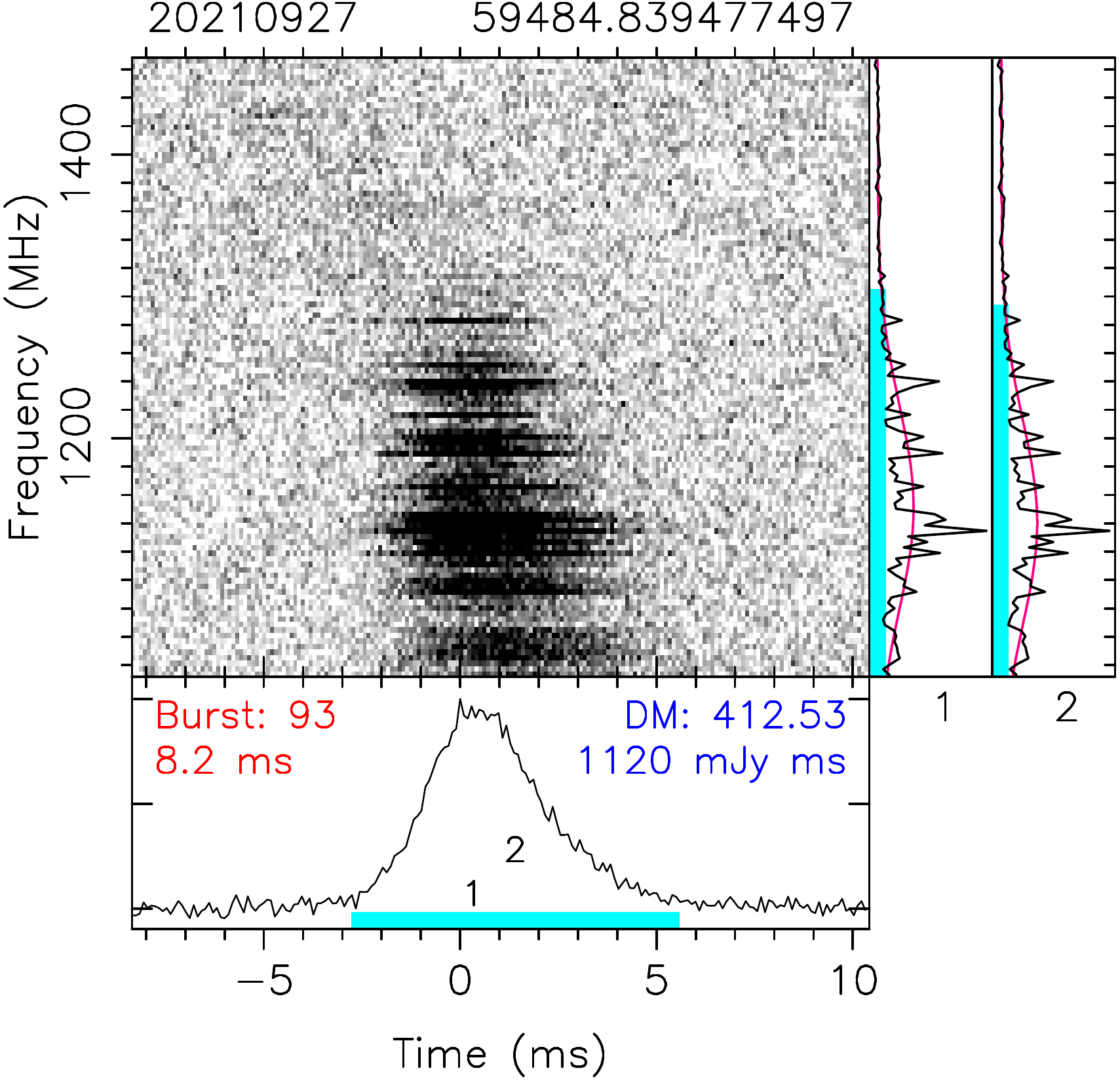}
    \includegraphics[height=37mm]{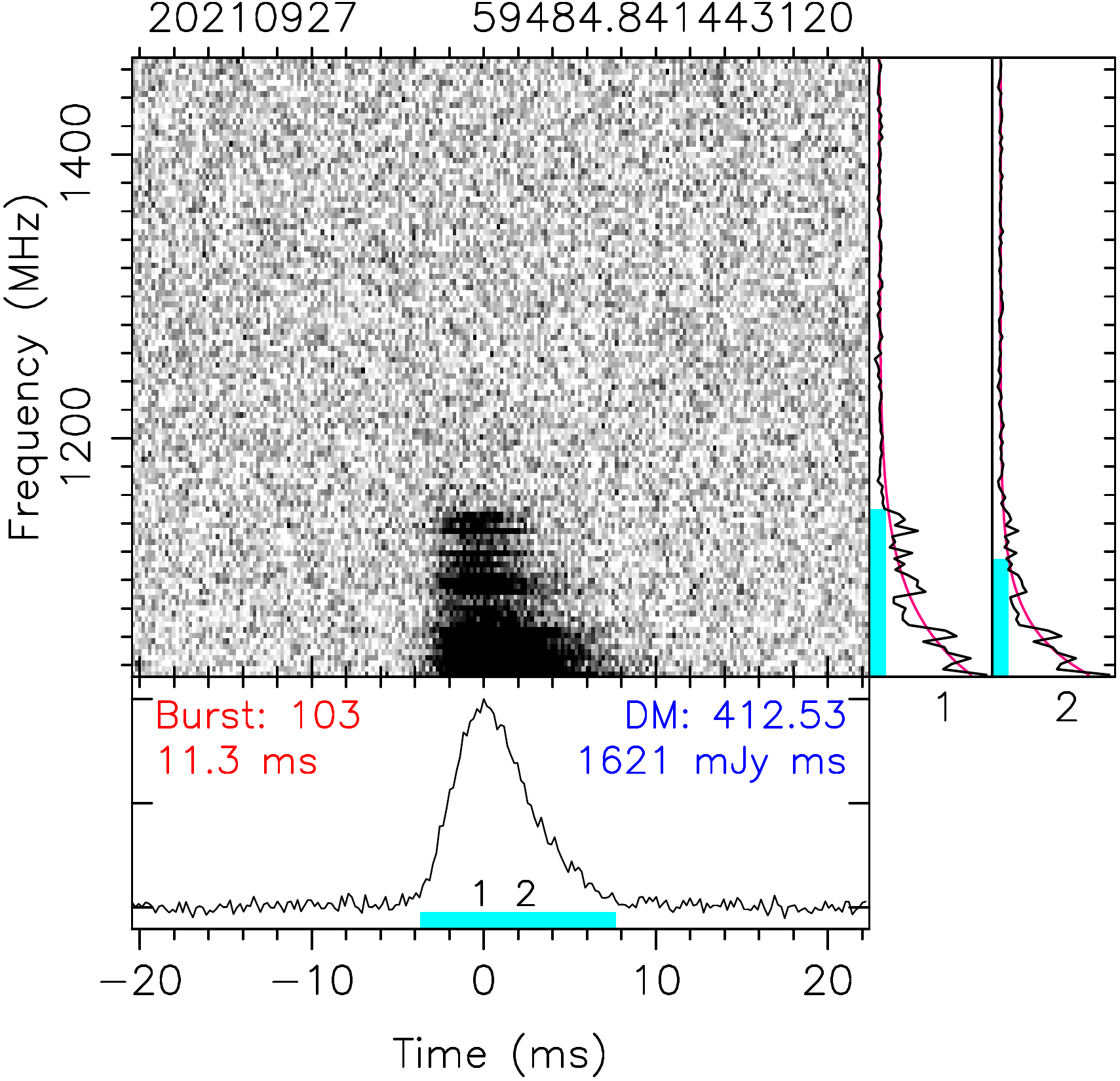}
\caption{The same as Figure~\ref{fig:appendix:D1W} but for bursts in D2-L.
}\label{fig:appendix:D2L}
\end{figure*}
\addtocounter{figure}{-1}
\begin{figure*}
\flushleft
    \includegraphics[height=37mm]{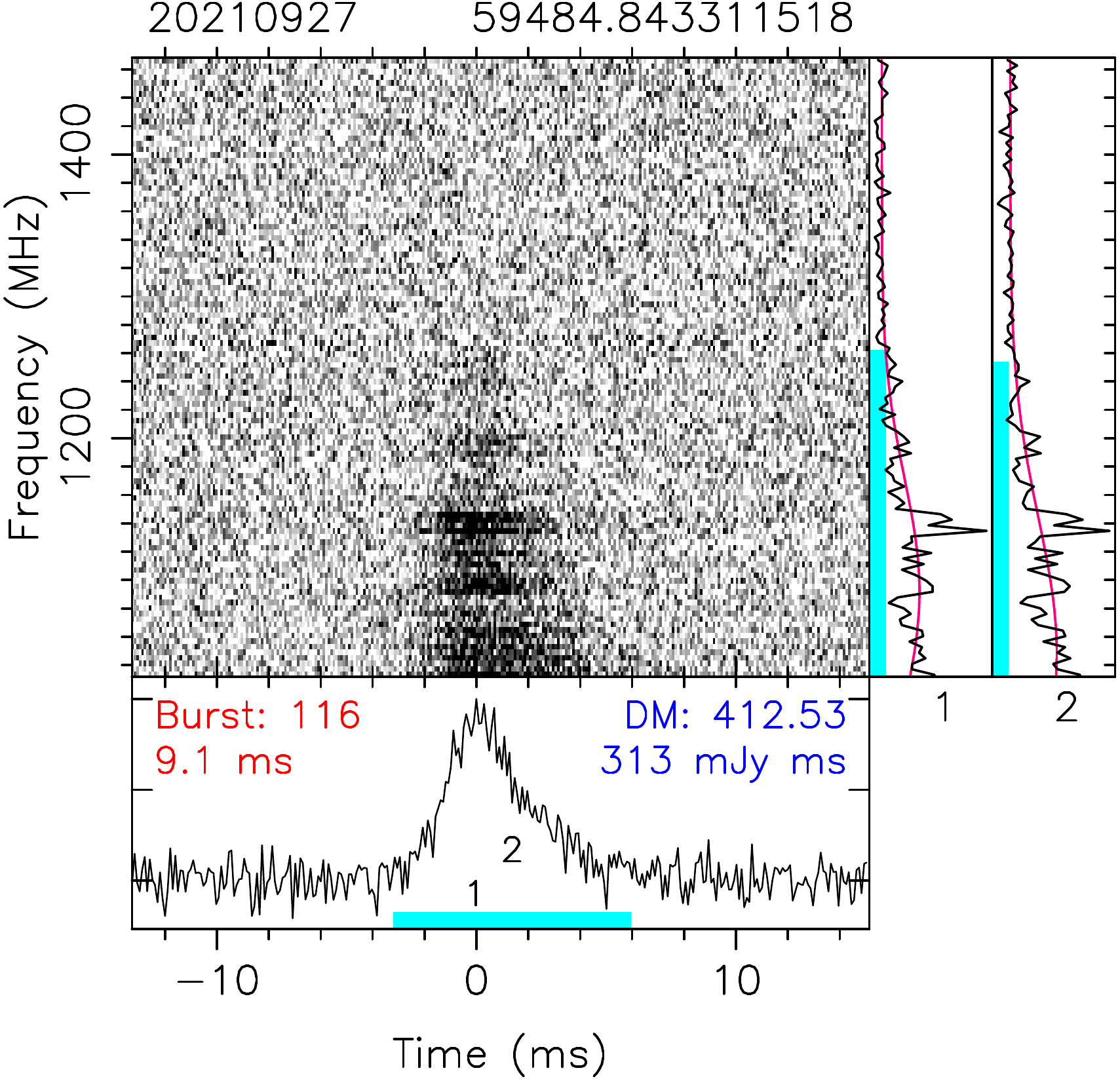}
    \includegraphics[height=37mm]{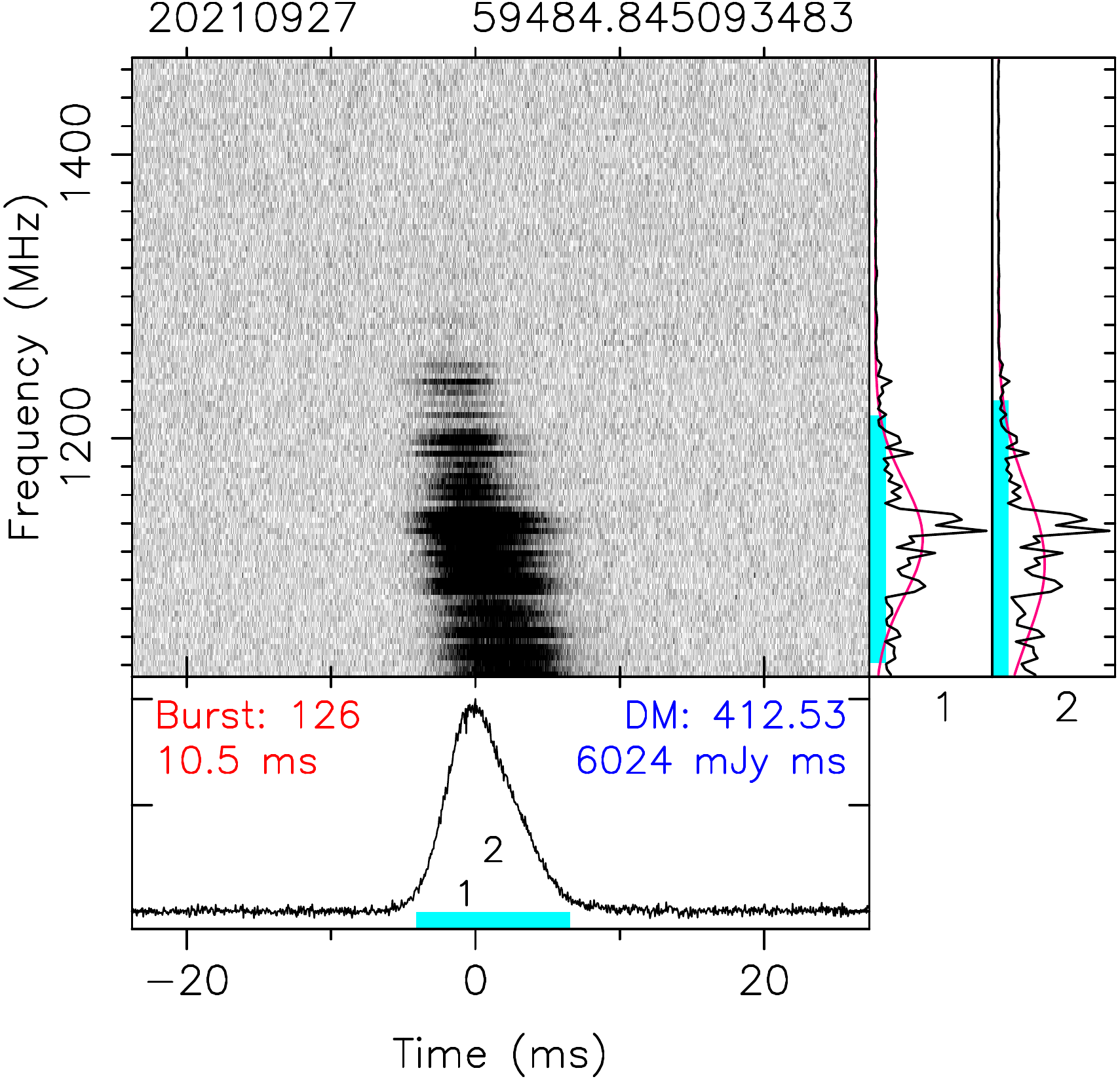}
    \includegraphics[height=37mm]{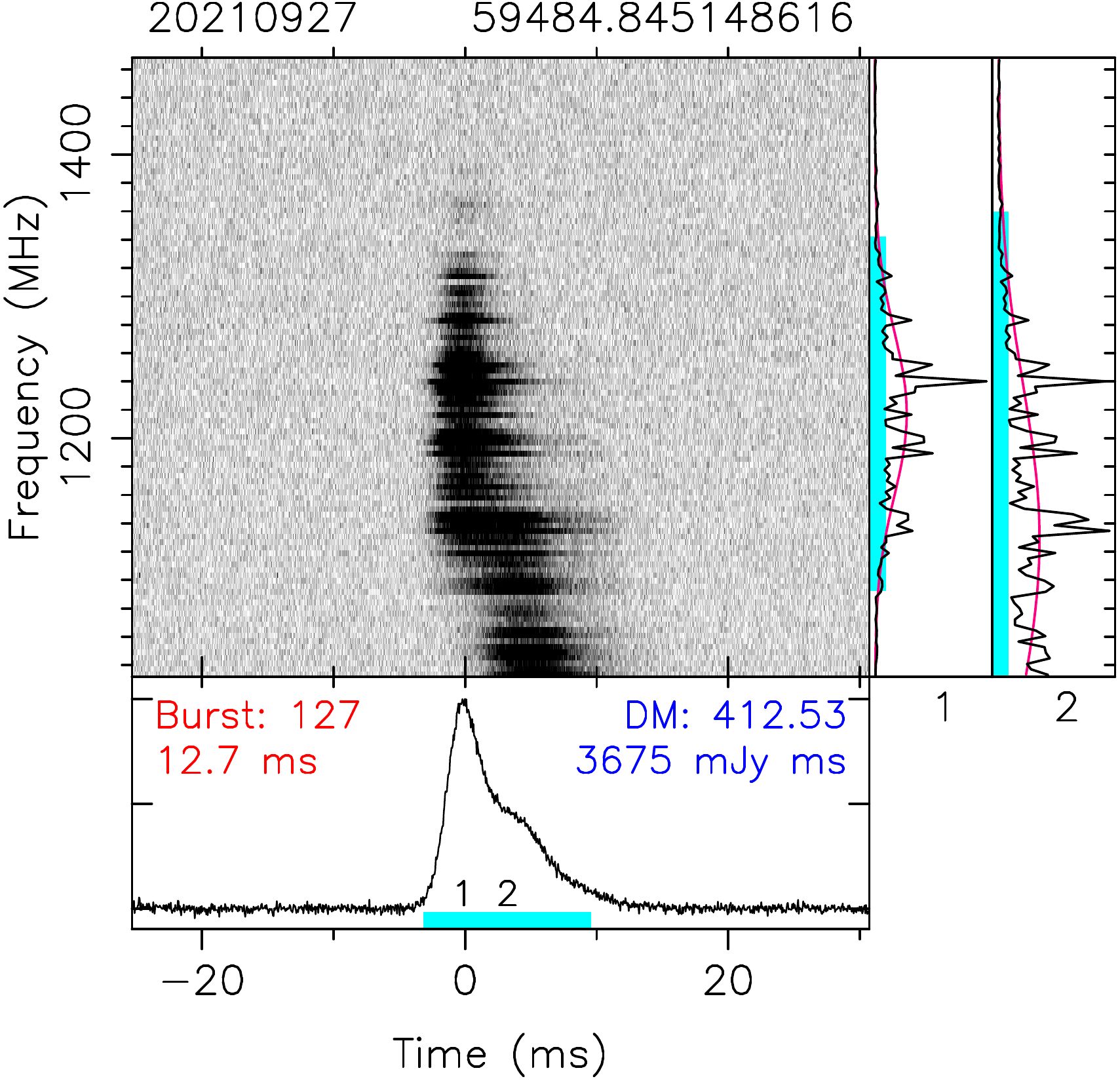}
    \includegraphics[height=37mm]{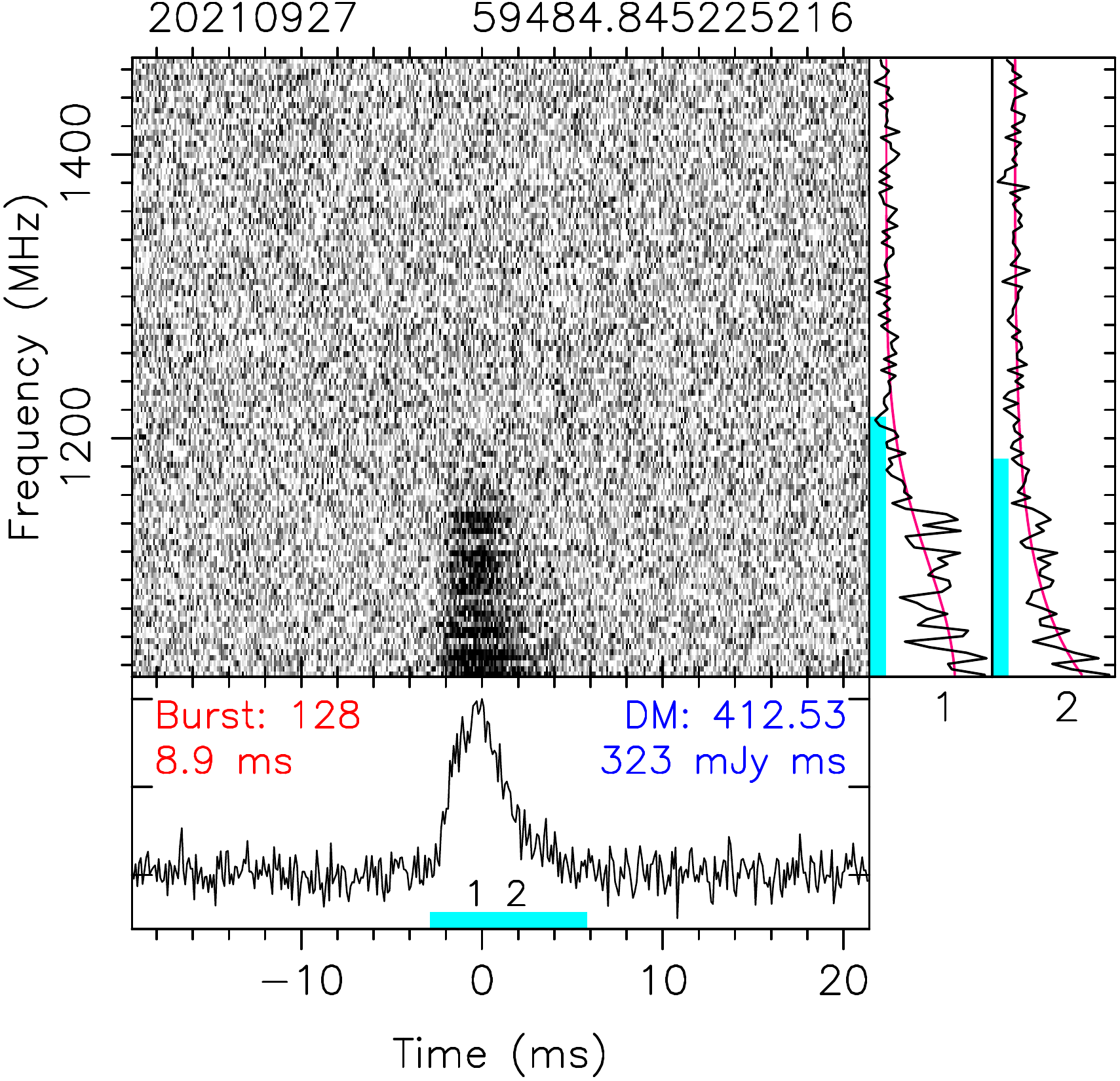}
    \includegraphics[height=37mm]{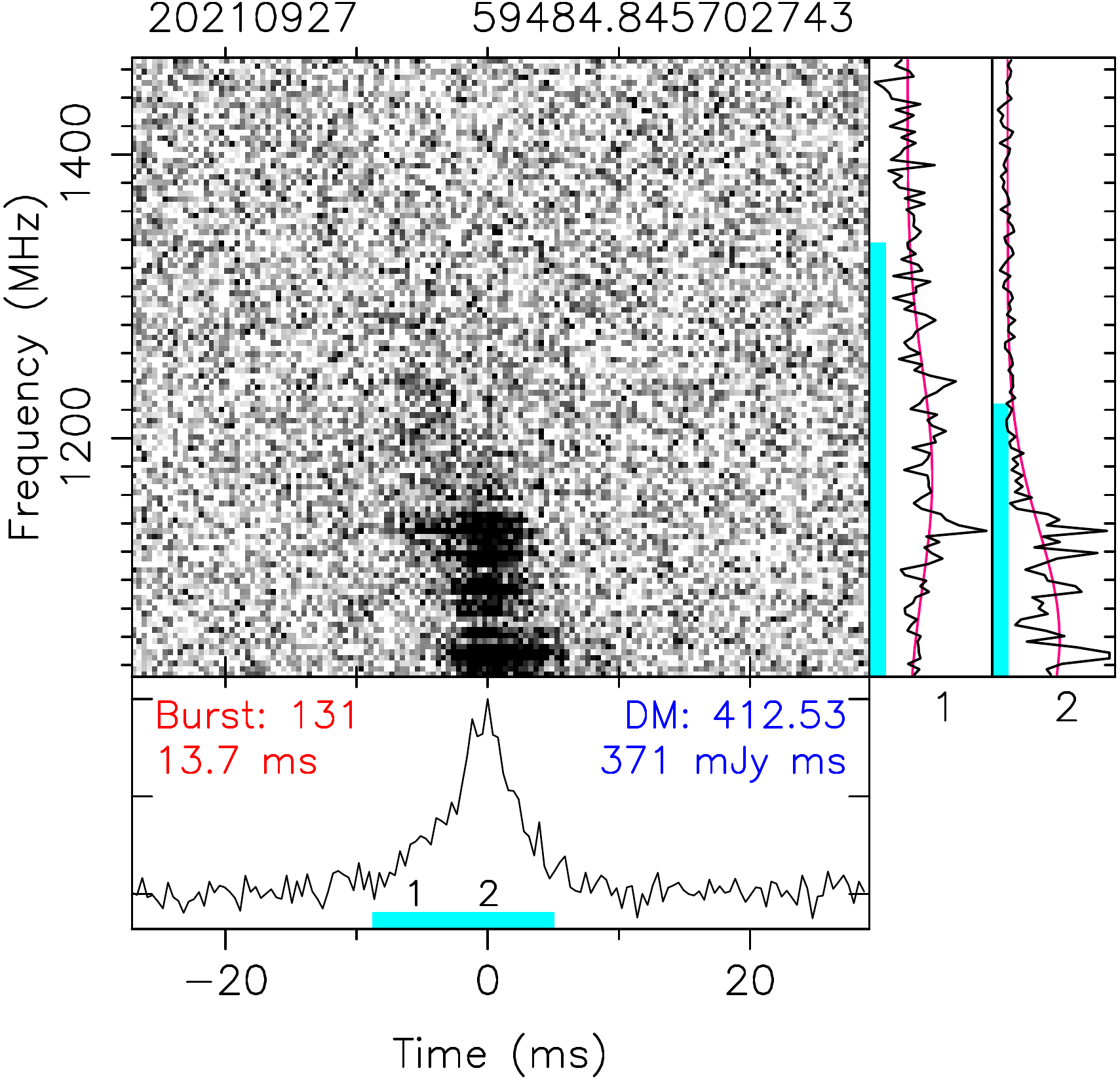}
    \includegraphics[height=37mm]{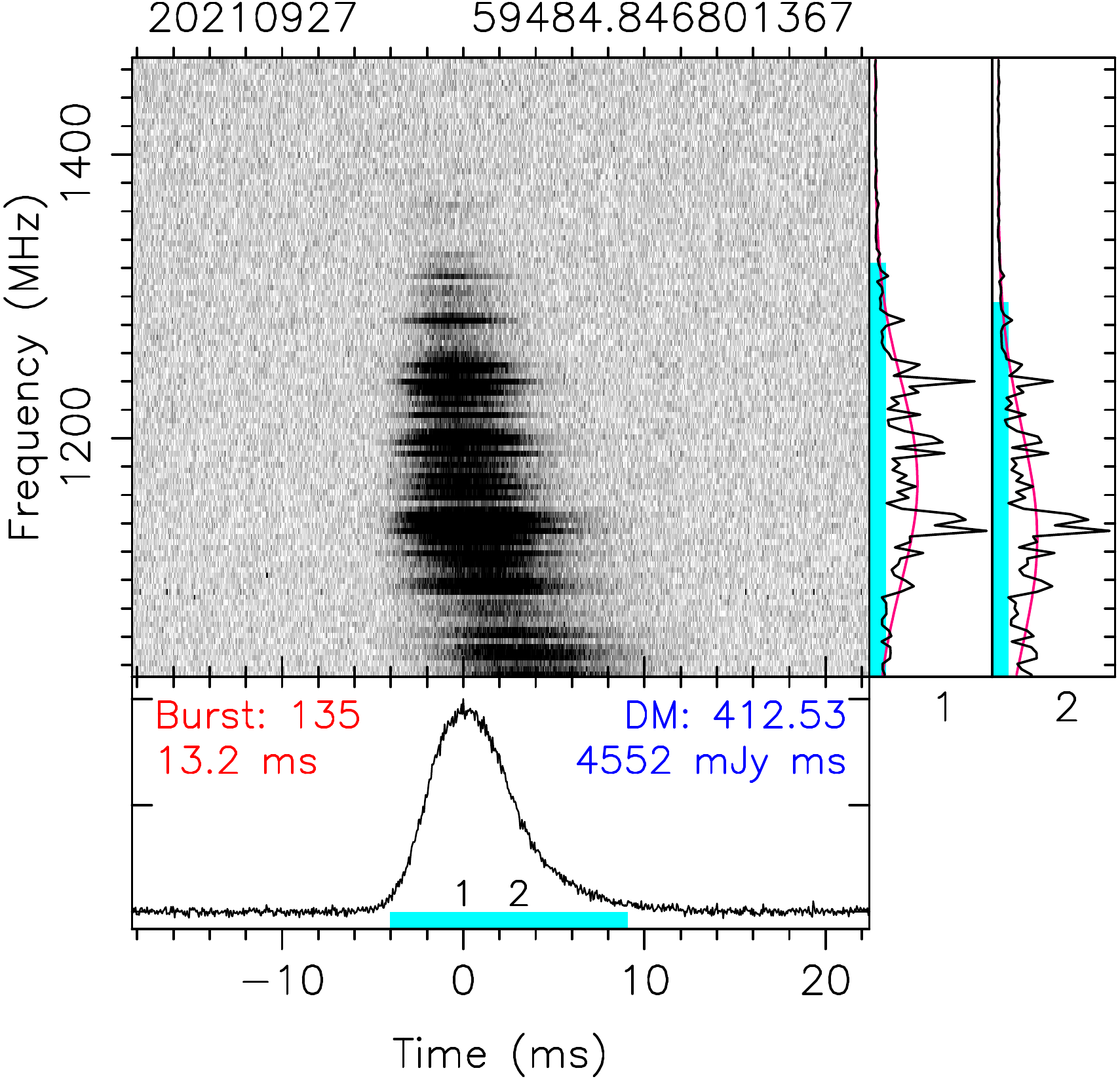}
    \includegraphics[height=37mm]{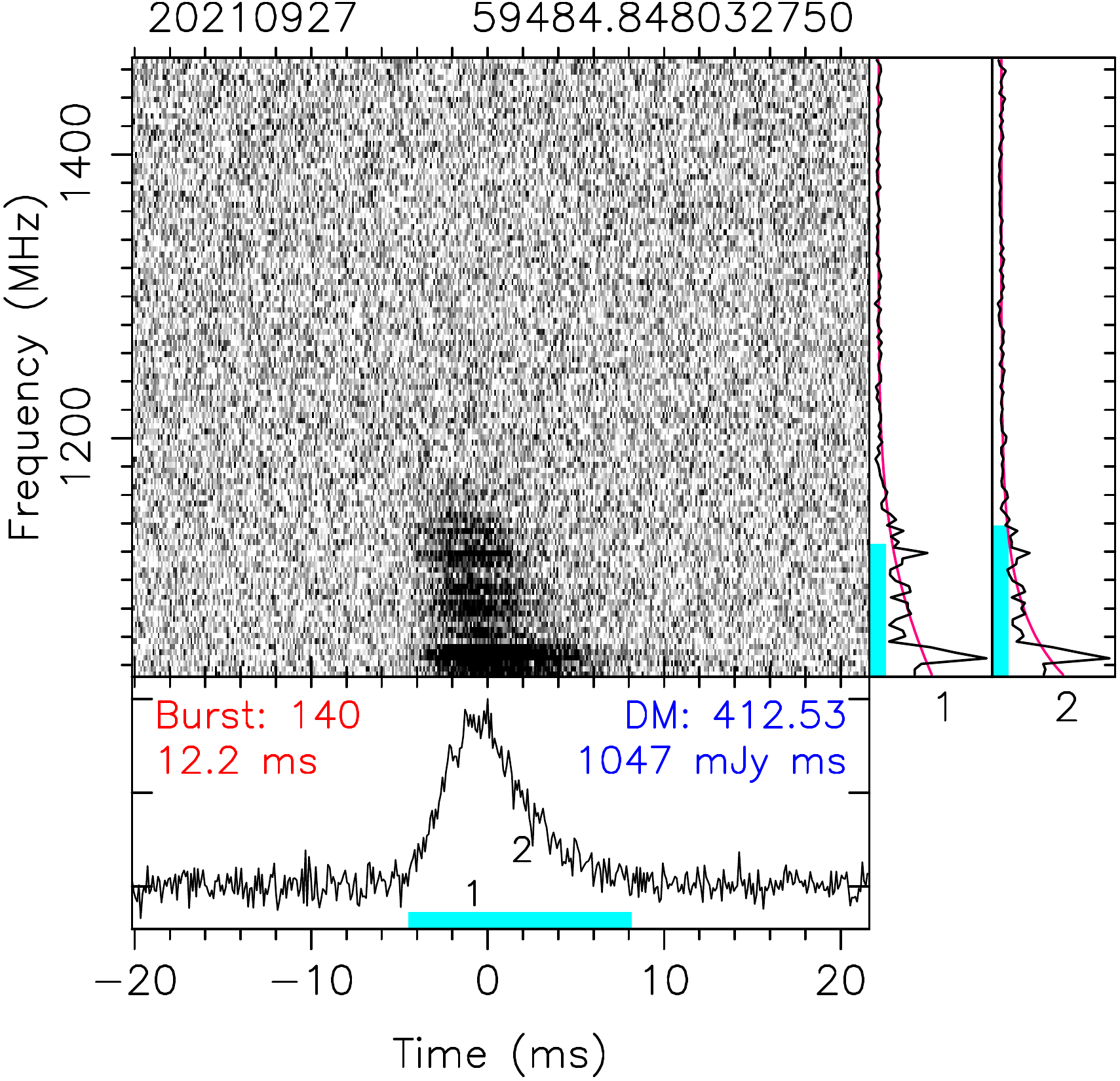}
    \includegraphics[height=37mm]{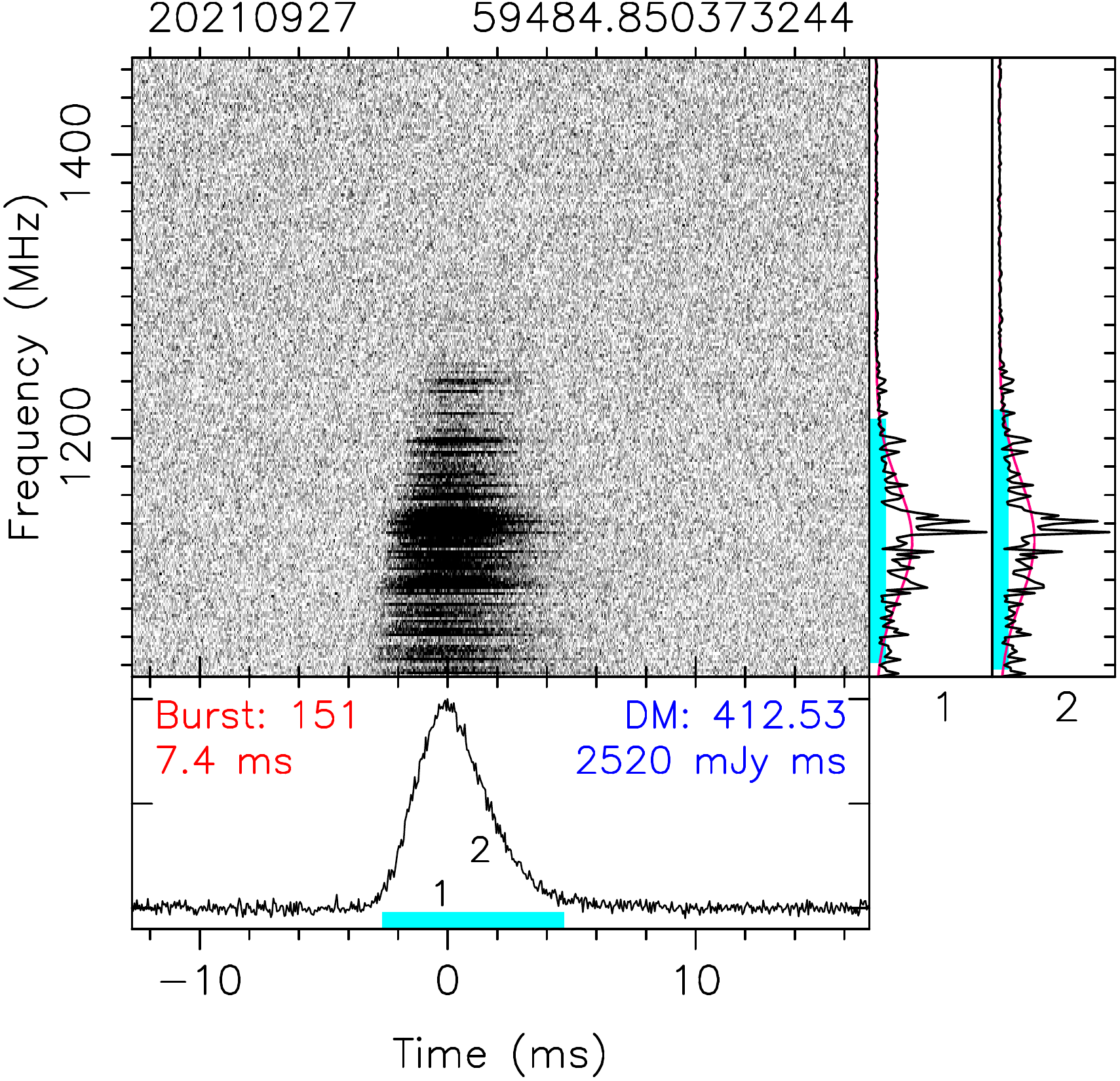}
    \includegraphics[height=37mm]{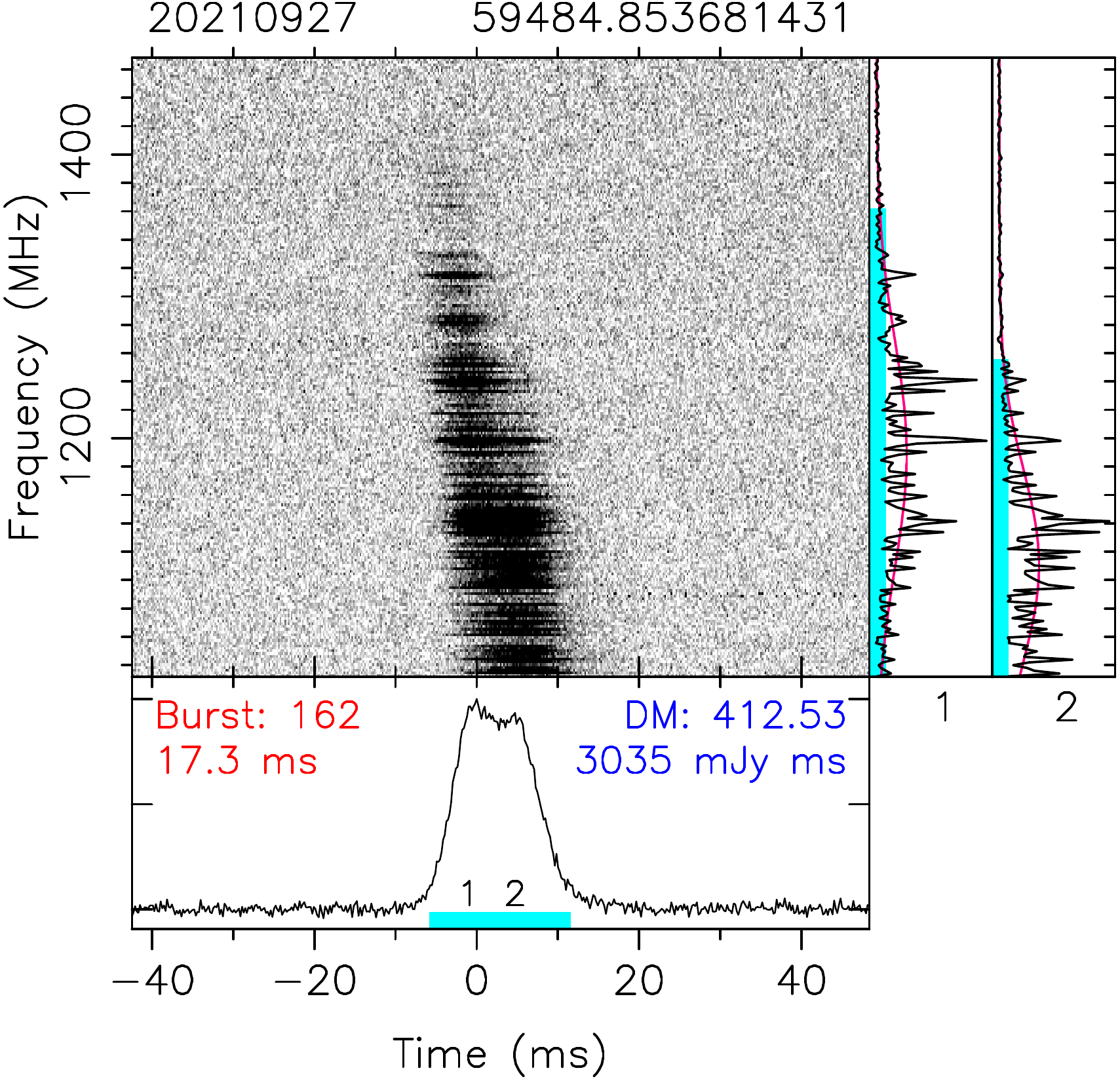}
    \includegraphics[height=37mm]{20210928/FRB20201124A_20210928_tracking-M01-P1-c512b1.fits-169-T-3535.657-3535.698-DM-412.5.pdf}
    \includegraphics[height=37mm]{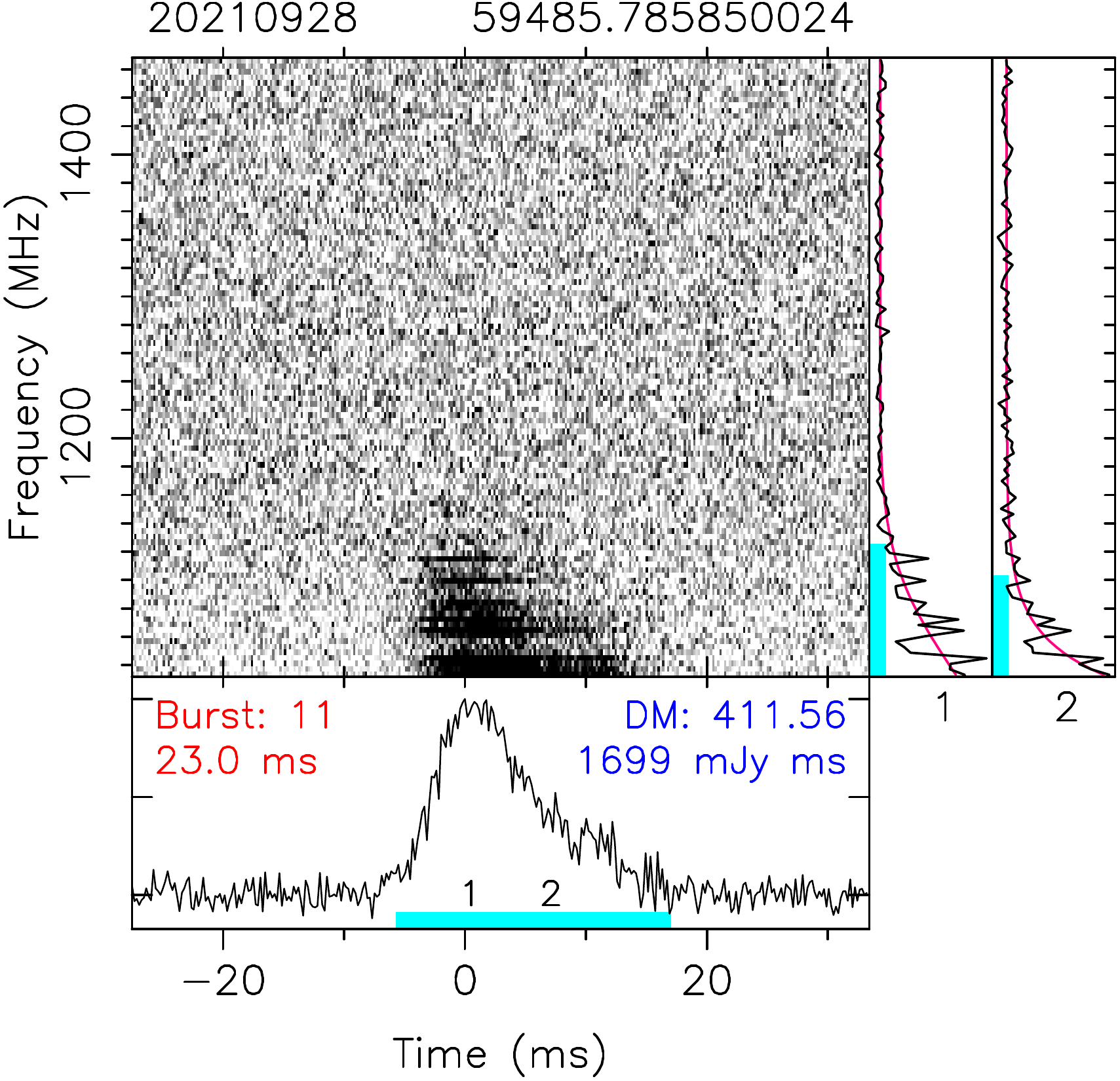}
    \includegraphics[height=37mm]{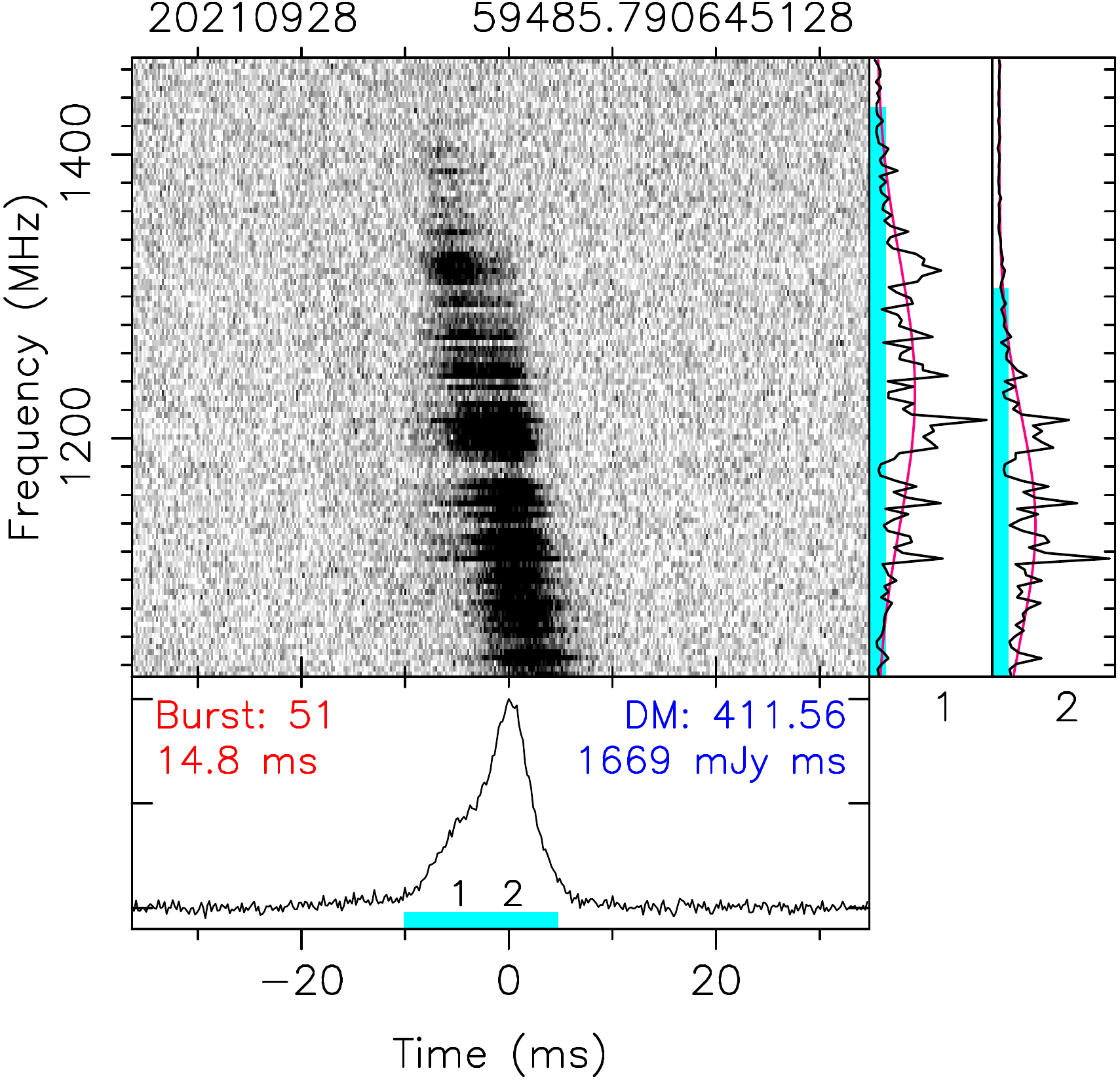}
    \includegraphics[height=37mm]{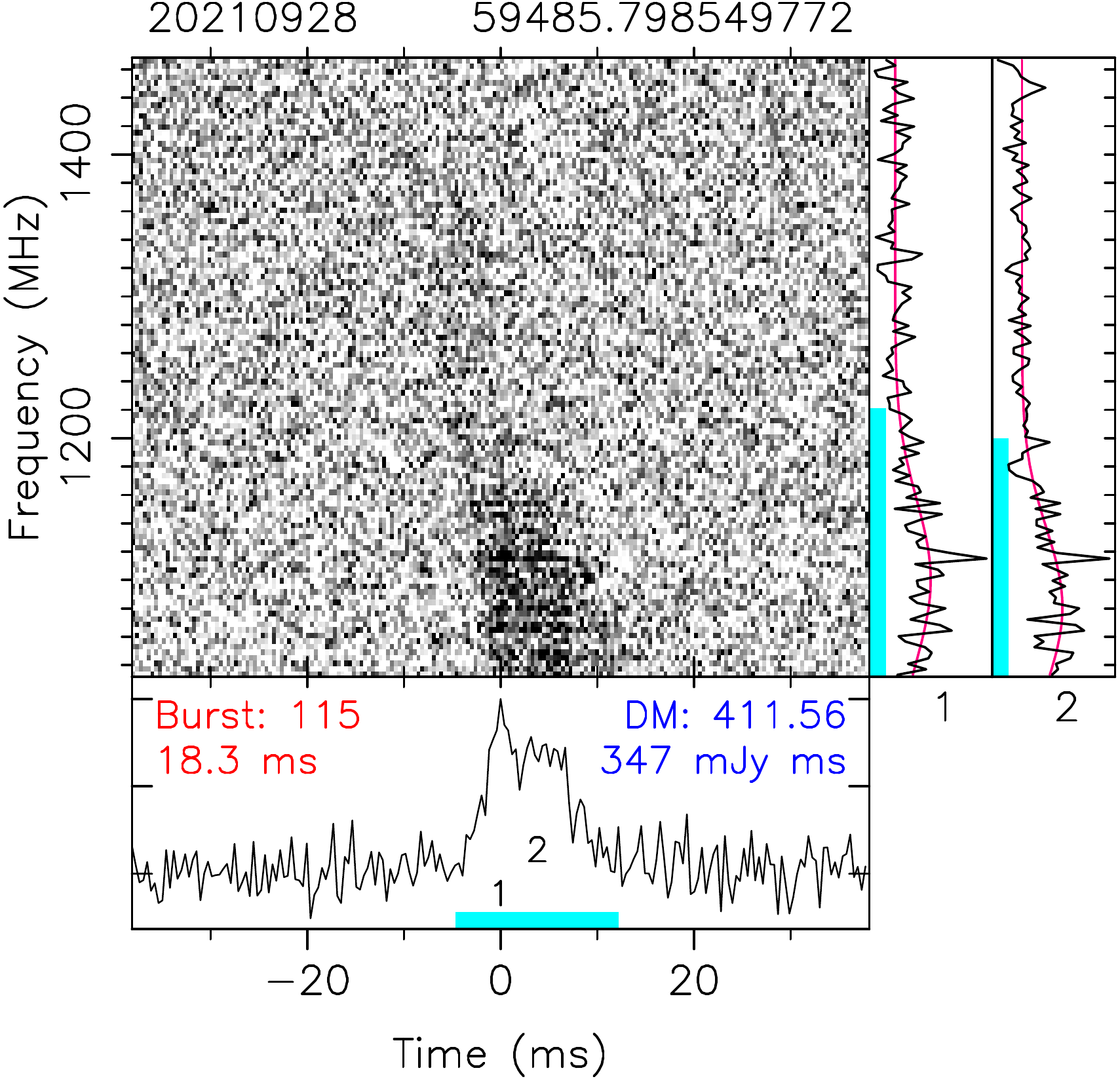}
    \includegraphics[height=37mm]{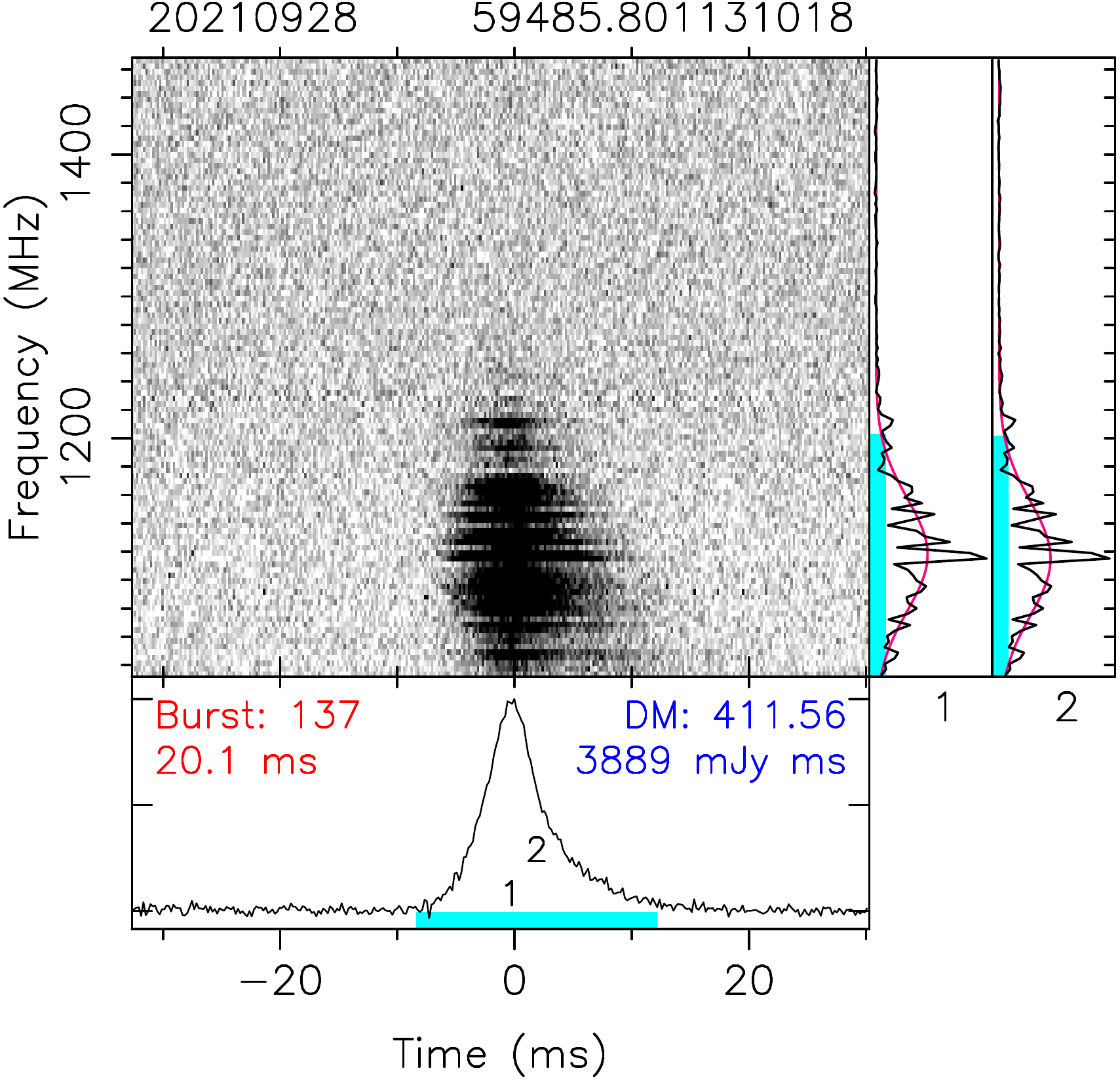}
    \includegraphics[height=37mm]{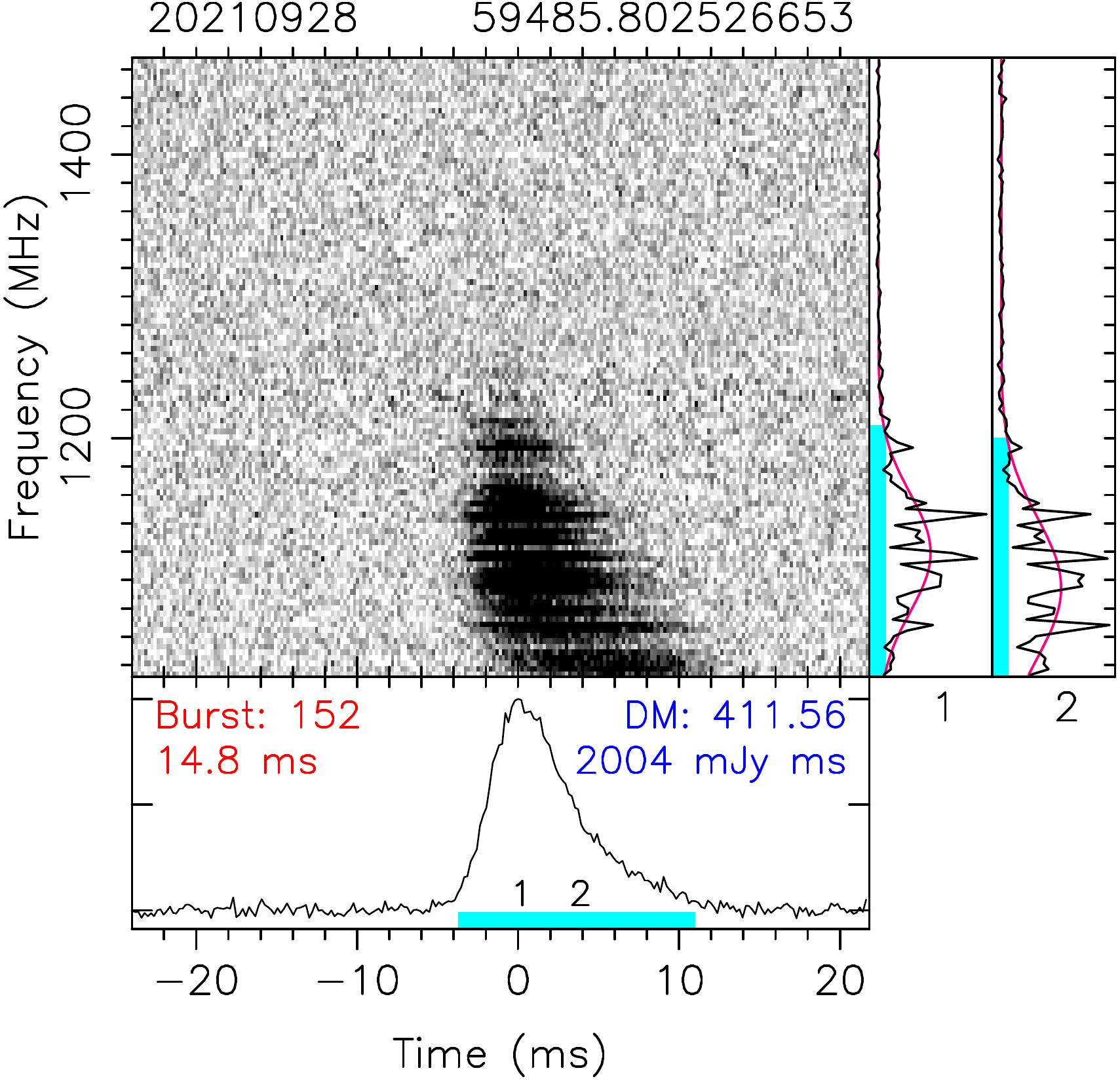}
    \includegraphics[height=37mm]{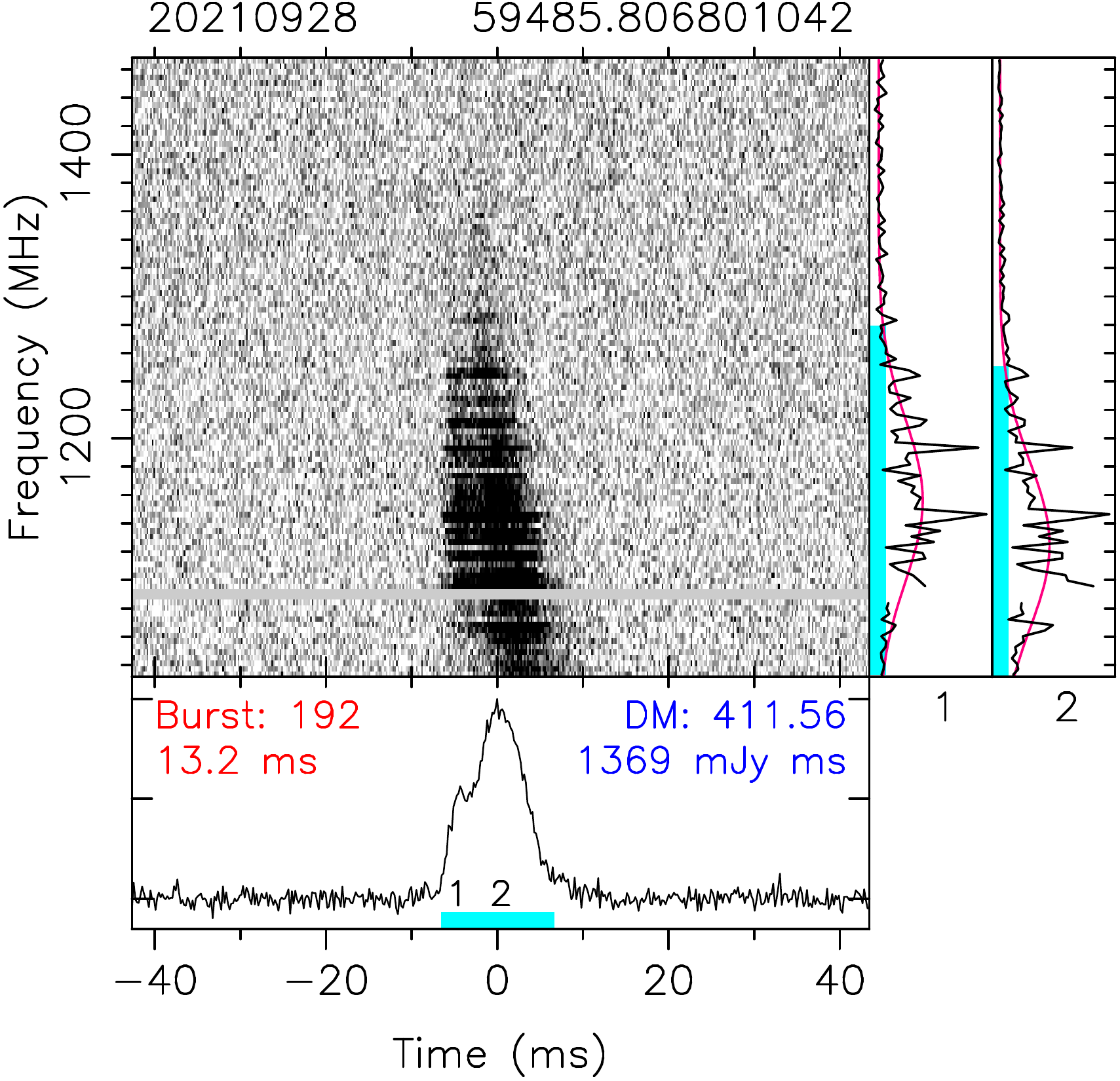}
    \includegraphics[height=37mm]{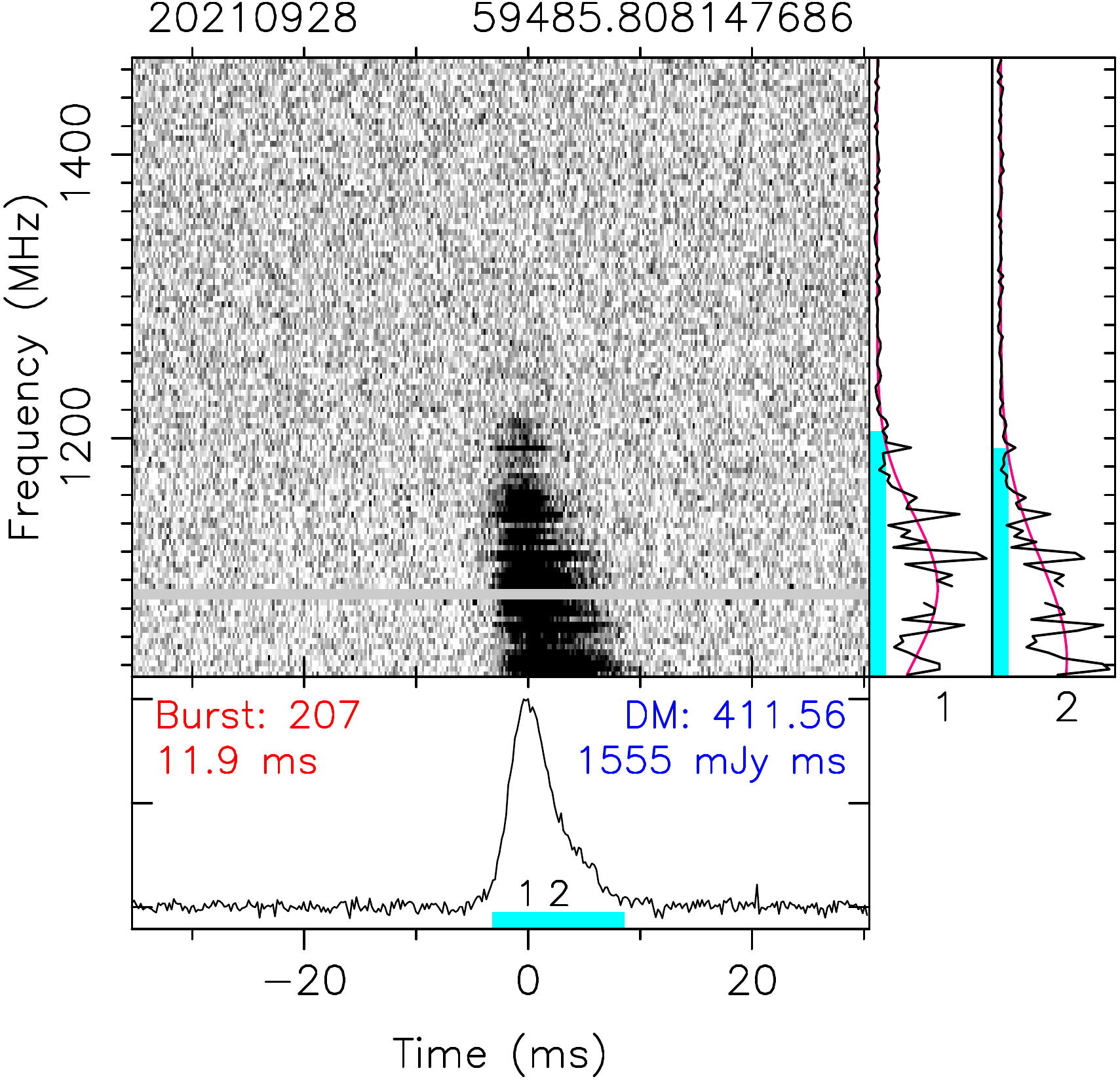}
    \includegraphics[height=37mm]{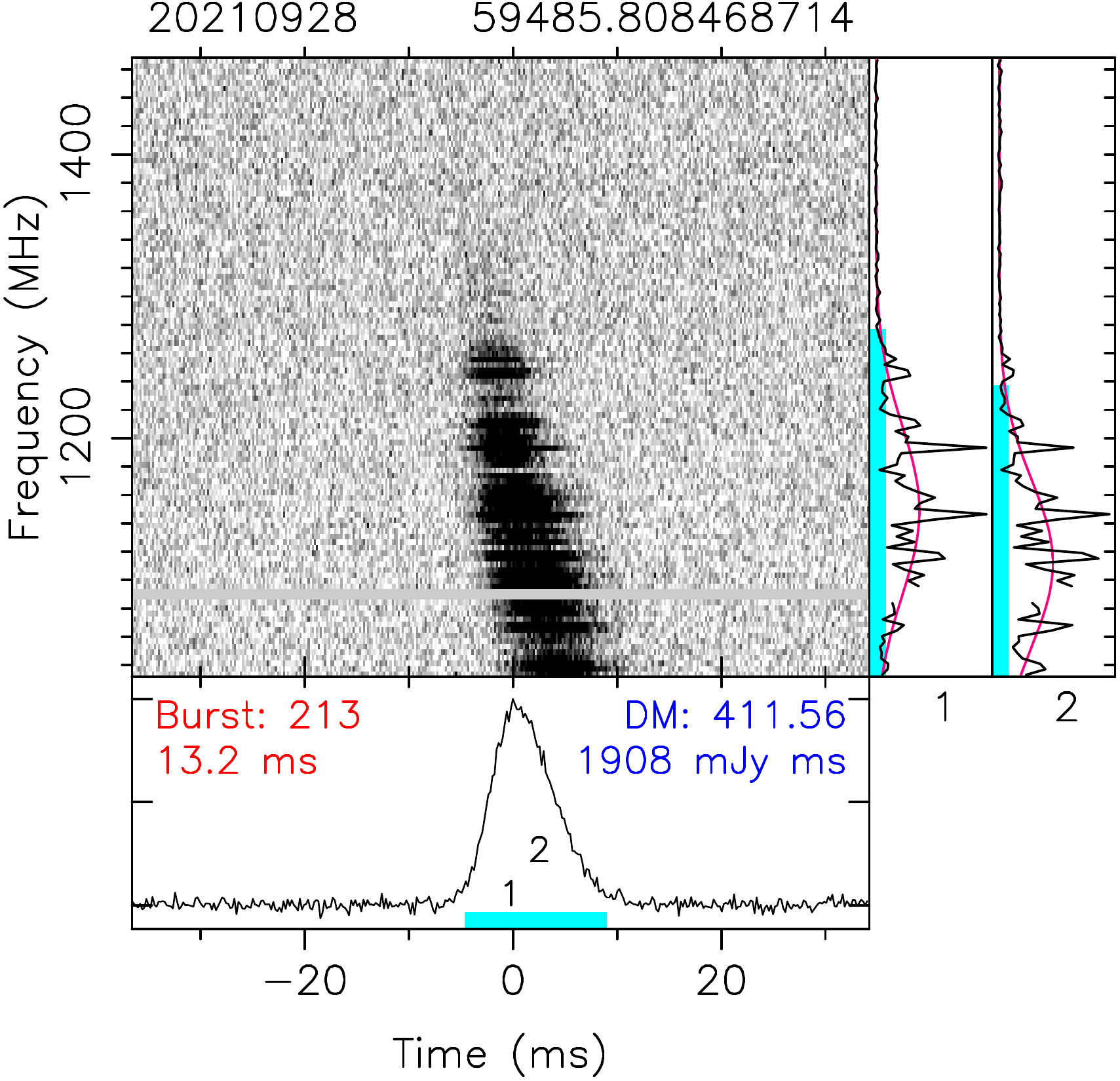}
    \includegraphics[height=37mm]{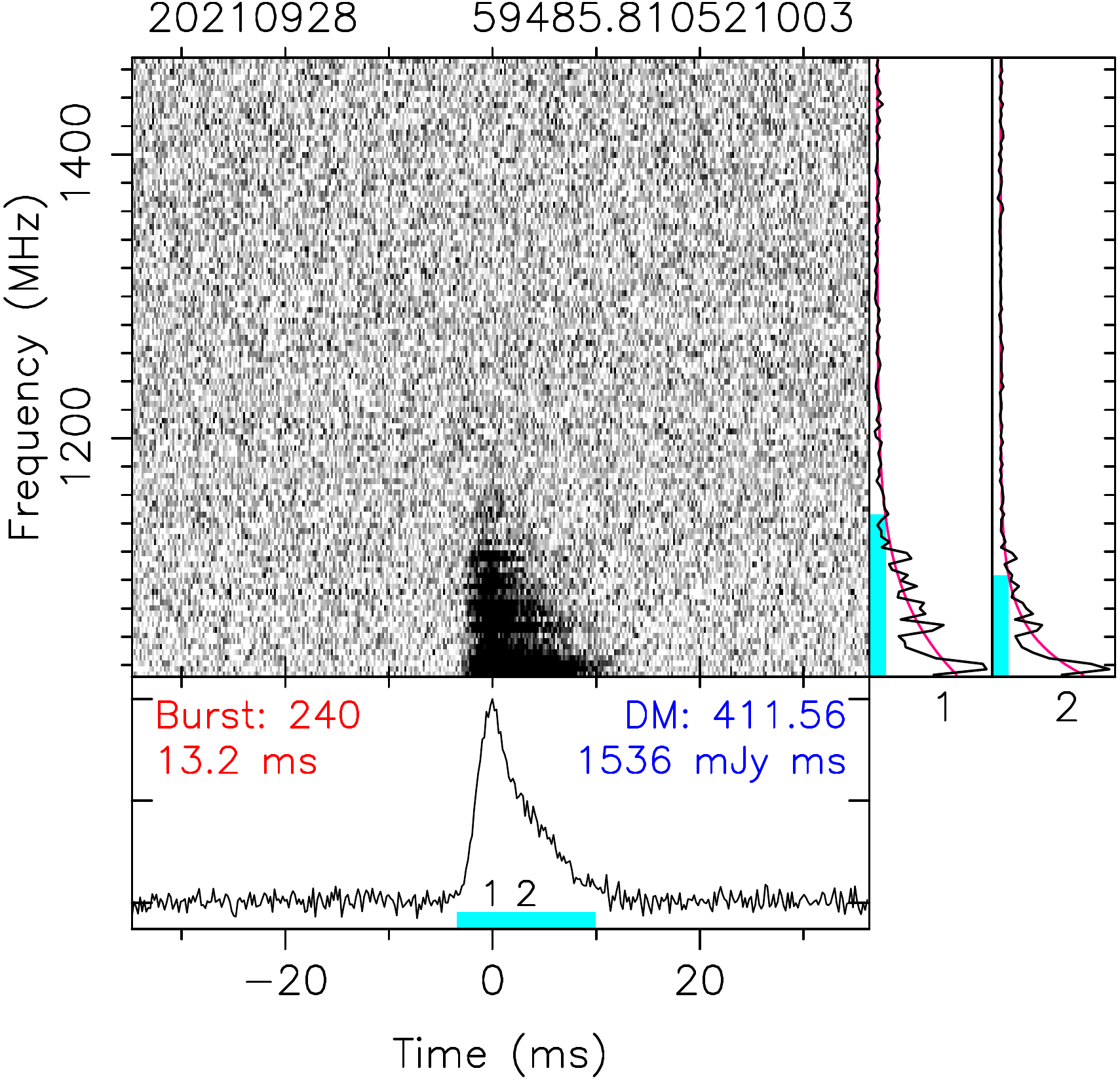}
    \includegraphics[height=37mm]{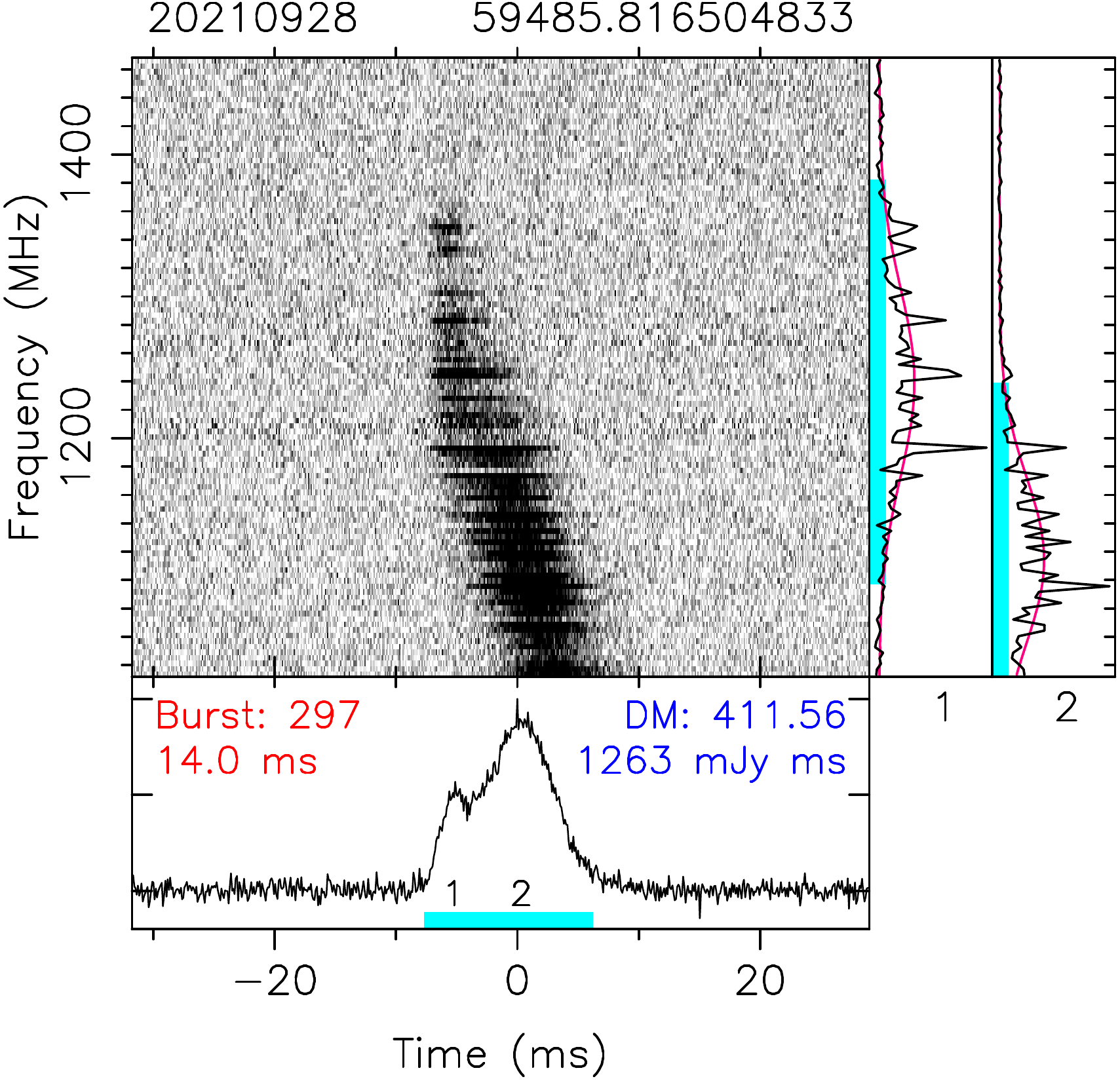}
    \includegraphics[height=37mm]{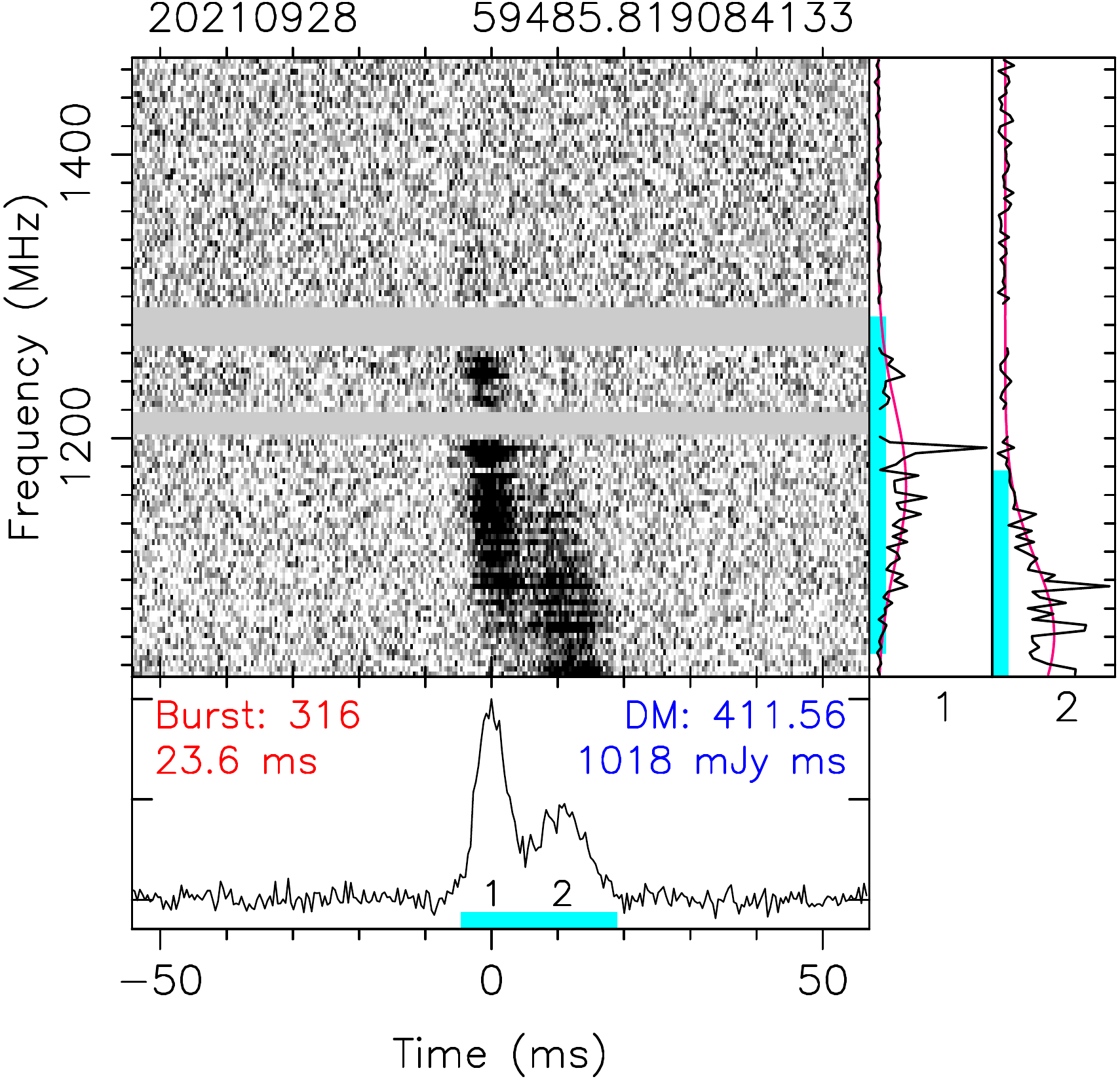}
    \includegraphics[height=37mm]{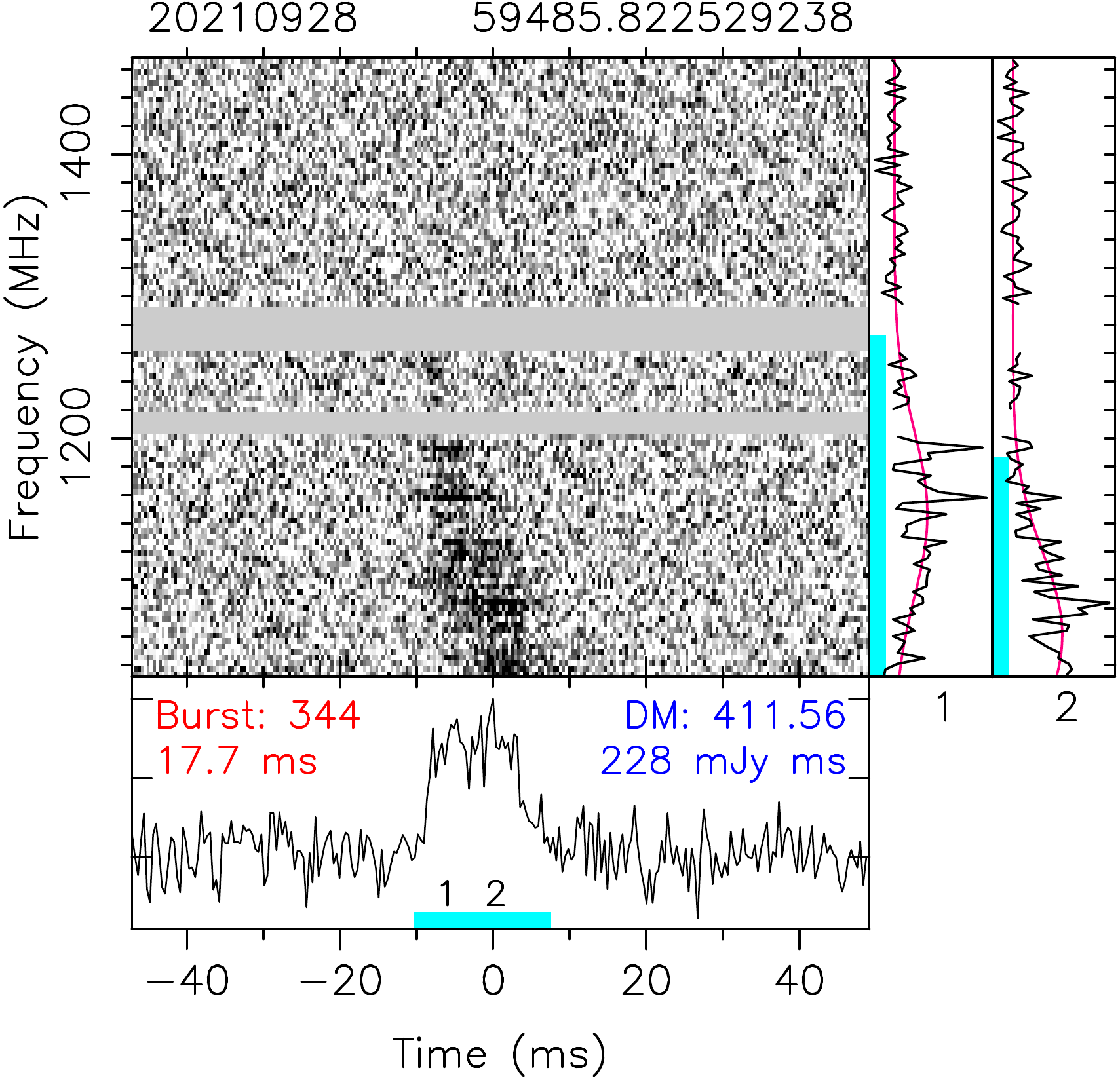}
    \includegraphics[height=37mm]{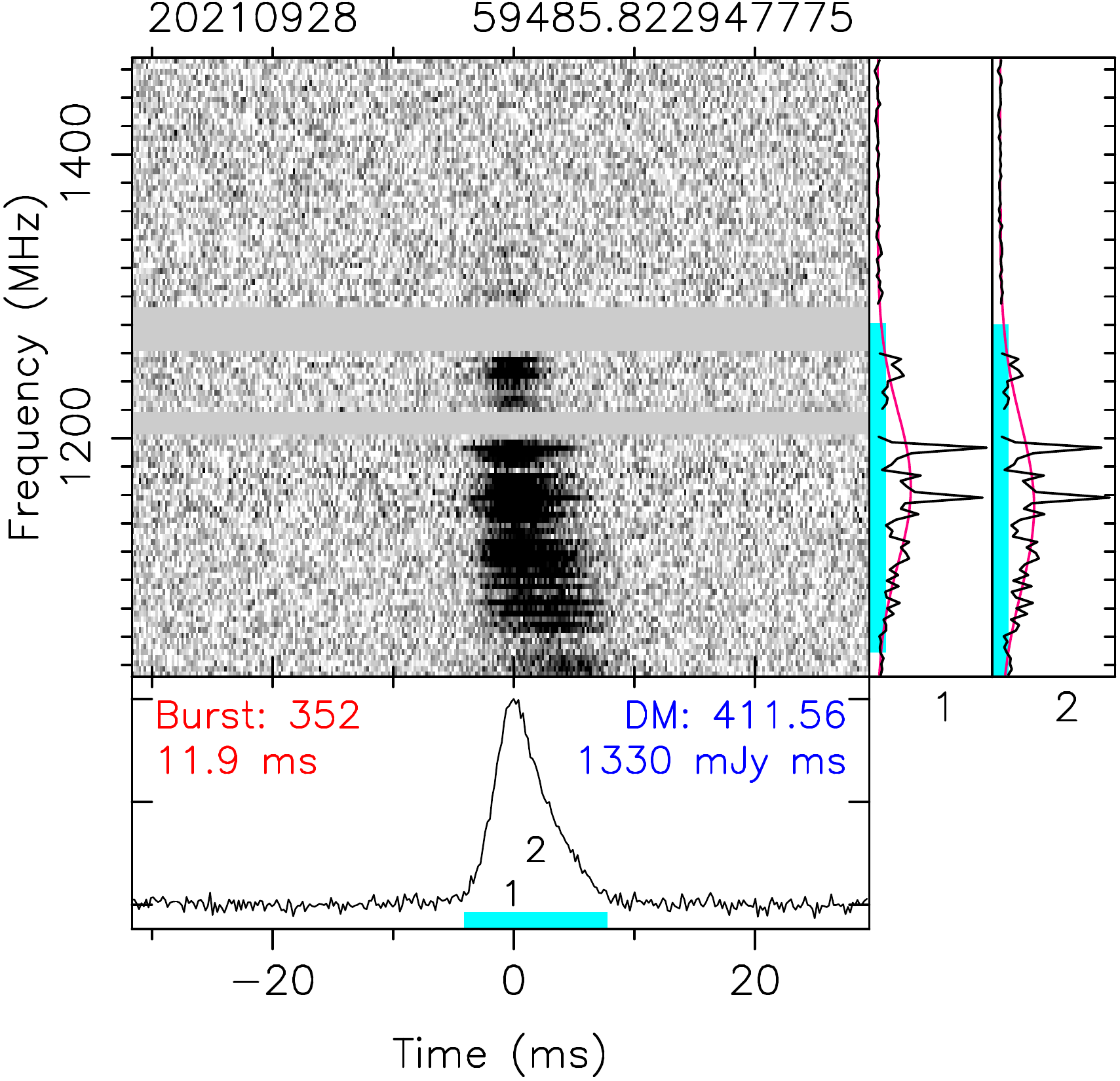}
\includegraphics[height=37mm]{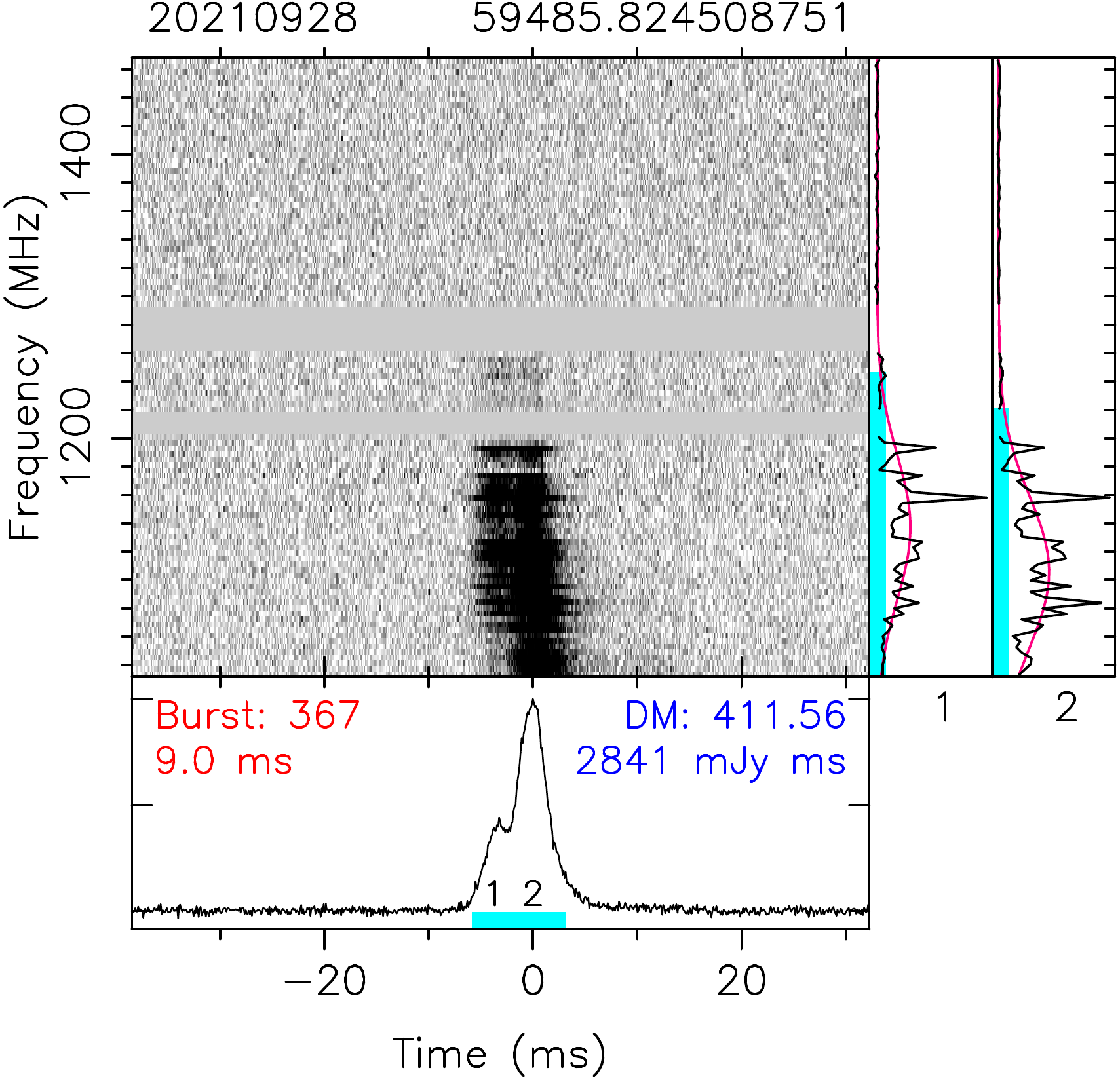}
\caption{\it{ -- continued and ended}.
    }
\end{figure*}

\begin{figure*}
    \flushleft
    \includegraphics[height=37mm]{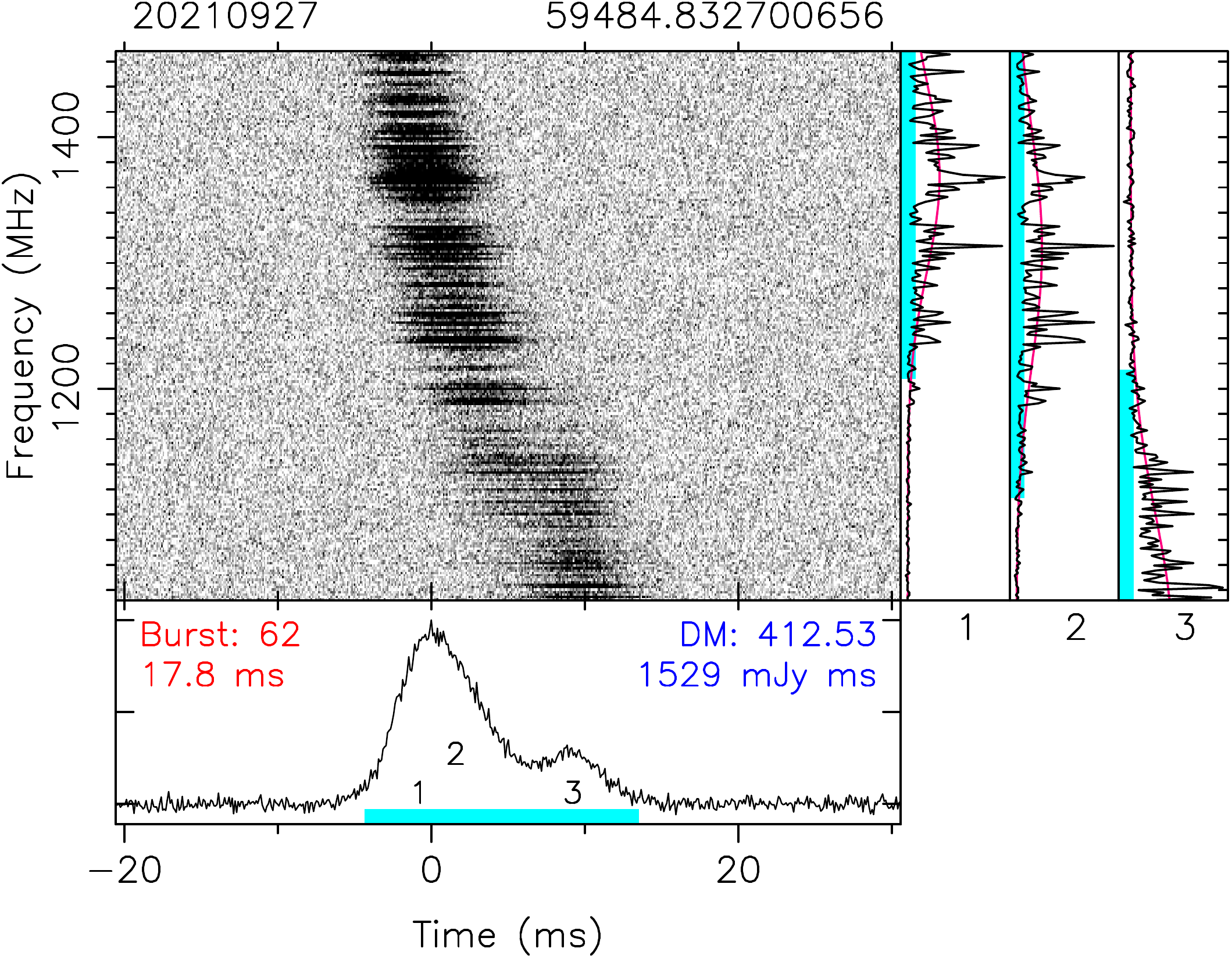}
    \includegraphics[height=37mm]{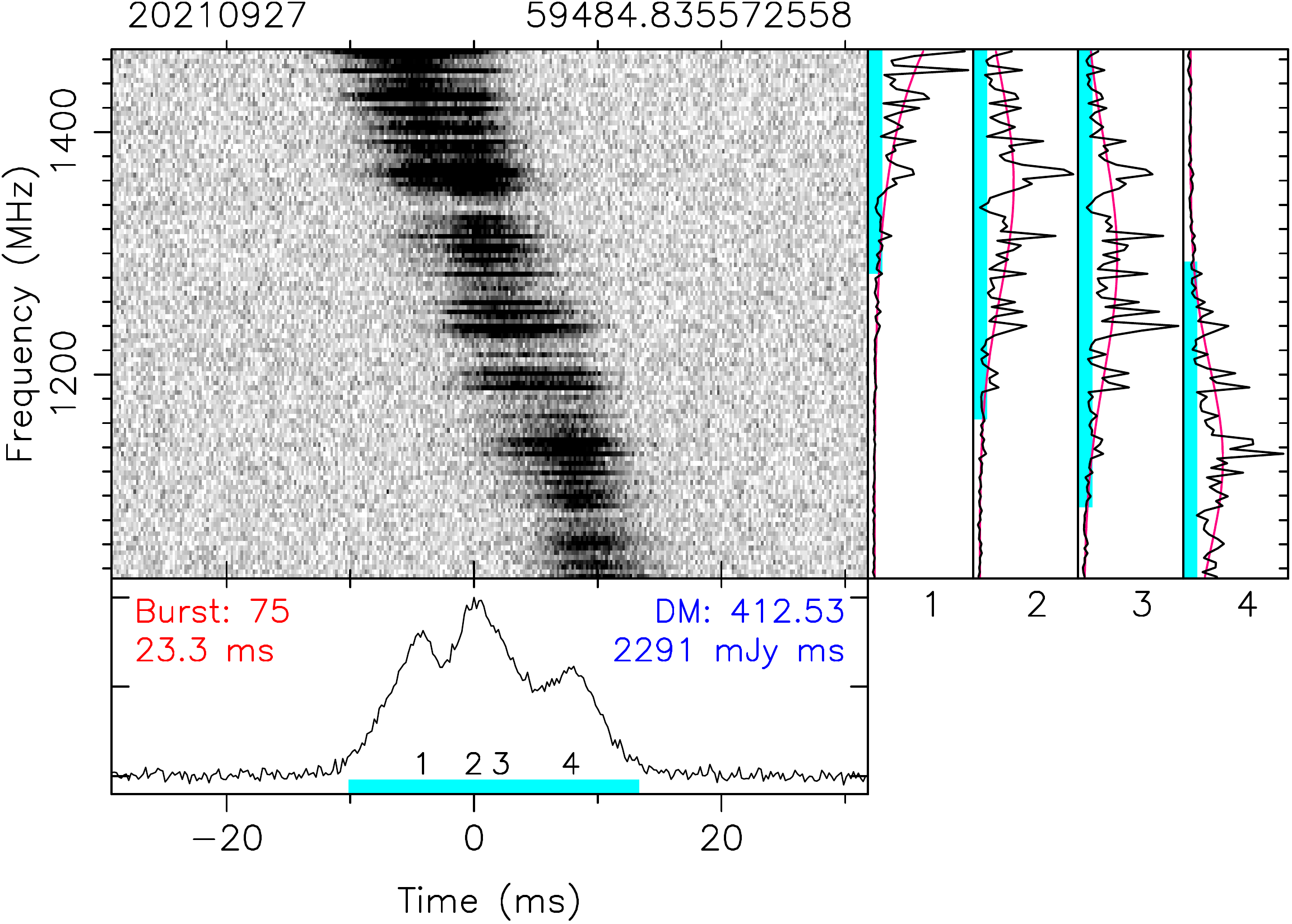}
    \includegraphics[height=37mm]{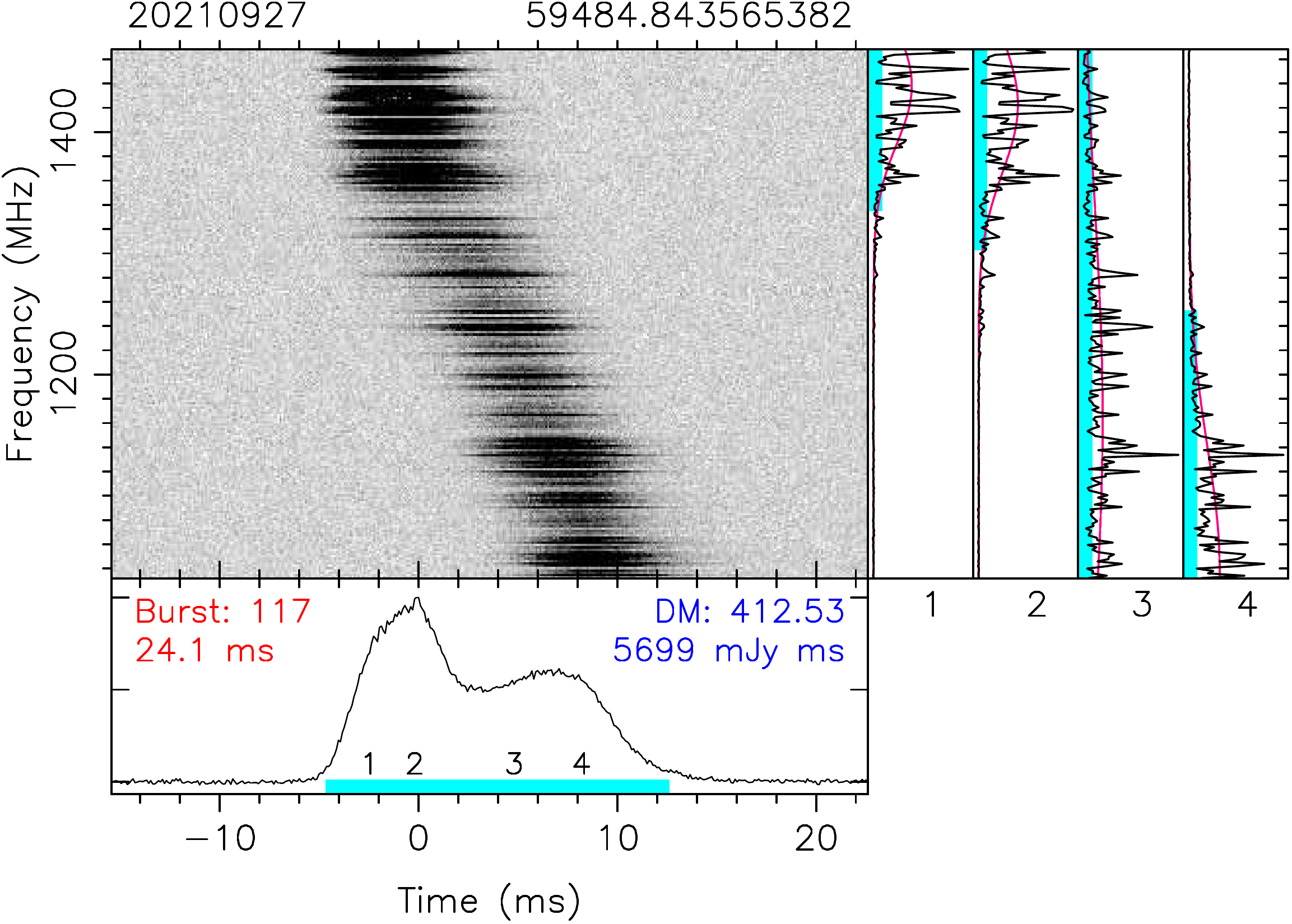}
    \includegraphics[height=37mm]{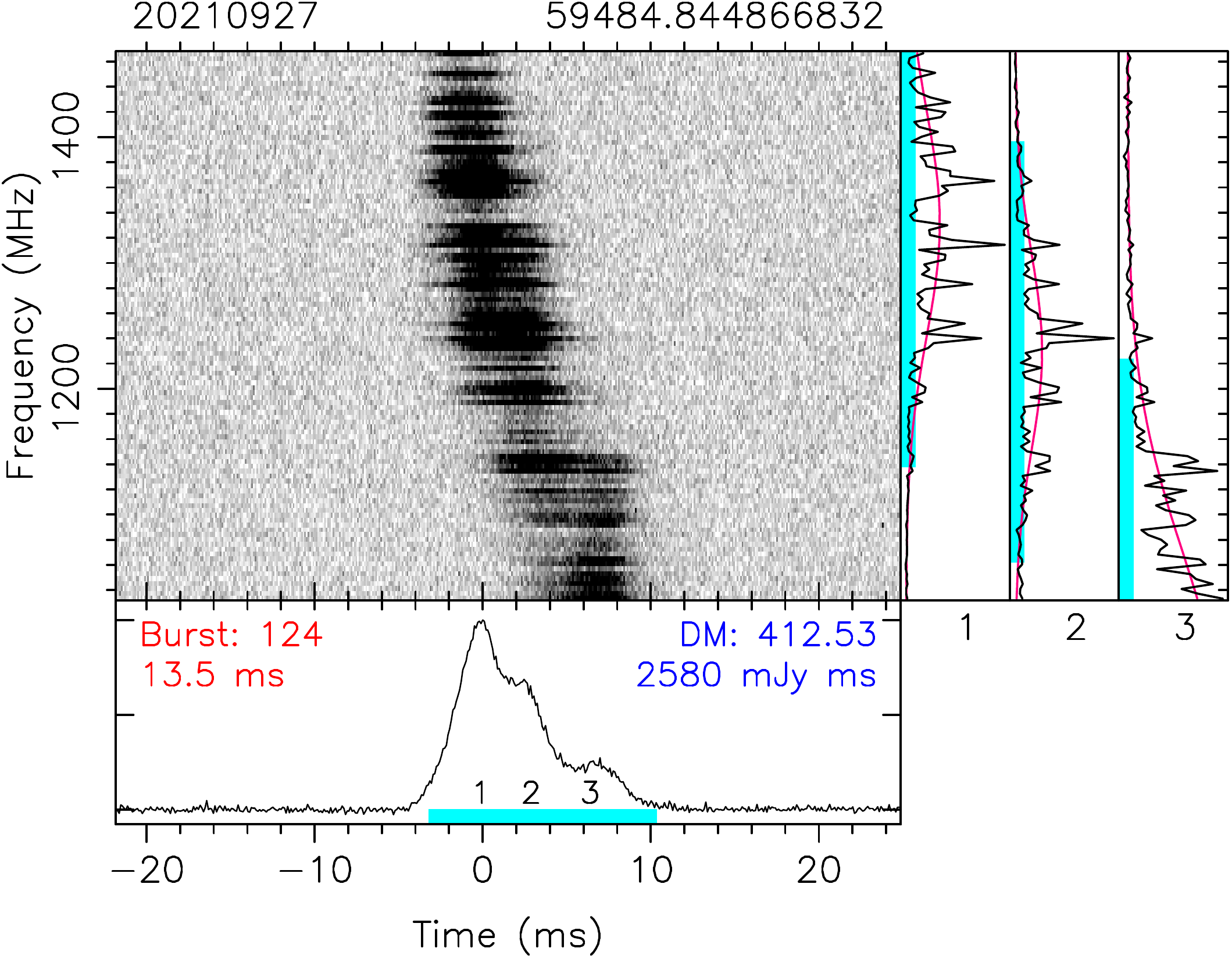}
    \includegraphics[height=37mm]{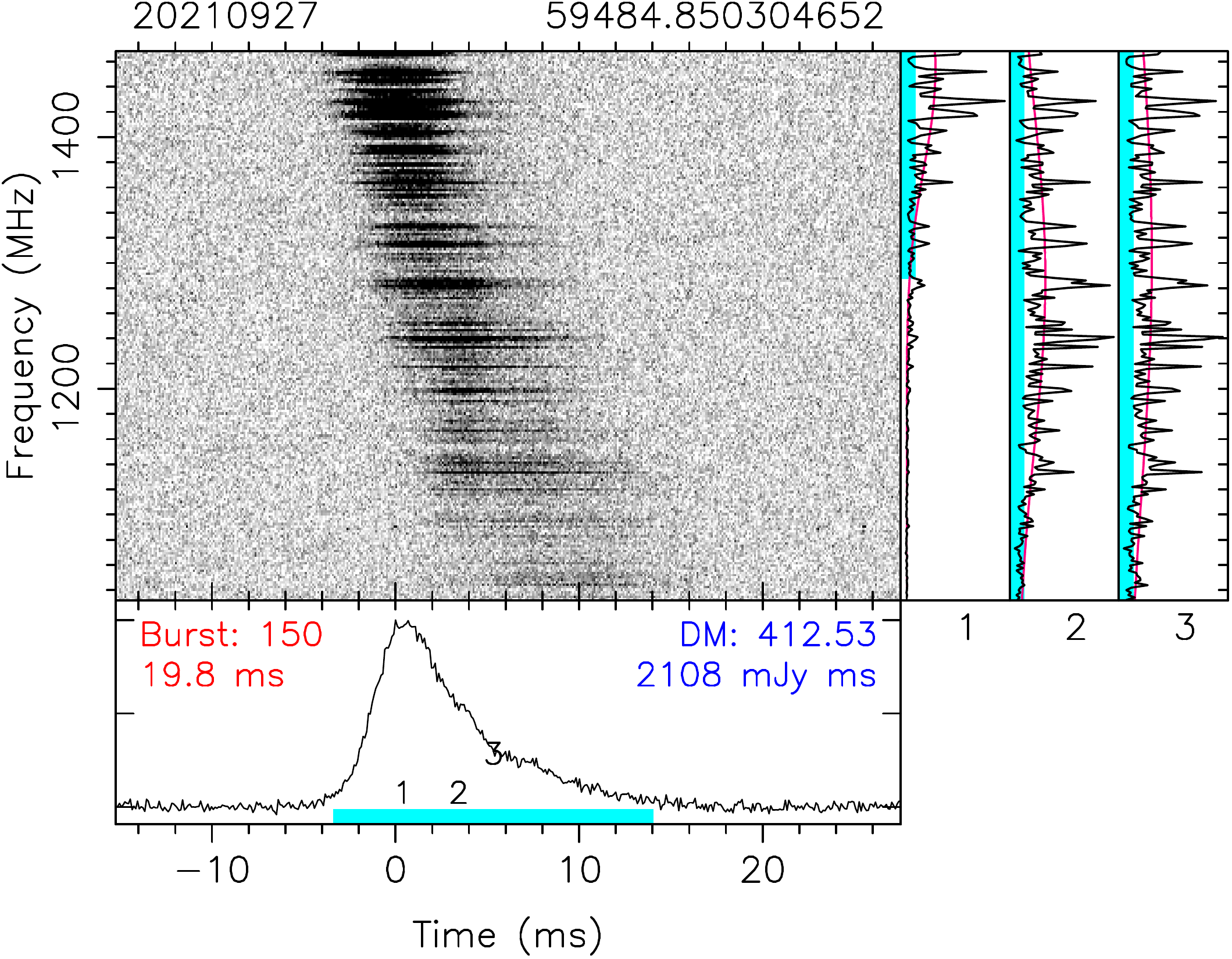}
    \includegraphics[height=37mm]{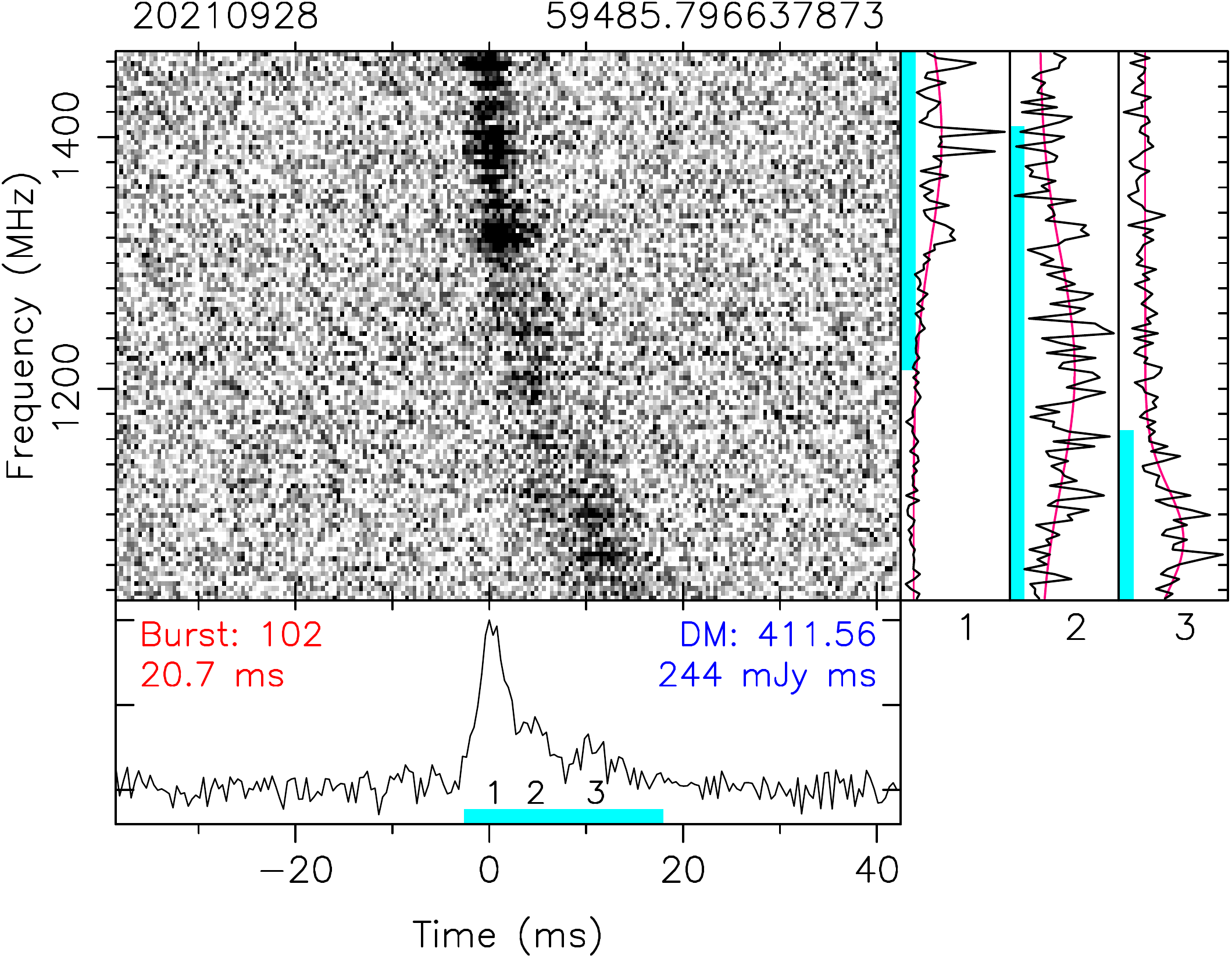}
    \includegraphics[height=37mm]{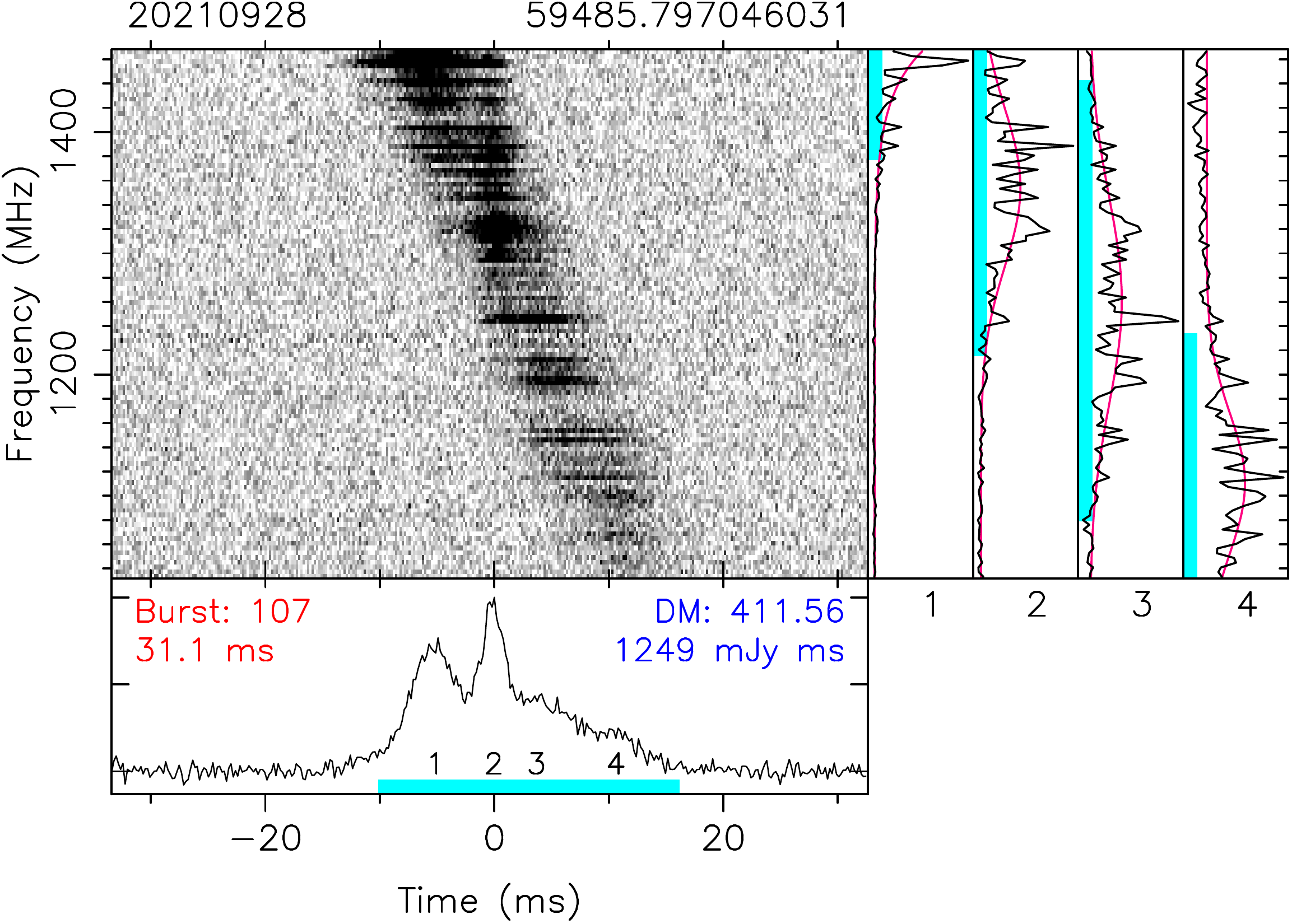}
    \includegraphics[height=37mm]{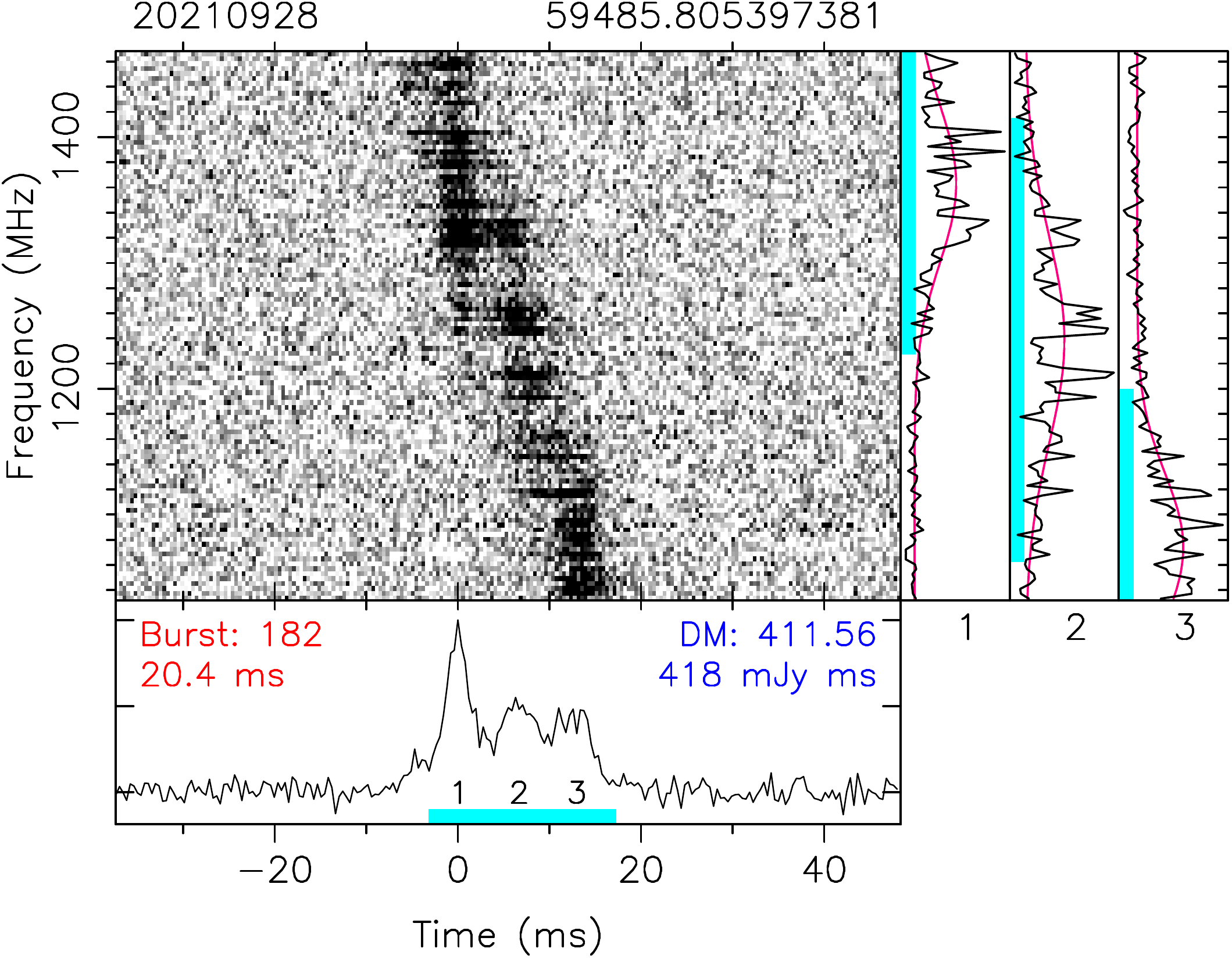}
    \includegraphics[height=37mm]{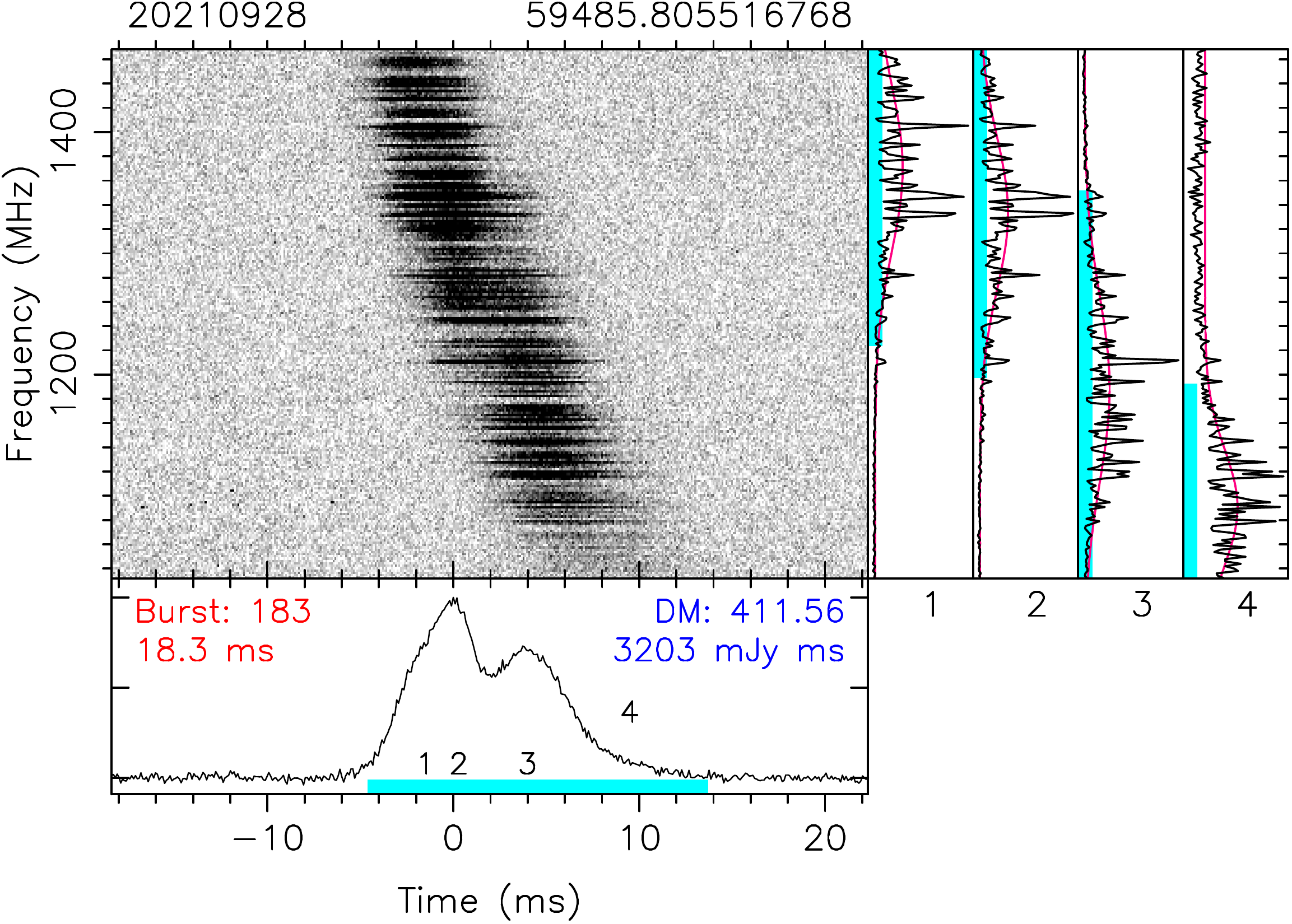}
    \includegraphics[height=37mm]{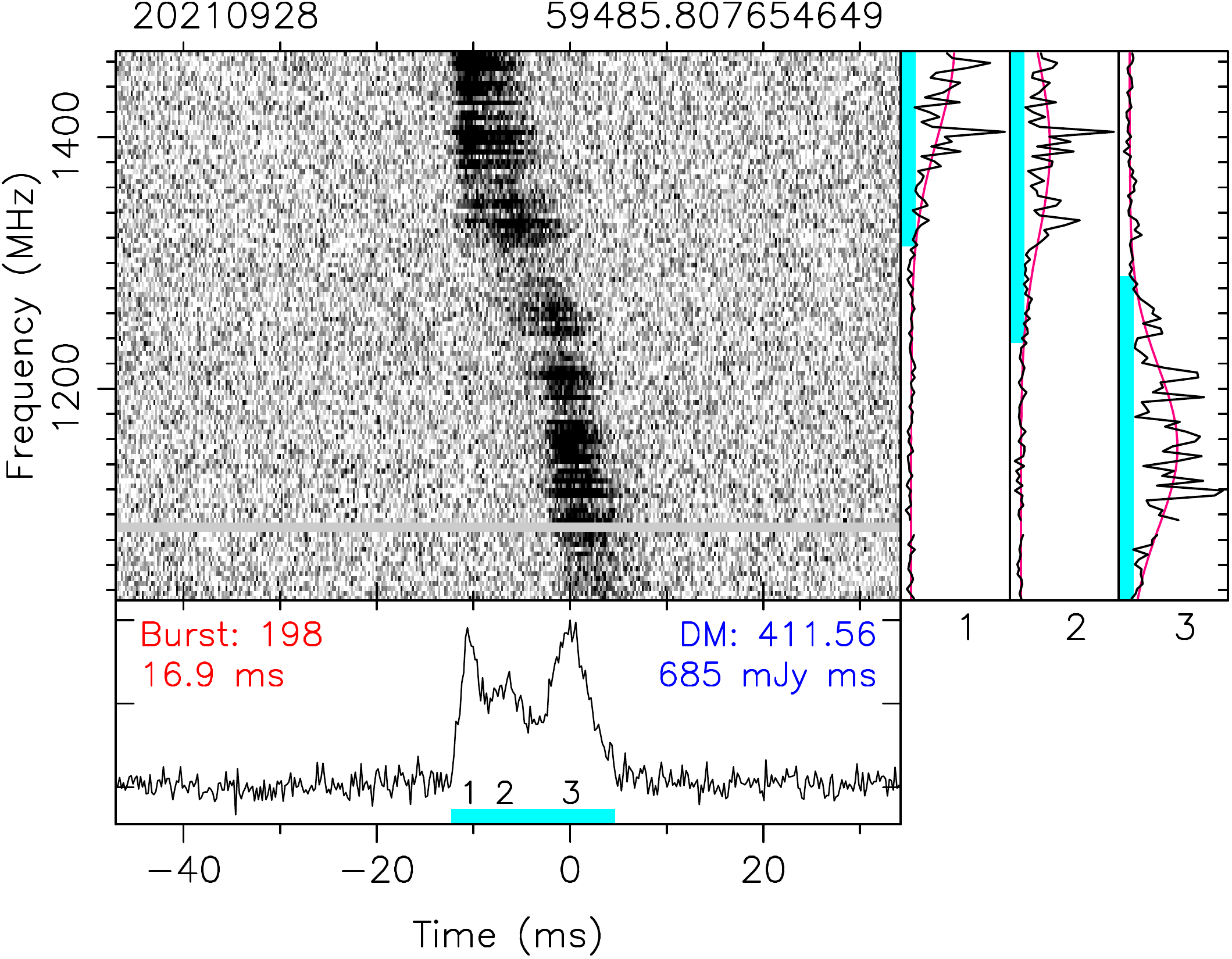}
    \includegraphics[height=37mm]{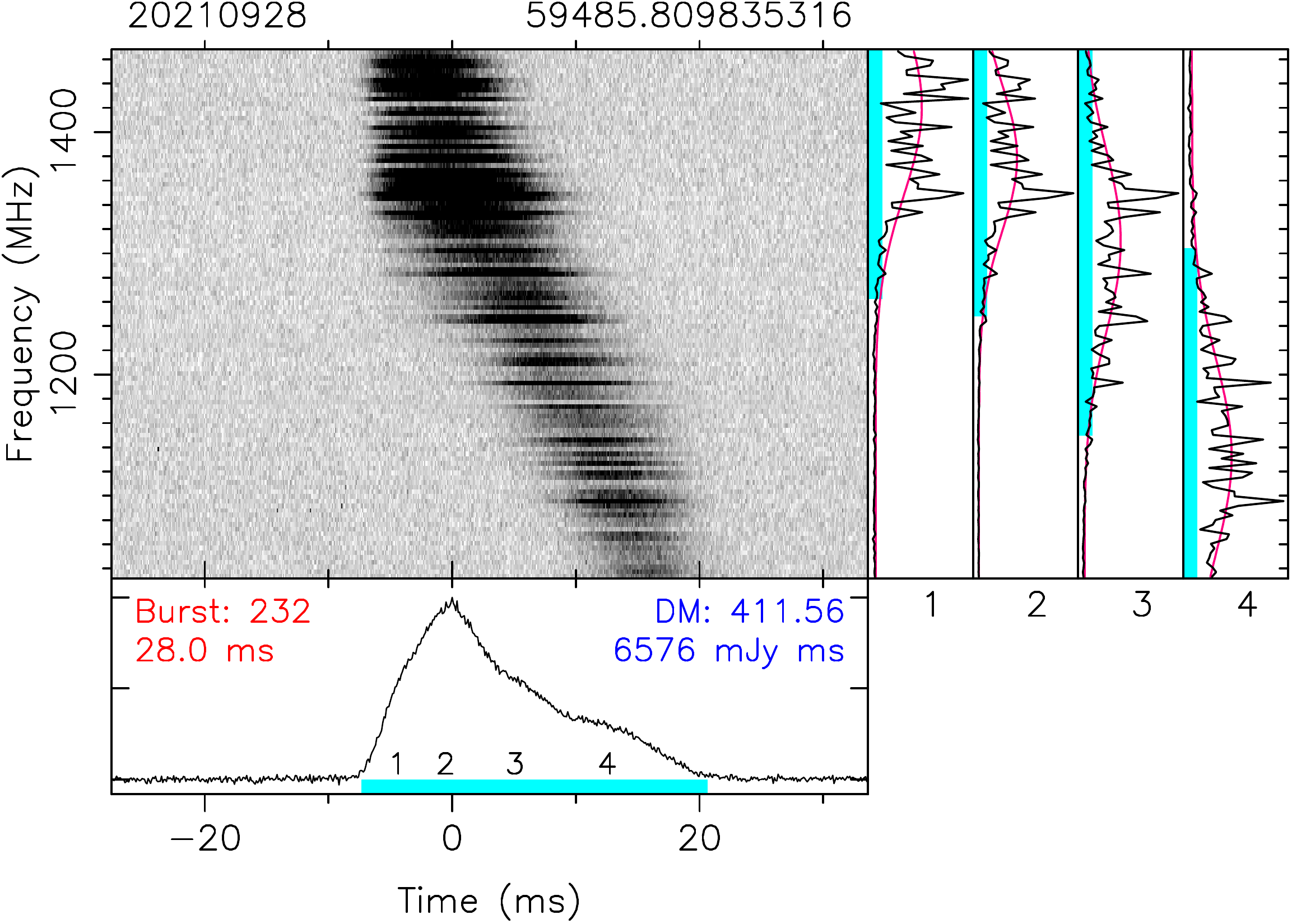}
    \includegraphics[height=37mm]{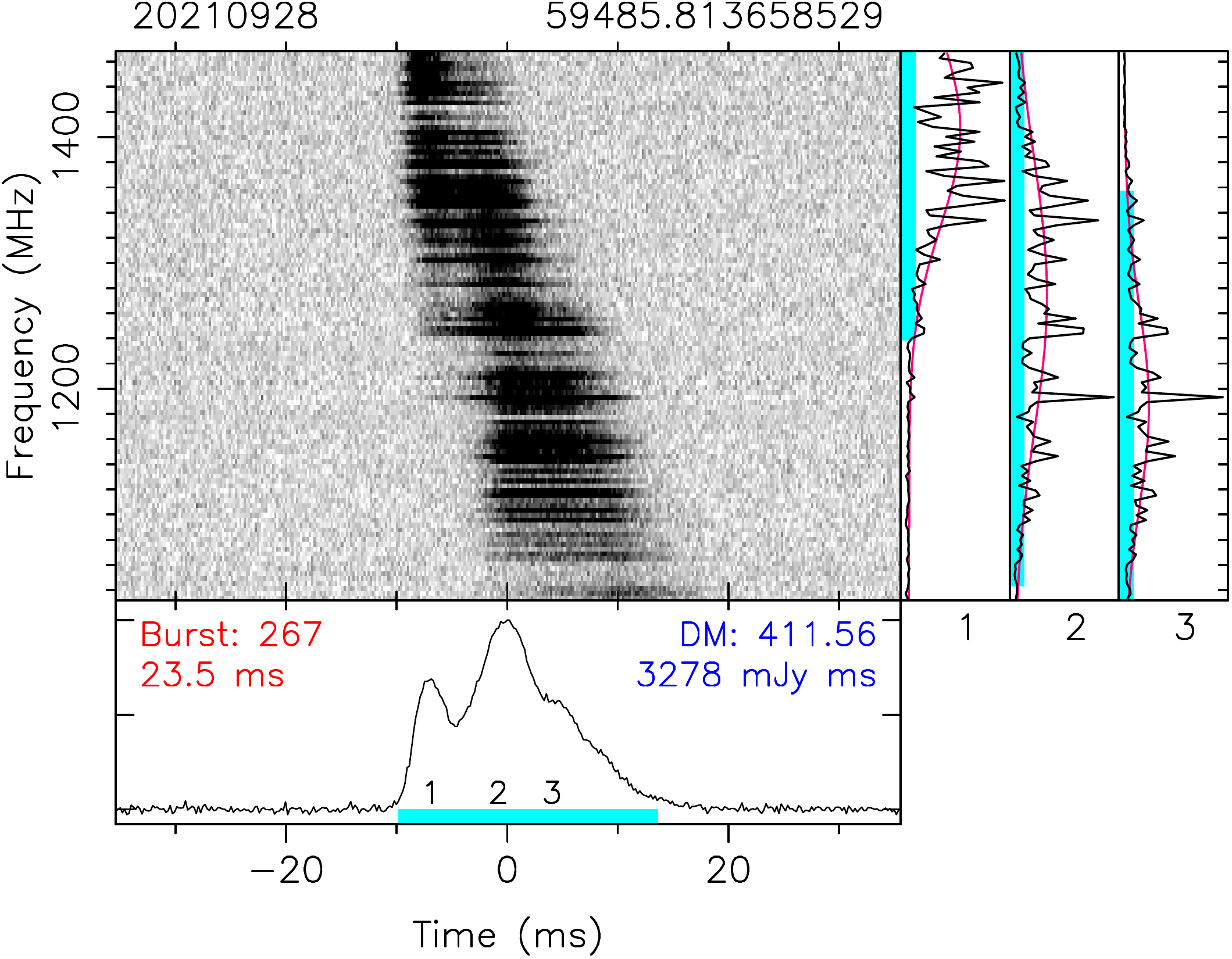}
    \includegraphics[height=37mm]{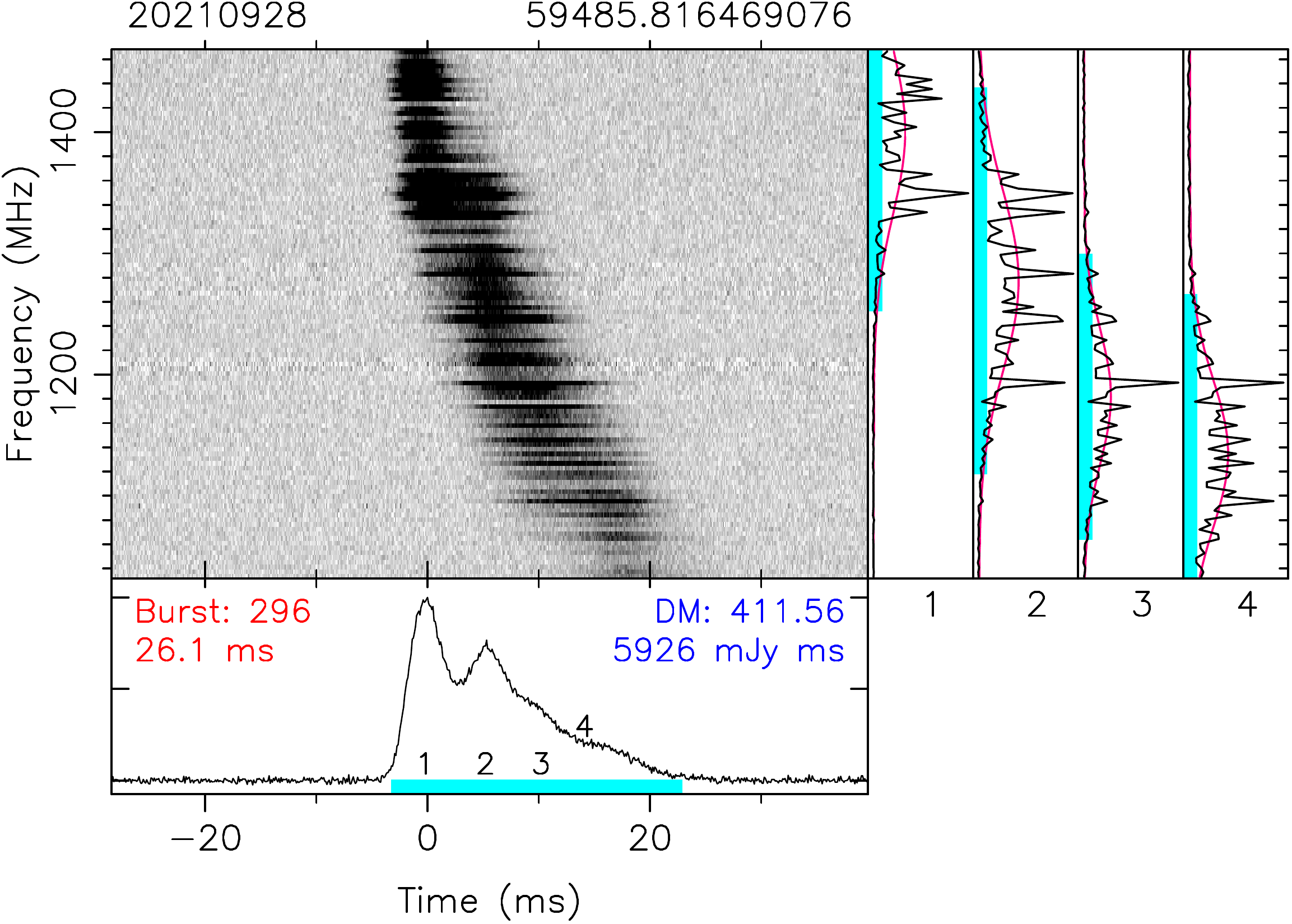}
    \includegraphics[height=37mm]{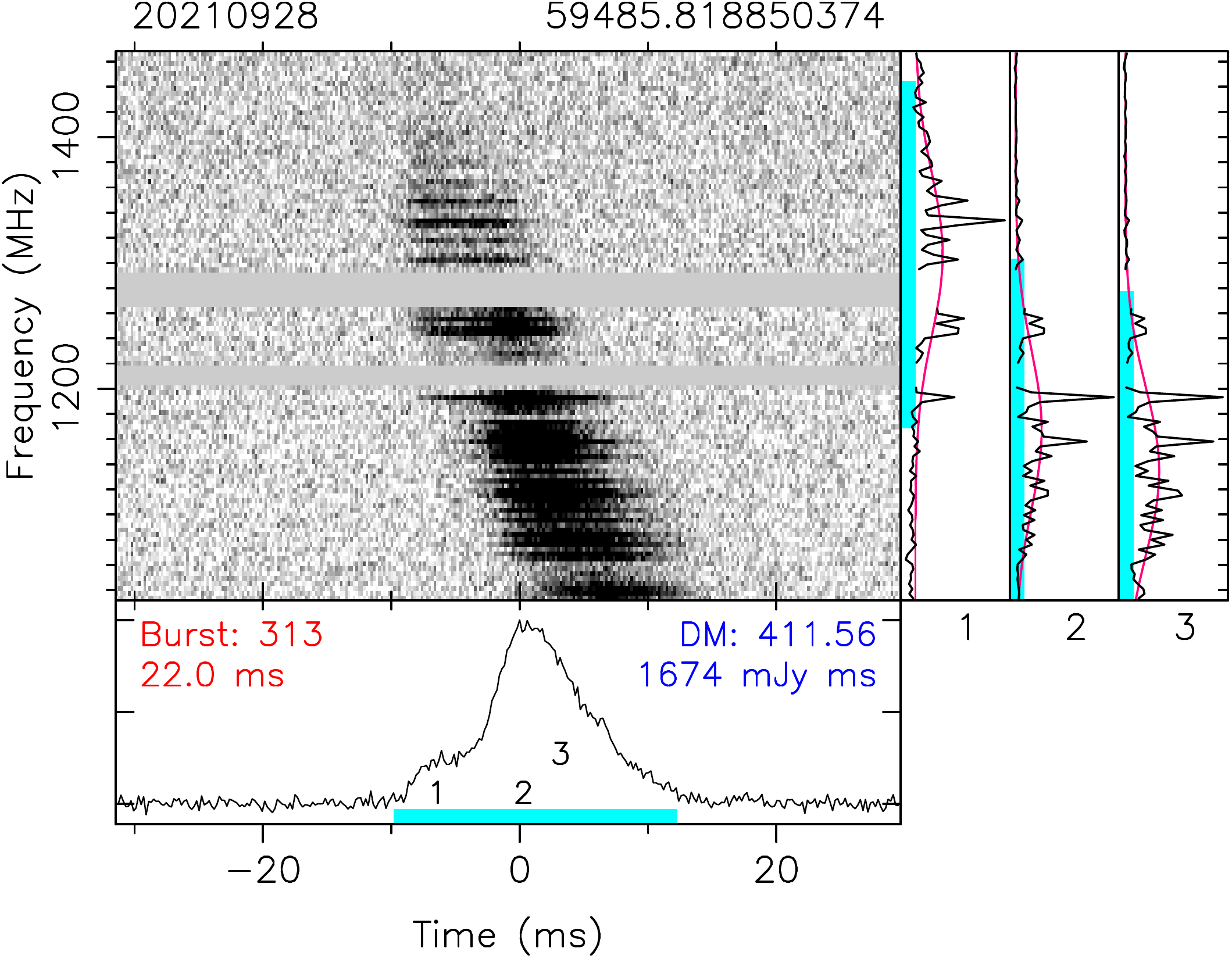}
    \includegraphics[height=37mm]{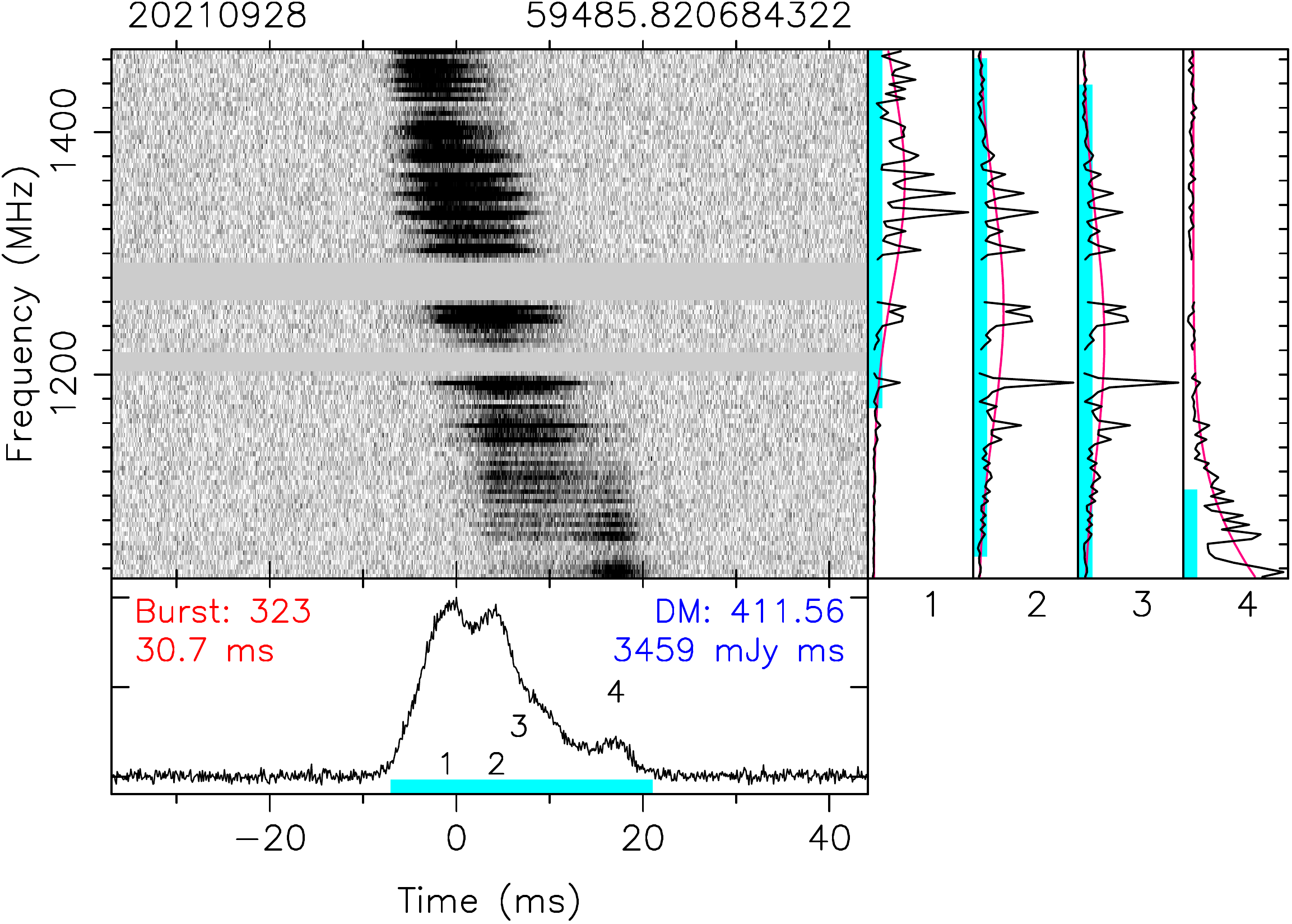}
    \includegraphics[height=37mm]{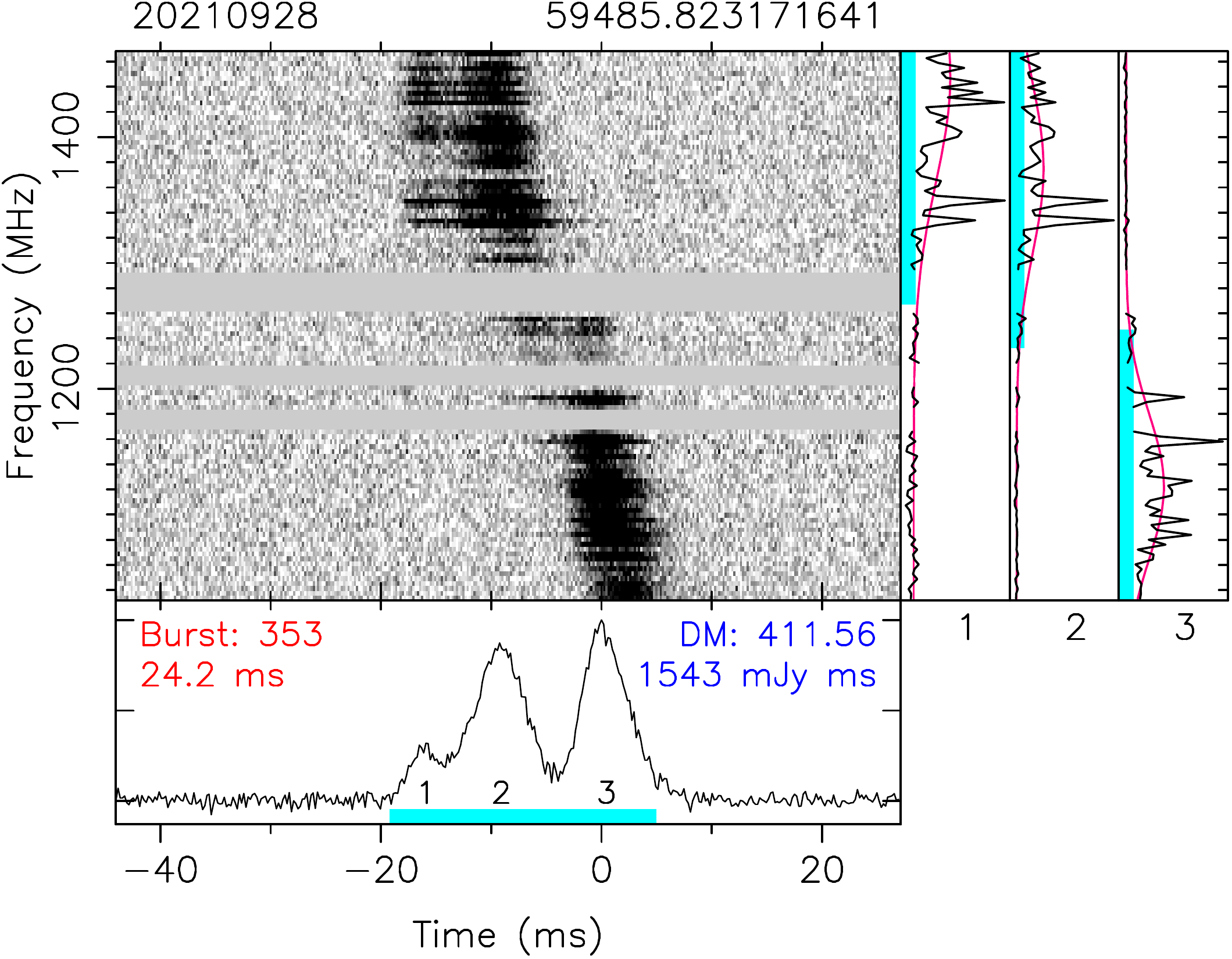}
\caption{The same as Figure~\ref{fig:appendix:D1W} but for bursts in Dm-W.
}\label{fig:appendix:DmW} 
\end{figure*}

\begin{figure*}
    \flushleft
    \includegraphics[height=37mm]{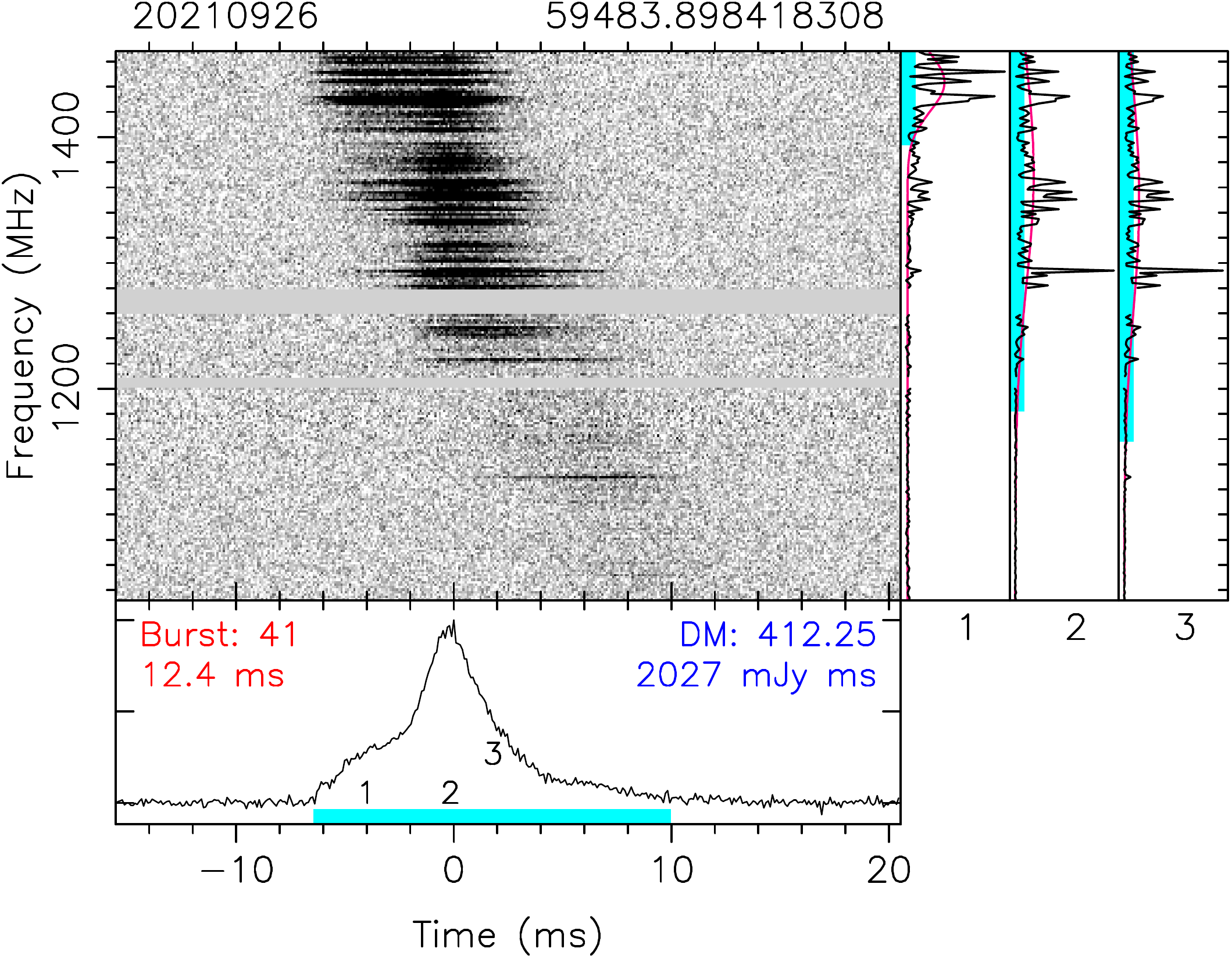}
    \includegraphics[height=37mm]{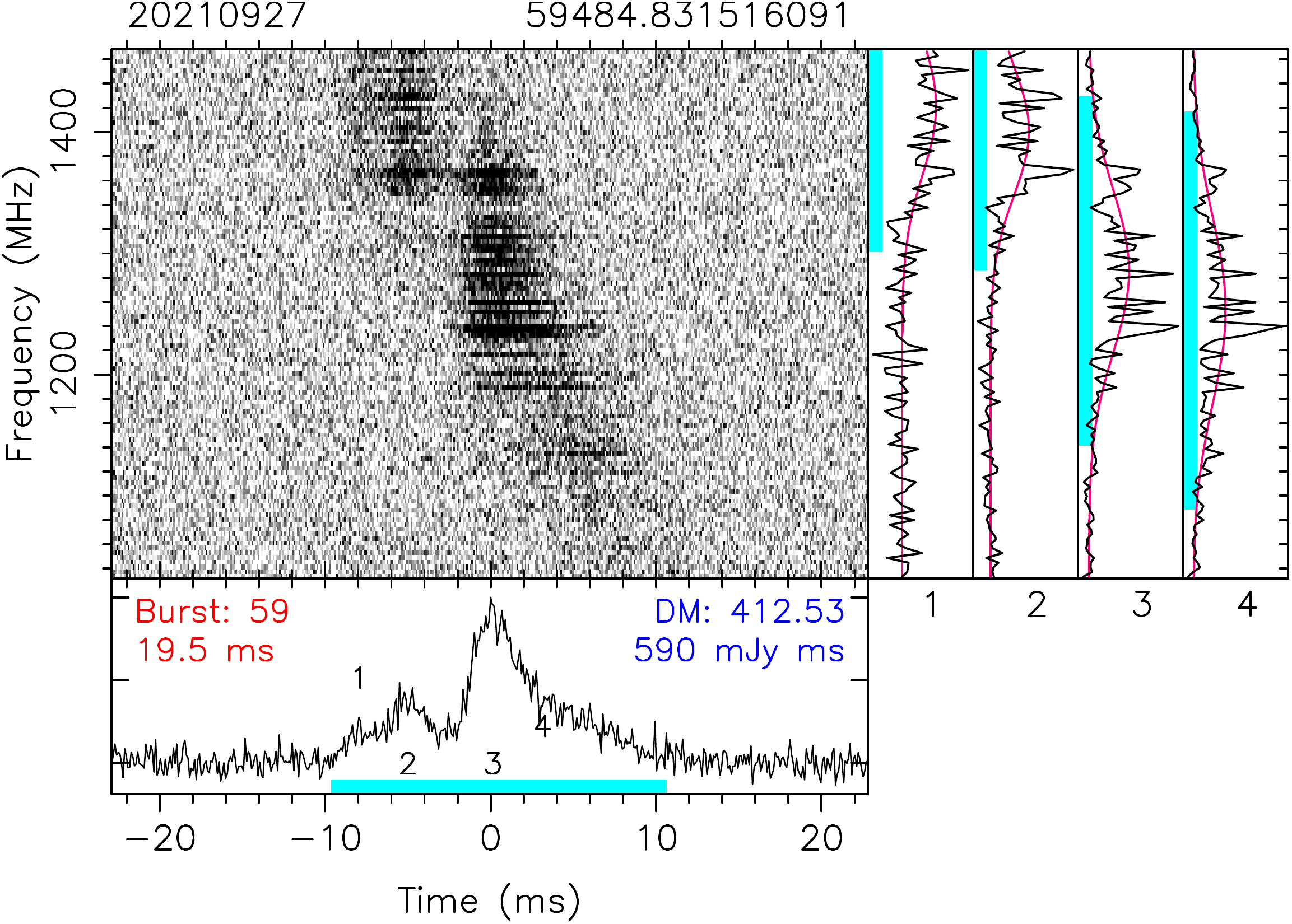}
    \includegraphics[height=37mm]{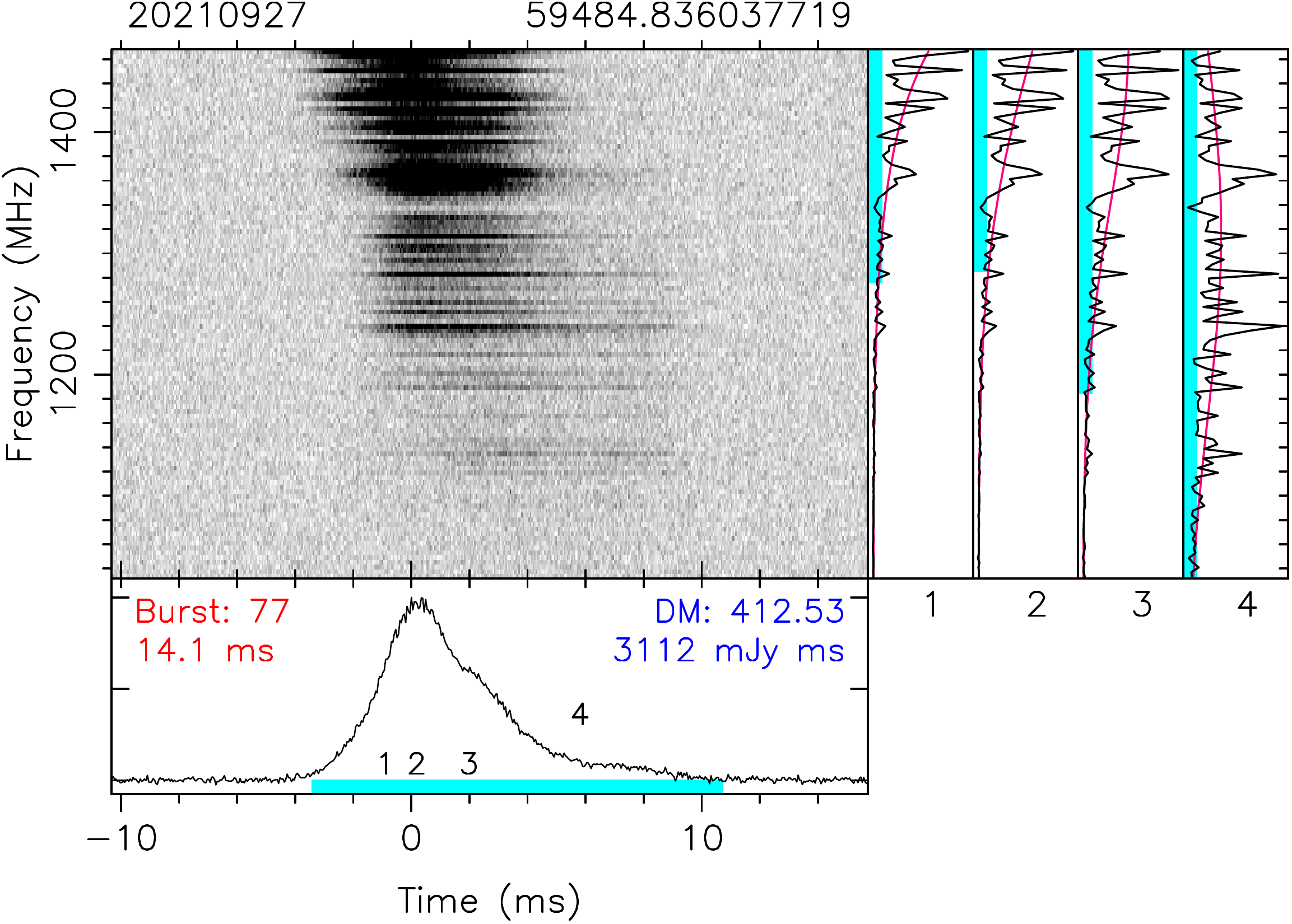}
    \includegraphics[height=37mm]{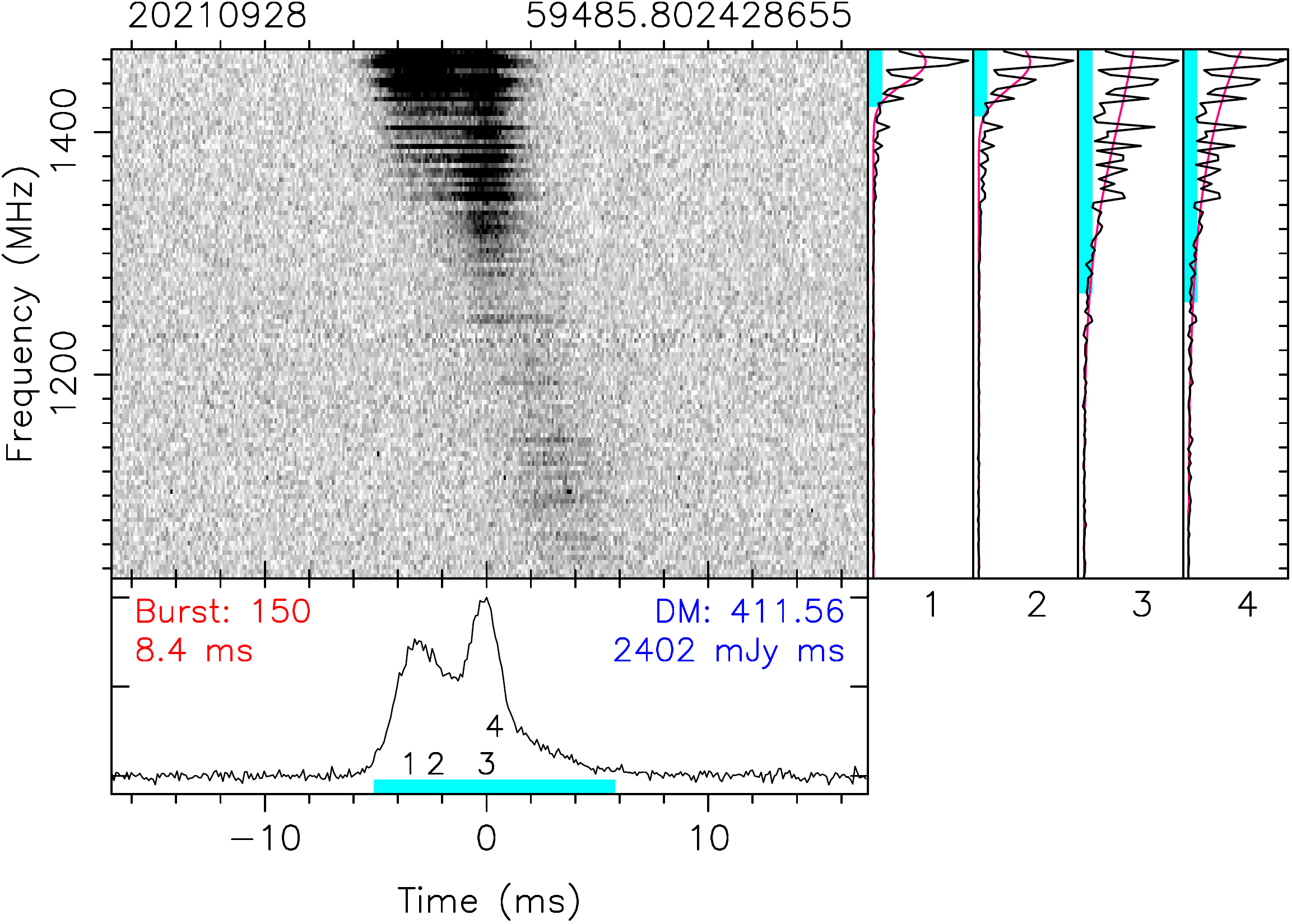}
    \includegraphics[height=37mm]{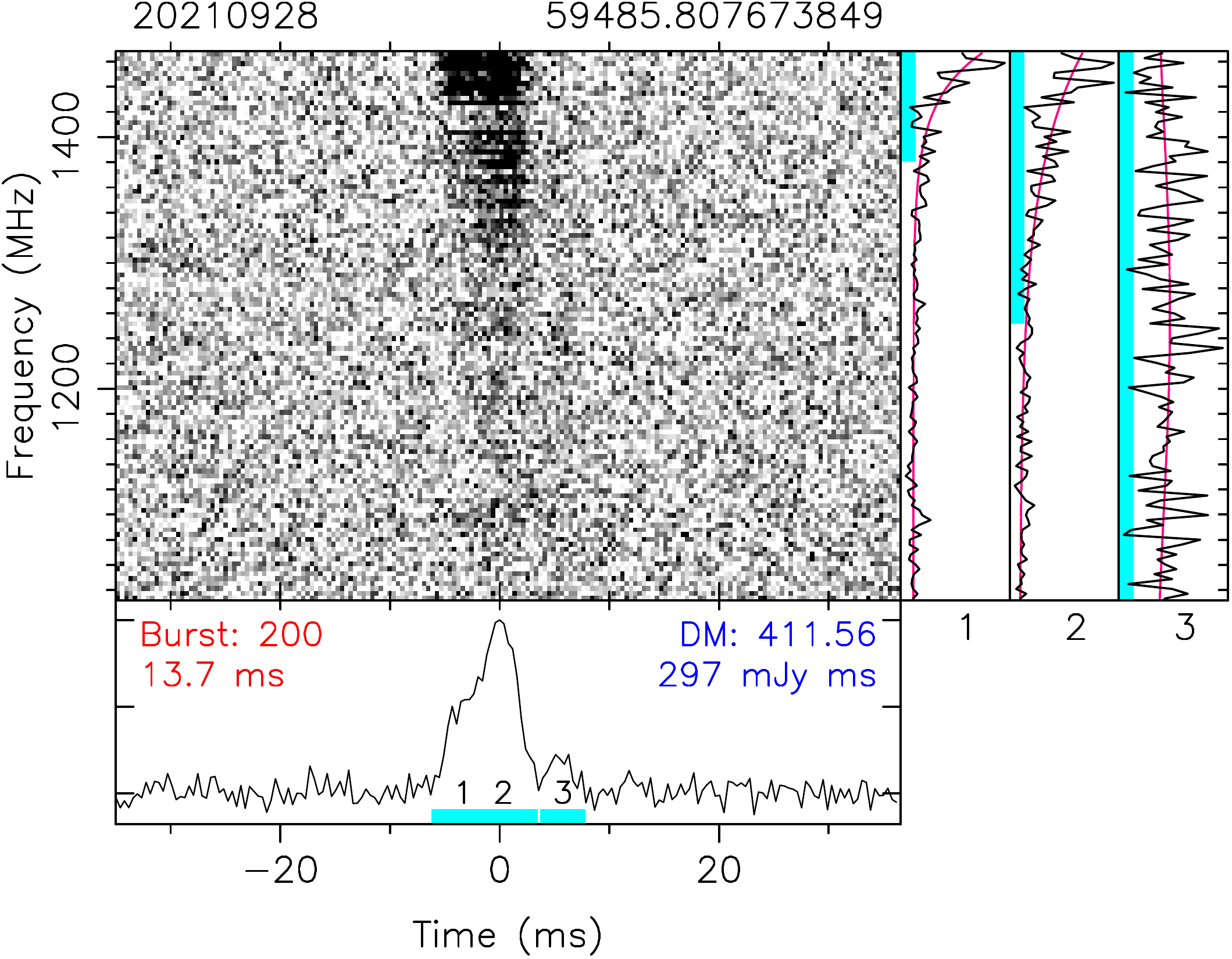}
    \includegraphics[height=37mm]{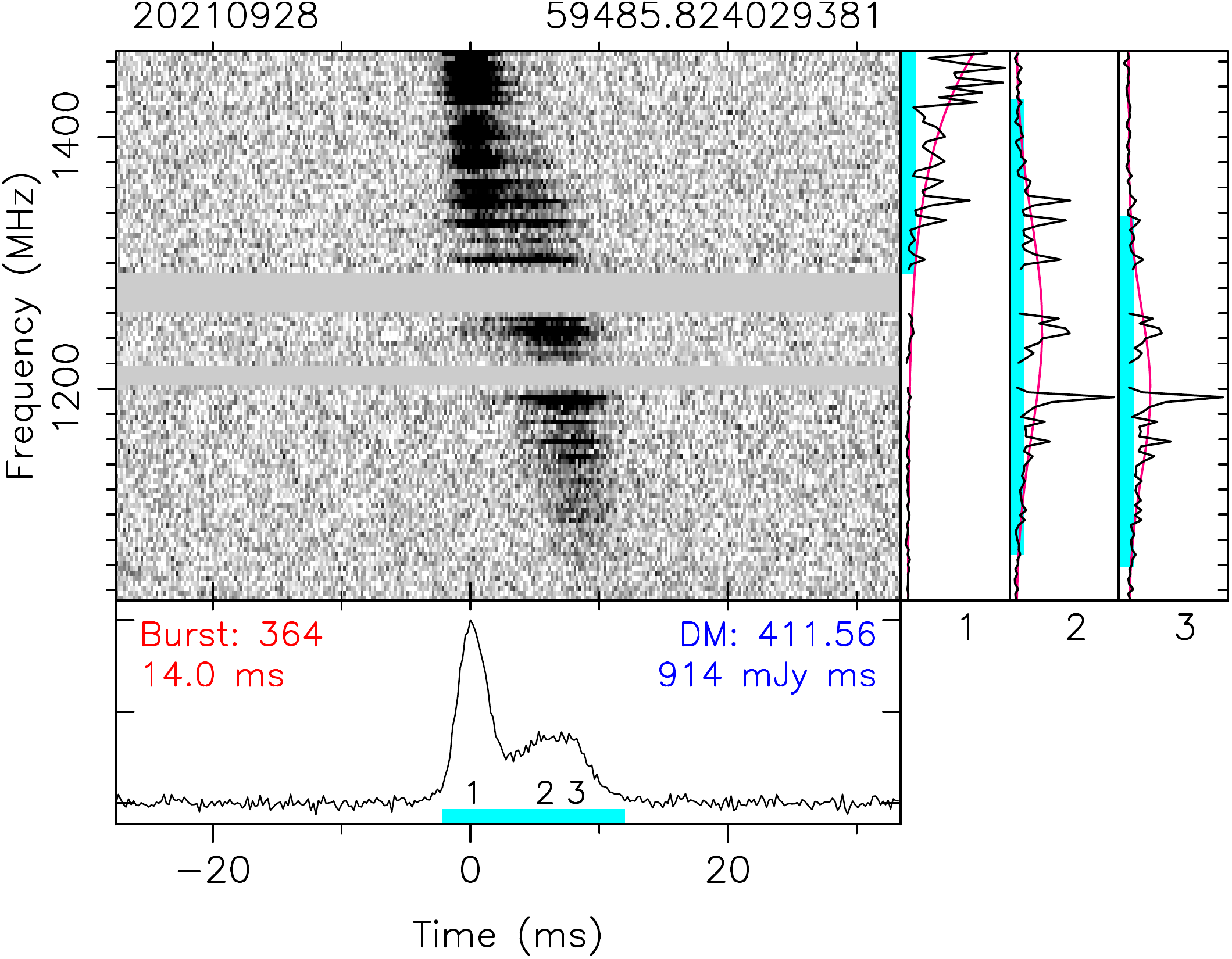}
\caption{The same as Figure~\ref{fig:appendix:D1W} but for bursts in Dm-H.
}\label{fig:appendix:DmH} 
\end{figure*}

\begin{figure*}
    \flushleft
    \includegraphics[height=37mm]{20210926/FRB20201124A_20210926_tracking-M01-P1-c512b1.fits-015-T-1567.217-1567.267-DM-412.4.pdf}
    \includegraphics[height=37mm]{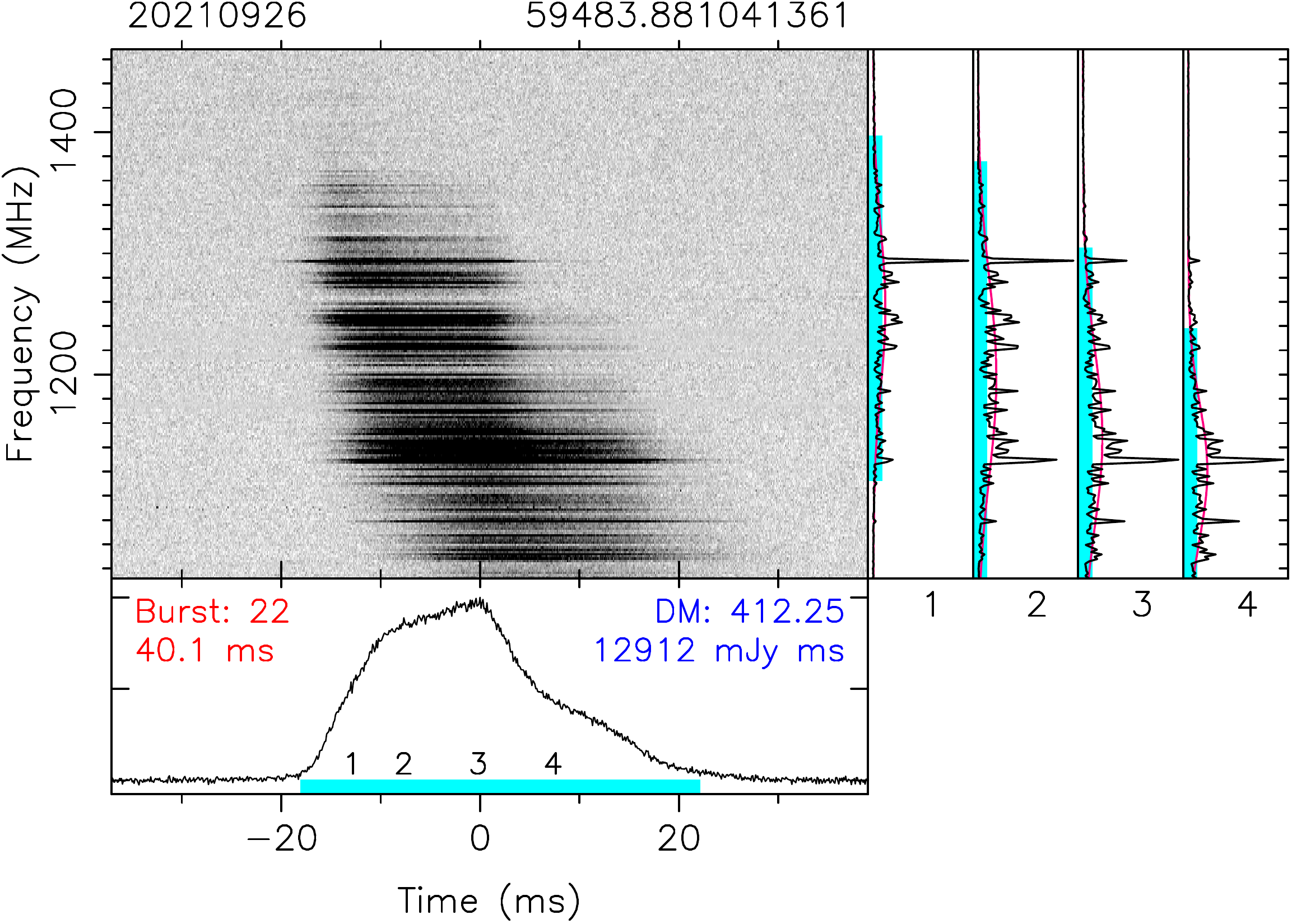}
    \includegraphics[height=37mm]{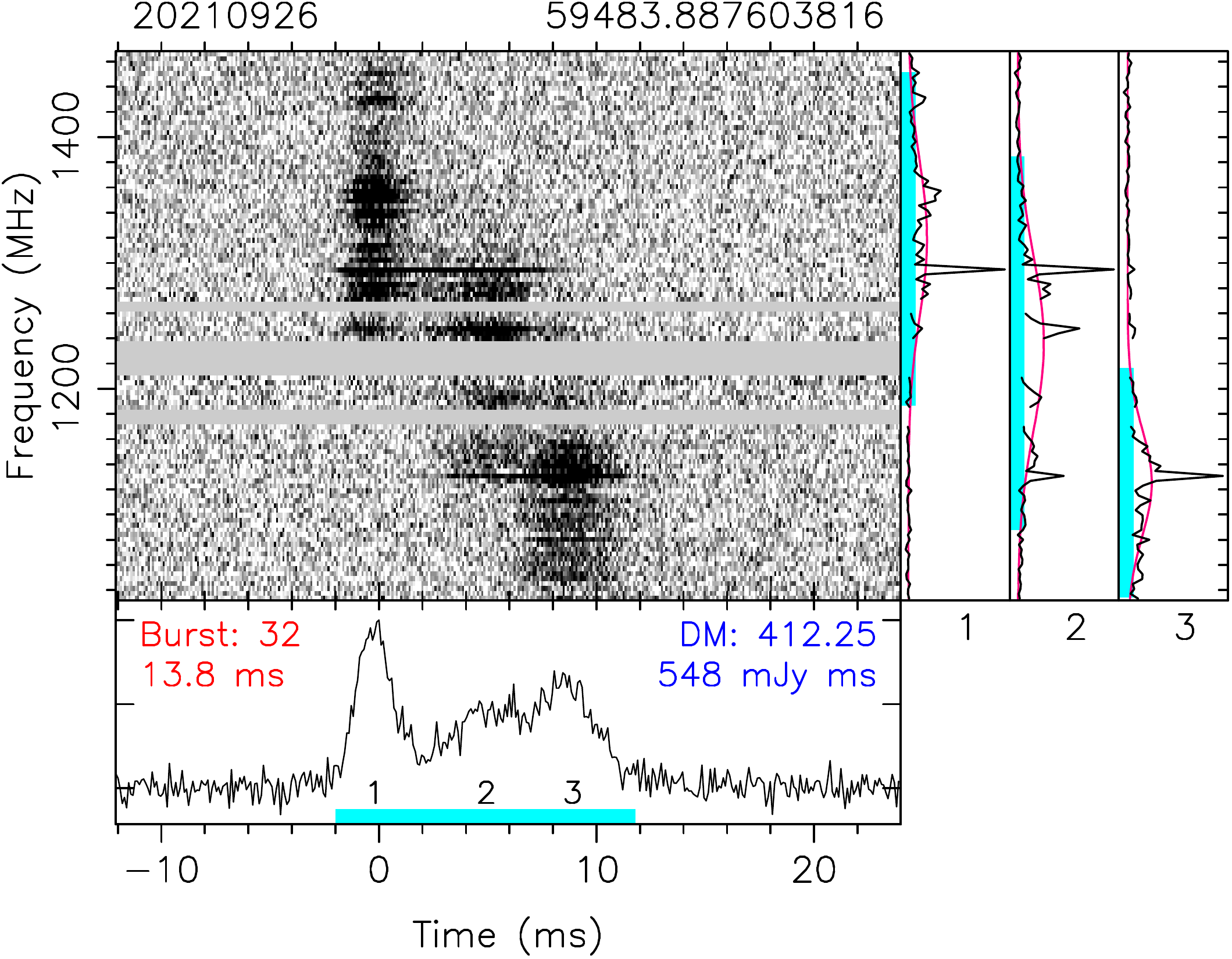}
    \includegraphics[height=37mm]{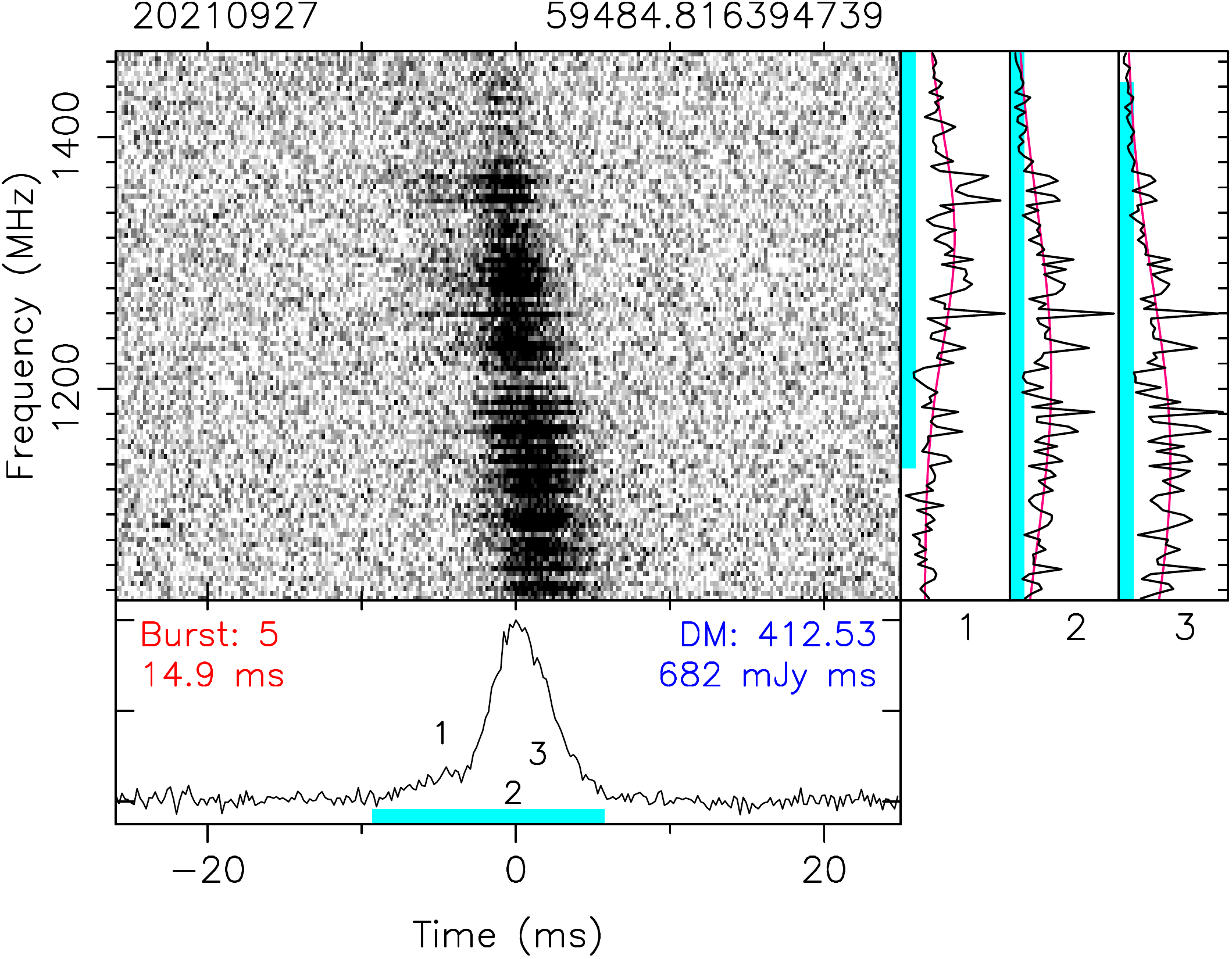}
    \includegraphics[height=37mm]{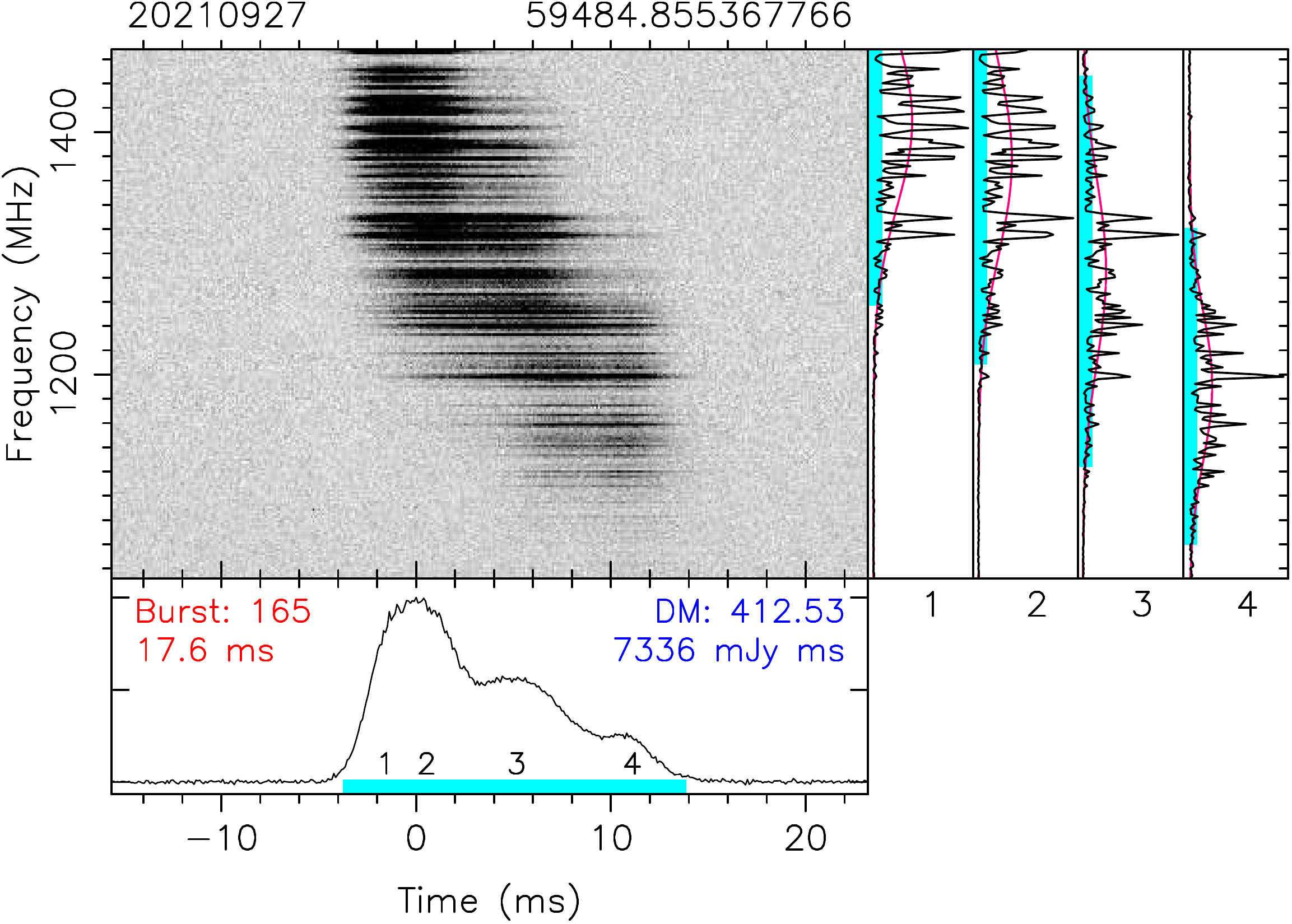}
    \includegraphics[height=37mm]{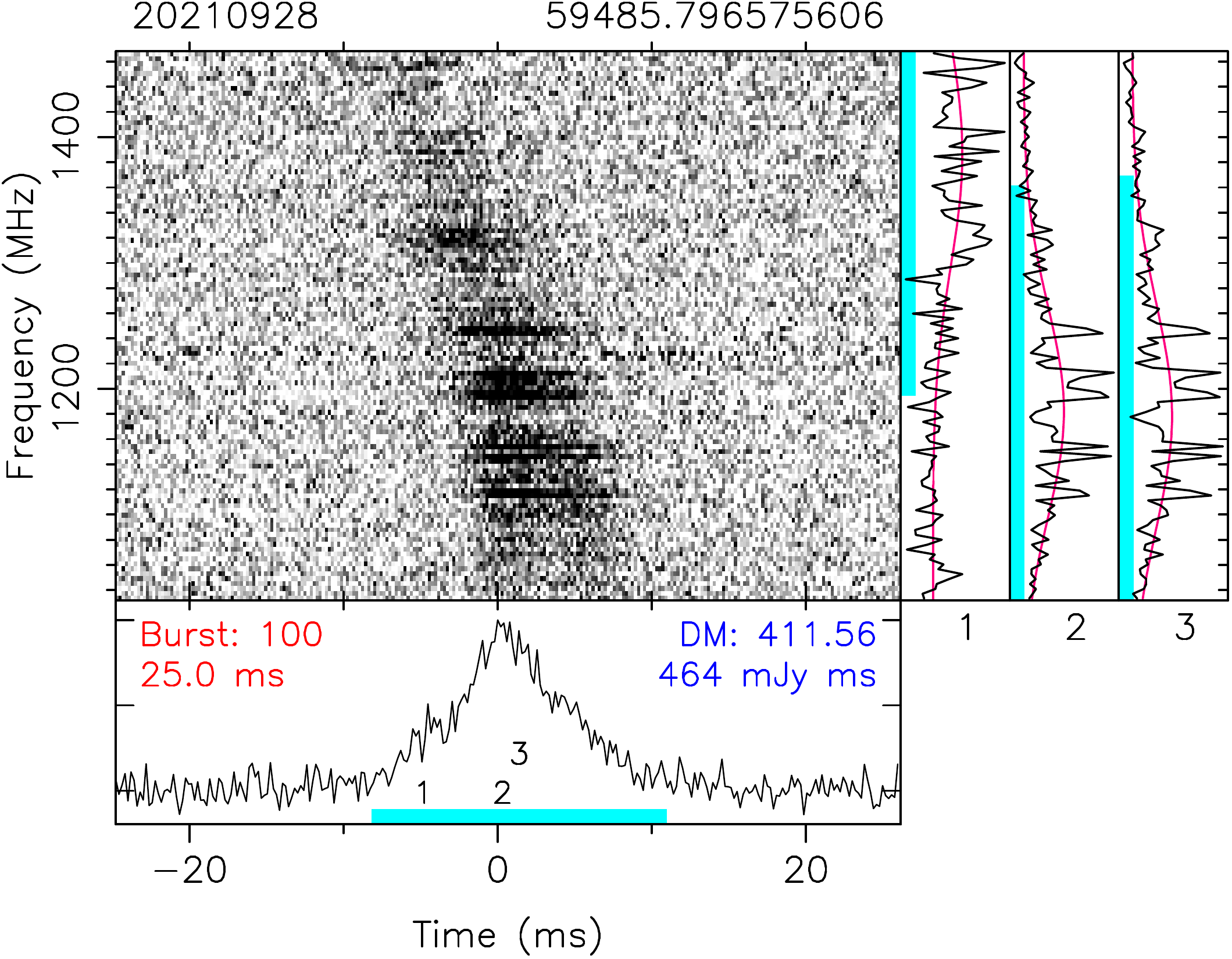}
    \includegraphics[height=37mm]{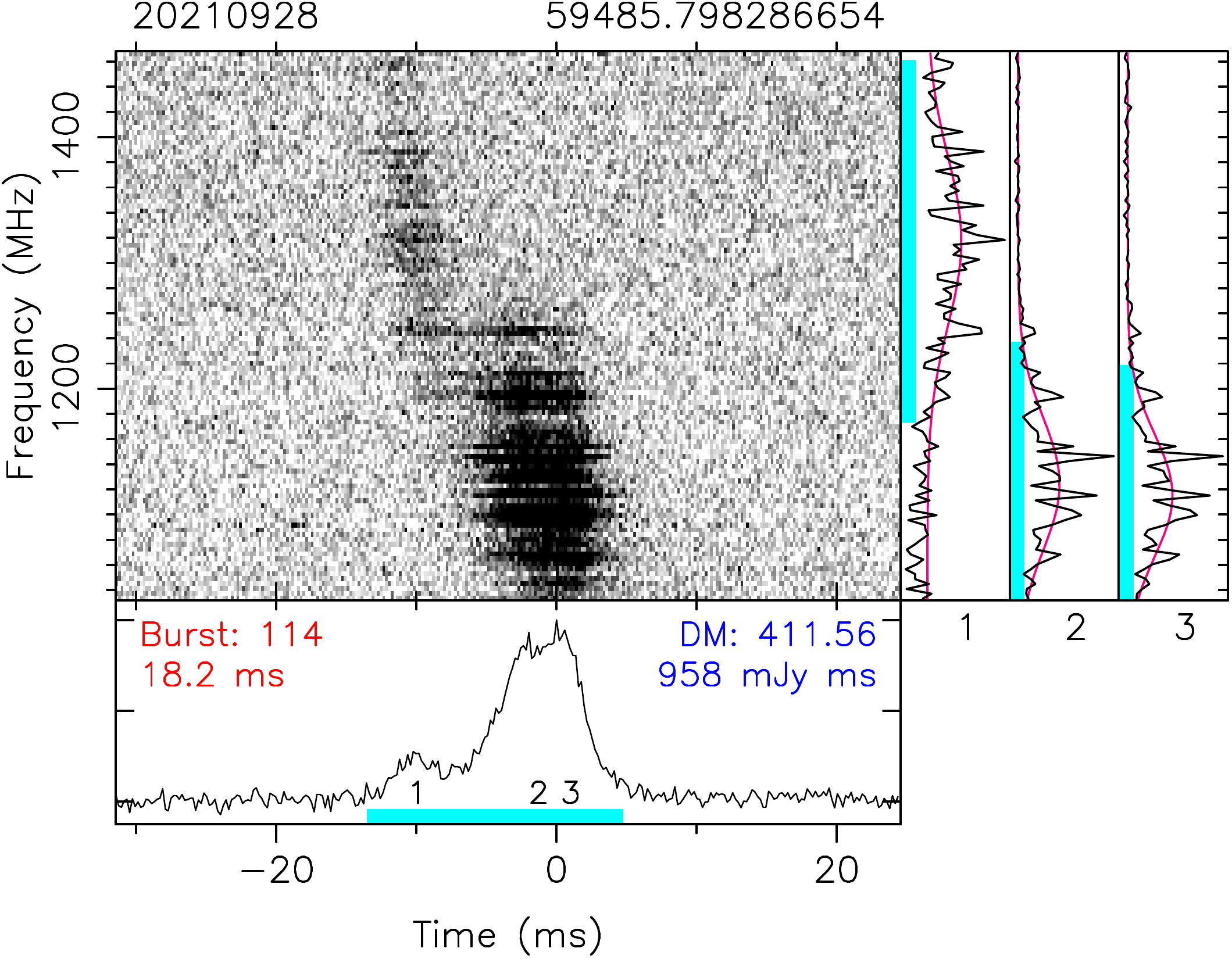}
    \includegraphics[height=37mm]{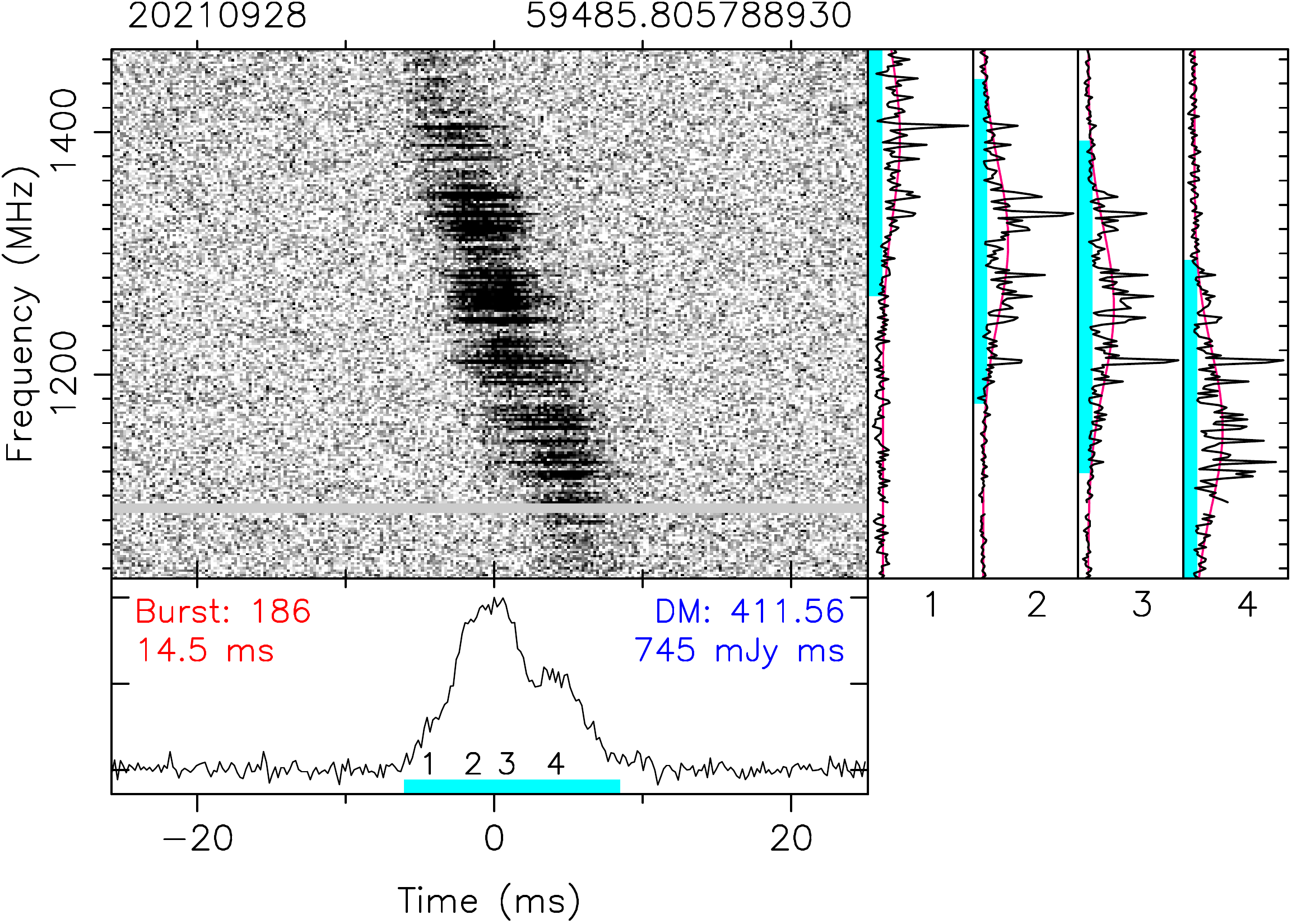}
\caption{The same as Figure~\ref{fig:appendix:D1W} but for bursts in Dm-M.
}\label{fig:appendix:DmM} 
\end{figure*}

\begin{figure*}
    \flushleft
    \includegraphics[height=37mm]{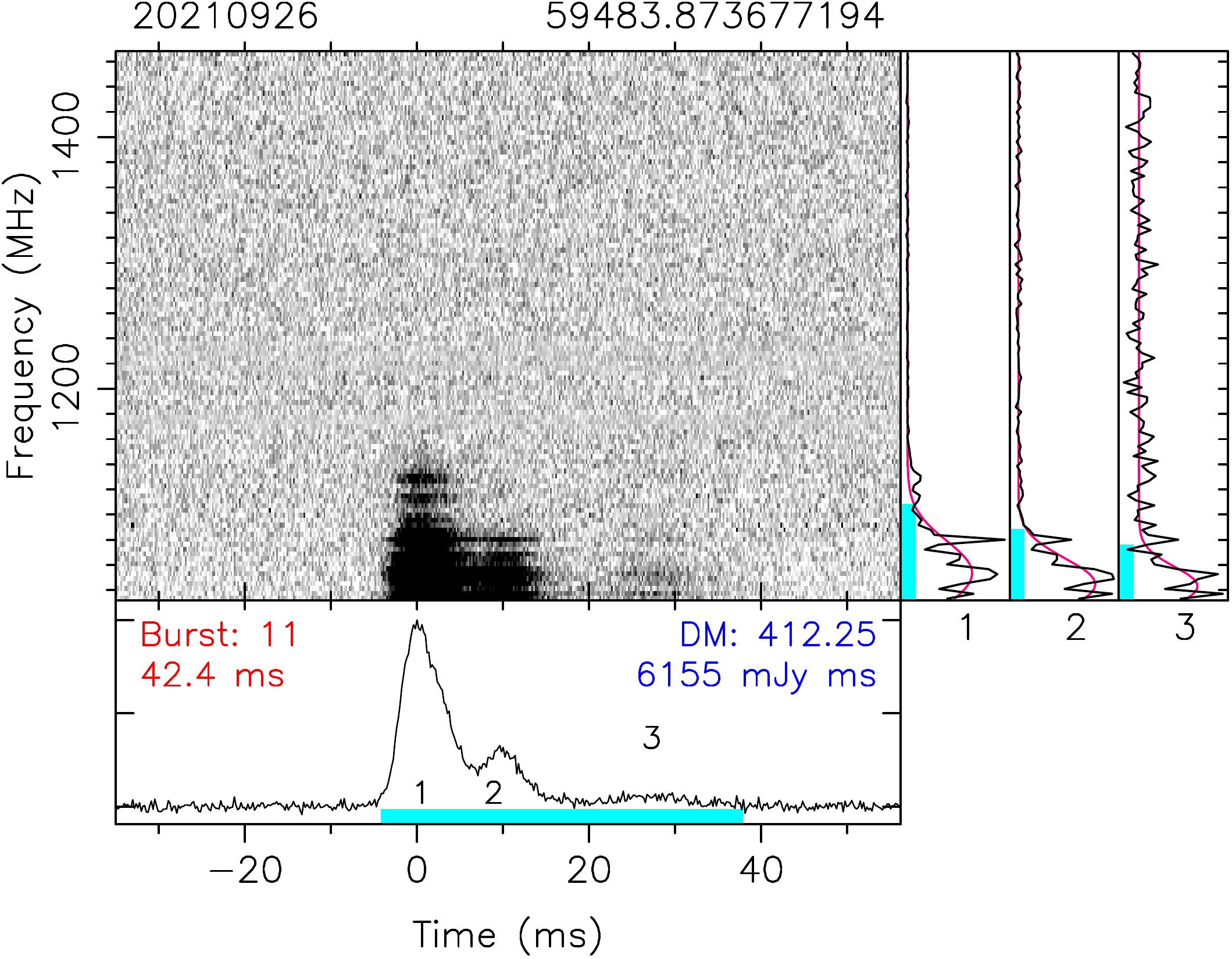}
    \includegraphics[height=37mm]{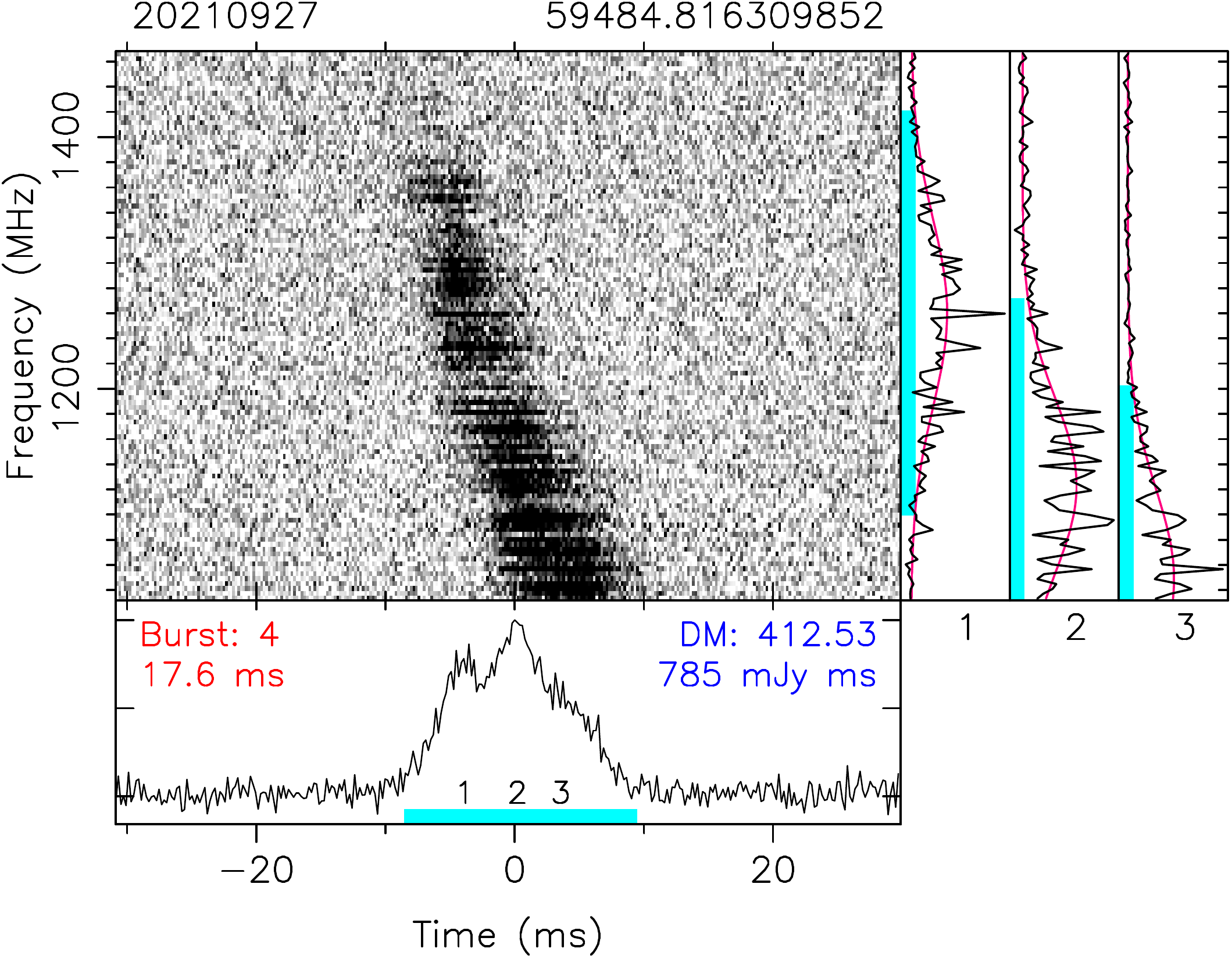}
    \includegraphics[height=37mm]{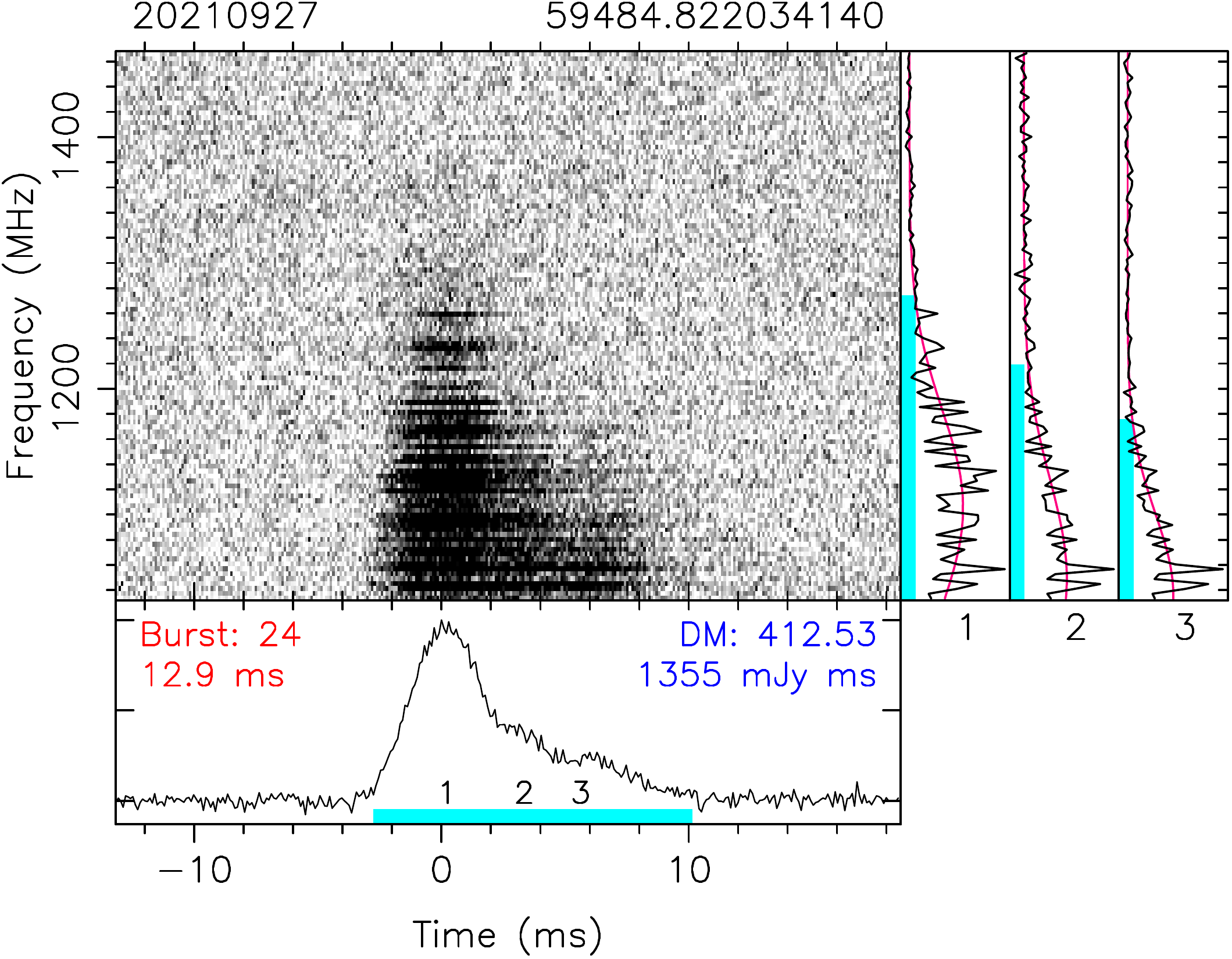}
    \includegraphics[height=37mm]{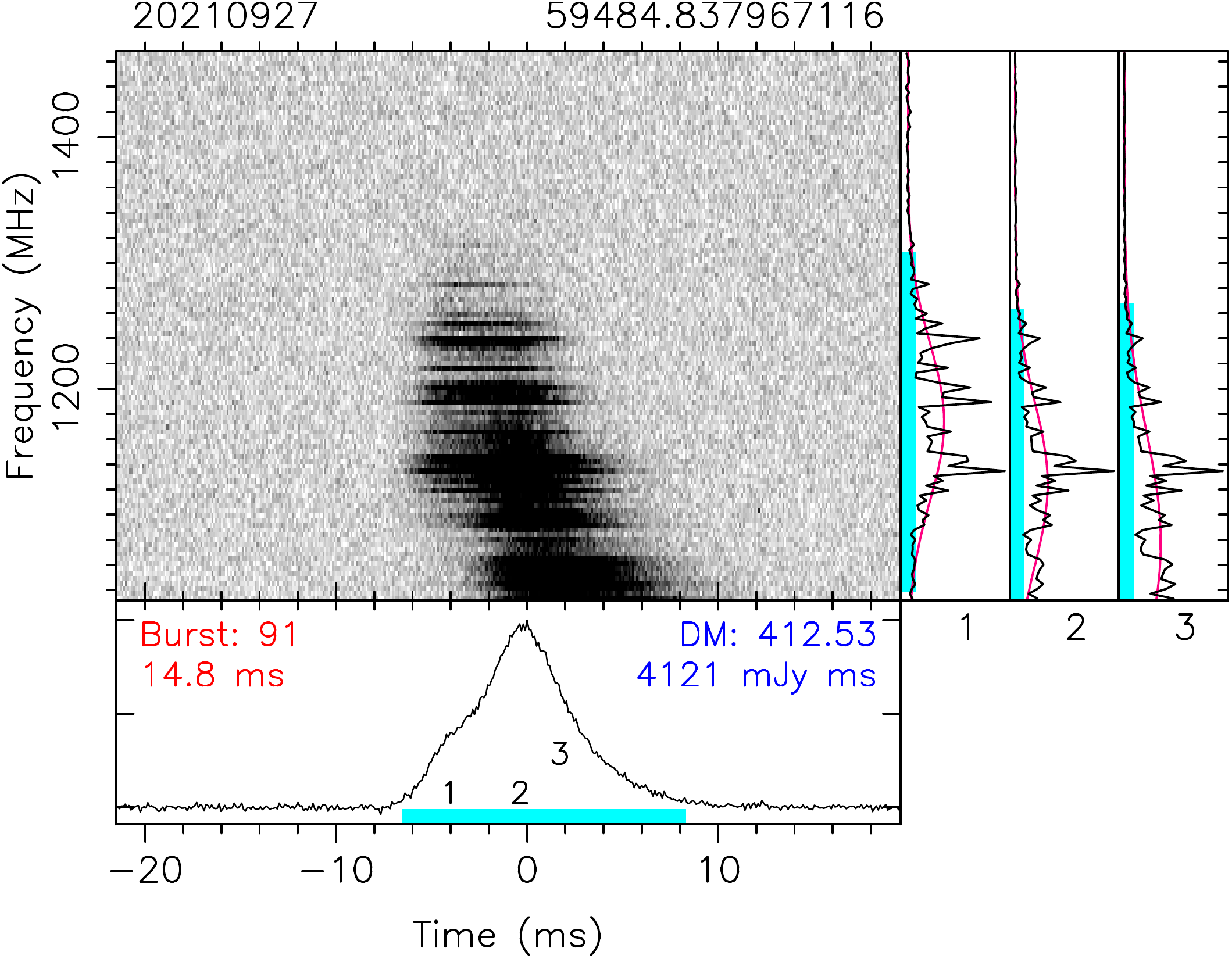}
    \includegraphics[height=37mm]{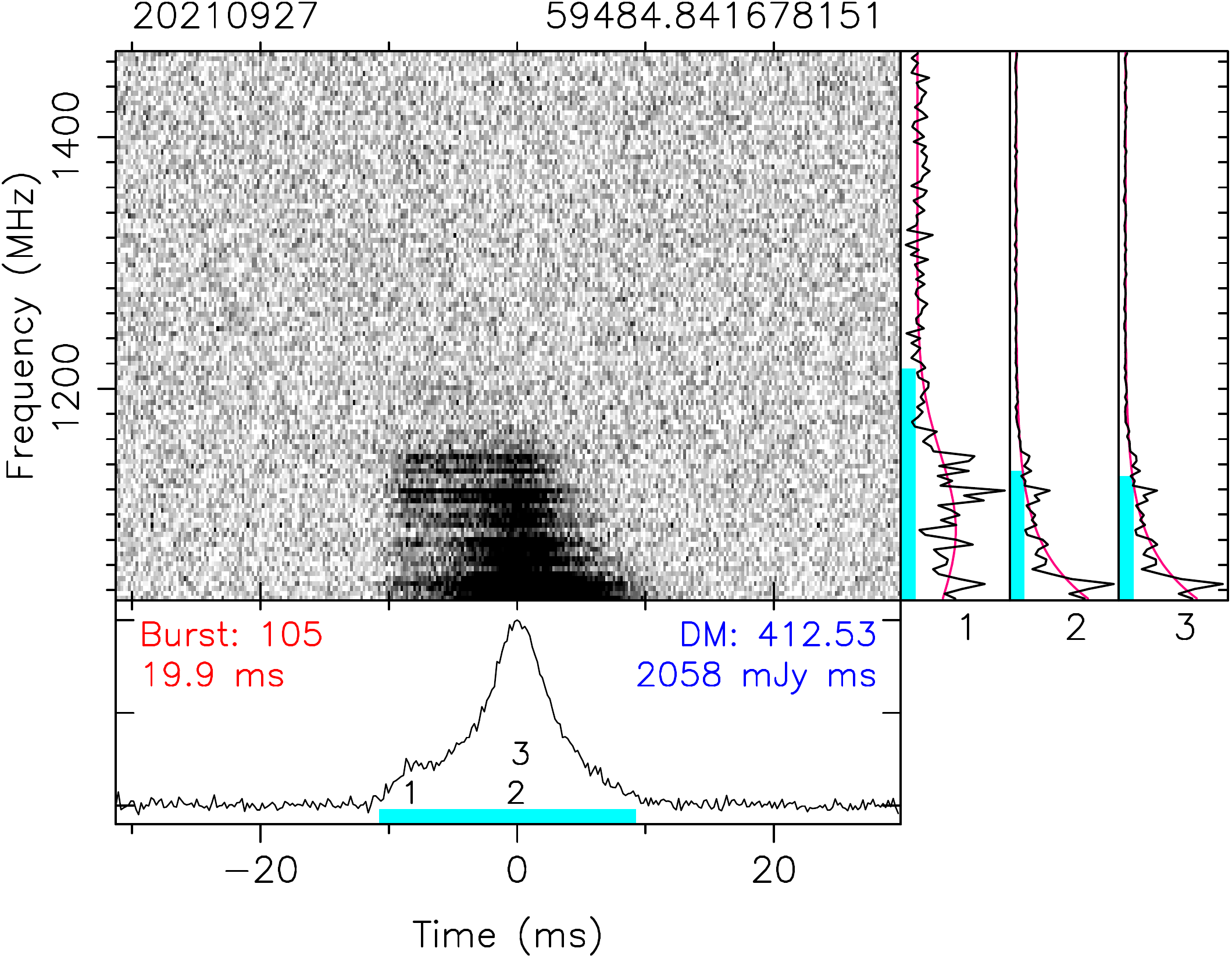}
    \includegraphics[height=37mm]{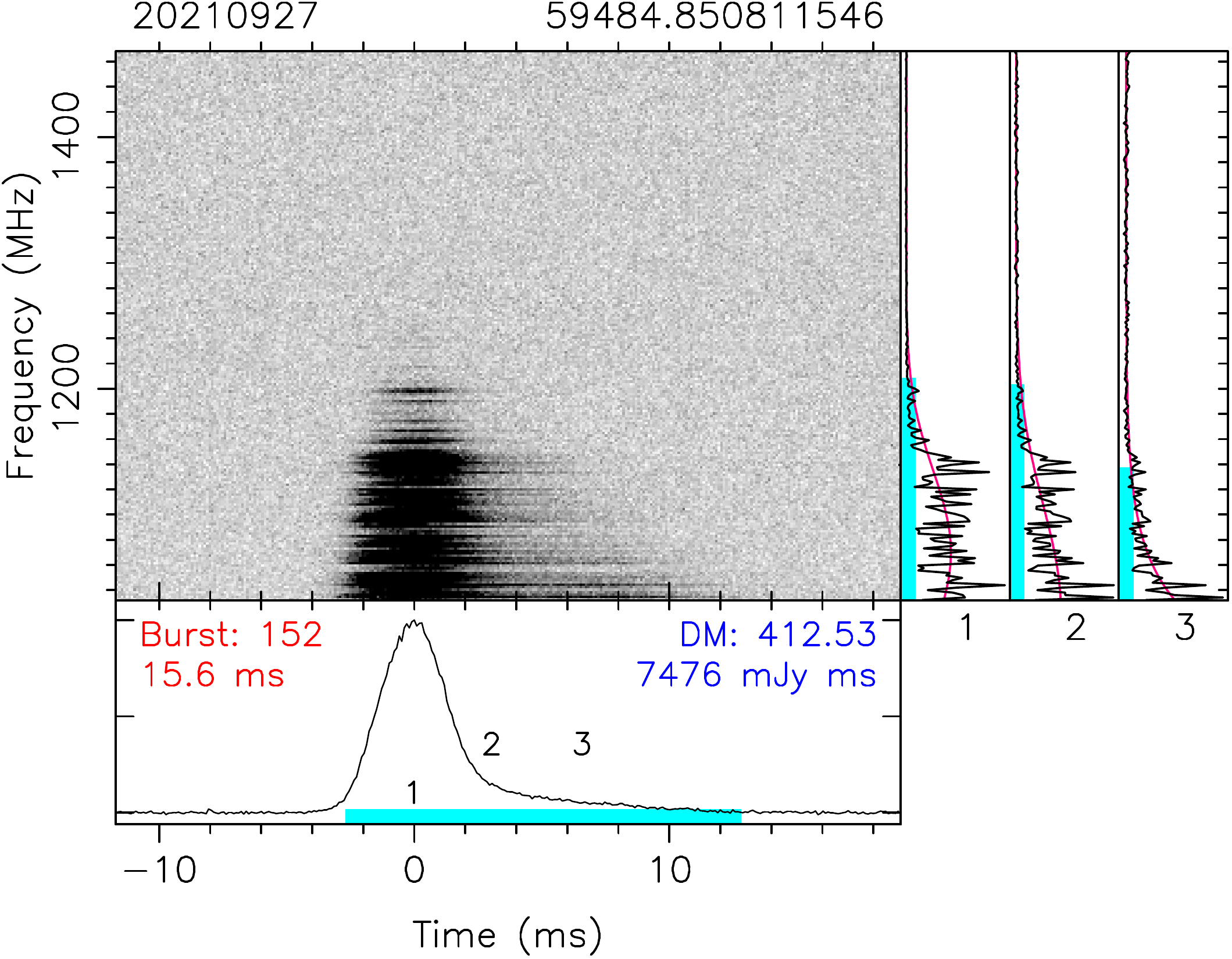}
    \includegraphics[height=37mm]{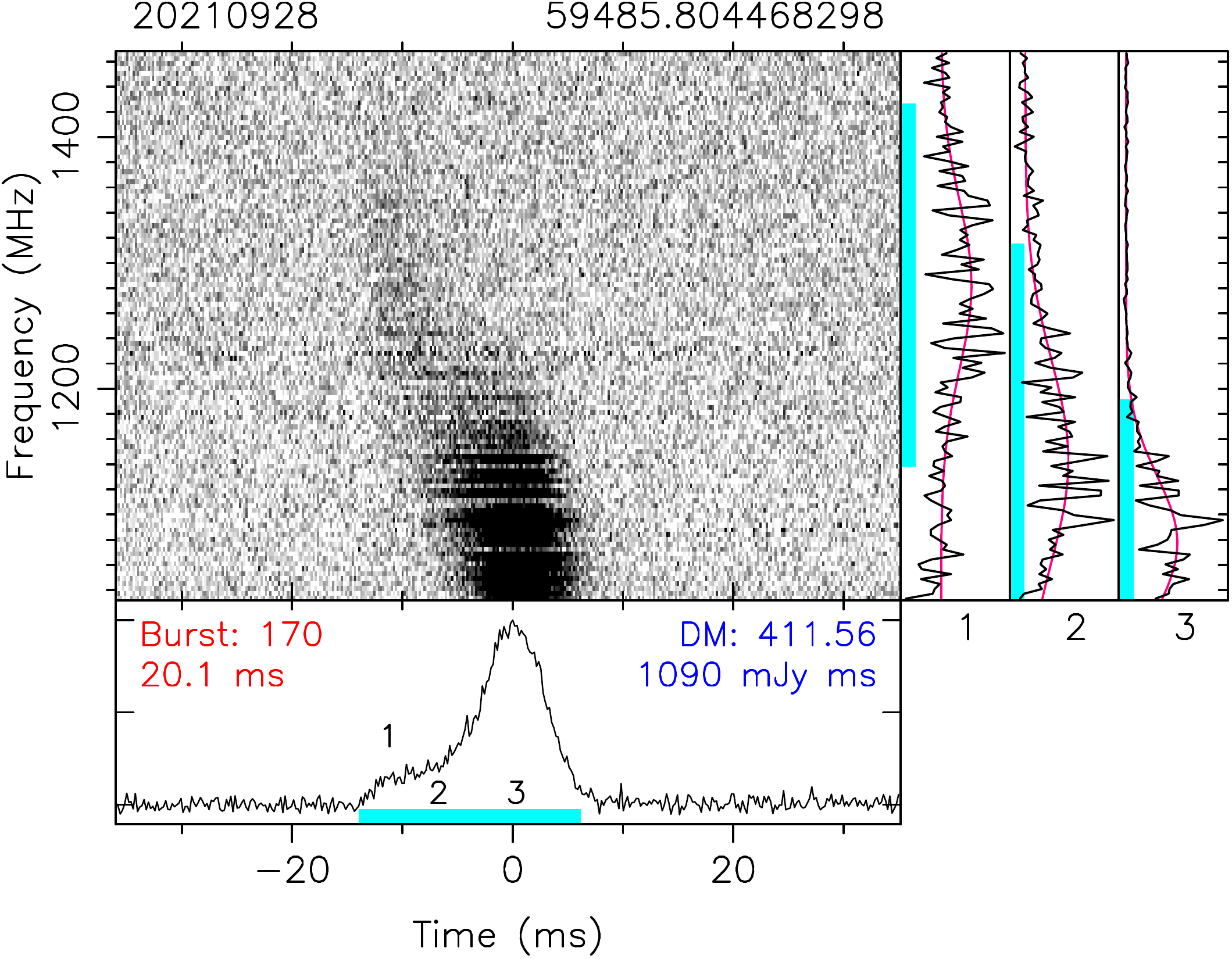}
    \includegraphics[height=37mm]{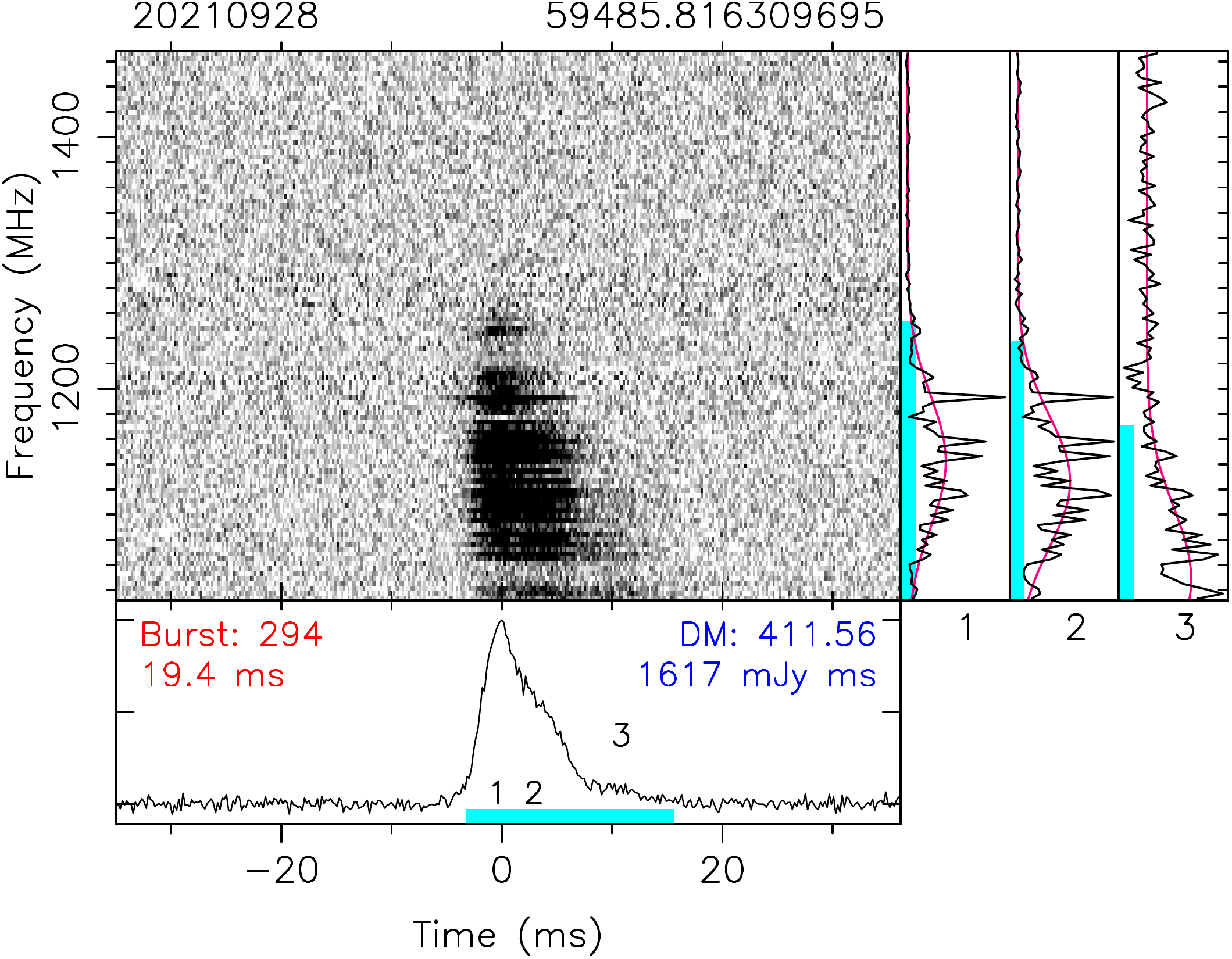}
\caption{The same as Figure~\ref{fig:appendix:D1W} but for bursts in Dm-L.
}\label{fig:appendix:DmL} 
\end{figure*}

\begin{figure*}
    \flushleft
    \includegraphics[height=37mm]{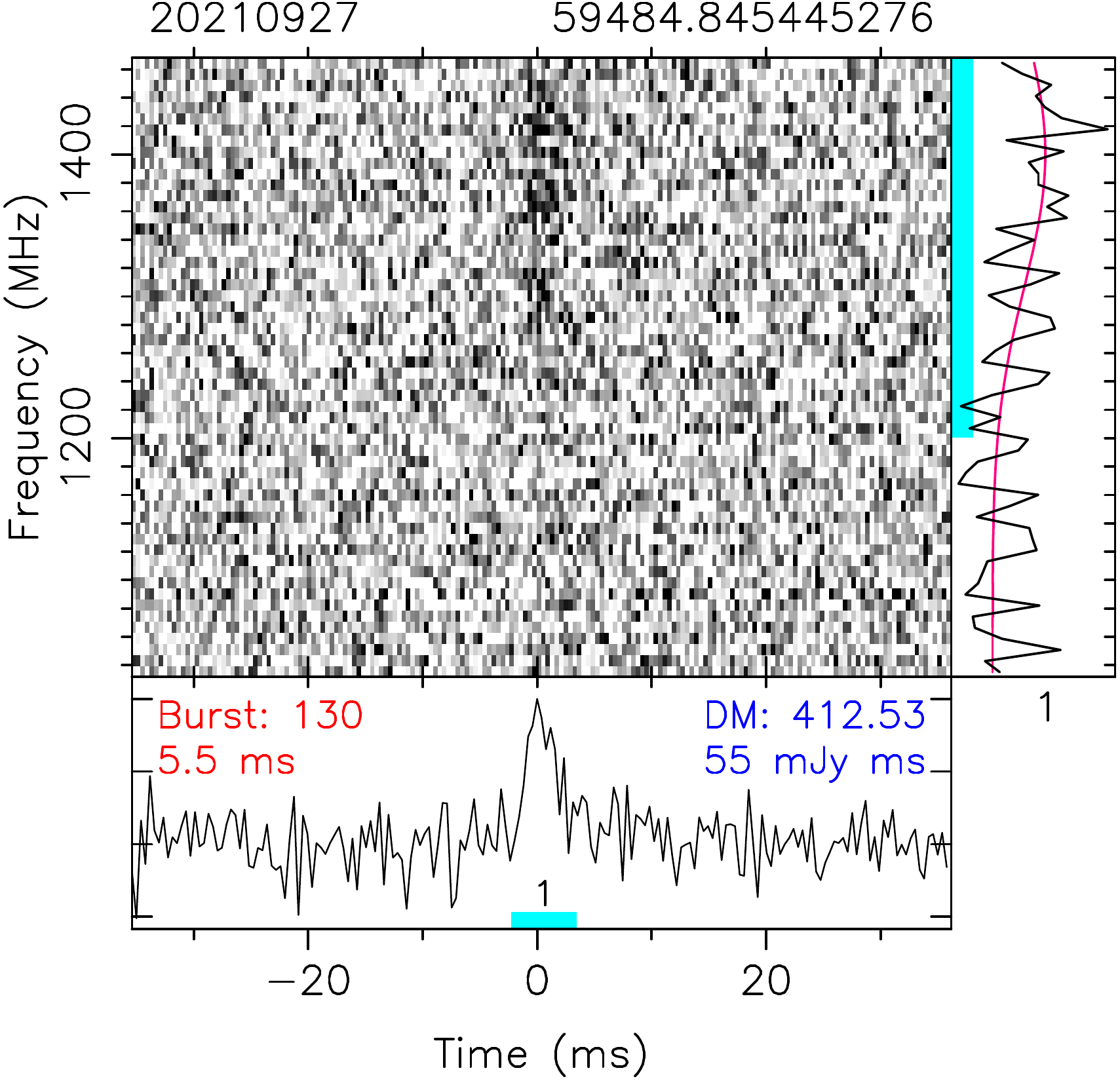}
    \includegraphics[height=37mm]{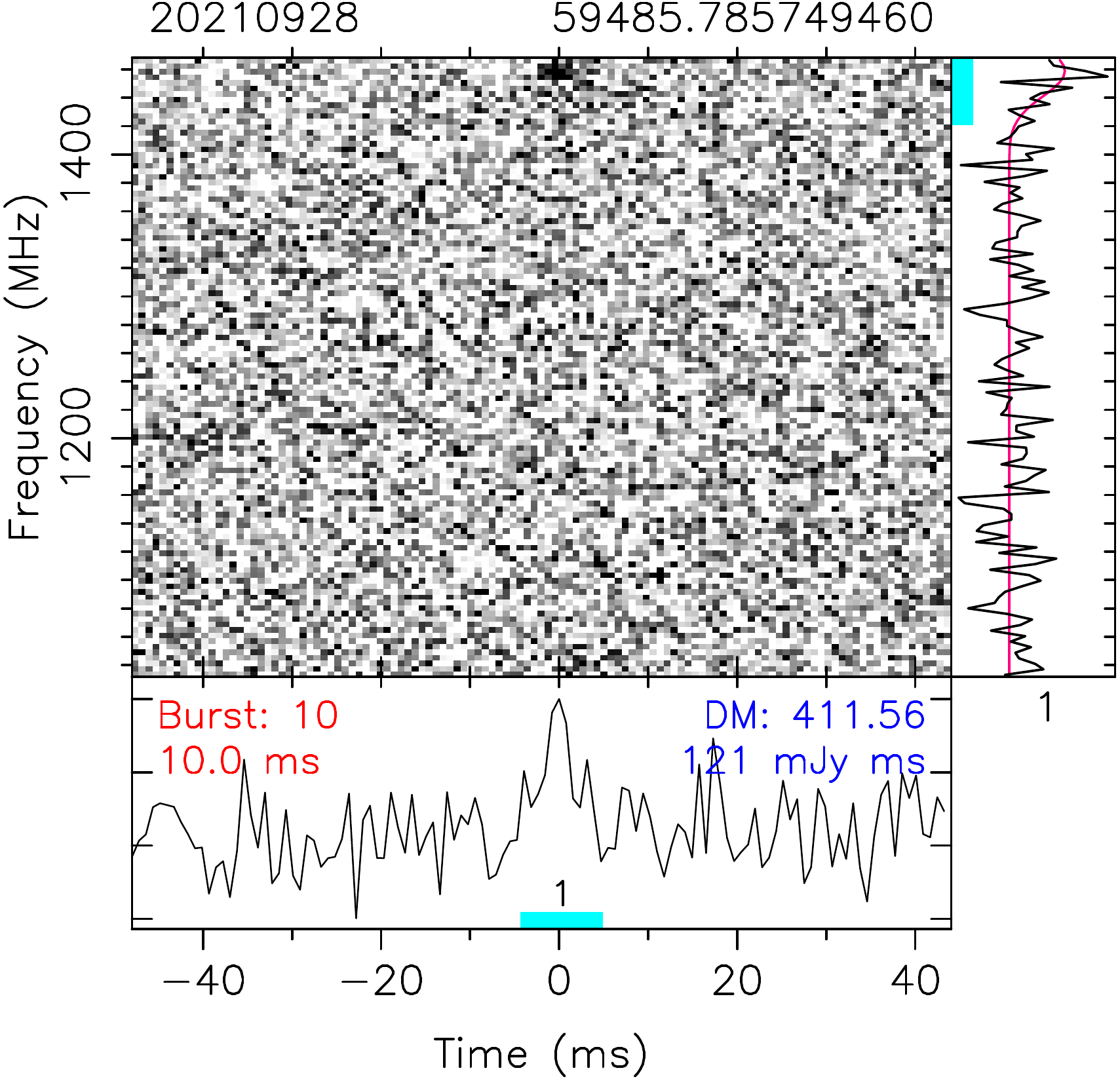}
    \includegraphics[height=37mm]{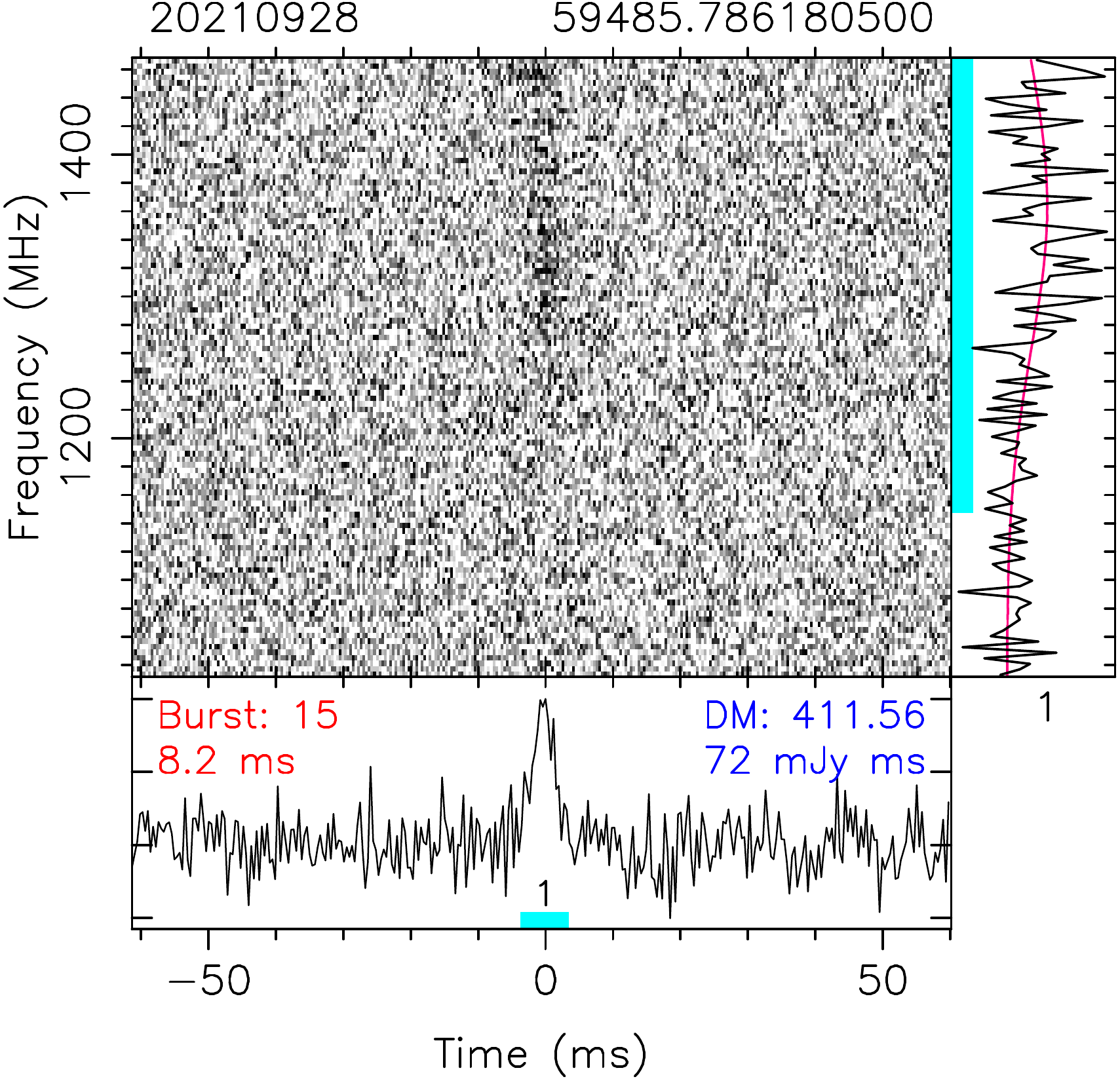}
    \includegraphics[height=37mm]{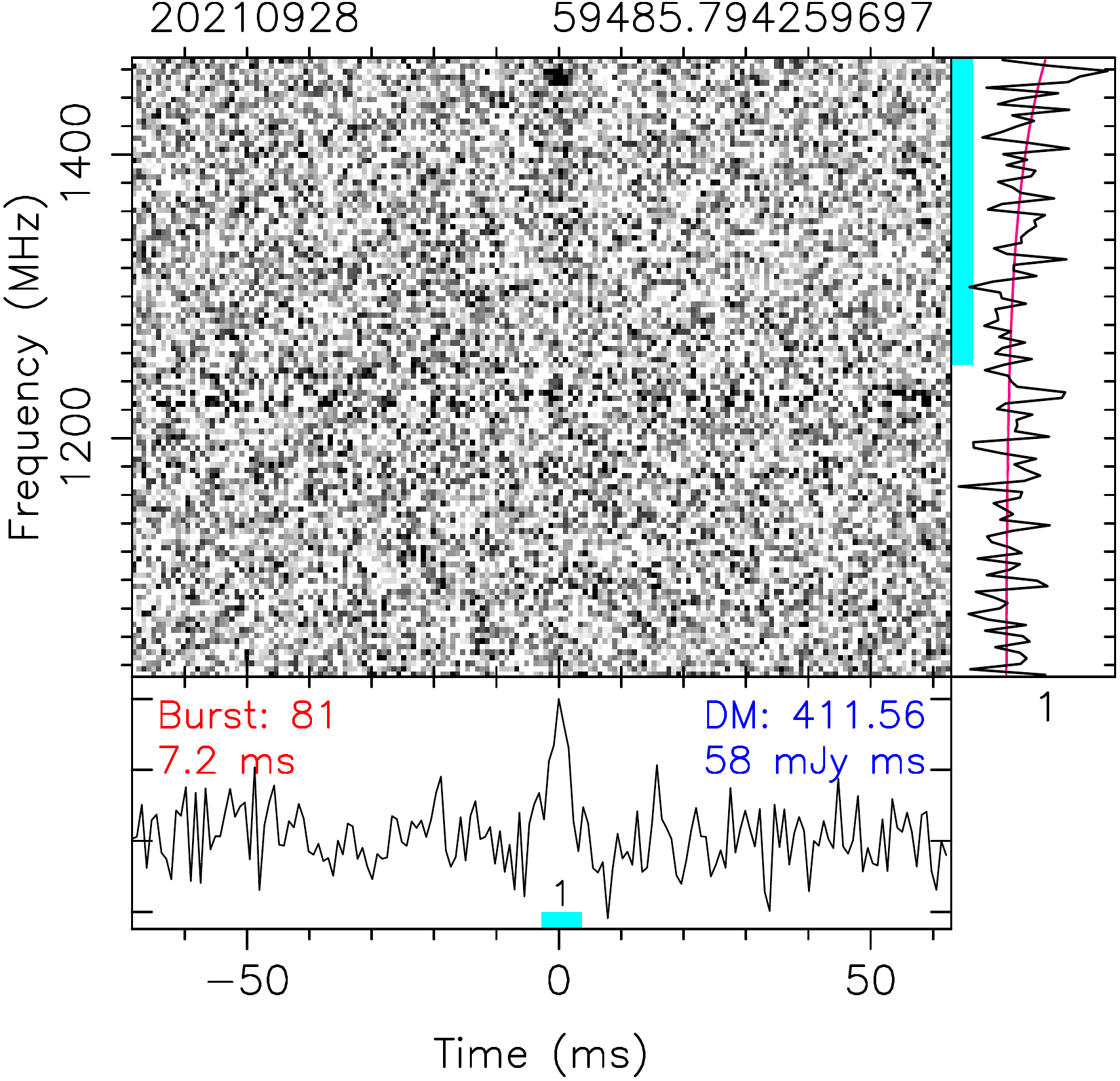}
    \includegraphics[height=37mm]{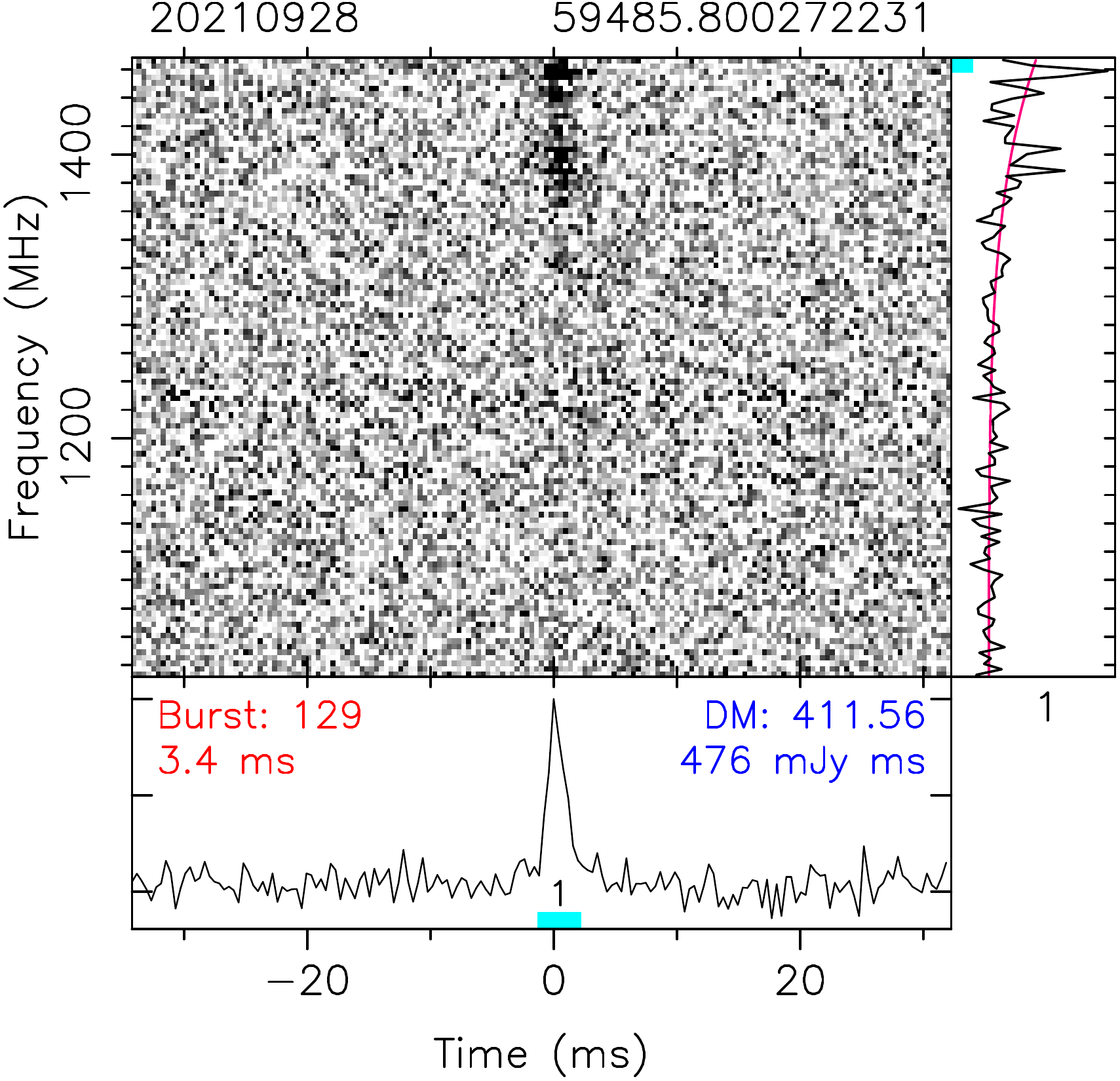}
    \includegraphics[height=37mm]{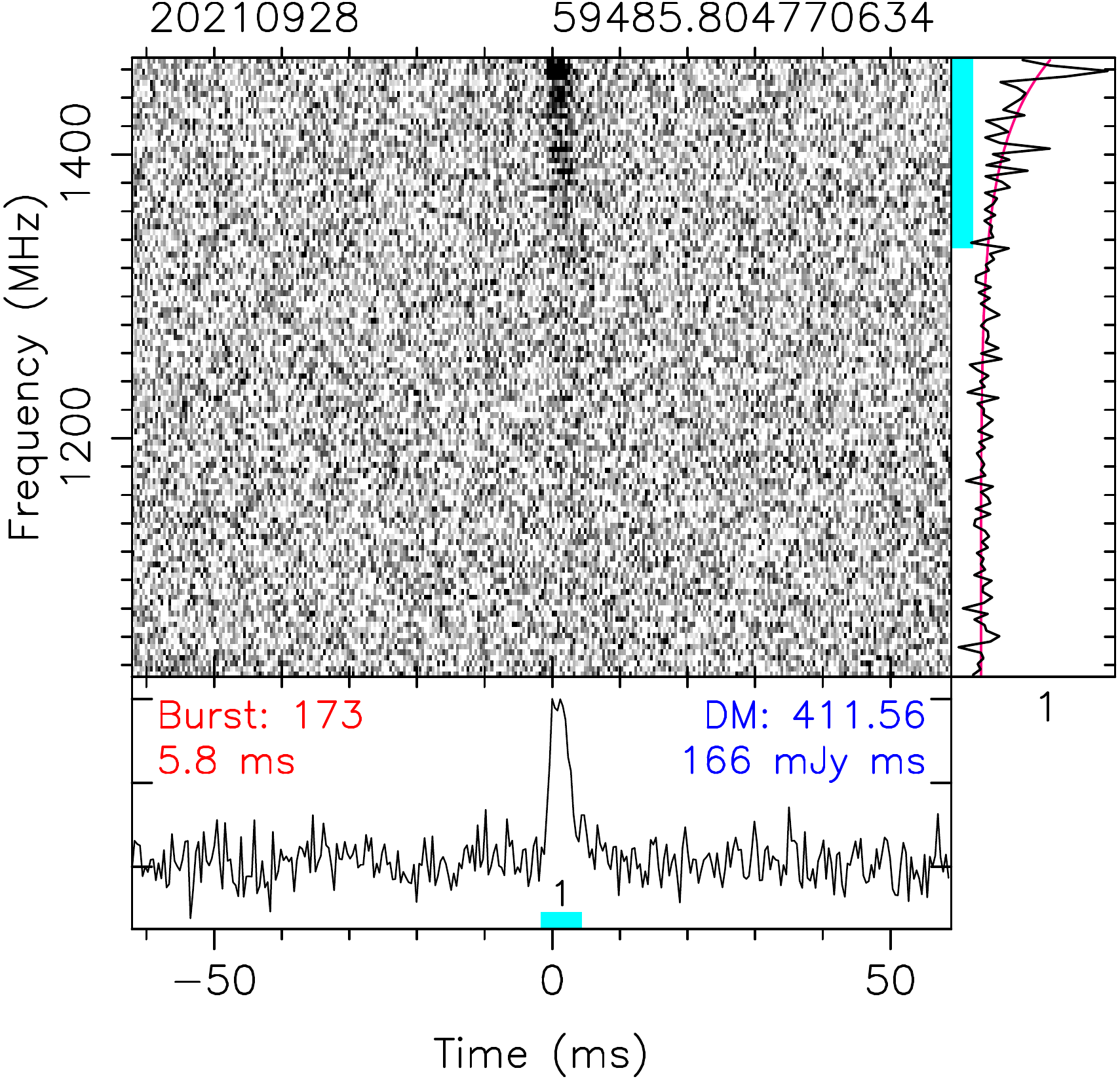}
    \includegraphics[height=37mm]{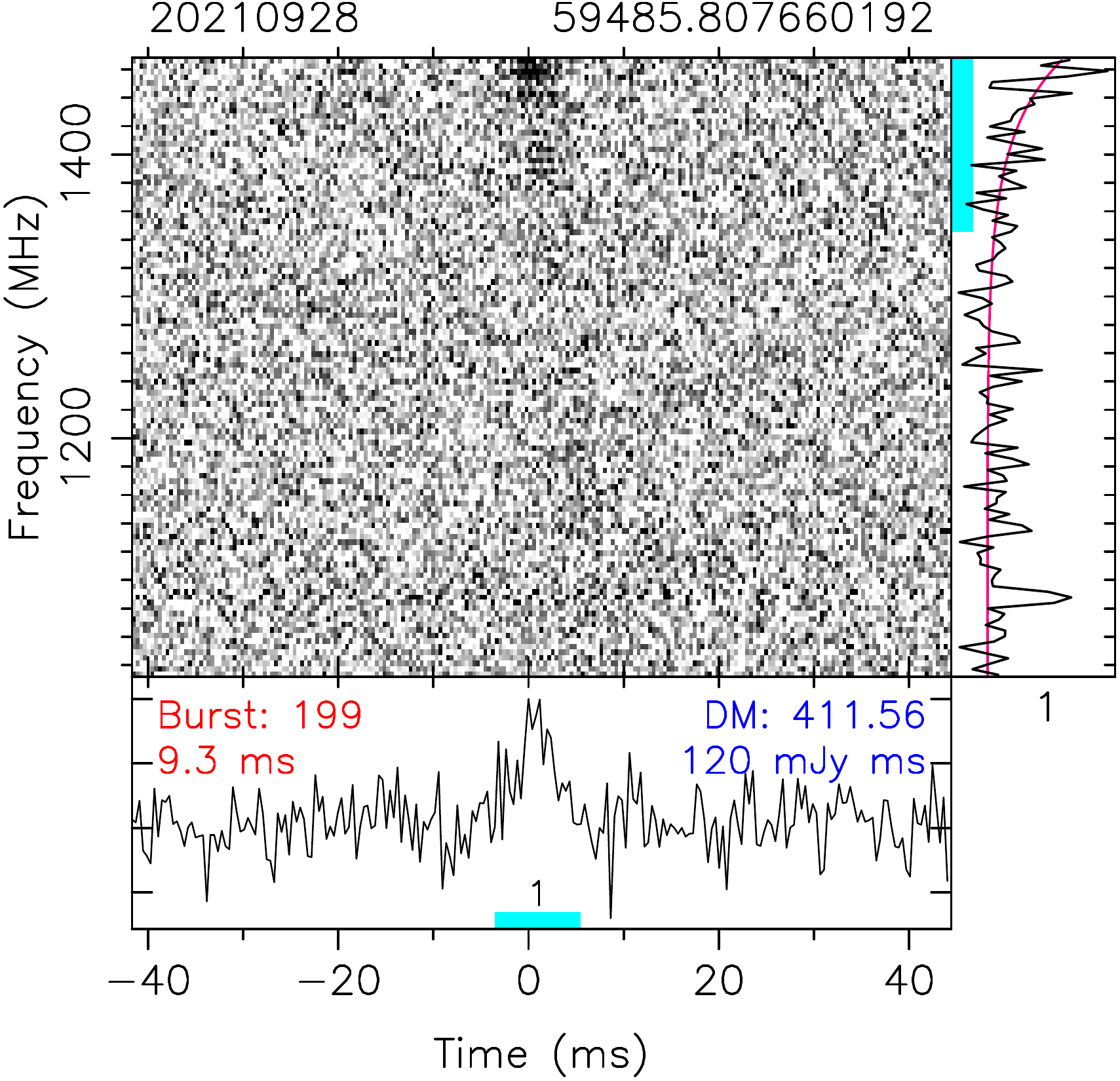}
    \includegraphics[height=37mm]{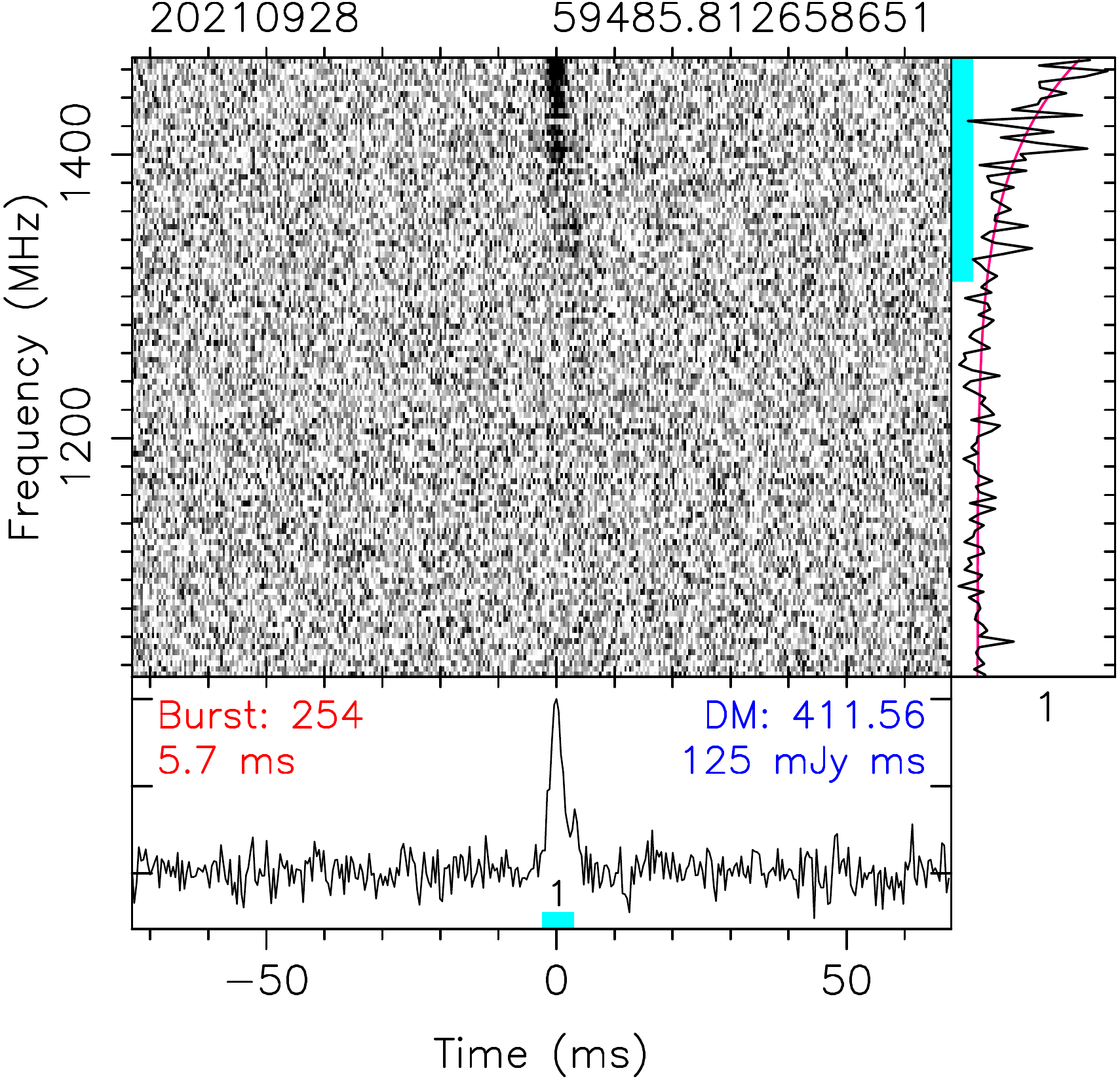}
    \includegraphics[height=37mm]{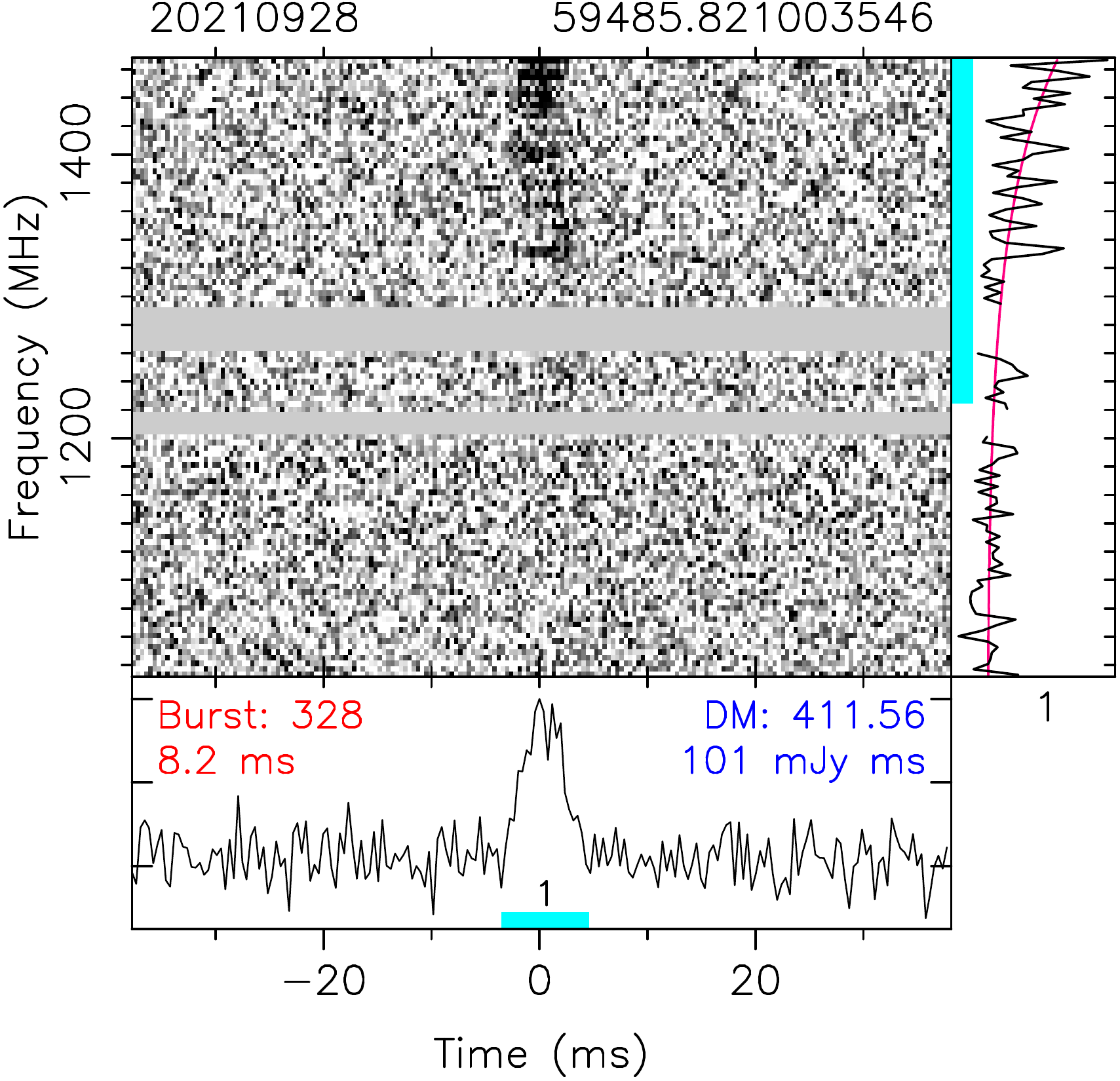}
    \includegraphics[height=37mm]{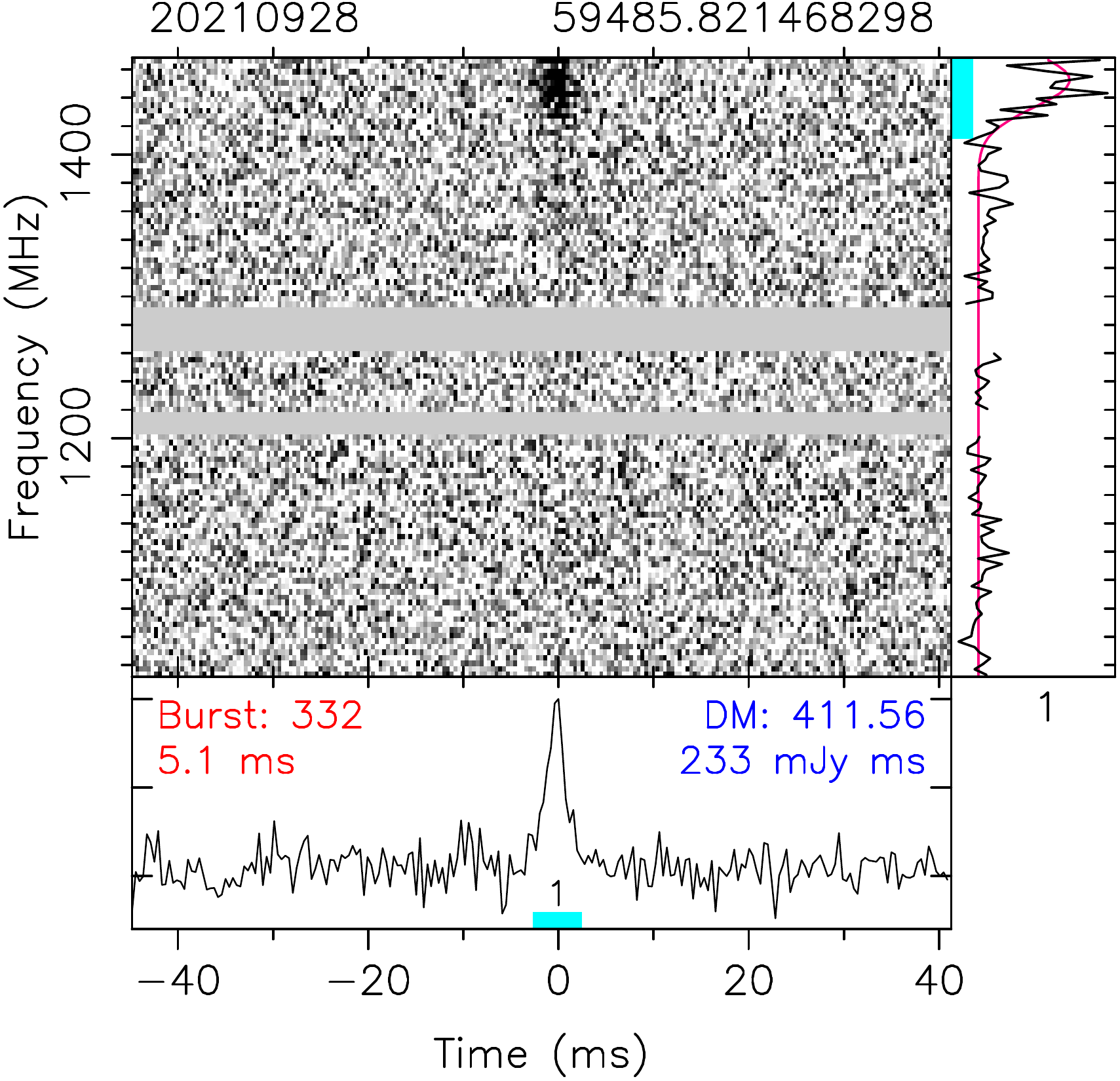}
\caption{The same as Figure~\ref{fig:appendix:D1W} but for bursts at higher band with no evidence of drifting (NE-H).
}\label{fig:appendix:NEH} 
\end{figure*}

\clearpage
\begin{figure*}
    \flushleft
    \includegraphics[height=37mm]{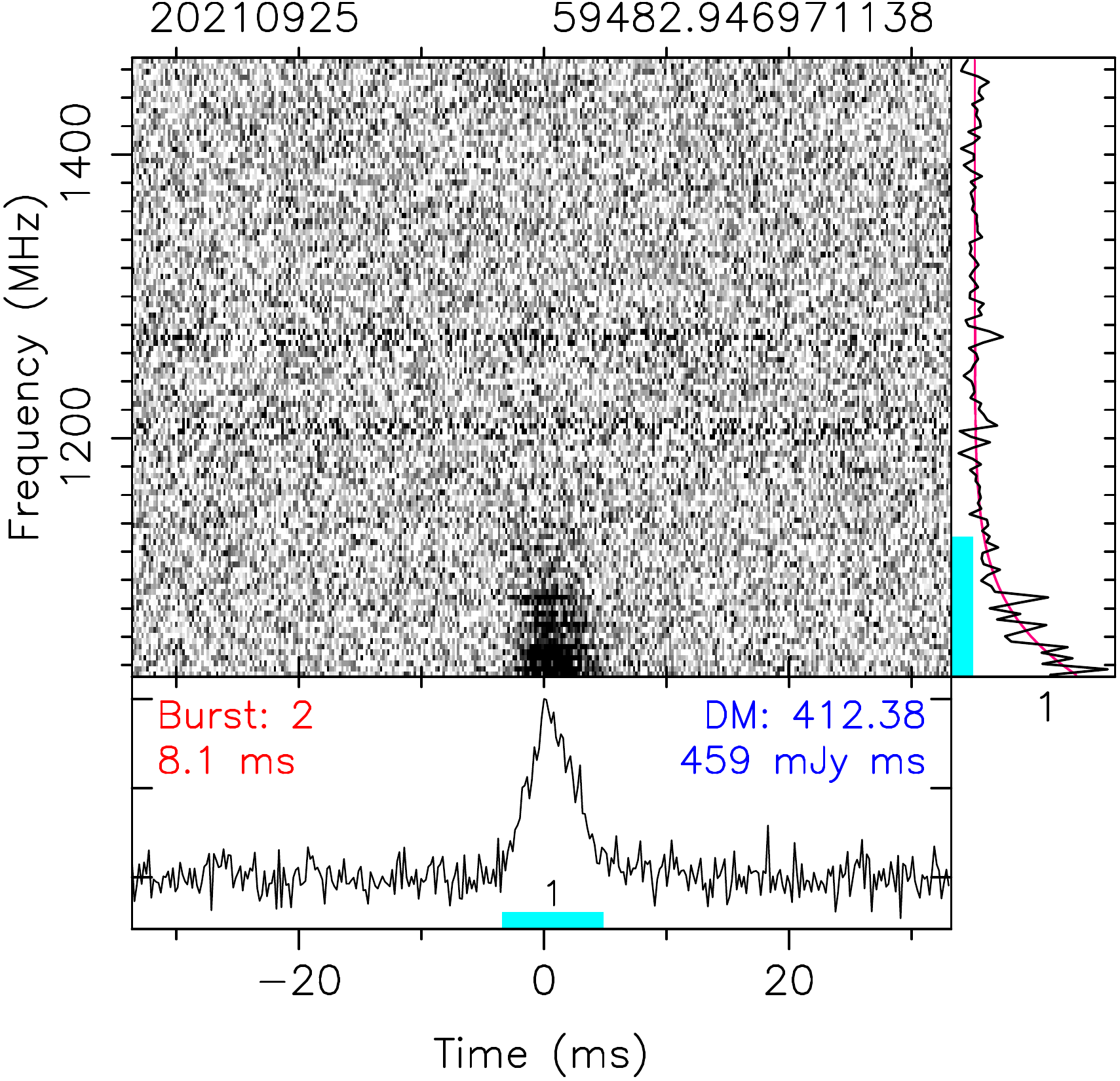}
    \includegraphics[height=37mm]{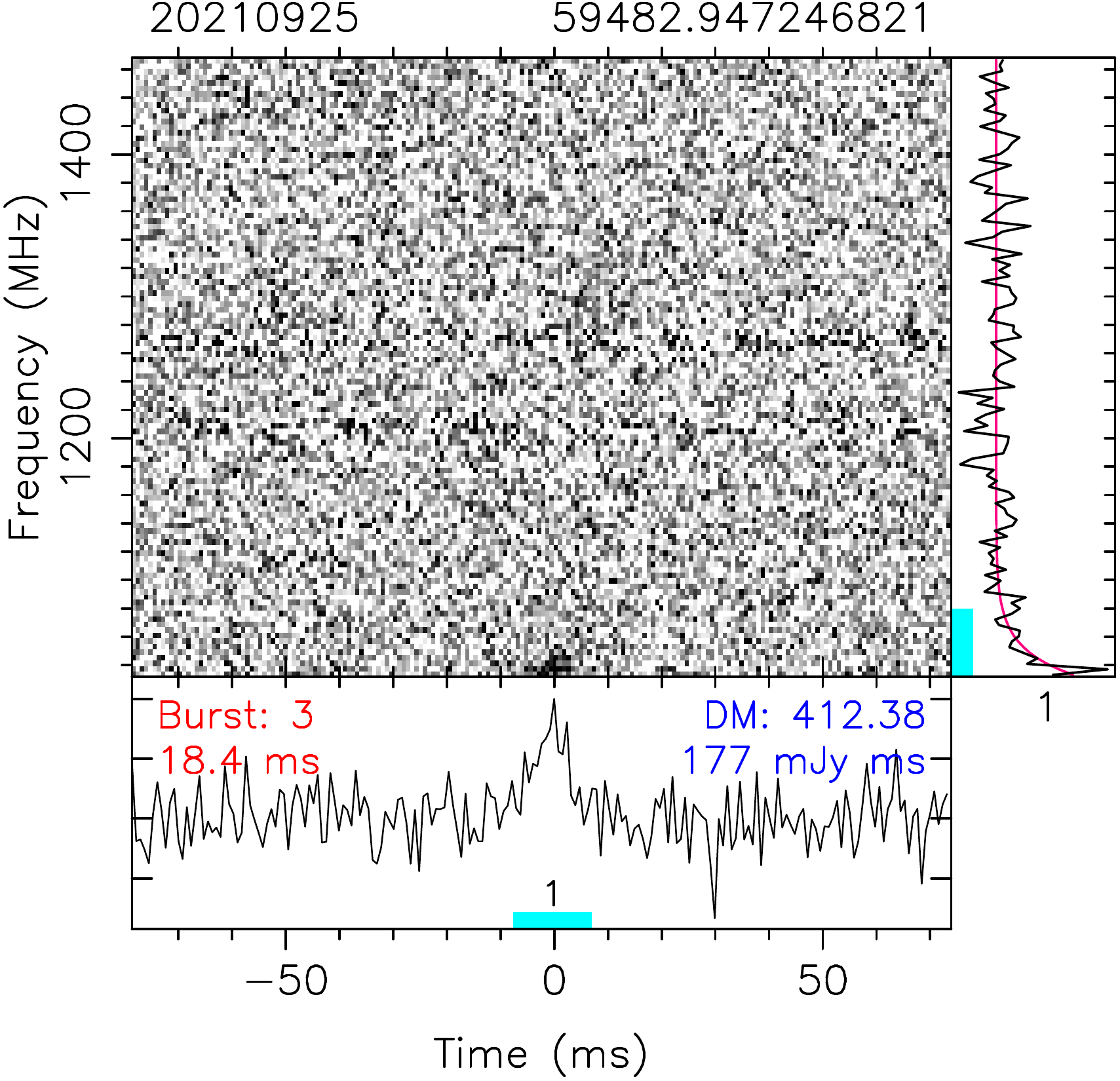}
    \includegraphics[height=37mm]{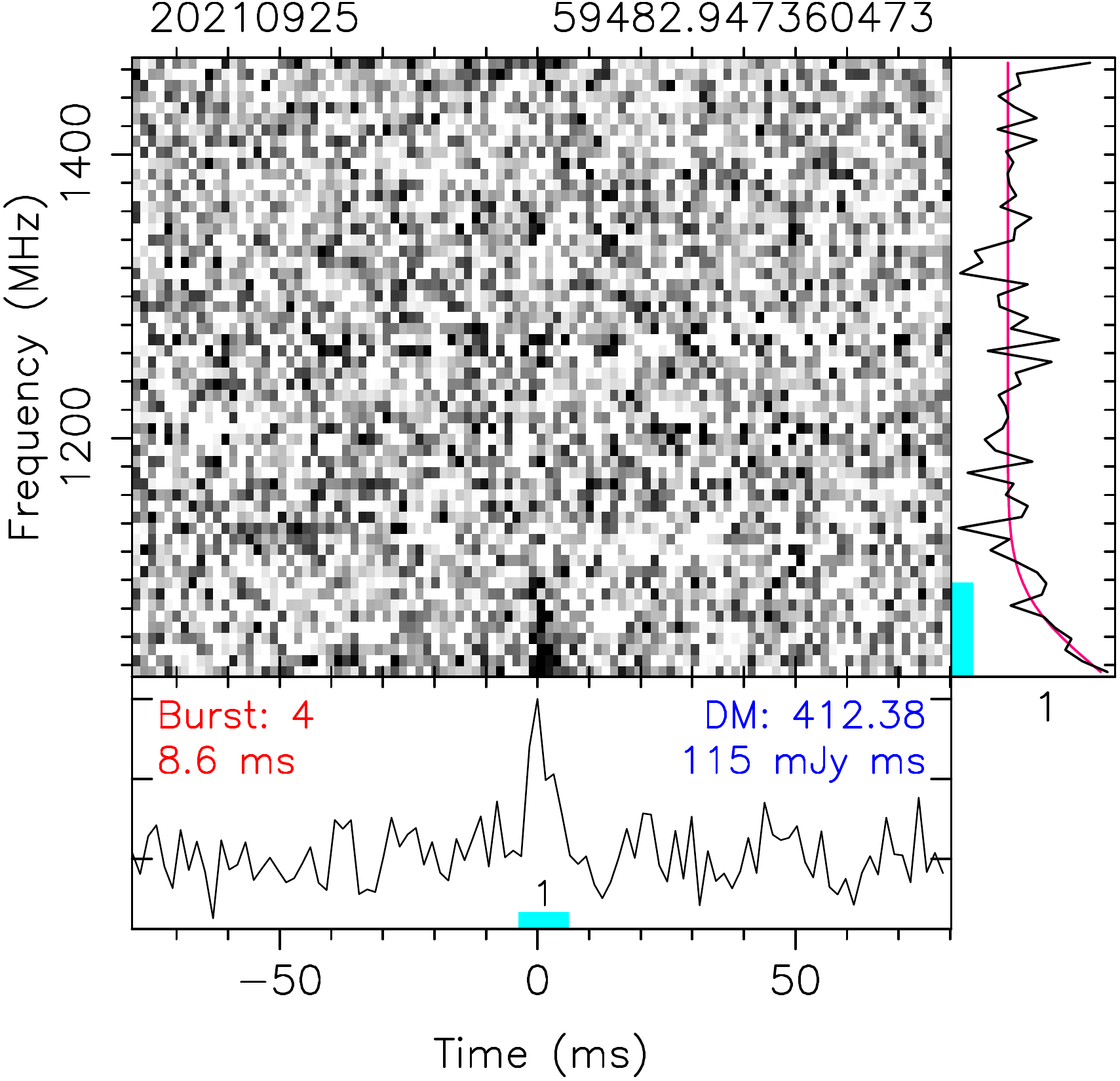}
    \includegraphics[height=37mm]{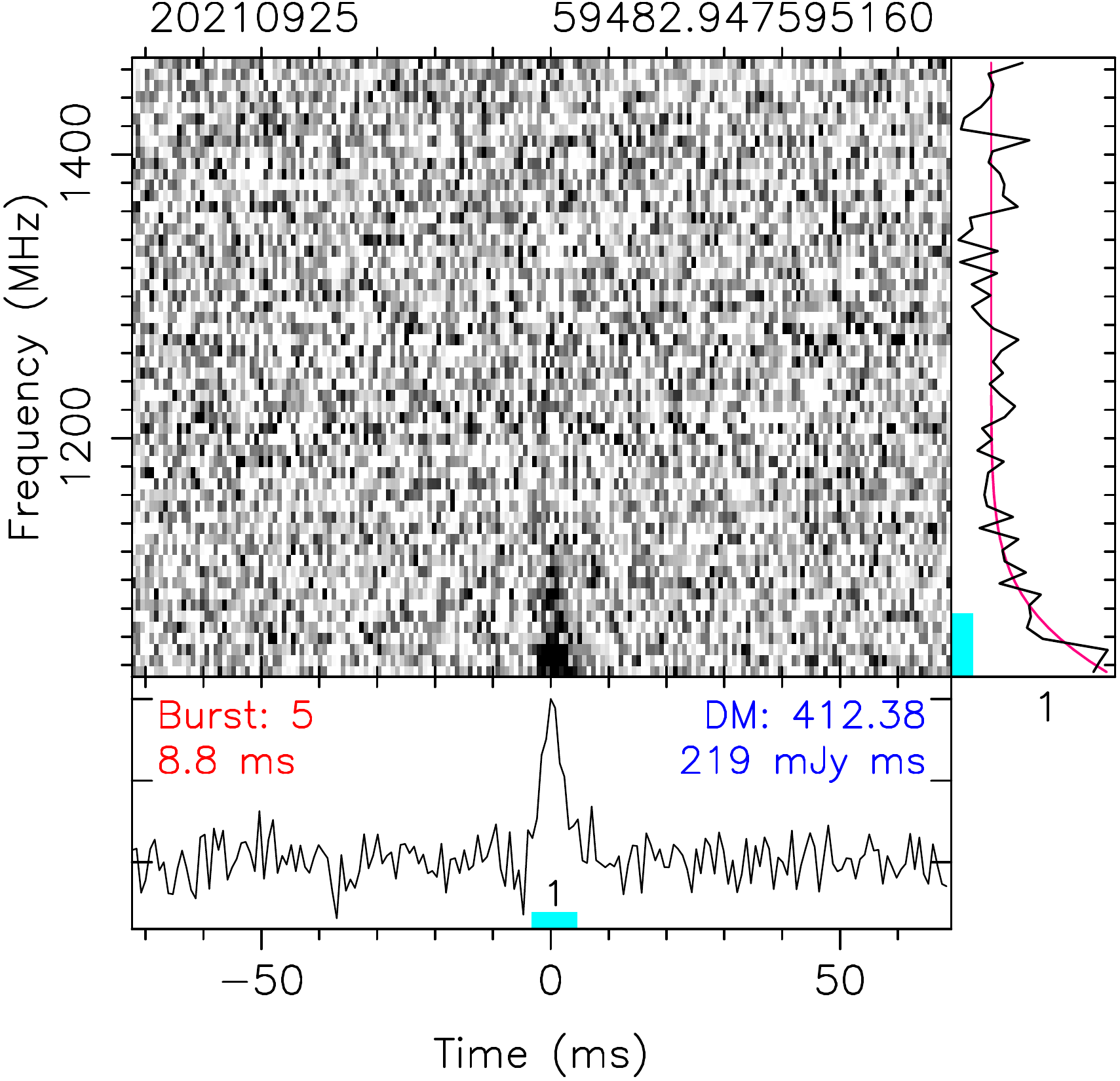}
    \includegraphics[height=37mm]{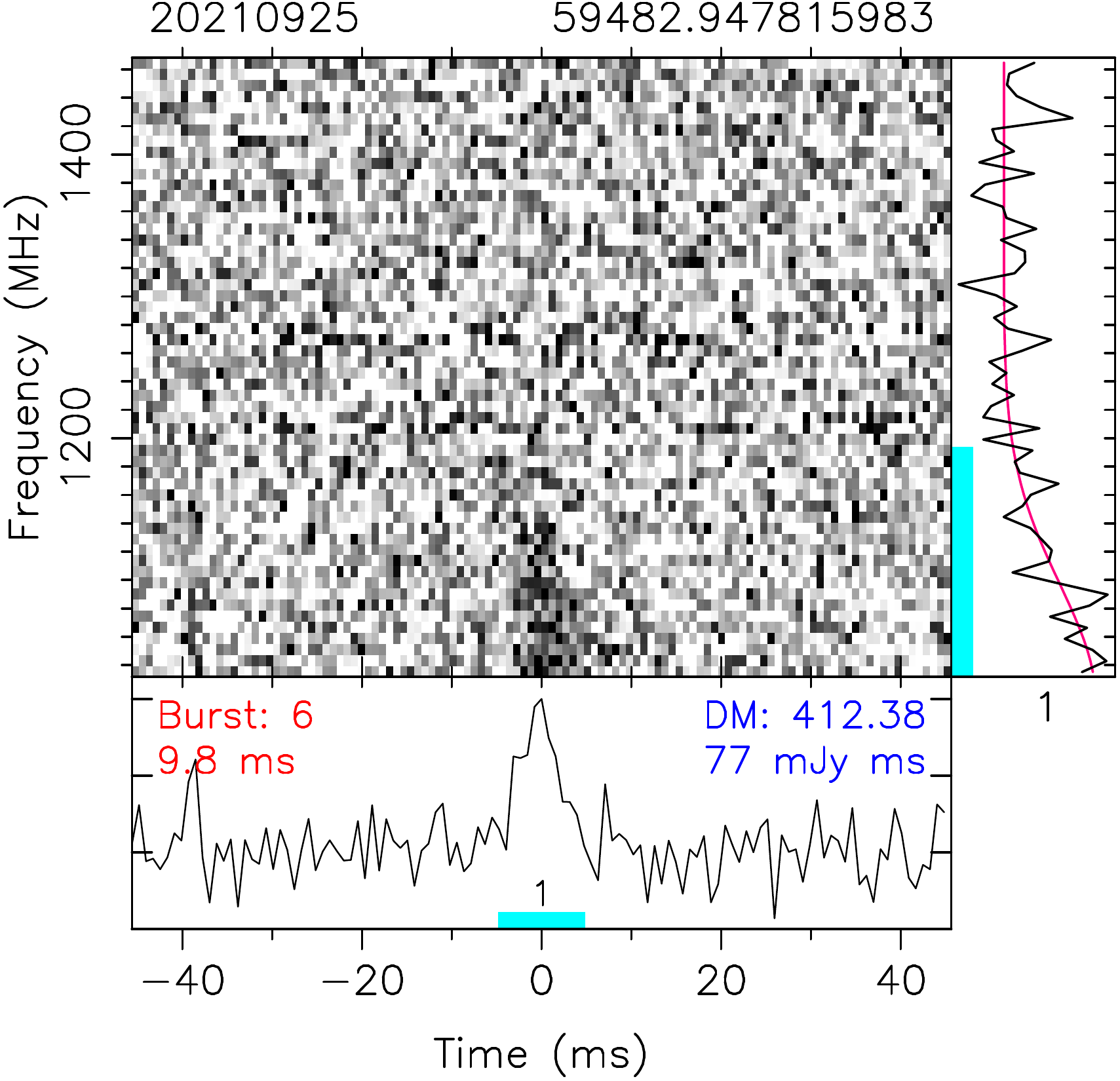}
    \includegraphics[height=37mm]{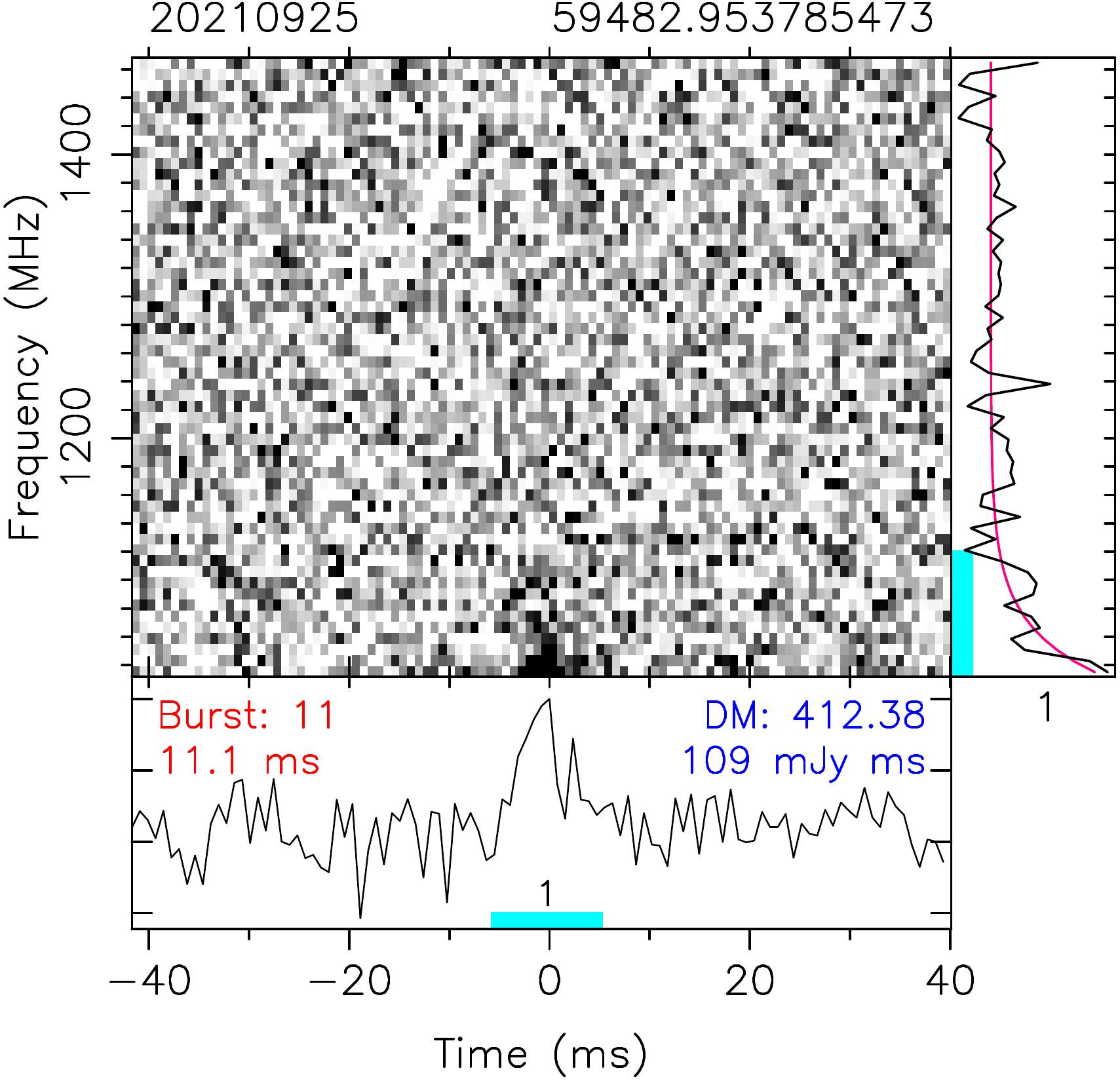}
    \includegraphics[height=37mm]{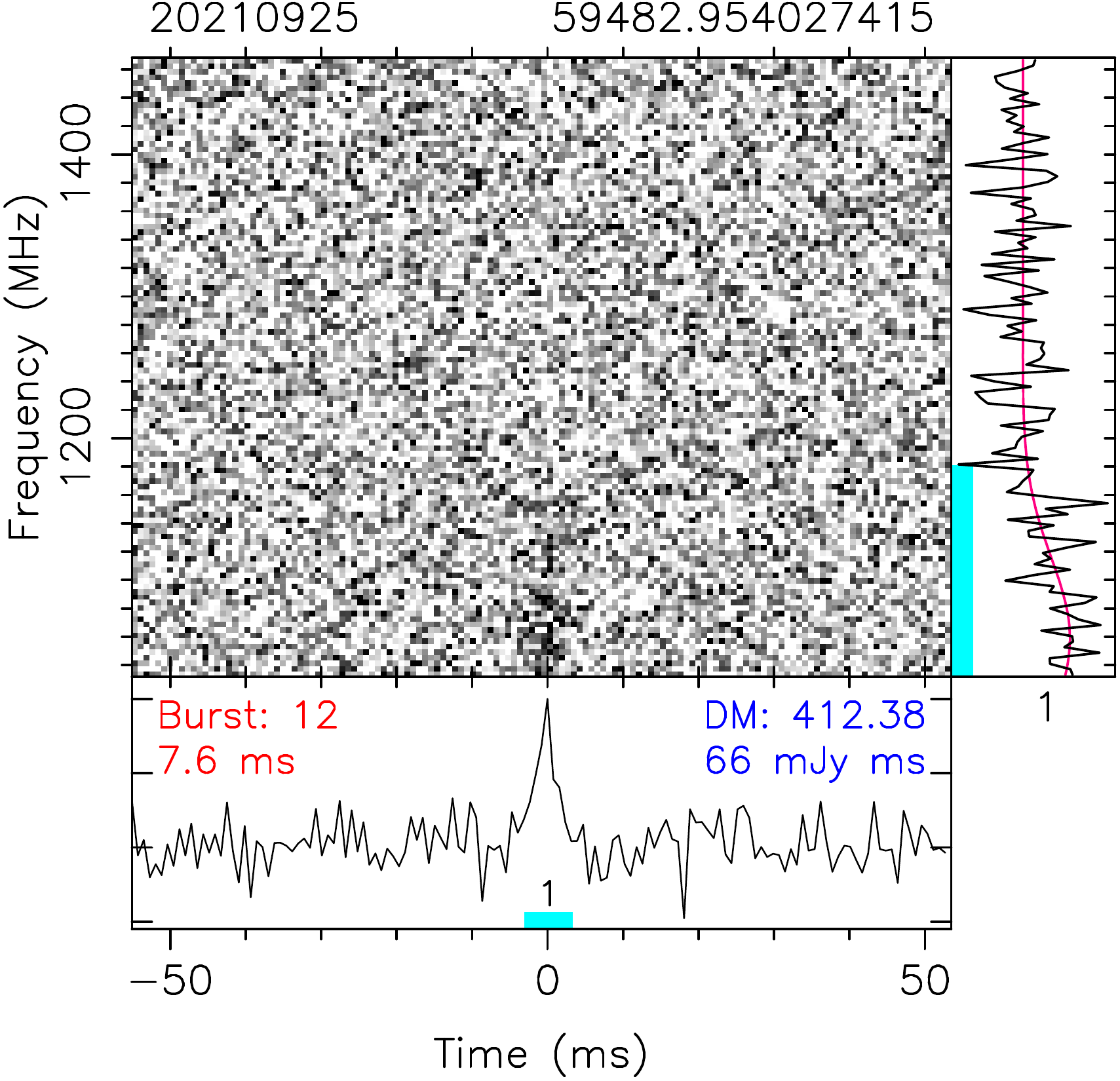}
    \includegraphics[height=37mm]{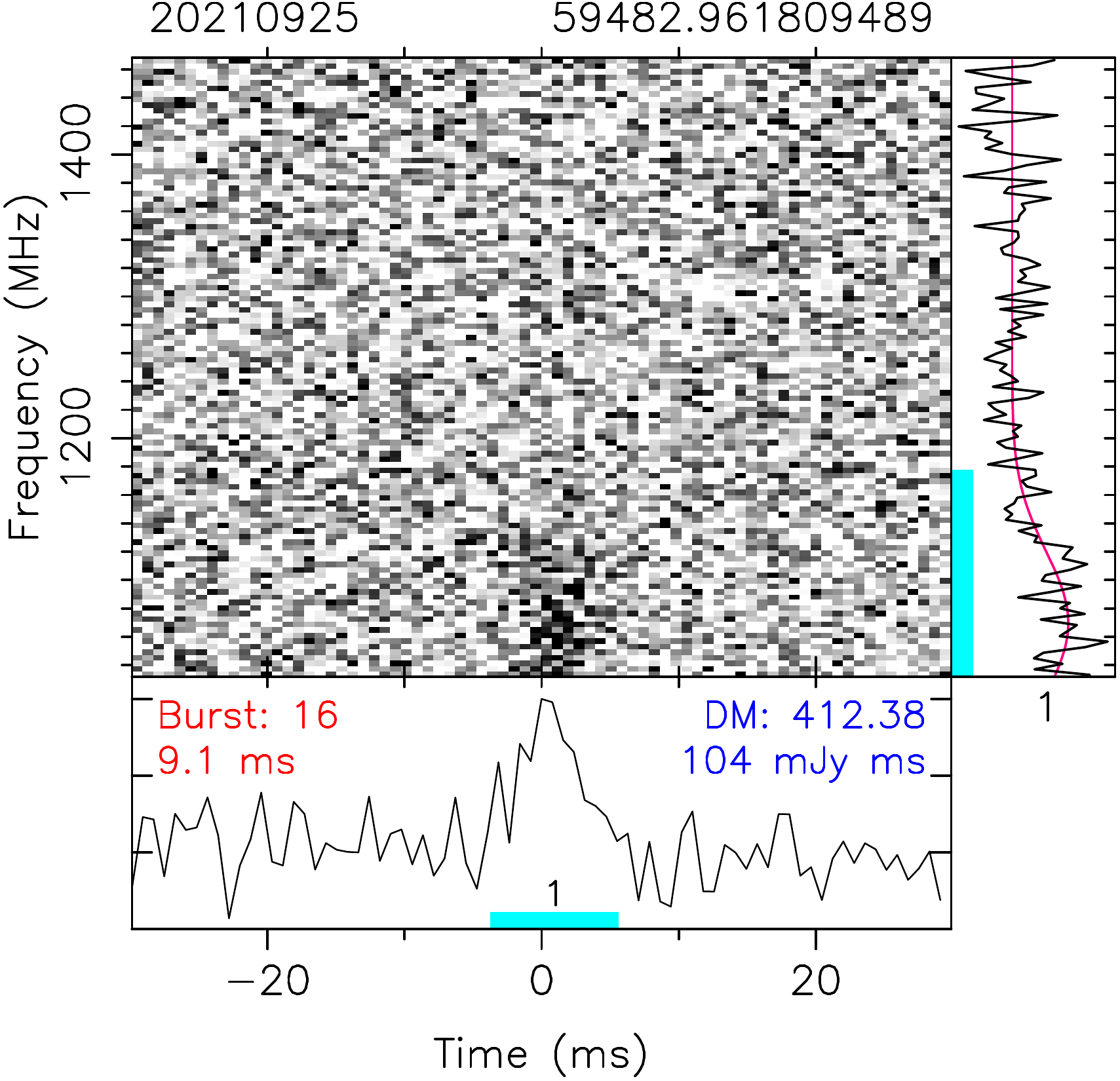}
    \includegraphics[height=37mm]{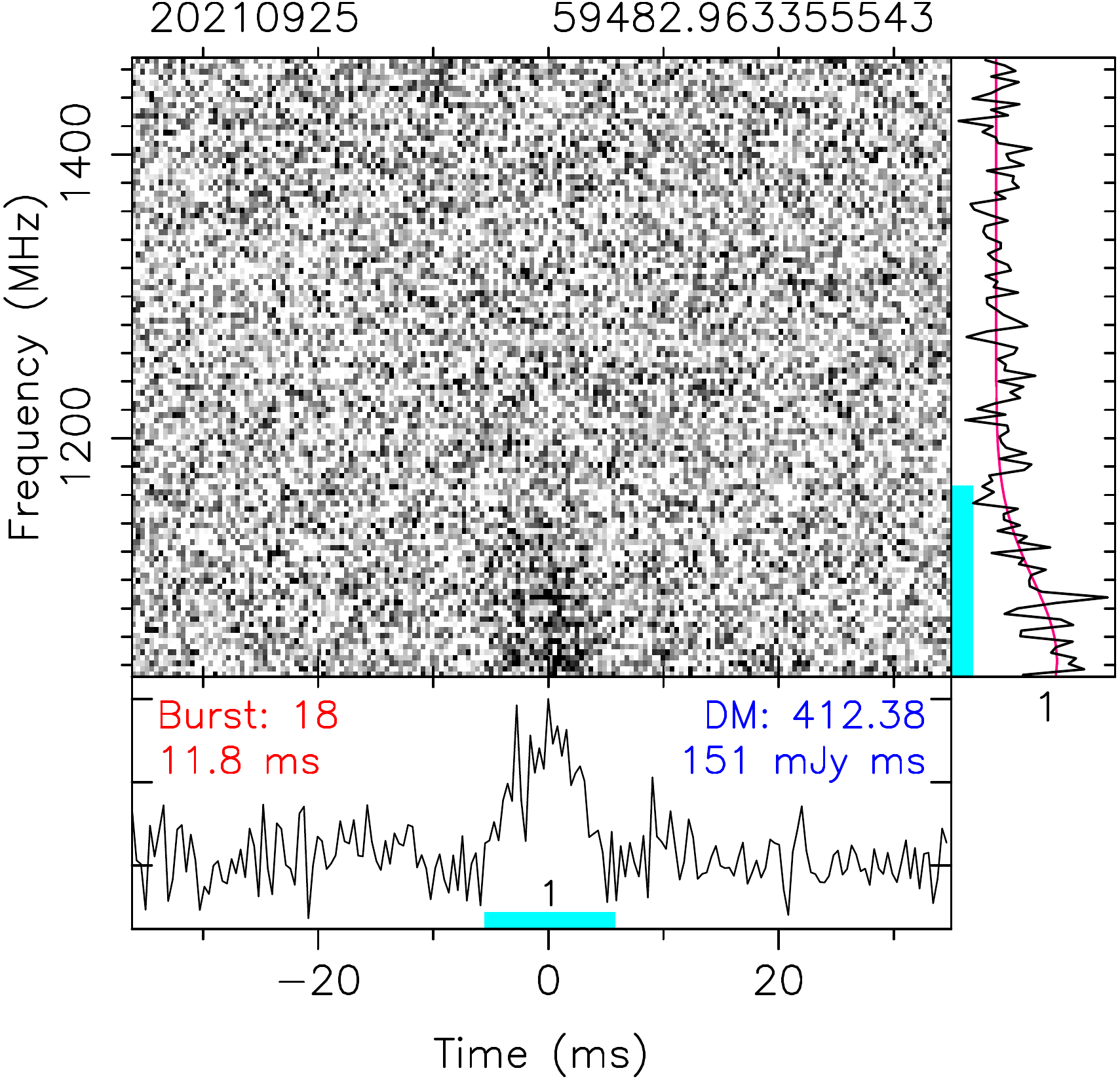}
    \includegraphics[height=37mm]{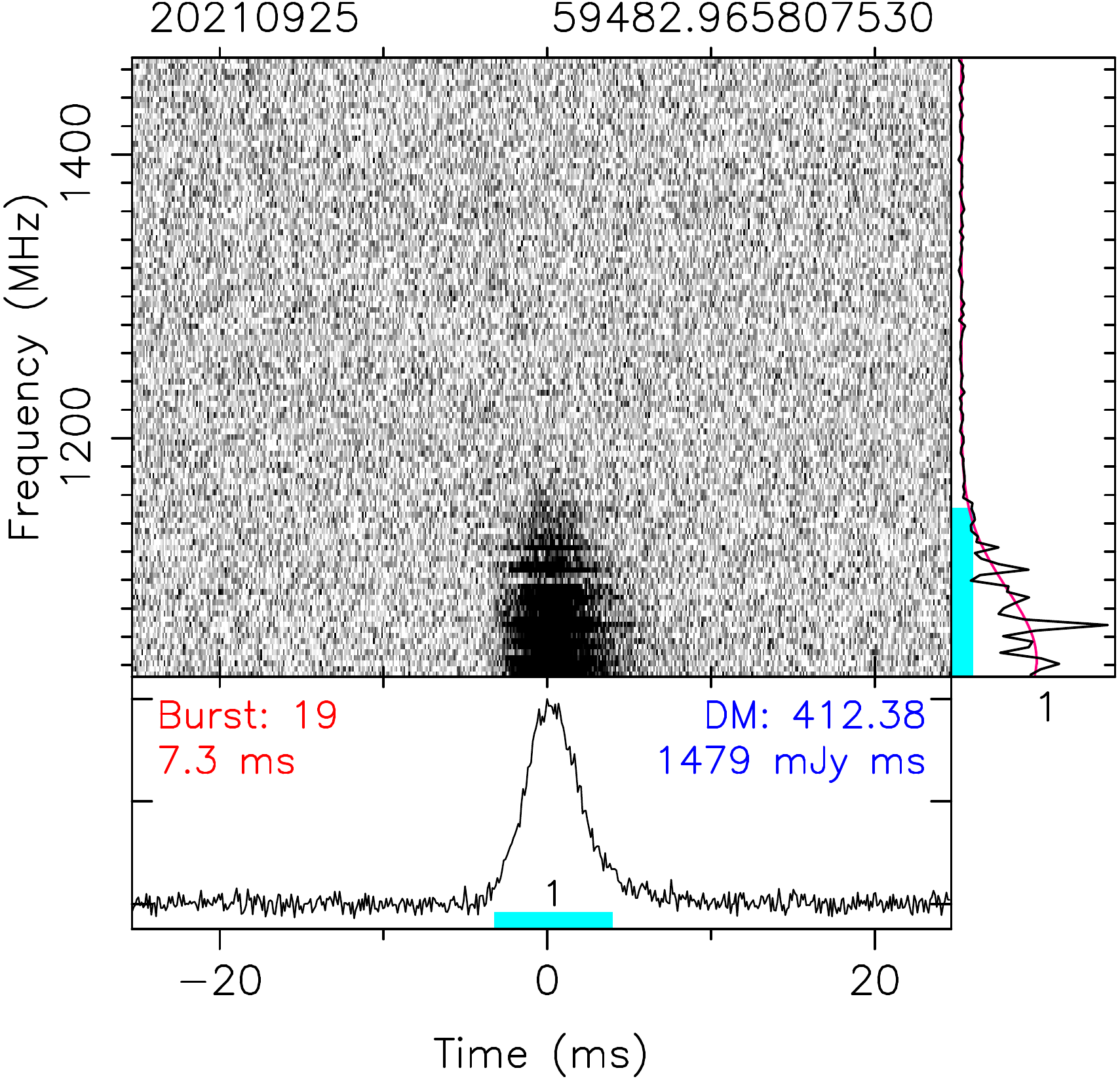}
    \includegraphics[height=37mm]{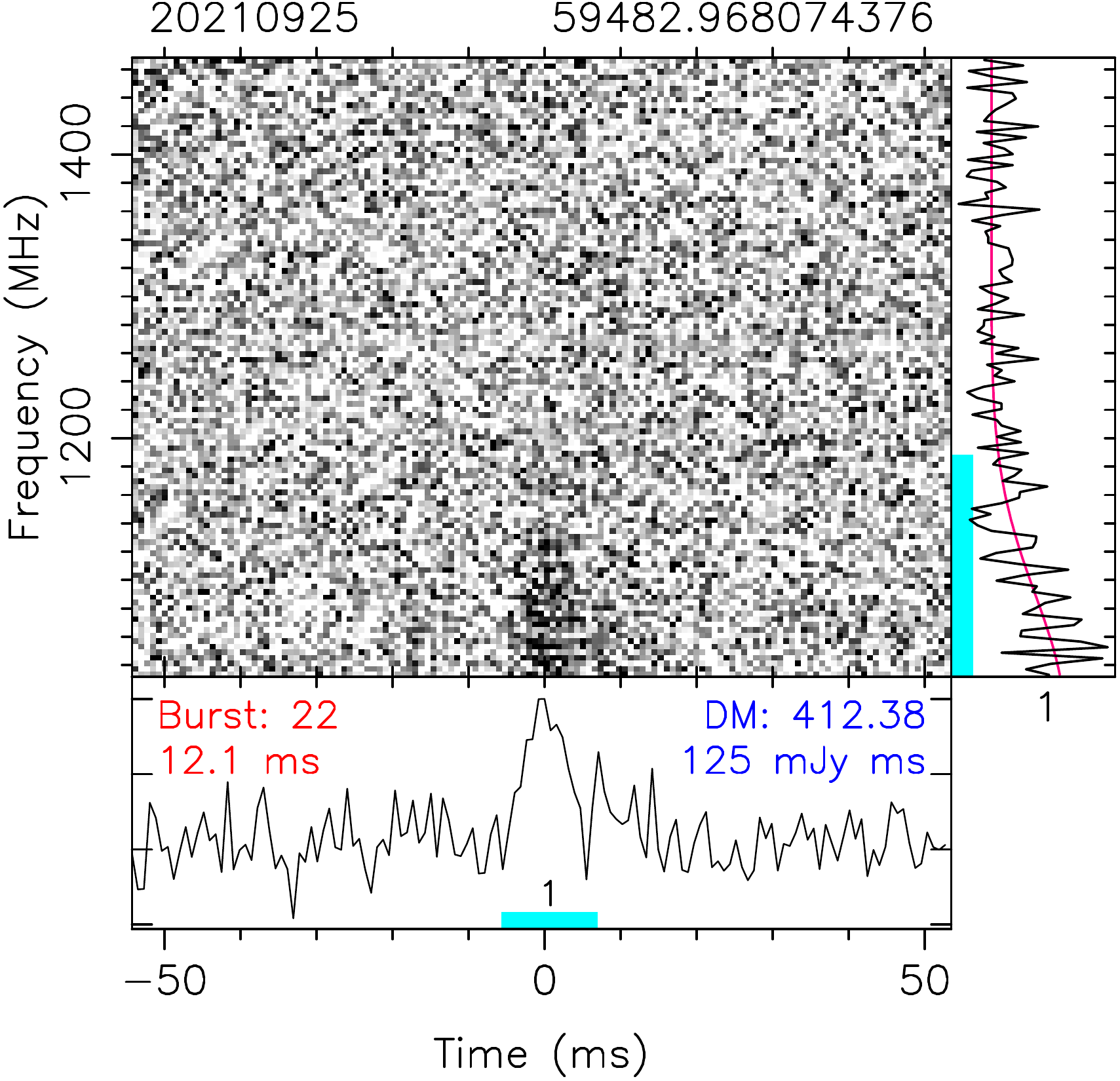}
    \includegraphics[height=37mm]{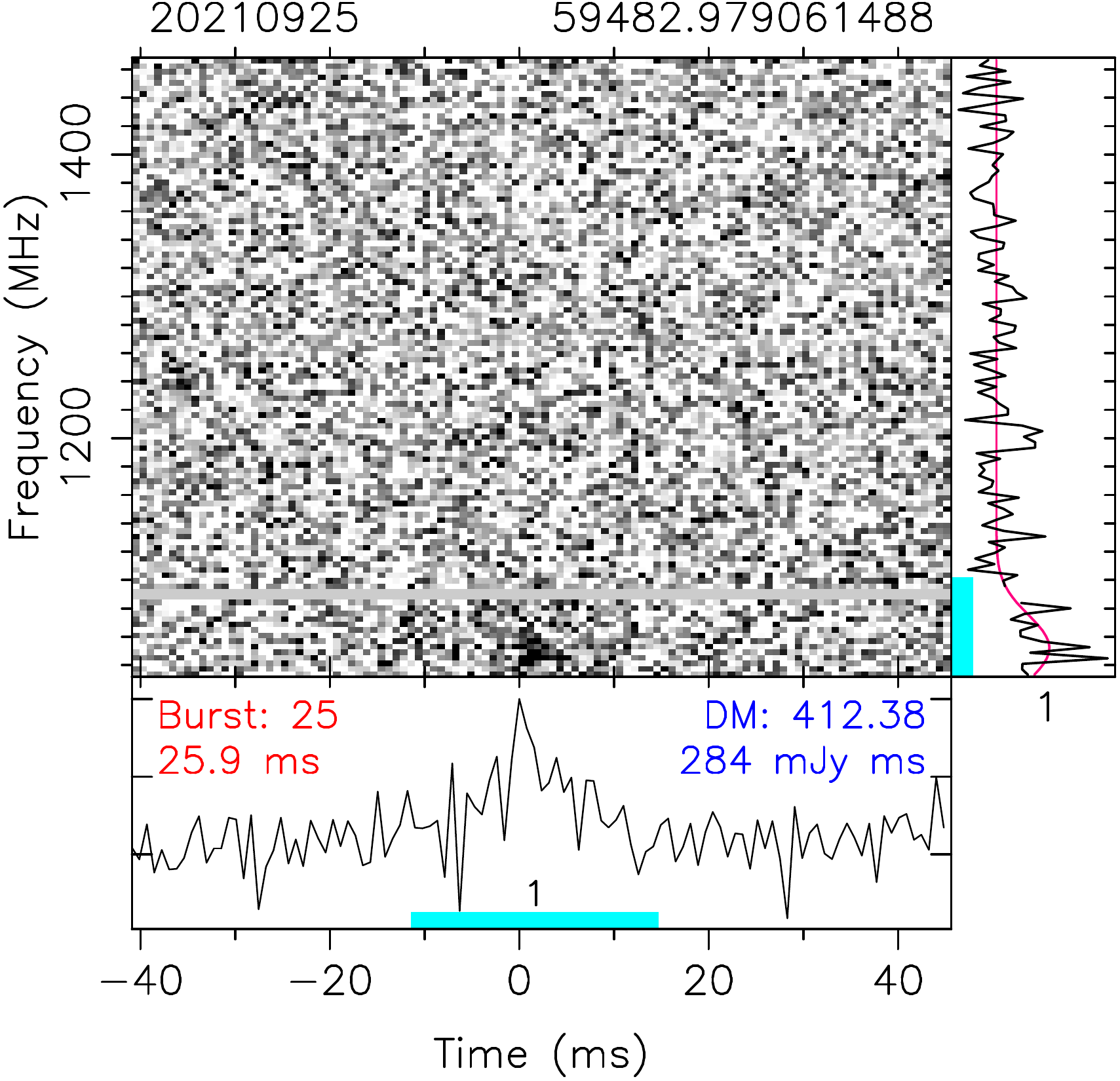}
    \includegraphics[height=37mm]{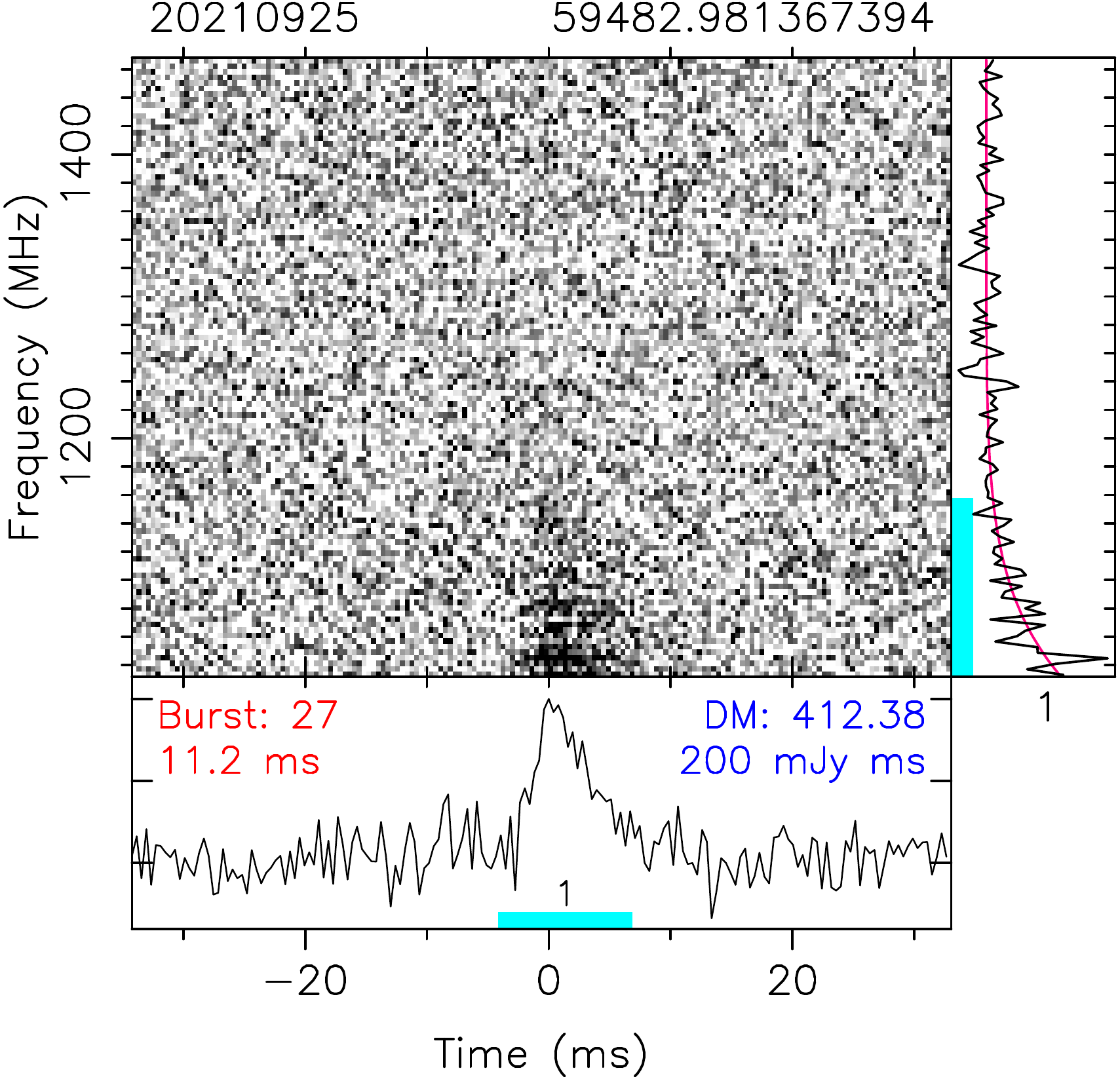}
    \includegraphics[height=37mm]{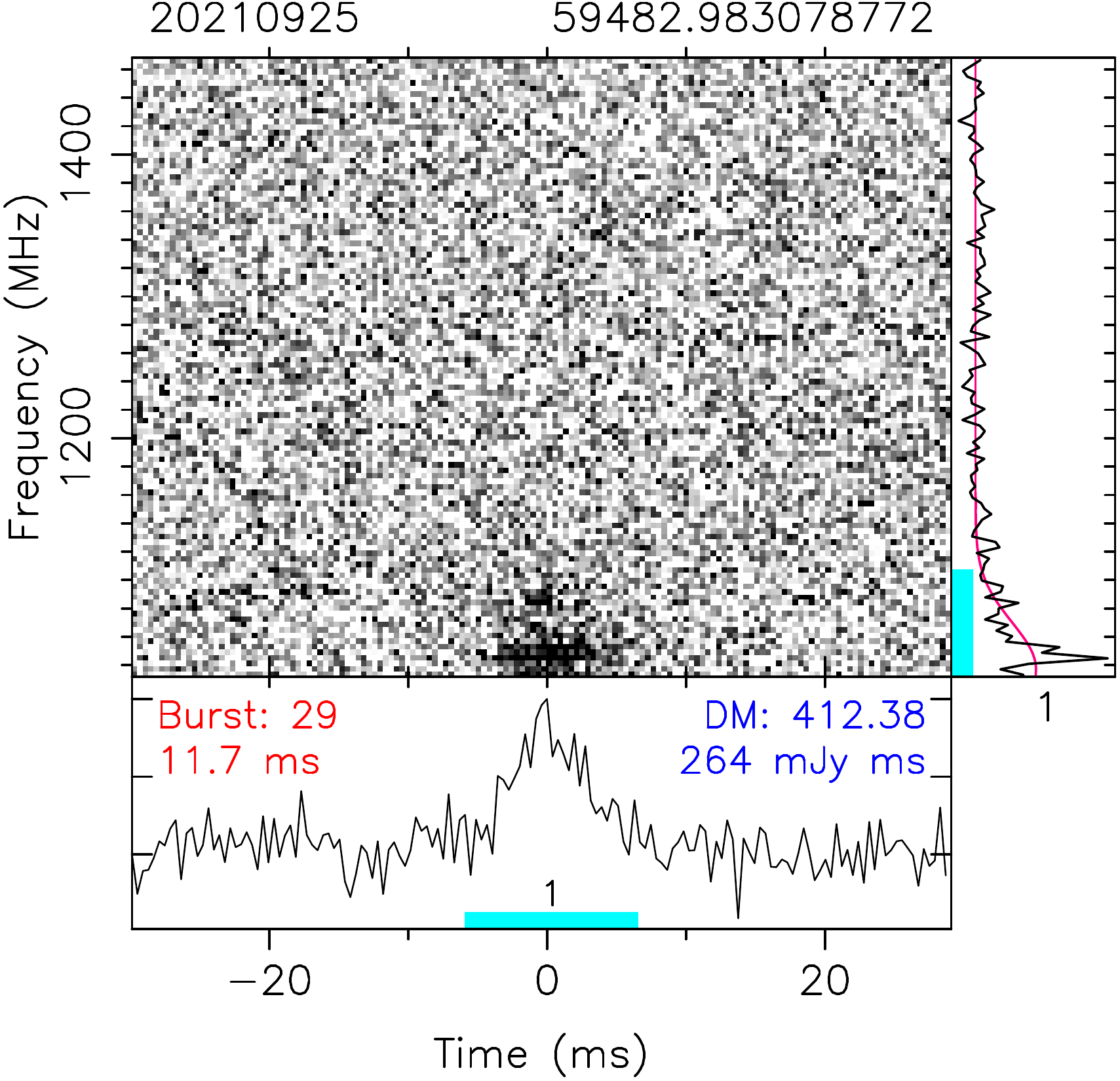}
    \includegraphics[height=37mm]{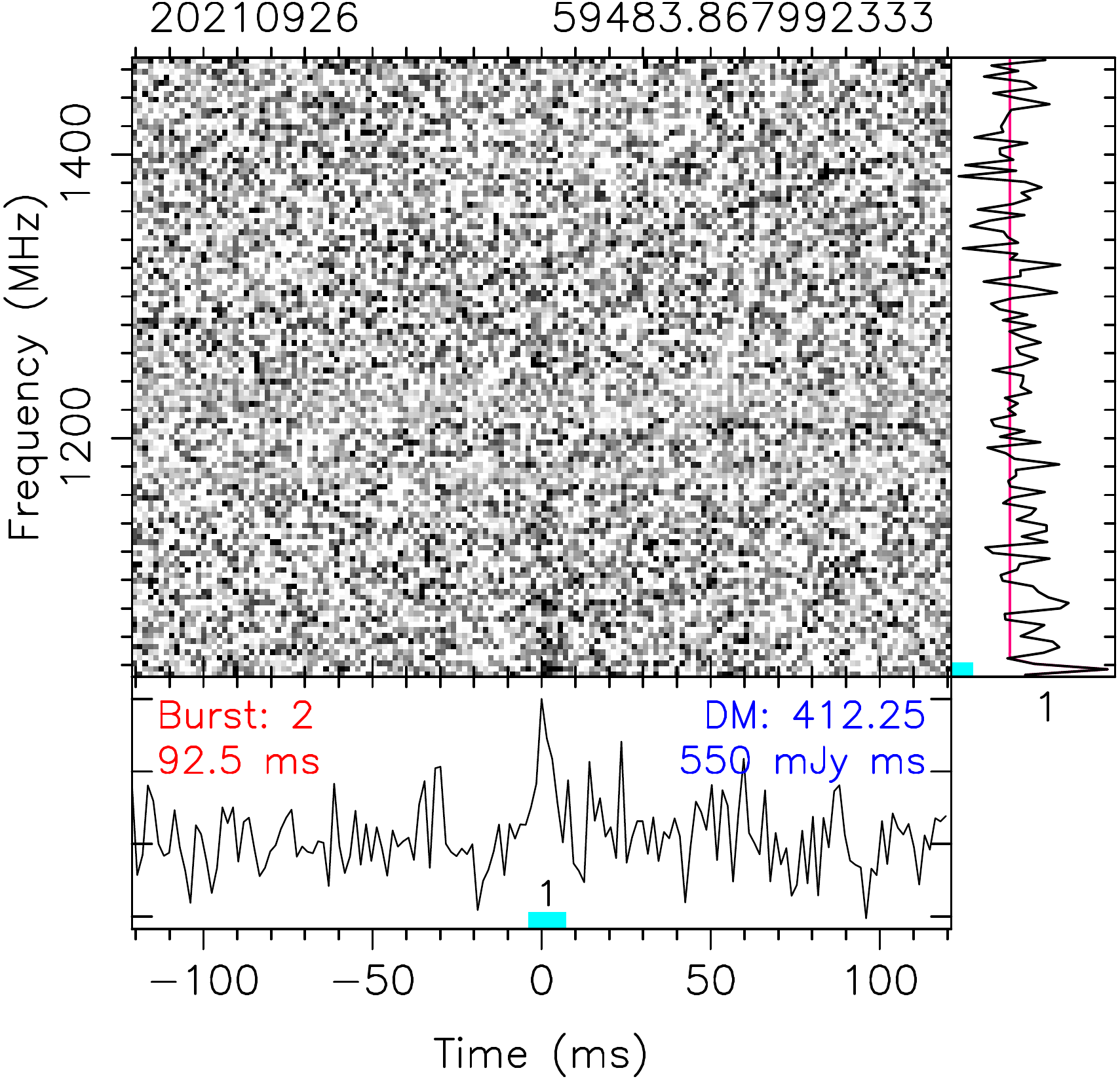}
    \includegraphics[height=37mm]{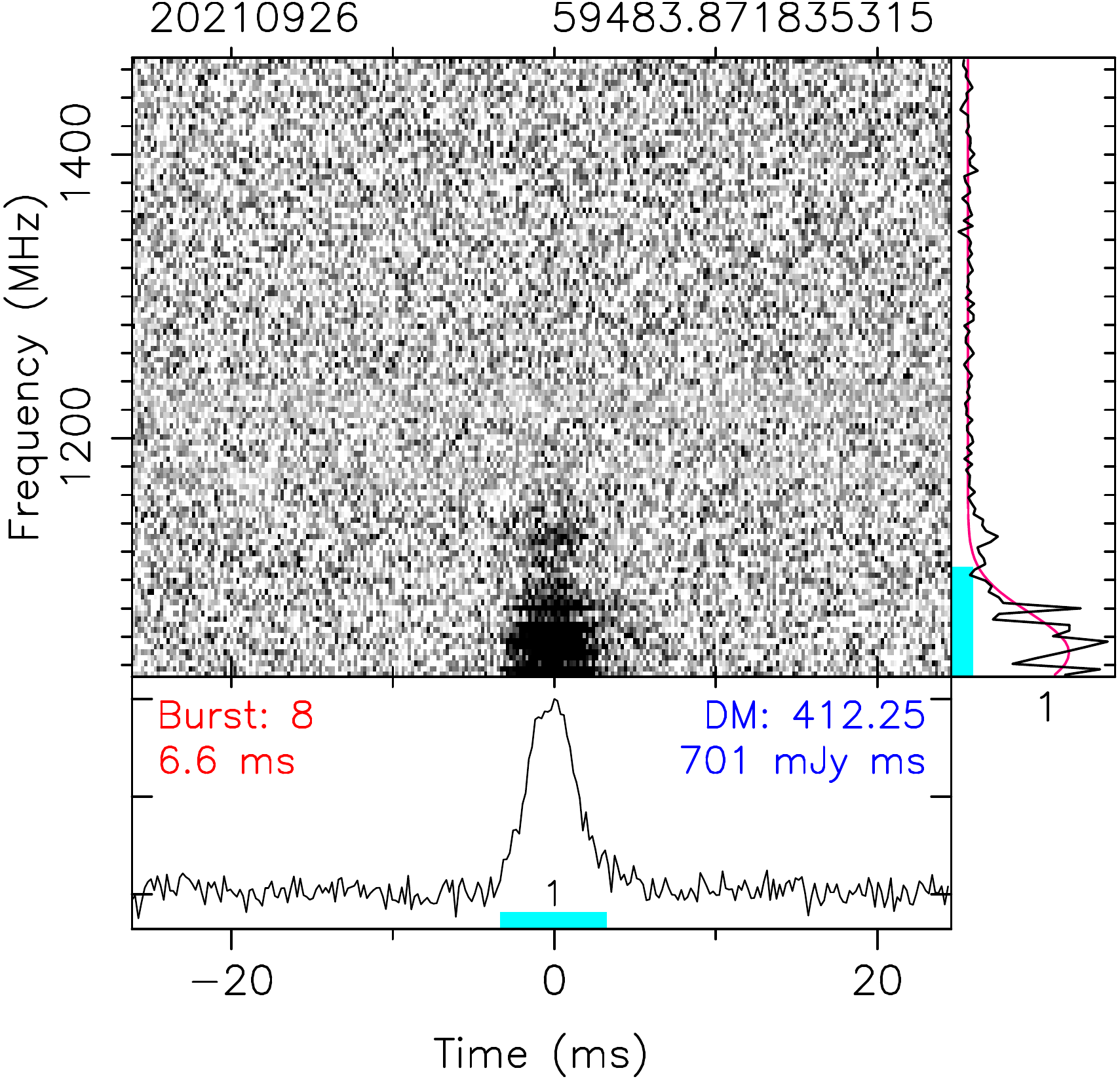}
    \includegraphics[height=37mm]{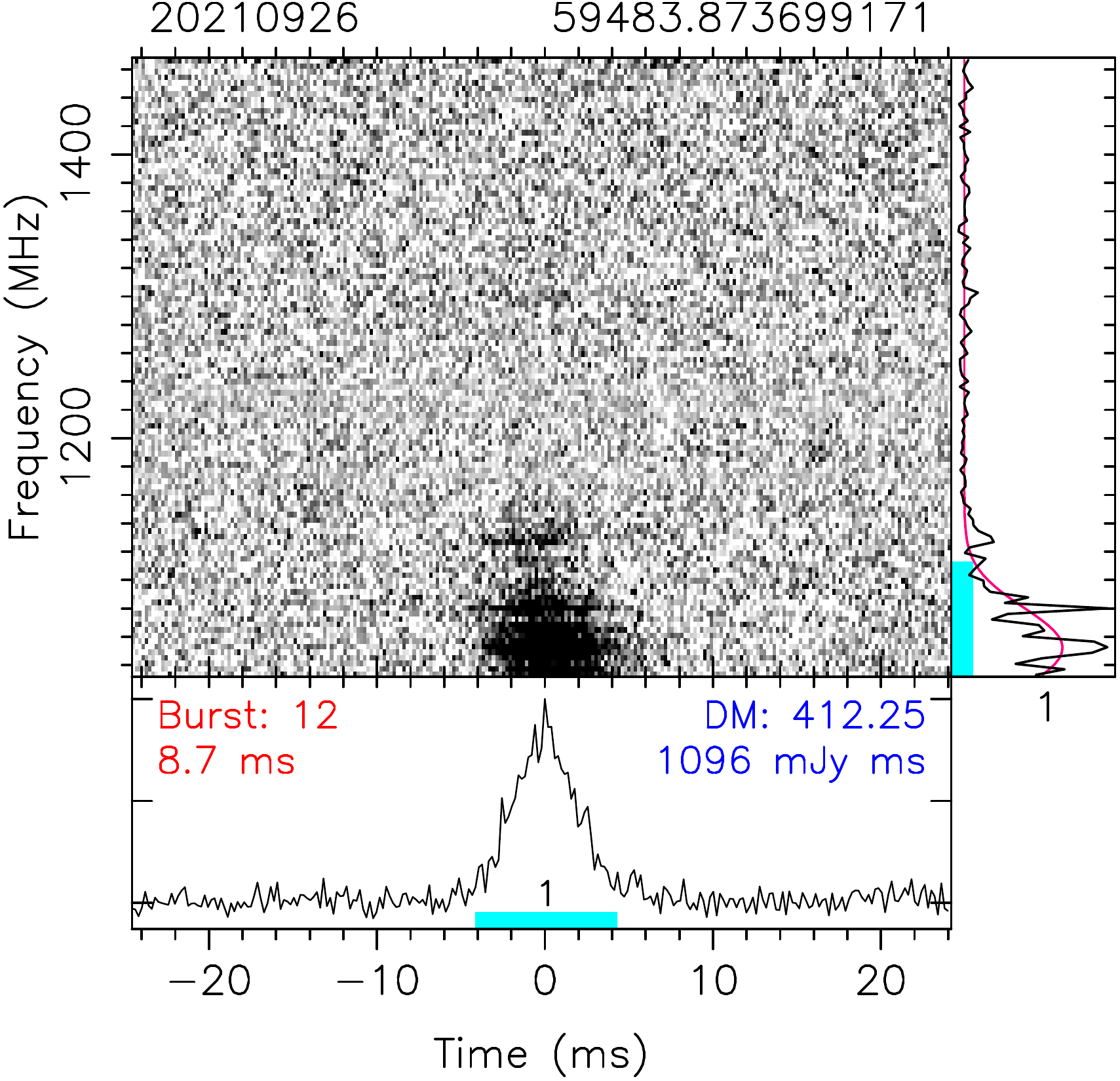}
    \includegraphics[height=37mm]{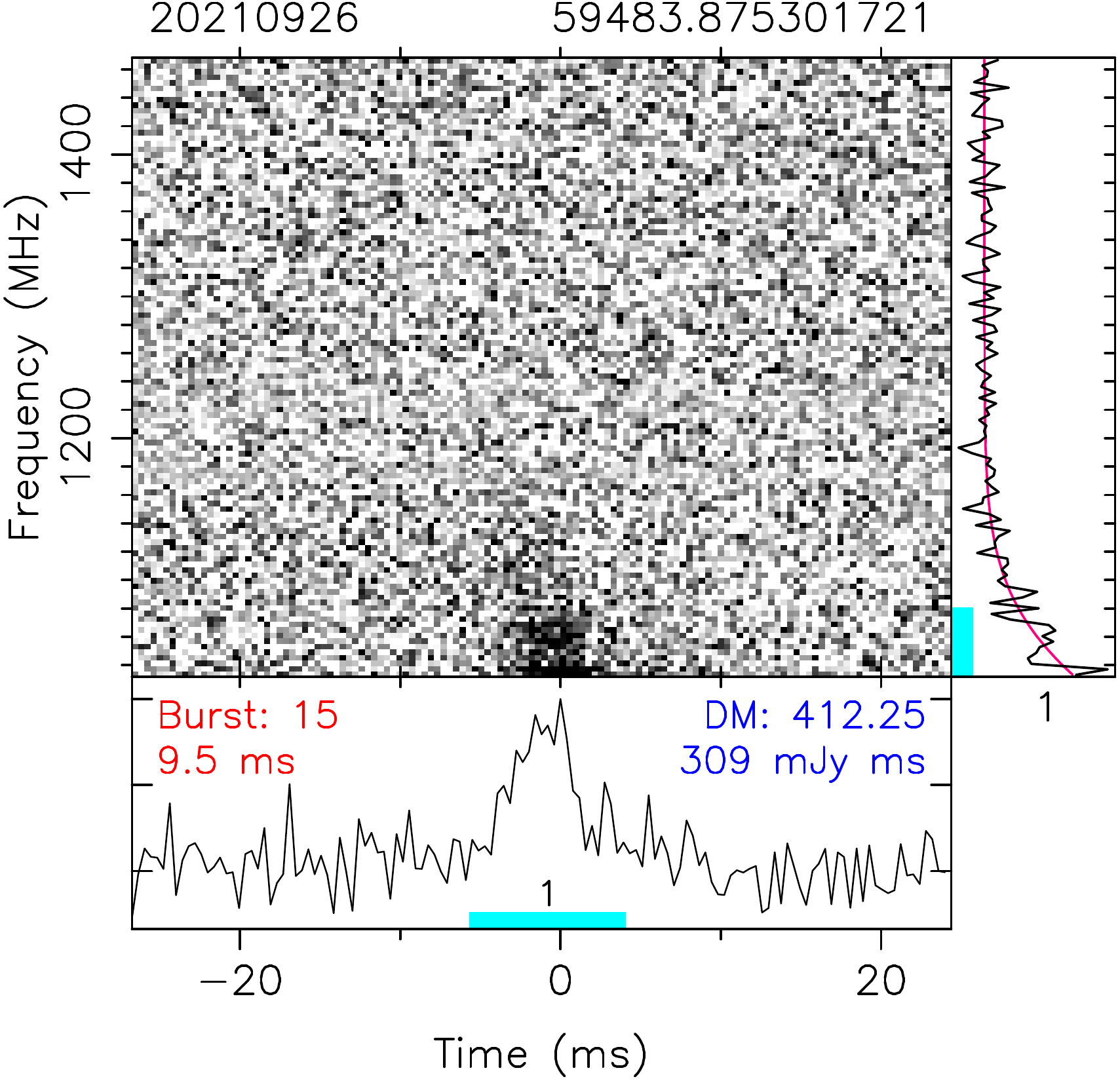}
    \includegraphics[height=37mm]{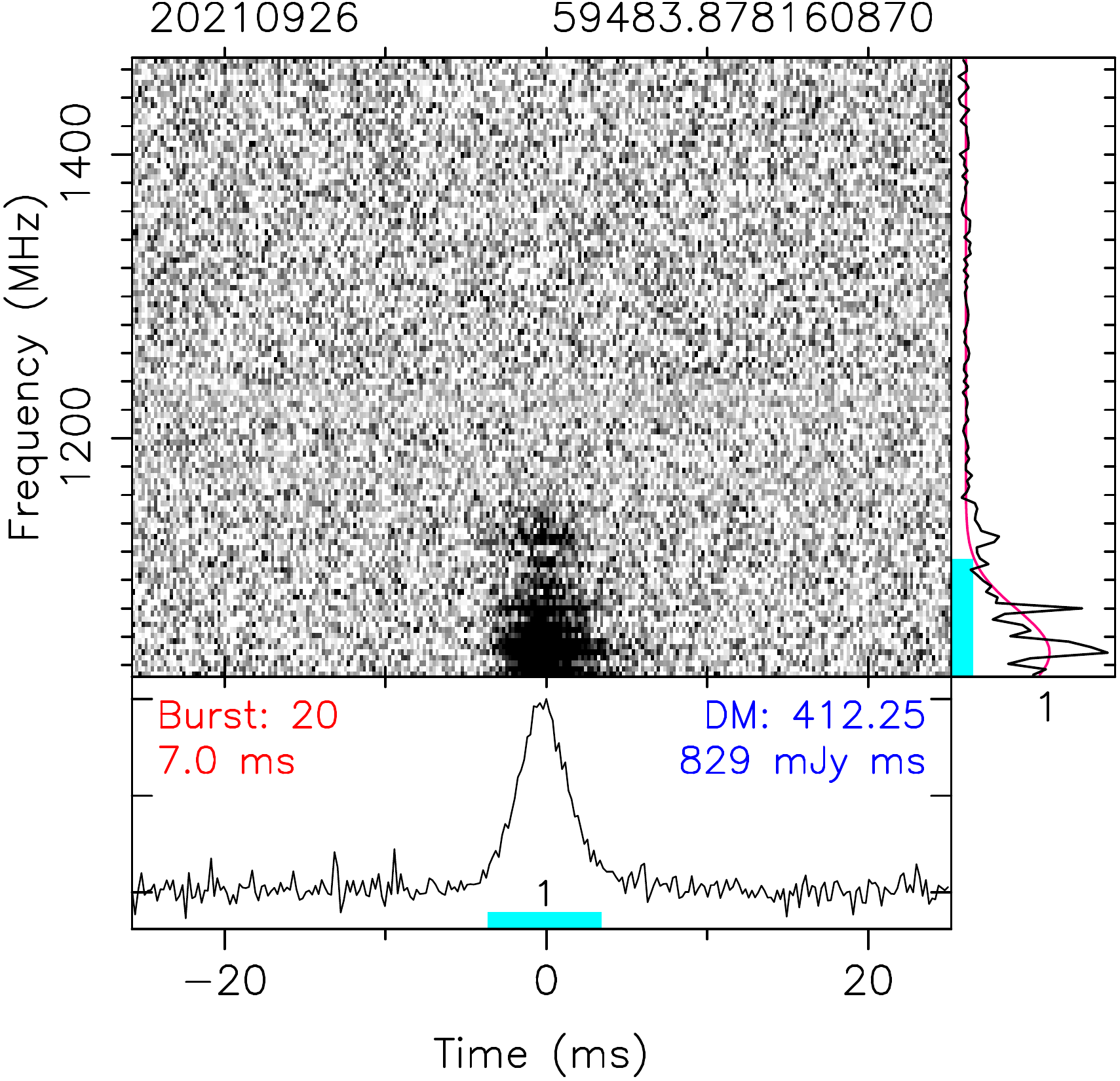}
    \includegraphics[height=37mm]{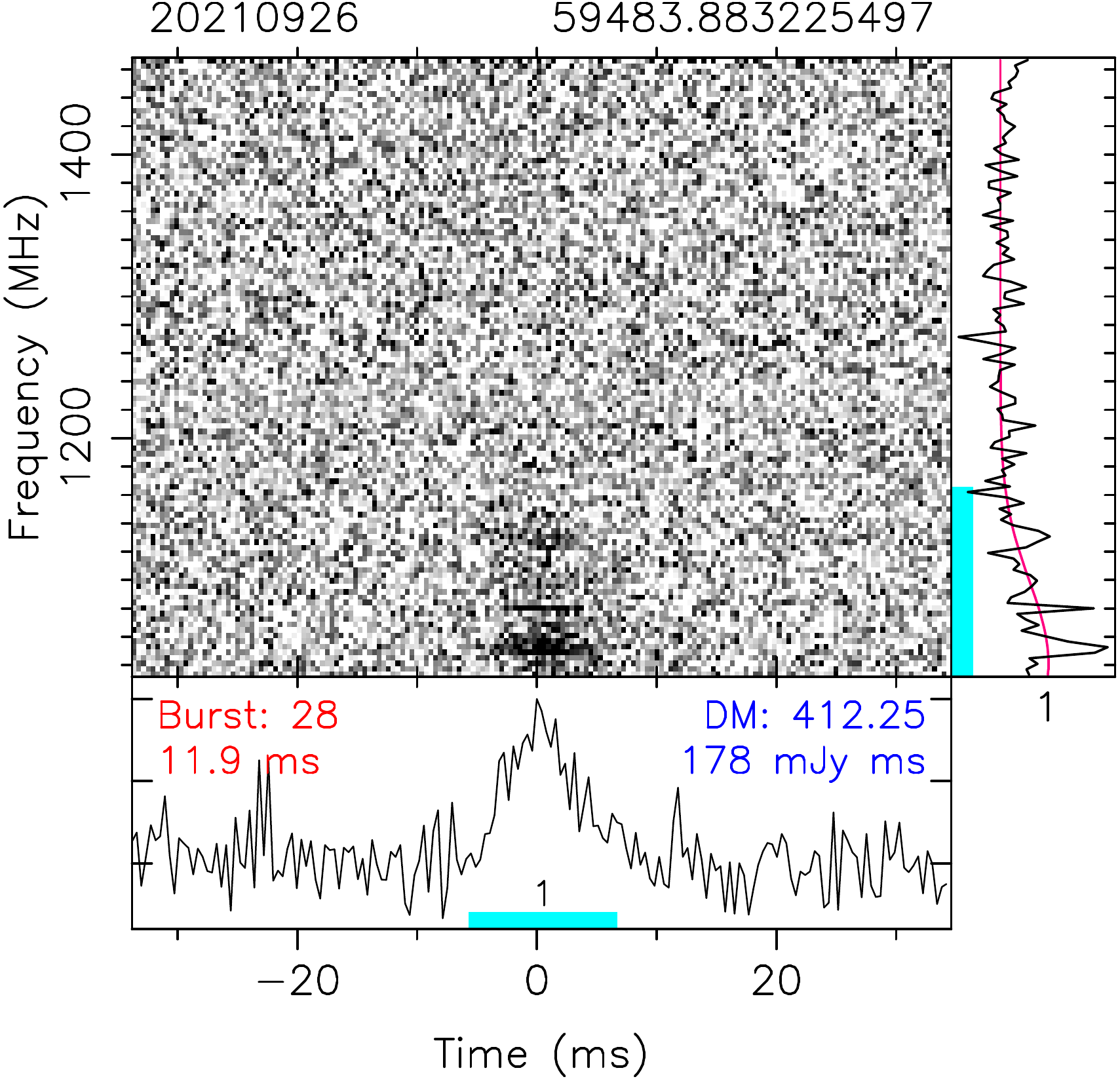}
    \includegraphics[height=37mm]{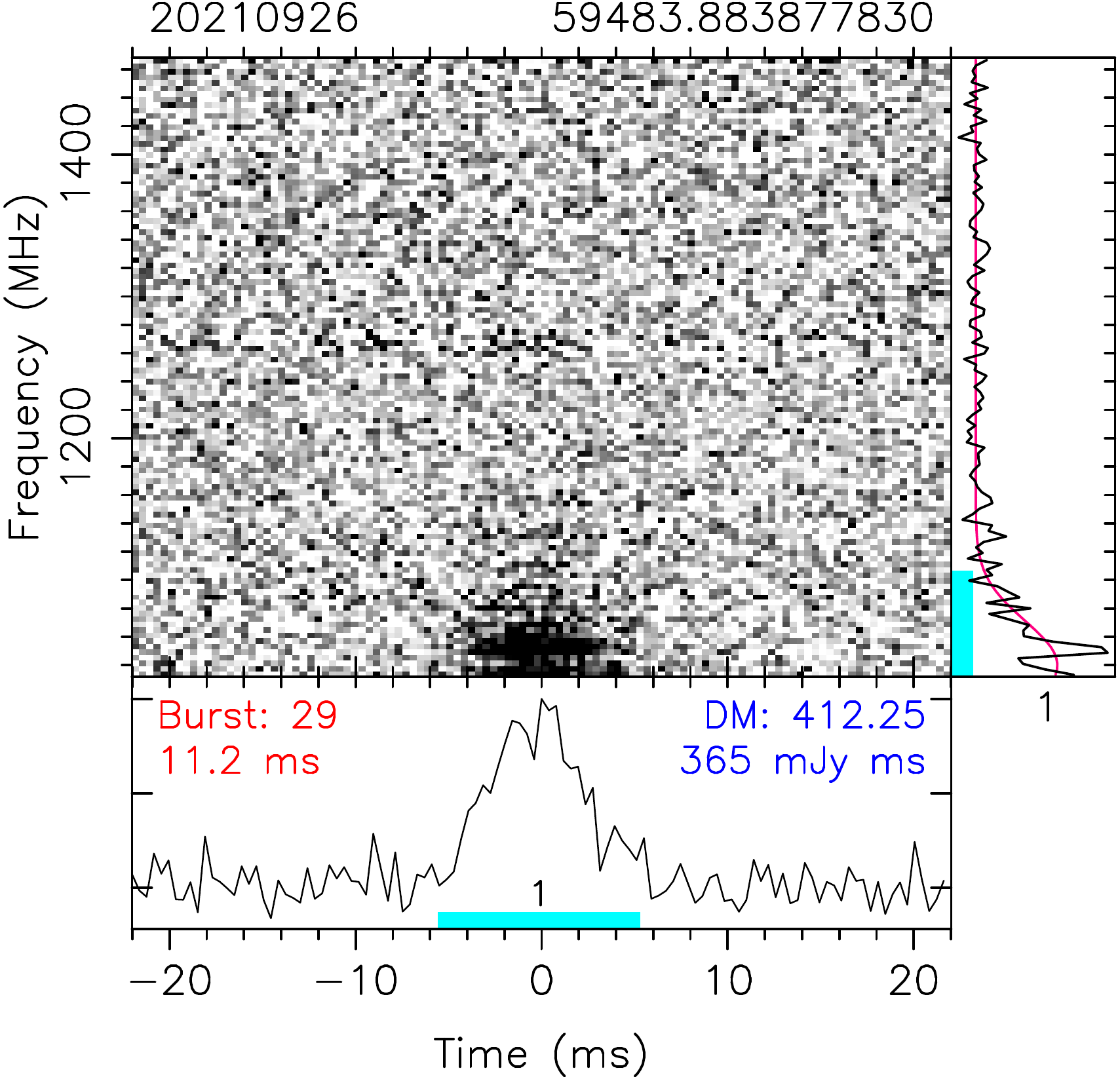}
    \includegraphics[height=37mm]{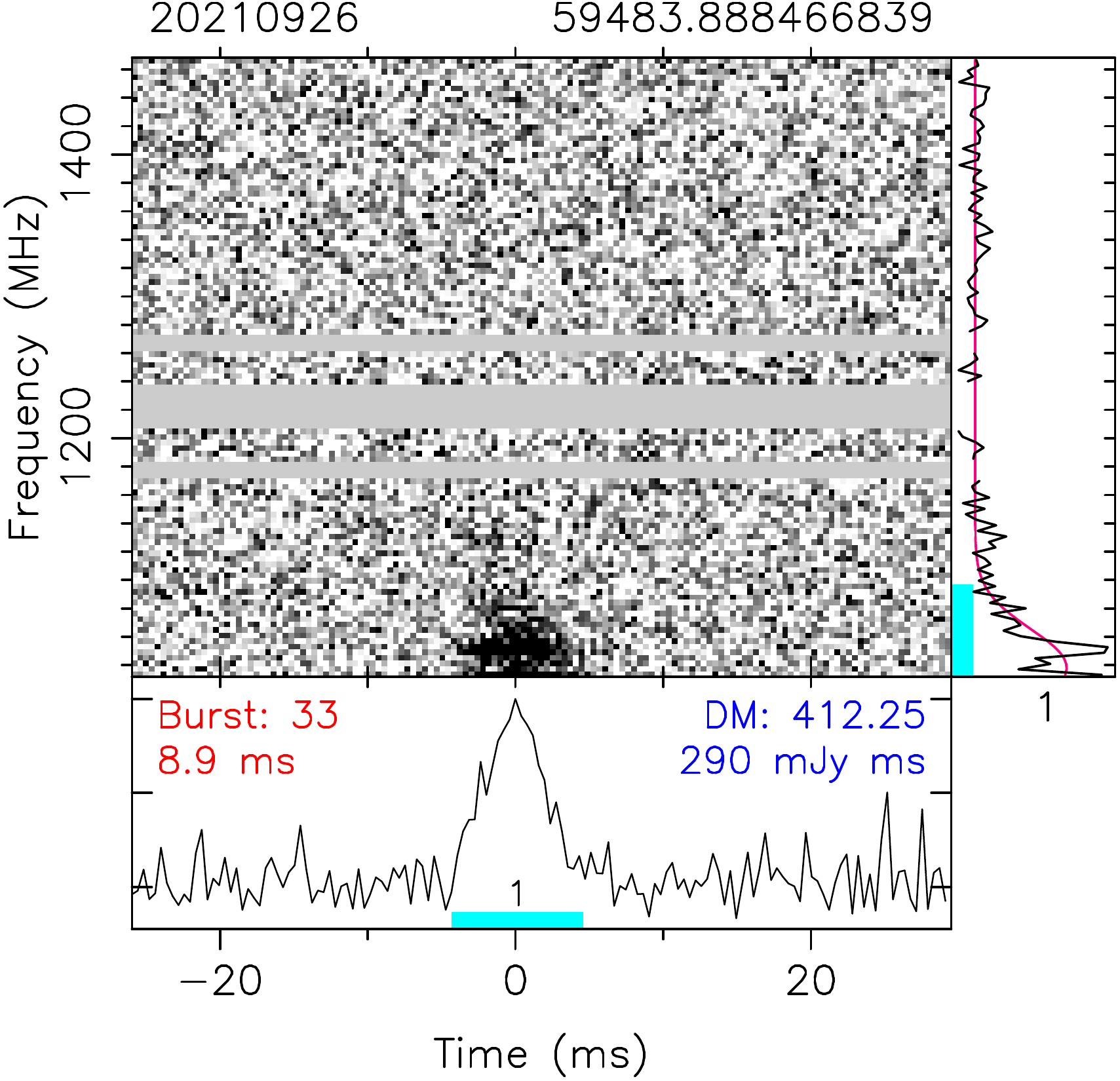}
    \includegraphics[height=37mm]{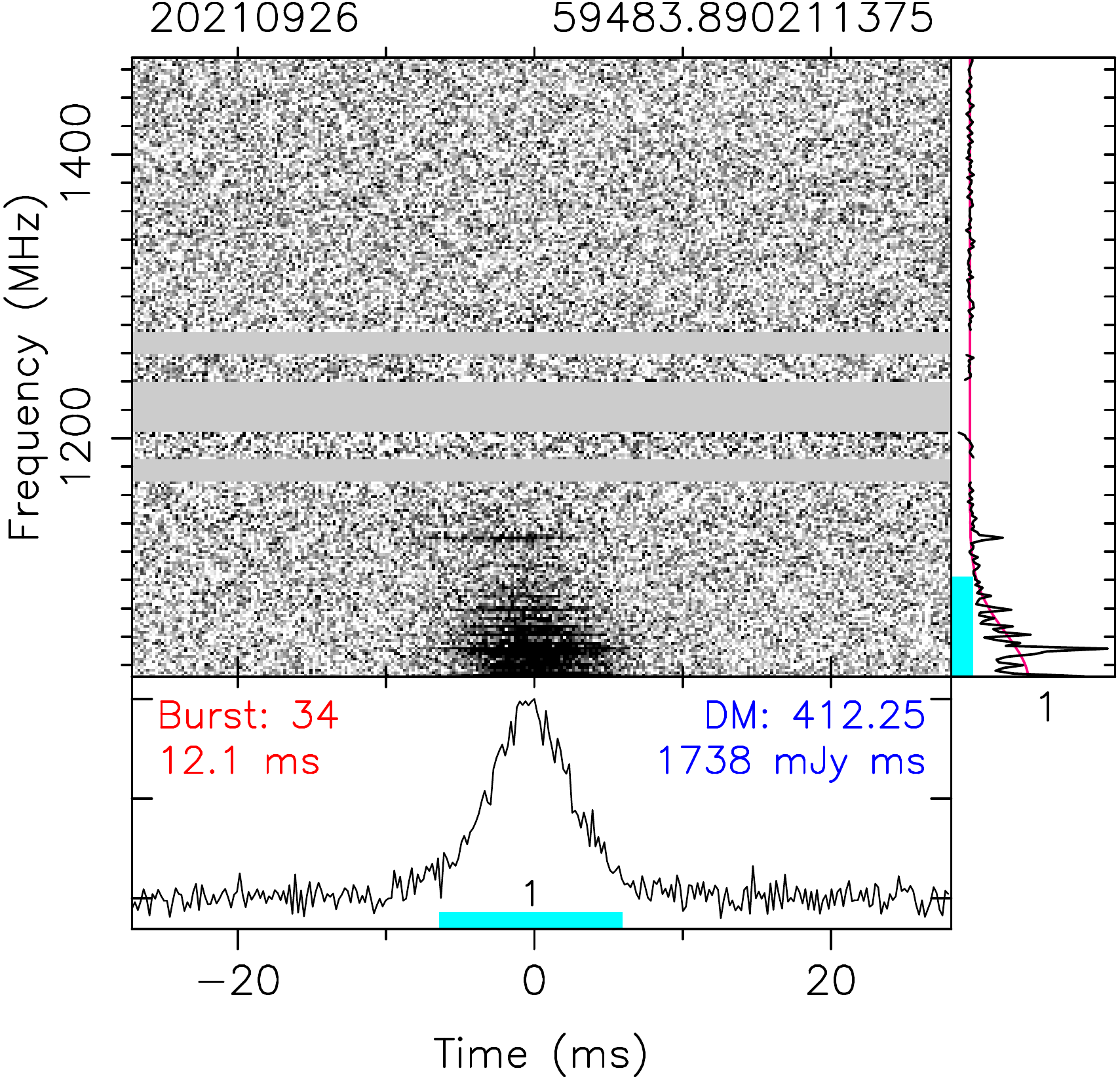}
    \includegraphics[height=37mm]{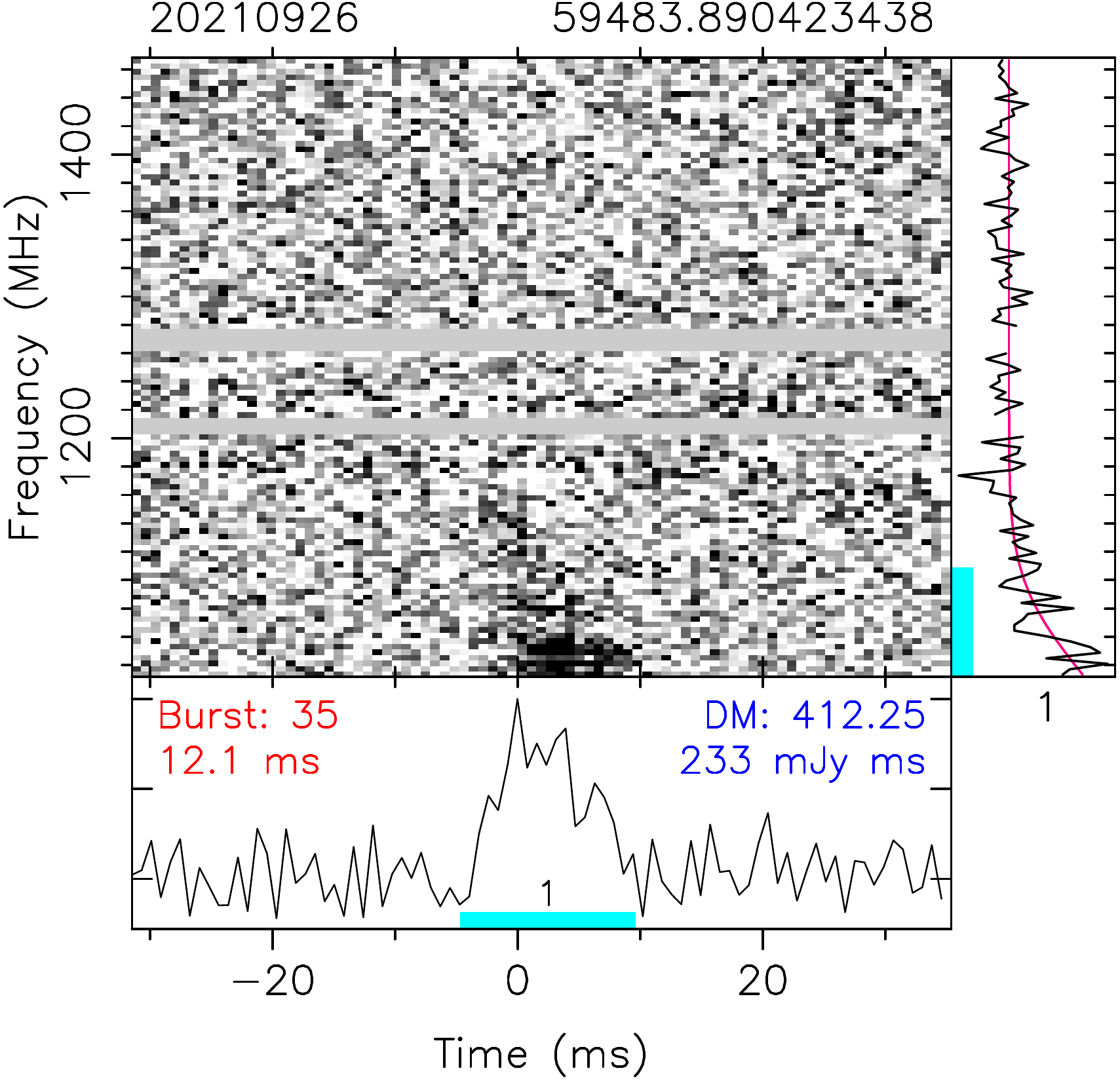}
\caption{The same as Figure~\ref{fig:appendix:D1W} but for bursts at lower band with no evidence of drifting (NE-L).
}\label{fig:appendix:NEL} 
\end{figure*}
\addtocounter{figure}{-1}
\begin{figure*}
    \flushleft
    \includegraphics[height=37mm]{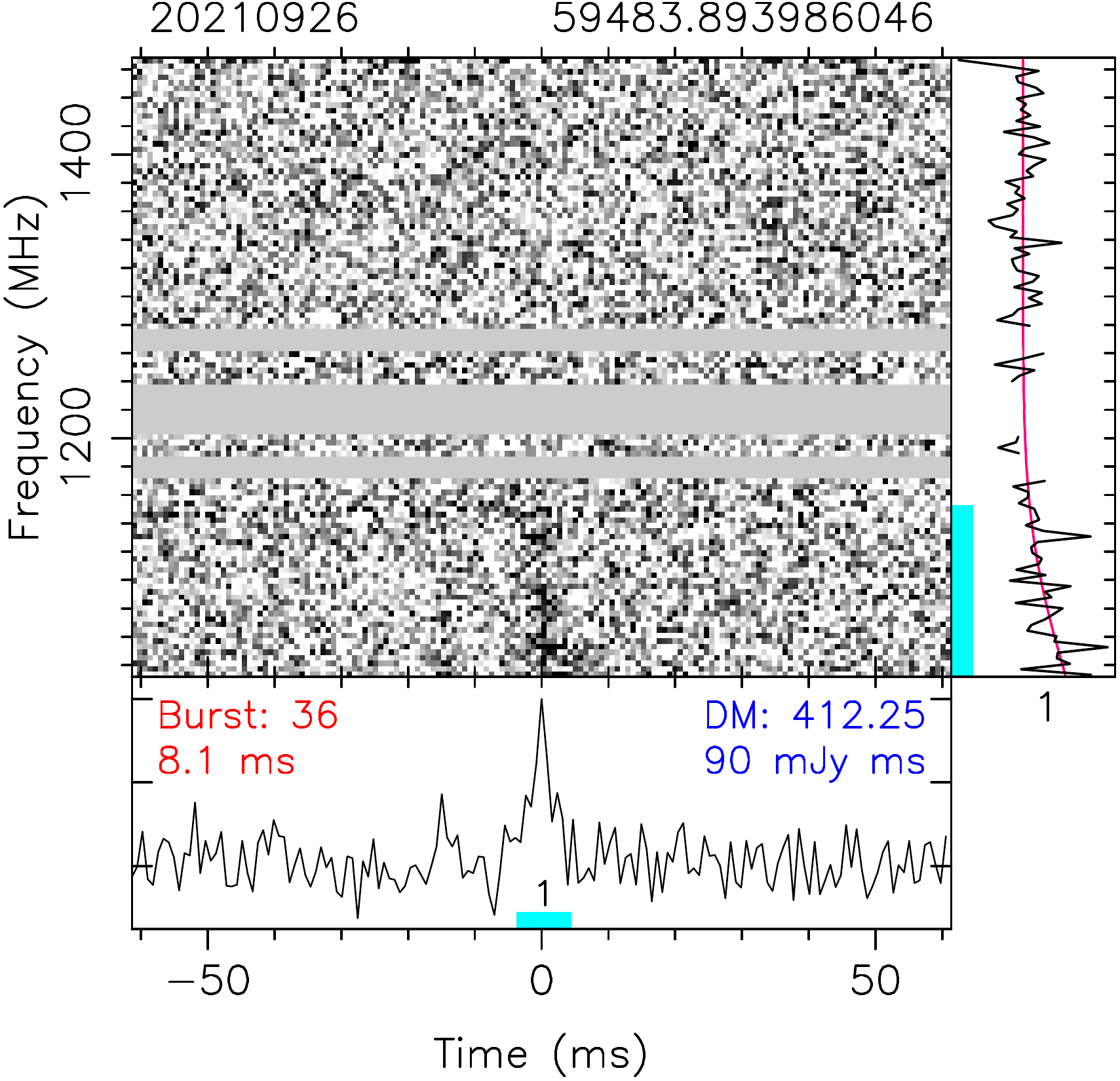}
    \includegraphics[height=37mm]{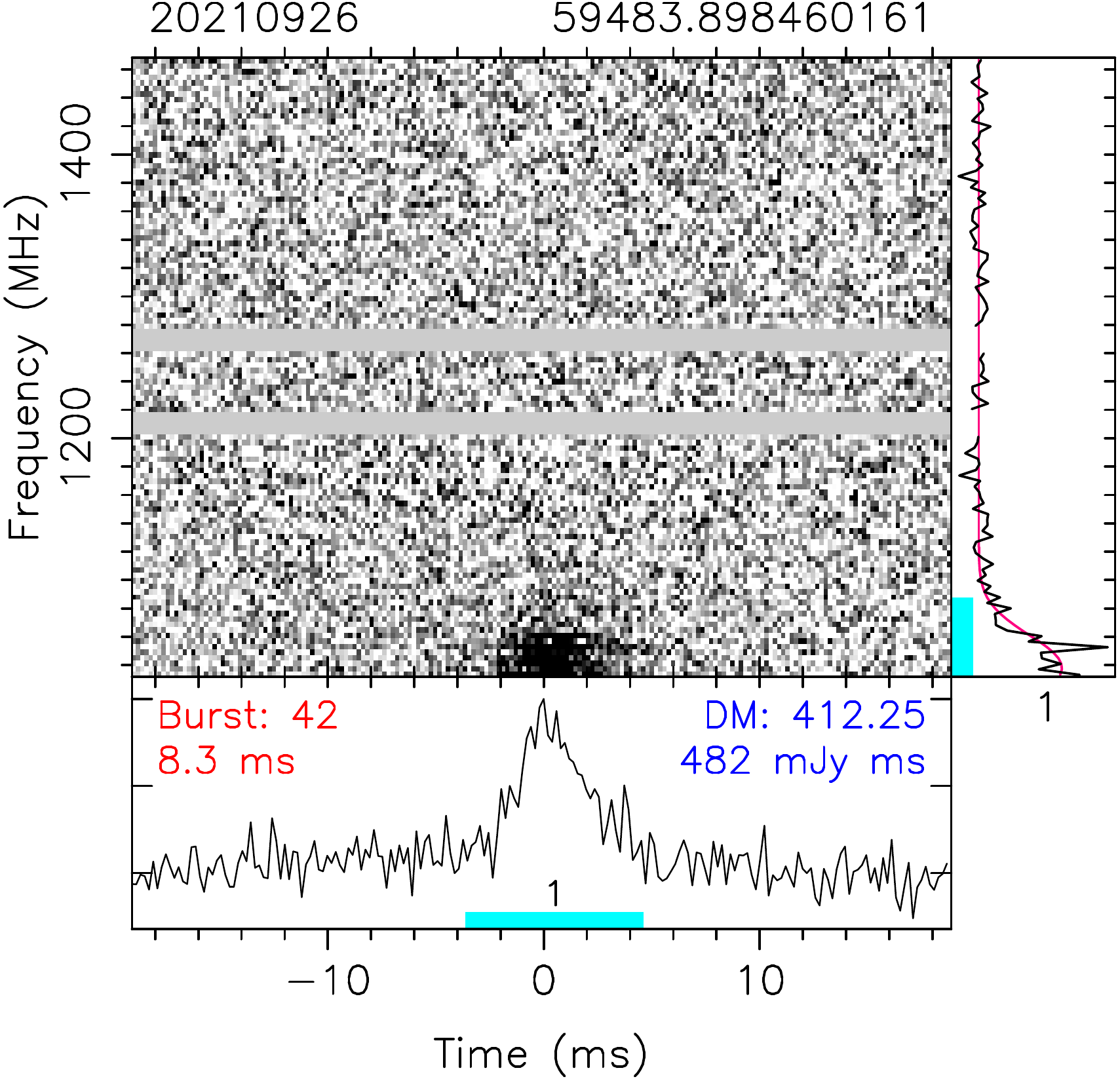}
    \includegraphics[height=37mm]{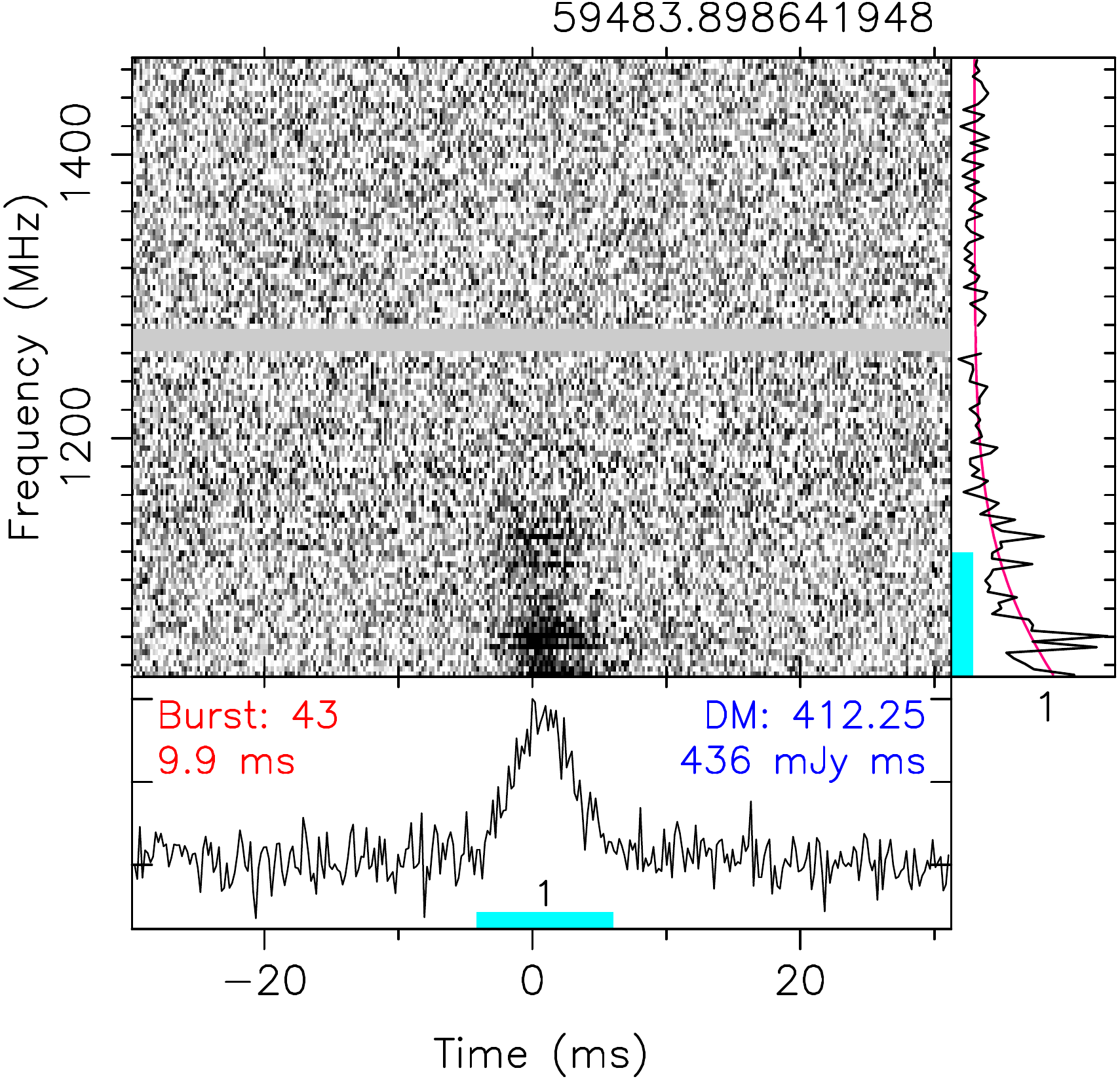}
    \includegraphics[height=37mm]{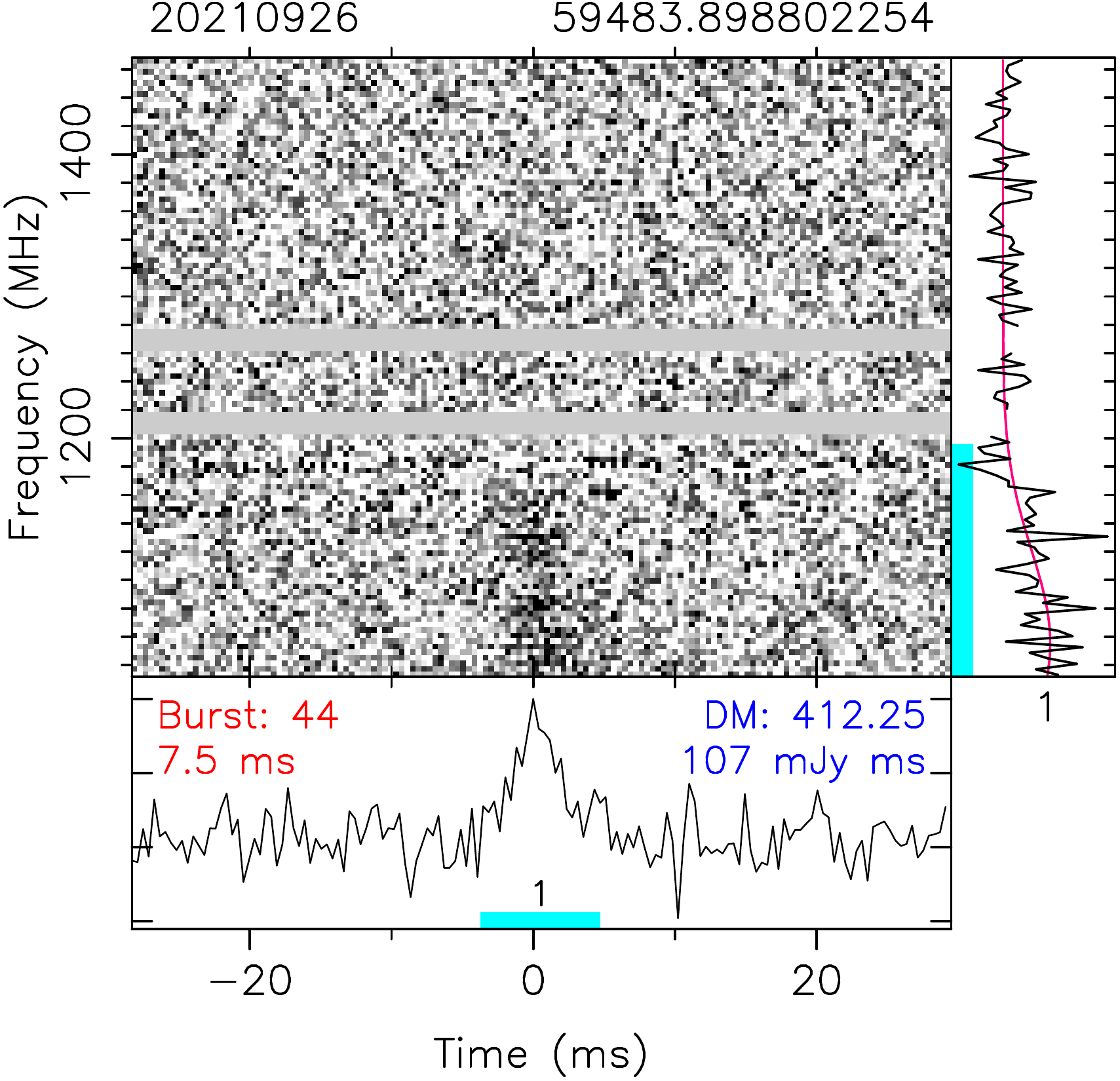}
    \includegraphics[height=37mm]{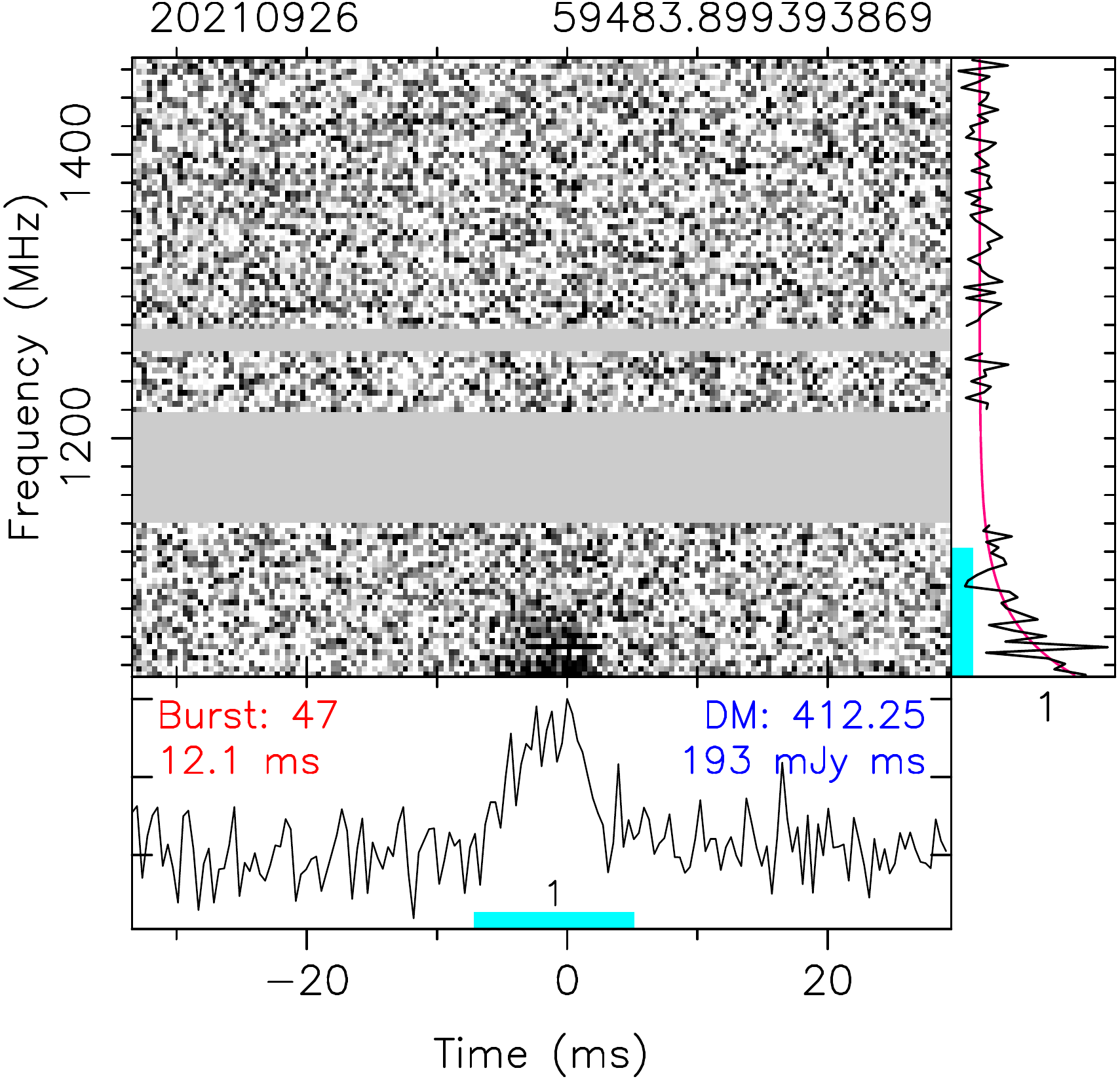}
    \includegraphics[height=37mm]{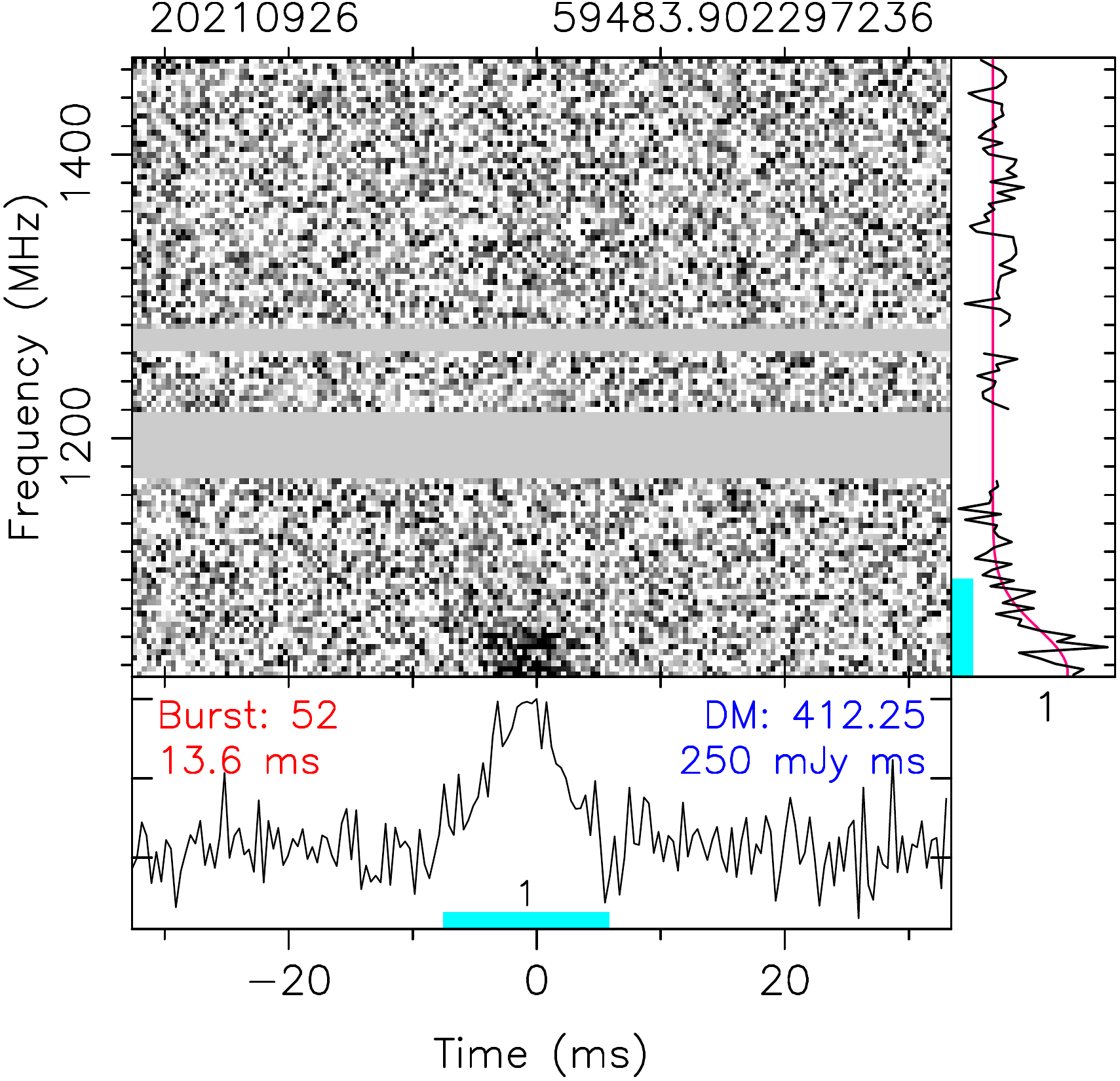}
    \includegraphics[height=37mm]{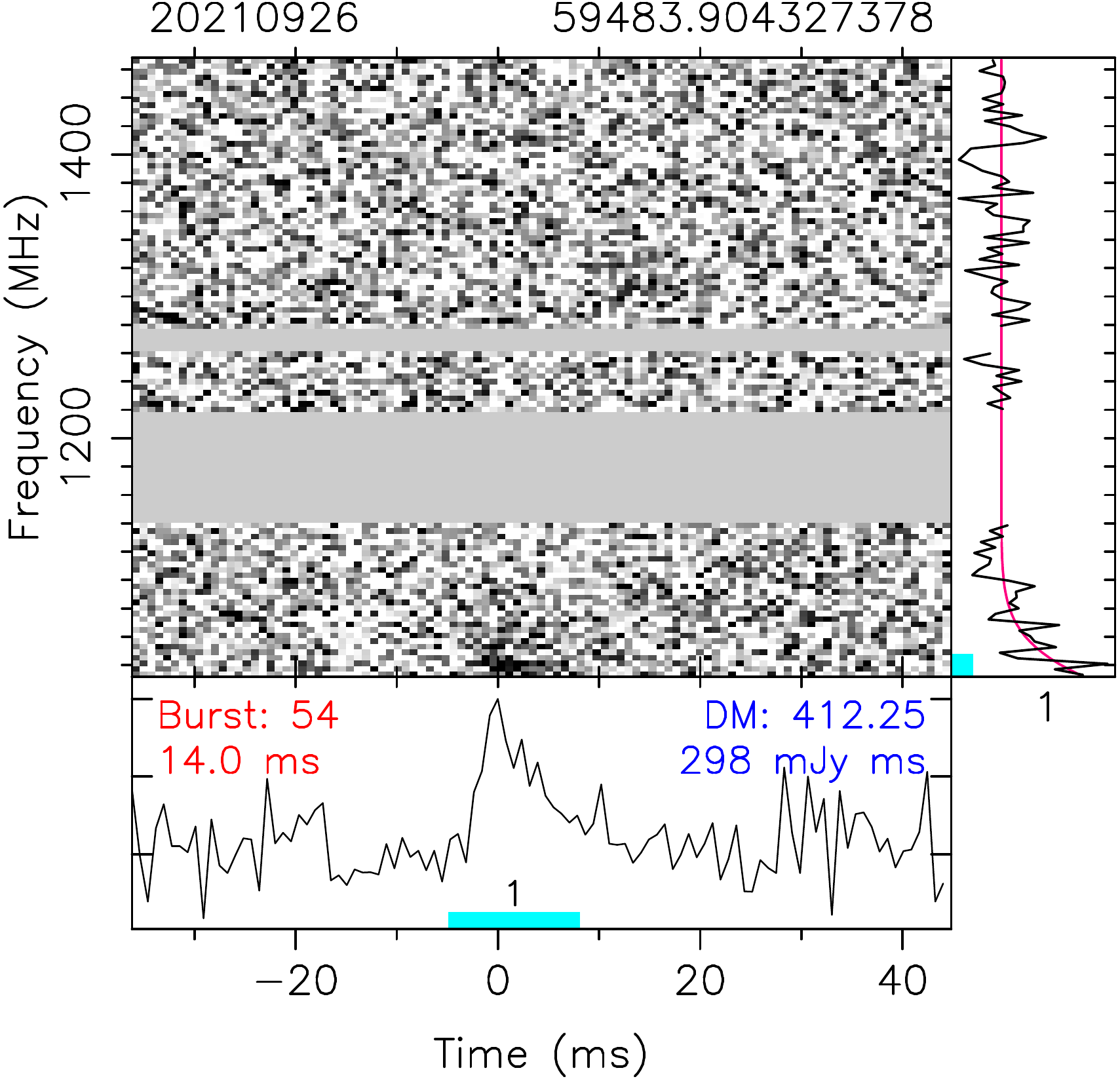}
    \includegraphics[height=37mm]{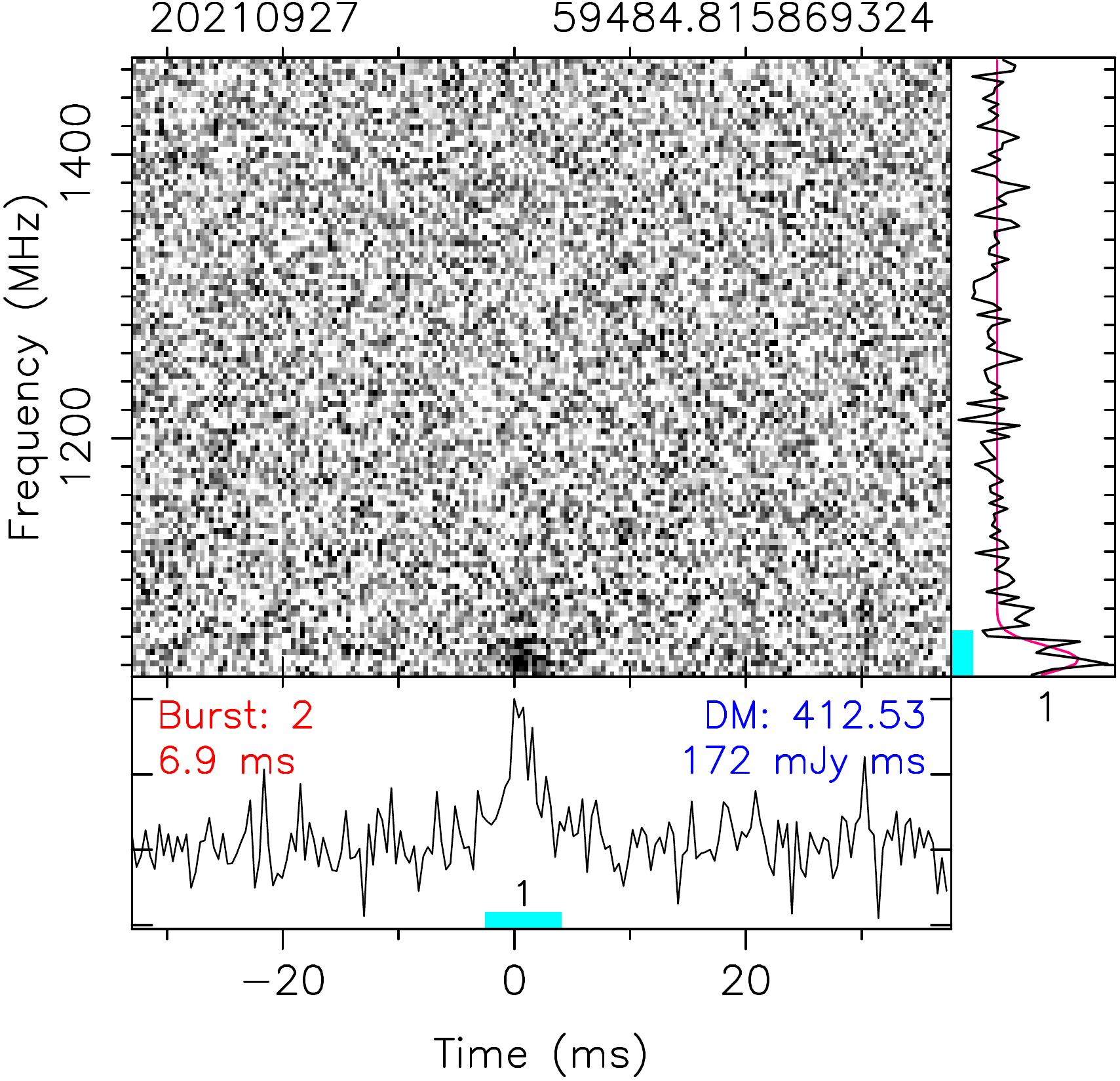}
    \includegraphics[height=37mm]{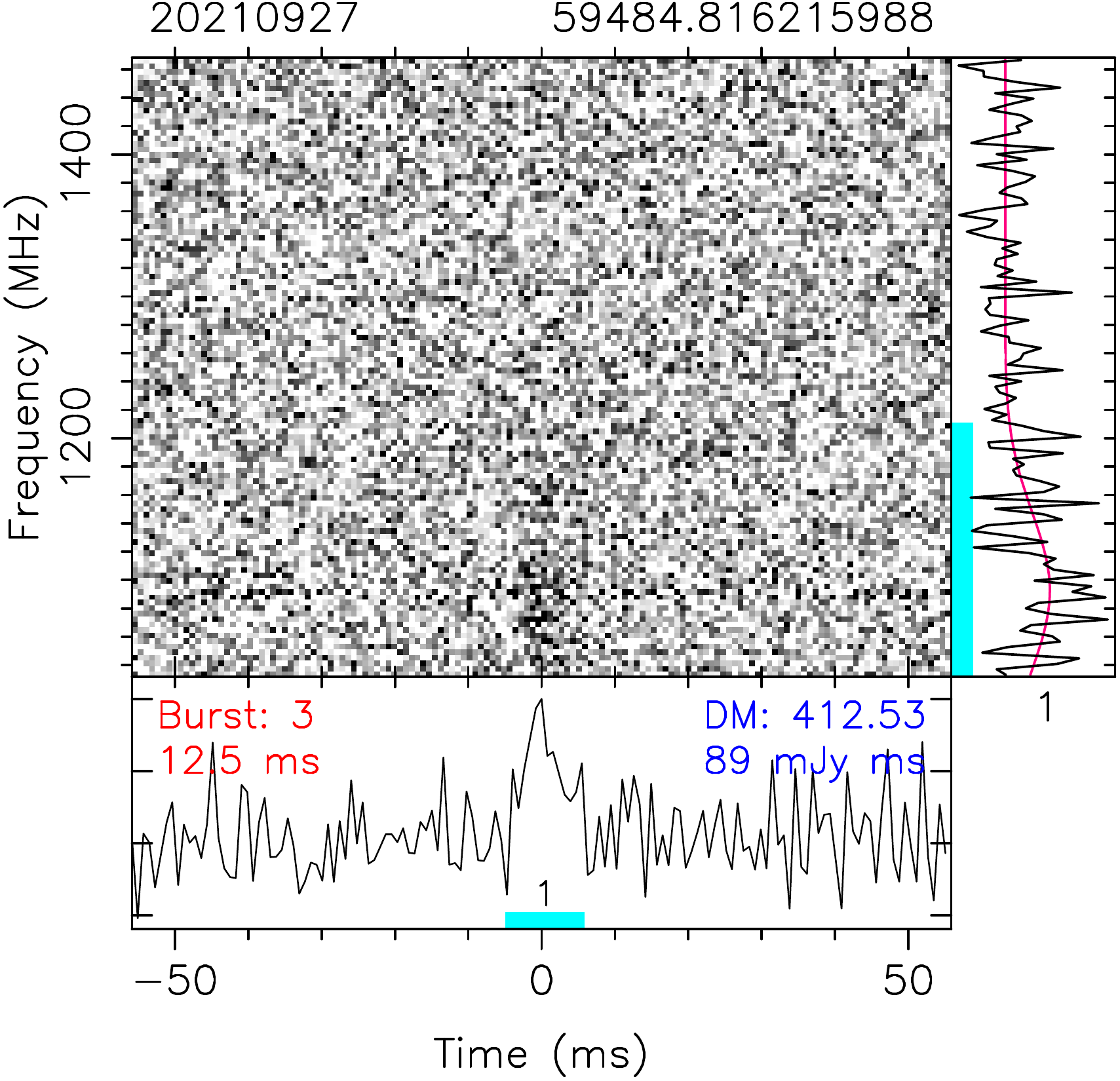}
    \includegraphics[height=37mm]{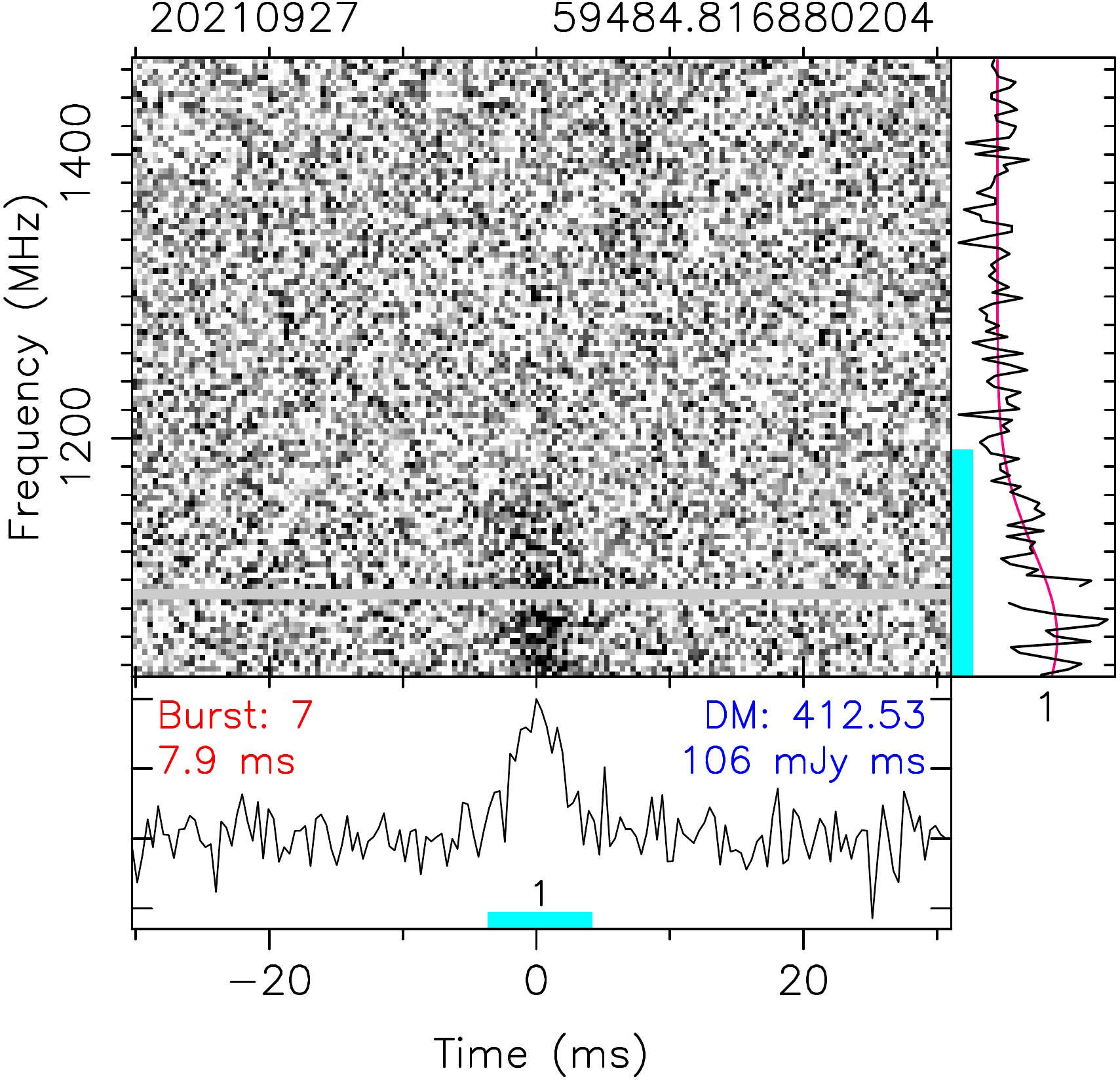}
    \includegraphics[height=37mm]{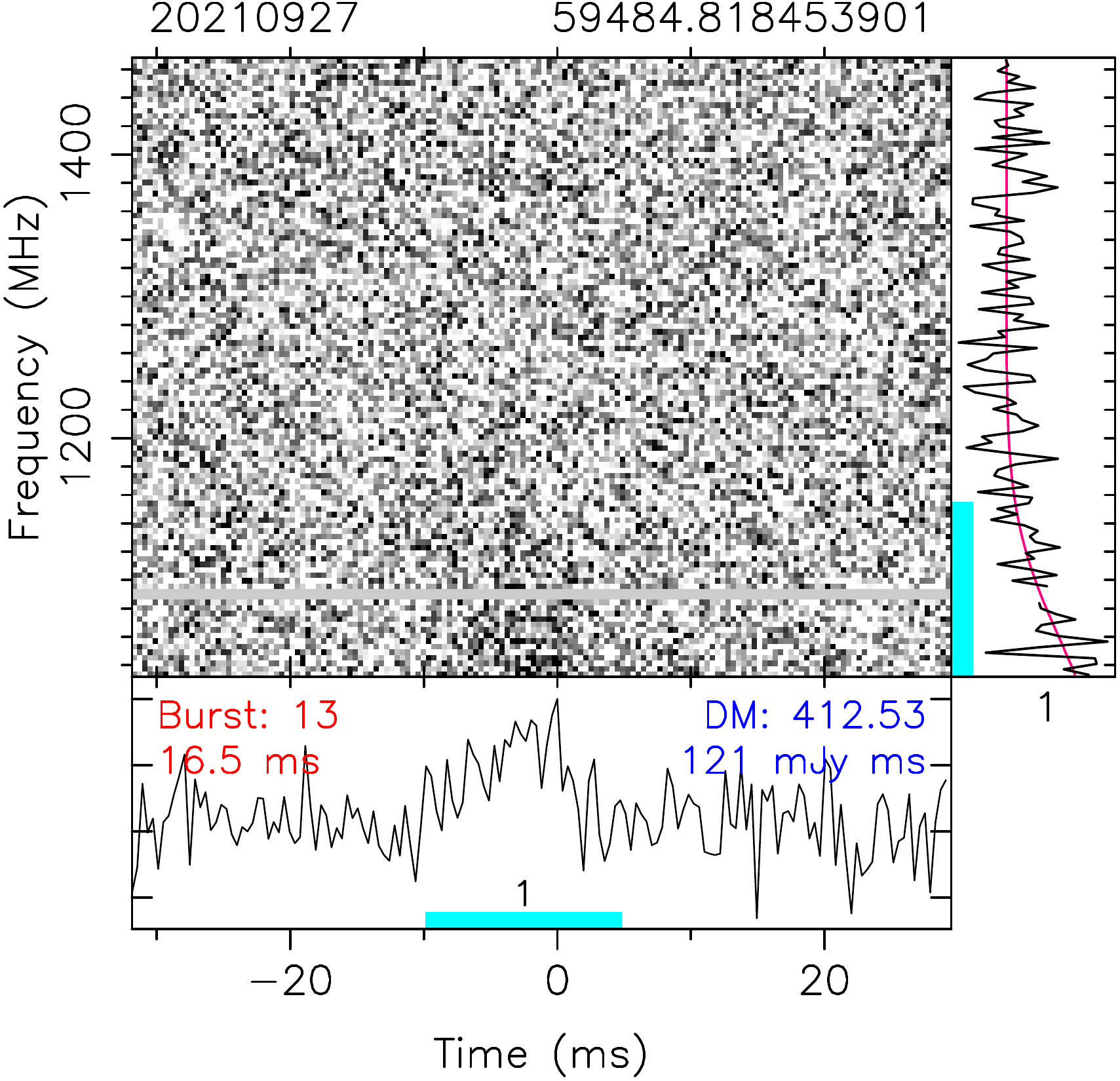}
    \includegraphics[height=37mm]{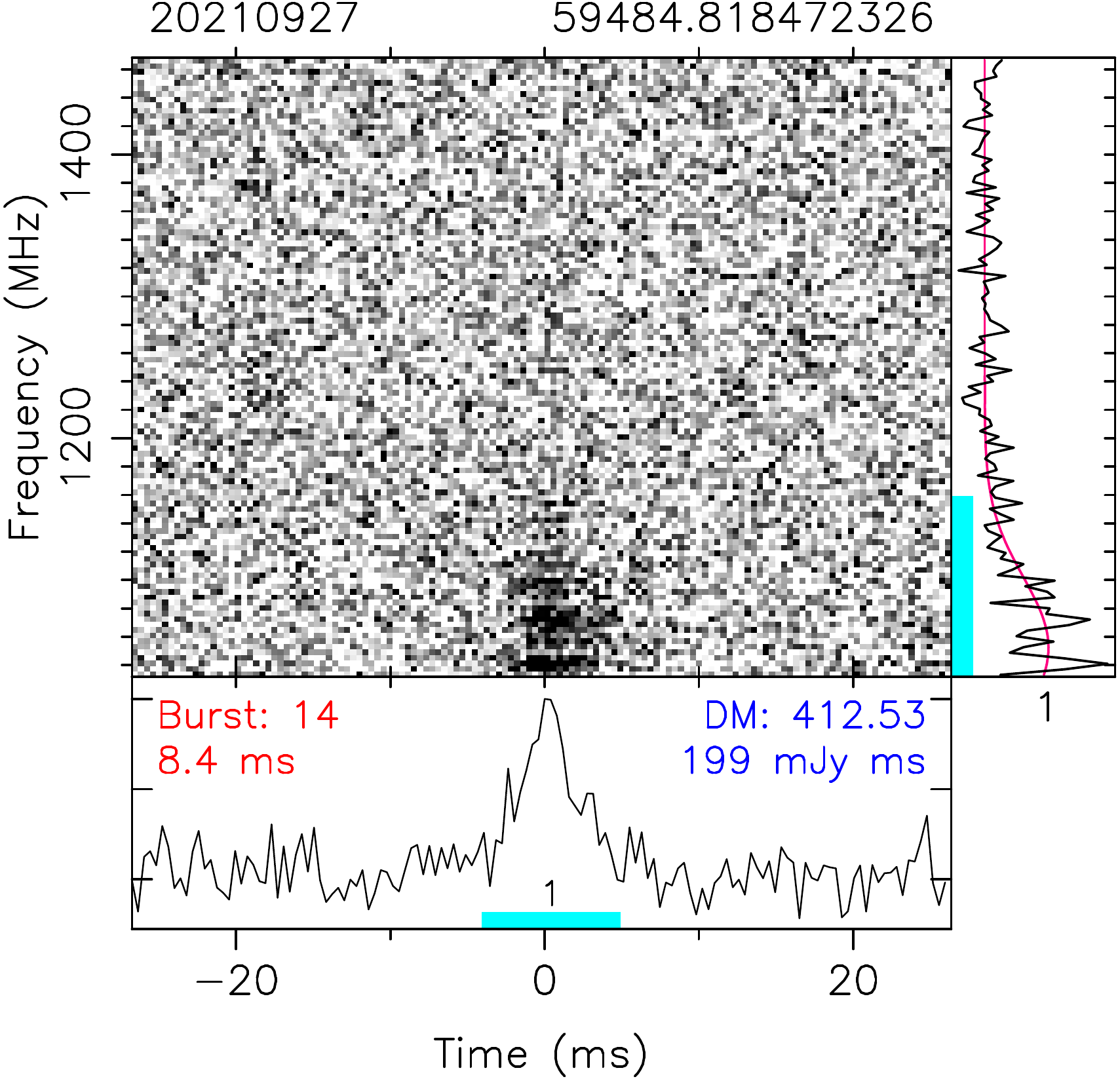}
    \includegraphics[height=37mm]{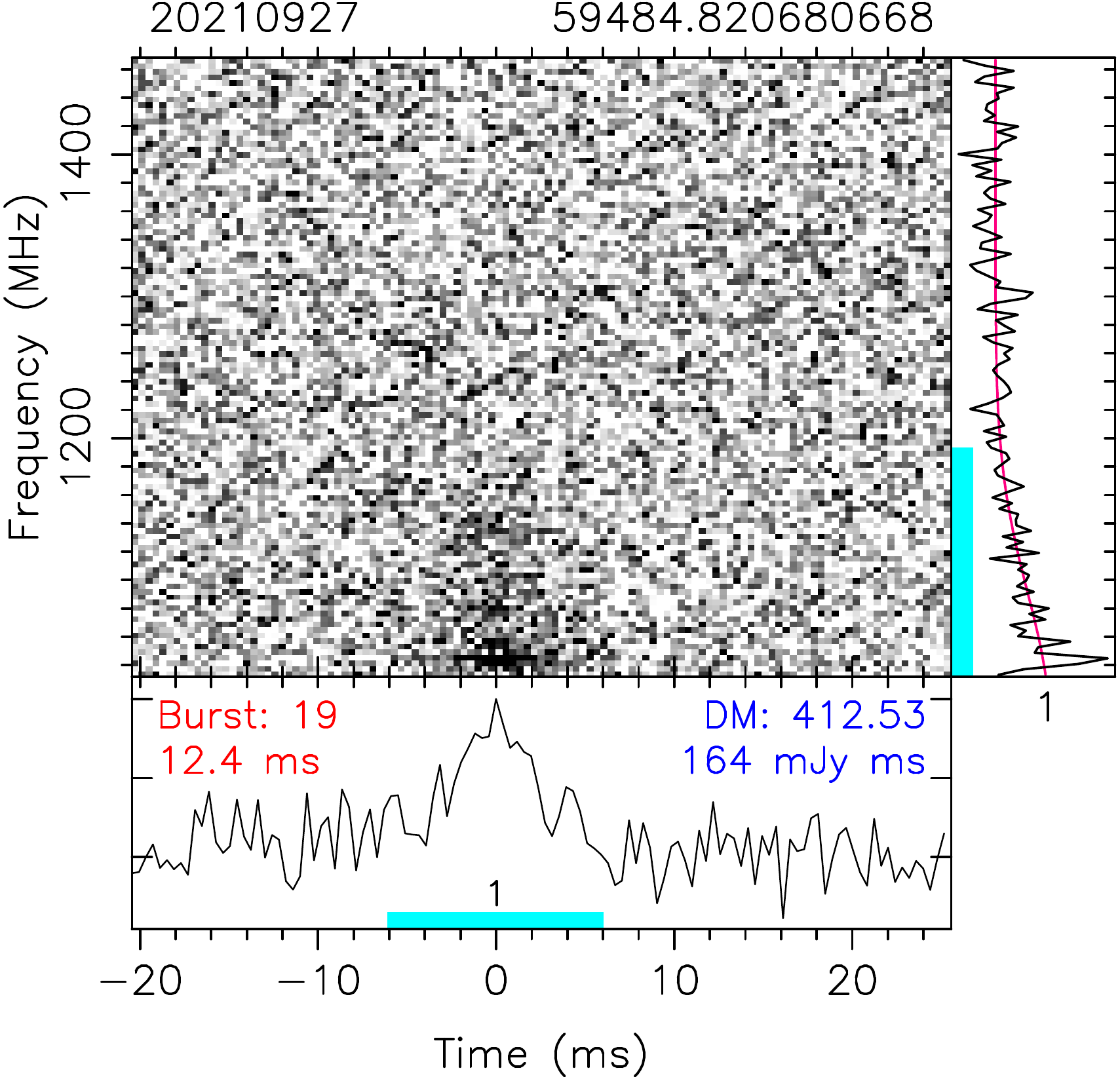}
    \includegraphics[height=37mm]{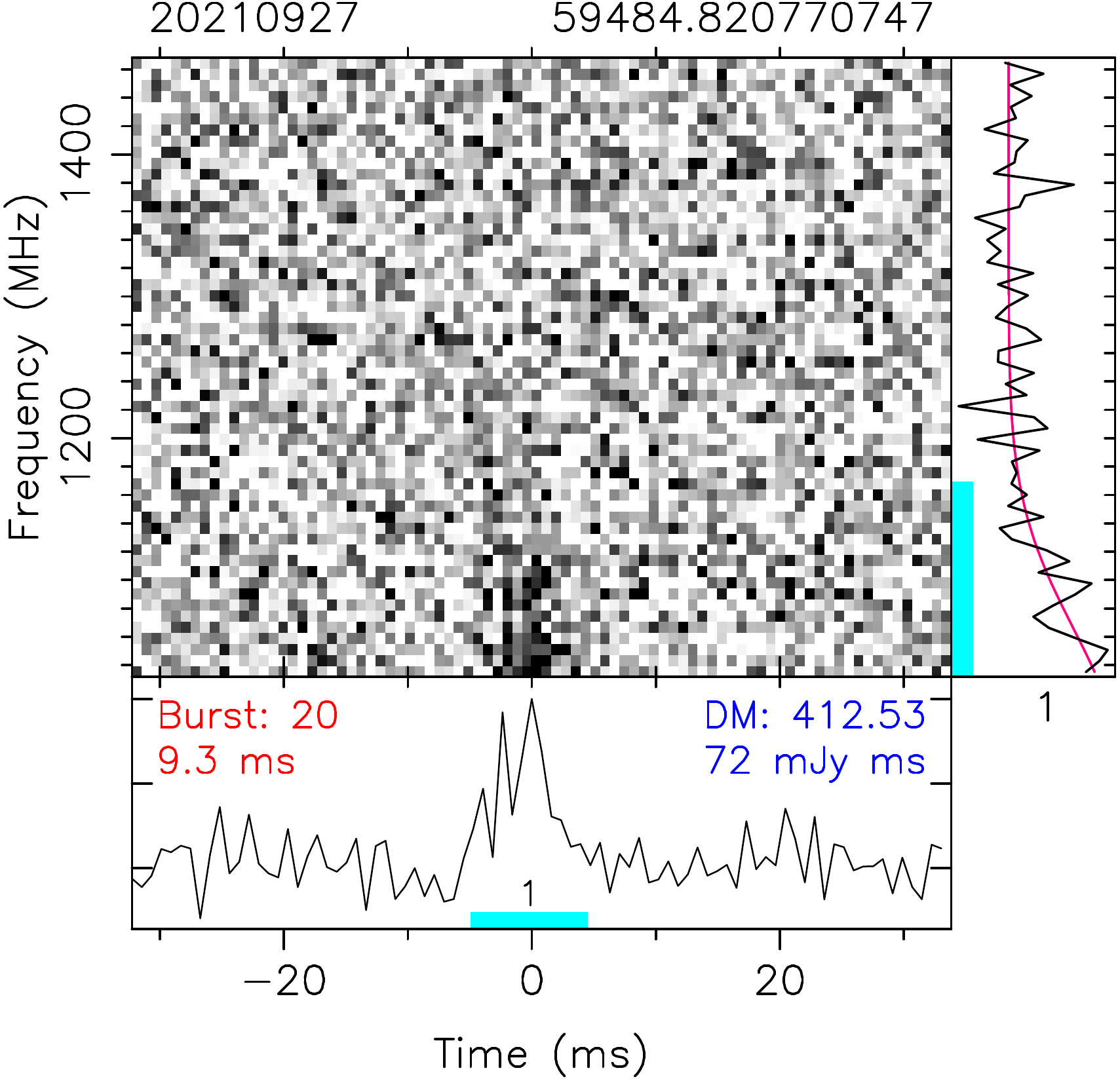}
    \includegraphics[height=37mm]{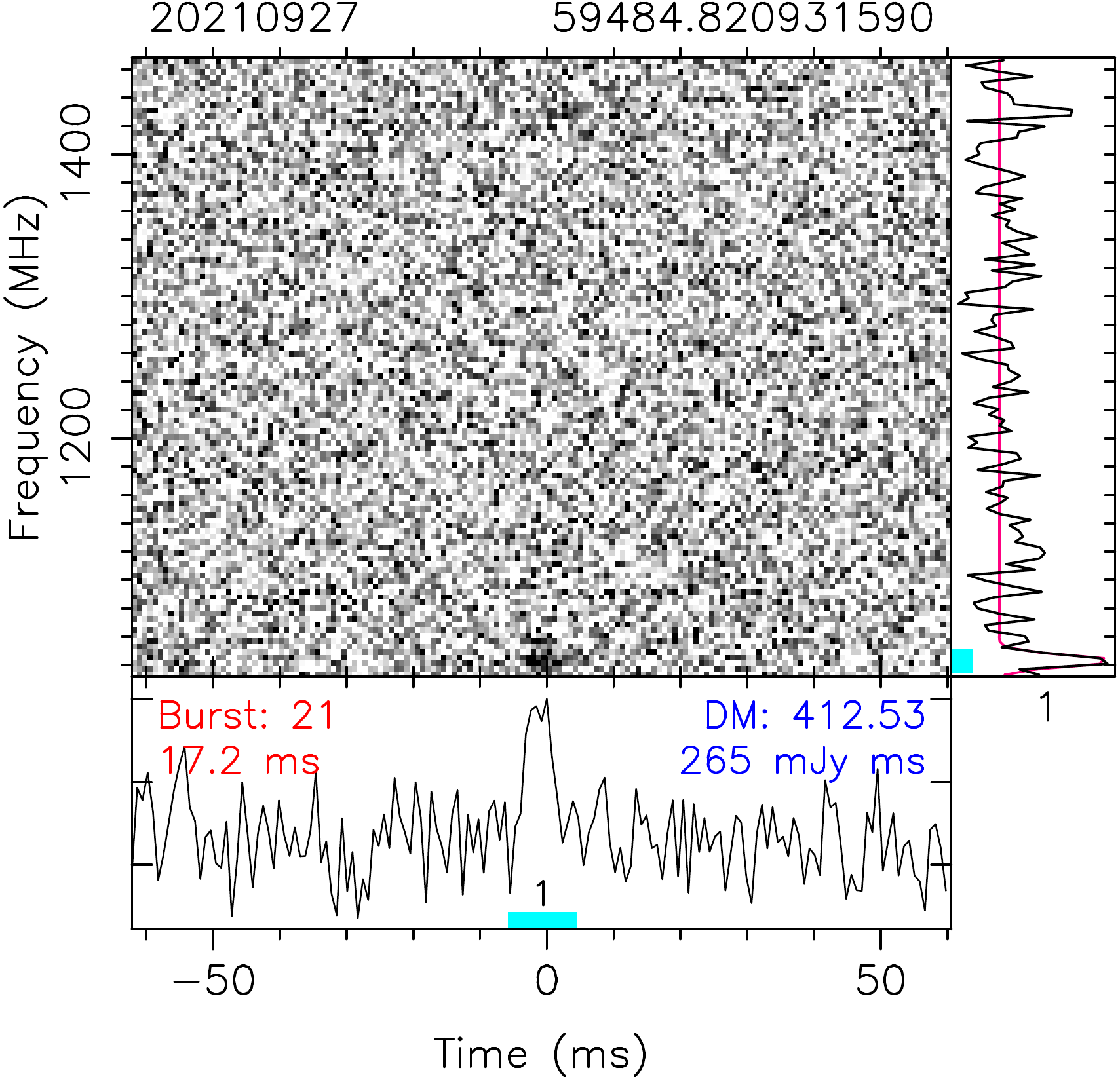}
    \includegraphics[height=37mm]{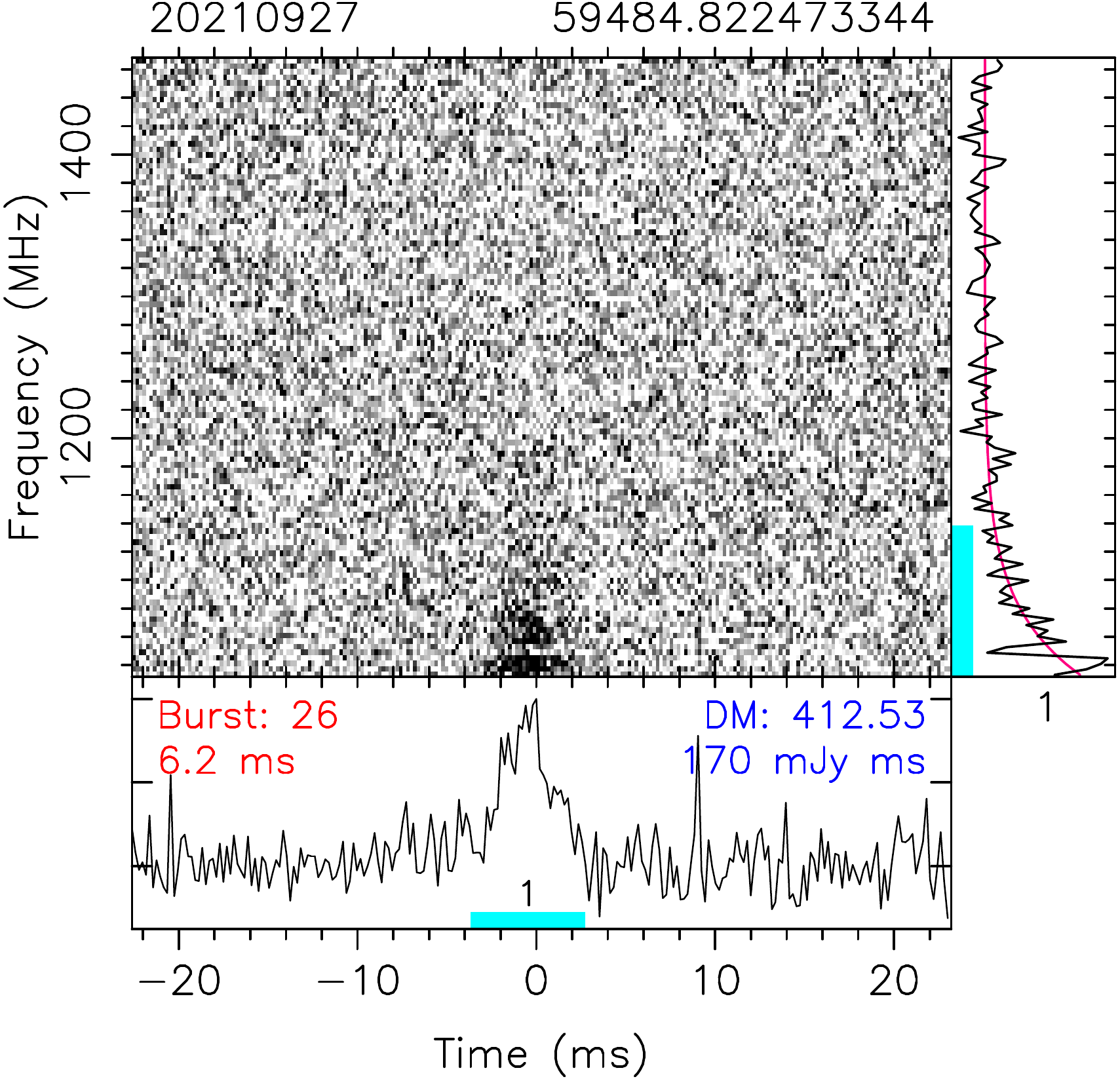}
    \includegraphics[height=37mm]{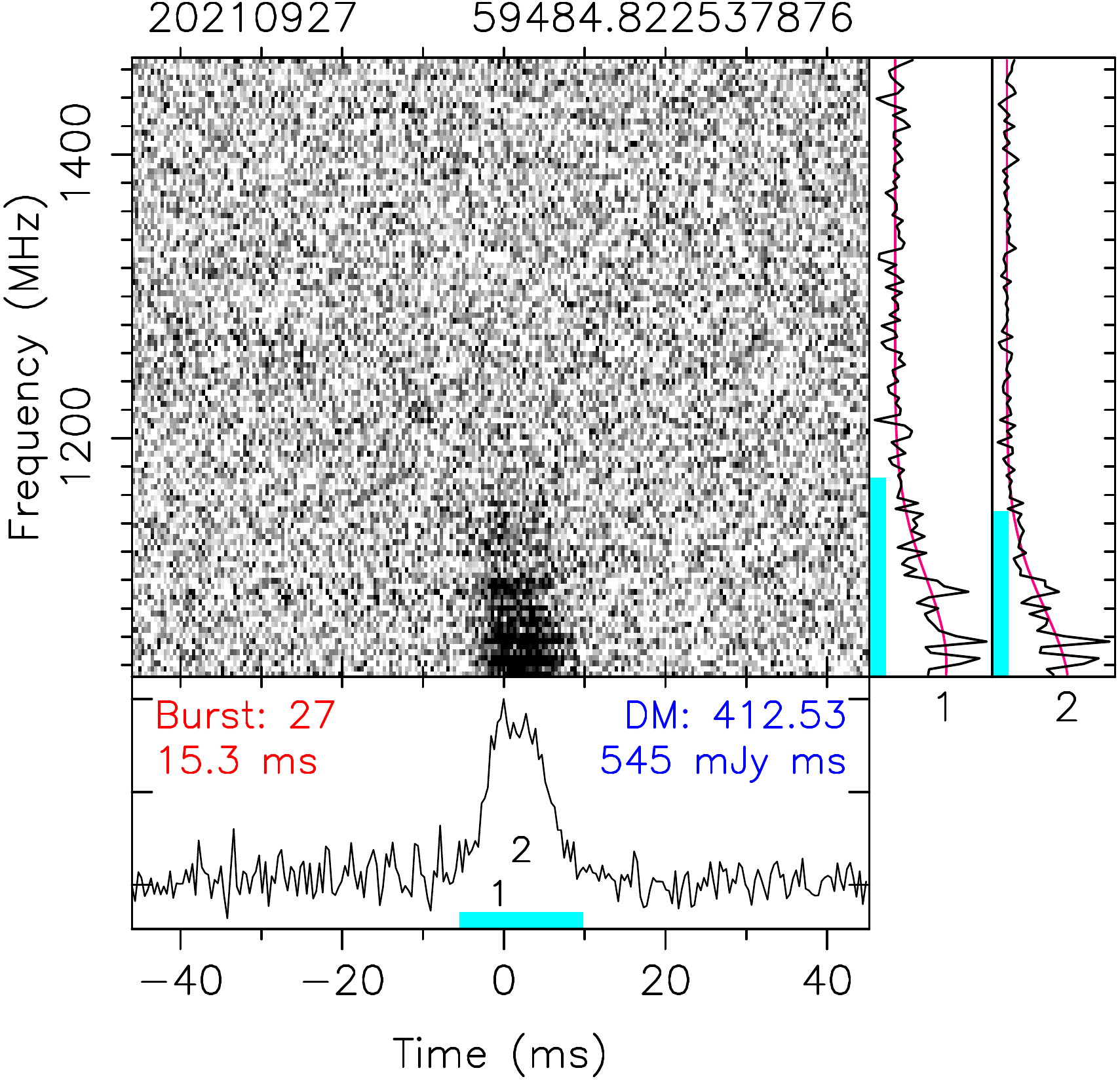}
    \includegraphics[height=37mm]{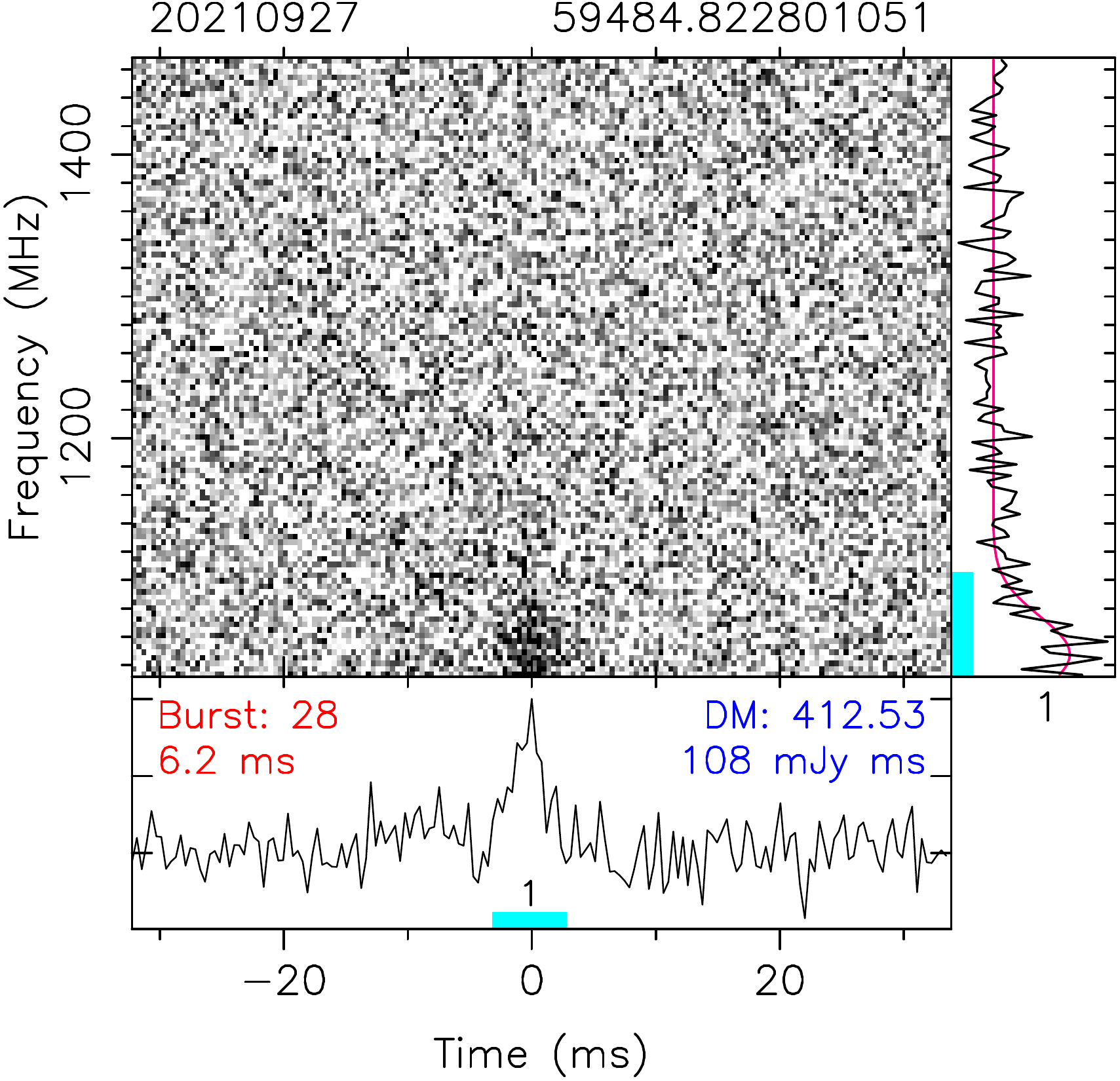}
    \includegraphics[height=37mm]{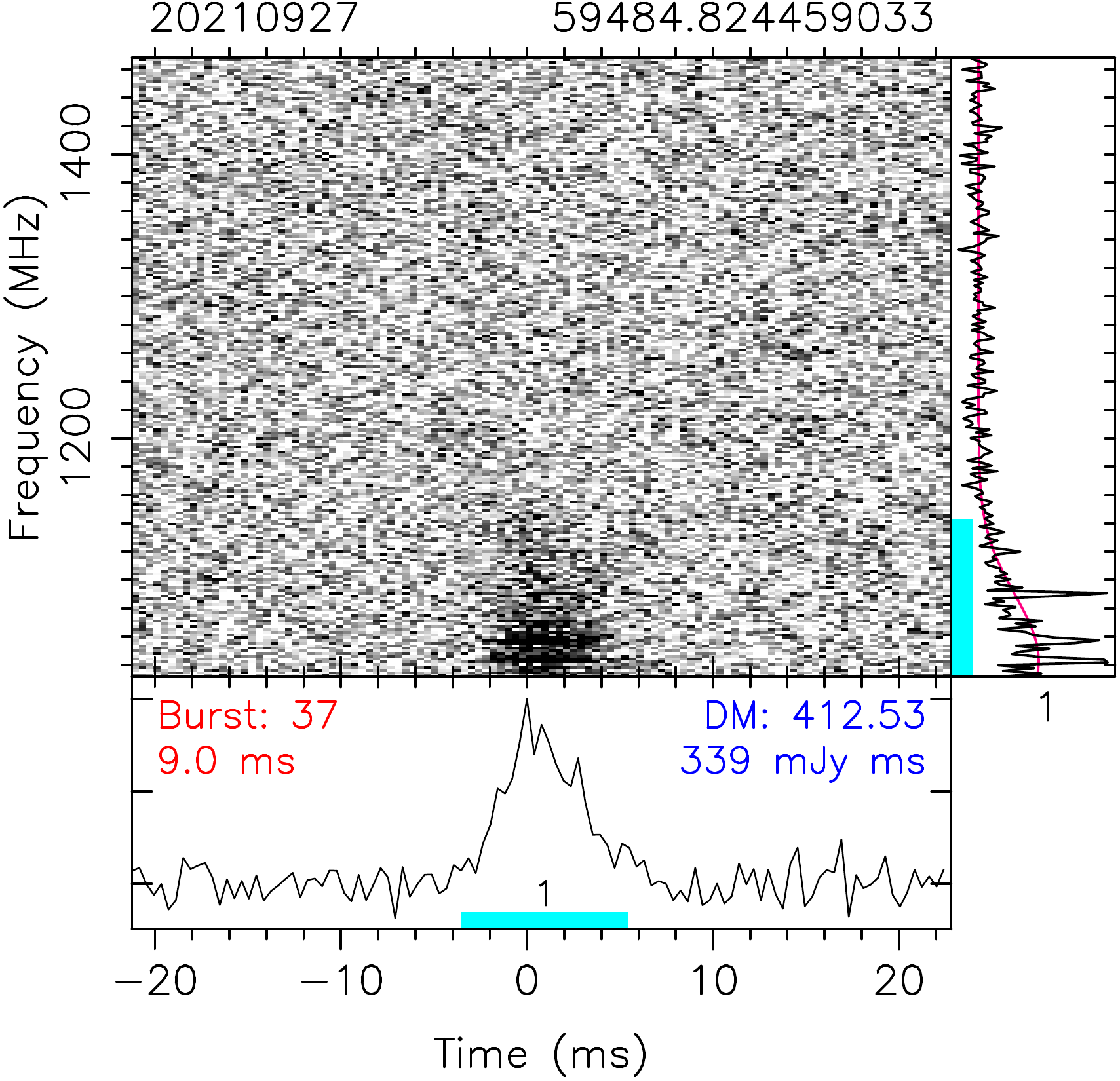}
    \includegraphics[height=37mm]{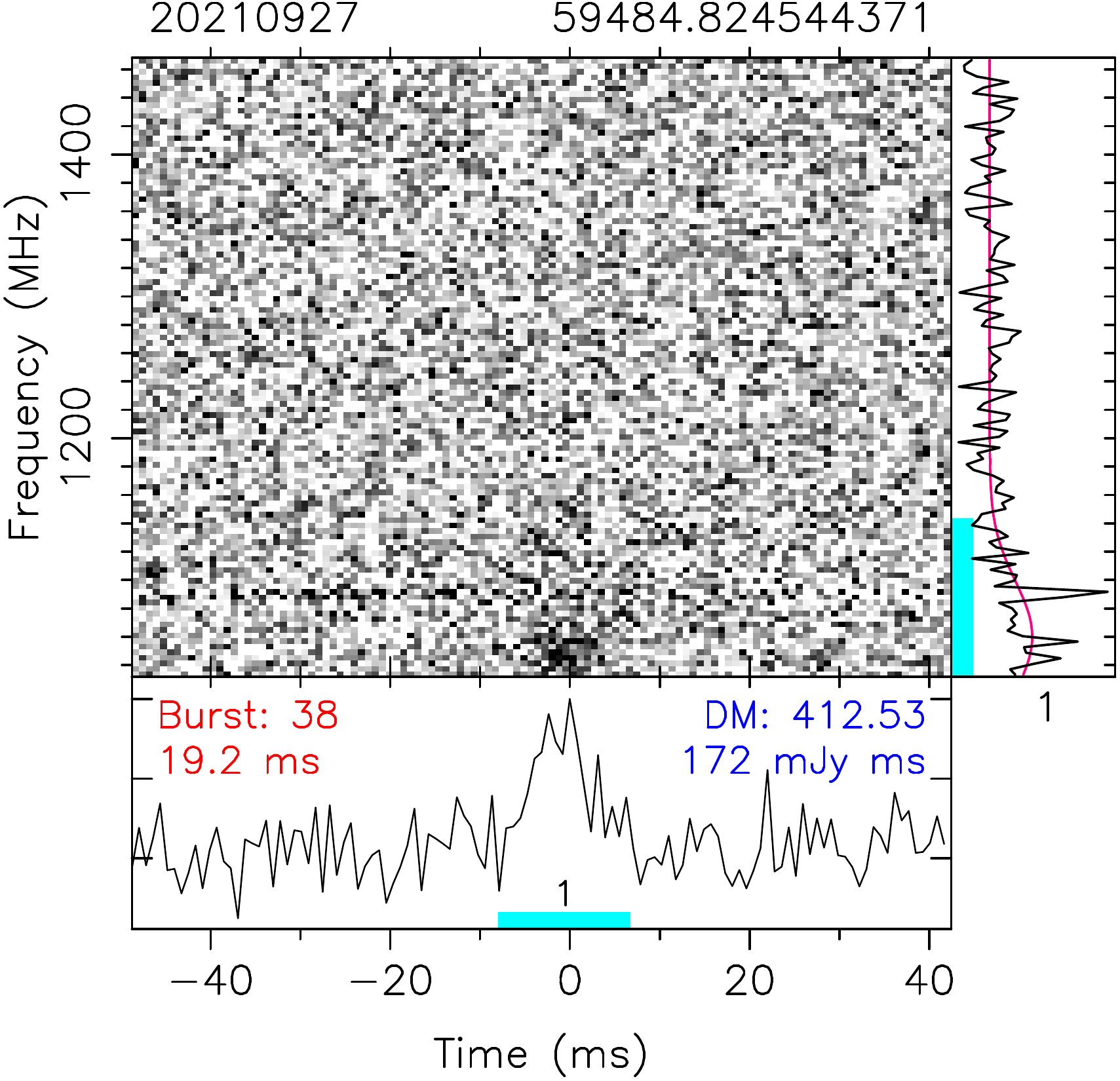}
    \includegraphics[height=37mm]{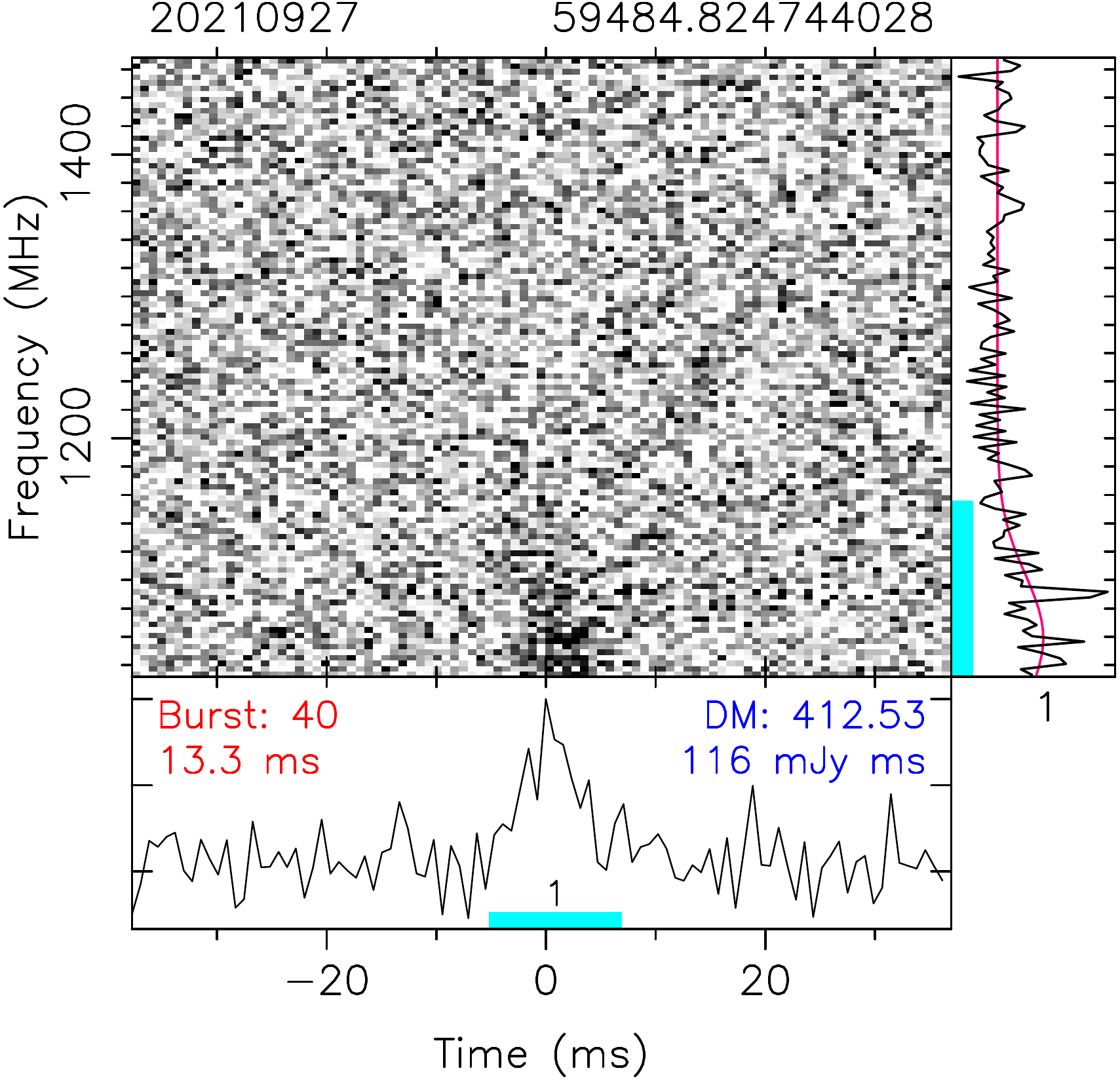}
    \includegraphics[height=37mm]{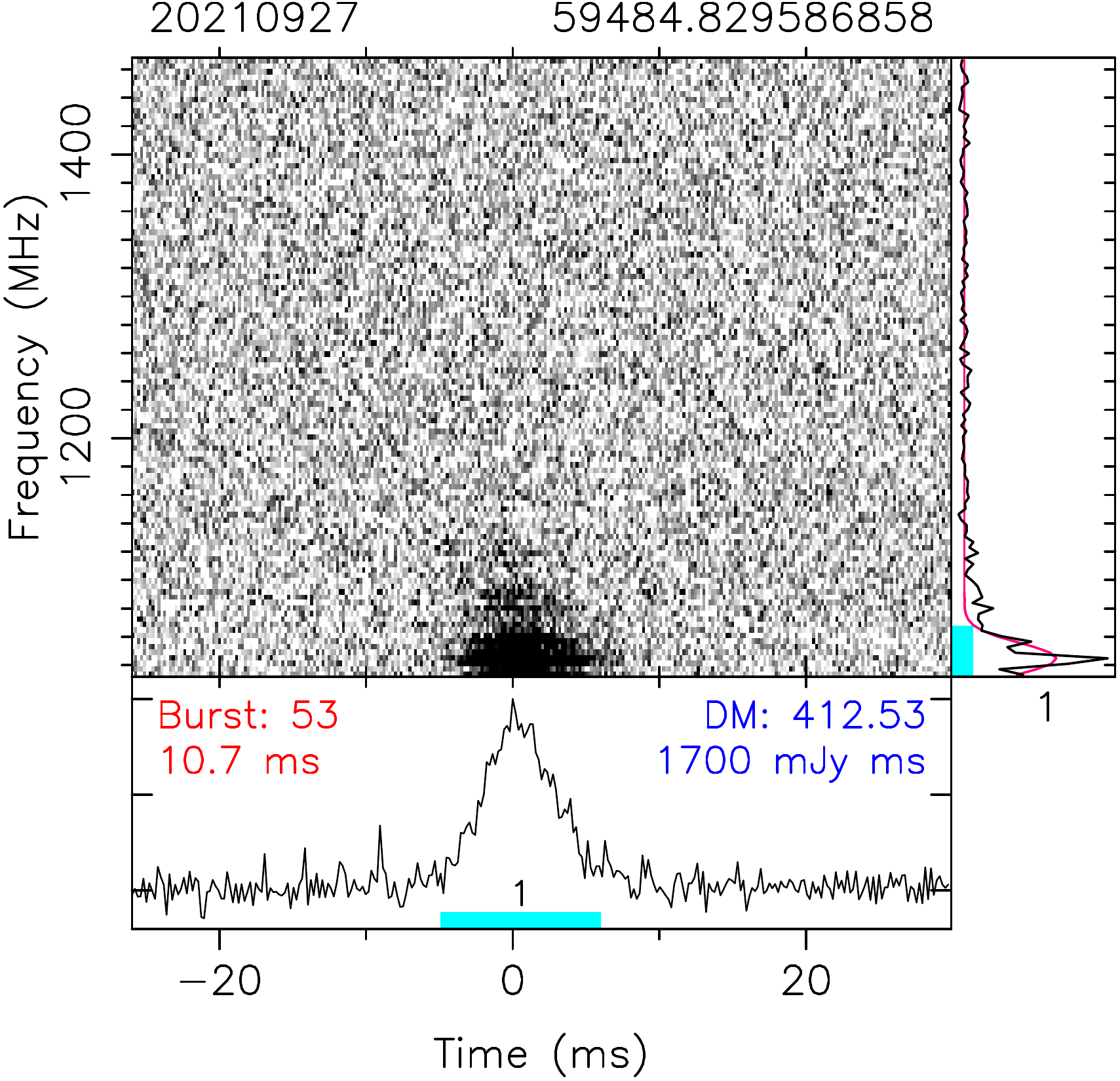}
    \includegraphics[height=37mm]{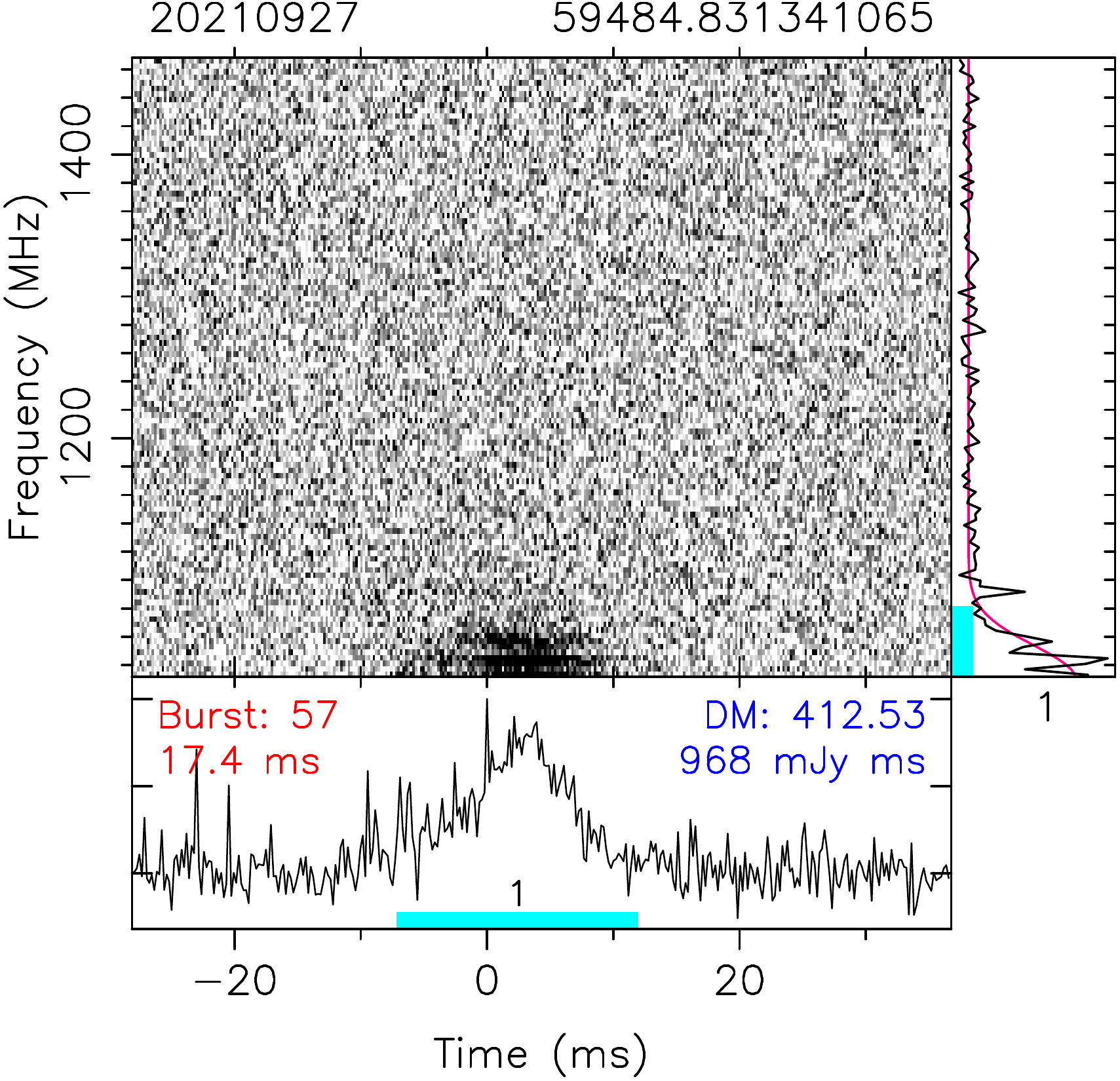}
    \includegraphics[height=37mm]{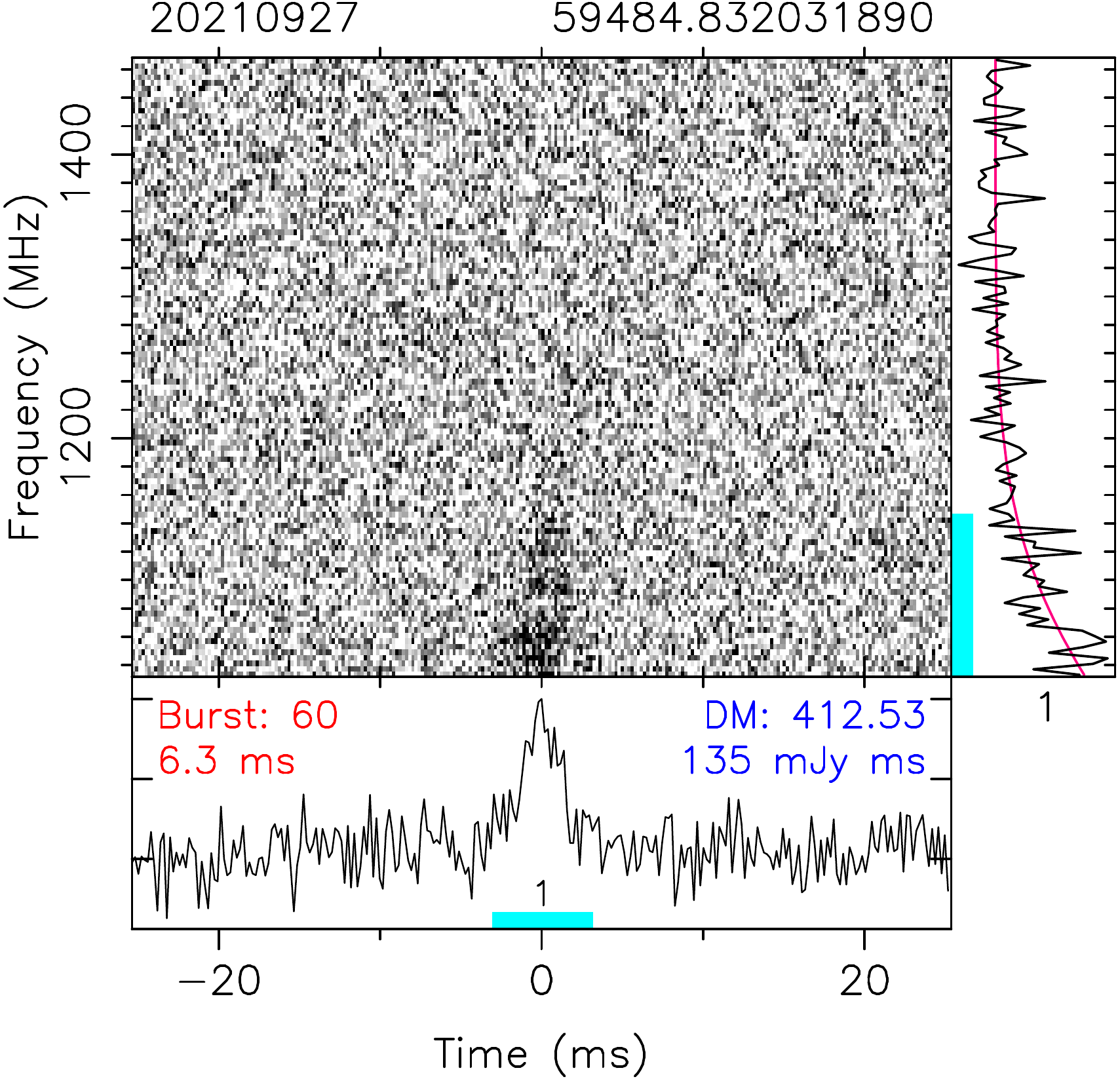}
    \caption{ \it{ -- continued}.
}
\end{figure*}
\addtocounter{figure}{-1}
\begin{figure*}
    \flushleft
    \includegraphics[height=37mm]{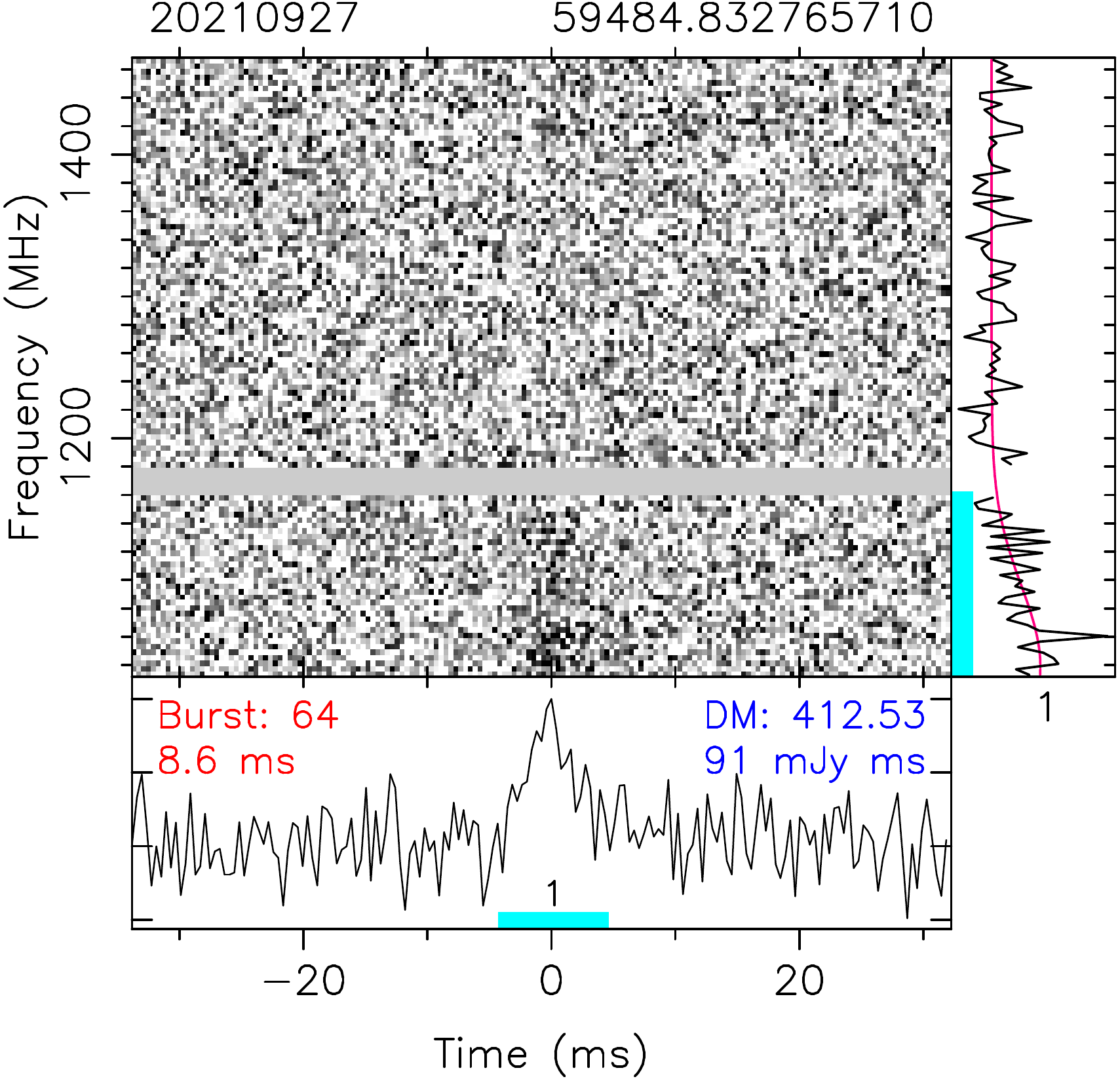}
    \includegraphics[height=37mm]{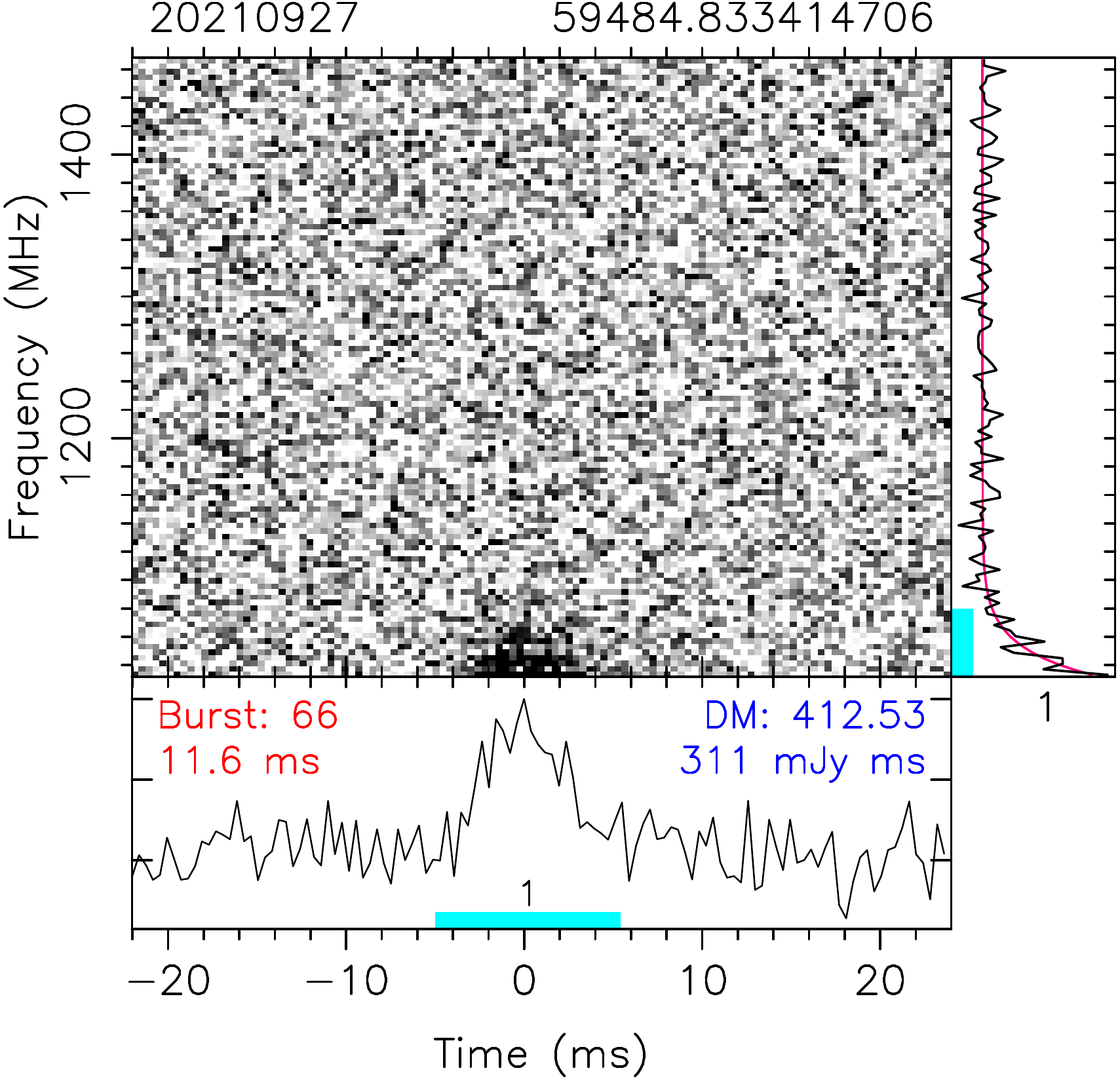}
    \includegraphics[height=37mm]{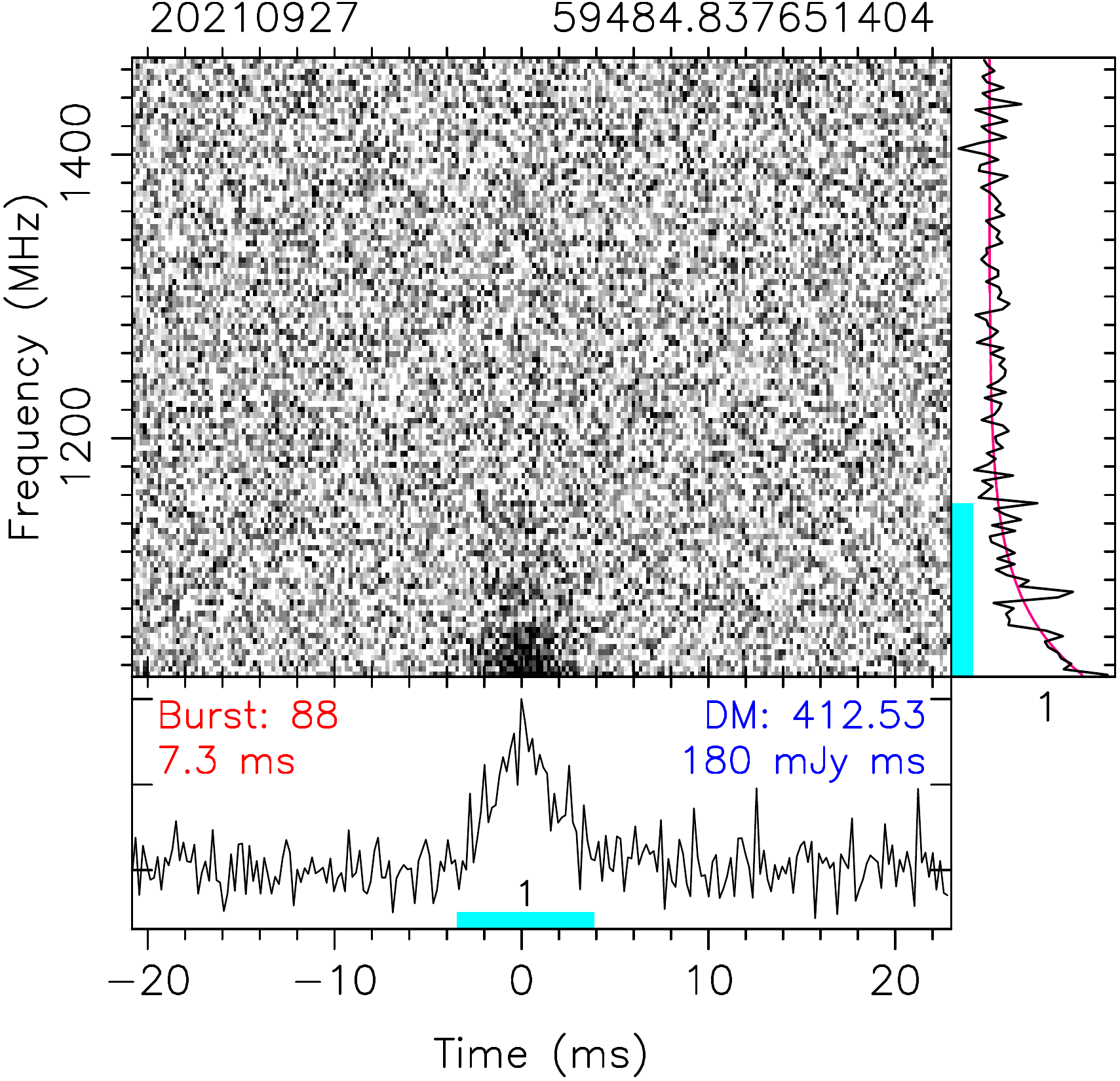}
    \includegraphics[height=37mm]{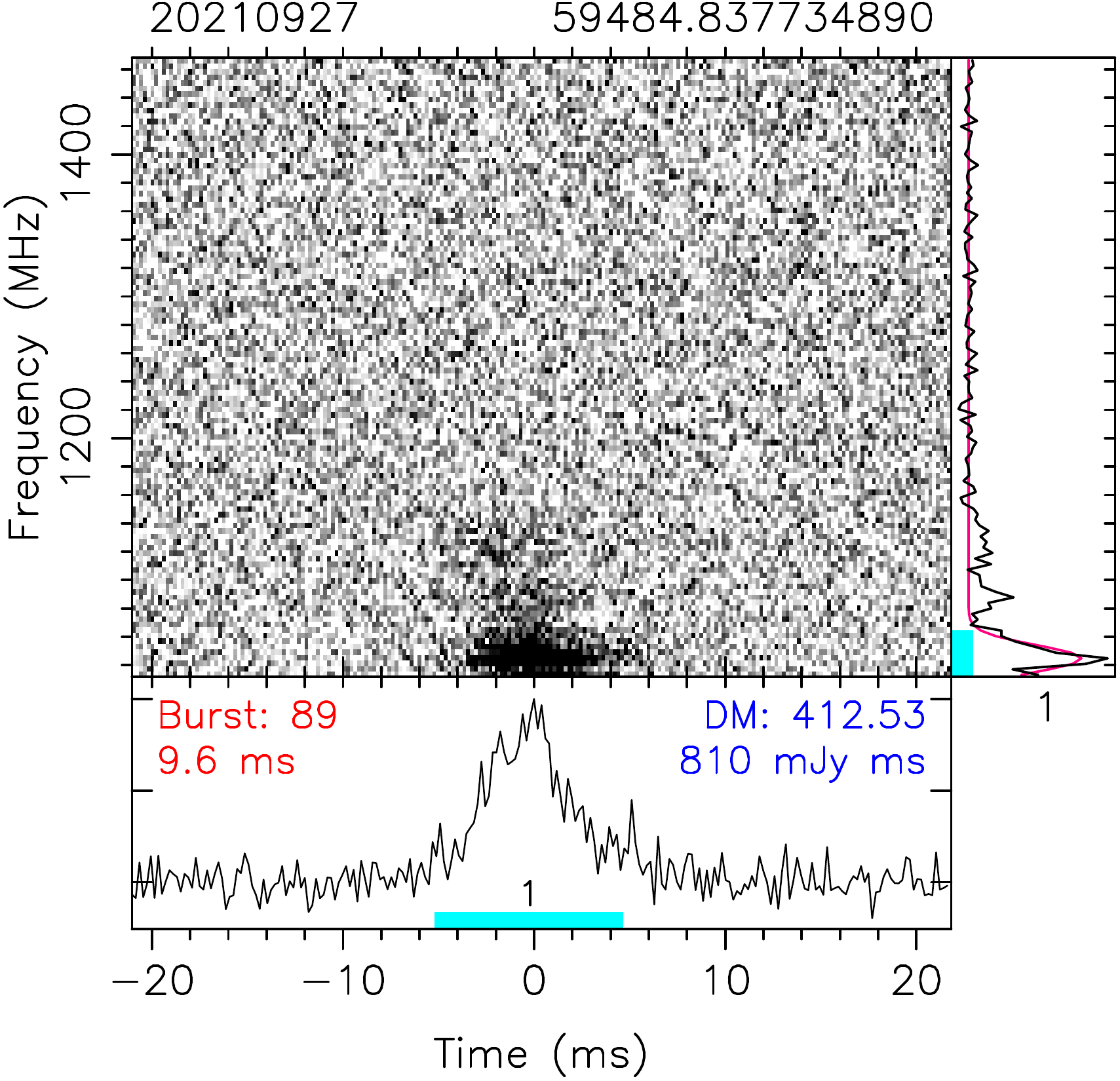}
    \includegraphics[height=37mm]{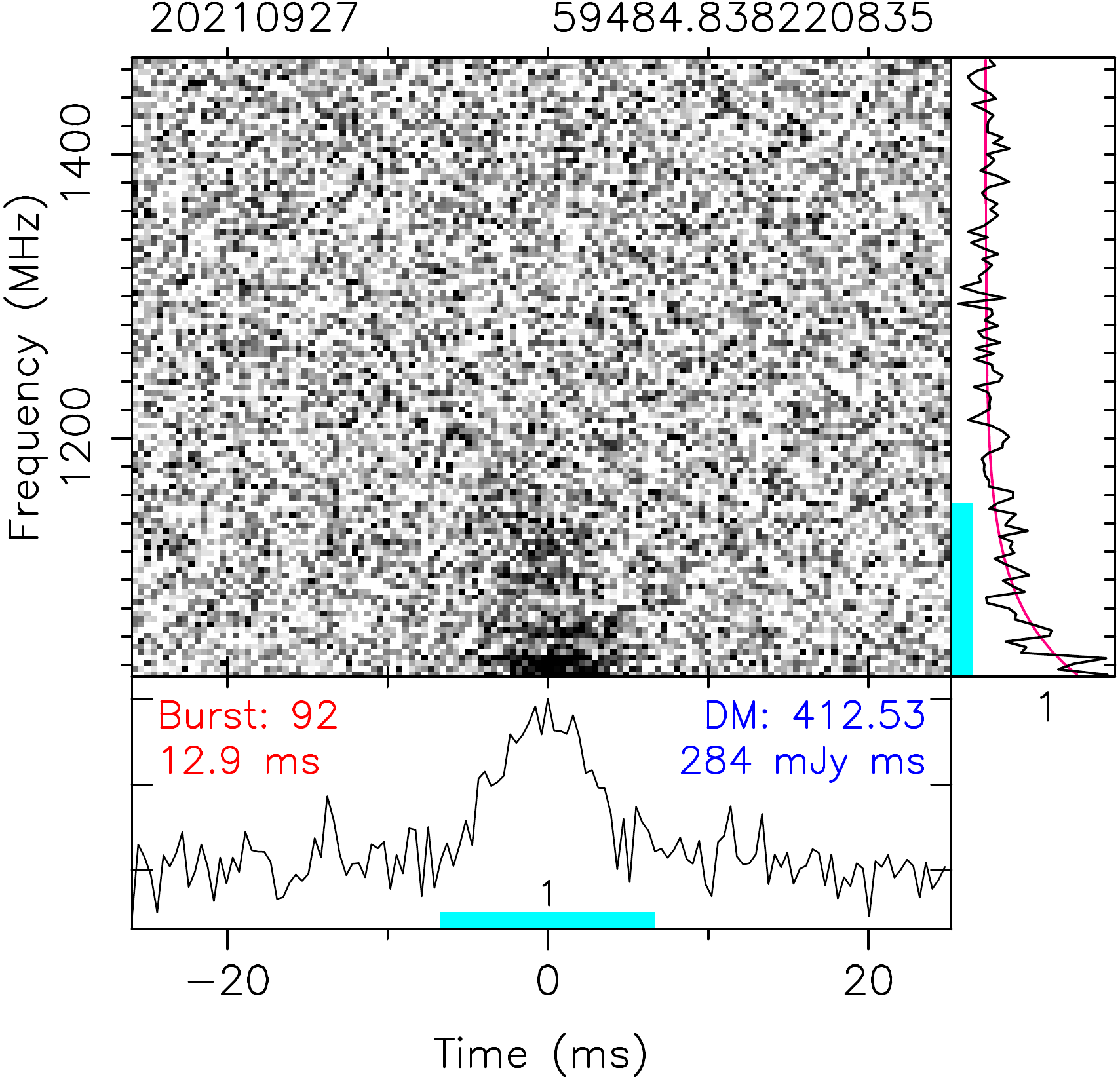}
    \includegraphics[height=37mm]{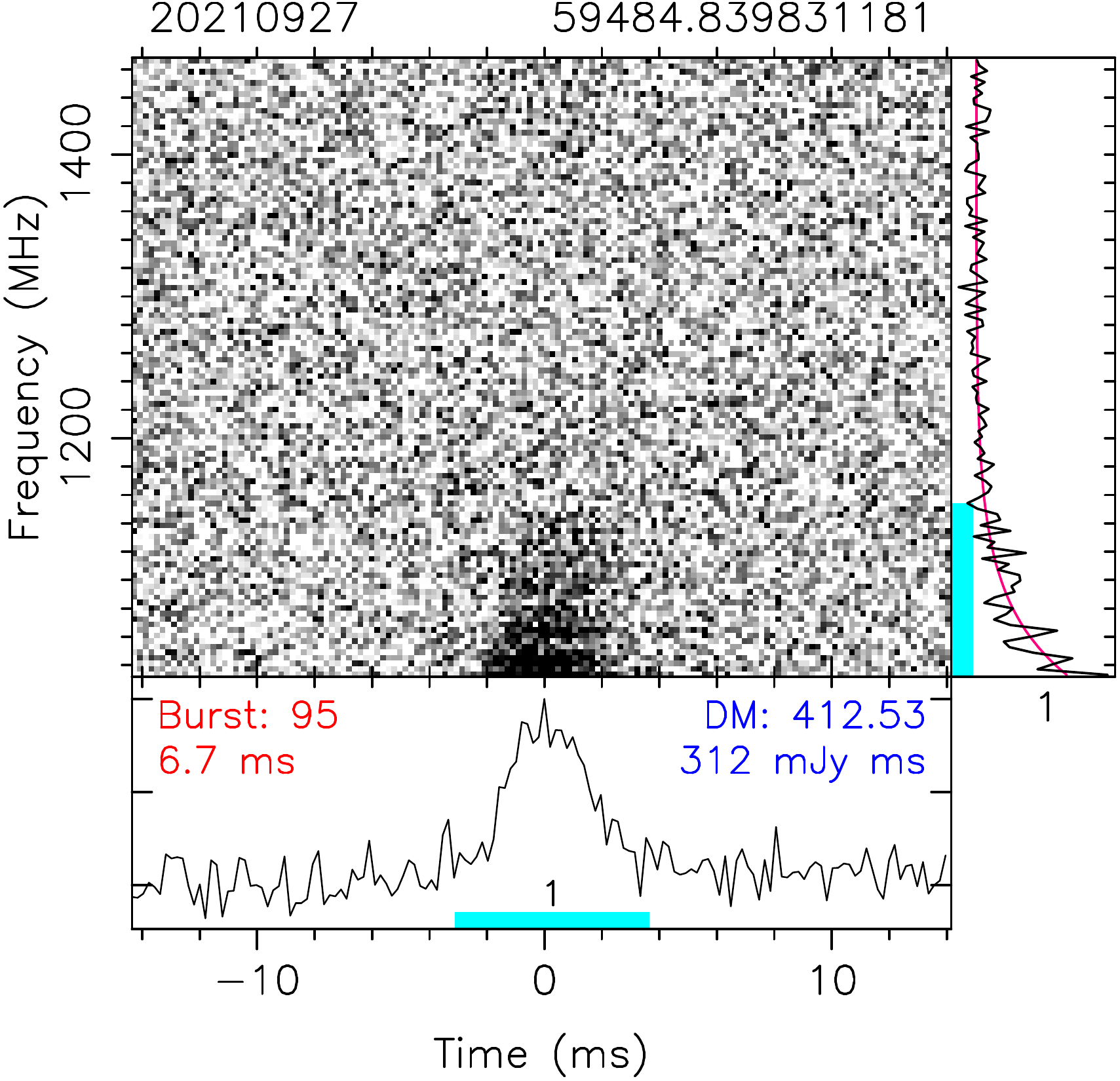}
    \includegraphics[height=37mm]{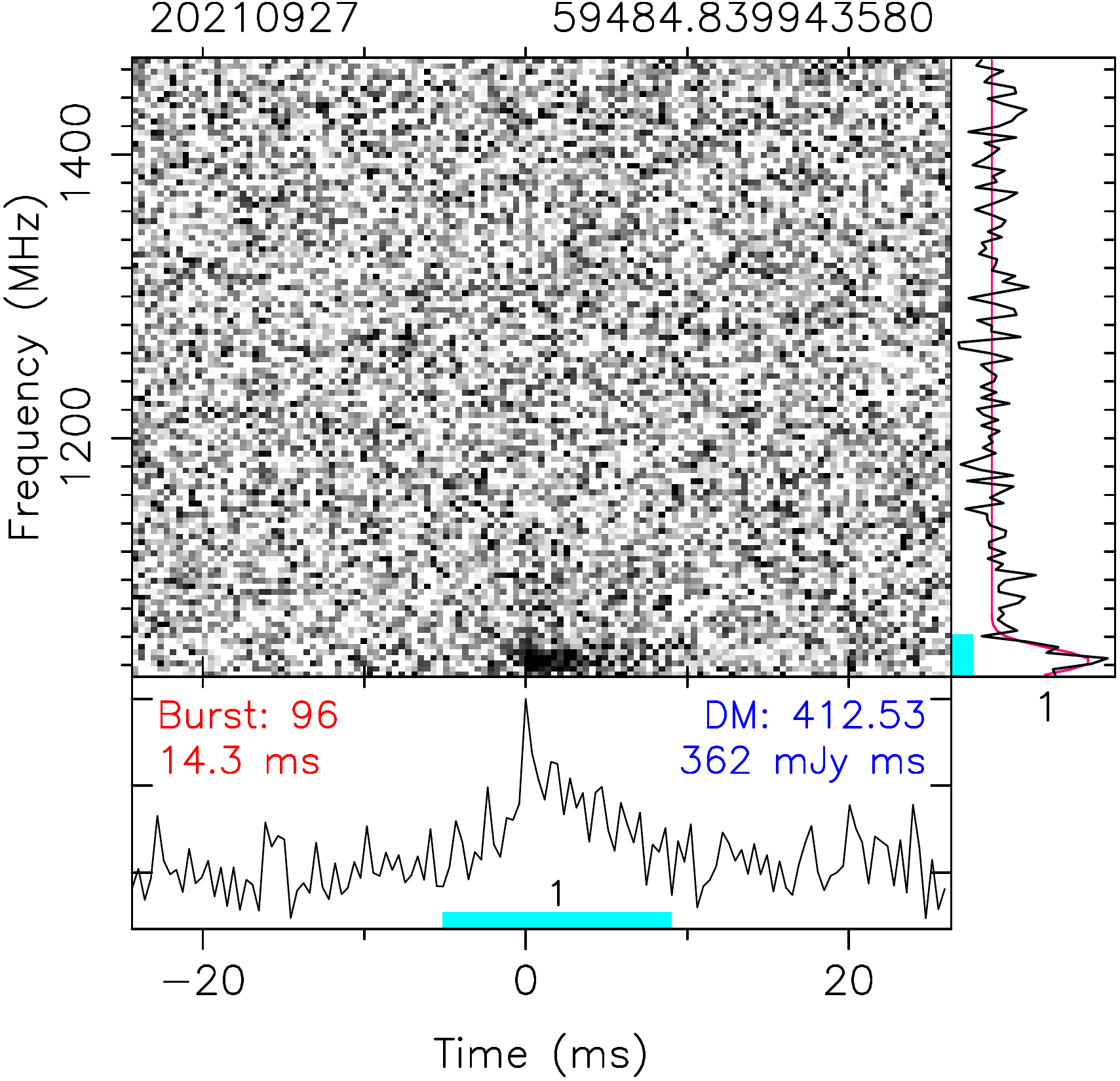}
    \includegraphics[height=37mm]{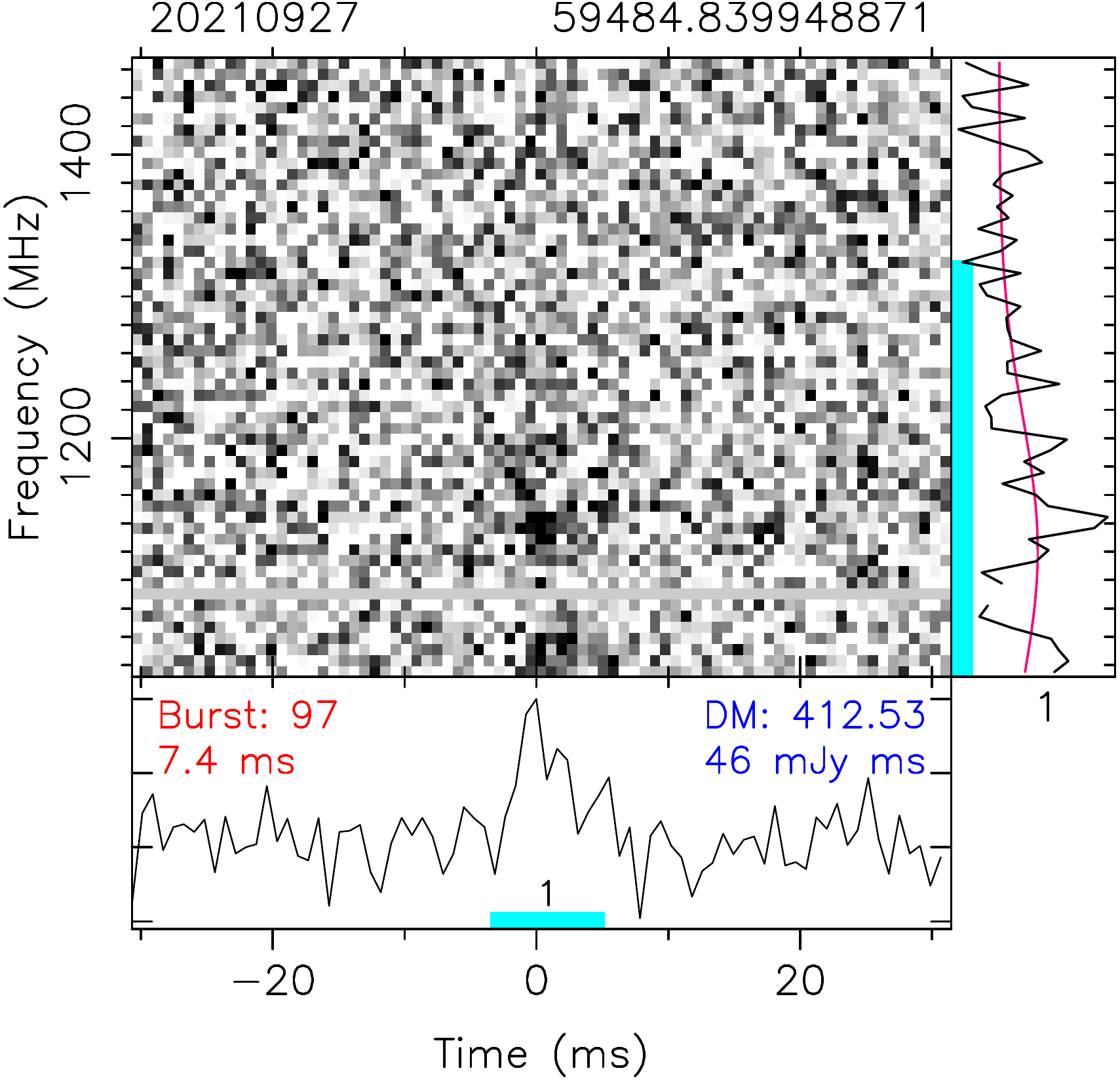}
    \includegraphics[height=37mm]{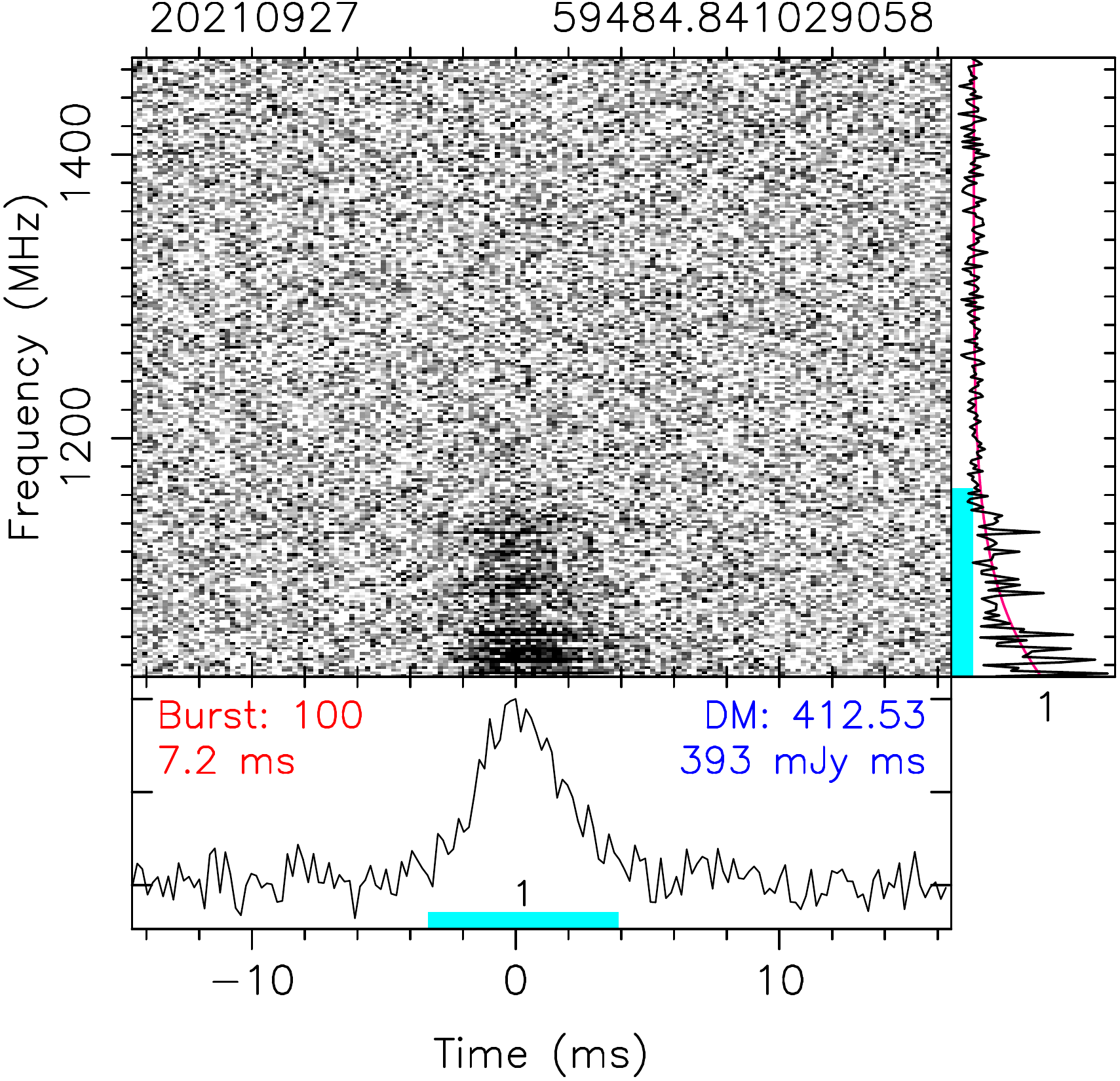}
    \includegraphics[height=37mm]{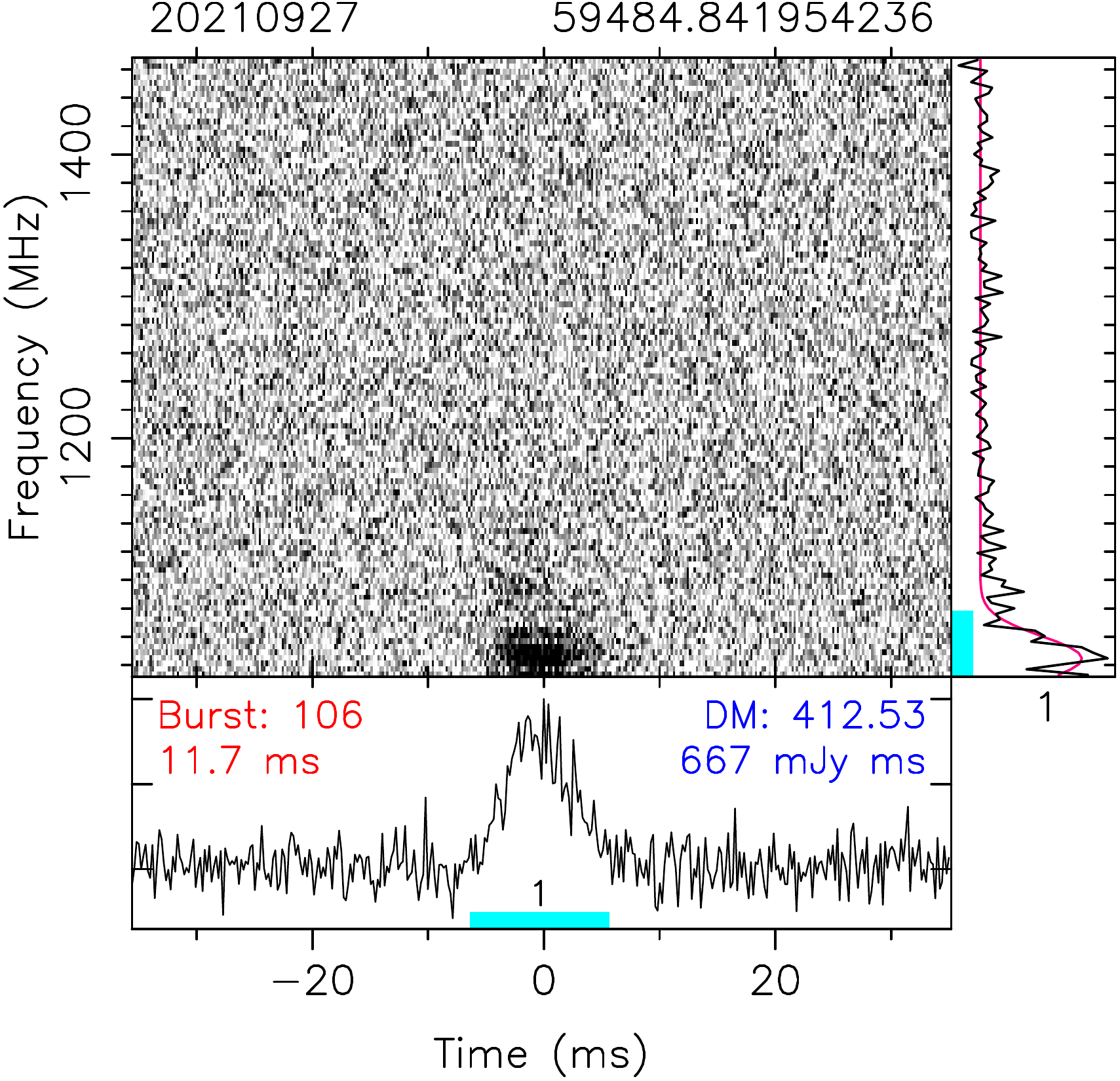}
    \includegraphics[height=37mm]{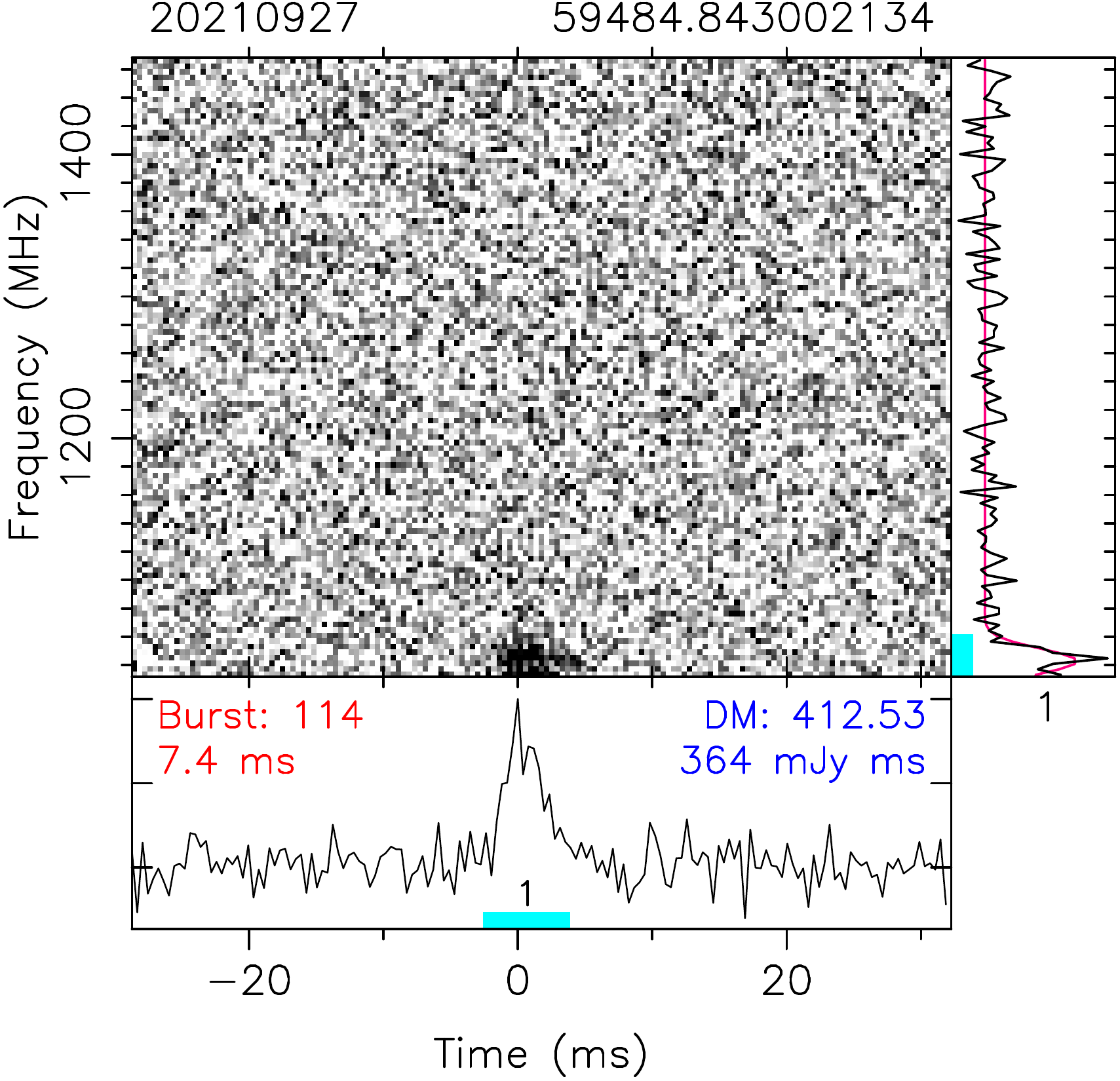}
    \includegraphics[height=37mm]{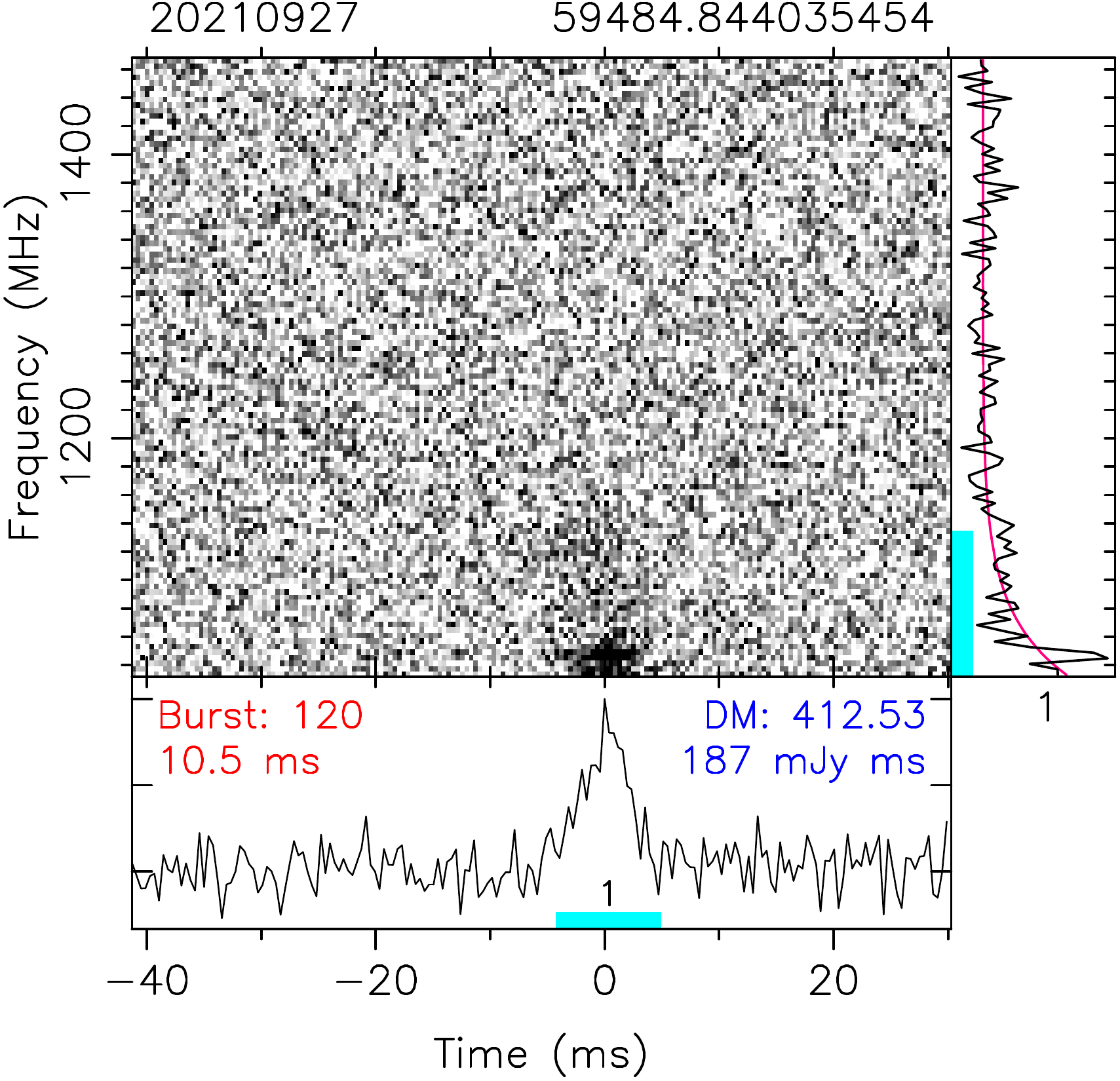}
    \includegraphics[height=37mm]{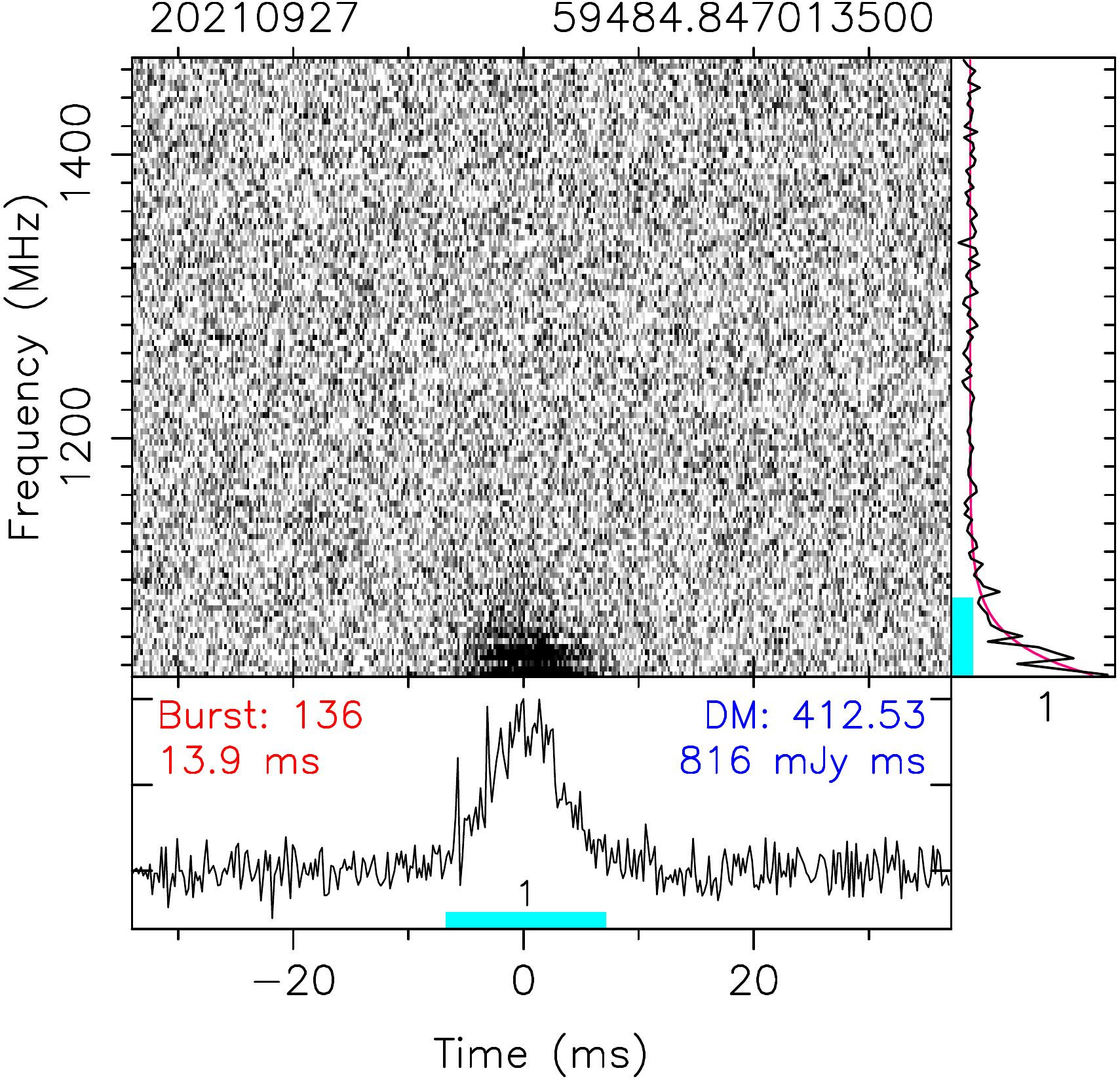}
    \includegraphics[height=37mm]{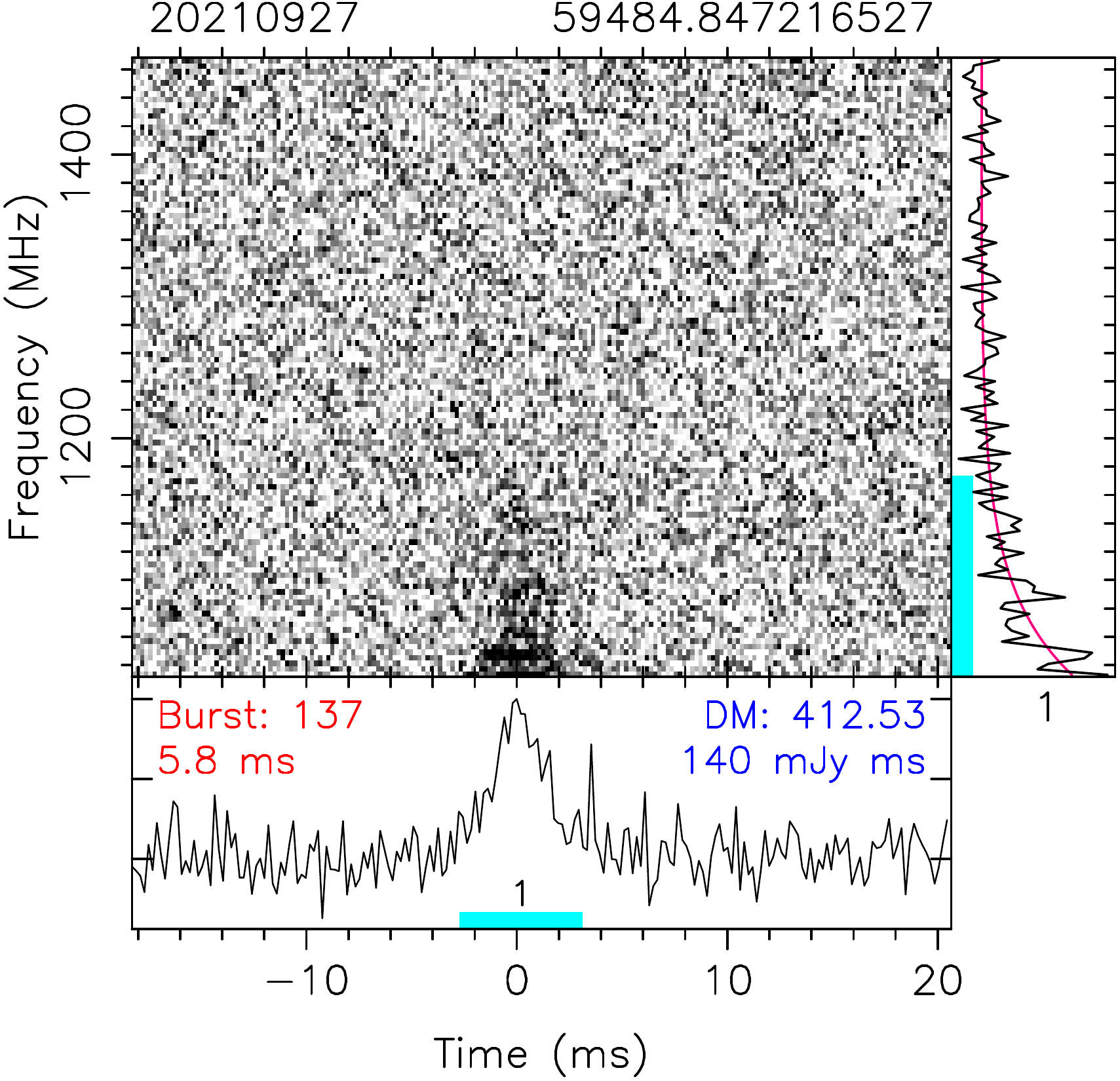}
    \includegraphics[height=37mm]{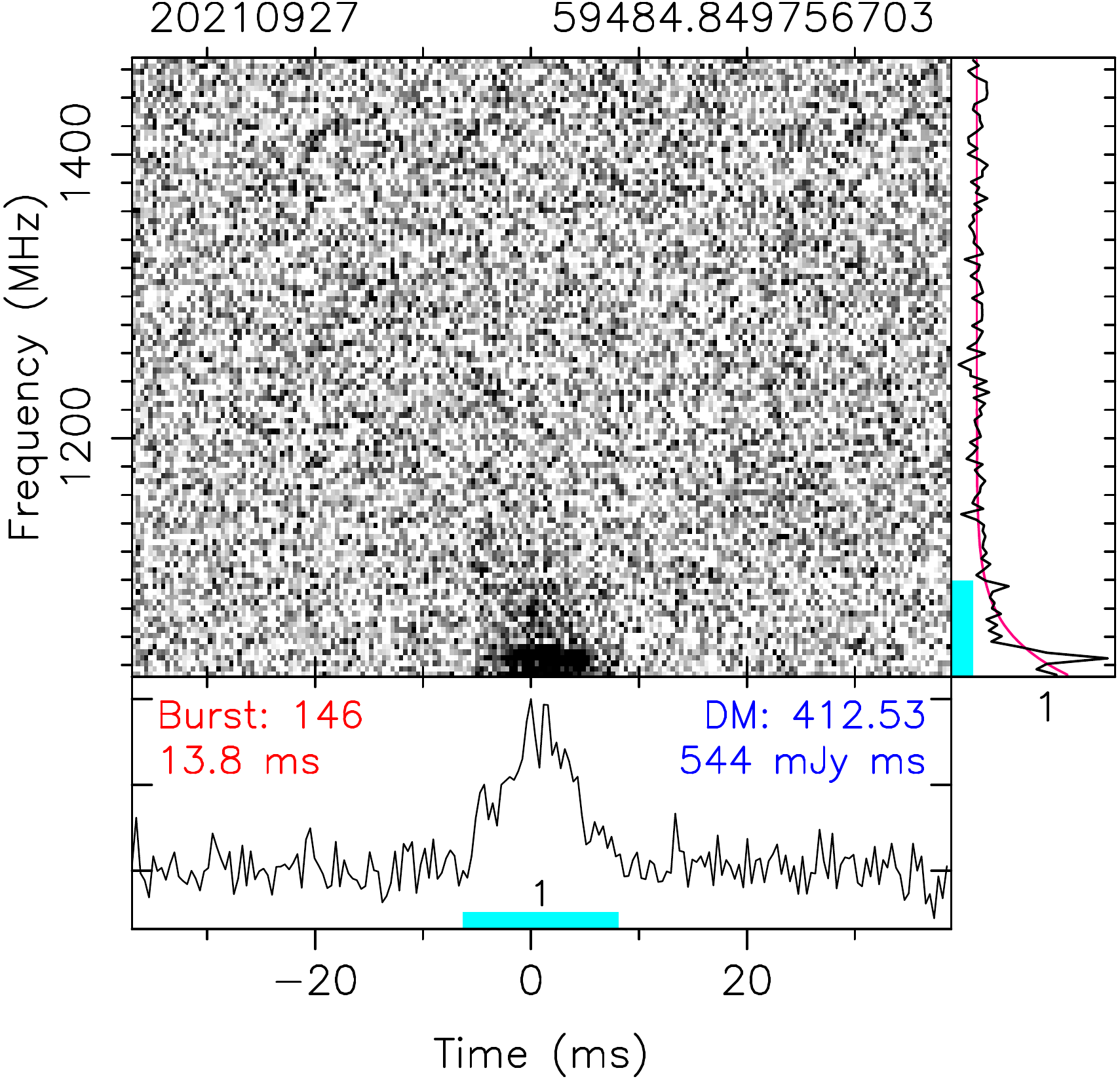}
    \includegraphics[height=37mm]{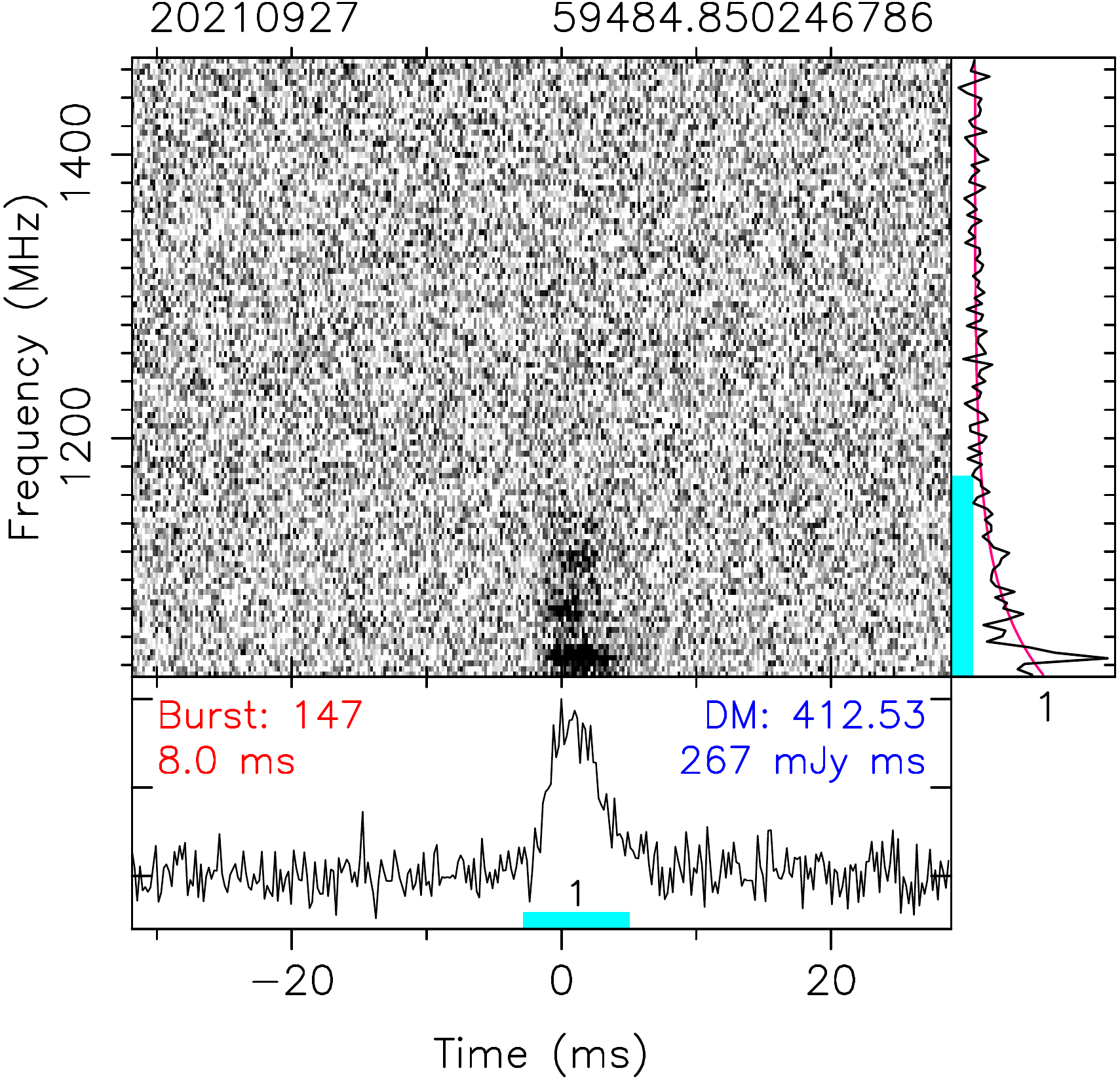}
    \includegraphics[height=37mm]{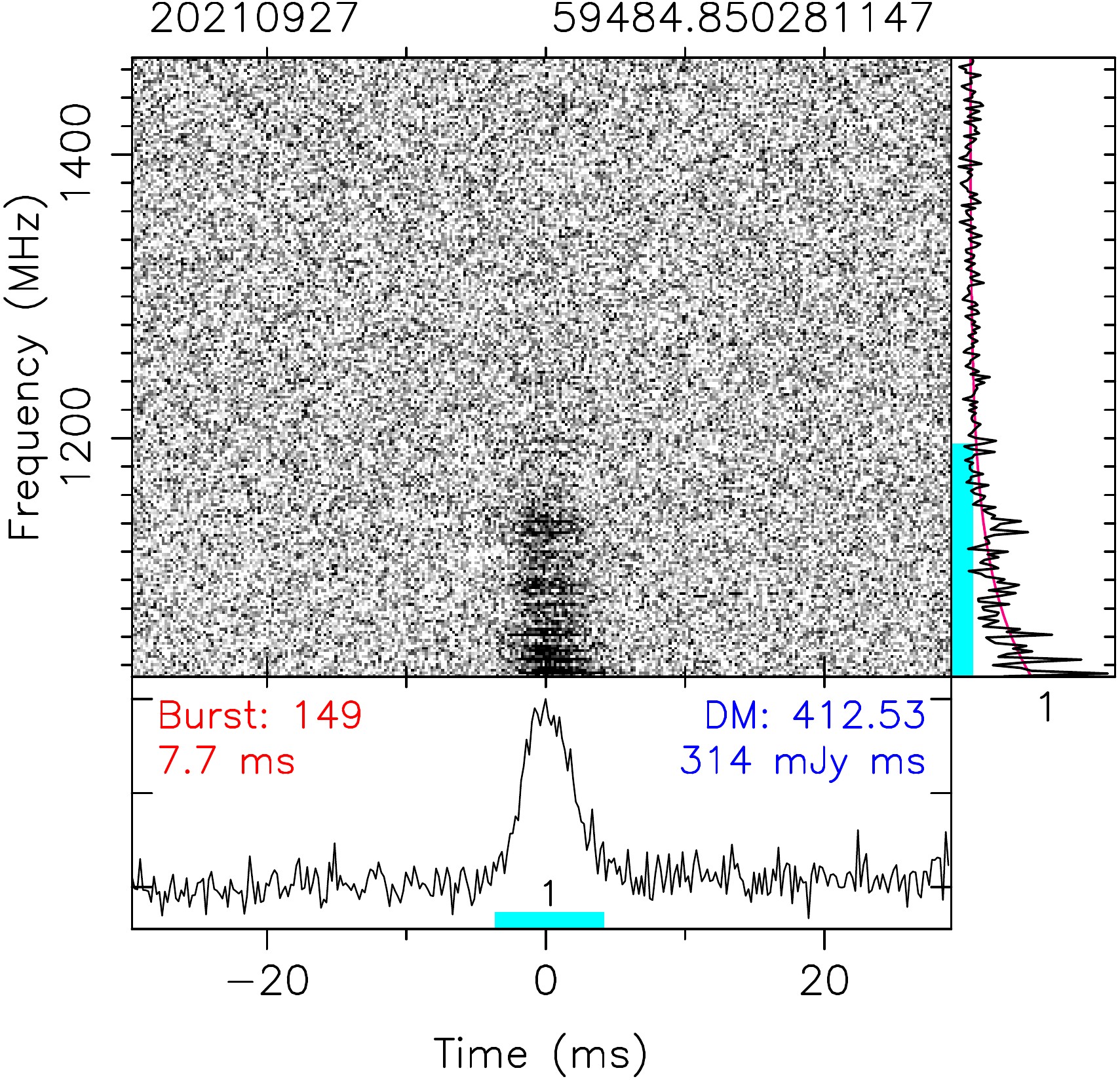}
    \includegraphics[height=37mm]{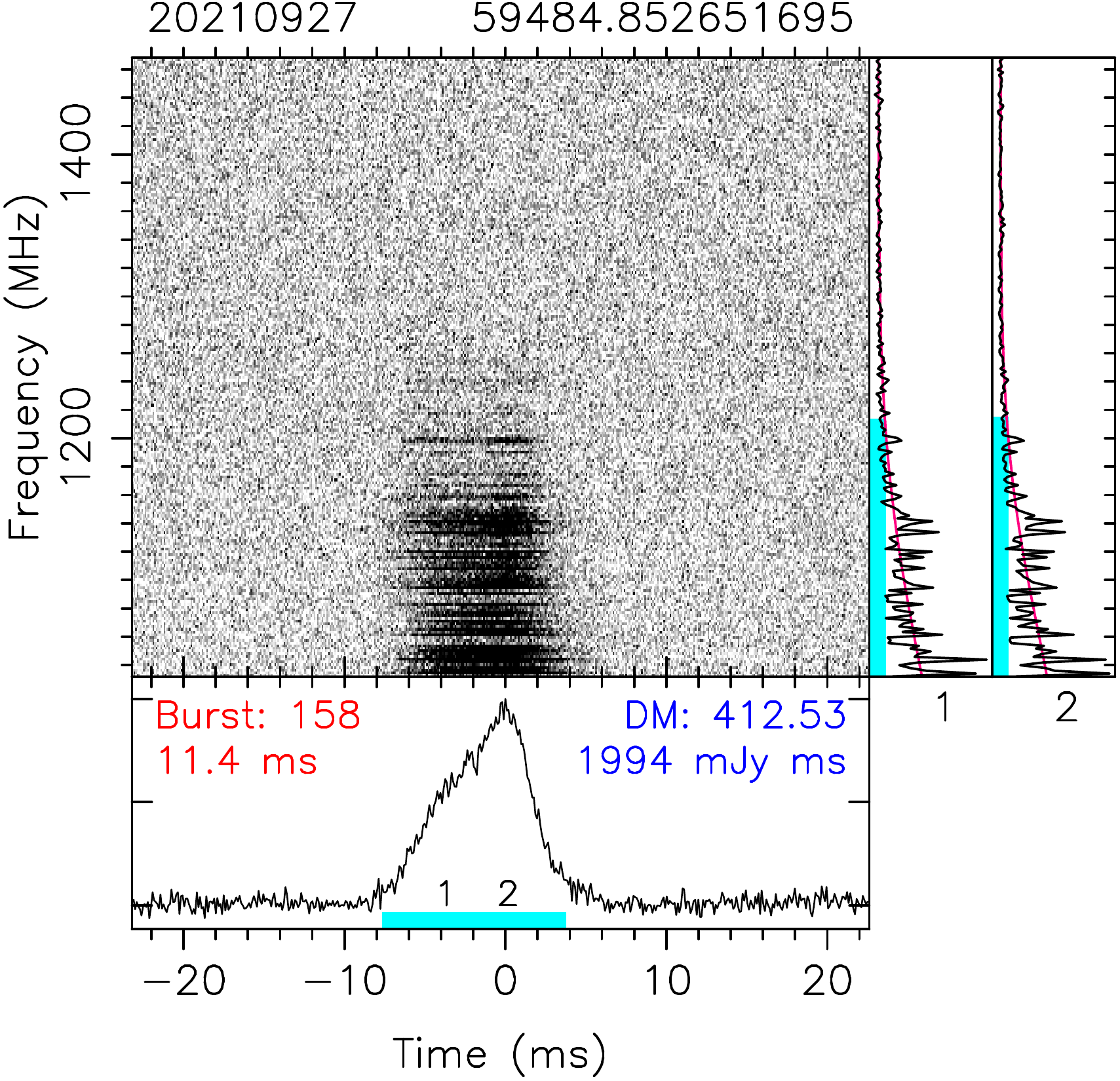}
    \includegraphics[height=37mm]{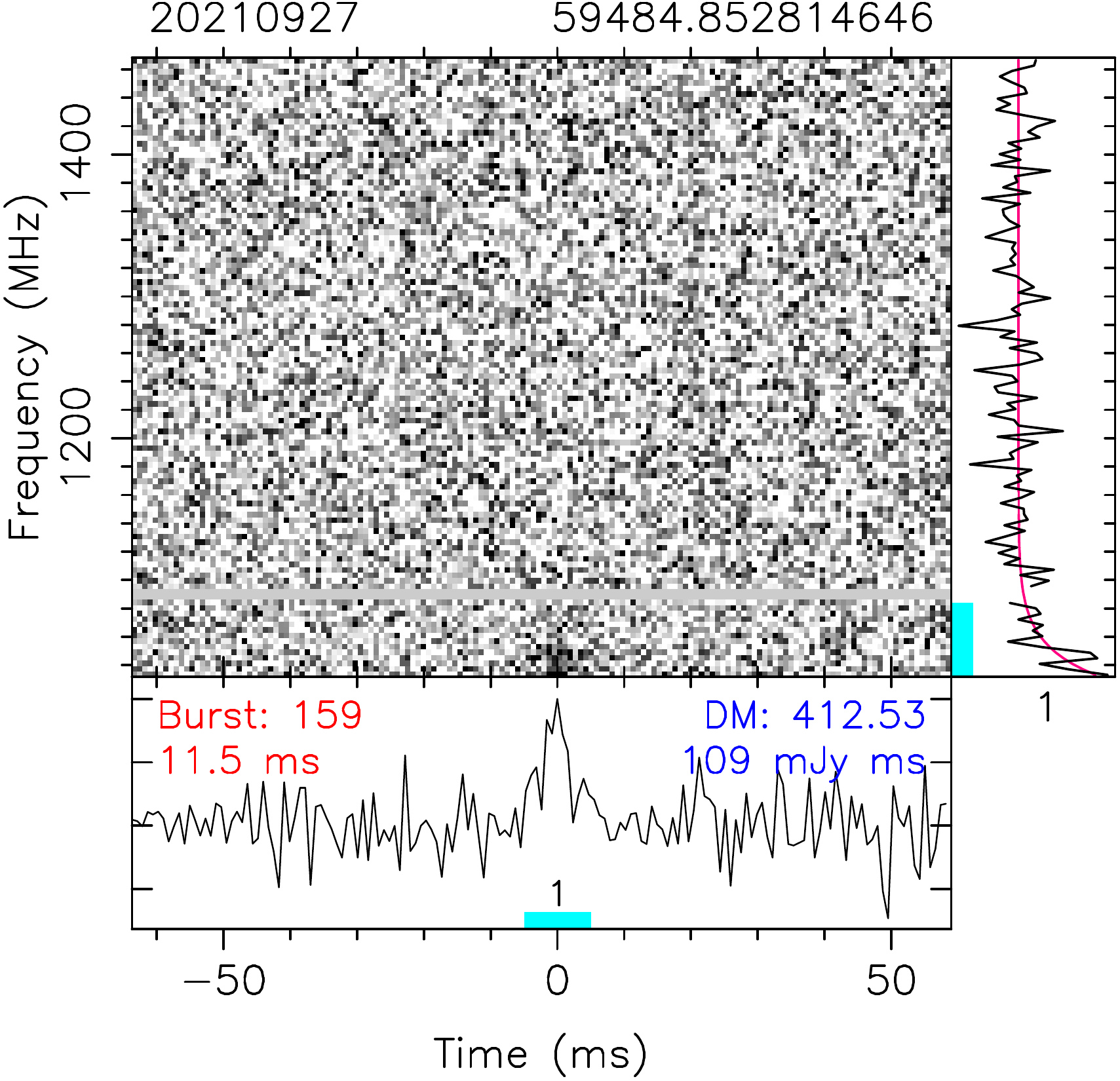}
    \includegraphics[height=37mm]{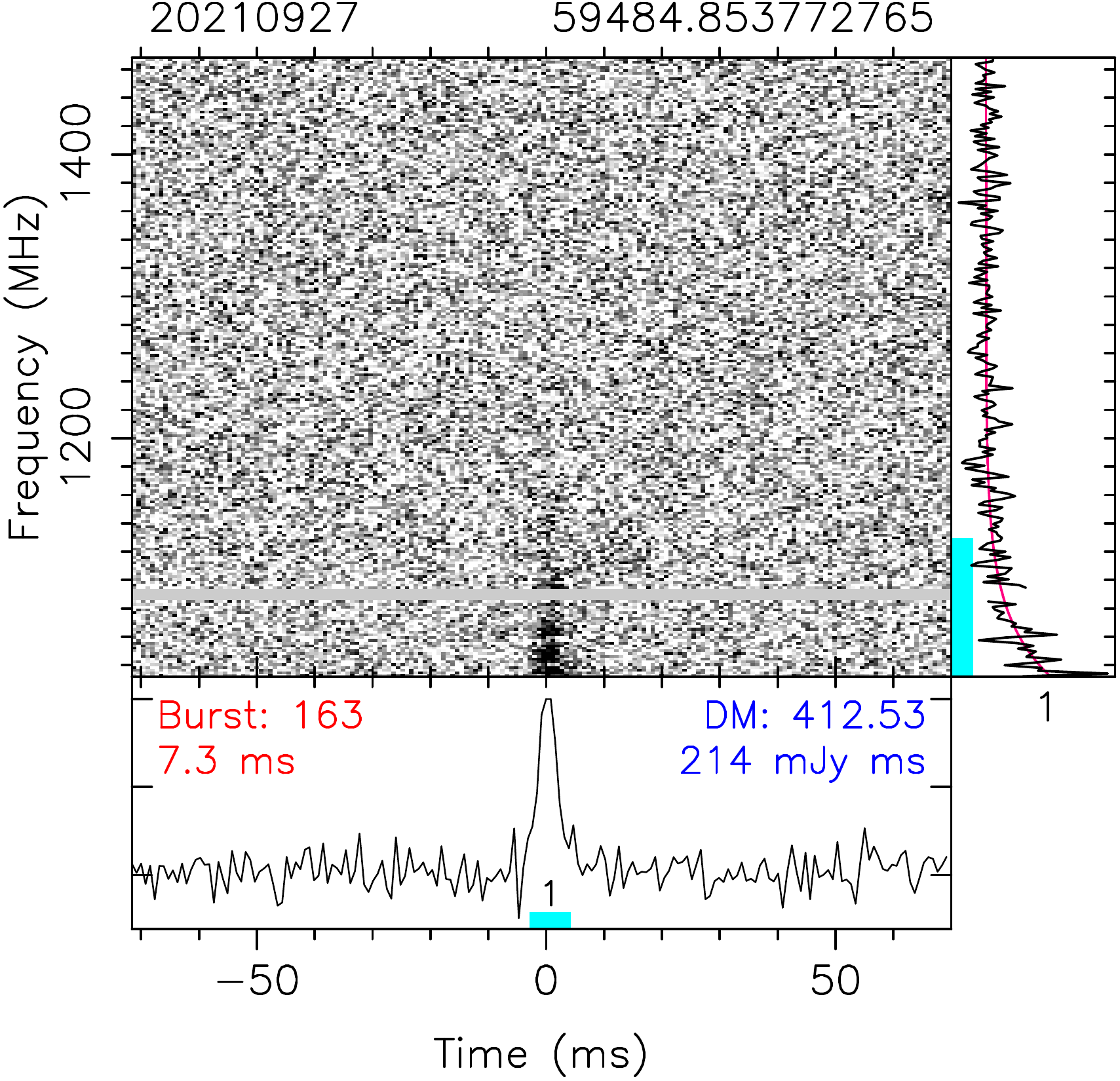}
    \includegraphics[height=37mm]{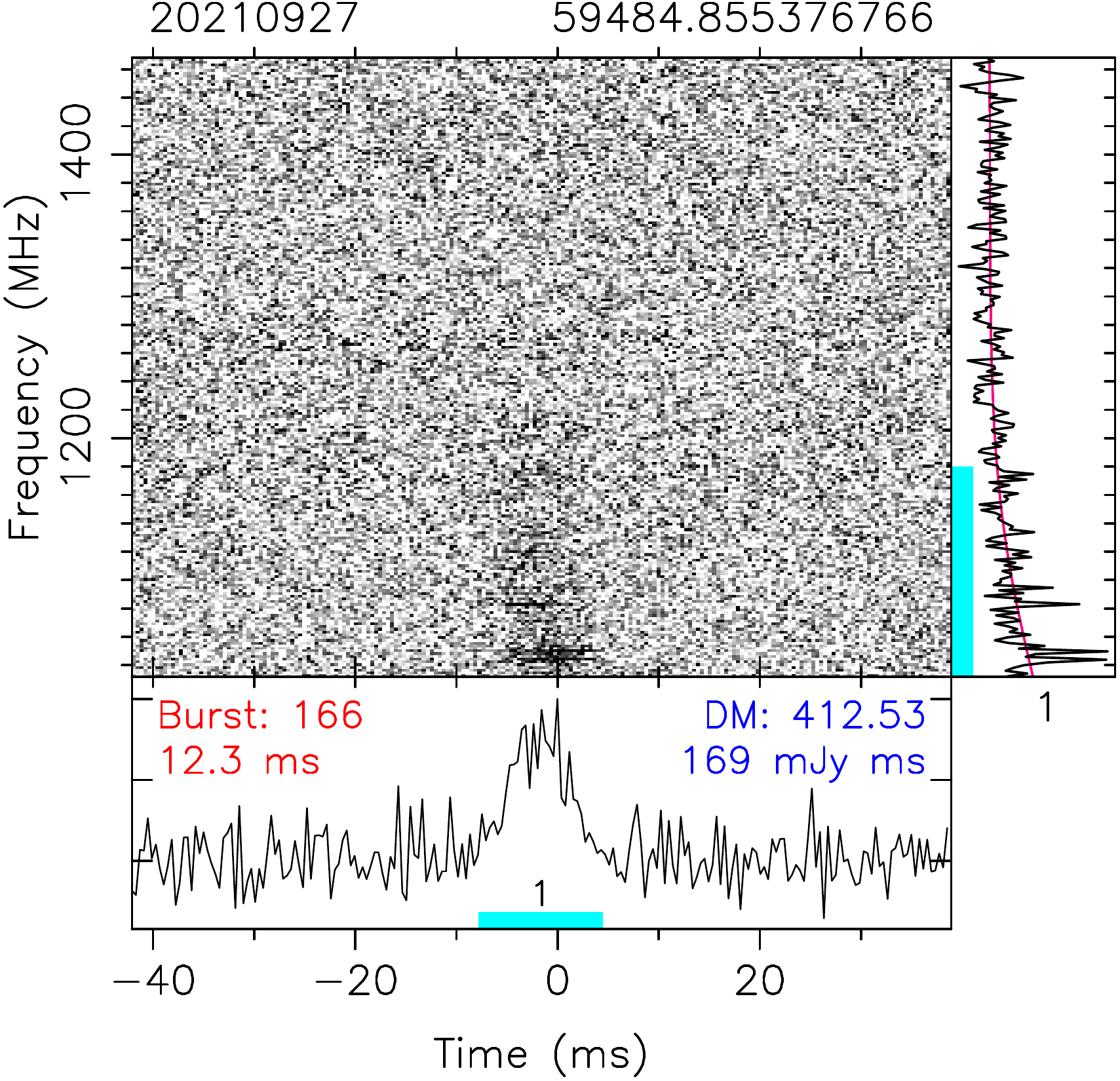}
    \includegraphics[height=37mm]{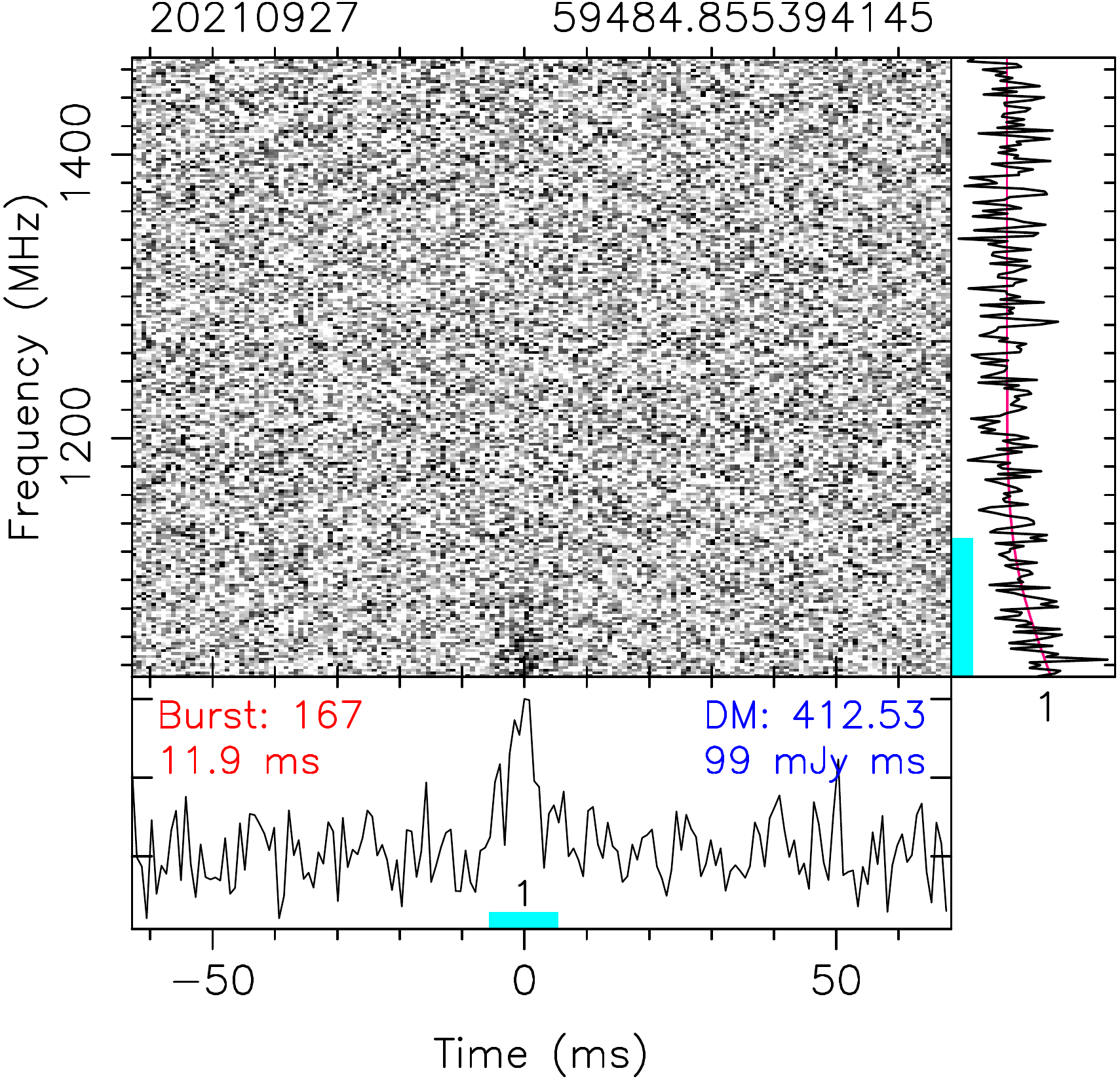}
    \includegraphics[height=37mm]{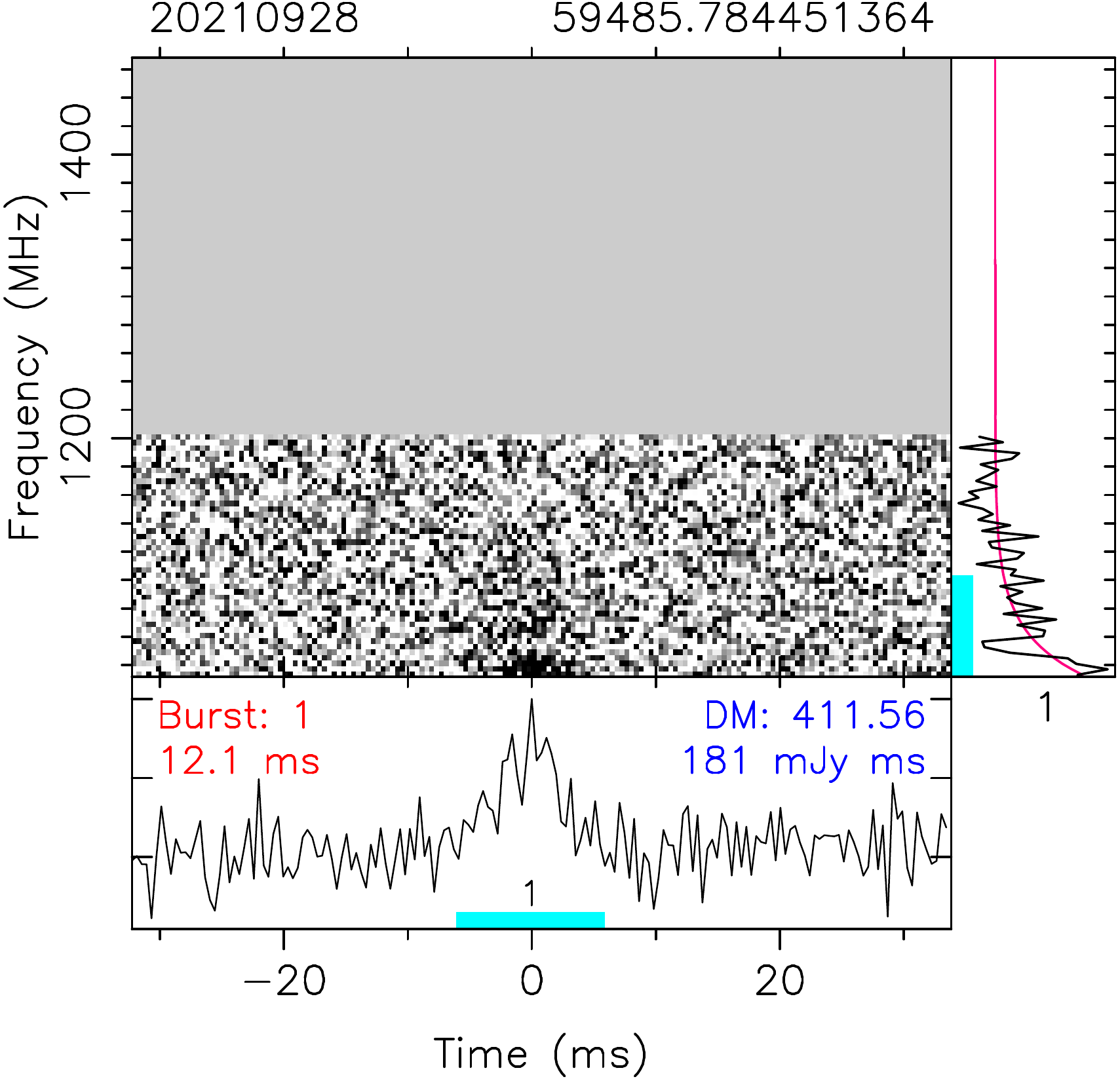}
    \includegraphics[height=37mm]{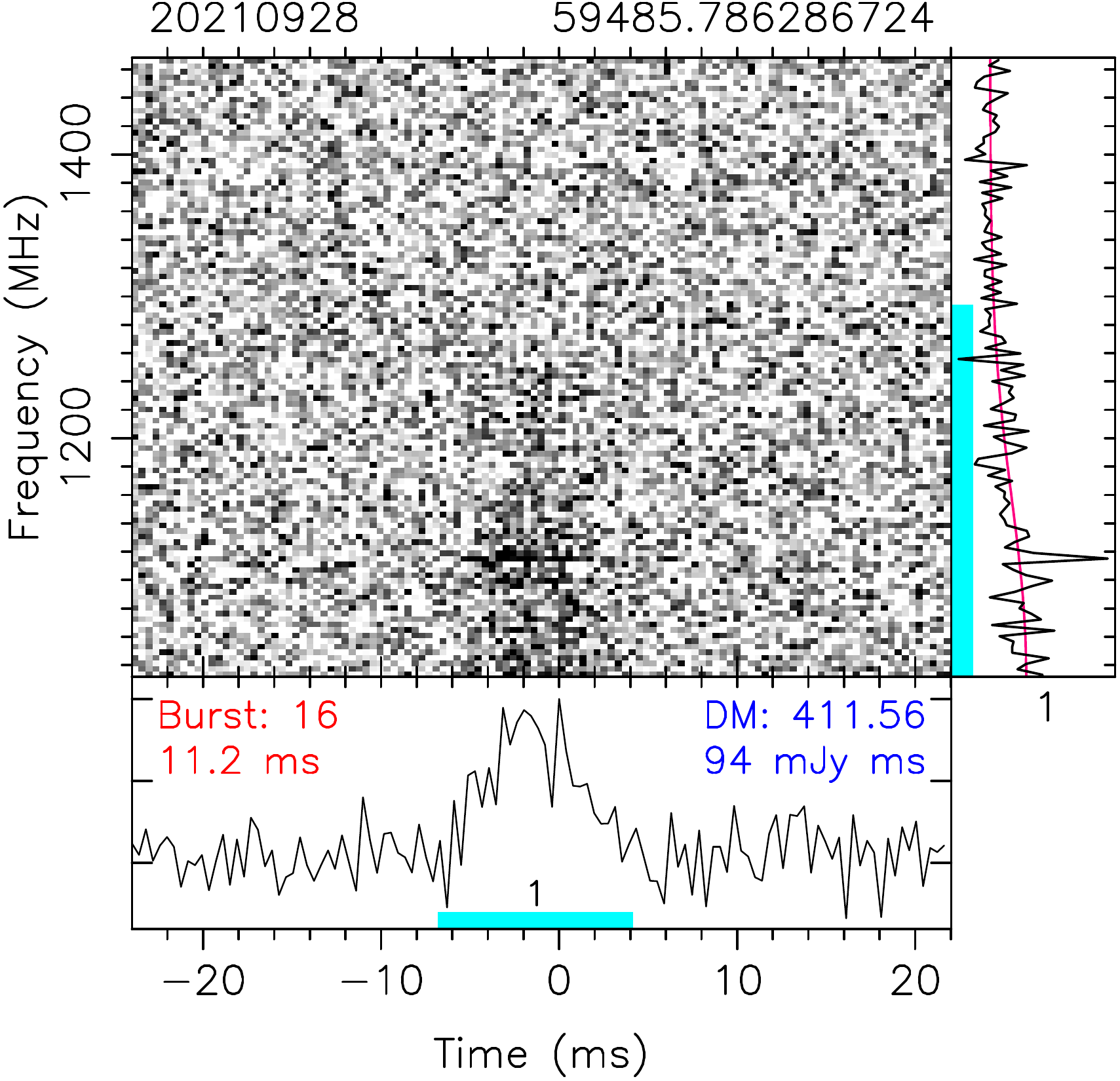}
    \caption{\it{ -- continued}.
}
\end{figure*}
\addtocounter{figure}{-1}
\begin{figure*}
    \flushleft
    \includegraphics[height=37mm]{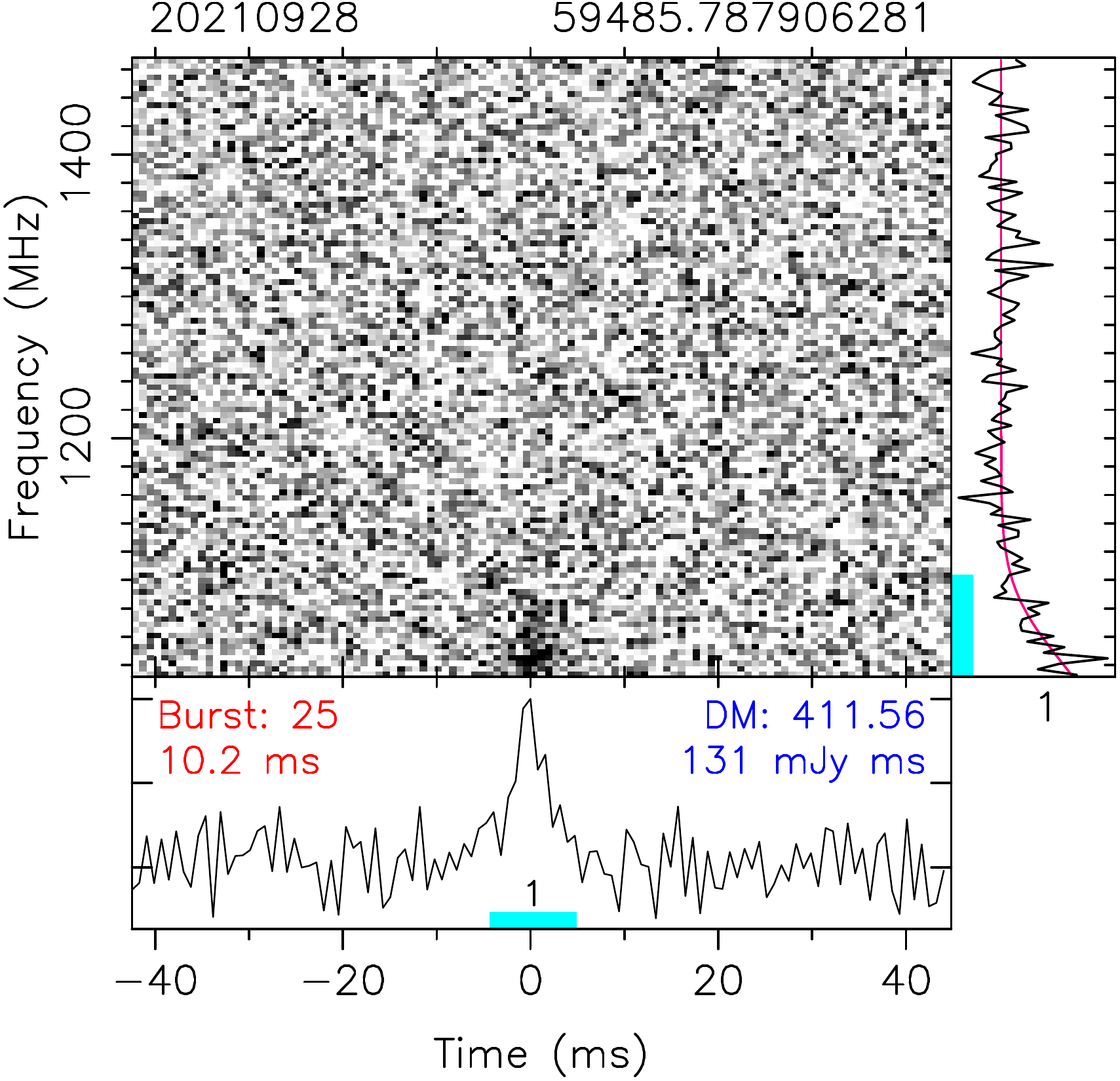}
    \includegraphics[height=37mm]{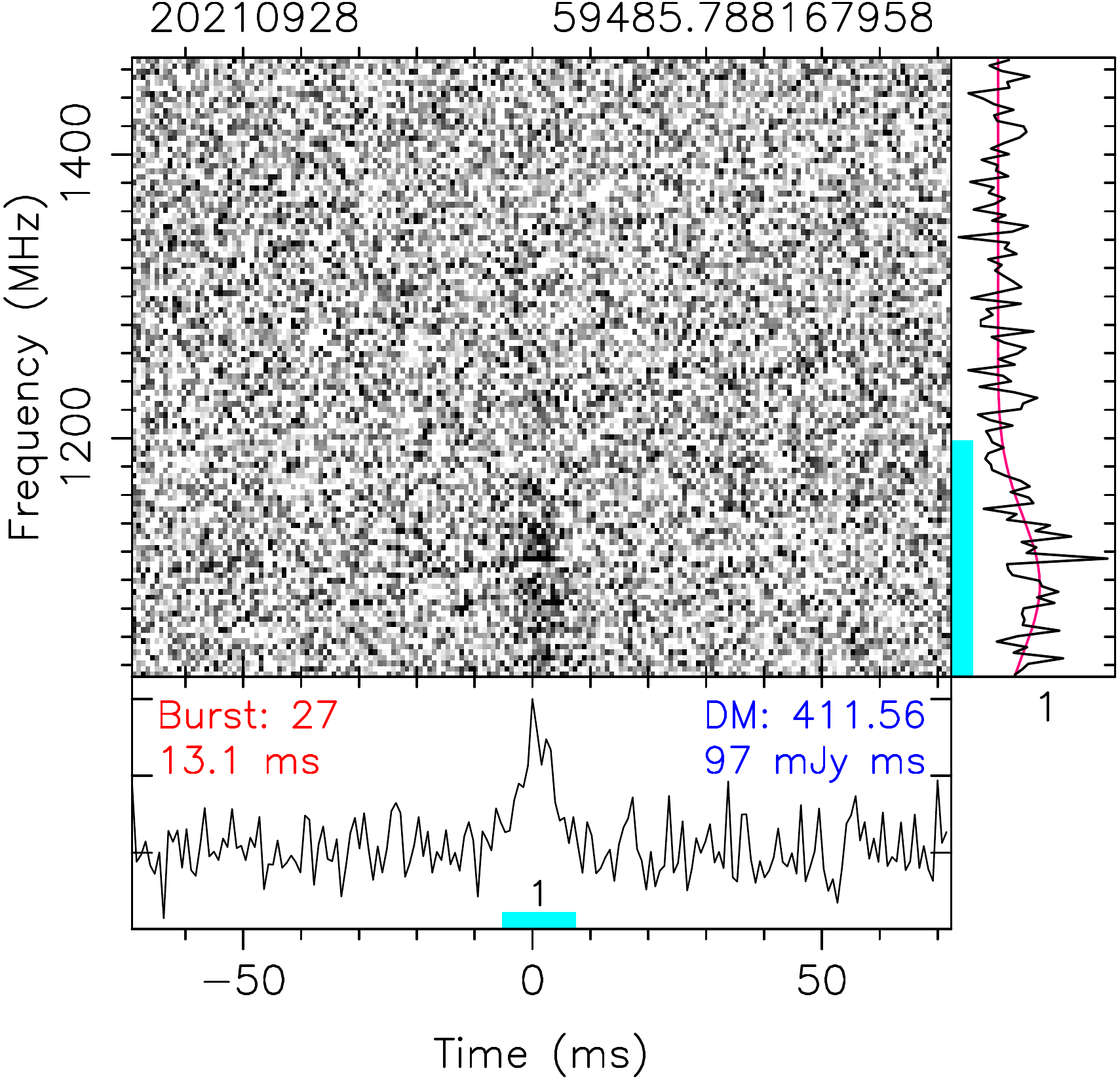}
    \includegraphics[height=37mm]{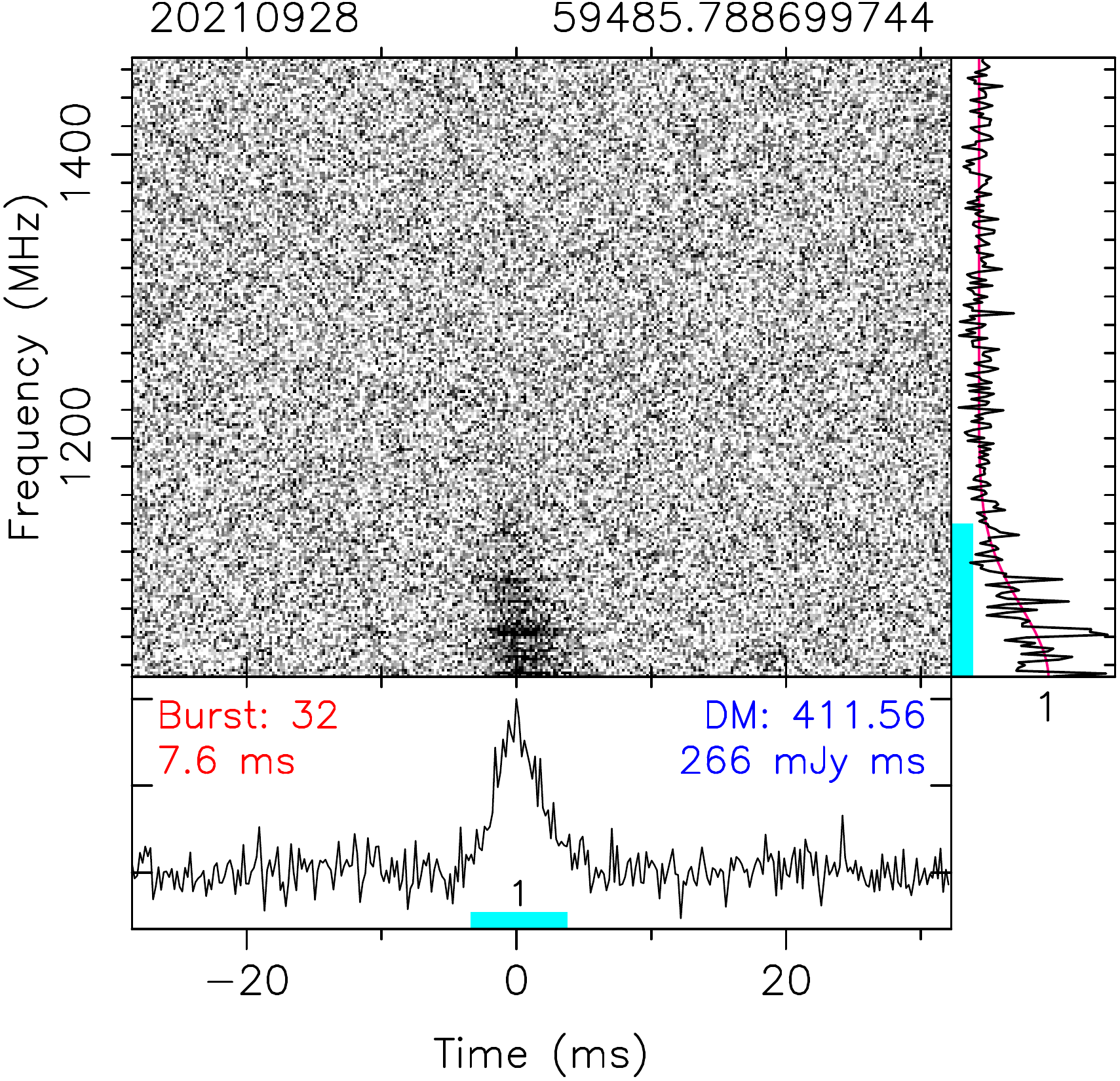}
    \includegraphics[height=37mm]{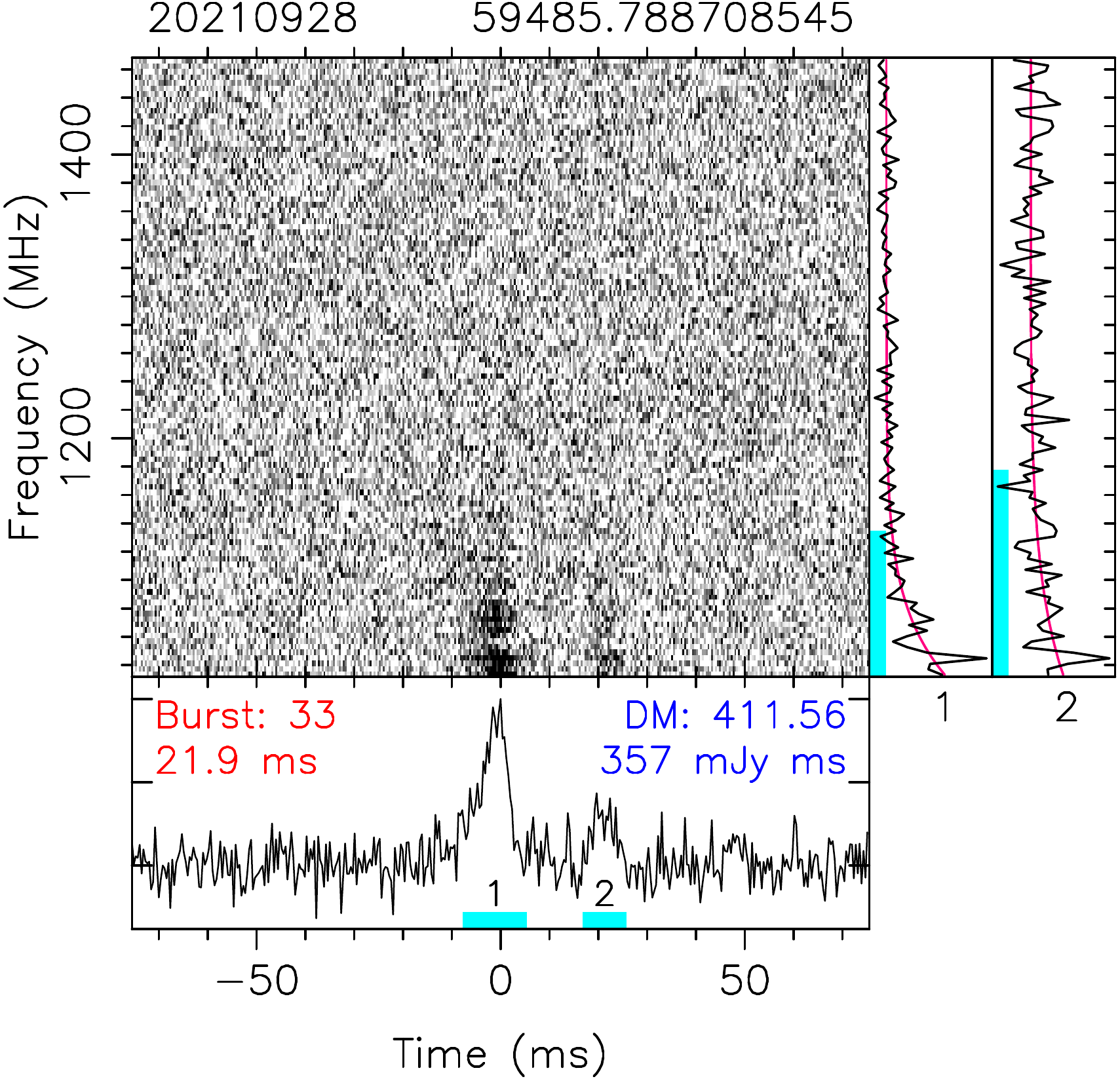}
    \includegraphics[height=37mm]{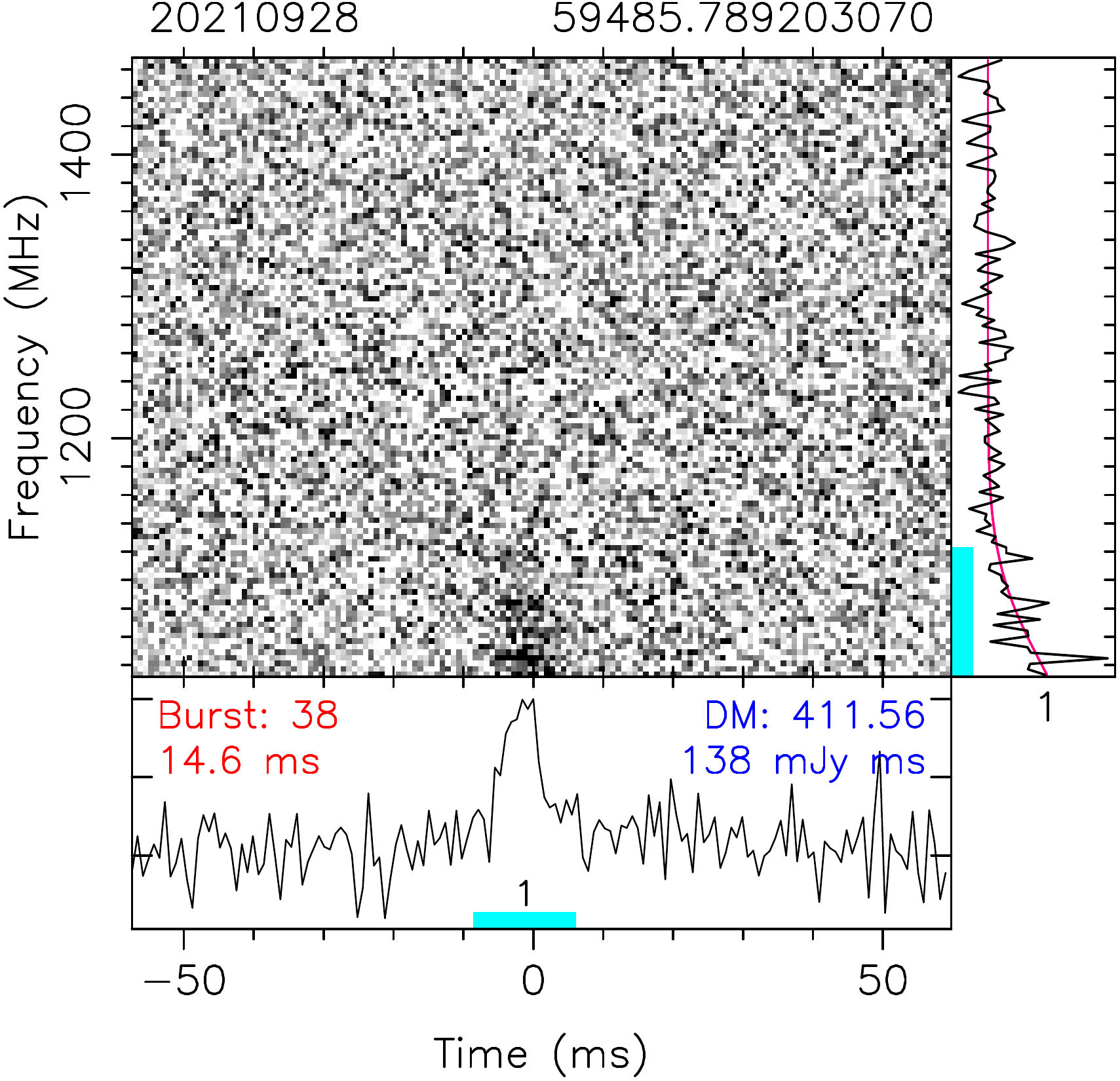}
    \includegraphics[height=37mm]{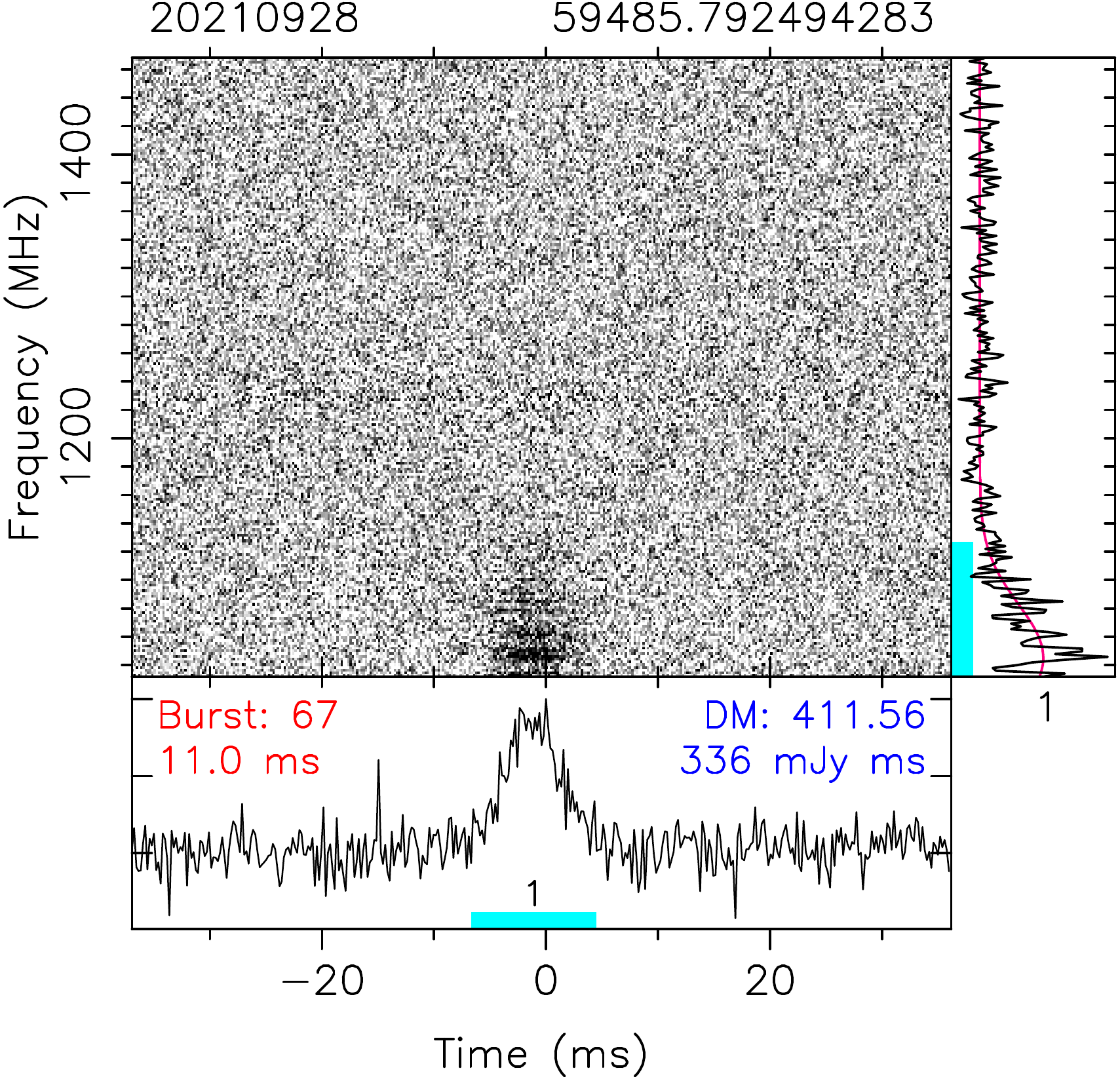}
    \includegraphics[height=37mm]{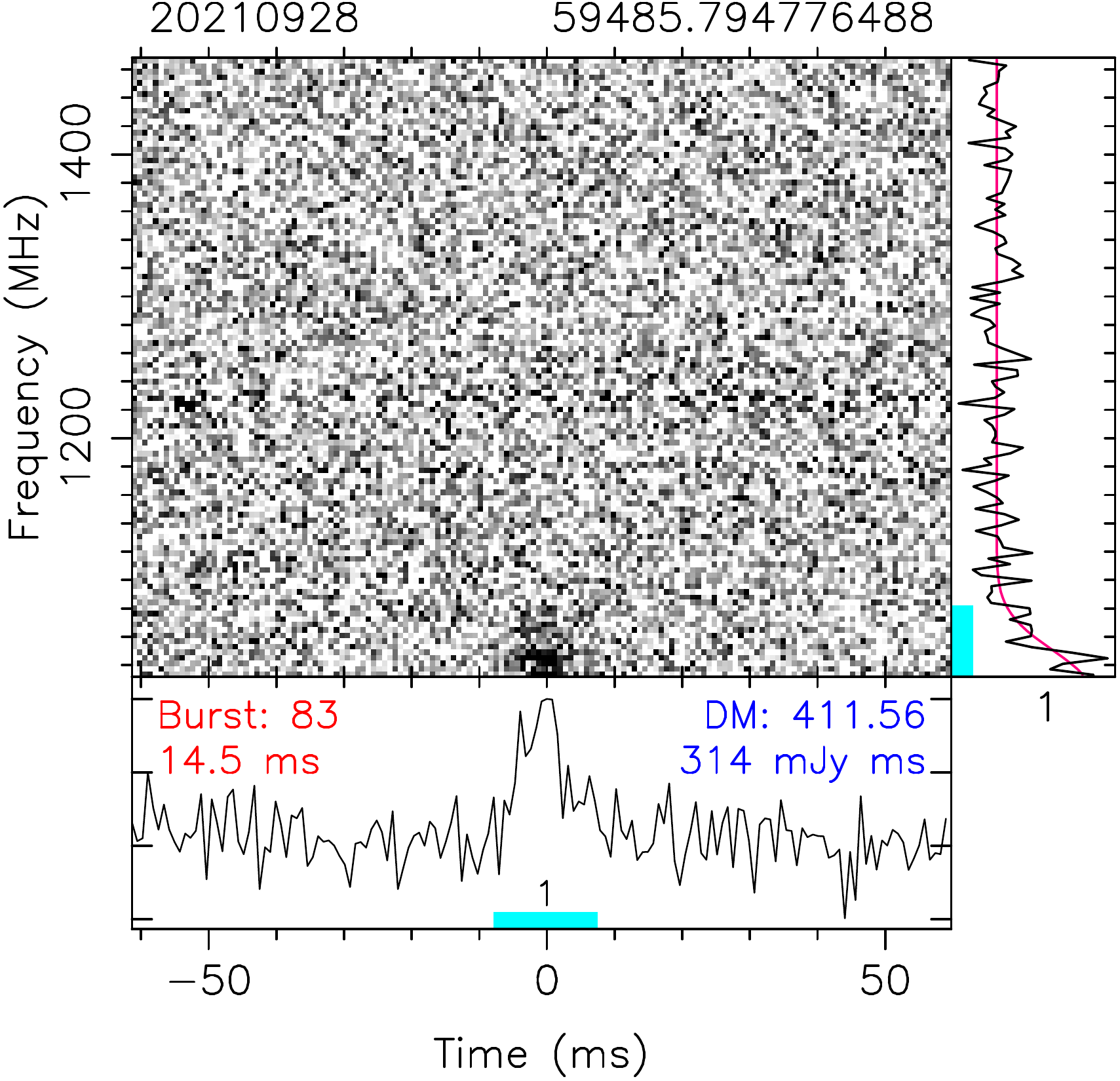}
    \includegraphics[height=37mm]{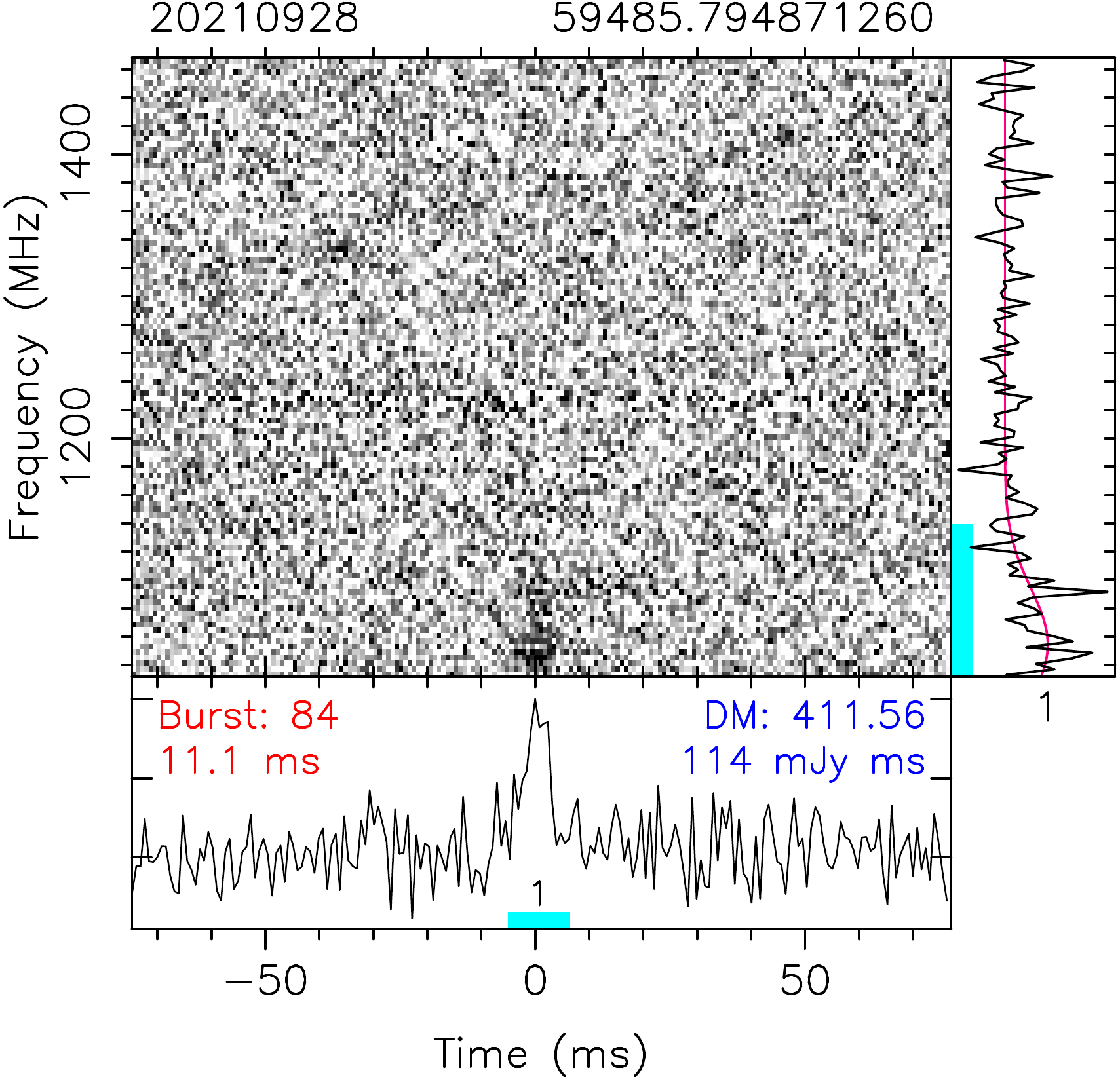}
    \includegraphics[height=37mm]{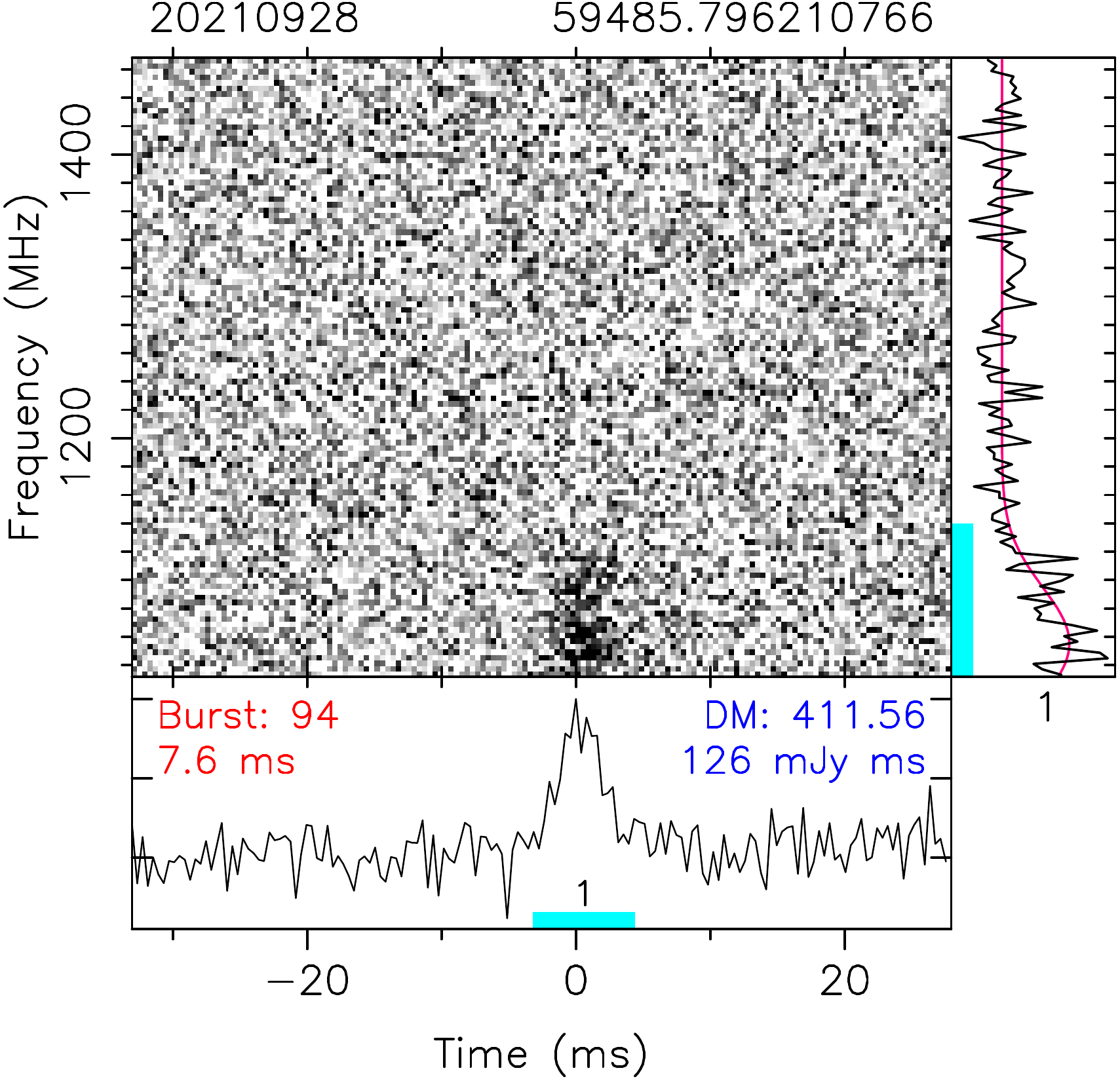}
    \includegraphics[height=37mm]{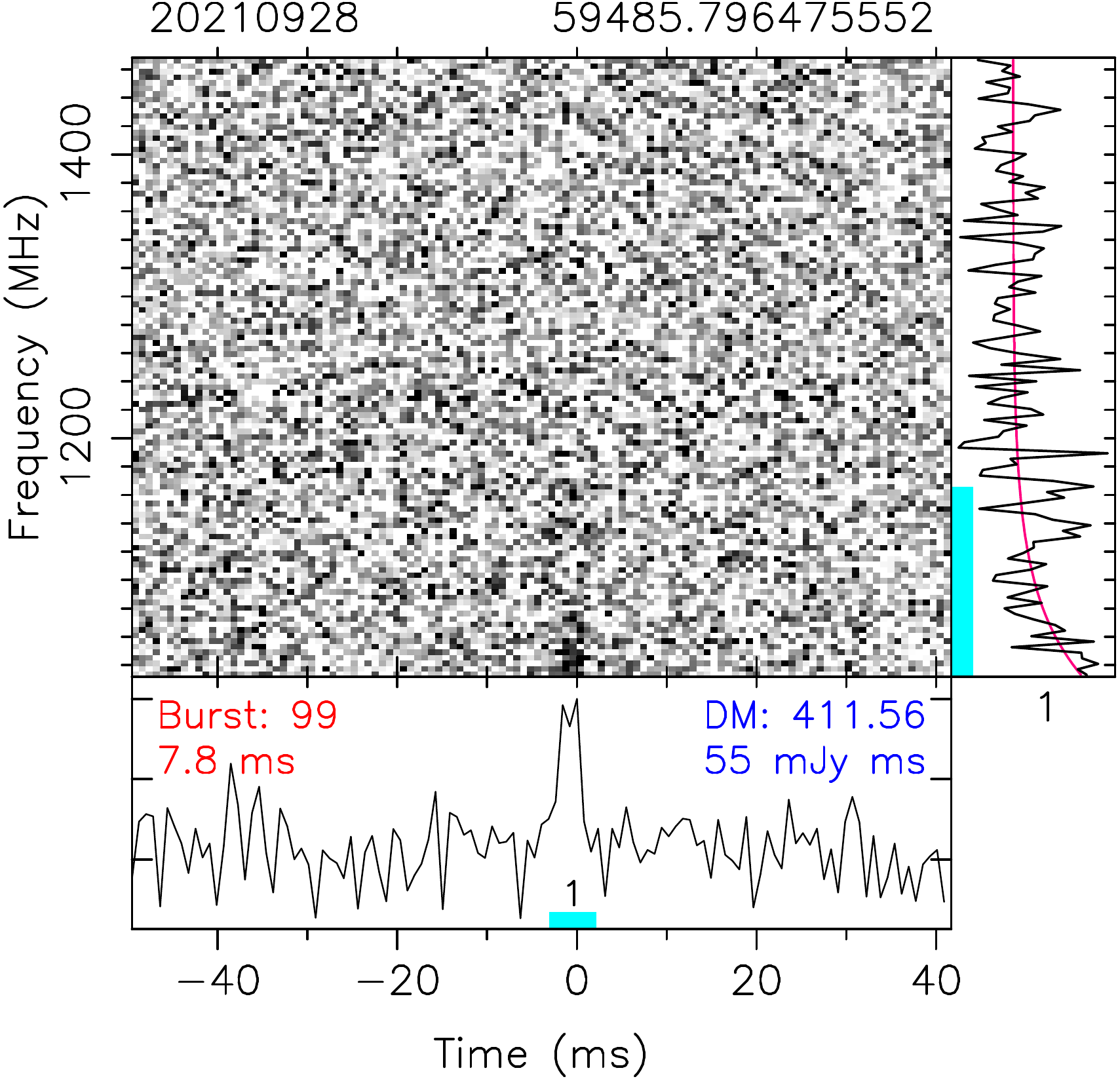}
    \includegraphics[height=37mm]{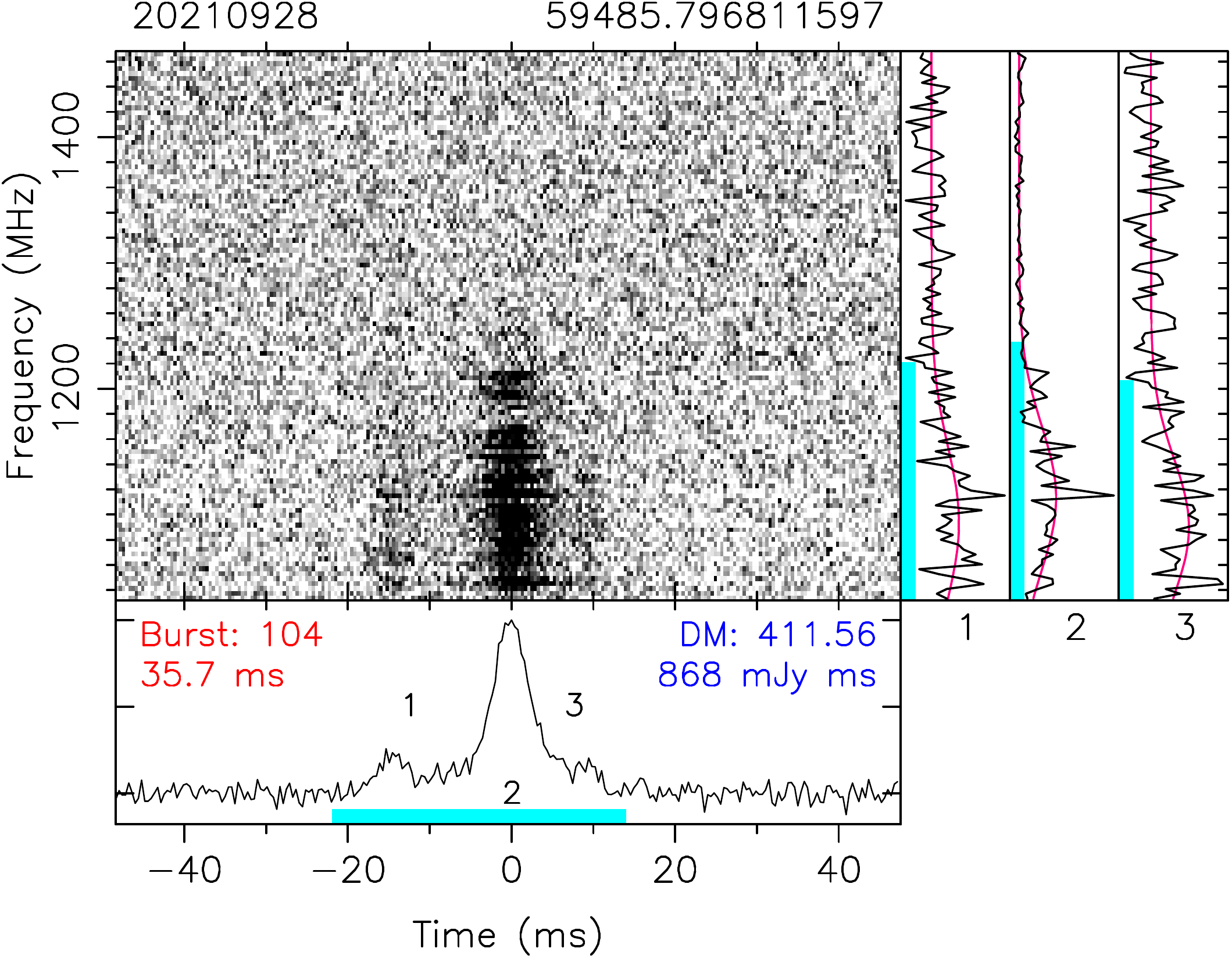}
    \includegraphics[height=37mm]{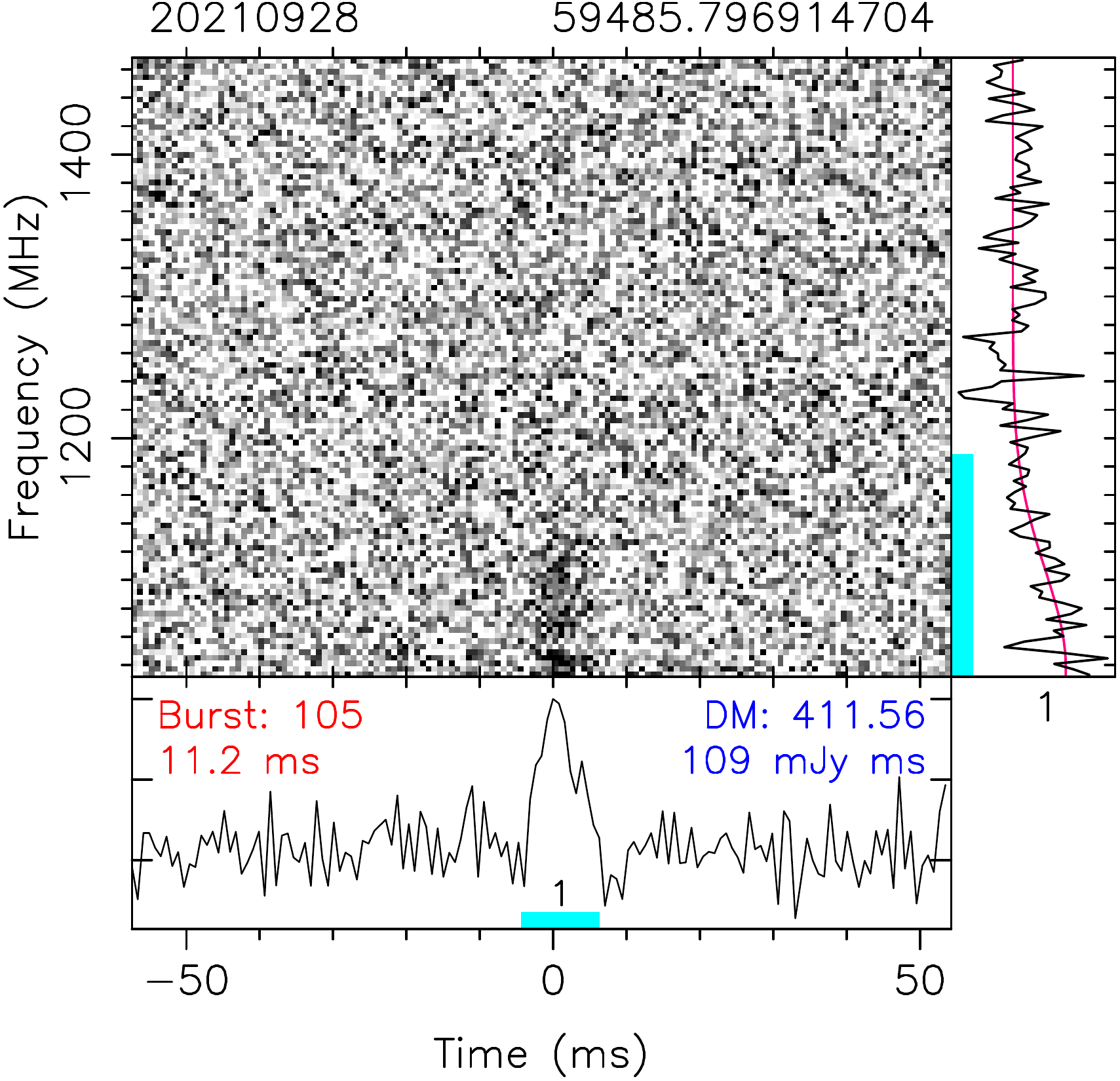}
    \includegraphics[height=37mm]{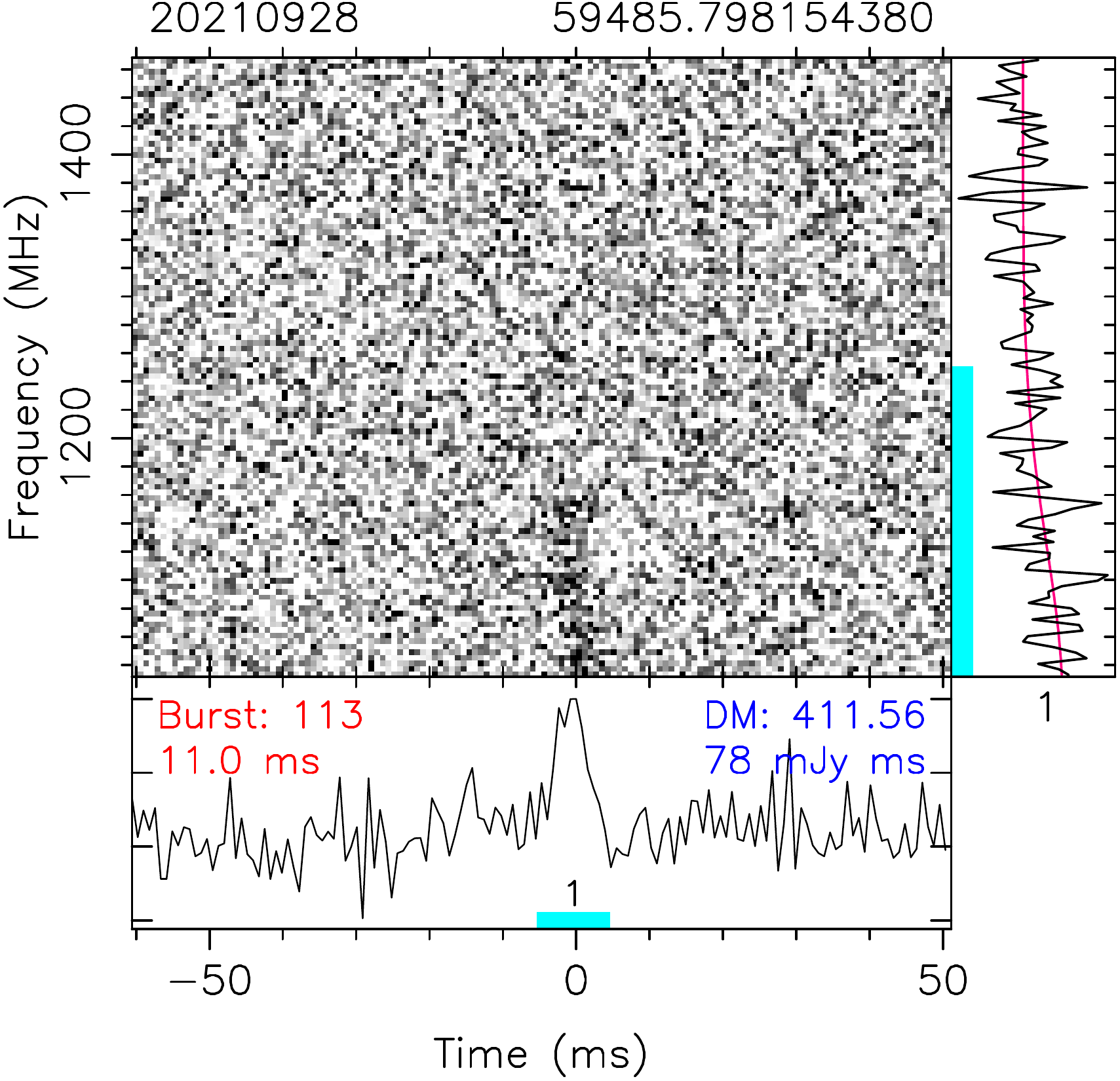}
    \includegraphics[height=37mm]{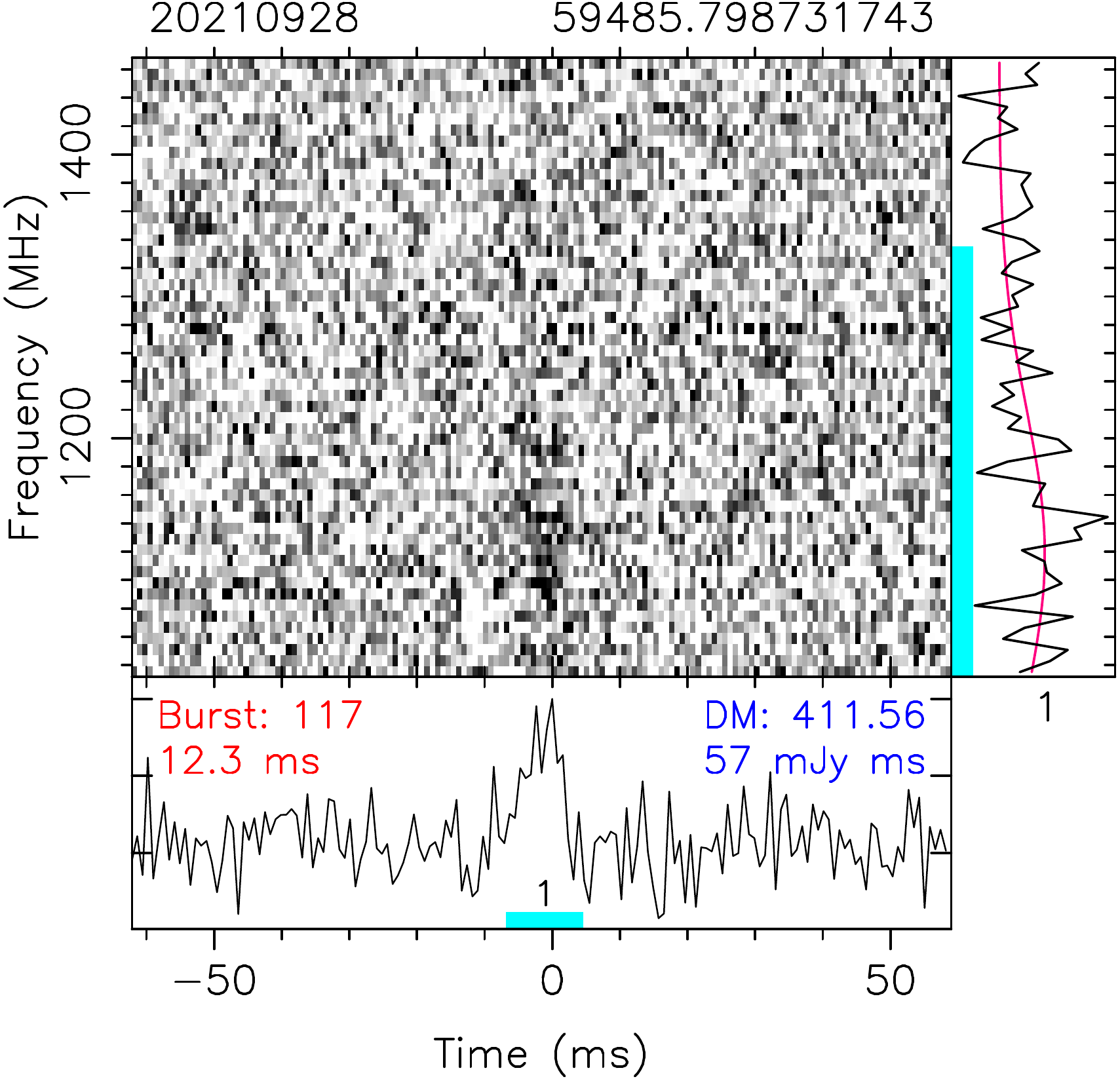}
    \includegraphics[height=37mm]{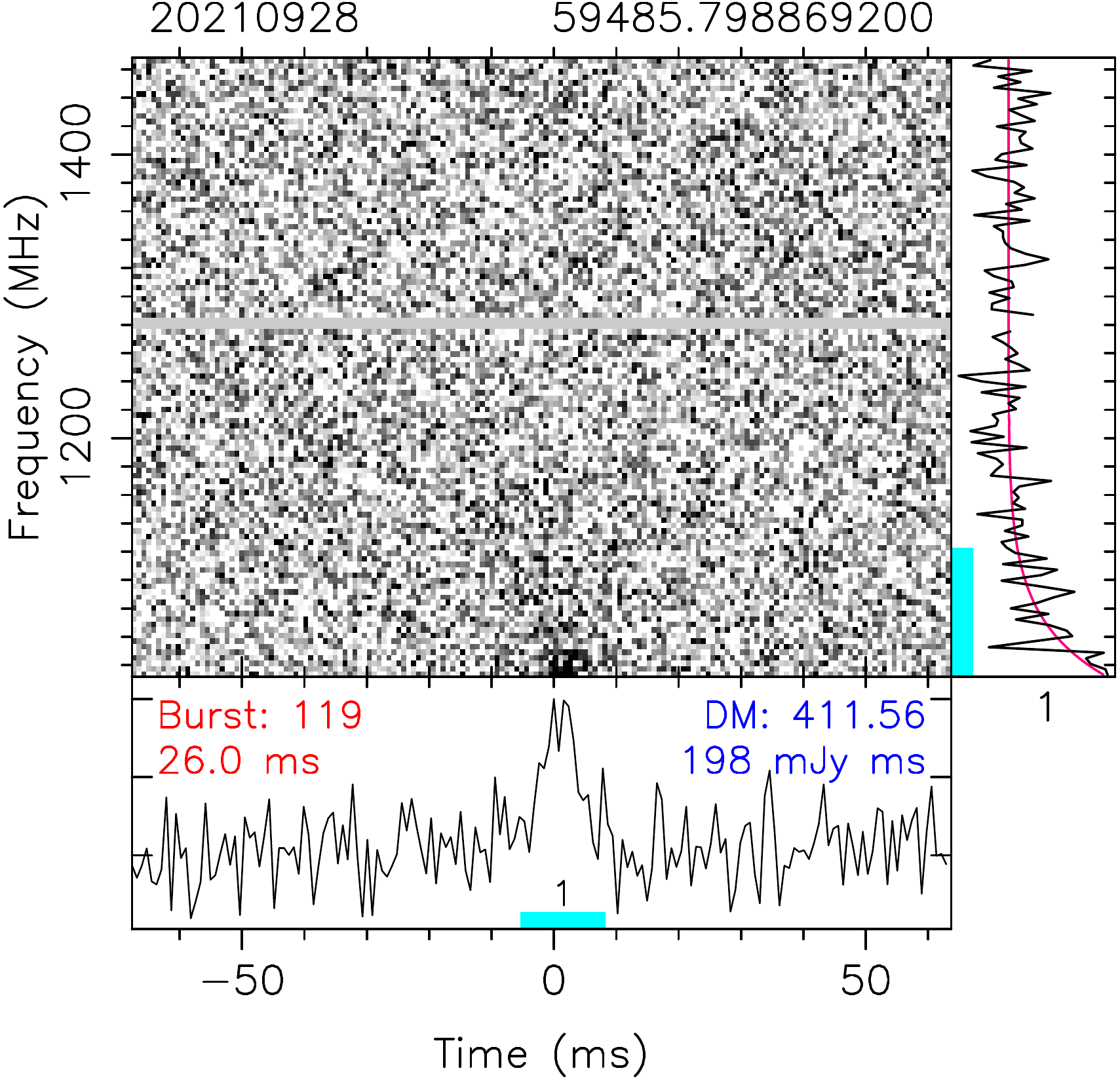}
    \includegraphics[height=37mm]{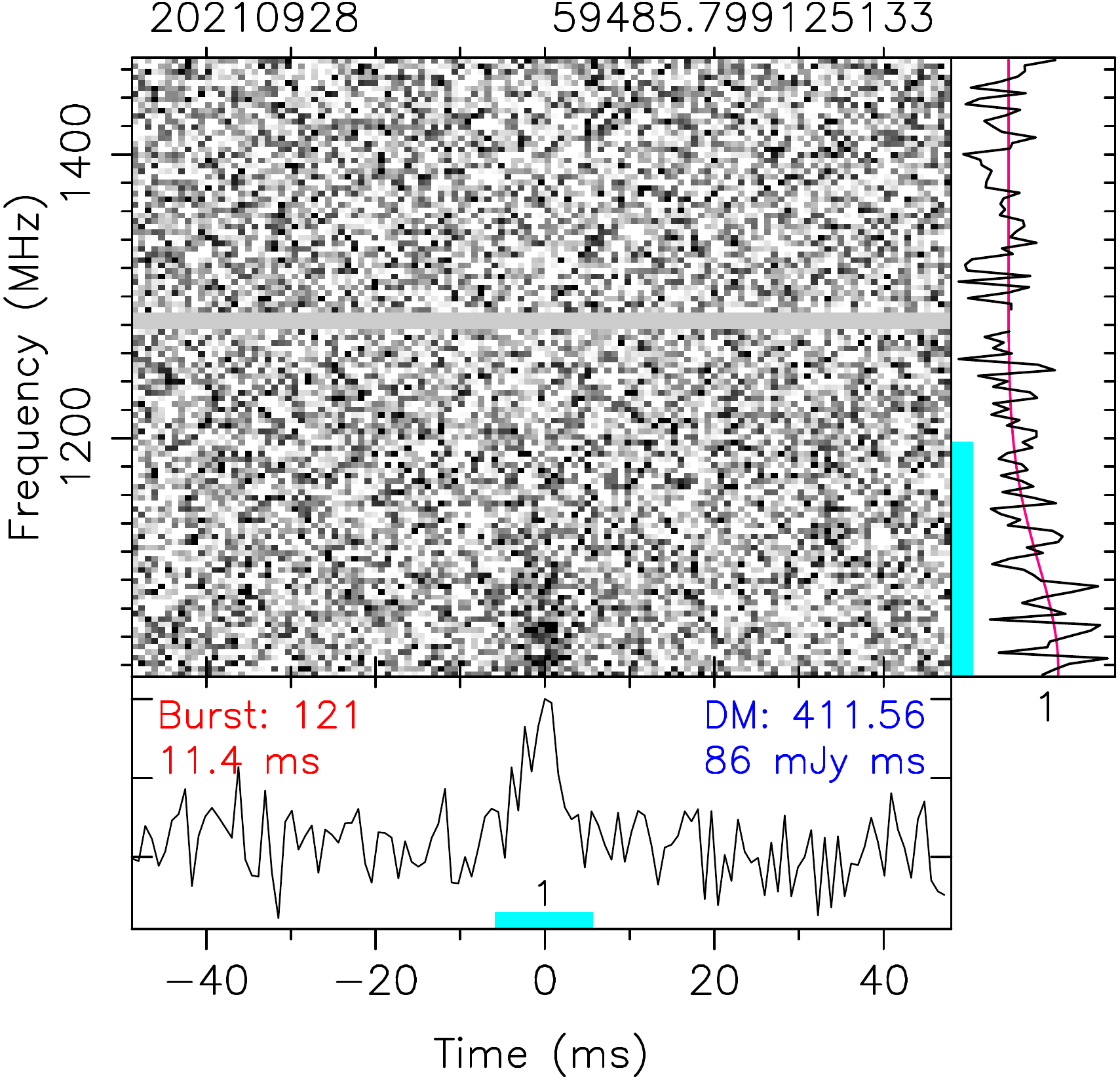}
    \includegraphics[height=37mm]{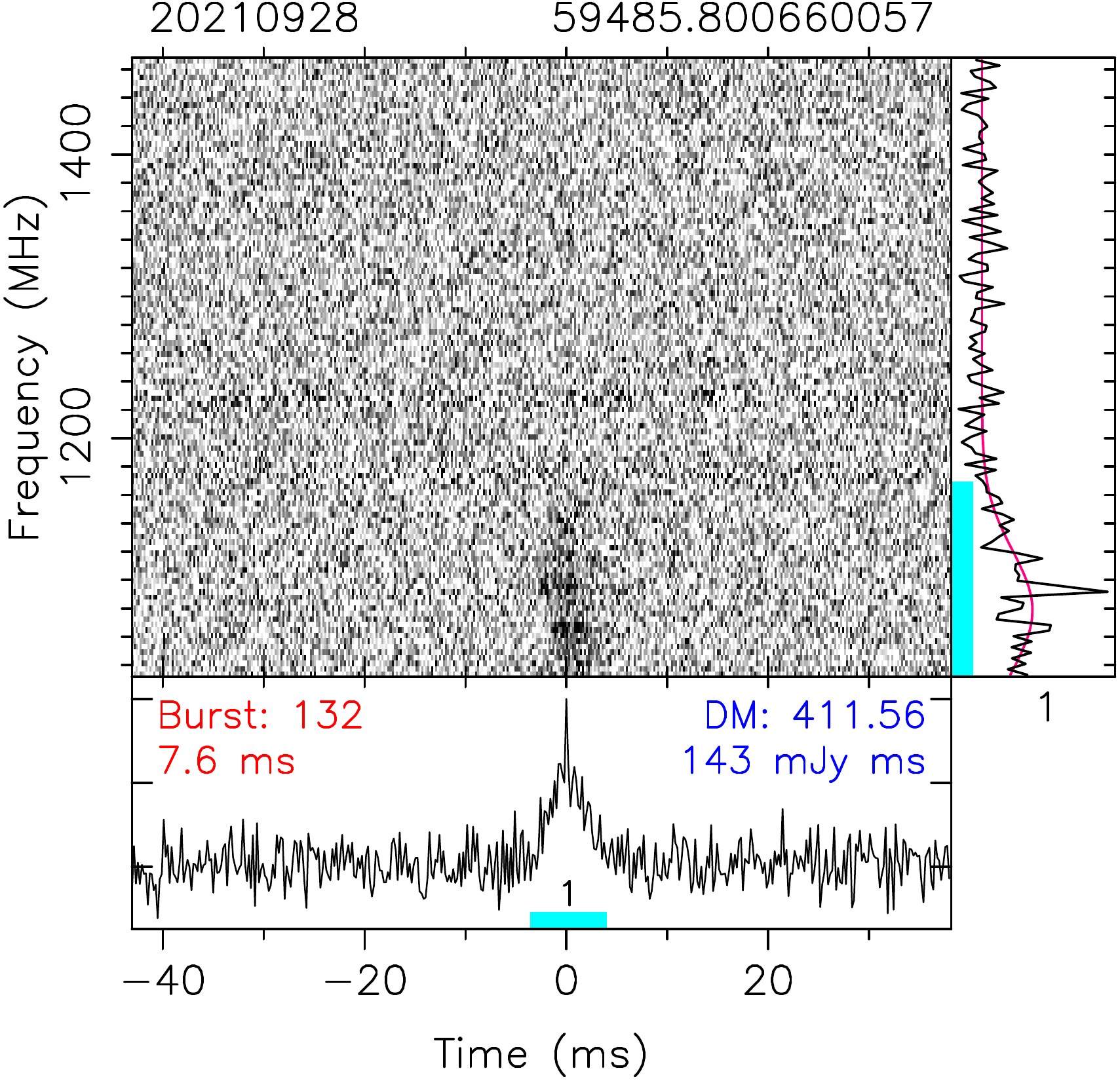}
    \includegraphics[height=37mm]{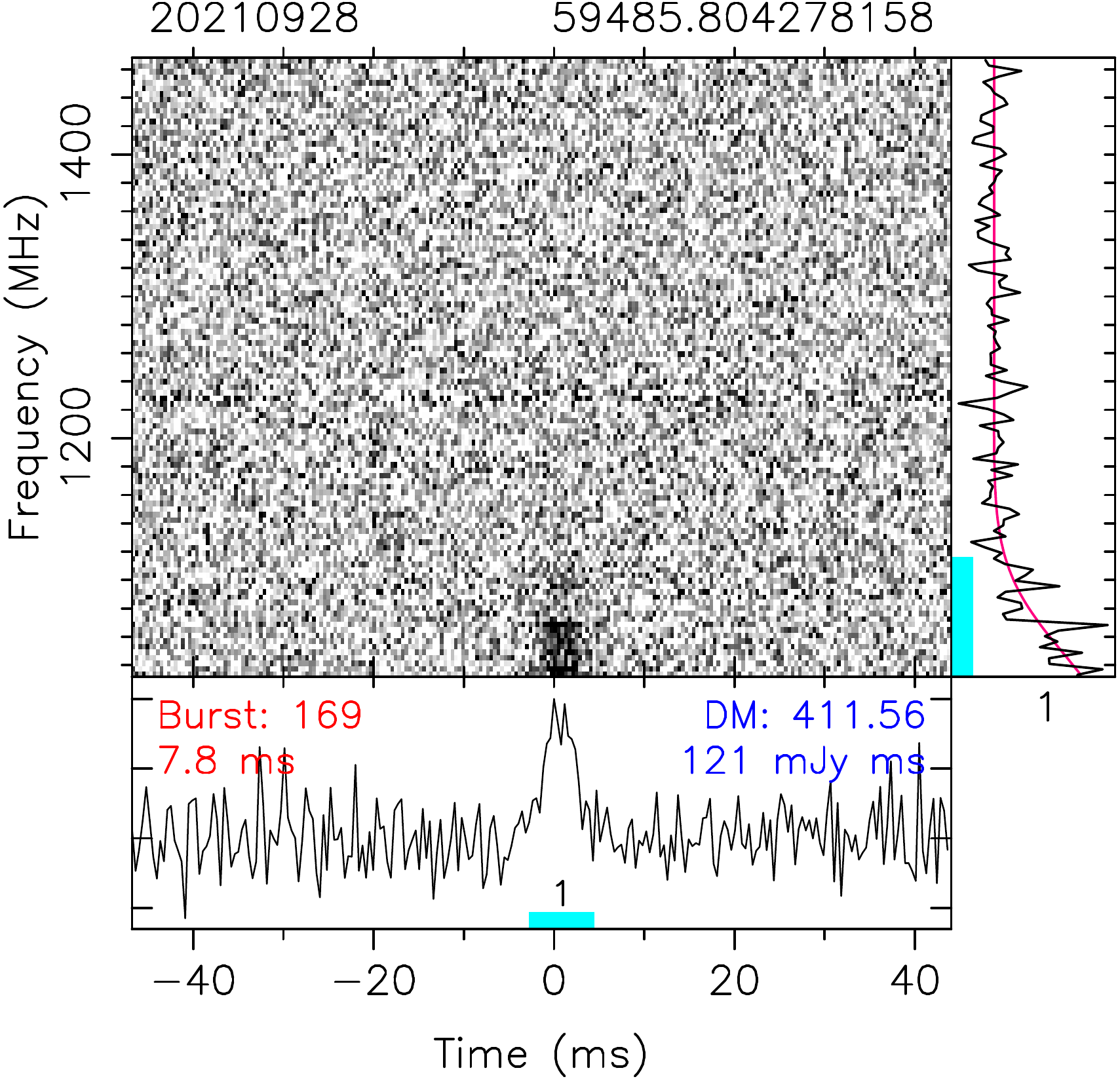}
    \includegraphics[height=37mm]{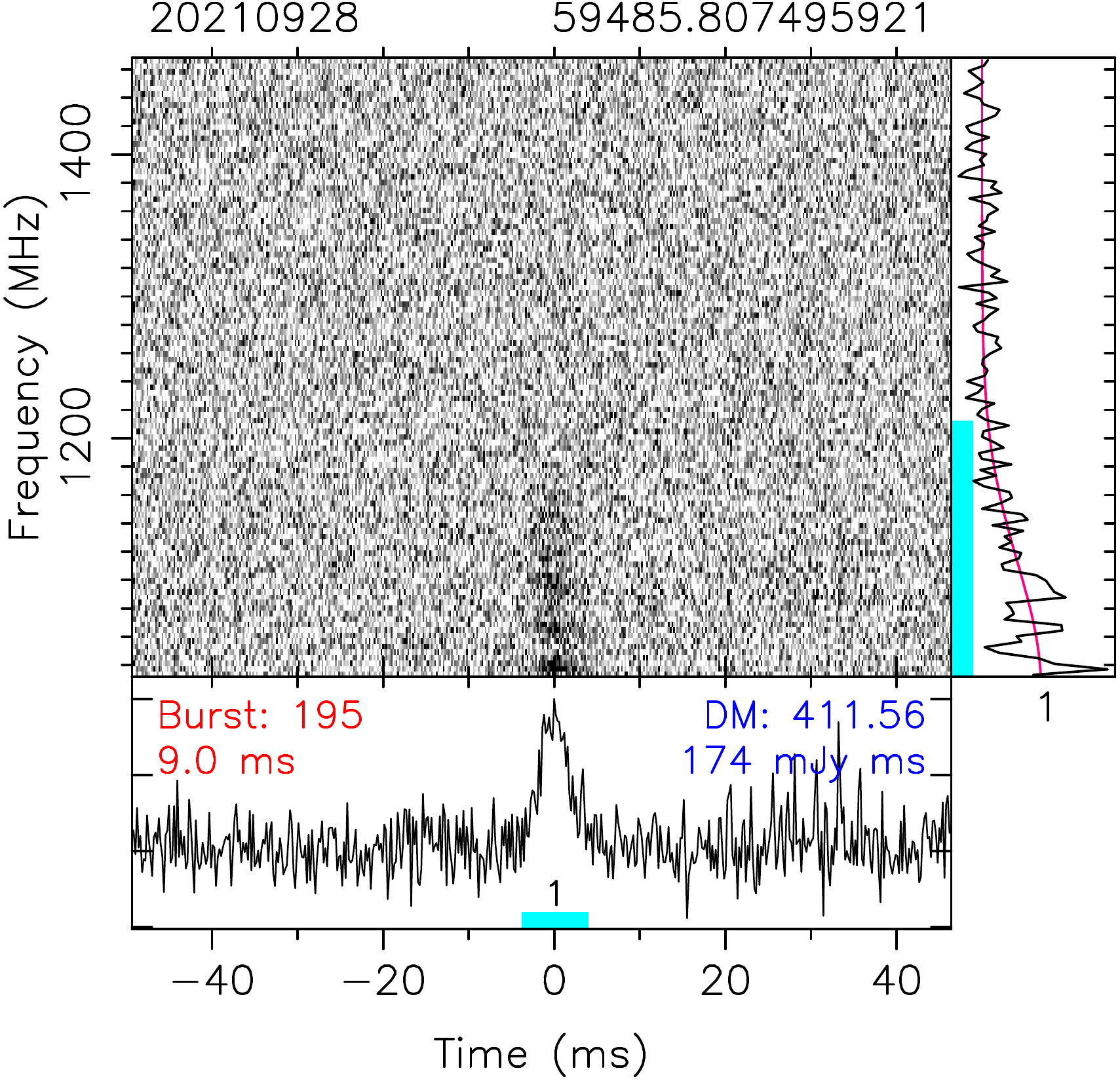}
    \includegraphics[height=37mm]{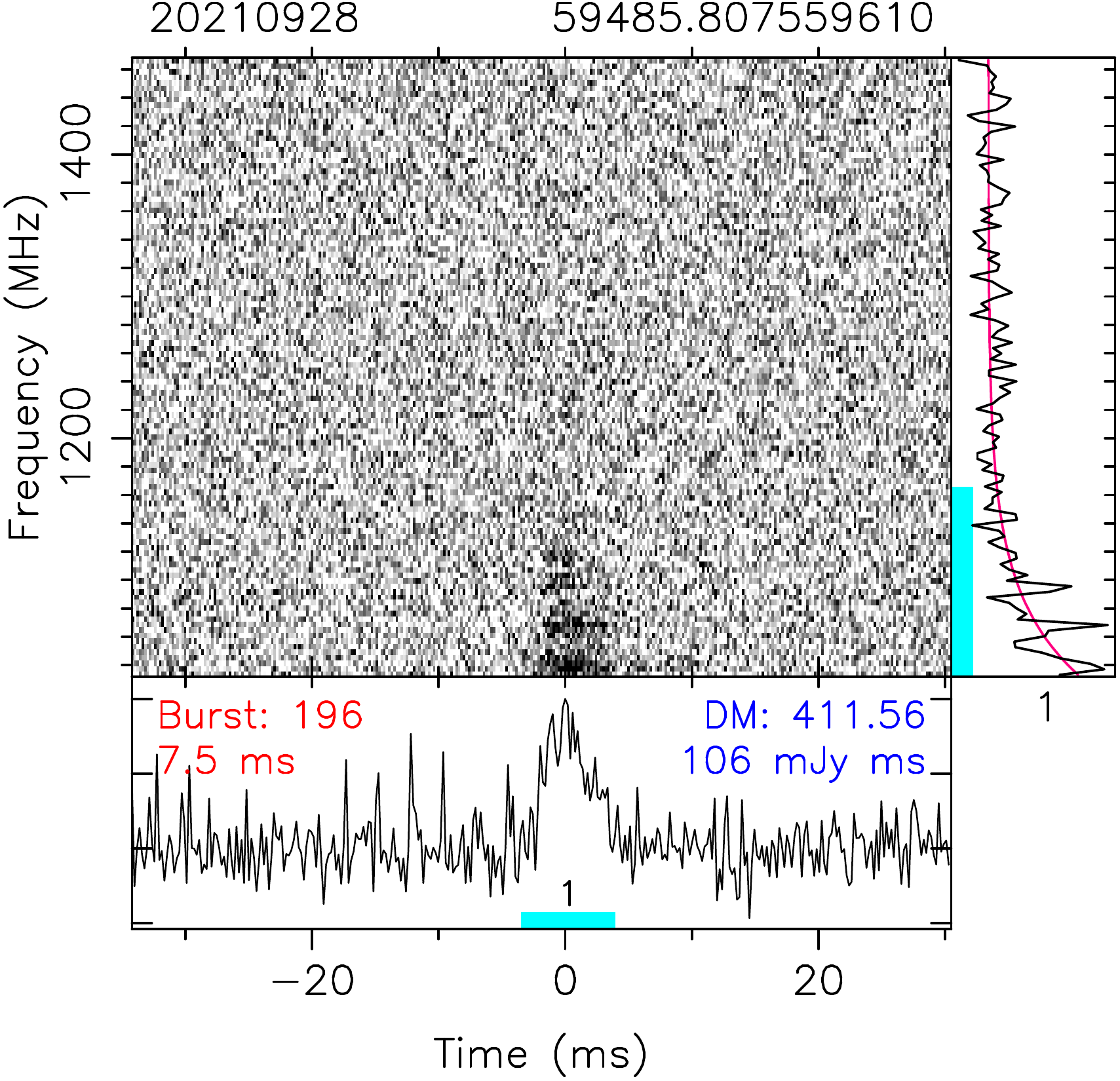}
    \includegraphics[height=37mm]{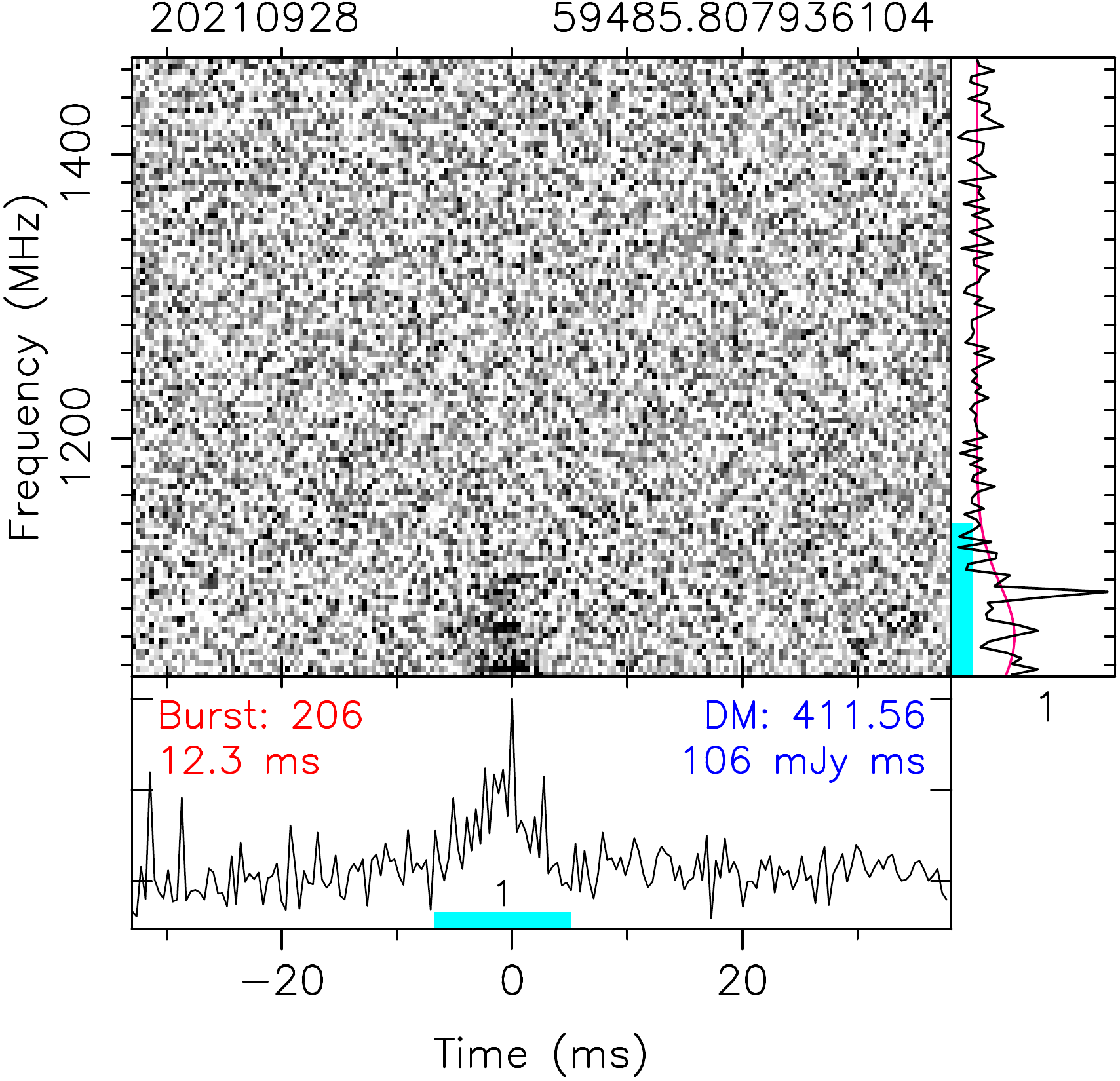}
    \includegraphics[height=37mm]{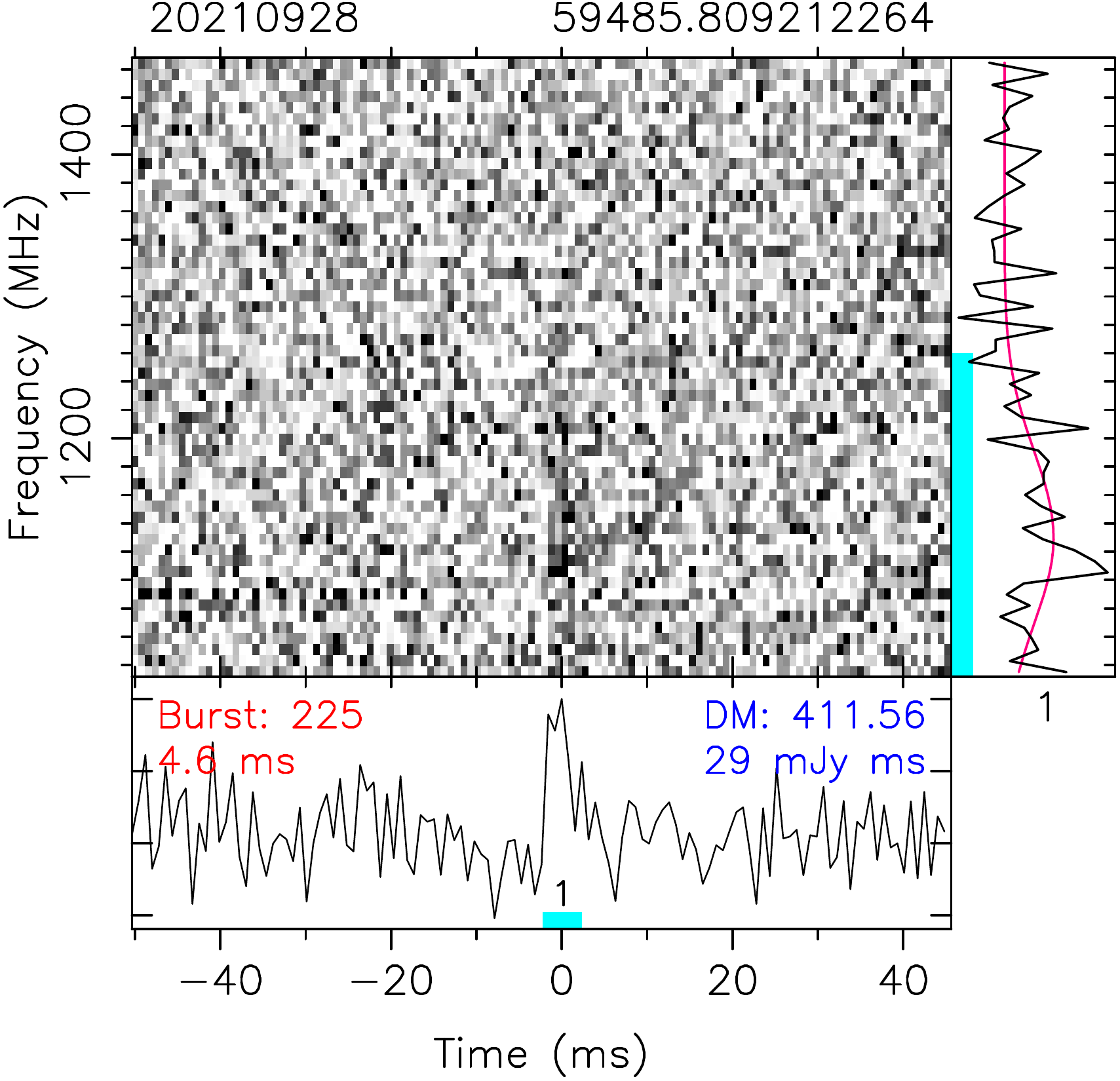}
    \includegraphics[height=37mm]{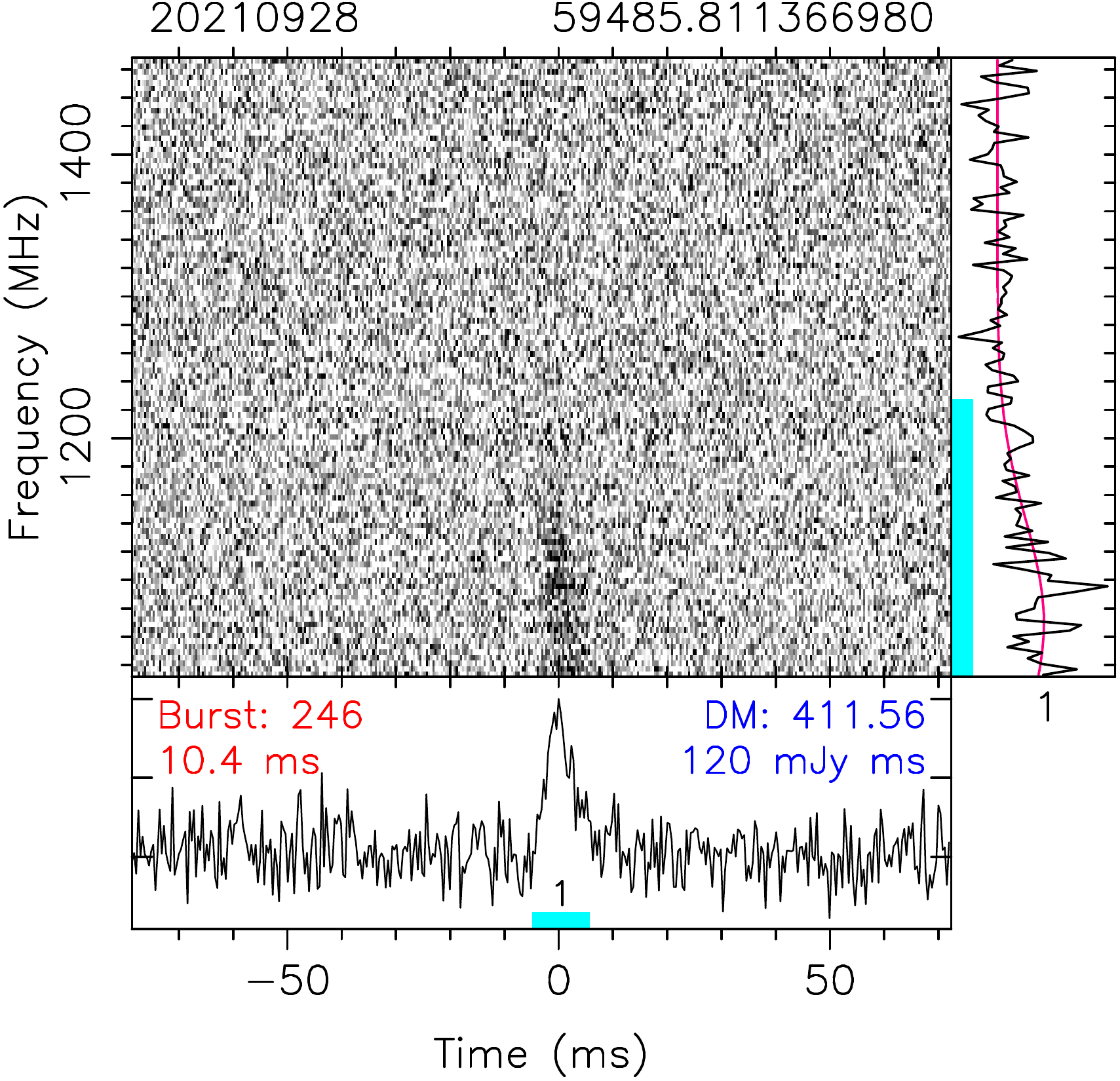}
    \includegraphics[height=37mm]{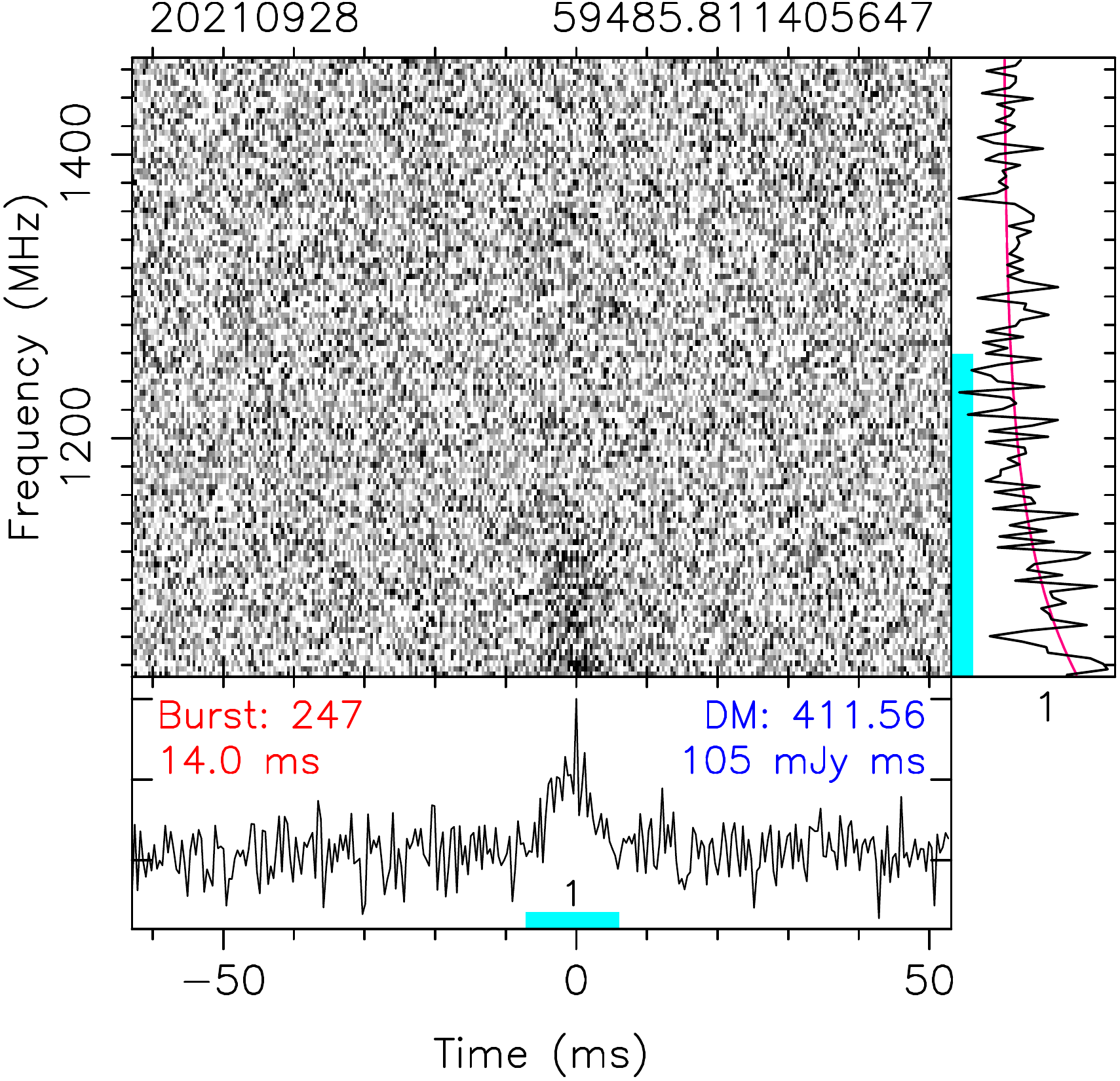}
\caption{\it{ -- continued}.
}
\end{figure*}
\addtocounter{figure}{-1}
\begin{figure*}
    \flushleft
    \includegraphics[height=37mm]{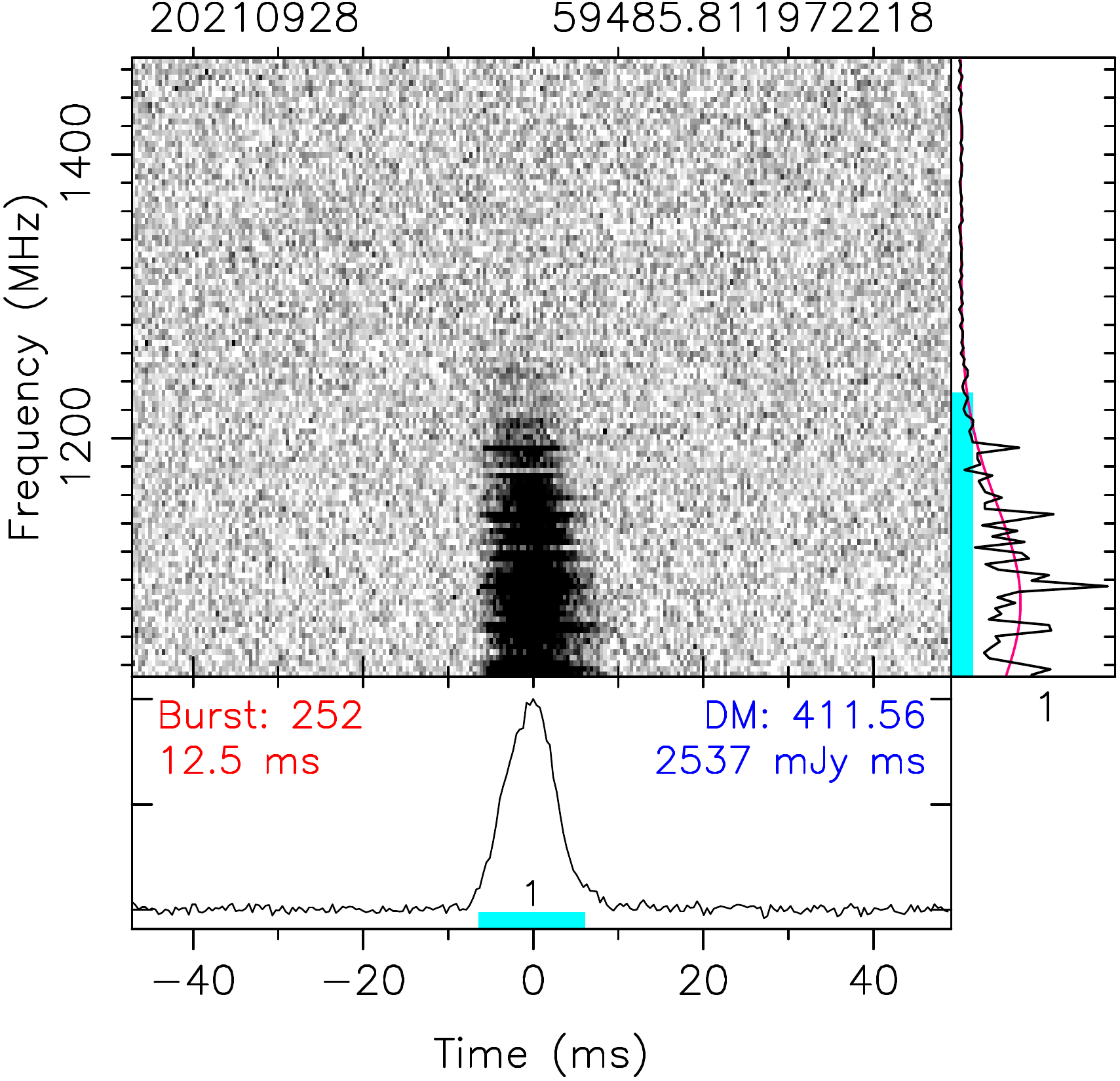}
    \includegraphics[height=37mm]{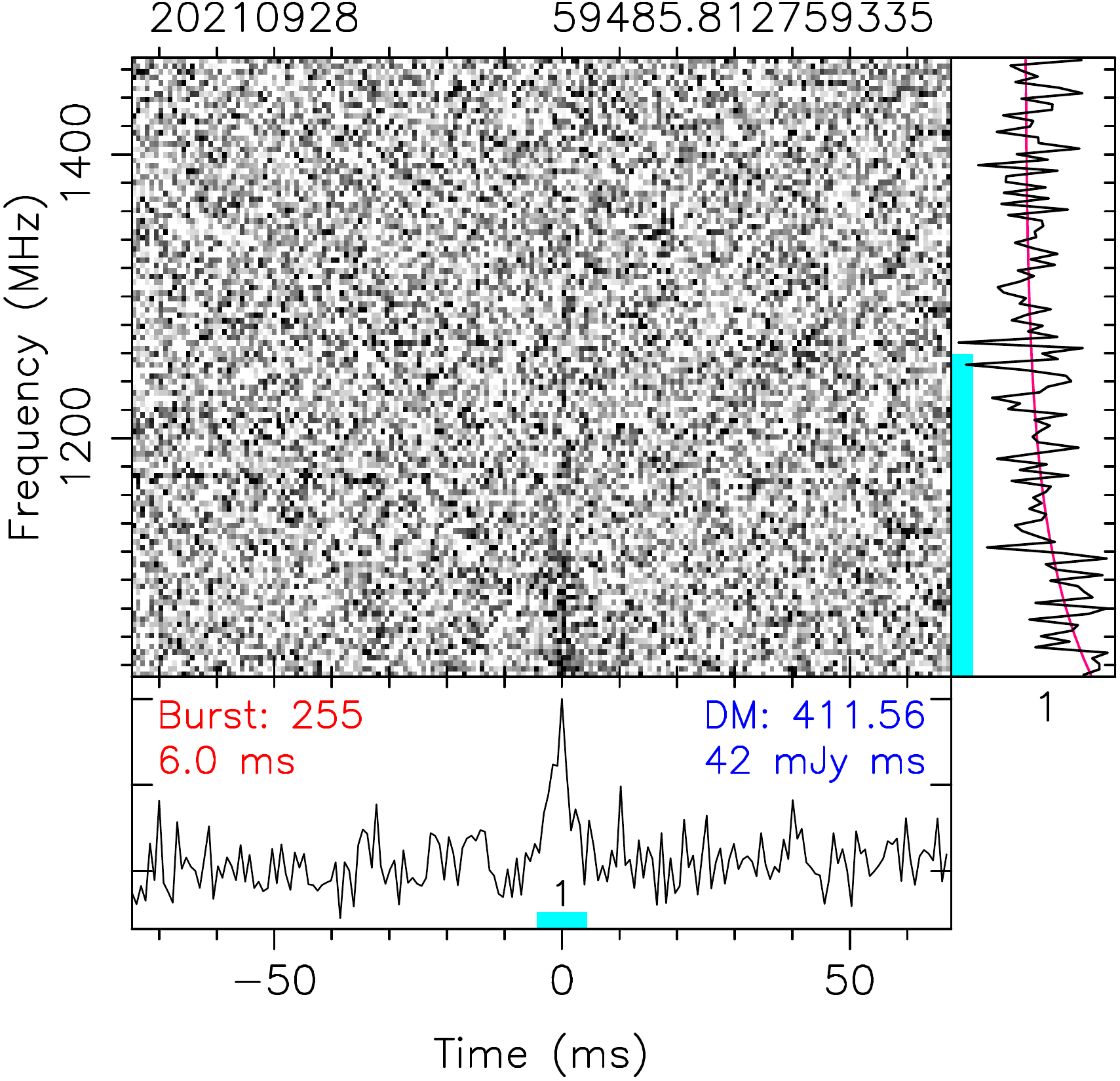}
    \includegraphics[height=37mm]{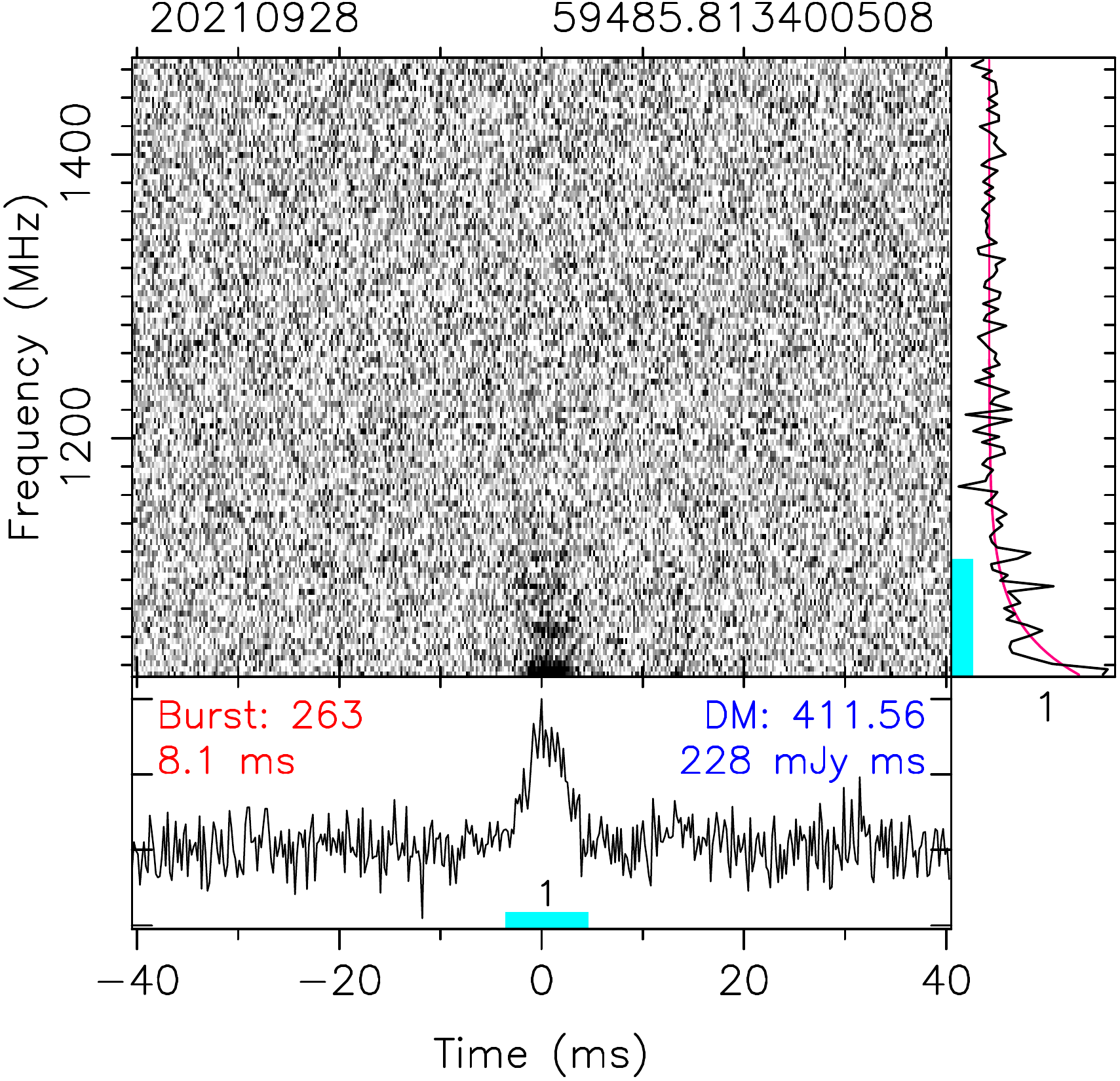}
    \includegraphics[height=37mm]{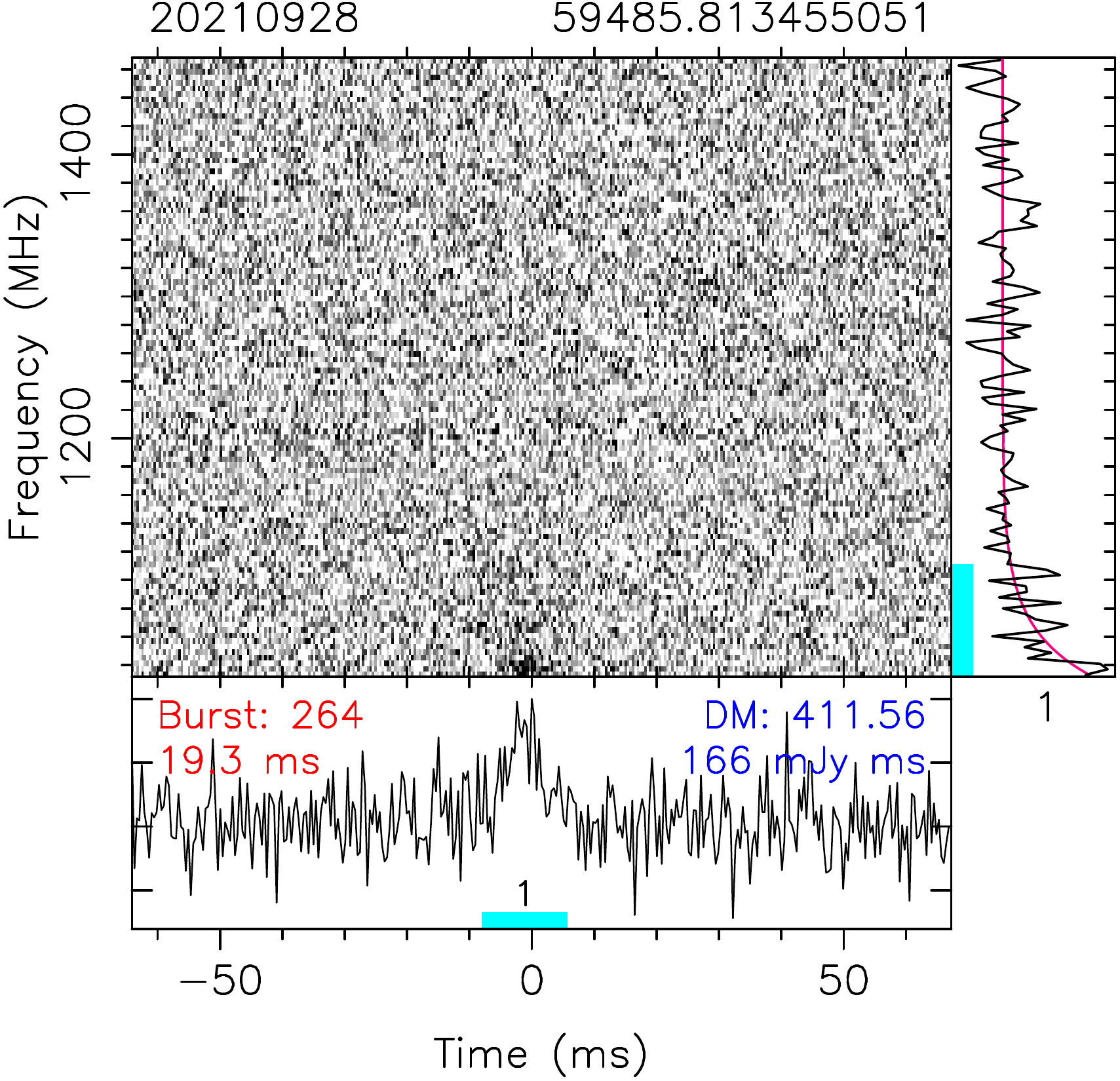}
    \includegraphics[height=37mm]{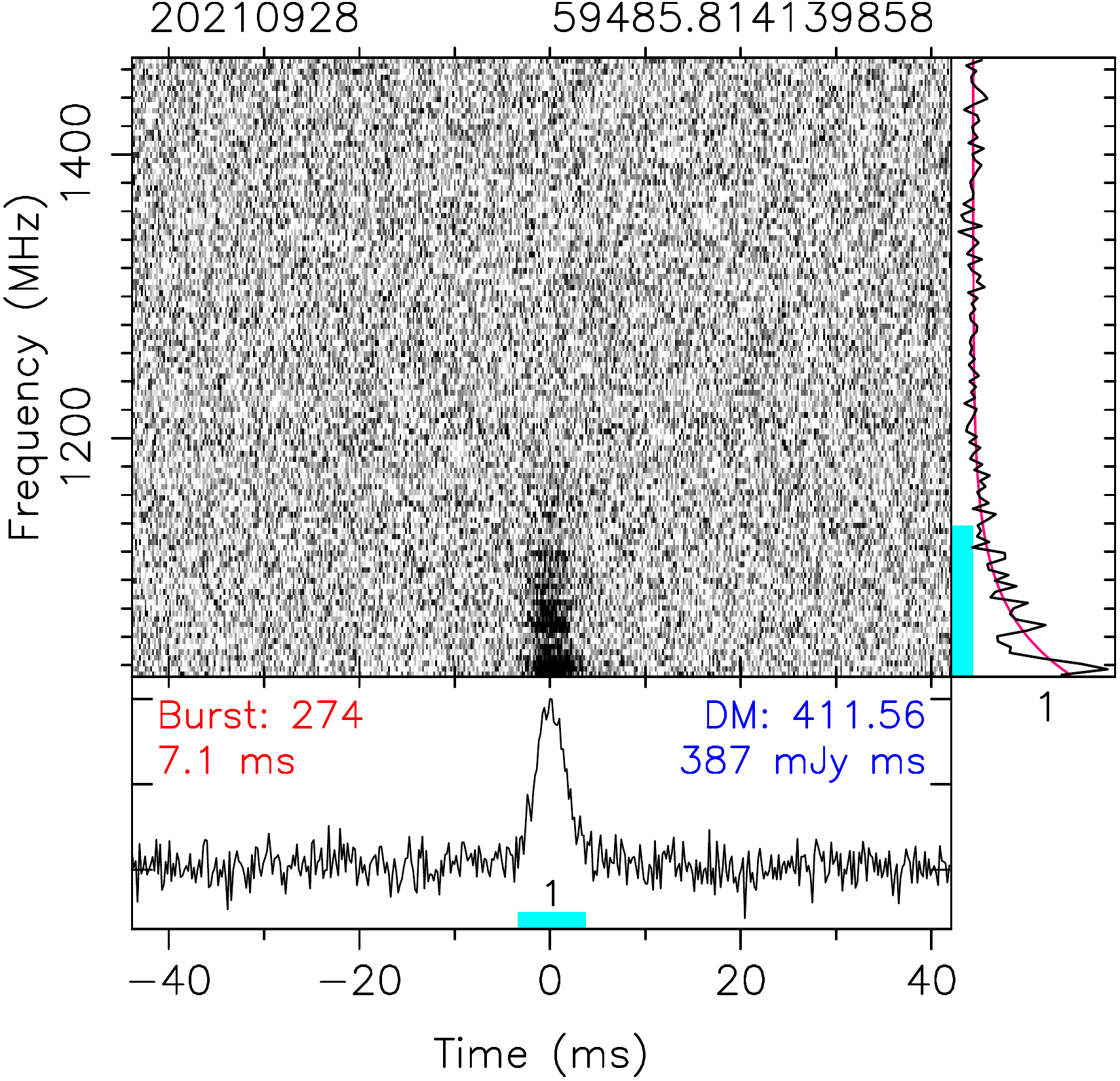}
    \includegraphics[height=37mm]{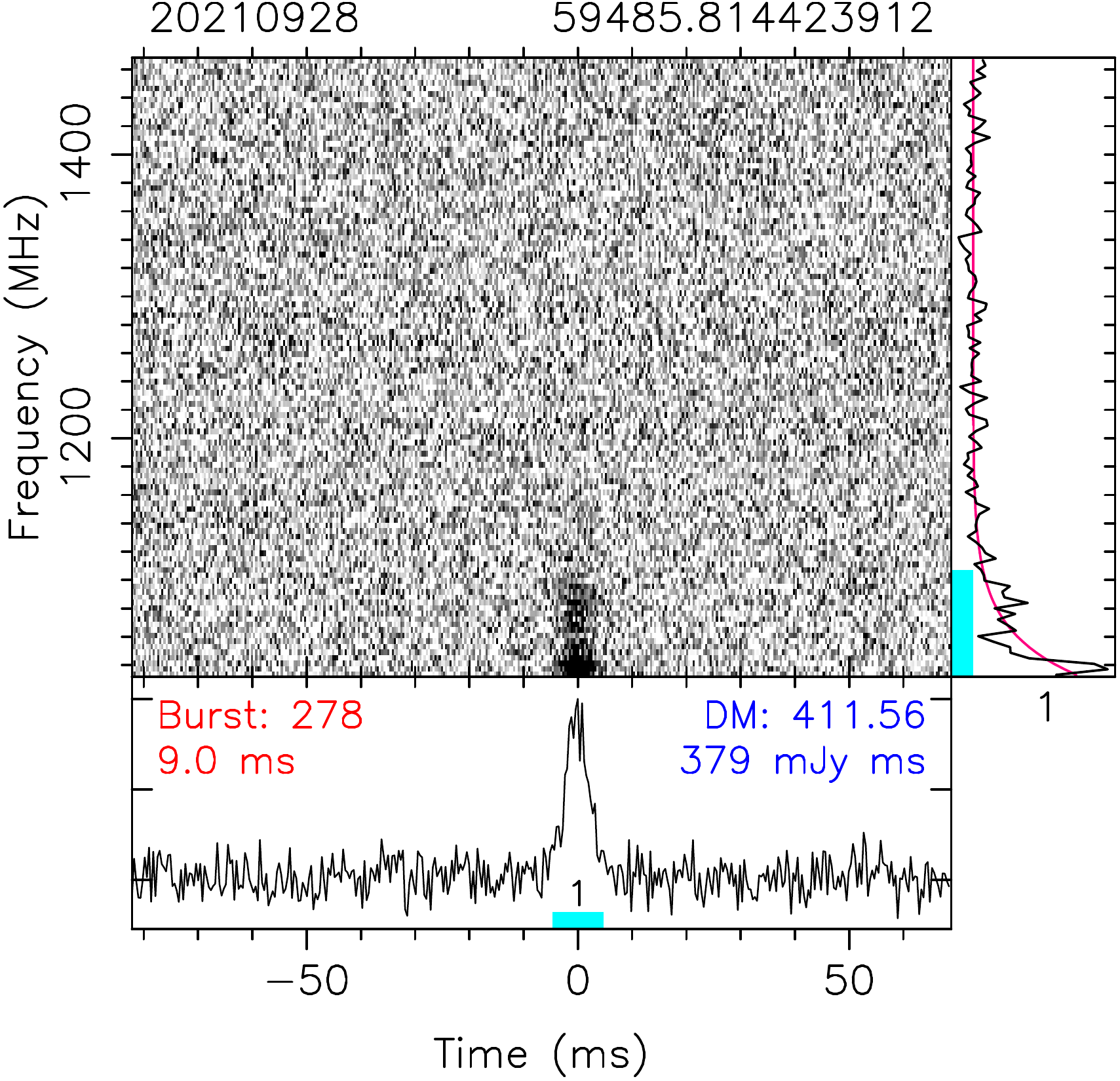}
    \includegraphics[height=37mm]{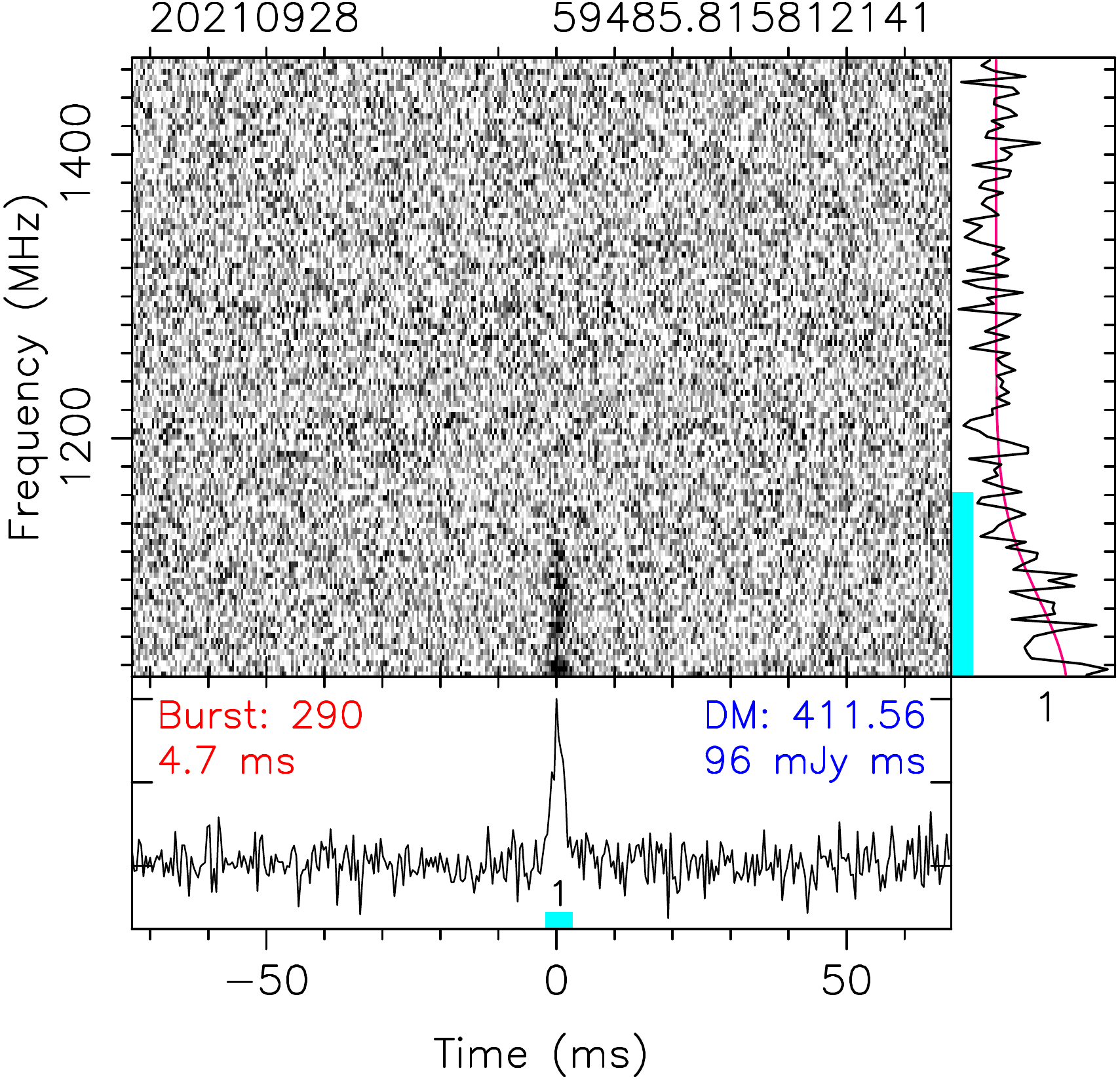}
    \includegraphics[height=37mm]{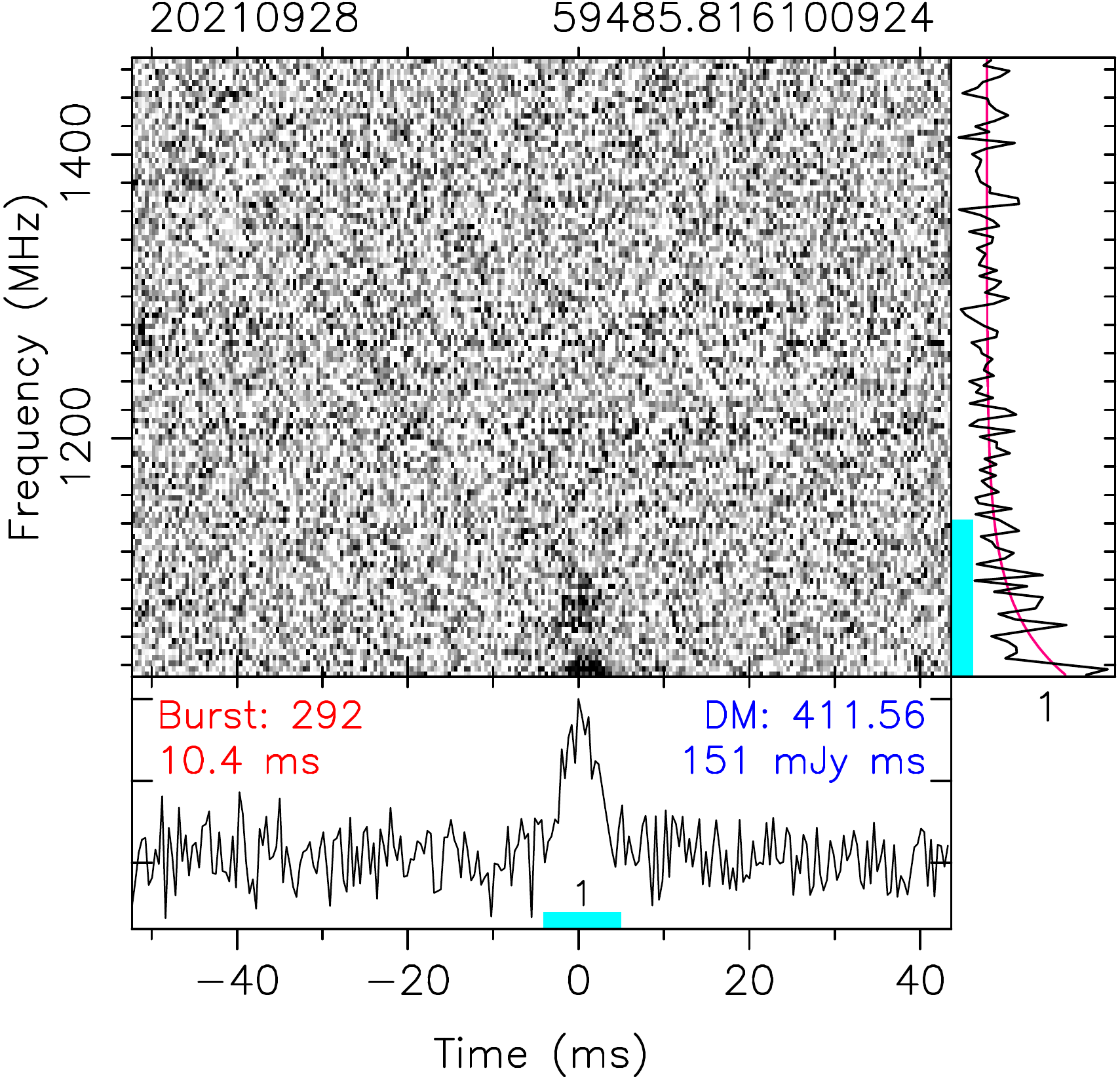}
    \includegraphics[height=37mm]{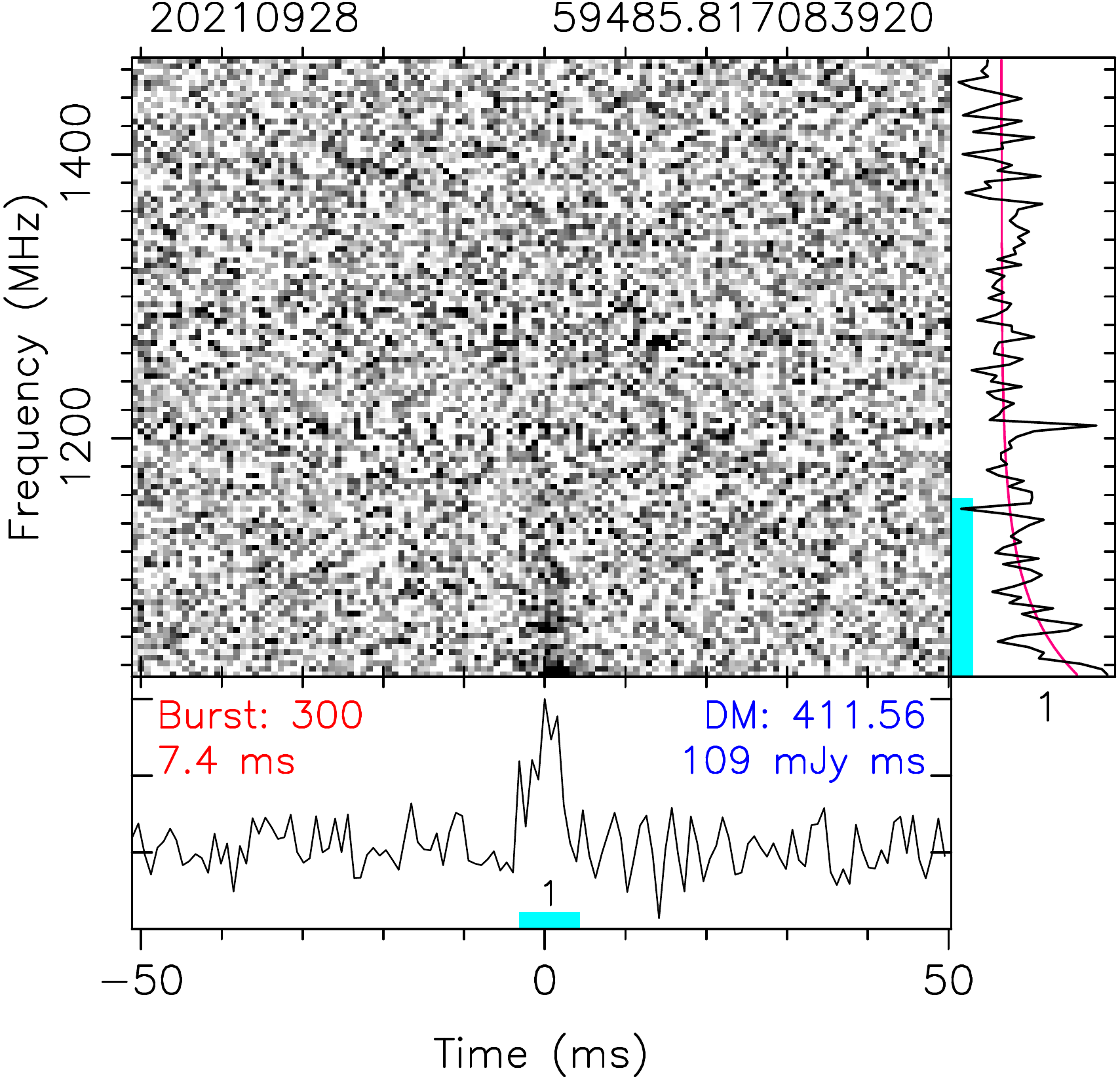}
    \includegraphics[height=37mm]{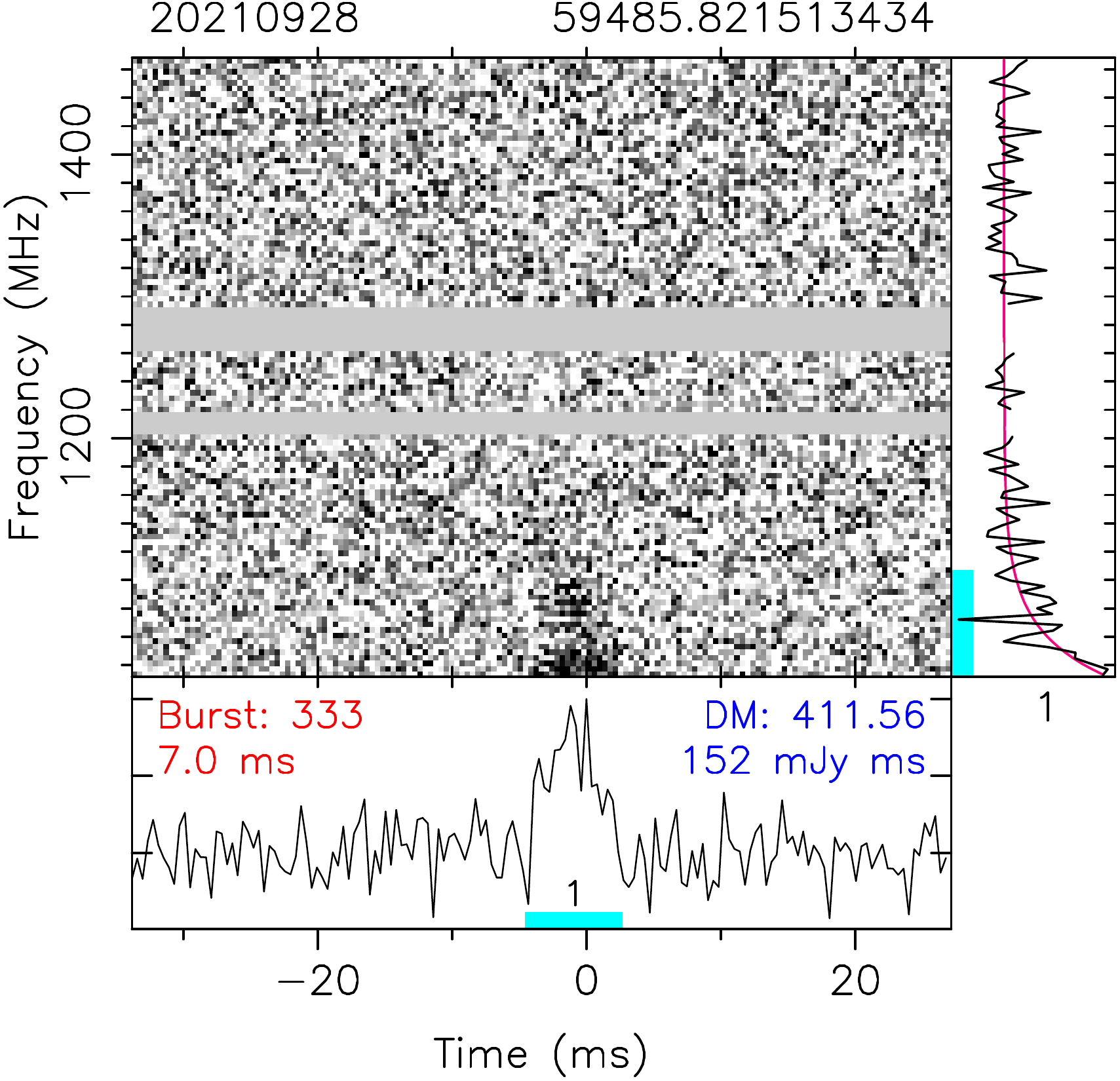}
    \includegraphics[height=37mm]{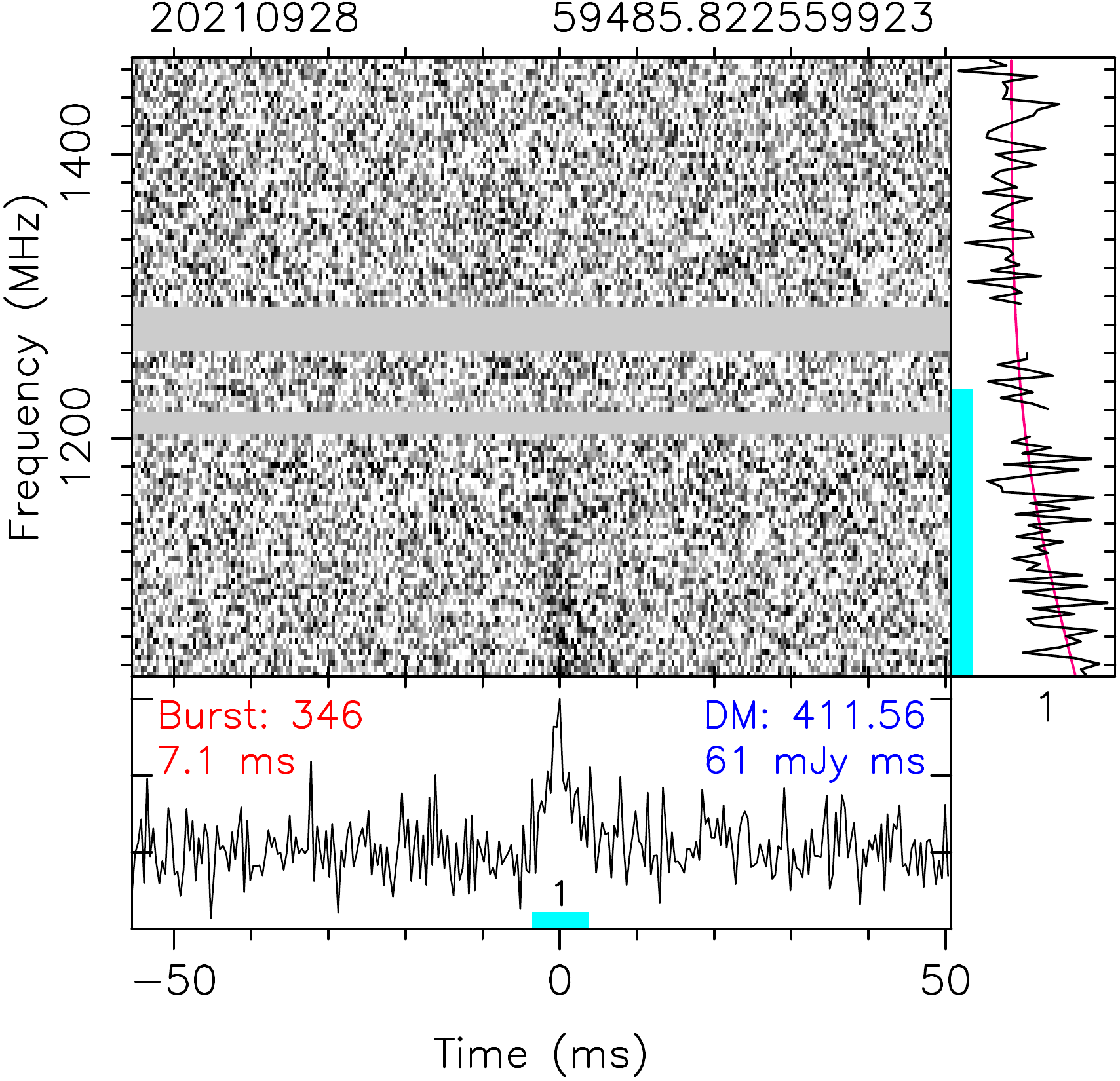}
\caption{\it{ -- continued and ended}.
}
\end{figure*}
\clearpage

\begin{figure*}
    \flushleft
    \includegraphics[height=37mm]{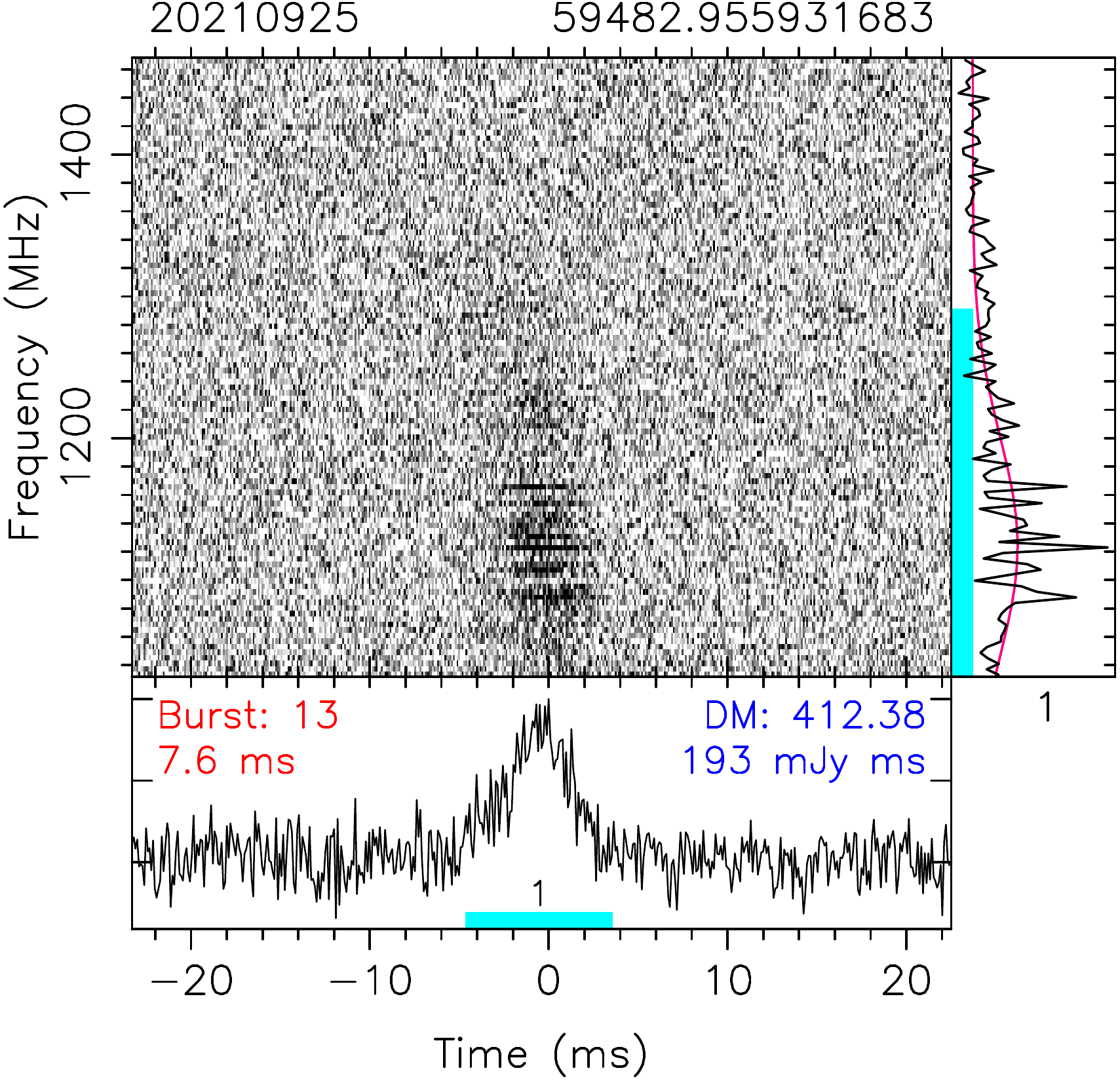}
    \includegraphics[height=37mm]{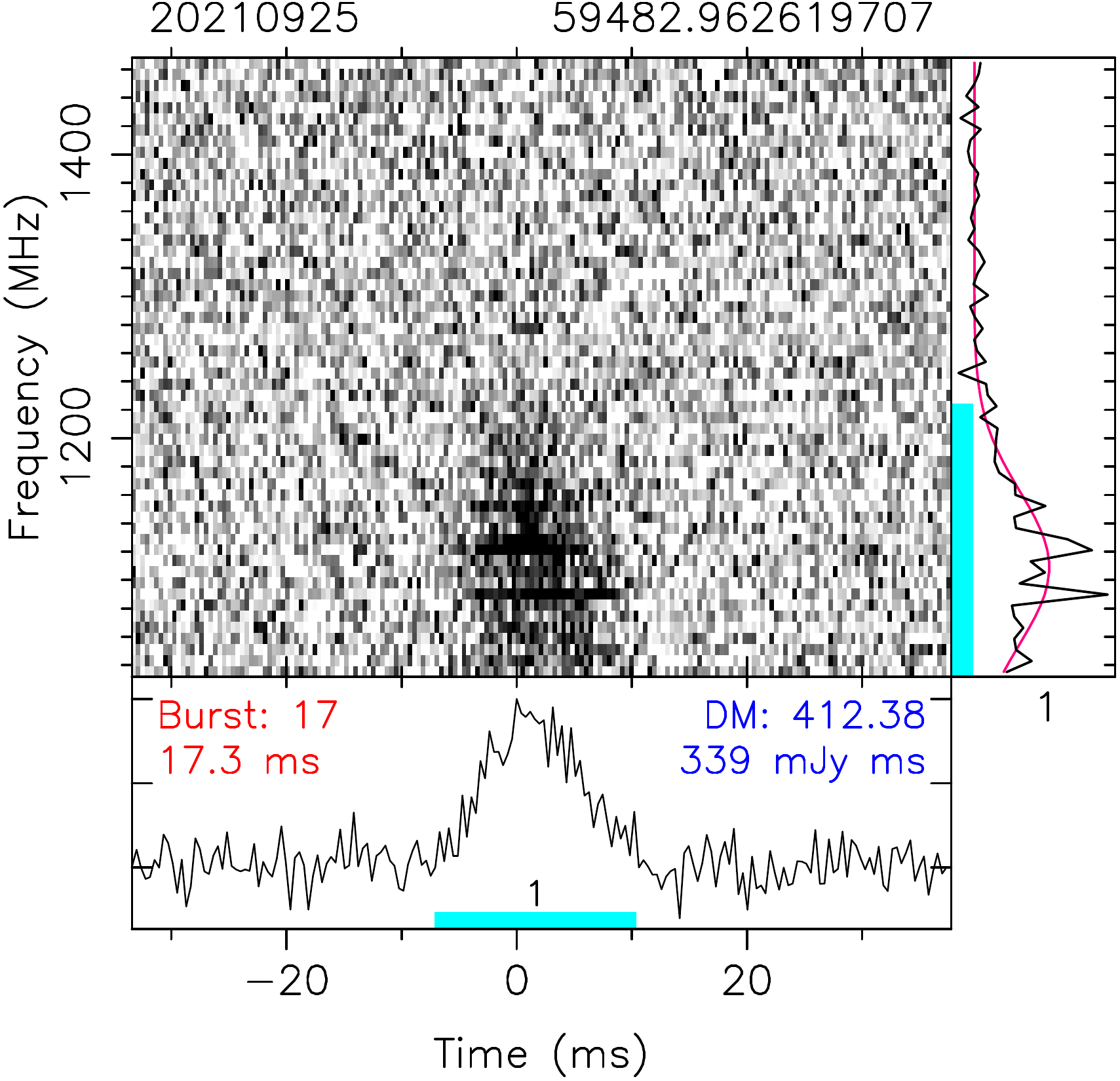}
    \includegraphics[height=37mm]{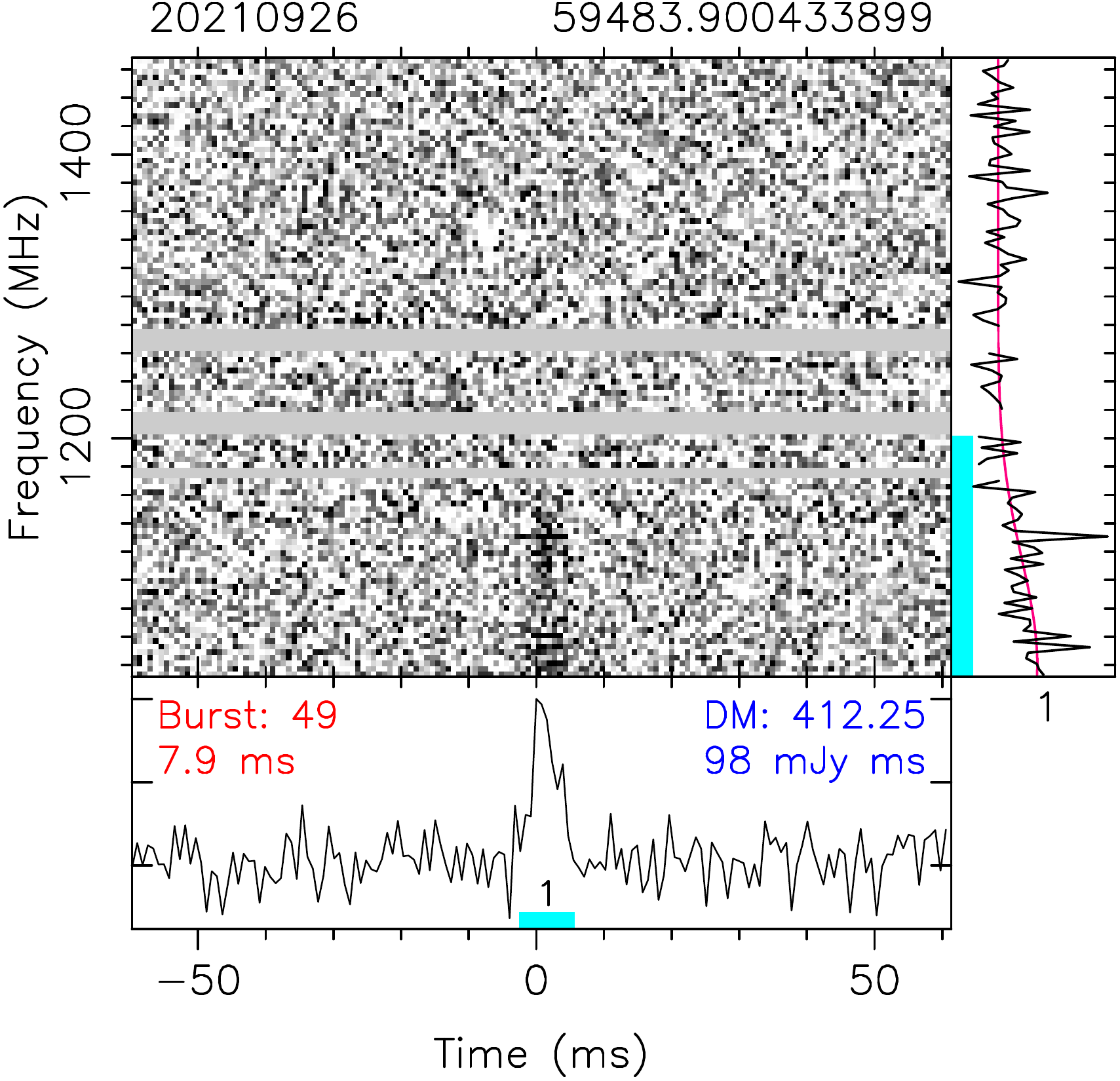}
    \includegraphics[height=37mm]{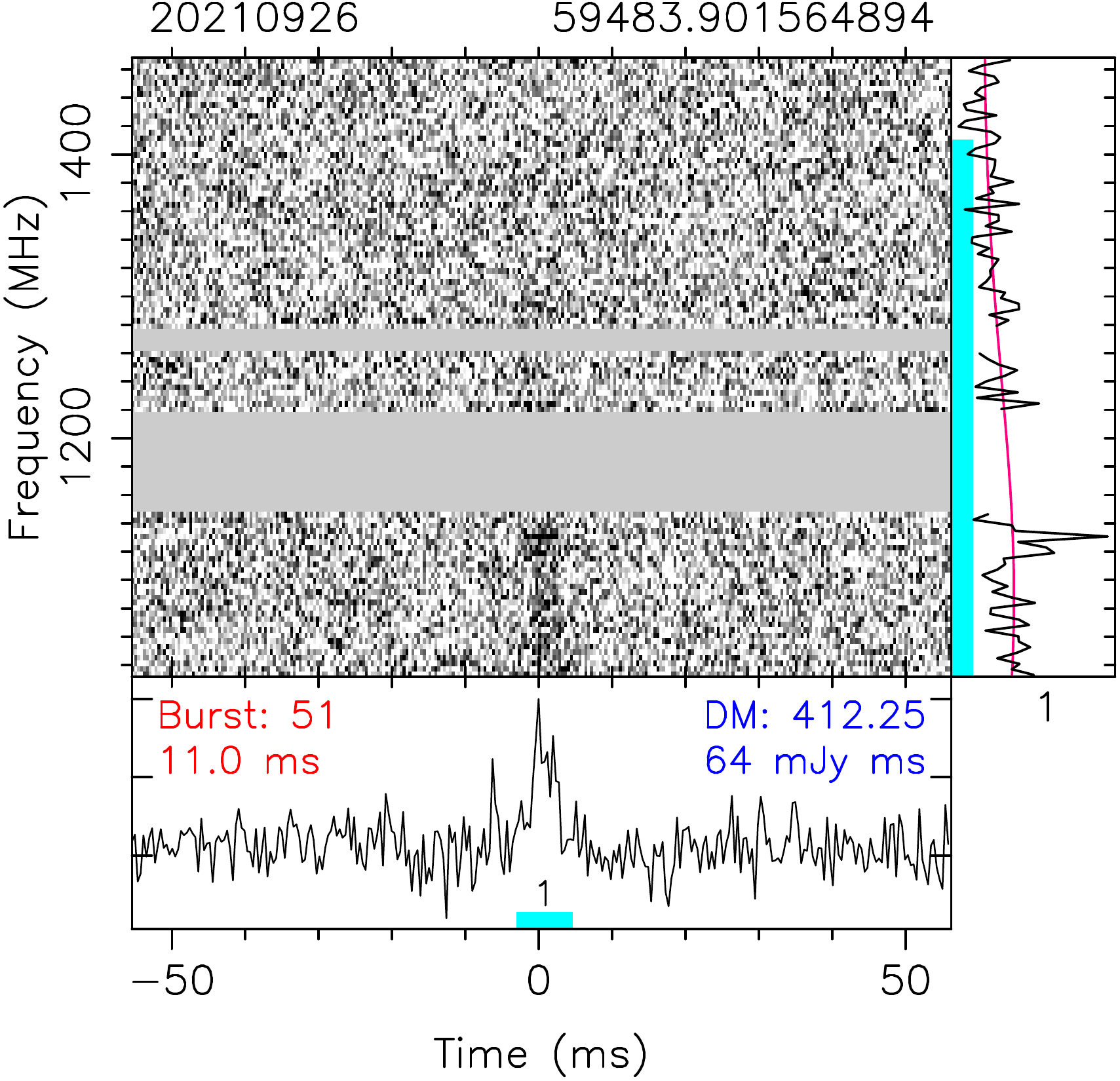}
    \includegraphics[height=37mm]{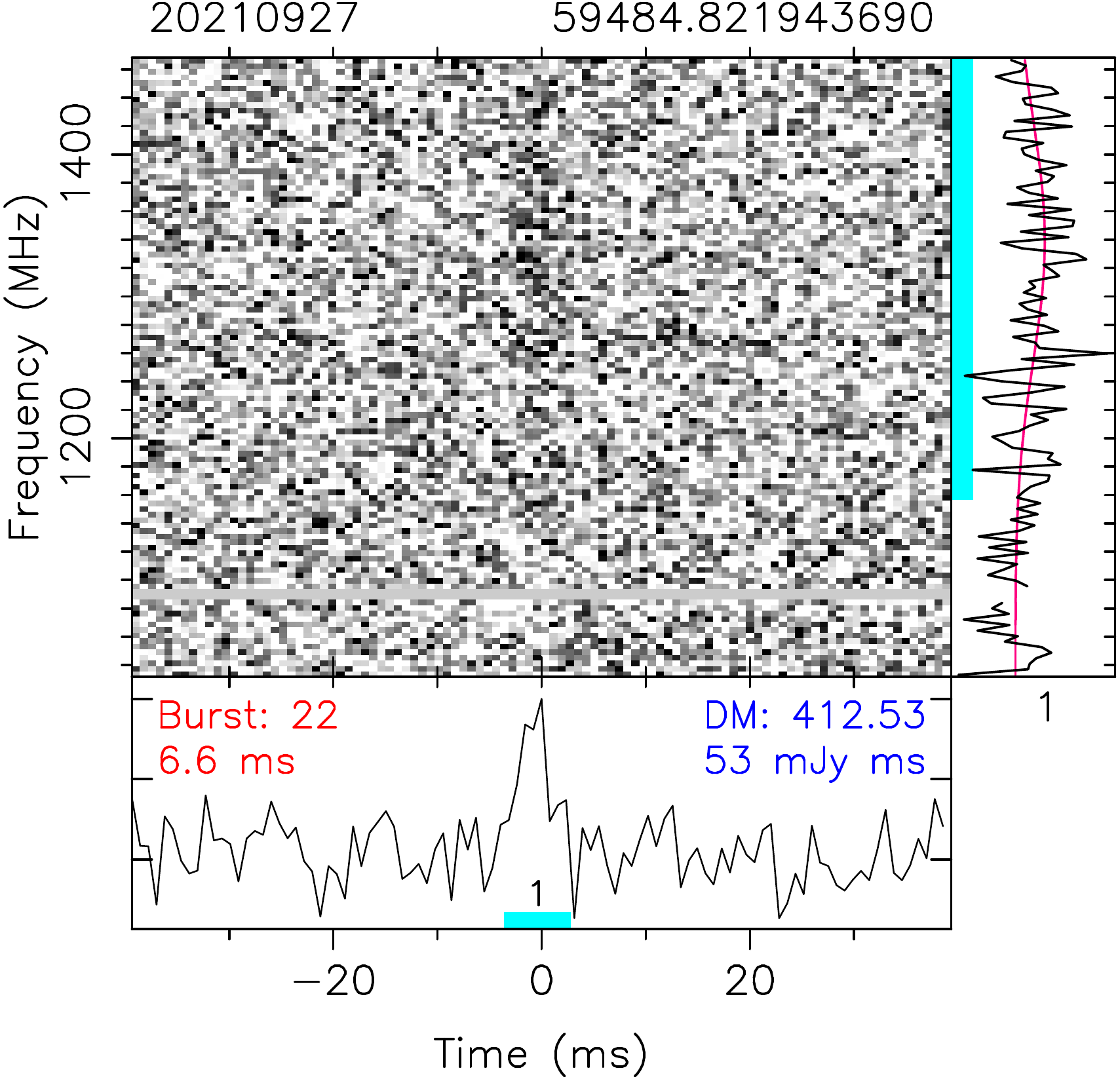}
    \includegraphics[height=37mm]{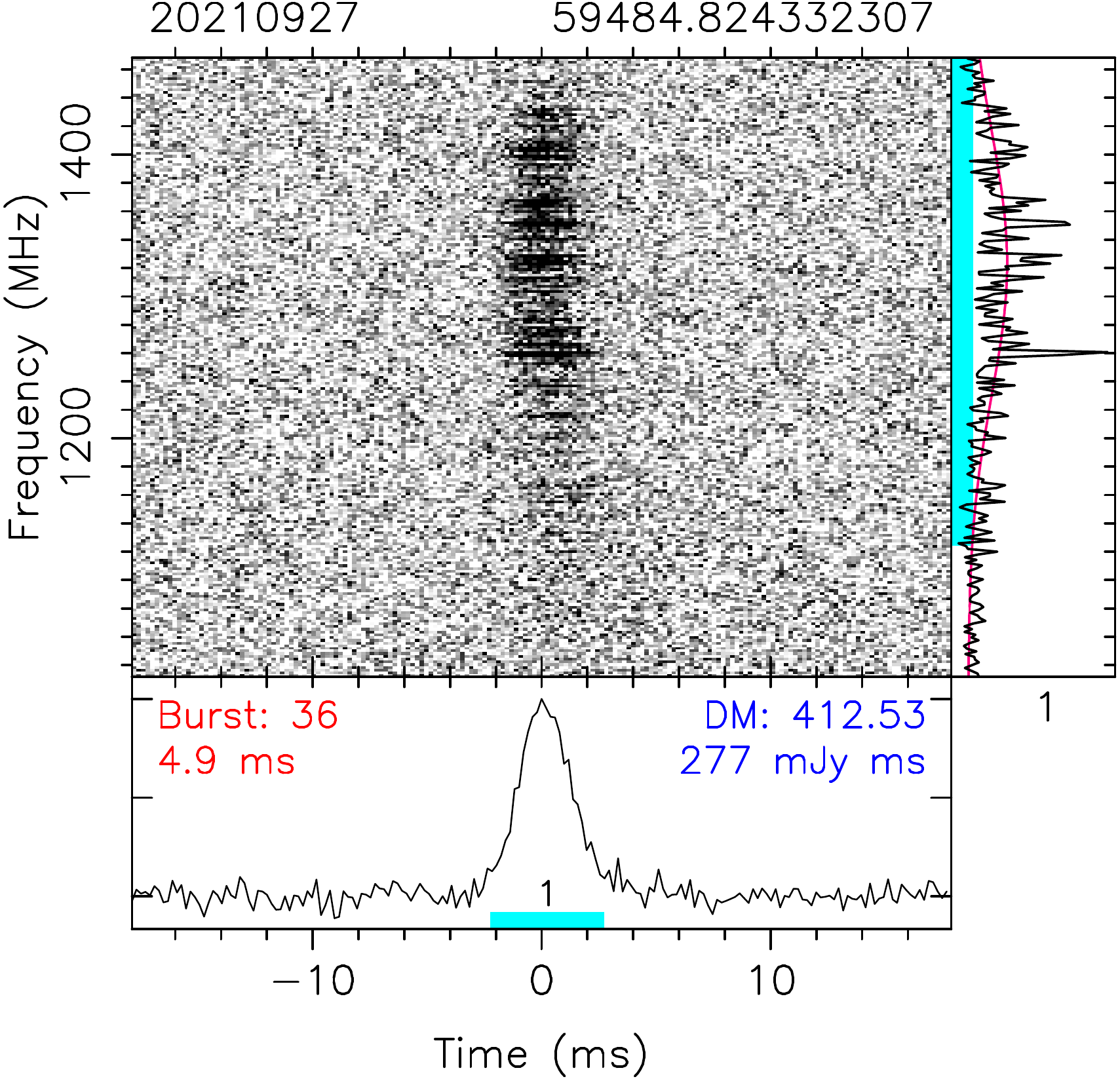}
    \includegraphics[height=37mm]{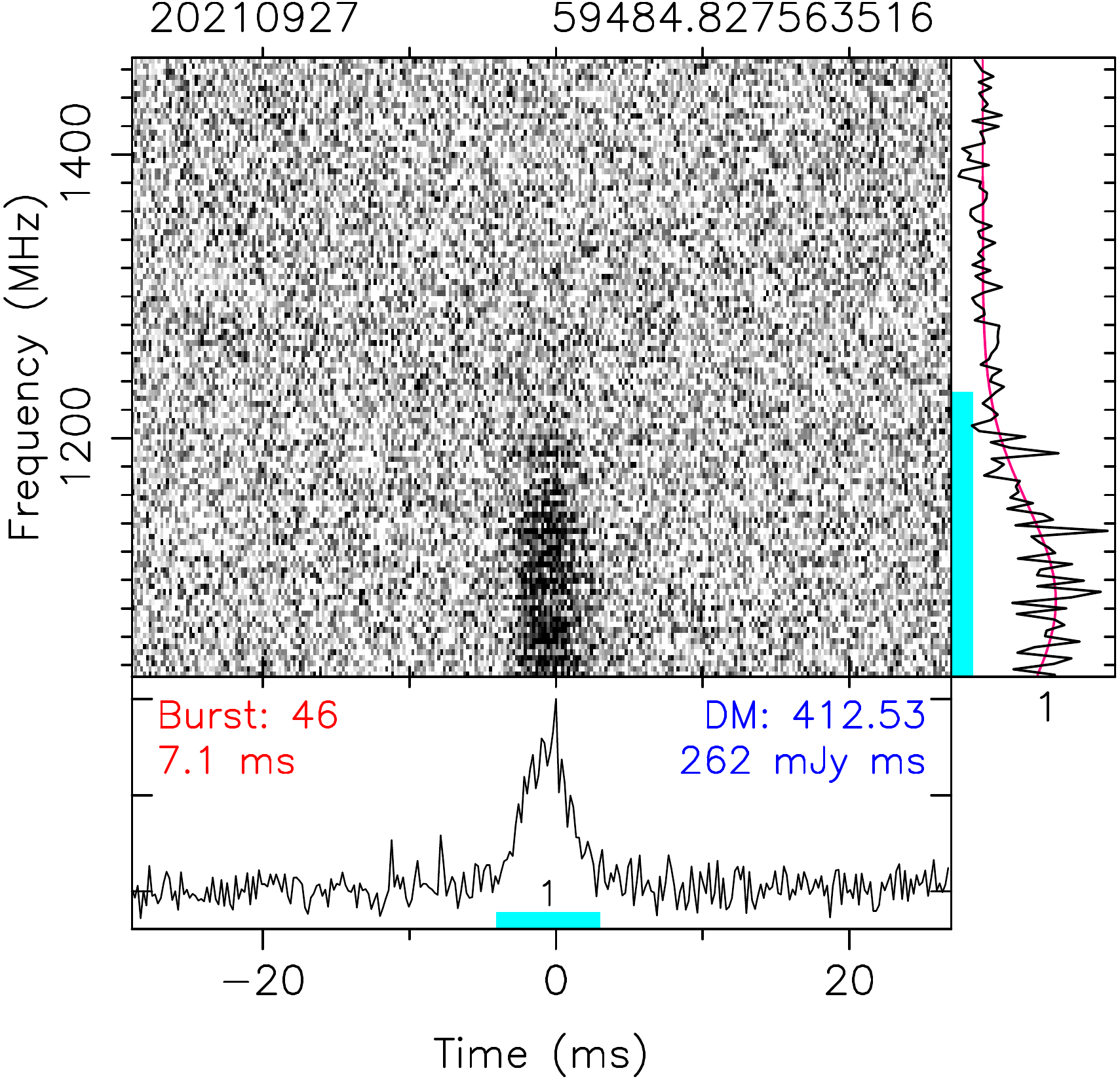}
    \includegraphics[height=37mm]{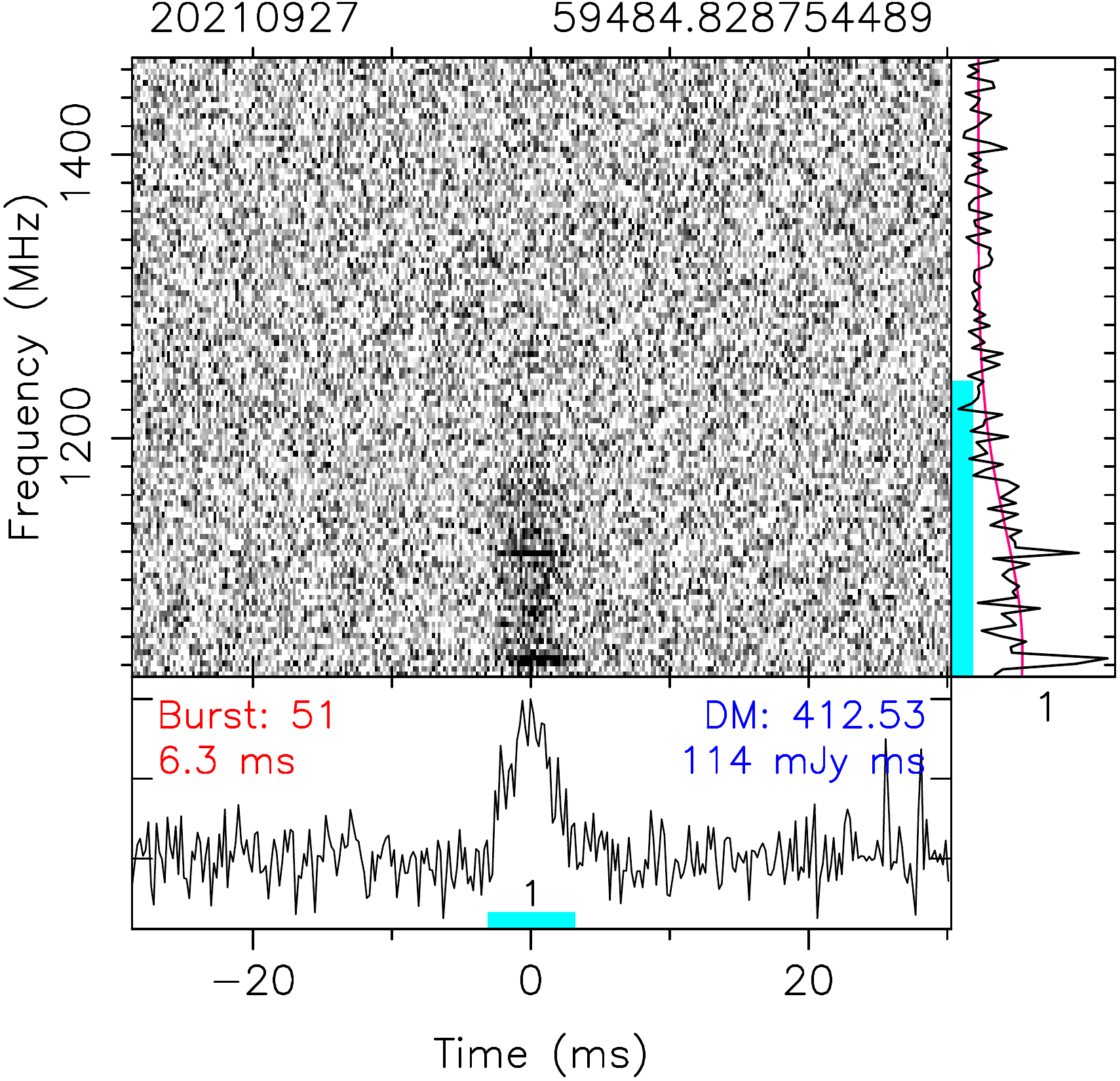}
    \includegraphics[height=37mm]{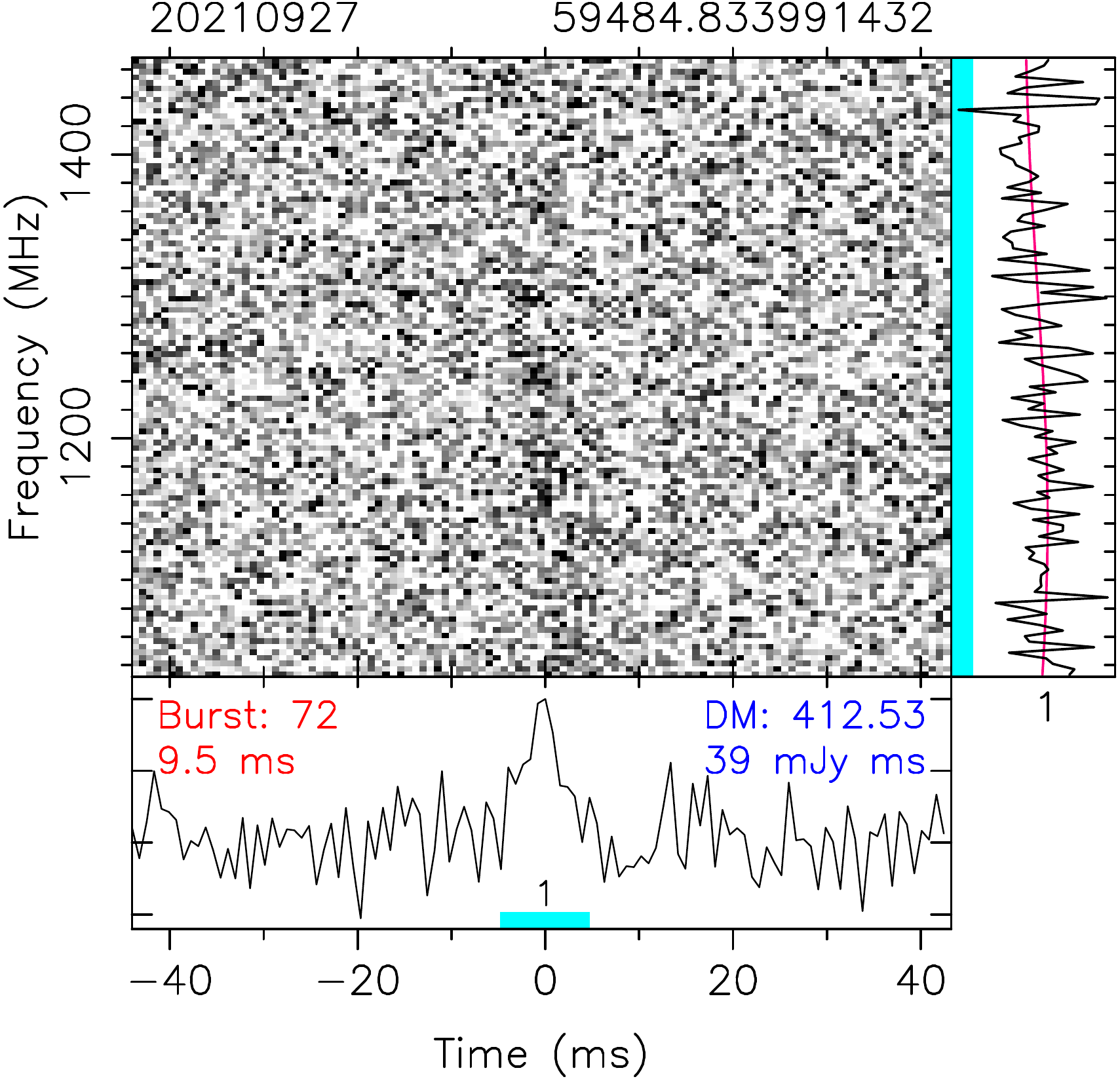}
    \includegraphics[height=37mm]{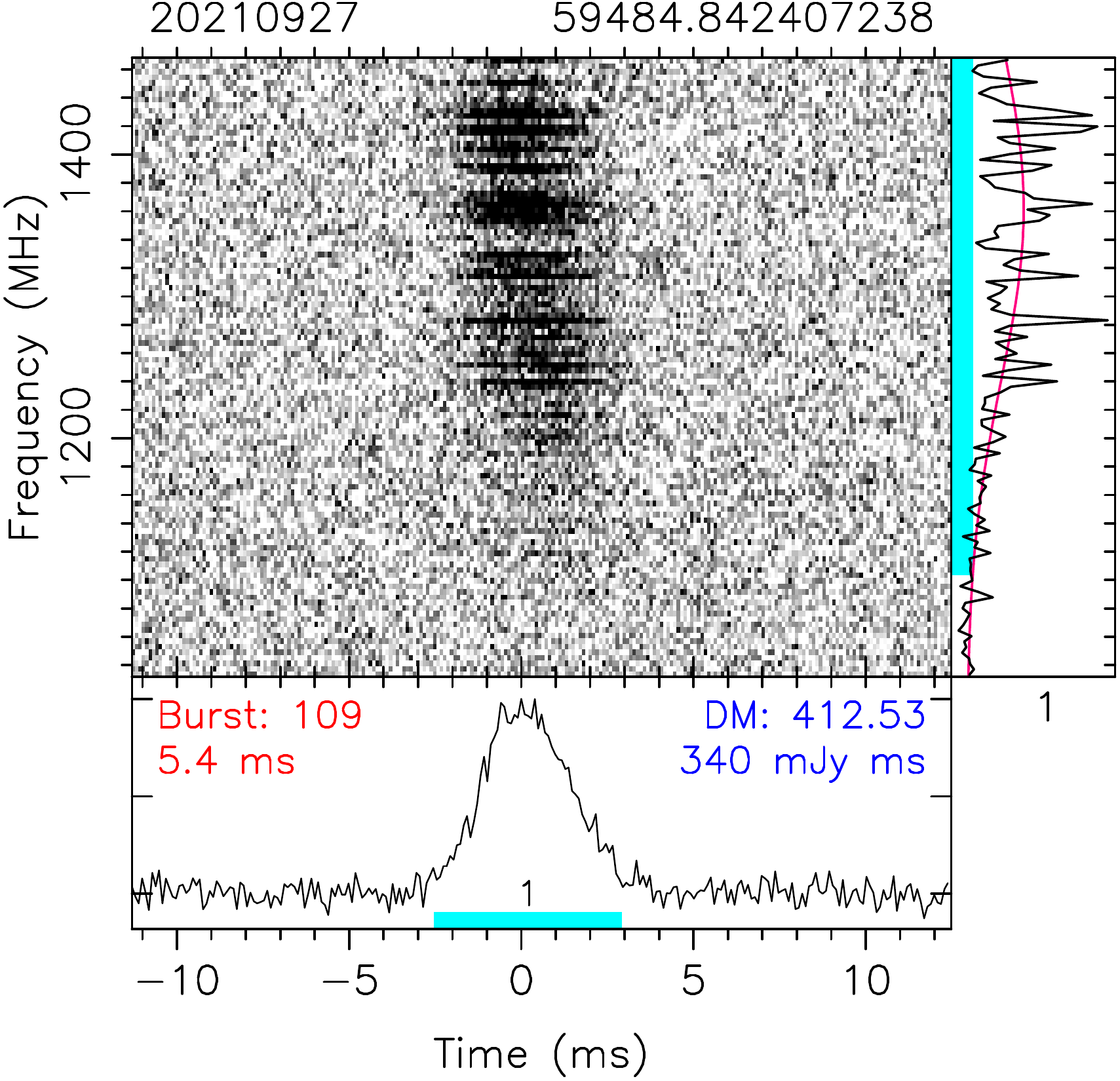}
    \includegraphics[height=37mm]{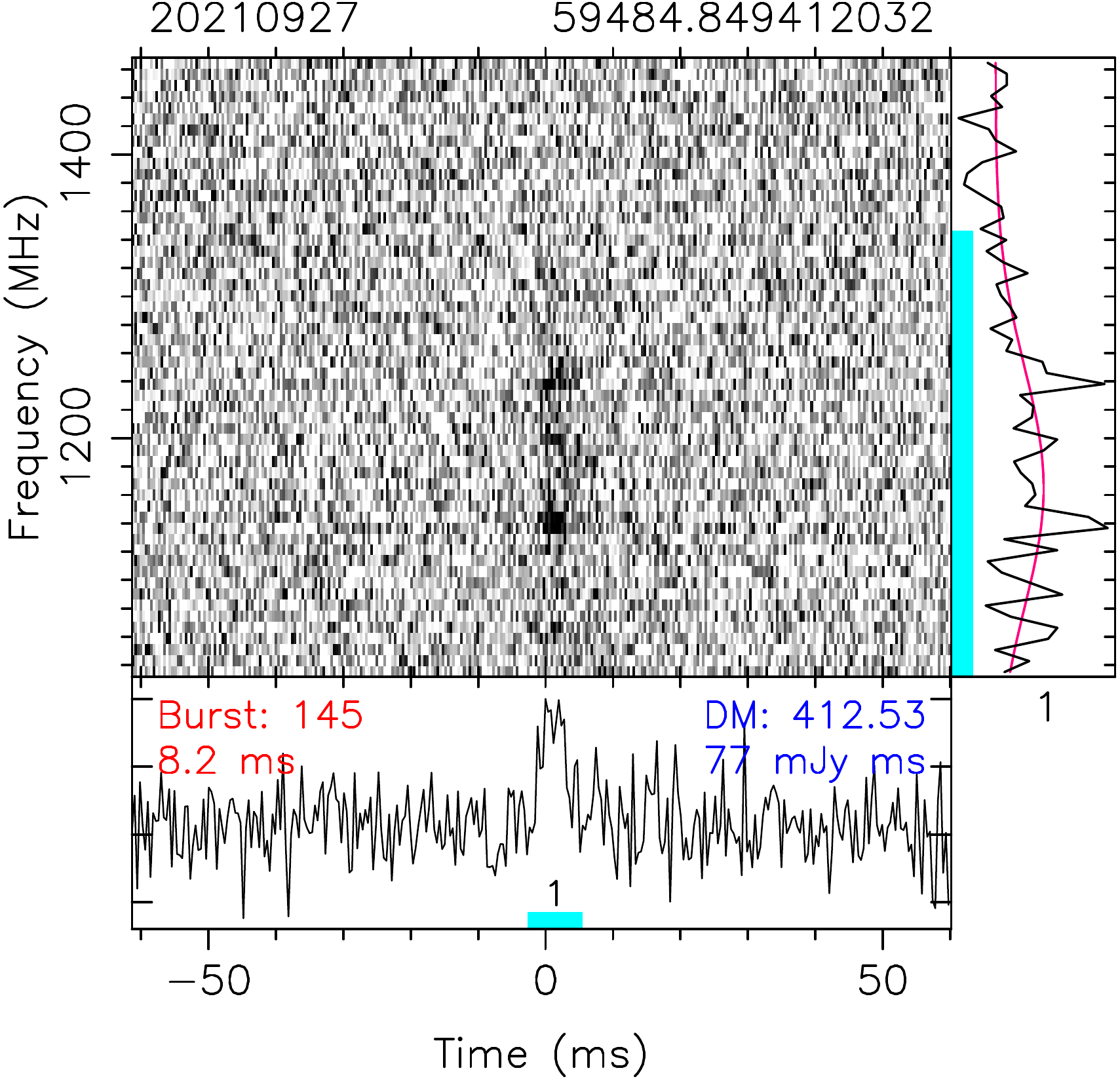}
    \includegraphics[height=37mm]{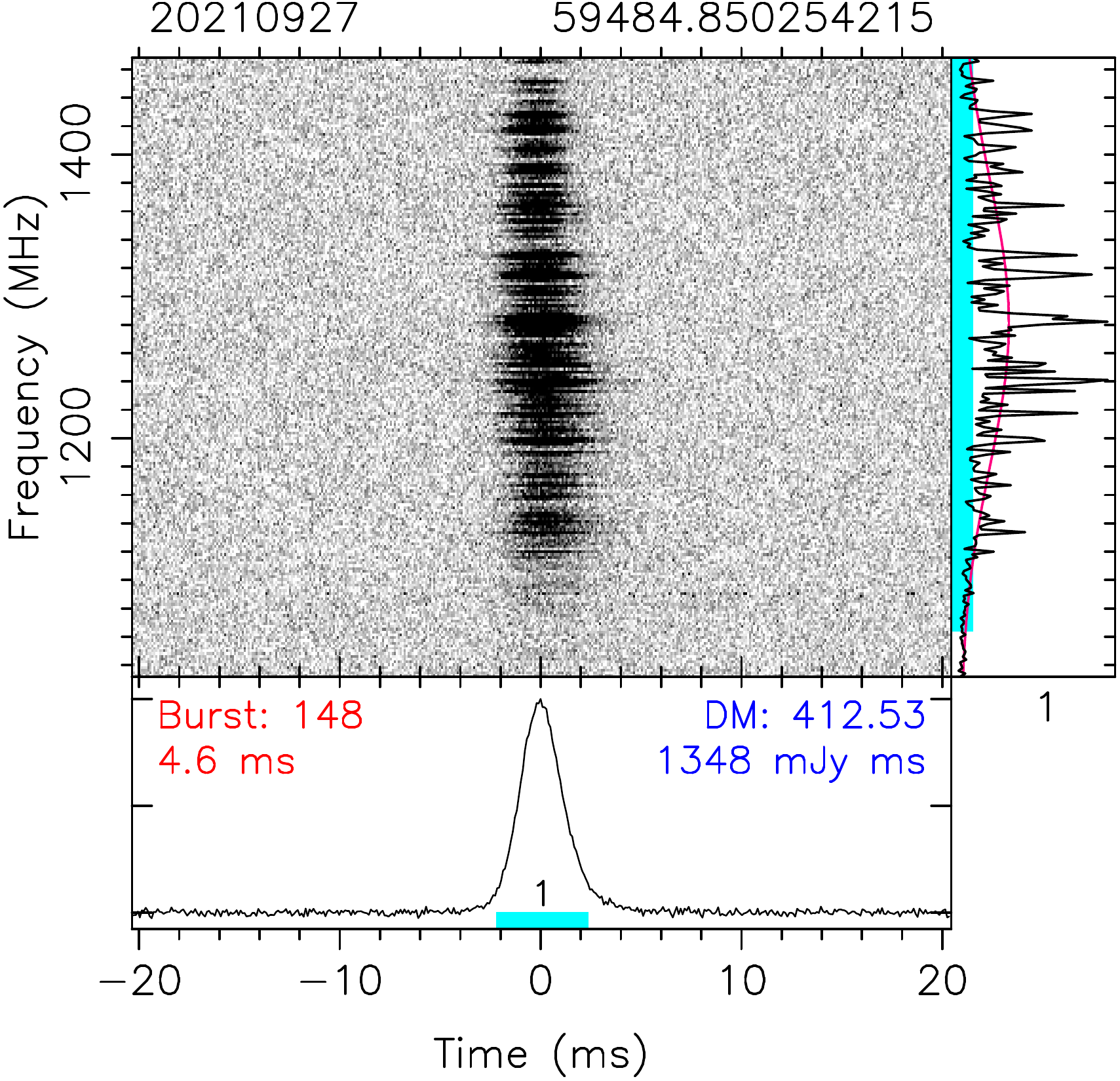}
    \includegraphics[height=37mm]{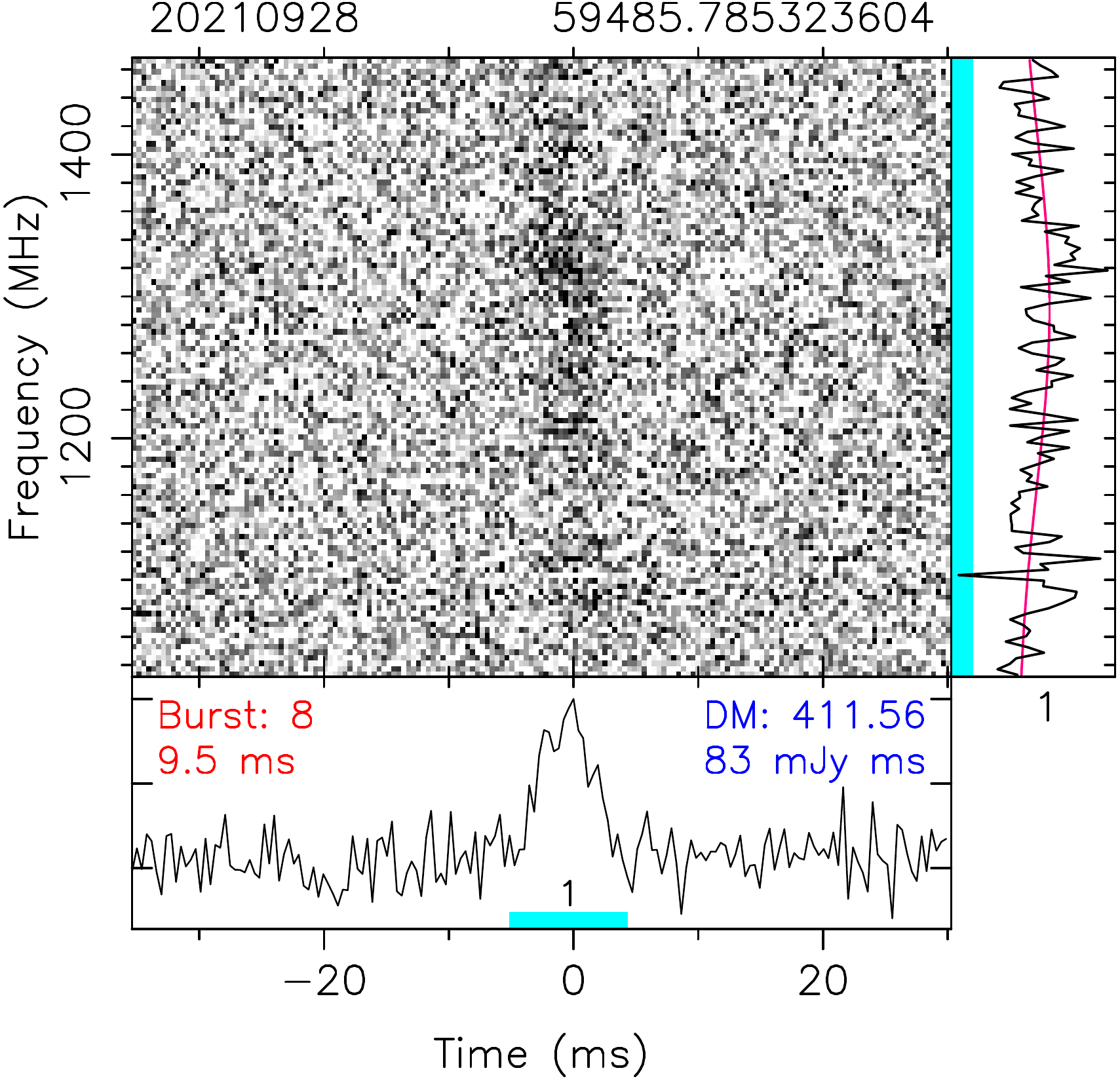}
    \includegraphics[height=37mm]{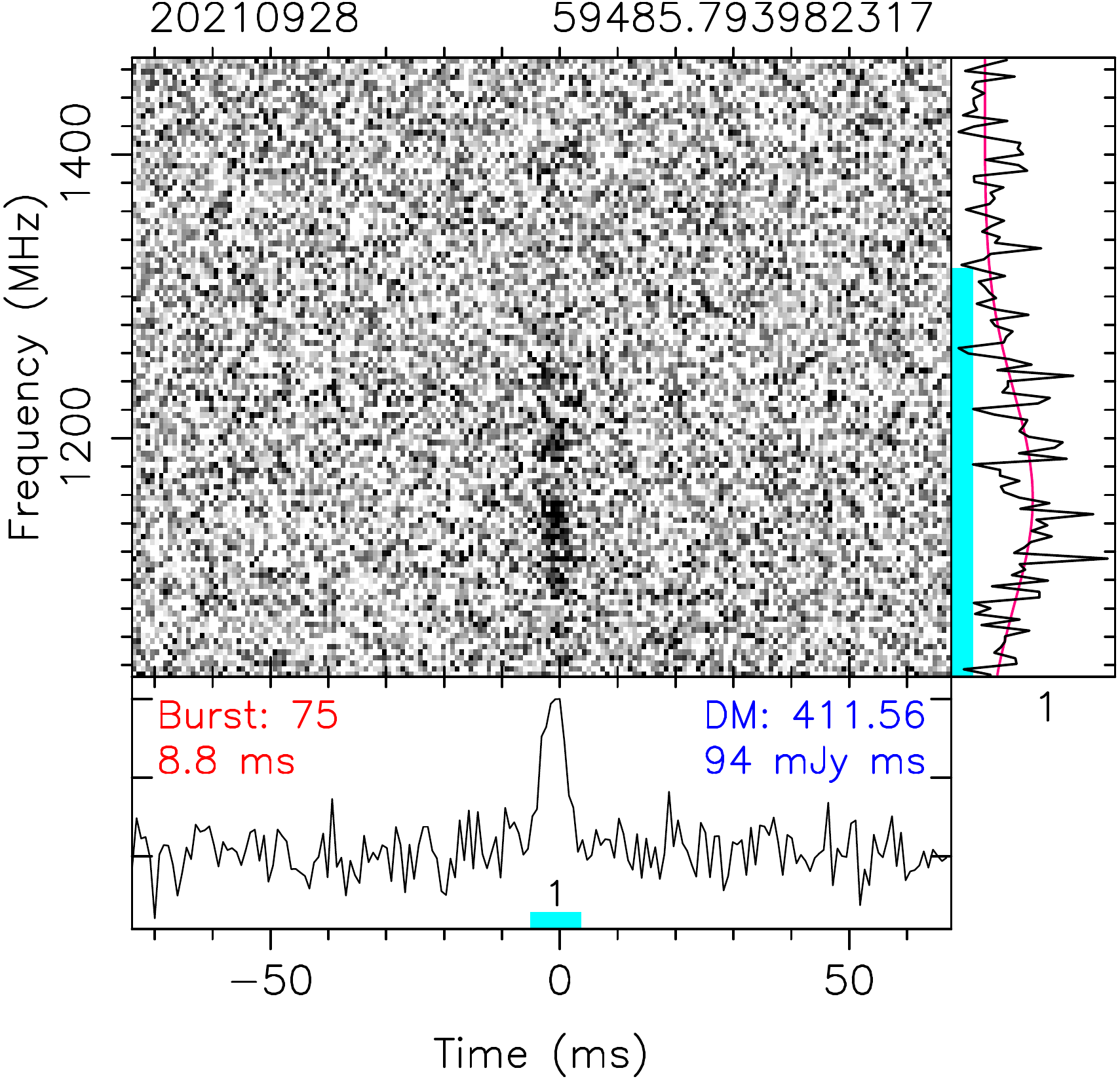}
    \includegraphics[height=37mm]{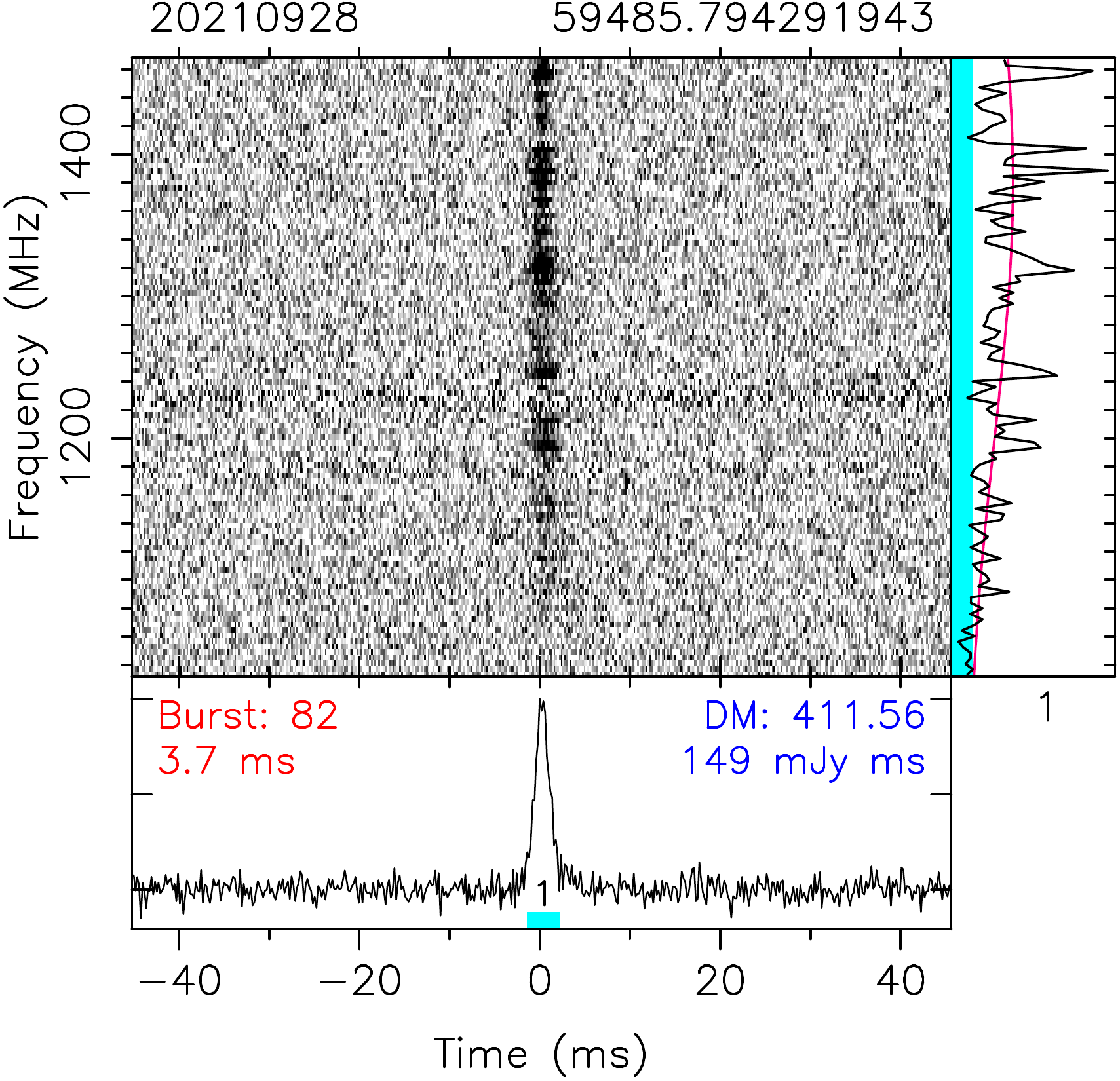}
    \includegraphics[height=37mm]{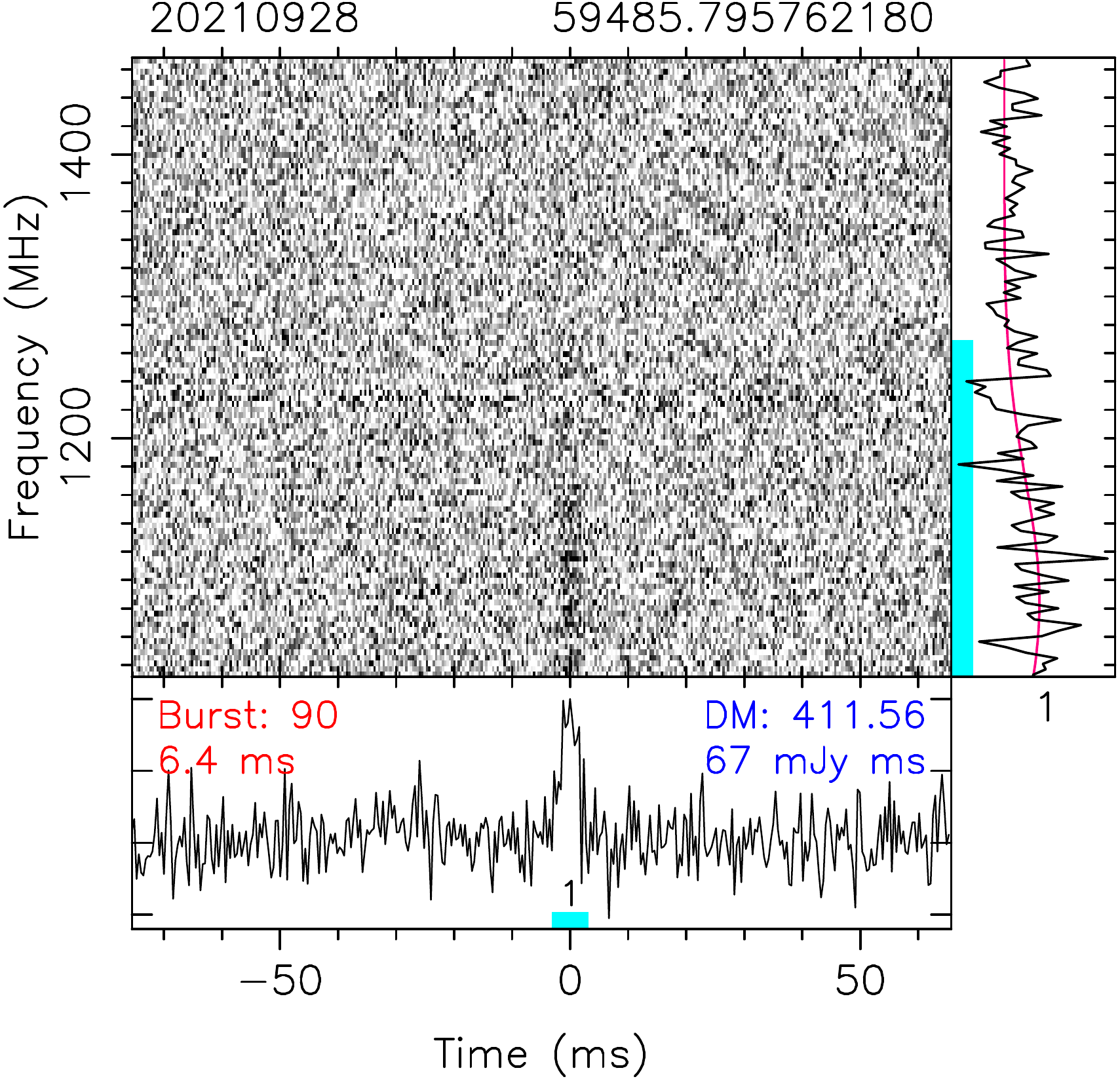}
    \includegraphics[height=37mm]{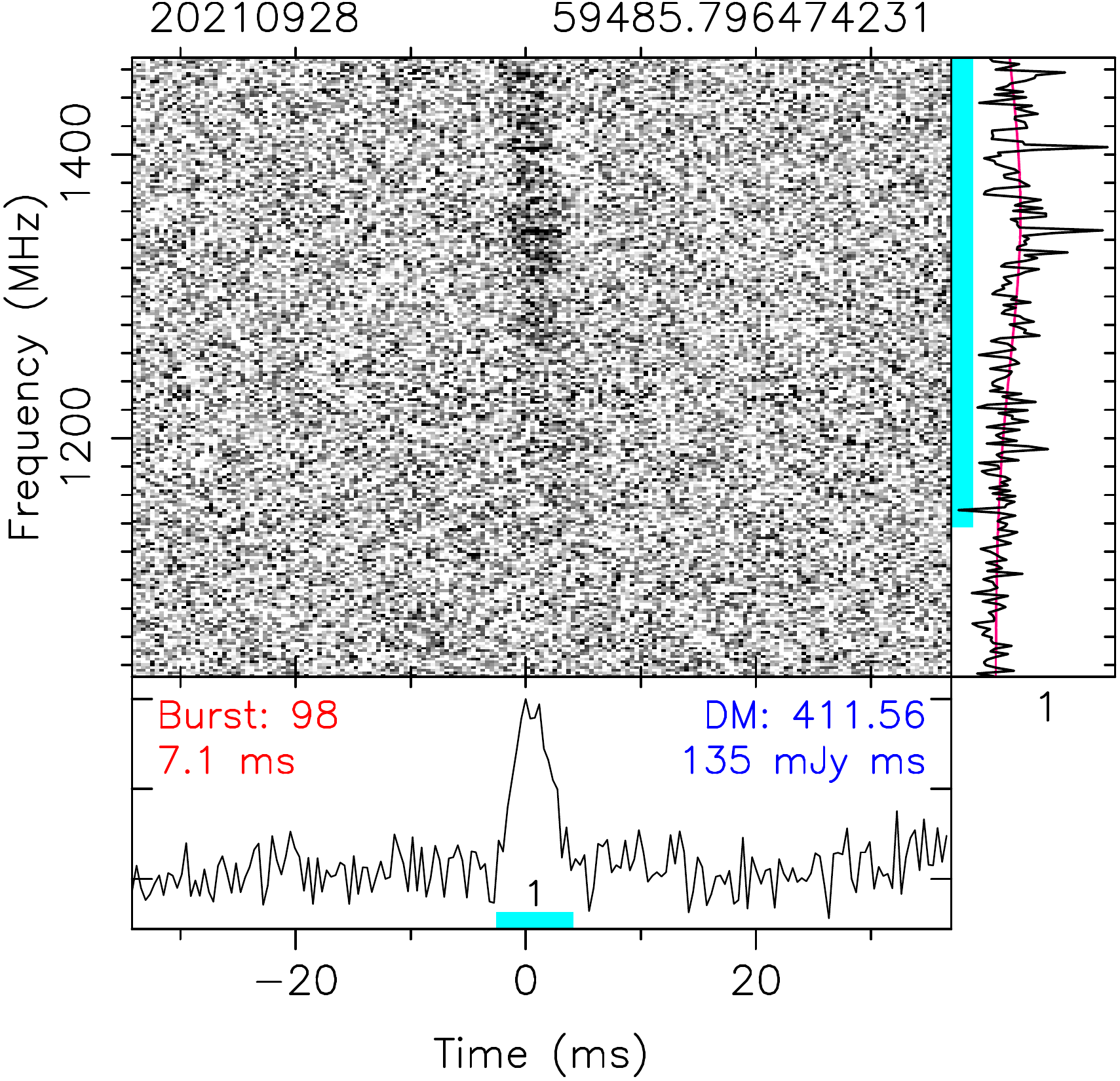}
    \includegraphics[height=37mm]{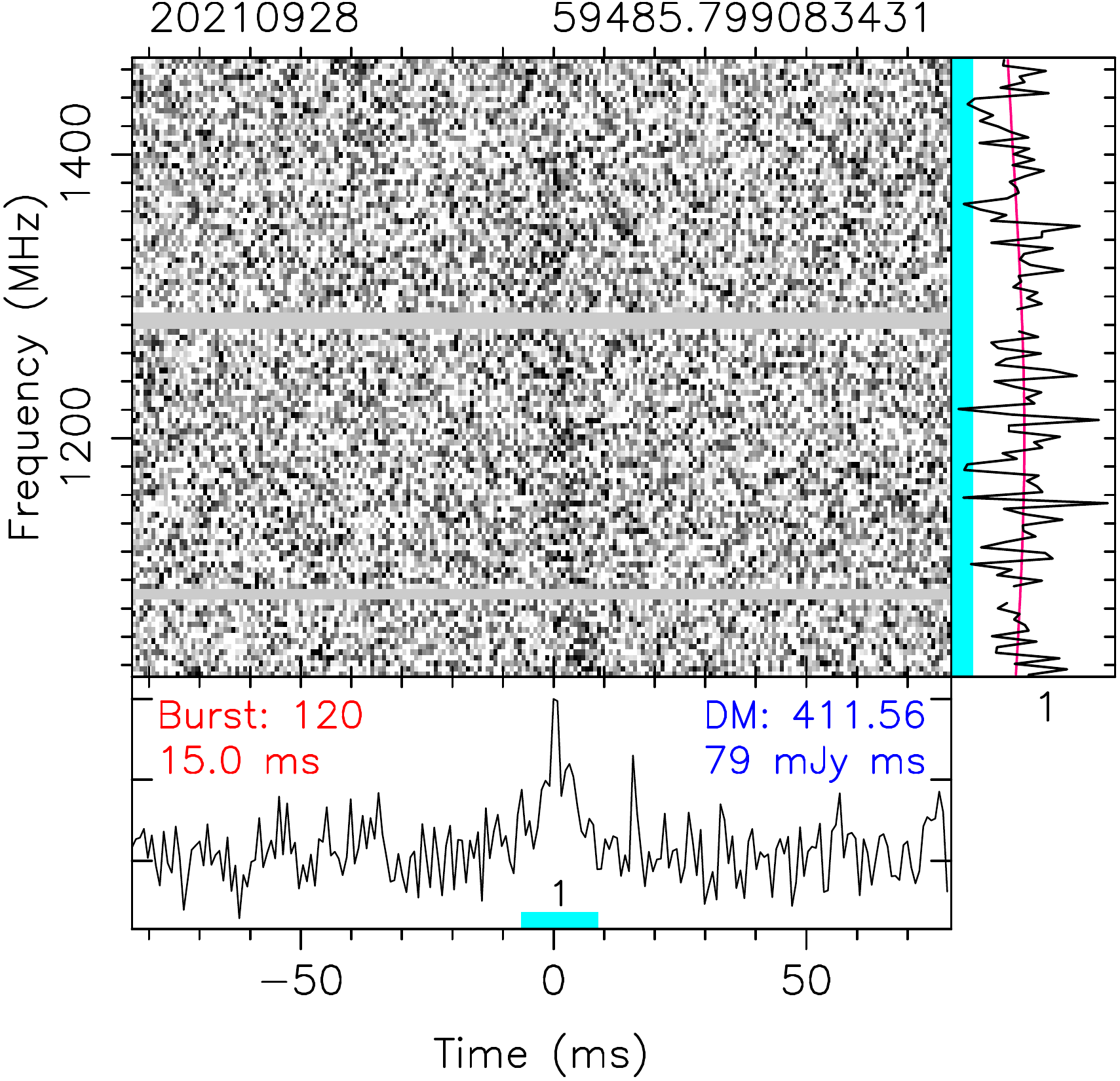}
    \includegraphics[height=37mm]{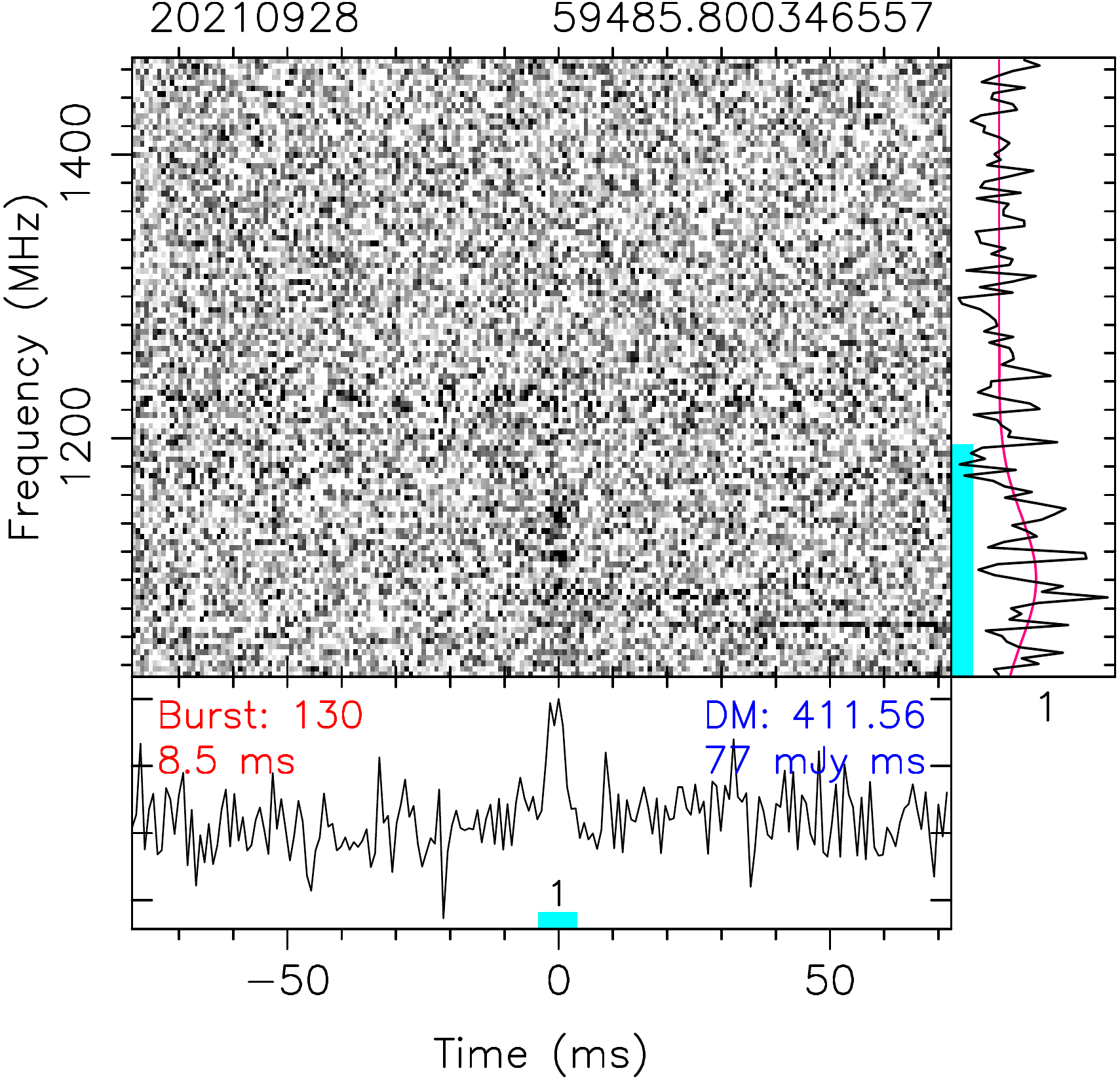}
    \includegraphics[height=37mm]{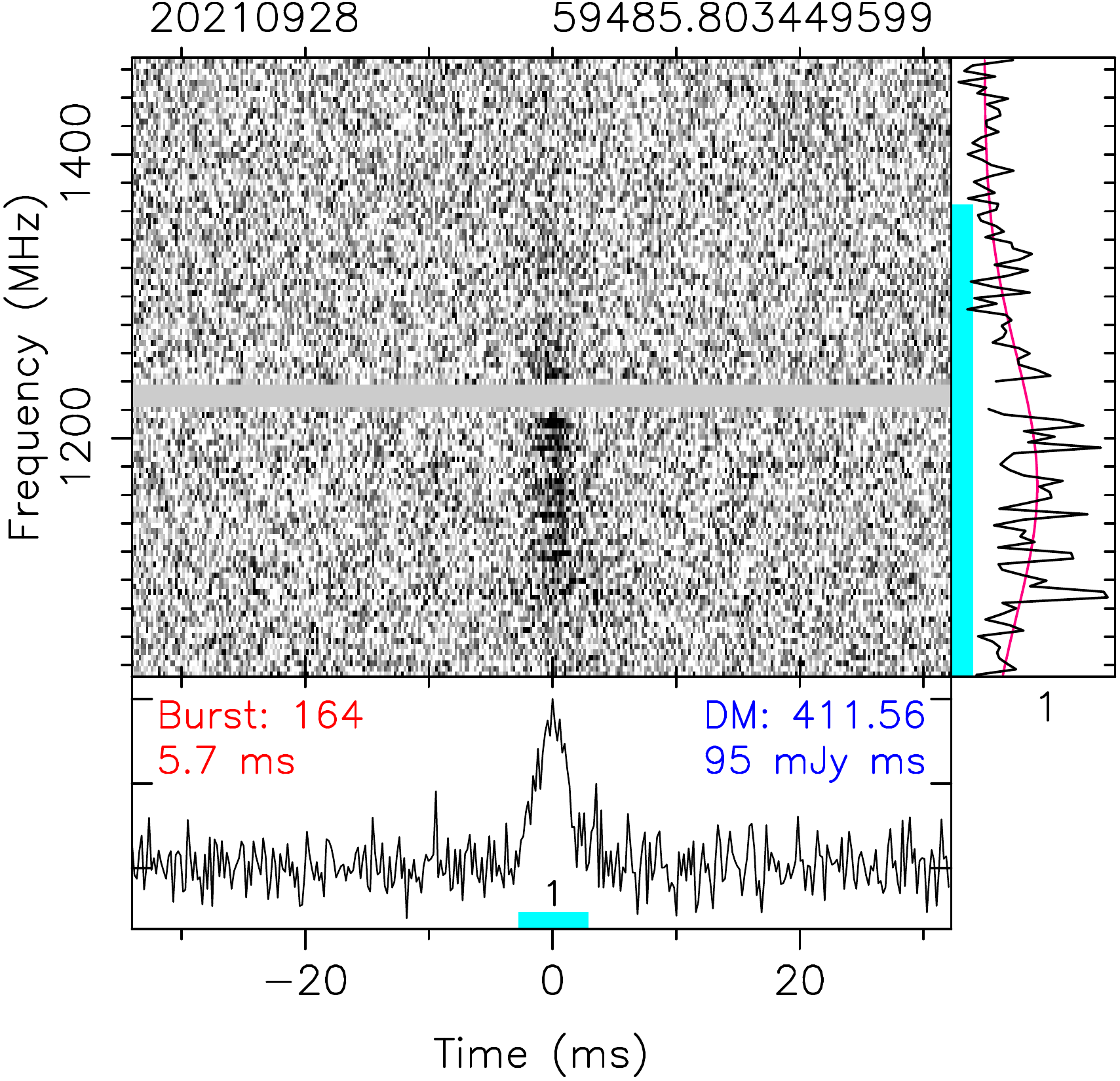}
    \includegraphics[height=37mm]{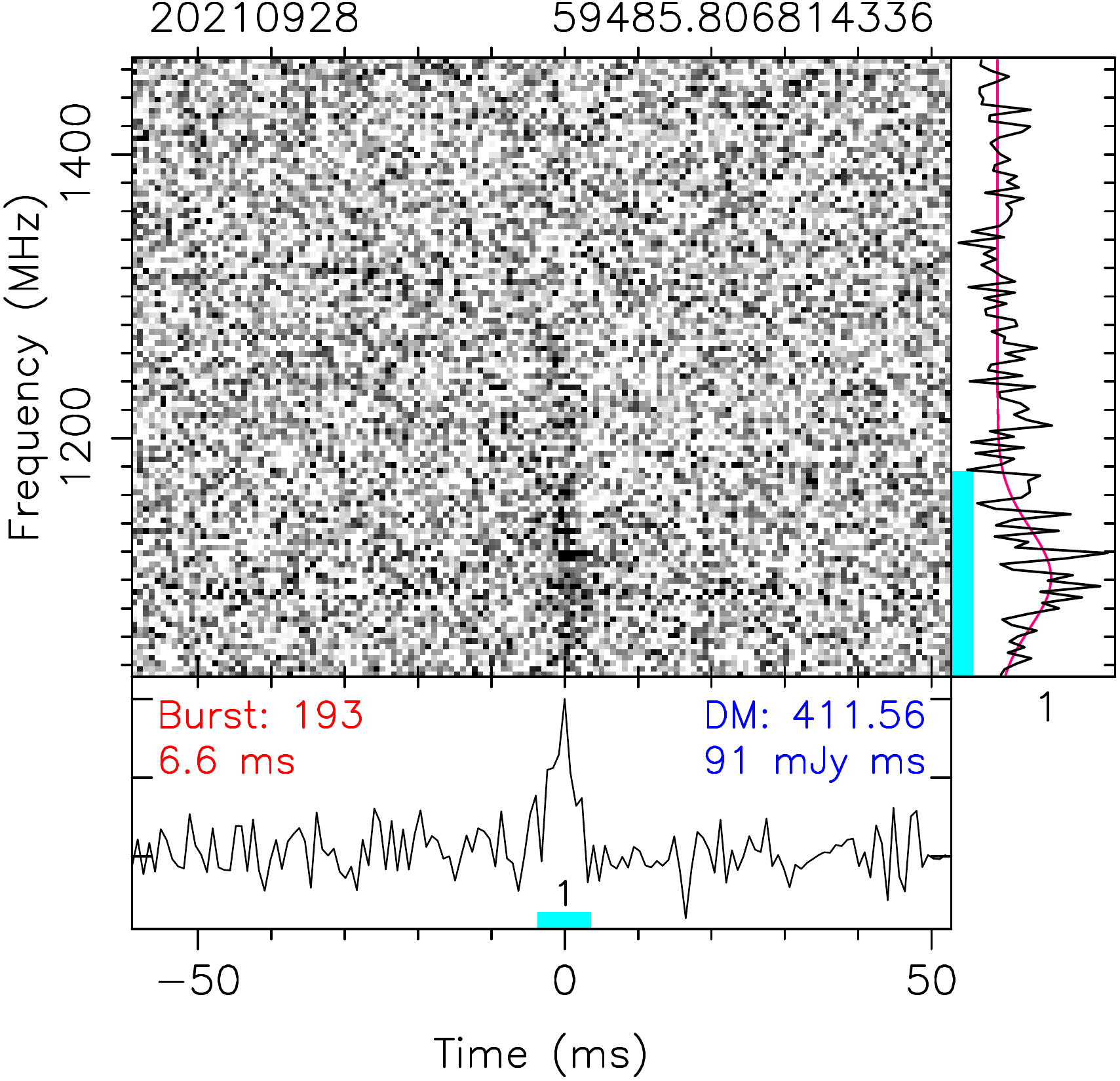}
    \includegraphics[height=37mm]{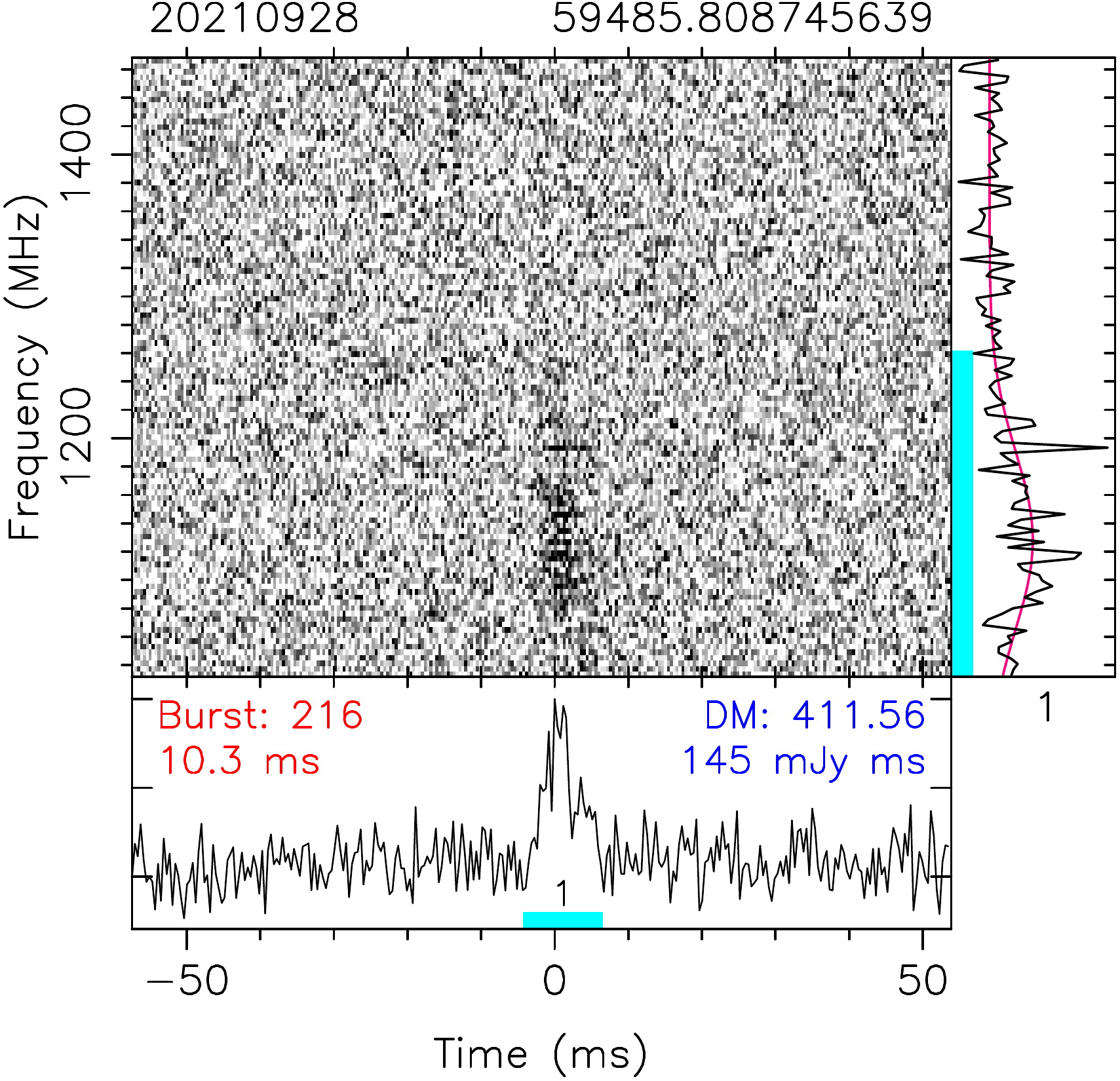}
    \includegraphics[height=37mm]{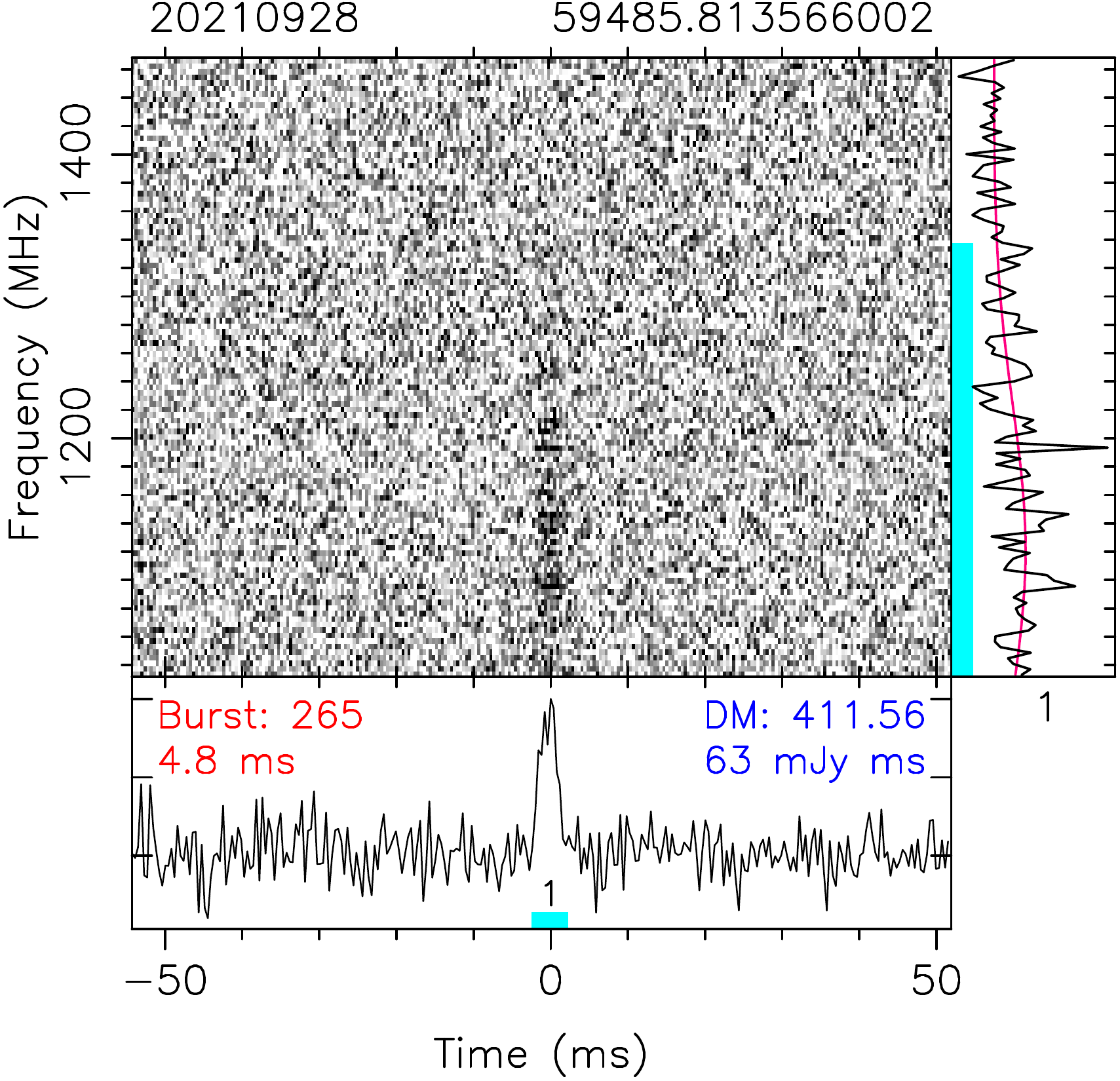}
    \includegraphics[height=37mm]{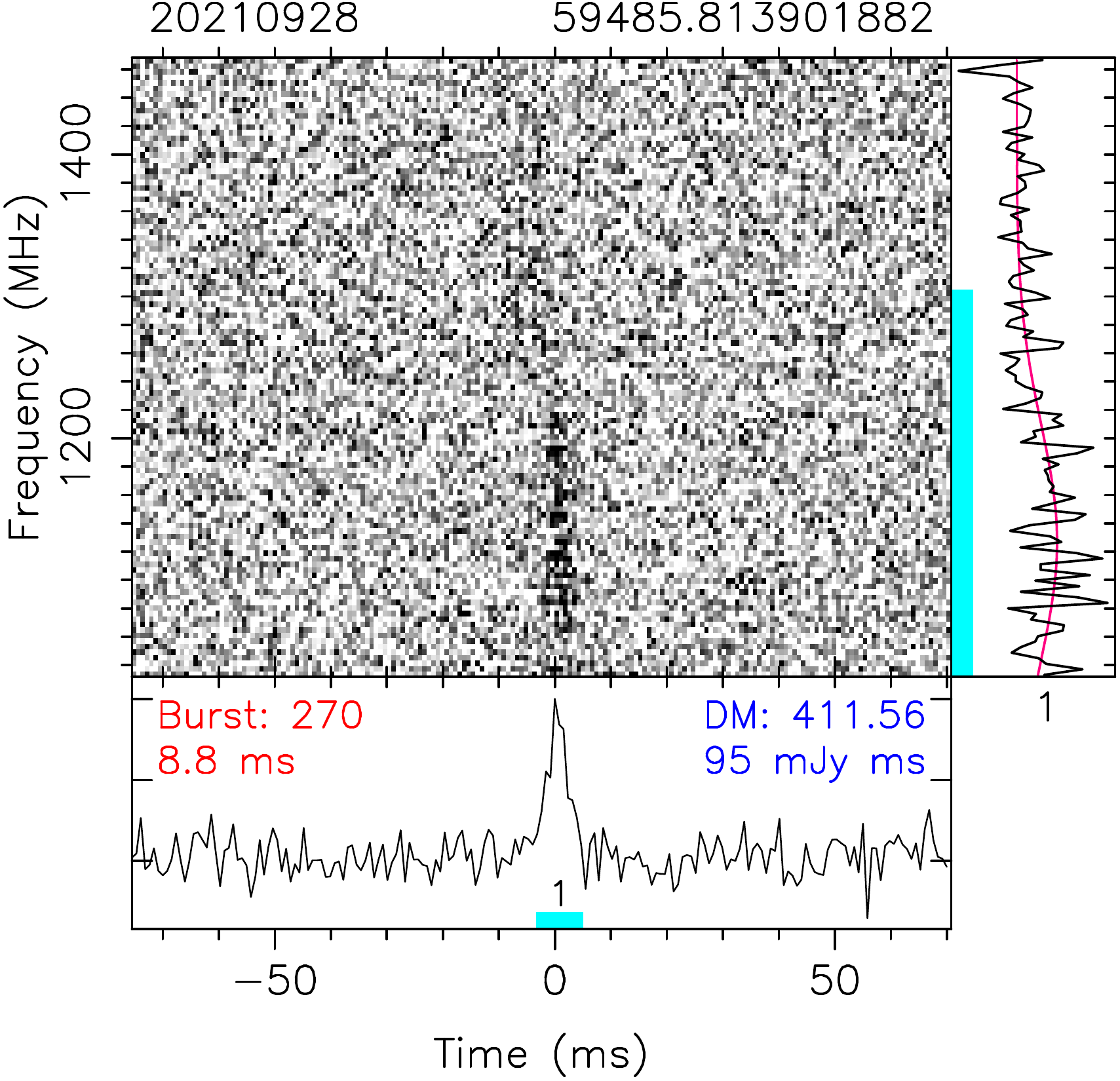}
\caption{The same as Figure~\ref{fig:appendix:ND} but for bursts with no drifting (ND).
}\label{fig:appendix:ND} 
\end{figure*}
\addtocounter{figure}{-1}
\begin{figure*}
    \flushleft
    \includegraphics[height=37mm]{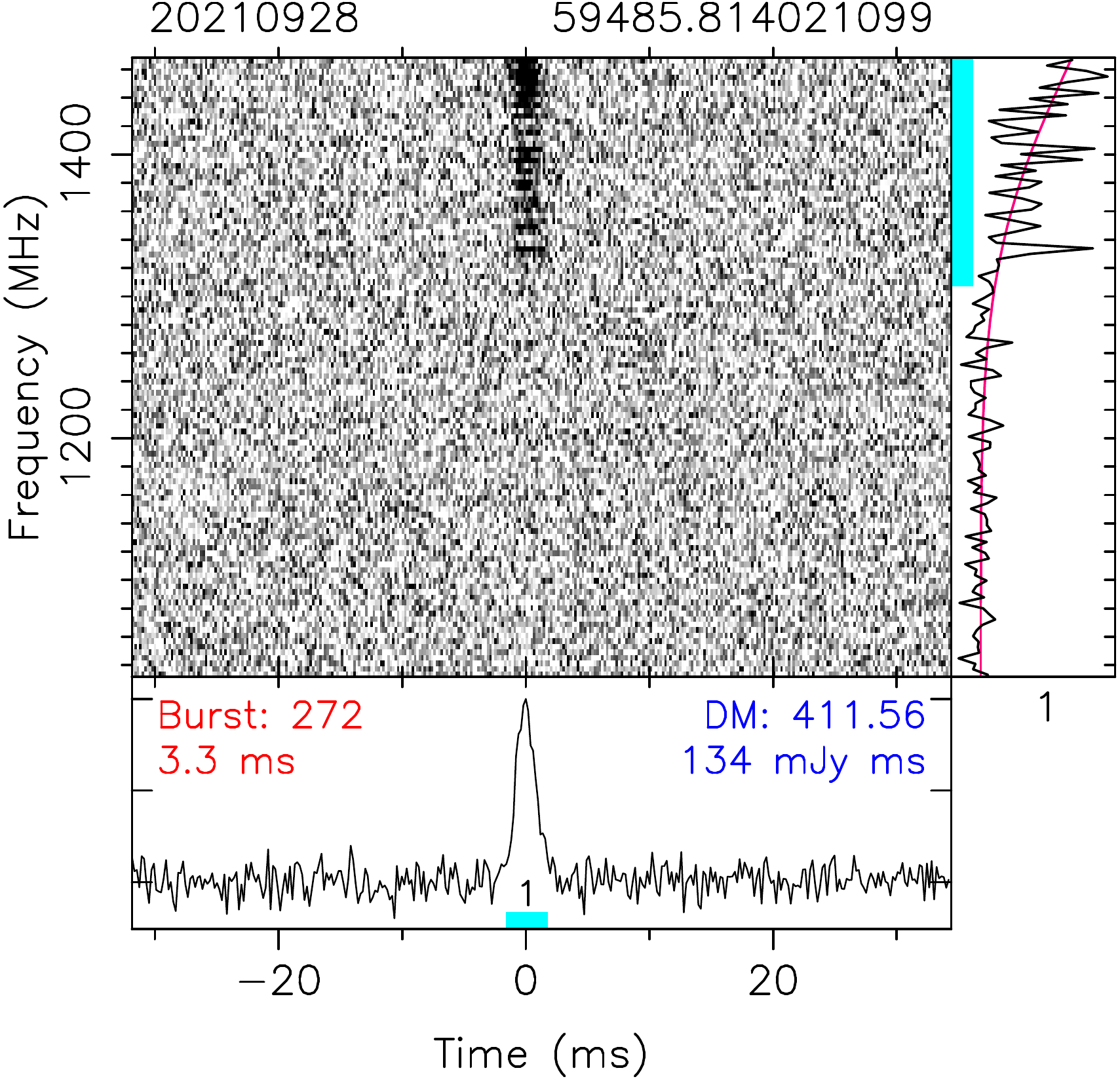}
    \includegraphics[height=37mm]{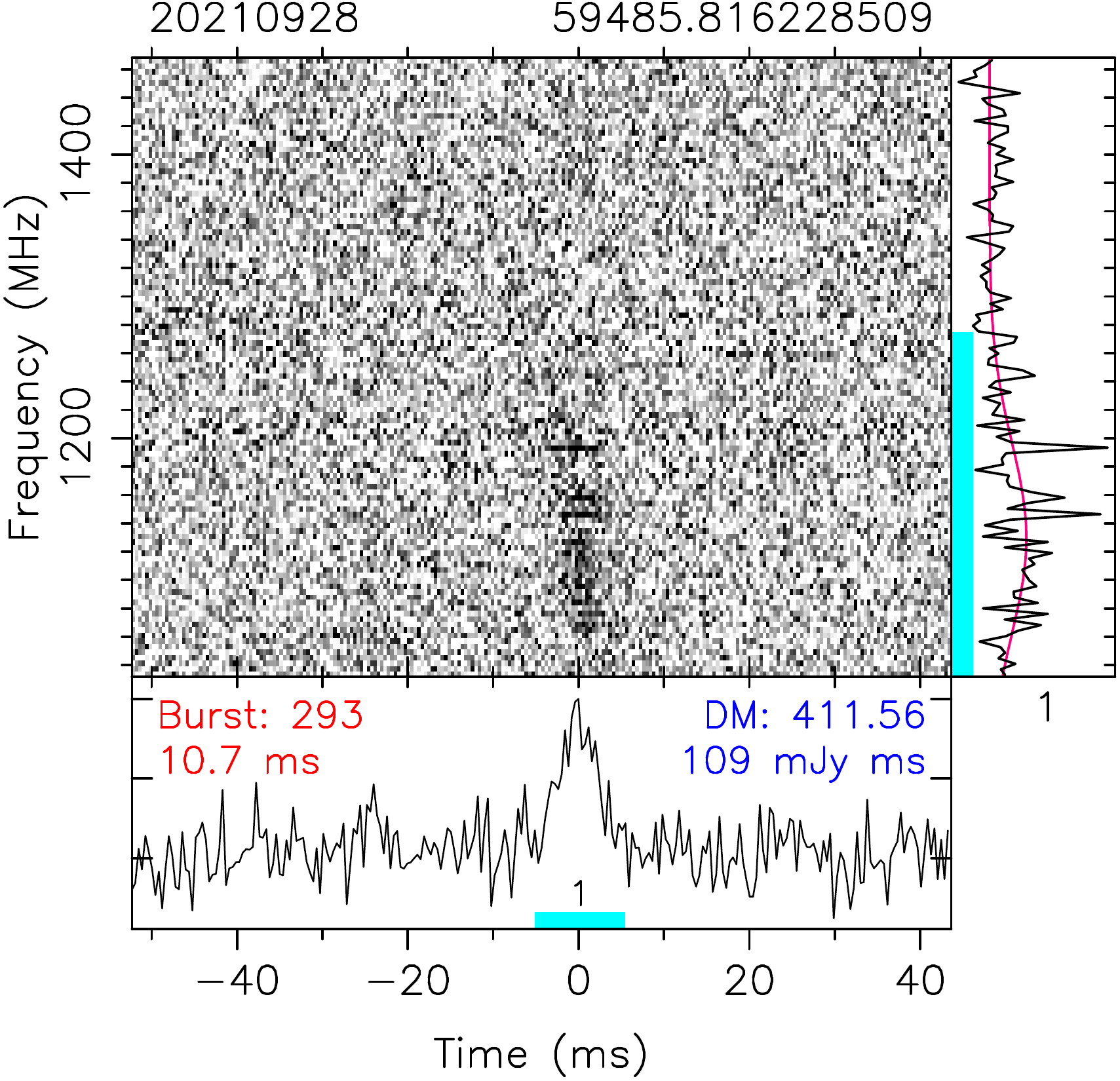}
    \includegraphics[height=37mm]{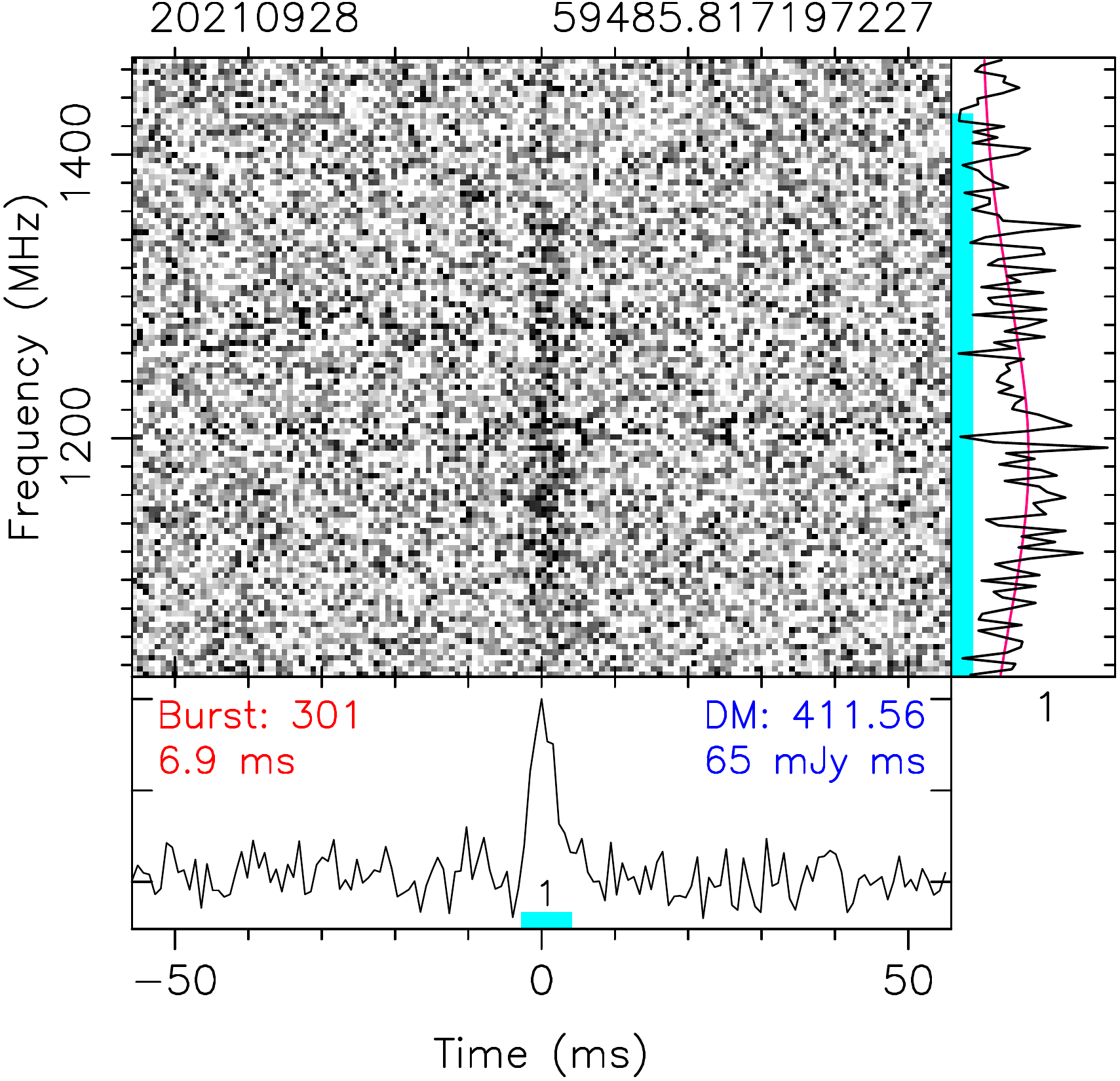}
    \includegraphics[height=37mm]{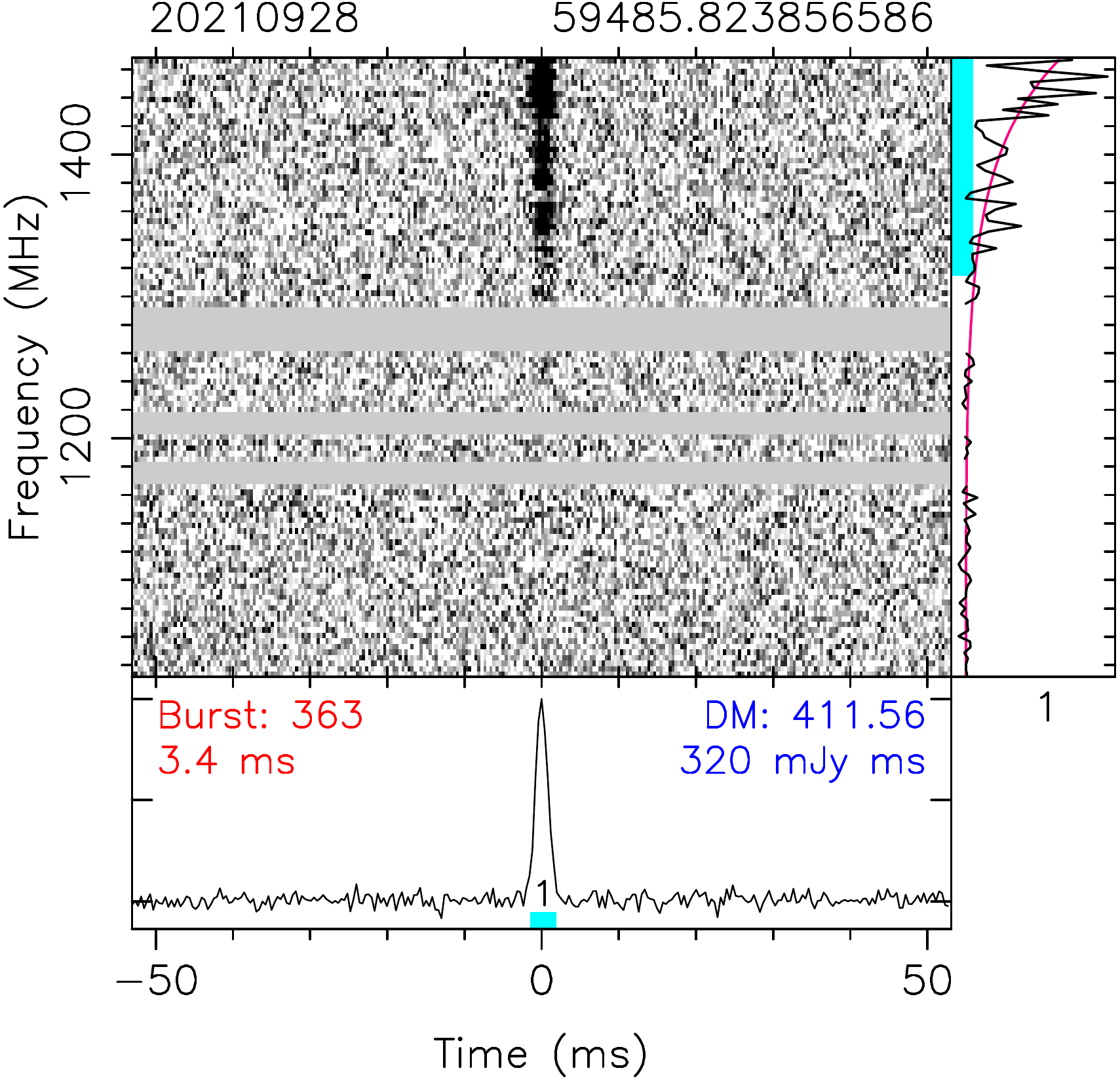}
    \caption{ \it{ -- continued and ended}.
}
\end{figure*}

\begin{figure*}
    \flushleft
    \includegraphics[height=37mm]{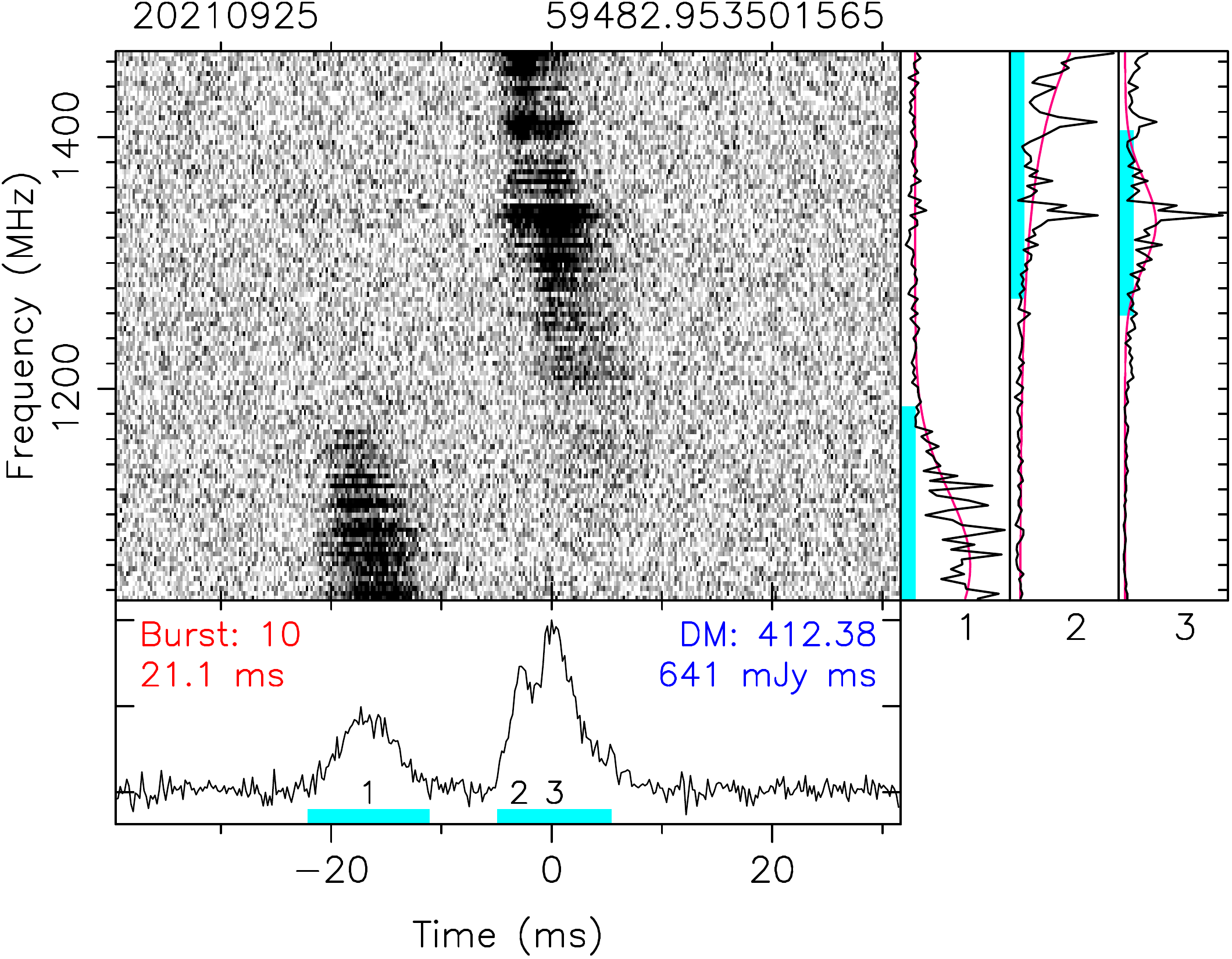}
    \includegraphics[height=37mm]{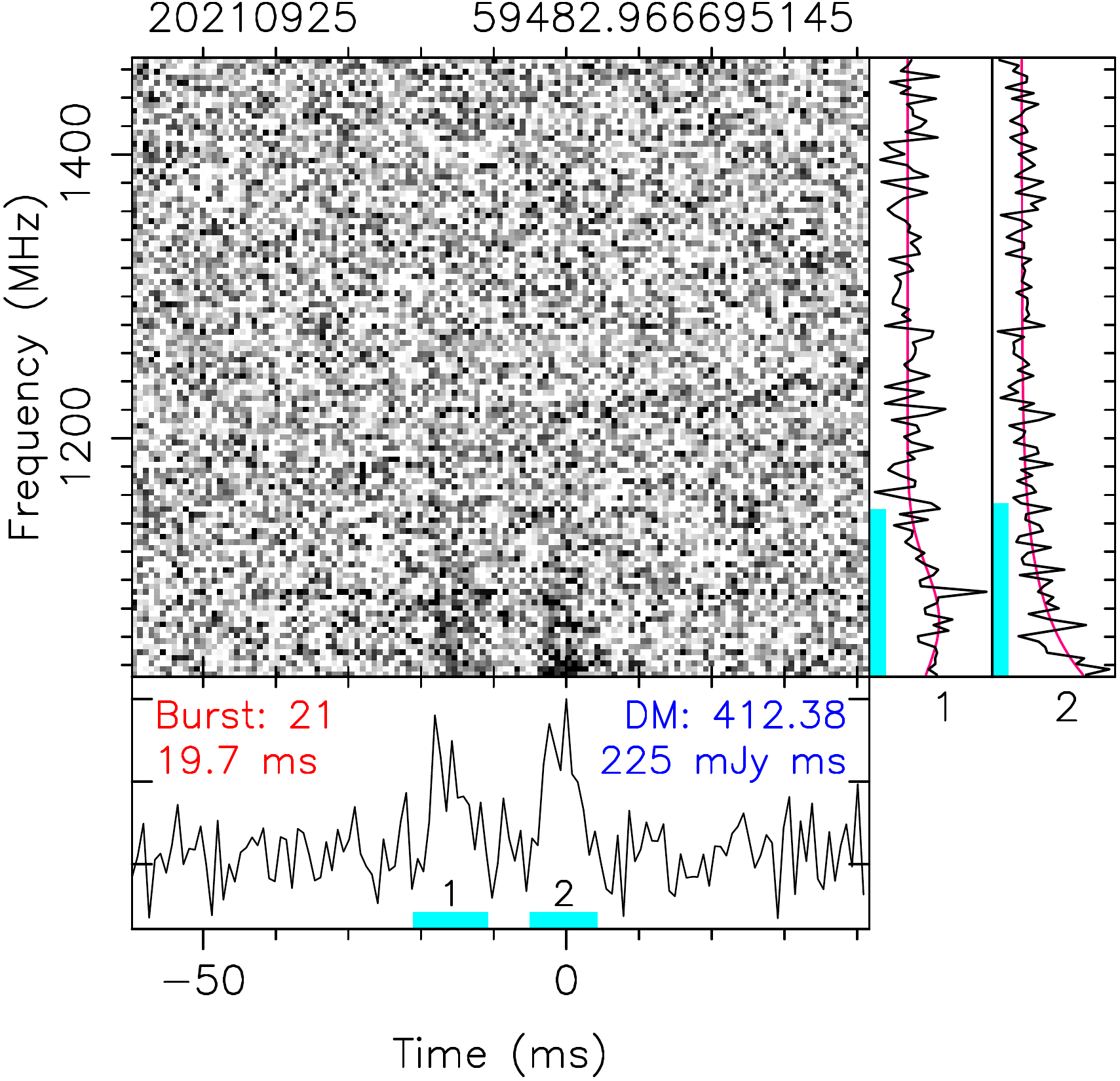}
    \includegraphics[height=37mm]{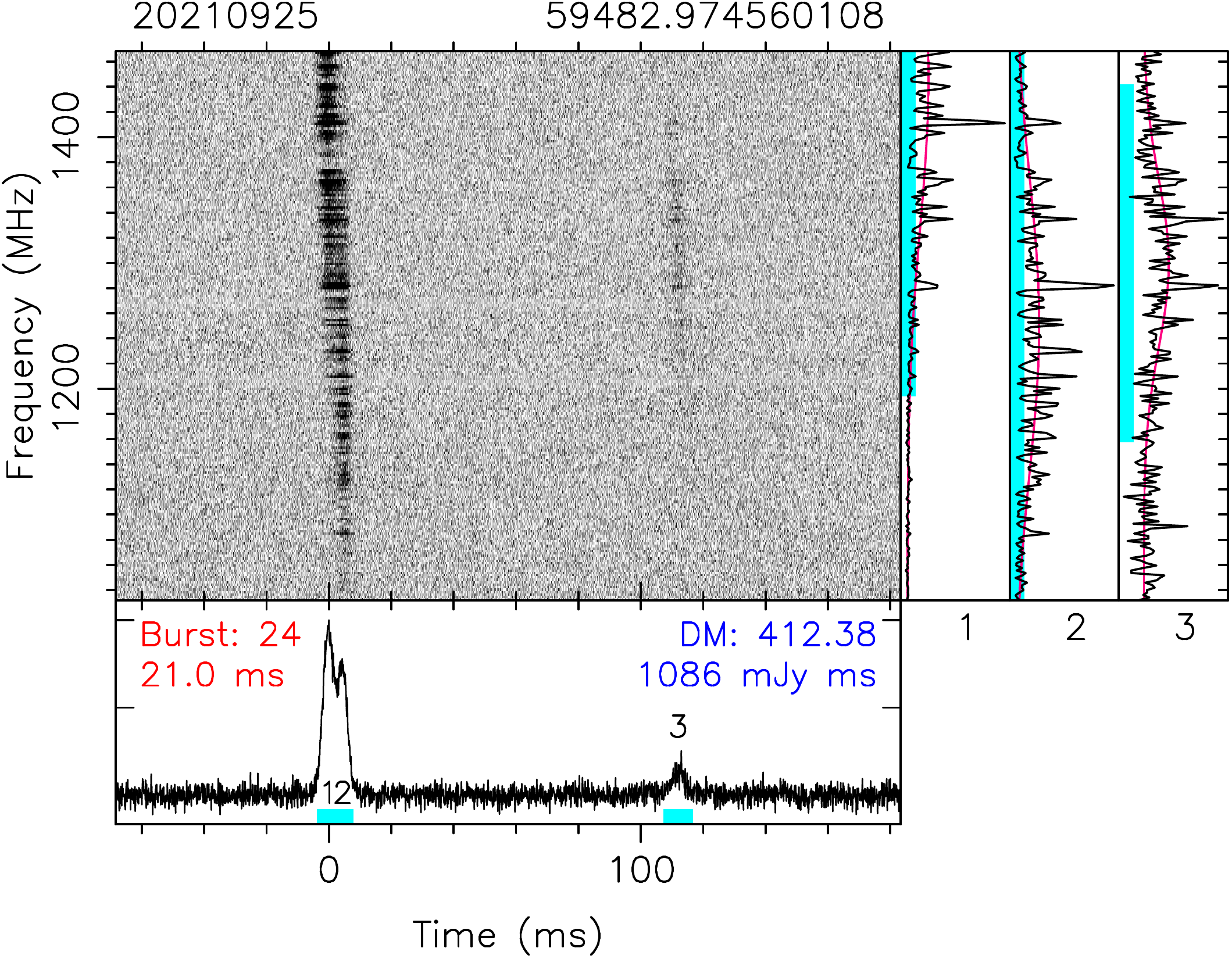} 
    \includegraphics[height=37mm]{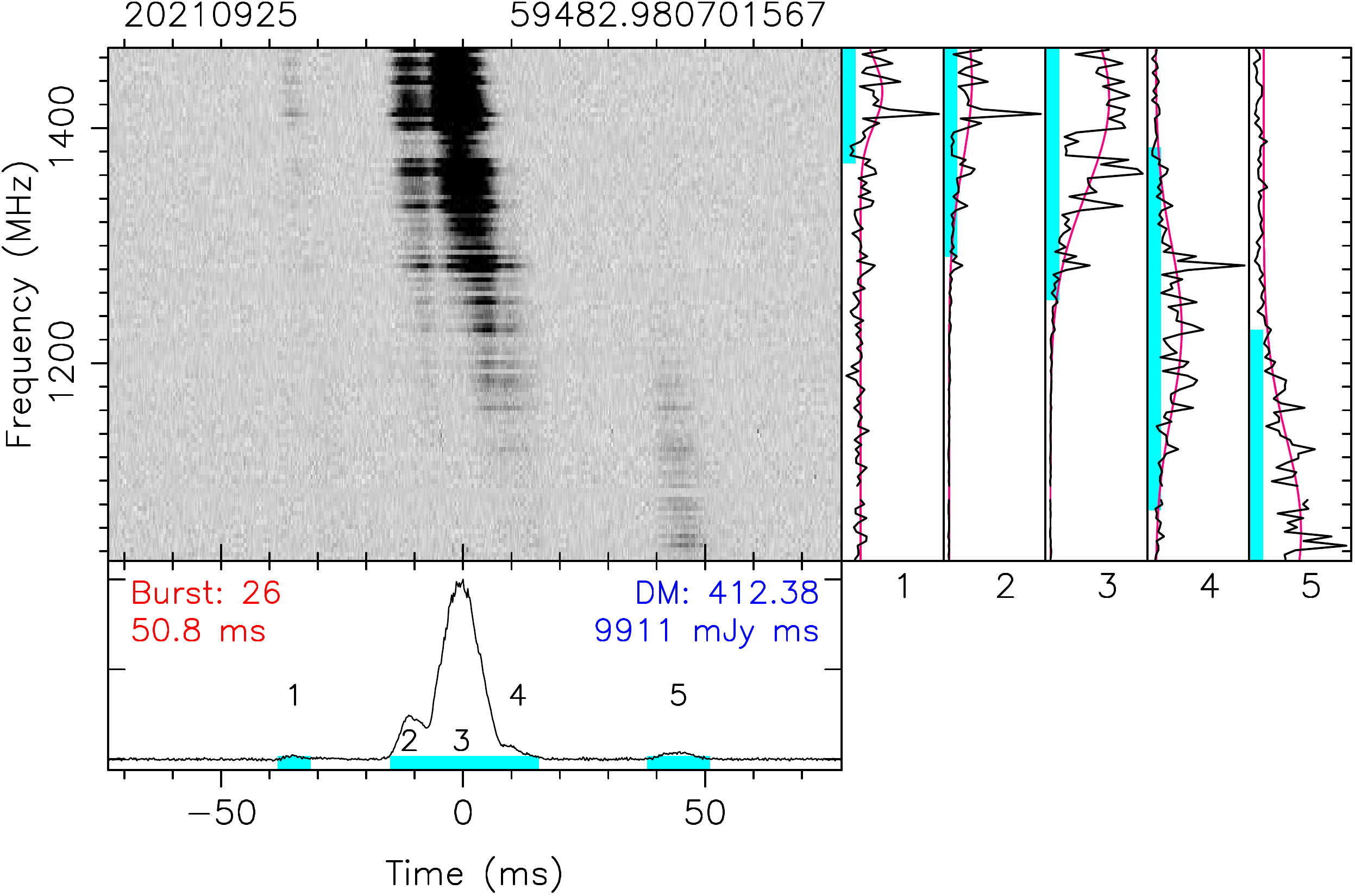}
    \includegraphics[height=37mm]{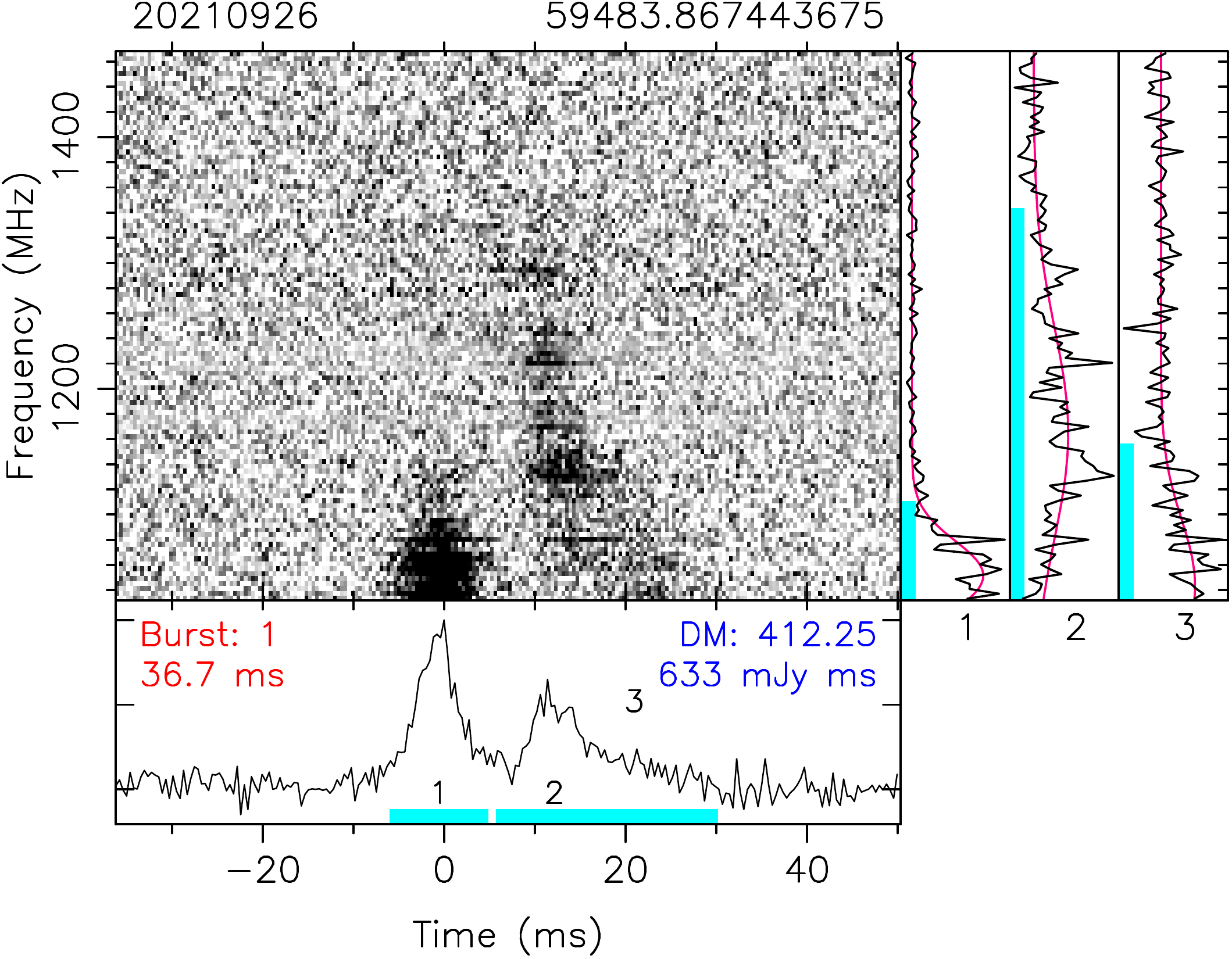}
    \includegraphics[height=37mm]{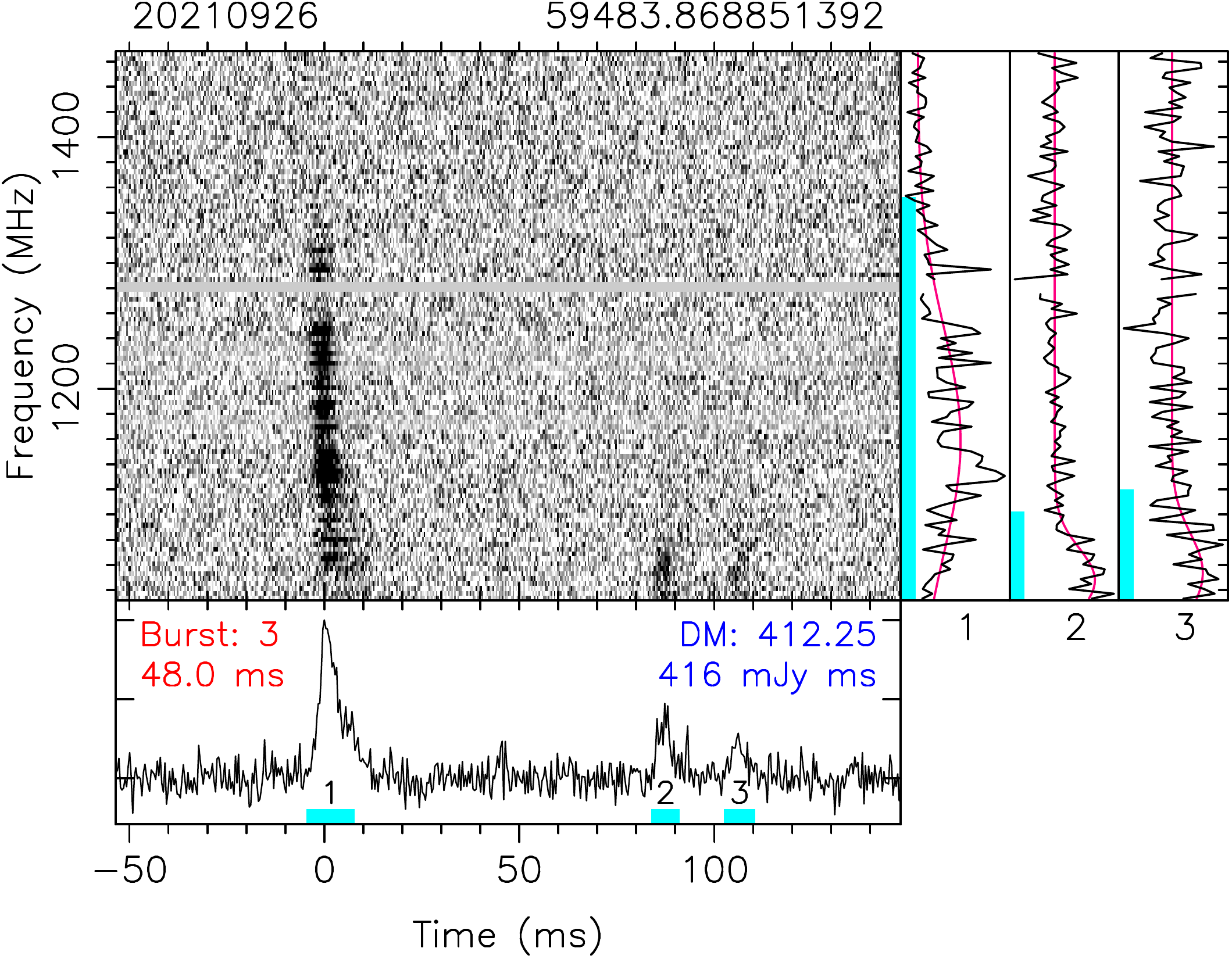}
    \includegraphics[height=37mm]{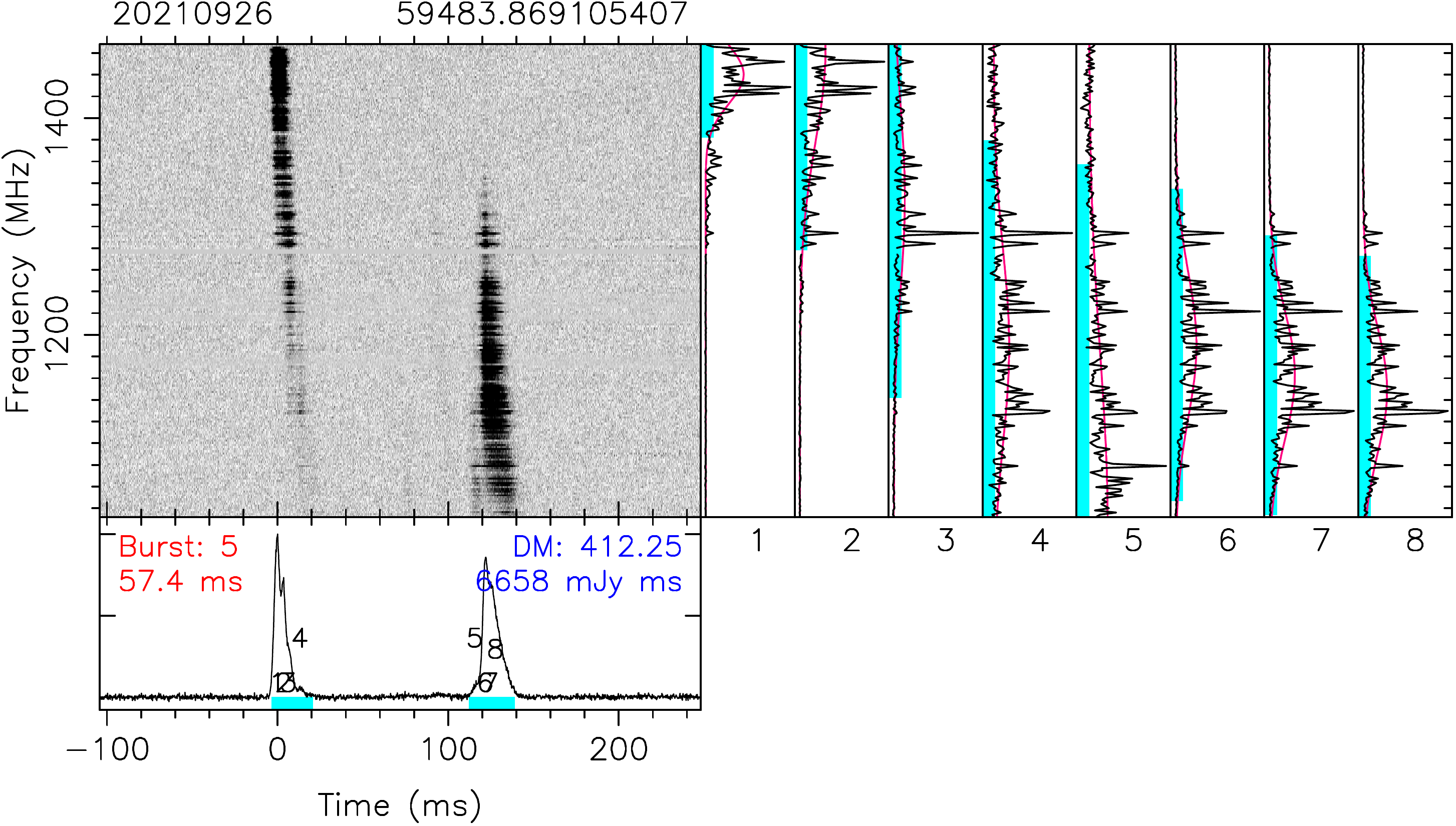}
    \includegraphics[height=37mm]{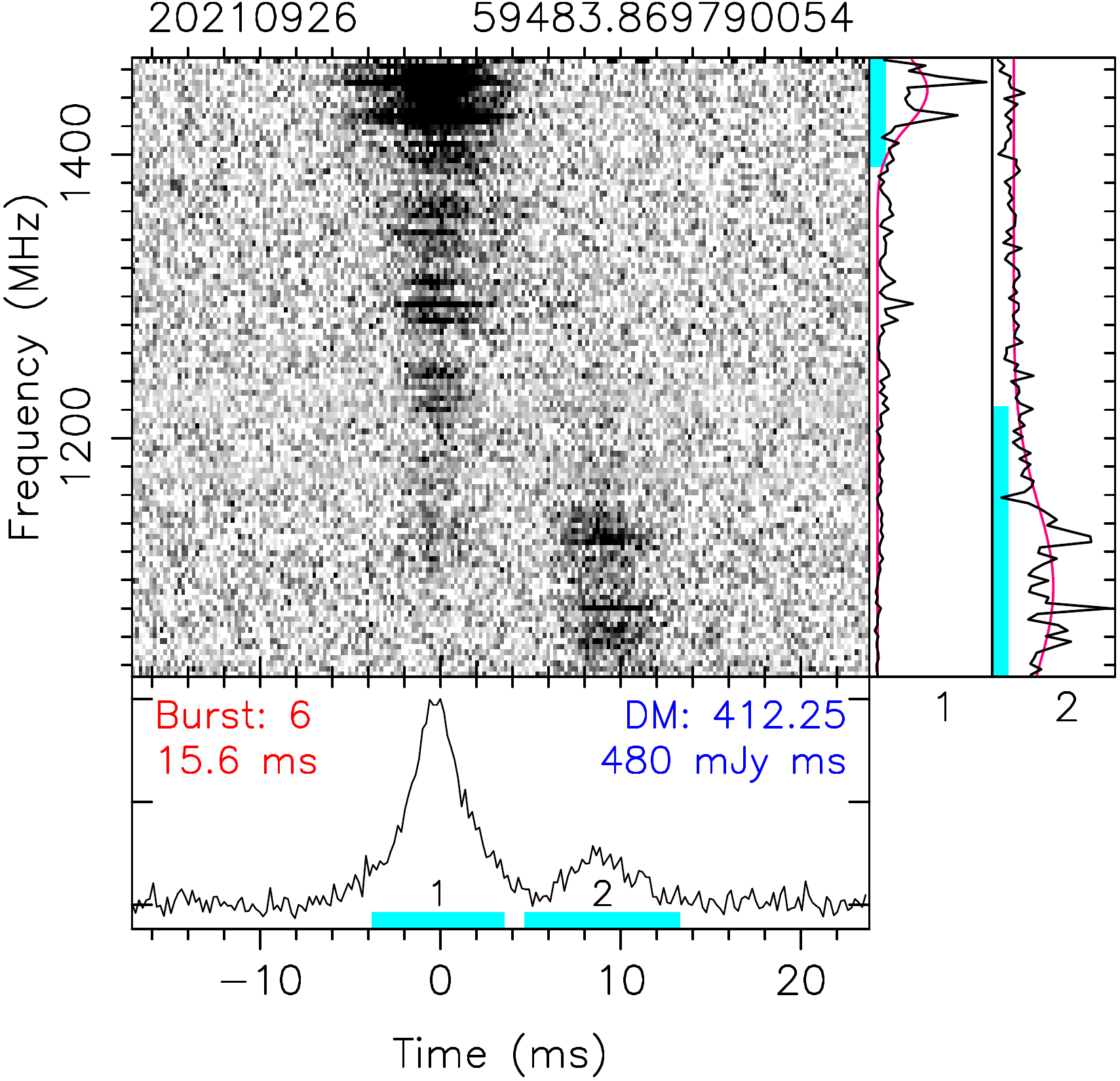}
    \includegraphics[height=37mm]{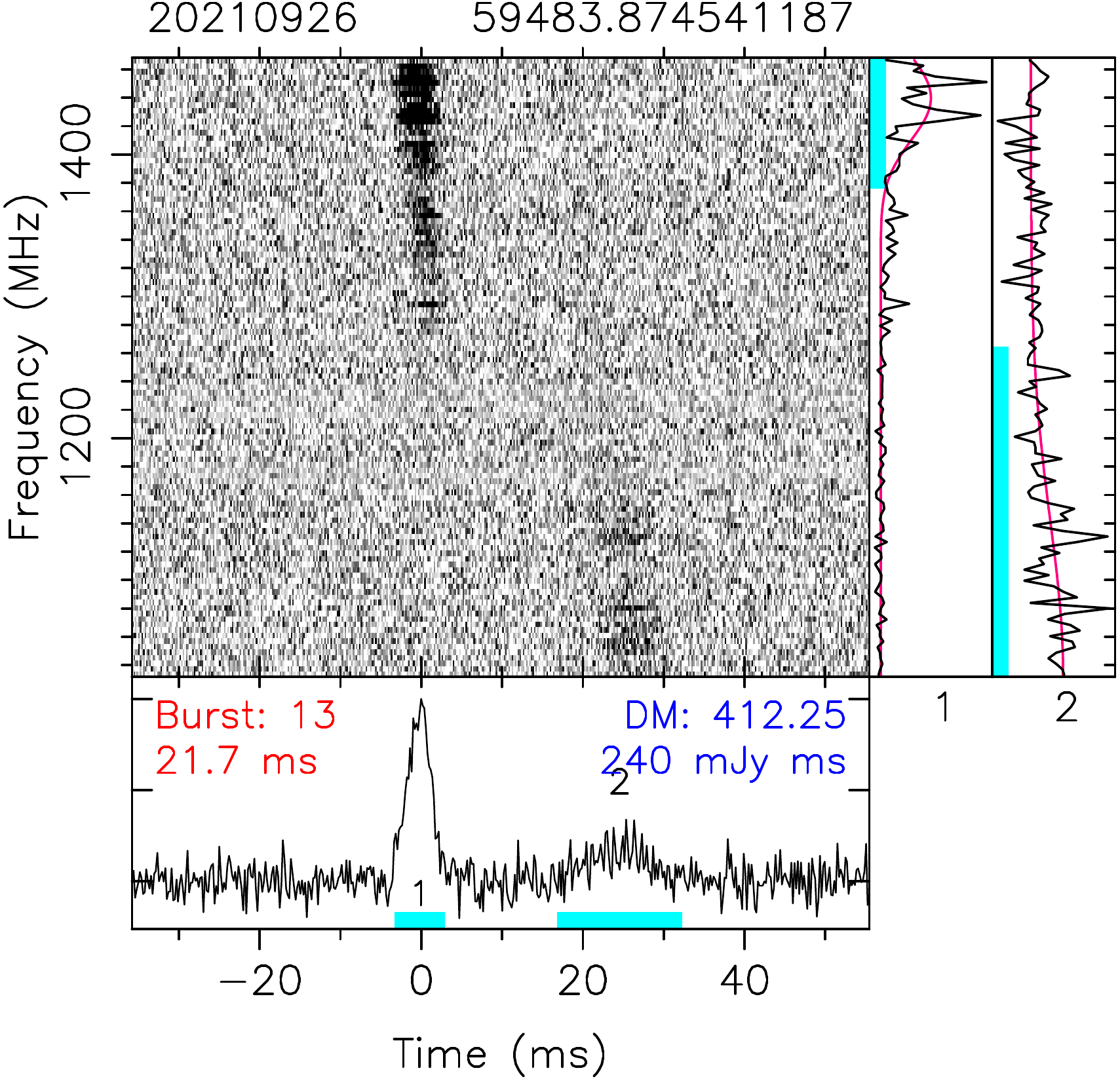}
    \includegraphics[height=37mm]{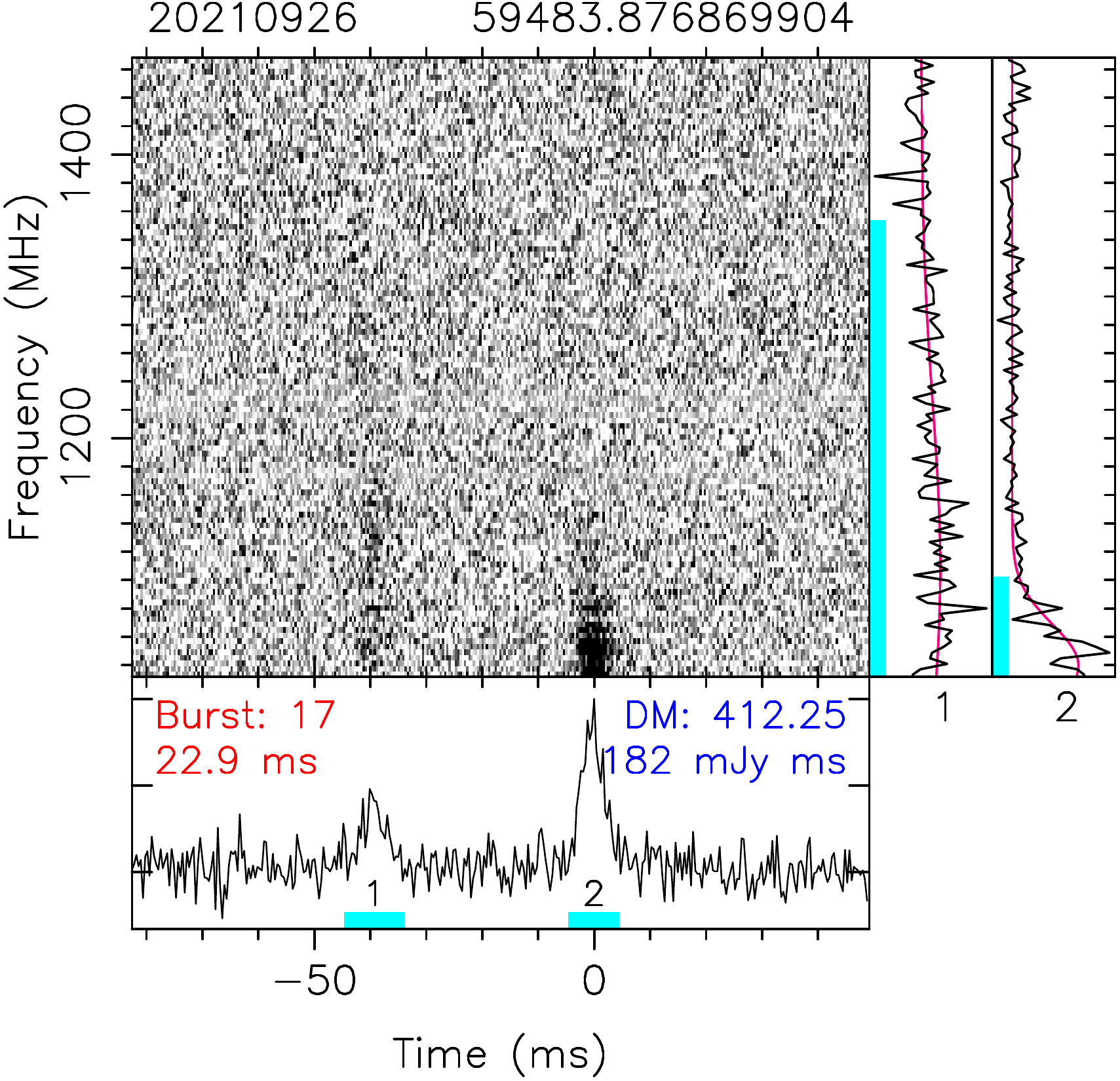}
    \includegraphics[height=37mm]{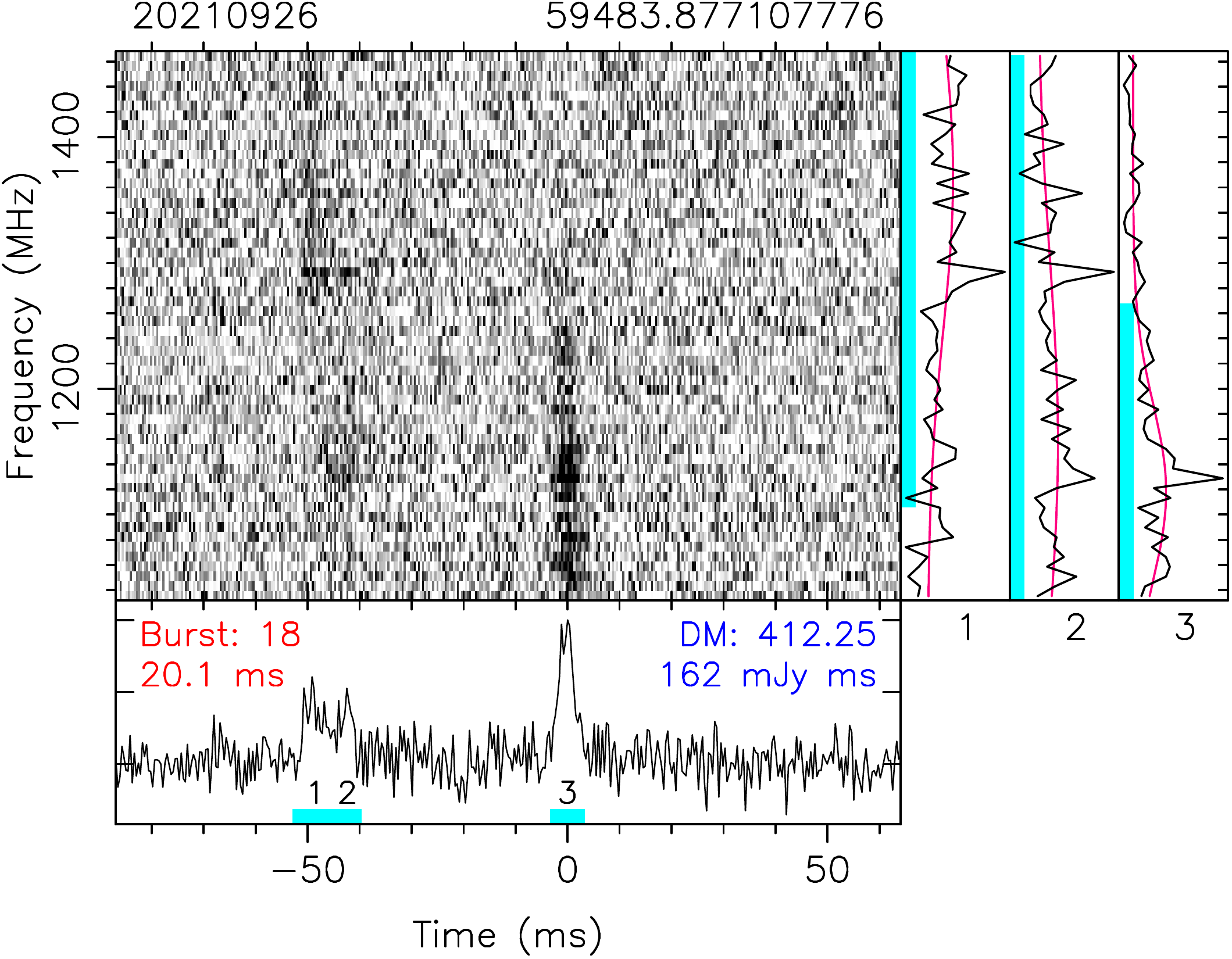}
    \includegraphics[height=37mm]{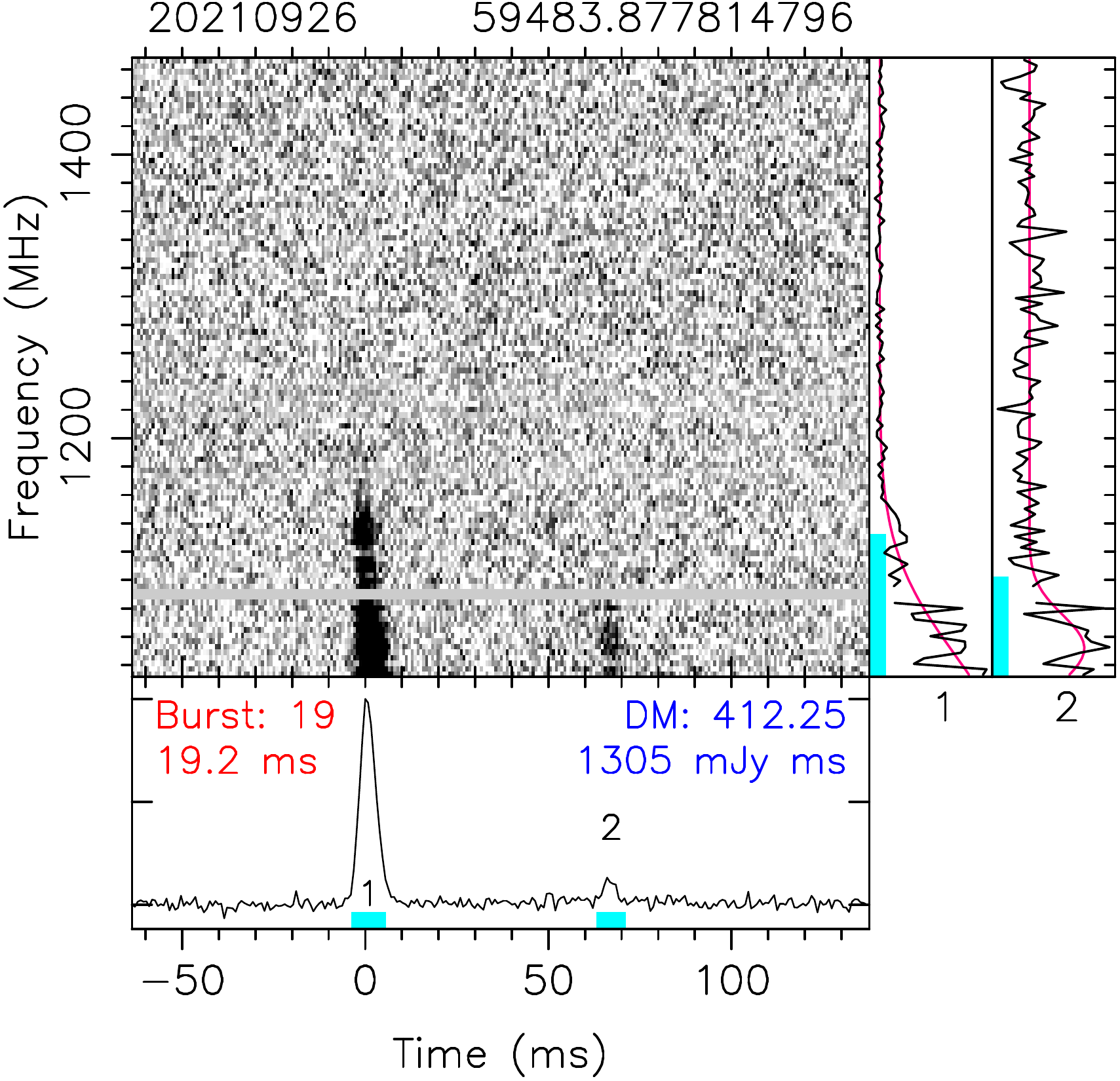}
    \includegraphics[height=37mm]{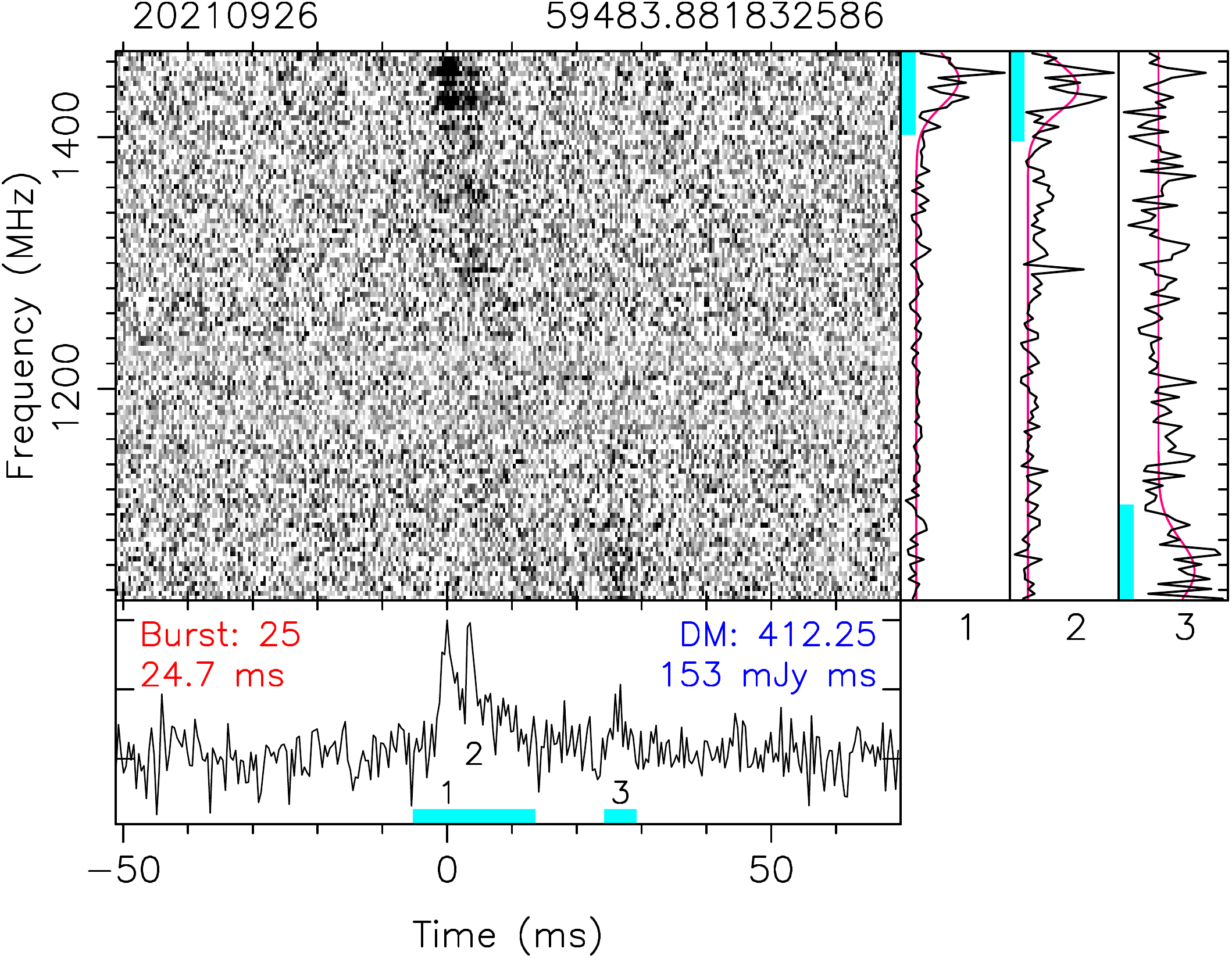}
    \includegraphics[height=37mm]{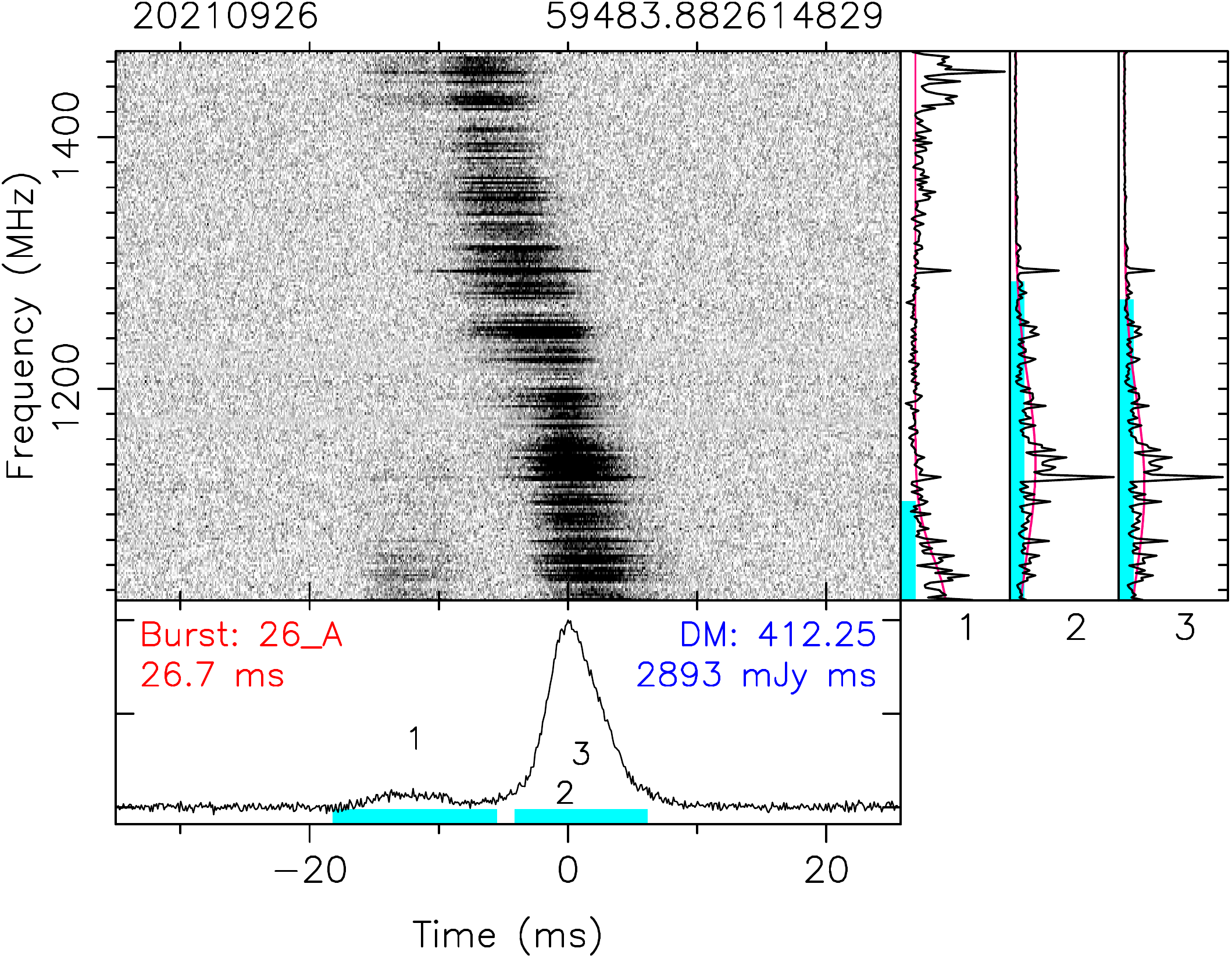}
    \includegraphics[height=37mm]{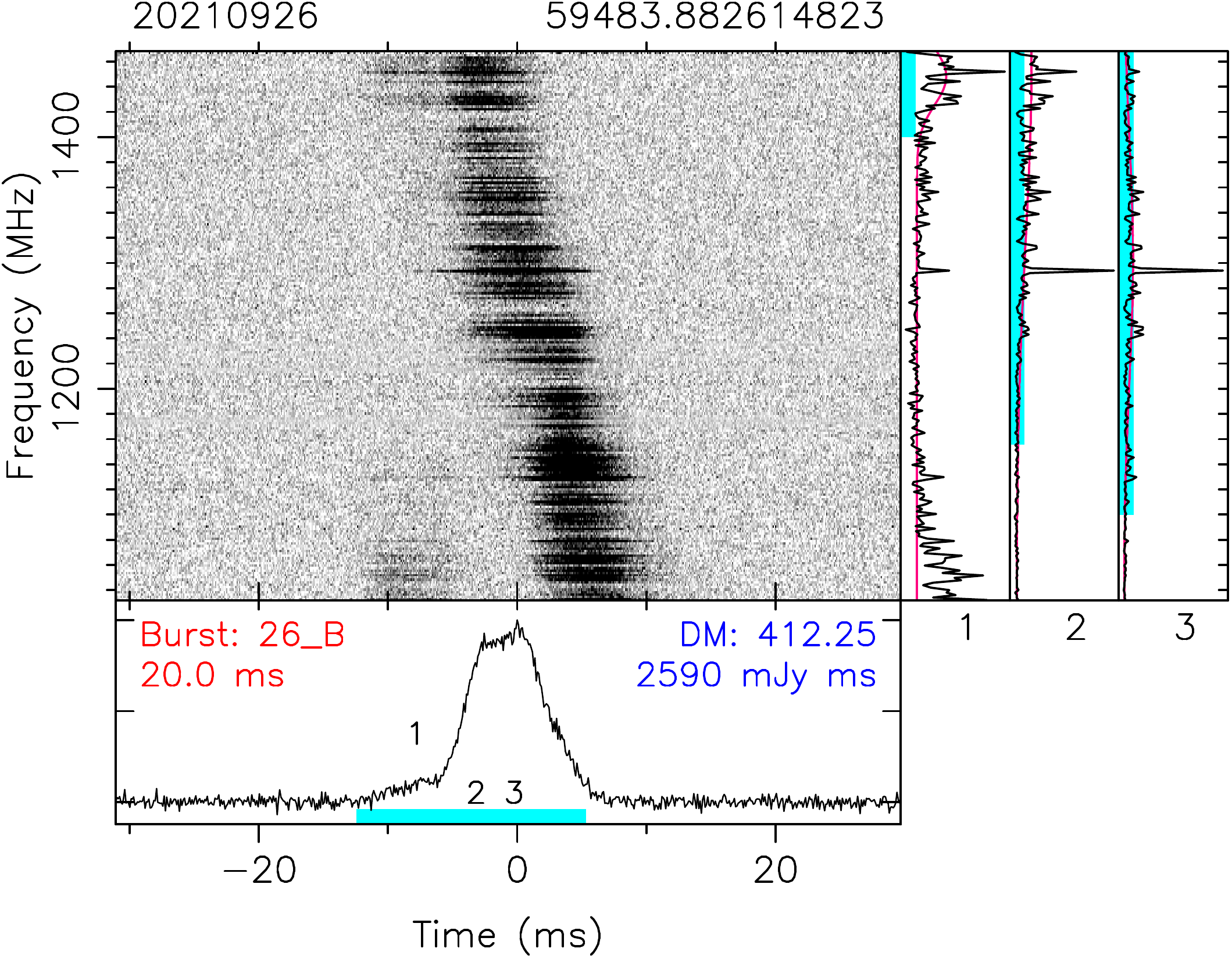}
    \includegraphics[height=37mm]{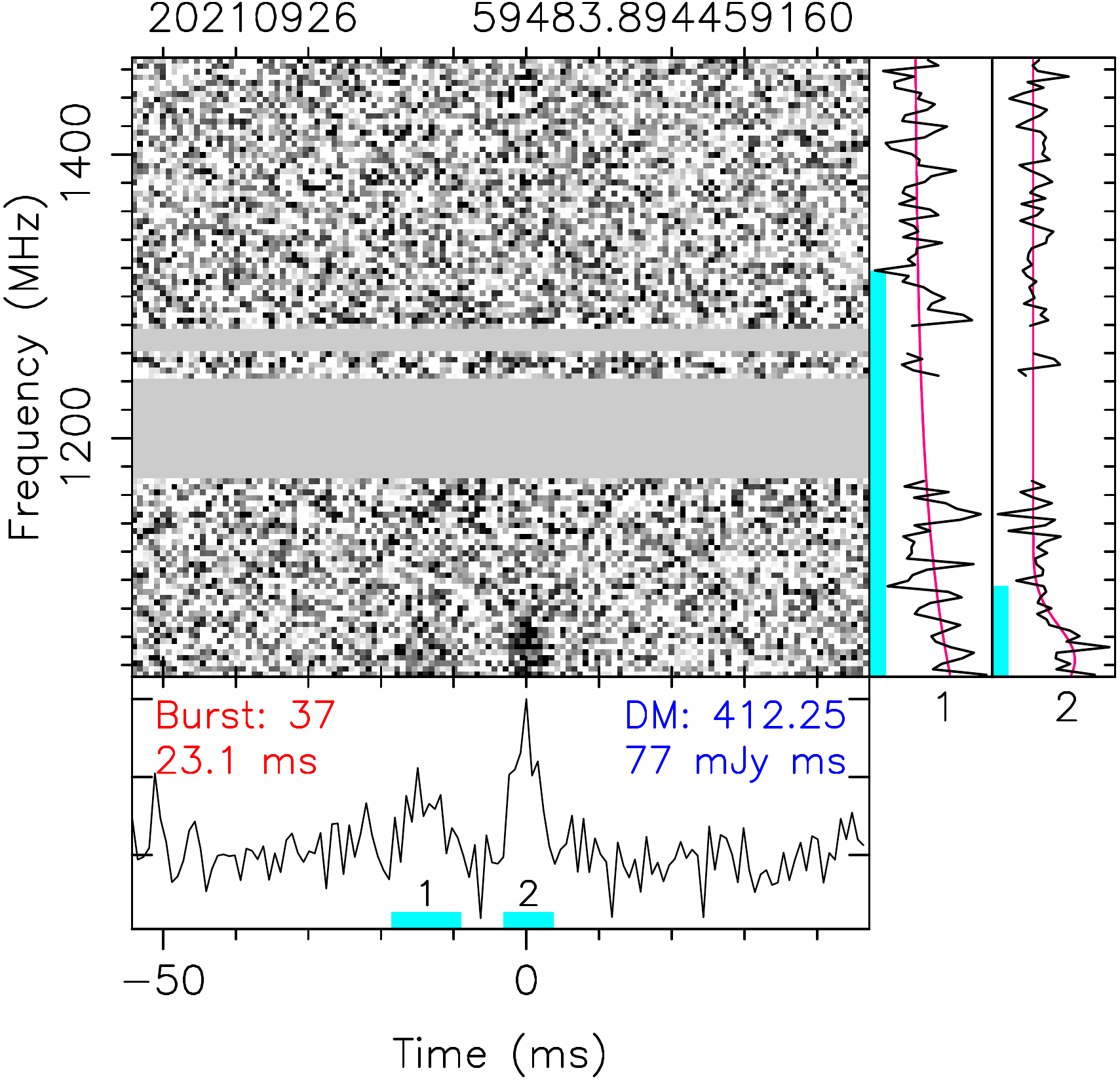}
    \includegraphics[height=37mm]{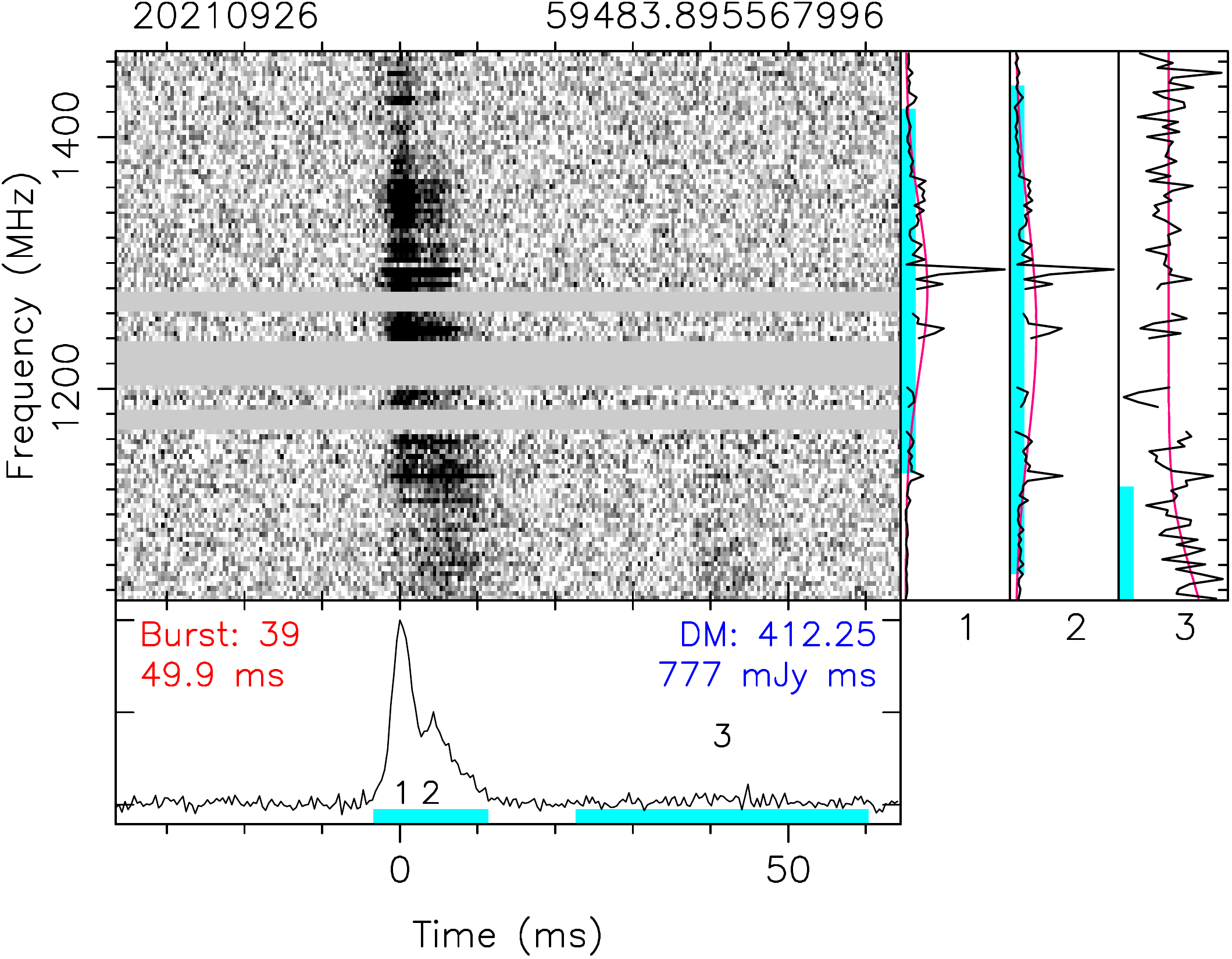}
    \includegraphics[height=37mm]{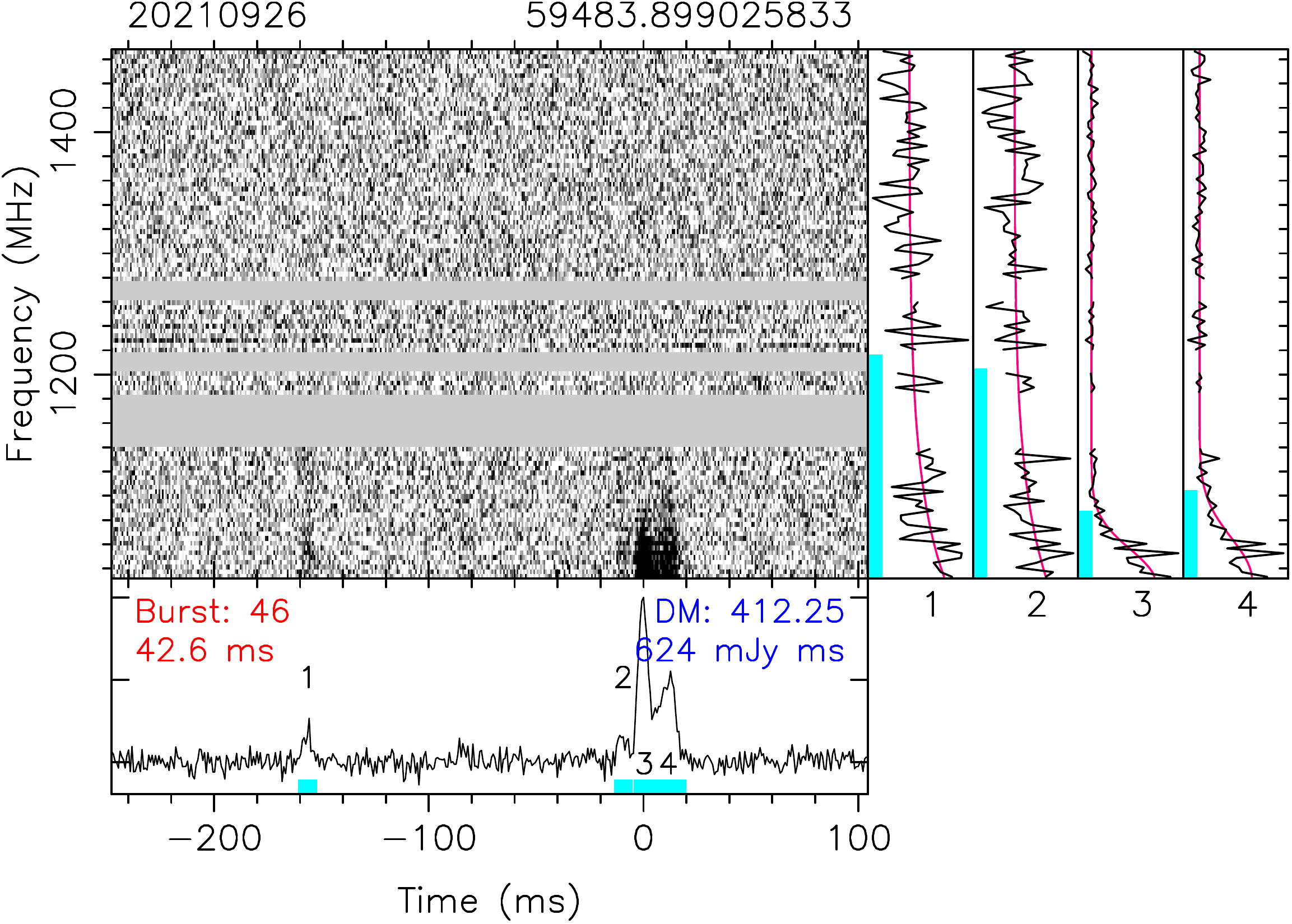}
\caption{The same as Figure~\ref{fig:appendix:D1W} but for complex bursts (C).
}\label{fig:appendix:C} 
\end{figure*}
\addtocounter{figure}{-1}
\begin{figure*}
    \flushleft
    \includegraphics[height=37mm]{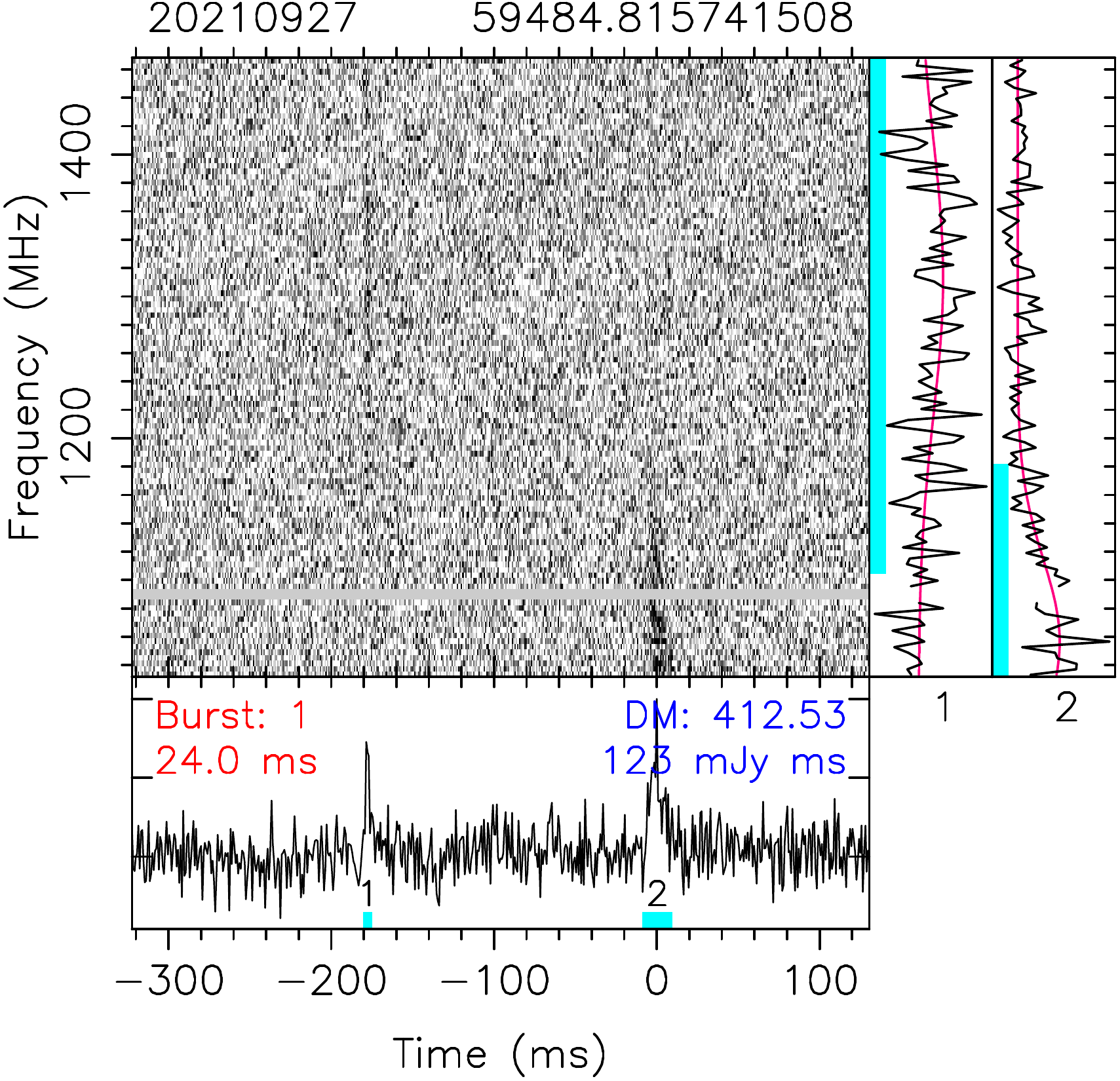} 
    \includegraphics[height=37mm]{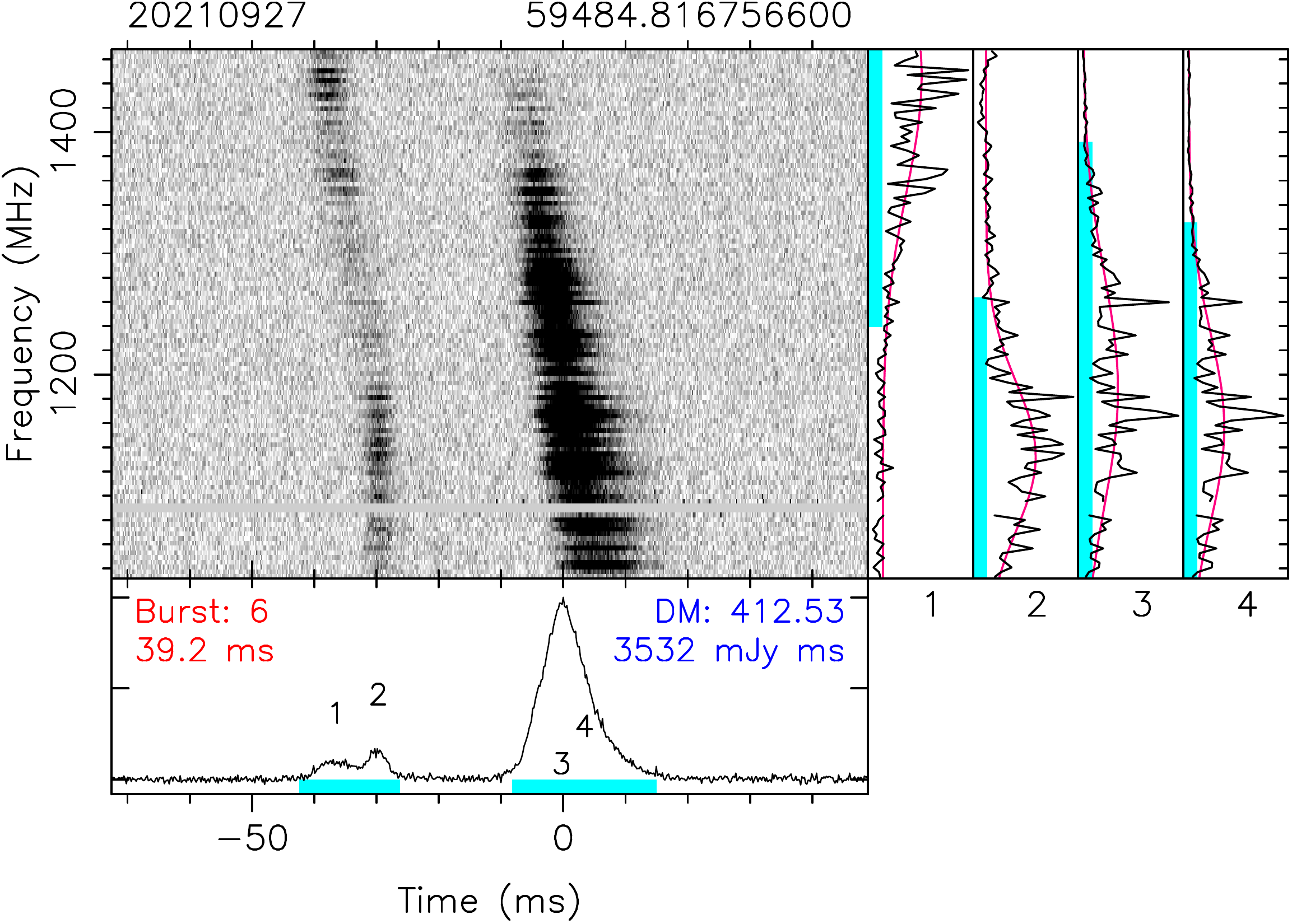}
    \includegraphics[height=37mm]{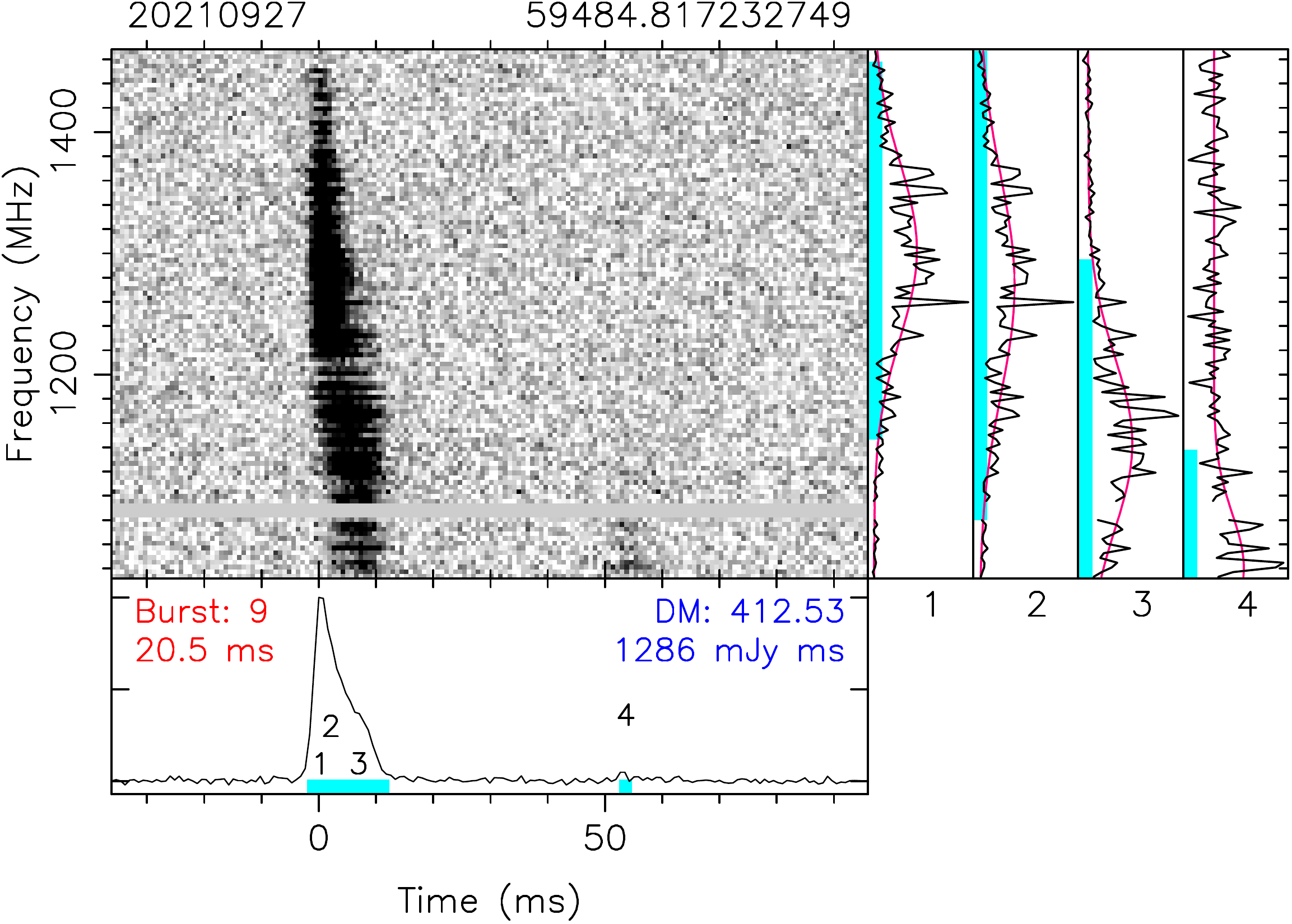} 
    \includegraphics[height=37mm]{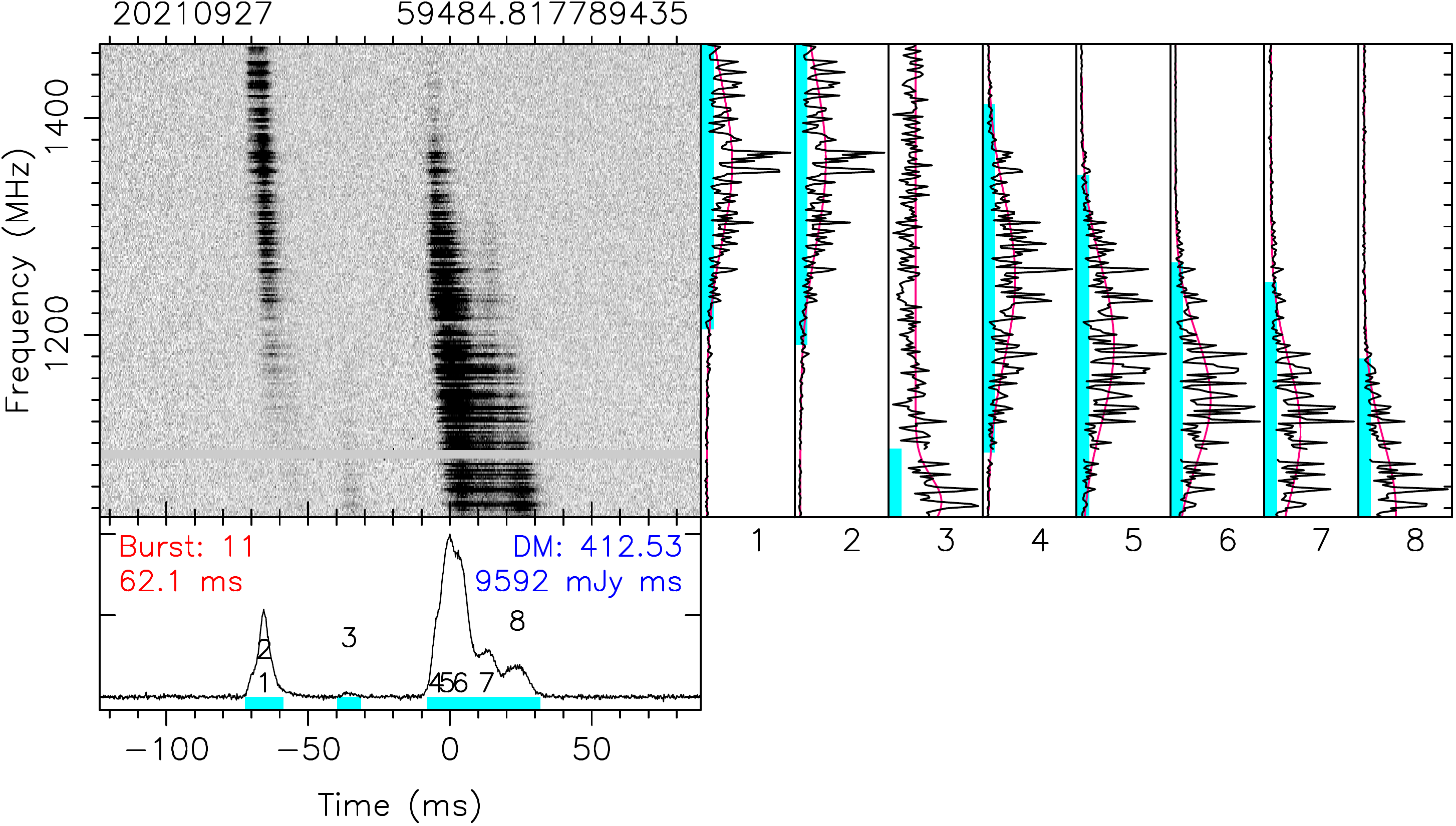} 
    \includegraphics[height=37mm]{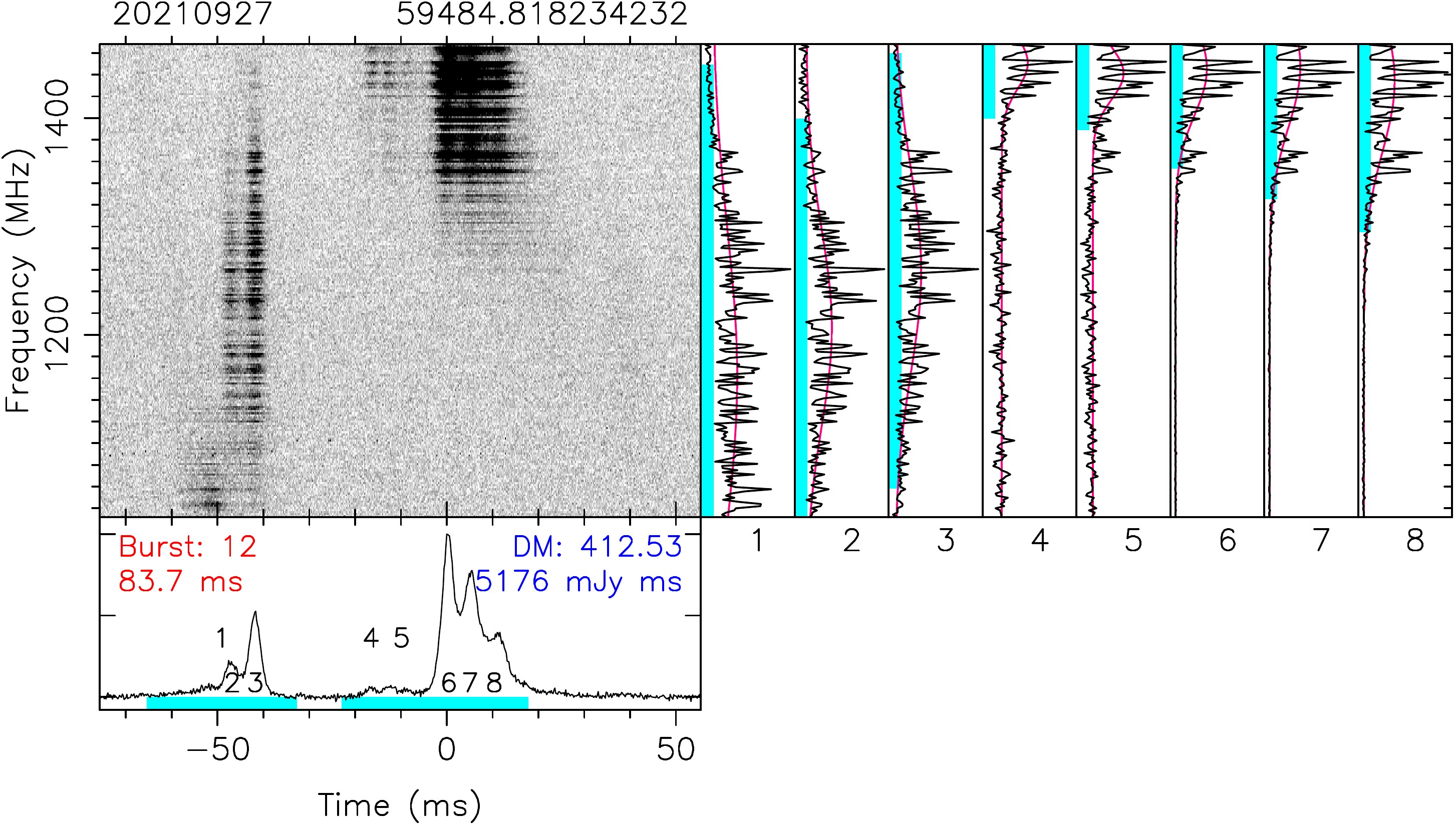} 
    \includegraphics[height=37mm]{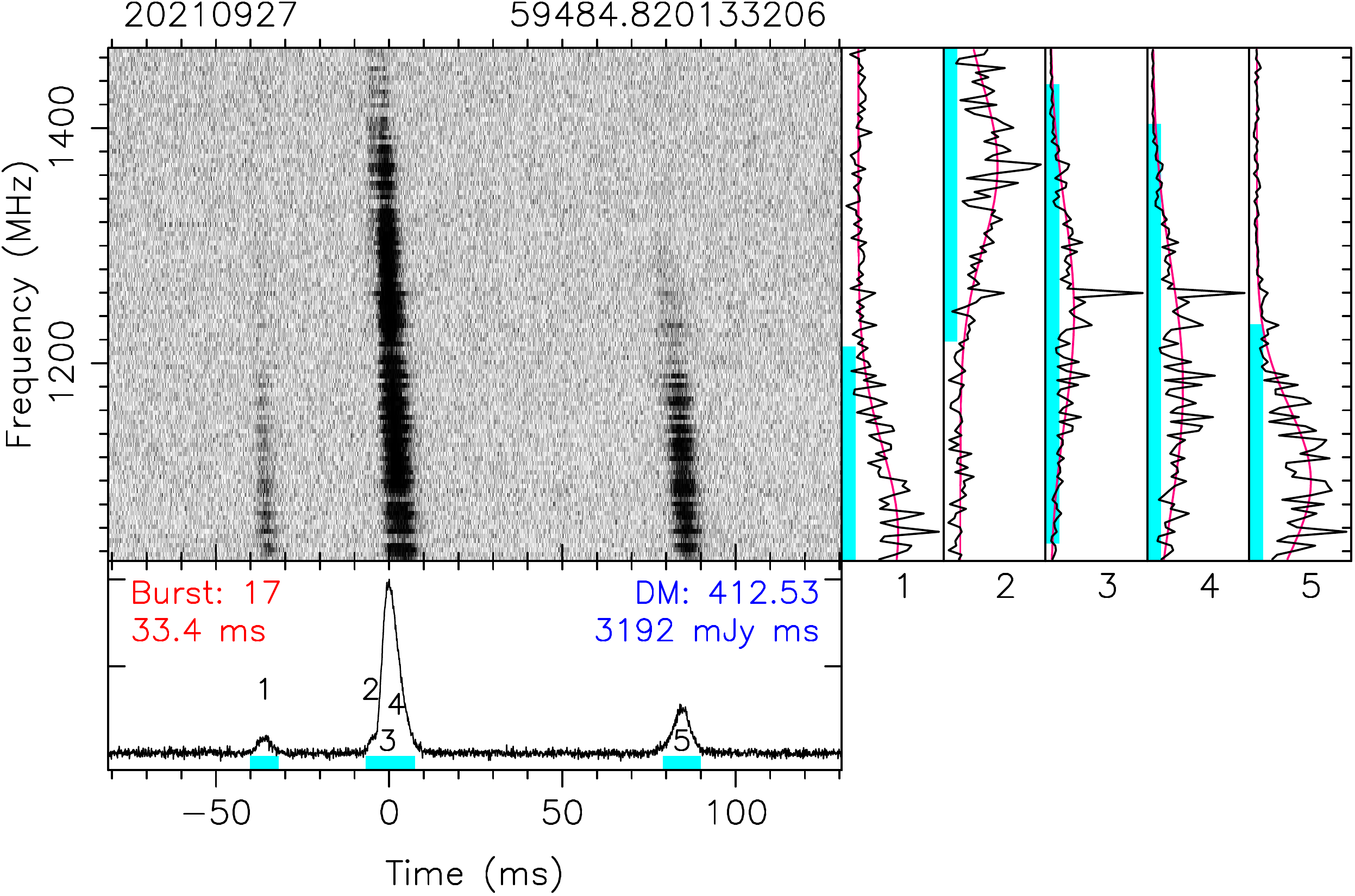} 
    \includegraphics[height=37mm]{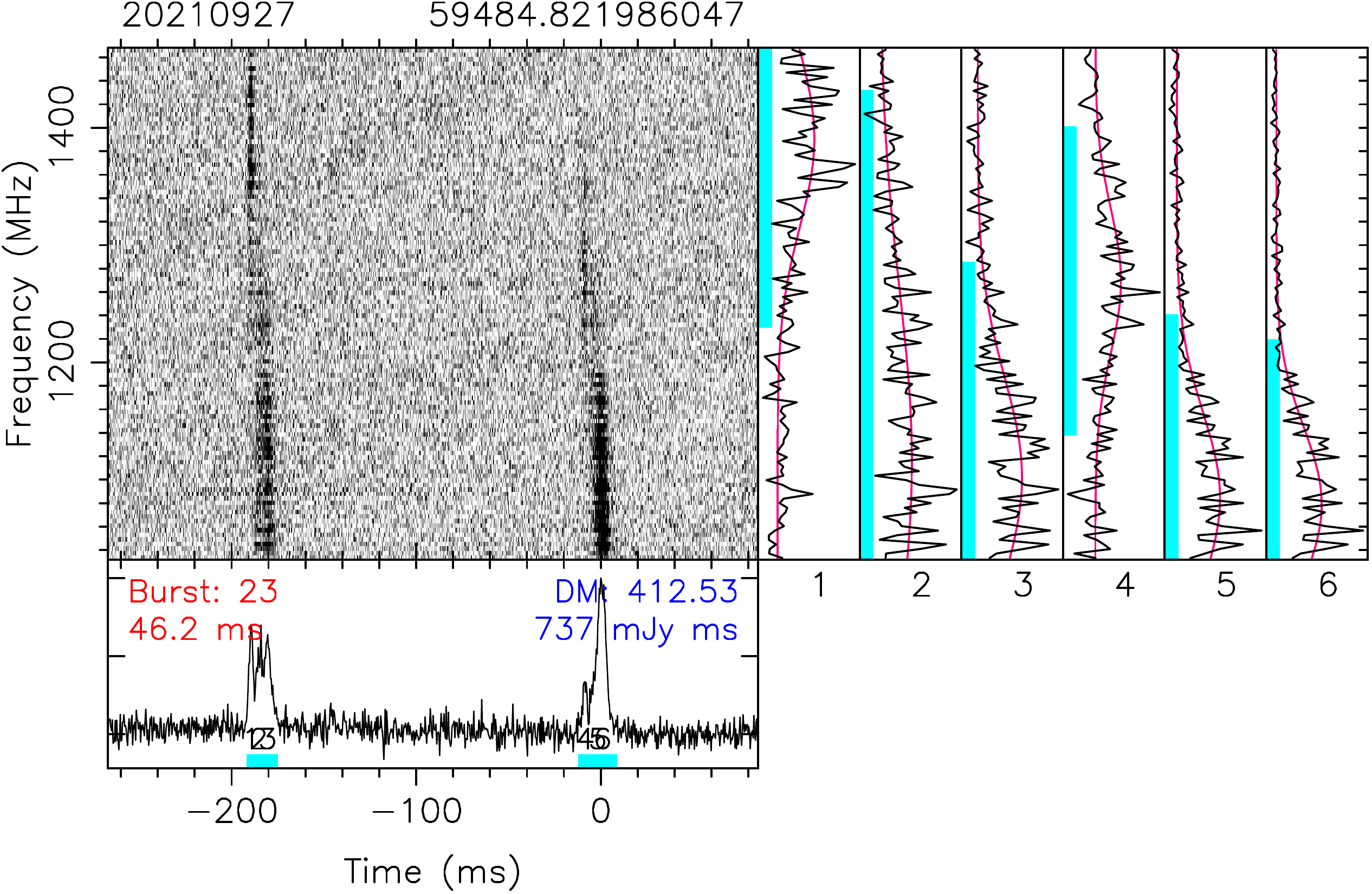} 
    \includegraphics[height=37mm]{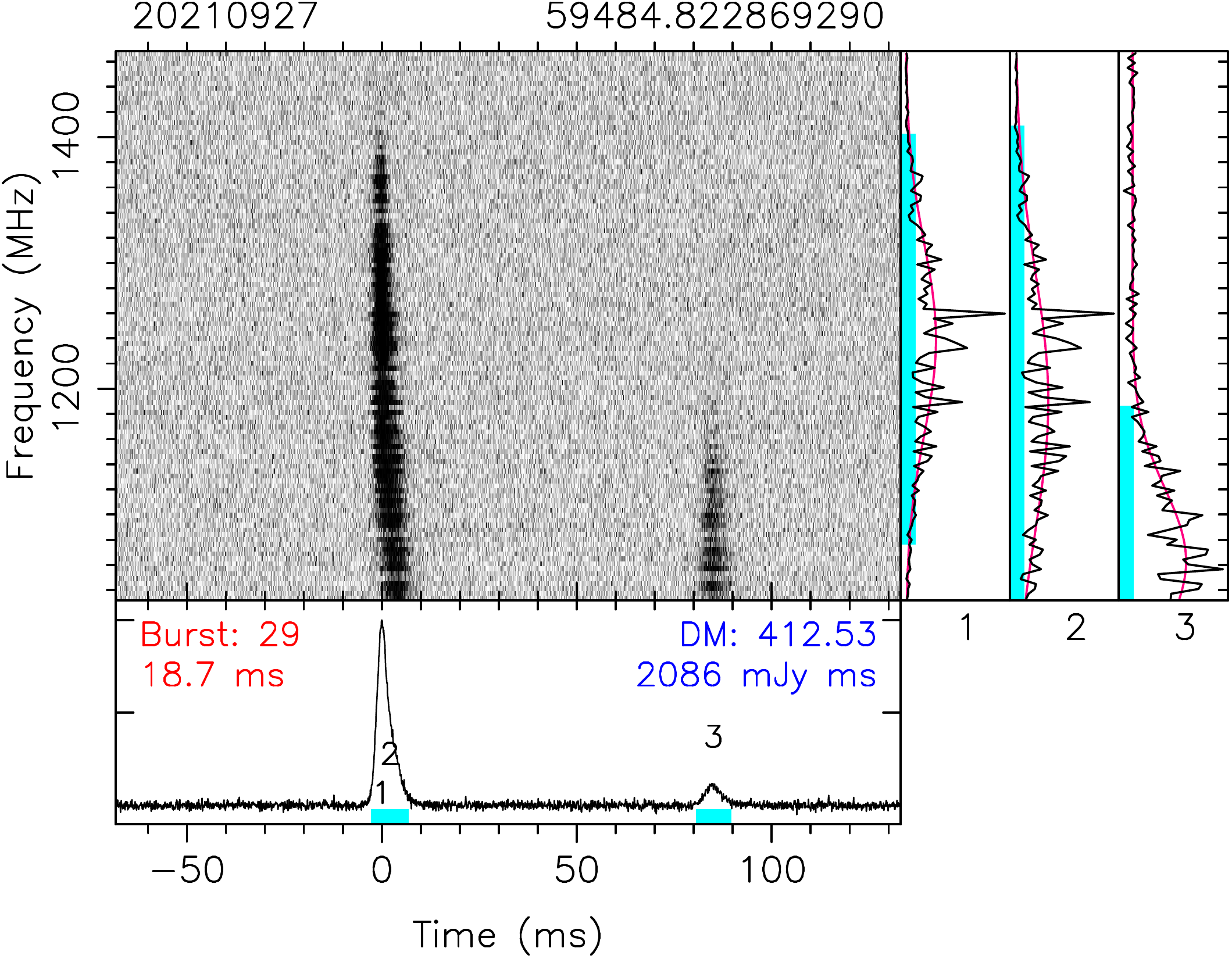} 
    \includegraphics[height=37mm]{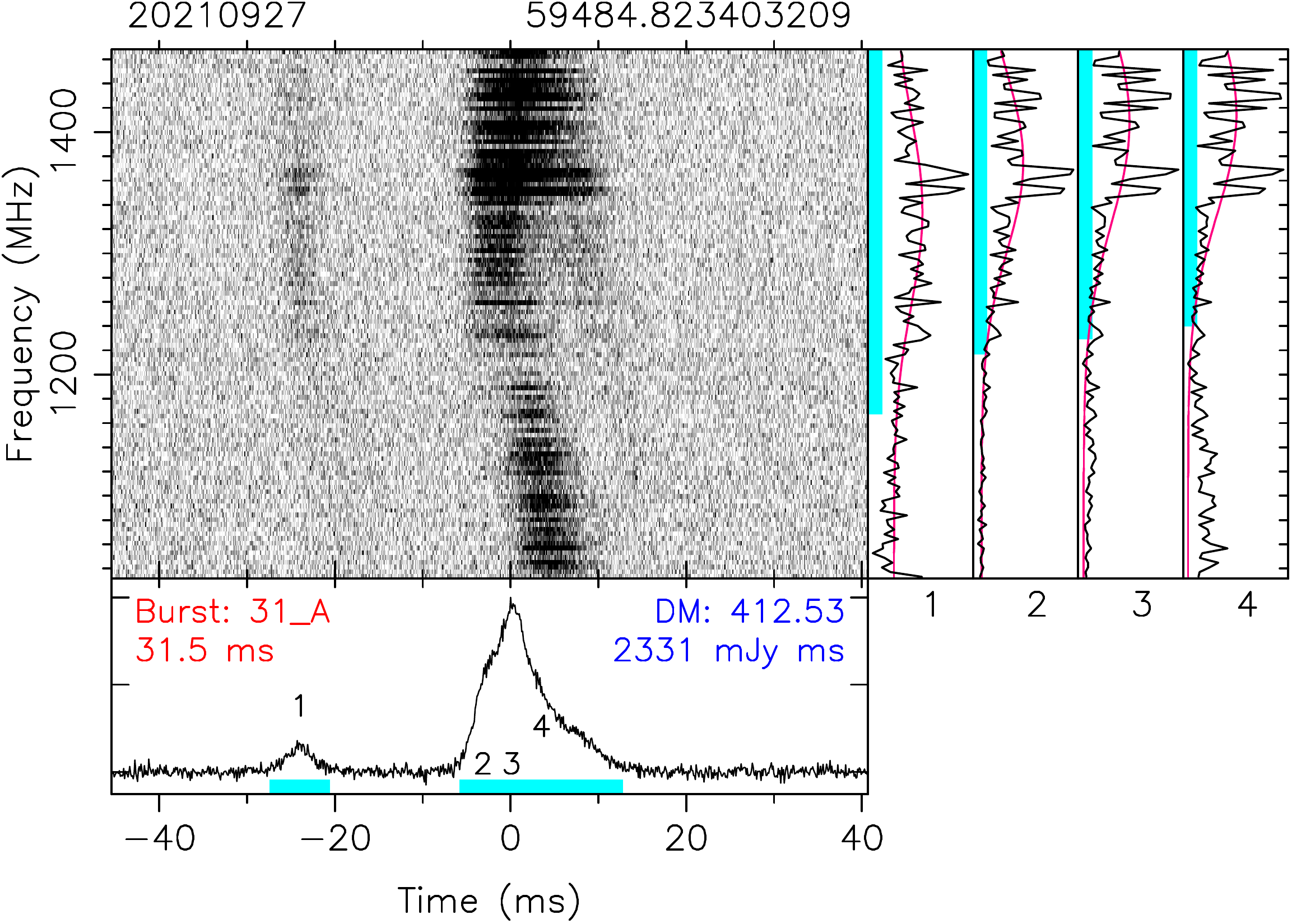}
    \includegraphics[height=37mm]{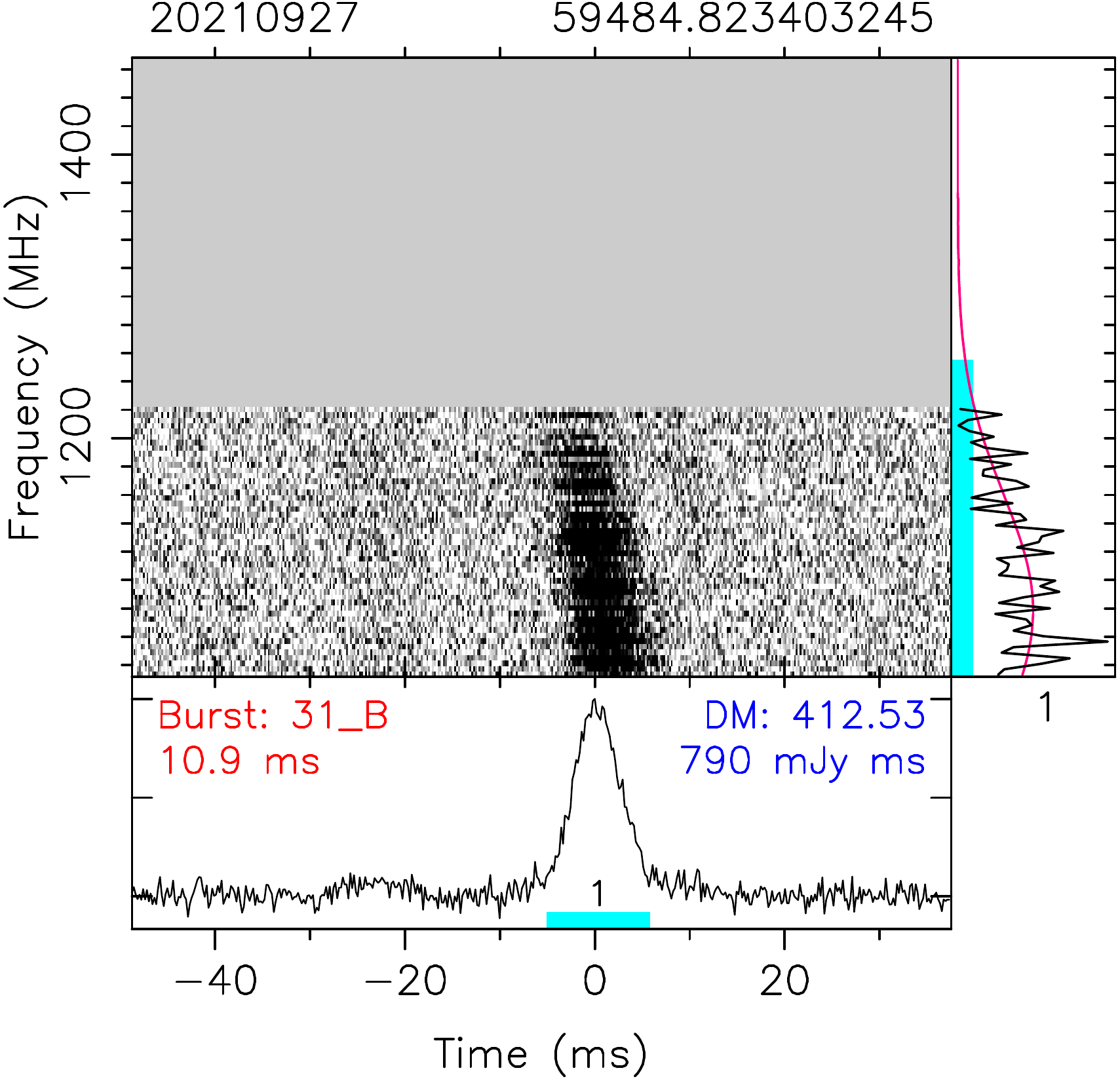}
    \includegraphics[height=37mm]{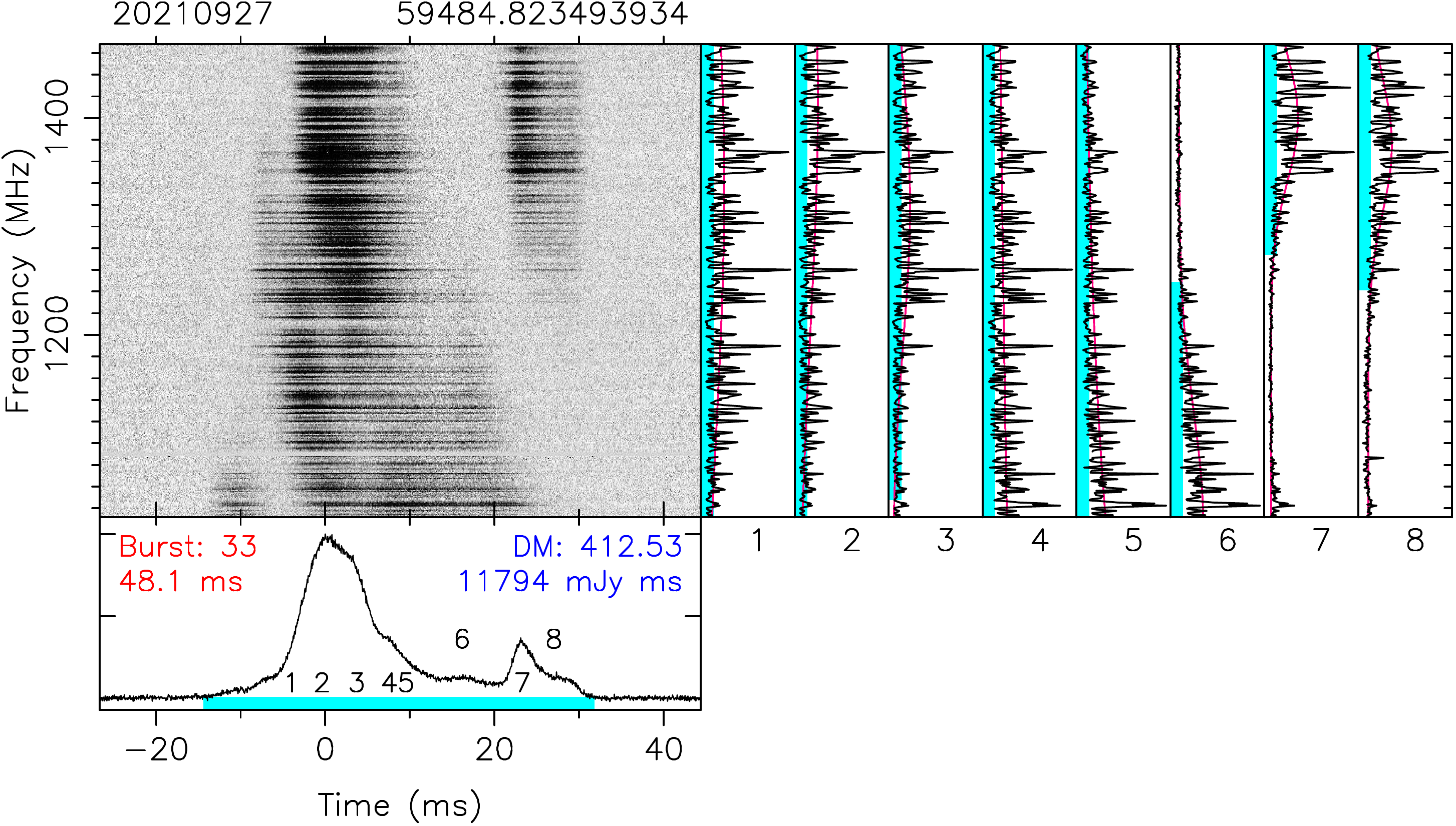}
    \includegraphics[height=37mm]{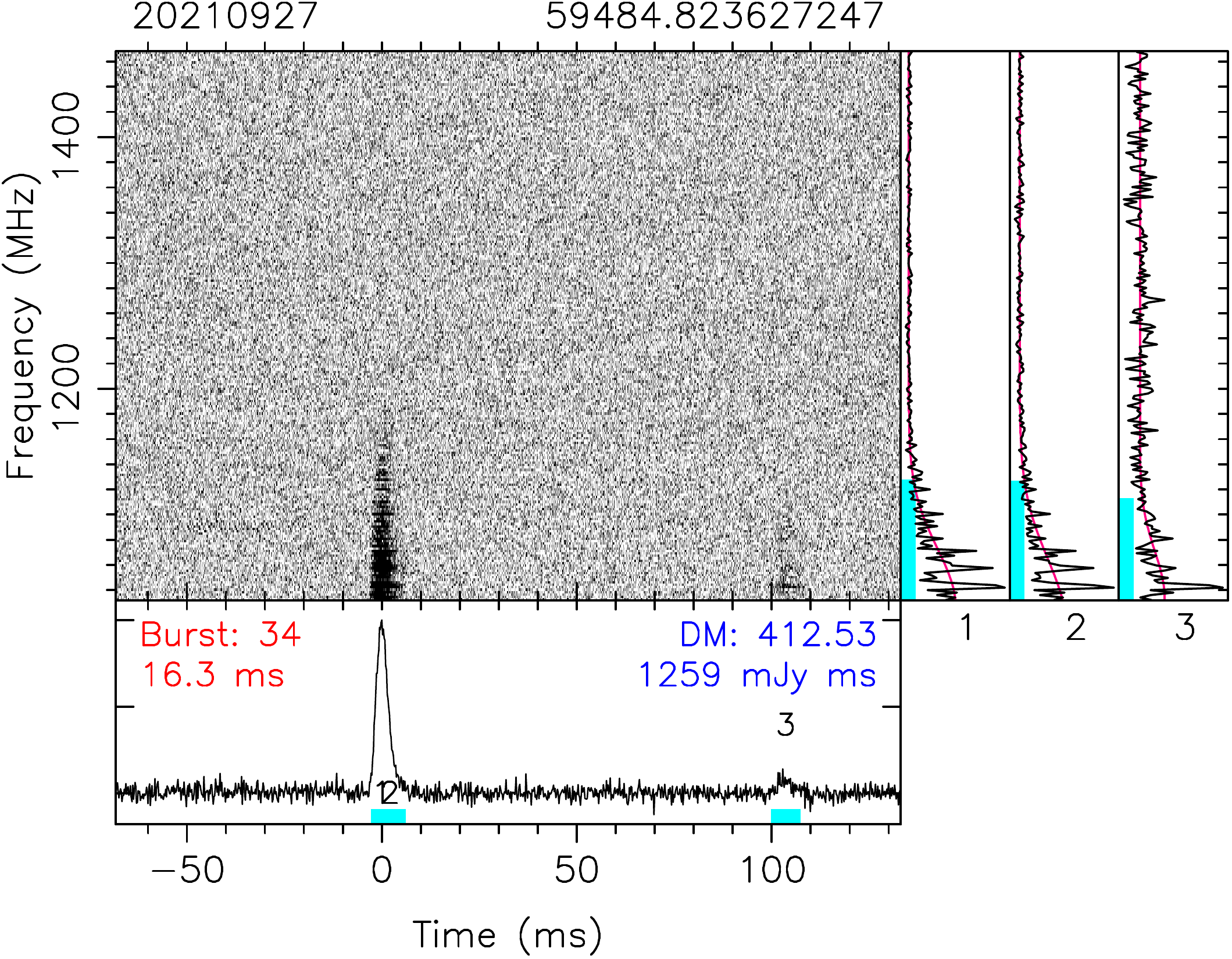} 
    \includegraphics[height=37mm]{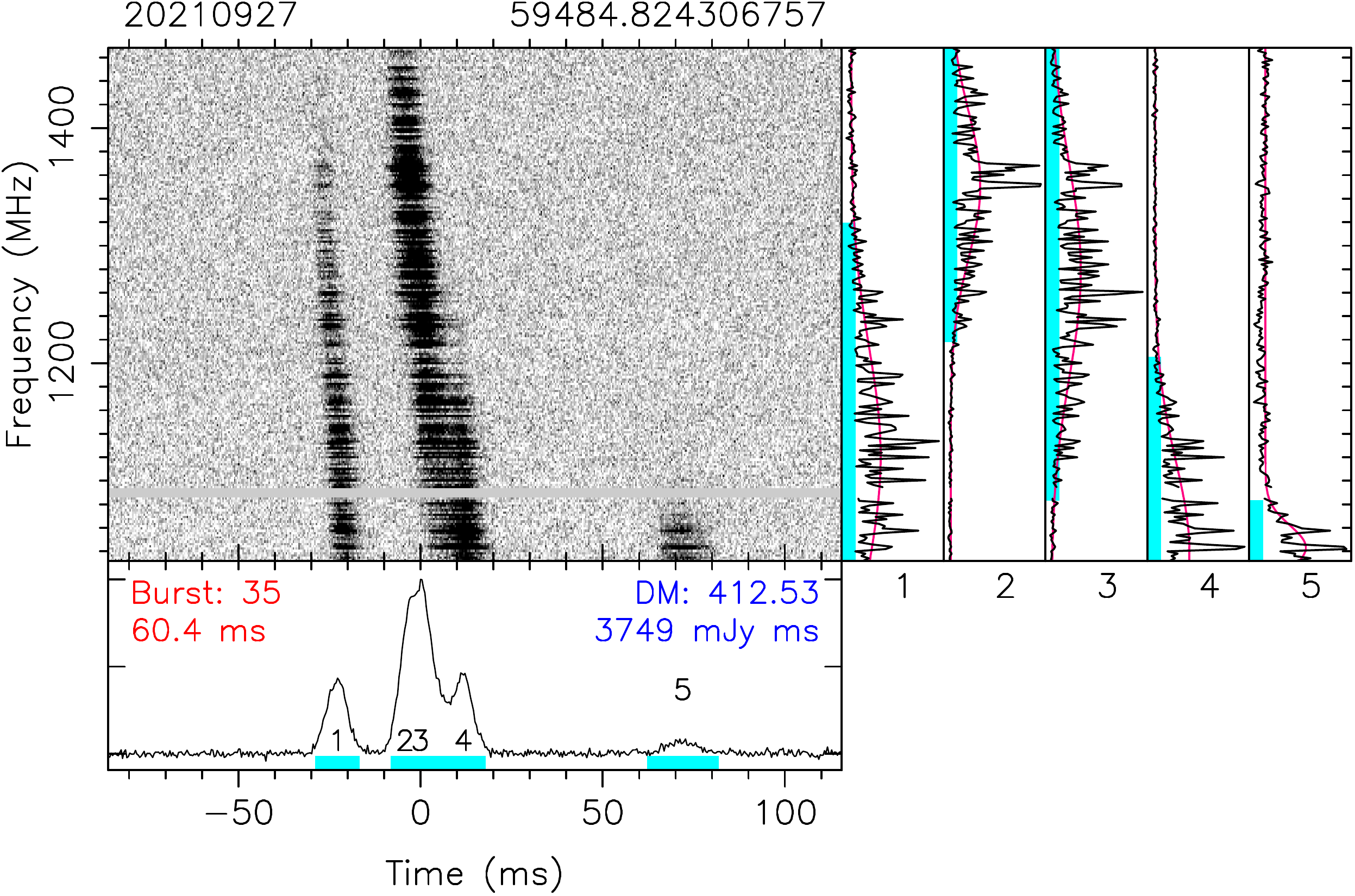} 
    \includegraphics[height=37mm]{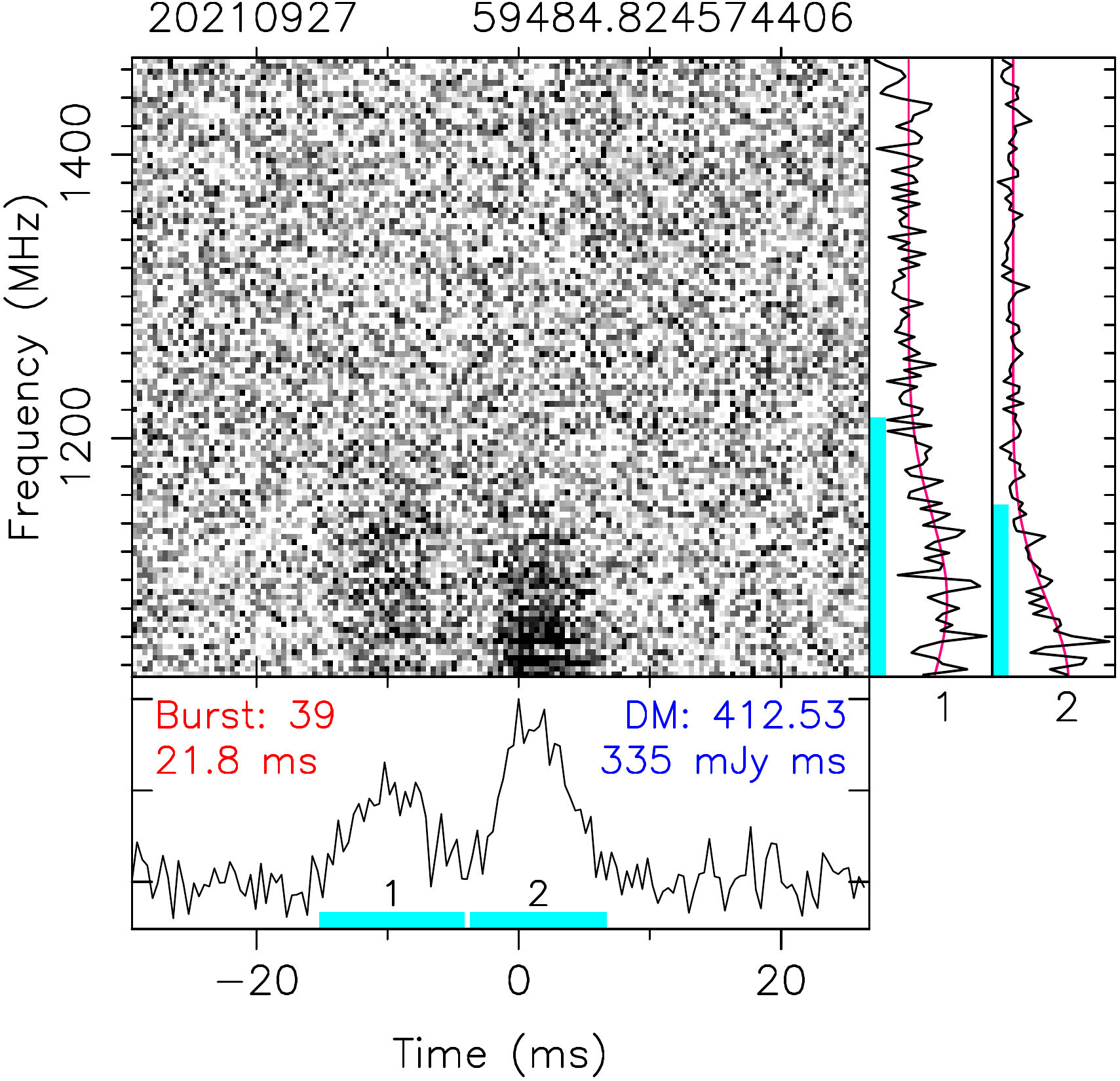}
    \includegraphics[height=37mm]{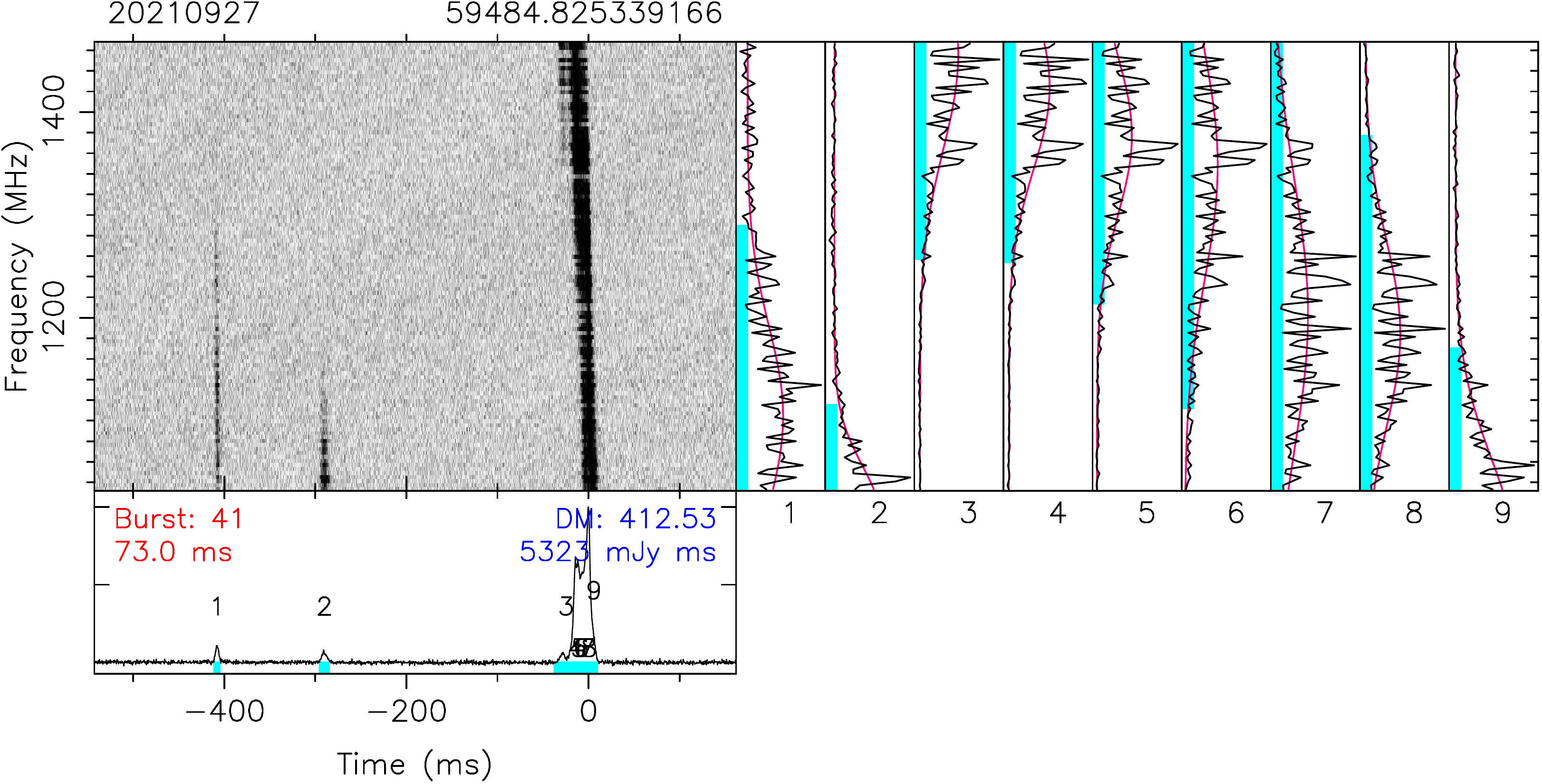} 
    \includegraphics[height=37mm]{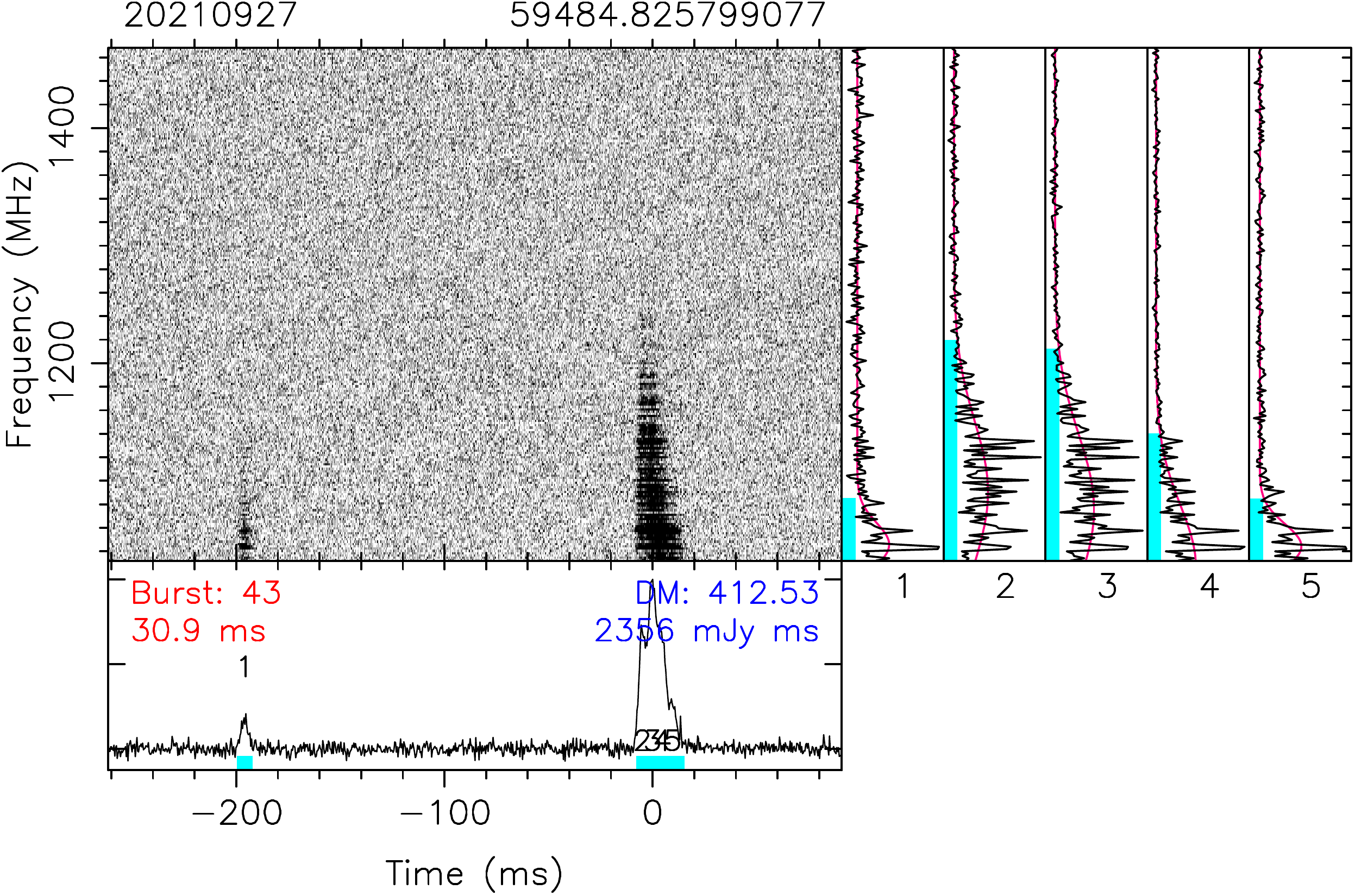} 
\caption{\it{ -- continued}.
}
\end{figure*}
\addtocounter{figure}{-1}
\begin{figure*}
    \flushleft
    \includegraphics[height=37mm]{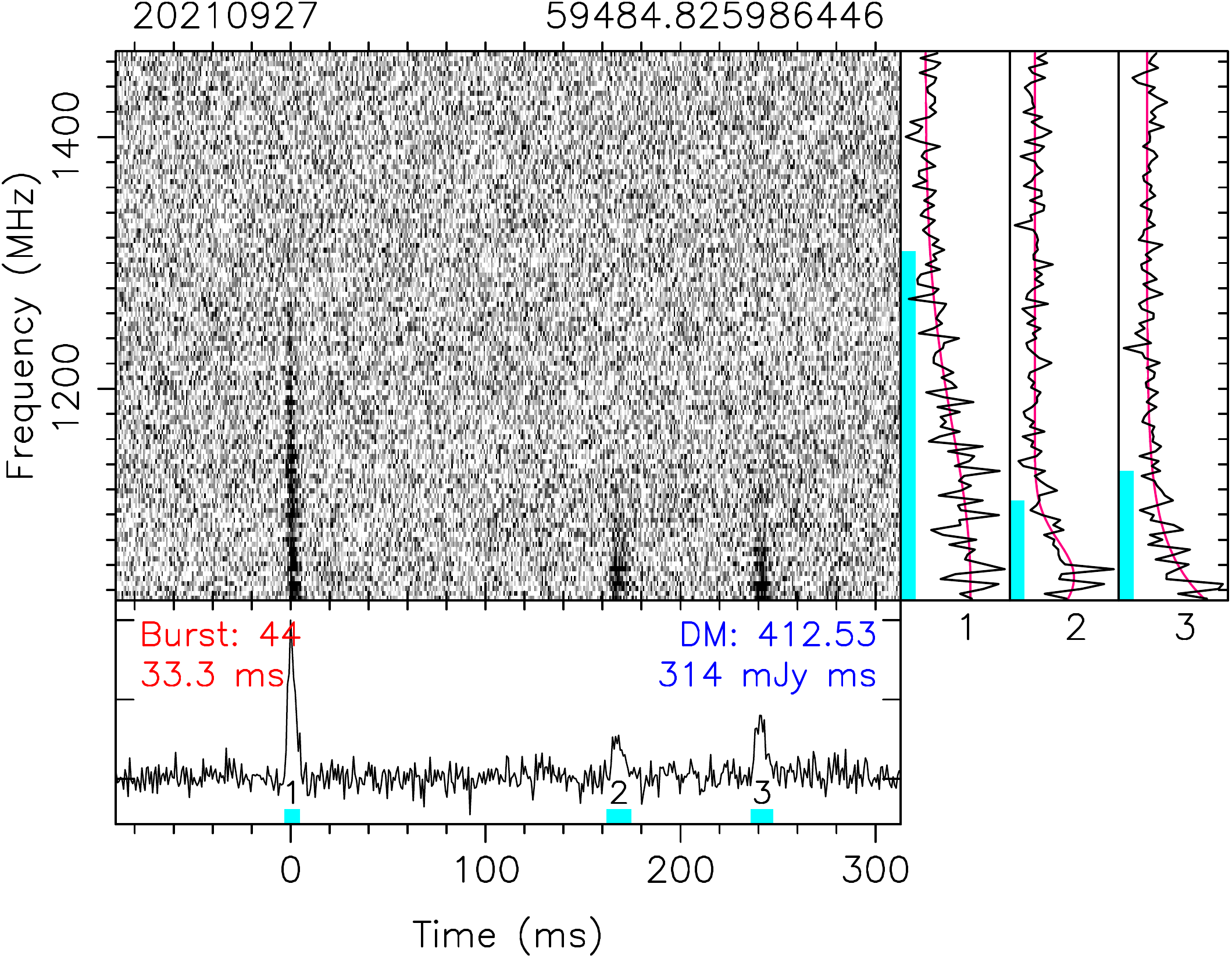} 
    \includegraphics[height=37mm]{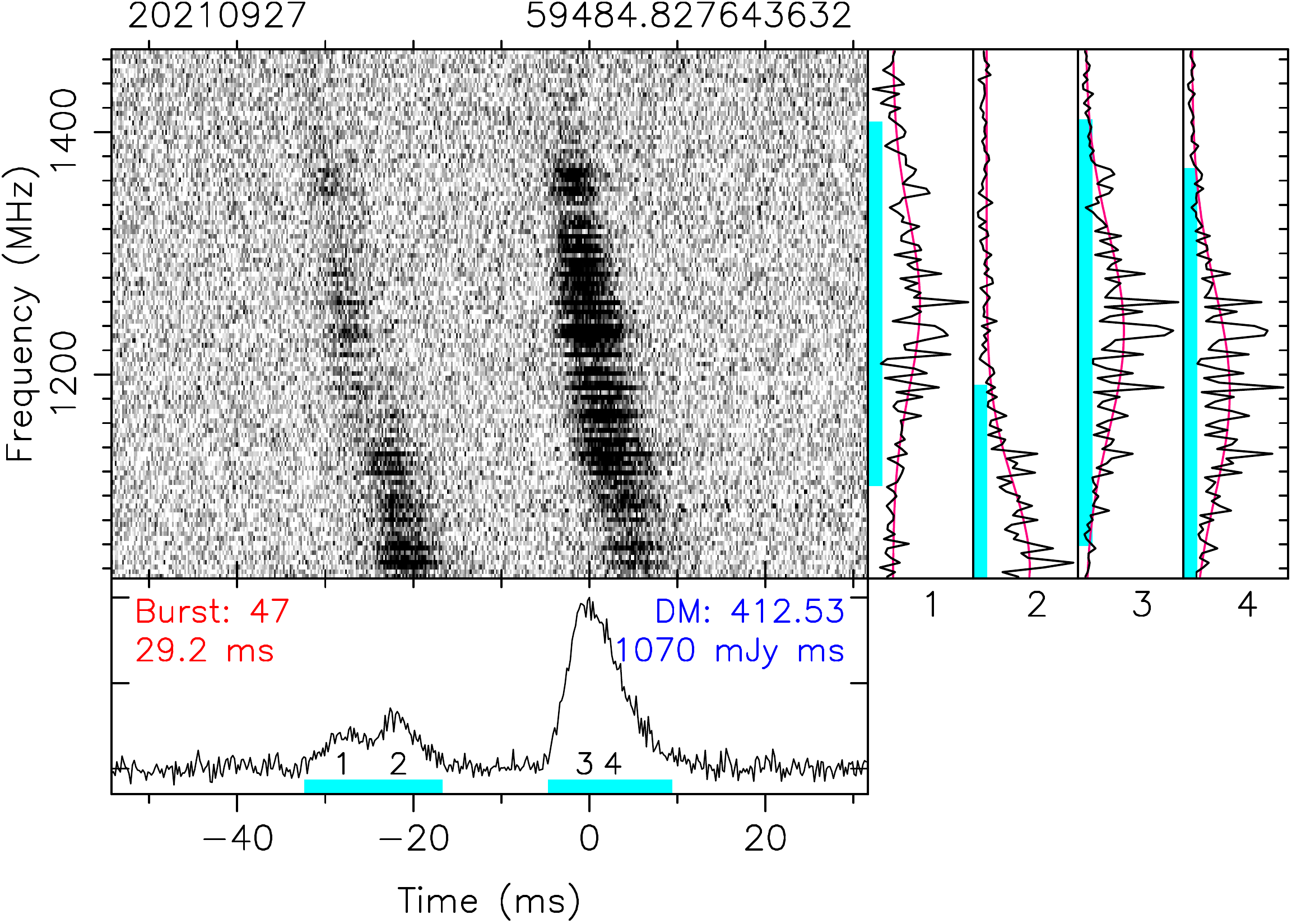}
    \includegraphics[height=37mm]{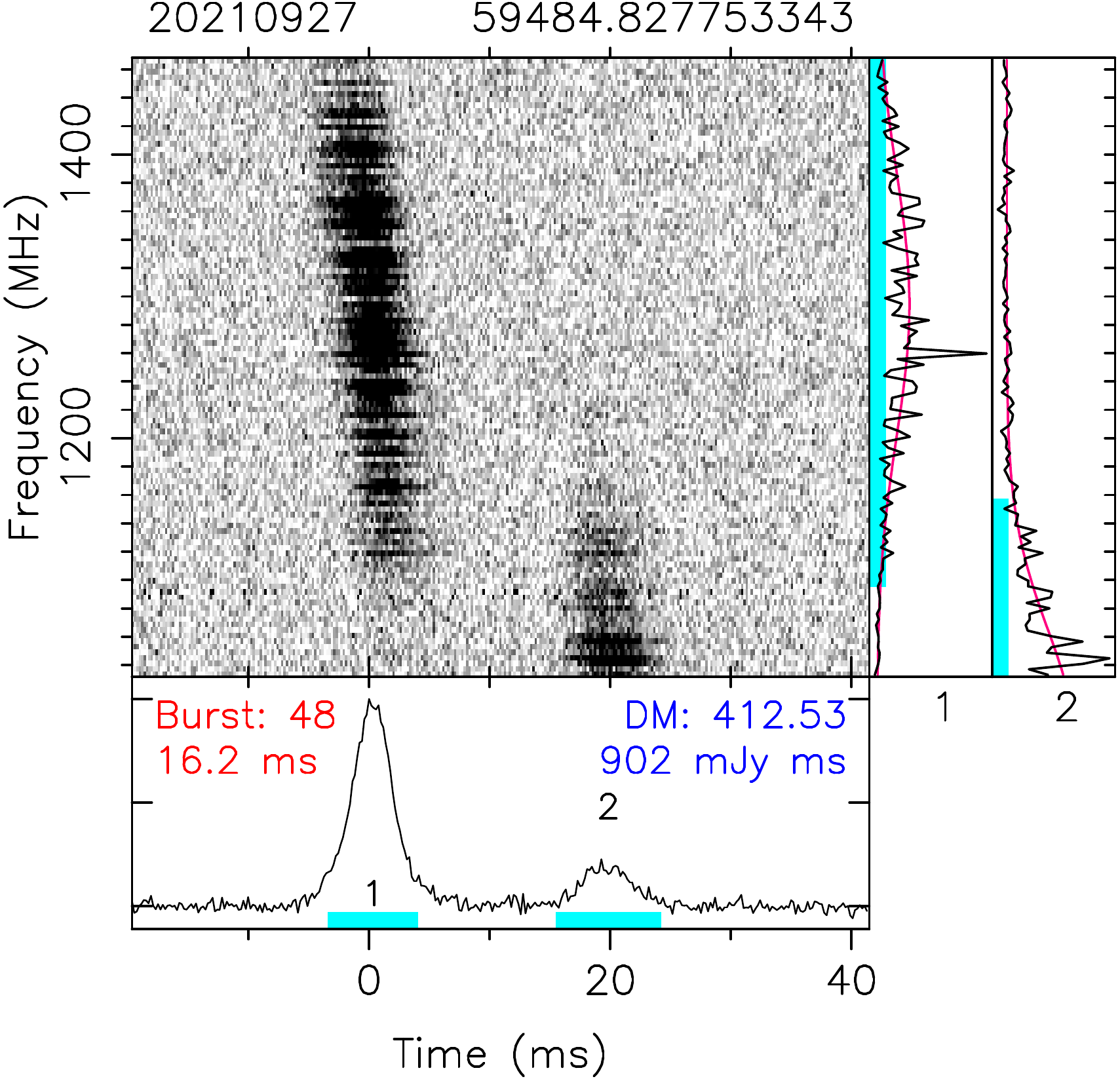}
    \includegraphics[height=37mm]{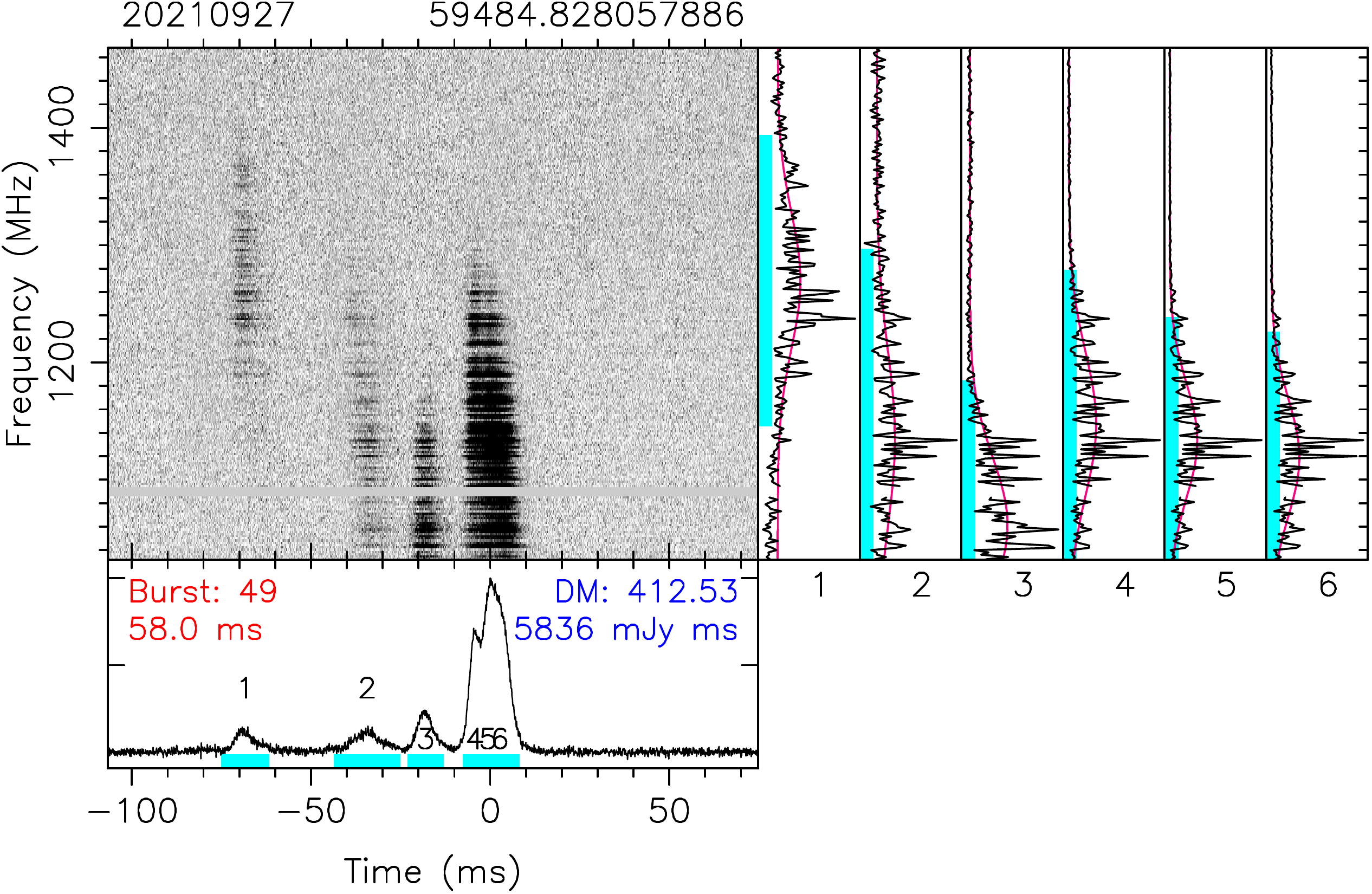}
    \includegraphics[height=37mm]{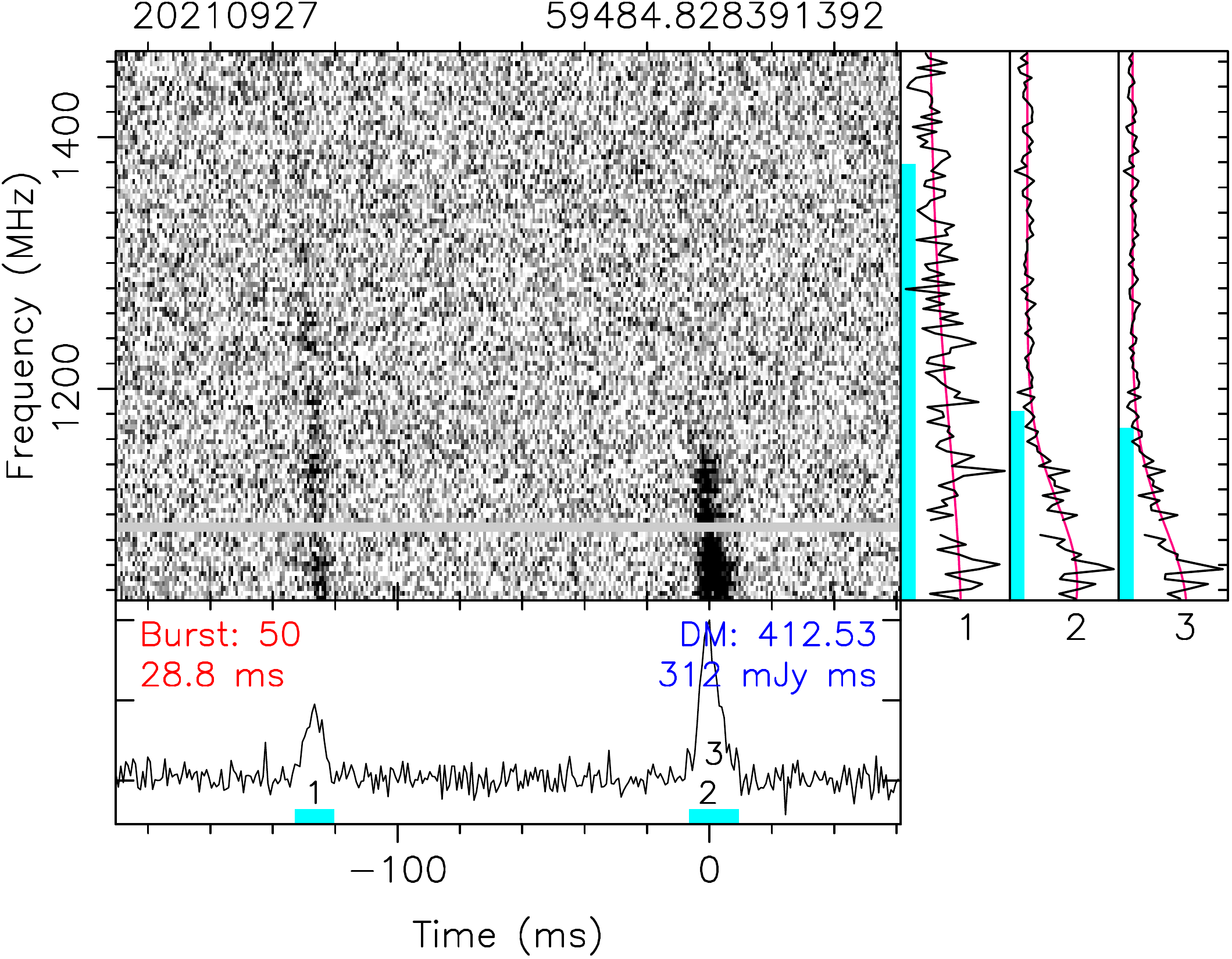} 
    \includegraphics[height=37mm]{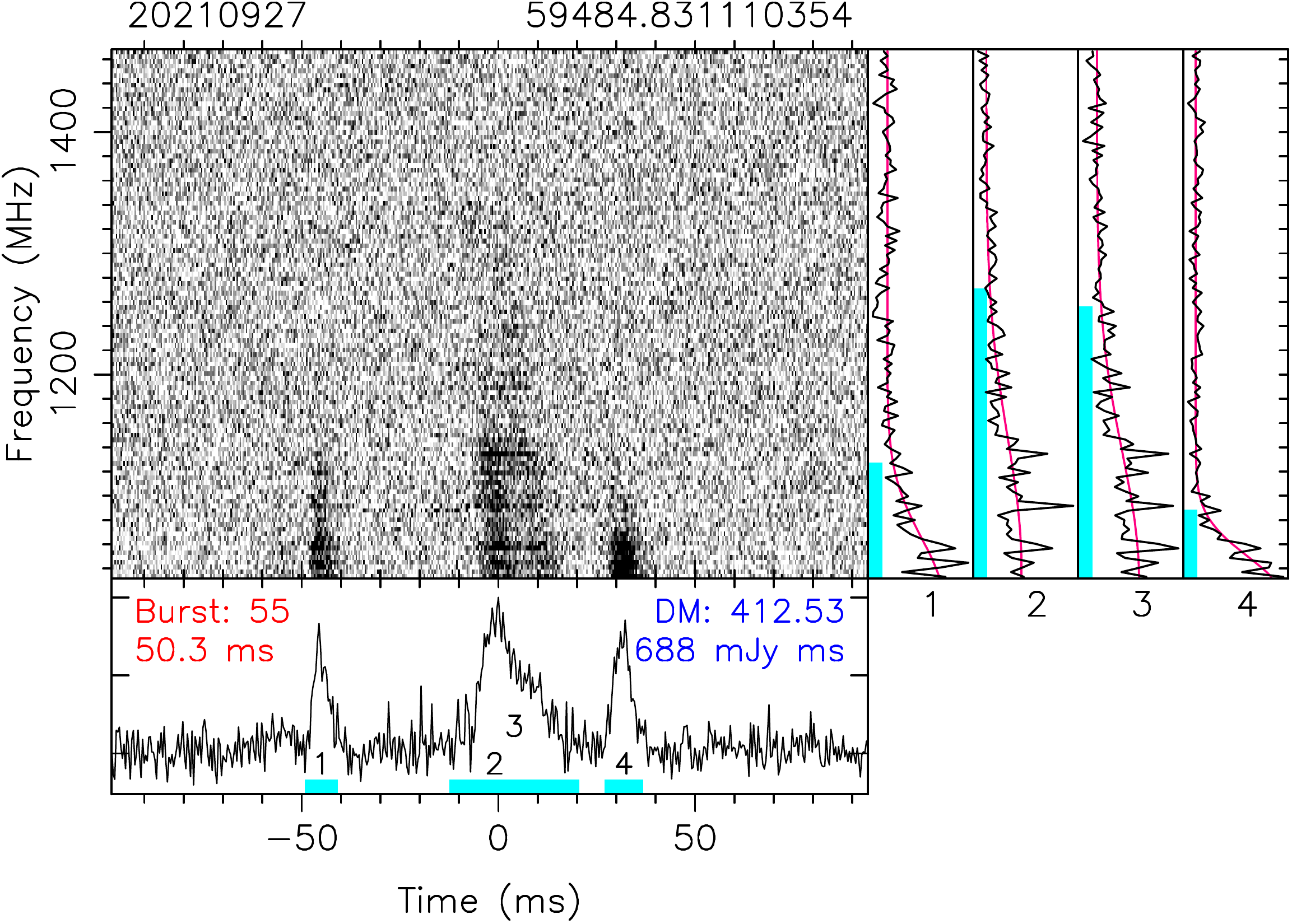}
    \includegraphics[height=37mm]{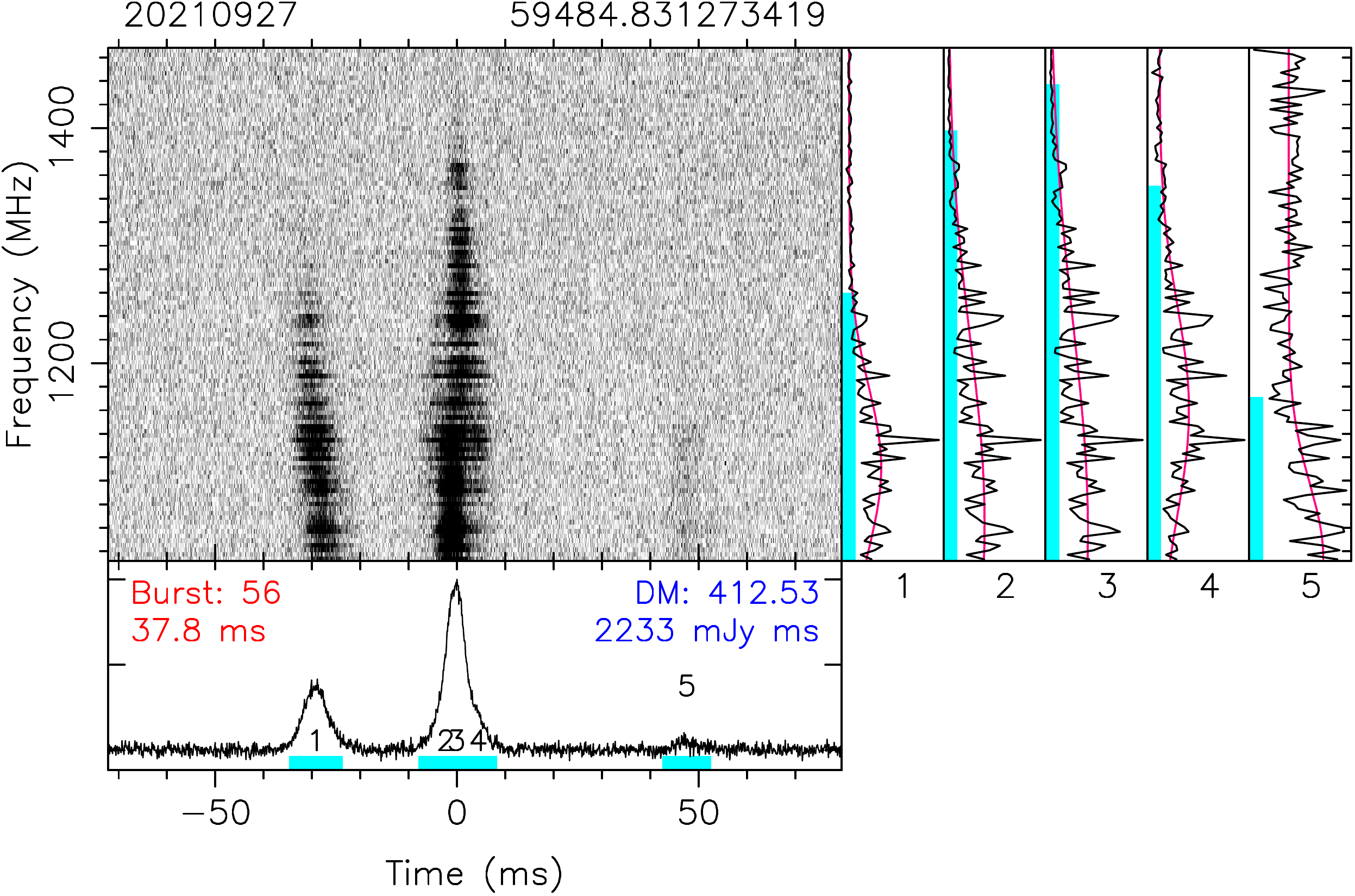}
    \includegraphics[height=37mm]{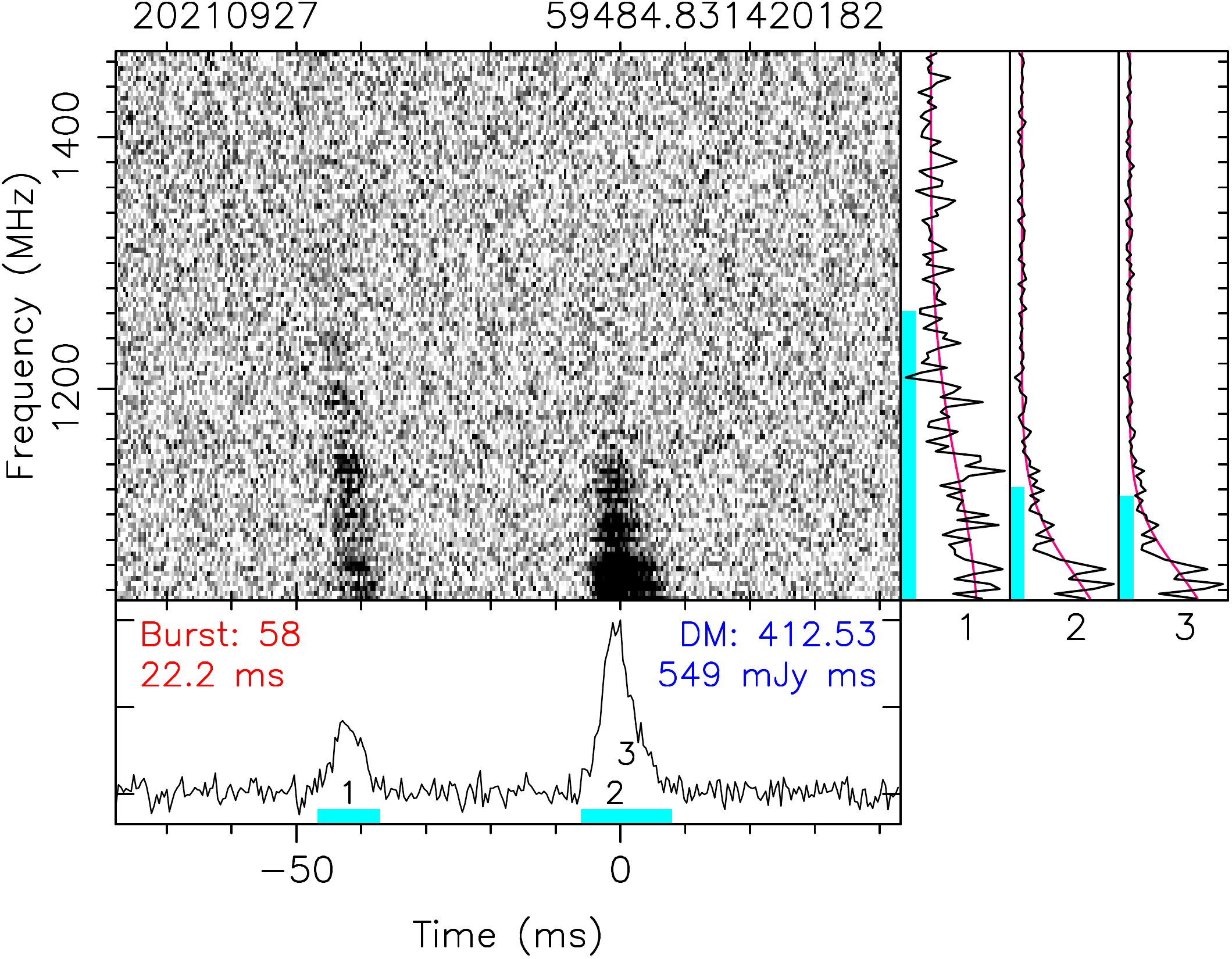} 
    \includegraphics[height=37mm]{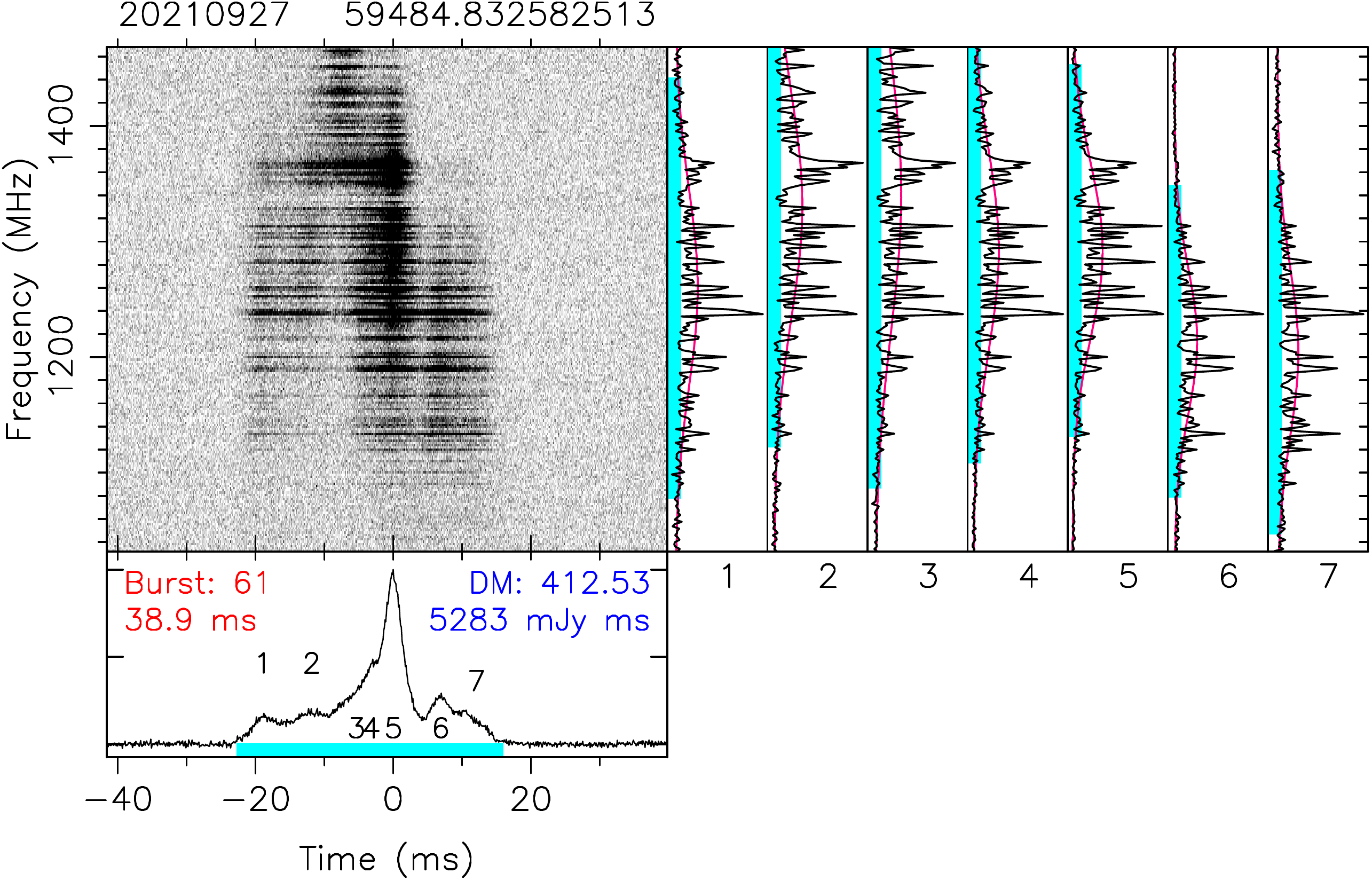}
    \includegraphics[height=37mm]{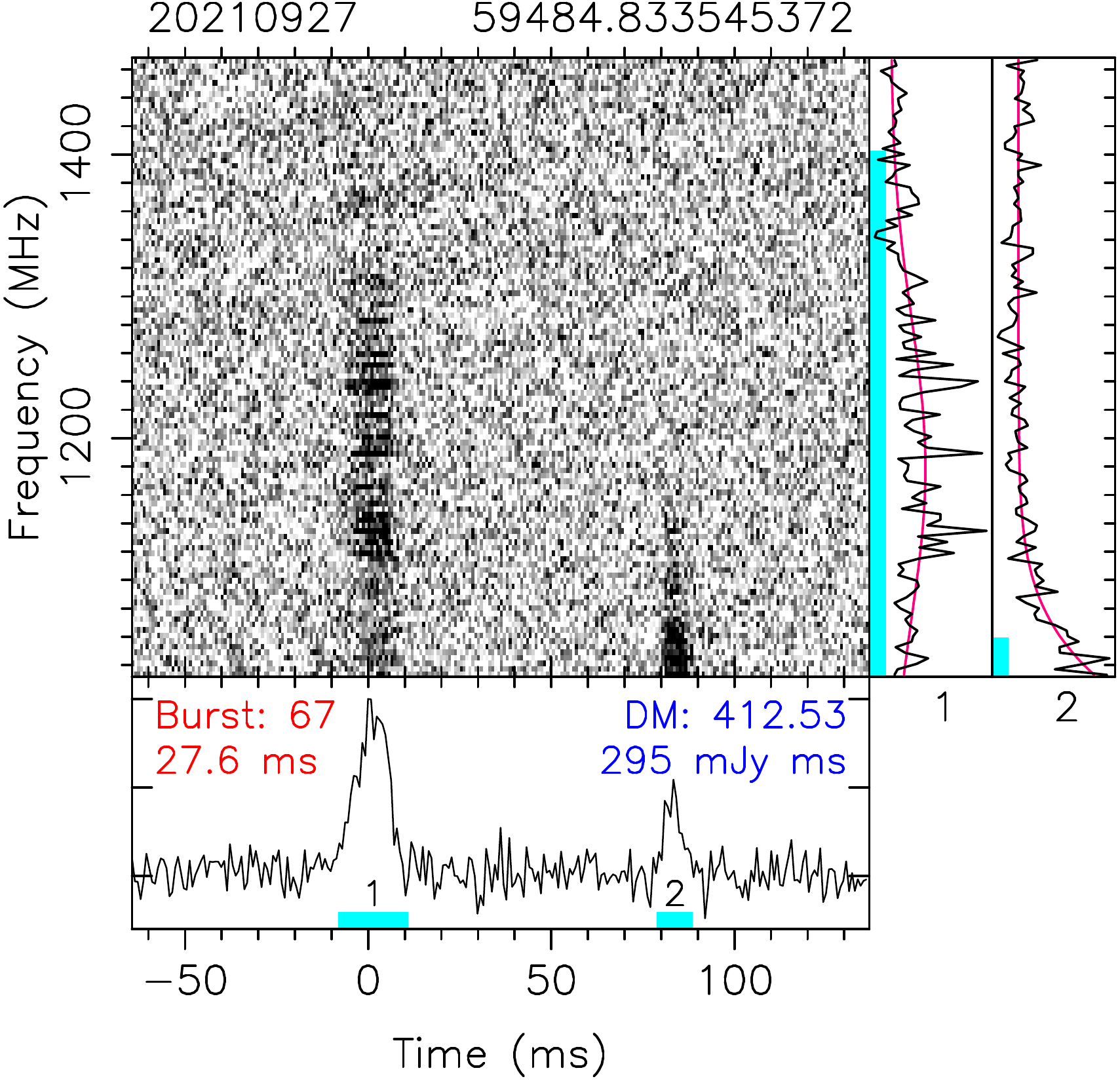} 
    \includegraphics[height=37mm]{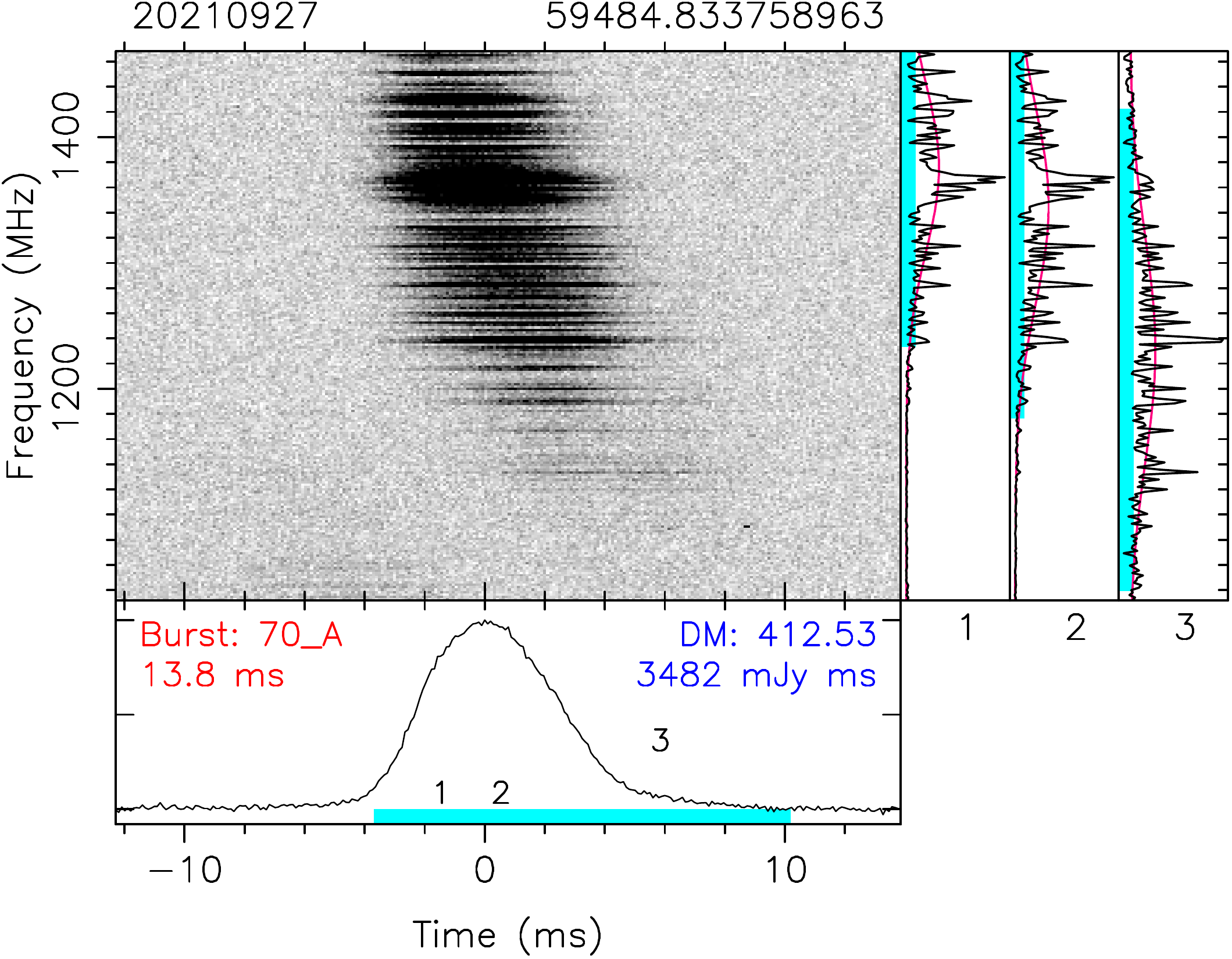}
    \includegraphics[height=37mm]{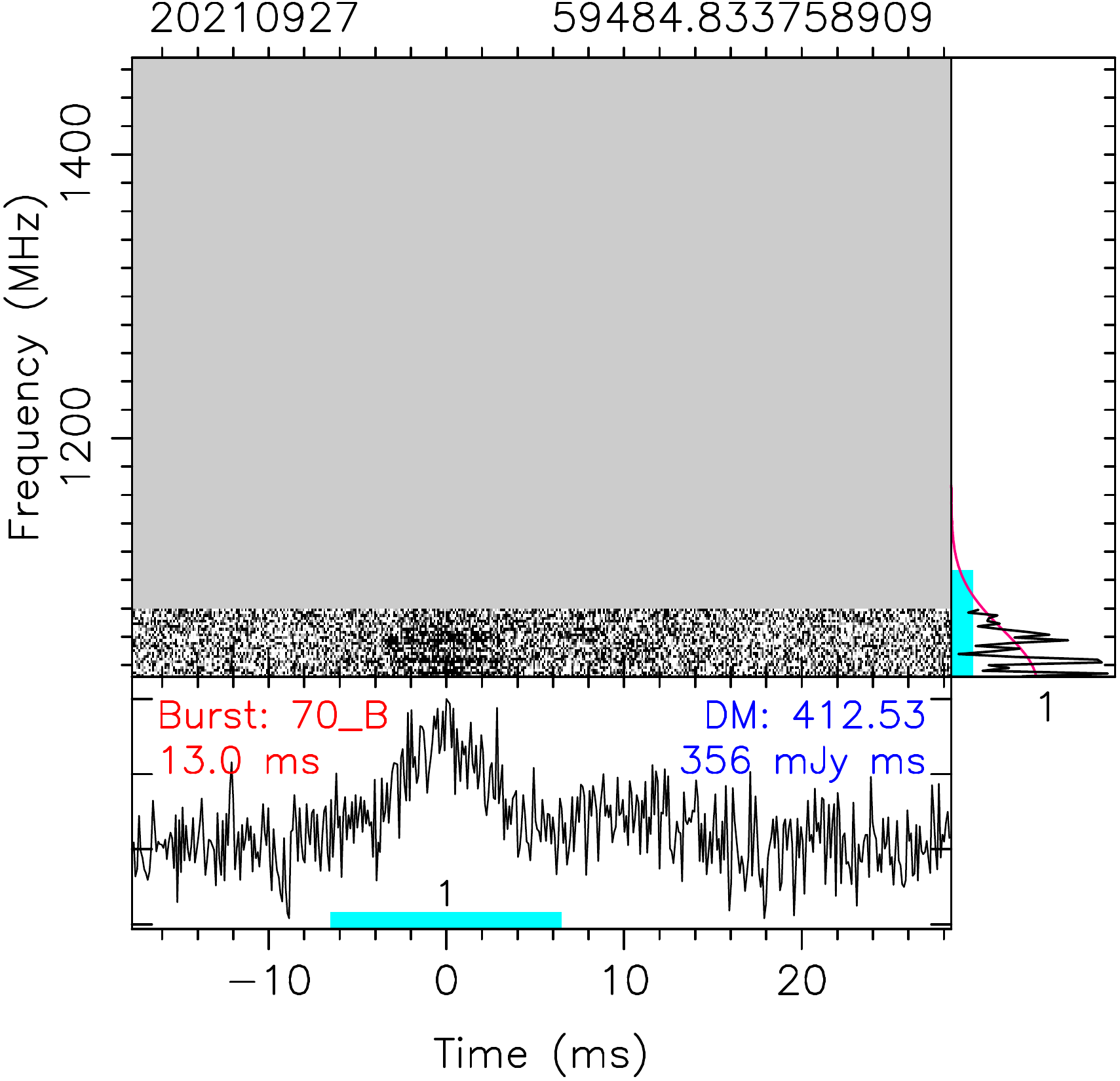}
    \includegraphics[height=37mm]{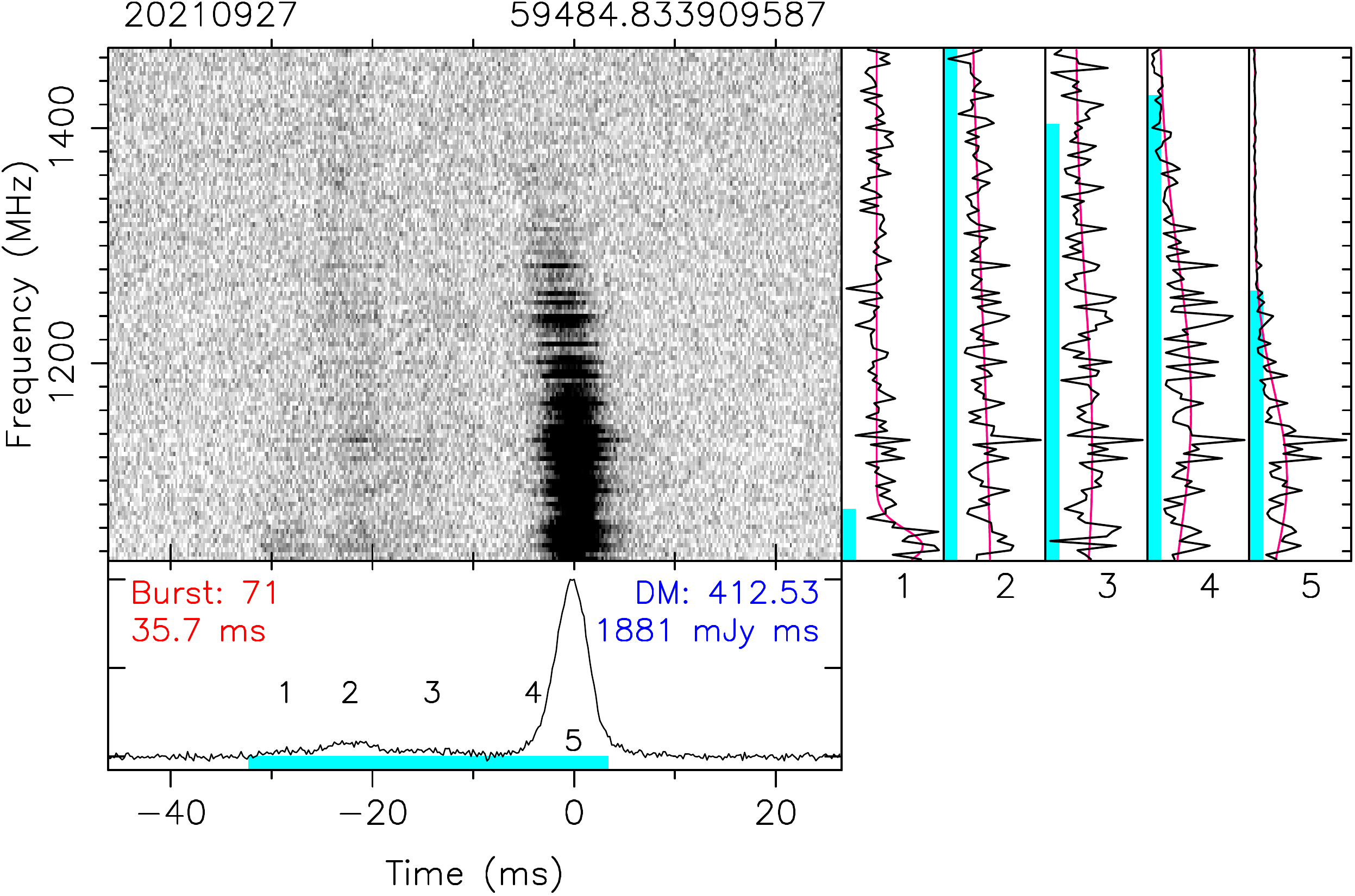}
    \includegraphics[height=37mm]{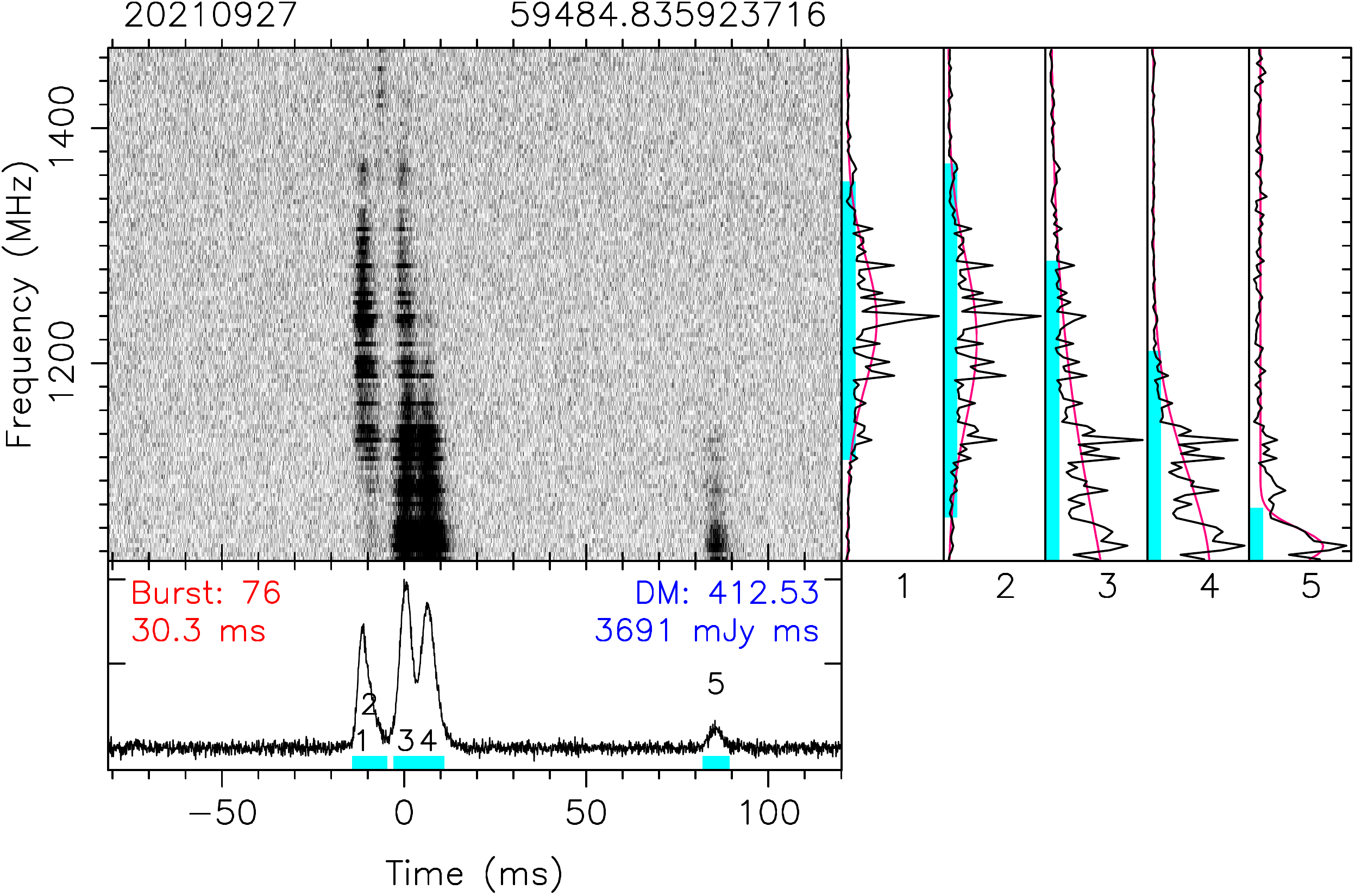} 
    \includegraphics[height=37mm]{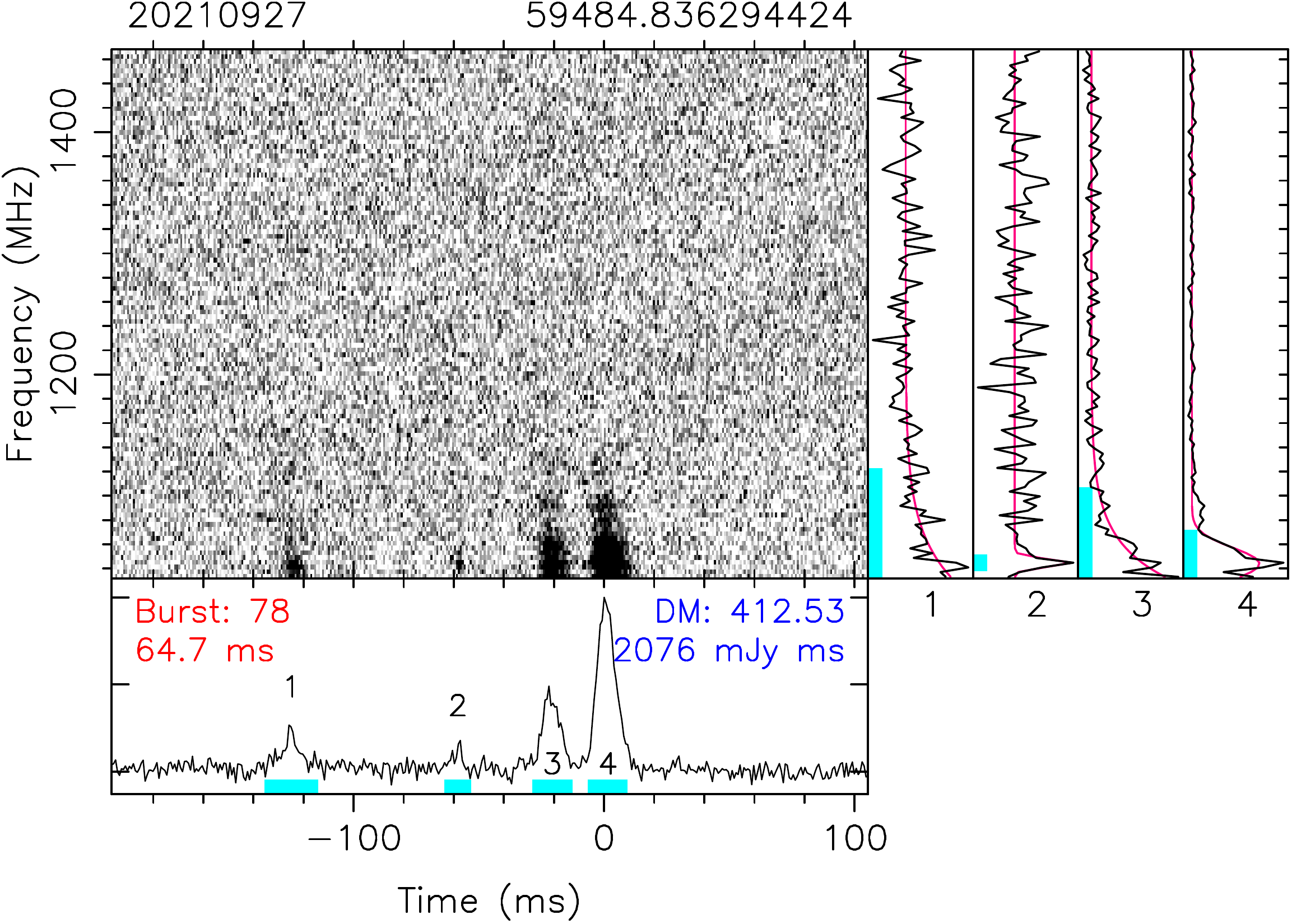} 
    \includegraphics[height=37mm]{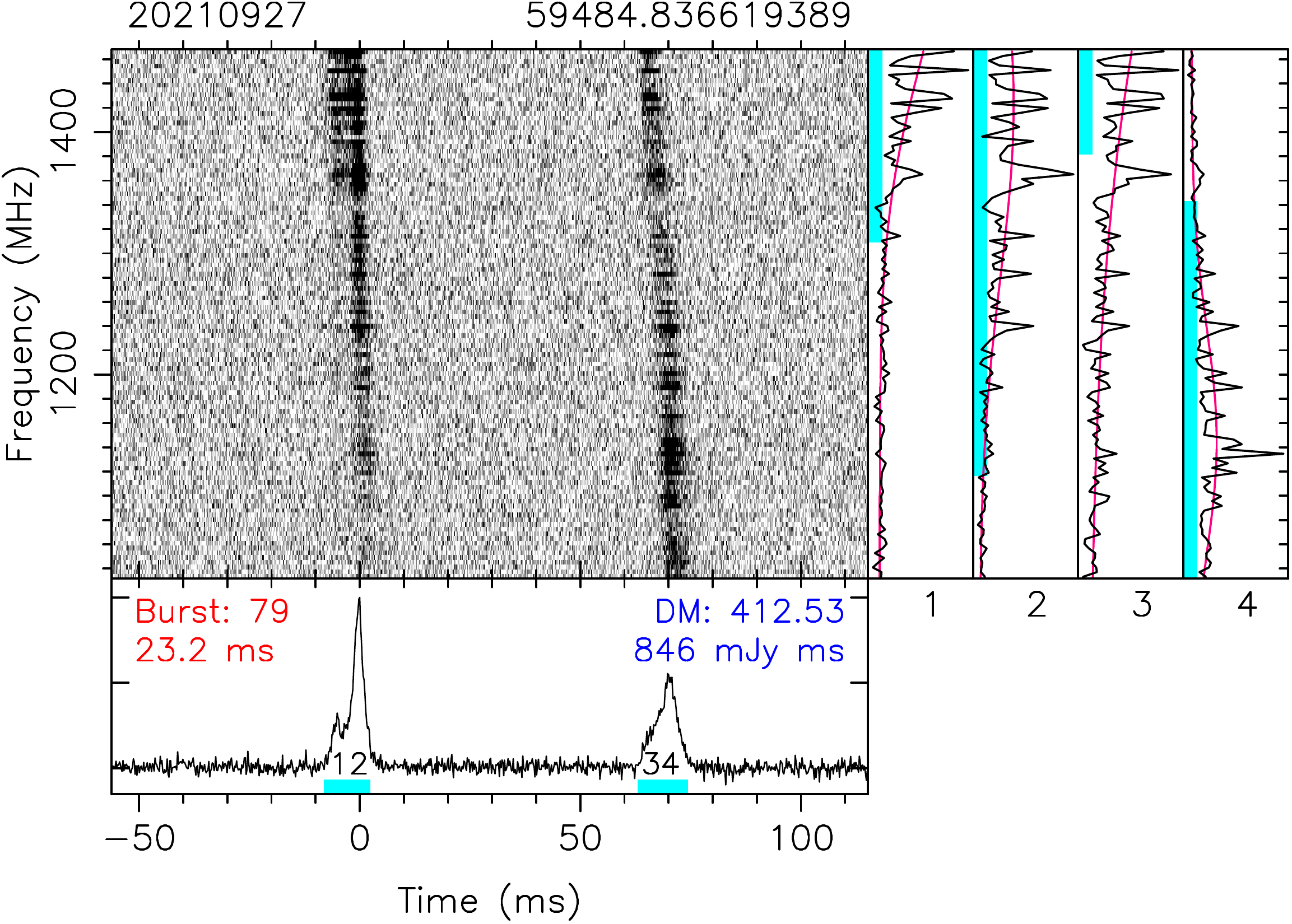} 
    \includegraphics[height=38mm]{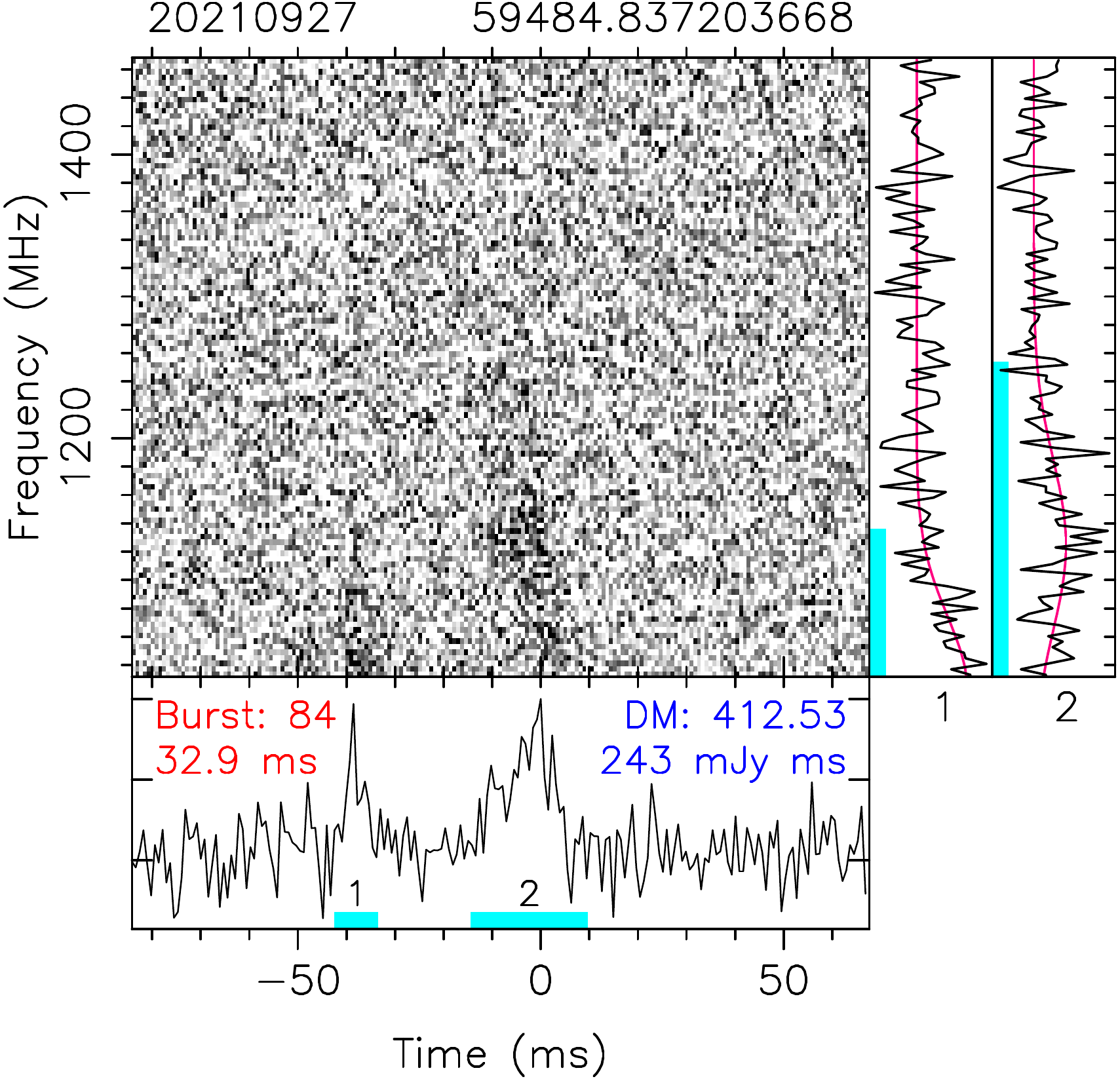}  
    \includegraphics[height=37mm]{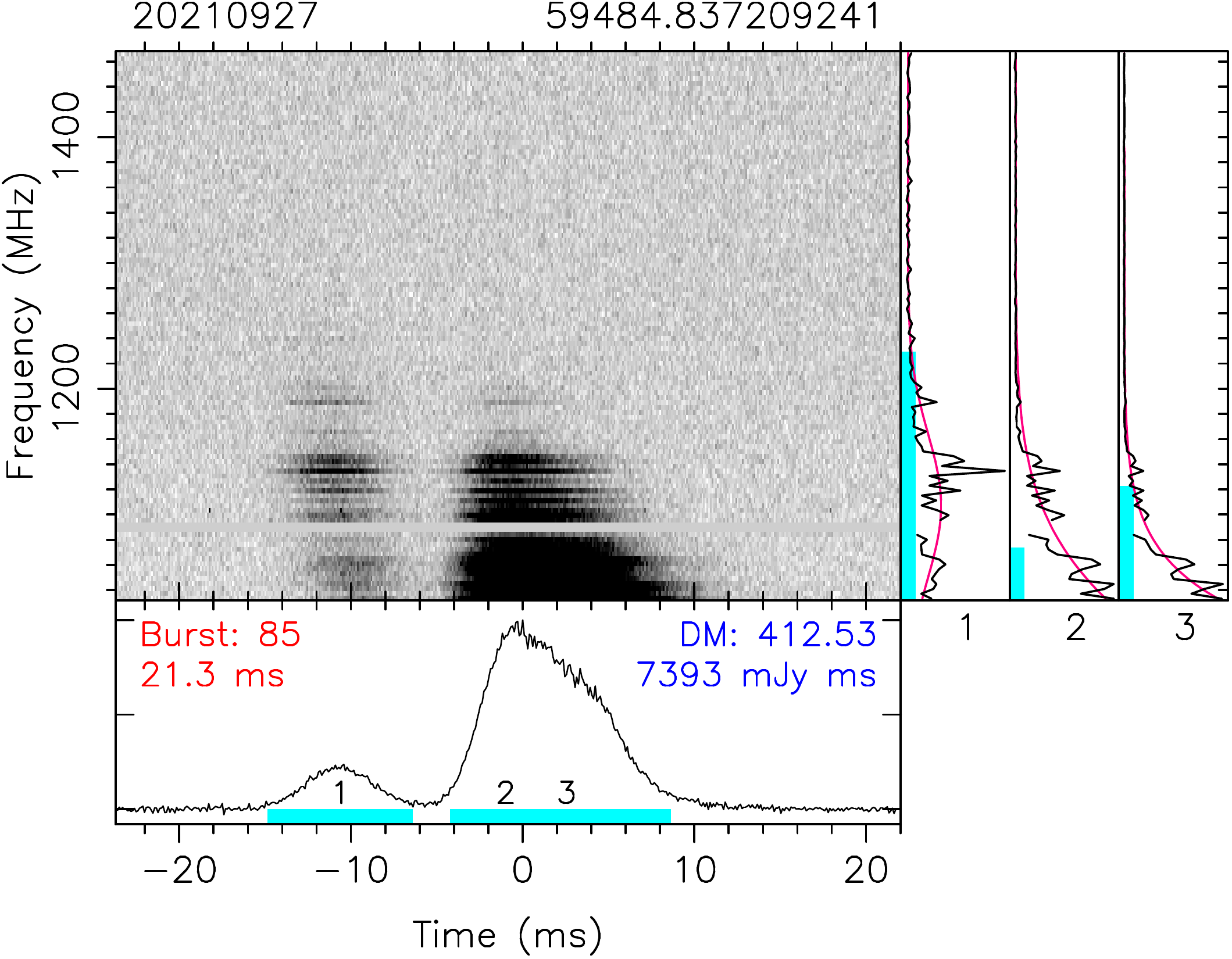}
\caption{\it{ -- continued}.
}
\end{figure*}
\addtocounter{figure}{-1}
\begin{figure*}
    \flushleft
    \includegraphics[height=37mm]{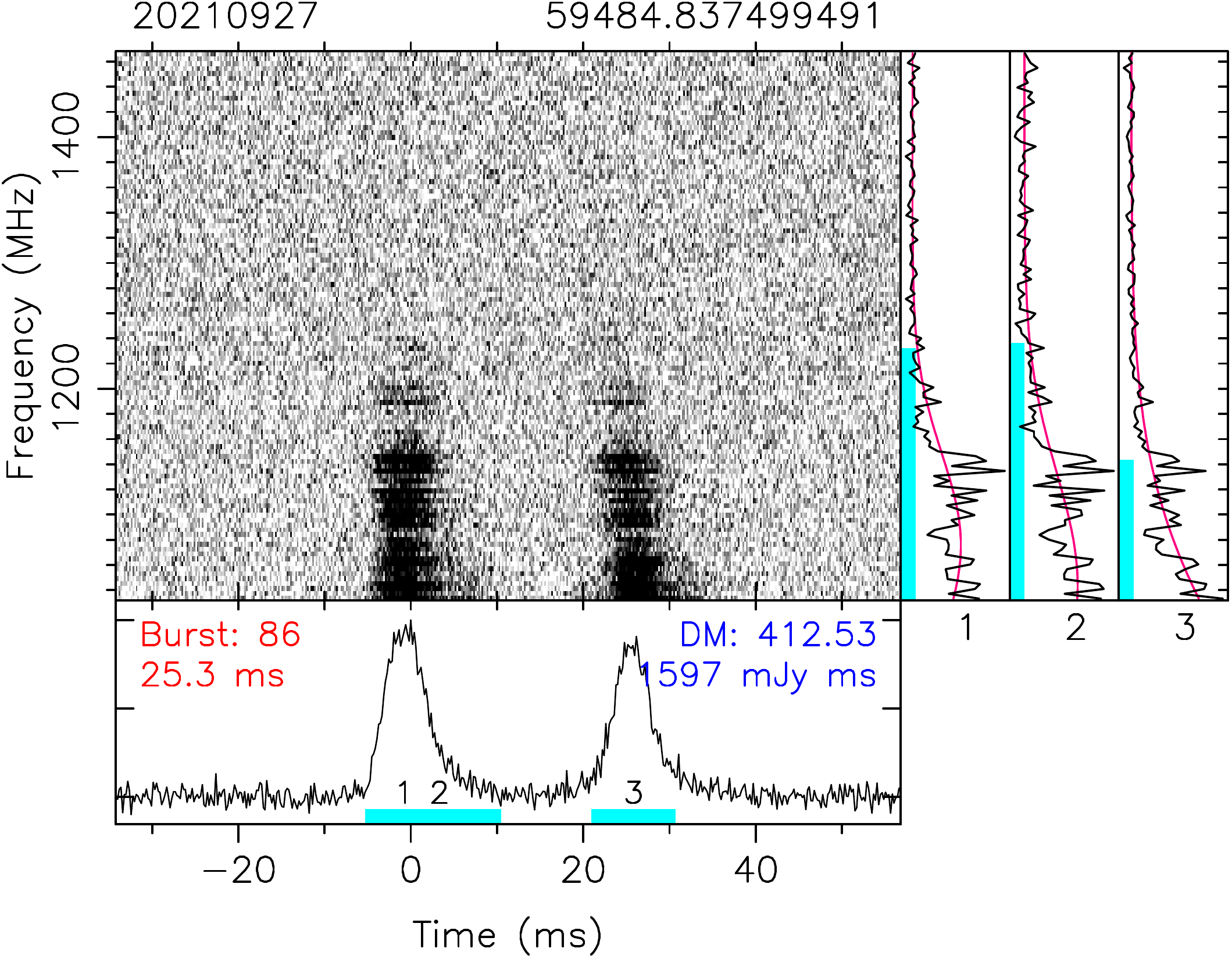}
    \includegraphics[height=37mm]{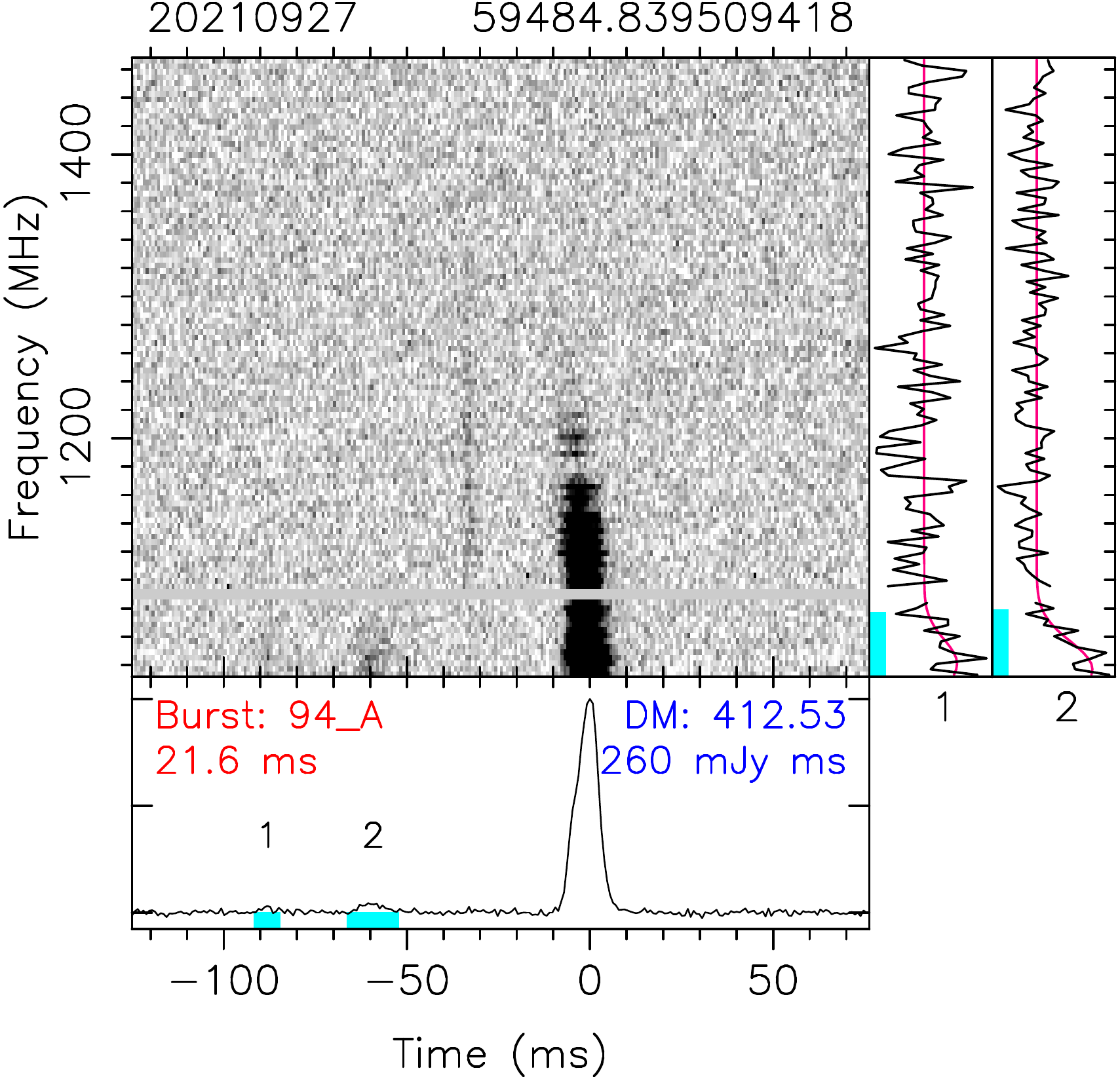} 
    \includegraphics[height=37mm]{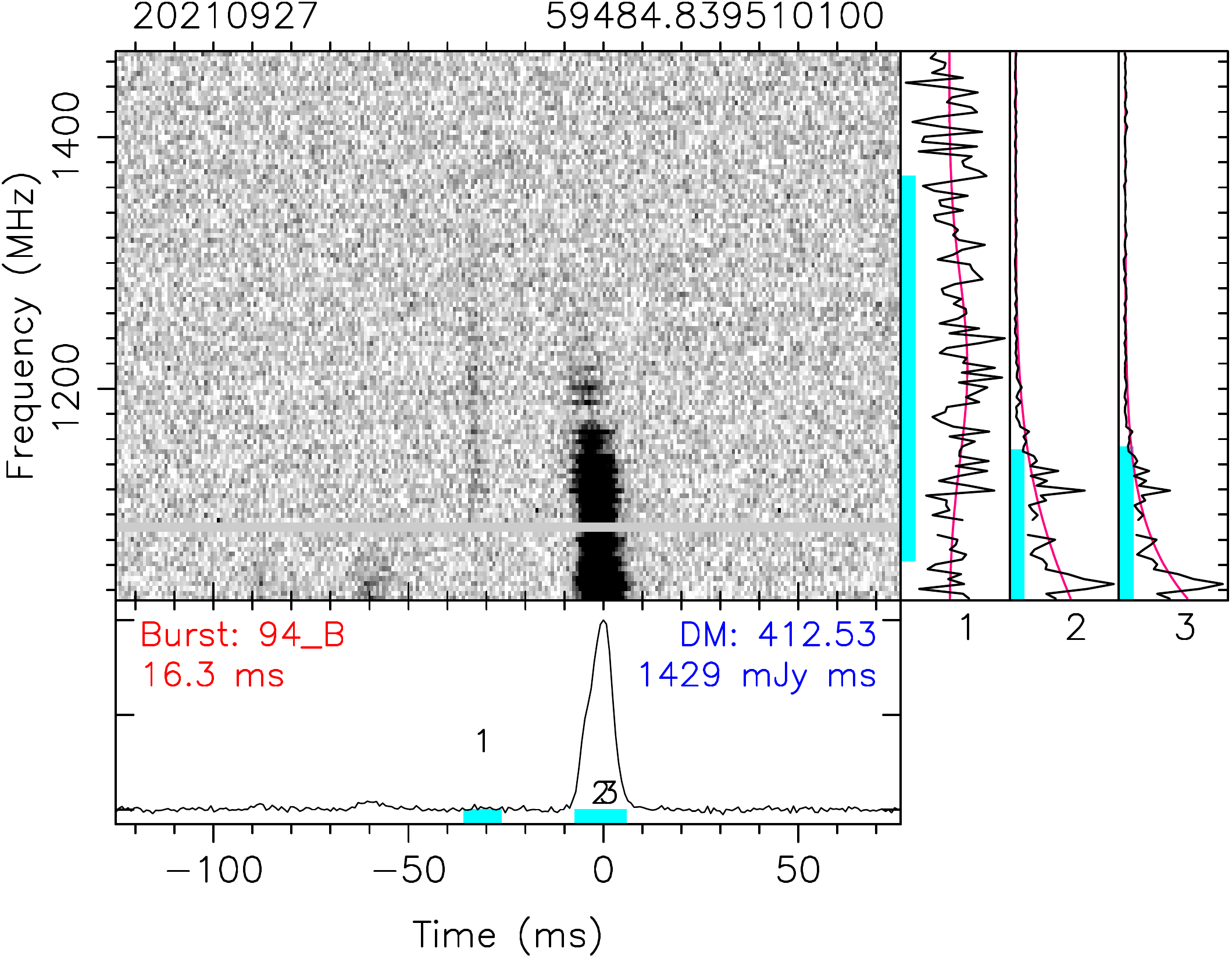}
    \includegraphics[height=37mm]{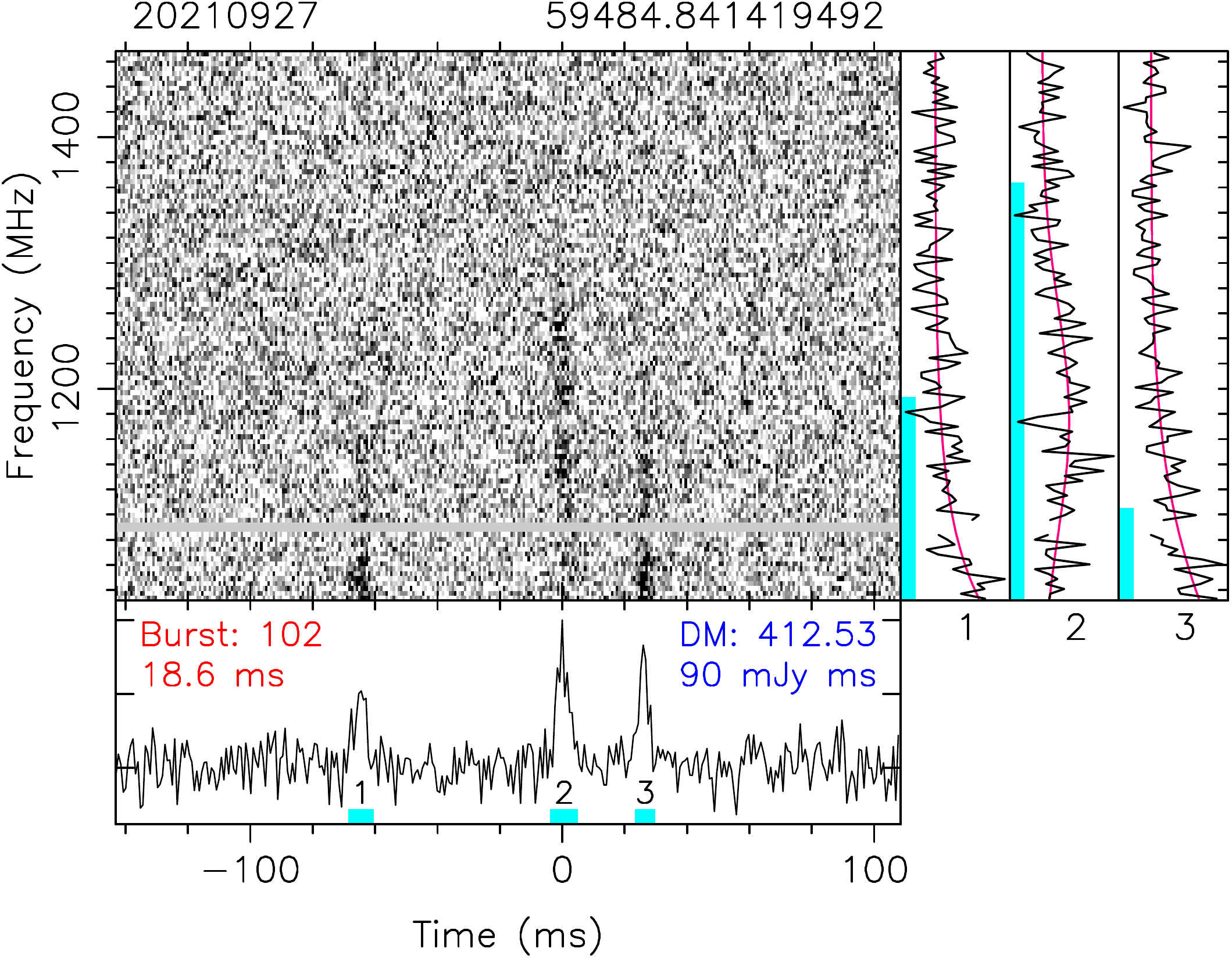} 
    \includegraphics[height=37mm]{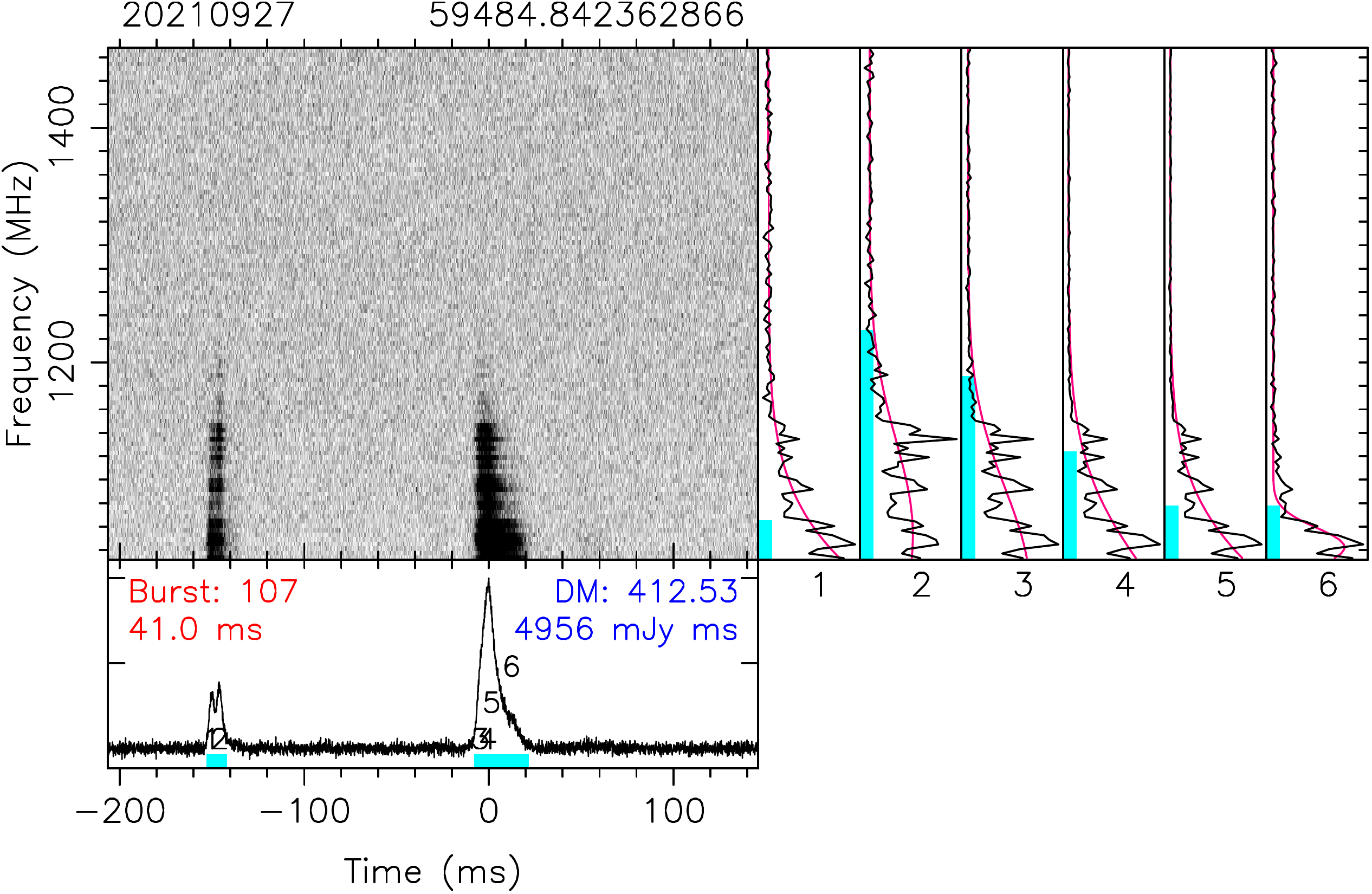} 
    \includegraphics[height=37mm]{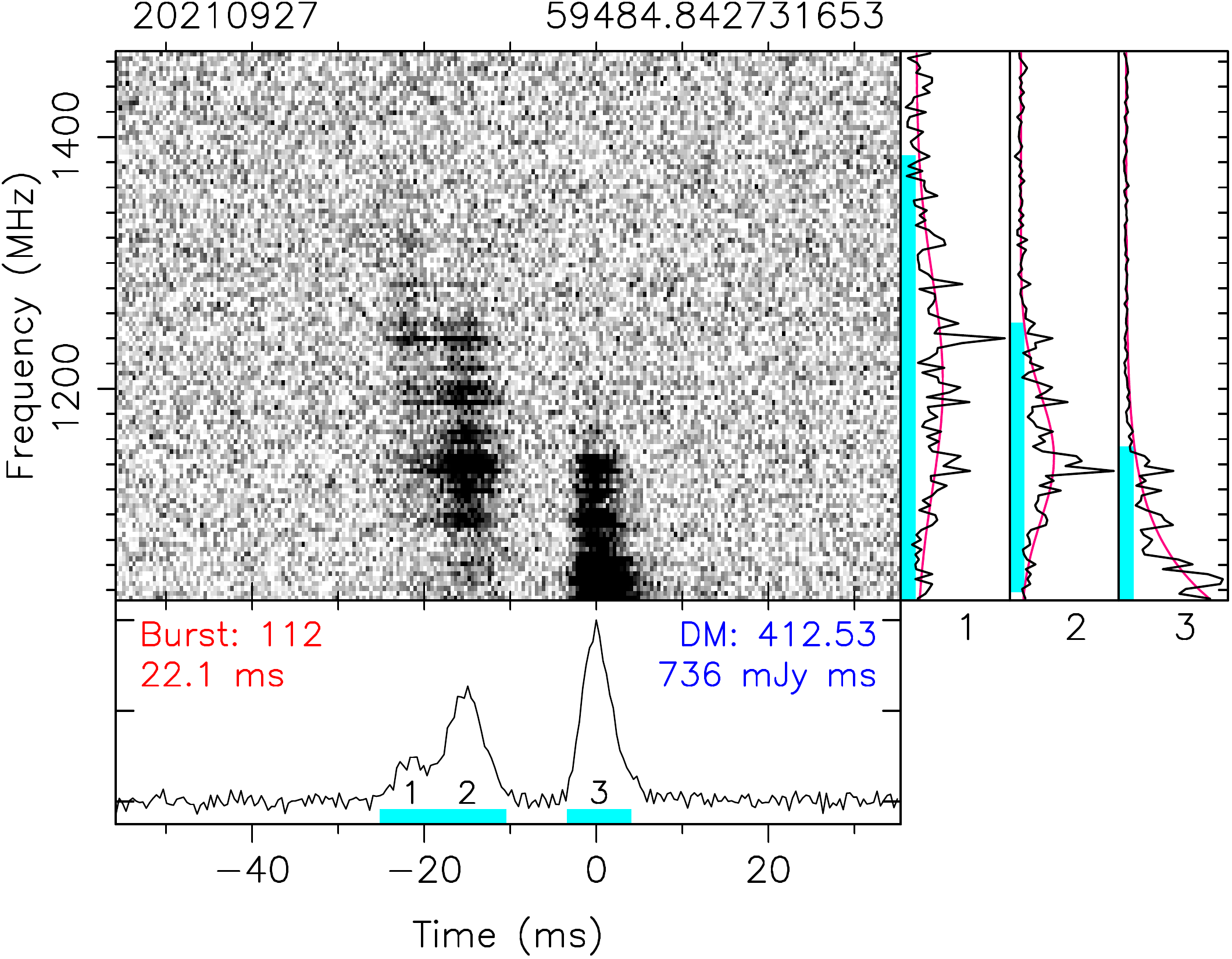}
    \includegraphics[height=37mm]{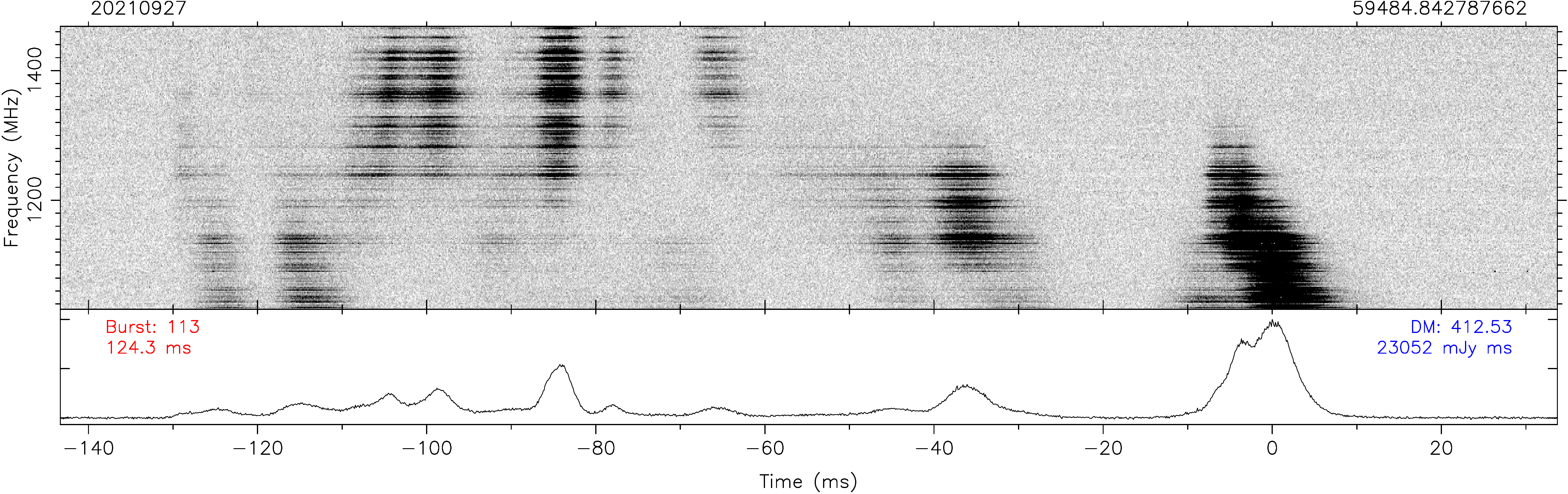}
    \includegraphics[height=37mm]{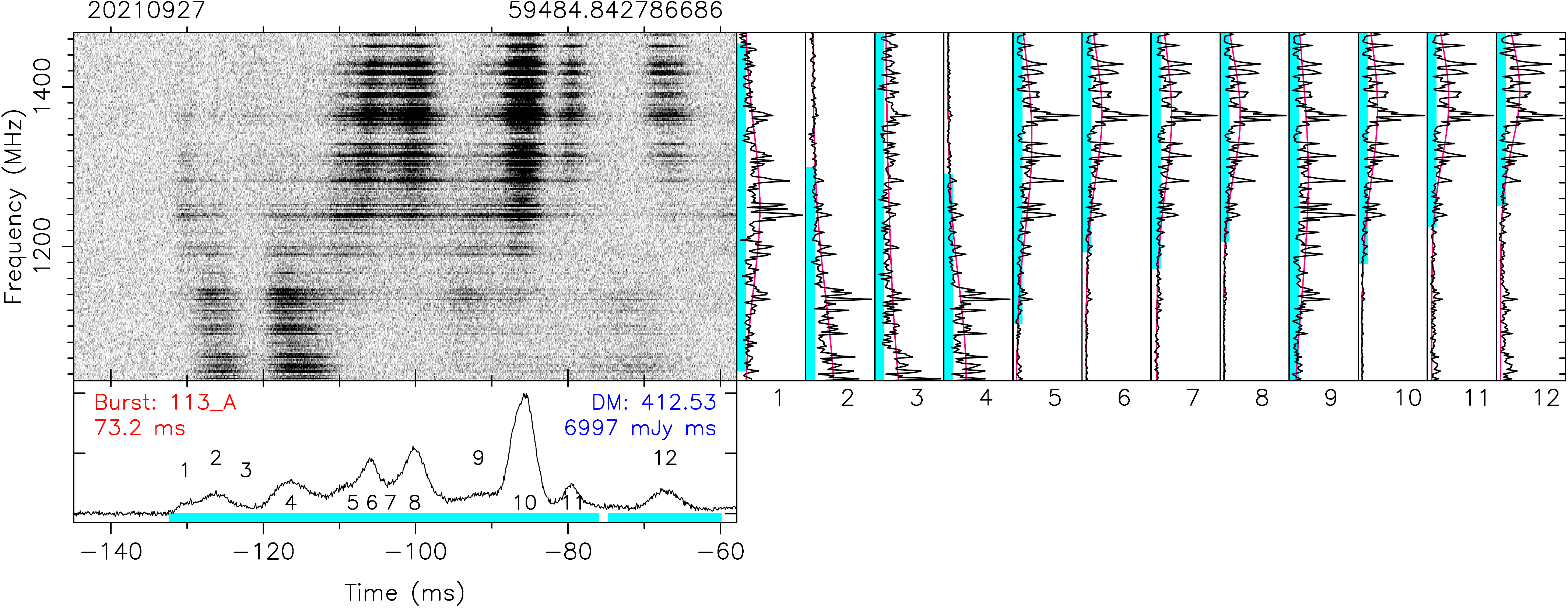}
    \includegraphics[height=37mm]{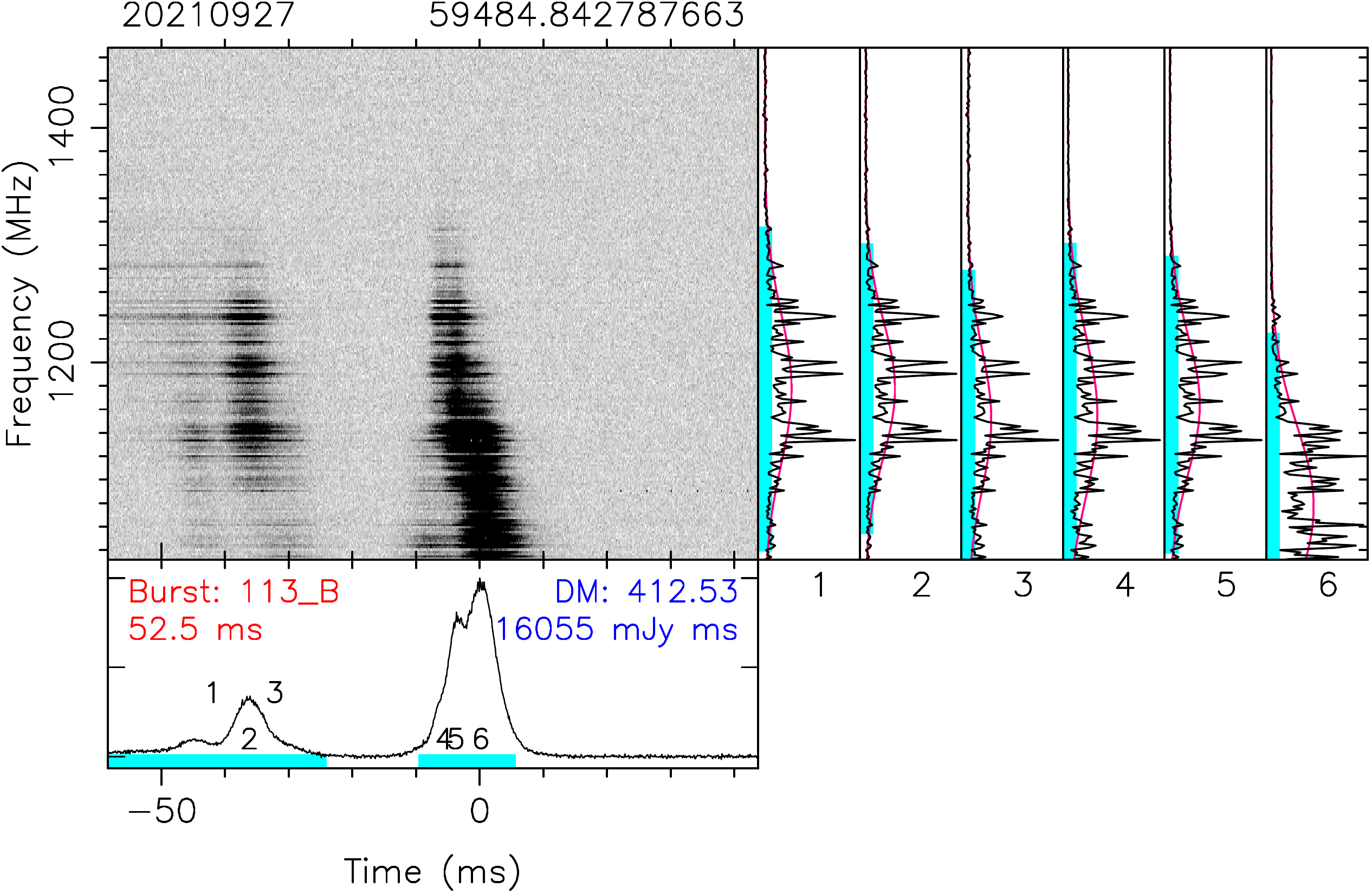}
    \includegraphics[height=37mm]{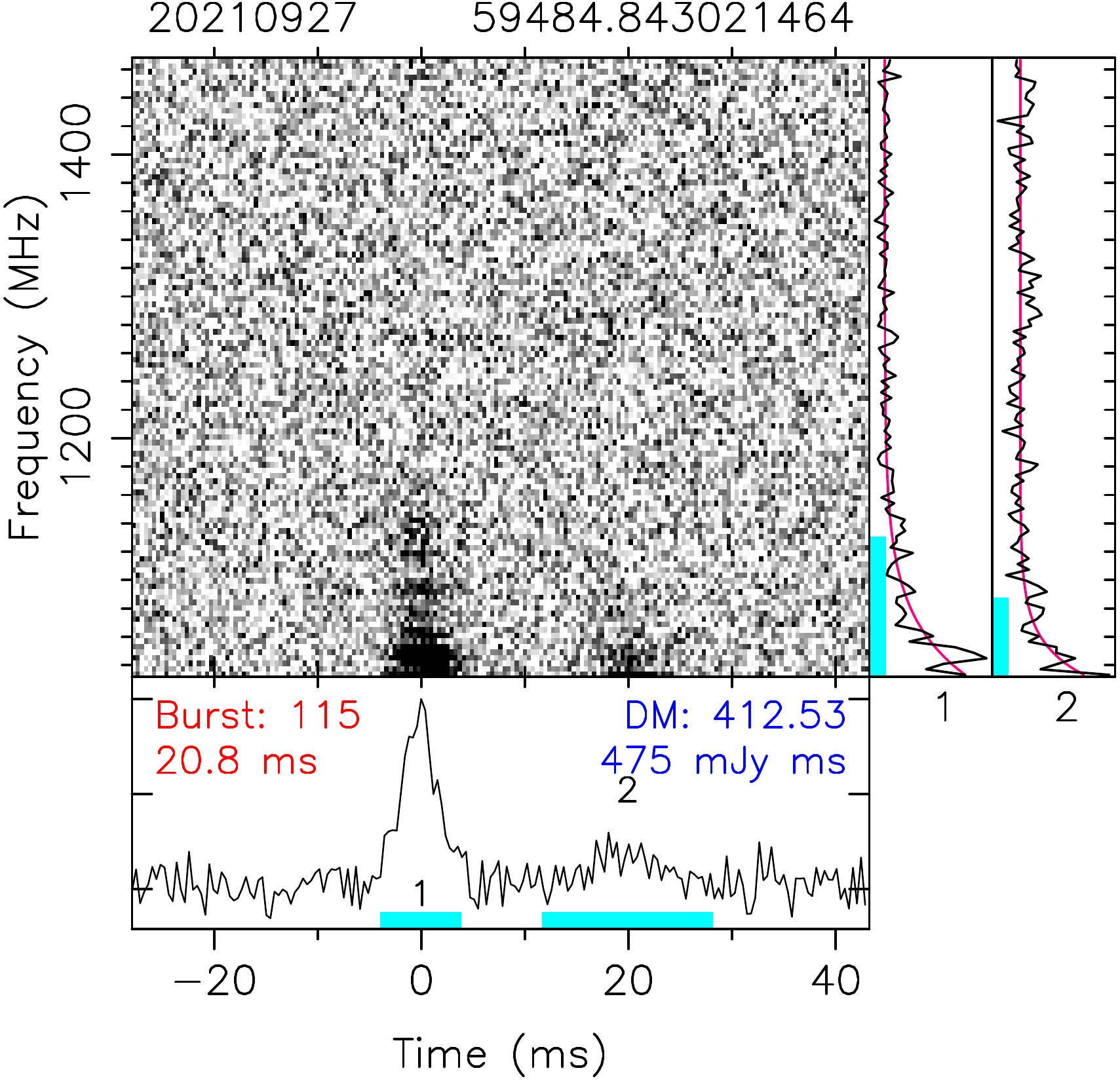}
    \includegraphics[height=37mm]{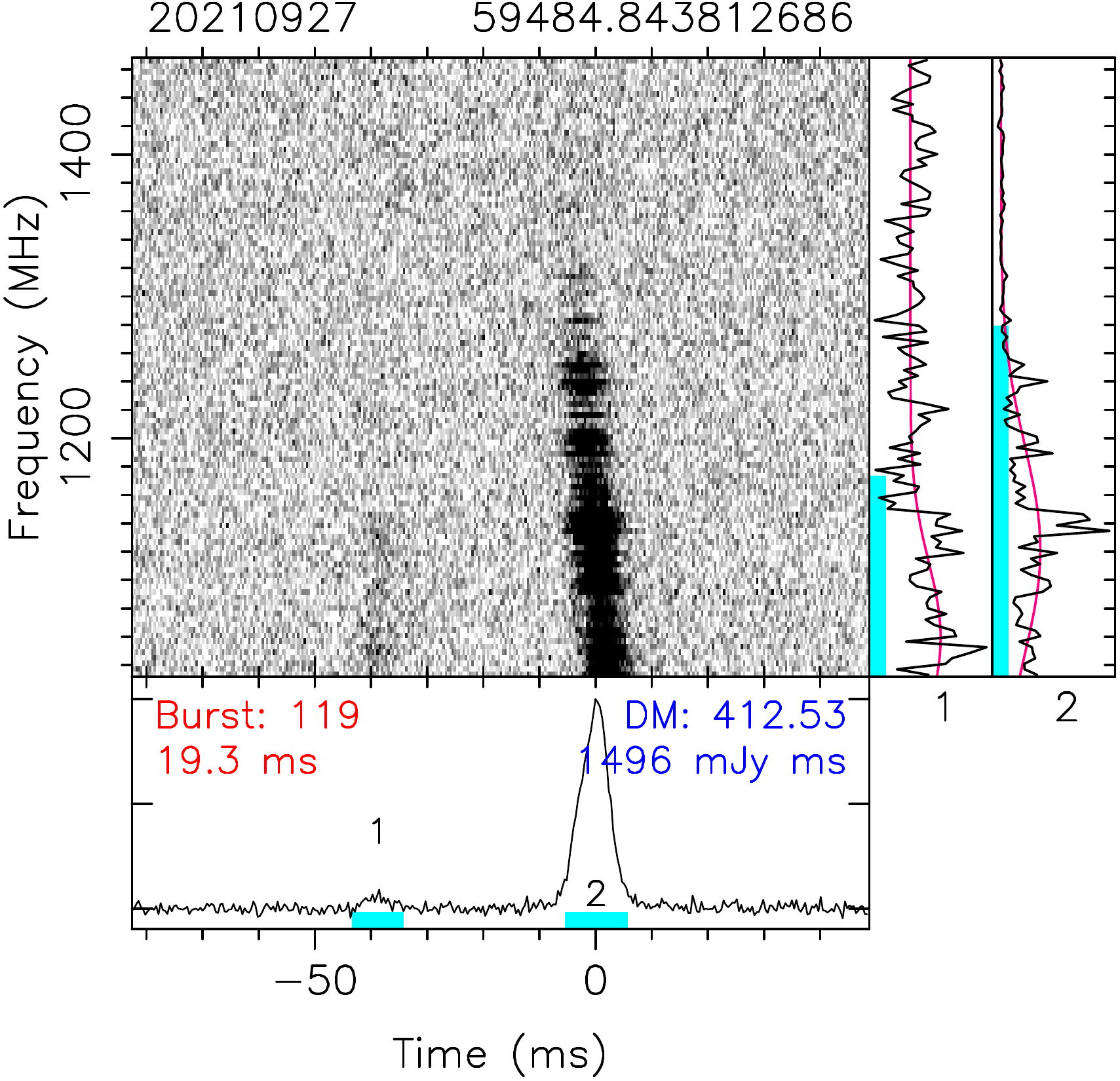}
    \includegraphics[height=37mm]{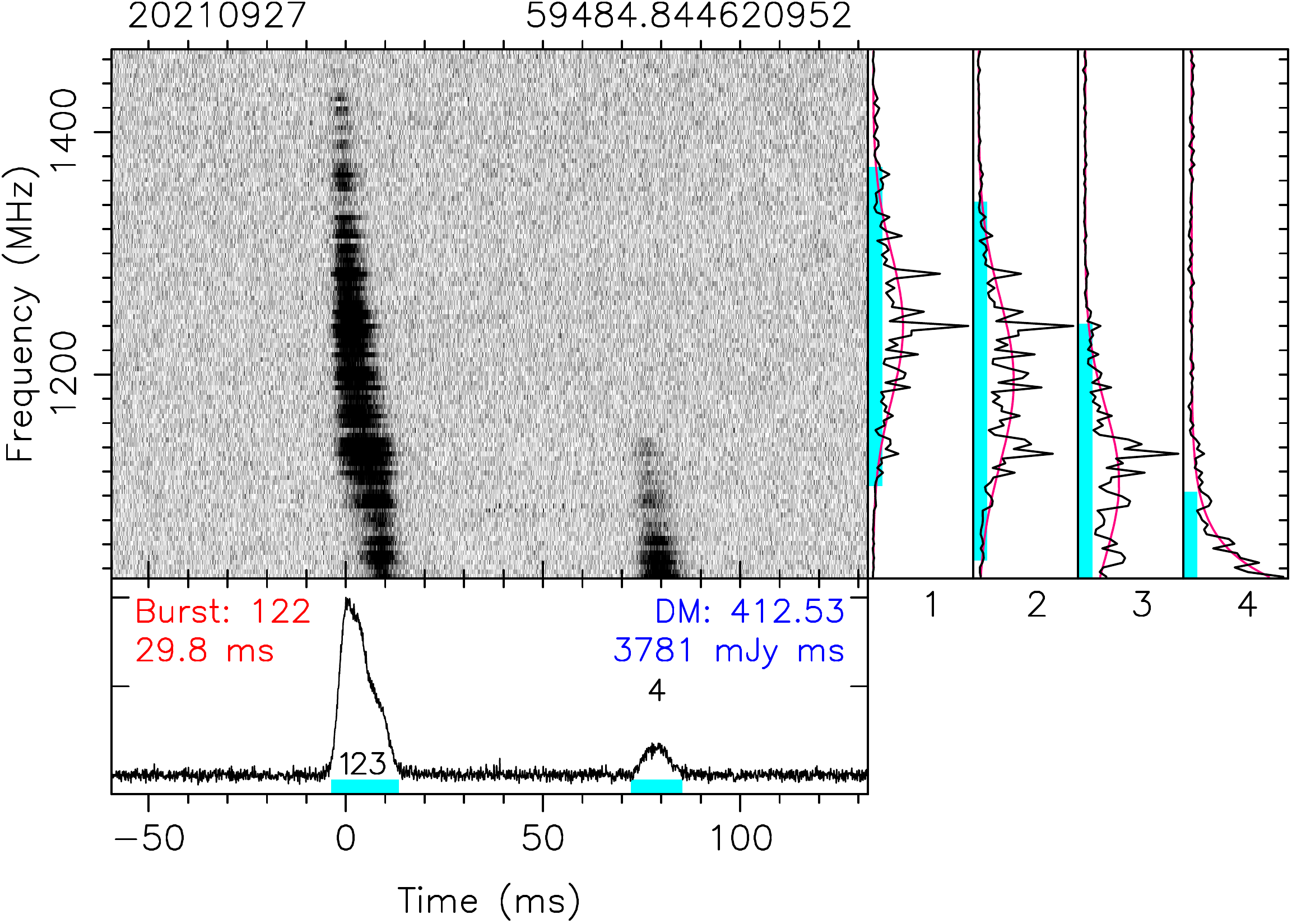} 
    \includegraphics[height=37mm]{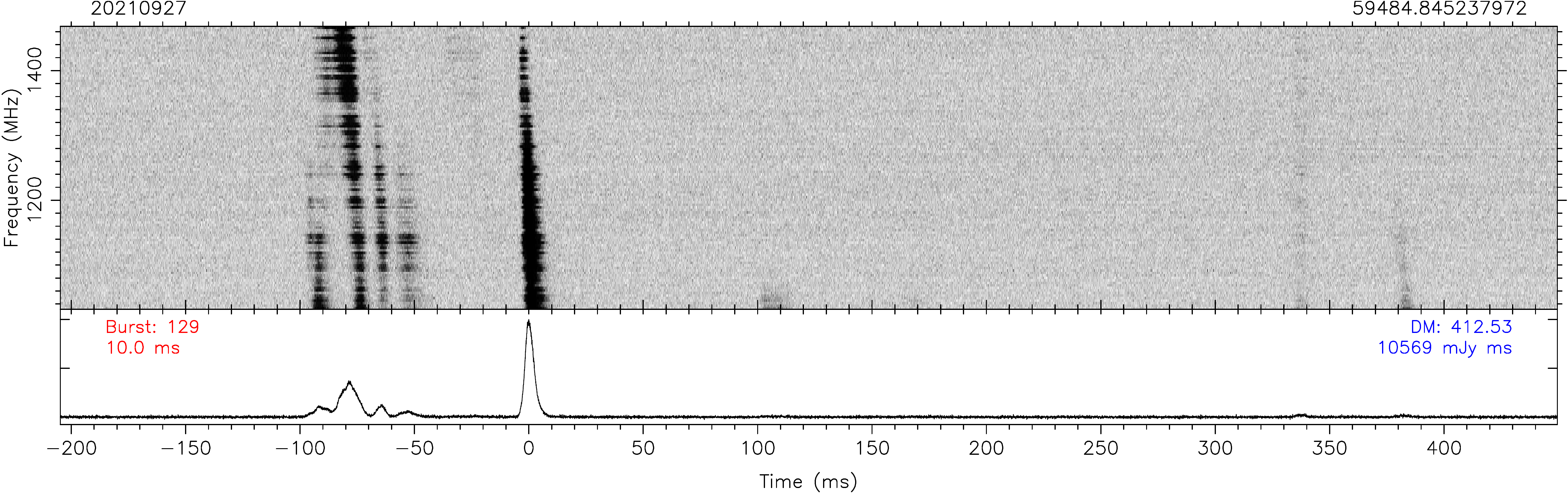} 
\caption{\it{ -- continued}.
}
\end{figure*}
\addtocounter{figure}{-1}
\begin{figure*}
    \flushleft
    \includegraphics[height=37mm]{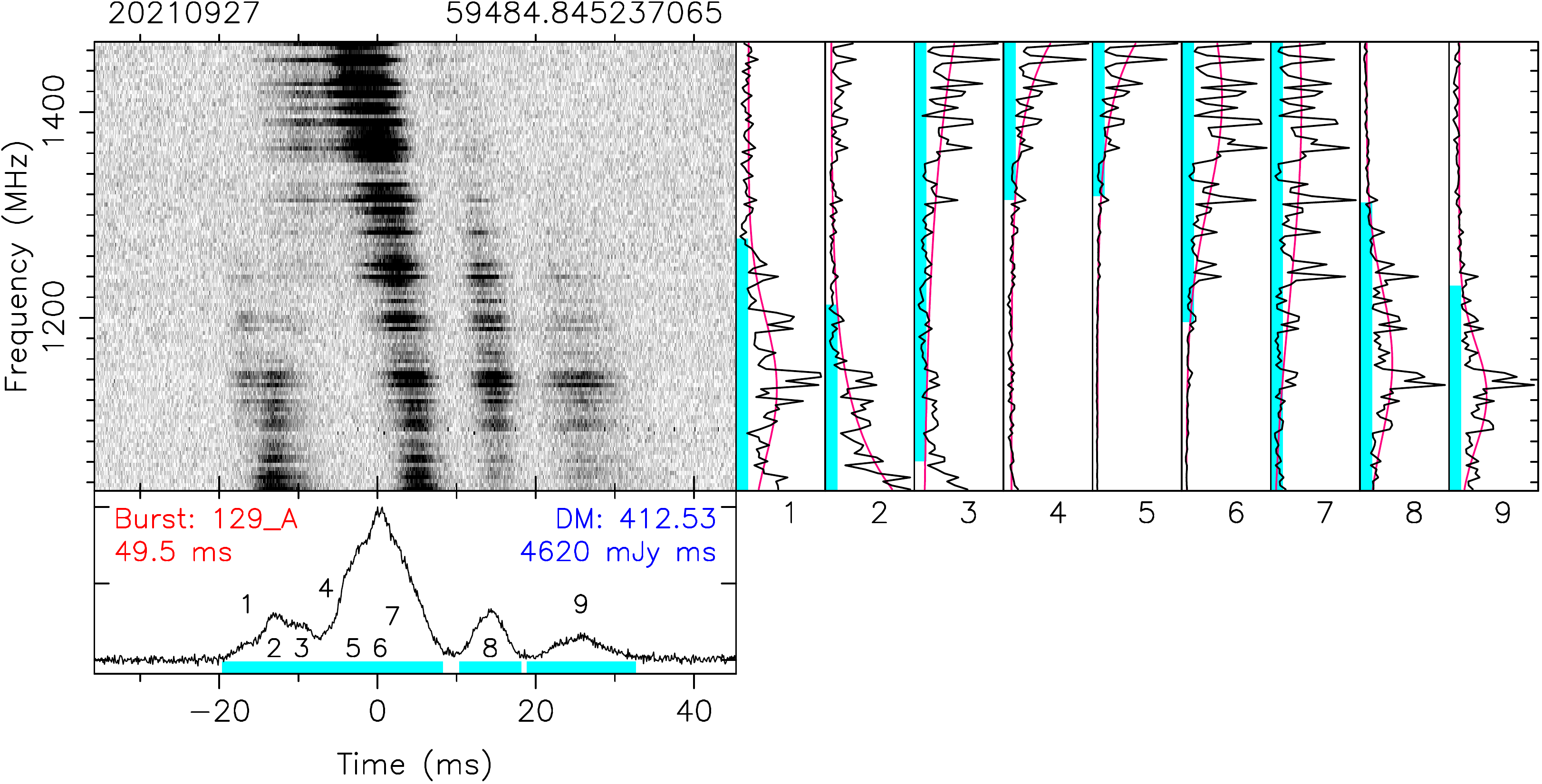} 
    \includegraphics[height=37mm]{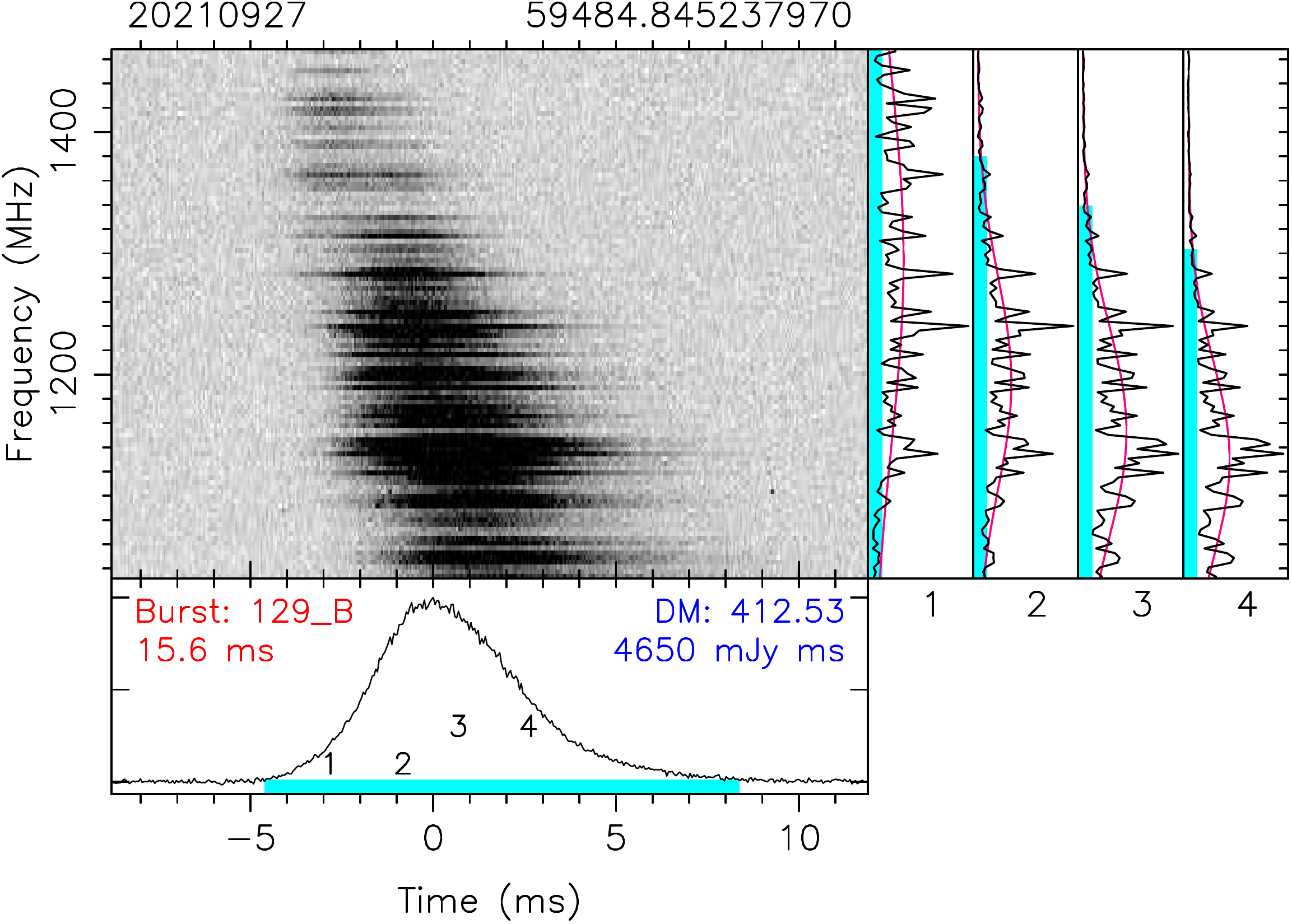} 
    \includegraphics[height=37mm]{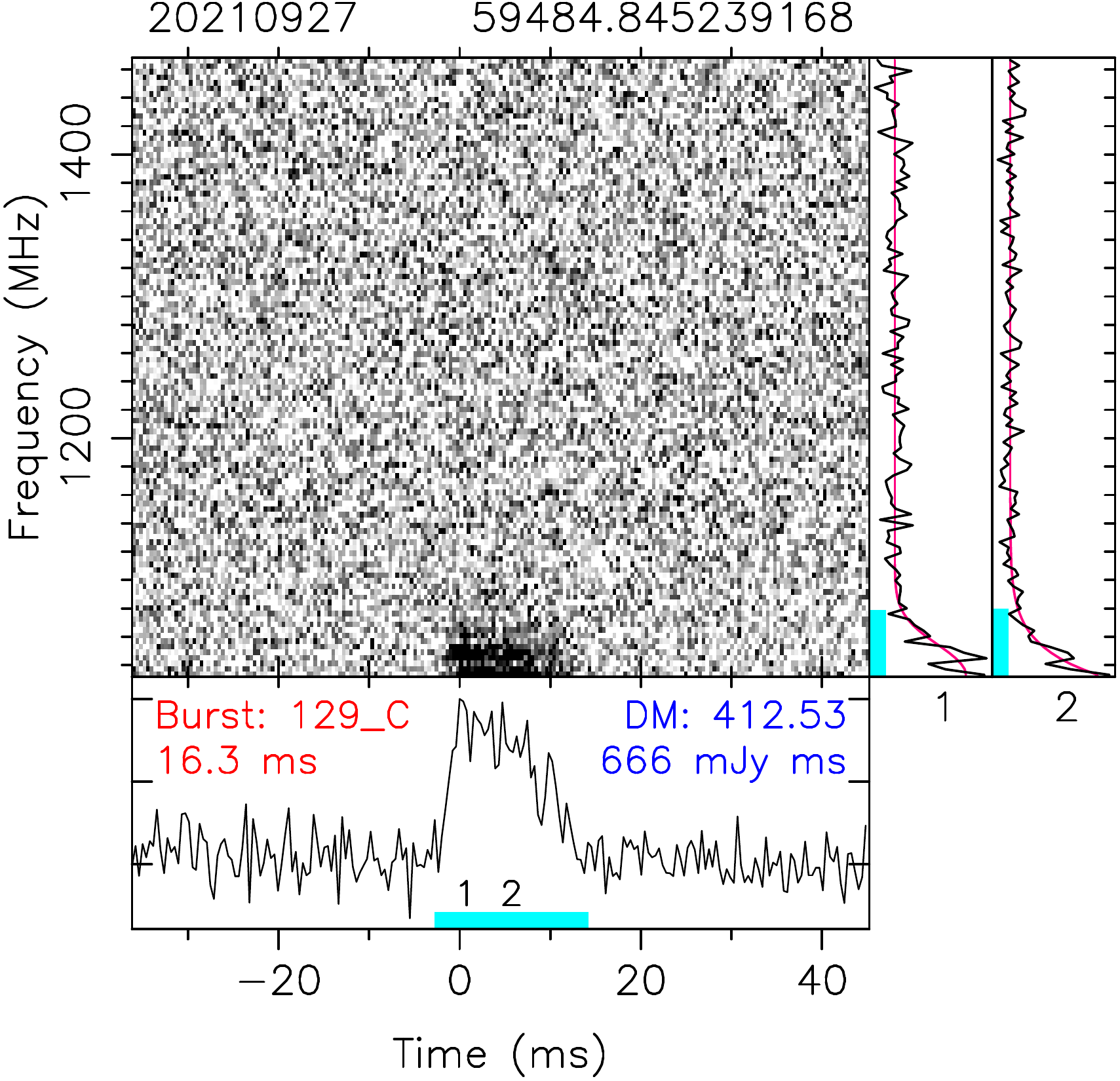} 
    \includegraphics[height=37mm]{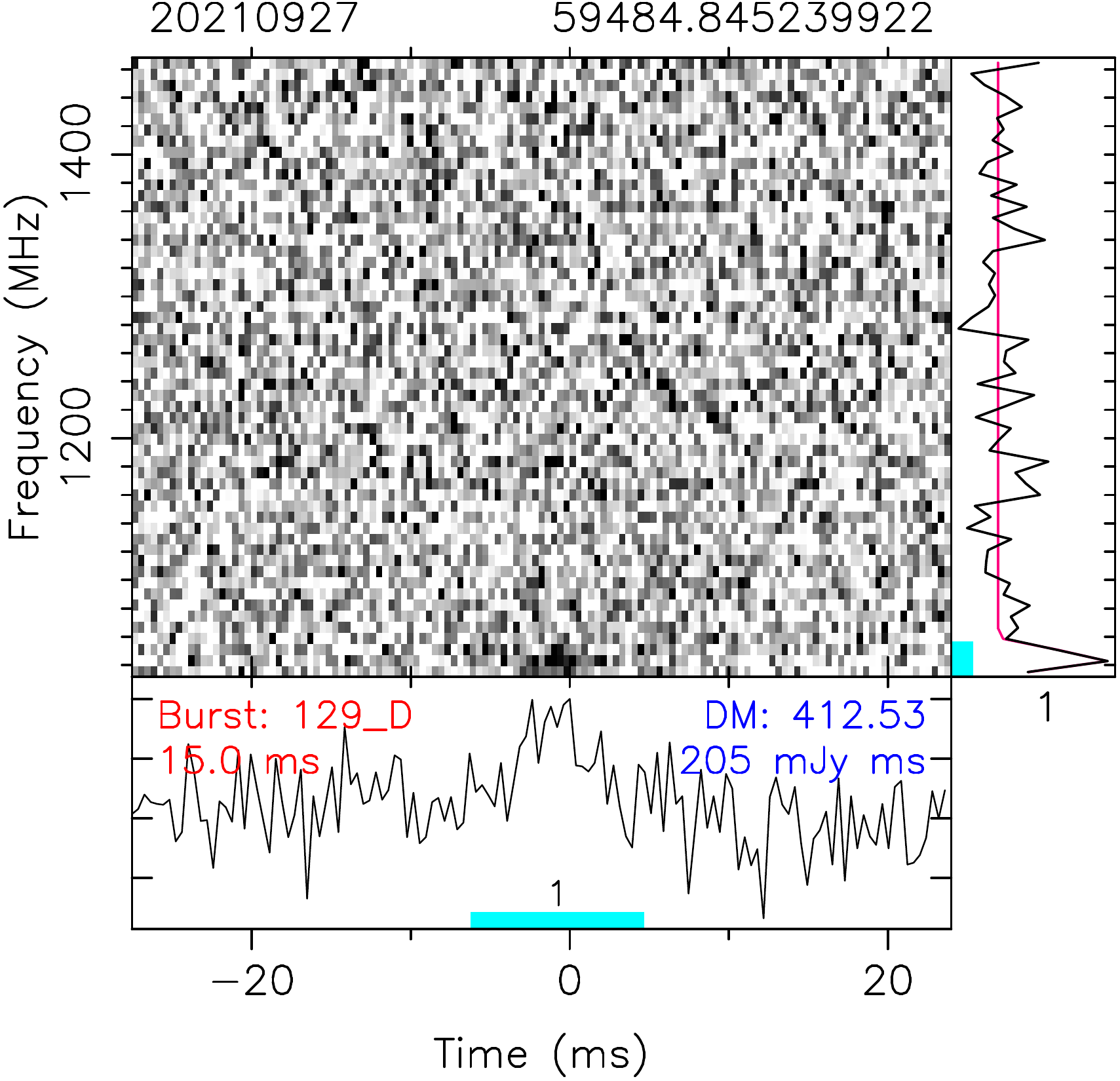} 
    \includegraphics[height=37mm]{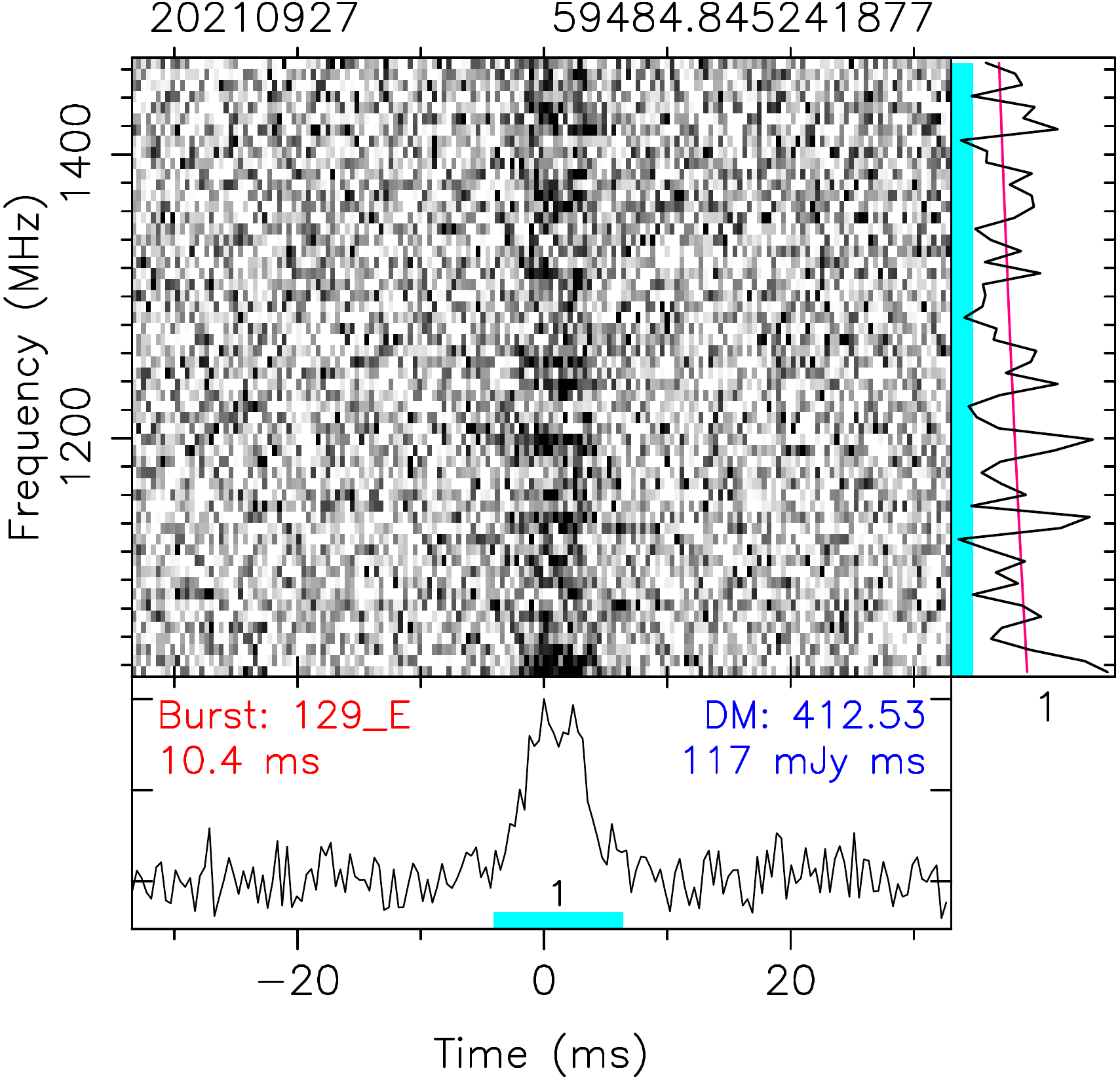} 
    \includegraphics[height=37mm]{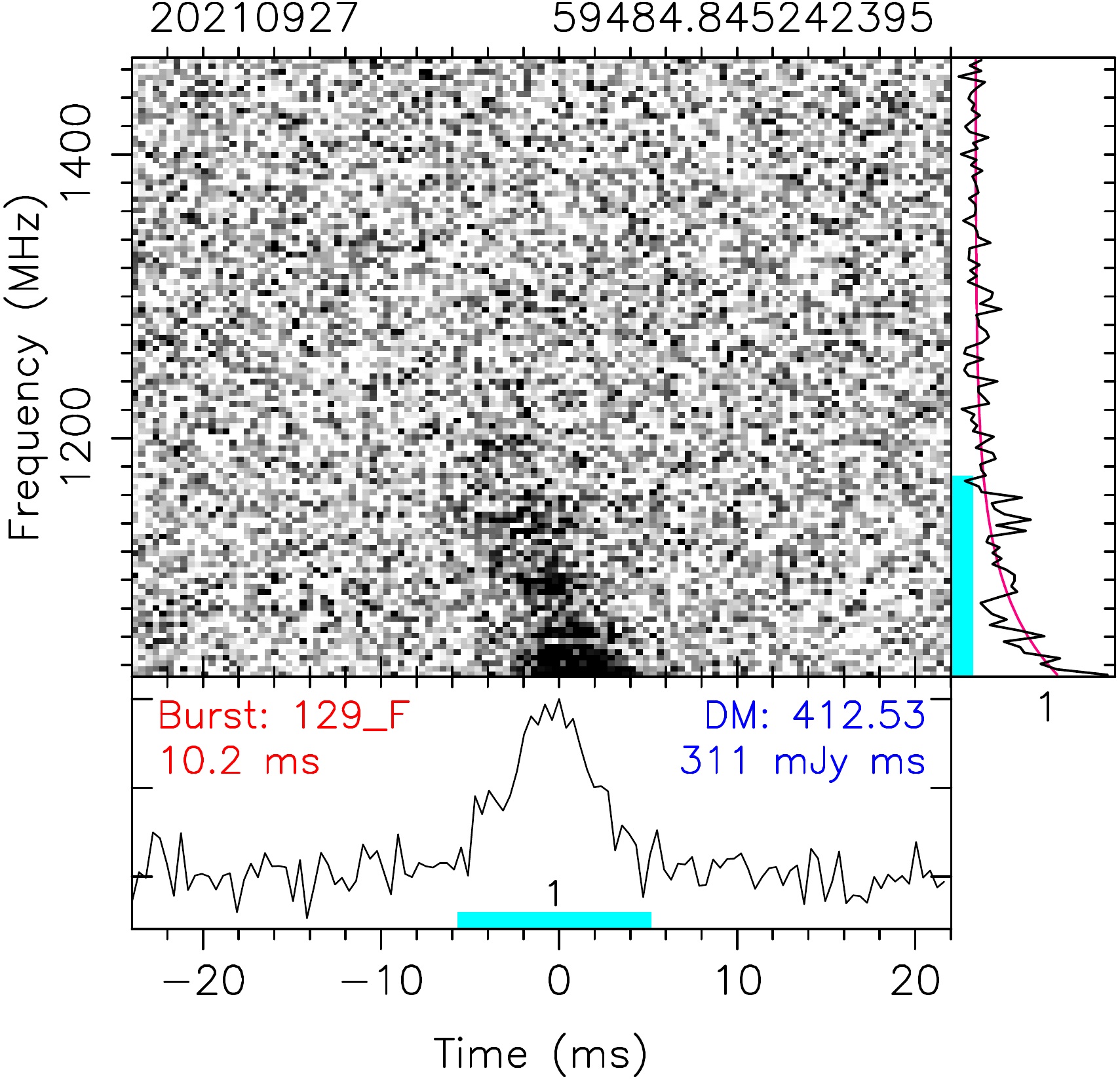} 
    \includegraphics[height=37mm]{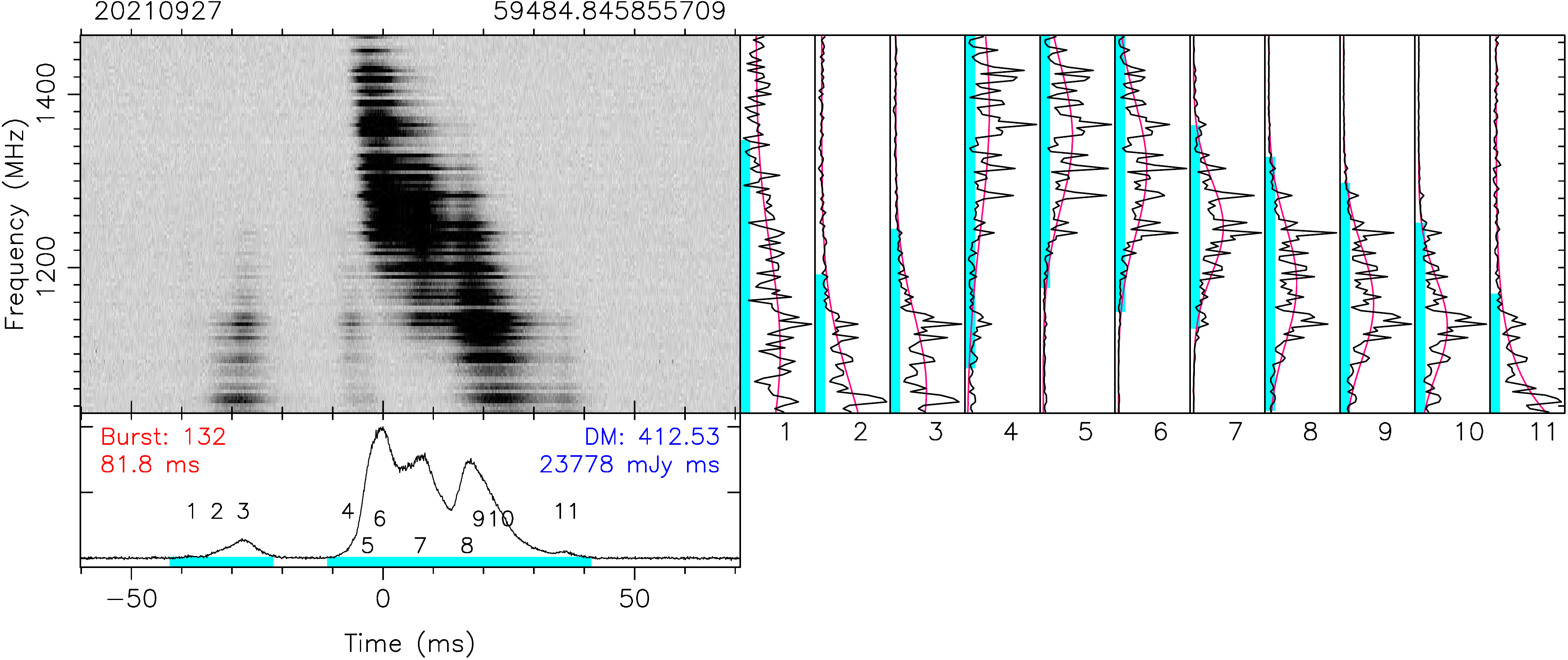} 
    \includegraphics[height=37mm]{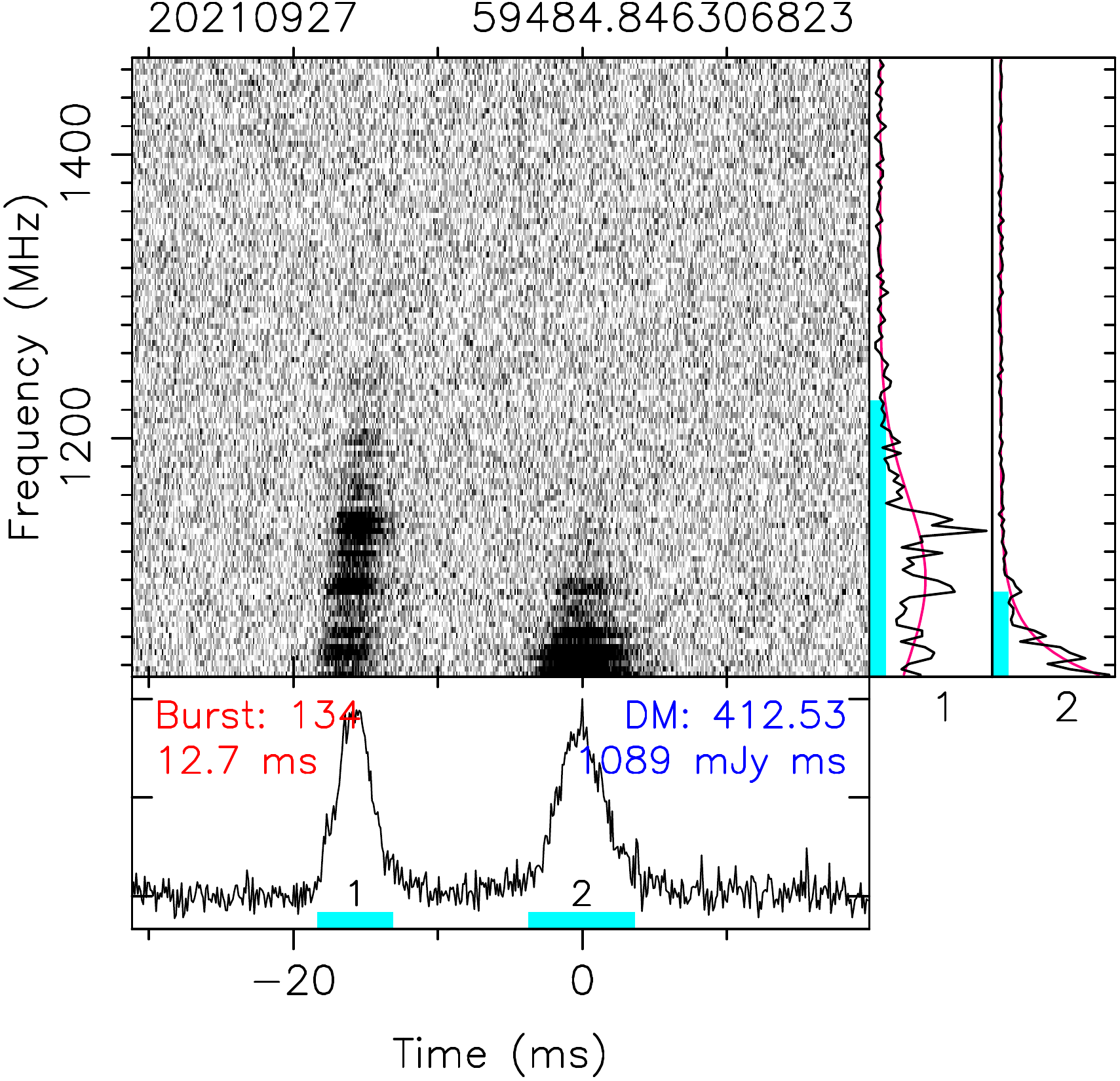}
    \includegraphics[height=37mm]{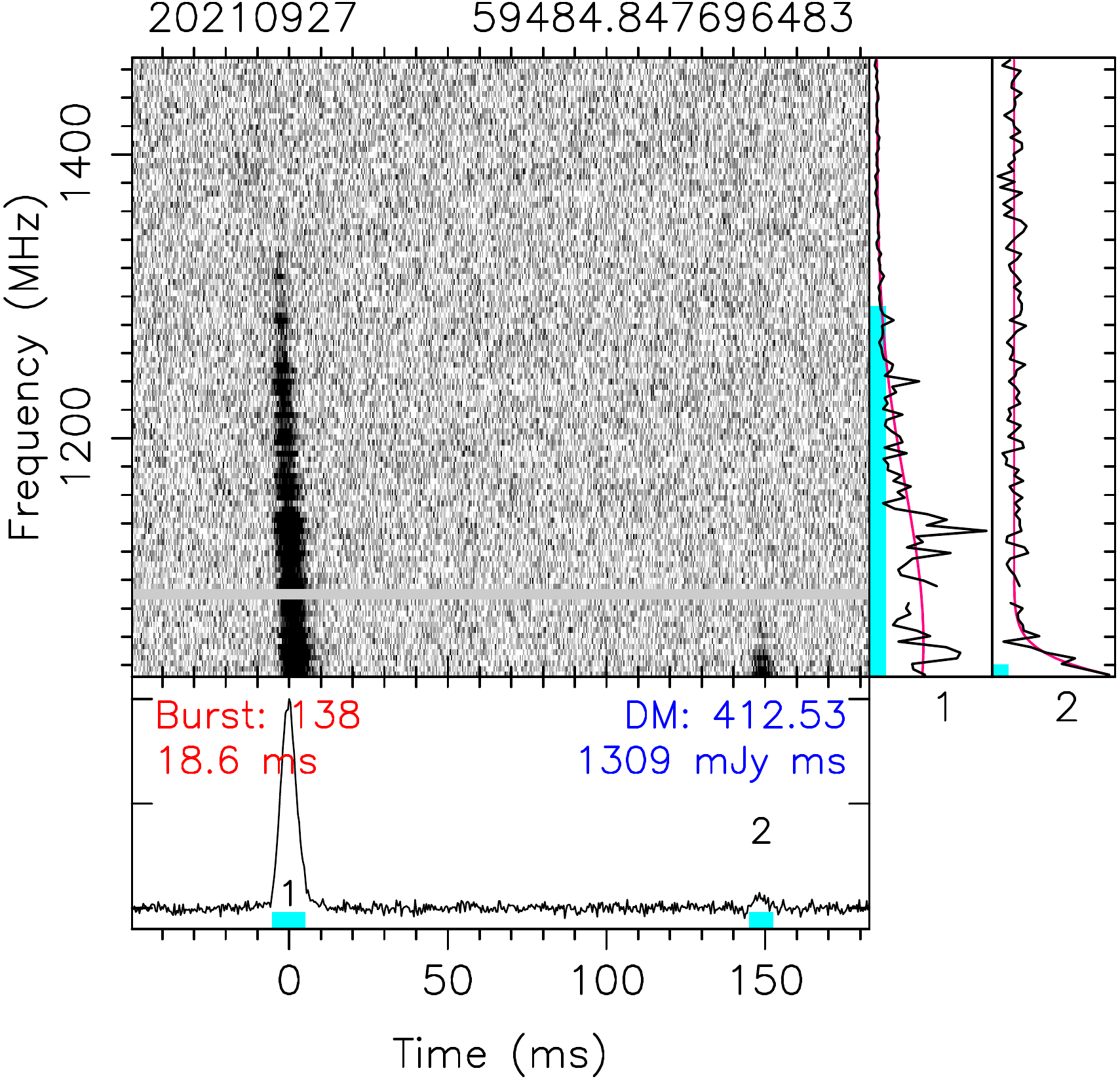} 
    \includegraphics[height=37mm]{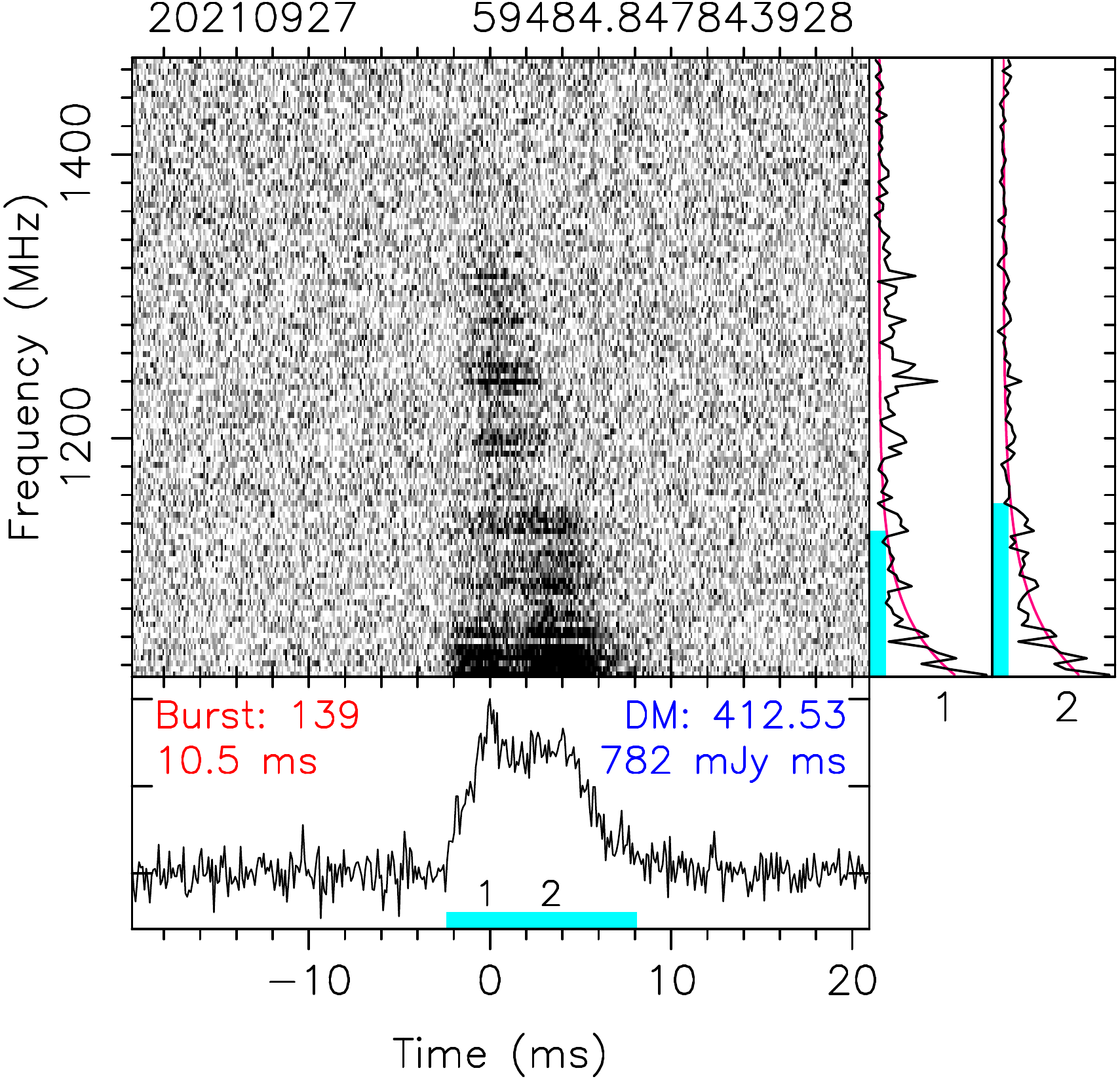}
    \includegraphics[height=37mm]{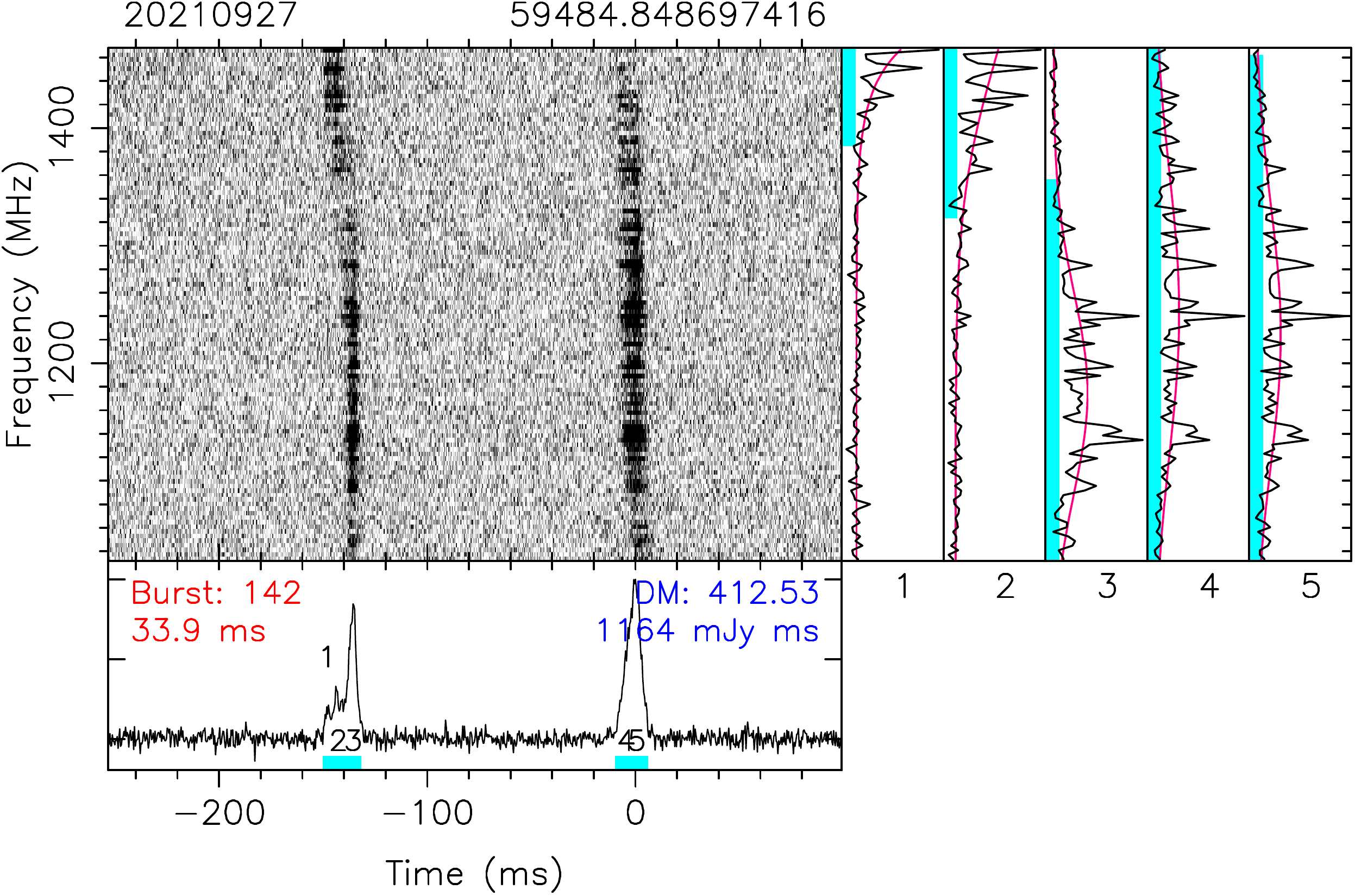} 
    \includegraphics[height=37mm]{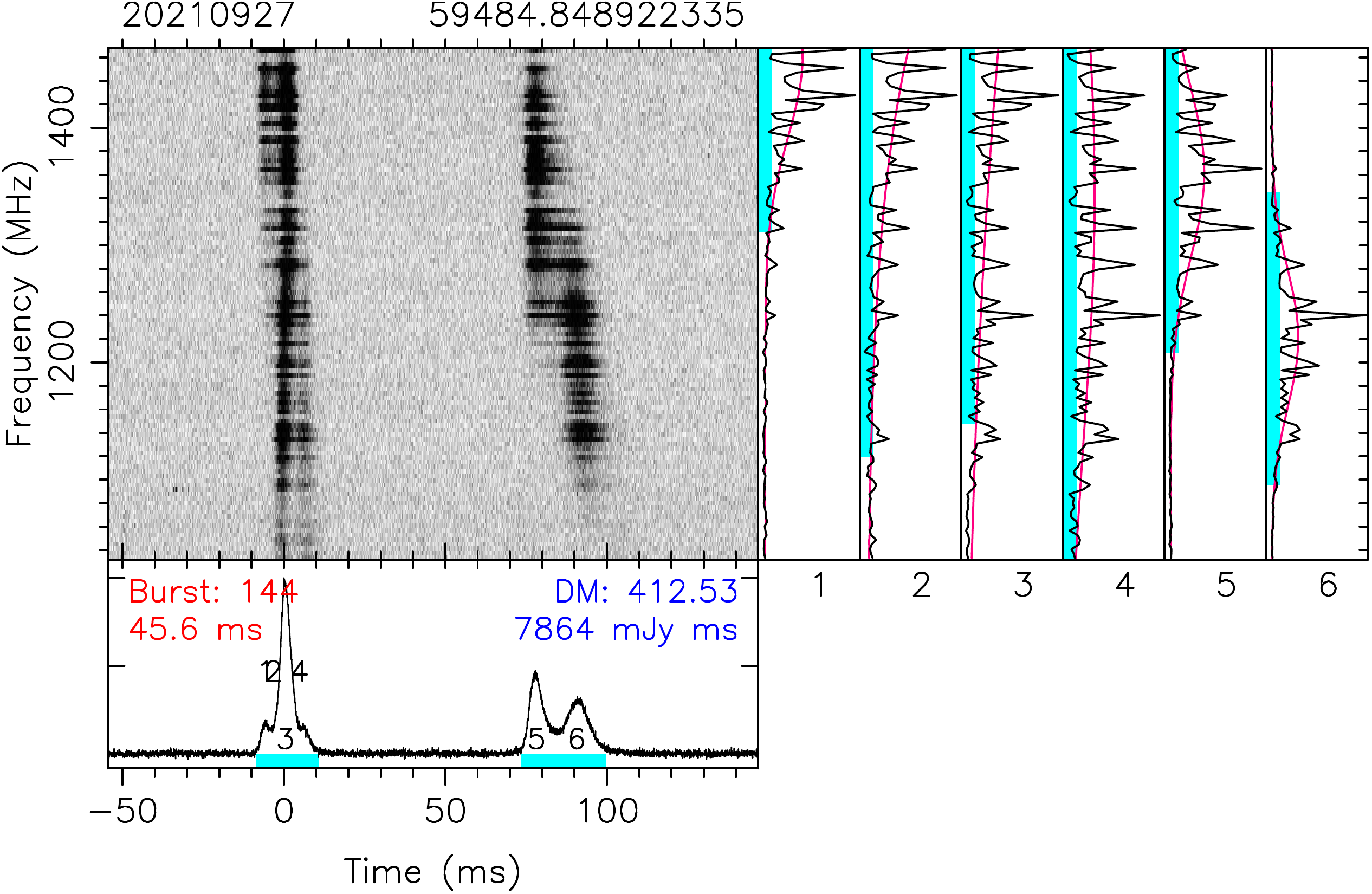} 
    \includegraphics[height=37mm]{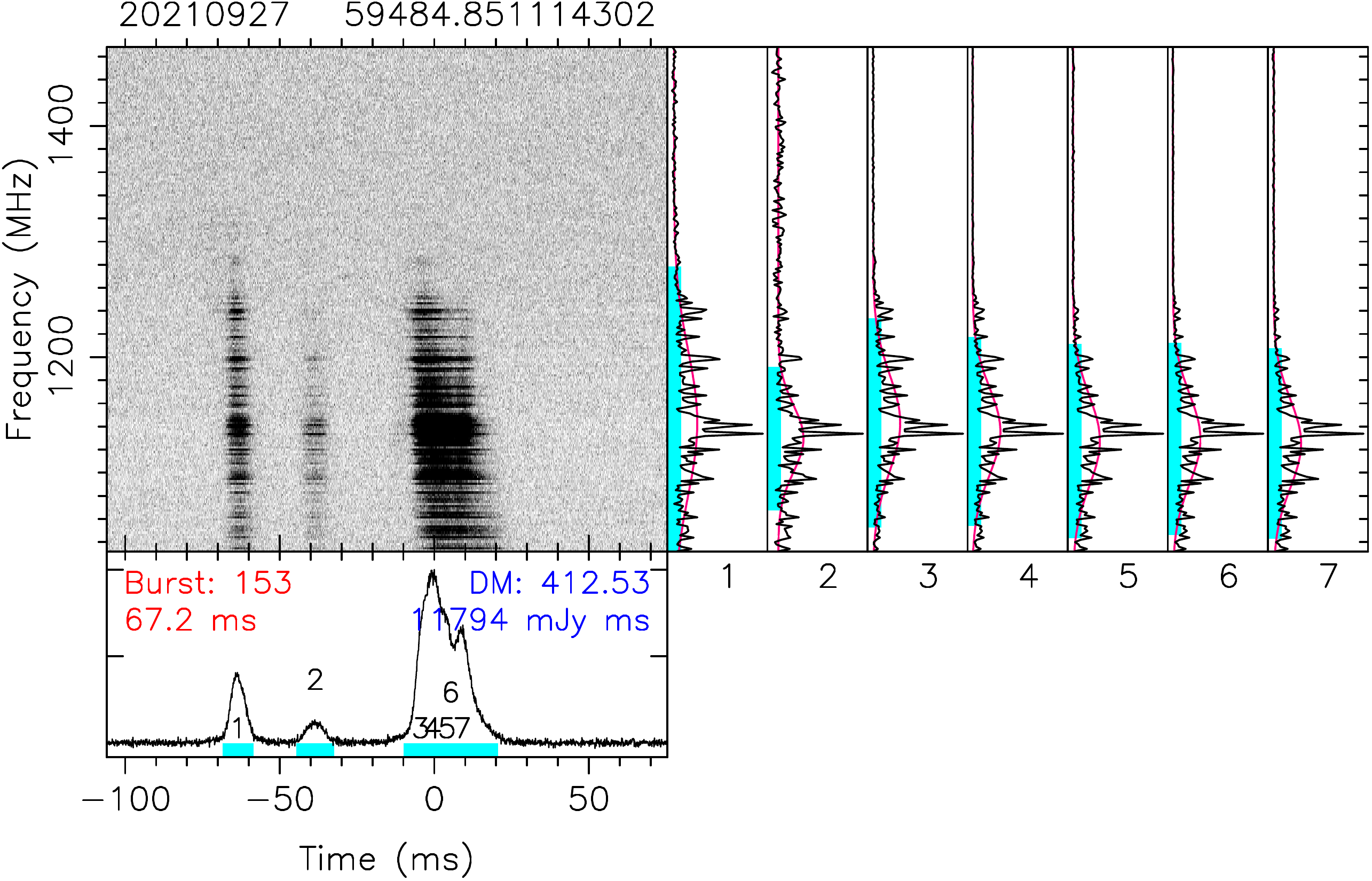} 
    \includegraphics[height=37mm]{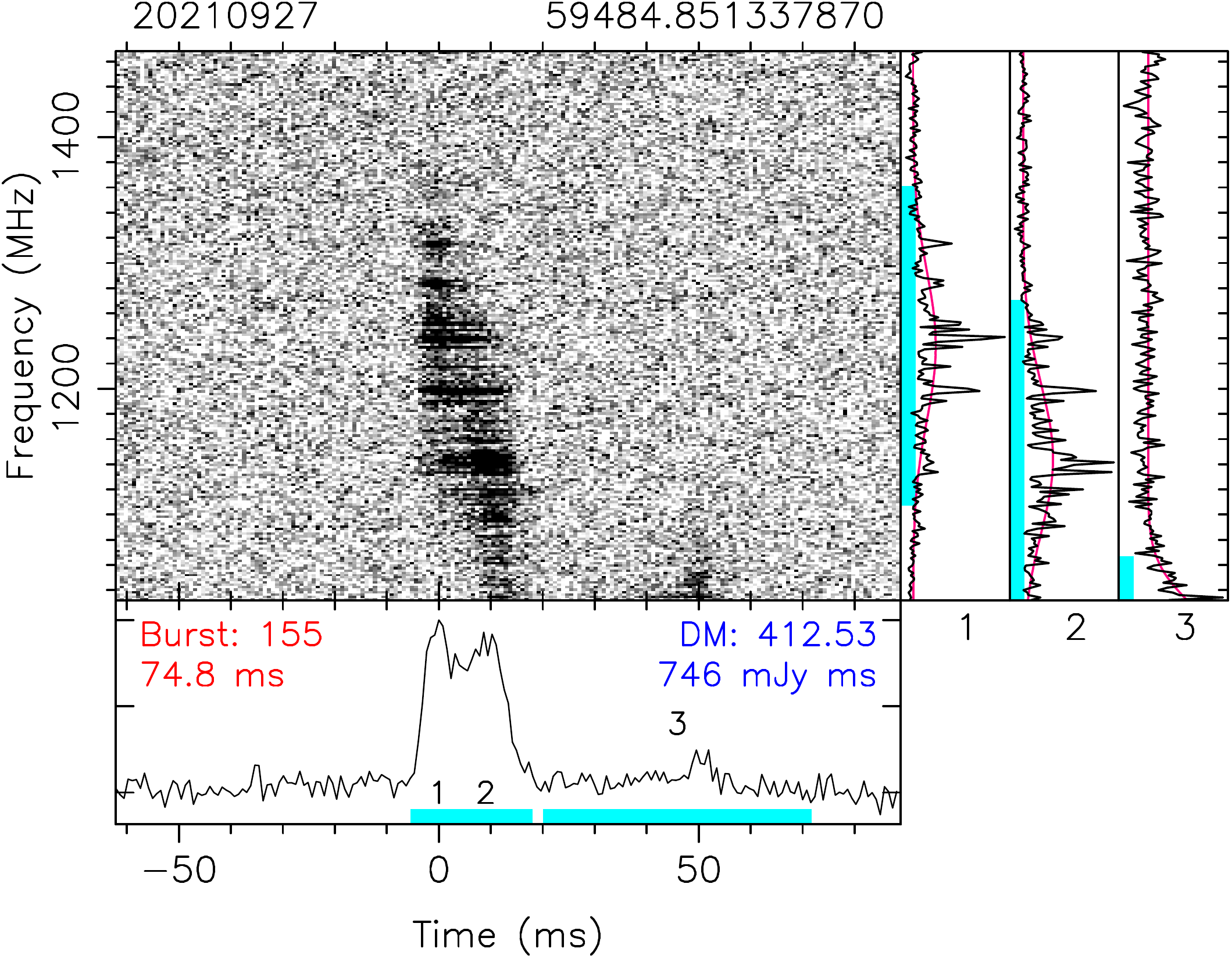}
    \includegraphics[height=37mm]{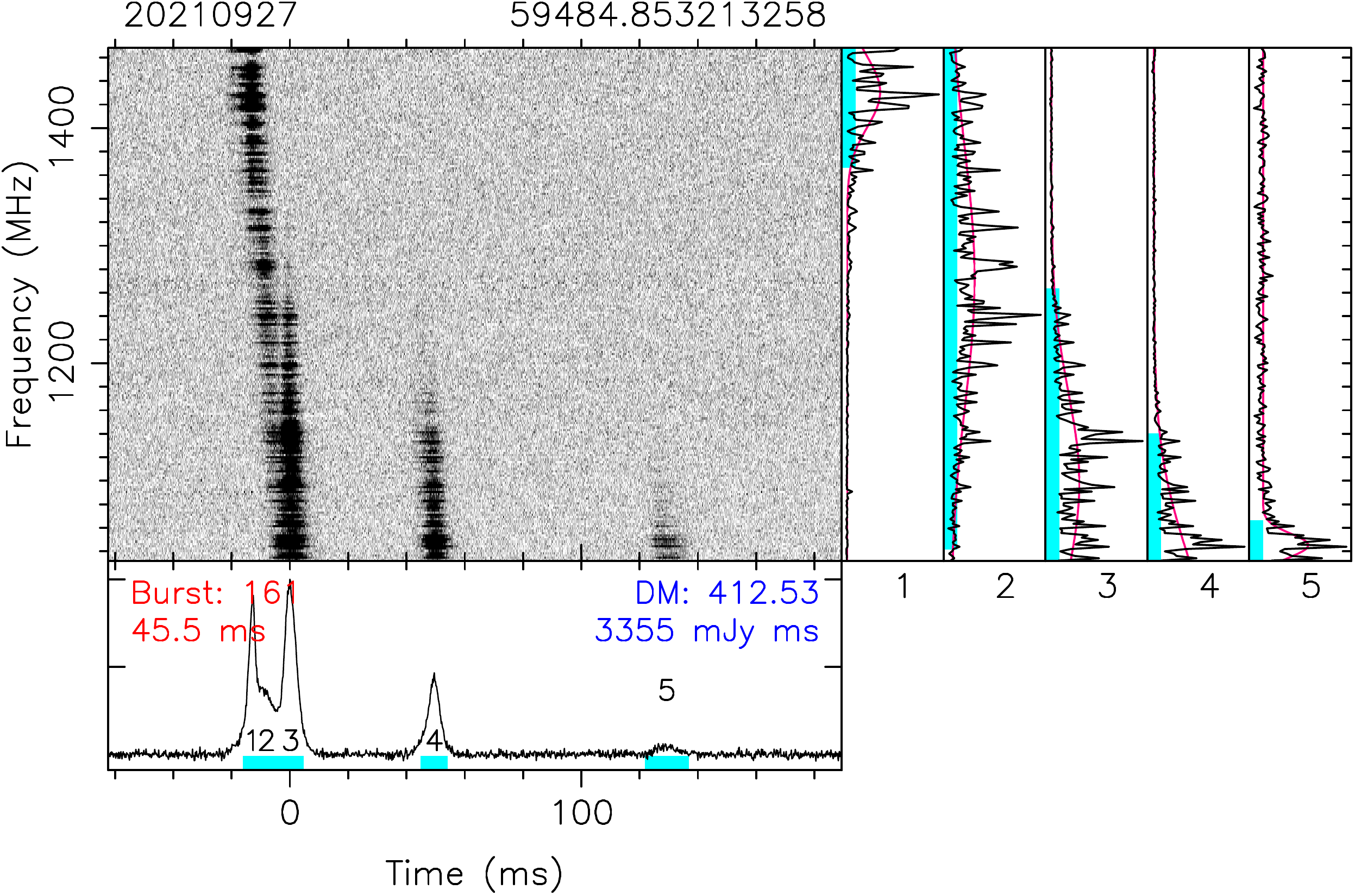} 
    \includegraphics[height=37mm]{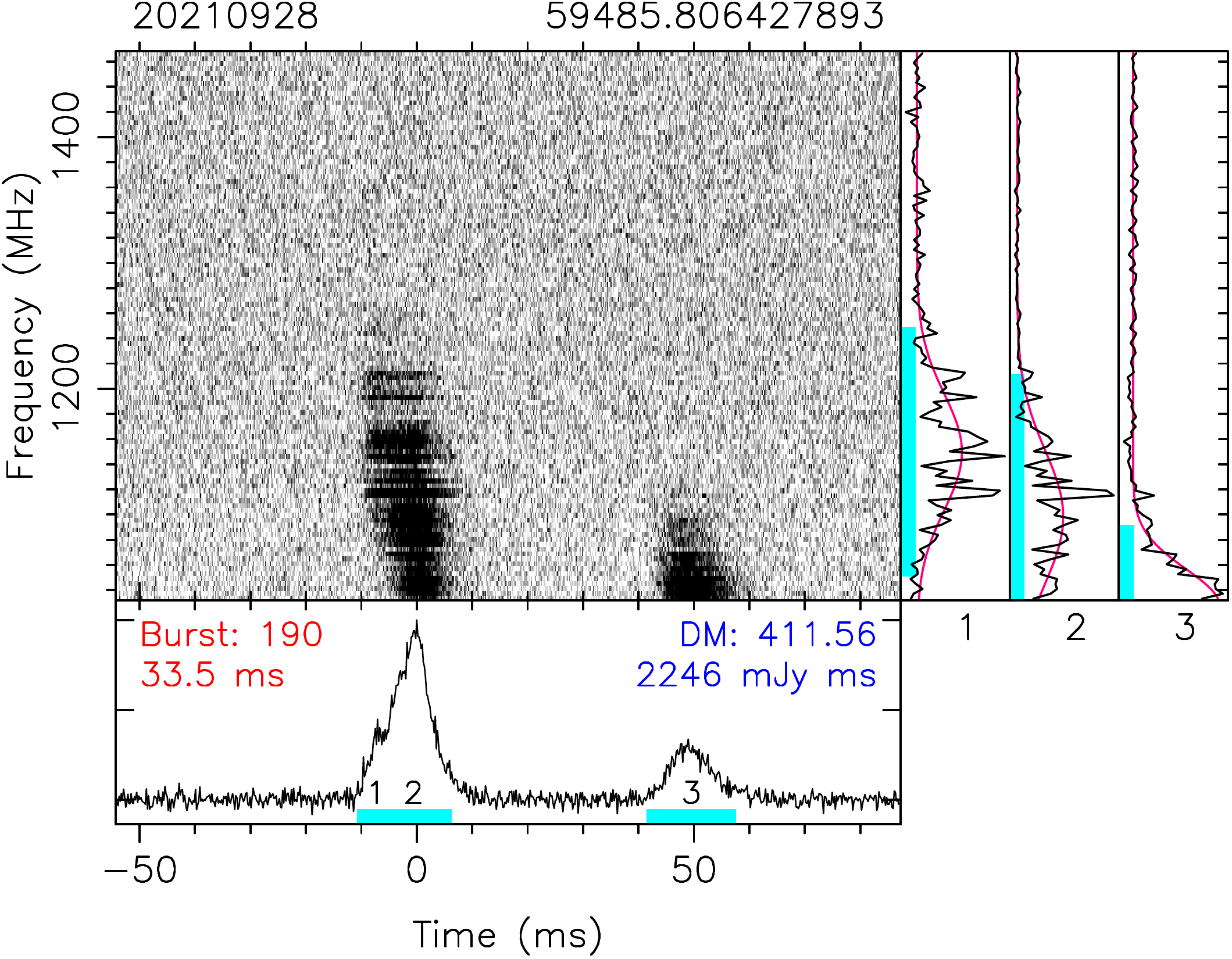}
    \includegraphics[height=37mm]{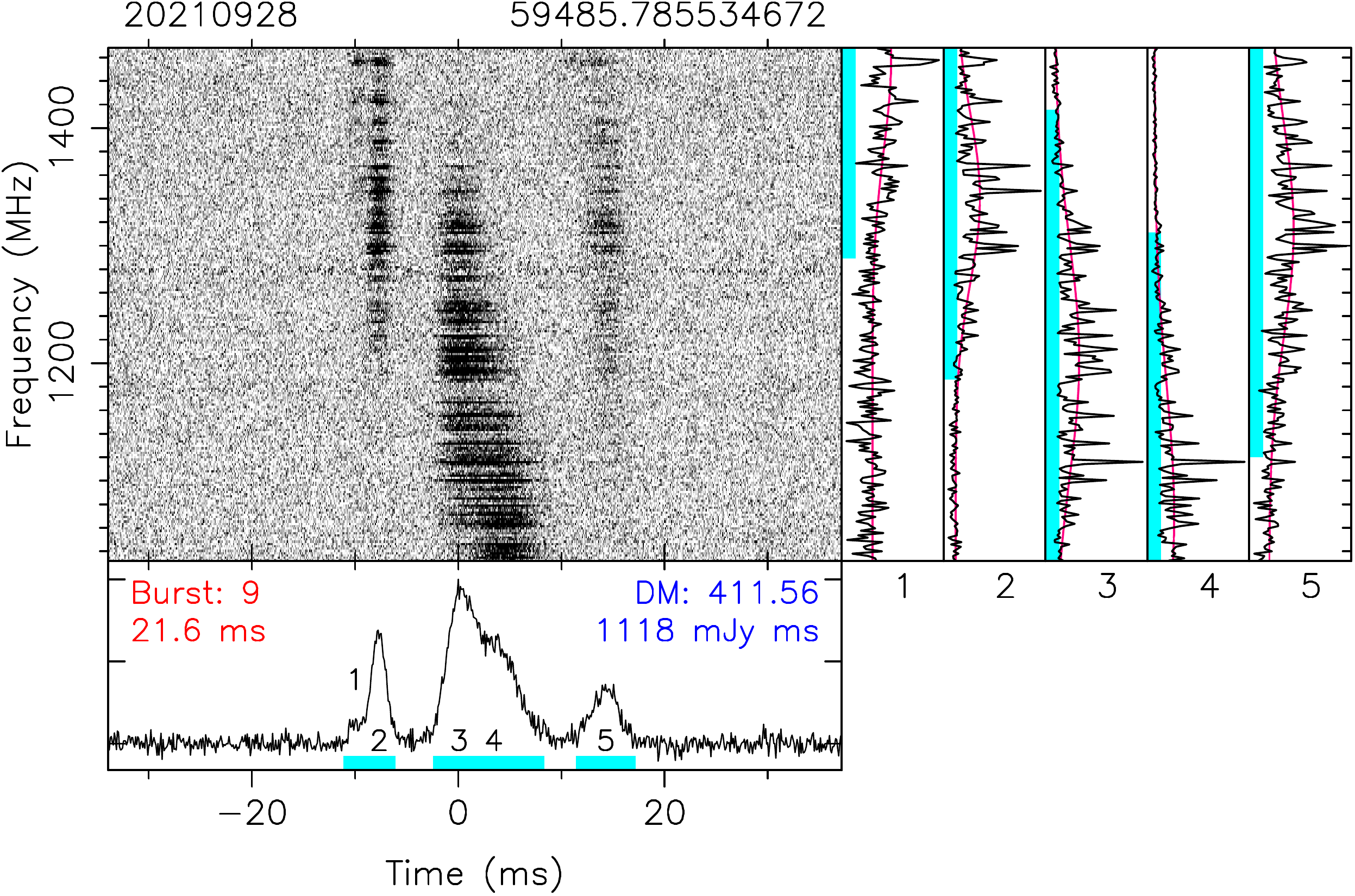}
    \includegraphics[height=37mm]{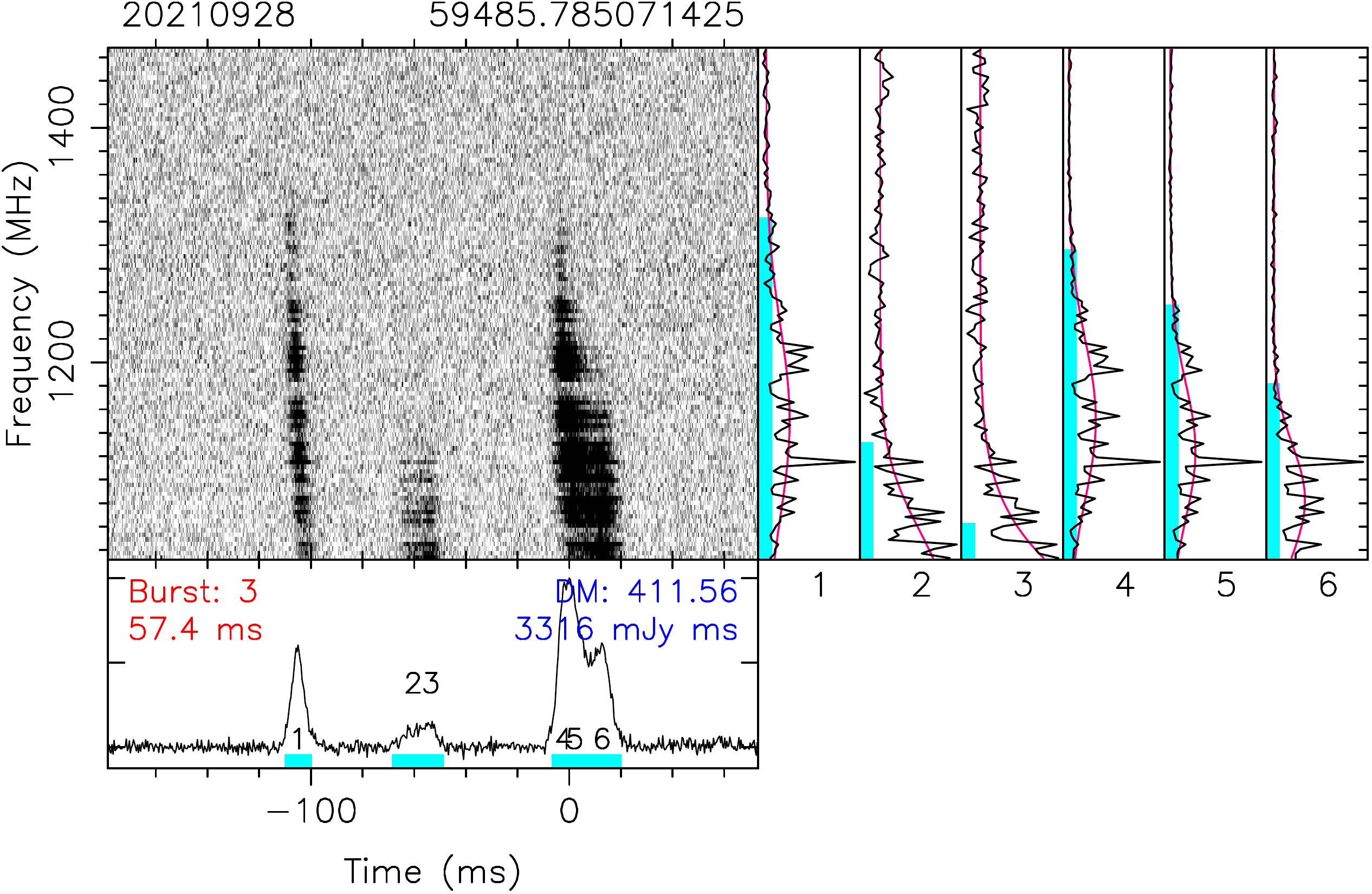}  
\caption{\it{ -- continued}.
}
\end{figure*}
\addtocounter{figure}{-1}
\begin{figure*}
    \flushleft
    \includegraphics[height=37mm]{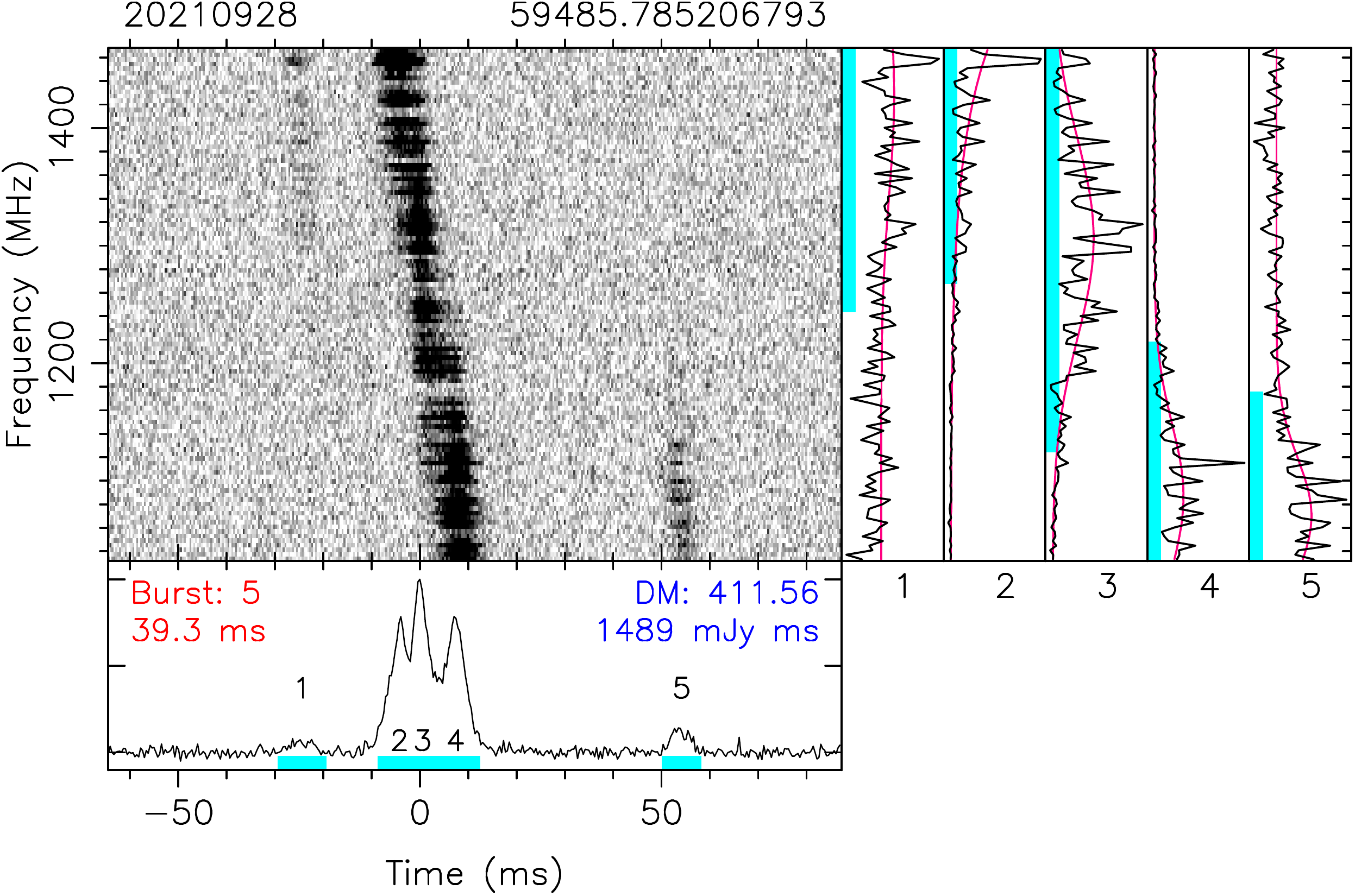}
    \includegraphics[height=37mm]{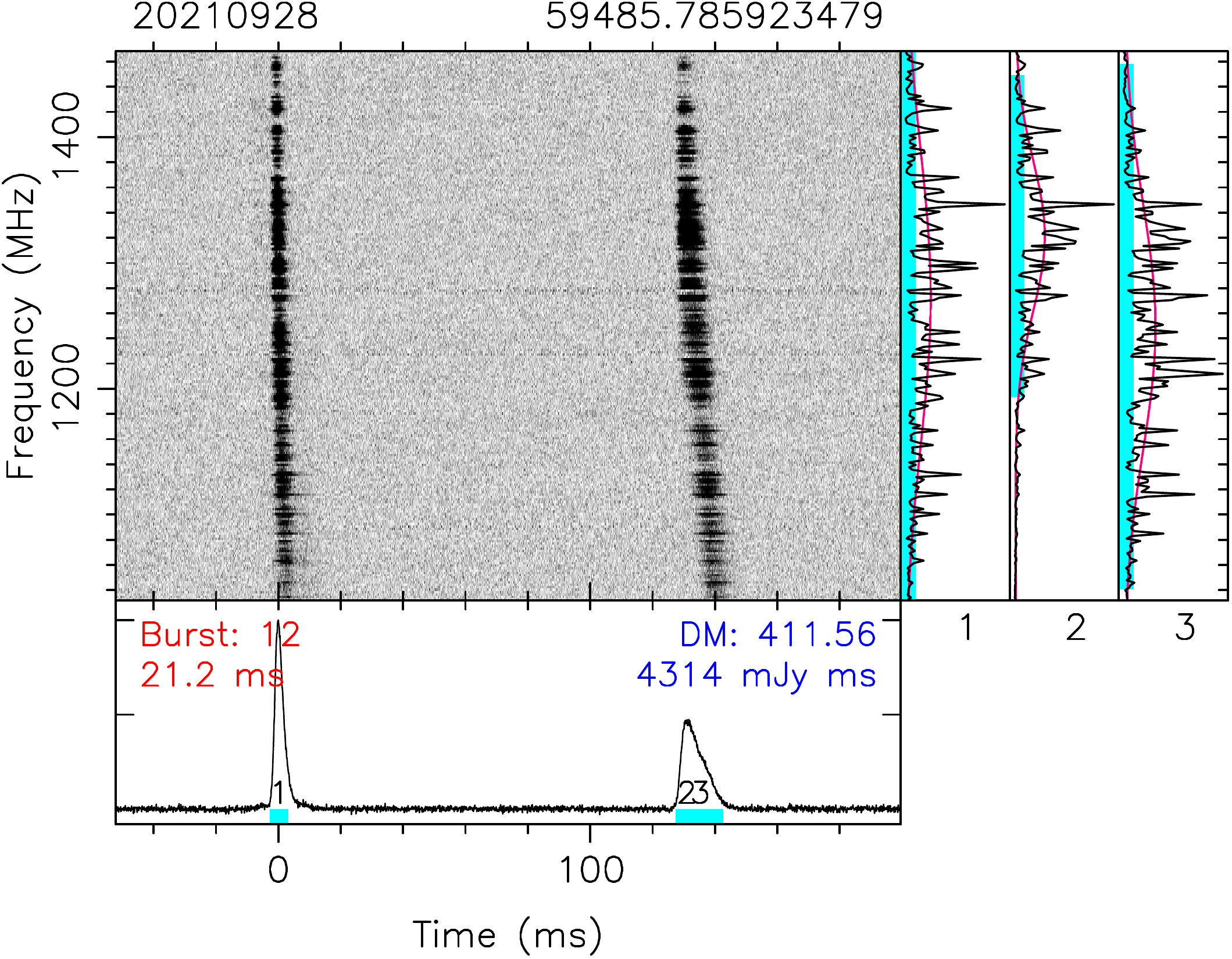}
    \includegraphics[height=37mm]{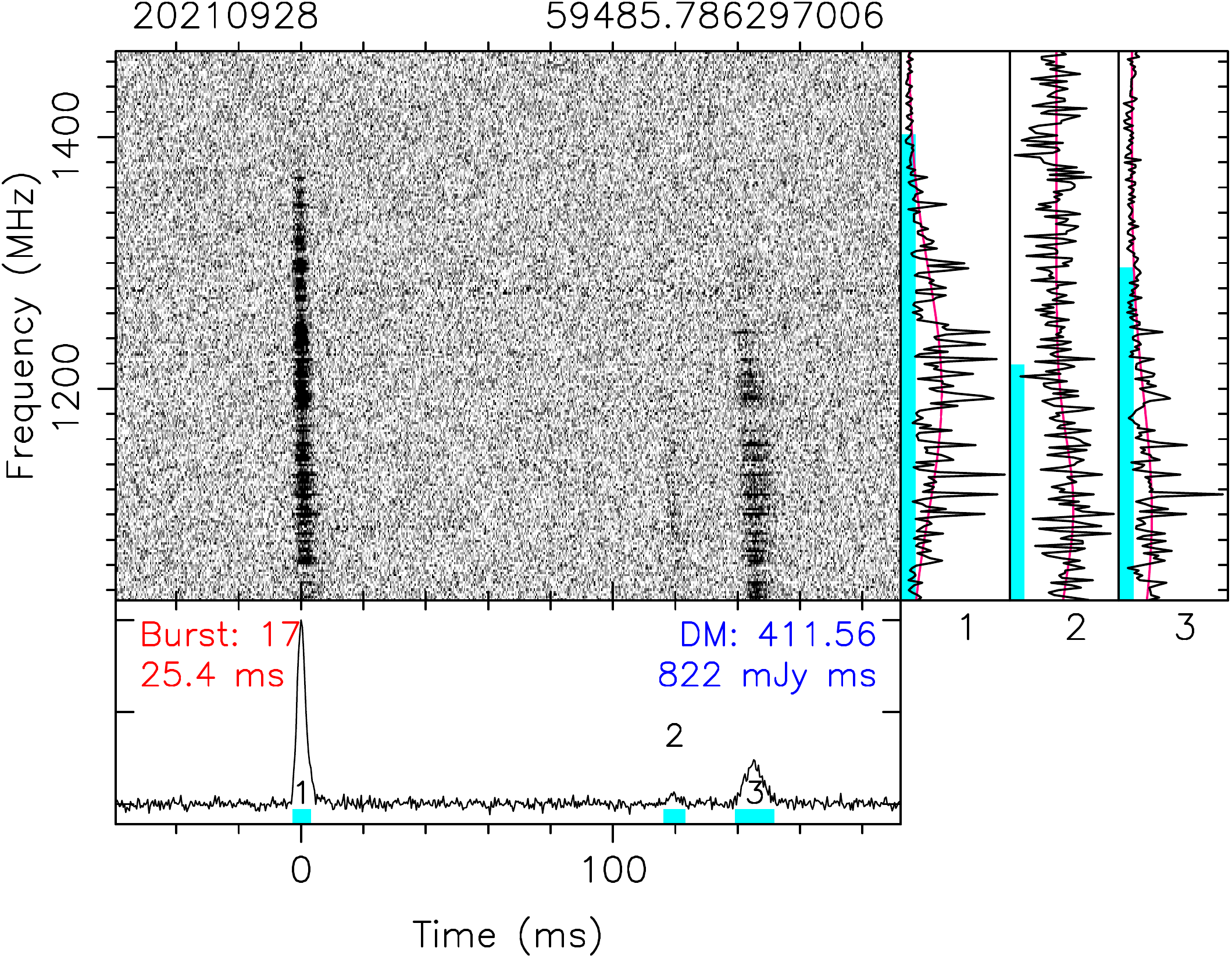}
    \includegraphics[height=37mm]{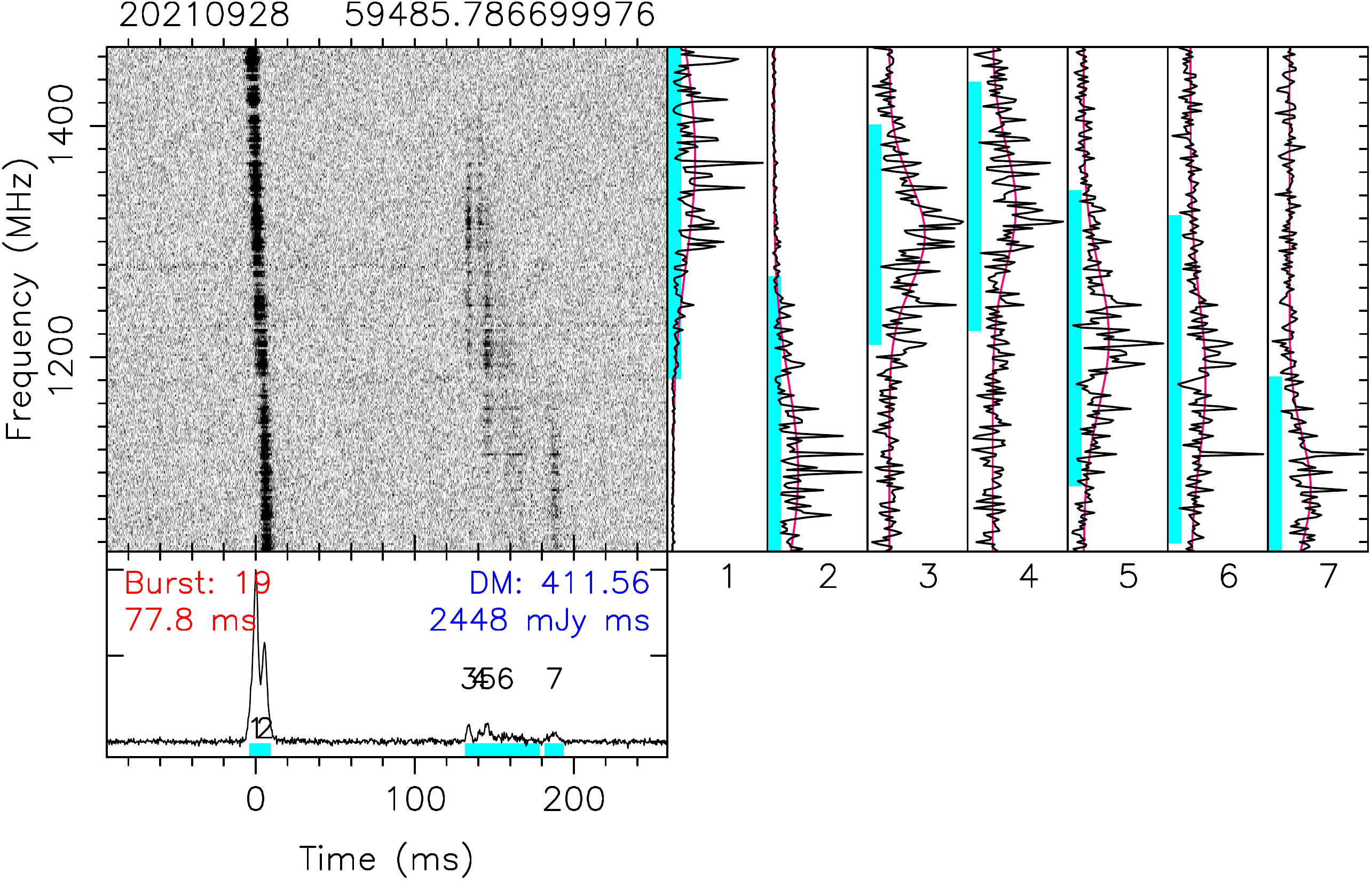}
    \includegraphics[height=37mm]{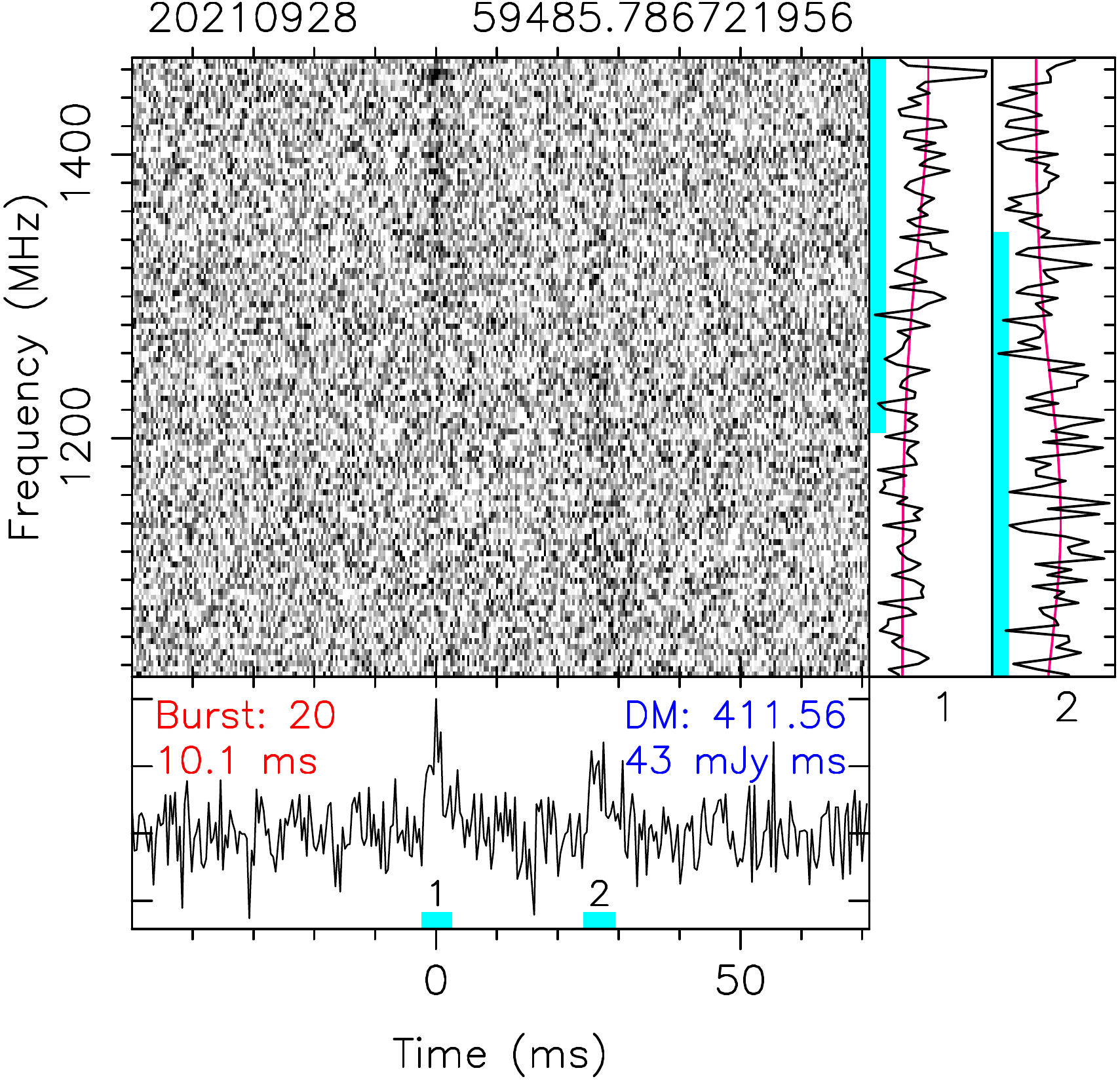}
    \includegraphics[height=37mm]{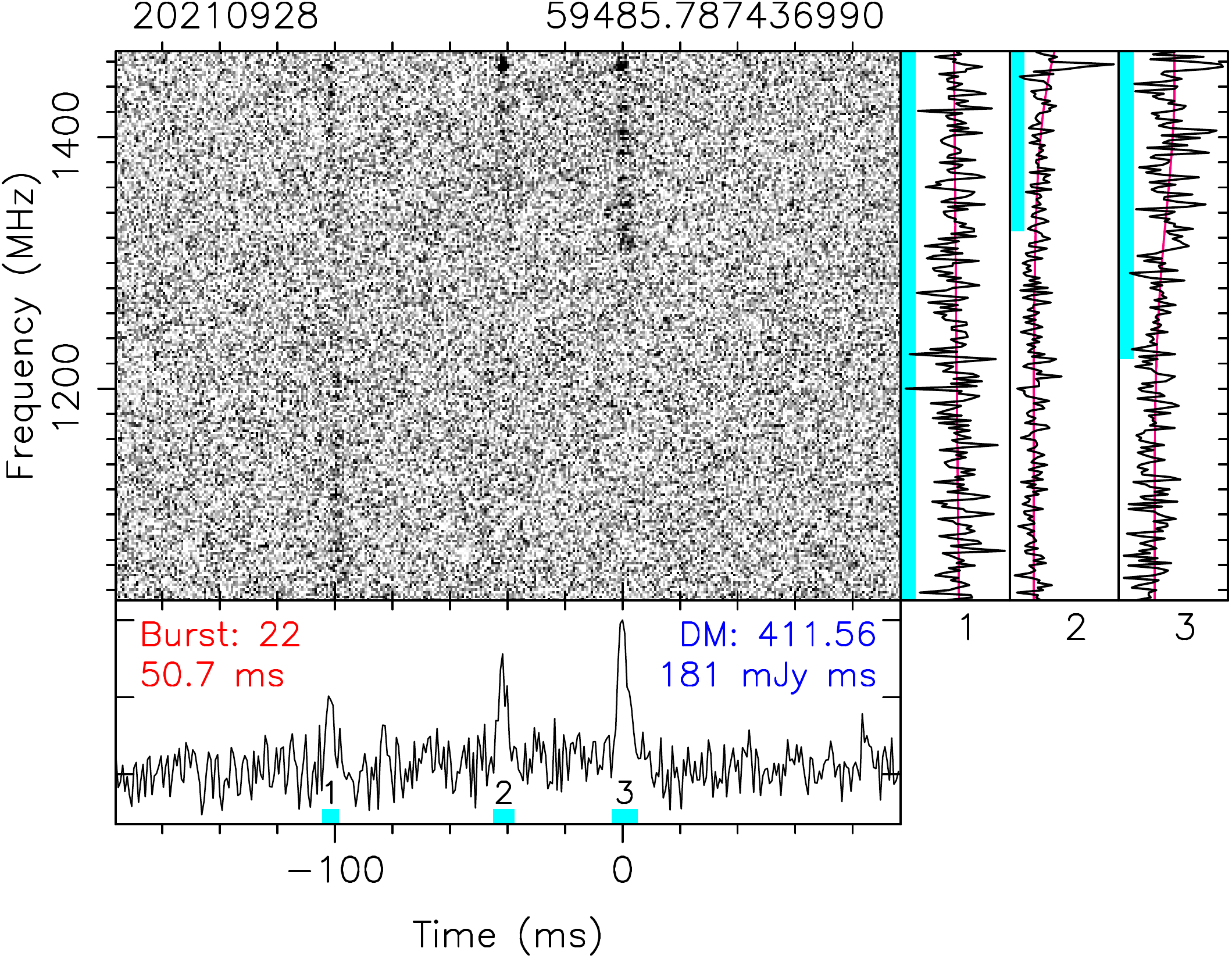}
    \includegraphics[height=37mm]{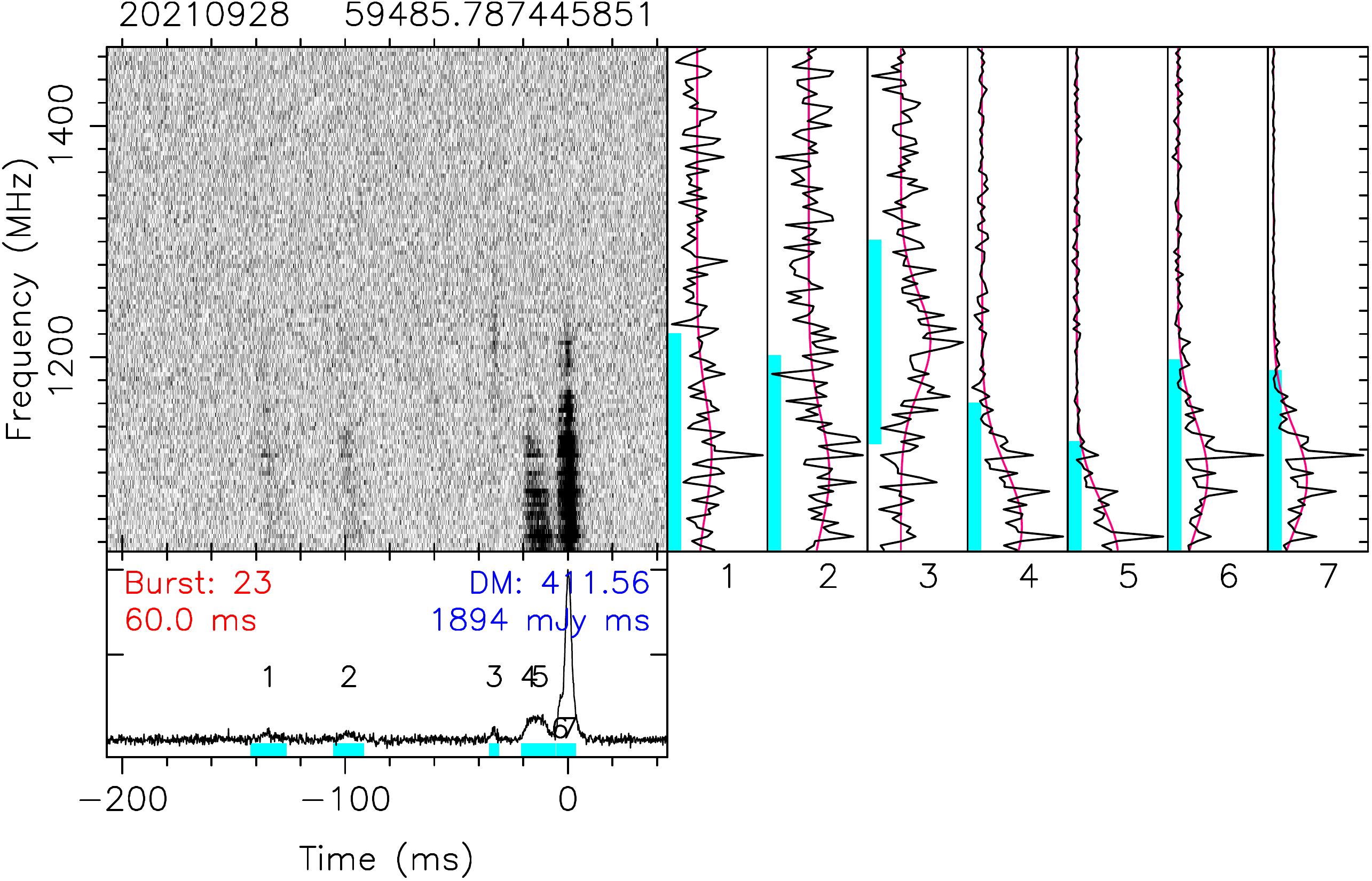}
    \includegraphics[height=37mm]{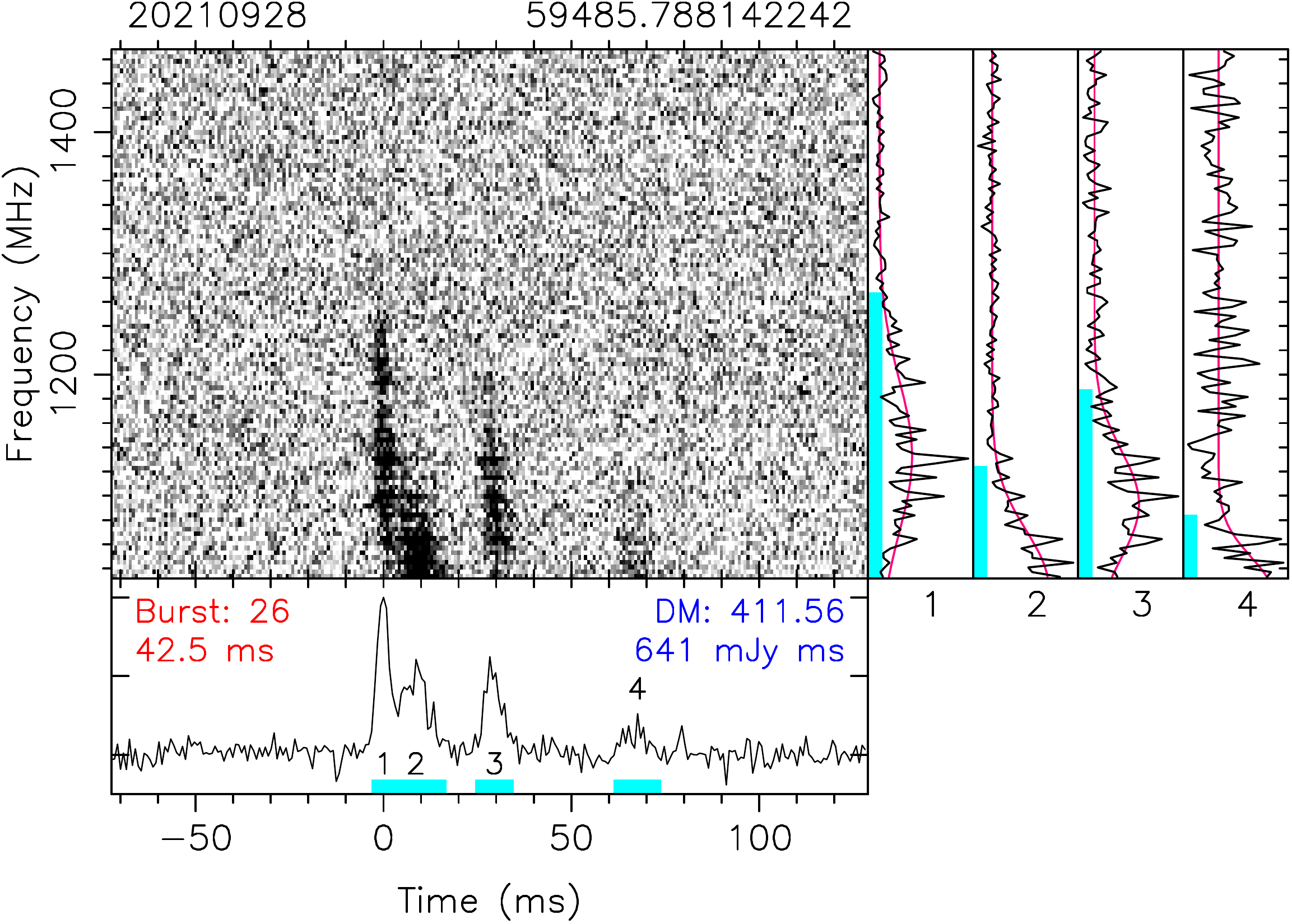}
    \includegraphics[height=37mm]{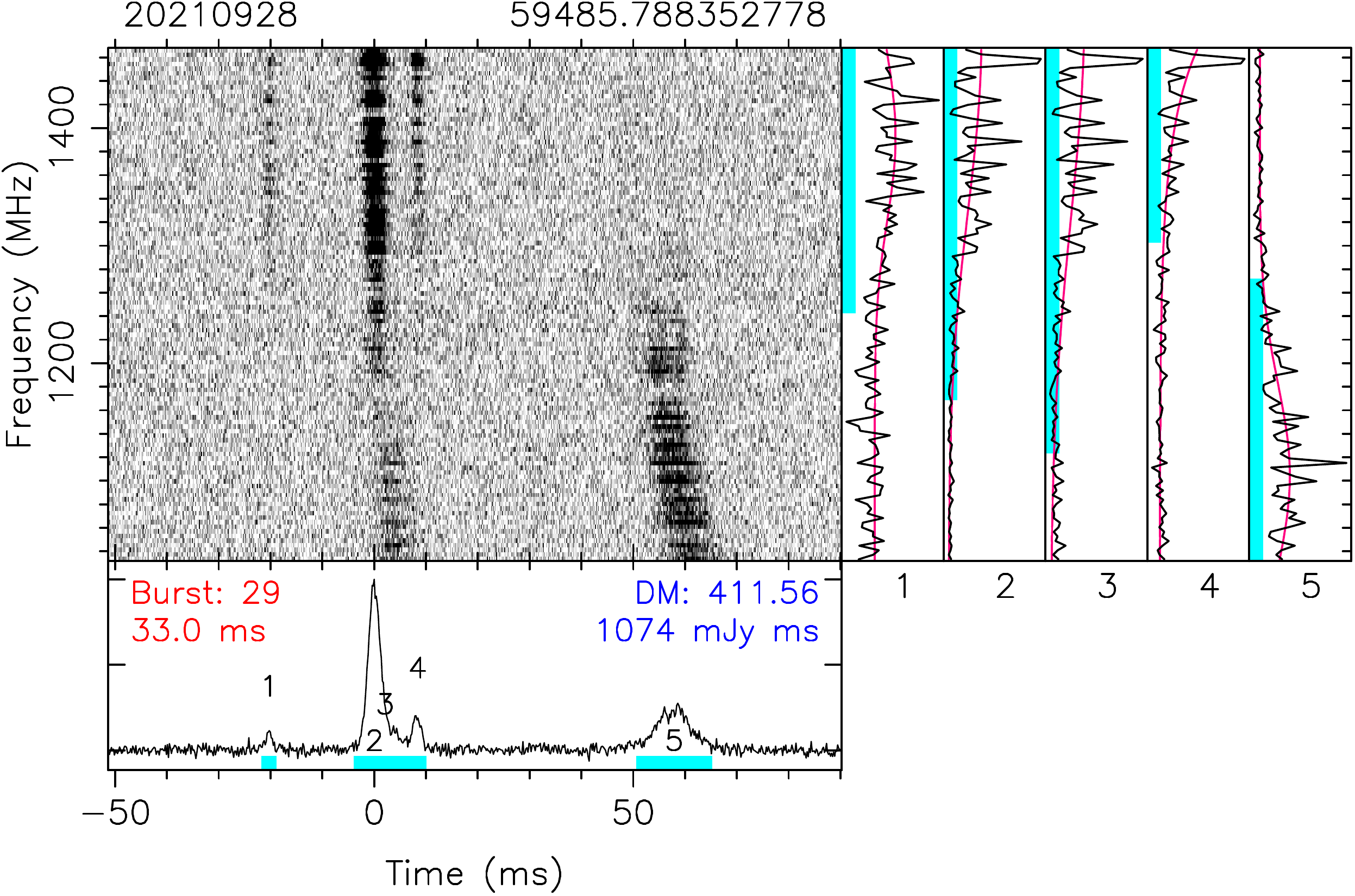}
    \includegraphics[height=37mm]{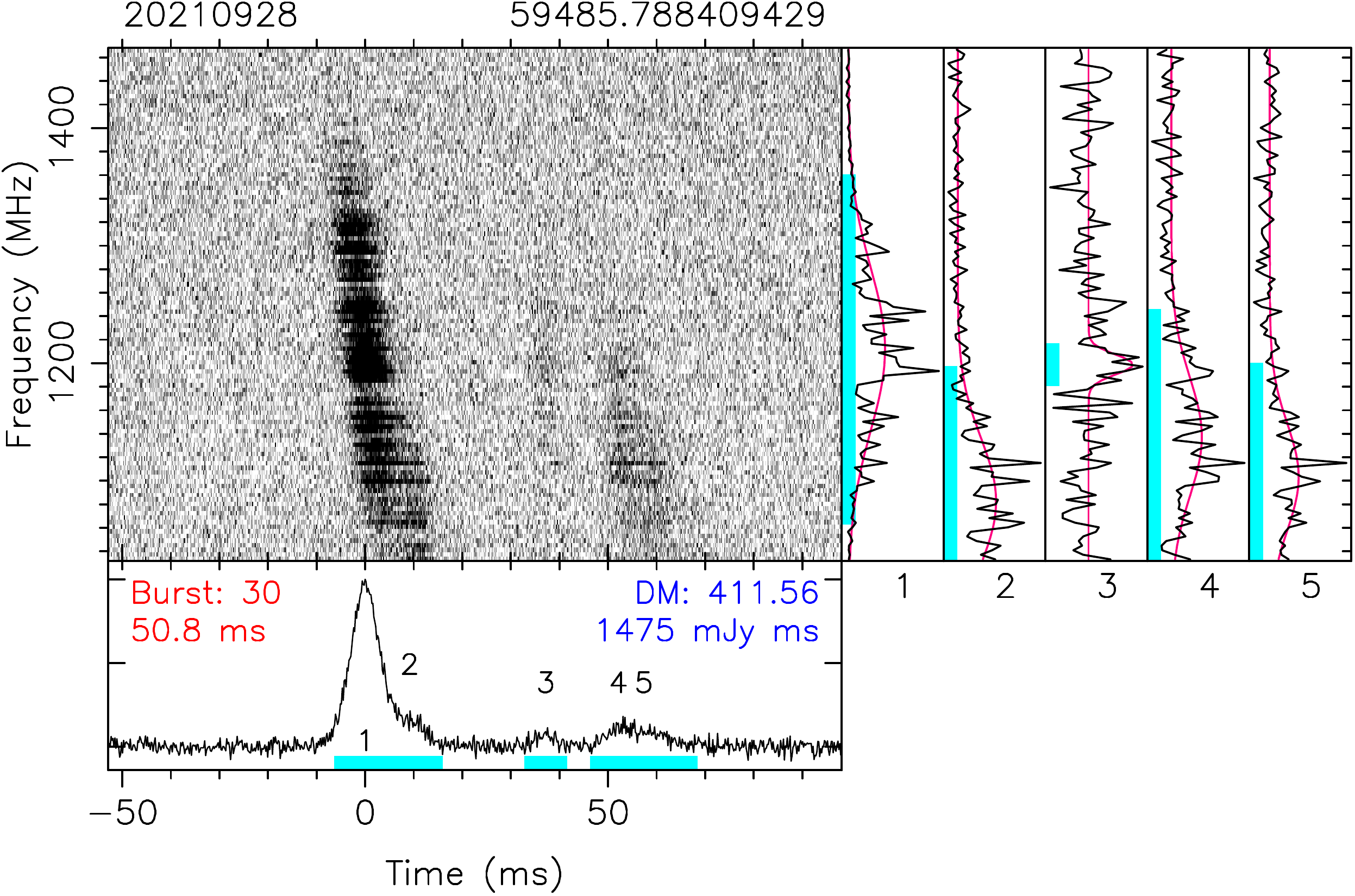}
    \includegraphics[height=37mm]{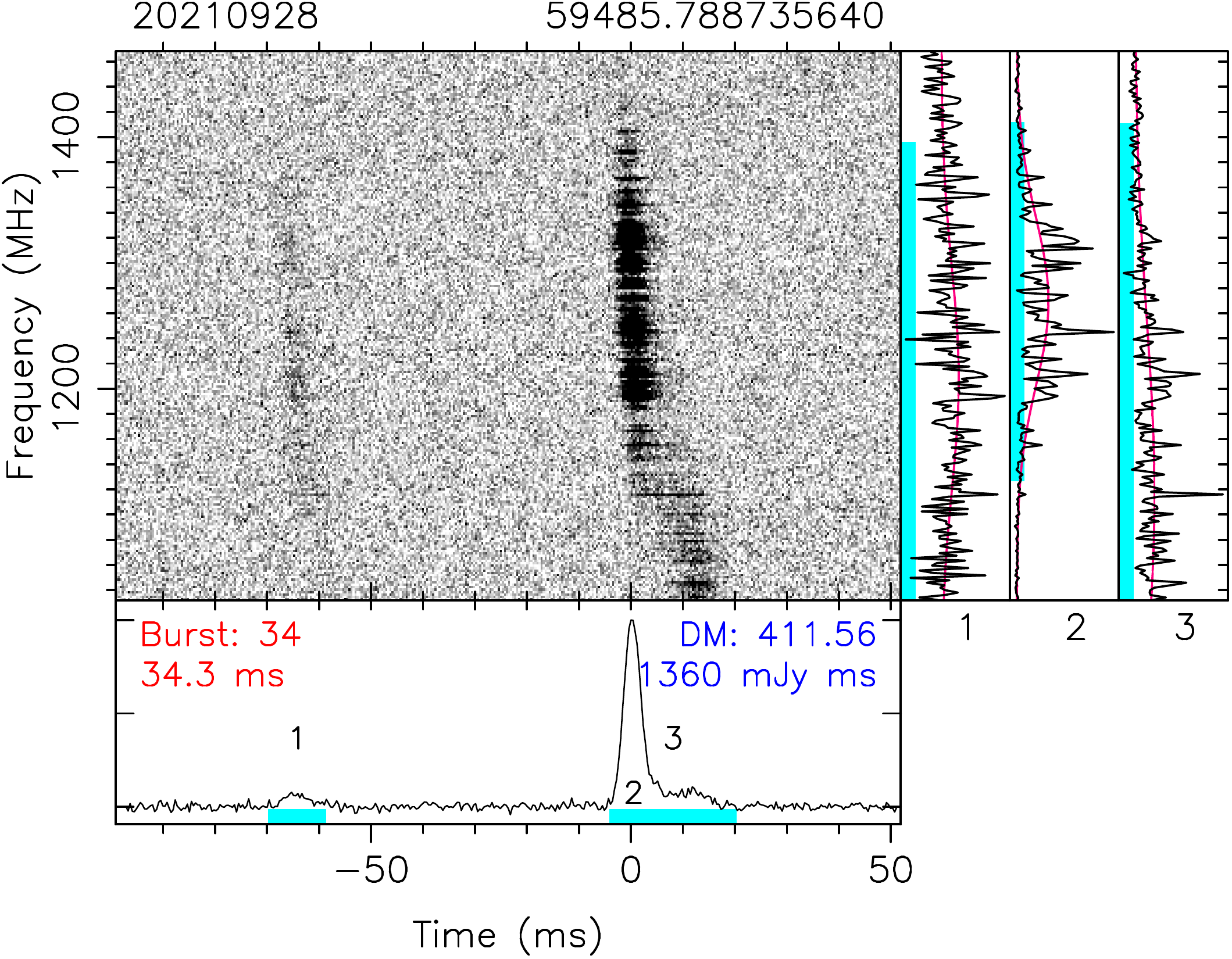}
    \includegraphics[height=37mm]{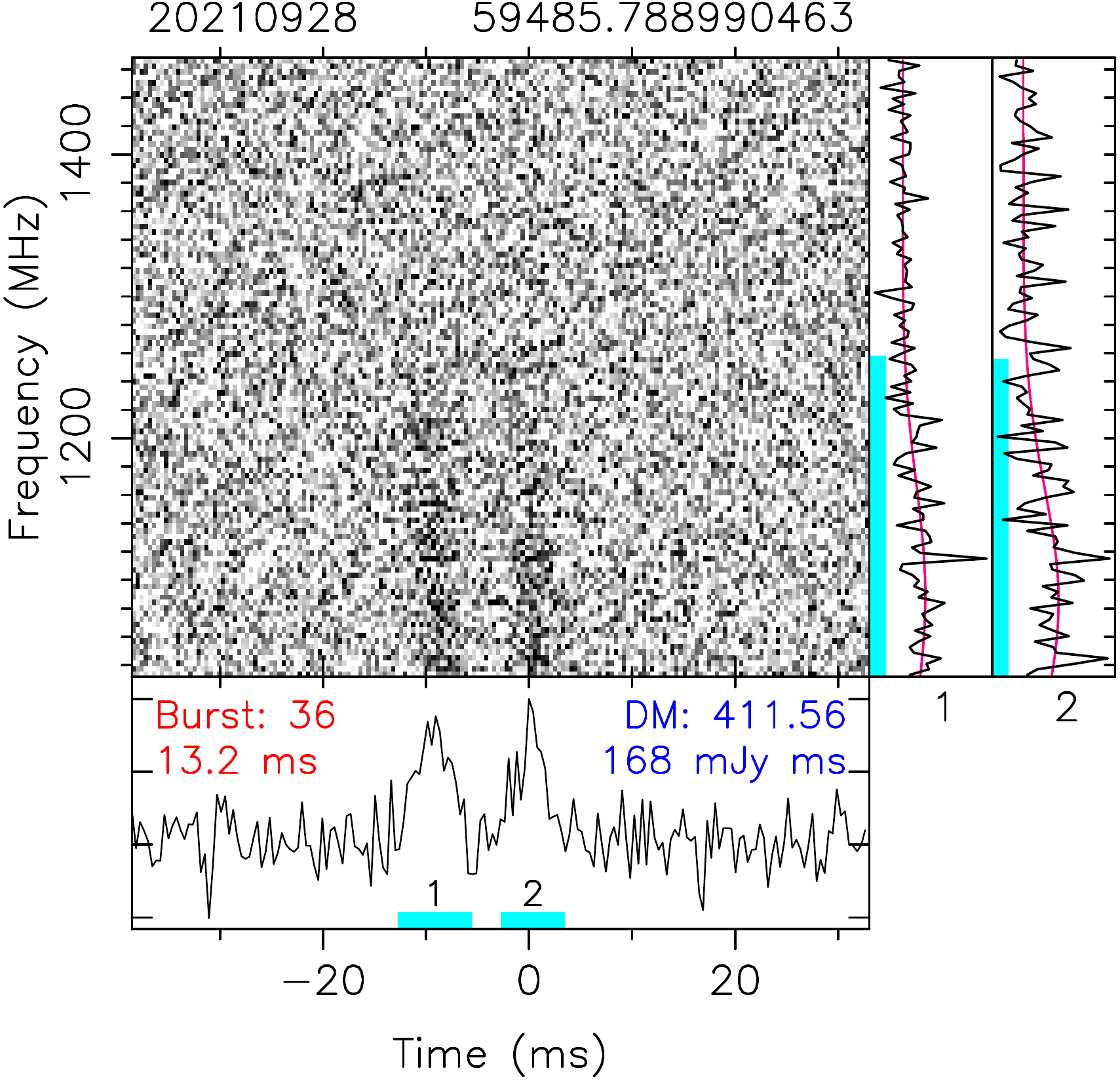}
    \includegraphics[height=37mm]{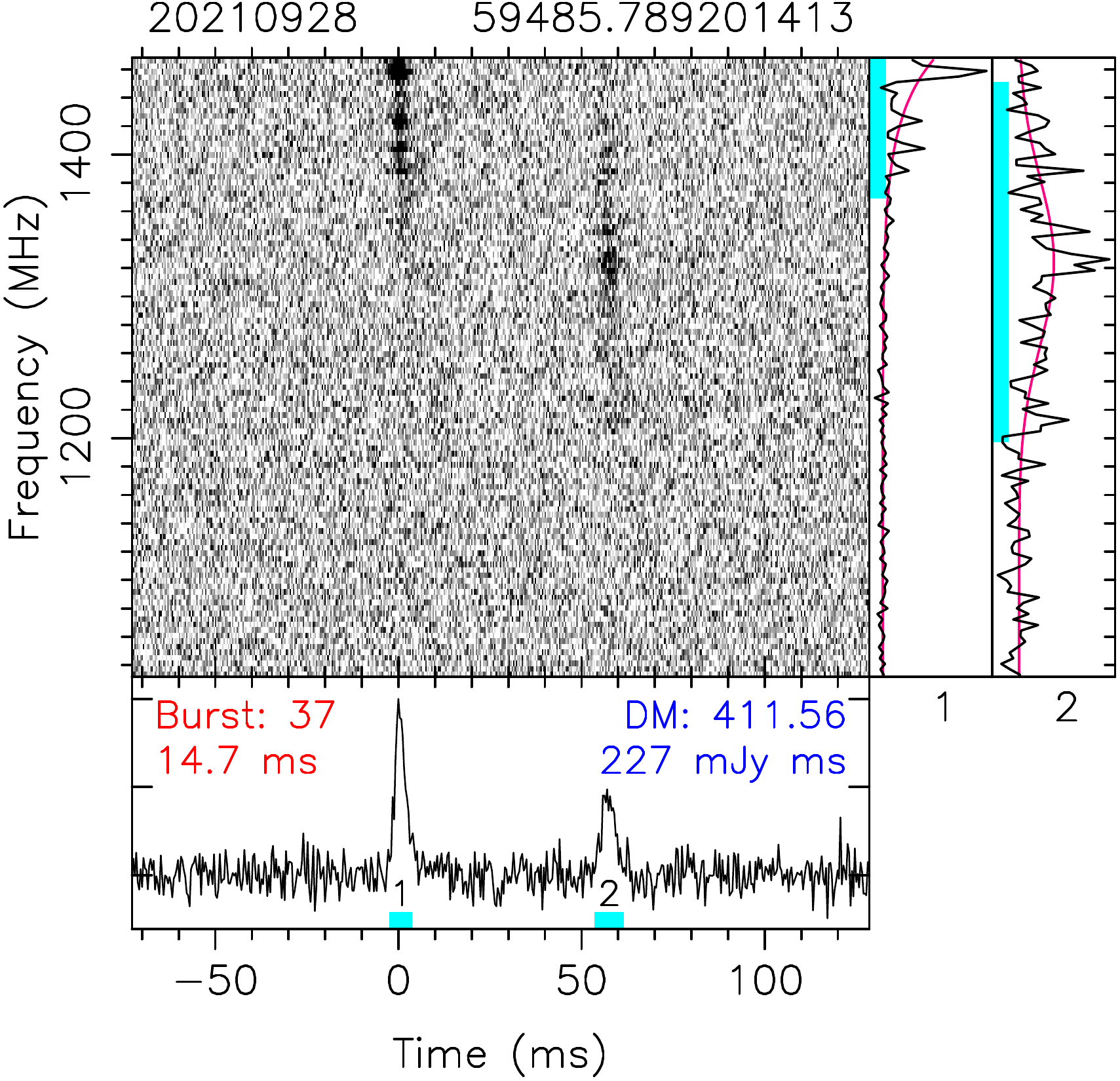}
    \includegraphics[height=37mm]{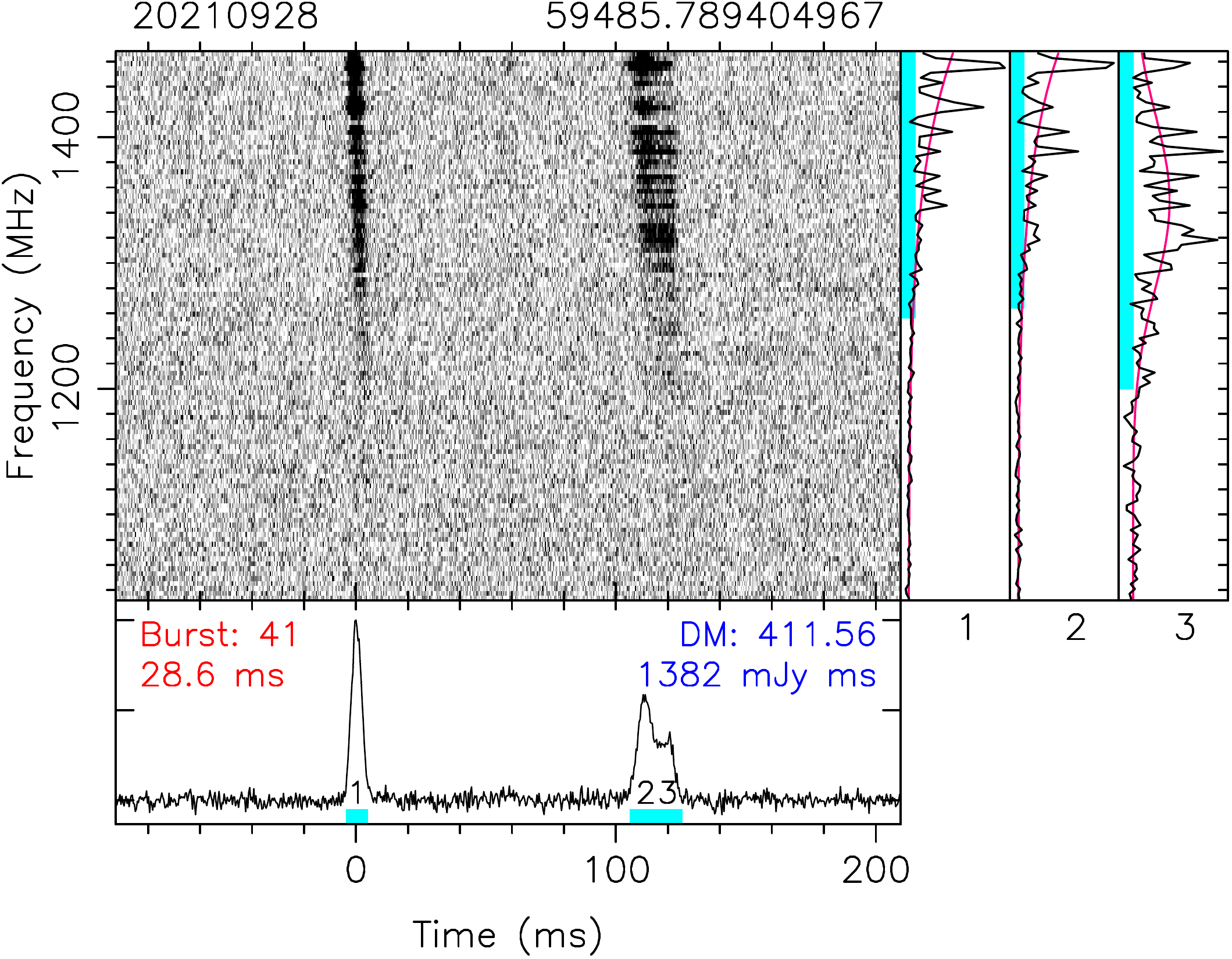}
    \includegraphics[height=37mm]{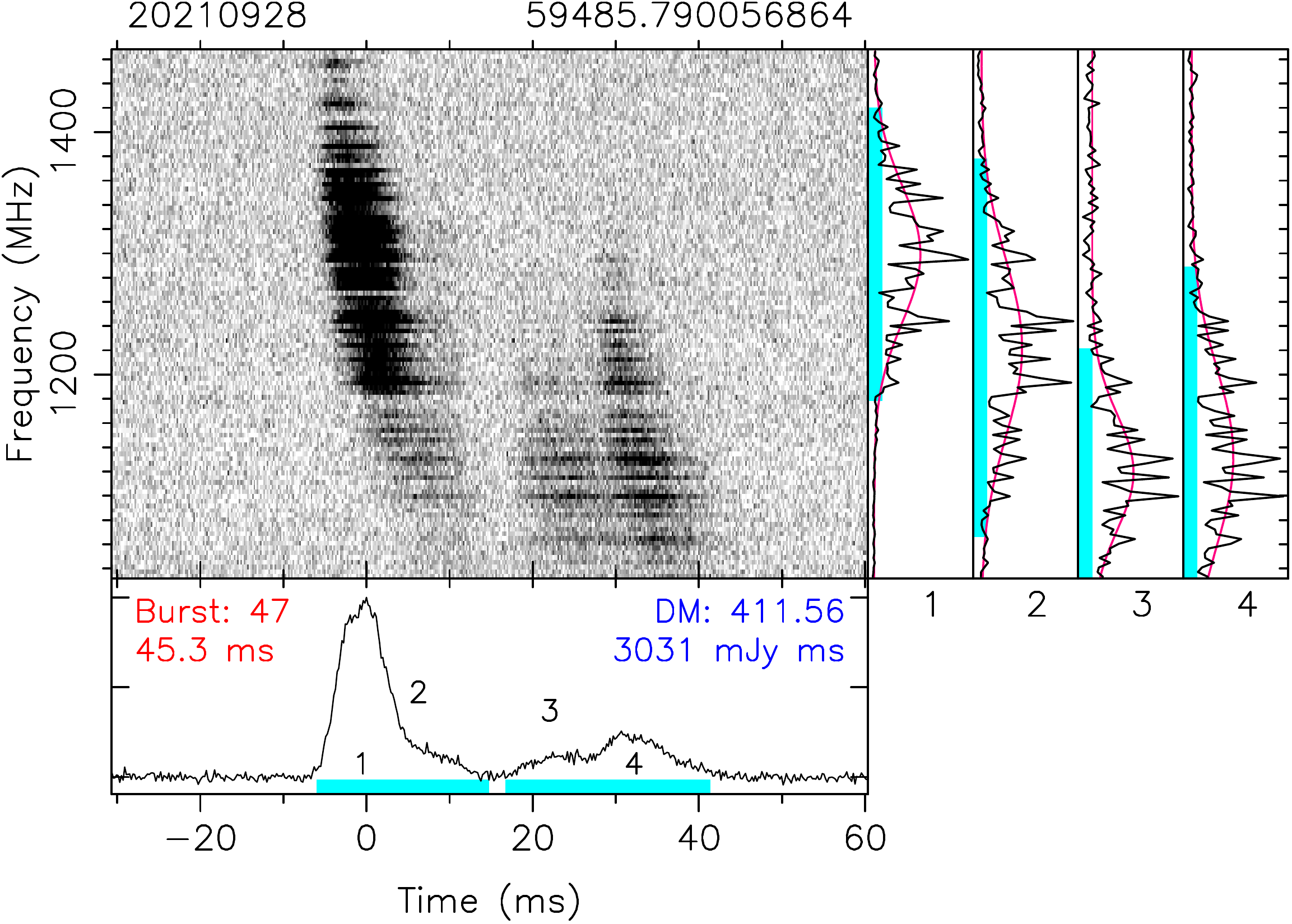}
    \includegraphics[height=37mm]{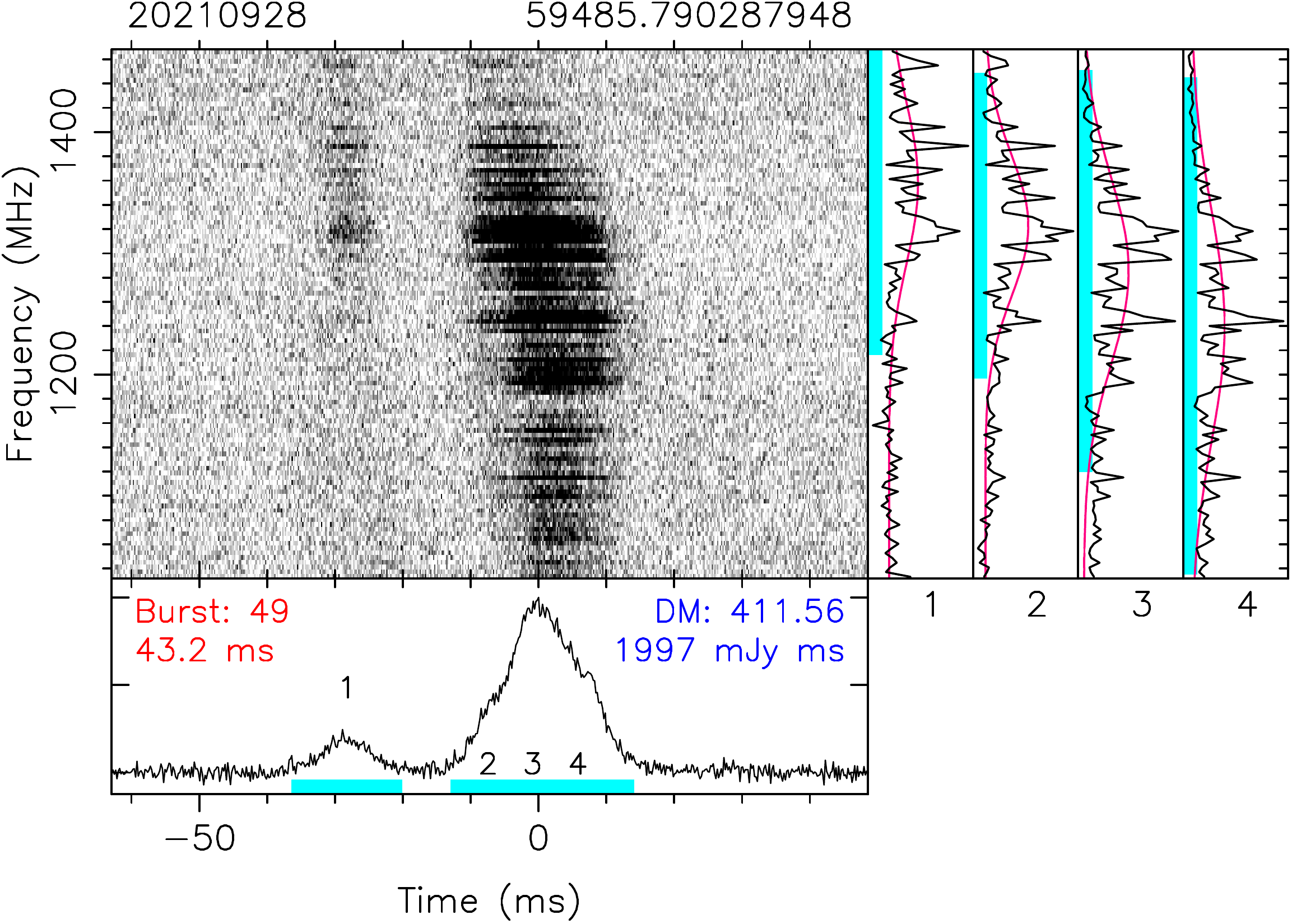}
    \includegraphics[height=37mm]{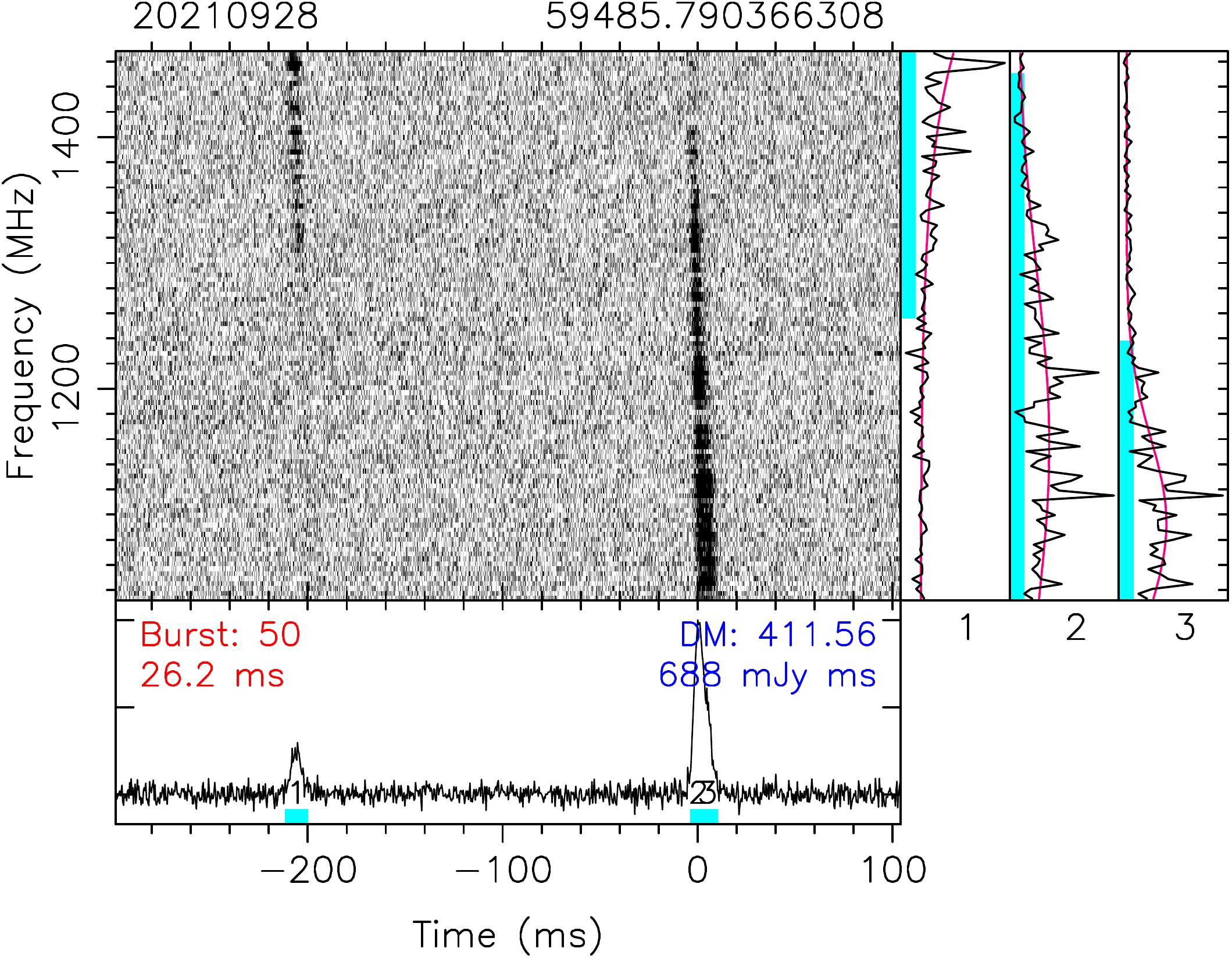}
    \includegraphics[height=37mm]{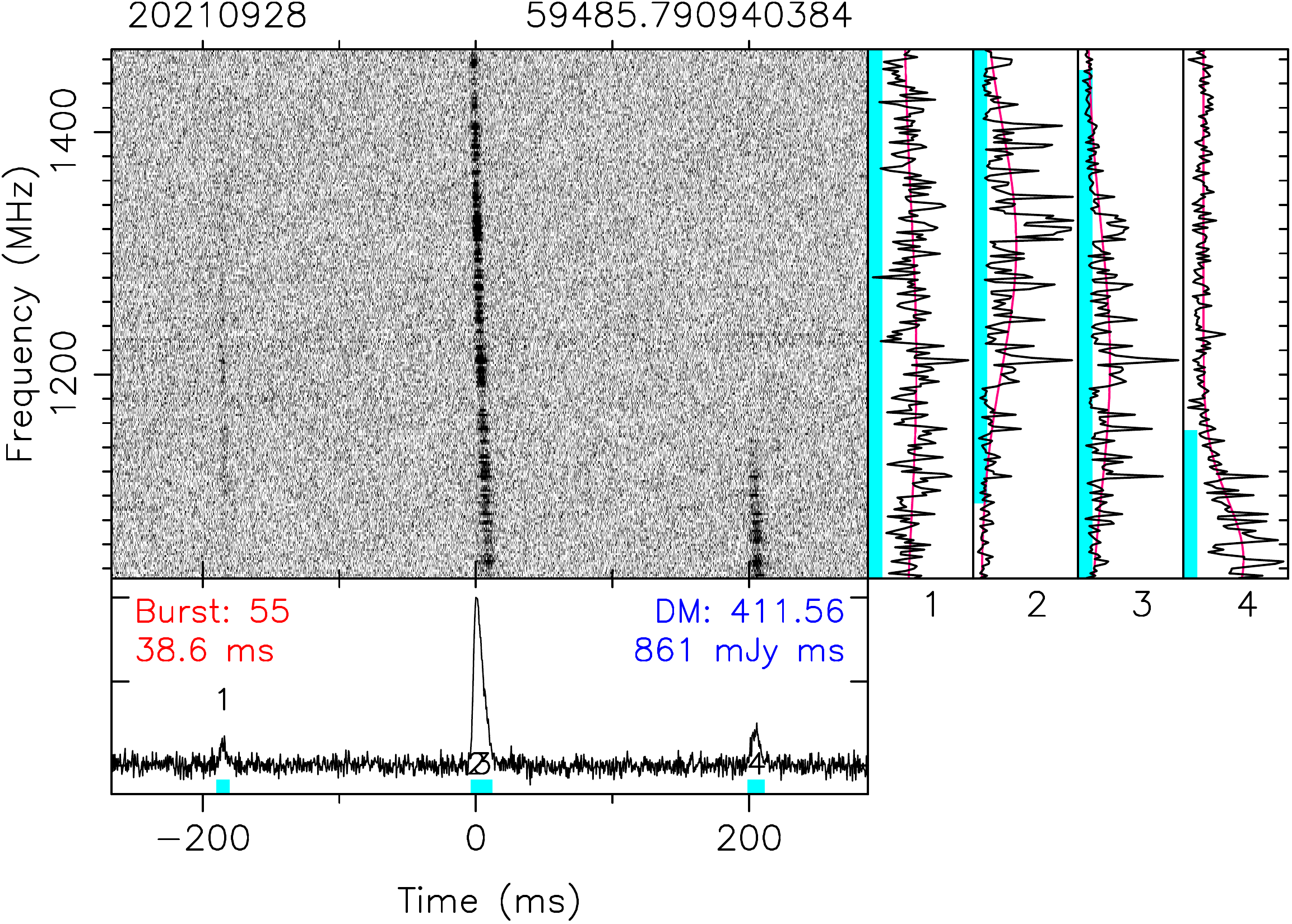}
\caption{\it{ -- continued}.
}
\end{figure*}
\addtocounter{figure}{-1}
\begin{figure*}
    \flushleft
    \includegraphics[height=37mm]{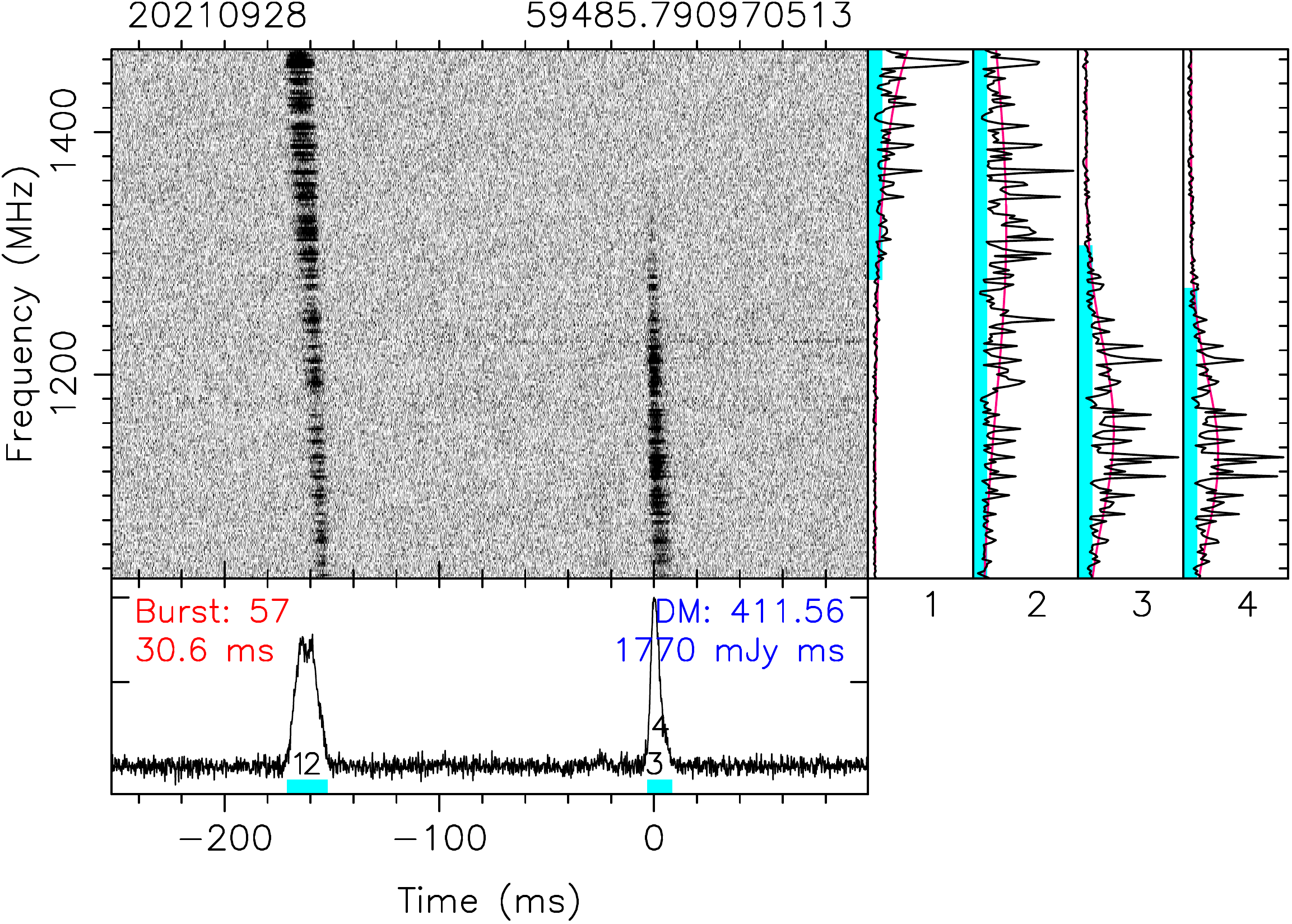}
    \includegraphics[height=37mm]{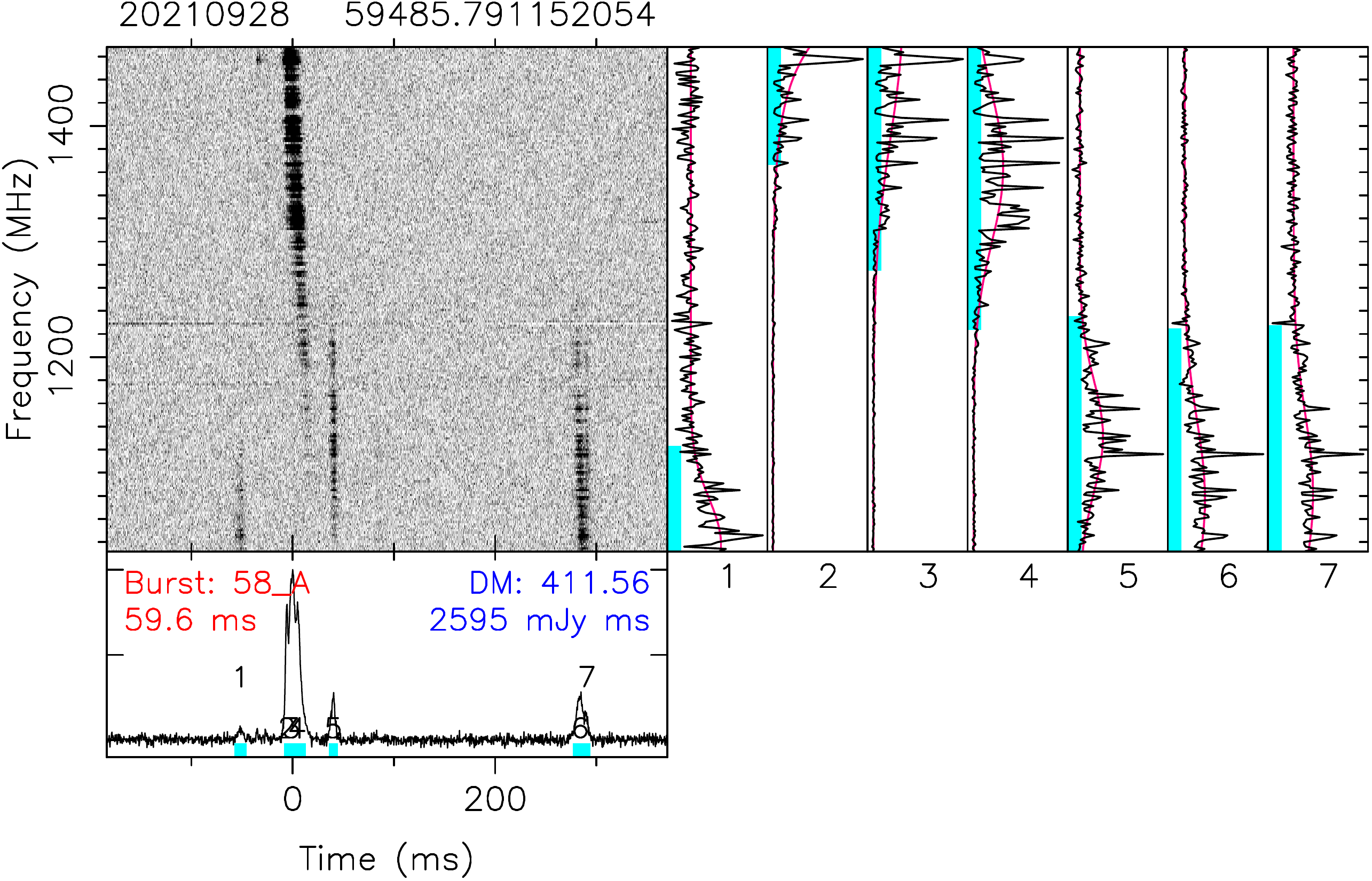}
    \includegraphics[height=37mm]{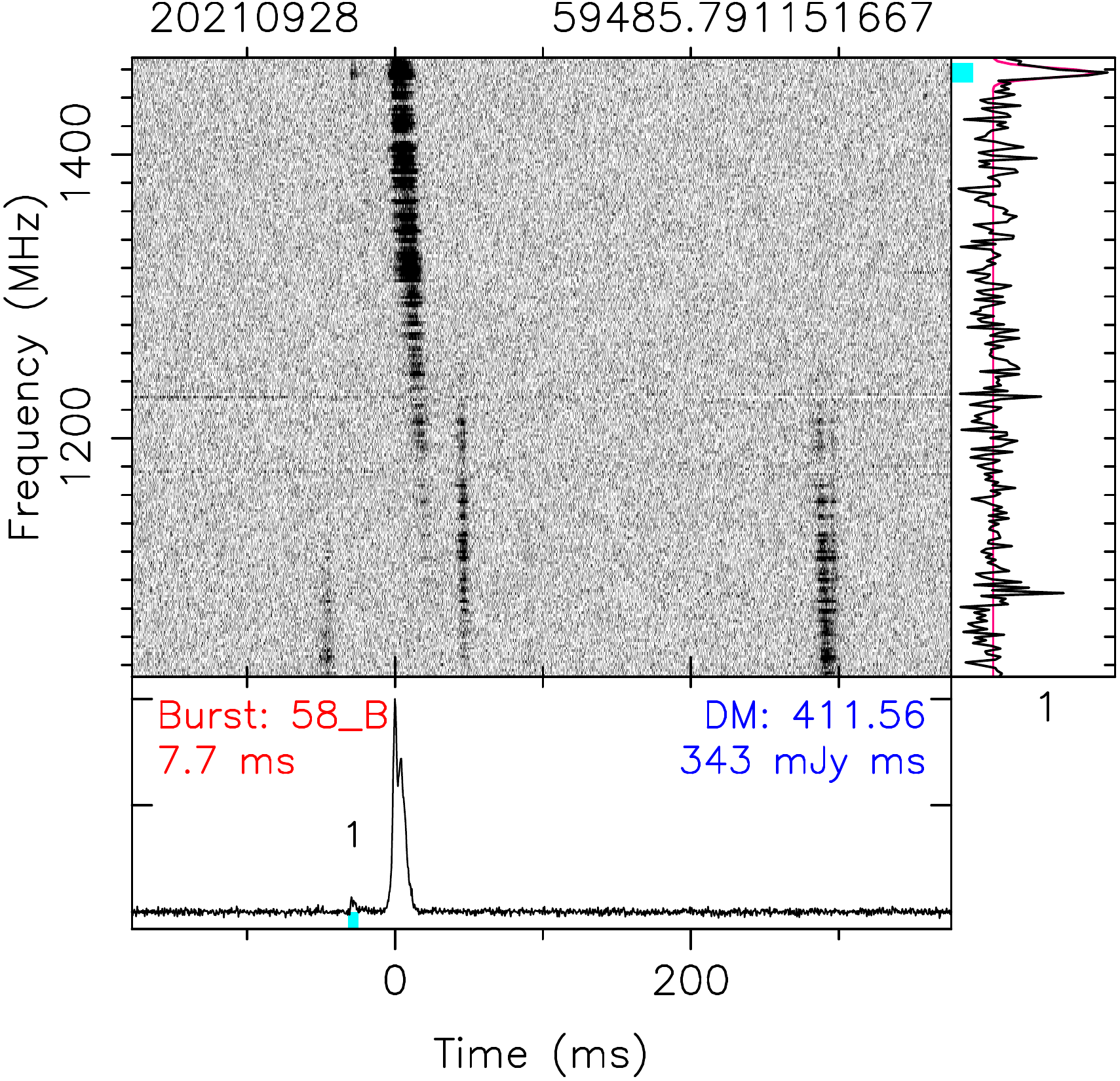}
    \includegraphics[height=37mm]{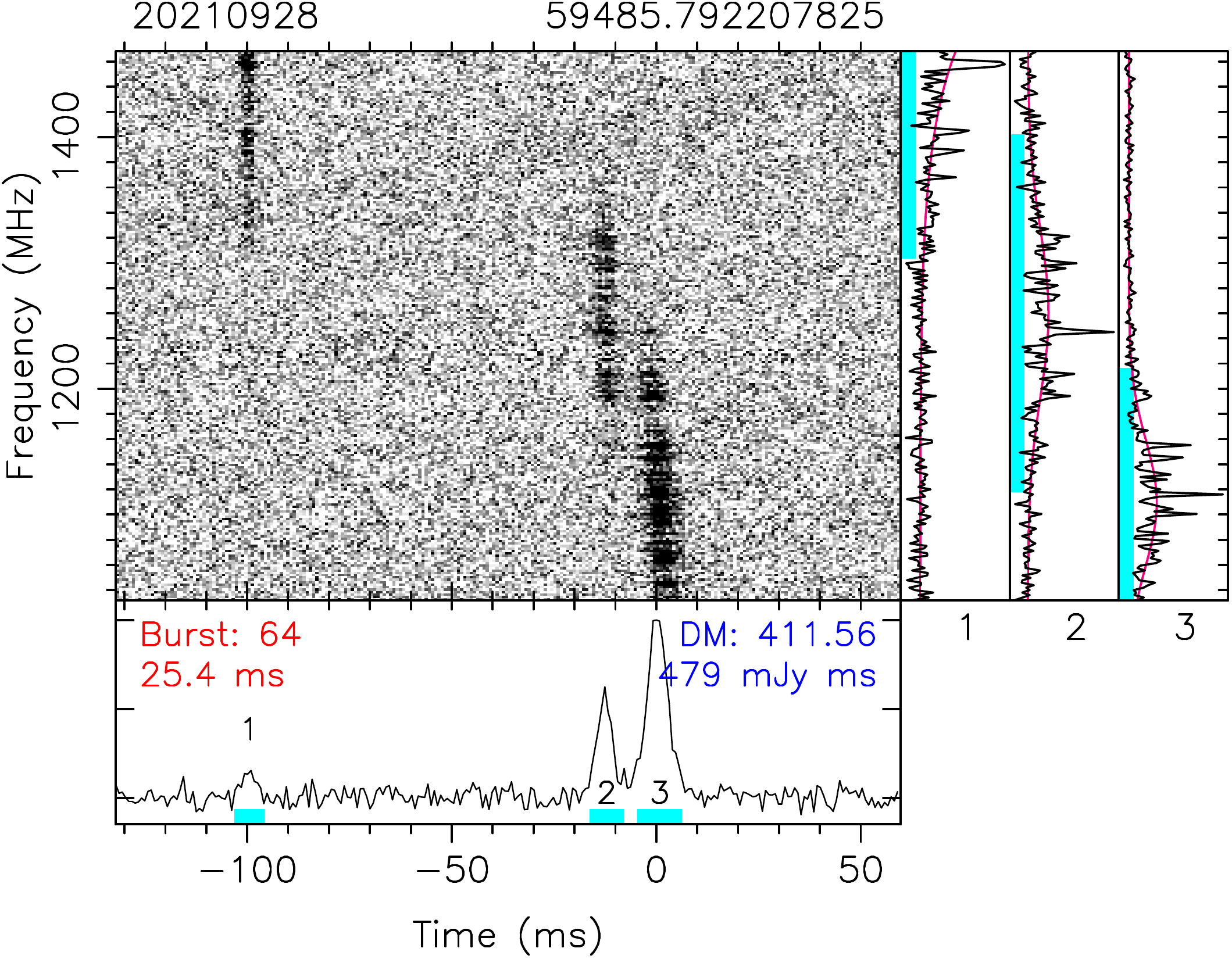}
    \includegraphics[height=37mm]{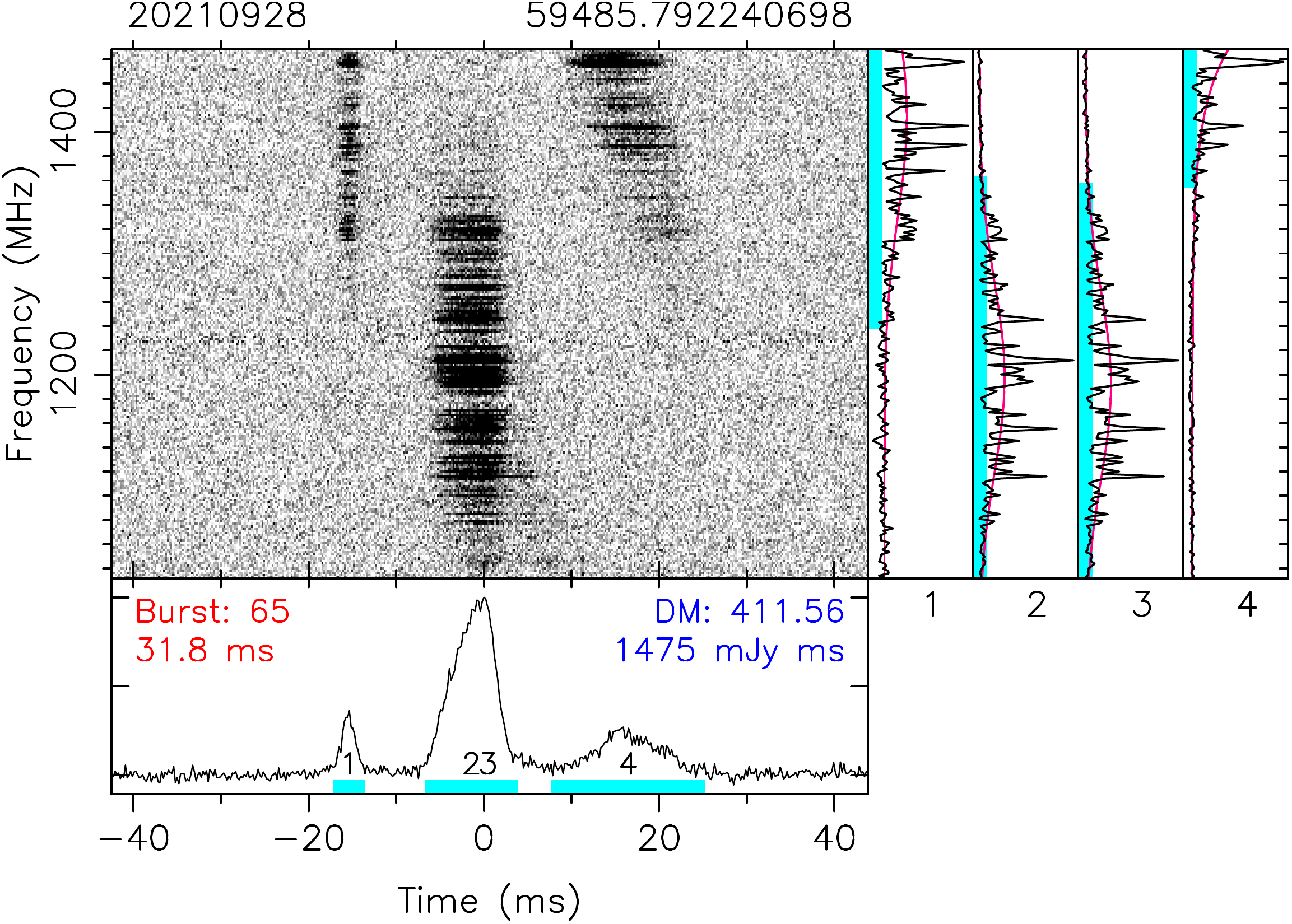}
    \includegraphics[height=37mm]{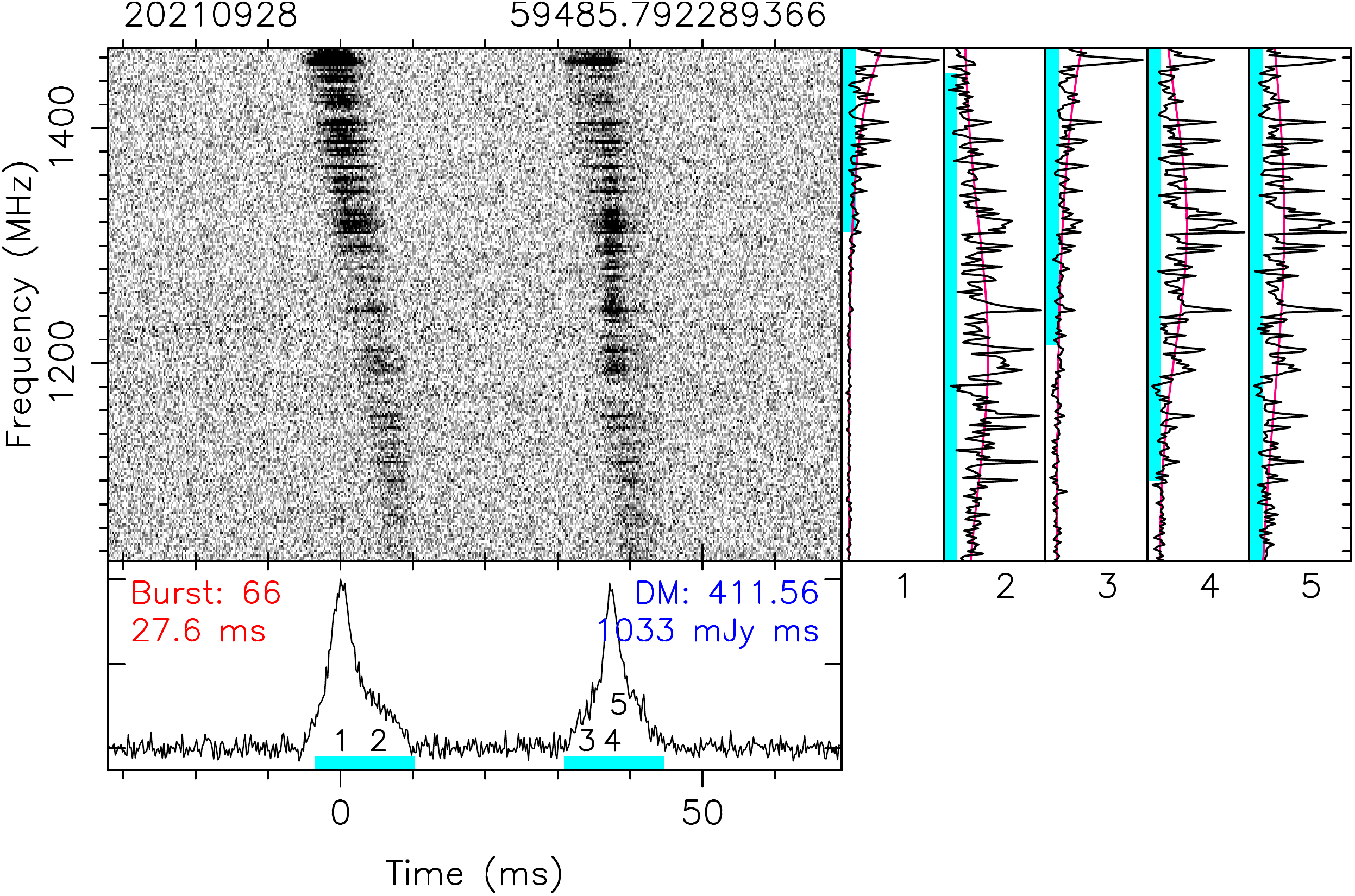}
    \includegraphics[height=37mm]{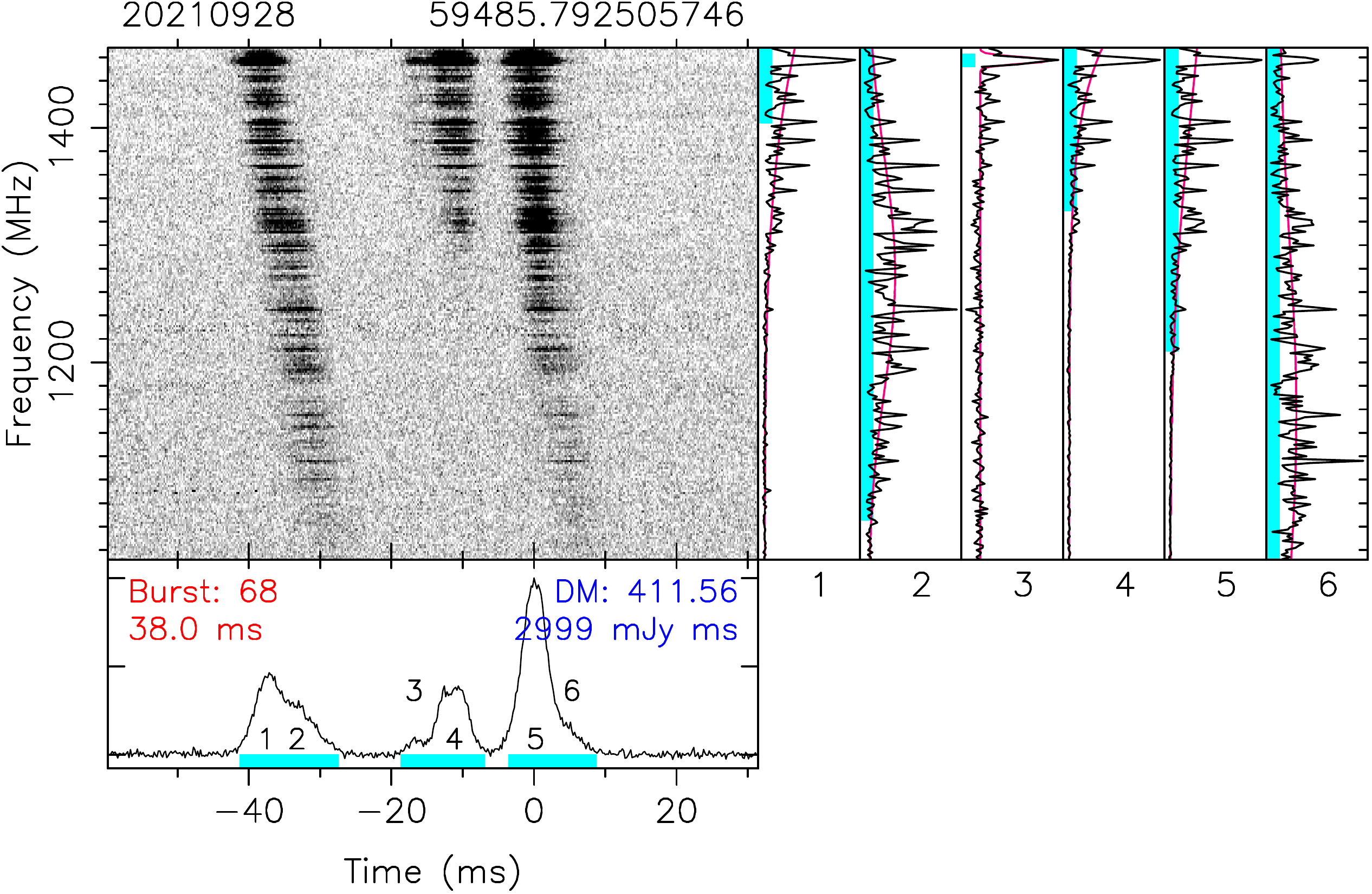}
    \includegraphics[height=37mm]{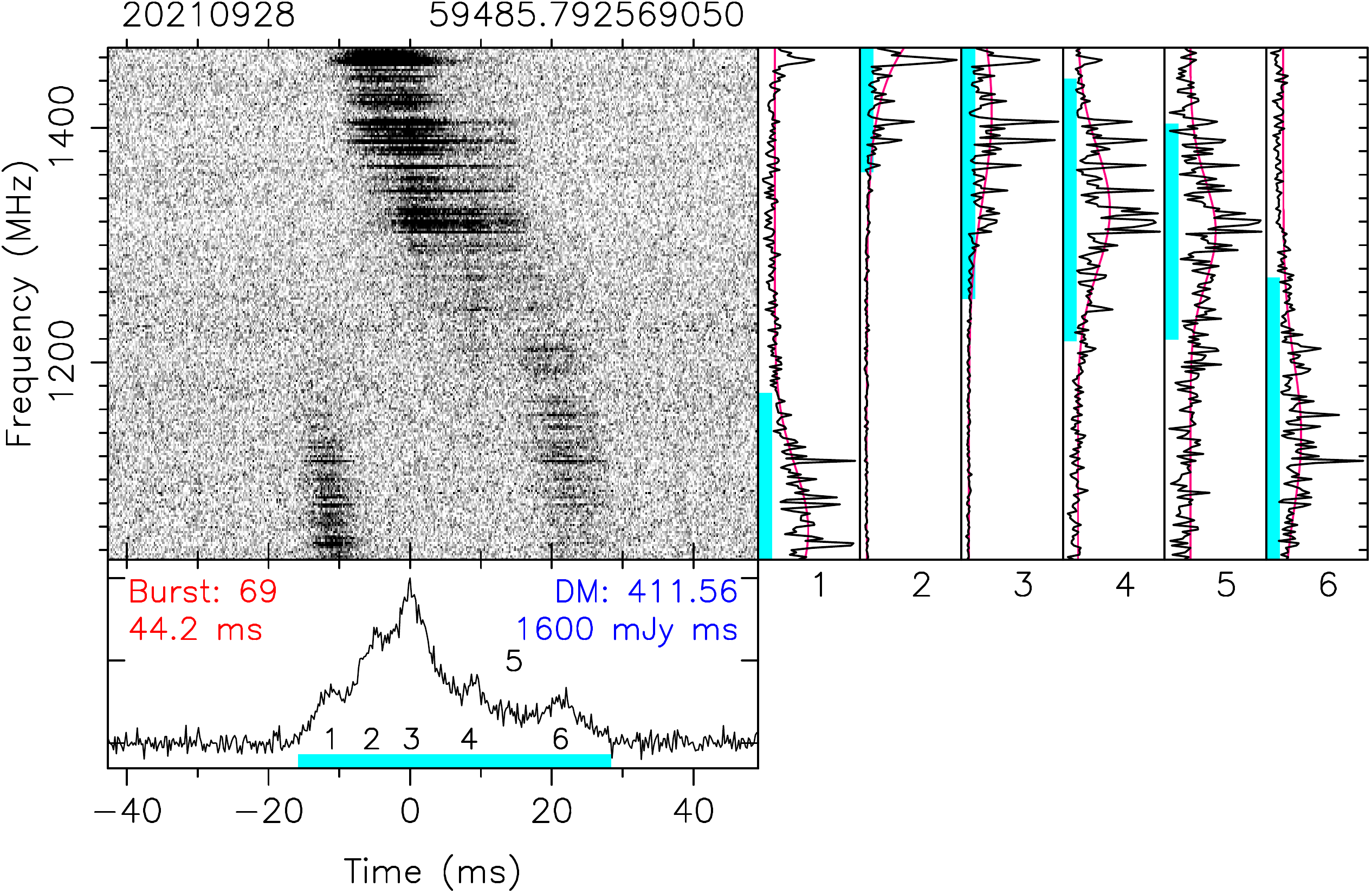}
    \includegraphics[height=37mm]{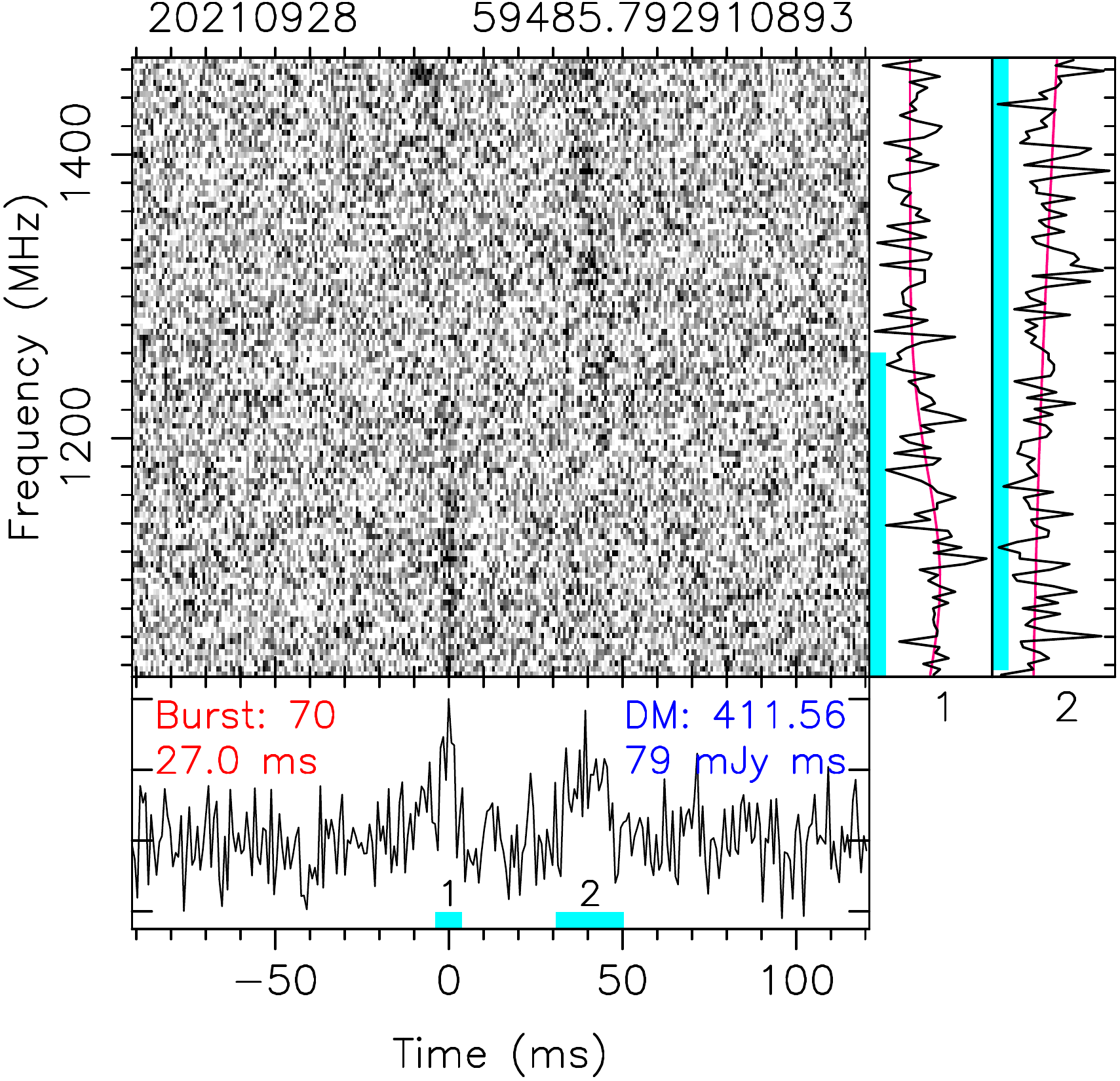}
    \includegraphics[height=37mm]{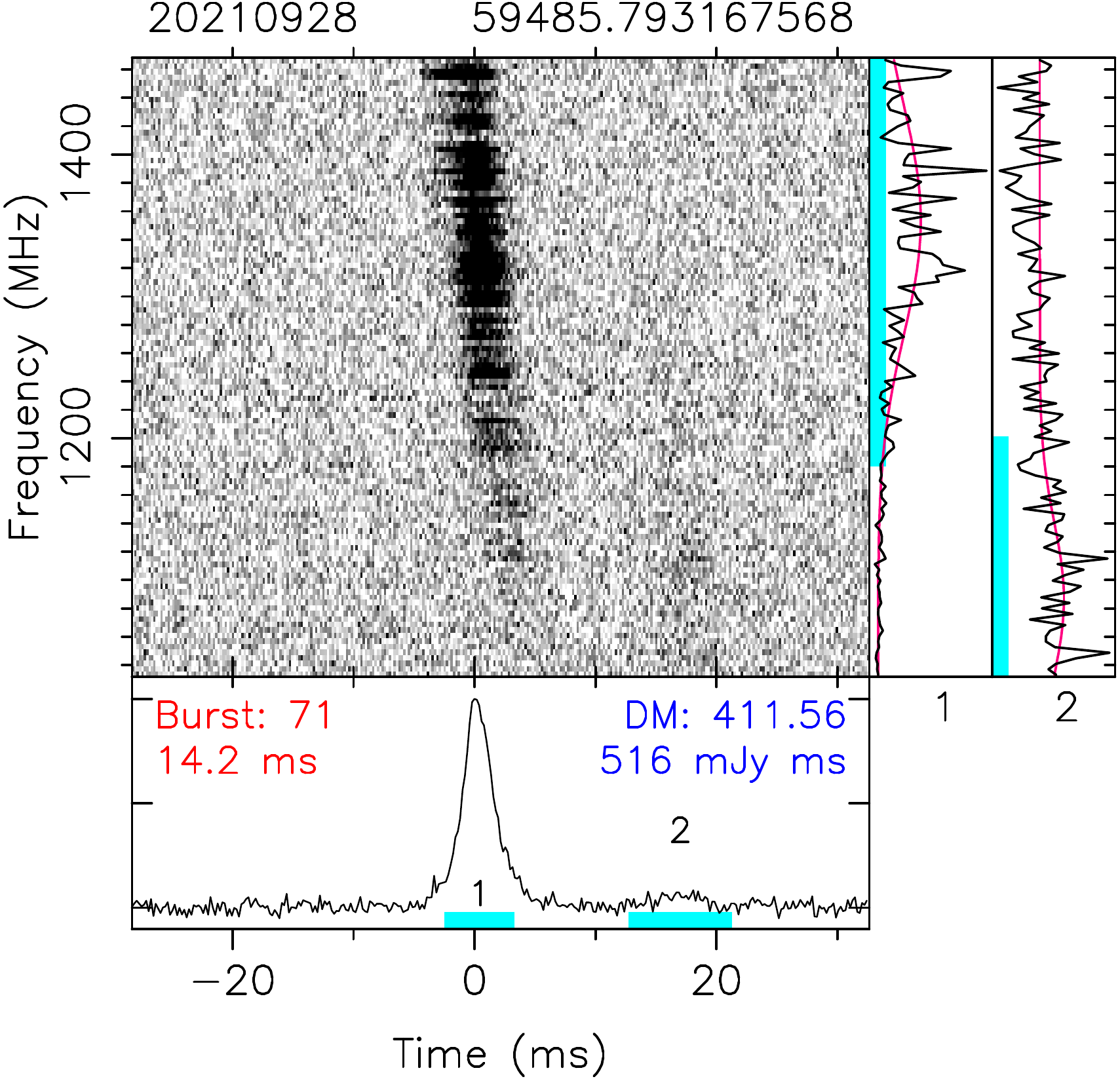}
    \includegraphics[height=37mm]{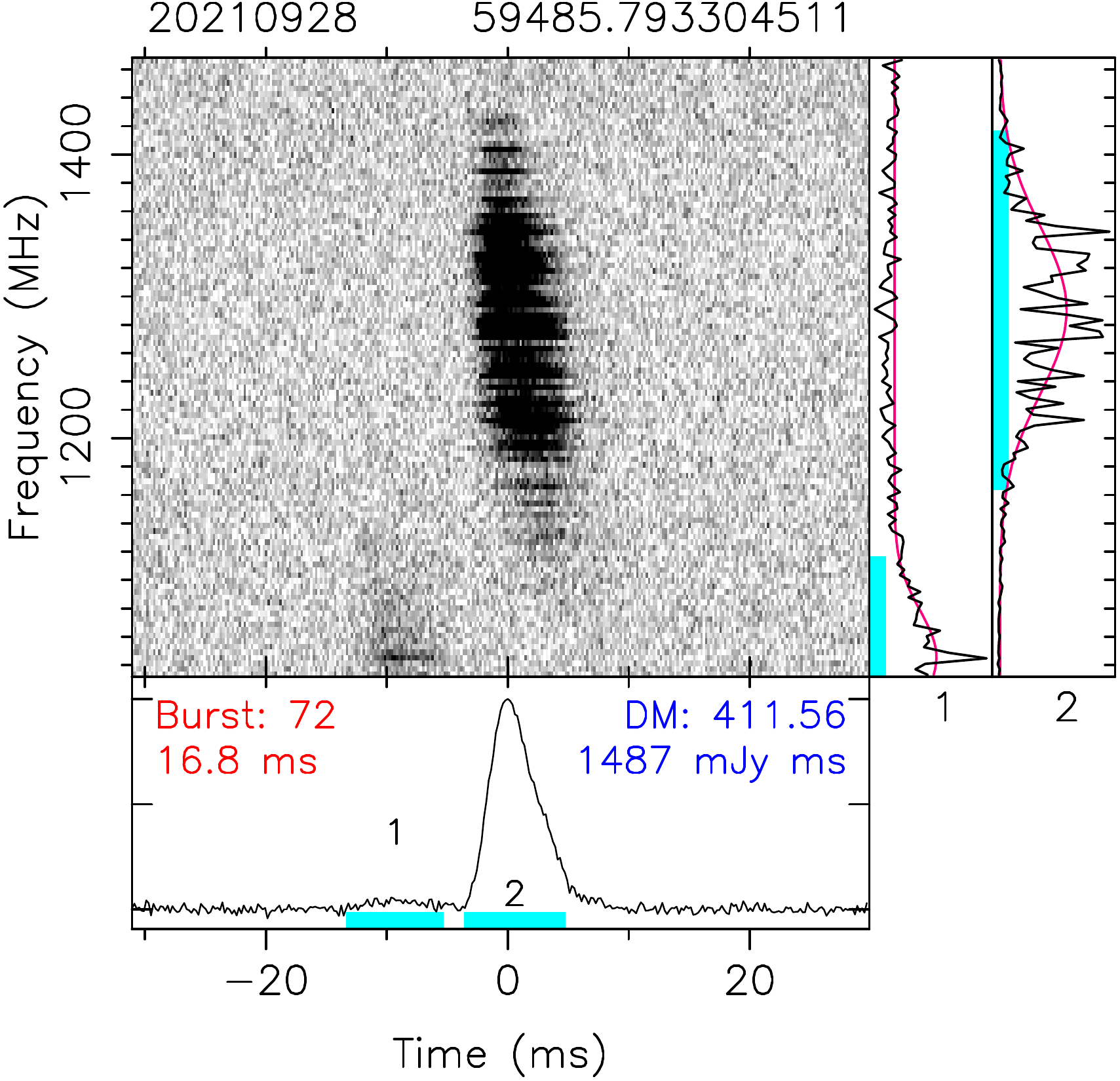}
    \includegraphics[height=37mm]{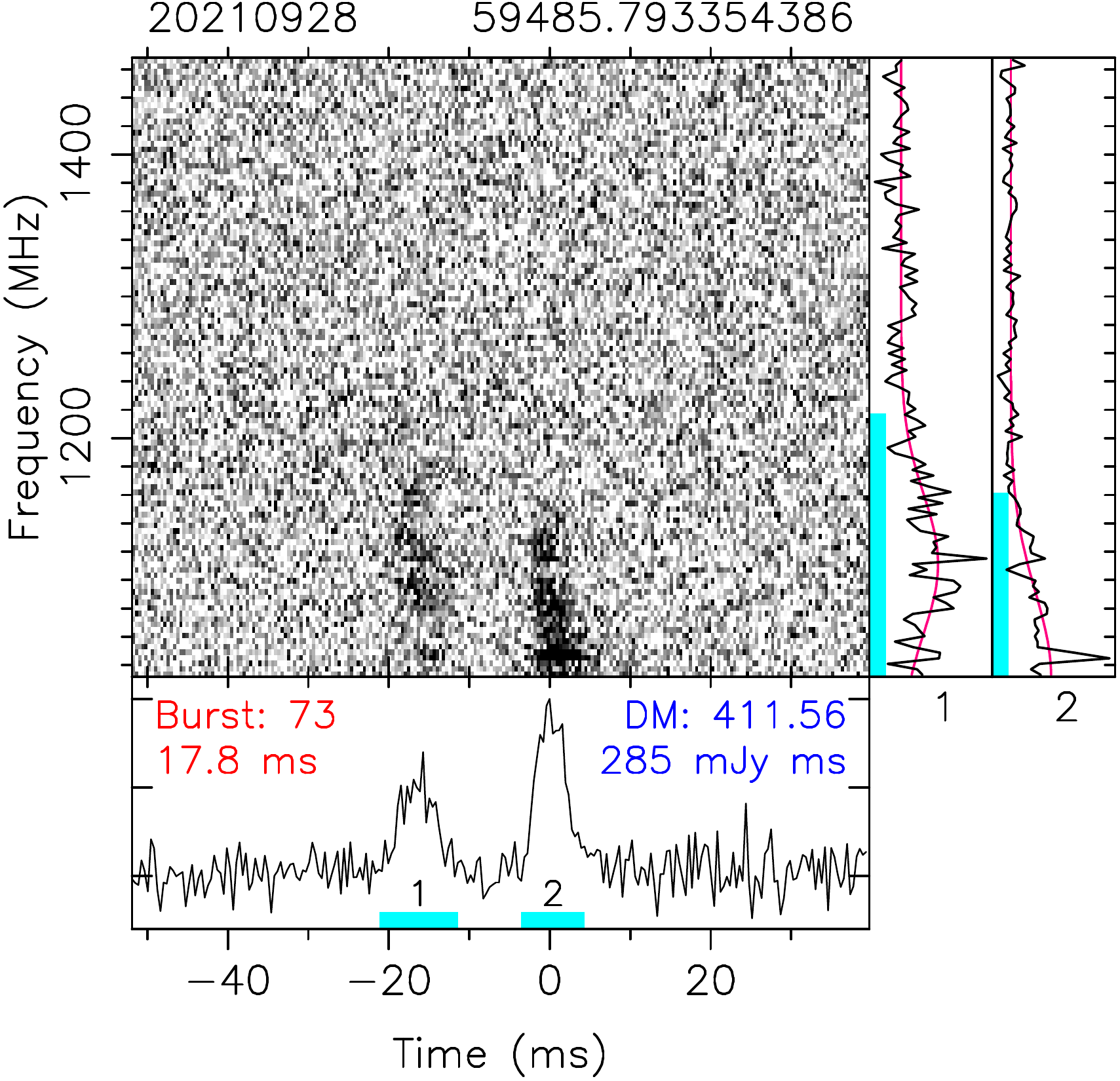}
    \includegraphics[height=37mm]{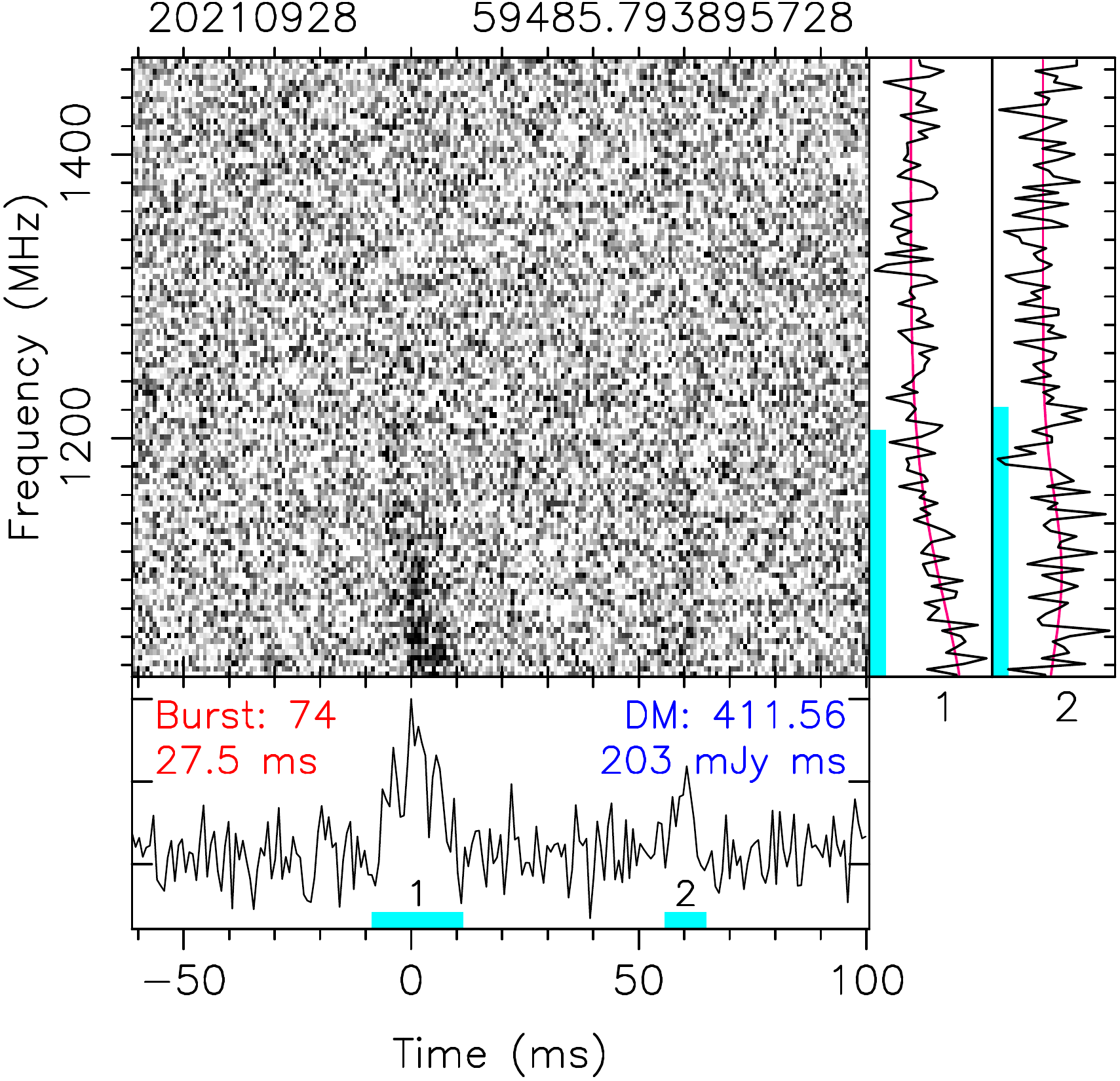}
    \includegraphics[height=37mm]{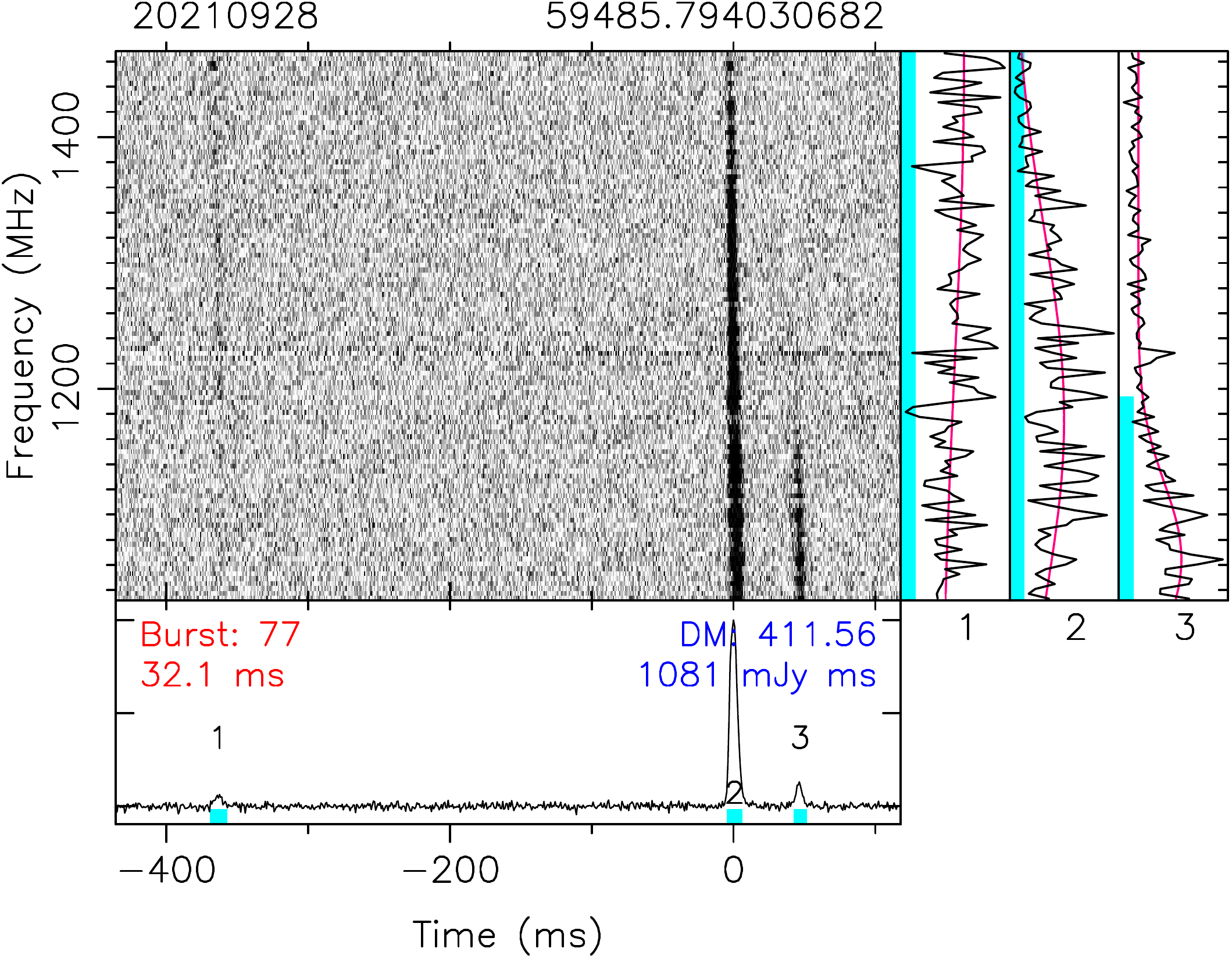}
    \includegraphics[height=37mm]{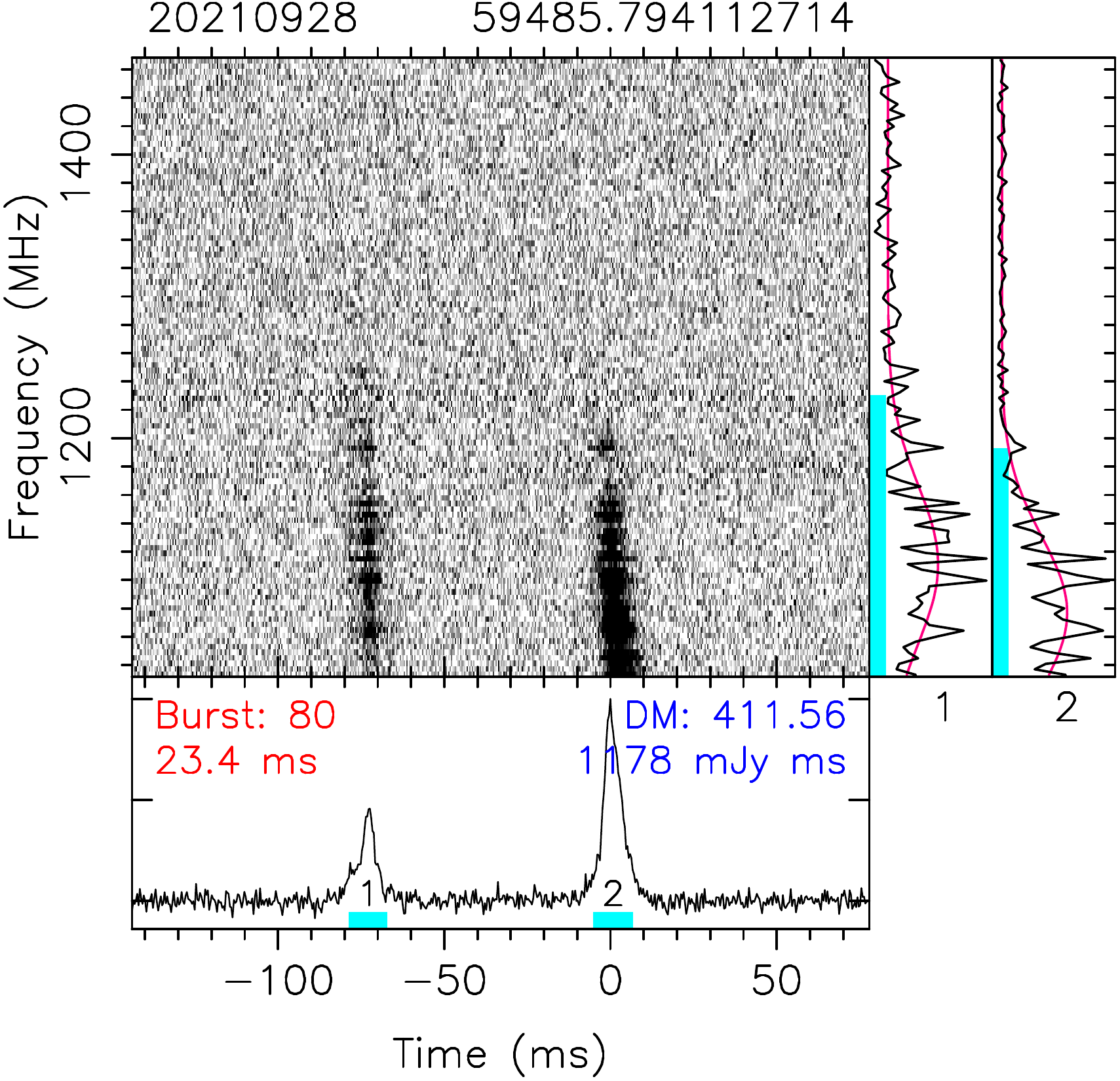}
    \includegraphics[height=37mm]{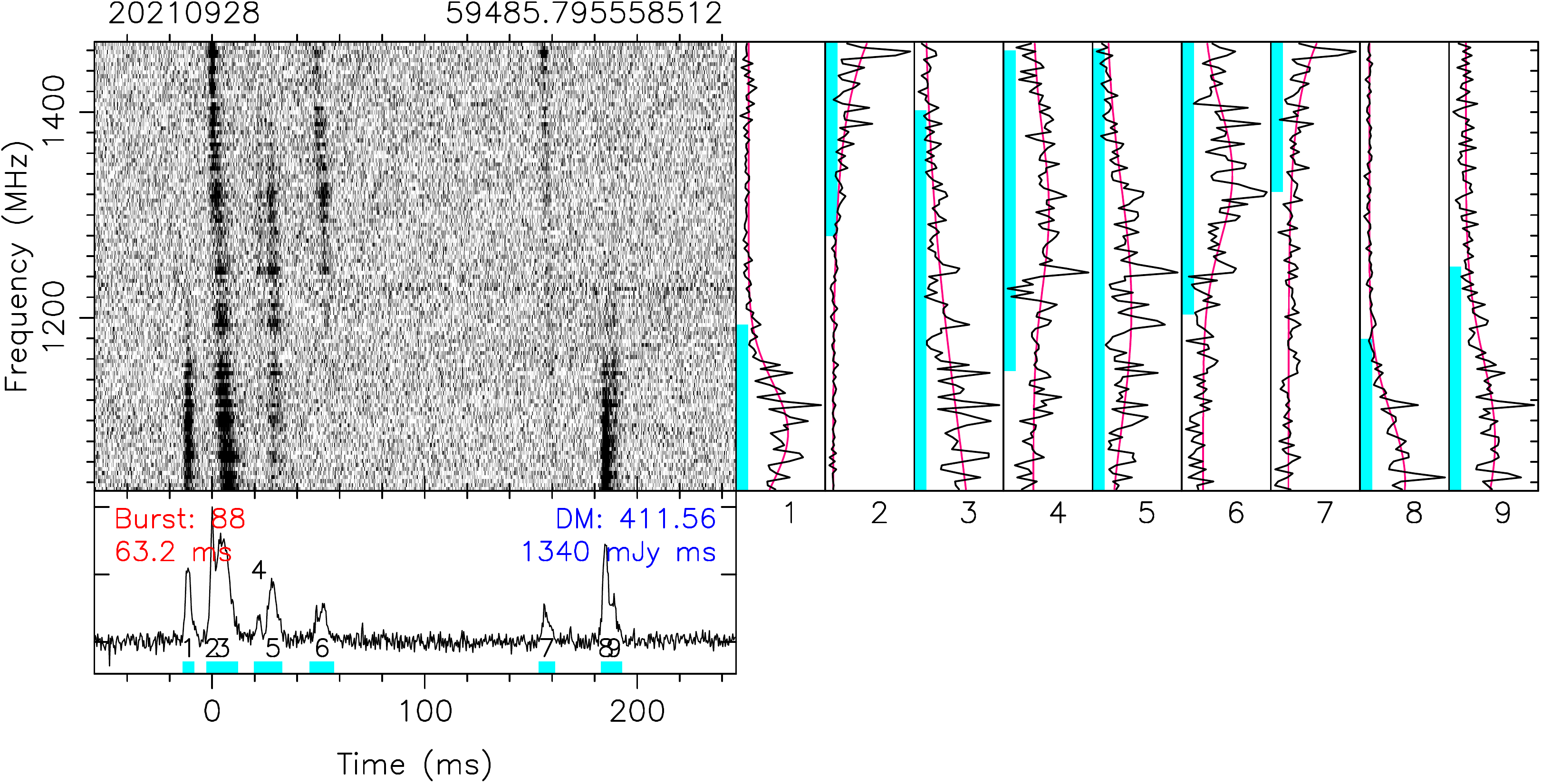}
    \includegraphics[height=37mm]{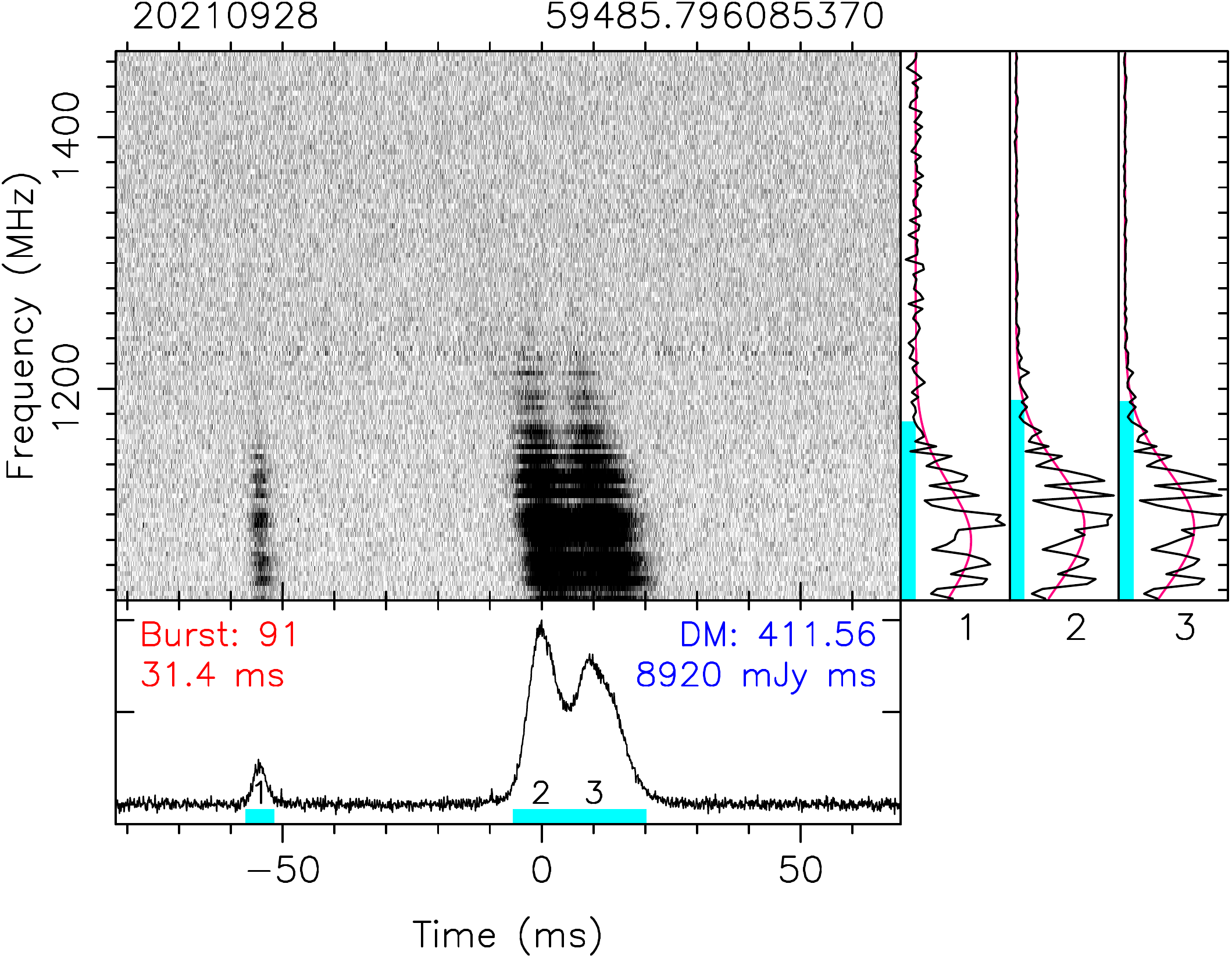}
    \includegraphics[height=37mm]{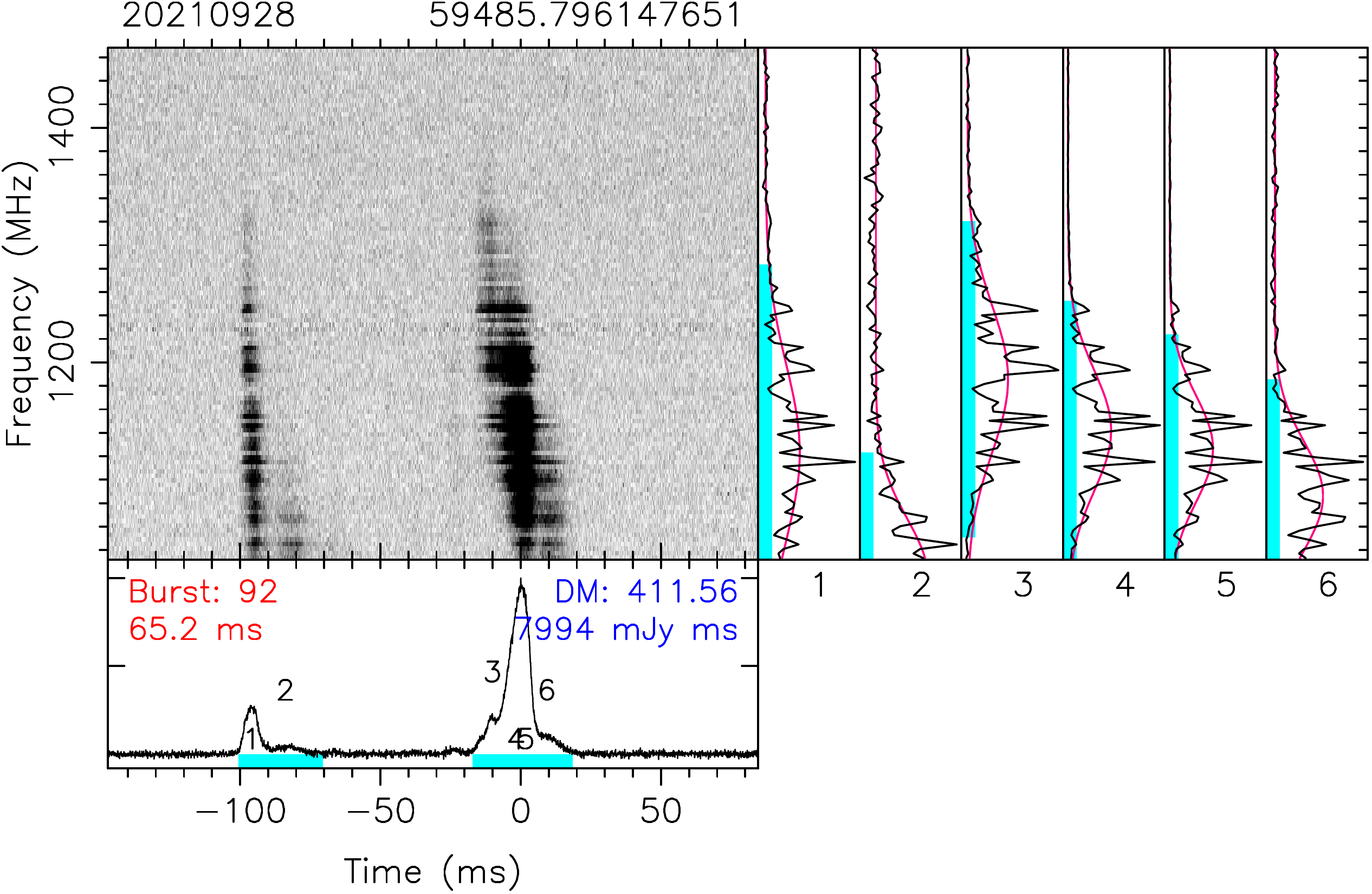}
    \includegraphics[height=37mm]{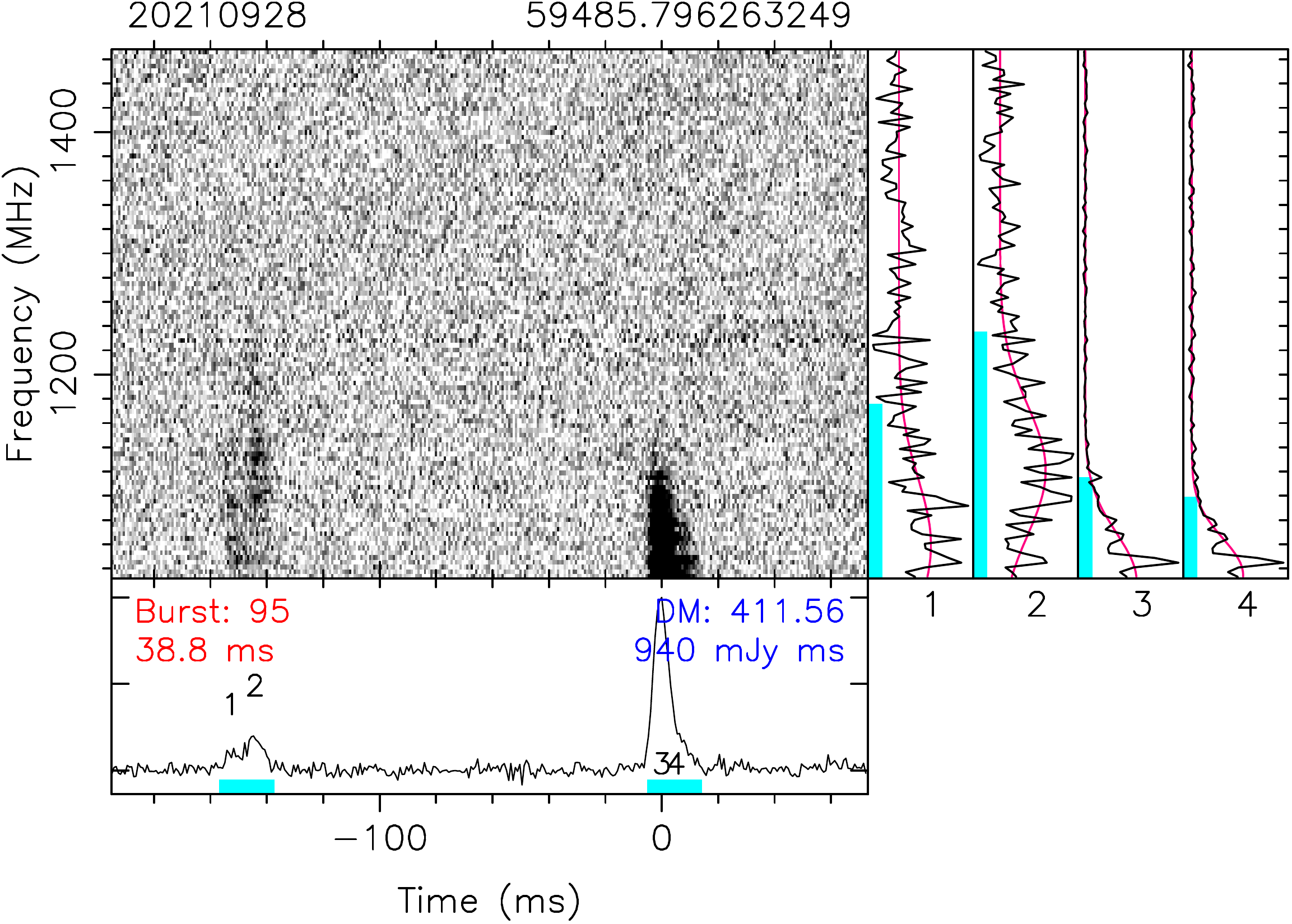}
\caption{\it{ -- continued}.
}
\end{figure*}
\addtocounter{figure}{-1}
\begin{figure*}
    \flushleft
    \includegraphics[height=37mm]{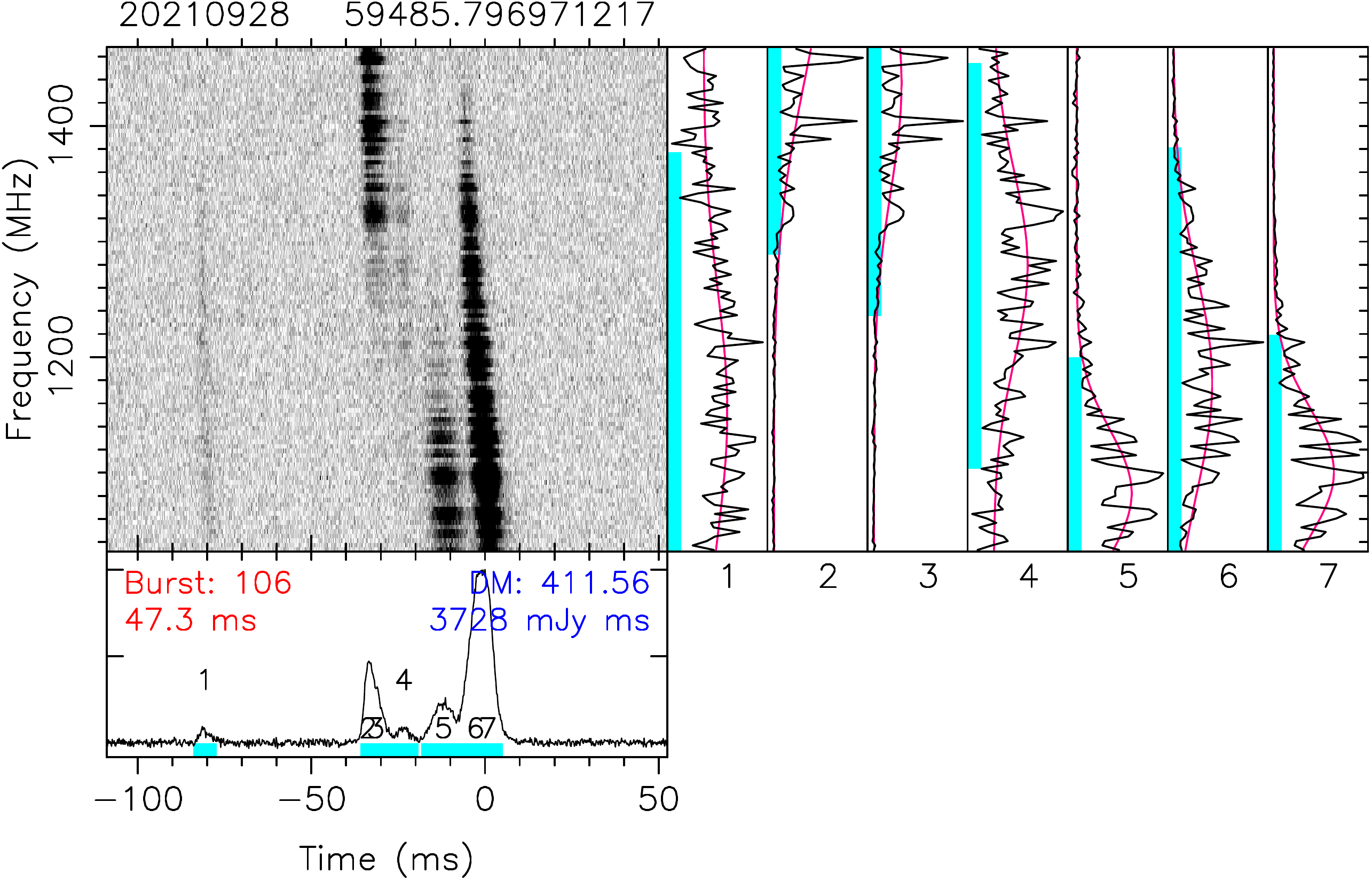}
    \includegraphics[height=37mm]{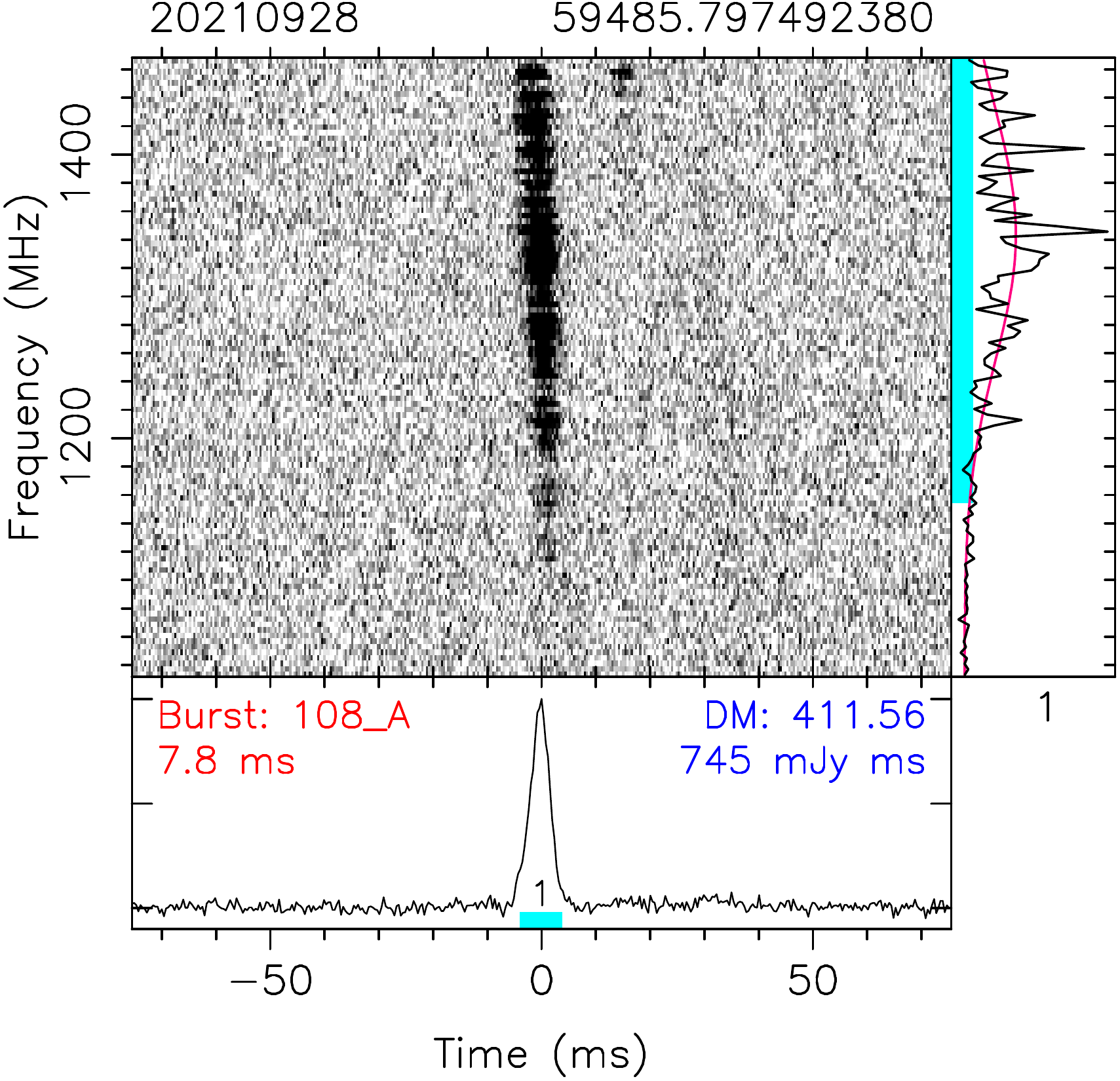}
    \includegraphics[height=37mm]{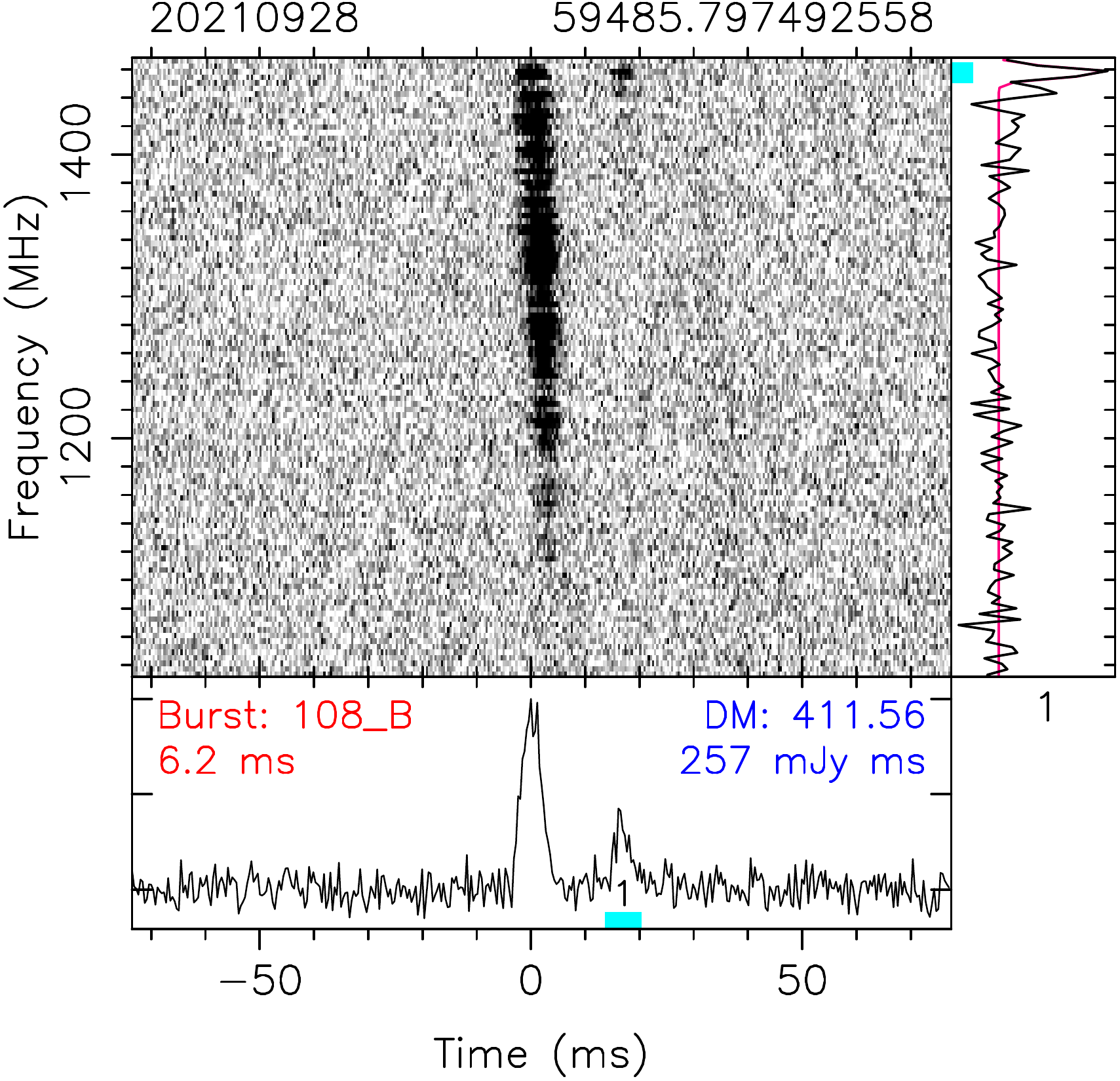}
    \includegraphics[height=37mm]{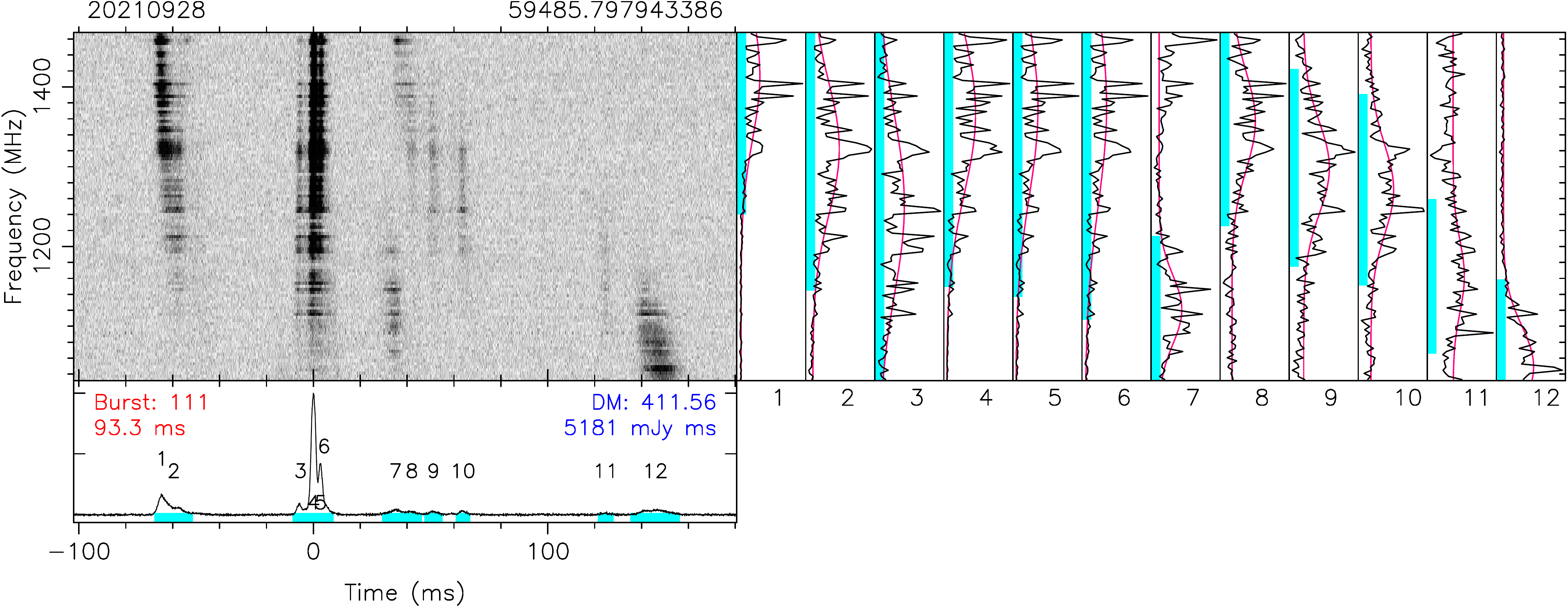}
    \includegraphics[height=37mm]{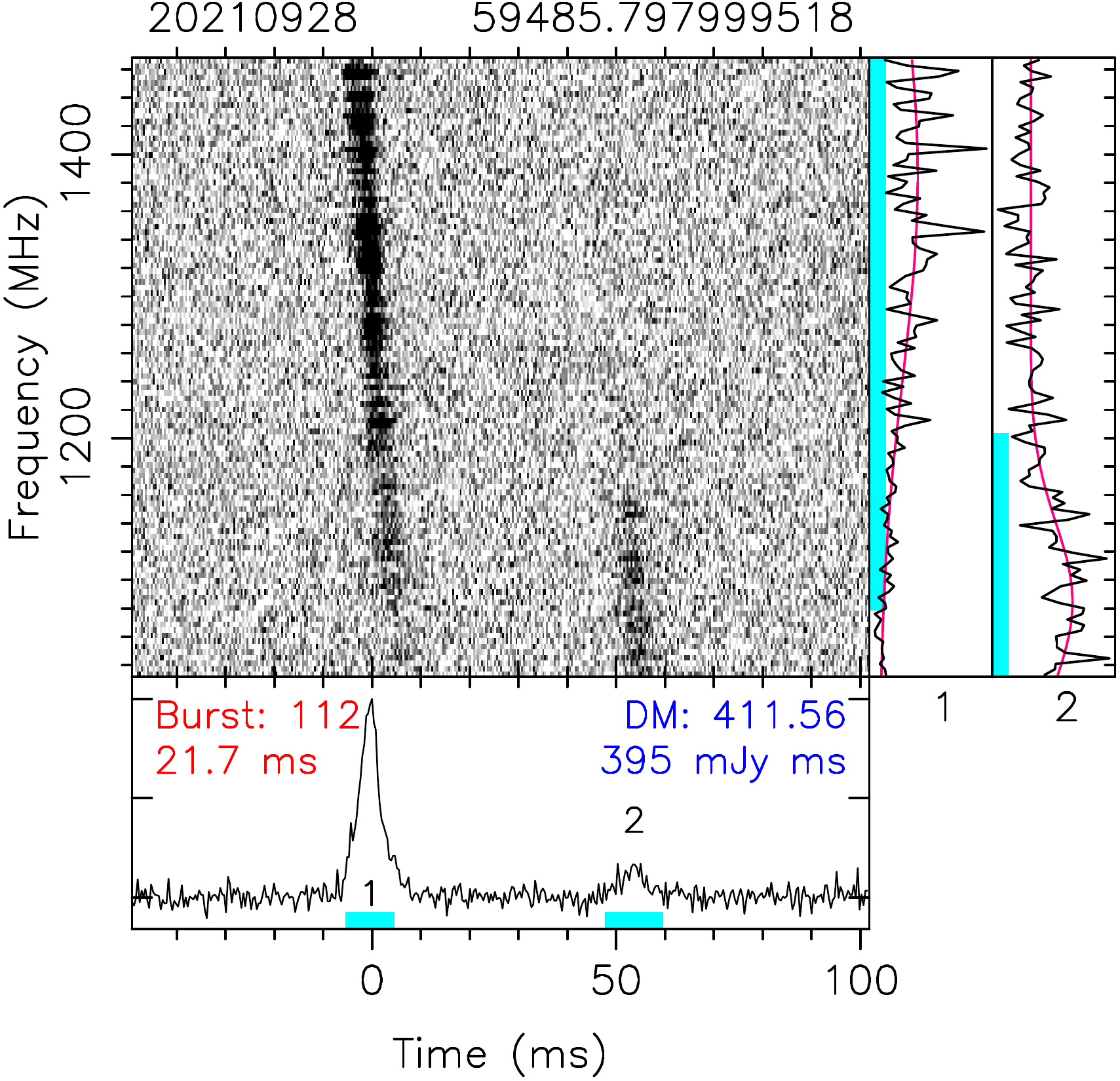}
    \includegraphics[height=37mm]{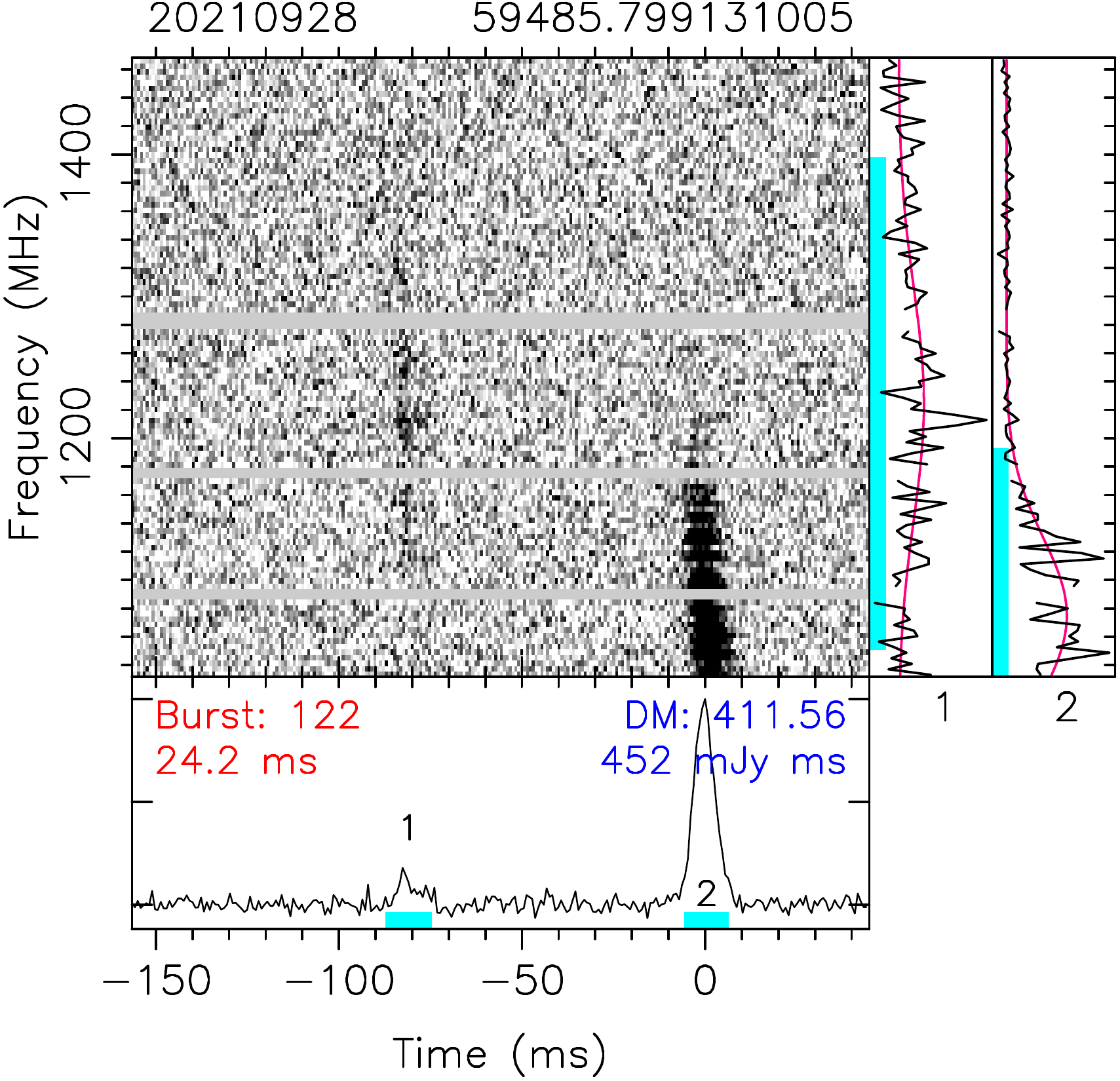}
    \includegraphics[height=37mm]{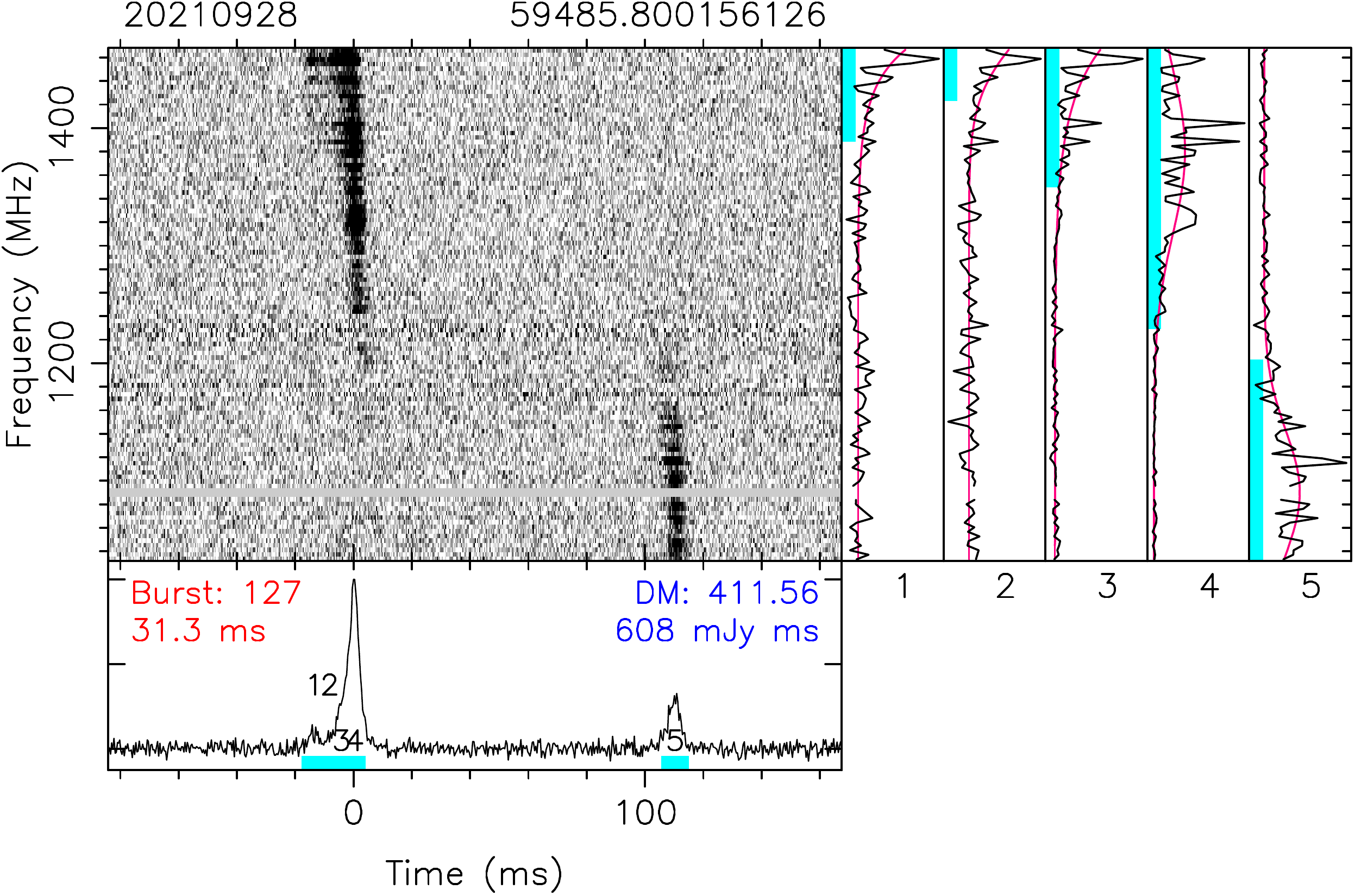}
    \includegraphics[height=37mm]{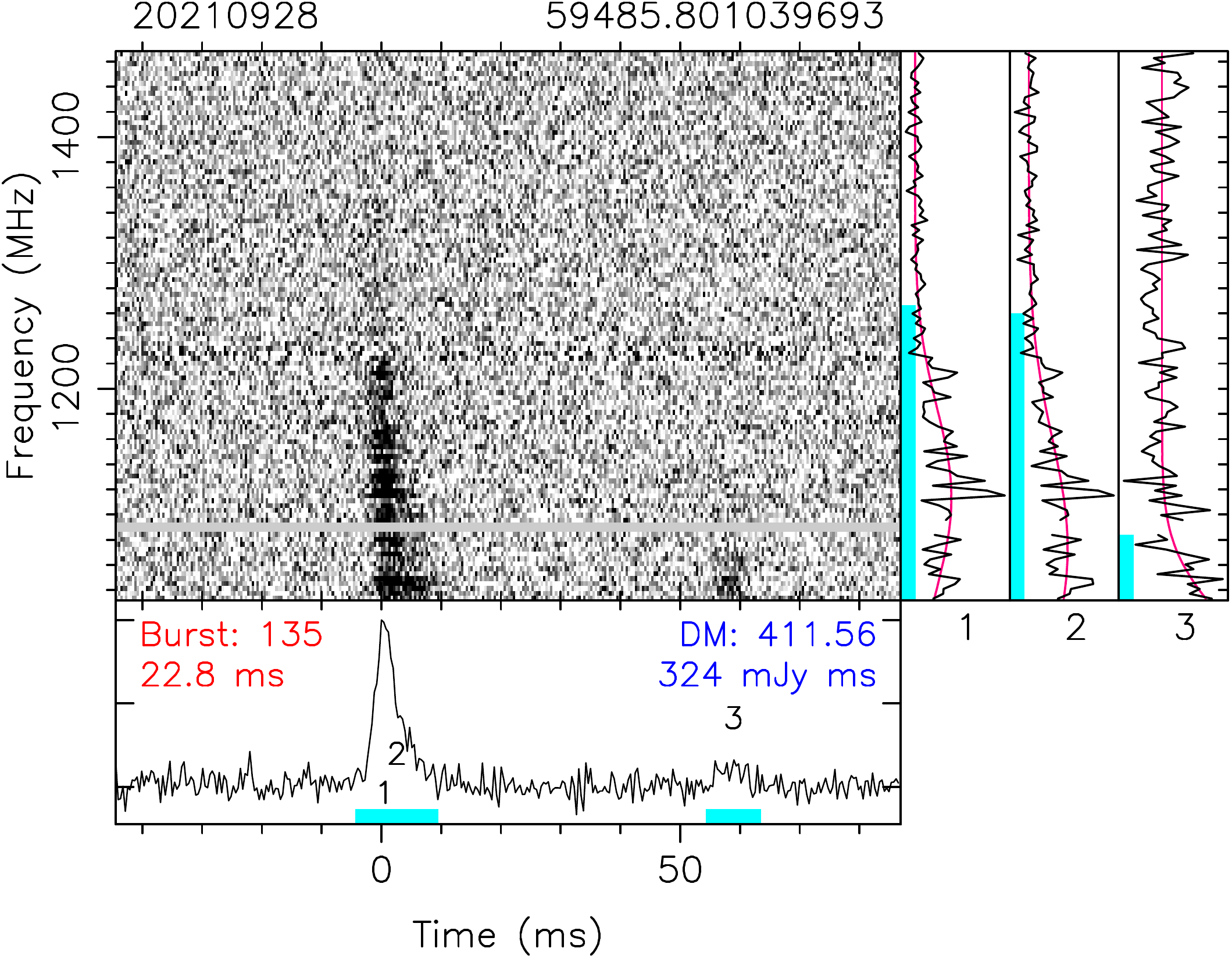}
    \includegraphics[height=37mm]{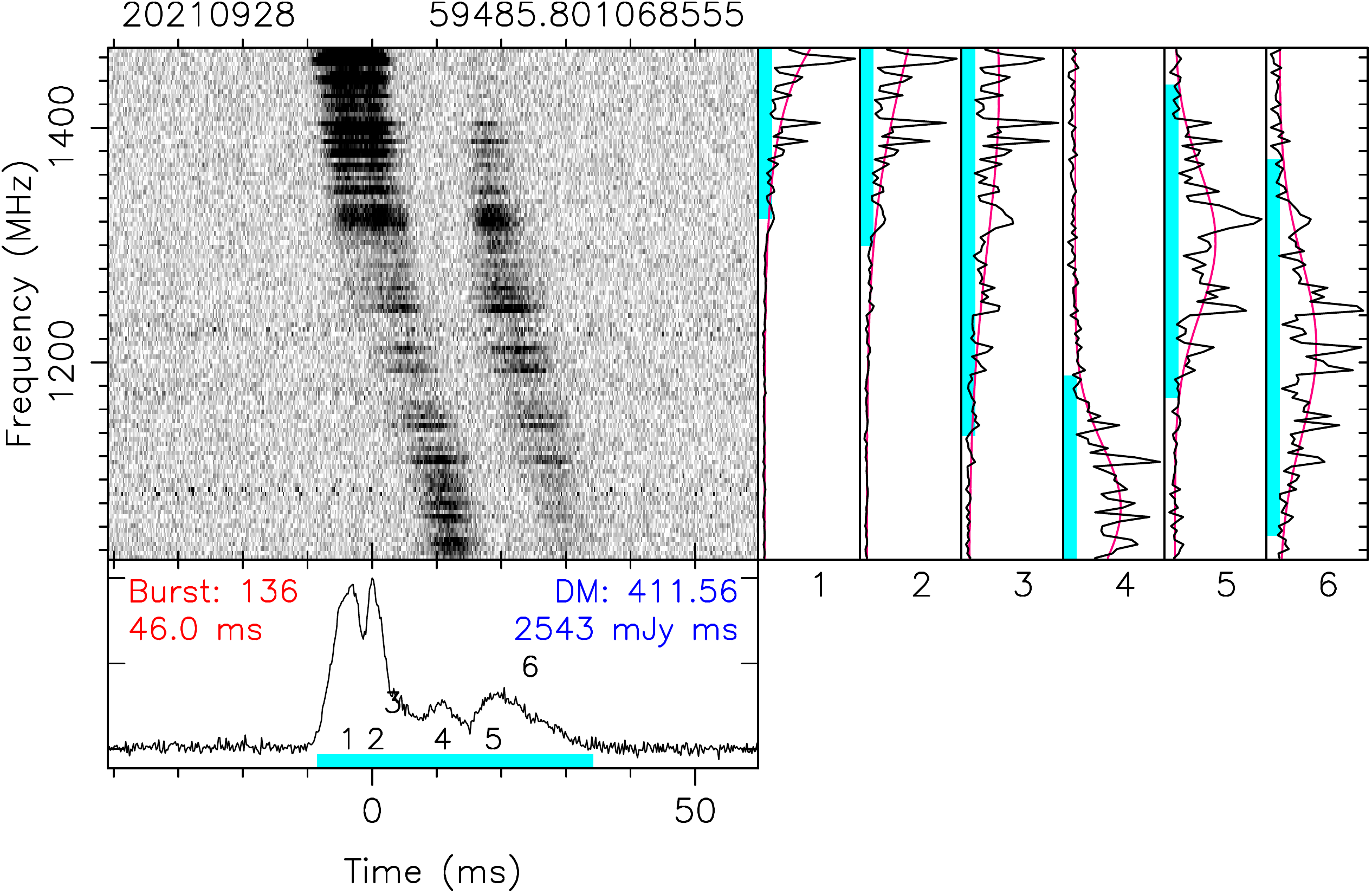}
    \includegraphics[height=37mm]{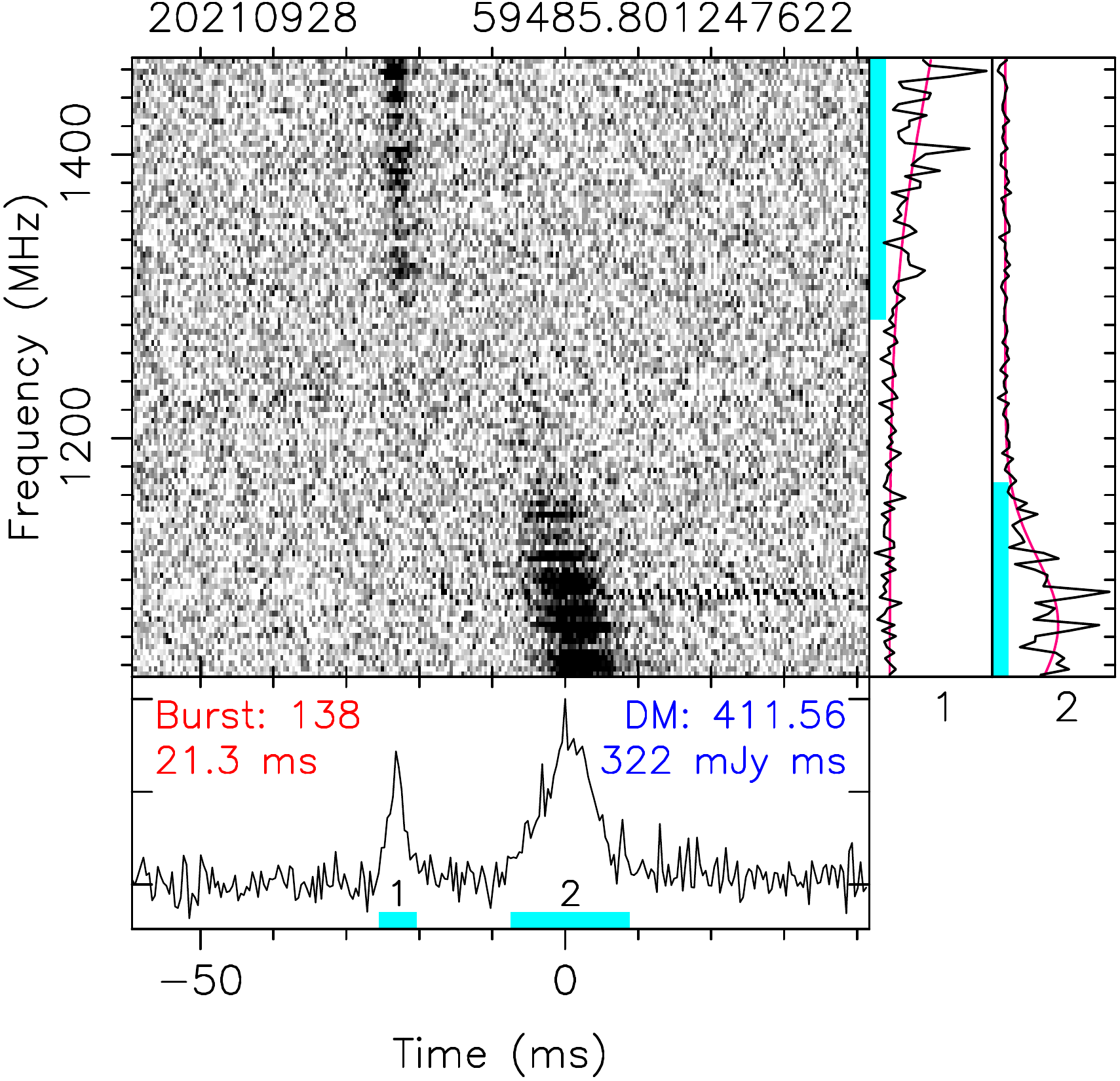}
    \includegraphics[height=37mm]{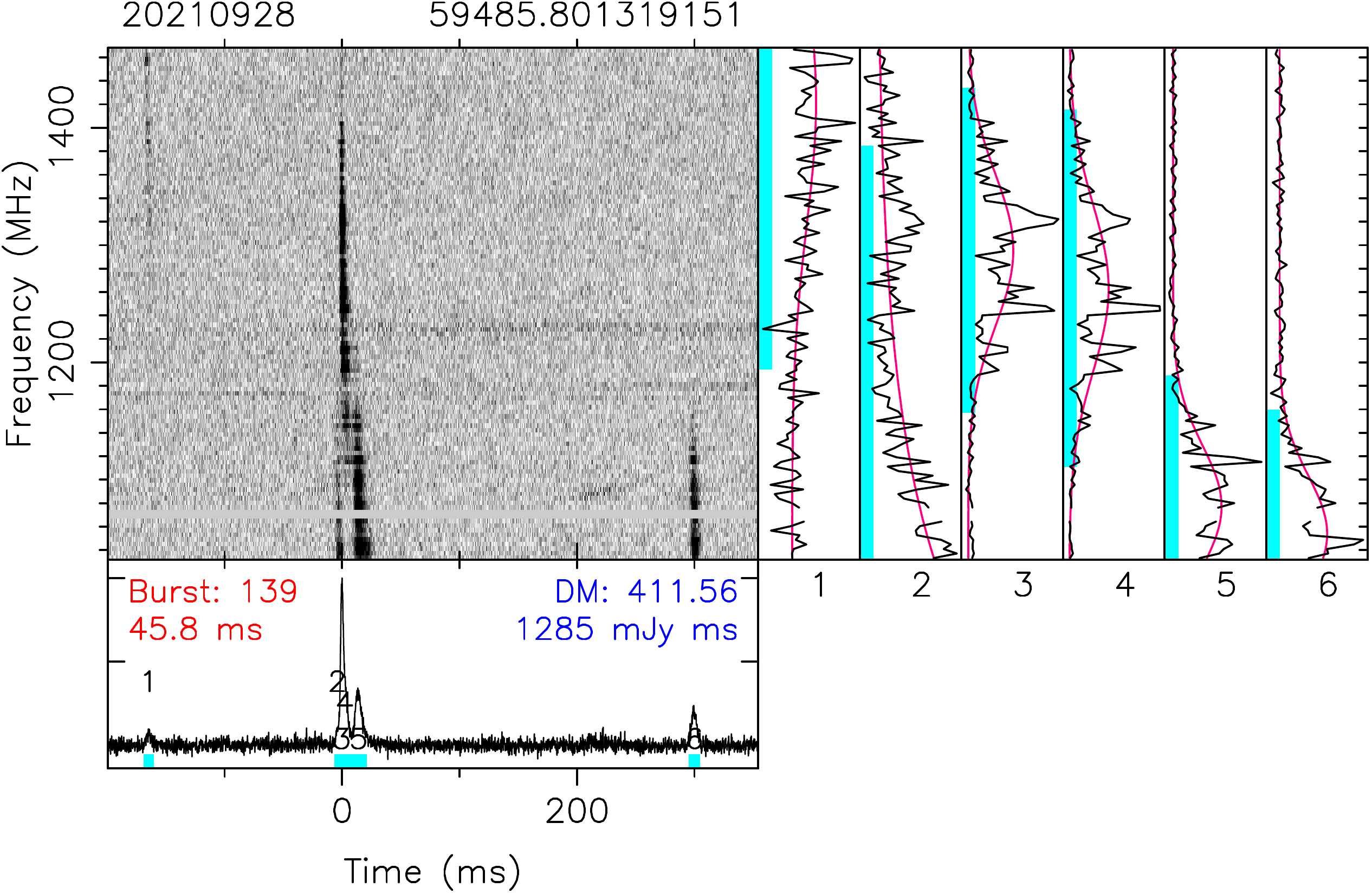}
    \includegraphics[height=37mm]{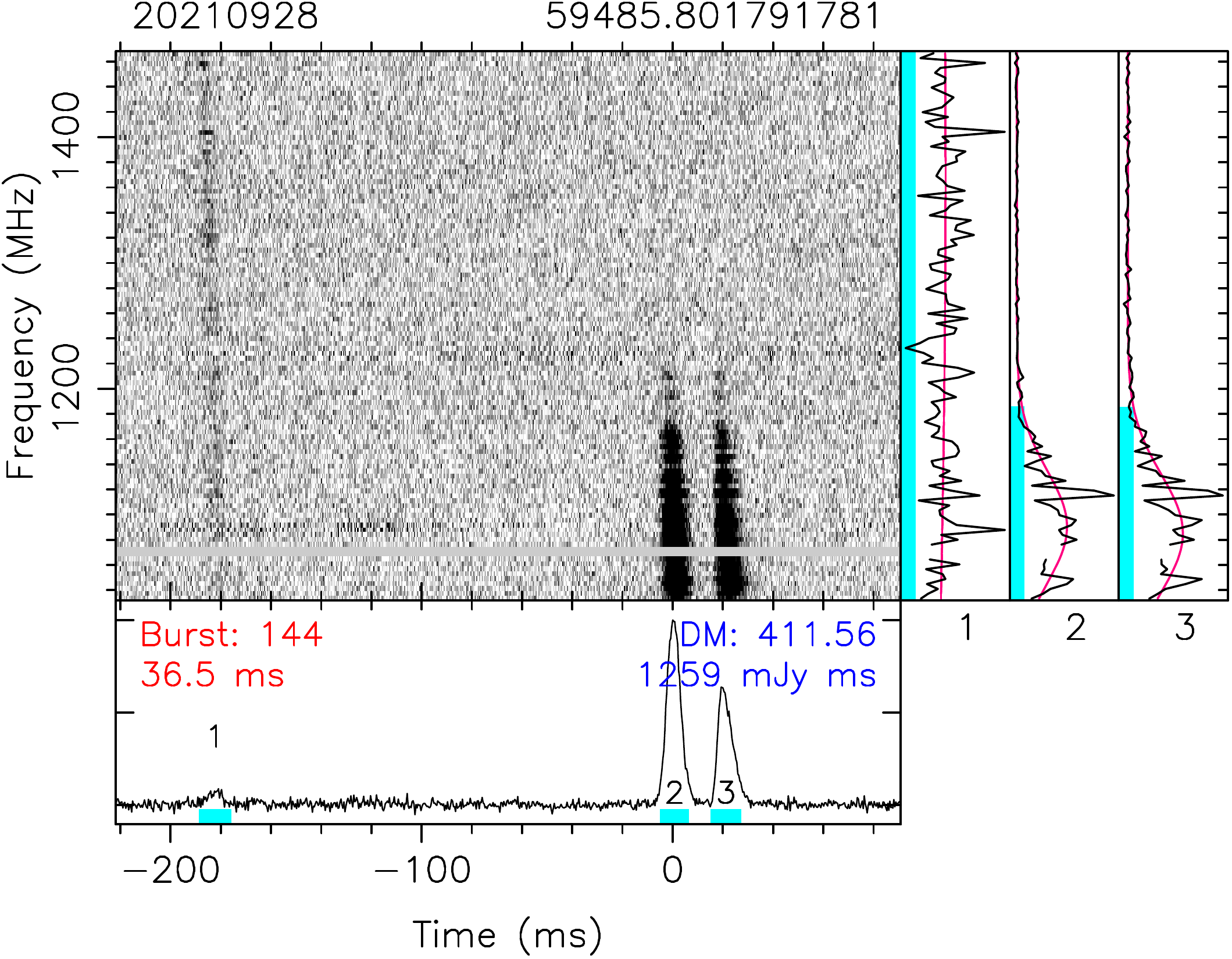}
    \includegraphics[height=37mm]{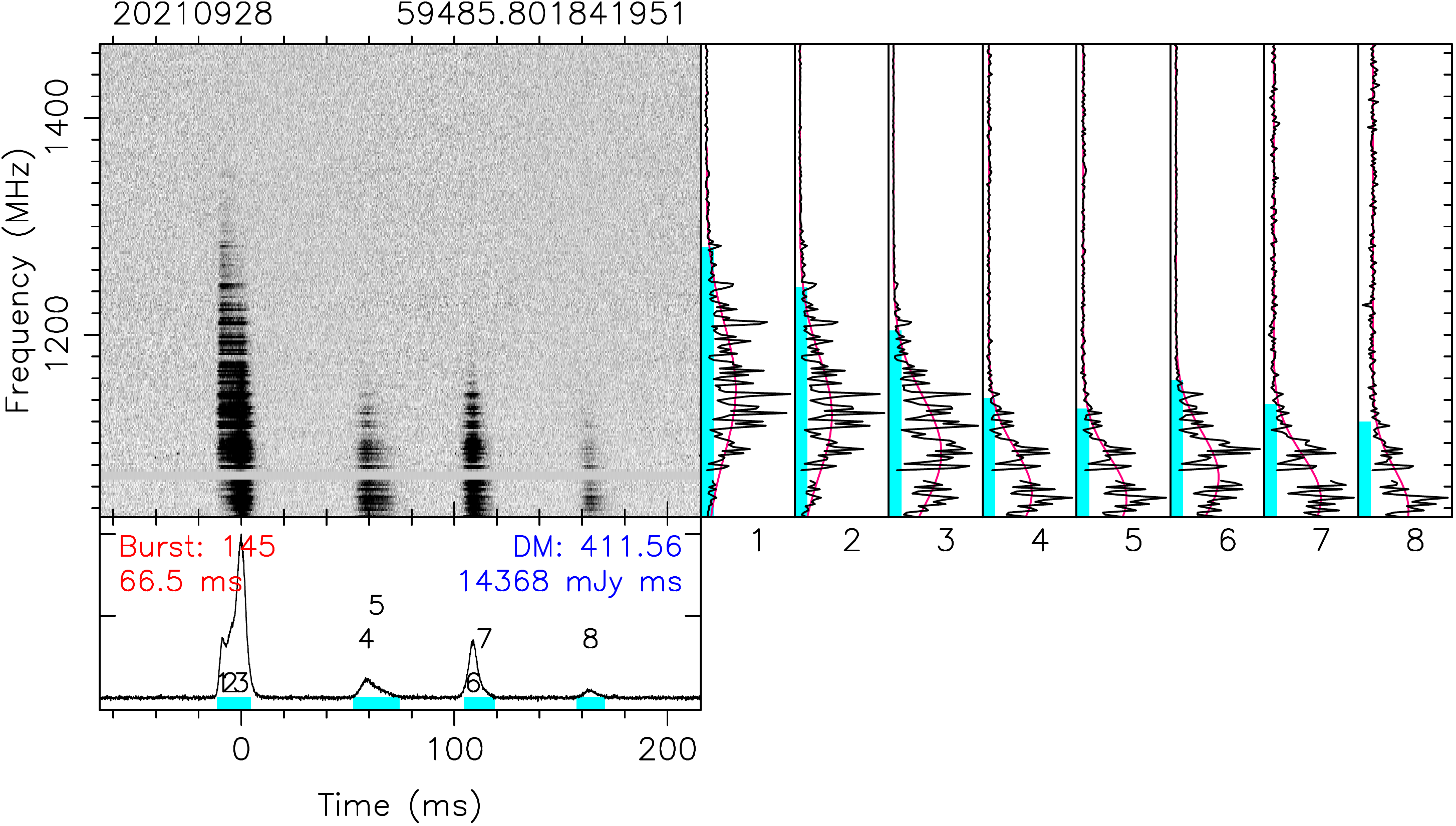}
    \includegraphics[height=37mm]{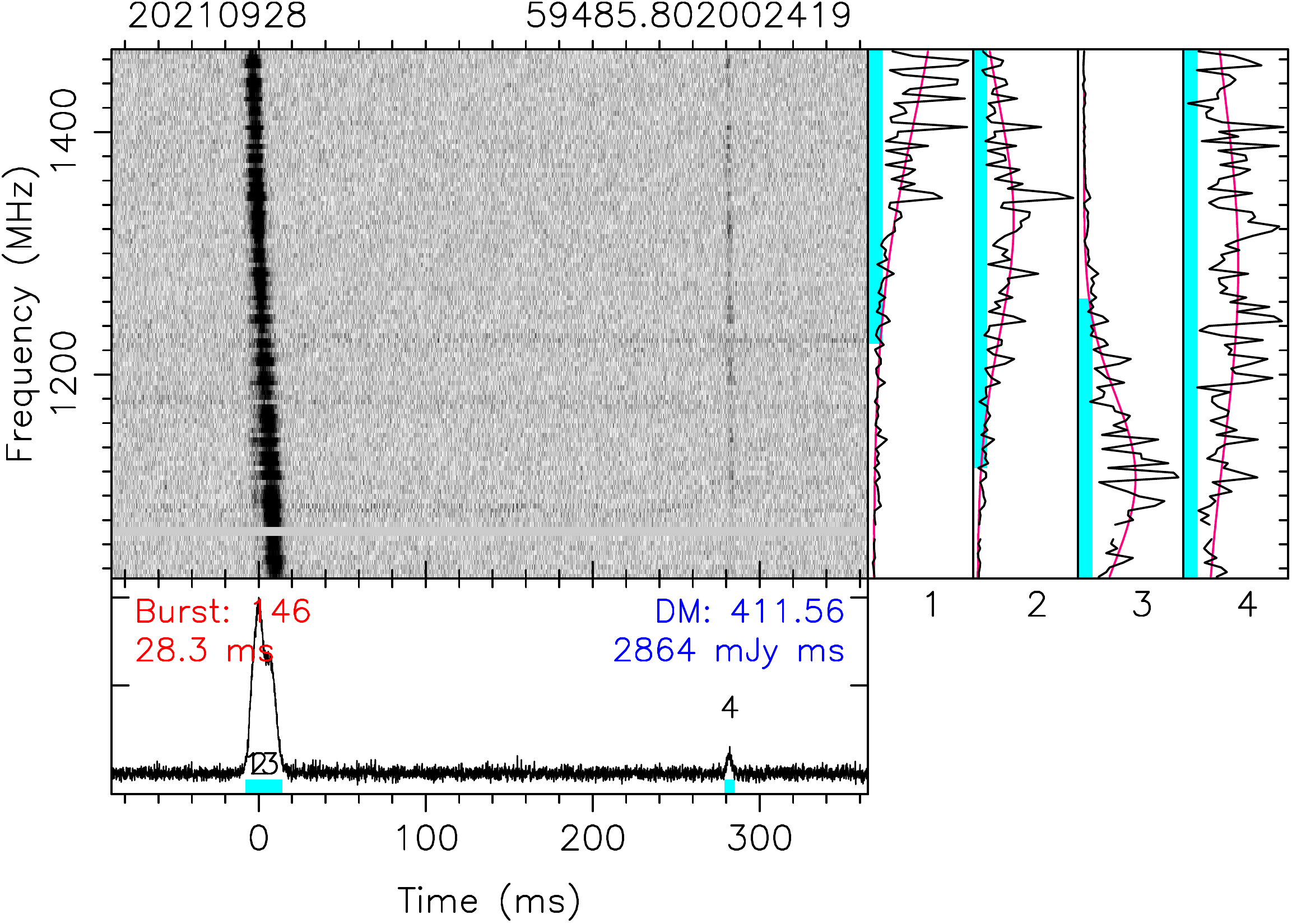}
    \includegraphics[height=37mm]{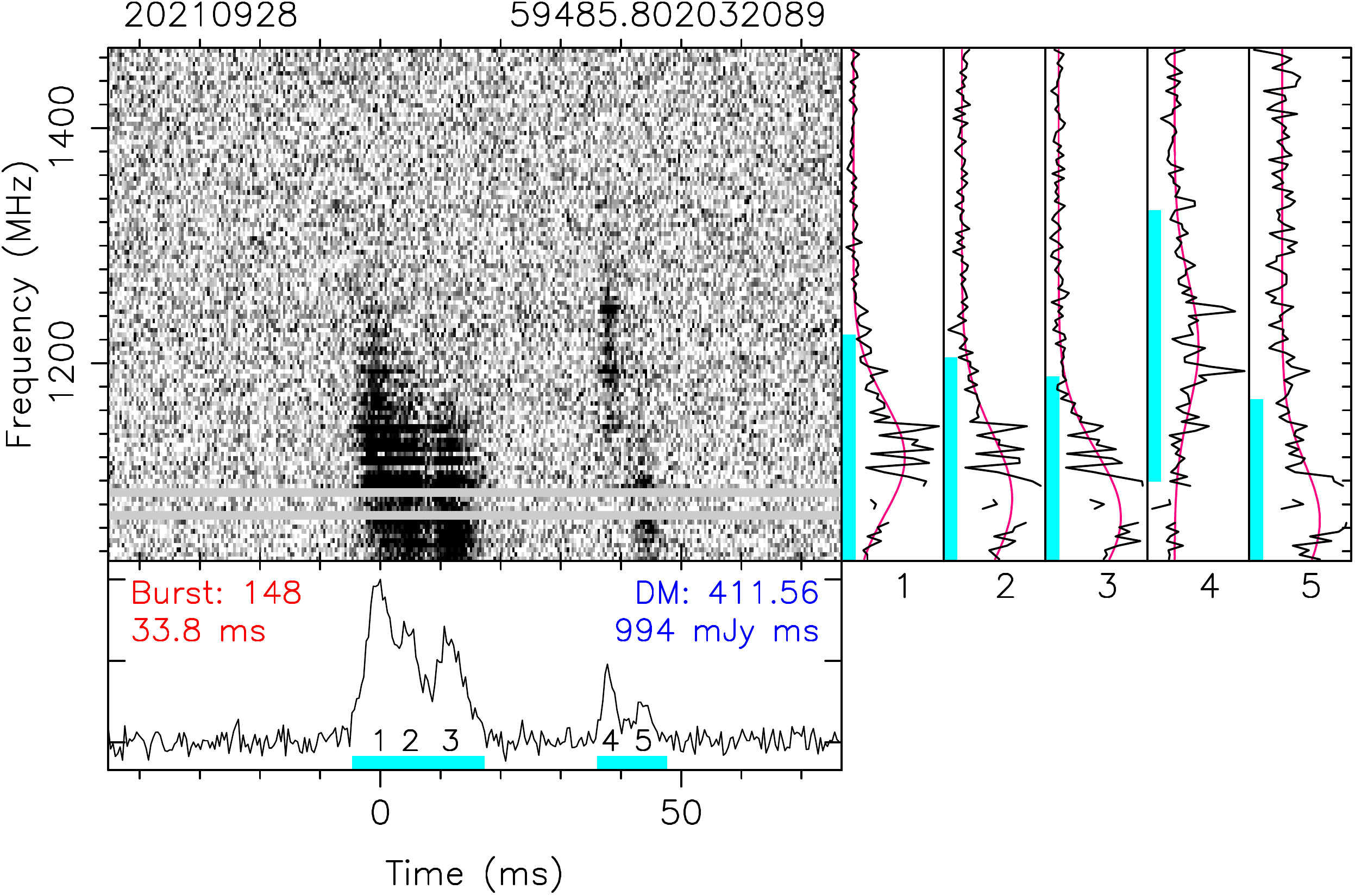}
    \includegraphics[height=37mm]{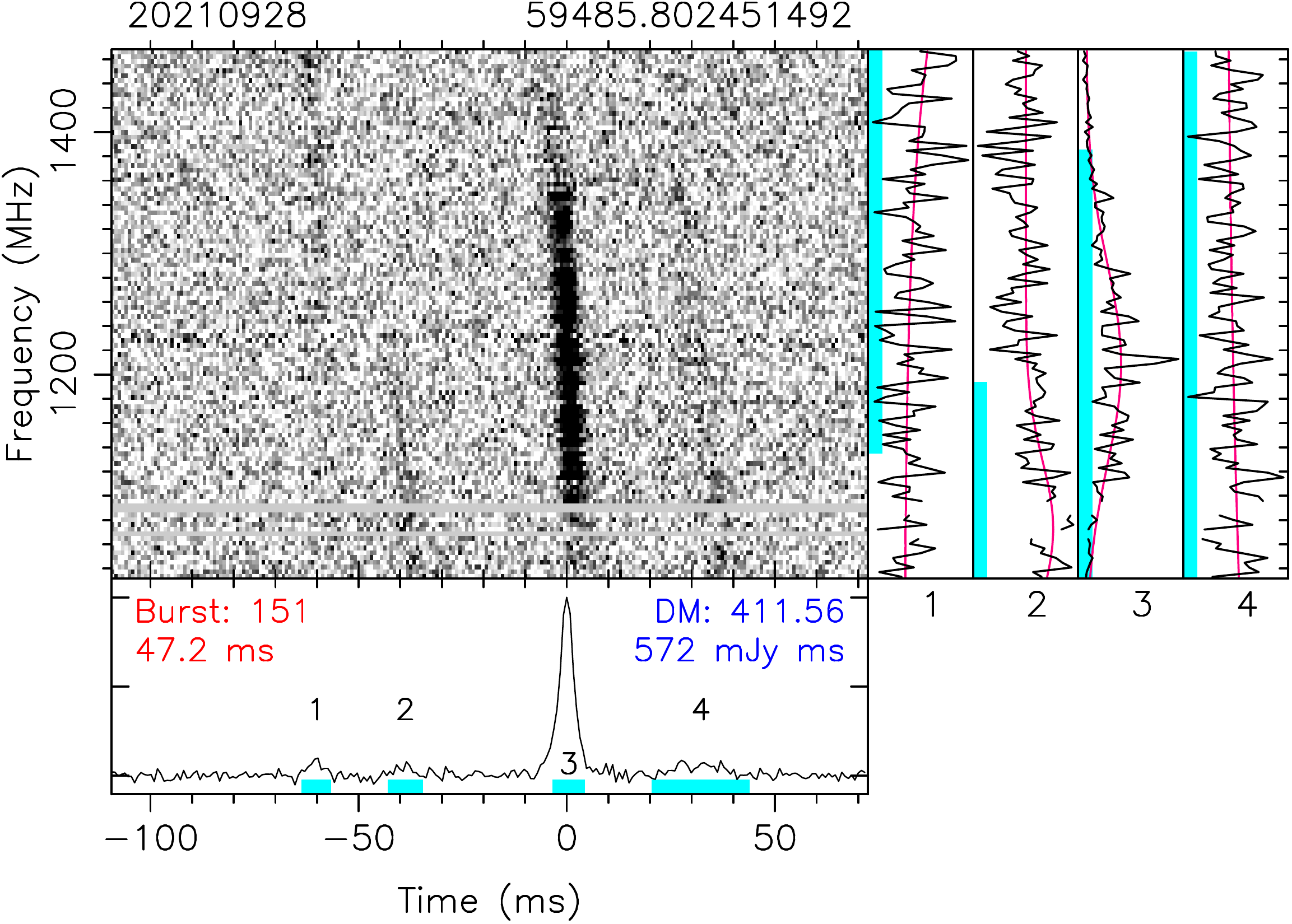}
    \includegraphics[height=37mm]{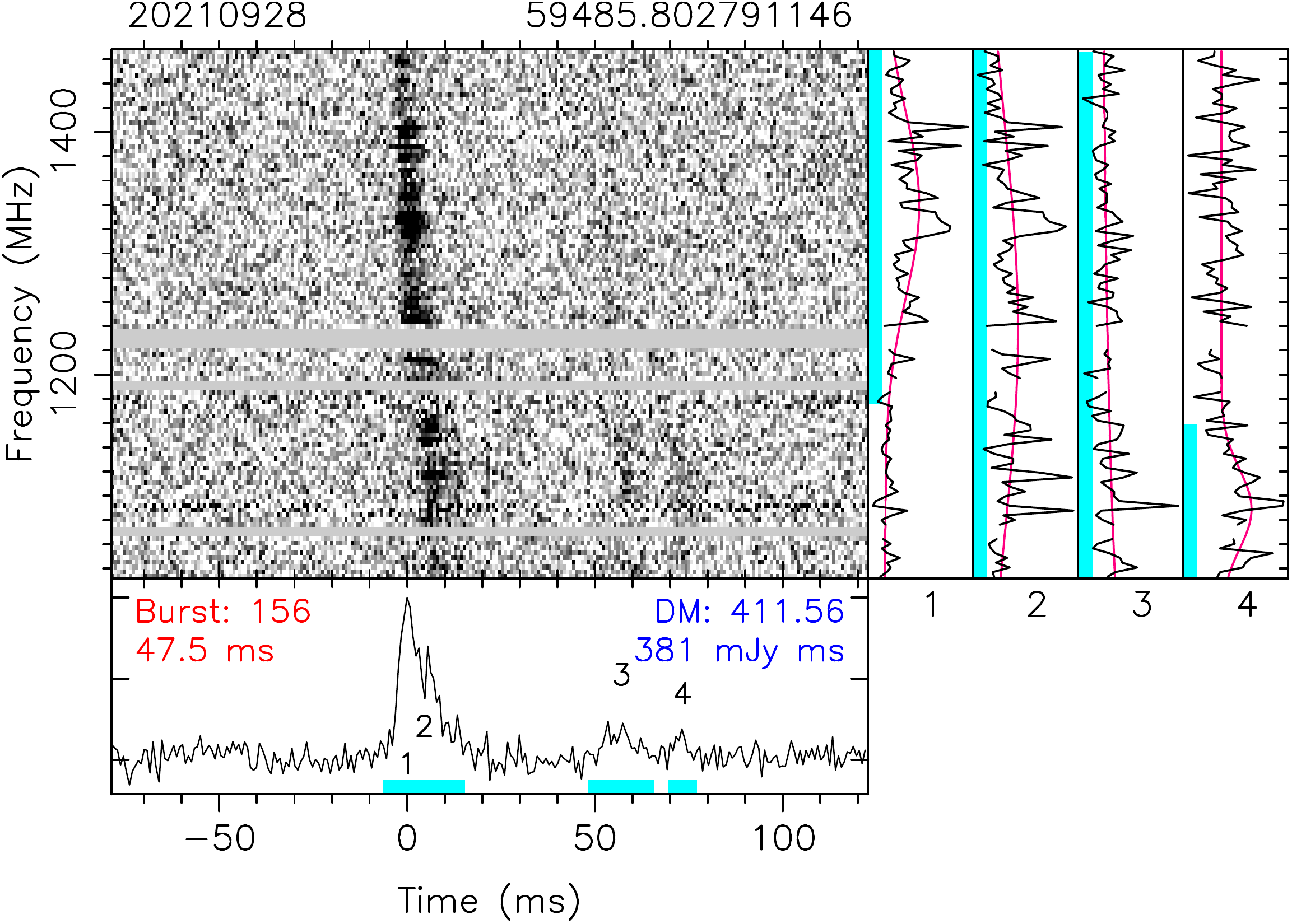}
\caption{\it{ -- continued}.
}
\end{figure*}
\addtocounter{figure}{-1}
\begin{figure*}
    \flushleft
    \includegraphics[height=37mm]{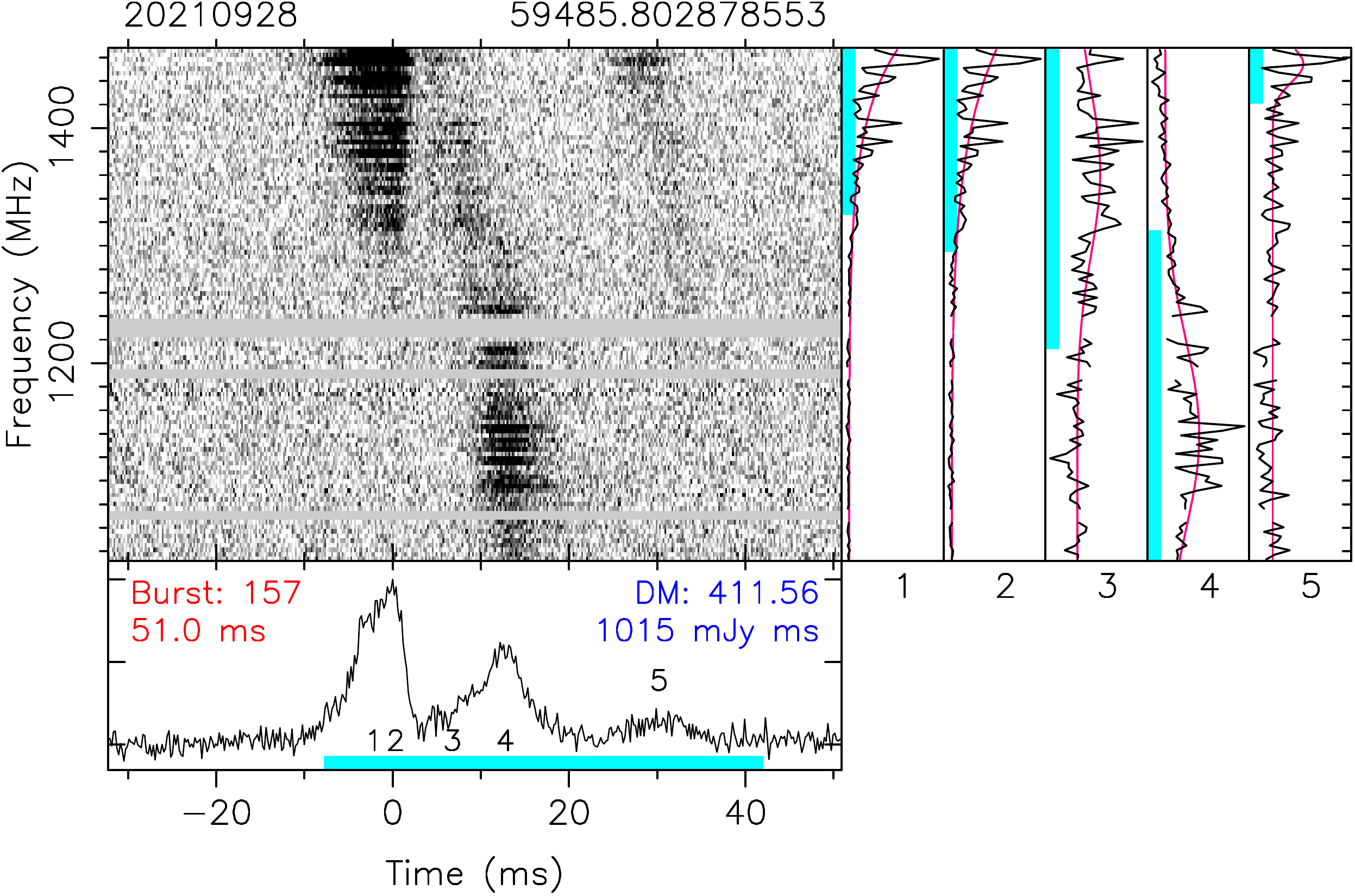}
    \includegraphics[height=37mm]{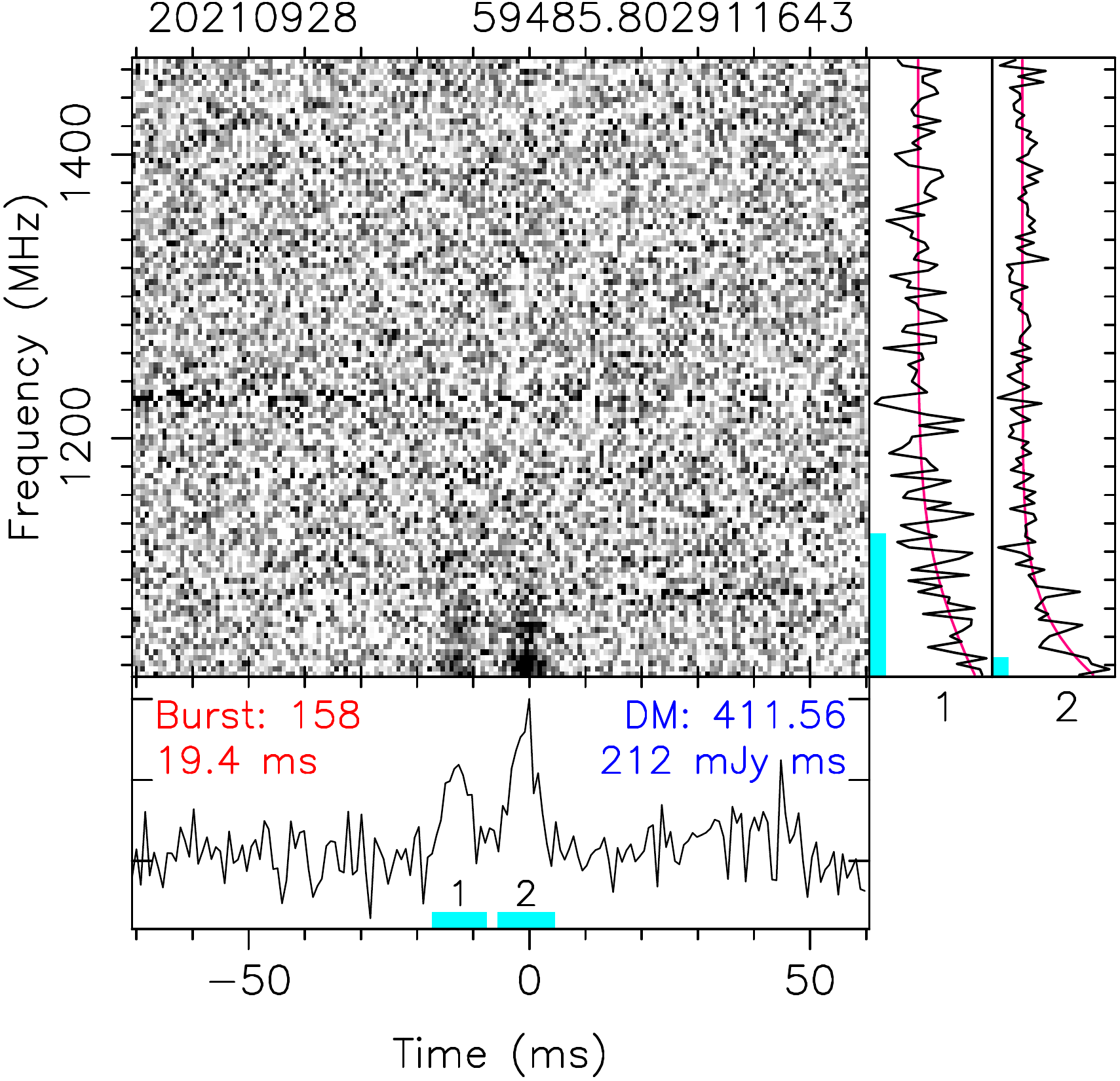}
    \includegraphics[height=37mm]{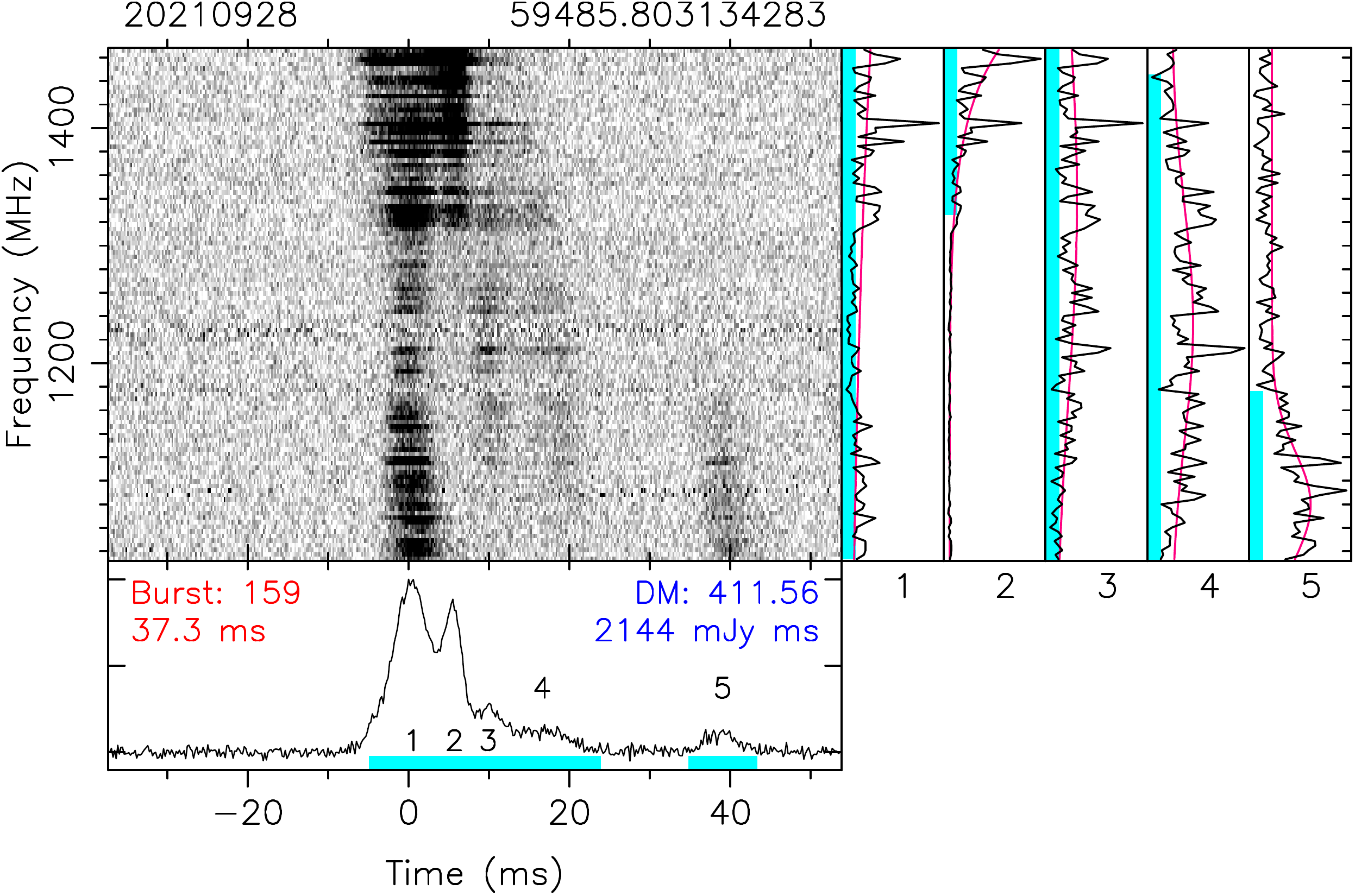}
    \includegraphics[height=37mm]{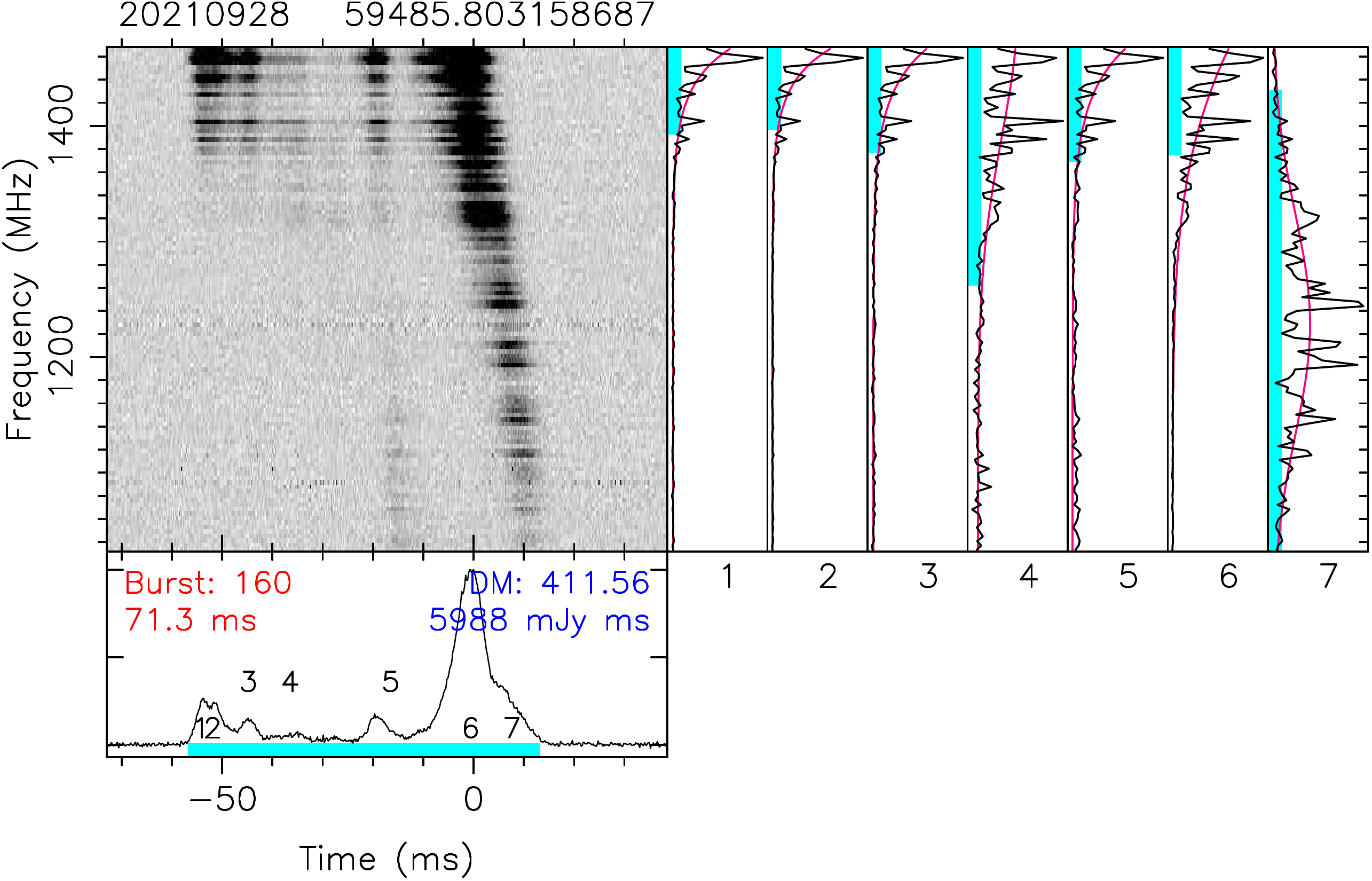}
    \includegraphics[height=37mm]{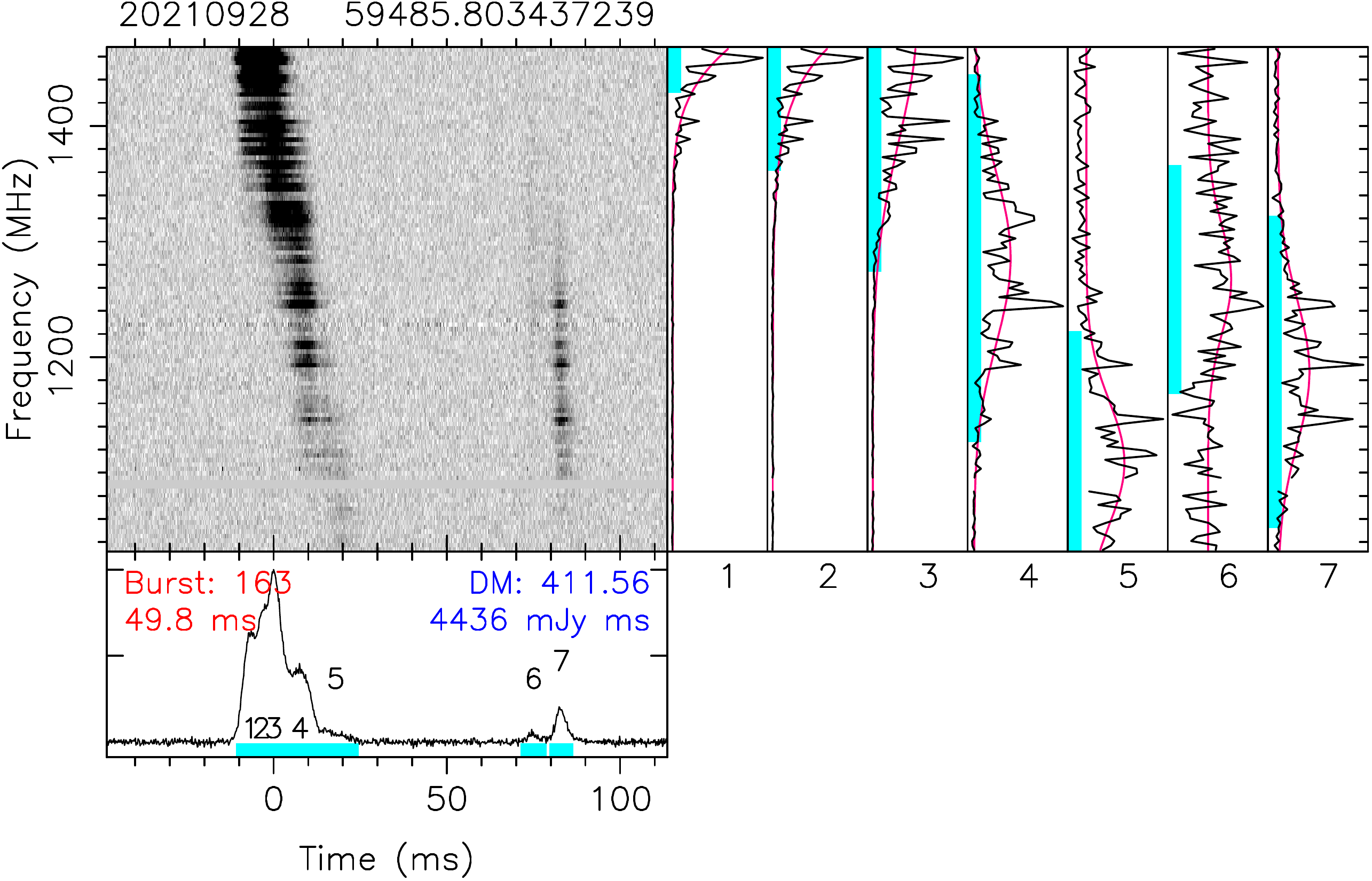}
    \includegraphics[height=37mm]{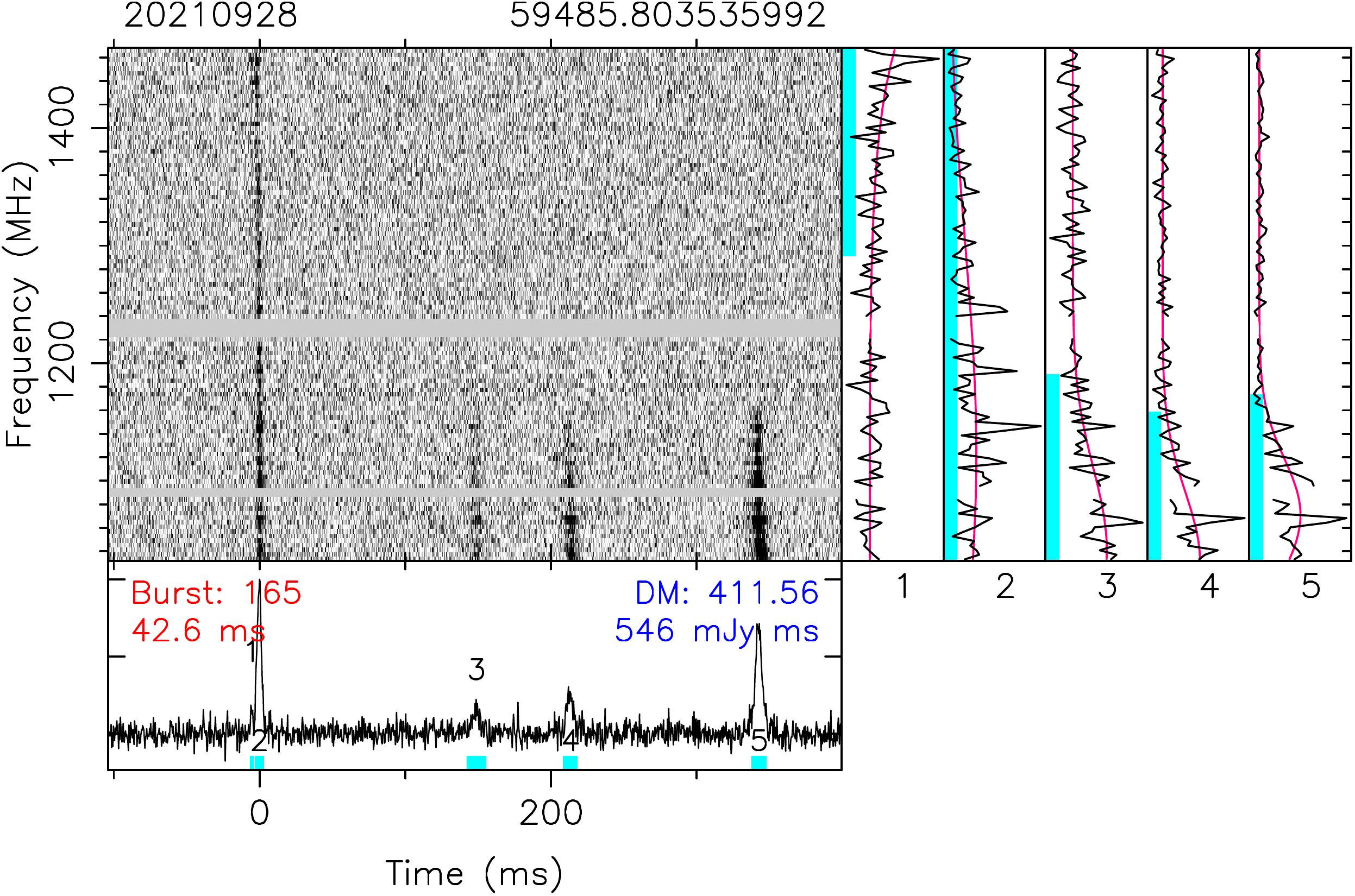}
    \includegraphics[height=37mm]{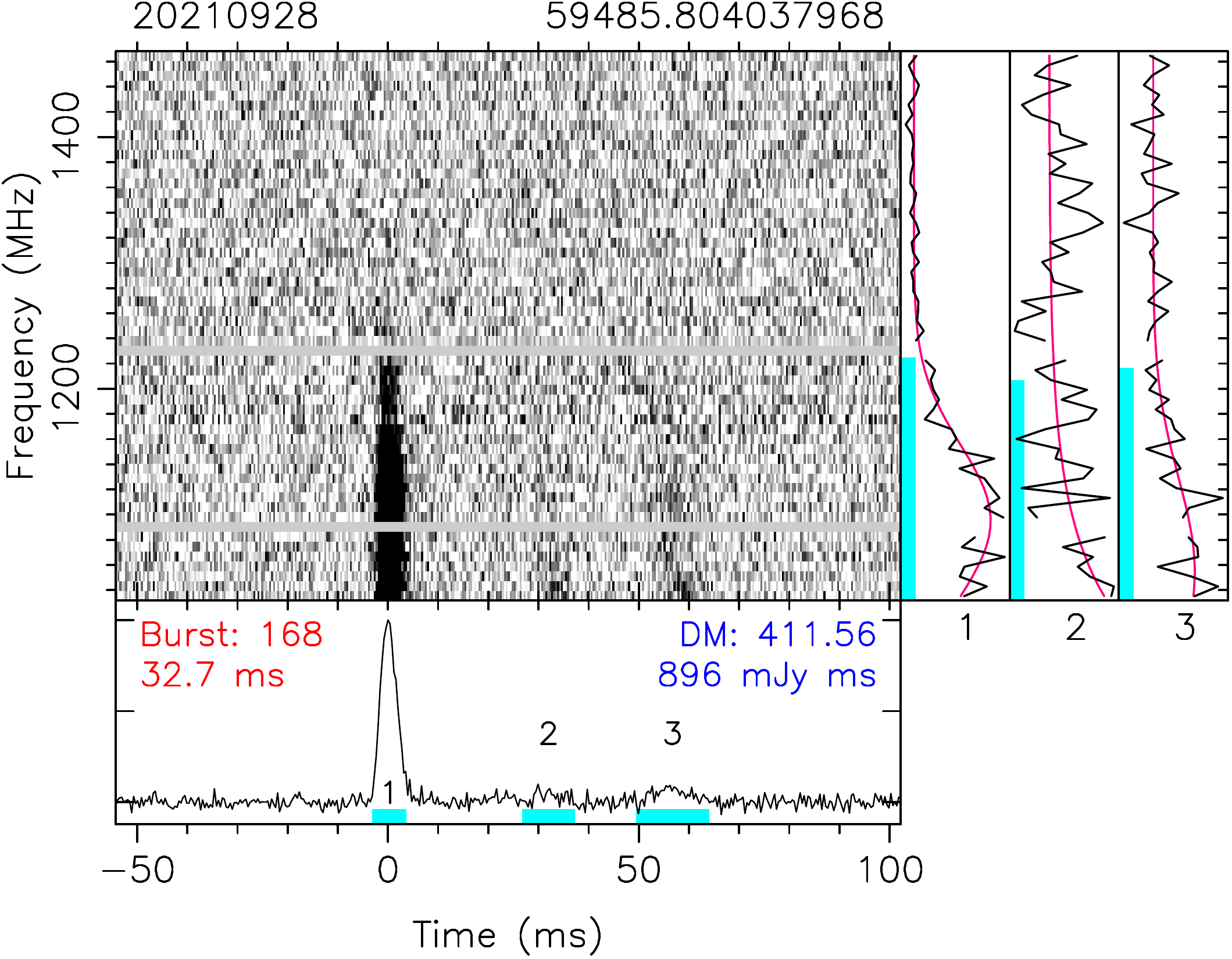}
    \includegraphics[height=37mm]{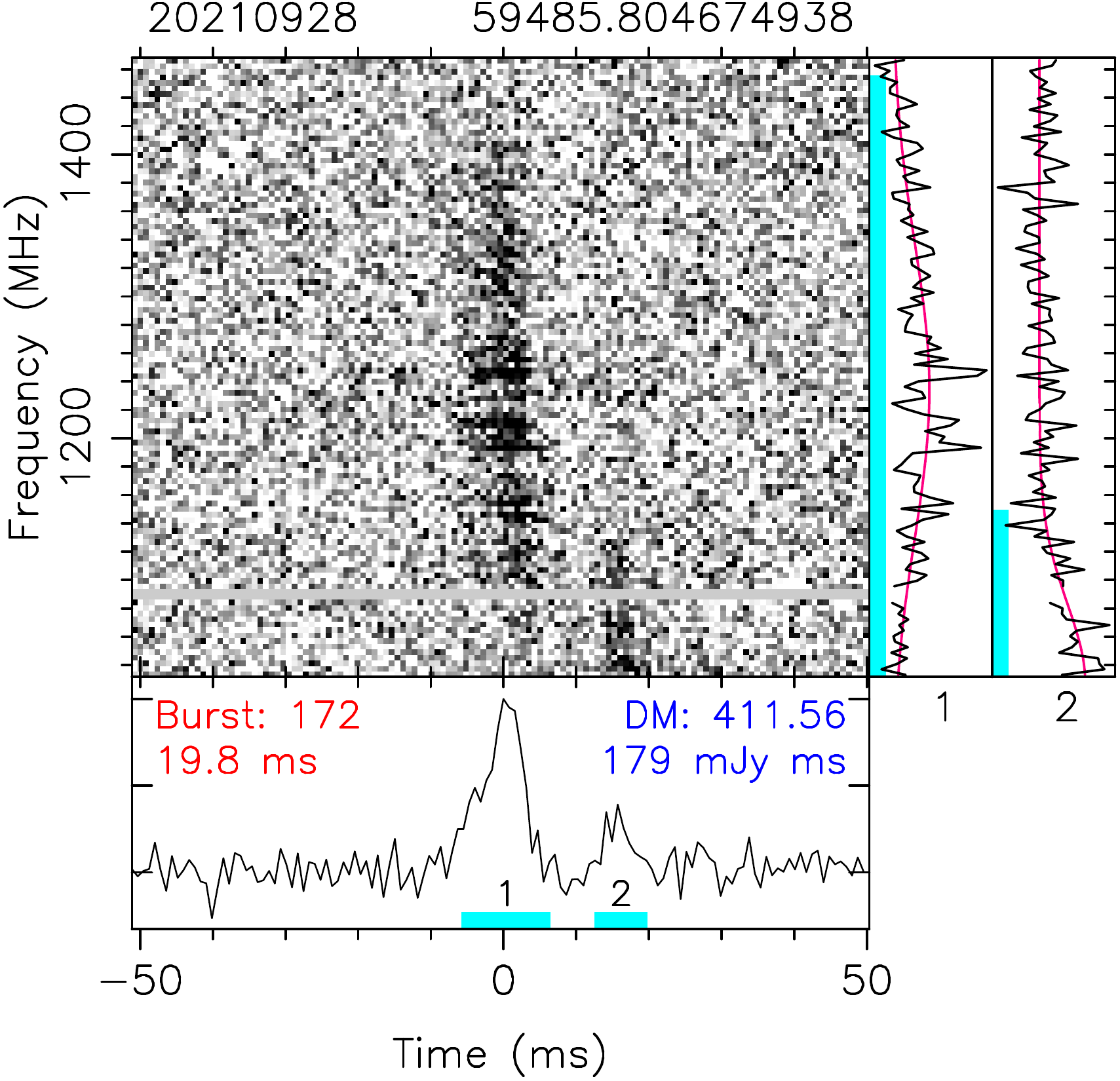}
    \includegraphics[height=37mm]{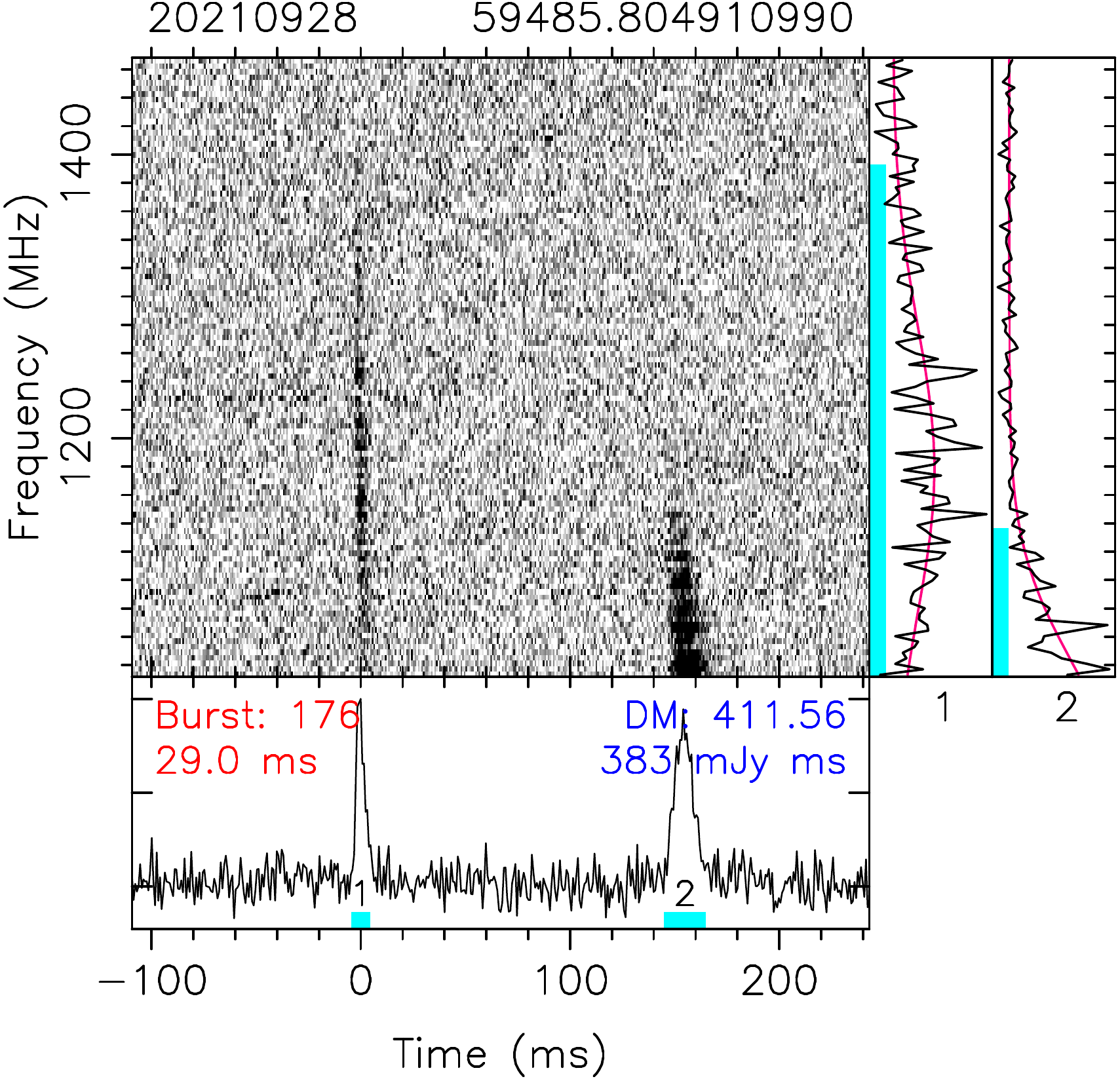}
    \includegraphics[height=37mm]{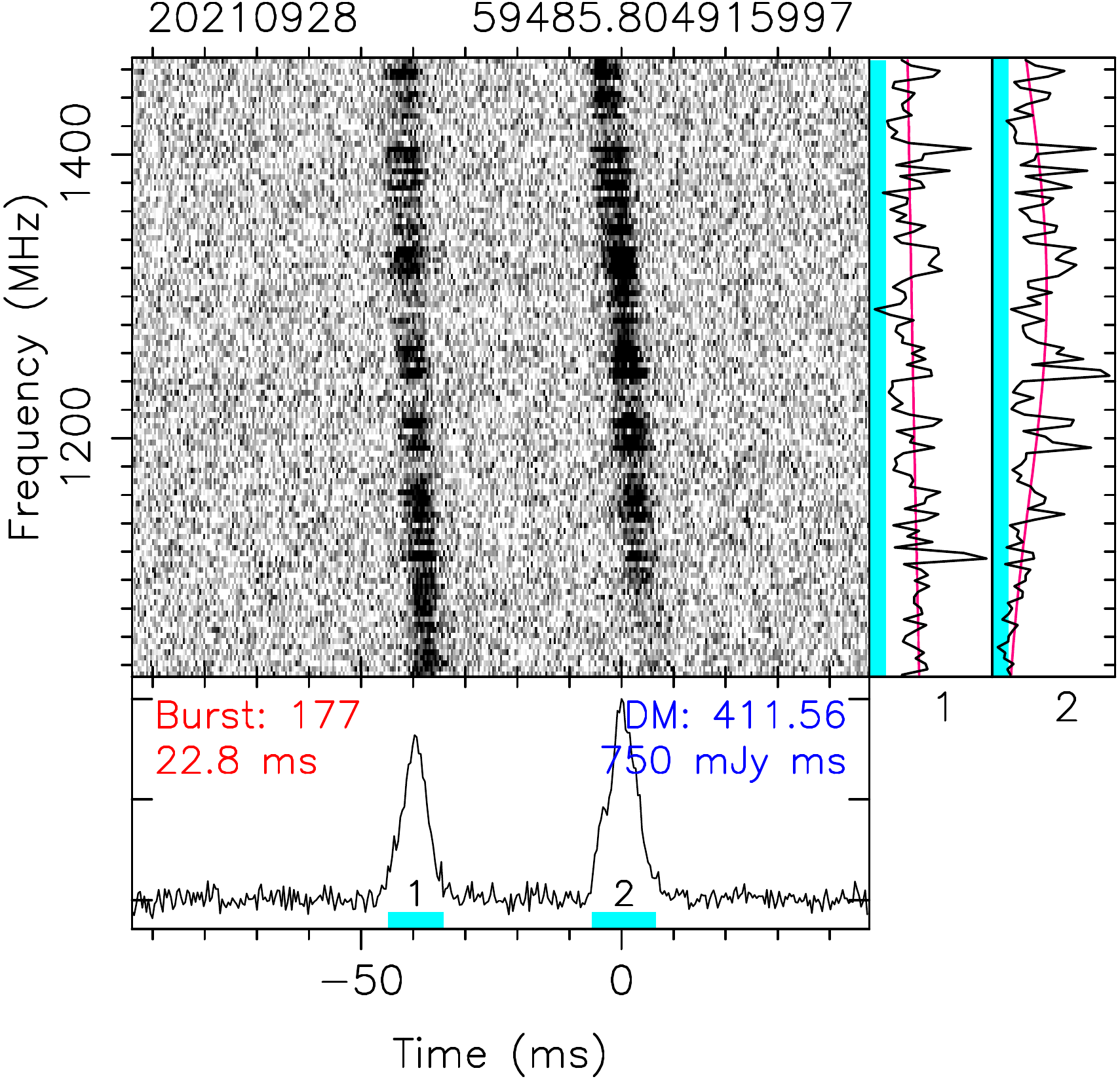}
    \includegraphics[height=37mm]{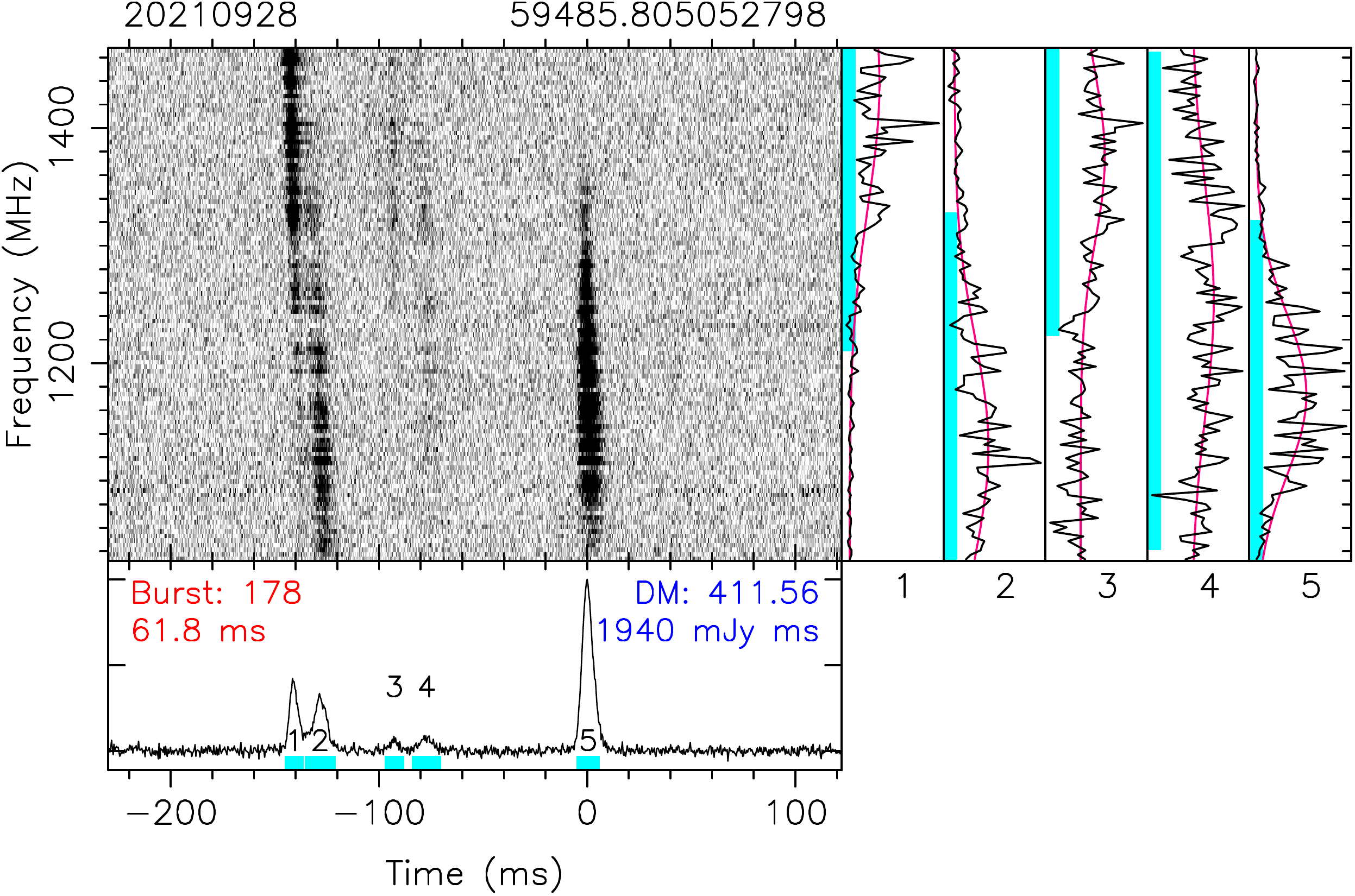}
    \includegraphics[height=37mm]{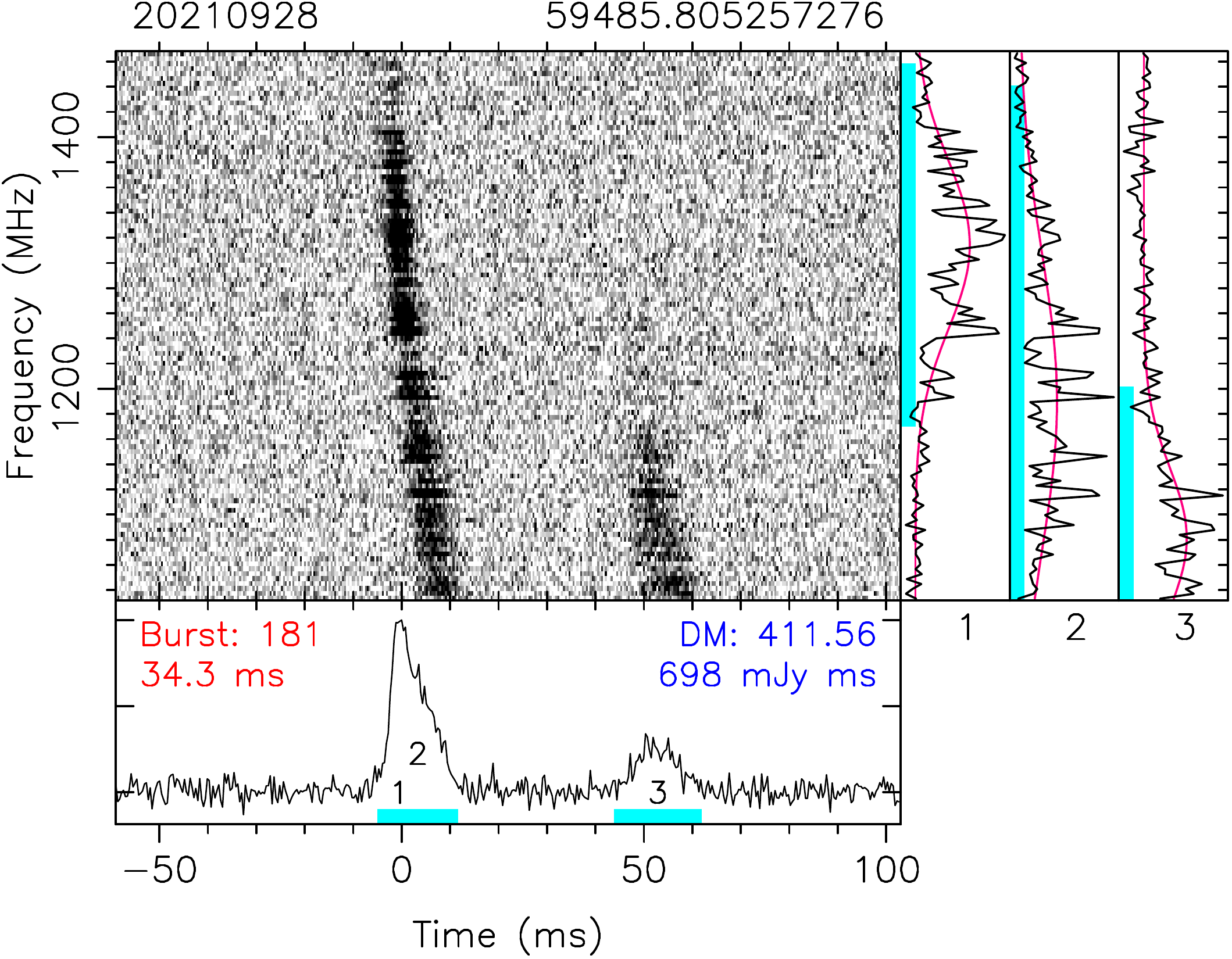}
    \includegraphics[height=37mm]{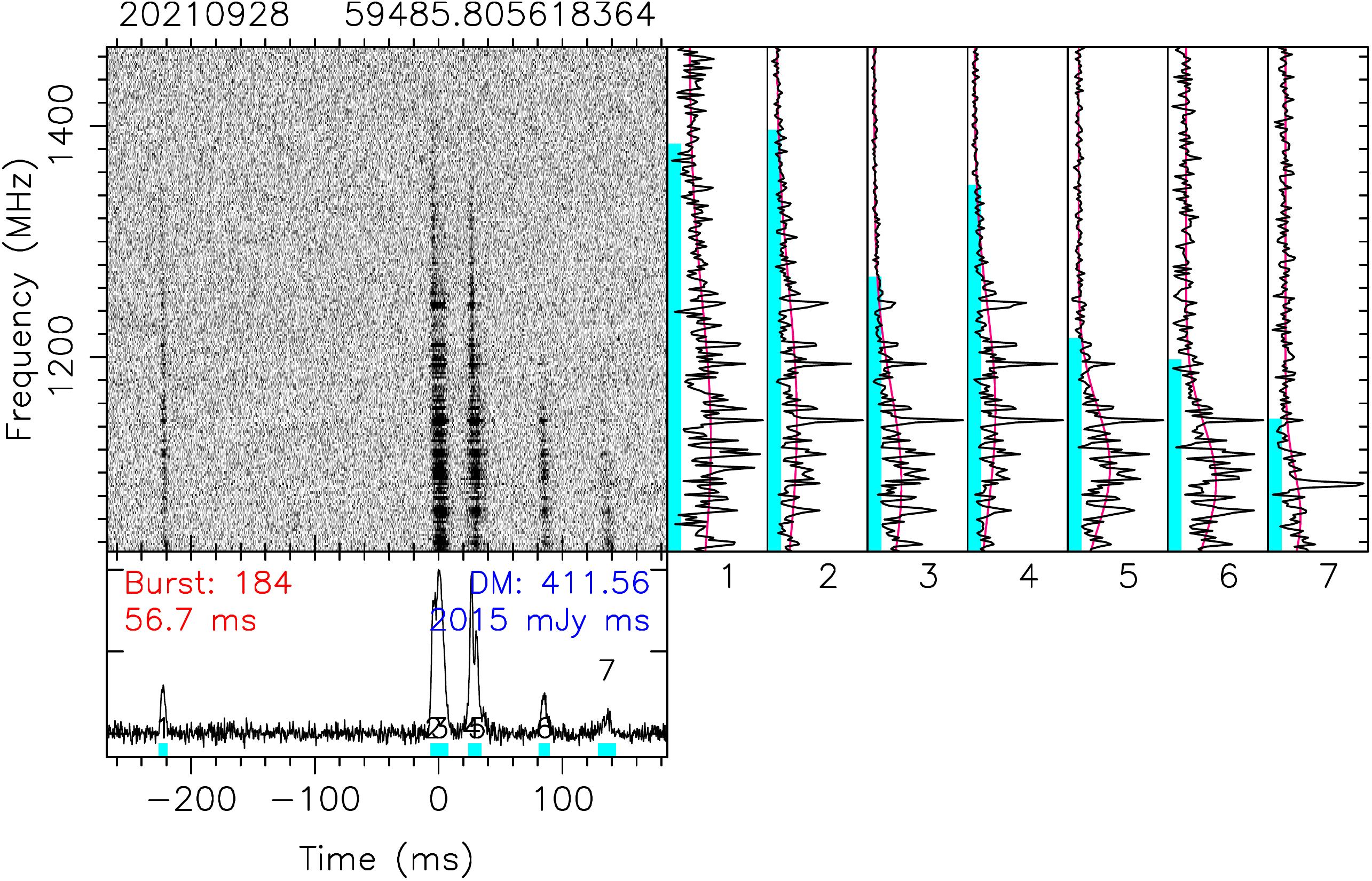}
    \includegraphics[height=37mm]{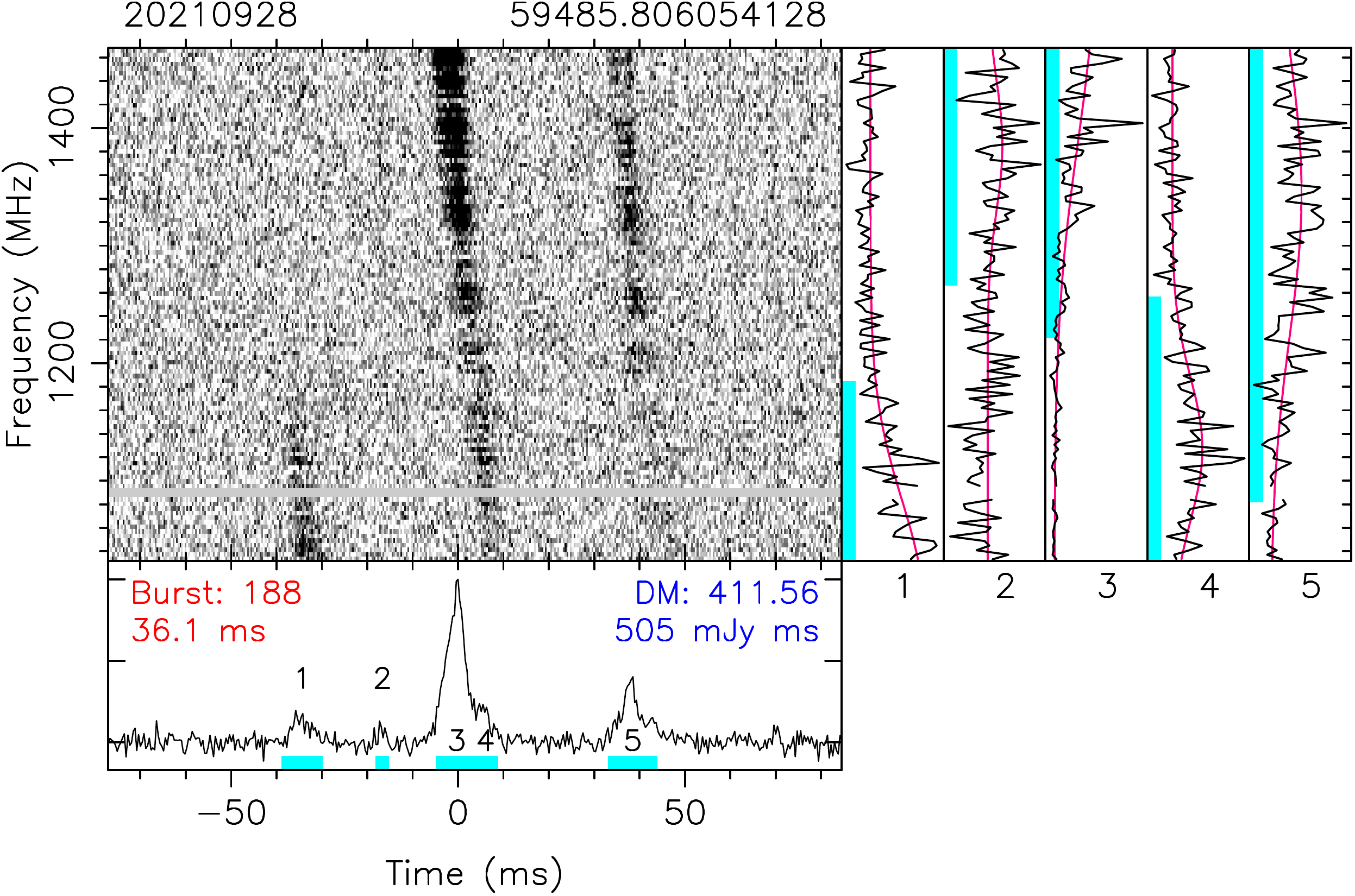}
    \includegraphics[height=37mm]{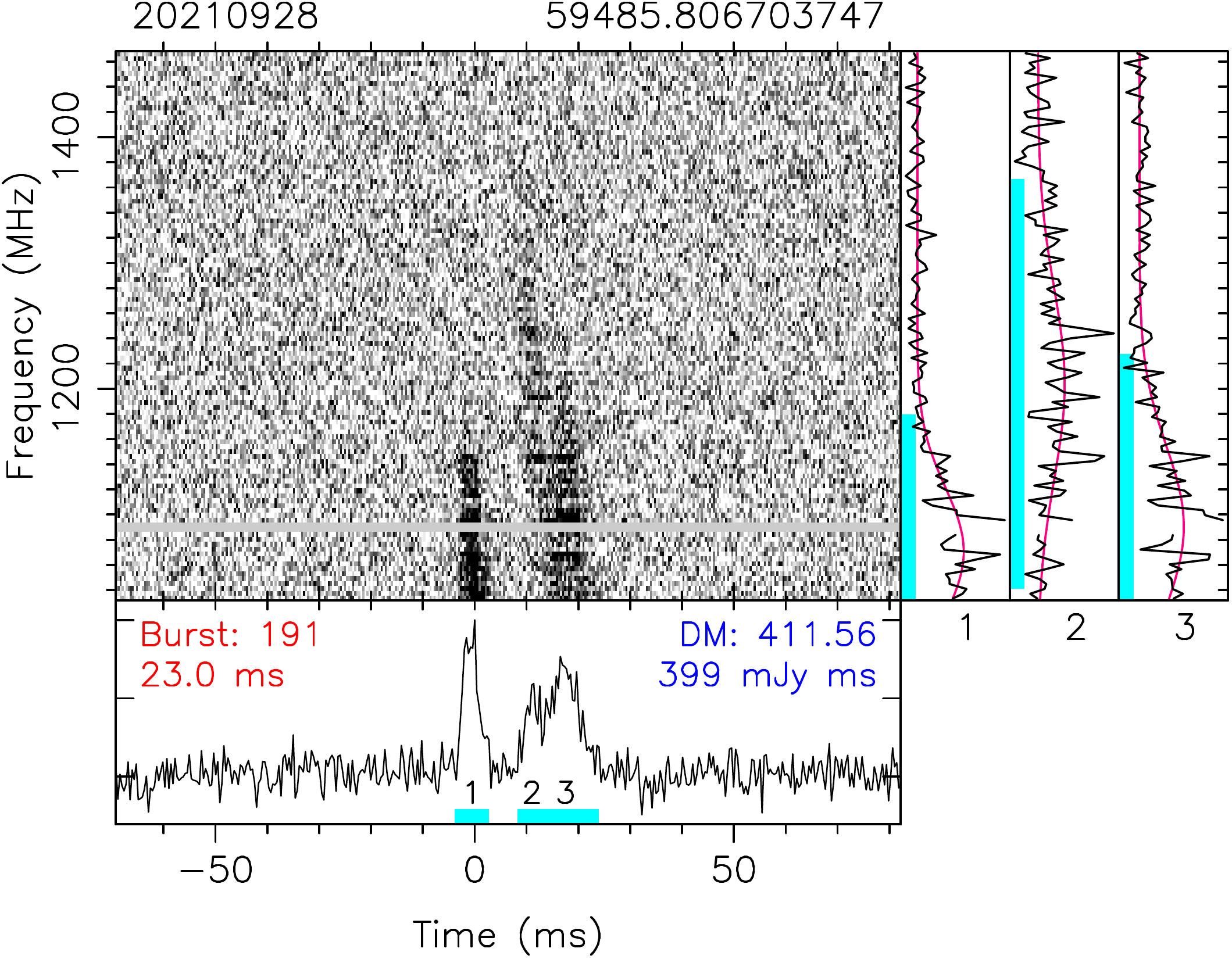}
    \includegraphics[height=37mm]{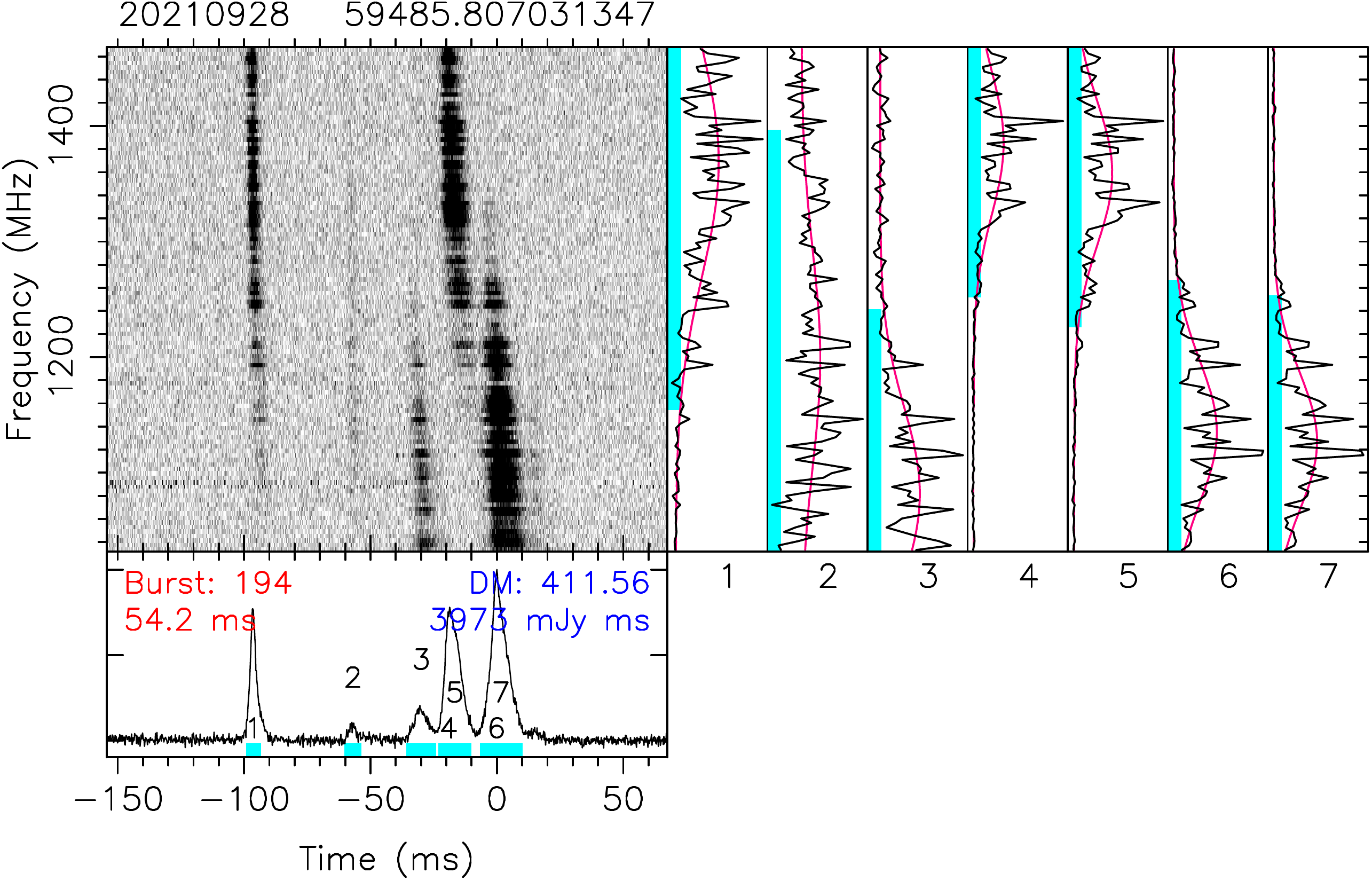}
    \includegraphics[height=37mm]{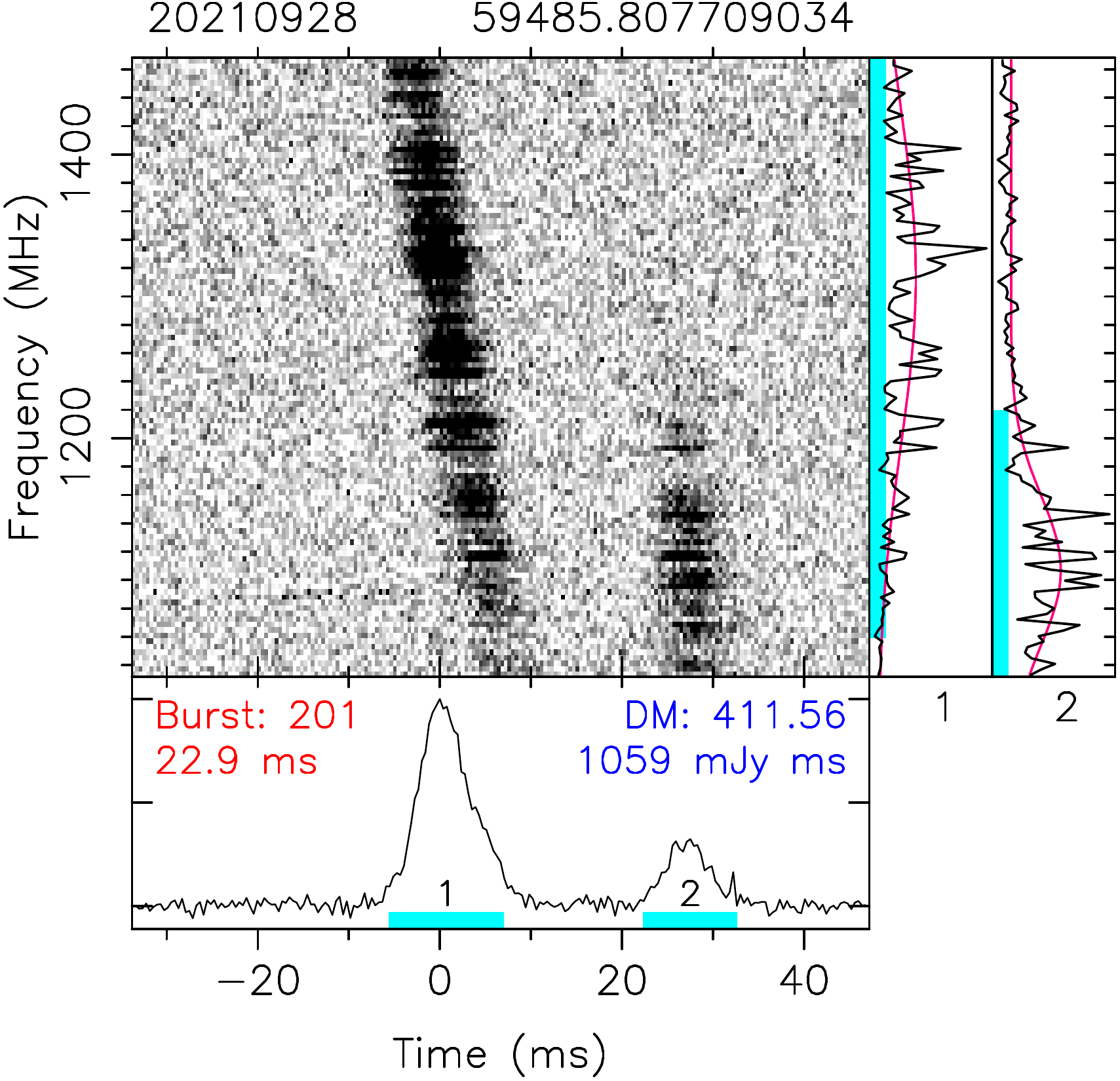}
\caption{\it{ -- continued}.
}
\end{figure*}
\addtocounter{figure}{-1}
\begin{figure*}
    \flushleft
    \includegraphics[height=37mm]{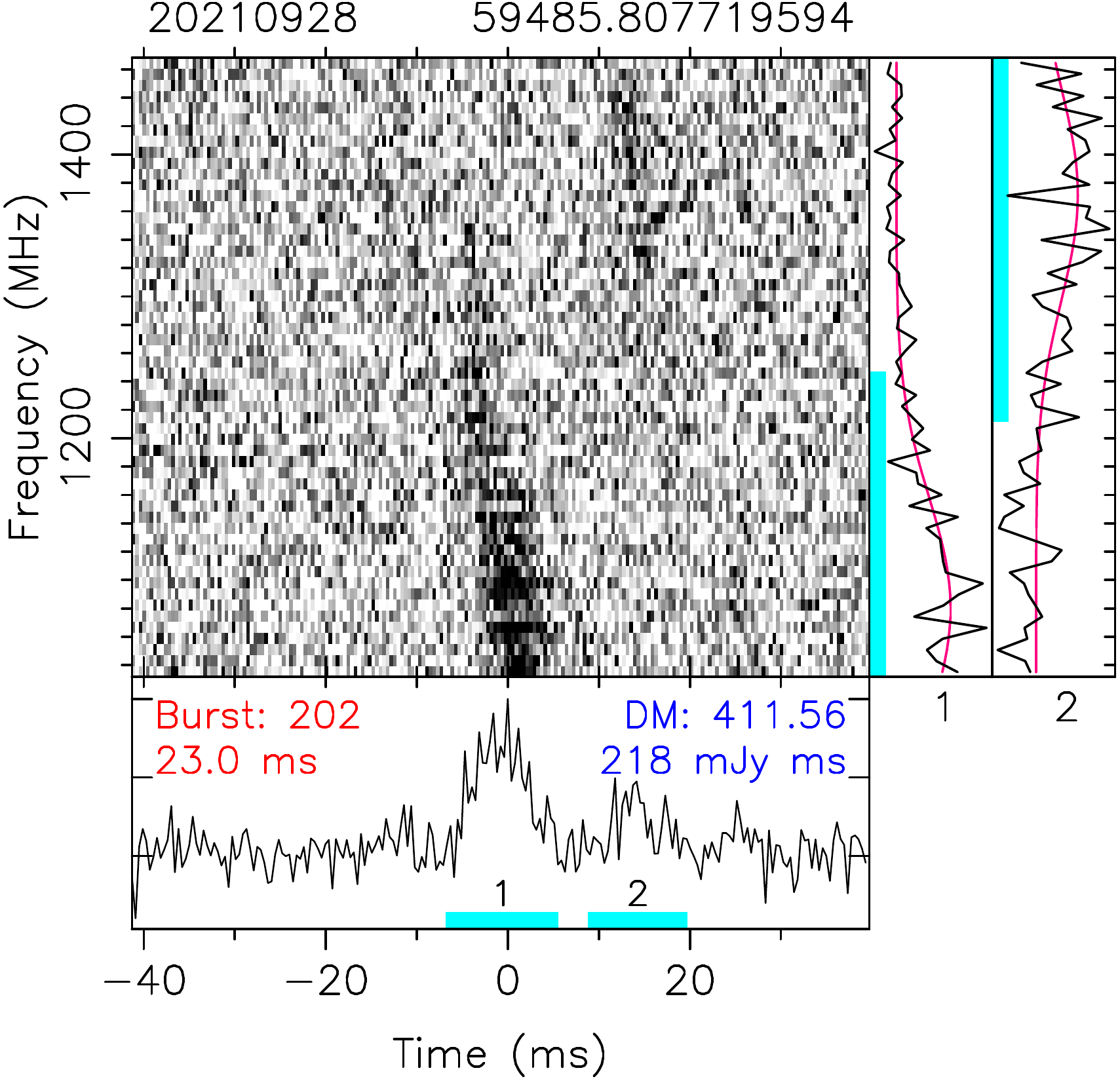}
    \includegraphics[height=37mm]{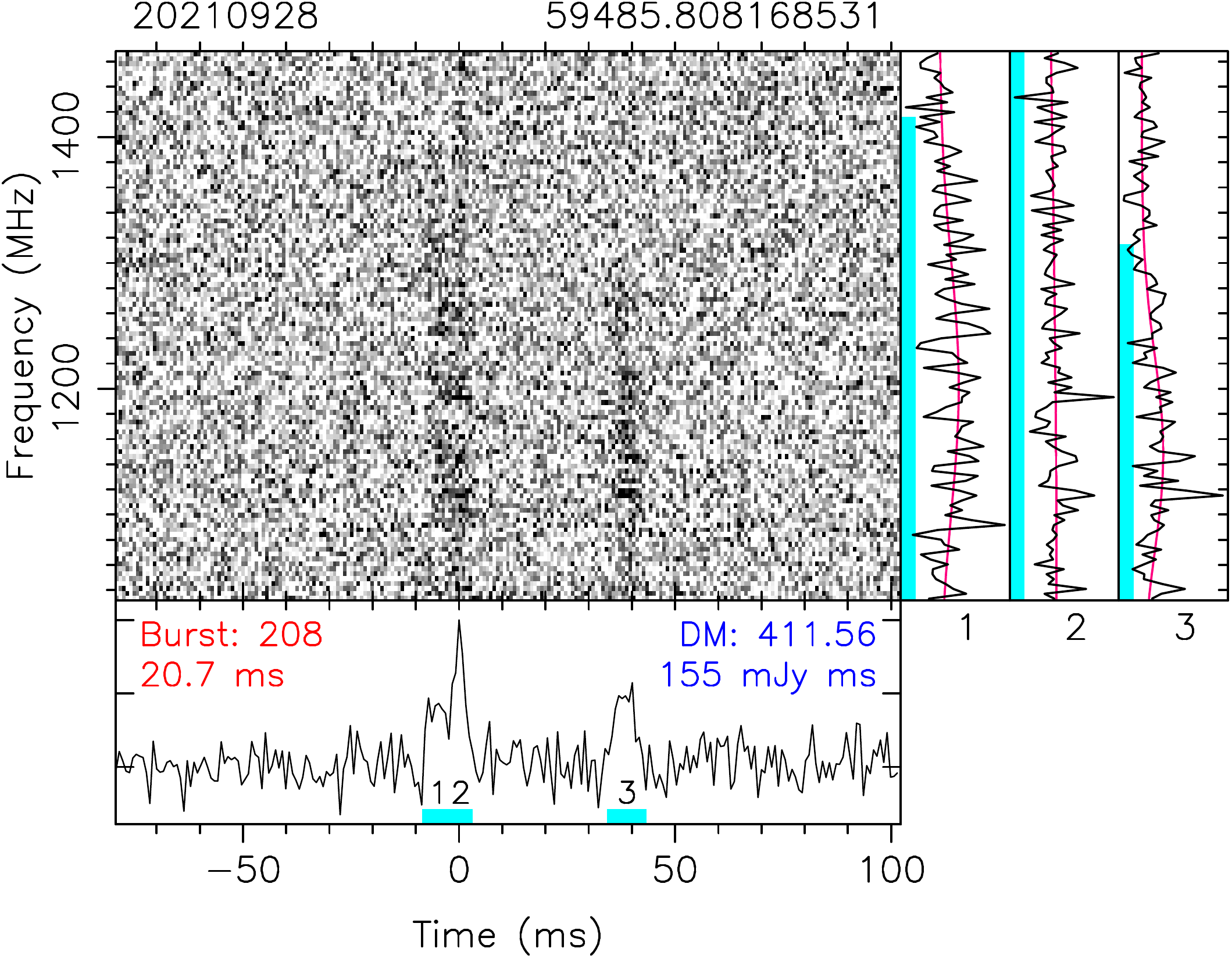}
    \includegraphics[height=37mm]{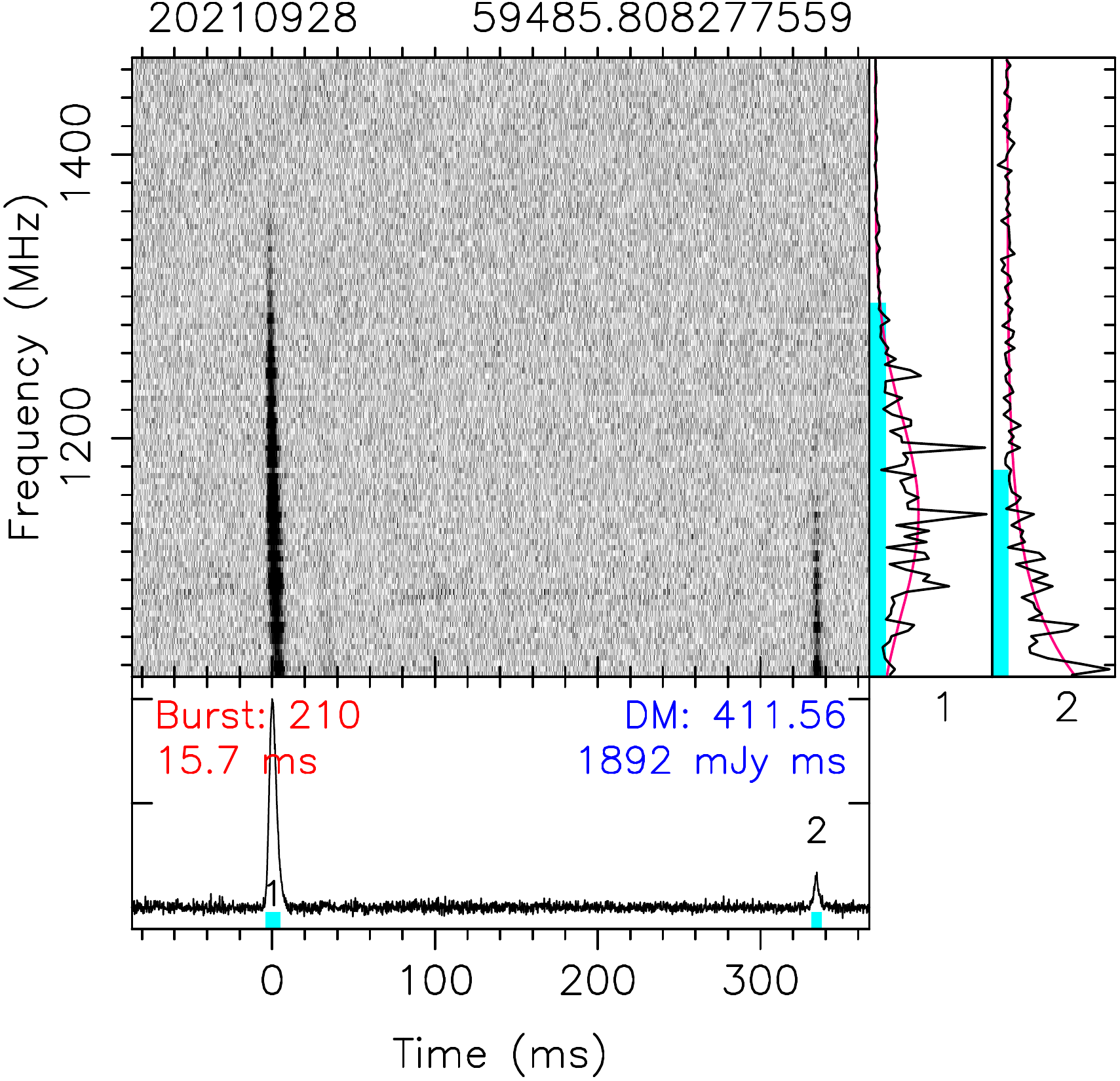}
    \includegraphics[height=37mm]{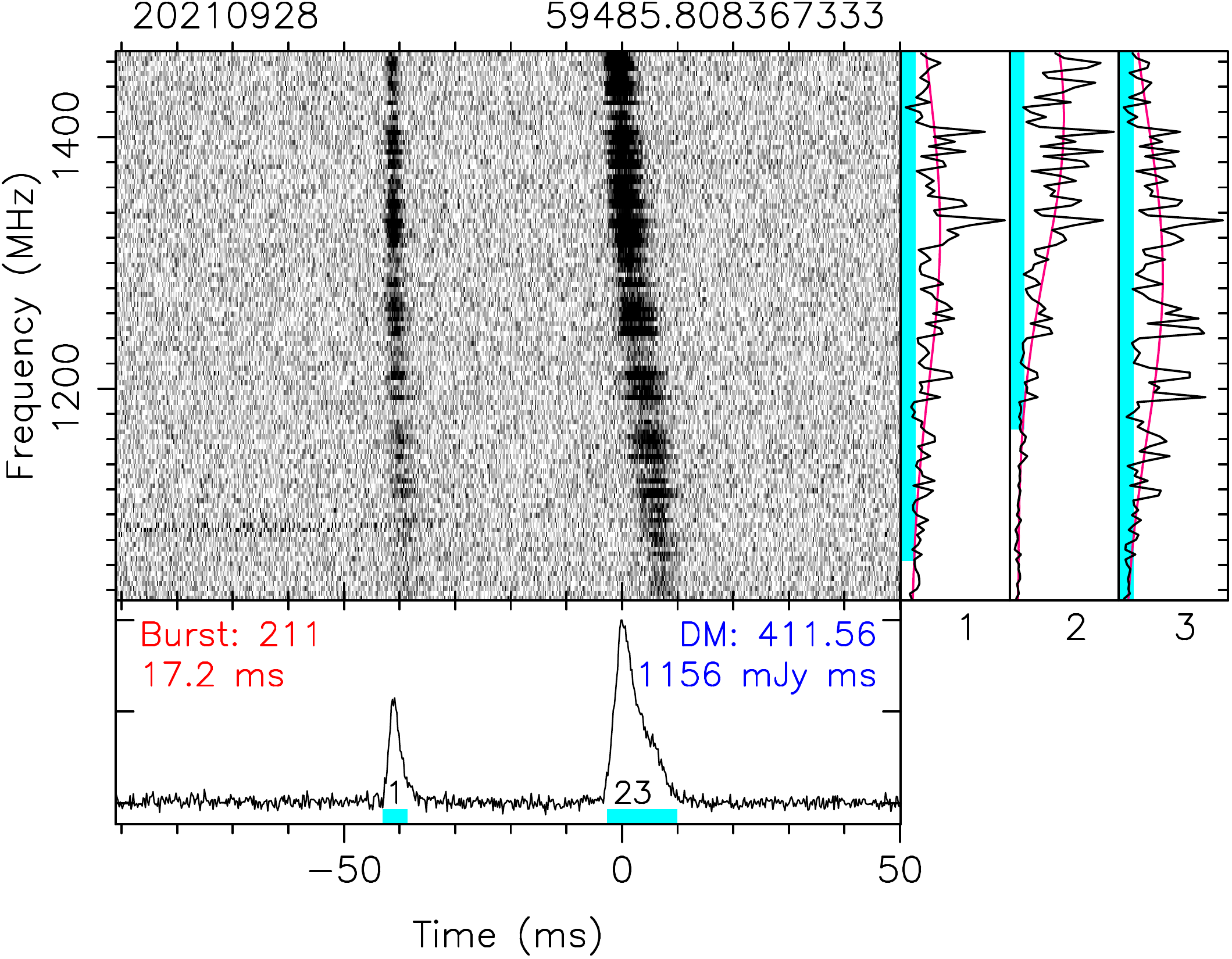}
    \includegraphics[height=37mm]{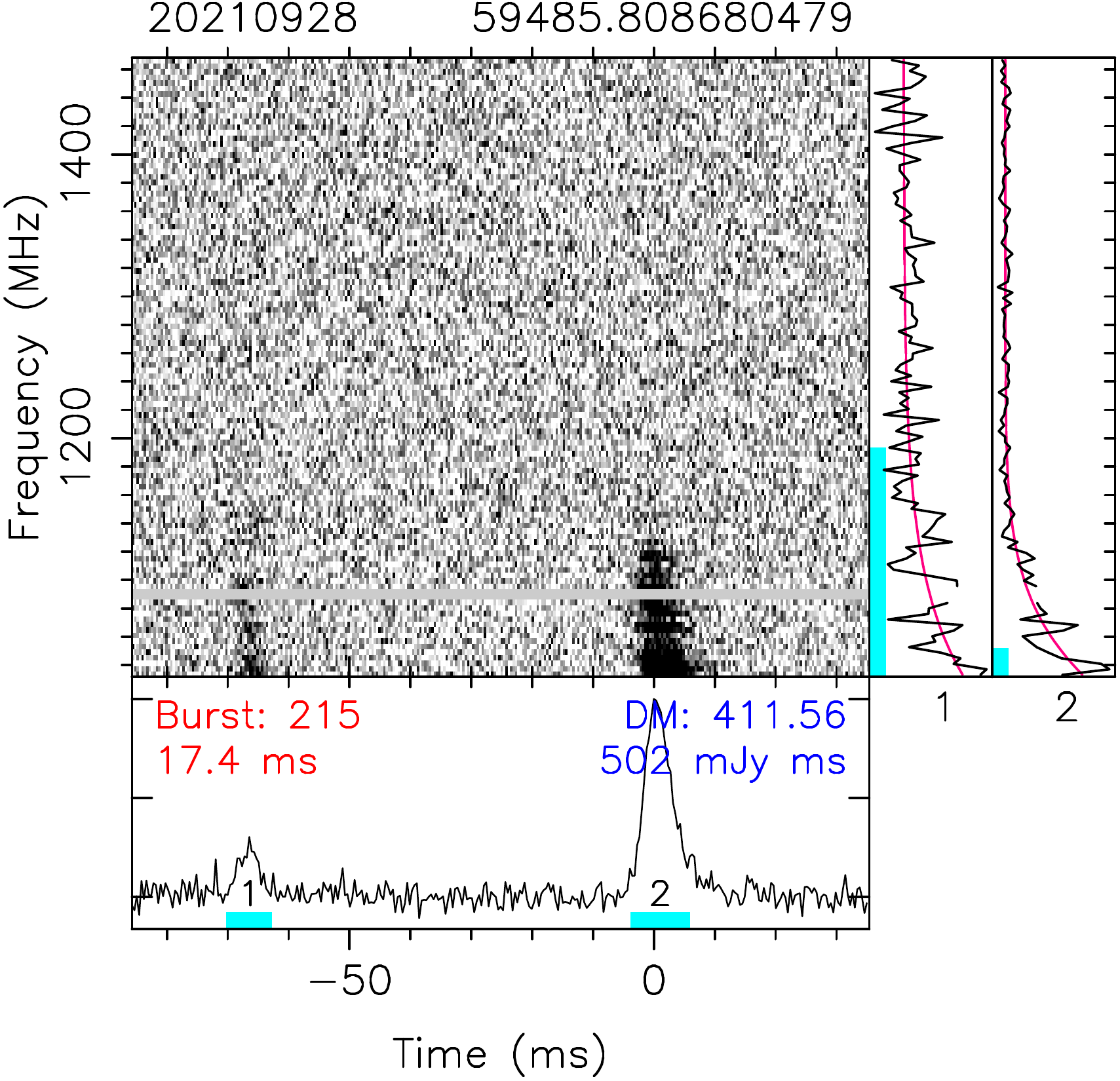}
    \includegraphics[height=37mm]{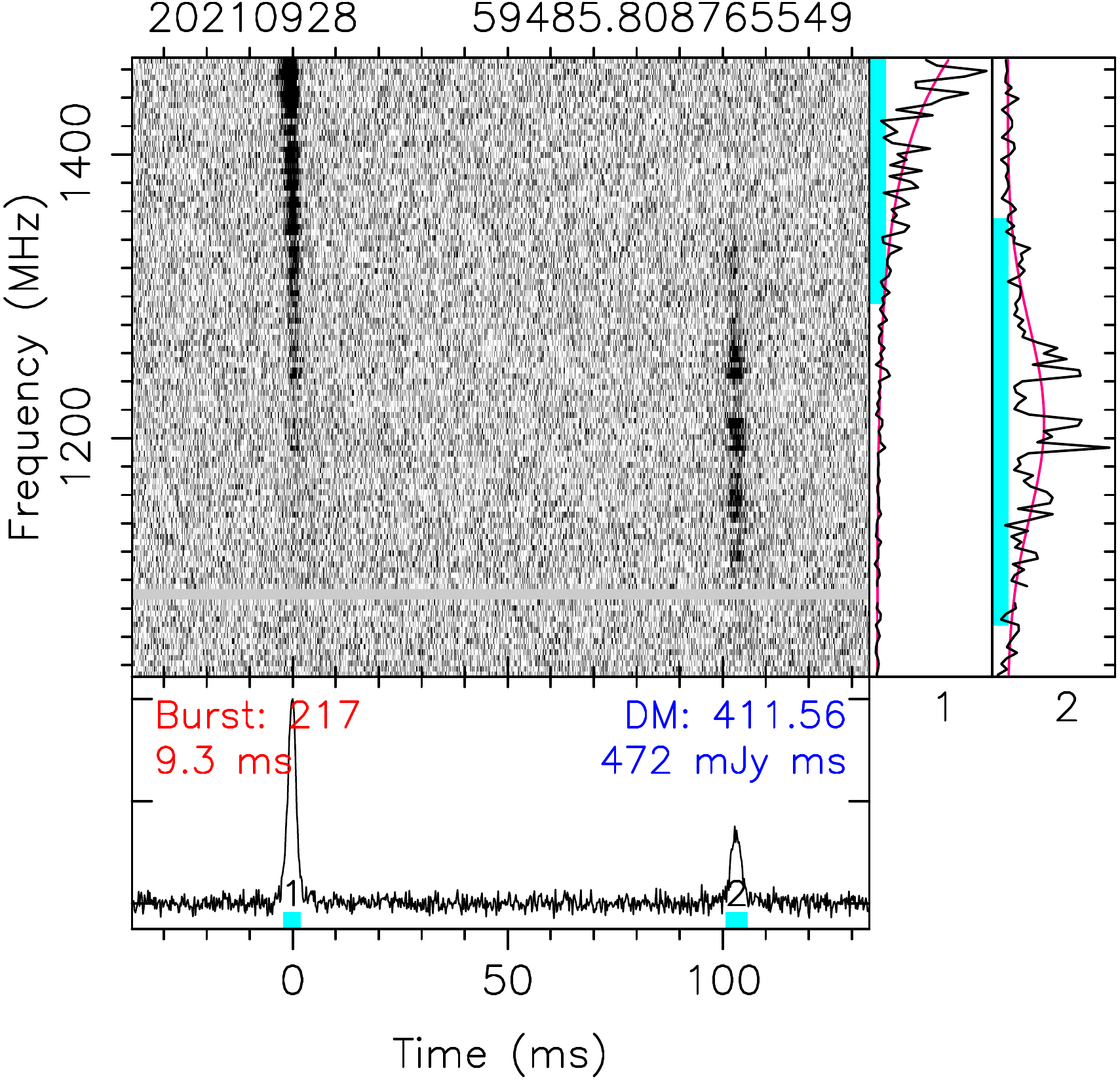}
    \includegraphics[height=37mm]{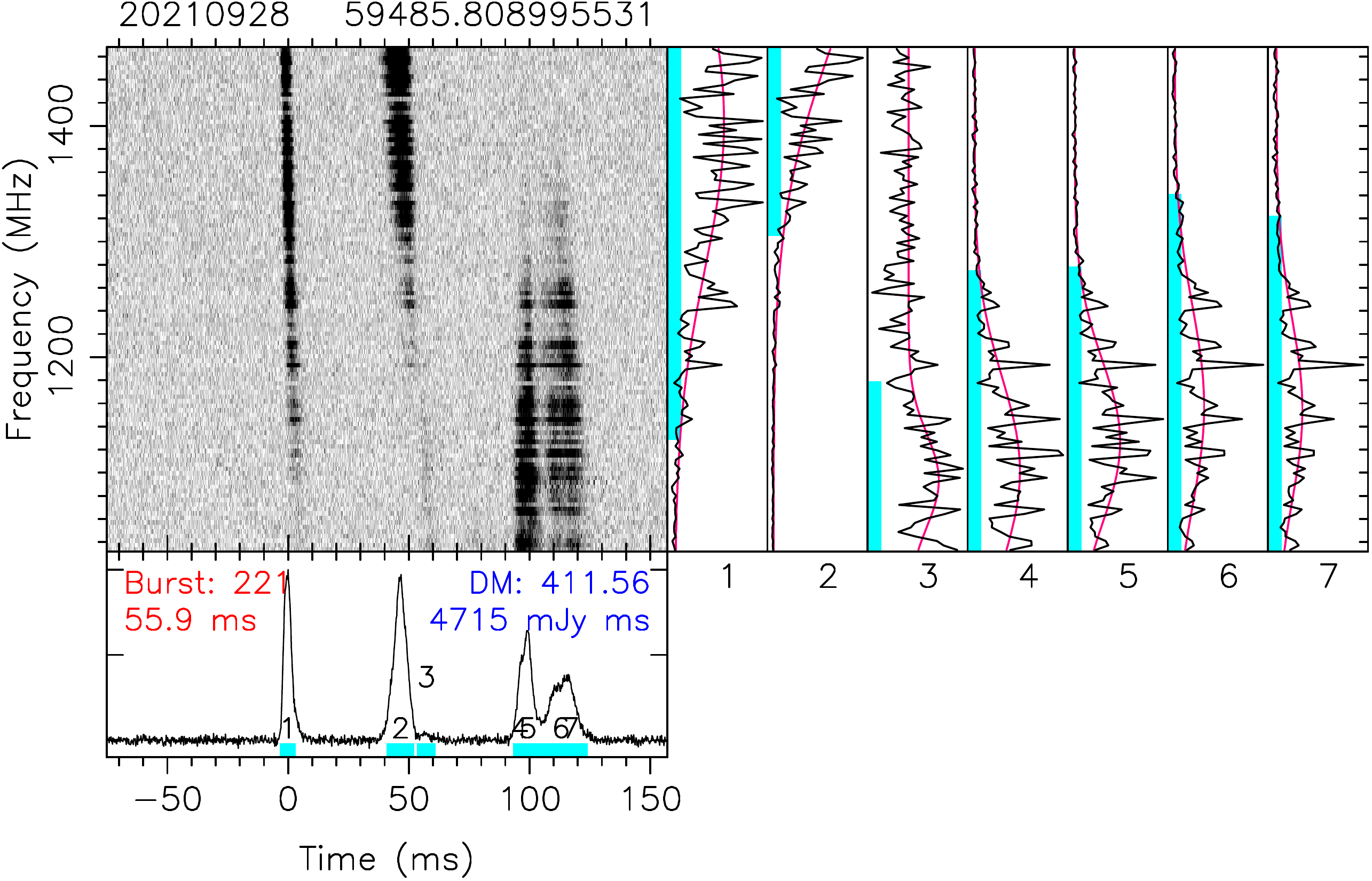}
    \includegraphics[height=37mm]{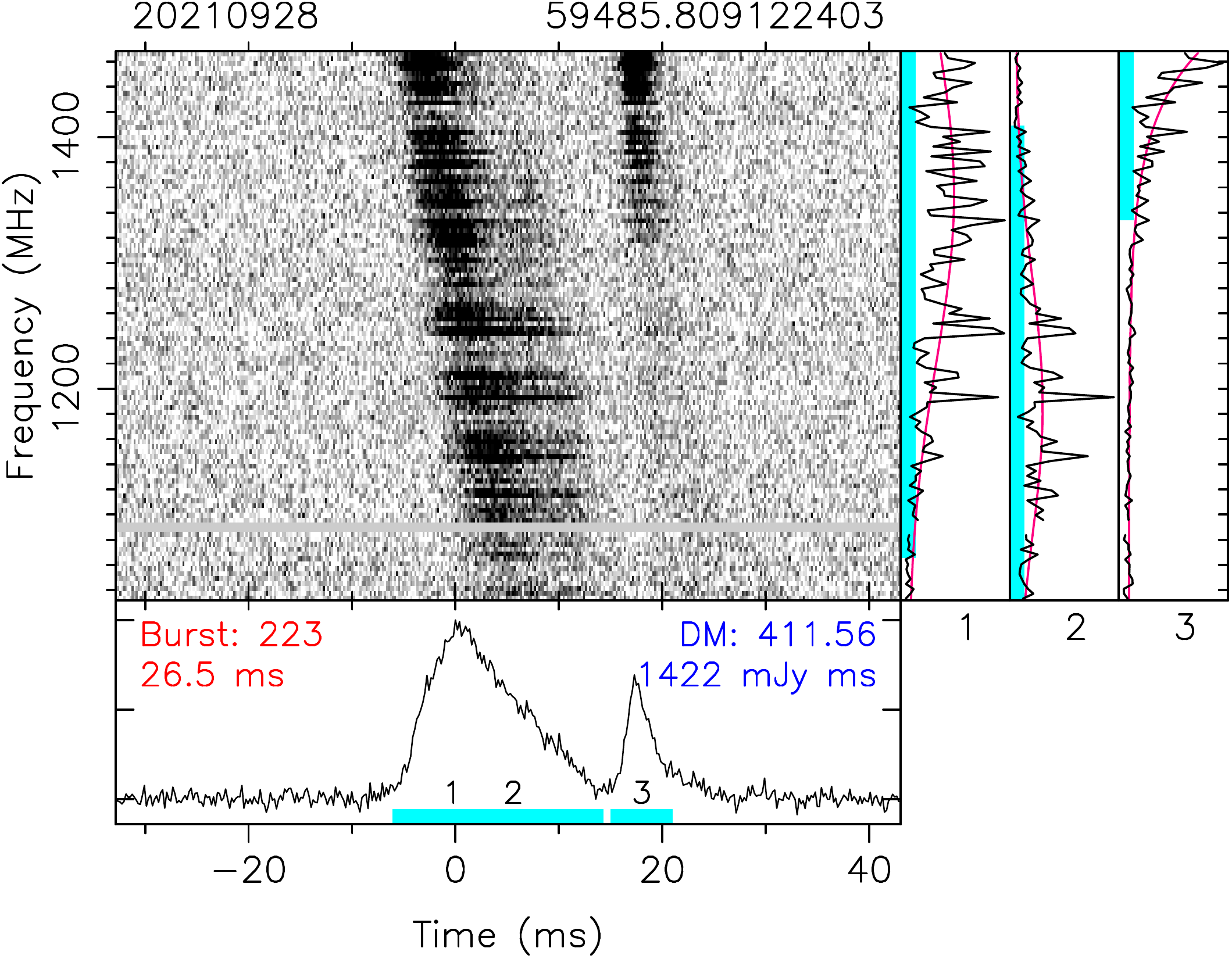}
    \includegraphics[height=37mm]{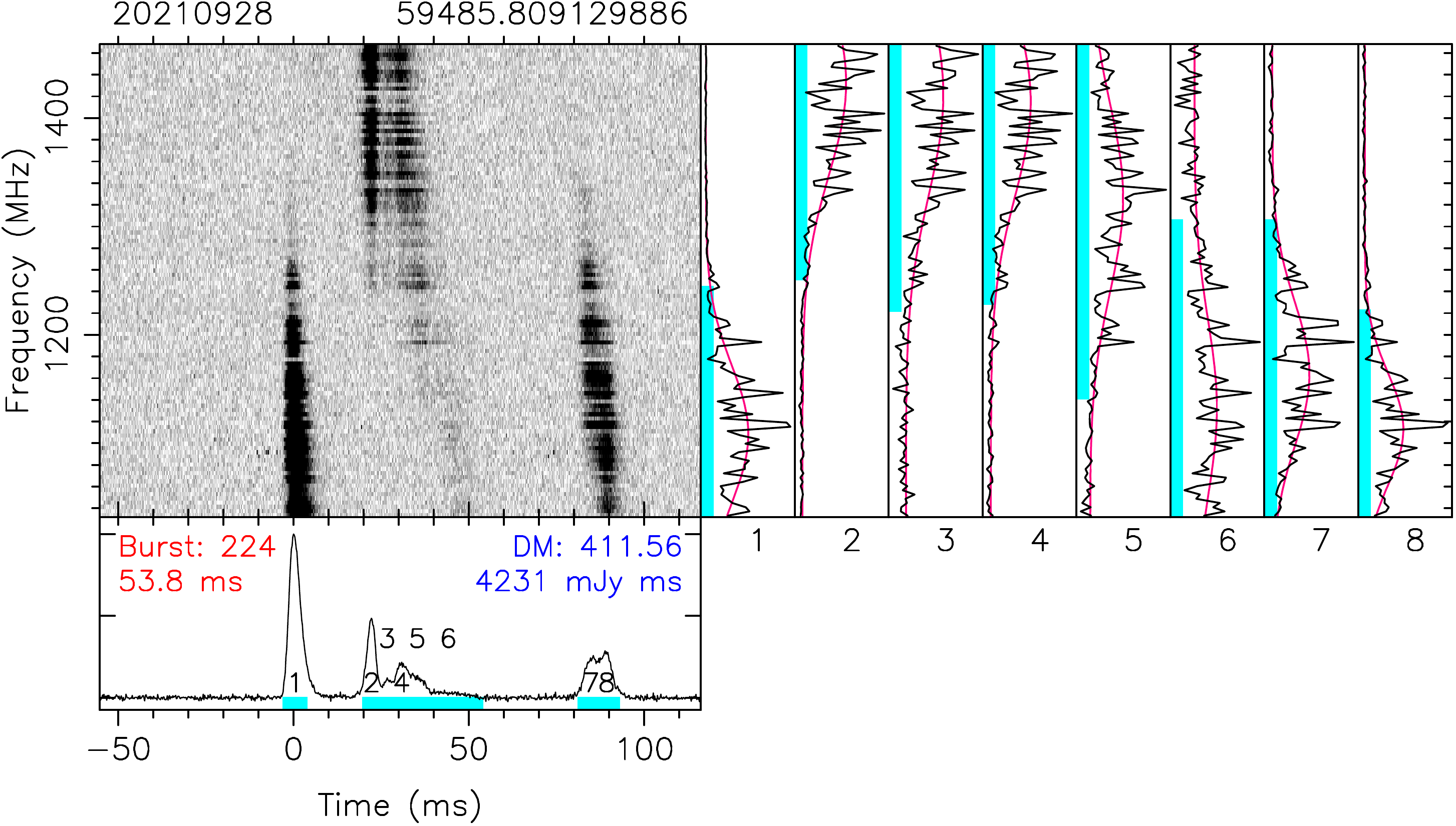}
    \includegraphics[height=37mm]{20210929/FRB20201124A_20210929_tracking-M01-P1-c512b1.fits-228-T-2219.777-2219.938-DM-411.6.pdf}
    \includegraphics[height=37mm]{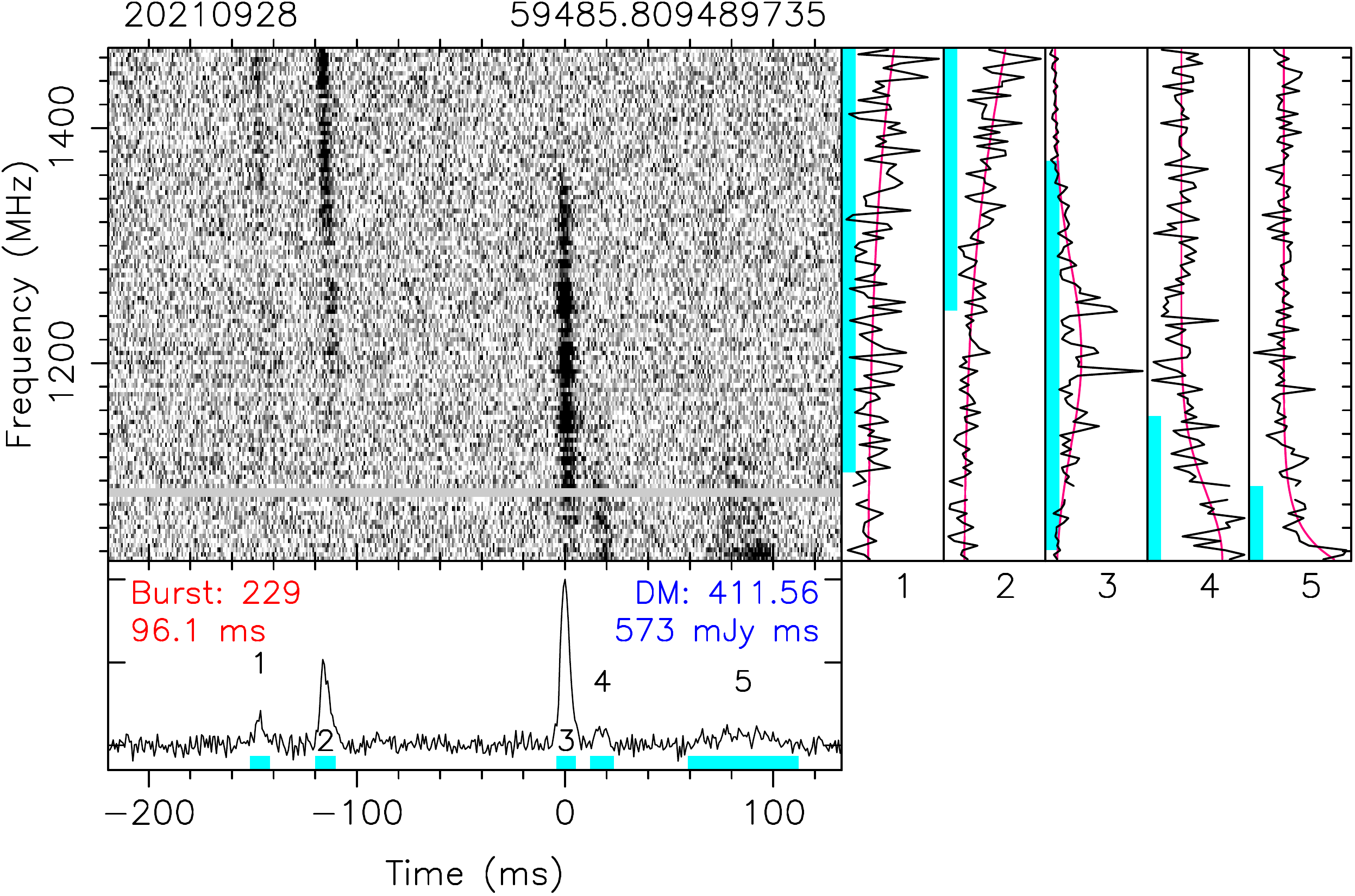}
    \includegraphics[height=37mm]{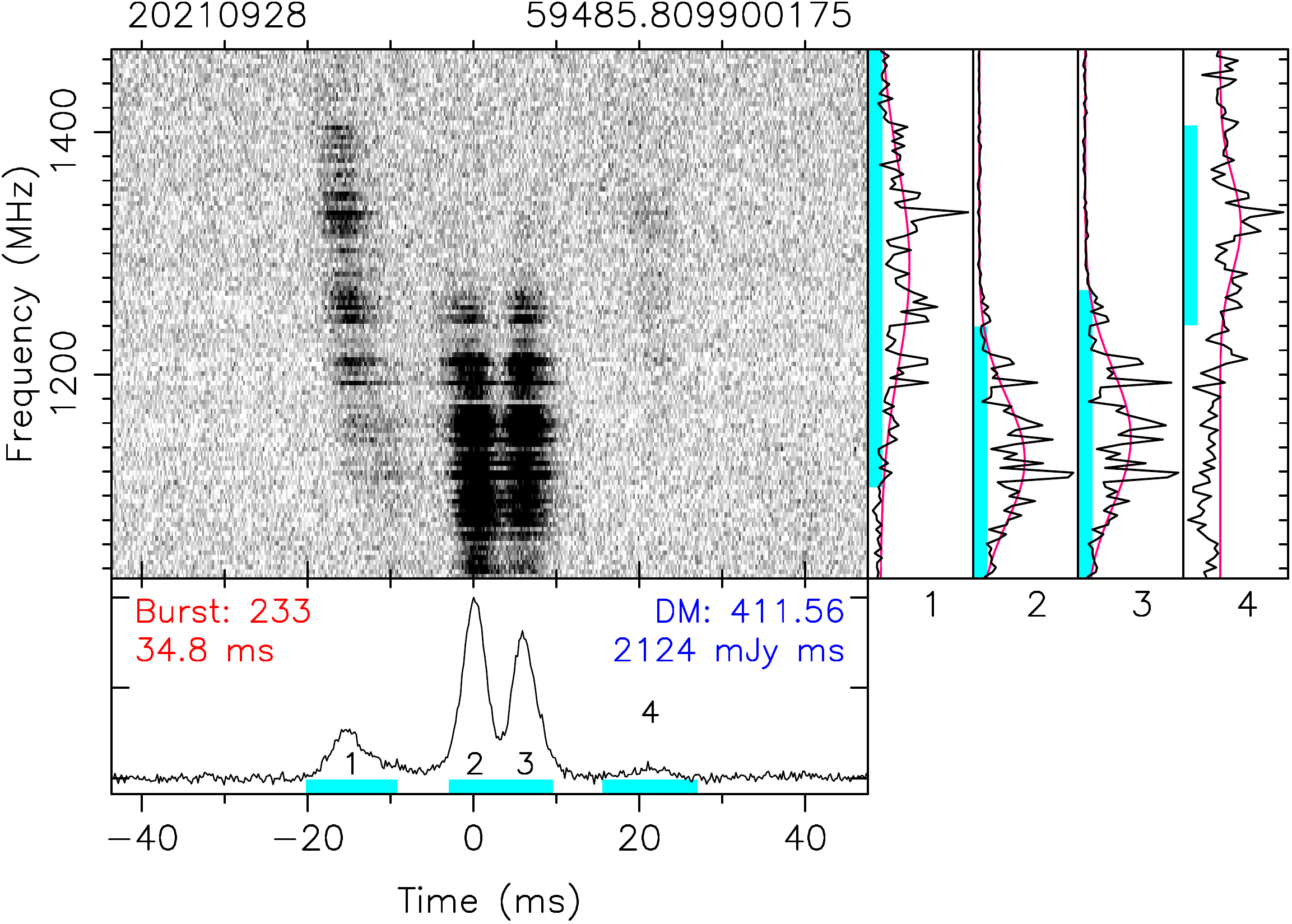}
    \includegraphics[height=37mm]{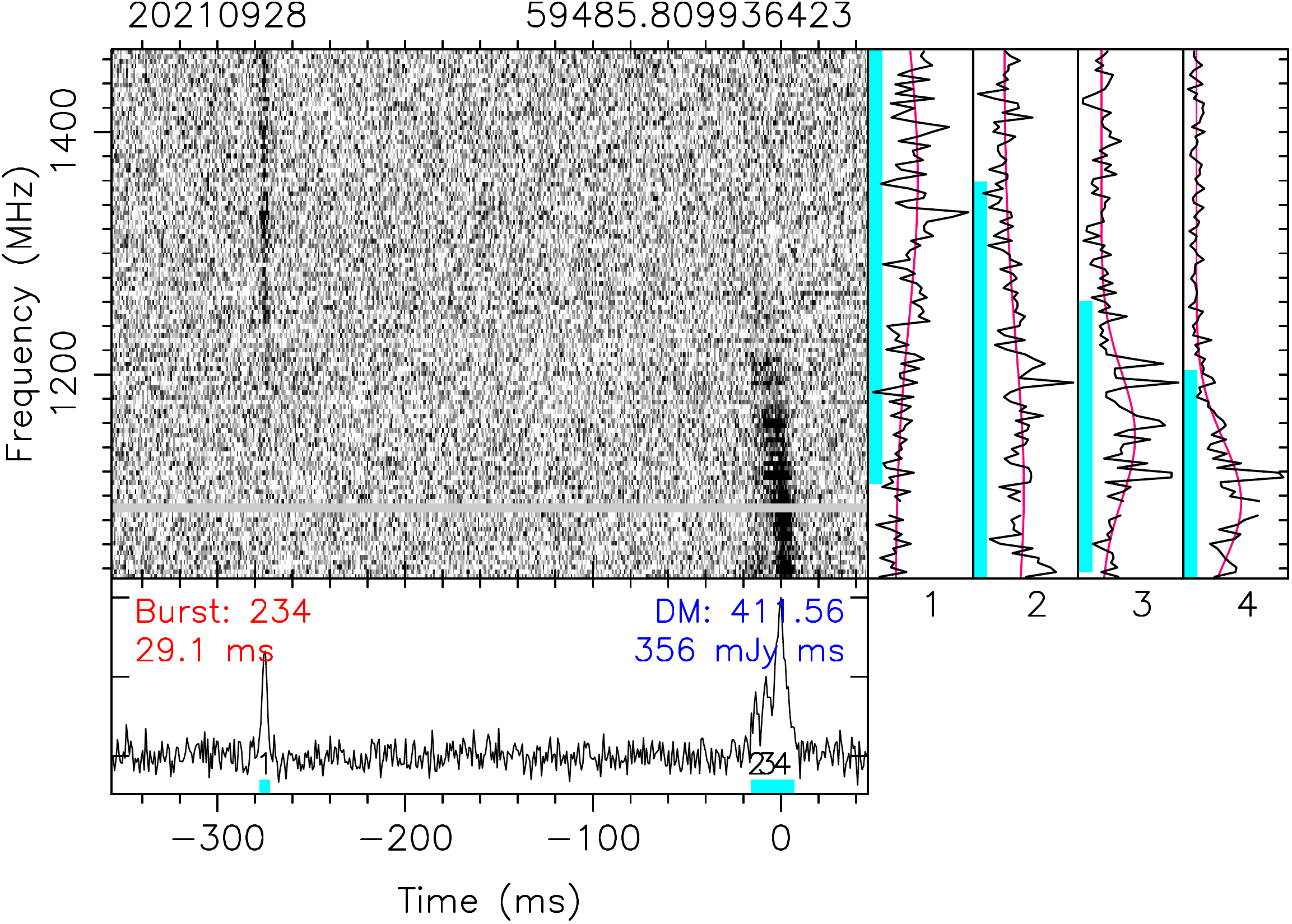}
    \includegraphics[height=37mm]{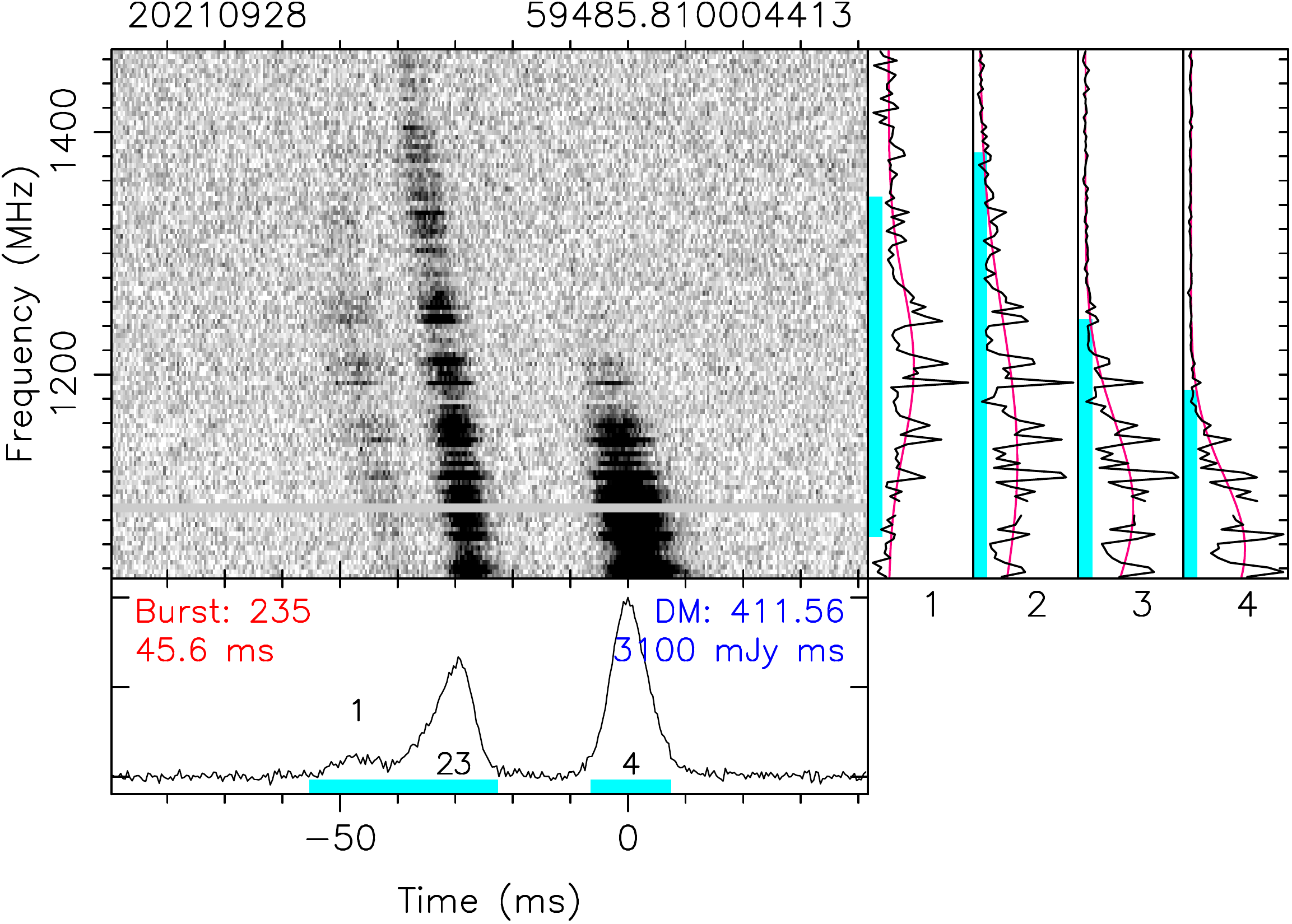}
    \includegraphics[height=37mm]{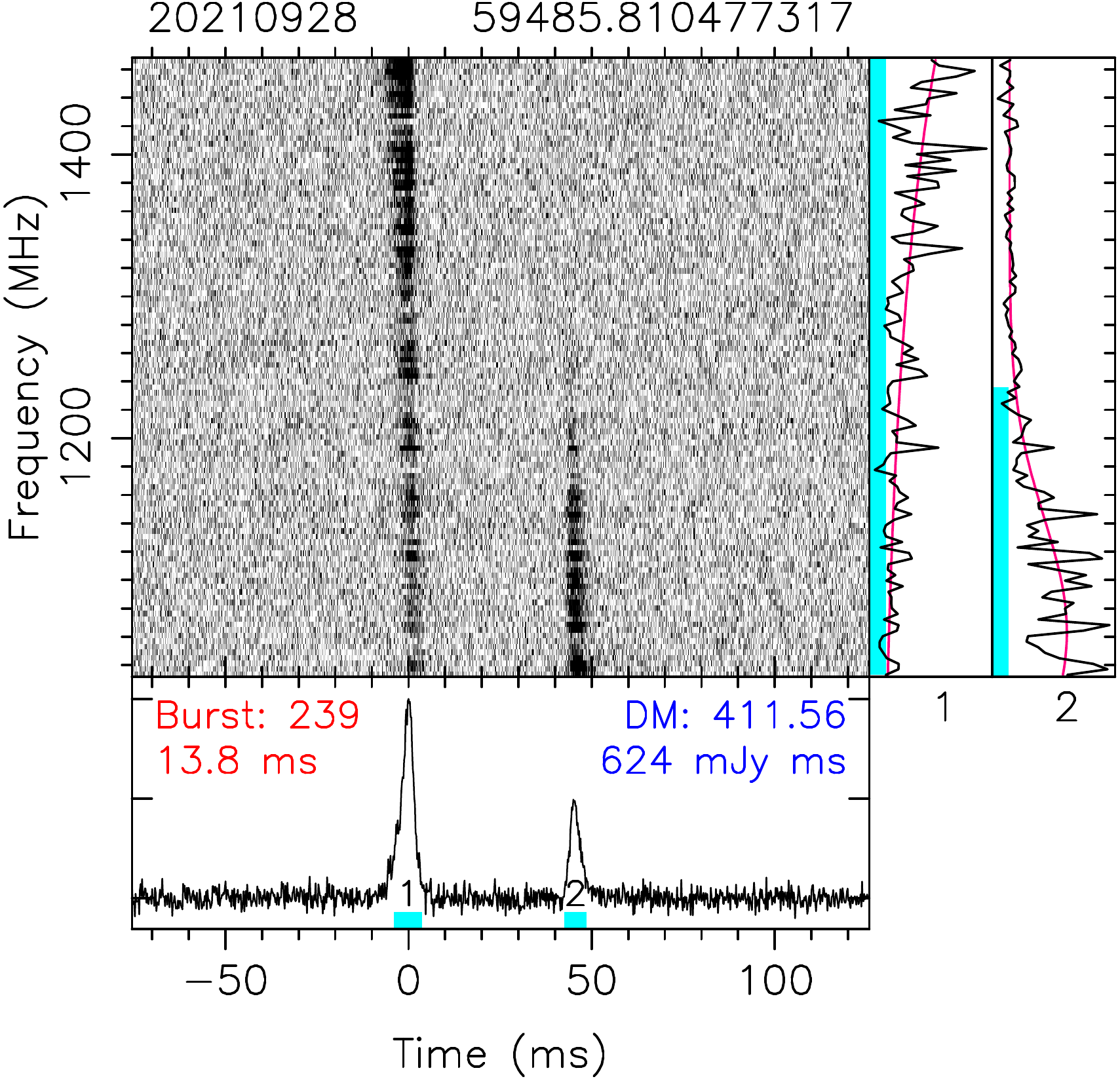}
    \includegraphics[height=37mm]{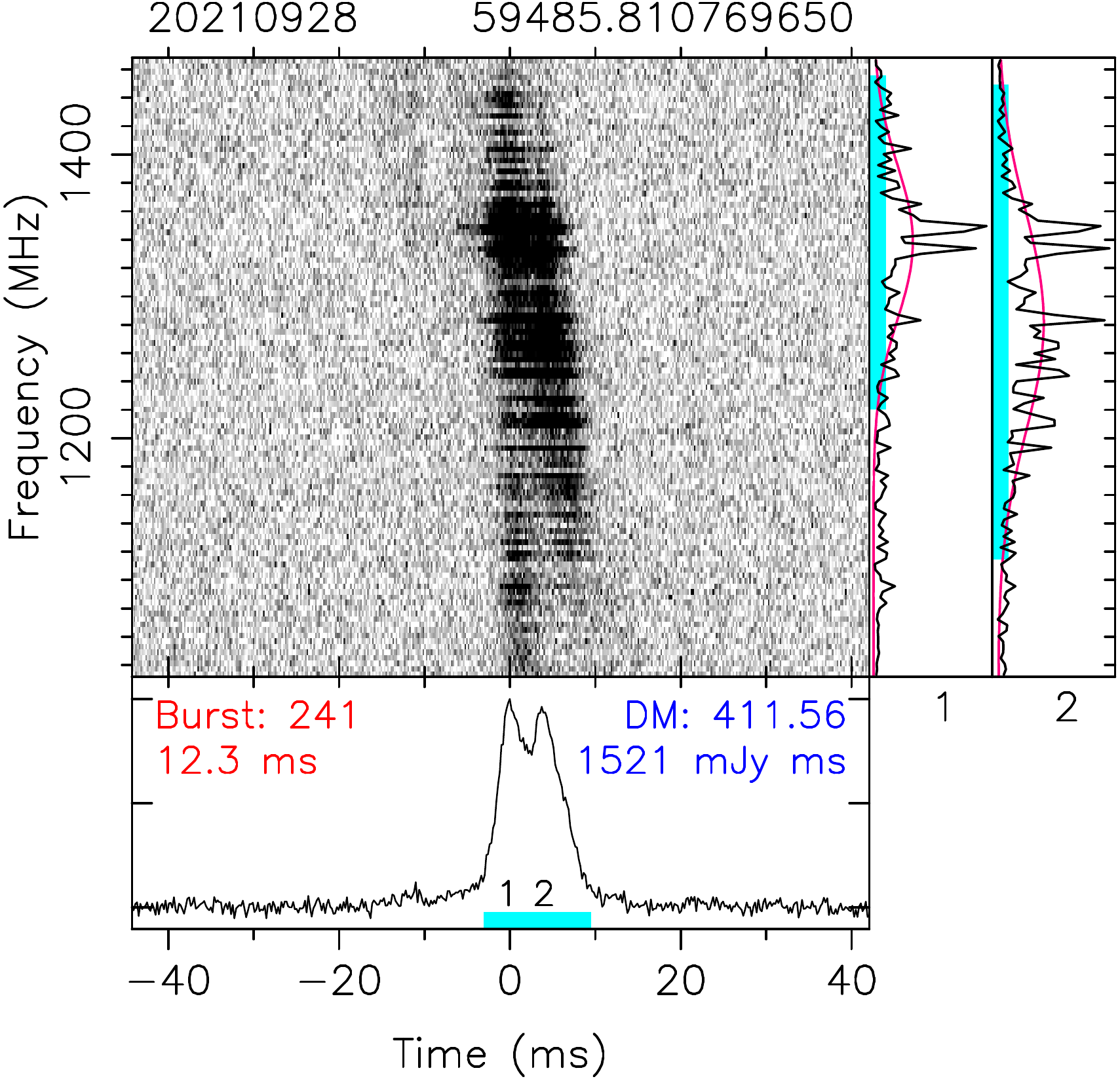}
\caption{\it{ -- continued}.
}
\end{figure*}
\addtocounter{figure}{-1}
\begin{figure*}
    \flushleft
    \includegraphics[height=37mm]{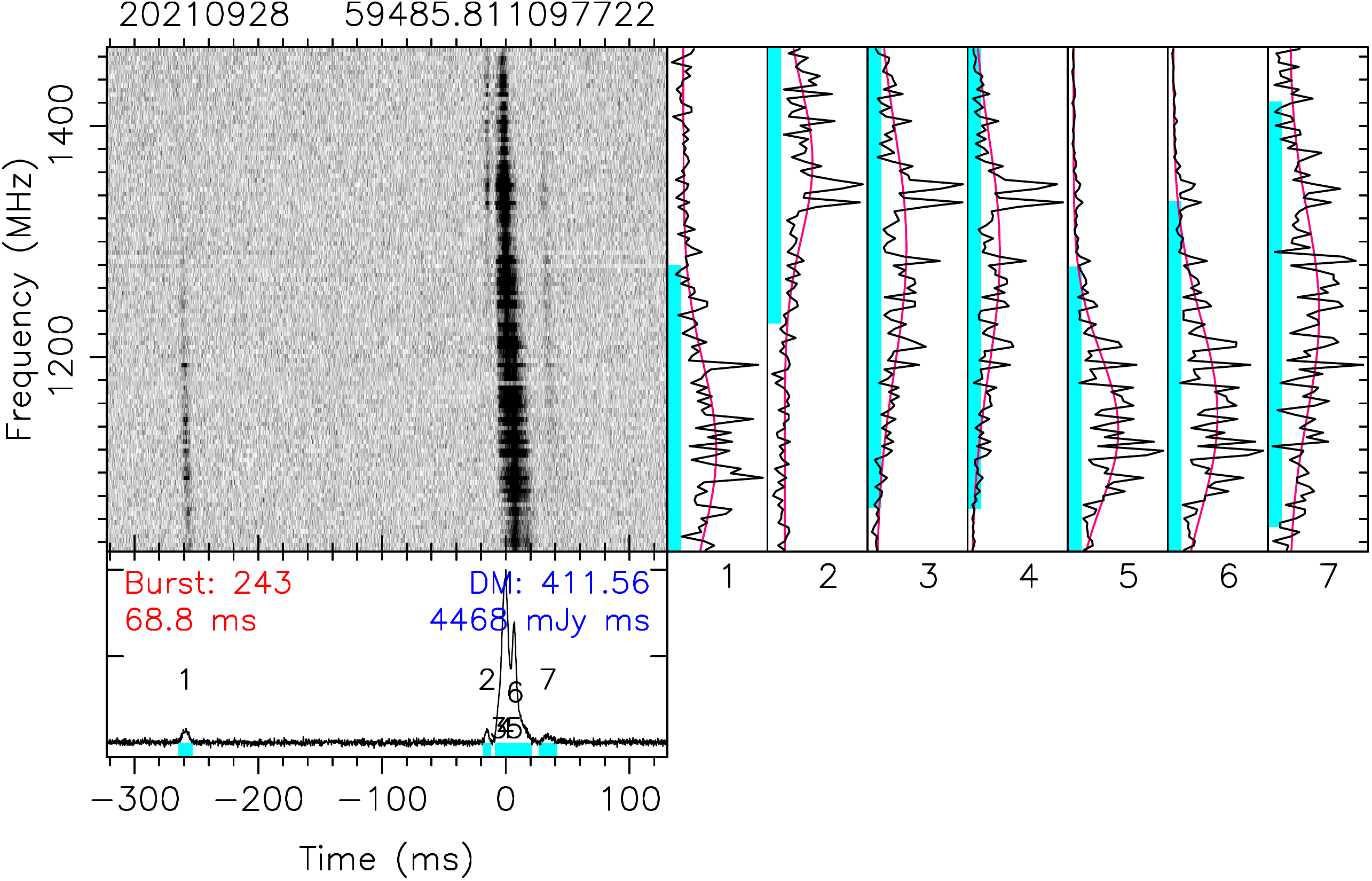}
    \includegraphics[height=37mm]{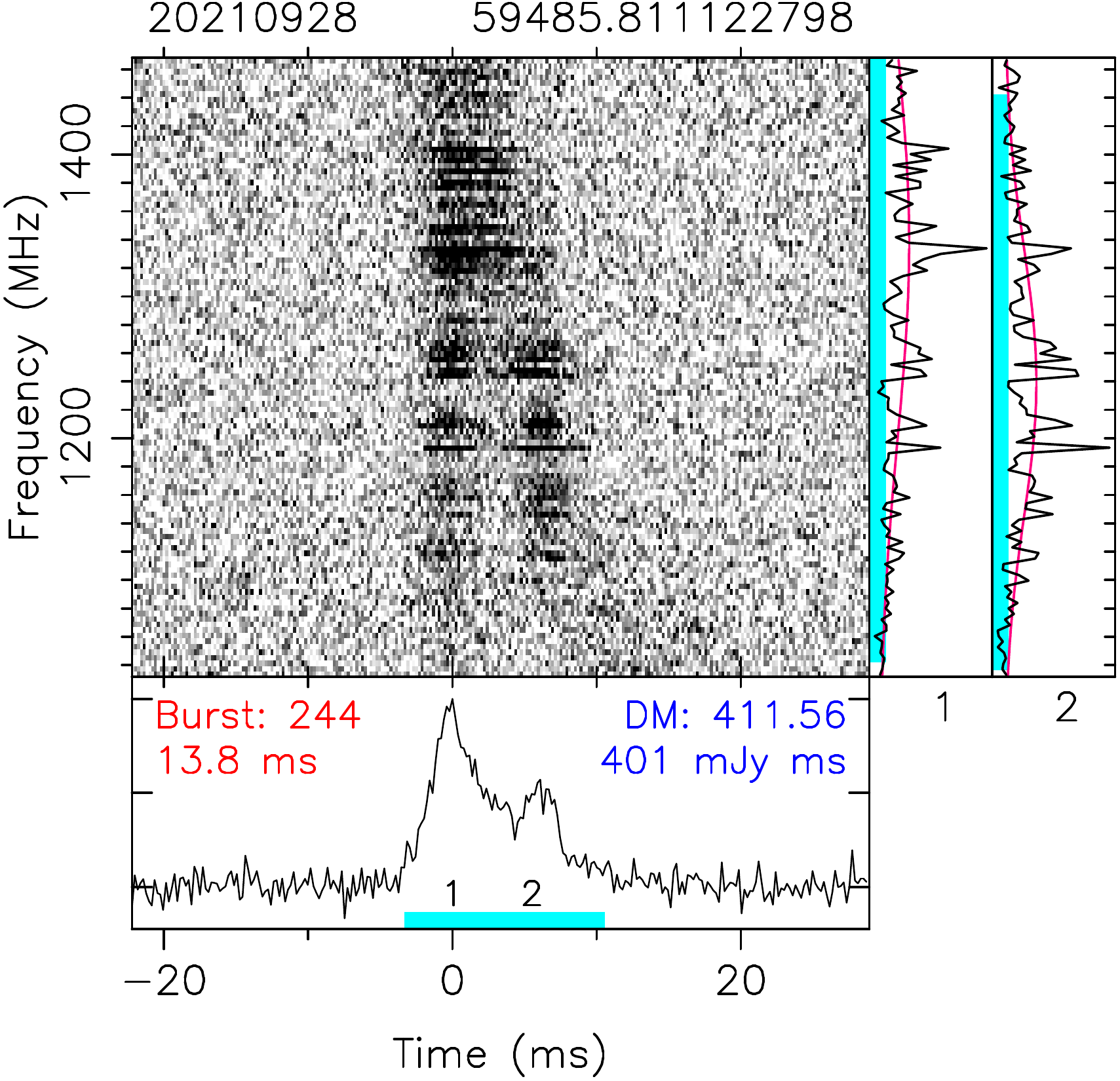}
    \includegraphics[height=37mm]{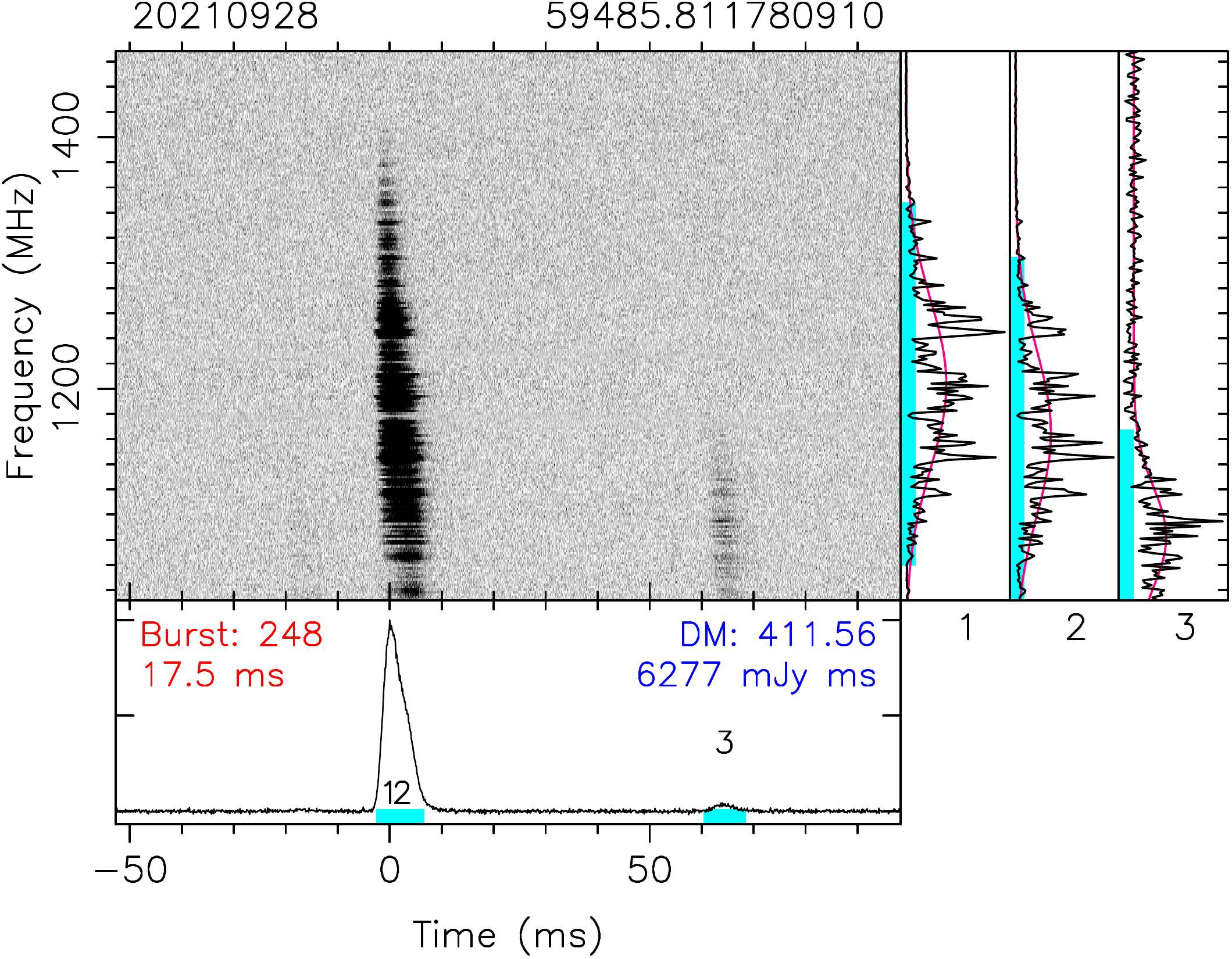}
    \includegraphics[height=37mm]{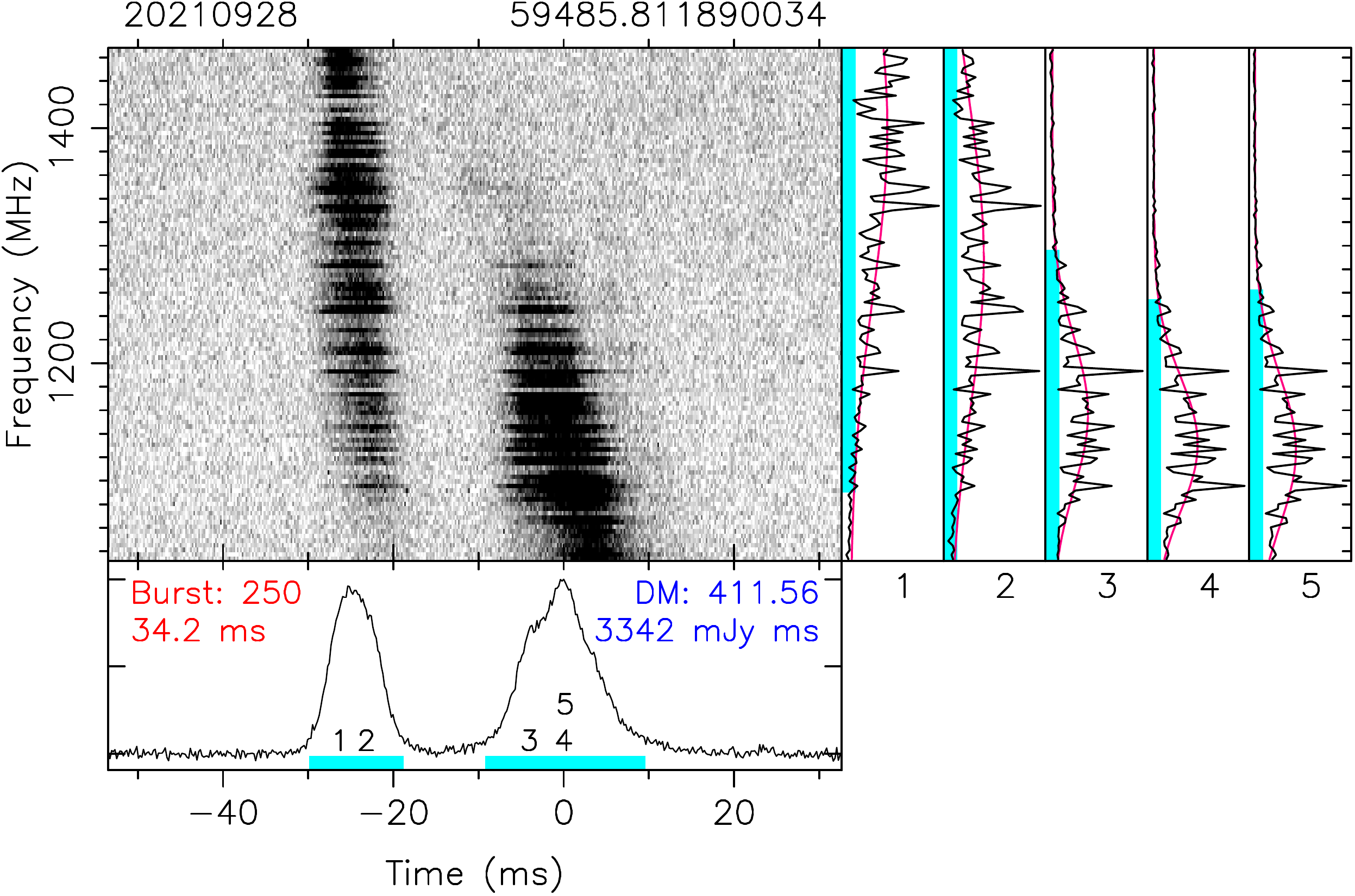}
    \includegraphics[height=37mm]{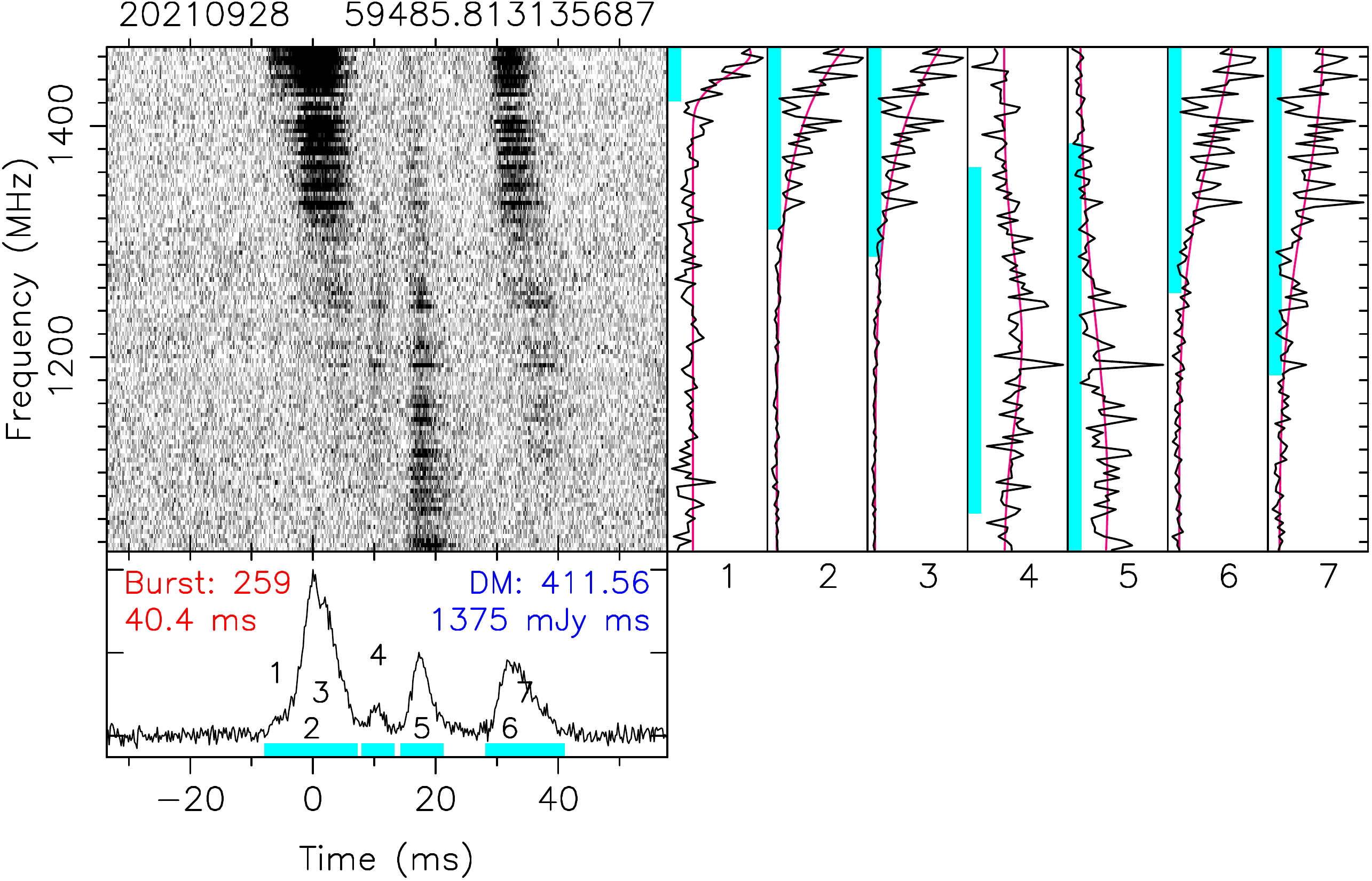}
    \includegraphics[height=37mm]{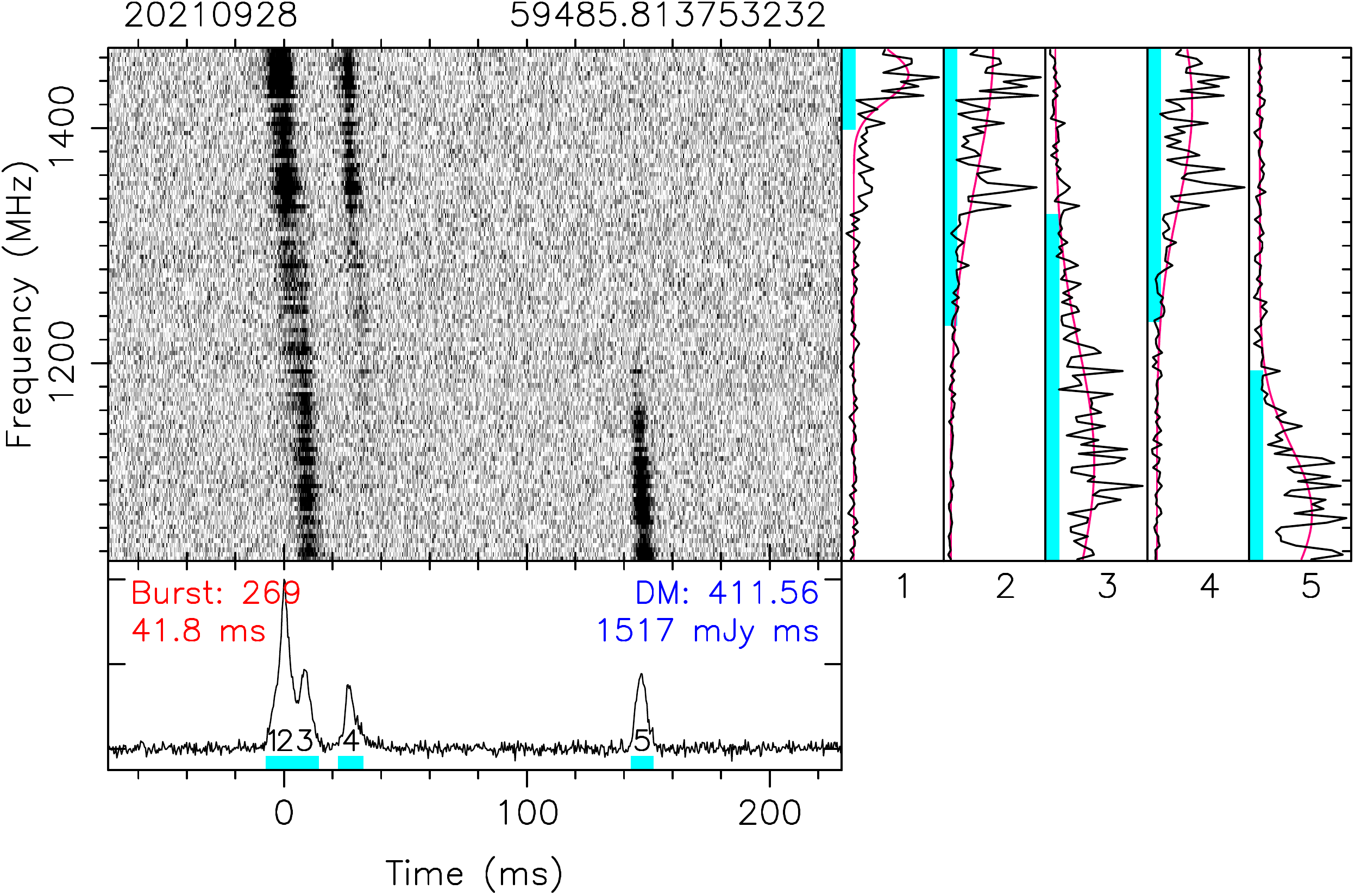}
    \includegraphics[height=37mm]{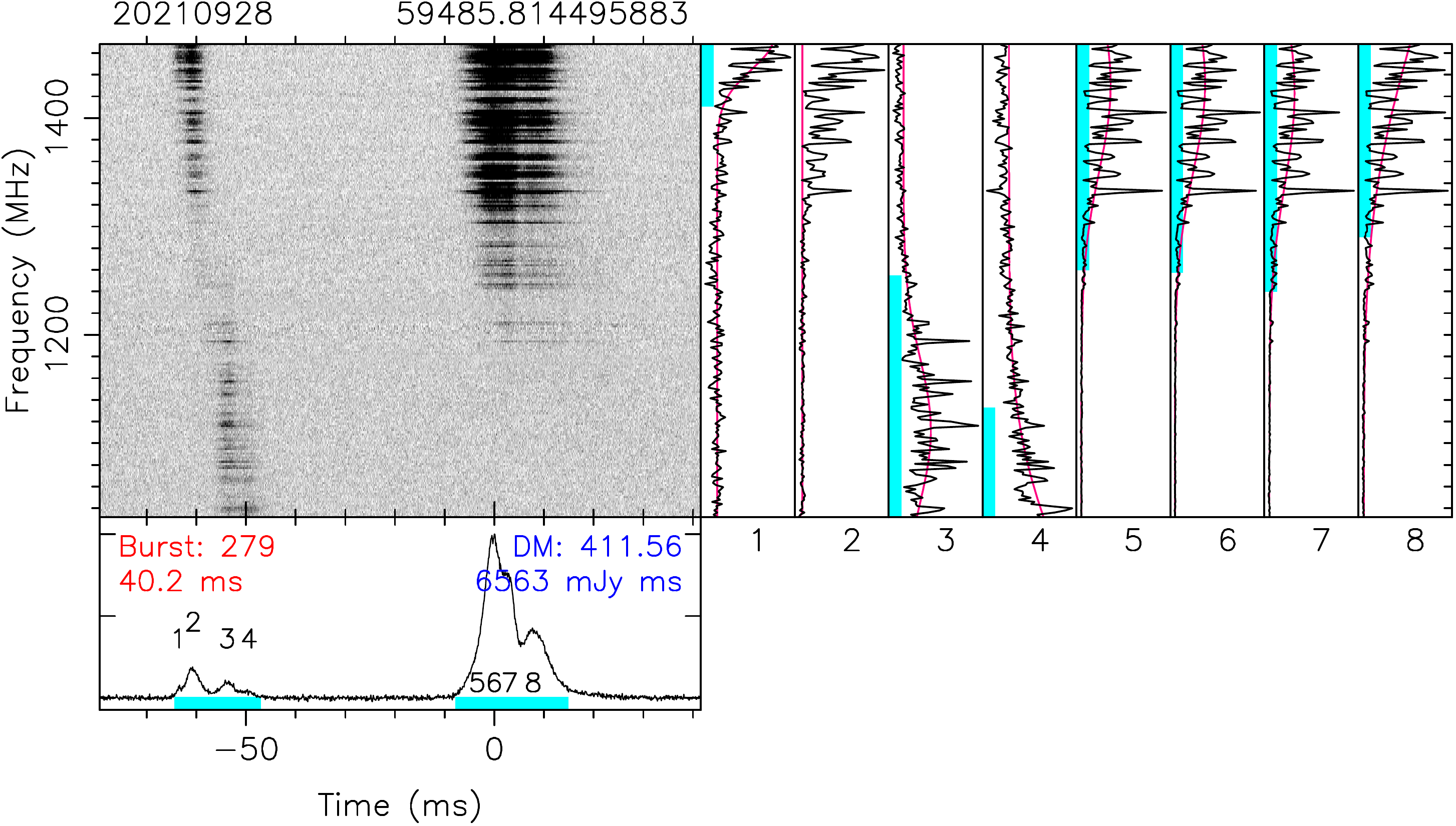}
    \includegraphics[height=37mm]{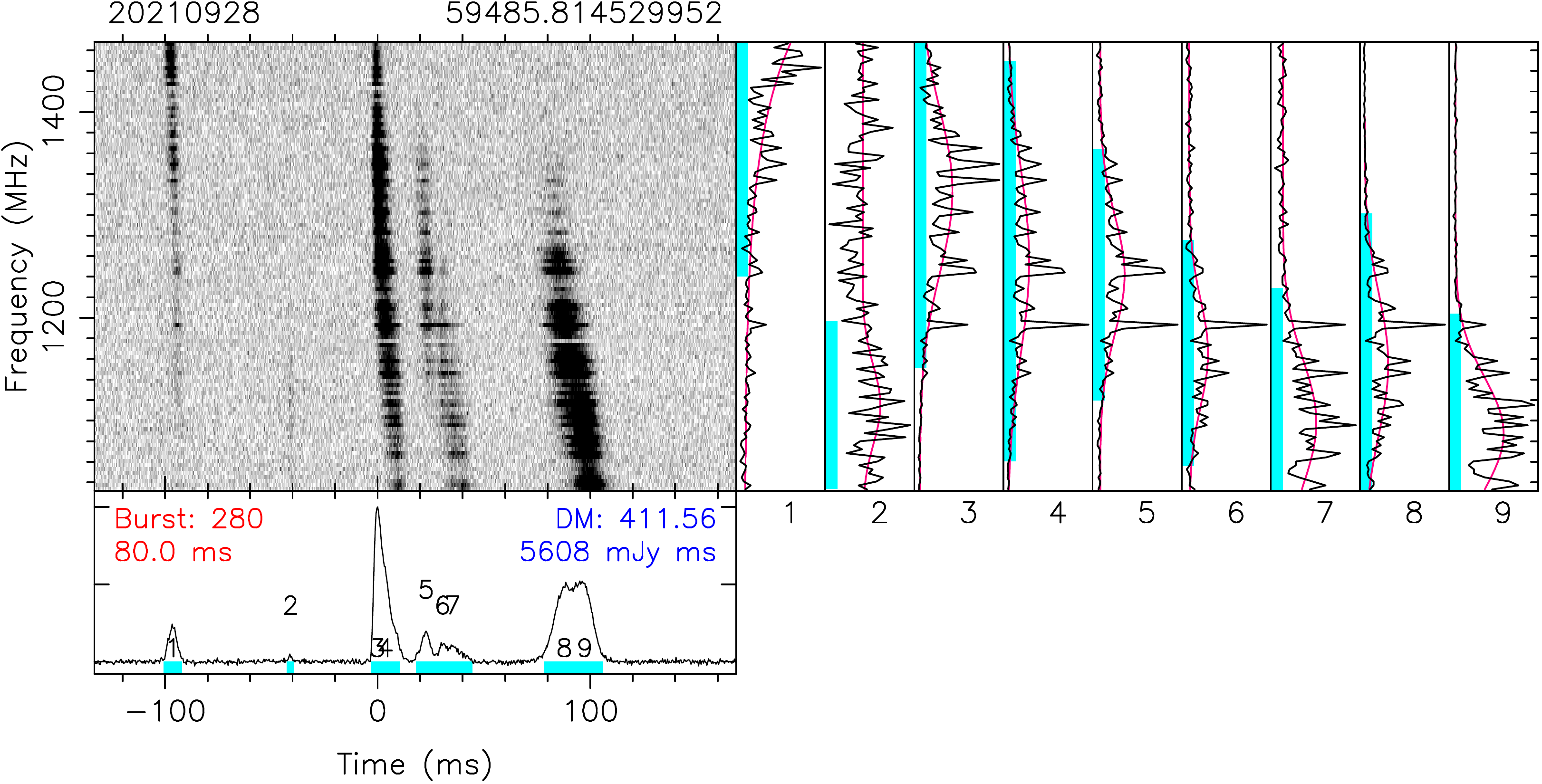}
    \includegraphics[height=37mm]{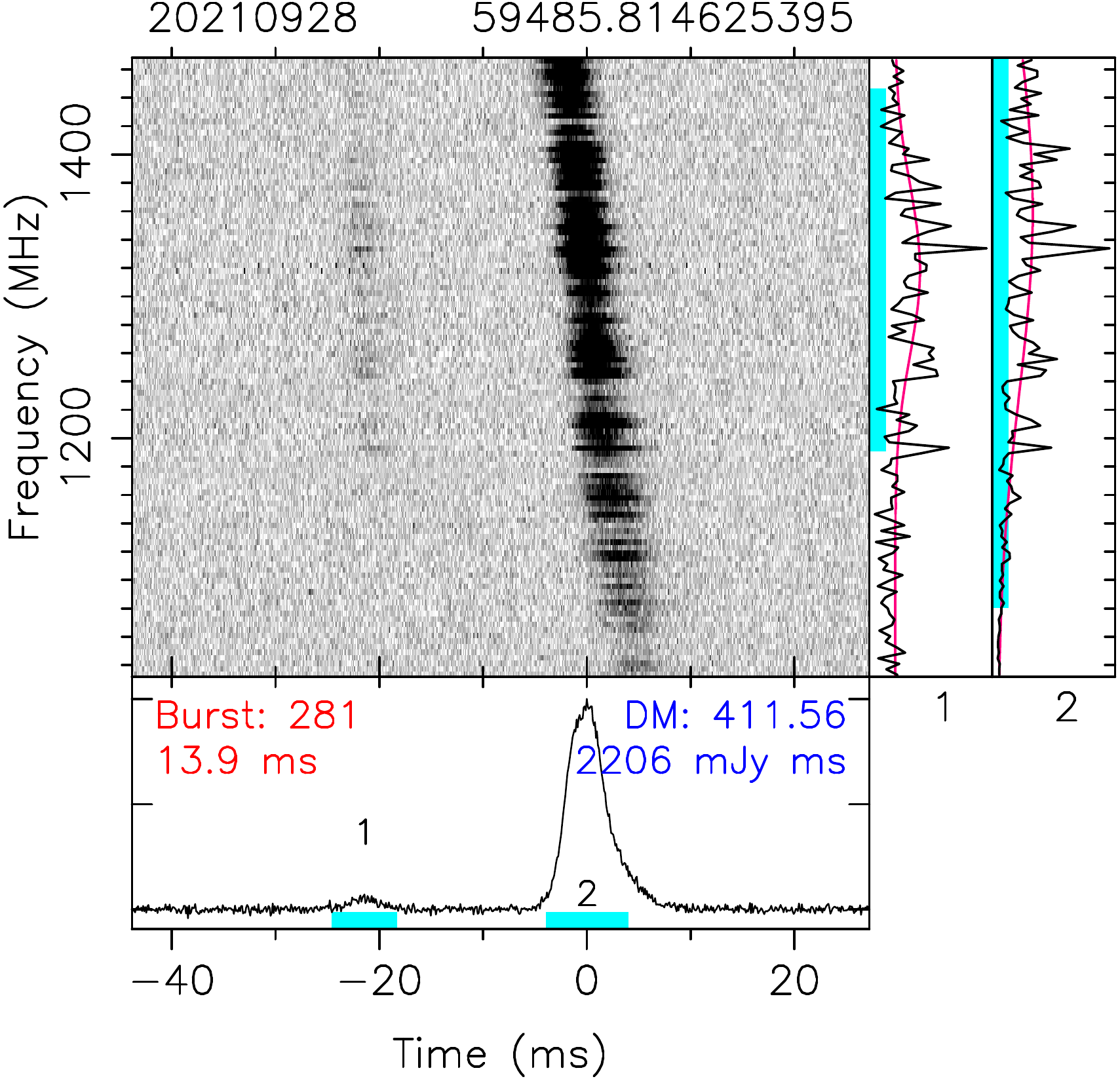}
    \includegraphics[height=37mm]{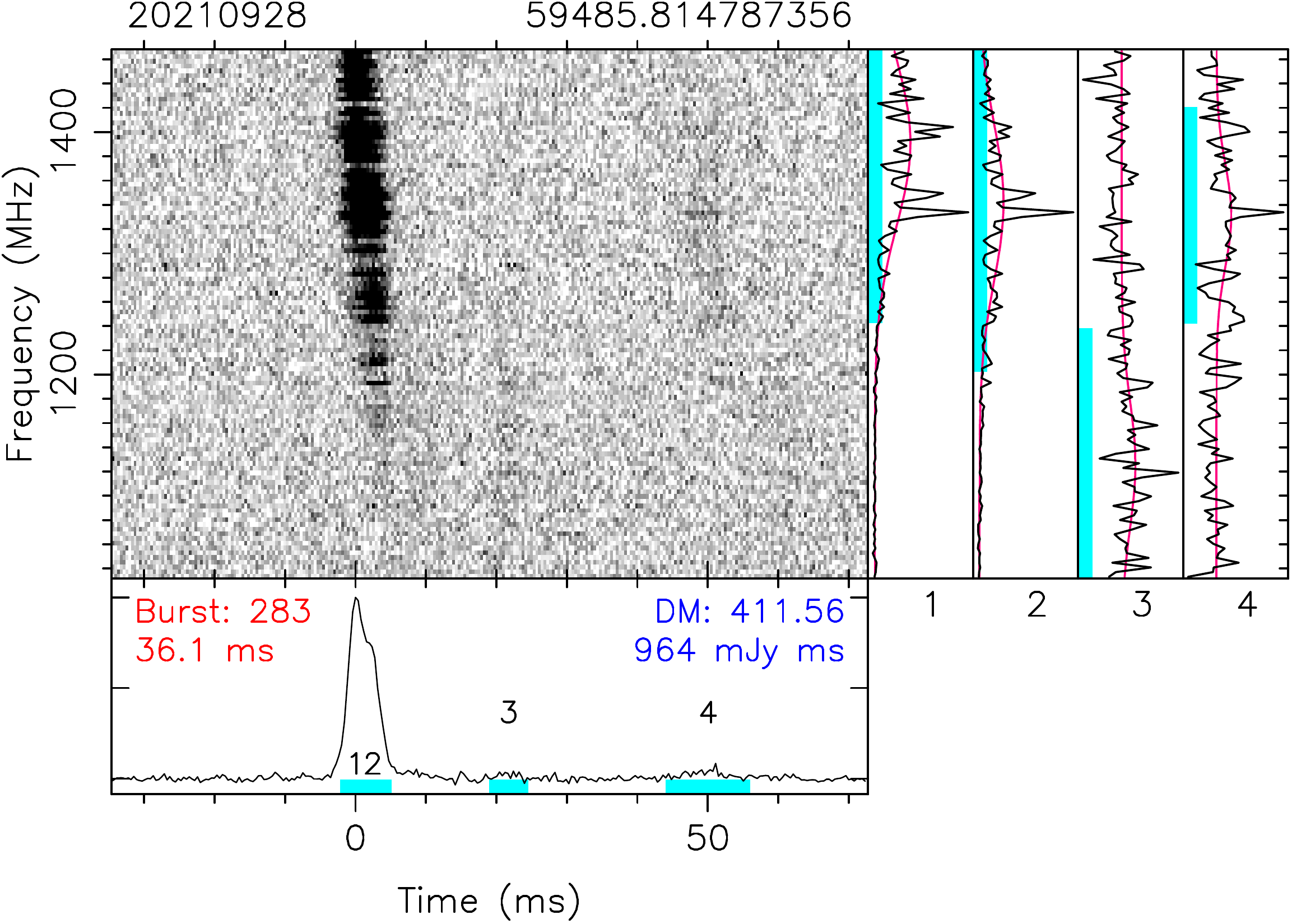}
    \includegraphics[height=37mm]{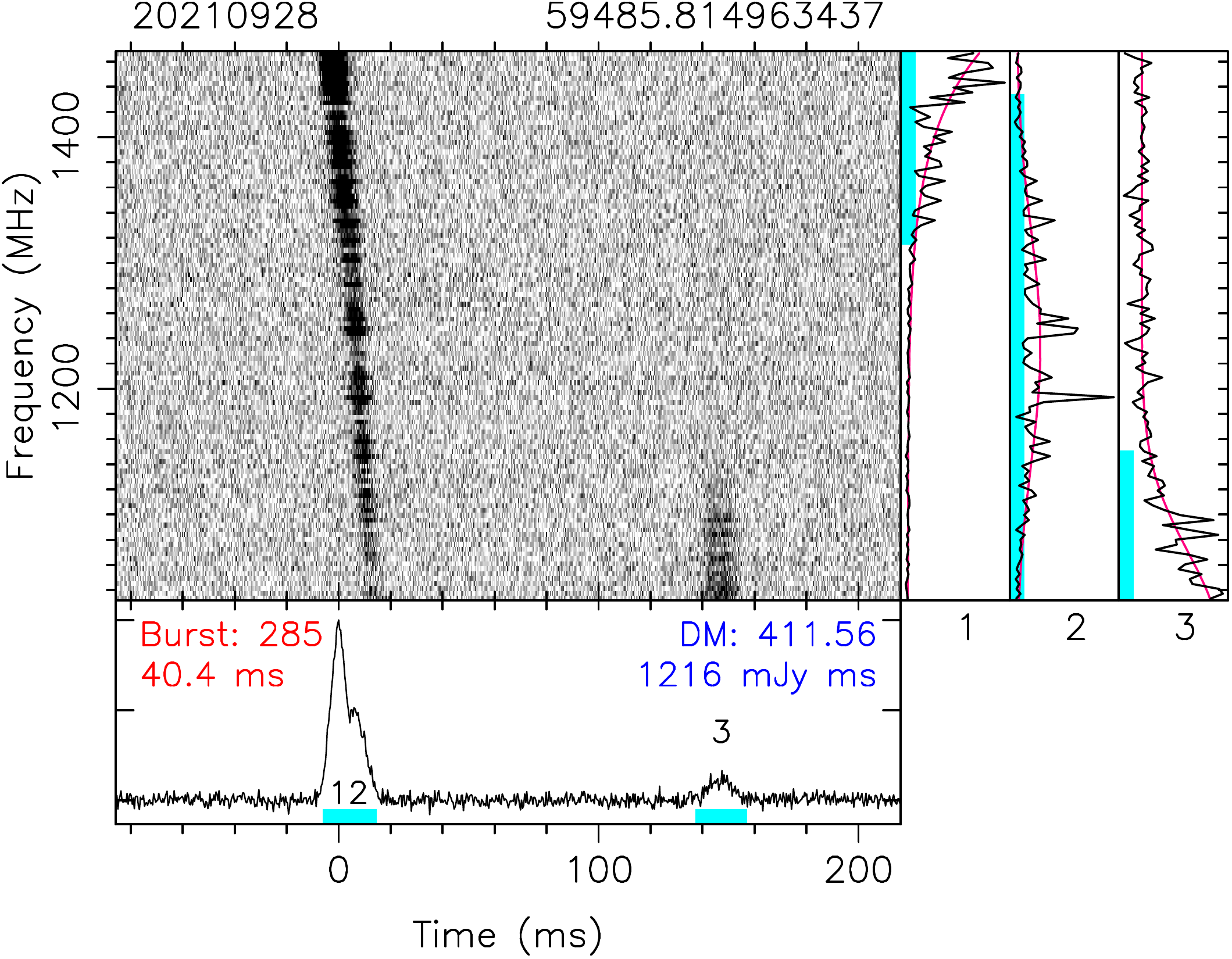}
    \includegraphics[height=37mm]{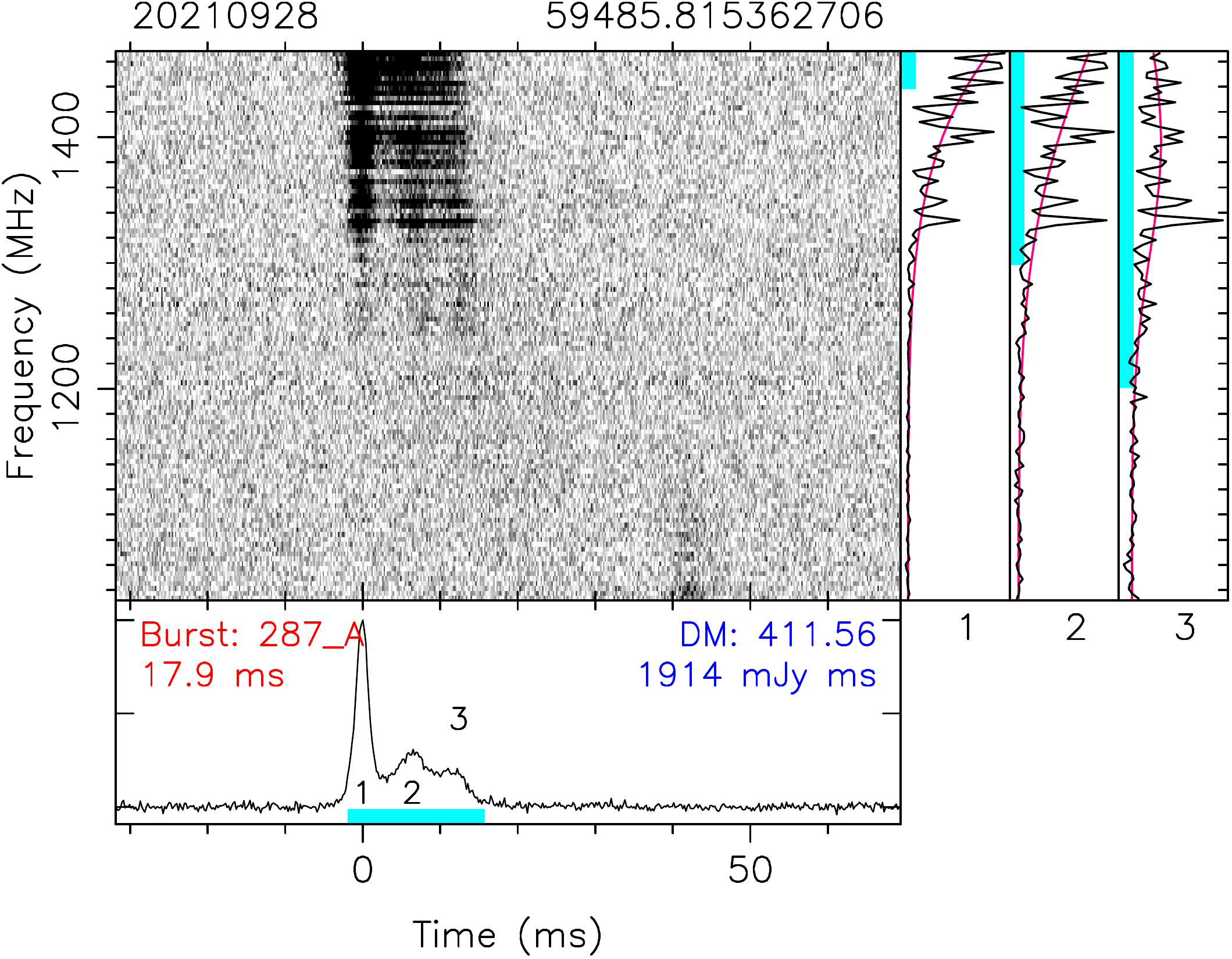}
    \includegraphics[height=37mm]{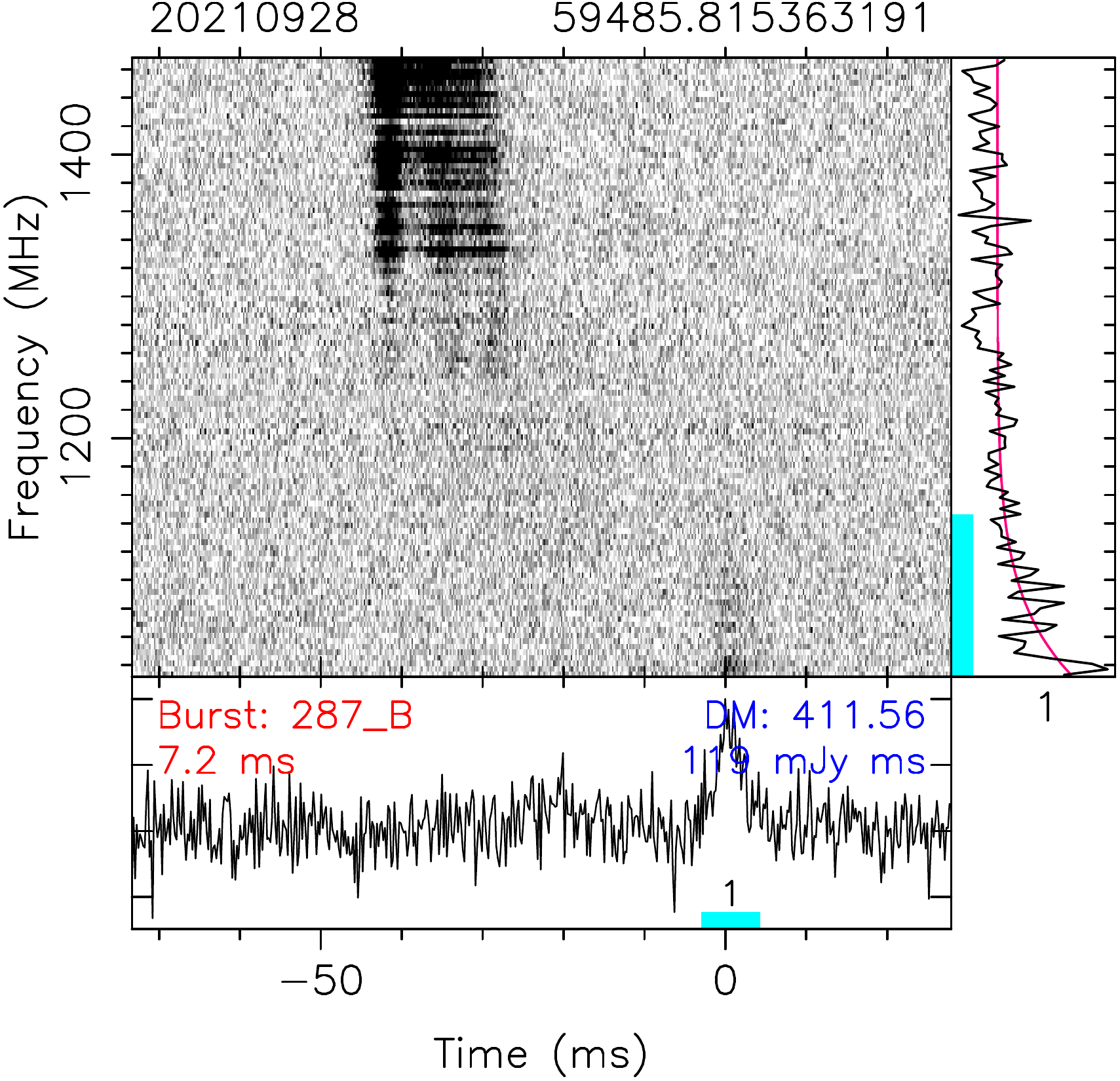}
    \includegraphics[height=37mm]{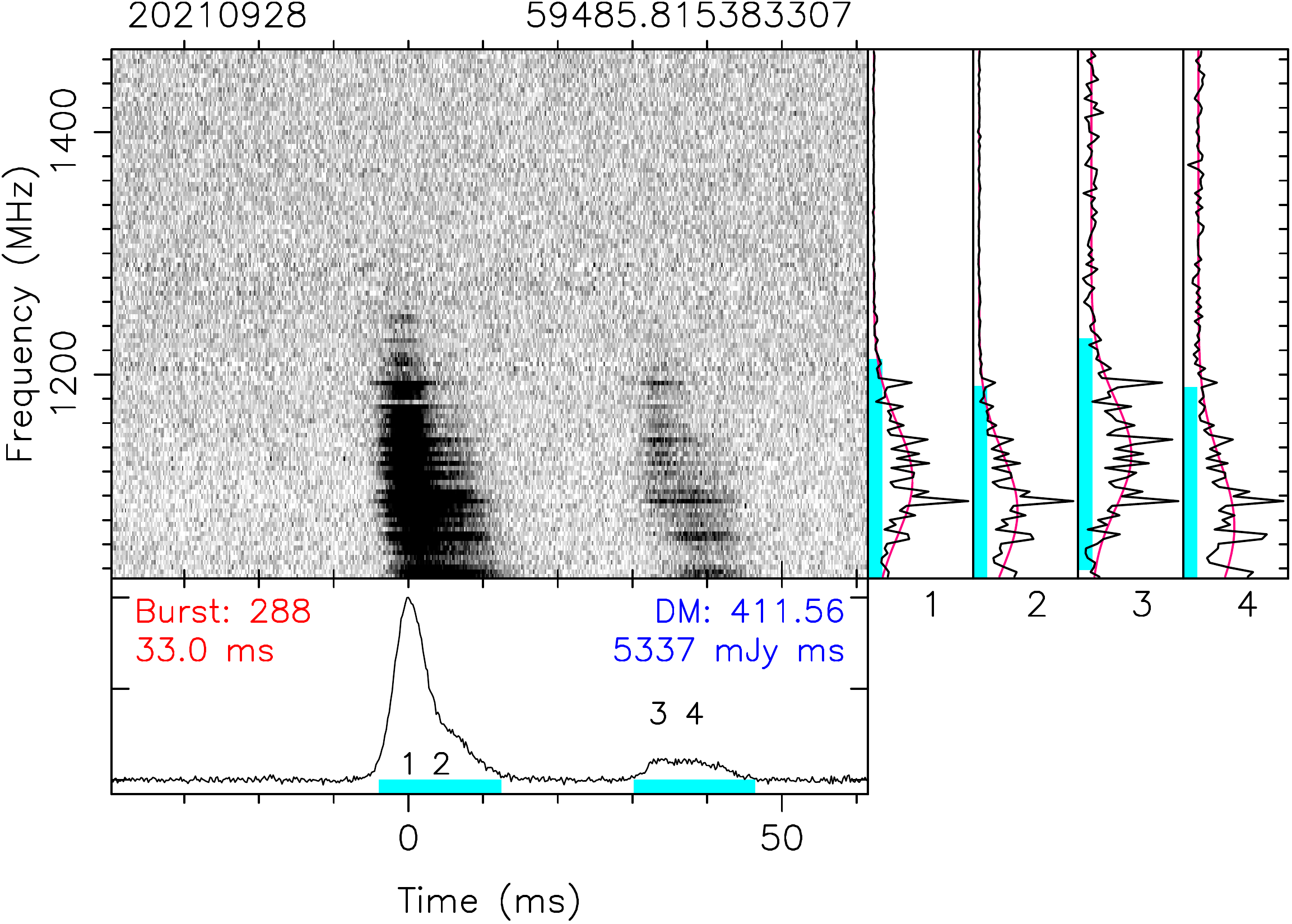}
    \includegraphics[height=37mm]{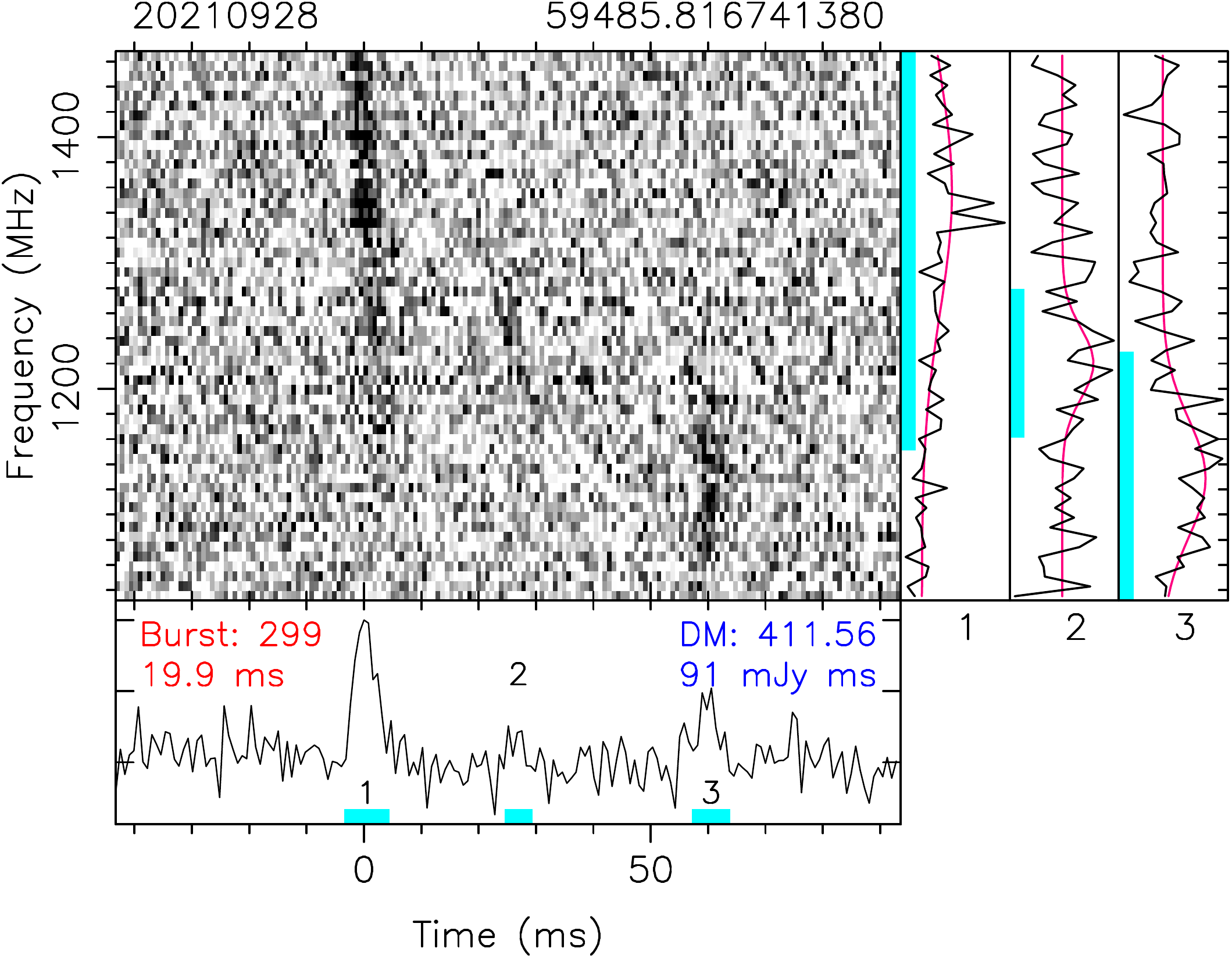}
    \includegraphics[height=37mm]{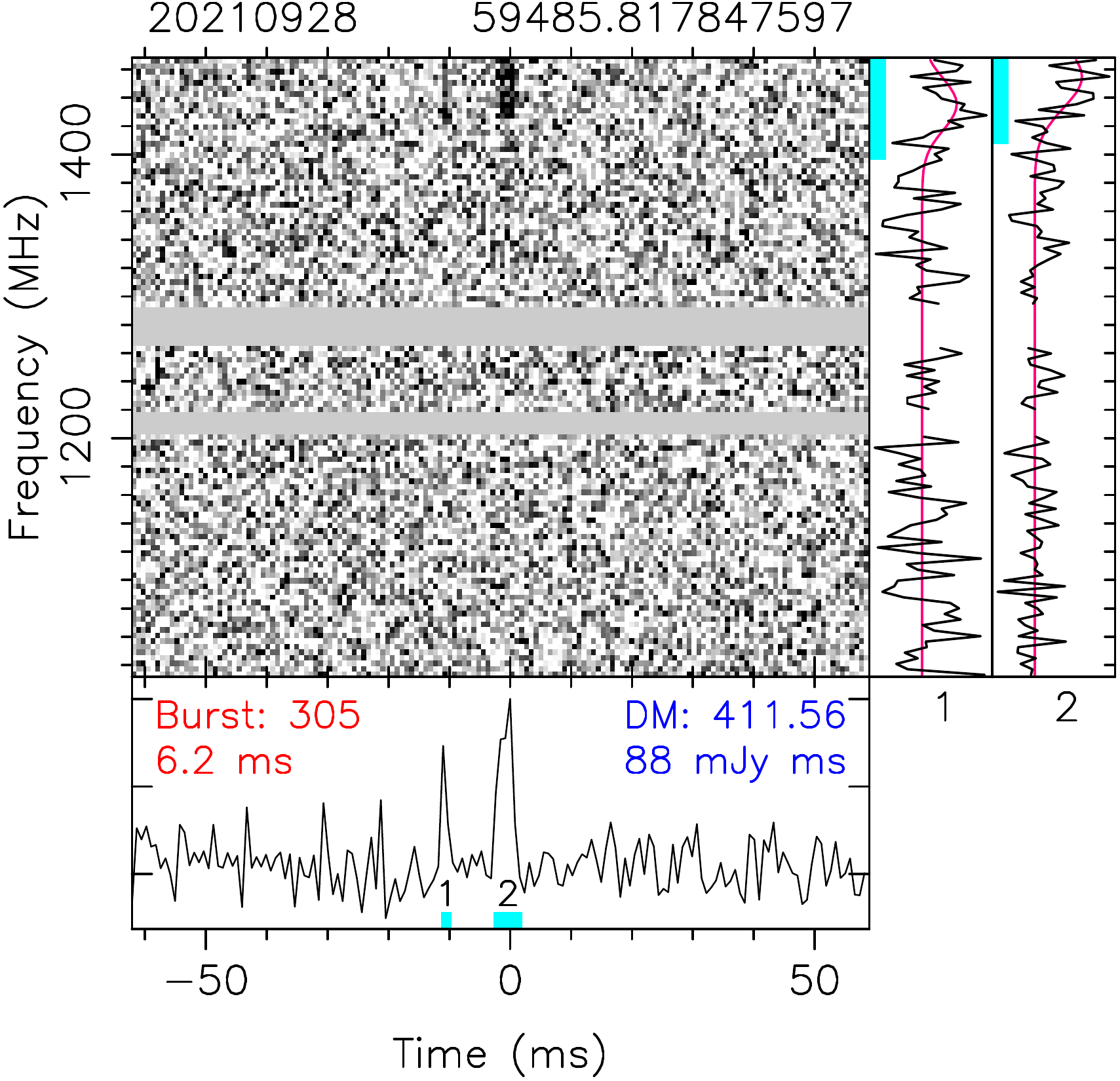}
    \caption{ \it{ -- continued and ended}.
}
\end{figure*}
\addtocounter{figure}{-1}
\begin{figure*}
    \flushleft
    \includegraphics[height=37mm]{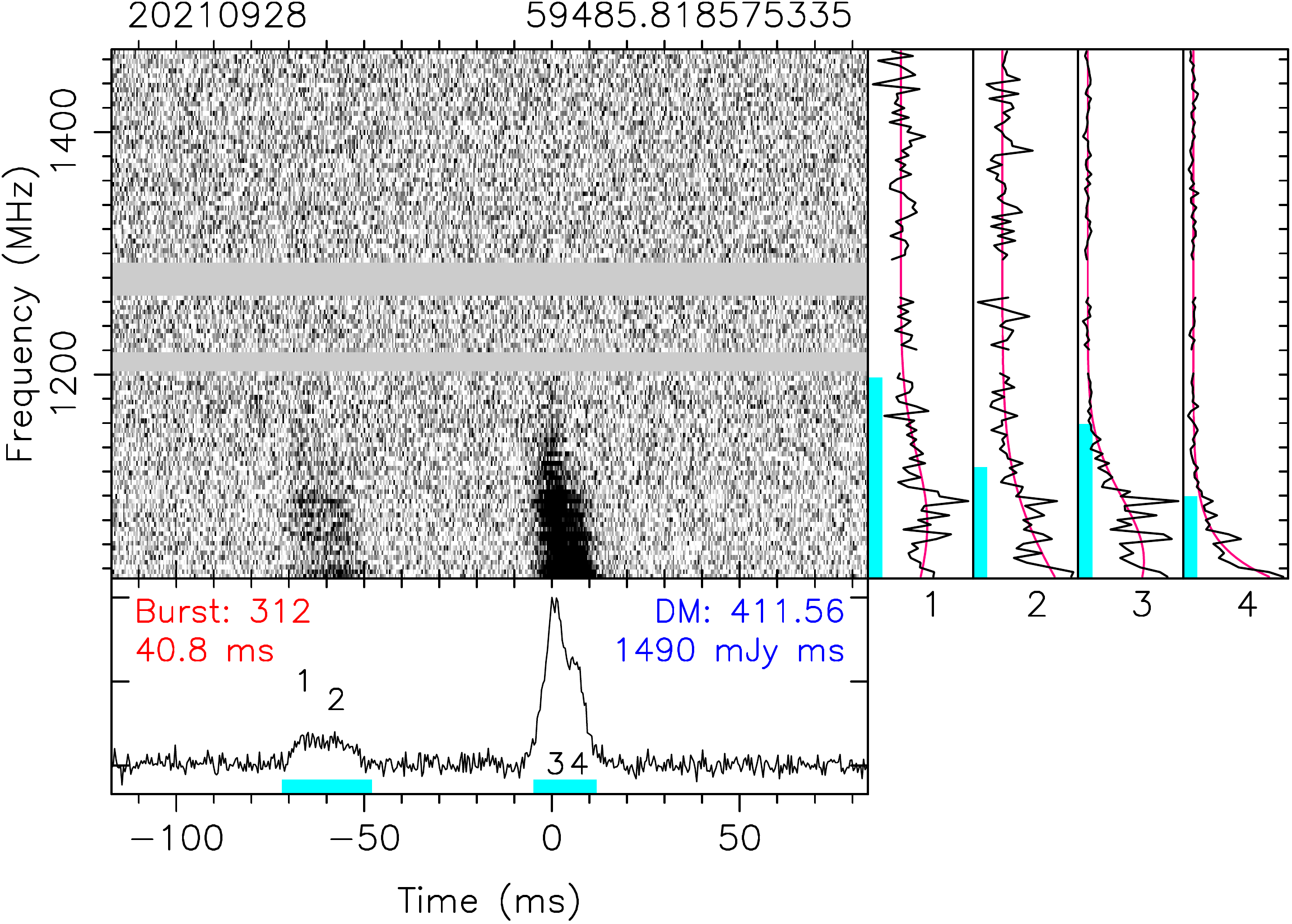}
    \includegraphics[height=37mm]{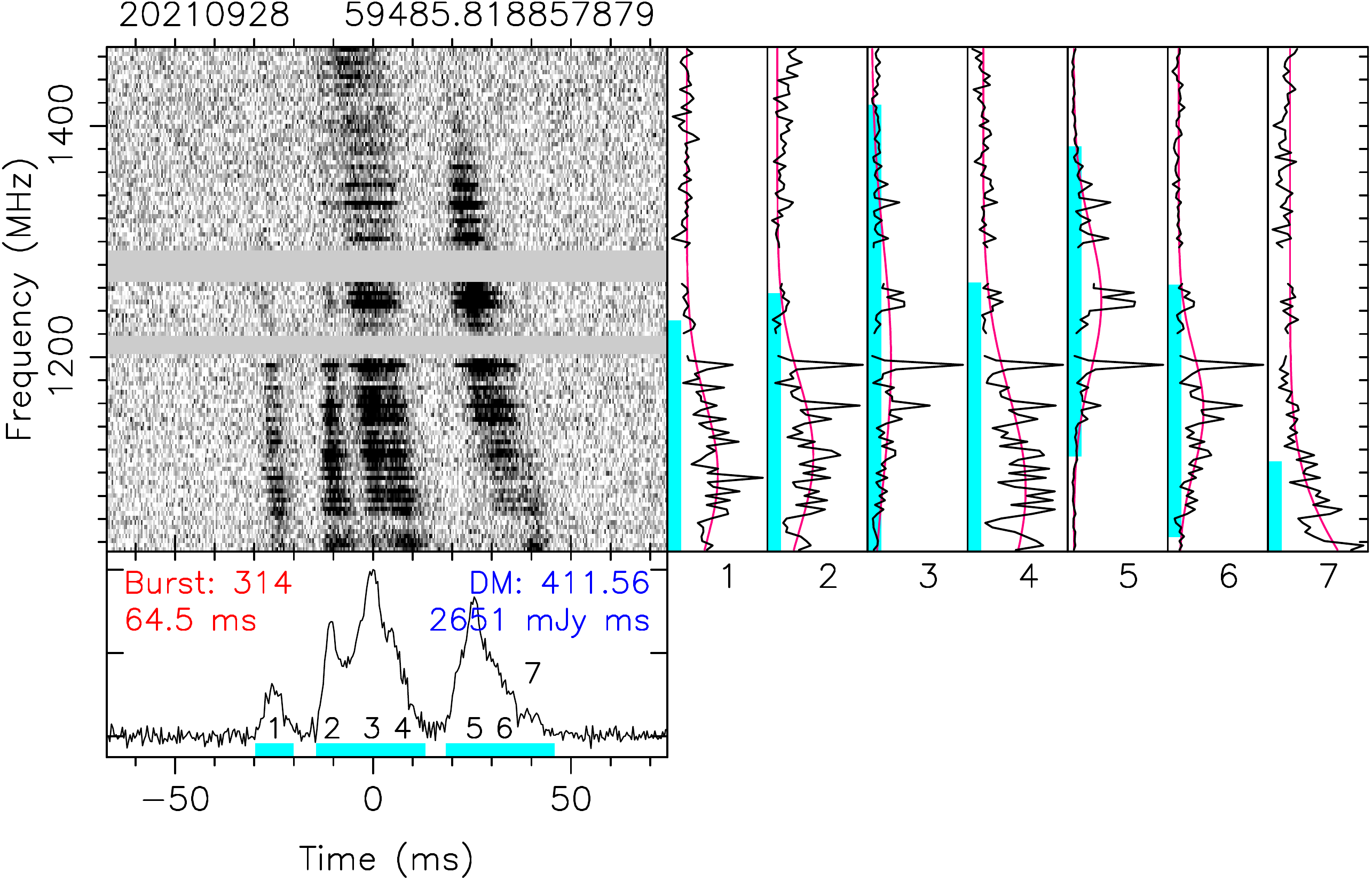}
    \includegraphics[height=37mm]{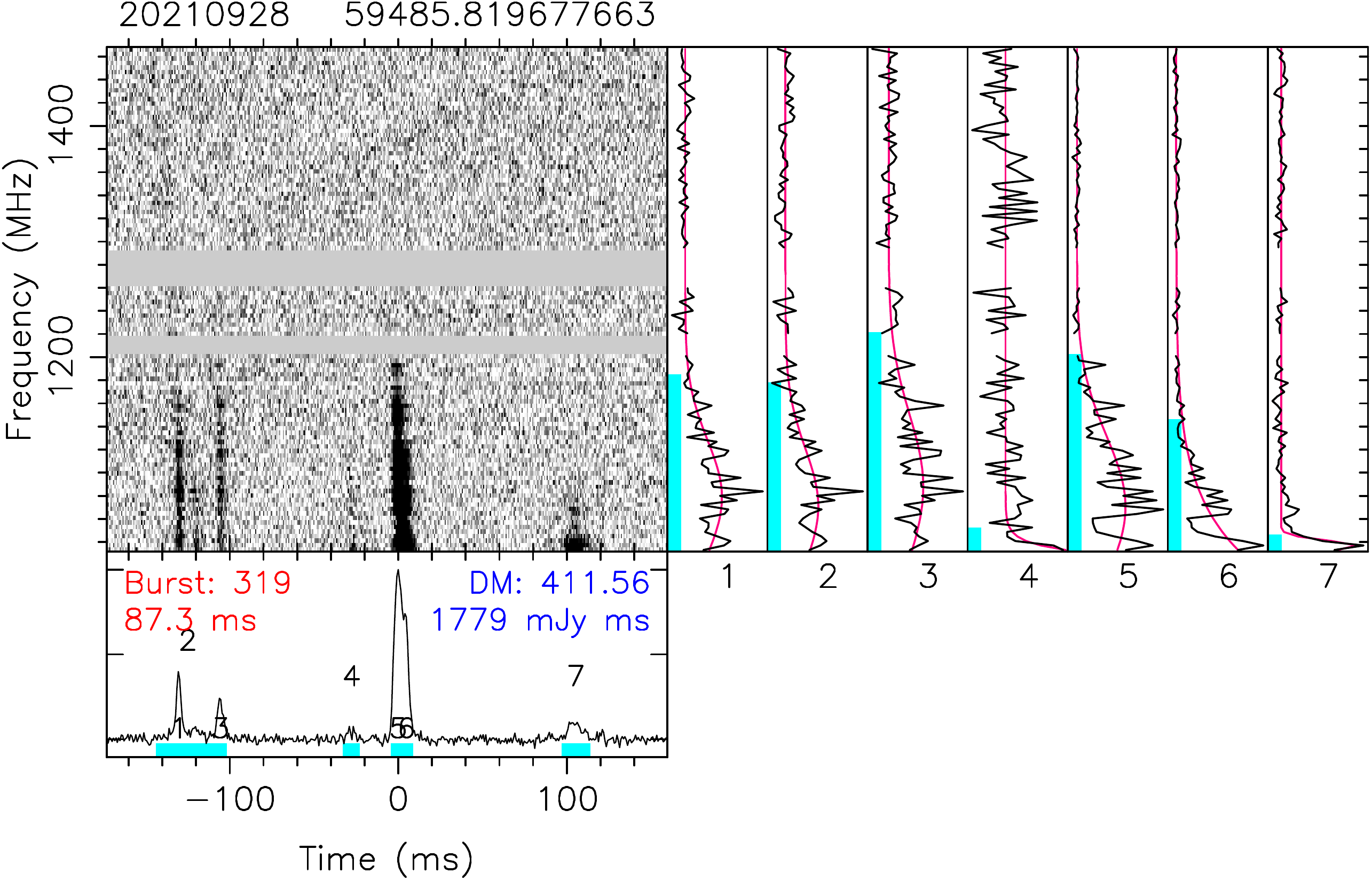}
    \includegraphics[height=37mm]{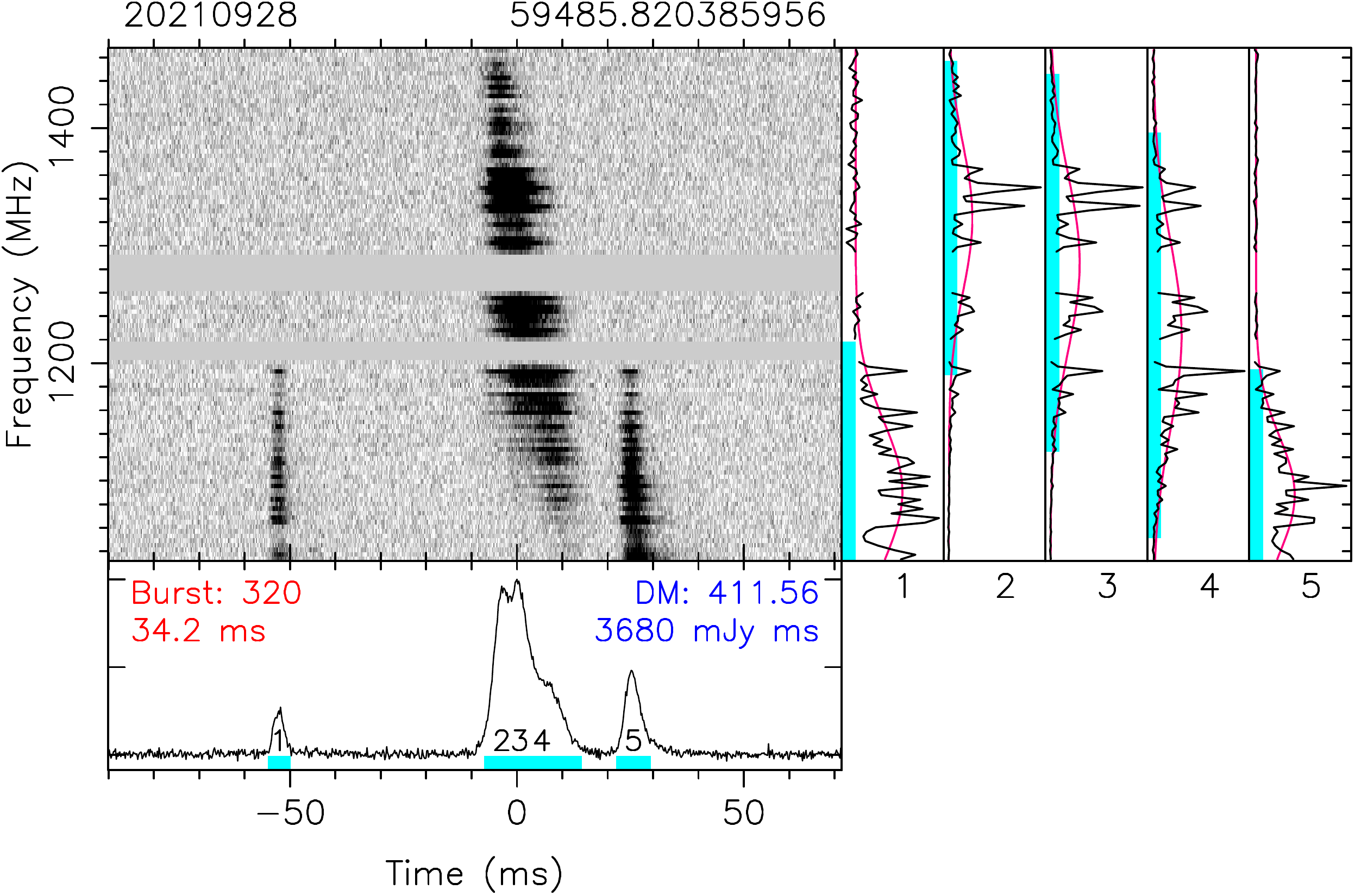}
    \includegraphics[height=37mm]{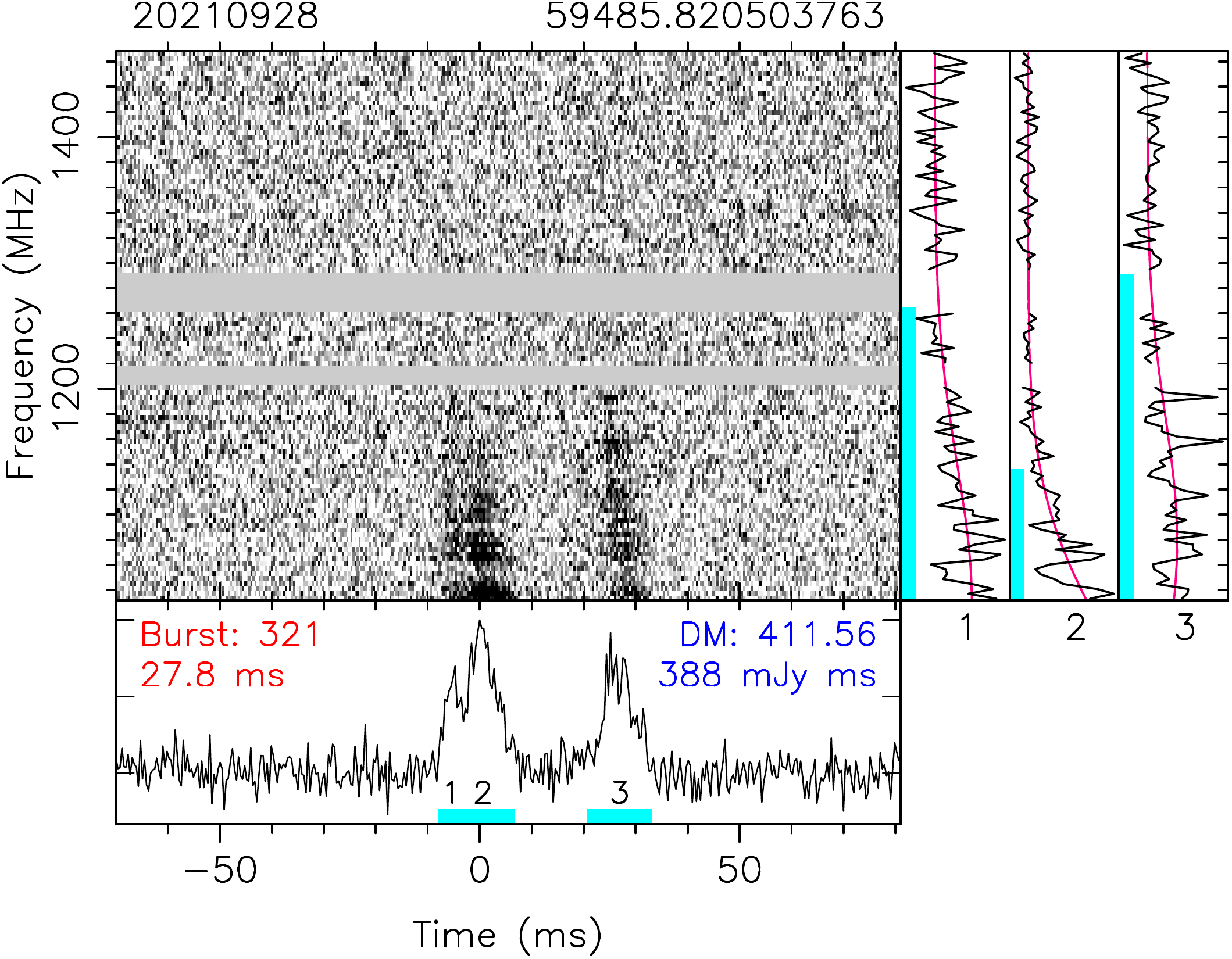}
    \includegraphics[height=37mm]{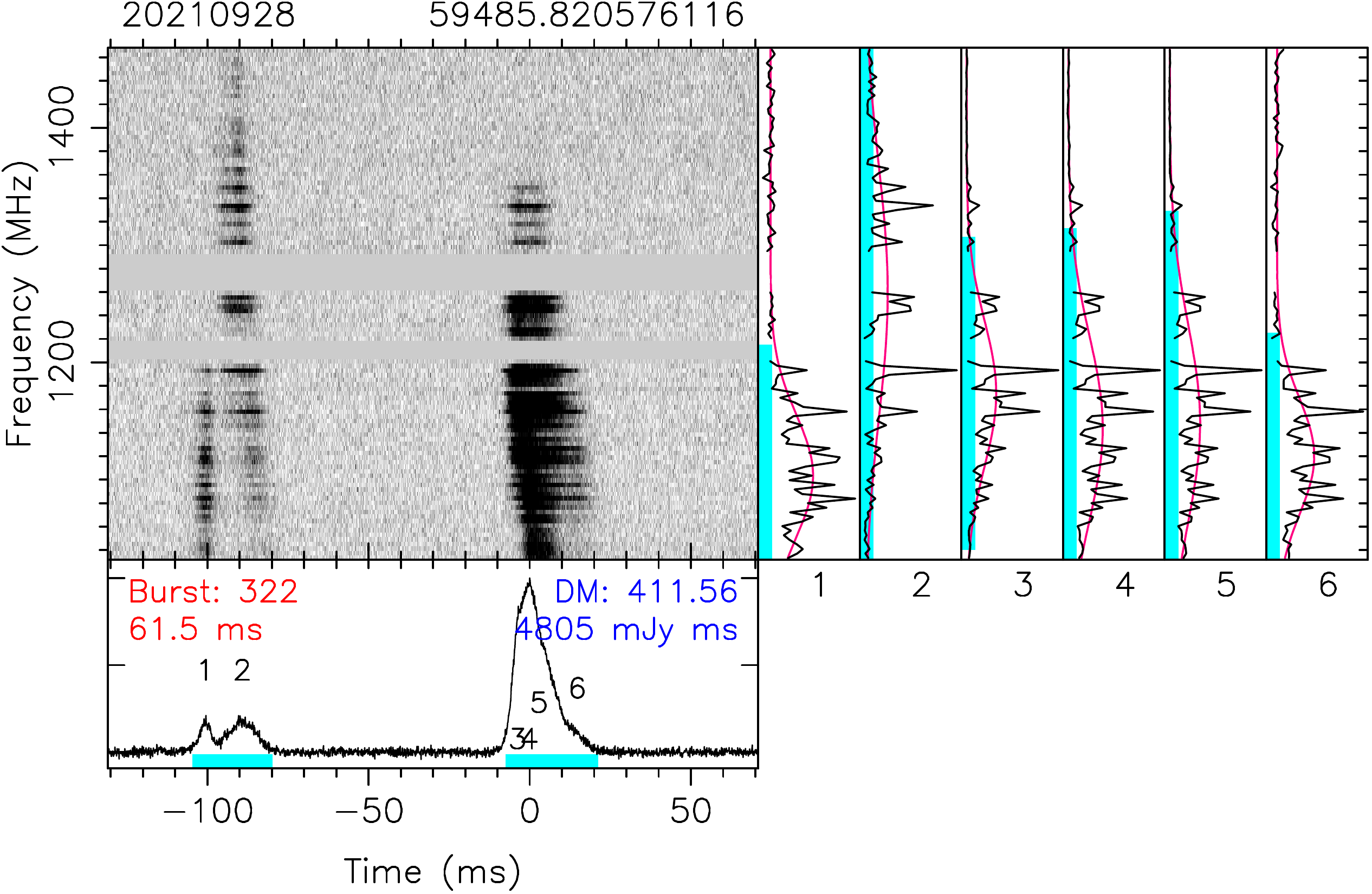}
    \includegraphics[height=37mm]{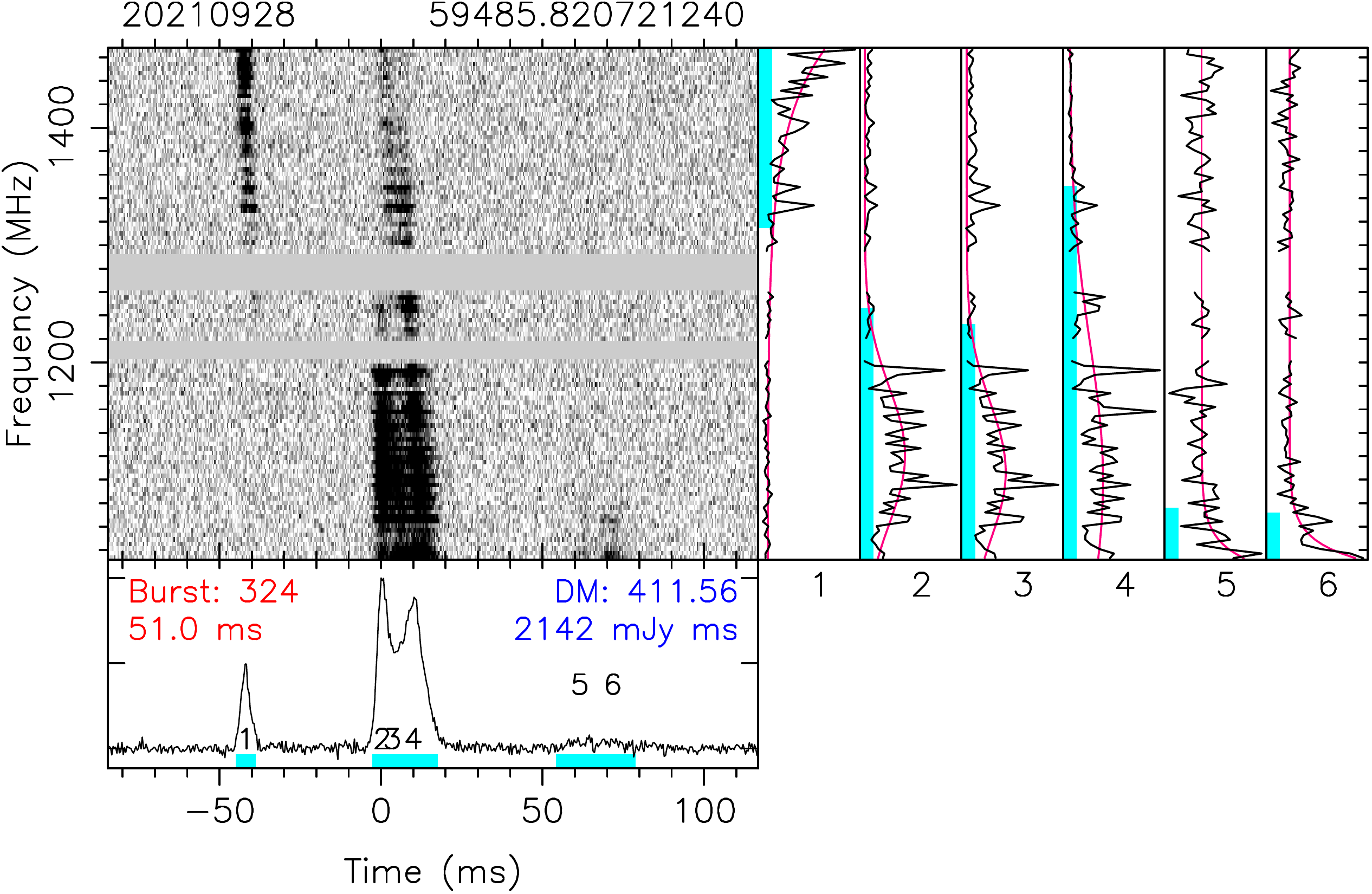}
    \includegraphics[height=37mm]{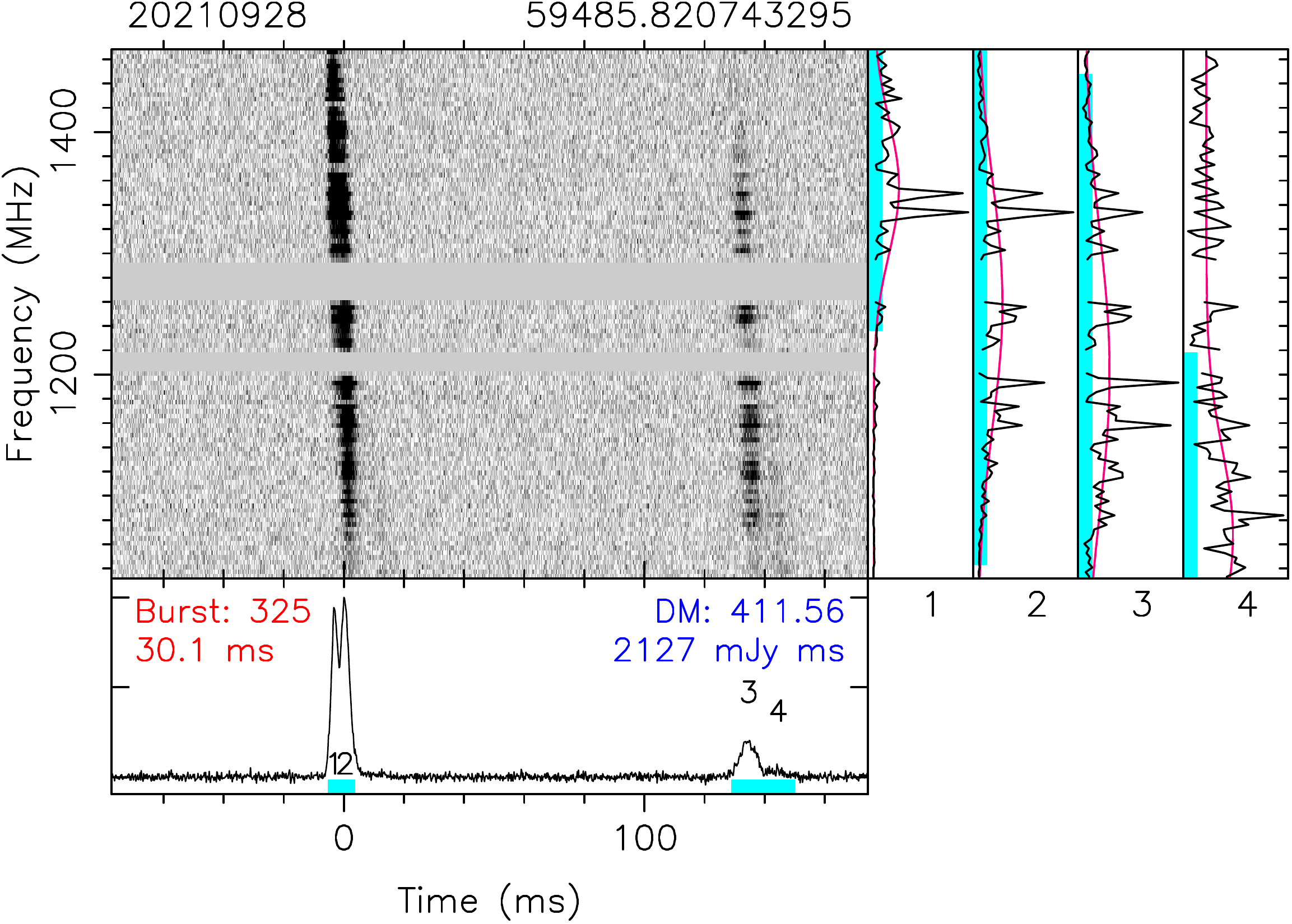}
    \includegraphics[height=37mm]{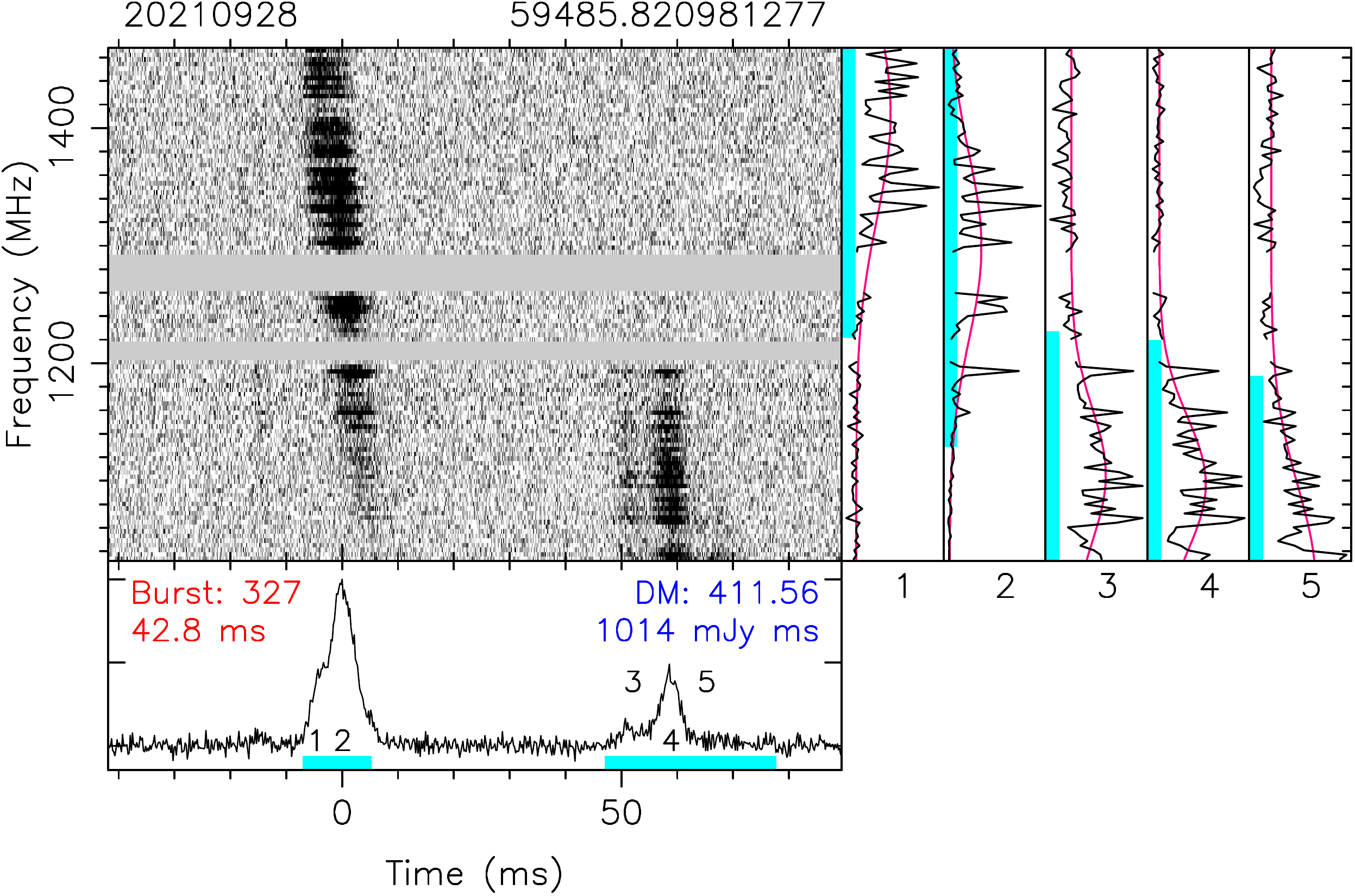}
    \includegraphics[height=37mm]{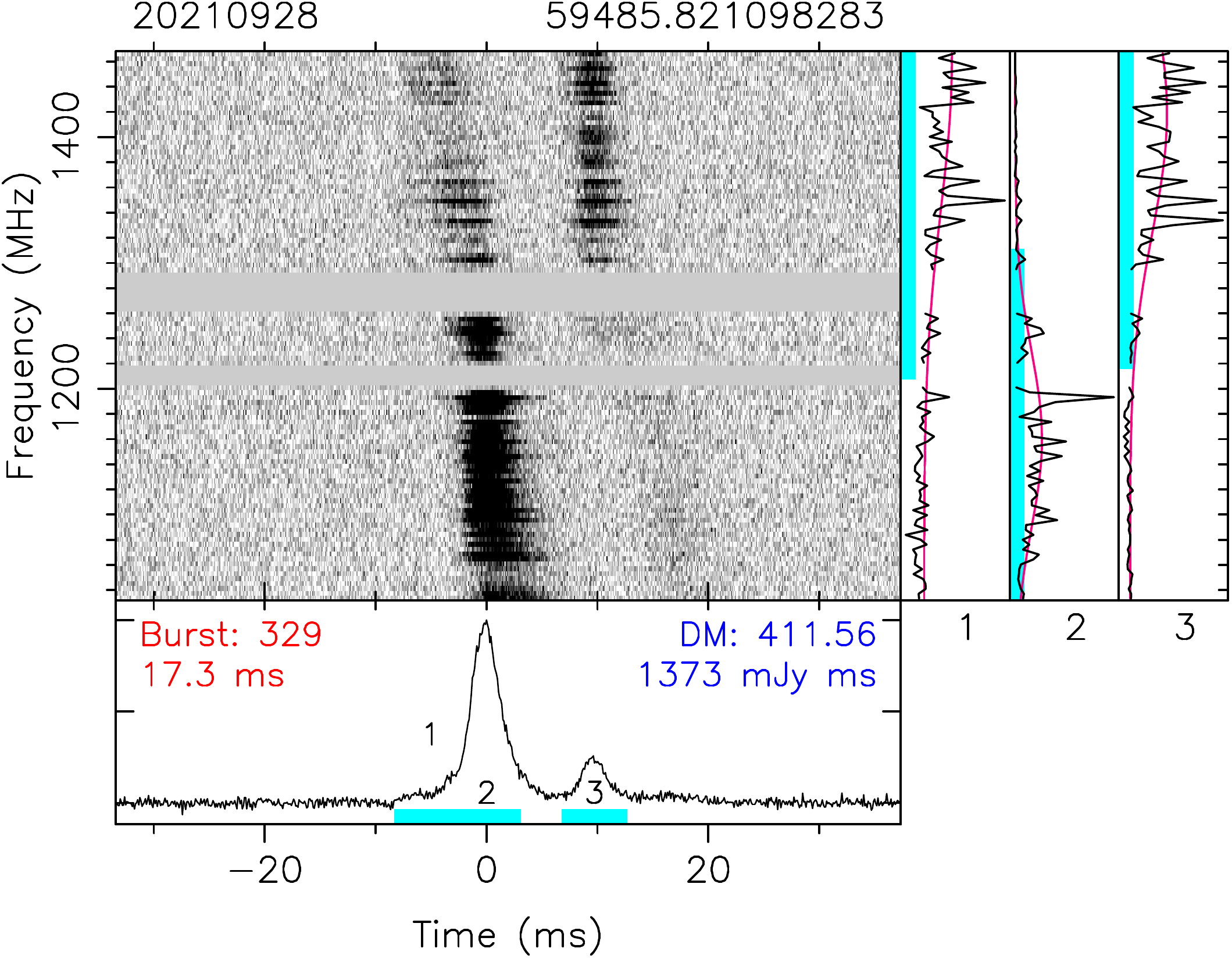}
    \includegraphics[height=37mm]{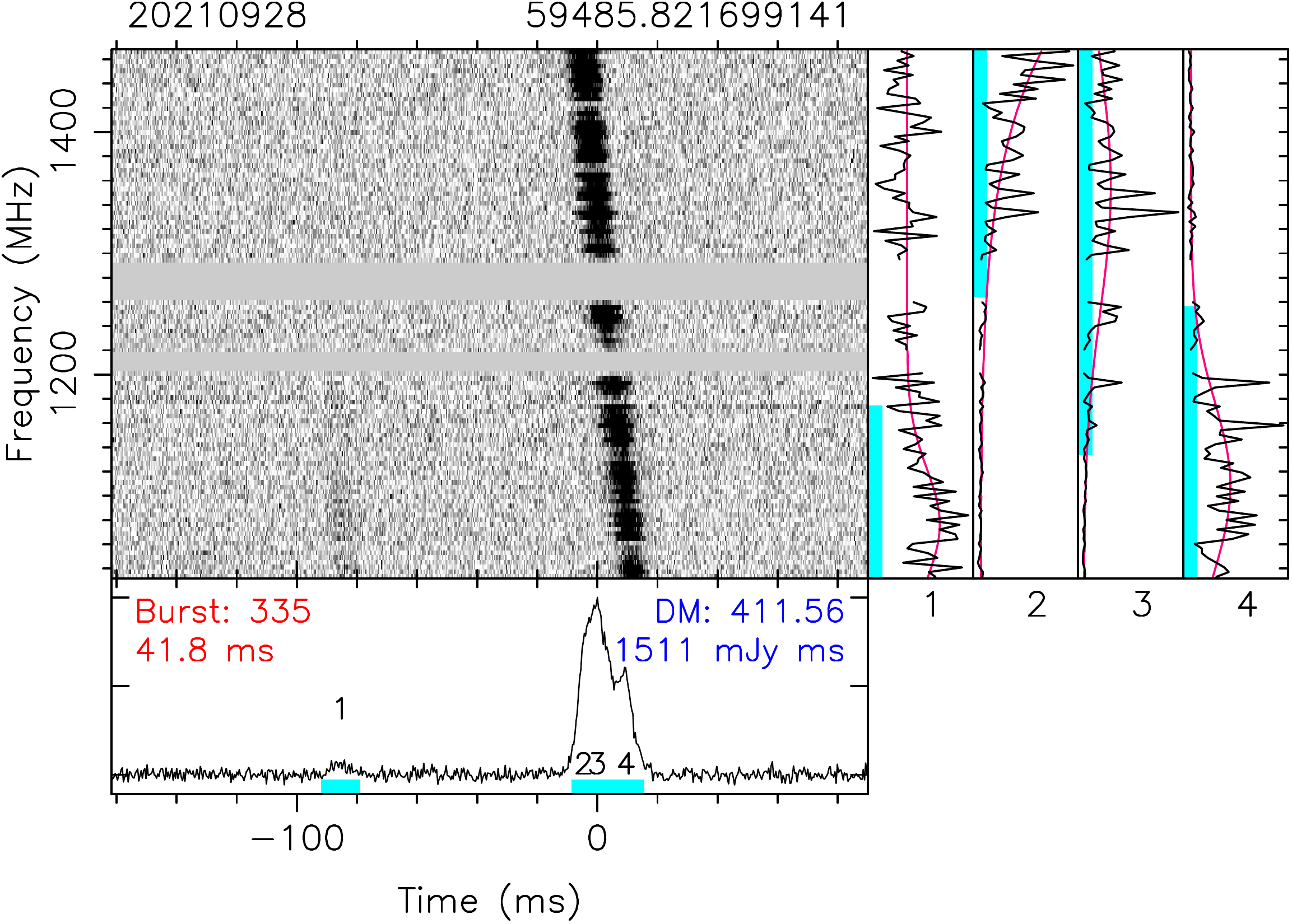}
    \includegraphics[height=37mm]{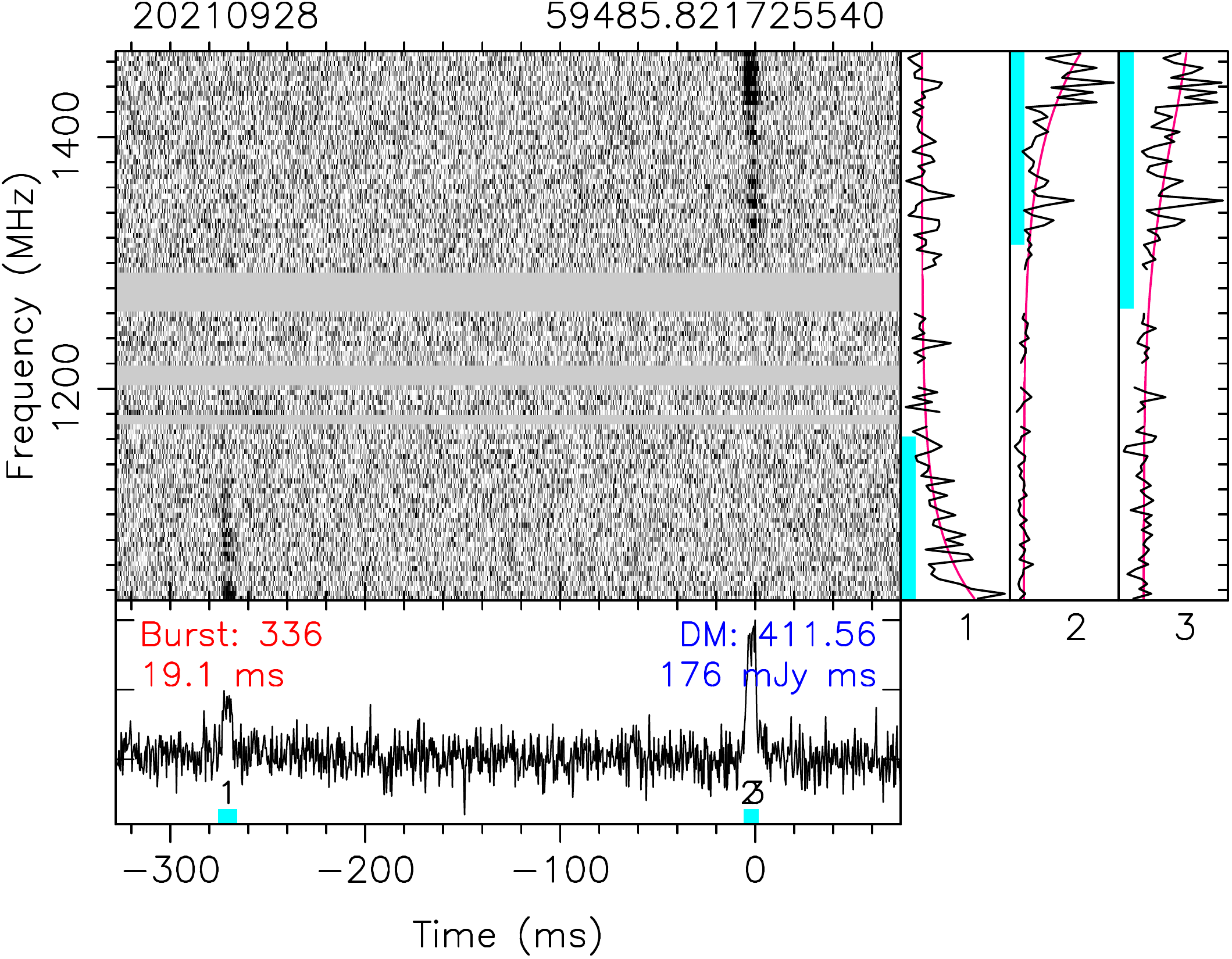}
    \includegraphics[height=37mm]{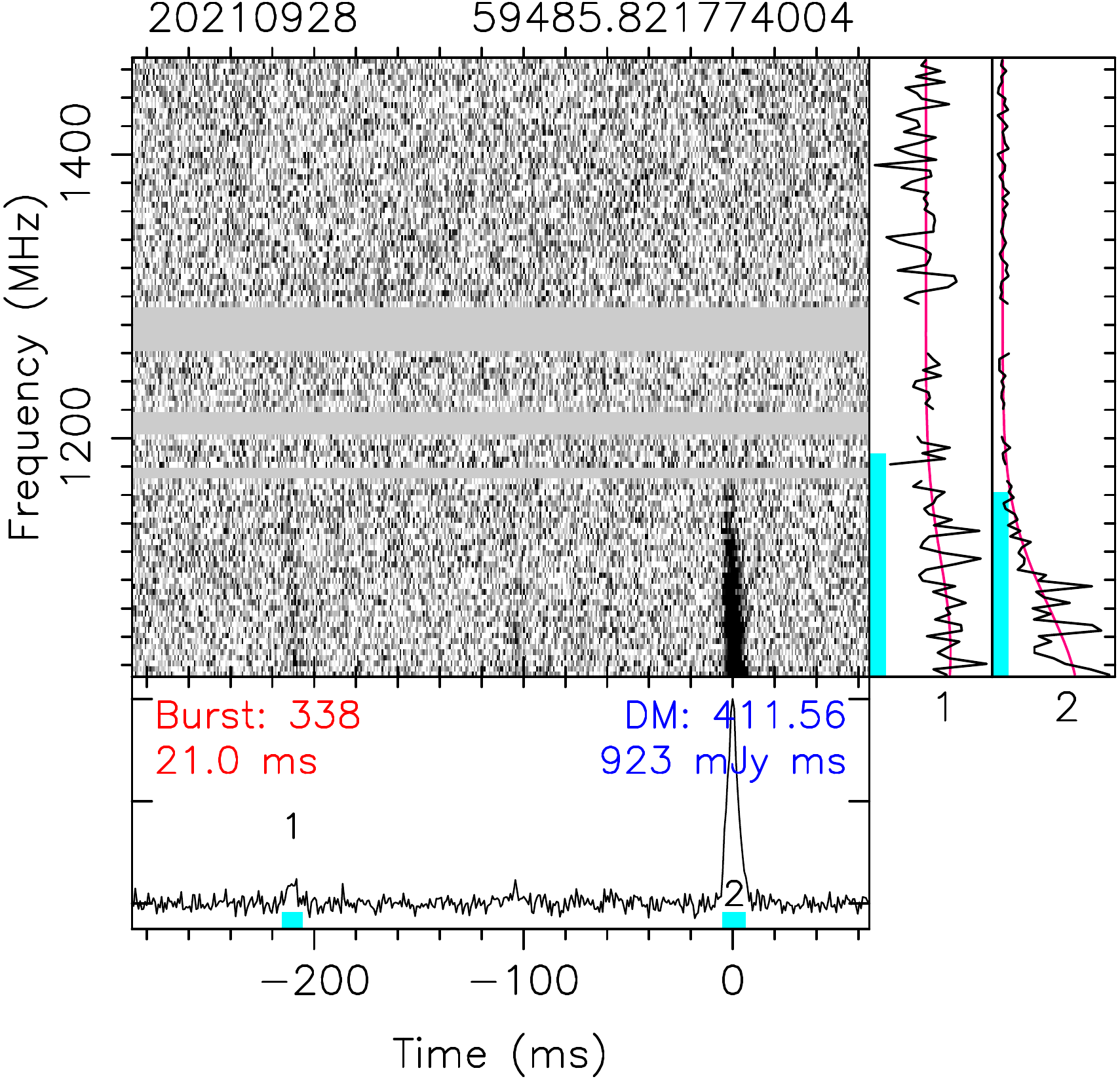}
    \includegraphics[height=37mm]{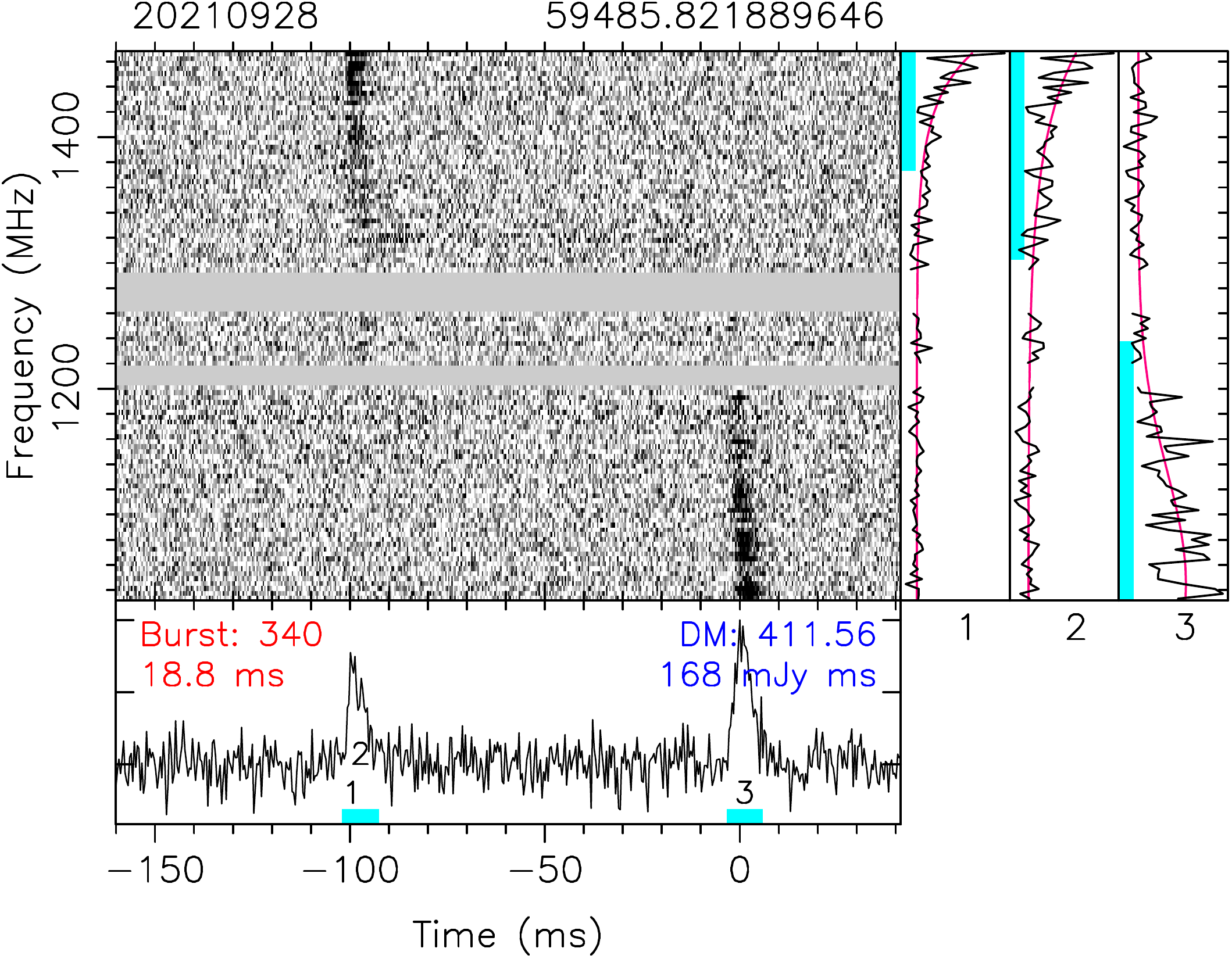}
    \includegraphics[height=37mm]{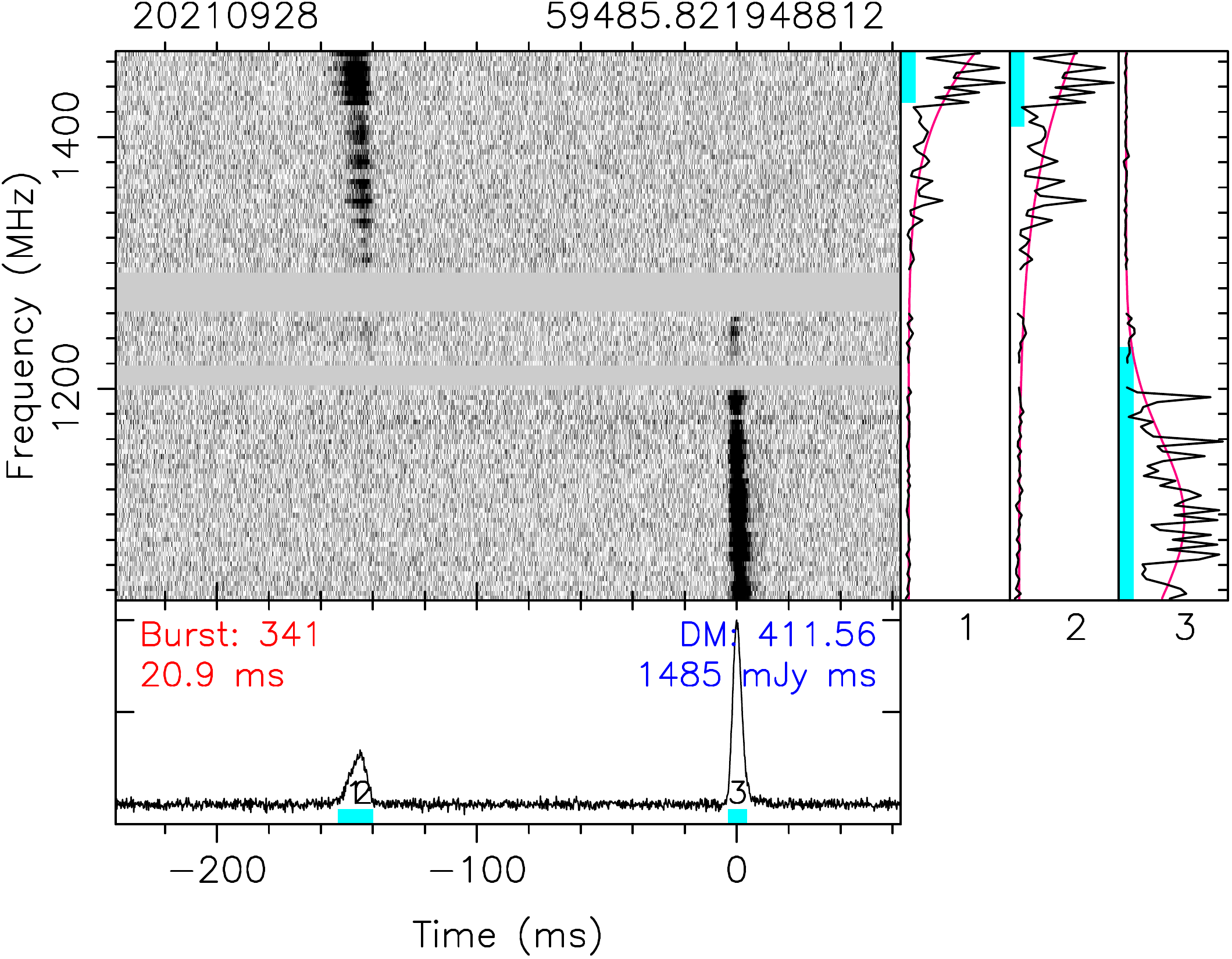}
    \includegraphics[height=37mm]{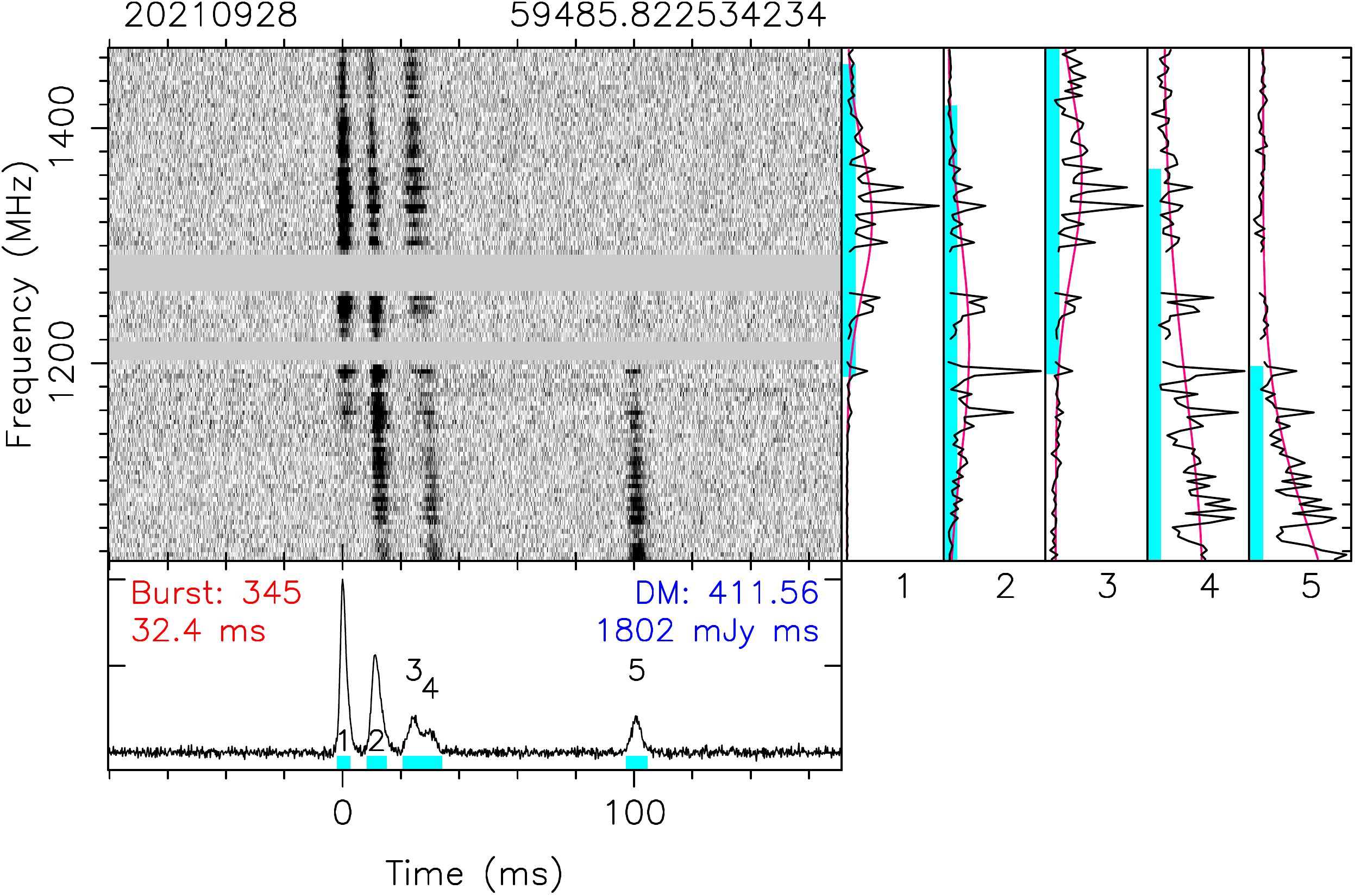}
    \includegraphics[height=38mm]{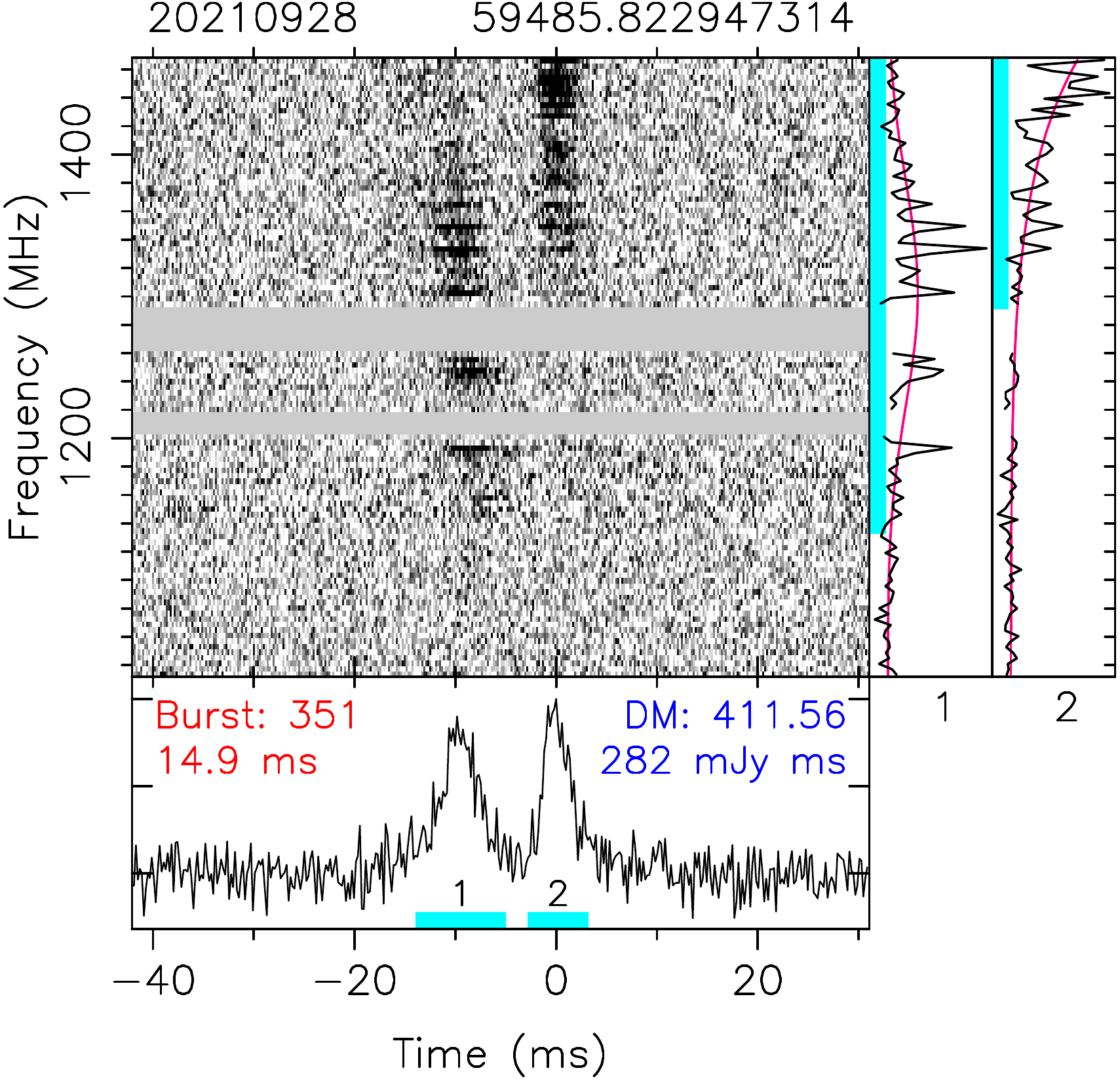} 
    \includegraphics[height=37mm]{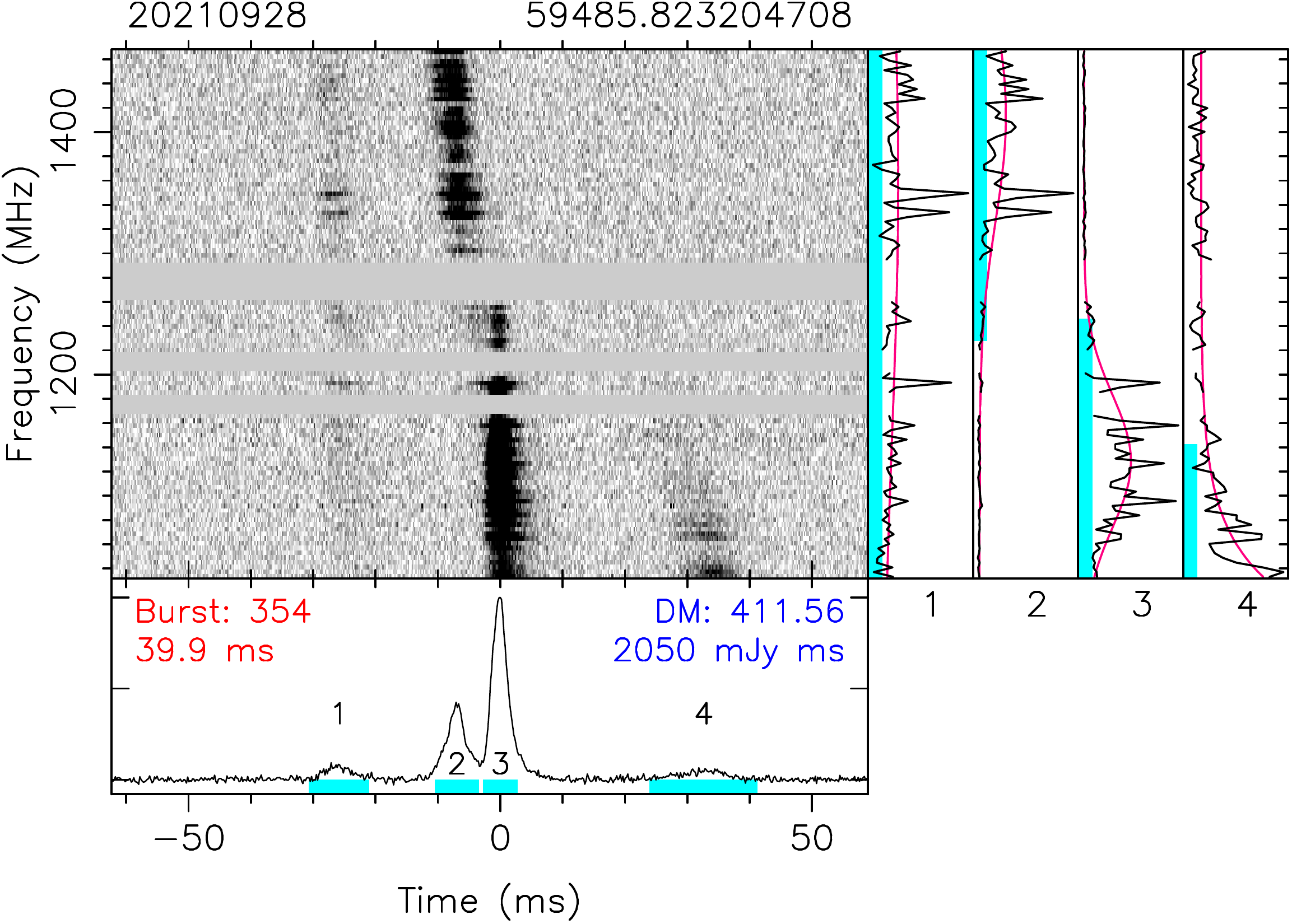}
    \caption{ \it{ -- continued and ended}.
}
\end{figure*}
\addtocounter{figure}{-1}
\begin{figure*}
    \flushleft
    \includegraphics[height=37mm]{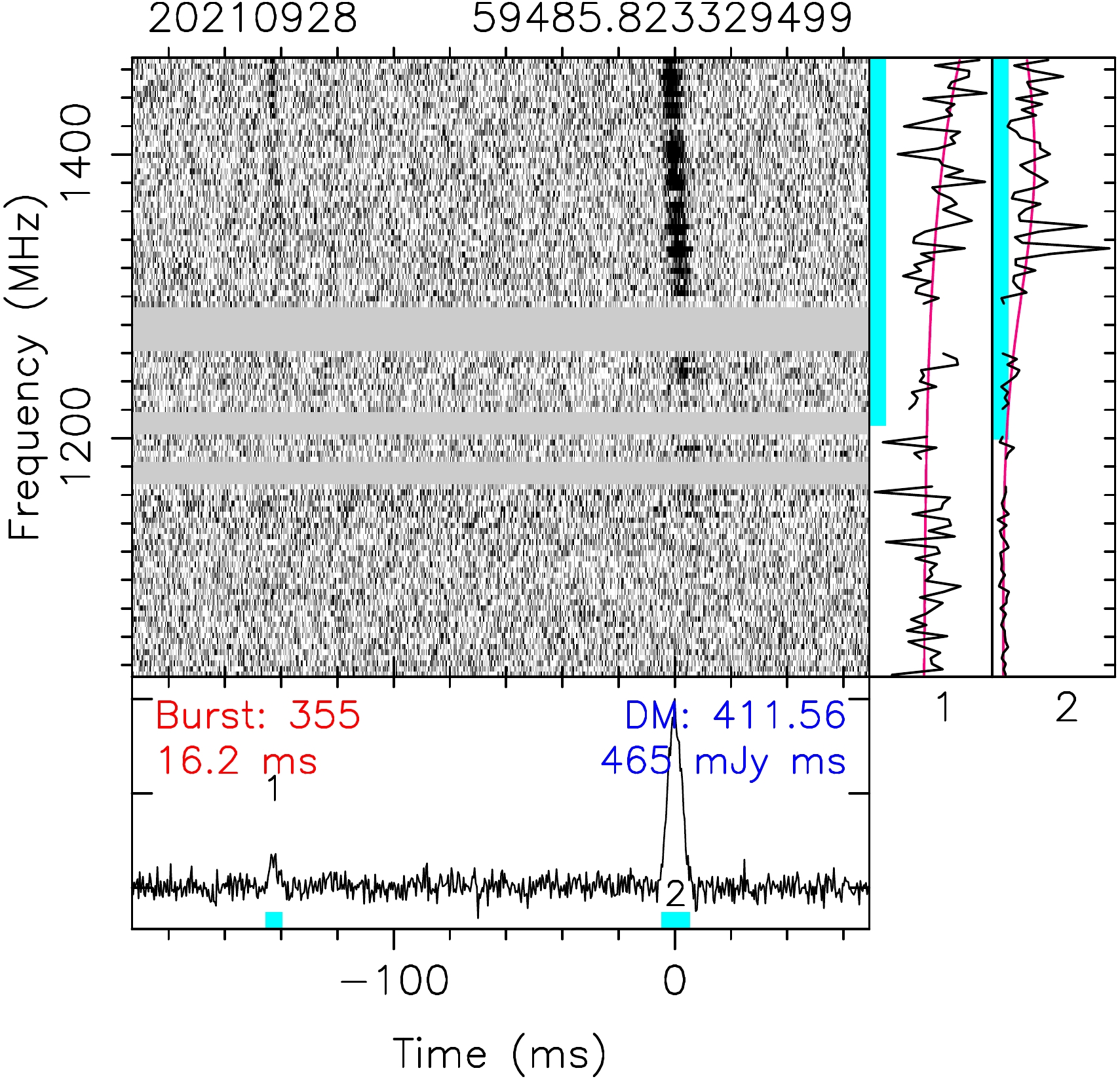}
    \includegraphics[height=37mm]{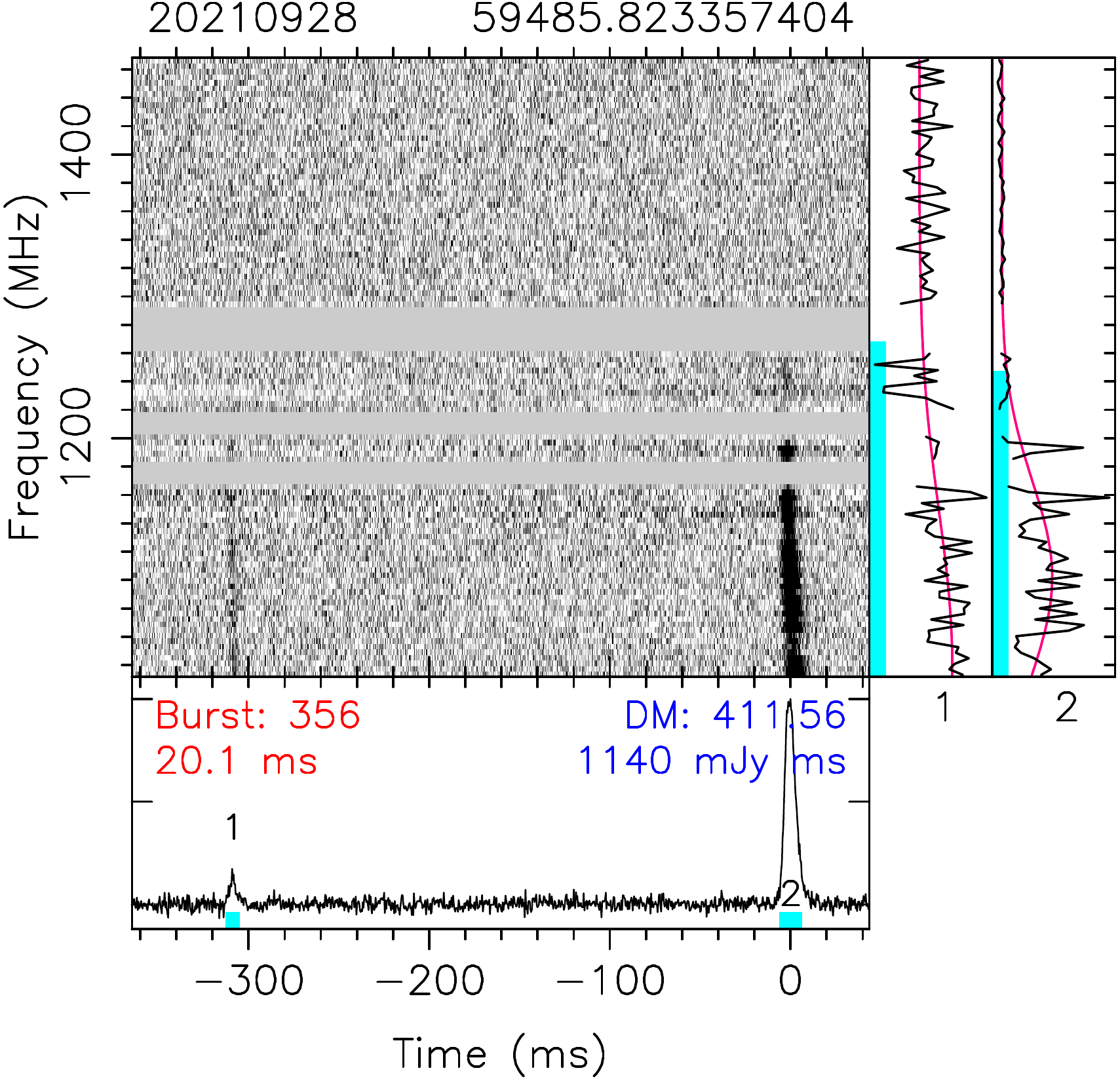}
    \includegraphics[height=37mm]{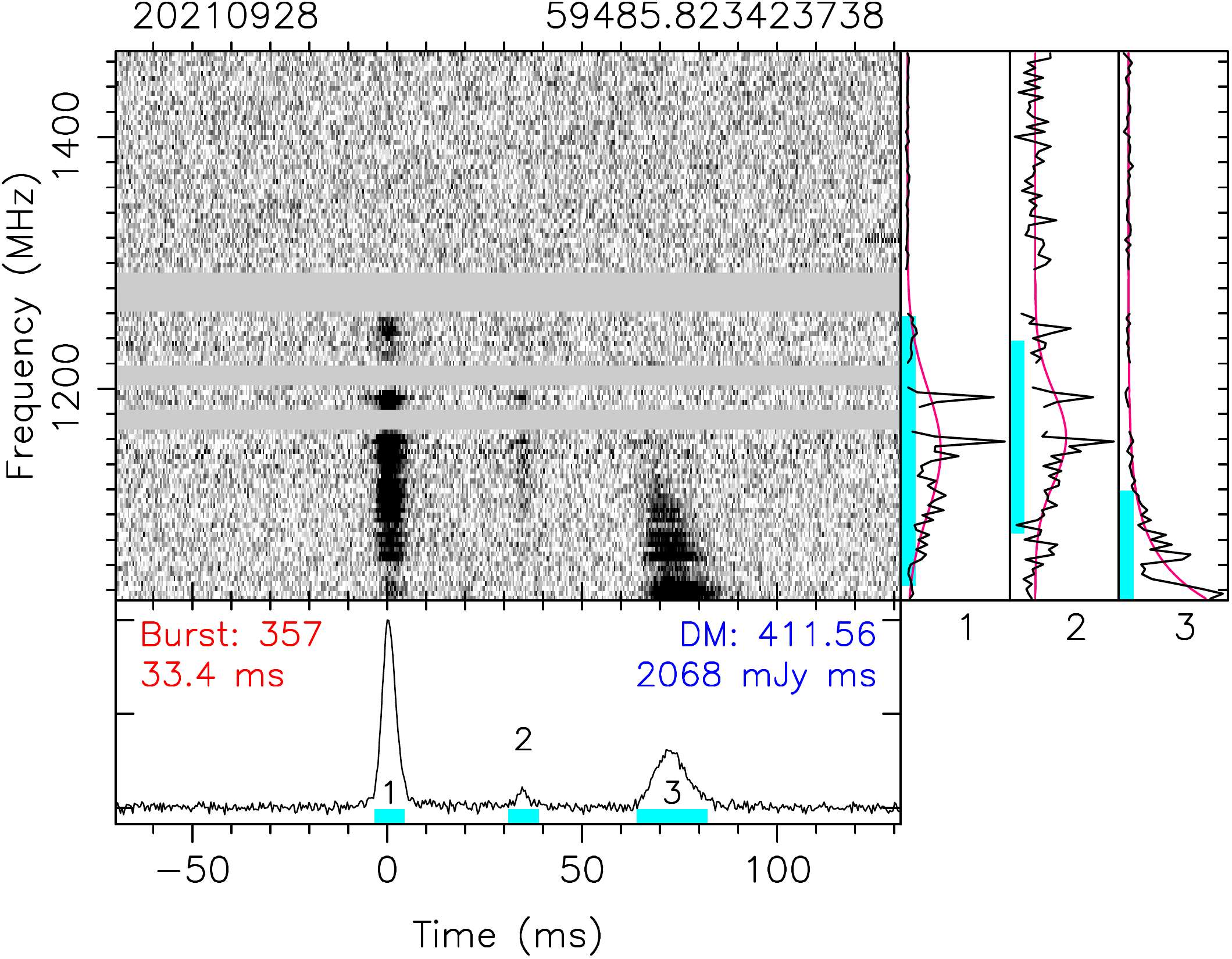}
    \includegraphics[height=37mm]{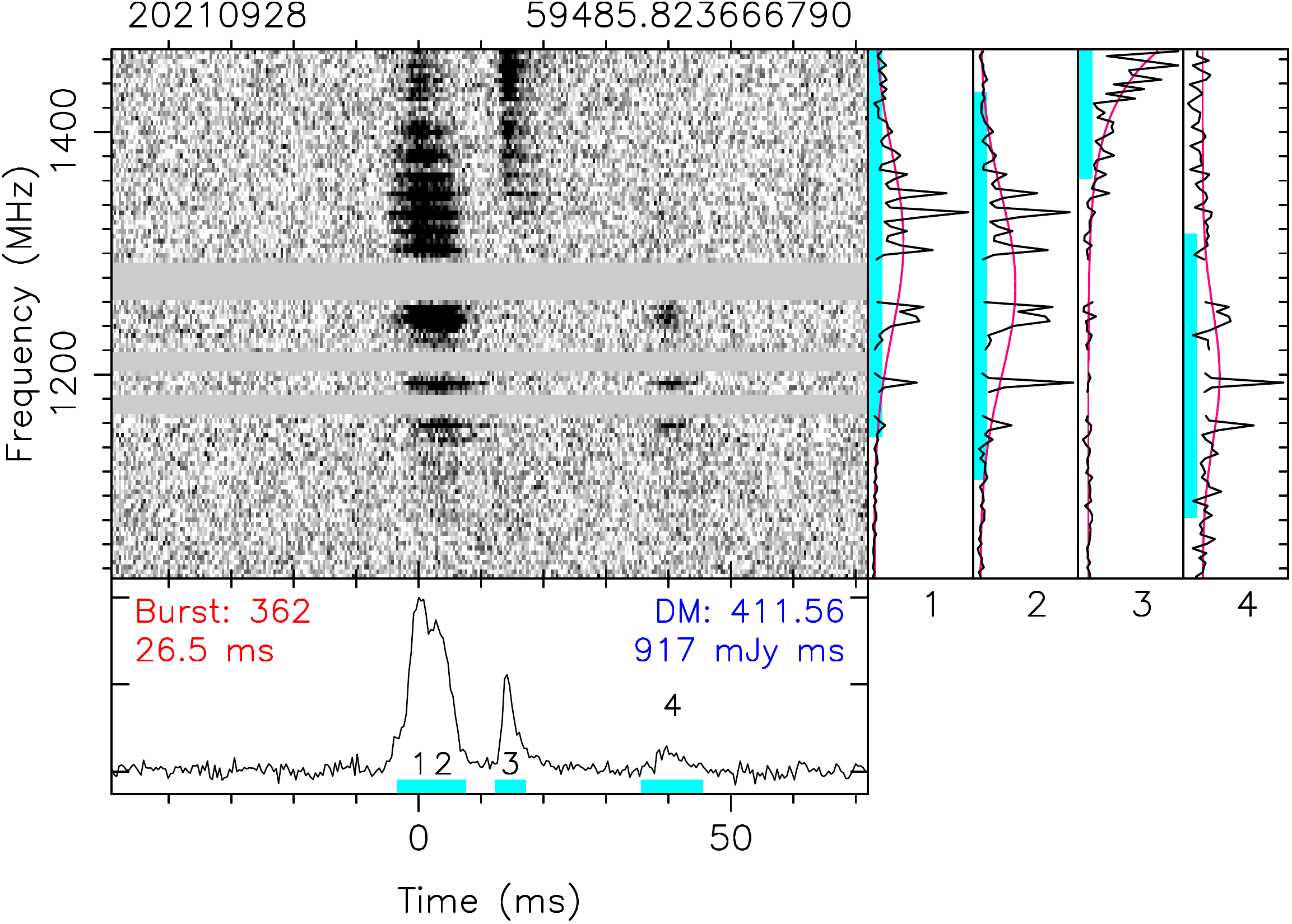}
    \includegraphics[height=38mm]{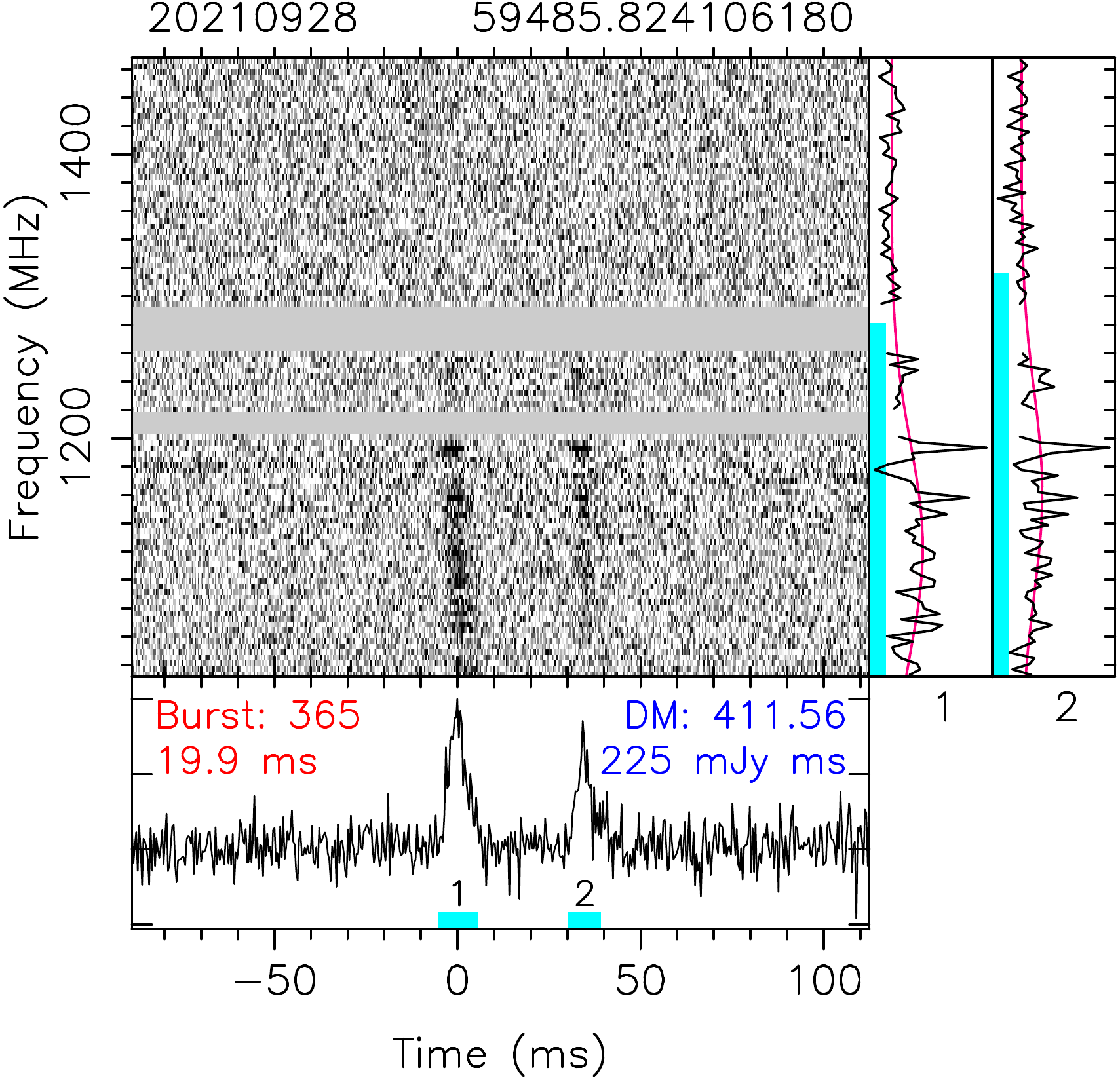}
    \includegraphics[height=37mm]{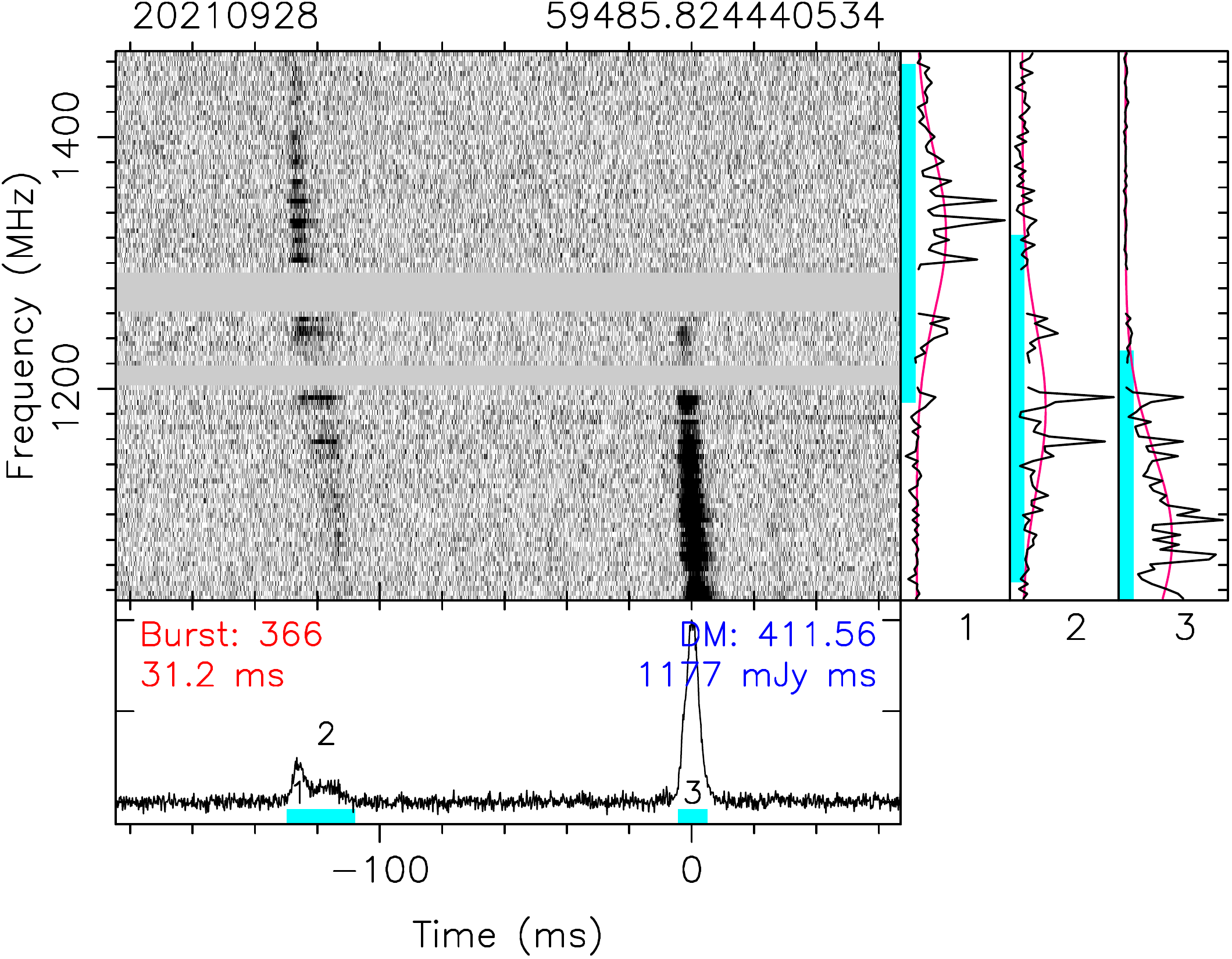}
    \includegraphics[height=37mm]{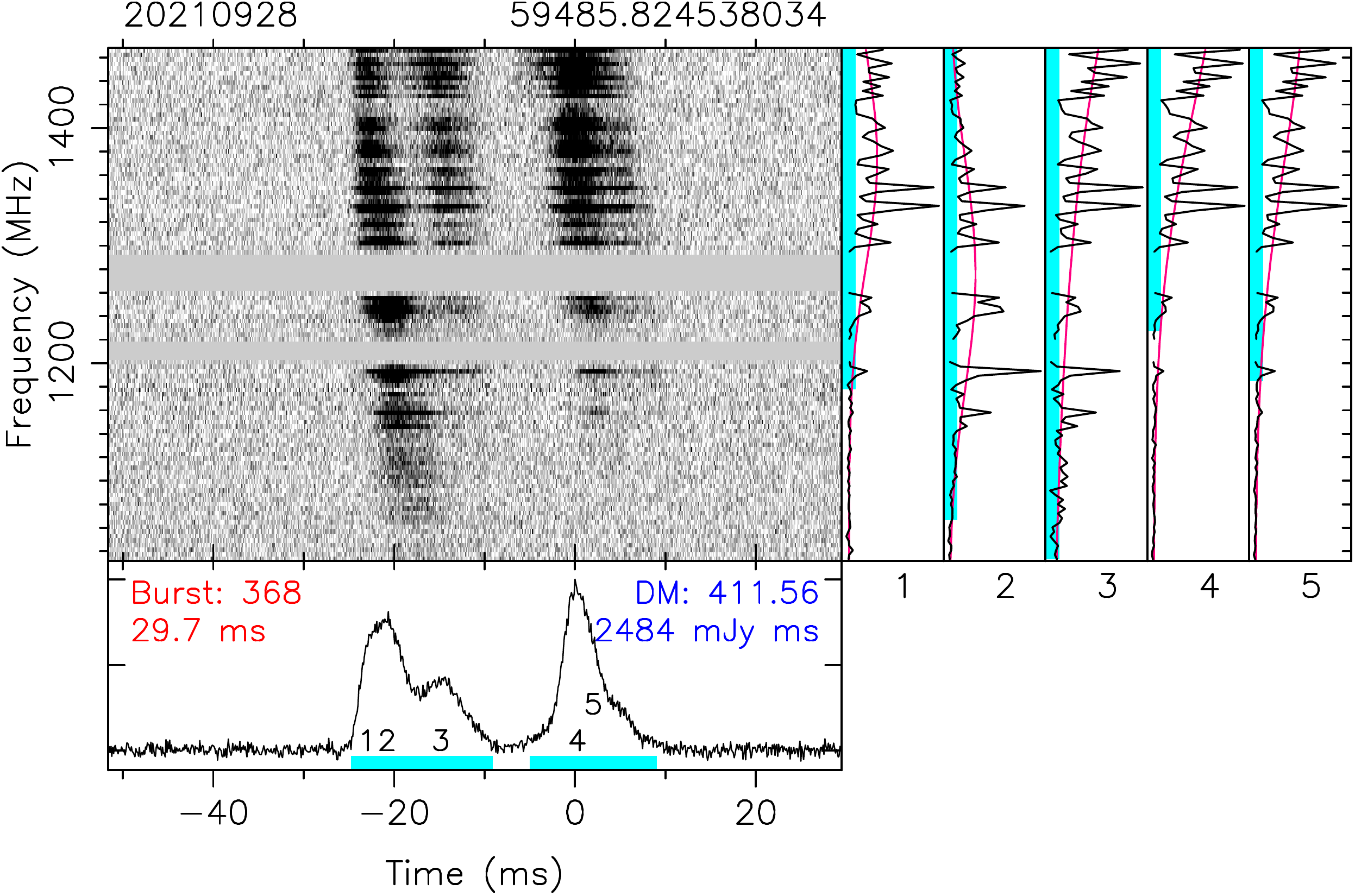}
    \caption{ \it{ -- continued and ended}.
}
\end{figure*}
\clearpage

\end{document}